
\documentclass[acmsmall,screen,nonacm]{acmart}
\AtBeginDocument{%
  }

\setcopyright{none}
\settopmatter{printacmref=false,printfolios=true}

\usepackage{mdframed}
\usepackage{framed}
\usepackage{amsthm, amsfonts, amscd}
\usepackage{mathpartir}
\usepackage{stmaryrd}
\usepackage{aliascnt}
\usepackage[nameinlink,capitalize]{cleveref}
\usepackage{multicol}
\usepackage{wrapfig} 
\usepackage{extarrows}
\usepackage{tikz}
\usetikzlibrary{calc}
\usetikzlibrary{arrows.meta}
\numberwithin{equation}{section}

\newtheorem{thm}{Theorem}[section]

\newaliascnt{lem}{thm}
\newtheorem{lem}[lem]{Lemma}
\aliascntresetthe{lem}

\newaliascnt{prop}{thm}
\newtheorem{prop}[prop]{Proposition}
\aliascntresetthe{prop}

\newaliascnt{cor}{thm}
\newtheorem{cor}[cor]{Corollary}
\aliascntresetthe{cor}

\theoremstyle{definition}
\newaliascnt{defn}{thm}
\newtheorem{defn}[defn]{Definition}
\aliascntresetthe{defn}

\newaliascnt{ex}{thm}

\aliascntresetthe{ex}

\theoremstyle{remark}
\newaliascnt{rem}{thm}

\aliascntresetthe{rem}

\crefname{lem}{Lemma}{Lemmas}
\Crefname{lem}{Lemma}{Lemmas}

\usepackage{color}
\usepackage{bm} %
\usepackage{listings}
\usepackage{booktabs}
\usepackage{hyperref}
\usepackage{subcaption}

\definecolor{math}{rgb}{.624,.251,.259}
\definecolor{lang}{rgb}{.224,.451,.686}
\definecolor{lange}{rgb}{.29,.14,.56}
\definecolor{meta}{rgb}{.581,.067,.0}
\newcommand{\GLang}[0]{Grassmann}
\newcommand{\NLang}[0]{NumPy}

\newcommand{\slabel}[1]{}

\DeclareMathOperator*{\lvadd}{\textcolor{lang}{+}}
\DeclareMathOperator*{\lvscale}{\textcolor{lang}{\ast}}
\newcommand{\lvzero}{\textcolor{lang}{0}}
\DeclareMathOperator*{\lbzero}{\textcolor{lang}{\{\}}}
\newcommand{\lbone}{\textcolor{lang}{\{1\}}}
\newcommand{\len}[1]{\textcolor{lang}t_{\textcolor{lang}#1}}

\newcommand{\lee}[0]{\textcolor{lang}t}

\newcommand{\lsn}[1]{\textcolor{lang}h_{\textcolor{lang}#1}}
\newcommand{\lprop}{\textcolor{lang}p}
\newcommand{\lsnp}[1]{\textcolor{lang}h_{\textcolor{lang}#1}^{\textcolor{lang}\prime}}
\newcommand{\ls}[0]{\textcolor{lang}h}

\newcommand{\els}[0]{\textcolor{lange}h}
\newcommand{\elty}[0]{\textcolor{lange}T}
\newcommand{\ecolon}[0]{\textcolor{lange}{\colon}}
\newcommand{\lbn}[1]{\textcolor{lang}b_{\textcolor{lang}#1}}
\newcommand{\lb}[0]{\textcolor{lang}b}
\newcommand{\lbp}[0]{\textcolor{lang}{b^{\prime}}}
\newcommand{\lenv}[1]{\textcolor{lang}v_{\textcolor{lang}#1}}

\newcommand{\leev}[0]{\textcolor{lang}v}
\newcommand{\leevp}[0]{\textcolor{lang}{v^{\prime}}}
\newcommand{\lsnv}[1]{\textcolor{lang}{hv}_{\textcolor{lang}#1}}
\newcommand{\lsv}[0]{\textcolor{lang}{hv}}
\newcommand{\lsvp}[0]{\textcolor{lang}{hv^{\prime}}}

\newcommand{\lsnvpp}[1]{\textcolor{lang}{hv_{#1}^{\prime\prime}}}

\newcommand{\lsempty}{\textcolor{lang}{\{0\}}}
\newcommand{\lR}{\textcolor{lang}{\mathbb{R}}}
\newcommand{\losum}{\mathbin{\textcolor{lang}{\oplus}}}
\newcommand{\loprod}{\mathbin{\textcolor{lang}{\otimes}}}
\newcommand{\lstar}{\textcolor{lang}{\star}}

\DeclareMathOperator*{\lbosum}{\textcolor{lang}{\oplus}}
\DeclareMathOperator*{\lboprod}{\textcolor{lang}{\otimes}}
\DeclareMathOperator*{\lbstar}{\textcolor{lang}{\star}}

\DeclareMathOperator*{\Kernel}{\textcolor{lang}{Kernel}}
\DeclareMathOperator*{\CoKernel}{\textcolor{lang}{CoKernel}}
\DeclareMathOperator*{\Image}{\textcolor{lang}{Image}}
\DeclareMathOperator*{\CoImage}{\textcolor{lang}{CoImage}}

\DeclareMathOperator*{\pBasis}{\textcolor{lang}{ProjectedBasis}}
\DeclareMathOperator*{\iBasis}{\textcolor{lang}{ImageBasis}}

\newcommand{\linner}[1]{\textcolor{lang}{\langle\cdot, \cdot\rangle_{#1}}}

\newcommand{\limp}{\mathbin{\multimap}}  %

\DeclareMathOperator*{\llambda}{\textcolor{lang}{\lambda}}
\newcommand{\lllambda}{\textcolor{lang}{\lambda^{\limp}}}
\DeclareMathOperator*{\ldotsp}{\textcolor{lang}{.\;}}
\newcommand{\lllambdaArg}[2]{\lllambda \llpl #1 \llpr \ldotsp #2}
\DeclareMathOperator*{\nllambda}{\textcolor{lang}{\lambda^{\rightarrow}}}
\newcommand{\nllambdaArg}[2]{\nllambda \llpl #1 \llpr \ldotsp #2}

\DeclareMathOperator*{\lrh}{\textcolor{lang}{\rightarrow}}
\DeclareMathOperator*{\lpseudo}{\textcolor{lang}{\dagger}}
\DeclareMathOperator*{\lsharp}{\textcolor{lang}{\sharp}}
\DeclareMathOperator*{\lflat}{\textcolor{lang}{\flat}}
\DeclareMathOperator*{\lembd}{\textcolor{lang}{\iota}}
\newcommand{\lproj}{\textcolor{lang}{\pi}}

\newcommand{\leval}[3]{\ensuremath{{#1}^{#2}\!\left(#3\right)}}
\newcommand{\lneval}[2]{\ensuremath{{#1}\!\left(#2\right)}}
\DeclareMathOperator*{\fnto}{\textcolor{lang}{\to}}

\newcommand{\ldelta}[2]{\textcolor{lang}{\delta}_{#1, #2}}
\newcommand{\lidelta}[1]{\textcolor{lang}{\delta}_{#1}}

\newcommand{\lkprojpp}[1]{\textcolor{lang}{\pi}_{#1}}
\newcommand{\lkembedpp}[1]{\textcolor{lang}{\iota}_{#1}}
\DeclareMathOperator{\luncurry}{\textcolor{lang}{uncurry}}
\DeclareMathOperator*{\lcurry}{\textcolor{lang}{curry}}
\DeclareMathOperator*{\ltr}{\textcolor{lang}{Tr}}

\newcommand{\fntype}[3]{(#1\colon #2\fnto #3)}
\newcommand{\fntypep}[3]{#1\colon #2\fnto #3}

\newcommand{\lty}{\textcolor{lang}{\mathrm{T}}}
\newcommand{\ltyn}[1]{\lty{}_{#1}}
\newcommand{\ltynx}[1]{\lty{}_{#1}}

\newcommand{\lky}{\textcolor{lang}{\mathrm{k}}}
\DeclareMathOperator*{\lkbase}{\textcolor{lang}{\star}}
\newcommand{\lkn}[1]{\lky_{#1}}
\newcommand{\lknto}[2]{#1 \fnto #2}

\DeclareMathOperator*{\ltyv}{\textcolor{lang}{Tv}}
\newcommand{\ltynv}[1]{\ltyv{}_{#1}}

\newcommand{\llpl}{\mathopen{\textcolor{lang}{(}}}
\newcommand{\llpr}{\mathclose{\textcolor{lang}{)}}}
\DeclareMathOperator*{\lctx}{\textcolor{lang}{(\Delta, \epsilon)}}

\newcommand{\ldisun}[0]{\mathop{\textcolor{lang}{\sqcup}}}

\DeclareMathOperator*{\lDelta}{\textcolor{lang}{\Delta}}
\DeclareMathOperator*{\lepsilon}{\textcolor{lang}{\epsilon}}

\DeclareMathOperator*{\lctxnle}{\textcolor{lang}{(\emptyset{}, \epsilon)}}

\DeclareMathOperator*{\lemptyset}{\textcolor{lang}{\emptyset{}}}

\newcommand{\fv}[1]{\textcolor{meta}{fv(}#1\textcolor{meta}{)}}
\newcommand{\sof}[1]{\textcolor{meta}{spaceOf(}#1\textcolor{meta}{)}}
\newcommand{\tof}[1]{\textcolor{meta}{typeOf(}#1\textcolor{meta}{)}}

\newcommand{\iof}[1]{\textcolor{meta}{innerOf(}#1\textcolor{meta}{)}}
\newcommand{\dof}[2]{\textcolor{meta}{dimOf(}#1, #2\textcolor{meta}{)}}
\newcommand{\shpOf}[2]{\textcolor{meta}{shapeOf(}#1, #2\textcolor{meta}{)}}

\DeclareMathOperator*{\svalid}{\lkbase{}}

\DeclareMathOperator*{\tvalid}{\lky{}}
\DeclareMathOperator*{\bvalid}{\textcolor{meta}{ValidBasis}}
\DeclareMathOperator*{\lbf}{\textcolor{lang}{\triangleleft}}
\DeclareMathOperator*{\lbfr}{\textcolor{lang}{\triangleright}}

\DeclareMathOperator{\SE}{\textcolor{meta}{\mathcal{E}}}
\DeclareMathOperator{\ST}{\textcolor{meta}{\mathcal{T}}}
\DeclareMathOperator{\SB}{\textcolor{meta}{\mathcal{B}}}
\newcommand{\Sem}[2]{\textcolor{meta}{\mathcal{E}(}#1\textcolor{meta}{, }#2\textcolor{meta}{)}}
\newcommand{\SemCheck}[2]{\textcolor{meta}{\mathcal{E}^{\mathrm{check}}(}#1\textcolor{meta}{, }#2\textcolor{meta}{)}}
\newcommand{\SemT}[2]{\textcolor{meta}{\mathcal{T}(}#1\textcolor{meta}{, }#2\textcolor{meta}{)}}
\newcommand{\SemBB}[2]{\textcolor{meta}{\mathcal{B}(}#1\textcolor{meta}{, }#2\textcolor{meta}{)}}

\newcommand{\SemL}[4]{\textcolor{meta}{\mathcal{E}(}#3\vdash #1\colon #4 \textcolor{meta}{, }#2\textcolor{meta}{)}}

\newcommand{\SemUS}[3]{\textcolor{meta}{\mathcal{T}(}#3\vdash #1 \colon\tvalid \textcolor{meta}{, }#2\textcolor{meta}{)}}
\newcommand{\SemLK}[3]{\SemL{#1}{#2}{\lctxnle}{#3}}

\newcommand{\SemLUS}[2]{\SemUS{#1}{#2}{\lctxnle}}

\newcommand{\SemA}[3]{{\textcolor{meta}{\mathcal{E}(}#1\textcolor{meta}{, }#2\textcolor{meta}{)_{#3}}}}

\newcommand{\isStdBasis}[1]{\textcolor{meta}{\mathrm{isStdBasis}(}#1\textcolor{meta}{)}}

\newcommand{\stdBasisCtx}[2]{\textcolor{meta}{\mathrm{stdBasis}(}#1, #2\textcolor{meta}{)}}
\newcommand{\stdBasisDCtx}[1]{\stdBasisCtx{#1}{\lctxnle}}

\newcommand{\subst}[2]{\textcolor{meta}{[}#2\textcolor{meta}{/}#1\textcolor{meta}{]}}

\DeclareMathOperator*{\contract}{Contract}
\newcommand{\contractN}[3]{\contract_{#1}(#2, #3)}
\DeclareMathOperator*{\TTV}{TTV}

\DeclareMathOperator*{\MTM}{MTM}
\DeclareMathOperator*{\Transpose}{Transpose}
\DeclareMathOperator*{\Reshape}{Reshape}
\DeclareMathOperator*{\Krn}{Kron}
\DeclareMathOperator*{\Blk}{BlockDiag}
\newcommand{\vkrn}[0]{\textcolor{lang}{\Krn}}
\newcommand{\vblk}[0]{\textcolor{lang}{\Blk}}

\newcommand{\LTTVPP}[3]{\textcolor{lang}{\TTV{}}_{#1}\llpl #2, #3 \llpr}

\newcommand{\LTTT}[3]{\textcolor{lang}{\MTM{}_{#1}}\llpl #2, #3\llpr}

\newcommand{\Trans}[0]{\textcolor{lang}{\Transpose}}
\newcommand{\Reshap}[0]{\textcolor{lang}{\Reshape}}
\DeclareMathOperator*{\argmin}{arg\,min}
\newcommand{\norm}[1]{\left\lVert#1\right\rVert}

\DeclareMathOperator*{\evalArrow}{\Rightarrow}

\DeclareMathOperator*{\evalCtx}{\ensuremath{\Psi}}
\DeclareMathOperator*{\evalCtxVars}{\ensuremath{\Psi}}
\DeclareMathOperator*{\evalCtxSpace}{\ensuremath{\epsilon}}
\DeclareMathOperator*{\valCheckCtxSym}{\ensuremath{\Gamma_v}}
\newcommand{\valCheckCtx}[0]{\textcolor{lang}{\valCheckCtxSym}}
\newcommand{\evalJudge}[4]{#1\vdash #2 \xlongrightarrow{#3} #4}
\newcommand{\evalBBJudge}[4]{#1\vdash #2 \xlongleftrightarrow{#3} #4}
\newcommand{\evalSJudge}[3]{#1\vdash #2 \evalArrow{} #3}
\newcommand{\evalJudgeK}[3]{\evalJudge{\evalCtx}{#1}{#2}{#3}}
\newcommand{\evalBBJudgeK}[3]{\evalBBJudge{\evalCtx}{#1}{#2}{#3}}
\newcommand{\evalSJudgeK}[2]{\evalSJudge{\evalCtx}{#1}{#2}}

\newcommand{\frameB}[3]{\llambda  (#1\lbf #2) \lbfr #3}
\newcommand{\iBasisC}[1]{\iBasis{}\llpl{}#1\llpr{}}
\newcommand{\pBasisC}[3]{\pBasis{}\llpl{}#1, #2, #3\llpr{}}

\DeclareMathOperator*{\lsum}{\textcolor{lang}{\sum}}

\DeclareMathOperator*{\GramSchdmit}{\textcolor{math}{GS}}
\DeclareMathOperator*{\LexInd}{\textcolor{math}{LexInd}}

\DeclareMathOperator*{\FSV}{\textcolor{math}{FilteredSingularVectors}}

\DeclareMathOperator*{\GSVDU}{{\operatorname{GSVDU}}}
\DeclareMathOperator*{\GSVDUI}{{\operatorname{GSVDUI}}}
\DeclareMathOperator*{\GSVDS}{{\operatorname{GSVDS}}}
\DeclareMathOperator*{\GSVDV}{{\operatorname{GSVDV}}}
\DeclareMathOperator*{\GSVDVI}{{\operatorname{GSVDVI}}}

\usepackage{generated/commands/glang-lean}
\providecommand{\glleanliterate}{}
\renewcommand{\glleanliterate}{\lstset{literate=%
  {∀}{{$\forall$}}1
  {∃}{{$\exists$}}1
  {¬}{{$\neg$}}1
  {→}{{$\to$}}1
  {←}{{$\leftarrow$}}1
  {↔}{{$\leftrightarrow$}}1
  {⇒}{{$\Rightarrow$}}1
  {⟹}{{$\Longrightarrow$}}1
  {↦}{{$\mapsto$}}1
  {∧}{{$\wedge$}}1
  {∨}{{$\vee$}}1
  {⊕}{{$\oplus$}}1
  {⊗}{{$\otimes$}}1
  {≤}{{$\leq$}}1
  {≥}{{$\geq$}}1
  {≠}{{$\neq$}}1
  {≡}{{$\equiv$}}1
  {≈}{{$\approx$}}1
  {∈}{{$\in$}}1
  {∉}{{$\notin$}}1
  {⊆}{{$\subseteq$}}1
  {∪}{{$\cup$}}1
  {∩}{{$\cap$}}1
  {∖}{{$\setminus$}}1
  {∅}{{$\emptyset$}}1
  {⊥}{{$\bot$}}1
  {⊤}{{$\top$}}1
  {·}{{$\cdot$}}1
  {×}{{$\times$}}1
  {∘}{{$\circ$}}1
  {∑}{{$\sum$}}1
  {⊢}{{$\vdash$}}1
  {⊨}{{$\models$}}1
  {⟦}{{$\llbracket$}}1
  {⟧}{{$\rrbracket$}}1
  {⟨}{{$\langle$}}1
  {⟩}{{$\rangle$}}1
  {…}{{$\ldots$}}1
  {†}{{$\dagger$}}1
  {ℝ}{{$\mathbb{R}$}}1
  {ℕ}{{$\mathbb{N}$}}1
  {α}{{$\alpha$}}1
  {β}{{$\beta$}}1
  {δ}{{$\delta$}}1
  {ε}{{$\varepsilon$}}1
  {η}{{$\eta$}}1
  {λ}{{$\lambda$}}1
  {μ}{{$\mu$}}1
  {π}{{$\pi$}}1
  {ρ}{{$\rho$}}1
  {σ}{{$\sigma$}}1
  {τ}{{$\tau$}}1
  {φ}{{$\varphi$}}1
  {Γ}{{$\Gamma$}}1
  {Δ}{{$\Delta$}}1
  {Π}{{$\Pi$}}1
  {Σ}{{$\Sigma$}}1
  {Ψ}{{$\Psi$}}1
  {₀}{{$_0$}}1
  {₁}{{$_1$}}1
  {₂}{{$_2$}}1
  {₃}{{$_3$}}1
  {ₙ}{{$_n$}}1
  {ᵢ}{{$_i$}}1
  {ₜ}{{$_t$}}1
  {ᗮ}{{$^{\perp}$}}1
  {−}{{$-$}}1
  {⁻}{{$^-$}}1
  {¹}{{$^1$}}1
  {⌊}{{$\lfloor$}}1
  {⌋}{{$\rfloor$}}1
  {—}{{---}}1
  {▸}{{$\blacktriangleright$}}1
  {ₗ}{{$_l$}}1
  {ˡ}{{$^l$}}1
  {″}{{$''$}}1
  {⋯}{{$\cdots$}}1
  {⊛}{{$\circledast$}}1
  {⊸}{{$\multimap$}}1
  {⟷}{{$\longleftrightarrow$}}1
  {§}{{\S}}1
  {①}{{(1)}}1
  {②}{{(2)}}1
  {④}{{(4)}}1
  {⑥}{{(6)}}1
  {✓}{{$\checkmark$}}1
  {♯}{{$\sharp$}}1
  {•}{{$\bullet$}}1
  {ᵀ}{{$^T$}}1
  {⁺}{{$^+$}}1
}}

\gldefsyntax{space.empty}{space}{\ls{}}{\lsempty{}}{Empty Space}{\texttt{GLang.Space.empty}}{GLang/Syntax.lean:28}
\gldefsyntax{space.real}{space}{\ls{}}{\lR{}}{Scalar Space}{\texttt{GLang.Space.real}}{GLang/Syntax.lean:29}
\gldefsyntax{space.dsum}{space}{\ls{}}{\lsn{0}\losum\lsn{1}}{Direct Sum Space}{\texttt{GLang.Space.dsum}}{GLang/Syntax.lean:31}
\gldefsyntax{space.dual}{space}{\ls{}}{\ls{}^{\lstar}}{Dual Space}{\texttt{GLang.Space.dual}}{GLang/Syntax.lean:30}
\gldefsyntax{space.tprod}{space}{\ls{}}{\lsn{0}\loprod\lsn{1}}{Tensor Product Space}{\texttt{GLang.Space.tprod}}{GLang/Syntax.lean:32}
\gldefsyntax{space.ker}{space}{\ls{}}{\Kernel{}(\lee)}{Kernel of a Linear Operator}{\texttt{GLang.Space.ker}}{GLang/Syntax.lean:39}
\gldefsyntax{space.coker}{space}{\ls{}}{\CoKernel{}(\lee)}{Cokernel of a Linear Operator}{\texttt{GLang.Space.coker}}{GLang/Syntax.lean:40}
\gldefsyntax{space.img}{space}{\ls{}}{\Image{}(\lee)}{Image of a Linear Operator}{\texttt{GLang.Space.img}}{GLang/Syntax.lean:41}
\gldefsyntax{space.coimg}{space}{\ls{}}{\CoImage{}(\lee)}{Coimage of a Linear Operator}{\texttt{GLang.Space.coimg}}{GLang/Syntax.lean:42}
\gldefsyntaxrows{space}{
$\ls{}$ & $::=$ & $\lsempty{}$ & Empty Space\\
 & $|$ & $\lR{}$ & Scalar Space\\
 & $|$ & $\lsn{0}\losum\lsn{1}$ & Direct Sum Space\\
 & $|$ & $\ls{}^{\lstar}$ & Dual Space\\
 & $|$ & $\lsn{0}\loprod\lsn{1}$ & Tensor Product Space\\
 & $|$ & $\Kernel{}(\lee)$ & Kernel of a Linear Operator\\
 & $|$ & $\CoKernel{}(\lee)$ & Cokernel of a Linear Operator\\
 & $|$ & $\Image{}(\lee)$ & Image of a Linear Operator\\
 & $|$ & $\CoImage{}(\lee)$ & Coimage of a Linear Operator\\
}
\gldefsyntax{ty.space}{ty}{\lty{}}{\ls{}}{Space as a Type}{\texttt{GLang.Ty.space}}{GLang/Syntax.lean:49}
\gldefsyntax{ty.piTy}{ty}{\lty{}}{\fntypep{x}{\ltyn{0}}{\ltyn{1}}}{Dependent Function Type}{\texttt{GLang.Ty.pi\allowbreak Ty}}{GLang/Syntax.lean:50}
\gldefsyntaxrows{ty}{
$\lty{}$ & $::=$ & $\ls{}$ & Space as a Type\\
 & $|$ & $\fntypep{x}{\ltyn{0}}{\ltyn{1}}$ & Dependent Function Type\\
}
\gldefsyntax{kind.base}{kind}{\lky{}}{\lkbase{}}{Space Kind}{\texttt{GLang.Kind.base}}{GLang/Syntax.lean:125}
\gldefsyntax{kind.arrow}{kind}{\lky{}}{\lknto{\lkn{0}}{\lkn{1}}}{Function Kind}{\texttt{GLang.Kind.arrow}}{GLang/Syntax.lean:126}
\gldefsyntaxrows{kind}{
$\lky{}$ & $::=$ & $\lkbase{}$ & Space Kind\\
 & $|$ & $\lknto{\lkn{0}}{\lkn{1}}$ & Function Kind\\
}
\gldefsyntax{term.zero}{term}{\lee}{\lvzero_{\els{}}}{Zero Vector}{\texttt{GLang.Term.zero}}{GLang/Syntax.lean:66}
\gldefsyntax{term.smul}{term}{\lee}{\len{0} \lvscale \len{1}}{Scalar-Vector Product}{\texttt{GLang.Term.smul}}{GLang/Syntax.lean:67}
\gldefsyntax{term.real}{term}{\lee}{r}{Real Number}{\texttt{GLang.Term.real}}{GLang/Syntax.lean:68}
\gldefsyntax{term.add}{term}{\lee}{\len{0} \lvadd \len{1}}{Vector Sum}{\texttt{GLang.Term.add}}{GLang/Syntax.lean:69}
\gldefsyntax{term.inner}{term}{\lee}{\linner{}_{\els{}}}{Inner Product}{\texttt{GLang.Term.inner}}{GLang/Syntax.lean:70}
\gldefsyntax{term.appL}{term}{\lee}{\leval{\len{0}}{}{\len{1}}}{Linear Application}{\texttt{GLang.Term.app\allowbreak L}}{GLang/Syntax.lean:71}
\gldefsyntax{term.appI}{term}{\lee}{\leval{\len{0}}{}{\len{1}}}{Inner-Product Application}{\texttt{GLang.Term.app\allowbreak I}}{GLang/Syntax.lean:72}
\gldefsyntax{term.appN}{term}{\lee}{\leval{\len{0}}{}{\len{1}}}{Non-linear Application}{\texttt{GLang.Term.app\allowbreak N}}{GLang/Syntax.lean:73}
\gldefsyntax{term.llam}{term}{\lee}{\lllambdaArg{x\ecolon{}\els{}}{\lee}}{Multi-linear Function}{\texttt{GLang.Term.llam}}{GLang/Syntax.lean:74}
\gldefsyntax{term.nllam}{term}{\lee}{\nllambdaArg{x\ecolon{}\elty{}}{\lee}}{Non-linear Function}{\texttt{GLang.Term.nllam}}{GLang/Syntax.lean:75}
\gldefsyntax{term.var}{term}{\lee}{x}{Variable}{\texttt{GLang.Term.var}}{GLang/Syntax.lean:76}
\gldefsyntax{term.pinvT}{term}{\lee}{\lee{}^{\lpseudo{}}}{Linear Pseudo-Inverse (tensor)}{\texttt{GLang.Term.pinv\allowbreak T}}{GLang/Syntax.lean:77}
\gldefsyntax{term.pinvD}{term}{\lee}{\lee{}^{\lpseudo{}}}{Linear Pseudo-Inverse (dual)}{\texttt{GLang.Term.pinv\allowbreak D}}{GLang/Syntax.lean:78}
\gldefsyntax{term.dsumE}{term}{\lee}{\len{0}\losum\len{1}}{Direct Sum}{\texttt{GLang.Term.dsum\allowbreak E}}{GLang/Syntax.lean:79}
\gldefsyntax{term.tprodE}{term}{\lee}{\len{0}\loprod\len{1}}{Tensor Product}{\texttt{GLang.Term.tprod\allowbreak E}}{GLang/Syntax.lean:80}
\gldefsyntax{term.embed}{term}{\lee}{\lembd{}_{\els}}{Embedding}{\texttt{GLang.Term.embed}}{GLang/Syntax.lean:81}
\gldefsyntaxrows{term}{
$\lee$ & $::=$ & $\lvzero_{\els{}}$ & Zero Vector\\
 & $|$ & $\len{0} \lvscale \len{1}$ & Scalar-Vector Product\\
 & $|$ & $r$ & Real Number\\
 & $|$ & $\len{0} \lvadd \len{1}$ & Vector Sum\\
 & $|$ & $\linner{}_{\els{}}$ & Inner Product\\
 & $|$ & $\leval{\len{0}}{}{\len{1}}$ & Linear Application\\
 & $|$ & $\leval{\len{0}}{}{\len{1}}$ & Inner-Product Application\\
 & $|$ & $\leval{\len{0}}{}{\len{1}}$ & Non-linear Application\\
 & $|$ & $\lllambdaArg{x\ecolon{}\els{}}{\lee}$ & Multi-linear Function\\
 & $|$ & $\nllambdaArg{x\ecolon{}\elty{}}{\lee}$ & Non-linear Function\\
 & $|$ & $x$ & Variable\\
 & $|$ & $\lee{}^{\lpseudo{}}$ & Linear Pseudo-Inverse (tensor)\\
 & $|$ & $\lee{}^{\lpseudo{}}$ & Linear Pseudo-Inverse (dual)\\
 & $|$ & $\len{0}\losum\len{1}$ & Direct Sum\\
 & $|$ & $\len{0}\loprod\len{1}$ & Tensor Product\\
 & $|$ & $\lembd{}_{\els}$ & Embedding\\
}
\gldefsyntax{basis.empty}{basis}{\lb{}}{\lbzero{}}{Basis of $\lsempty$}{\texttt{GLang.Basis.empty}}{GLang/Syntax.lean:111}
\gldefsyntax{basis.one}{basis}{\lb{}}{\lbone{}}{Basis of $\lR$}{\texttt{GLang.Basis.one}}{GLang/Syntax.lean:112}
\gldefsyntax{basis.dual}{basis}{\lb{}}{\lb{}^{\lbstar}}{Dual Basis}{\texttt{GLang.Basis.dual}}{GLang/Syntax.lean:113}
\gldefsyntax{basis.dsum}{basis}{\lb{}}{\lbn{0}\lbosum\lbn{1}}{Direct Sum Basis}{\texttt{GLang.Basis.dsum}}{GLang/Syntax.lean:114}
\gldefsyntax{basis.tprod}{basis}{\lb{}}{\lbn{0}\lboprod\lbn{1}}{Tensor Product Basis}{\texttt{GLang.Basis.tprod}}{GLang/Syntax.lean:115}
\gldefsyntax{basis.proj}{basis}{\lb{}}{\pBasisC{\ls}{\lbn{0}}{\lbn{1}}}{Projected Basis}{\texttt{GLang.Basis.proj}}{GLang/Syntax.lean:116}
\gldefsyntax{basis.image}{basis}{\lb{}}{\iBasisC{\lb{}}}{Image Basis}{\texttt{GLang.Basis.image}}{GLang/Syntax.lean:117}
\gldefsyntax{basis.terms}{basis}{\lb{}}{\llpl{}\len{0},\dots,\len{n}\llpr{}_{\lb{}}}{Term-sequence Basis}{\texttt{GLang.Basis.terms}}{GLang/Syntax.lean:118}
\gldefsyntax{basis.frame}{basis}{\lb{}}{\frameB{x}{\lbn{0}}{\lbn{1}}}{Frame (Dependent Basis)}{\texttt{GLang.Basis.frame}}{GLang/Syntax.lean:121}
\gldefsyntaxrows{basis}{
$\lb{}$ & $::=$ & $\lbzero{}$ & Basis of $\lsempty$\\
 & $|$ & $\lbone{}$ & Basis of $\lR$\\
 & $|$ & $\lb{}^{\lbstar}$ & Dual Basis\\
 & $|$ & $\lbn{0}\lbosum\lbn{1}$ & Direct Sum Basis\\
 & $|$ & $\lbn{0}\lboprod\lbn{1}$ & Tensor Product Basis\\
 & $|$ & $\pBasisC{\ls}{\lbn{0}}{\lbn{1}}$ & Projected Basis\\
 & $|$ & $\iBasisC{\lb{}}$ & Image Basis\\
 & $|$ & $\llpl{}\len{0},\dots,\len{n}\llpr{}_{\lb{}}$ & Term-sequence Basis\\
 & $|$ & $\frameB{x}{\lbn{0}}{\lbn{1}}$ & Frame (Dependent Basis)\\
}
\gldefsyntax{valFFSProp.kernel}{valFFSProp}{p}{\Kernel{}}{\texttt{kernel}}{\texttt{GLang.Val\allowbreak FFSProp.kernel}}{GLang/Values.lean:57}
\gldefsyntax{valFFSProp.cokernel}{valFFSProp}{p}{\CoKernel{}}{\texttt{cokernel}}{\texttt{GLang.Val\allowbreak FFSProp.cokernel}}{GLang/Values.lean:58}
\gldefsyntax{valFFSProp.image}{valFFSProp}{p}{\Image{}}{\texttt{image}}{\texttt{GLang.Val\allowbreak FFSProp.image}}{GLang/Values.lean:59}
\gldefsyntax{valFFSProp.coimage}{valFFSProp}{p}{\CoImage{}}{\texttt{coimage}}{\texttt{GLang.Val\allowbreak FFSProp.coimage}}{GLang/Values.lean:60}
\gldefsyntaxrows{valFFSProp}{
$p$ & $::=$ & $\Kernel{}$ & \texttt{kernel}\\
 & $|$ & $\CoKernel{}$ & \texttt{cokernel}\\
 & $|$ & $\Image{}$ & \texttt{image}\\
 & $|$ & $\CoImage{}$ & \texttt{coimage}\\
}
\gldefsyntax{valSpace.empty}{valSpace}{\lsv{}}{\lsempty{}}{\texttt{empty}}{\texttt{GLang.Val\allowbreak Space.empty}}{GLang/Values.lean:76}
\gldefsyntax{valSpace.real}{valSpace}{\lsv{}}{\lR{}}{\texttt{real}}{\texttt{GLang.Val\allowbreak Space.real}}{GLang/Values.lean:77}
\gldefsyntax{valSpace.dsum}{valSpace}{\lsv{}}{\lsnv{0} \losum \lsnv{1}}{\texttt{dsum}}{\texttt{GLang.Val\allowbreak Space.dsum}}{GLang/Values.lean:78}
\gldefsyntax{valSpace.tprod}{valSpace}{\lsv{}}{\lsnv{0} \loprod \lsnv{1}}{\texttt{tprod}}{\texttt{GLang.Val\allowbreak Space.tprod}}{GLang/Values.lean:79}
\gldefsyntax{valSpace.ffs}{valSpace}{\lsv{}}{\lR^{\leev{},\lprop{},\lsv{}}}{Four fundamental subspaces of a value linear map}{\texttt{GLang.Val\allowbreak Space.ffs}}{GLang/Values.lean:84}
\gldefsyntaxrows{valSpace}{
$\lsv{}$ & $::=$ & $\lsempty{}$ & \texttt{empty}\\
 & $|$ & $\lR{}$ & \texttt{real}\\
 & $|$ & $\lsnv{0} \losum \lsnv{1}$ & \texttt{dsum}\\
 & $|$ & $\lsnv{0} \loprod \lsnv{1}$ & \texttt{tprod}\\
 & $|$ & $\lR^{\leev{},\lprop{},\lsv{}}$ & Four fundamental subspaces of a value linear map\\
}
\gldefsyntax{valTy.space}{valTy}{\ltyv{}}{\lsv{}}{\texttt{space}}{\texttt{GLang.Val\allowbreak Ty.space}}{GLang/Values.lean:88}
\gldefsyntax{valTy.piTy}{valTy}{\ltyv{}}{(x\colon \ltynv{0}\fnto \ltynv{1})}{\texttt{pi\allowbreak Ty}}{\texttt{GLang.Val\allowbreak Ty.pi\allowbreak Ty}}{GLang/Values.lean:89}
\gldefsyntaxrows{valTy}{
$\ltyv{}$ & $::=$ & $\lsv{}$ & \texttt{space}\\
 & $|$ & $(x\colon \ltynv{0}\fnto \ltynv{1})$ & \texttt{pi\allowbreak Ty}\\
}
\gldefsyntax{valExpr.zero}{valExpr}{\leev{}}{\lvzero{}_{\lsv{}}}{Vector-space generic}{\texttt{GLang.Val\allowbreak Expr.zero}}{GLang/Values.lean:94}
\gldefsyntax{valExpr.rlit}{valExpr}{\leev{}}{r}{\texttt{rlit}}{\texttt{GLang.Val\allowbreak Expr.rlit}}{GLang/Values.lean:95}
\gldefsyntax{valExpr.scale}{valExpr}{\leev{}}{\lenv{0} \lvscale \lenv{1}}{\texttt{scale}}{\texttt{GLang.Val\allowbreak Expr.scale}}{GLang/Values.lean:96}
\gldefsyntax{valExpr.vadd}{valExpr}{\leev{}}{\lenv{0} \lvadd_{\lsv{}} \lenv{1}}{\texttt{vadd}}{\texttt{GLang.Val\allowbreak Expr.vadd}}{GLang/Values.lean:97}
\gldefsyntax{valExpr.dsumE}{valExpr}{\leev{}}{\lenv{0}\losum \lenv{1}}{\texttt{dsum\allowbreak E}}{\texttt{GLang.Val\allowbreak Expr.dsum\allowbreak E}}{GLang/Values.lean:98}
\gldefsyntax{valExpr.tprodE}{valExpr}{\leev{}}{\lenv{0}\loprod \lenv{1}}{\texttt{tprod\allowbreak E}}{\texttt{GLang.Val\allowbreak Expr.tprod\allowbreak E}}{GLang/Values.lean:99}
\gldefsyntax{valExpr.delta}{valExpr}{\leev{}}{\ldelta{n}{\lsv{}}}{Basis vector / identity tensor}{\texttt{GLang.Val\allowbreak Expr.delta}}{GLang/Values.lean:101}
\gldefsyntax{valExpr.idelta}{valExpr}{\leev{}}{\lidelta{\lsv{}}}{\texttt{idelta}}{\texttt{GLang.Val\allowbreak Expr.idelta}}{GLang/Values.lean:102}
\gldefsyntax{valExpr.vsum}{valExpr}{\leev{}}{\lsum^{\lsv{}}_{v\in \mathit{vs}}v}{Sum over a value-space index}{\texttt{GLang.Val\allowbreak Expr.vsum}}{GLang/Values.lean:109}
\gldefsyntax{valExpr.vlam}{valExpr}{\leev{}}{\llambda(x\colon \ltynv{0})\lrh \ltynv{1}\colon \leev{}}{Lambdas}{\texttt{GLang.Val\allowbreak Expr.vlam}}{GLang/Values.lean:114}
\gldefsyntax{valExpr.var}{valExpr}{\leev{}}{x_{\ltyv{}}}{Free variable}{\texttt{GLang.Val\allowbreak Expr.var}}{GLang/Values.lean:125}
\gldefsyntax{valExpr.ttv}{valExpr}{\leev{}}{\LTTVPP{\lsv{}}{\lenv{0}}{\lenv{1}}}{Tensor-vector partial contraction (paper:}{\texttt{GLang.Val\allowbreak Expr.ttv}}{GLang/Values.lean:129}
\gldefsyntax{valExpr.mtm}{valExpr}{\leev{}}{\glvmtm{\lsnv{c}}{\lsnv{1}}{\lenv{0}}{\lenv{1}}}{Matrix-tensor-matrix / "MTM" (paper:}{\texttt{GLang.Val\allowbreak Expr.mtm}}{GLang/Values.lean:134}
\gldefsyntax{valExpr.appV}{valExpr}{\leev{}}{\glvapp{\lenv{0}}{\lenv{1}}}{Application (linear / non-linear share one constructor in the value layer; typing distinguishes)}{\texttt{GLang.Val\allowbreak Expr.app\allowbreak V}}{GLang/Values.lean:139}
\gldefsyntax{valExpr.vkrn}{valExpr}{\leev{}}{\vkrn(\lenv{0}, \lenv{1})}{Kronecker / block products on values}{\texttt{GLang.Val\allowbreak Expr.vkrn}}{GLang/Values.lean:141}
\gldefsyntax{valExpr.vblk}{valExpr}{\leev{}}{\vblk(\lenv{0}, \lenv{1})}{\texttt{vblk}}{\texttt{GLang.Val\allowbreak Expr.vblk}}{GLang/Values.lean:142}
\gldefsyntax{valExpr.reshap}{valExpr}{\leev{}}{\glvreshape{\leev{}}{\ltynv{t}}{\ltynv{s}}}{Reshape between value types}{\texttt{GLang.Val\allowbreak Expr.reshap}}{GLang/Values.lean:146}
\gldefsyntax{valExpr.vtrans}{valExpr}{\leev{}}{\Trans(\leev{})}{Transpose of a tensor-shaped value}{\texttt{GLang.Val\allowbreak Expr.vtrans}}{GLang/Values.lean:148}
\gldefsyntax{valExpr.gsvdS}{valExpr}{\leev{}}{\glvop{GSVDS}(\lenv{0},\lenv{1},\lenv{2})}{(G)SVD}{\texttt{GLang.Val\allowbreak Expr.gsvd\allowbreak S}}{GLang/Values.lean:150}
\gldefsyntax{valExpr.gsvdU}{valExpr}{\leev{}}{\glvop{GSVDU}(\lenv{0},\lenv{1},\lenv{2})}{\texttt{gsvd\allowbreak U}}{\texttt{GLang.Val\allowbreak Expr.gsvd\allowbreak U}}{GLang/Values.lean:151}
\gldefsyntax{valExpr.gsvdUI}{valExpr}{\leev{}}{\glvop{GSVDUI}(\lenv{0},\lenv{1},\lenv{2})}{\texttt{gsvd\allowbreak UI}}{\texttt{GLang.Val\allowbreak Expr.gsvd\allowbreak UI}}{GLang/Values.lean:152}
\gldefsyntax{valExpr.gsvdV}{valExpr}{\leev{}}{\glvop{GSVDV}(\lenv{0},\lenv{1},\lenv{2})}{\texttt{gsvd\allowbreak V}}{\texttt{GLang.Val\allowbreak Expr.gsvd\allowbreak V}}{GLang/Values.lean:153}
\gldefsyntax{valExpr.gsvdVI}{valExpr}{\leev{}}{\glvop{GSVDVI}(\lenv{0},\lenv{1},\lenv{2})}{\texttt{gsvd\allowbreak VI}}{\texttt{GLang.Val\allowbreak Expr.gsvd\allowbreak VI}}{GLang/Values.lean:154}
\gldefsyntax{valExpr.pinv}{valExpr}{\leev{}}{\leev{}^{+}}{Pseudo-inverse of a value (paper writes}{\texttt{GLang.Val\allowbreak Expr.pinv}}{GLang/Values.lean:157}
\gldefsyntax{valExpr.gpinv}{valExpr}{\leev{}}{\glvop{GPINV}(\lenv{0},\lenv{1},\lenv{2})}{Generalized pseudo-inverse with explicit inner-product operands}{\texttt{GLang.Val\allowbreak Expr.gpinv}}{GLang/Values.lean:159}
\gldefsyntax{valExpr.rrefRep}{valExpr}{\leev{}}{\glvop{RREFRep}(\lenv{0},\lenv{1})}{RREF representation}{\texttt{GLang.Val\allowbreak Expr.rref\allowbreak Rep}}{GLang/Values.lean:161}
\gldefsyntax{valExpr.inv}{valExpr}{\leev{}}{\leev{}^{-1}}{Inverse-form}{\texttt{GLang.Val\allowbreak Expr.inv}}{GLang/Values.lean:169}
\gldefsyntax{valExpr.svds}{valExpr}{\leev{}}{\glvop{SVDS}(\leev{})}{Standard-basis singular-value block}{\texttt{GLang.Val\allowbreak Expr.svds}}{GLang/Values.lean:175}
\gldefsyntaxrows{valExpr}{
$\leev{}$ & $::=$ & $\lvzero{}_{\lsv{}}$ & Vector-space generic\\
 & $|$ & $r$ & \texttt{rlit}\\
 & $|$ & $\lenv{0} \lvscale \lenv{1}$ & \texttt{scale}\\
 & $|$ & $\lenv{0} \lvadd_{\lsv{}} \lenv{1}$ & \texttt{vadd}\\
 & $|$ & $\lenv{0}\losum \lenv{1}$ & \texttt{dsum\allowbreak E}\\
 & $|$ & $\lenv{0}\loprod \lenv{1}$ & \texttt{tprod\allowbreak E}\\
 & $|$ & $\ldelta{n}{\lsv{}}$ & Basis vector / identity tensor\\
 & $|$ & $\lidelta{\lsv{}}$ & \texttt{idelta}\\
 & $|$ & $\lsum^{\lsv{}}_{v\in \mathit{vs}}v$ & Sum over a value-space index\\
 & $|$ & $\llambda(x\colon \ltynv{0})\lrh \ltynv{1}\colon \leev{}$ & Lambdas\\
 & $|$ & $x_{\ltyv{}}$ & Free variable\\
 & $|$ & $\LTTVPP{\lsv{}}{\lenv{0}}{\lenv{1}}$ & Tensor-vector partial contraction (paper:\\
 & $|$ & $\glvmtm{\lsnv{c}}{\lsnv{1}}{\lenv{0}}{\lenv{1}}$ & Matrix-tensor-matrix / "MTM" (paper:\\
 & $|$ & $\glvapp{\lenv{0}}{\lenv{1}}$ & Application (linear / non-linear share one constructor in the value layer; typing distinguishes)\\
 & $|$ & $\vkrn(\lenv{0}, \lenv{1})$ & Kronecker / block products on values\\
 & $|$ & $\vblk(\lenv{0}, \lenv{1})$ & \texttt{vblk}\\
 & $|$ & $\glvreshape{\leev{}}{\ltynv{t}}{\ltynv{s}}$ & Reshape between value types\\
 & $|$ & $\Trans(\leev{})$ & Transpose of a tensor-shaped value\\
 & $|$ & $\glvop{GSVDS}(\lenv{0},\lenv{1},\lenv{2})$ & (G)SVD\\
 & $|$ & $\glvop{GSVDU}(\lenv{0},\lenv{1},\lenv{2})$ & \texttt{gsvd\allowbreak U}\\
 & $|$ & $\glvop{GSVDUI}(\lenv{0},\lenv{1},\lenv{2})$ & \texttt{gsvd\allowbreak UI}\\
 & $|$ & $\glvop{GSVDV}(\lenv{0},\lenv{1},\lenv{2})$ & \texttt{gsvd\allowbreak V}\\
 & $|$ & $\glvop{GSVDVI}(\lenv{0},\lenv{1},\lenv{2})$ & \texttt{gsvd\allowbreak VI}\\
 & $|$ & $\leev{}^{+}$ & Pseudo-inverse of a value (paper writes\\
 & $|$ & $\glvop{GPINV}(\lenv{0},\lenv{1},\lenv{2})$ & Generalized pseudo-inverse with explicit inner-product operands\\
 & $|$ & $\glvop{RREFRep}(\lenv{0},\lenv{1})$ & RREF representation\\
 & $|$ & $\leev{}^{-1}$ & Inverse-form\\
 & $|$ & $\glvop{SVDS}(\leev{})$ & Standard-basis singular-value block\\
}
\gldefsyntax{valKind.base}{valKind}{\lky{}}{\lkbase{}}{\texttt{base}}{\texttt{GLang.Val\allowbreak Kind.base}}{GLang/Values.lean:210}
\gldefsyntax{valKind.arrow}{valKind}{\lky{}}{\lkn{0} \fnto \lkn{1}}{\texttt{arrow}}{\texttt{GLang.Val\allowbreak Kind.arrow}}{GLang/Values.lean:211}
\gldefsyntaxrows{valKind}{
$\lky{}$ & $::=$ & $\lkbase{}$ & \texttt{base}\\
 & $|$ & $\lkn{0} \fnto \lkn{1}$ & \texttt{arrow}\\
}
\gldefsyntax{meta.lctx}{meta}{\lDelta{}}{\lDelta{}}{Linear context}{\texttt{GLang.LCtx}}{GLang/Typing.lean:29}
\gldefsyntax{meta.nlctx}{meta}{\lepsilon{}}{\lepsilon{}}{Non-linear context}{\texttt{GLang.NLCtx}}{GLang/Typing.lean:30}
\gldefsyntax{meta.evalCtx}{meta}{\evalCtx{}}{\evalCtx{}}{Evaluation context (the store)}{\texttt{GLang.Eval\allowbreak Ctx}}{GLang/ValueOperational.lean:133}
\gldefsyntax{meta.denv}{meta}{\sigma}{\sigma}{Denotational environment}{\texttt{GLang.DEnv}}{GLang/Denot.lean:580}
\gldefsyntax{meta.valCtx}{meta}{\valCheckCtx{}}{\valCheckCtx{}}{Value typing context}{\texttt{GLang.Val\allowbreak Ctx}}{GLang/ValueTyping.lean:43}
\gldefsyntax{meta.valEnv}{meta}{\glvEnv}{\glvEnv}{Value semantic environment}{\texttt{GLang.Val\allowbreak Env\allowbreak Sem}}{GLang/ValueDenotSem.lean:425}
\gldefsyntax{meta.directOracle}{meta}{\mathcal{O}(\glvEnv;\Gamma;\Delta)}{\mathcal{O}(\glvEnv;\Gamma;\Delta)}{Constructive decomposition oracle at a named environment and local type/value stacks}{\texttt{GLang.direct\allowbreak Oracle}}{GLang/Adequacy/DirectOracle.lean:14}
\gldefsyntax{meta.tof}{meta}{\tof{\lb{}}}{\tof{\lb{}}}{Type of a basis}{\texttt{GLang.Basis.tof}}{GLang/Basis.lean:74}
\gldefsyntax{meta.sof}{meta}{\sof{\lb{}}}{\sof{\lb{}}}{Space of a basis}{\texttt{GLang.Basis.sof}}{GLang/Basis.lean:60}
\gldefsyntax{meta.stdBasis}{meta}{\stdBasisCtx{\lty{}}{\lctxnle}}{\stdBasisCtx{\lty{}}{\lctxnle}}{Syntax-directed standard basis of a source type}{\texttt{GLang.Basis.std}}{GLang/Basis.lean:523}
\gldefsyntax{meta.stdBasisSpace}{meta}{\stdBasisCtx{\ls{}}{\lctxnle}}{\stdBasisCtx{\ls{}}{\lctxnle}}{Space-level recursion used by the standard-basis construction}{\texttt{GLang.Basis.std\allowbreak Space}}{GLang/Basis.lean:506}
\gldefsyntax{meta.iof}{meta}{\iof{\ls{}}}{\iof{\ls{}}}{Derived source term denoting the canonical inner product on a space}{\texttt{GLang.Space.iof}}{GLang/InnerProduct.lean:193}
\gldefsyntax{meta.dof}{meta}{\dof{\lsv{}}{\glvEnv}}{\dof{\lsv{}}{\glvEnv}}{Degrees of freedom of a value space}{\texttt{GLang.Val\allowbreak Space.dof}}{GLang/ValueDenot.lean:135}
\gldefsyntax{meta.shpOf}{meta}{\shpOf{\lsv{}}{\glvEnv}}{\shpOf{\lsv{}}{\glvEnv}}{Shape of a value space}{\texttt{GLang.Val\allowbreak Space.shp\allowbreak Of}}{GLang/ValueDenot.lean:116}
\gldefsyntax{meta.inferType}{meta}{\operatorname{inferType}(\leev{})}{\operatorname{inferType}(\leev{})}{Inferred type of a value}{\texttt{GLang.Val\allowbreak Expr.infer\allowbreak Type}}{GLang/ValueDenot.lean:790}
\gldefsyntax{meta.inferSpace}{meta}{\operatorname{inferSpace}(\leev{})}{\operatorname{inferSpace}(\leev{})}{Inferred result space of a value}{\texttt{GLang.Val\allowbreak Expr.infer\allowbreak Space}}{GLang/ValueDenot.lean:1020}
\gldefsyntax{meta.denotTerm}{meta}{\Sem{\lee{}}{\sigma}}{\Sem{\lee{}}{\sigma}}{Denotation of a source term}{\texttt{GLang.denot\allowbreak Term}}{GLang/Denot.lean:877}
\gldefsyntax{meta.denotTy}{meta}{\SemT{\lty{}}{\sigma}}{\SemT{\lty{}}{\sigma}}{Denotation of a source type}{\texttt{GLang.denot\allowbreak Ty}}{GLang/Denot.lean:874}
\gldefsyntax{meta.denotBasis}{meta}{\SemBB{\lb{}}{\sigma}}{\SemBB{\lb{}}{\sigma}}{Denotation of a basis}{\texttt{GLang.denot\allowbreak Basis}}{GLang/Basis.lean:886}
\gldefsyntax{meta.denotVal}{meta}{\SemA{\leev{}}{\glvEnv}{}}{\SemA{\leev{}}{\glvEnv}{}}{Denotation of a value expression}{\texttt{GLang.Adequacy.Partial\allowbreak Direct.eval\allowbreak Val}}{GLang/Adequacy/PartialSourcePublic.lean:12}
\gldefmetarows{
$\lDelta{}$ & Linear context & \texttt{\footnotesize GLang.LCtx}\\
$\lepsilon{}$ & Non-linear context & \texttt{\footnotesize GLang.NLCtx}\\
$\evalCtx{}$ & Evaluation context & \texttt{\footnotesize GLang.Eval\allowbreak Ctx}\\
$\sigma$ & Denotational environment & \texttt{\footnotesize GLang.DEnv}\\
$\valCheckCtx{}$ & Value typing context & \texttt{\footnotesize GLang.Val\allowbreak Ctx}\\
$\glvEnv$ & Value semantic environment & \texttt{\footnotesize GLang.Val\allowbreak Env\allowbreak Sem}\\
$\mathcal{O}(\glvEnv;\Gamma;\Delta)$ & Constructive decomposition oracle at a named environment and local type/value stacks & \texttt{\footnotesize GLang.direct\allowbreak Oracle}\\
$\tof{\lb{}}$ & Type of a basis & \texttt{\footnotesize GLang.Basis.tof}\\
$\sof{\lb{}}$ & Space of a basis & \texttt{\footnotesize GLang.Basis.sof}\\
$\stdBasisCtx{\lty{}}{\lctxnle}$ & Syntax-directed standard basis of a source type & \texttt{\footnotesize GLang.Basis.std}\\
$\stdBasisCtx{\ls{}}{\lctxnle}$ & Space-level recursion used by the standard-basis construction & \texttt{\footnotesize GLang.Basis.std\allowbreak Space}\\
$\iof{\ls{}}$ & Derived source term denoting the canonical inner product on a space & \texttt{\footnotesize GLang.Space.iof}\\
$\dof{\lsv{}}{\glvEnv}$ & Degrees of freedom of a value space & \texttt{\footnotesize GLang.Val\allowbreak Space.dof}\\
$\shpOf{\lsv{}}{\glvEnv}$ & Shape of a value space & \texttt{\footnotesize GLang.Val\allowbreak Space.shp\allowbreak Of}\\
$\operatorname{inferType}(\leev{})$ & Inferred type of a value & \texttt{\footnotesize GLang.Val\allowbreak Expr.infer\allowbreak Type}\\
$\operatorname{inferSpace}(\leev{})$ & Inferred result space of a value & \texttt{\footnotesize GLang.Val\allowbreak Expr.infer\allowbreak Space}\\
$\Sem{\lee{}}{\sigma}$ & Denotation of a source term & \texttt{\footnotesize GLang.denot\allowbreak Term}\\
$\SemT{\lty{}}{\sigma}$ & Denotation of a source type & \texttt{\footnotesize GLang.denot\allowbreak Ty}\\
$\SemBB{\lb{}}{\sigma}$ & Denotation of a basis & \texttt{\footnotesize GLang.denot\allowbreak Basis}\\
$\SemA{\leev{}}{\glvEnv}{}$ & Denotation of a value expression & \texttt{\footnotesize GLang.Adequacy.Partial\allowbreak Direct.eval\allowbreak Val}\\
}
\gldefjudgment{inductiveCtxWF}{\vdash(\lDelta{},\lepsilon{})}{Inductive well-formed source contexts.}{fig:gl-context-wf}
\gldefrule{inductiveCtxWF.empty}{inductiveCtxWF}{\textsc{Ctx-Empty}}{\phantom{.}}{\vdash (\lemptyset{}, \lemptyset{})}{\texttt{GLang.Inductive\allowbreak Ctx\allowbreak WF.empty}}{GLang/SourceContextInductive.lean:29}
\gldefruleexplanation{inductiveCtxWF.empty}{The empty linear and nonlinear contexts are well formed.}
\gldefrule{inductiveCtxWF.linear}{inductiveCtxWF}{\textsc{Ctx-Linear}}{\vdash (\lDelta{}, \lepsilon{}) \\ \neg(\mathit{x} \in \lDelta{}) \\ \neg(\mathit{x} \in \lepsilon{}) \\ \forall \mathit{y}.\; \mathit{y} \in \mathsf{fv}(\ls{}) \Rightarrow \mathit{y} \in \lepsilon{} \\ (\lemptyset, \lepsilon{})\vdash \ls{}\colon\svalid}{\vdash ((\lDelta{}, \mathit{x}\colon \ls{}), \lepsilon{})}{\texttt{GLang.Inductive\allowbreak Ctx\allowbreak WF.linear}}{GLang/SourceContextInductive.lean:30}
\gldefruleexplanation{inductiveCtxWF.linear}{A fresh linear variable may be added at a valid space whose free variables already belong to $\lepsilon$.}
\gldefrule{inductiveCtxWF.nonlinear}{inductiveCtxWF}{\textsc{Ctx-Nonlinear}}{\vdash (\lDelta{}, \lepsilon{}) \\ \neg(\mathit{x} \in \lDelta{}) \\ \neg(\mathit{x} \in \lepsilon{}) \\ \forall \mathit{y}.\; \mathit{y} \in \mathsf{fv}(\lty{}) \Rightarrow \mathit{y} \in \lepsilon{} \\ (\lemptyset, \lepsilon{})\vdash \lty{}\colon \lky{}}{\vdash (\lDelta{}, (\lepsilon{}, \mathit{x}\colon \lty{}))}{\texttt{GLang.Inductive\allowbreak Ctx\allowbreak WF.nonlinear}}{GLang/SourceContextInductive.lean:36}
\gldefruleexplanation{inductiveCtxWF.nonlinear}{A fresh nonlinear variable may be added at a well-kinded type whose free variables already belong to $\lepsilon$.}
\gldefrulerows{inductiveCtxWF}{
\textsc{Ctx-Empty} & $\vdash (\lemptyset{}, \lemptyset{})$ & \texttt{\footnotesize GLang/Source\allowbreak Context\allowbreak Inductive.lean:29}\\
\textsc{Ctx-Linear} & $\vdash ((\lDelta{}, \mathit{x}\colon \ls{}), \lepsilon{})$ & \texttt{\footnotesize GLang/Source\allowbreak Context\allowbreak Inductive.lean:30}\\
\textsc{Ctx-Nonlinear} & $\vdash (\lDelta{}, (\lepsilon{}, \mathit{x}\colon \lty{}))$ & \texttt{\footnotesize GLang/Source\allowbreak Context\allowbreak Inductive.lean:36}\\
}{
\gpfit{\inferrule*[]{\phantom{.}}{\vdash (\lemptyset{}, \lemptyset{})}}
\and
\gpfit{\inferrule*[]{\vdash (\lDelta{}, \lepsilon{}) \\ \neg(\mathit{x} \in \lDelta{}) \\ \neg(\mathit{x} \in \lepsilon{}) \\ \forall \mathit{y}.\; \mathit{y} \in \mathsf{fv}(\ls{}) \Rightarrow \mathit{y} \in \lepsilon{} \\ (\lemptyset, \lepsilon{})\vdash \ls{}\colon\svalid}{\vdash ((\lDelta{}, \mathit{x}\colon \ls{}), \lepsilon{})}}
\and
\gpfit{\inferrule*[]{\vdash (\lDelta{}, \lepsilon{}) \\ \neg(\mathit{x} \in \lDelta{}) \\ \neg(\mathit{x} \in \lepsilon{}) \\ \forall \mathit{y}.\; \mathit{y} \in \mathsf{fv}(\lty{}) \Rightarrow \mathit{y} \in \lepsilon{} \\ (\lemptyset, \lepsilon{})\vdash \lty{}\colon \lky{}}{\vdash (\lDelta{}, (\lepsilon{}, \mathit{x}\colon \lty{}))}}
}{
\gpfit{\inferrule*[lab=\textsc{Ctx-Empty}]{\phantom{.}}{\vdash (\lemptyset{}, \lemptyset{})}}
\and
\gpfit{\inferrule*[lab=\textsc{Ctx-Linear}]{\vdash (\lDelta{}, \lepsilon{}) \\ \neg(\mathit{x} \in \lDelta{}) \\ \neg(\mathit{x} \in \lepsilon{}) \\ \forall \mathit{y}.\; \mathit{y} \in \mathsf{fv}(\ls{}) \Rightarrow \mathit{y} \in \lepsilon{} \\ (\lemptyset, \lepsilon{})\vdash \ls{}\colon\svalid}{\vdash ((\lDelta{}, \mathit{x}\colon \ls{}), \lepsilon{})}}
\and
\gpfit{\inferrule*[lab=\textsc{Ctx-Nonlinear}]{\vdash (\lDelta{}, \lepsilon{}) \\ \neg(\mathit{x} \in \lDelta{}) \\ \neg(\mathit{x} \in \lepsilon{}) \\ \forall \mathit{y}.\; \mathit{y} \in \mathsf{fv}(\lty{}) \Rightarrow \mathit{y} \in \lepsilon{} \\ (\lemptyset, \lepsilon{})\vdash \lty{}\colon \lky{}}{\vdash (\lDelta{}, (\lepsilon{}, \mathit{x}\colon \lty{}))}}
}
\gldefjudgment{sValid}{\lctxnle\vdash\ls\colon\svalid}{Well-formed spaces (generated from \texttt{GLang.SValid}).}{fig:gl-svalid}
\gldefrule{sValid.empty}{sValid}{\textsc{S-Empty}}{\phantom{.}}{(\lemptyset, \lepsilon{})\vdash \lsempty{}\colon\svalid}{\texttt{GLang.SValid.empty}}{GLang/Typing.lean:133}
\gldefruleexplanation{sValid.empty}{The zero-dimensional space is well formed in every nonlinear context.}
\gldefrule{sValid.real}{sValid}{\textsc{S-Real}}{\phantom{.}}{(\lemptyset, \lepsilon{})\vdash \lR{}\colon\svalid}{\texttt{GLang.SValid.real}}{GLang/Typing.lean:135}
\gldefruleexplanation{sValid.real}{The scalar space $\mathbb R$ is well formed in every nonlinear context.}
\gldefrule{sValid.dual}{sValid}{\textsc{S-Dual}}{(\lemptyset, \lepsilon{})\vdash \ls{}\colon\svalid}{(\lemptyset, \lepsilon{})\vdash \ls{}^{\lstar}\colon\svalid}{\texttt{GLang.SValid.dual}}{GLang/Typing.lean:137}
\gldefruleexplanation{sValid.dual}{The dual of a well-formed space is well formed.}
\gldefrule{sValid.dsum}{sValid}{\textsc{S-Dsum}}{\forall i.\; (\lemptyset, \lepsilon{})\vdash \lsn{i}\colon\svalid}{(\lemptyset, \lepsilon{})\vdash \lsn{0} \losum \lsn{1}\colon\svalid}{\texttt{GLang.SValid.dsum}}{GLang/Typing.lean:139}
\gldefruleexplanation{sValid.dsum}{Two well-formed spaces determine a well-formed direct sum.}
\gldefrule{sValid.tprod}{sValid}{\textsc{S-Tprod}}{\forall i.\; (\lemptyset, \lepsilon{})\vdash \lsn{i}\colon\svalid}{(\lemptyset, \lepsilon{})\vdash \lsn{0} \loprod \lsn{1}\colon\svalid}{\texttt{GLang.SValid.tprod}}{GLang/Typing.lean:141}
\gldefruleexplanation{sValid.tprod}{Two well-formed spaces determine a well-formed tensor product.}
\gldefrule{sValid.ker}{sValid}{\textsc{S-Ker}}{(\lemptyset{}, \lepsilon{})\vdash \lee{}\colon \lsn{0}^{\lstar} \loprod \lsn{1}}{(\lemptyset, \lepsilon{})\vdash \Kernel(\lee{})\colon\svalid}{\texttt{GLang.SValid.ker}}{GLang/Typing.lean:149}
\gldefruleexplanation{sValid.ker}{A well-typed linear map with no linear free variables determines a well-formed kernel.}
\gldefrule{sValid.coker}{sValid}{\textsc{S-Coker}}{(\lemptyset{}, \lepsilon{})\vdash \lee{}\colon \lsn{0}^{\lstar} \loprod \lsn{1}}{(\lemptyset, \lepsilon{})\vdash \CoKernel(\lee{})\colon\svalid}{\texttt{GLang.SValid.coker}}{GLang/Typing.lean:152}
\gldefruleexplanation{sValid.coker}{A well-typed linear map with no linear free variables determines a well-formed cokernel.}
\gldefrule{sValid.img}{sValid}{\textsc{S-Img}}{(\lemptyset{}, \lepsilon{})\vdash \lee{}\colon \lsn{0}^{\lstar} \loprod \lsn{1}}{(\lemptyset, \lepsilon{})\vdash \Image(\lee{})\colon\svalid}{\texttt{GLang.SValid.img}}{GLang/Typing.lean:155}
\gldefruleexplanation{sValid.img}{A well-typed linear map with no linear free variables determines a well-formed image.}
\gldefrule{sValid.coimg}{sValid}{\textsc{S-Coimg}}{(\lemptyset{}, \lepsilon{})\vdash \lee{}\colon \lsn{0}^{\lstar} \loprod \lsn{1}}{(\lemptyset, \lepsilon{})\vdash \CoImage(\lee{})\colon\svalid}{\texttt{GLang.SValid.coimg}}{GLang/Typing.lean:158}
\gldefruleexplanation{sValid.coimg}{A well-typed linear map with no linear free variables determines a well-formed coimage.}
\gldefrulerows{sValid}{
\textsc{S-Empty} & $(\lemptyset, \lepsilon{})\vdash \lsempty{}\colon\svalid$ & \texttt{\footnotesize GLang/Typing.lean:133}\\
\textsc{S-Real} & $(\lemptyset, \lepsilon{})\vdash \lR{}\colon\svalid$ & \texttt{\footnotesize GLang/Typing.lean:135}\\
\textsc{S-Dual} & $(\lemptyset, \lepsilon{})\vdash \ls{}^{\lstar}\colon\svalid$ & \texttt{\footnotesize GLang/Typing.lean:137}\\
\textsc{S-Dsum} & $(\lemptyset, \lepsilon{})\vdash \lsn{0} \losum \lsn{1}\colon\svalid$ & \texttt{\footnotesize GLang/Typing.lean:139}\\
\textsc{S-Tprod} & $(\lemptyset, \lepsilon{})\vdash \lsn{0} \loprod \lsn{1}\colon\svalid$ & \texttt{\footnotesize GLang/Typing.lean:141}\\
\textsc{S-Ker} & $(\lemptyset, \lepsilon{})\vdash \Kernel(\lee{})\colon\svalid$ & \texttt{\footnotesize GLang/Typing.lean:149}\\
\textsc{S-Coker} & $(\lemptyset, \lepsilon{})\vdash \CoKernel(\lee{})\colon\svalid$ & \texttt{\footnotesize GLang/Typing.lean:152}\\
\textsc{S-Img} & $(\lemptyset, \lepsilon{})\vdash \Image(\lee{})\colon\svalid$ & \texttt{\footnotesize GLang/Typing.lean:155}\\
\textsc{S-Coimg} & $(\lemptyset, \lepsilon{})\vdash \CoImage(\lee{})\colon\svalid$ & \texttt{\footnotesize GLang/Typing.lean:158}\\
}{
\gpfit{\inferrule*[]{\phantom{.}}{(\lemptyset, \lepsilon{})\vdash \lsempty{}\colon\svalid}}
\and
\gpfit{\inferrule*[]{\phantom{.}}{(\lemptyset, \lepsilon{})\vdash \lR{}\colon\svalid}}
\and
\gpfit{\inferrule*[]{(\lemptyset, \lepsilon{})\vdash \ls{}\colon\svalid}{(\lemptyset, \lepsilon{})\vdash \ls{}^{\lstar}\colon\svalid}}
\and
\gpfit{\inferrule*[]{\forall i.\; (\lemptyset, \lepsilon{})\vdash \lsn{i}\colon\svalid}{(\lemptyset, \lepsilon{})\vdash \lsn{0} \losum \lsn{1}\colon\svalid}}
\and
\gpfit{\inferrule*[]{\forall i.\; (\lemptyset, \lepsilon{})\vdash \lsn{i}\colon\svalid}{(\lemptyset, \lepsilon{})\vdash \lsn{0} \loprod \lsn{1}\colon\svalid}}
\and
\gpfit{\inferrule*[]{(\lemptyset{}, \lepsilon{})\vdash \lee{}\colon \lsn{0}^{\lstar} \loprod \lsn{1}}{(\lemptyset, \lepsilon{})\vdash \Kernel(\lee{})\colon\svalid}}
\and
\gpfit{\inferrule*[]{(\lemptyset{}, \lepsilon{})\vdash \lee{}\colon \lsn{0}^{\lstar} \loprod \lsn{1}}{(\lemptyset, \lepsilon{})\vdash \CoKernel(\lee{})\colon\svalid}}
\and
\gpfit{\inferrule*[]{(\lemptyset{}, \lepsilon{})\vdash \lee{}\colon \lsn{0}^{\lstar} \loprod \lsn{1}}{(\lemptyset, \lepsilon{})\vdash \Image(\lee{})\colon\svalid}}
\and
\gpfit{\inferrule*[]{(\lemptyset{}, \lepsilon{})\vdash \lee{}\colon \lsn{0}^{\lstar} \loprod \lsn{1}}{(\lemptyset, \lepsilon{})\vdash \CoImage(\lee{})\colon\svalid}}
}{
\gpfit{\inferrule*[lab=\textsc{S-Empty}]{\phantom{.}}{(\lemptyset, \lepsilon{})\vdash \lsempty{}\colon\svalid}}
\and
\gpfit{\inferrule*[lab=\textsc{S-Real}]{\phantom{.}}{(\lemptyset, \lepsilon{})\vdash \lR{}\colon\svalid}}
\and
\gpfit{\inferrule*[lab=\textsc{S-Dual}]{(\lemptyset, \lepsilon{})\vdash \ls{}\colon\svalid}{(\lemptyset, \lepsilon{})\vdash \ls{}^{\lstar}\colon\svalid}}
\and
\gpfit{\inferrule*[lab=\textsc{S-Dsum}]{\forall i.\; (\lemptyset, \lepsilon{})\vdash \lsn{i}\colon\svalid}{(\lemptyset, \lepsilon{})\vdash \lsn{0} \losum \lsn{1}\colon\svalid}}
\and
\gpfit{\inferrule*[lab=\textsc{S-Tprod}]{\forall i.\; (\lemptyset, \lepsilon{})\vdash \lsn{i}\colon\svalid}{(\lemptyset, \lepsilon{})\vdash \lsn{0} \loprod \lsn{1}\colon\svalid}}
\and
\gpfit{\inferrule*[lab=\textsc{S-Ker}]{(\lemptyset{}, \lepsilon{})\vdash \lee{}\colon \lsn{0}^{\lstar} \loprod \lsn{1}}{(\lemptyset, \lepsilon{})\vdash \Kernel(\lee{})\colon\svalid}}
\and
\gpfit{\inferrule*[lab=\textsc{S-Coker}]{(\lemptyset{}, \lepsilon{})\vdash \lee{}\colon \lsn{0}^{\lstar} \loprod \lsn{1}}{(\lemptyset, \lepsilon{})\vdash \CoKernel(\lee{})\colon\svalid}}
\and
\gpfit{\inferrule*[lab=\textsc{S-Img}]{(\lemptyset{}, \lepsilon{})\vdash \lee{}\colon \lsn{0}^{\lstar} \loprod \lsn{1}}{(\lemptyset, \lepsilon{})\vdash \Image(\lee{})\colon\svalid}}
\and
\gpfit{\inferrule*[lab=\textsc{S-Coimg}]{(\lemptyset{}, \lepsilon{})\vdash \lee{}\colon \lsn{0}^{\lstar} \loprod \lsn{1}}{(\lemptyset, \lepsilon{})\vdash \CoImage(\lee{})\colon\svalid}}
}
\gldefjudgment{tValid}{\lctxnle\vdash\lty\colon\tvalid}{Well-formed types (generated from \texttt{GLang.TValid}).}{fig:gl-tvalid}
\gldefrule{tValid.space}{tValid}{\textsc{K-Space}}{(\lemptyset, \lepsilon{})\vdash \ls{}\colon\svalid}{(\lemptyset, \lepsilon{})\vdash \ls{}\colon \lkbase{}}{\texttt{GLang.TValid.space}}{GLang/Typing.lean:165}
\gldefruleexplanation{tValid.space}{A well-formed space is a type of base kind.}
\gldefrule{tValid.piTy}{tValid}{\textsc{K-Pi\allowbreak Ty}}{(\lemptyset, \lepsilon{})\vdash \ltyn{0}\colon \lkn{0} \\ (\lemptyset, (\lepsilon{}, \mathit{x}\colon \ltyn{0}))\vdash \ltyn{1}\colon \lkn{1}}{(\lemptyset, \lepsilon{})\vdash (x\colon \ltyn{0}\fnto \ltyn{1})\colon \lkn{0} \fnto \lkn{1}}{\texttt{GLang.TValid.pi\allowbreak Ty}}{GLang/Typing.lean:170}
\gldefruleexplanation{tValid.piTy}{A well-kinded domain and a codomain well kinded under a fresh domain variable form a dependent function type of arrow kind.}
\gldefrulerows{tValid}{
\textsc{K-Space} & $(\lemptyset, \lepsilon{})\vdash \ls{}\colon \lkbase{}$ & \texttt{\footnotesize GLang/Typing.lean:165}\\
\textsc{K-Pi\allowbreak Ty} & $(\lemptyset, \lepsilon{})\vdash (x\colon \ltyn{0}\fnto \ltyn{1})\colon \lkn{0} \fnto \lkn{1}$ & \texttt{\footnotesize GLang/Typing.lean:170}\\
}{
\gpfit{\inferrule*[]{(\lemptyset, \lepsilon{})\vdash \ls{}\colon\svalid}{(\lemptyset, \lepsilon{})\vdash \ls{}\colon \lkbase{}}}
\and
\gpfit{\inferrule*[]{(\lemptyset, \lepsilon{})\vdash \ltyn{0}\colon \lkn{0} \\ (\lemptyset, (\lepsilon{}, \mathit{x}\colon \ltyn{0}))\vdash \ltyn{1}\colon \lkn{1}}{(\lemptyset, \lepsilon{})\vdash (x\colon \ltyn{0}\fnto \ltyn{1})\colon \lkn{0} \fnto \lkn{1}}}
}{
\gpfit{\inferrule*[lab=\textsc{K-Space}]{(\lemptyset, \lepsilon{})\vdash \ls{}\colon\svalid}{(\lemptyset, \lepsilon{})\vdash \ls{}\colon \lkbase{}}}
\and
\gpfit{\inferrule*[lab=\textsc{K-Pi\allowbreak Ty}]{(\lemptyset, \lepsilon{})\vdash \ltyn{0}\colon \lkn{0} \\ (\lemptyset, (\lepsilon{}, \mathit{x}\colon \ltyn{0}))\vdash \ltyn{1}\colon \lkn{1}}{(\lemptyset, \lepsilon{})\vdash (x\colon \ltyn{0}\fnto \ltyn{1})\colon \lkn{0} \fnto \lkn{1}}}
}
\gldefjudgment{subType}{\lctxnle\vdash\lsn{0}\leq\lsn{1}}{Subspace relation (generated from \texttt{GLang.SubType}).}{fig:gl-subtype}
\gldefrule{subType.subKer}{subType}{\textsc{Sub-Sub\allowbreak Ker}}{(\lemptyset{}, \lepsilon{})\vdash \lee{}\colon \lsn{0}^{\lstar} \loprod \lsn{1}}{(\lemptyset, \lepsilon{})\vdash \Kernel(\lee{})\leq \lsn{0}}{\texttt{GLang.Sub\allowbreak Type.sub\_\allowbreak ker}}{GLang/Typing.lean:181}
\gldefruleexplanation{subType.subKer}{The kernel of a well-typed linear map is a subspace of its domain.}
\gldefrule{subType.subCoker}{subType}{\textsc{Sub-Sub\allowbreak Coker}}{(\lemptyset{}, \lepsilon{})\vdash \lee{}\colon \lsn{0}^{\lstar} \loprod \lsn{1}}{(\lemptyset, \lepsilon{})\vdash \CoKernel(\lee{})\leq \lsn{0}}{\texttt{GLang.Sub\allowbreak Type.sub\_\allowbreak coker}}{GLang/Typing.lean:184}
\gldefruleexplanation{subType.subCoker}{The cokernel of a well-typed linear map is a subspace of its domain.}
\gldefrule{subType.subImg}{subType}{\textsc{Sub-Sub\allowbreak Img}}{(\lemptyset{}, \lepsilon{})\vdash \lee{}\colon \lsn{0}^{\lstar} \loprod \lsn{1}}{(\lemptyset, \lepsilon{})\vdash \Image(\lee{})\leq \lsn{1}}{\texttt{GLang.Sub\allowbreak Type.sub\_\allowbreak img}}{GLang/Typing.lean:187}
\gldefruleexplanation{subType.subImg}{The image of a well-typed linear map is a subspace of its codomain.}
\gldefrule{subType.subCoimg}{subType}{\textsc{Sub-Sub\allowbreak Coimg}}{(\lemptyset{}, \lepsilon{})\vdash \lee{}\colon \lsn{0}^{\lstar} \loprod \lsn{1}}{(\lemptyset, \lepsilon{})\vdash \CoImage(\lee{})\leq \lsn{1}}{\texttt{GLang.Sub\allowbreak Type.sub\_\allowbreak coimg}}{GLang/Typing.lean:190}
\gldefruleexplanation{subType.subCoimg}{The coimage of a well-typed linear map is a subspace of its codomain.}
\gldefrulerows{subType}{
\textsc{Sub-Sub\allowbreak Ker} & $(\lemptyset, \lepsilon{})\vdash \Kernel(\lee{})\leq \lsn{0}$ & \texttt{\footnotesize GLang/Typing.lean:181}\\
\textsc{Sub-Sub\allowbreak Coker} & $(\lemptyset, \lepsilon{})\vdash \CoKernel(\lee{})\leq \lsn{0}$ & \texttt{\footnotesize GLang/Typing.lean:184}\\
\textsc{Sub-Sub\allowbreak Img} & $(\lemptyset, \lepsilon{})\vdash \Image(\lee{})\leq \lsn{1}$ & \texttt{\footnotesize GLang/Typing.lean:187}\\
\textsc{Sub-Sub\allowbreak Coimg} & $(\lemptyset, \lepsilon{})\vdash \CoImage(\lee{})\leq \lsn{1}$ & \texttt{\footnotesize GLang/Typing.lean:190}\\
}{
\gpfit{\inferrule*[]{(\lemptyset{}, \lepsilon{})\vdash \lee{}\colon \lsn{0}^{\lstar} \loprod \lsn{1}}{(\lemptyset, \lepsilon{})\vdash \Kernel(\lee{})\leq \lsn{0}}}
\and
\gpfit{\inferrule*[]{(\lemptyset{}, \lepsilon{})\vdash \lee{}\colon \lsn{0}^{\lstar} \loprod \lsn{1}}{(\lemptyset, \lepsilon{})\vdash \CoKernel(\lee{})\leq \lsn{0}}}
\and
\gpfit{\inferrule*[]{(\lemptyset{}, \lepsilon{})\vdash \lee{}\colon \lsn{0}^{\lstar} \loprod \lsn{1}}{(\lemptyset, \lepsilon{})\vdash \Image(\lee{})\leq \lsn{1}}}
\and
\gpfit{\inferrule*[]{(\lemptyset{}, \lepsilon{})\vdash \lee{}\colon \lsn{0}^{\lstar} \loprod \lsn{1}}{(\lemptyset, \lepsilon{})\vdash \CoImage(\lee{})\leq \lsn{1}}}
}{
\gpfit{\inferrule*[lab=\textsc{Sub-Sub\allowbreak Ker}]{(\lemptyset{}, \lepsilon{})\vdash \lee{}\colon \lsn{0}^{\lstar} \loprod \lsn{1}}{(\lemptyset, \lepsilon{})\vdash \Kernel(\lee{})\leq \lsn{0}}}
\and
\gpfit{\inferrule*[lab=\textsc{Sub-Sub\allowbreak Coker}]{(\lemptyset{}, \lepsilon{})\vdash \lee{}\colon \lsn{0}^{\lstar} \loprod \lsn{1}}{(\lemptyset, \lepsilon{})\vdash \CoKernel(\lee{})\leq \lsn{0}}}
\and
\gpfit{\inferrule*[lab=\textsc{Sub-Sub\allowbreak Img}]{(\lemptyset{}, \lepsilon{})\vdash \lee{}\colon \lsn{0}^{\lstar} \loprod \lsn{1}}{(\lemptyset, \lepsilon{})\vdash \Image(\lee{})\leq \lsn{1}}}
\and
\gpfit{\inferrule*[lab=\textsc{Sub-Sub\allowbreak Coimg}]{(\lemptyset{}, \lepsilon{})\vdash \lee{}\colon \lsn{0}^{\lstar} \loprod \lsn{1}}{(\lemptyset, \lepsilon{})\vdash \CoImage(\lee{})\leq \lsn{1}}}
}
\gldefjudgment{hasType}{\lctx\vdash\lee\colon\lty}{Term typing (generated from \texttt{GLang.HasType}).}{fig:gl-hastype}
\gldefrule{hasType.zero}{hasType}{\textsc{T-Zero}}{(\lemptyset, \lepsilon{})\vdash \ls{}\colon\svalid}{(\lDelta{}, \lepsilon{})\vdash \lvzero{}_{\ls{}}\colon \ls{}}{\texttt{GLang.Has\allowbreak Type.zero}}{GLang/Typing.lean:198}
\gldefruleexplanation{hasType.zero}{The zero vector inhabits any well-formed space under any linear context.}
\gldefrule{hasType.smul}{hasType}{\textsc{T-Smul}}{({\lDelta{}_{0}}, \lepsilon{})\vdash \len{0}\colon \lR{} \\ ({\lDelta{}_{1}}, \lepsilon{})\vdash \len{1}\colon \ls{}}{({\lDelta{}_{0}} \ldisun {\lDelta{}_{1}}, \lepsilon{})\vdash \len{0} \lvscale \len{1}\colon \ls{}}{\texttt{GLang.Has\allowbreak Type.smul}}{GLang/Typing.lean:205}
\gldefruleexplanation{hasType.smul}{Scaling a vector by a scalar preserves its space and combines disjoint linear contexts.}
\gldefrule{hasType.add}{hasType}{\textsc{T-Add}}{\forall i.\; (\lDelta{}, \lepsilon{})\vdash \len{i}\colon \ls{}}{(\lDelta{}, \lepsilon{})\vdash \len{0} \lvadd \len{1}\colon \ls{}}{\texttt{GLang.Has\allowbreak Type.add}}{GLang/Typing.lean:214}
\gldefruleexplanation{hasType.add}{Two vectors in the same space and linear context may be added in that space.}
\gldefrule{hasType.inner}{hasType}{\textsc{T-Inner}}{(\lemptyset, \lepsilon{})\vdash \ls{}\colon\svalid}{(\lemptyset{}, \lepsilon{})\vdash \linner{}_{\ls{}}\colon \ls{}^{\lstar} \loprod \ls{}^{\lstar}}{\texttt{GLang.Has\allowbreak Type.inner}}{GLang/Typing.lean:221}
\gldefruleexplanation{hasType.inner}{The canonical inner product on $\ls$ has type $\ls^*\loprod\ls^*$ with no linear free variables.}
\gldefrule{hasType.realLit}{hasType}{\textsc{T-Real\allowbreak Lit}}{\phantom{.}}{(\lemptyset{}, \lepsilon{})\vdash \mathit{r}\colon \lR{}}{\texttt{GLang.Has\allowbreak Type.real\_\allowbreak lit}}{GLang/Typing.lean:226}
\gldefruleexplanation{hasType.realLit}{A real literal has scalar type with an empty linear context.}
\gldefrule{hasType.appLinear}{hasType}{\textsc{T-App\allowbreak Linear}}{({\lDelta{}_{0}}, \lepsilon{})\vdash \len{0}\colon \lsn{0}^{\lstar} \loprod \lsn{1} \\ ({\lDelta{}_{1}}, \lepsilon{})\vdash \len{1}\colon \lsn{0}}{({\lDelta{}_{0}} \ldisun {\lDelta{}_{1}}, \lepsilon{})\vdash \leval{\len{0}}{}{\len{1}}\colon \lsn{1}}{\texttt{GLang.Has\allowbreak Type.app\_\allowbreak linear}}{GLang/Typing.lean:233}
\gldefruleexplanation{hasType.appLinear}{Applying a linear map to a vector in its domain yields its codomain and combines disjoint linear contexts.}
\gldefrule{hasType.appInner}{hasType}{\textsc{T-App\allowbreak Inner}}{({\lDelta{}_{0}}, \lepsilon{})\vdash \len{0}\colon \ls{}^{\lstar} \\ ({\lDelta{}_{1}}, \lepsilon{})\vdash \len{1}\colon \ls{}}{({\lDelta{}_{0}} \ldisun {\lDelta{}_{1}}, \lepsilon{})\vdash \leval{\len{0}}{}{\len{1}}\colon \lR{}}{\texttt{GLang.Has\allowbreak Type.app\_\allowbreak inner}}{GLang/Typing.lean:247}
\gldefruleexplanation{hasType.appInner}{Applying a covector to a vector in its space yields a scalar and combines disjoint linear contexts.}
\gldefrule{hasType.appNonlinear}{hasType}{\textsc{T-App\allowbreak Nonlinear}}{(\lemptyset{}, \lepsilon{})\vdash \len{0}\colon (x\colon \ltyn{0}\fnto \ltyn{1}) \\ (\lemptyset{}, \lepsilon{})\vdash \len{1}\colon \ltyn{0}}{(\lemptyset{}, \lepsilon{})\vdash \leval{\len{0}}{}{\len{1}}\colon \ltyn{1}\subst{x}{\len{1}}}{\texttt{GLang.Has\allowbreak Type.app\_\allowbreak nonlinear}}{GLang/Typing.lean:260}
\gldefruleexplanation{hasType.appNonlinear}{Dependent application substitutes the argument into the result type, with both operands typed in the empty linear context.}
\gldefrule{hasType.varLin}{hasType}{\textsc{T-Var\allowbreak Lin}}{\phantom{.}}{(\{(\mathit{x}, \ls{})\}, \lepsilon{})\vdash \mathit{x}\colon \ls{}}{\texttt{GLang.Has\allowbreak Type.var\_\allowbreak lin}}{GLang/Typing.lean:267}
\gldefruleexplanation{hasType.varLin}{A linear variable has its declared space under the singleton linear context containing that variable.}
\gldefrule{hasType.varNlin}{hasType}{\textsc{T-Var\allowbreak Nlin}}{(\mathit{x}\colon \lty{}) \in \lepsilon{}}{(\lemptyset{}, \lepsilon{})\vdash \mathit{x}\colon \lty{}}{\texttt{GLang.Has\allowbreak Type.var\_\allowbreak nlin}}{GLang/Typing.lean:274}
\gldefruleexplanation{hasType.varNlin}{A variable declared in $\lepsilon$ has its declared type under the empty linear context.}
\gldefrule{hasType.pinvTensor}{hasType}{\textsc{T-Pinv\allowbreak Tensor}}{(\lemptyset{}, \lepsilon{})\vdash \lee{}\colon \lsn{0}^{\lstar} \loprod \lsn{1}}{(\lemptyset{}, \lepsilon{})\vdash \lee{}^{\lpseudo}\colon \lsn{1}^{\lstar} \loprod \lsn{0}}{\texttt{GLang.Has\allowbreak Type.pinv\_\allowbreak tensor}}{GLang/Typing.lean:281}
\gldefruleexplanation{hasType.pinvTensor}{The pseudoinverse reverses the domain and codomain of a linear map typed with no linear free variables.}
\gldefrule{hasType.pinvDual}{hasType}{\textsc{T-Pinv\allowbreak Dual}}{(\lemptyset{}, \lepsilon{})\vdash \lee{}\colon \ls{}^{\lstar}}{(\lemptyset{}, \lepsilon{})\vdash \lee{}^{\lpseudo}\colon \lR{}^{\lstar} \loprod \ls{}}{\texttt{GLang.Has\allowbreak Type.pinv\_\allowbreak dual}}{GLang/Typing.lean:288}
\gldefruleexplanation{hasType.pinvDual}{The pseudoinverse of a covector on $\ls$ is a linear map from $\mathbb R$ to $\ls$, with no linear free variables.}
\gldefrule{hasType.promote}{hasType}{\textsc{T-Promote}}{\forall y \in \mathit{lin}.\; y \notin \bigcup_{(\cdot\colon s) \in \lDelta{},\, \mathit{lin}} \mathrm{fv}(s) \;\cup\; \bigcup_{(\cdot\colon \tau) \in \lepsilon{}'} \mathrm{fv}(\tau) \\ (\mathit{lin} \ldisun \lDelta{}, \lepsilon{}')\vdash \lee{}\colon \lty{}}{(\lDelta{}, \mathit{lin} \ldisun \lepsilon{}')\vdash \lee{}\colon \lty{}}{\texttt{GLang.Has\allowbreak Type.promote}}{GLang/Typing.lean:303}
\gldefruleexplanation{hasType.promote}{The linear context may be moved into the nonlinear context without changing the term or its type unless a variable is used to define a space, as these must depend non-linearly.}
\gldefrule{hasType.dsumE}{hasType}{\textsc{T-Dsum\allowbreak E}}{\forall i.\; (\lDelta{}, \lepsilon{})\vdash \len{i}\colon \lsn{i}}{(\lDelta{}, \lepsilon{})\vdash \len{0}\losum \len{1}\colon \lsn{0} \losum \lsn{1}}{\texttt{GLang.Has\allowbreak Type.dsum\allowbreak E}}{GLang/Typing.lean:318}
\gldefruleexplanation{hasType.dsumE}{Two vectors typed under the same linear context form a vector in the direct sum of their spaces.}
\gldefrule{hasType.tprodE}{hasType}{\textsc{T-Tprod\allowbreak E}}{\forall i.\; ({\lDelta{}_{i}}, \lepsilon{})\vdash \len{i}\colon \lsn{i}}{({\lDelta{}_{0}} \ldisun {\lDelta{}_{1}}, \lepsilon{})\vdash \len{0}\loprod \len{1}\colon \lsn{0} \loprod \lsn{1}}{\texttt{GLang.Has\allowbreak Type.tprod\allowbreak E}}{GLang/Typing.lean:327}
\gldefruleexplanation{hasType.tprodE}{Tensoring two vectors forms a vector in the tensor product of their spaces and combines disjoint linear contexts.}
\gldefrule{hasType.llamReal}{hasType}{\textsc{T-Llam\allowbreak Real}}{(\lemptyset, \lepsilon{})\vdash \ls{}\colon\svalid \\ ((\lDelta{}, \mathit{x}\colon \ls{}), \lepsilon{})\vdash \lee{}\colon \lR{}}{(\lDelta{}, \lepsilon{})\vdash (\lllambda \llpl x\colon \ls{}\llpr \ldotsp \lee{})\colon \ls{}^{\lstar}}{\texttt{GLang.Has\allowbreak Type.llam\_\allowbreak real}}{GLang/Typing.lean:340}
\gldefruleexplanation{hasType.llamReal}{Abstracting a fresh linear variable from a scalar-valued body produces a covector on the variable's space.}
\gldefrule{hasType.llamTensor}{hasType}{\textsc{T-Llam\allowbreak Tensor}}{\ls{} \neq \lR{} \\ (\lemptyset, \lepsilon{})\vdash \lsn{0}\colon\svalid \\ (\lemptyset, \lepsilon{})\vdash \ls{}\colon\svalid \\ ((\lDelta{}, \mathit{x}\colon \lsn{0}), \lepsilon{})\vdash \lee{}\colon \ls{}}{(\lDelta{}, \lepsilon{})\vdash (\lllambda \llpl x\colon \lsn{0}\llpr \ldotsp \lee{})\colon \lsn{0}^{\lstar} \loprod \ls{}}{\texttt{GLang.Has\allowbreak Type.llam\_\allowbreak tensor}}{GLang/Typing.lean:347}
\gldefruleexplanation{hasType.llamTensor}{Abstracting a fresh linear variable from a body in a non-scalar space produces the corresponding linear-map tensor.}
\gldefrule{hasType.embed}{hasType}{\textsc{T-Embed}}{(\lemptyset, \lepsilon{})\vdash \lsn{0}\leq \lsn{1}}{(\lemptyset{}, \lepsilon{})\vdash \lembd{}_{\lsn{0}}\colon \lsn{0}^{\lstar} \loprod \lsn{1}}{\texttt{GLang.Has\allowbreak Type.embed}}{GLang/Typing.lean:355}
\gldefruleexplanation{hasType.embed}{A subspace inclusion is a linear map from the subspace to its ambient space with no linear free variables.}
\gldefrule{hasType.nllam}{hasType}{\textsc{T-Nllam}}{(\lemptyset, \lepsilon{})\vdash (x\colon \ltyn{0}\fnto \ltyn{1})\colon \lkn{0} \fnto \lkn{1} \\ (\lemptyset{}, (\lepsilon{}, \mathit{x}\colon \ltyn{0}))\vdash \lee{}\colon \ltyn{1}}{(\lemptyset{}, \lepsilon{})\vdash (\nllambda \llpl x\colon \ltyn{0}\llpr \ldotsp \lee{})\colon (x\colon \ltyn{0}\fnto \ltyn{1})}{\texttt{GLang.Has\allowbreak Type.nllam}}{GLang/Typing.lean:364}
\gldefruleexplanation{hasType.nllam}{A body typed under a fresh nonlinear argument forms a dependent function when the complete function type is well kinded.}
\gldefrulerows{hasType}{
\textsc{T-Zero} & $(\lDelta{}, \lepsilon{})\vdash \lvzero{}_{\ls{}}\colon \ls{}$ & \texttt{\footnotesize GLang/Typing.lean:198}\\
\textsc{T-Smul} & $({\lDelta{}_{0}} \ldisun {\lDelta{}_{1}}, \lepsilon{})\vdash \len{0} \lvscale \len{1}\colon \ls{}$ & \texttt{\footnotesize GLang/Typing.lean:205}\\
\textsc{T-Add} & $(\lDelta{}, \lepsilon{})\vdash \len{0} \lvadd \len{1}\colon \ls{}$ & \texttt{\footnotesize GLang/Typing.lean:214}\\
\textsc{T-Inner} & $(\lemptyset{}, \lepsilon{})\vdash \linner{}_{\ls{}}\colon \ls{}^{\lstar} \loprod \ls{}^{\lstar}$ & \texttt{\footnotesize GLang/Typing.lean:221}\\
\textsc{T-Real\allowbreak Lit} & $(\lemptyset{}, \lepsilon{})\vdash \mathit{r}\colon \lR{}$ & \texttt{\footnotesize GLang/Typing.lean:226}\\
\textsc{T-App\allowbreak Linear} & $({\lDelta{}_{0}} \ldisun {\lDelta{}_{1}}, \lepsilon{})\vdash \leval{\len{0}}{}{\len{1}}\colon \lsn{1}$ & \texttt{\footnotesize GLang/Typing.lean:233}\\
\textsc{T-App\allowbreak Inner} & $({\lDelta{}_{0}} \ldisun {\lDelta{}_{1}}, \lepsilon{})\vdash \leval{\len{0}}{}{\len{1}}\colon \lR{}$ & \texttt{\footnotesize GLang/Typing.lean:247}\\
\textsc{T-App\allowbreak Nonlinear} & $(\lemptyset{}, \lepsilon{})\vdash \leval{\len{0}}{}{\len{1}}\colon \ltyn{1}\subst{x}{\len{1}}$ & \texttt{\footnotesize GLang/Typing.lean:260}\\
\textsc{T-Var\allowbreak Lin} & $(\{(\mathit{x}, \ls{})\}, \lepsilon{})\vdash \mathit{x}\colon \ls{}$ & \texttt{\footnotesize GLang/Typing.lean:267}\\
\textsc{T-Var\allowbreak Nlin} & $(\lemptyset{}, \lepsilon{})\vdash \mathit{x}\colon \lty{}$ & \texttt{\footnotesize GLang/Typing.lean:274}\\
\textsc{T-Pinv\allowbreak Tensor} & $(\lemptyset{}, \lepsilon{})\vdash \lee{}^{\lpseudo}\colon \lsn{1}^{\lstar} \loprod \lsn{0}$ & \texttt{\footnotesize GLang/Typing.lean:281}\\
\textsc{T-Pinv\allowbreak Dual} & $(\lemptyset{}, \lepsilon{})\vdash \lee{}^{\lpseudo}\colon \lR{}^{\lstar} \loprod \ls{}$ & \texttt{\footnotesize GLang/Typing.lean:288}\\
\textsc{T-Promote} & $(\lDelta{}, \mathit{lin} \ldisun \lepsilon{}')\vdash \lee{}\colon \lty{}$ & \texttt{\footnotesize GLang/Typing.lean:303}\\
\textsc{T-Dsum\allowbreak E} & $(\lDelta{}, \lepsilon{})\vdash \len{0}\losum \len{1}\colon \lsn{0} \losum \lsn{1}$ & \texttt{\footnotesize GLang/Typing.lean:318}\\
\textsc{T-Tprod\allowbreak E} & $({\lDelta{}_{0}} \ldisun {\lDelta{}_{1}}, \lepsilon{})\vdash \len{0}\loprod \len{1}\colon \lsn{0} \loprod \lsn{1}$ & \texttt{\footnotesize GLang/Typing.lean:327}\\
\textsc{T-Llam\allowbreak Real} & $(\lDelta{}, \lepsilon{})\vdash (\lllambda \llpl x\colon \ls{}\llpr \ldotsp \lee{})\colon \ls{}^{\lstar}$ & \texttt{\footnotesize GLang/Typing.lean:340}\\
\textsc{T-Llam\allowbreak Tensor} & $(\lDelta{}, \lepsilon{})\vdash (\lllambda \llpl x\colon \lsn{0}\llpr \ldotsp \lee{})\colon \lsn{0}^{\lstar} \loprod \ls{}$ & \texttt{\footnotesize GLang/Typing.lean:347}\\
\textsc{T-Embed} & $(\lemptyset{}, \lepsilon{})\vdash \lembd{}_{\lsn{0}}\colon \lsn{0}^{\lstar} \loprod \lsn{1}$ & \texttt{\footnotesize GLang/Typing.lean:355}\\
\textsc{T-Nllam} & $(\lemptyset{}, \lepsilon{})\vdash (\nllambda \llpl x\colon \ltyn{0}\llpr \ldotsp \lee{})\colon (x\colon \ltyn{0}\fnto \ltyn{1})$ & \texttt{\footnotesize GLang/Typing.lean:364}\\
}{
\gpfit{\inferrule*[]{(\lemptyset, \lepsilon{})\vdash \ls{}\colon\svalid}{(\lDelta{}, \lepsilon{})\vdash \lvzero{}_{\ls{}}\colon \ls{}}}
\and
\gpfit{\inferrule*[]{({\lDelta{}_{0}}, \lepsilon{})\vdash \len{0}\colon \lR{} \\ ({\lDelta{}_{1}}, \lepsilon{})\vdash \len{1}\colon \ls{}}{({\lDelta{}_{0}} \ldisun {\lDelta{}_{1}}, \lepsilon{})\vdash \len{0} \lvscale \len{1}\colon \ls{}}}
\and
\gpfit{\inferrule*[]{\forall i.\; (\lDelta{}, \lepsilon{})\vdash \len{i}\colon \ls{}}{(\lDelta{}, \lepsilon{})\vdash \len{0} \lvadd \len{1}\colon \ls{}}}
\and
\gpfit{\inferrule*[]{(\lemptyset, \lepsilon{})\vdash \ls{}\colon\svalid}{(\lemptyset{}, \lepsilon{})\vdash \linner{}_{\ls{}}\colon \ls{}^{\lstar} \loprod \ls{}^{\lstar}}}
\and
\gpfit{\inferrule*[]{\phantom{.}}{(\lemptyset{}, \lepsilon{})\vdash \mathit{r}\colon \lR{}}}
\and
\gpfit{\inferrule*[]{({\lDelta{}_{0}}, \lepsilon{})\vdash \len{0}\colon \lsn{0}^{\lstar} \loprod \lsn{1} \\ ({\lDelta{}_{1}}, \lepsilon{})\vdash \len{1}\colon \lsn{0}}{({\lDelta{}_{0}} \ldisun {\lDelta{}_{1}}, \lepsilon{})\vdash \leval{\len{0}}{}{\len{1}}\colon \lsn{1}}}
\and
\gpfit{\inferrule*[]{({\lDelta{}_{0}}, \lepsilon{})\vdash \len{0}\colon \ls{}^{\lstar} \\ ({\lDelta{}_{1}}, \lepsilon{})\vdash \len{1}\colon \ls{}}{({\lDelta{}_{0}} \ldisun {\lDelta{}_{1}}, \lepsilon{})\vdash \leval{\len{0}}{}{\len{1}}\colon \lR{}}}
\and
\gpfit{\inferrule*[]{(\lemptyset{}, \lepsilon{})\vdash \len{0}\colon (x\colon \ltyn{0}\fnto \ltyn{1}) \\ (\lemptyset{}, \lepsilon{})\vdash \len{1}\colon \ltyn{0}}{(\lemptyset{}, \lepsilon{})\vdash \leval{\len{0}}{}{\len{1}}\colon \ltyn{1}\subst{x}{\len{1}}}}
\and
\gpfit{\inferrule*[]{\phantom{.}}{(\{(\mathit{x}, \ls{})\}, \lepsilon{})\vdash \mathit{x}\colon \ls{}}}
\and
\gpfit{\inferrule*[]{(\mathit{x}\colon \lty{}) \in \lepsilon{}}{(\lemptyset{}, \lepsilon{})\vdash \mathit{x}\colon \lty{}}}
\and
\gpfit{\inferrule*[]{(\lemptyset{}, \lepsilon{})\vdash \lee{}\colon \lsn{0}^{\lstar} \loprod \lsn{1}}{(\lemptyset{}, \lepsilon{})\vdash \lee{}^{\lpseudo}\colon \lsn{1}^{\lstar} \loprod \lsn{0}}}
\and
\gpfit{\inferrule*[]{(\lemptyset{}, \lepsilon{})\vdash \lee{}\colon \ls{}^{\lstar}}{(\lemptyset{}, \lepsilon{})\vdash \lee{}^{\lpseudo}\colon \lR{}^{\lstar} \loprod \ls{}}}
\and
\gpfit{\inferrule*[]{\forall y \in \mathit{lin}.\; y \notin \bigcup_{(\cdot\colon s) \in \lDelta{},\, \mathit{lin}} \mathrm{fv}(s) \;\cup\; \bigcup_{(\cdot\colon \tau) \in \lepsilon{}'} \mathrm{fv}(\tau) \\ (\mathit{lin} \ldisun \lDelta{}, \lepsilon{}')\vdash \lee{}\colon \lty{}}{(\lDelta{}, \mathit{lin} \ldisun \lepsilon{}')\vdash \lee{}\colon \lty{}}}
\and
\gpfit{\inferrule*[]{\forall i.\; (\lDelta{}, \lepsilon{})\vdash \len{i}\colon \lsn{i}}{(\lDelta{}, \lepsilon{})\vdash \len{0}\losum \len{1}\colon \lsn{0} \losum \lsn{1}}}
\and
\gpfit{\inferrule*[]{\forall i.\; ({\lDelta{}_{i}}, \lepsilon{})\vdash \len{i}\colon \lsn{i}}{({\lDelta{}_{0}} \ldisun {\lDelta{}_{1}}, \lepsilon{})\vdash \len{0}\loprod \len{1}\colon \lsn{0} \loprod \lsn{1}}}
\and
\gpfit{\inferrule*[]{(\lemptyset, \lepsilon{})\vdash \ls{}\colon\svalid \\ ((\lDelta{}, \mathit{x}\colon \ls{}), \lepsilon{})\vdash \lee{}\colon \lR{}}{(\lDelta{}, \lepsilon{})\vdash (\lllambda \llpl x\colon \ls{}\llpr \ldotsp \lee{})\colon \ls{}^{\lstar}}}
\and
\gpfit{\inferrule*[]{\ls{} \neq \lR{} \\ (\lemptyset, \lepsilon{})\vdash \lsn{0}\colon\svalid \\ (\lemptyset, \lepsilon{})\vdash \ls{}\colon\svalid \\ ((\lDelta{}, \mathit{x}\colon \lsn{0}), \lepsilon{})\vdash \lee{}\colon \ls{}}{(\lDelta{}, \lepsilon{})\vdash (\lllambda \llpl x\colon \lsn{0}\llpr \ldotsp \lee{})\colon \lsn{0}^{\lstar} \loprod \ls{}}}
\and
\gpfit{\inferrule*[]{(\lemptyset, \lepsilon{})\vdash \lsn{0}\leq \lsn{1}}{(\lemptyset{}, \lepsilon{})\vdash \lembd{}_{\lsn{0}}\colon \lsn{0}^{\lstar} \loprod \lsn{1}}}
\and
\gpfit{\inferrule*[]{(\lemptyset, \lepsilon{})\vdash (x\colon \ltyn{0}\fnto \ltyn{1})\colon \lkn{0} \fnto \lkn{1} \\ (\lemptyset{}, (\lepsilon{}, \mathit{x}\colon \ltyn{0}))\vdash \lee{}\colon \ltyn{1}}{(\lemptyset{}, \lepsilon{})\vdash (\nllambda \llpl x\colon \ltyn{0}\llpr \ldotsp \lee{})\colon (x\colon \ltyn{0}\fnto \ltyn{1})}}
}{
\gpfit{\inferrule*[lab=\textsc{T-Zero}]{(\lemptyset, \lepsilon{})\vdash \ls{}\colon\svalid}{(\lDelta{}, \lepsilon{})\vdash \lvzero{}_{\ls{}}\colon \ls{}}}
\and
\gpfit{\inferrule*[lab=\textsc{T-Smul}]{({\lDelta{}_{0}}, \lepsilon{})\vdash \len{0}\colon \lR{} \\ ({\lDelta{}_{1}}, \lepsilon{})\vdash \len{1}\colon \ls{}}{({\lDelta{}_{0}} \ldisun {\lDelta{}_{1}}, \lepsilon{})\vdash \len{0} \lvscale \len{1}\colon \ls{}}}
\and
\gpfit{\inferrule*[lab=\textsc{T-Add}]{\forall i.\; (\lDelta{}, \lepsilon{})\vdash \len{i}\colon \ls{}}{(\lDelta{}, \lepsilon{})\vdash \len{0} \lvadd \len{1}\colon \ls{}}}
\and
\gpfit{\inferrule*[lab=\textsc{T-Inner}]{(\lemptyset, \lepsilon{})\vdash \ls{}\colon\svalid}{(\lemptyset{}, \lepsilon{})\vdash \linner{}_{\ls{}}\colon \ls{}^{\lstar} \loprod \ls{}^{\lstar}}}
\and
\gpfit{\inferrule*[lab=\textsc{T-Real\allowbreak Lit}]{\phantom{.}}{(\lemptyset{}, \lepsilon{})\vdash \mathit{r}\colon \lR{}}}
\and
\gpfit{\inferrule*[lab=\textsc{T-App\allowbreak Linear}]{({\lDelta{}_{0}}, \lepsilon{})\vdash \len{0}\colon \lsn{0}^{\lstar} \loprod \lsn{1} \\ ({\lDelta{}_{1}}, \lepsilon{})\vdash \len{1}\colon \lsn{0}}{({\lDelta{}_{0}} \ldisun {\lDelta{}_{1}}, \lepsilon{})\vdash \leval{\len{0}}{}{\len{1}}\colon \lsn{1}}}
\and
\gpfit{\inferrule*[lab=\textsc{T-App\allowbreak Inner}]{({\lDelta{}_{0}}, \lepsilon{})\vdash \len{0}\colon \ls{}^{\lstar} \\ ({\lDelta{}_{1}}, \lepsilon{})\vdash \len{1}\colon \ls{}}{({\lDelta{}_{0}} \ldisun {\lDelta{}_{1}}, \lepsilon{})\vdash \leval{\len{0}}{}{\len{1}}\colon \lR{}}}
\and
\gpfit{\inferrule*[lab=\textsc{T-App\allowbreak Nonlinear}]{(\lemptyset{}, \lepsilon{})\vdash \len{0}\colon (x\colon \ltyn{0}\fnto \ltyn{1}) \\ (\lemptyset{}, \lepsilon{})\vdash \len{1}\colon \ltyn{0}}{(\lemptyset{}, \lepsilon{})\vdash \leval{\len{0}}{}{\len{1}}\colon \ltyn{1}\subst{x}{\len{1}}}}
\and
\gpfit{\inferrule*[lab=\textsc{T-Var\allowbreak Lin}]{\phantom{.}}{(\{(\mathit{x}, \ls{})\}, \lepsilon{})\vdash \mathit{x}\colon \ls{}}}
\and
\gpfit{\inferrule*[lab=\textsc{T-Var\allowbreak Nlin}]{(\mathit{x}\colon \lty{}) \in \lepsilon{}}{(\lemptyset{}, \lepsilon{})\vdash \mathit{x}\colon \lty{}}}
\and
\gpfit{\inferrule*[lab=\textsc{T-Pinv\allowbreak Tensor}]{(\lemptyset{}, \lepsilon{})\vdash \lee{}\colon \lsn{0}^{\lstar} \loprod \lsn{1}}{(\lemptyset{}, \lepsilon{})\vdash \lee{}^{\lpseudo}\colon \lsn{1}^{\lstar} \loprod \lsn{0}}}
\and
\gpfit{\inferrule*[lab=\textsc{T-Pinv\allowbreak Dual}]{(\lemptyset{}, \lepsilon{})\vdash \lee{}\colon \ls{}^{\lstar}}{(\lemptyset{}, \lepsilon{})\vdash \lee{}^{\lpseudo}\colon \lR{}^{\lstar} \loprod \ls{}}}
\and
\gpfit{\inferrule*[lab=\textsc{T-Promote}]{\forall y \in \mathit{lin}.\; y \notin \bigcup_{(\cdot\colon s) \in \lDelta{},\, \mathit{lin}} \mathrm{fv}(s) \;\cup\; \bigcup_{(\cdot\colon \tau) \in \lepsilon{}'} \mathrm{fv}(\tau) \\ (\mathit{lin} \ldisun \lDelta{}, \lepsilon{}')\vdash \lee{}\colon \lty{}}{(\lDelta{}, \mathit{lin} \ldisun \lepsilon{}')\vdash \lee{}\colon \lty{}}}
\and
\gpfit{\inferrule*[lab=\textsc{T-Dsum\allowbreak E}]{\forall i.\; (\lDelta{}, \lepsilon{})\vdash \len{i}\colon \lsn{i}}{(\lDelta{}, \lepsilon{})\vdash \len{0}\losum \len{1}\colon \lsn{0} \losum \lsn{1}}}
\and
\gpfit{\inferrule*[lab=\textsc{T-Tprod\allowbreak E}]{\forall i.\; ({\lDelta{}_{i}}, \lepsilon{})\vdash \len{i}\colon \lsn{i}}{({\lDelta{}_{0}} \ldisun {\lDelta{}_{1}}, \lepsilon{})\vdash \len{0}\loprod \len{1}\colon \lsn{0} \loprod \lsn{1}}}
\and
\gpfit{\inferrule*[lab=\textsc{T-Llam\allowbreak Real}]{(\lemptyset, \lepsilon{})\vdash \ls{}\colon\svalid \\ ((\lDelta{}, \mathit{x}\colon \ls{}), \lepsilon{})\vdash \lee{}\colon \lR{}}{(\lDelta{}, \lepsilon{})\vdash (\lllambda \llpl x\colon \ls{}\llpr \ldotsp \lee{})\colon \ls{}^{\lstar}}}
\and
\gpfit{\inferrule*[lab=\textsc{T-Llam\allowbreak Tensor}]{\ls{} \neq \lR{} \\ (\lemptyset, \lepsilon{})\vdash \lsn{0}\colon\svalid \\ (\lemptyset, \lepsilon{})\vdash \ls{}\colon\svalid \\ ((\lDelta{}, \mathit{x}\colon \lsn{0}), \lepsilon{})\vdash \lee{}\colon \ls{}}{(\lDelta{}, \lepsilon{})\vdash (\lllambda \llpl x\colon \lsn{0}\llpr \ldotsp \lee{})\colon \lsn{0}^{\lstar} \loprod \ls{}}}
\and
\gpfit{\inferrule*[lab=\textsc{T-Embed}]{(\lemptyset, \lepsilon{})\vdash \lsn{0}\leq \lsn{1}}{(\lemptyset{}, \lepsilon{})\vdash \lembd{}_{\lsn{0}}\colon \lsn{0}^{\lstar} \loprod \lsn{1}}}
\and
\gpfit{\inferrule*[lab=\textsc{T-Nllam}]{(\lemptyset, \lepsilon{})\vdash (x\colon \ltyn{0}\fnto \ltyn{1})\colon \lkn{0} \fnto \lkn{1} \\ (\lemptyset{}, (\lepsilon{}, \mathit{x}\colon \ltyn{0}))\vdash \lee{}\colon \ltyn{1}}{(\lemptyset{}, \lepsilon{})\vdash (\nllambda \llpl x\colon \ltyn{0}\llpr \ldotsp \lee{})\colon (x\colon \ltyn{0}\fnto \ltyn{1})}}
}
\gldefjudgment{basisValid}{\lctxnle\vdash\lb{}\colon\bvalid{}}{Basis validity (generated from \texttt{GLang.Basis.BasisValid}).}{fig:gl-basisvalid}
\gldefrule{basisValid.empty}{basisValid}{\textsc{B-Empty}}{\phantom{.}}{(\emptyset,\lepsilon{})\vdash \lbzero{}\colon \bvalid{}}{\texttt{GLang.Basis.Basis\allowbreak Valid.empty}}{GLang/Basis.lean:162}
\gldefruleexplanation{basisValid.empty}{The empty basis is valid for the zero-dimensional space.}
\gldefrule{basisValid.one}{basisValid}{\textsc{B-One}}{\phantom{.}}{(\emptyset,\lepsilon{})\vdash \lbone{}\colon \bvalid{}}{\texttt{GLang.Basis.Basis\allowbreak Valid.one}}{GLang/Basis.lean:163}
\gldefruleexplanation{basisValid.one}{The unit basis is valid for $\mathbb R$.}
\gldefrule{basisValid.dual}{basisValid}{\textsc{B-Dual}}{(\emptyset,\lepsilon{})\vdash \lb{}\colon \bvalid{} \\ \tof{\lb{}} = \sof{\lb{}}}{(\emptyset,\lepsilon{})\vdash \lb{}^{\lstar}\colon \bvalid{}}{\texttt{GLang.Basis.Basis\allowbreak Valid.dual}}{GLang/Basis.lean:164}
\gldefruleexplanation{basisValid.dual}{The dual of a valid space basis is valid.}
\gldefrule{basisValid.dsum}{basisValid}{\textsc{B-Dsum}}{\forall i.\; (\emptyset,\lepsilon{})\vdash \lbn{i}\colon \bvalid{} \\ \forall i.\; \tof{\lbn{i}} = \sof{\lbn{i}}}{(\emptyset,\lepsilon{})\vdash \lbn{0} \losum \lbn{1}\colon \bvalid{}}{\texttt{GLang.Basis.Basis\allowbreak Valid.dsum}}{GLang/Basis.lean:167}
\gldefruleexplanation{basisValid.dsum}{The direct sum of two valid space bases is valid.}
\gldefrule{basisValid.tprod}{basisValid}{\textsc{B-Tprod}}{\forall i.\; (\emptyset,\lepsilon{})\vdash \lbn{i}\colon \bvalid{} \\ \forall i.\; \tof{\lbn{i}} = \sof{\lbn{i}}}{(\emptyset,\lepsilon{})\vdash \lbn{0} \loprod \lbn{1}\colon \bvalid{}}{\texttt{GLang.Basis.Basis\allowbreak Valid.tprod}}{GLang/Basis.lean:171}
\gldefruleexplanation{basisValid.tprod}{The tensor product of two valid space bases is valid.}
\gldefrule{basisValid.projKer}{basisValid}{\textsc{B-Proj\allowbreak Ker}}{\forall i.\; (\emptyset,\lepsilon{})\vdash \lbn{i}\colon \bvalid{} \\ \forall i.\; \tof{\lbn{i}} = \lsn{i} \\ (\lemptyset{}, \lepsilon{})\vdash \lee{}\colon \lsn{0}^{\lstar} \loprod \lsn{1}}{(\emptyset,\lepsilon{})\vdash \pBasisC{\Kernel(\lee{})}{\lbn{0}}{\lbn{1}}\colon \bvalid{}}{\texttt{GLang.Basis.Basis\allowbreak Valid.proj\_\allowbreak ker}}{GLang/Basis.lean:180}
\gldefruleexplanation{basisValid.projKer}{Valid bases aligned with a well-typed map's domain and codomain determine a valid projected kernel basis.}
\gldefrule{basisValid.projCoker}{basisValid}{\textsc{B-Proj\allowbreak Coker}}{\forall i.\; (\emptyset,\lepsilon{})\vdash \lbn{i}\colon \bvalid{} \\ \forall i.\; \tof{\lbn{i}} = \lsn{i} \\ (\lemptyset{}, \lepsilon{})\vdash \lee{}\colon \lsn{0}^{\lstar} \loprod \lsn{1}}{(\emptyset,\lepsilon{})\vdash \pBasisC{\CoKernel(\lee{})}{\lbn{0}}{\lbn{1}}\colon \bvalid{}}{\texttt{GLang.Basis.Basis\allowbreak Valid.proj\_\allowbreak coker}}{GLang/Basis.lean:184}
\gldefruleexplanation{basisValid.projCoker}{Valid bases aligned with a well-typed map's domain and codomain determine a valid projected cokernel basis.}
\gldefrule{basisValid.projImg}{basisValid}{\textsc{B-Proj\allowbreak Img}}{\forall i.\; (\emptyset,\lepsilon{})\vdash \lbn{i}\colon \bvalid{} \\ \forall i.\; \tof{\lbn{i}} = \lsn{i} \\ (\lemptyset{}, \lepsilon{})\vdash \lee{}\colon \lsn{0}^{\lstar} \loprod \lsn{1}}{(\emptyset,\lepsilon{})\vdash \pBasisC{\Image(\lee{})}{\lbn{0}}{\lbn{1}}\colon \bvalid{}}{\texttt{GLang.Basis.Basis\allowbreak Valid.proj\_\allowbreak img}}{GLang/Basis.lean:188}
\gldefruleexplanation{basisValid.projImg}{Valid bases aligned with a well-typed map's domain and codomain determine a valid projected image basis.}
\gldefrule{basisValid.projCoimg}{basisValid}{\textsc{B-Proj\allowbreak Coimg}}{\forall i.\; (\emptyset,\lepsilon{})\vdash \lbn{i}\colon \bvalid{} \\ \forall i.\; \tof{\lbn{i}} = \lsn{i} \\ (\lemptyset{}, \lepsilon{})\vdash \lee{}\colon \lsn{0}^{\lstar} \loprod \lsn{1}}{(\emptyset,\lepsilon{})\vdash \pBasisC{\CoImage(\lee{})}{\lbn{0}}{\lbn{1}}\colon \bvalid{}}{\texttt{GLang.Basis.Basis\allowbreak Valid.proj\_\allowbreak coimg}}{GLang/Basis.lean:192}
\gldefruleexplanation{basisValid.projCoimg}{Valid bases aligned with a well-typed map's domain and codomain determine a valid projected coimage basis.}
\gldefrule{basisValid.image}{basisValid}{\textsc{B-Image}}{(\emptyset,\lepsilon{})\vdash \lb{}\colon \bvalid{} \\ \sof{\lb{}} = \CoKernel(\lee{}) \\ (\lemptyset{}, \lepsilon{})\vdash \lee{}\colon \lsn{0}^{\lstar} \loprod \lsn{1}}{(\emptyset,\lepsilon{})\vdash \iBasisC{\lb{}}\colon \bvalid{}}{\texttt{GLang.Basis.Basis\allowbreak Valid.image}}{GLang/Basis.lean:198}
\gldefruleexplanation{basisValid.image}{A valid cokernel basis and its well-typed defining map determine a valid image basis.}
\gldefrule{basisValid.terms}{basisValid}{\textsc{B-Terms}}{(\emptyset,\lepsilon{})\vdash \lb{}\colon \bvalid{} \\ \tof{\lb{}} = \ls{} \\ (\lemptyset, \lepsilon{})\vdash \ls{}\colon\svalid \\ \forall i.\; (\lemptyset{}, \lepsilon{})\vdash \len{i}\colon \ls{} \\ n \geq 0}{(\emptyset,\lepsilon{})\vdash \llpl{} \len{0},\dots,\len{n}\llpr{}_{\lb{}}\colon \bvalid{}}{\texttt{GLang.Basis.Basis\allowbreak Valid.terms}}{GLang/Basis.lean:204}
\gldefruleexplanation{basisValid.terms}{A nonempty family of vectors in a well-formed space, together with a valid fallback basis of that space, determines a valid term basis.}
\gldefrule{basisValid.frame}{basisValid}{\textsc{B-Frame}}{(\emptyset,\lepsilon{})\vdash \lbn{0}\colon \bvalid{} \\ (\emptyset,(\lepsilon{}, \mathit{x}\colon \tof{\lbn{0}}))\vdash \lbn{1}\colon \bvalid{}}{(\emptyset,\lepsilon{})\vdash \frameB{x}{\lbn{0}}{\lbn{1}}\colon \bvalid{}}{\texttt{GLang.Basis.Basis\allowbreak Valid.frame}}{GLang/Basis.lean:213}
\gldefruleexplanation{basisValid.frame}{A valid domain basis and a valid fiber basis under a fresh domain variable determine a valid dependent frame.}
\gldefrulerows{basisValid}{
\textsc{B-Empty} & $(\emptyset,\lepsilon{})\vdash \lbzero{}\colon \bvalid{}$ & \texttt{\footnotesize GLang/Basis.lean:162}\\
\textsc{B-One} & $(\emptyset,\lepsilon{})\vdash \lbone{}\colon \bvalid{}$ & \texttt{\footnotesize GLang/Basis.lean:163}\\
\textsc{B-Dual} & $(\emptyset,\lepsilon{})\vdash \lb{}^{\lstar}\colon \bvalid{}$ & \texttt{\footnotesize GLang/Basis.lean:164}\\
\textsc{B-Dsum} & $(\emptyset,\lepsilon{})\vdash \lbn{0} \losum \lbn{1}\colon \bvalid{}$ & \texttt{\footnotesize GLang/Basis.lean:167}\\
\textsc{B-Tprod} & $(\emptyset,\lepsilon{})\vdash \lbn{0} \loprod \lbn{1}\colon \bvalid{}$ & \texttt{\footnotesize GLang/Basis.lean:171}\\
\textsc{B-Proj\allowbreak Ker} & $(\emptyset,\lepsilon{})\vdash \pBasisC{\Kernel(\lee{})}{\lbn{0}}{\lbn{1}}\colon \bvalid{}$ & \texttt{\footnotesize GLang/Basis.lean:180}\\
\textsc{B-Proj\allowbreak Coker} & $(\emptyset,\lepsilon{})\vdash \pBasisC{\CoKernel(\lee{})}{\lbn{0}}{\lbn{1}}\colon \bvalid{}$ & \texttt{\footnotesize GLang/Basis.lean:184}\\
\textsc{B-Proj\allowbreak Img} & $(\emptyset,\lepsilon{})\vdash \pBasisC{\Image(\lee{})}{\lbn{0}}{\lbn{1}}\colon \bvalid{}$ & \texttt{\footnotesize GLang/Basis.lean:188}\\
\textsc{B-Proj\allowbreak Coimg} & $(\emptyset,\lepsilon{})\vdash \pBasisC{\CoImage(\lee{})}{\lbn{0}}{\lbn{1}}\colon \bvalid{}$ & \texttt{\footnotesize GLang/Basis.lean:192}\\
\textsc{B-Image} & $(\emptyset,\lepsilon{})\vdash \iBasisC{\lb{}}\colon \bvalid{}$ & \texttt{\footnotesize GLang/Basis.lean:198}\\
\textsc{B-Terms} & $(\emptyset,\lepsilon{})\vdash \llpl{} \len{0},\dots,\len{n}\llpr{}_{\lb{}}\colon \bvalid{}$ & \texttt{\footnotesize GLang/Basis.lean:204}\\
\textsc{B-Frame} & $(\emptyset,\lepsilon{})\vdash \frameB{x}{\lbn{0}}{\lbn{1}}\colon \bvalid{}$ & \texttt{\footnotesize GLang/Basis.lean:213}\\
}{
\gpfit{\inferrule*[]{\phantom{.}}{(\emptyset,\lepsilon{})\vdash \lbzero{}\colon \bvalid{}}}
\and
\gpfit{\inferrule*[]{\phantom{.}}{(\emptyset,\lepsilon{})\vdash \lbone{}\colon \bvalid{}}}
\and
\gpfit{\inferrule*[]{(\emptyset,\lepsilon{})\vdash \lb{}\colon \bvalid{} \\ \tof{\lb{}} = \sof{\lb{}}}{(\emptyset,\lepsilon{})\vdash \lb{}^{\lstar}\colon \bvalid{}}}
\and
\gpfit{\inferrule*[]{\forall i.\; (\emptyset,\lepsilon{})\vdash \lbn{i}\colon \bvalid{} \\ \forall i.\; \tof{\lbn{i}} = \sof{\lbn{i}}}{(\emptyset,\lepsilon{})\vdash \lbn{0} \losum \lbn{1}\colon \bvalid{}}}
\and
\gpfit{\inferrule*[]{\forall i.\; (\emptyset,\lepsilon{})\vdash \lbn{i}\colon \bvalid{} \\ \forall i.\; \tof{\lbn{i}} = \sof{\lbn{i}}}{(\emptyset,\lepsilon{})\vdash \lbn{0} \loprod \lbn{1}\colon \bvalid{}}}
\and
\gpfit{\inferrule*[]{\forall i.\; (\emptyset,\lepsilon{})\vdash \lbn{i}\colon \bvalid{} \\ \forall i.\; \tof{\lbn{i}} = \lsn{i} \\ (\lemptyset{}, \lepsilon{})\vdash \lee{}\colon \lsn{0}^{\lstar} \loprod \lsn{1}}{(\emptyset,\lepsilon{})\vdash \pBasisC{\Kernel(\lee{})}{\lbn{0}}{\lbn{1}}\colon \bvalid{}}}
\and
\gpfit{\inferrule*[]{\forall i.\; (\emptyset,\lepsilon{})\vdash \lbn{i}\colon \bvalid{} \\ \forall i.\; \tof{\lbn{i}} = \lsn{i} \\ (\lemptyset{}, \lepsilon{})\vdash \lee{}\colon \lsn{0}^{\lstar} \loprod \lsn{1}}{(\emptyset,\lepsilon{})\vdash \pBasisC{\CoKernel(\lee{})}{\lbn{0}}{\lbn{1}}\colon \bvalid{}}}
\and
\gpfit{\inferrule*[]{\forall i.\; (\emptyset,\lepsilon{})\vdash \lbn{i}\colon \bvalid{} \\ \forall i.\; \tof{\lbn{i}} = \lsn{i} \\ (\lemptyset{}, \lepsilon{})\vdash \lee{}\colon \lsn{0}^{\lstar} \loprod \lsn{1}}{(\emptyset,\lepsilon{})\vdash \pBasisC{\Image(\lee{})}{\lbn{0}}{\lbn{1}}\colon \bvalid{}}}
\and
\gpfit{\inferrule*[]{\forall i.\; (\emptyset,\lepsilon{})\vdash \lbn{i}\colon \bvalid{} \\ \forall i.\; \tof{\lbn{i}} = \lsn{i} \\ (\lemptyset{}, \lepsilon{})\vdash \lee{}\colon \lsn{0}^{\lstar} \loprod \lsn{1}}{(\emptyset,\lepsilon{})\vdash \pBasisC{\CoImage(\lee{})}{\lbn{0}}{\lbn{1}}\colon \bvalid{}}}
\and
\gpfit{\inferrule*[]{(\emptyset,\lepsilon{})\vdash \lb{}\colon \bvalid{} \\ \sof{\lb{}} = \CoKernel(\lee{}) \\ (\lemptyset{}, \lepsilon{})\vdash \lee{}\colon \lsn{0}^{\lstar} \loprod \lsn{1}}{(\emptyset,\lepsilon{})\vdash \iBasisC{\lb{}}\colon \bvalid{}}}
\and
\gpfit{\inferrule*[]{(\emptyset,\lepsilon{})\vdash \lb{}\colon \bvalid{} \\ \tof{\lb{}} = \ls{} \\ (\lemptyset, \lepsilon{})\vdash \ls{}\colon\svalid \\ \forall i.\; (\lemptyset{}, \lepsilon{})\vdash \len{i}\colon \ls{} \\ n \geq 0}{(\emptyset,\lepsilon{})\vdash \llpl{} \len{0},\dots,\len{n}\llpr{}_{\lb{}}\colon \bvalid{}}}
\and
\gpfit{\inferrule*[]{(\emptyset,\lepsilon{})\vdash \lbn{0}\colon \bvalid{} \\ (\emptyset,(\lepsilon{}, \mathit{x}\colon \tof{\lbn{0}}))\vdash \lbn{1}\colon \bvalid{}}{(\emptyset,\lepsilon{})\vdash \frameB{x}{\lbn{0}}{\lbn{1}}\colon \bvalid{}}}
}{
\gpfit{\inferrule*[lab=\textsc{B-Empty}]{\phantom{.}}{(\emptyset,\lepsilon{})\vdash \lbzero{}\colon \bvalid{}}}
\and
\gpfit{\inferrule*[lab=\textsc{B-One}]{\phantom{.}}{(\emptyset,\lepsilon{})\vdash \lbone{}\colon \bvalid{}}}
\and
\gpfit{\inferrule*[lab=\textsc{B-Dual}]{(\emptyset,\lepsilon{})\vdash \lb{}\colon \bvalid{} \\ \tof{\lb{}} = \sof{\lb{}}}{(\emptyset,\lepsilon{})\vdash \lb{}^{\lstar}\colon \bvalid{}}}
\and
\gpfit{\inferrule*[lab=\textsc{B-Dsum}]{\forall i.\; (\emptyset,\lepsilon{})\vdash \lbn{i}\colon \bvalid{} \\ \forall i.\; \tof{\lbn{i}} = \sof{\lbn{i}}}{(\emptyset,\lepsilon{})\vdash \lbn{0} \losum \lbn{1}\colon \bvalid{}}}
\and
\gpfit{\inferrule*[lab=\textsc{B-Tprod}]{\forall i.\; (\emptyset,\lepsilon{})\vdash \lbn{i}\colon \bvalid{} \\ \forall i.\; \tof{\lbn{i}} = \sof{\lbn{i}}}{(\emptyset,\lepsilon{})\vdash \lbn{0} \loprod \lbn{1}\colon \bvalid{}}}
\and
\gpfit{\inferrule*[lab=\textsc{B-Proj\allowbreak Ker}]{\forall i.\; (\emptyset,\lepsilon{})\vdash \lbn{i}\colon \bvalid{} \\ \forall i.\; \tof{\lbn{i}} = \lsn{i} \\ (\lemptyset{}, \lepsilon{})\vdash \lee{}\colon \lsn{0}^{\lstar} \loprod \lsn{1}}{(\emptyset,\lepsilon{})\vdash \pBasisC{\Kernel(\lee{})}{\lbn{0}}{\lbn{1}}\colon \bvalid{}}}
\and
\gpfit{\inferrule*[lab=\textsc{B-Proj\allowbreak Coker}]{\forall i.\; (\emptyset,\lepsilon{})\vdash \lbn{i}\colon \bvalid{} \\ \forall i.\; \tof{\lbn{i}} = \lsn{i} \\ (\lemptyset{}, \lepsilon{})\vdash \lee{}\colon \lsn{0}^{\lstar} \loprod \lsn{1}}{(\emptyset,\lepsilon{})\vdash \pBasisC{\CoKernel(\lee{})}{\lbn{0}}{\lbn{1}}\colon \bvalid{}}}
\and
\gpfit{\inferrule*[lab=\textsc{B-Proj\allowbreak Img}]{\forall i.\; (\emptyset,\lepsilon{})\vdash \lbn{i}\colon \bvalid{} \\ \forall i.\; \tof{\lbn{i}} = \lsn{i} \\ (\lemptyset{}, \lepsilon{})\vdash \lee{}\colon \lsn{0}^{\lstar} \loprod \lsn{1}}{(\emptyset,\lepsilon{})\vdash \pBasisC{\Image(\lee{})}{\lbn{0}}{\lbn{1}}\colon \bvalid{}}}
\and
\gpfit{\inferrule*[lab=\textsc{B-Proj\allowbreak Coimg}]{\forall i.\; (\emptyset,\lepsilon{})\vdash \lbn{i}\colon \bvalid{} \\ \forall i.\; \tof{\lbn{i}} = \lsn{i} \\ (\lemptyset{}, \lepsilon{})\vdash \lee{}\colon \lsn{0}^{\lstar} \loprod \lsn{1}}{(\emptyset,\lepsilon{})\vdash \pBasisC{\CoImage(\lee{})}{\lbn{0}}{\lbn{1}}\colon \bvalid{}}}
\and
\gpfit{\inferrule*[lab=\textsc{B-Image}]{(\emptyset,\lepsilon{})\vdash \lb{}\colon \bvalid{} \\ \sof{\lb{}} = \CoKernel(\lee{}) \\ (\lemptyset{}, \lepsilon{})\vdash \lee{}\colon \lsn{0}^{\lstar} \loprod \lsn{1}}{(\emptyset,\lepsilon{})\vdash \iBasisC{\lb{}}\colon \bvalid{}}}
\and
\gpfit{\inferrule*[lab=\textsc{B-Terms}]{(\emptyset,\lepsilon{})\vdash \lb{}\colon \bvalid{} \\ \tof{\lb{}} = \ls{} \\ (\lemptyset, \lepsilon{})\vdash \ls{}\colon\svalid \\ \forall i.\; (\lemptyset{}, \lepsilon{})\vdash \len{i}\colon \ls{} \\ n \geq 0}{(\emptyset,\lepsilon{})\vdash \llpl{} \len{0},\dots,\len{n}\llpr{}_{\lb{}}\colon \bvalid{}}}
\and
\gpfit{\inferrule*[lab=\textsc{B-Frame}]{(\emptyset,\lepsilon{})\vdash \lbn{0}\colon \bvalid{} \\ (\emptyset,(\lepsilon{}, \mathit{x}\colon \tof{\lbn{0}}))\vdash \lbn{1}\colon \bvalid{}}{(\emptyset,\lepsilon{})\vdash \frameB{x}{\lbn{0}}{\lbn{1}}\colon \bvalid{}}}
}
\gldefjudgment{lc.space}{\mathsf{lc}(\ls{})}{Local closure of spaces (generated from \texttt{GLang.Space.lc}).}{fig:gl-lc-space}
\gldefrule{lc.space.empty}{lc.space}{\textsc{LC-S-Empty}}{\phantom{.}}{\mathsf{lc}(\lsempty{})}{\texttt{GLang.Space.lc.empty}}{GLang/SourceLN.lean:179}
\gldefrule{lc.space.real}{lc.space}{\textsc{LC-S-Real}}{\phantom{.}}{\mathsf{lc}(\lR{})}{\texttt{GLang.Space.lc.real}}{GLang/SourceLN.lean:180}
\gldefrule{lc.space.dual}{lc.space}{\textsc{LC-S-Dual}}{\mathsf{lc}(\ls{})}{\mathsf{lc}(\mathsf{dual}(\ls{}))}{\texttt{GLang.Space.lc.dual}}{GLang/SourceLN.lean:181}
\gldefrule{lc.space.dsum}{lc.space}{\textsc{LC-S-Dsum}}{\mathsf{lc}(\lsn{0}) \\ \mathsf{lc}(\lsn{1})}{\mathsf{lc}(\mathsf{dsum}(\lsn{0}, \lsn{1}))}{\texttt{GLang.Space.lc.dsum}}{GLang/SourceLN.lean:182}
\gldefrule{lc.space.tprod}{lc.space}{\textsc{LC-S-Tprod}}{\mathsf{lc}(\lsn{0}) \\ \mathsf{lc}(\lsn{1})}{\mathsf{lc}(\mathsf{tprod}(\lsn{0}, \lsn{1}))}{\texttt{GLang.Space.lc.tprod}}{GLang/SourceLN.lean:183}
\gldefrule{lc.space.ker}{lc.space}{\textsc{LC-S-Ker}}{\mathsf{lc}(\lsn{0}) \\ \mathsf{lc}(\lsn{1}) \\ \mathsf{lc}(\lee{})}{\mathsf{lc}(\mathsf{ker}(\lsn{0}, \lsn{1}, \lee{}))}{\texttt{GLang.Space.lc.ker}}{GLang/SourceLN.lean:184}
\gldefrule{lc.space.coker}{lc.space}{\textsc{LC-S-Coker}}{\mathsf{lc}(\lsn{0}) \\ \mathsf{lc}(\lsn{1}) \\ \mathsf{lc}(\lee{})}{\mathsf{lc}(\mathsf{coker}(\lsn{0}, \lsn{1}, \lee{}))}{\texttt{GLang.Space.lc.coker}}{GLang/SourceLN.lean:185}
\gldefrule{lc.space.img}{lc.space}{\textsc{LC-S-Img}}{\mathsf{lc}(\lsn{0}) \\ \mathsf{lc}(\lsn{1}) \\ \mathsf{lc}(\lee{})}{\mathsf{lc}(\mathsf{img}(\lsn{0}, \lsn{1}, \lee{}))}{\texttt{GLang.Space.lc.img}}{GLang/SourceLN.lean:186}
\gldefrule{lc.space.coimg}{lc.space}{\textsc{LC-S-Coimg}}{\mathsf{lc}(\lsn{0}) \\ \mathsf{lc}(\lsn{1}) \\ \mathsf{lc}(\lee{})}{\mathsf{lc}(\mathsf{coimg}(\lsn{0}, \lsn{1}, \lee{}))}{\texttt{GLang.Space.lc.coimg}}{GLang/SourceLN.lean:187}
\gldefrulerows{lc.space}{
\textsc{LC-S-Empty} & $\mathsf{lc}(\lsempty{})$ & \texttt{\footnotesize GLang/Source\allowbreak LN.lean:179}\\
\textsc{LC-S-Real} & $\mathsf{lc}(\lR{})$ & \texttt{\footnotesize GLang/Source\allowbreak LN.lean:180}\\
\textsc{LC-S-Dual} & $\mathsf{lc}(\mathsf{dual}(\ls{}))$ & \texttt{\footnotesize GLang/Source\allowbreak LN.lean:181}\\
\textsc{LC-S-Dsum} & $\mathsf{lc}(\mathsf{dsum}(\lsn{0}, \lsn{1}))$ & \texttt{\footnotesize GLang/Source\allowbreak LN.lean:182}\\
\textsc{LC-S-Tprod} & $\mathsf{lc}(\mathsf{tprod}(\lsn{0}, \lsn{1}))$ & \texttt{\footnotesize GLang/Source\allowbreak LN.lean:183}\\
\textsc{LC-S-Ker} & $\mathsf{lc}(\mathsf{ker}(\lsn{0}, \lsn{1}, \lee{}))$ & \texttt{\footnotesize GLang/Source\allowbreak LN.lean:184}\\
\textsc{LC-S-Coker} & $\mathsf{lc}(\mathsf{coker}(\lsn{0}, \lsn{1}, \lee{}))$ & \texttt{\footnotesize GLang/Source\allowbreak LN.lean:185}\\
\textsc{LC-S-Img} & $\mathsf{lc}(\mathsf{img}(\lsn{0}, \lsn{1}, \lee{}))$ & \texttt{\footnotesize GLang/Source\allowbreak LN.lean:186}\\
\textsc{LC-S-Coimg} & $\mathsf{lc}(\mathsf{coimg}(\lsn{0}, \lsn{1}, \lee{}))$ & \texttt{\footnotesize GLang/Source\allowbreak LN.lean:187}\\
}{
\gpfit{\inferrule*[]{\phantom{.}}{\mathsf{lc}(\lsempty{})}}
\and
\gpfit{\inferrule*[]{\phantom{.}}{\mathsf{lc}(\lR{})}}
\and
\gpfit{\inferrule*[]{\mathsf{lc}(\ls{})}{\mathsf{lc}(\mathsf{dual}(\ls{}))}}
\and
\gpfit{\inferrule*[]{\mathsf{lc}(\lsn{0}) \\ \mathsf{lc}(\lsn{1})}{\mathsf{lc}(\mathsf{dsum}(\lsn{0}, \lsn{1}))}}
\and
\gpfit{\inferrule*[]{\mathsf{lc}(\lsn{0}) \\ \mathsf{lc}(\lsn{1})}{\mathsf{lc}(\mathsf{tprod}(\lsn{0}, \lsn{1}))}}
\and
\gpfit{\inferrule*[]{\mathsf{lc}(\lsn{0}) \\ \mathsf{lc}(\lsn{1}) \\ \mathsf{lc}(\lee{})}{\mathsf{lc}(\mathsf{ker}(\lsn{0}, \lsn{1}, \lee{}))}}
\and
\gpfit{\inferrule*[]{\mathsf{lc}(\lsn{0}) \\ \mathsf{lc}(\lsn{1}) \\ \mathsf{lc}(\lee{})}{\mathsf{lc}(\mathsf{coker}(\lsn{0}, \lsn{1}, \lee{}))}}
\and
\gpfit{\inferrule*[]{\mathsf{lc}(\lsn{0}) \\ \mathsf{lc}(\lsn{1}) \\ \mathsf{lc}(\lee{})}{\mathsf{lc}(\mathsf{img}(\lsn{0}, \lsn{1}, \lee{}))}}
\and
\gpfit{\inferrule*[]{\mathsf{lc}(\lsn{0}) \\ \mathsf{lc}(\lsn{1}) \\ \mathsf{lc}(\lee{})}{\mathsf{lc}(\mathsf{coimg}(\lsn{0}, \lsn{1}, \lee{}))}}
}{
\gpfit{\inferrule*[lab=\textsc{LC-S-Empty}]{\phantom{.}}{\mathsf{lc}(\lsempty{})}}
\and
\gpfit{\inferrule*[lab=\textsc{LC-S-Real}]{\phantom{.}}{\mathsf{lc}(\lR{})}}
\and
\gpfit{\inferrule*[lab=\textsc{LC-S-Dual}]{\mathsf{lc}(\ls{})}{\mathsf{lc}(\mathsf{dual}(\ls{}))}}
\and
\gpfit{\inferrule*[lab=\textsc{LC-S-Dsum}]{\mathsf{lc}(\lsn{0}) \\ \mathsf{lc}(\lsn{1})}{\mathsf{lc}(\mathsf{dsum}(\lsn{0}, \lsn{1}))}}
\and
\gpfit{\inferrule*[lab=\textsc{LC-S-Tprod}]{\mathsf{lc}(\lsn{0}) \\ \mathsf{lc}(\lsn{1})}{\mathsf{lc}(\mathsf{tprod}(\lsn{0}, \lsn{1}))}}
\and
\gpfit{\inferrule*[lab=\textsc{LC-S-Ker}]{\mathsf{lc}(\lsn{0}) \\ \mathsf{lc}(\lsn{1}) \\ \mathsf{lc}(\lee{})}{\mathsf{lc}(\mathsf{ker}(\lsn{0}, \lsn{1}, \lee{}))}}
\and
\gpfit{\inferrule*[lab=\textsc{LC-S-Coker}]{\mathsf{lc}(\lsn{0}) \\ \mathsf{lc}(\lsn{1}) \\ \mathsf{lc}(\lee{})}{\mathsf{lc}(\mathsf{coker}(\lsn{0}, \lsn{1}, \lee{}))}}
\and
\gpfit{\inferrule*[lab=\textsc{LC-S-Img}]{\mathsf{lc}(\lsn{0}) \\ \mathsf{lc}(\lsn{1}) \\ \mathsf{lc}(\lee{})}{\mathsf{lc}(\mathsf{img}(\lsn{0}, \lsn{1}, \lee{}))}}
\and
\gpfit{\inferrule*[lab=\textsc{LC-S-Coimg}]{\mathsf{lc}(\lsn{0}) \\ \mathsf{lc}(\lsn{1}) \\ \mathsf{lc}(\lee{})}{\mathsf{lc}(\mathsf{coimg}(\lsn{0}, \lsn{1}, \lee{}))}}
}
\gldefjudgment{lc.ty}{\mathsf{lc}(\lty{})}{Local closure of types (generated from \texttt{GLang.Ty.lc}).}{fig:gl-lc-ty}
\gldefrule{lc.ty.space}{lc.ty}{\textsc{LC-T-Space}}{\mathsf{lc}(\ls{})}{\mathsf{lc}(\mathsf{space}(\ls{}))}{\texttt{GLang.Ty.lc.space}}{GLang/SourceLN.lean:190}
\gldefrule{lc.ty.piTy}{lc.ty}{\textsc{LC-T-Pi\allowbreak Ty}}{\mathsf{lc}(\ltyn{0}) \\ \forall \mathit{x}.\; \neg(\mathit{x} \in \mathit{L}) \Rightarrow \mathsf{lc}(\mathsf{openVar}(\mathit{x}, \ltyn{1}))}{\mathsf{lc}(\mathsf{piTy}(\ltyn{0}, \ltyn{1}))}{\texttt{GLang.Ty.lc.pi\allowbreak Ty}}{GLang/SourceLN.lean:191}
\gldefrulerows{lc.ty}{
\textsc{LC-T-Space} & $\mathsf{lc}(\mathsf{space}(\ls{}))$ & \texttt{\footnotesize GLang/Source\allowbreak LN.lean:190}\\
\textsc{LC-T-Pi\allowbreak Ty} & $\mathsf{lc}(\mathsf{piTy}(\ltyn{0}, \ltyn{1}))$ & \texttt{\footnotesize GLang/Source\allowbreak LN.lean:191}\\
}{
\gpfit{\inferrule*[]{\mathsf{lc}(\ls{})}{\mathsf{lc}(\mathsf{space}(\ls{}))}}
\and
\gpfit{\inferrule*[]{\mathsf{lc}(\ltyn{0}) \\ \forall \mathit{x}.\; \neg(\mathit{x} \in \mathit{L}) \Rightarrow \mathsf{lc}(\mathsf{openVar}(\mathit{x}, \ltyn{1}))}{\mathsf{lc}(\mathsf{piTy}(\ltyn{0}, \ltyn{1}))}}
}{
\gpfit{\inferrule*[lab=\textsc{LC-T-Space}]{\mathsf{lc}(\ls{})}{\mathsf{lc}(\mathsf{space}(\ls{}))}}
\and
\gpfit{\inferrule*[lab=\textsc{LC-T-Pi\allowbreak Ty}]{\mathsf{lc}(\ltyn{0}) \\ \forall \mathit{x}.\; \neg(\mathit{x} \in \mathit{L}) \Rightarrow \mathsf{lc}(\mathsf{openVar}(\mathit{x}, \ltyn{1}))}{\mathsf{lc}(\mathsf{piTy}(\ltyn{0}, \ltyn{1}))}}
}
\gldefjudgment{lc.term}{\mathsf{lc}(\lee{})}{Local closure of terms (generated from \texttt{GLang.Term.lc}).}{fig:gl-lc-term}
\gldefrule{lc.term.var}{lc.term}{\textsc{LC-E-Var}}{\phantom{.}}{\mathsf{lc}(\mathsf{var}(\mathit{x}))}{\texttt{GLang.Term.lc.var}}{GLang/SourceLN.lean:195}
\gldefrule{lc.term.real}{lc.term}{\textsc{LC-E-Real}}{\phantom{.}}{\mathsf{lc}(\mathsf{real}(\mathit{r}))}{\texttt{GLang.Term.lc.real}}{GLang/SourceLN.lean:196}
\gldefrule{lc.term.zero}{lc.term}{\textsc{LC-E-Zero}}{\mathsf{lc}(\ls{})}{\mathsf{lc}(\mathsf{zero}(\ls{}))}{\texttt{GLang.Term.lc.zero}}{GLang/SourceLN.lean:197}
\gldefrule{lc.term.smul}{lc.term}{\textsc{LC-E-Smul}}{\mathsf{lc}(\len{0}) \\ \mathsf{lc}(\len{1})}{\mathsf{lc}(\mathsf{smul}(\len{0}, \len{1}))}{\texttt{GLang.Term.lc.smul}}{GLang/SourceLN.lean:198}
\gldefrule{lc.term.add}{lc.term}{\textsc{LC-E-Add}}{\mathsf{lc}(\len{0}) \\ \mathsf{lc}(\len{1})}{\mathsf{lc}(\mathsf{add}(\len{0}, \len{1}))}{\texttt{GLang.Term.lc.add}}{GLang/SourceLN.lean:199}
\gldefrule{lc.term.dsumE}{lc.term}{\textsc{LC-E-Dsum\allowbreak E}}{\mathsf{lc}(\len{0}) \\ \mathsf{lc}(\len{1})}{\mathsf{lc}(\mathsf{dsumE}(\len{0}, \len{1}))}{\texttt{GLang.Term.lc.dsum\allowbreak E}}{GLang/SourceLN.lean:200}
\gldefrule{lc.term.tprodE}{lc.term}{\textsc{LC-E-Tprod\allowbreak E}}{\mathsf{lc}(\len{0}) \\ \mathsf{lc}(\len{1})}{\mathsf{lc}(\mathsf{tprodE}(\len{0}, \len{1}))}{\texttt{GLang.Term.lc.tprod\allowbreak E}}{GLang/SourceLN.lean:201}
\gldefrule{lc.term.inner}{lc.term}{\textsc{LC-E-Inner}}{\mathsf{lc}(\ls{})}{\mathsf{lc}(\mathsf{inner}(\ls{}))}{\texttt{GLang.Term.lc.inner}}{GLang/SourceLN.lean:202}
\gldefrule{lc.term.embed}{lc.term}{\textsc{LC-E-Embed}}{\mathsf{lc}(\ls{})}{\mathsf{lc}(\mathsf{embed}(\ls{}))}{\texttt{GLang.Term.lc.embed}}{GLang/SourceLN.lean:203}
\gldefrule{lc.term.pinvT}{lc.term}{\textsc{LC-E-Pinv\allowbreak T}}{\mathsf{lc}(\lsn{0}) \\ \mathsf{lc}(\lsn{1}) \\ \mathsf{lc}(\lee{})}{\mathsf{lc}(\mathsf{pinvT}(\lsn{0}, \lsn{1}, \lee{}))}{\texttt{GLang.Term.lc.pinv\allowbreak T}}{GLang/SourceLN.lean:204}
\gldefrule{lc.term.pinvD}{lc.term}{\textsc{LC-E-Pinv\allowbreak D}}{\mathsf{lc}(\ls{}) \\ \mathsf{lc}(\lee{})}{\mathsf{lc}(\mathsf{pinvD}(\ls{}, \lee{}))}{\texttt{GLang.Term.lc.pinv\allowbreak D}}{GLang/SourceLN.lean:205}
\gldefrule{lc.term.llam}{lc.term}{\textsc{LC-E-Llam}}{\mathsf{lc}(\lsn{0}) \\ \mathsf{lc}(\ls{}) \\ \forall \mathit{x}.\; \neg(\mathit{x} \in \mathit{L}) \Rightarrow \mathsf{lc}(\mathsf{openVar}(\mathit{x}, \lee{}))}{\mathsf{lc}(\mathsf{llam}(\lsn{0}, \ls{}, \lee{}))}{\texttt{GLang.Term.lc.llam}}{GLang/SourceLN.lean:206}
\gldefrule{lc.term.nllam}{lc.term}{\textsc{LC-E-Nllam}}{\mathsf{lc}(\lty{}) \\ \forall \mathit{x}.\; \neg(\mathit{x} \in \mathit{L}) \Rightarrow \mathsf{lc}(\mathsf{openVar}(\mathit{x}, \lee{}))}{\mathsf{lc}(\mathsf{nllam}(\lty{}, \lee{}))}{\texttt{GLang.Term.lc.nllam}}{GLang/SourceLN.lean:209}
\gldefrule{lc.term.appL}{lc.term}{\textsc{LC-E-App\allowbreak L}}{\mathsf{lc}(\lsn{0}) \\ \mathsf{lc}(\lsn{1}) \\ \mathsf{lc}(\len{0}) \\ \mathsf{lc}(\len{1})}{\mathsf{lc}(\mathsf{appL}(\lsn{0}, \lsn{1}, \len{0}, \len{1}))}{\texttt{GLang.Term.lc.app\allowbreak L}}{GLang/SourceLN.lean:211}
\gldefrule{lc.term.appI}{lc.term}{\textsc{LC-E-App\allowbreak I}}{\mathsf{lc}(\ls{}) \\ \mathsf{lc}(\len{0}) \\ \mathsf{lc}(\len{1})}{\mathsf{lc}(\mathsf{appI}(\ls{}, \len{0}, \len{1}))}{\texttt{GLang.Term.lc.app\allowbreak I}}{GLang/SourceLN.lean:213}
\gldefrule{lc.term.appN}{lc.term}{\textsc{LC-E-App\allowbreak N}}{\mathsf{lc}(\len{0}) \\ \mathsf{lc}(\len{1})}{\mathsf{lc}(\mathsf{appN}(\len{0}, \len{1}))}{\texttt{GLang.Term.lc.app\allowbreak N}}{GLang/SourceLN.lean:214}
\gldefrulerows{lc.term}{
\textsc{LC-E-Var} & $\mathsf{lc}(\mathsf{var}(\mathit{x}))$ & \texttt{\footnotesize GLang/Source\allowbreak LN.lean:195}\\
\textsc{LC-E-Real} & $\mathsf{lc}(\mathsf{real}(\mathit{r}))$ & \texttt{\footnotesize GLang/Source\allowbreak LN.lean:196}\\
\textsc{LC-E-Zero} & $\mathsf{lc}(\mathsf{zero}(\ls{}))$ & \texttt{\footnotesize GLang/Source\allowbreak LN.lean:197}\\
\textsc{LC-E-Smul} & $\mathsf{lc}(\mathsf{smul}(\len{0}, \len{1}))$ & \texttt{\footnotesize GLang/Source\allowbreak LN.lean:198}\\
\textsc{LC-E-Add} & $\mathsf{lc}(\mathsf{add}(\len{0}, \len{1}))$ & \texttt{\footnotesize GLang/Source\allowbreak LN.lean:199}\\
\textsc{LC-E-Dsum\allowbreak E} & $\mathsf{lc}(\mathsf{dsumE}(\len{0}, \len{1}))$ & \texttt{\footnotesize GLang/Source\allowbreak LN.lean:200}\\
\textsc{LC-E-Tprod\allowbreak E} & $\mathsf{lc}(\mathsf{tprodE}(\len{0}, \len{1}))$ & \texttt{\footnotesize GLang/Source\allowbreak LN.lean:201}\\
\textsc{LC-E-Inner} & $\mathsf{lc}(\mathsf{inner}(\ls{}))$ & \texttt{\footnotesize GLang/Source\allowbreak LN.lean:202}\\
\textsc{LC-E-Embed} & $\mathsf{lc}(\mathsf{embed}(\ls{}))$ & \texttt{\footnotesize GLang/Source\allowbreak LN.lean:203}\\
\textsc{LC-E-Pinv\allowbreak T} & $\mathsf{lc}(\mathsf{pinvT}(\lsn{0}, \lsn{1}, \lee{}))$ & \texttt{\footnotesize GLang/Source\allowbreak LN.lean:204}\\
\textsc{LC-E-Pinv\allowbreak D} & $\mathsf{lc}(\mathsf{pinvD}(\ls{}, \lee{}))$ & \texttt{\footnotesize GLang/Source\allowbreak LN.lean:205}\\
\textsc{LC-E-Llam} & $\mathsf{lc}(\mathsf{llam}(\lsn{0}, \ls{}, \lee{}))$ & \texttt{\footnotesize GLang/Source\allowbreak LN.lean:206}\\
\textsc{LC-E-Nllam} & $\mathsf{lc}(\mathsf{nllam}(\lty{}, \lee{}))$ & \texttt{\footnotesize GLang/Source\allowbreak LN.lean:209}\\
\textsc{LC-E-App\allowbreak L} & $\mathsf{lc}(\mathsf{appL}(\lsn{0}, \lsn{1}, \len{0}, \len{1}))$ & \texttt{\footnotesize GLang/Source\allowbreak LN.lean:211}\\
\textsc{LC-E-App\allowbreak I} & $\mathsf{lc}(\mathsf{appI}(\ls{}, \len{0}, \len{1}))$ & \texttt{\footnotesize GLang/Source\allowbreak LN.lean:213}\\
\textsc{LC-E-App\allowbreak N} & $\mathsf{lc}(\mathsf{appN}(\len{0}, \len{1}))$ & \texttt{\footnotesize GLang/Source\allowbreak LN.lean:214}\\
}{
\gpfit{\inferrule*[]{\phantom{.}}{\mathsf{lc}(\mathsf{var}(\mathit{x}))}}
\and
\gpfit{\inferrule*[]{\phantom{.}}{\mathsf{lc}(\mathsf{real}(\mathit{r}))}}
\and
\gpfit{\inferrule*[]{\mathsf{lc}(\ls{})}{\mathsf{lc}(\mathsf{zero}(\ls{}))}}
\and
\gpfit{\inferrule*[]{\mathsf{lc}(\len{0}) \\ \mathsf{lc}(\len{1})}{\mathsf{lc}(\mathsf{smul}(\len{0}, \len{1}))}}
\and
\gpfit{\inferrule*[]{\mathsf{lc}(\len{0}) \\ \mathsf{lc}(\len{1})}{\mathsf{lc}(\mathsf{add}(\len{0}, \len{1}))}}
\and
\gpfit{\inferrule*[]{\mathsf{lc}(\len{0}) \\ \mathsf{lc}(\len{1})}{\mathsf{lc}(\mathsf{dsumE}(\len{0}, \len{1}))}}
\and
\gpfit{\inferrule*[]{\mathsf{lc}(\len{0}) \\ \mathsf{lc}(\len{1})}{\mathsf{lc}(\mathsf{tprodE}(\len{0}, \len{1}))}}
\and
\gpfit{\inferrule*[]{\mathsf{lc}(\ls{})}{\mathsf{lc}(\mathsf{inner}(\ls{}))}}
\and
\gpfit{\inferrule*[]{\mathsf{lc}(\ls{})}{\mathsf{lc}(\mathsf{embed}(\ls{}))}}
\and
\gpfit{\inferrule*[]{\mathsf{lc}(\lsn{0}) \\ \mathsf{lc}(\lsn{1}) \\ \mathsf{lc}(\lee{})}{\mathsf{lc}(\mathsf{pinvT}(\lsn{0}, \lsn{1}, \lee{}))}}
\and
\gpfit{\inferrule*[]{\mathsf{lc}(\ls{}) \\ \mathsf{lc}(\lee{})}{\mathsf{lc}(\mathsf{pinvD}(\ls{}, \lee{}))}}
\and
\gpfit{\inferrule*[]{\mathsf{lc}(\lsn{0}) \\ \mathsf{lc}(\ls{}) \\ \forall \mathit{x}.\; \neg(\mathit{x} \in \mathit{L}) \Rightarrow \mathsf{lc}(\mathsf{openVar}(\mathit{x}, \lee{}))}{\mathsf{lc}(\mathsf{llam}(\lsn{0}, \ls{}, \lee{}))}}
\and
\gpfit{\inferrule*[]{\mathsf{lc}(\lty{}) \\ \forall \mathit{x}.\; \neg(\mathit{x} \in \mathit{L}) \Rightarrow \mathsf{lc}(\mathsf{openVar}(\mathit{x}, \lee{}))}{\mathsf{lc}(\mathsf{nllam}(\lty{}, \lee{}))}}
\and
\gpfit{\inferrule*[]{\mathsf{lc}(\lsn{0}) \\ \mathsf{lc}(\lsn{1}) \\ \mathsf{lc}(\len{0}) \\ \mathsf{lc}(\len{1})}{\mathsf{lc}(\mathsf{appL}(\lsn{0}, \lsn{1}, \len{0}, \len{1}))}}
\and
\gpfit{\inferrule*[]{\mathsf{lc}(\ls{}) \\ \mathsf{lc}(\len{0}) \\ \mathsf{lc}(\len{1})}{\mathsf{lc}(\mathsf{appI}(\ls{}, \len{0}, \len{1}))}}
\and
\gpfit{\inferrule*[]{\mathsf{lc}(\len{0}) \\ \mathsf{lc}(\len{1})}{\mathsf{lc}(\mathsf{appN}(\len{0}, \len{1}))}}
}{
\gpfit{\inferrule*[lab=\textsc{LC-E-Var}]{\phantom{.}}{\mathsf{lc}(\mathsf{var}(\mathit{x}))}}
\and
\gpfit{\inferrule*[lab=\textsc{LC-E-Real}]{\phantom{.}}{\mathsf{lc}(\mathsf{real}(\mathit{r}))}}
\and
\gpfit{\inferrule*[lab=\textsc{LC-E-Zero}]{\mathsf{lc}(\ls{})}{\mathsf{lc}(\mathsf{zero}(\ls{}))}}
\and
\gpfit{\inferrule*[lab=\textsc{LC-E-Smul}]{\mathsf{lc}(\len{0}) \\ \mathsf{lc}(\len{1})}{\mathsf{lc}(\mathsf{smul}(\len{0}, \len{1}))}}
\and
\gpfit{\inferrule*[lab=\textsc{LC-E-Add}]{\mathsf{lc}(\len{0}) \\ \mathsf{lc}(\len{1})}{\mathsf{lc}(\mathsf{add}(\len{0}, \len{1}))}}
\and
\gpfit{\inferrule*[lab=\textsc{LC-E-Dsum\allowbreak E}]{\mathsf{lc}(\len{0}) \\ \mathsf{lc}(\len{1})}{\mathsf{lc}(\mathsf{dsumE}(\len{0}, \len{1}))}}
\and
\gpfit{\inferrule*[lab=\textsc{LC-E-Tprod\allowbreak E}]{\mathsf{lc}(\len{0}) \\ \mathsf{lc}(\len{1})}{\mathsf{lc}(\mathsf{tprodE}(\len{0}, \len{1}))}}
\and
\gpfit{\inferrule*[lab=\textsc{LC-E-Inner}]{\mathsf{lc}(\ls{})}{\mathsf{lc}(\mathsf{inner}(\ls{}))}}
\and
\gpfit{\inferrule*[lab=\textsc{LC-E-Embed}]{\mathsf{lc}(\ls{})}{\mathsf{lc}(\mathsf{embed}(\ls{}))}}
\and
\gpfit{\inferrule*[lab=\textsc{LC-E-Pinv\allowbreak T}]{\mathsf{lc}(\lsn{0}) \\ \mathsf{lc}(\lsn{1}) \\ \mathsf{lc}(\lee{})}{\mathsf{lc}(\mathsf{pinvT}(\lsn{0}, \lsn{1}, \lee{}))}}
\and
\gpfit{\inferrule*[lab=\textsc{LC-E-Pinv\allowbreak D}]{\mathsf{lc}(\ls{}) \\ \mathsf{lc}(\lee{})}{\mathsf{lc}(\mathsf{pinvD}(\ls{}, \lee{}))}}
\and
\gpfit{\inferrule*[lab=\textsc{LC-E-Llam}]{\mathsf{lc}(\lsn{0}) \\ \mathsf{lc}(\ls{}) \\ \forall \mathit{x}.\; \neg(\mathit{x} \in \mathit{L}) \Rightarrow \mathsf{lc}(\mathsf{openVar}(\mathit{x}, \lee{}))}{\mathsf{lc}(\mathsf{llam}(\lsn{0}, \ls{}, \lee{}))}}
\and
\gpfit{\inferrule*[lab=\textsc{LC-E-Nllam}]{\mathsf{lc}(\lty{}) \\ \forall \mathit{x}.\; \neg(\mathit{x} \in \mathit{L}) \Rightarrow \mathsf{lc}(\mathsf{openVar}(\mathit{x}, \lee{}))}{\mathsf{lc}(\mathsf{nllam}(\lty{}, \lee{}))}}
\and
\gpfit{\inferrule*[lab=\textsc{LC-E-App\allowbreak L}]{\mathsf{lc}(\lsn{0}) \\ \mathsf{lc}(\lsn{1}) \\ \mathsf{lc}(\len{0}) \\ \mathsf{lc}(\len{1})}{\mathsf{lc}(\mathsf{appL}(\lsn{0}, \lsn{1}, \len{0}, \len{1}))}}
\and
\gpfit{\inferrule*[lab=\textsc{LC-E-App\allowbreak I}]{\mathsf{lc}(\ls{}) \\ \mathsf{lc}(\len{0}) \\ \mathsf{lc}(\len{1})}{\mathsf{lc}(\mathsf{appI}(\ls{}, \len{0}, \len{1}))}}
\and
\gpfit{\inferrule*[lab=\textsc{LC-E-App\allowbreak N}]{\mathsf{lc}(\len{0}) \\ \mathsf{lc}(\len{1})}{\mathsf{lc}(\mathsf{appN}(\len{0}, \len{1}))}}
}
\gldefjudgment{lc.basis}{\mathsf{lc}(\lb{})}{Local closure of bases (generated from \texttt{GLang.Basis.lc}).}{fig:gl-lc-basis}
\gldefrule{lc.basis.empty}{lc.basis}{\textsc{LC-B-Empty}}{\phantom{.}}{\mathsf{lc}(\lbzero{})}{\texttt{GLang.Basis.lc.empty}}{GLang/SourceLN.lean:219}
\gldefrule{lc.basis.one}{lc.basis}{\textsc{LC-B-One}}{\phantom{.}}{\mathsf{lc}(\lbone{})}{\texttt{GLang.Basis.lc.one}}{GLang/SourceLN.lean:220}
\gldefrule{lc.basis.dual}{lc.basis}{\textsc{LC-B-Dual}}{\mathsf{lc}(\lb{})}{\mathsf{lc}(\mathsf{dual}(\lb{}))}{\texttt{GLang.Basis.lc.dual}}{GLang/SourceLN.lean:221}
\gldefrule{lc.basis.dsum}{lc.basis}{\textsc{LC-B-Dsum}}{\mathsf{lc}(\lbn{0}) \\ \mathsf{lc}(\lbn{1})}{\mathsf{lc}(\mathsf{dsum}(\lbn{0}, \lbn{1}))}{\texttt{GLang.Basis.lc.dsum}}{GLang/SourceLN.lean:222}
\gldefrule{lc.basis.tprod}{lc.basis}{\textsc{LC-B-Tprod}}{\mathsf{lc}(\lbn{0}) \\ \mathsf{lc}(\lbn{1})}{\mathsf{lc}(\mathsf{tprod}(\lbn{0}, \lbn{1}))}{\texttt{GLang.Basis.lc.tprod}}{GLang/SourceLN.lean:223}
\gldefrule{lc.basis.proj}{lc.basis}{\textsc{LC-B-Proj}}{\mathsf{lc}(\ls{}) \\ \mathsf{lc}(\lbn{0}) \\ \mathsf{lc}(\lbn{1})}{\mathsf{lc}(\mathsf{proj}(\ls{}, \lbn{0}, \lbn{1}))}{\texttt{GLang.Basis.lc.proj}}{GLang/SourceLN.lean:224}
\gldefrule{lc.basis.image}{lc.basis}{\textsc{LC-B-Image}}{\mathsf{lc}(\lb{})}{\mathsf{lc}(\mathsf{image}(\lb{}))}{\texttt{GLang.Basis.lc.image}}{GLang/SourceLN.lean:225}
\gldefrule{lc.basis.terms}{lc.basis}{\textsc{LC-B-Terms}}{\mathsf{lc}(\ls{}) \\ \forall \lee{}.\; \lee{} \in \mathit{es} \Rightarrow \mathsf{lc}(\lee{}) \\ \mathsf{lc}(\lb{})}{\mathsf{lc}(\mathsf{terms}(\ls{}, \mathit{es}, \lb{}))}{\texttt{GLang.Basis.lc.terms}}{GLang/SourceLN.lean:226}
\gldefrule{lc.basis.frame}{lc.basis}{\textsc{LC-B-Frame}}{\mathsf{lc}(\lbn{0}) \\ \forall \mathit{x}.\; \neg(\mathit{x} \in \mathit{L}) \Rightarrow \mathsf{lc}(\mathsf{openVar}(\mathit{x}, \lbn{1}))}{\mathsf{lc}(\mathsf{frame}(\lbn{0}, \lbn{1}))}{\texttt{GLang.Basis.lc.frame}}{GLang/SourceLN.lean:227}
\gldefrulerows{lc.basis}{
\textsc{LC-B-Empty} & $\mathsf{lc}(\lbzero{})$ & \texttt{\footnotesize GLang/Source\allowbreak LN.lean:219}\\
\textsc{LC-B-One} & $\mathsf{lc}(\lbone{})$ & \texttt{\footnotesize GLang/Source\allowbreak LN.lean:220}\\
\textsc{LC-B-Dual} & $\mathsf{lc}(\mathsf{dual}(\lb{}))$ & \texttt{\footnotesize GLang/Source\allowbreak LN.lean:221}\\
\textsc{LC-B-Dsum} & $\mathsf{lc}(\mathsf{dsum}(\lbn{0}, \lbn{1}))$ & \texttt{\footnotesize GLang/Source\allowbreak LN.lean:222}\\
\textsc{LC-B-Tprod} & $\mathsf{lc}(\mathsf{tprod}(\lbn{0}, \lbn{1}))$ & \texttt{\footnotesize GLang/Source\allowbreak LN.lean:223}\\
\textsc{LC-B-Proj} & $\mathsf{lc}(\mathsf{proj}(\ls{}, \lbn{0}, \lbn{1}))$ & \texttt{\footnotesize GLang/Source\allowbreak LN.lean:224}\\
\textsc{LC-B-Image} & $\mathsf{lc}(\mathsf{image}(\lb{}))$ & \texttt{\footnotesize GLang/Source\allowbreak LN.lean:225}\\
\textsc{LC-B-Terms} & $\mathsf{lc}(\mathsf{terms}(\ls{}, \mathit{es}, \lb{}))$ & \texttt{\footnotesize GLang/Source\allowbreak LN.lean:226}\\
\textsc{LC-B-Frame} & $\mathsf{lc}(\mathsf{frame}(\lbn{0}, \lbn{1}))$ & \texttt{\footnotesize GLang/Source\allowbreak LN.lean:227}\\
}{
\gpfit{\inferrule*[]{\phantom{.}}{\mathsf{lc}(\lbzero{})}}
\and
\gpfit{\inferrule*[]{\phantom{.}}{\mathsf{lc}(\lbone{})}}
\and
\gpfit{\inferrule*[]{\mathsf{lc}(\lb{})}{\mathsf{lc}(\mathsf{dual}(\lb{}))}}
\and
\gpfit{\inferrule*[]{\mathsf{lc}(\lbn{0}) \\ \mathsf{lc}(\lbn{1})}{\mathsf{lc}(\mathsf{dsum}(\lbn{0}, \lbn{1}))}}
\and
\gpfit{\inferrule*[]{\mathsf{lc}(\lbn{0}) \\ \mathsf{lc}(\lbn{1})}{\mathsf{lc}(\mathsf{tprod}(\lbn{0}, \lbn{1}))}}
\and
\gpfit{\inferrule*[]{\mathsf{lc}(\ls{}) \\ \mathsf{lc}(\lbn{0}) \\ \mathsf{lc}(\lbn{1})}{\mathsf{lc}(\mathsf{proj}(\ls{}, \lbn{0}, \lbn{1}))}}
\and
\gpfit{\inferrule*[]{\mathsf{lc}(\lb{})}{\mathsf{lc}(\mathsf{image}(\lb{}))}}
\and
\gpfit{\inferrule*[]{\mathsf{lc}(\ls{}) \\ \forall \lee{}.\; \lee{} \in \mathit{es} \Rightarrow \mathsf{lc}(\lee{}) \\ \mathsf{lc}(\lb{})}{\mathsf{lc}(\mathsf{terms}(\ls{}, \mathit{es}, \lb{}))}}
\and
\gpfit{\inferrule*[]{\mathsf{lc}(\lbn{0}) \\ \forall \mathit{x}.\; \neg(\mathit{x} \in \mathit{L}) \Rightarrow \mathsf{lc}(\mathsf{openVar}(\mathit{x}, \lbn{1}))}{\mathsf{lc}(\mathsf{frame}(\lbn{0}, \lbn{1}))}}
}{
\gpfit{\inferrule*[lab=\textsc{LC-B-Empty}]{\phantom{.}}{\mathsf{lc}(\lbzero{})}}
\and
\gpfit{\inferrule*[lab=\textsc{LC-B-One}]{\phantom{.}}{\mathsf{lc}(\lbone{})}}
\and
\gpfit{\inferrule*[lab=\textsc{LC-B-Dual}]{\mathsf{lc}(\lb{})}{\mathsf{lc}(\mathsf{dual}(\lb{}))}}
\and
\gpfit{\inferrule*[lab=\textsc{LC-B-Dsum}]{\mathsf{lc}(\lbn{0}) \\ \mathsf{lc}(\lbn{1})}{\mathsf{lc}(\mathsf{dsum}(\lbn{0}, \lbn{1}))}}
\and
\gpfit{\inferrule*[lab=\textsc{LC-B-Tprod}]{\mathsf{lc}(\lbn{0}) \\ \mathsf{lc}(\lbn{1})}{\mathsf{lc}(\mathsf{tprod}(\lbn{0}, \lbn{1}))}}
\and
\gpfit{\inferrule*[lab=\textsc{LC-B-Proj}]{\mathsf{lc}(\ls{}) \\ \mathsf{lc}(\lbn{0}) \\ \mathsf{lc}(\lbn{1})}{\mathsf{lc}(\mathsf{proj}(\ls{}, \lbn{0}, \lbn{1}))}}
\and
\gpfit{\inferrule*[lab=\textsc{LC-B-Image}]{\mathsf{lc}(\lb{})}{\mathsf{lc}(\mathsf{image}(\lb{}))}}
\and
\gpfit{\inferrule*[lab=\textsc{LC-B-Terms}]{\mathsf{lc}(\ls{}) \\ \forall \lee{}.\; \lee{} \in \mathit{es} \Rightarrow \mathsf{lc}(\lee{}) \\ \mathsf{lc}(\lb{})}{\mathsf{lc}(\mathsf{terms}(\ls{}, \mathit{es}, \lb{}))}}
\and
\gpfit{\inferrule*[lab=\textsc{LC-B-Frame}]{\mathsf{lc}(\lbn{0}) \\ \forall \mathit{x}.\; \neg(\mathit{x} \in \mathit{L}) \Rightarrow \mathsf{lc}(\mathsf{openVar}(\mathit{x}, \lbn{1}))}{\mathsf{lc}(\mathsf{frame}(\lbn{0}, \lbn{1}))}}
}
\gldefjudgment{lc.valSpace}{\mathsf{lc}(\lsv{})}{Local closure of value spaces (generated from \texttt{GLang.ValSpace.lc}).}{fig:gl-lc-valspace}
\gldefrule{lc.valSpace.empty}{lc.valSpace}{\textsc{LC-VS-Empty}}{\phantom{.}}{\mathsf{lc}(\lsempty{})}{\texttt{GLang.Val\allowbreak Space.lc.empty}}{GLang/ValueLN.lean:528}
\gldefrule{lc.valSpace.real}{lc.valSpace}{\textsc{LC-VS-Real}}{\phantom{.}}{\mathsf{lc}(\lR{})}{\texttt{GLang.Val\allowbreak Space.lc.real}}{GLang/ValueLN.lean:529}
\gldefrule{lc.valSpace.dsum}{lc.valSpace}{\textsc{LC-VS-Dsum}}{\mathsf{lc}(\lsnv{0}) \\ \mathsf{lc}(\lsnv{1})}{\mathsf{lc}(\mathsf{dsum}(\lsnv{0}, \lsnv{1}))}{\texttt{GLang.Val\allowbreak Space.lc.dsum}}{GLang/ValueLN.lean:530}
\gldefrule{lc.valSpace.tprod}{lc.valSpace}{\textsc{LC-VS-Tprod}}{\mathsf{lc}(\lsnv{0}) \\ \mathsf{lc}(\lsnv{1})}{\mathsf{lc}(\mathsf{tprod}(\lsnv{0}, \lsnv{1}))}{\texttt{GLang.Val\allowbreak Space.lc.tprod}}{GLang/ValueLN.lean:531}
\gldefrule{lc.valSpace.ffs}{lc.valSpace}{\textsc{LC-VS-Ffs}}{\mathsf{lc}(\lsv{}) \\ \mathsf{lc}(\leev{})}{\mathsf{lc}(\mathsf{ffs}(\lprop{}, \lsv{}, \leev{}))}{\texttt{GLang.Val\allowbreak Space.lc.ffs}}{GLang/ValueLN.lean:532}
\gldefrulerows{lc.valSpace}{
\textsc{LC-VS-Empty} & $\mathsf{lc}(\lsempty{})$ & \texttt{\footnotesize GLang/Value\allowbreak LN.lean:528}\\
\textsc{LC-VS-Real} & $\mathsf{lc}(\lR{})$ & \texttt{\footnotesize GLang/Value\allowbreak LN.lean:529}\\
\textsc{LC-VS-Dsum} & $\mathsf{lc}(\mathsf{dsum}(\lsnv{0}, \lsnv{1}))$ & \texttt{\footnotesize GLang/Value\allowbreak LN.lean:530}\\
\textsc{LC-VS-Tprod} & $\mathsf{lc}(\mathsf{tprod}(\lsnv{0}, \lsnv{1}))$ & \texttt{\footnotesize GLang/Value\allowbreak LN.lean:531}\\
\textsc{LC-VS-Ffs} & $\mathsf{lc}(\mathsf{ffs}(\lprop{}, \lsv{}, \leev{}))$ & \texttt{\footnotesize GLang/Value\allowbreak LN.lean:532}\\
}{
\gpfit{\inferrule*[]{\phantom{.}}{\mathsf{lc}(\lsempty{})}}
\and
\gpfit{\inferrule*[]{\phantom{.}}{\mathsf{lc}(\lR{})}}
\and
\gpfit{\inferrule*[]{\mathsf{lc}(\lsnv{0}) \\ \mathsf{lc}(\lsnv{1})}{\mathsf{lc}(\mathsf{dsum}(\lsnv{0}, \lsnv{1}))}}
\and
\gpfit{\inferrule*[]{\mathsf{lc}(\lsnv{0}) \\ \mathsf{lc}(\lsnv{1})}{\mathsf{lc}(\mathsf{tprod}(\lsnv{0}, \lsnv{1}))}}
\and
\gpfit{\inferrule*[]{\mathsf{lc}(\lsv{}) \\ \mathsf{lc}(\leev{})}{\mathsf{lc}(\mathsf{ffs}(\lprop{}, \lsv{}, \leev{}))}}
}{
\gpfit{\inferrule*[lab=\textsc{LC-VS-Empty}]{\phantom{.}}{\mathsf{lc}(\lsempty{})}}
\and
\gpfit{\inferrule*[lab=\textsc{LC-VS-Real}]{\phantom{.}}{\mathsf{lc}(\lR{})}}
\and
\gpfit{\inferrule*[lab=\textsc{LC-VS-Dsum}]{\mathsf{lc}(\lsnv{0}) \\ \mathsf{lc}(\lsnv{1})}{\mathsf{lc}(\mathsf{dsum}(\lsnv{0}, \lsnv{1}))}}
\and
\gpfit{\inferrule*[lab=\textsc{LC-VS-Tprod}]{\mathsf{lc}(\lsnv{0}) \\ \mathsf{lc}(\lsnv{1})}{\mathsf{lc}(\mathsf{tprod}(\lsnv{0}, \lsnv{1}))}}
\and
\gpfit{\inferrule*[lab=\textsc{LC-VS-Ffs}]{\mathsf{lc}(\lsv{}) \\ \mathsf{lc}(\leev{})}{\mathsf{lc}(\mathsf{ffs}(\lprop{}, \lsv{}, \leev{}))}}
}
\gldefjudgment{lc.valTy}{\mathsf{lc}(\ltyv{})}{Local closure of value types (generated from \texttt{GLang.ValTy.lc}).}{fig:gl-lc-valty}
\gldefrule{lc.valTy.space}{lc.valTy}{\textsc{LC-VT-Space}}{\mathsf{lc}(\lsv{})}{\mathsf{lc}(\mathsf{space}(\lsv{}))}{\texttt{GLang.Val\allowbreak Ty.lc.space}}{GLang/ValueLN.lean:535}
\gldefrule{lc.valTy.piTy}{lc.valTy}{\textsc{LC-VT-Pi\allowbreak Ty}}{\mathsf{lc}(\ltynv{0}) \\ \forall \mathit{x}.\; \neg(\mathit{x} \in \mathit{L}) \Rightarrow \mathsf{lc}(\mathsf{openVar}(\mathit{x}, \ltynv{0}, \ltynv{1}))}{\mathsf{lc}(\mathsf{piTy}(\ltynv{0}, \ltynv{1}))}{\texttt{GLang.Val\allowbreak Ty.lc.pi\allowbreak Ty}}{GLang/ValueLN.lean:536}
\gldefrulerows{lc.valTy}{
\textsc{LC-VT-Space} & $\mathsf{lc}(\mathsf{space}(\lsv{}))$ & \texttt{\footnotesize GLang/Value\allowbreak LN.lean:535}\\
\textsc{LC-VT-Pi\allowbreak Ty} & $\mathsf{lc}(\mathsf{piTy}(\ltynv{0}, \ltynv{1}))$ & \texttt{\footnotesize GLang/Value\allowbreak LN.lean:536}\\
}{
\gpfit{\inferrule*[]{\mathsf{lc}(\lsv{})}{\mathsf{lc}(\mathsf{space}(\lsv{}))}}
\and
\gpfit{\inferrule*[]{\mathsf{lc}(\ltynv{0}) \\ \forall \mathit{x}.\; \neg(\mathit{x} \in \mathit{L}) \Rightarrow \mathsf{lc}(\mathsf{openVar}(\mathit{x}, \ltynv{0}, \ltynv{1}))}{\mathsf{lc}(\mathsf{piTy}(\ltynv{0}, \ltynv{1}))}}
}{
\gpfit{\inferrule*[lab=\textsc{LC-VT-Space}]{\mathsf{lc}(\lsv{})}{\mathsf{lc}(\mathsf{space}(\lsv{}))}}
\and
\gpfit{\inferrule*[lab=\textsc{LC-VT-Pi\allowbreak Ty}]{\mathsf{lc}(\ltynv{0}) \\ \forall \mathit{x}.\; \neg(\mathit{x} \in \mathit{L}) \Rightarrow \mathsf{lc}(\mathsf{openVar}(\mathit{x}, \ltynv{0}, \ltynv{1}))}{\mathsf{lc}(\mathsf{piTy}(\ltynv{0}, \ltynv{1}))}}
}
\gldefjudgment{lc.valExpr}{\mathsf{lc}(\leev{})}{Local closure of value expressions (generated from \texttt{GLang.ValExpr.lc}).}{fig:gl-lc-valexpr}
\gldefrule{lc.valExpr.var}{lc.valExpr}{\textsc{LC-VE-Var}}{\mathsf{lc}(\ltyv{})}{\mathsf{lc}(\mathsf{var}(\mathit{x}, \ltyv{}))}{\texttt{GLang.Val\allowbreak Expr.lc.var}}{GLang/ValueLN.lean:541}
\gldefrule{lc.valExpr.vlam}{lc.valExpr}{\textsc{LC-VE-Vlam}}{\mathsf{lc}(\ltynv{0}) \\ \forall \mathit{x}.\; \neg(\mathit{x} \in \mathit{L}) \Rightarrow \mathsf{lc}(\mathsf{openVar}(\mathit{x}, \ltynv{0}, \ltynv{1})) \\ \forall \mathit{x}.\; \neg(\mathit{x} \in \mathit{L}) \Rightarrow \mathsf{lc}(\mathsf{openVar}(\mathit{x}, \ltynv{0}, \leev{}))}{\mathsf{lc}(\mathsf{vlam}(\ltynv{0}, \ltynv{1}, \leev{}))}{\texttt{GLang.Val\allowbreak Expr.lc.vlam}}{GLang/ValueLN.lean:542}
\gldefrule{lc.valExpr.zero}{lc.valExpr}{\textsc{LC-VE-Zero}}{\mathsf{lc}(\lsv{})}{\mathsf{lc}(\mathsf{zero}(\lsv{}))}{\texttt{GLang.Val\allowbreak Expr.lc.zero}}{GLang/ValueLN.lean:547}
\gldefrule{lc.valExpr.rlit}{lc.valExpr}{\textsc{LC-VE-Rlit}}{\phantom{.}}{\mathsf{lc}(\mathsf{rlit}(\mathit{r}))}{\texttt{GLang.Val\allowbreak Expr.lc.rlit}}{GLang/ValueLN.lean:548}
\gldefrule{lc.valExpr.scale}{lc.valExpr}{\textsc{LC-VE-Scale}}{\mathsf{lc}(\lenv{0}) \\ \mathsf{lc}(\lenv{1})}{\mathsf{lc}(\mathsf{scale}(\lenv{0}, \lenv{1}))}{\texttt{GLang.Val\allowbreak Expr.lc.scale}}{GLang/ValueLN.lean:549}
\gldefrule{lc.valExpr.vadd}{lc.valExpr}{\textsc{LC-VE-Vadd}}{\mathsf{lc}(\lsv{}) \\ \mathsf{lc}(\lenv{0}) \\ \mathsf{lc}(\lenv{1})}{\mathsf{lc}(\mathsf{vadd}(\lsv{}, \lenv{0}, \lenv{1}))}{\texttt{GLang.Val\allowbreak Expr.lc.vadd}}{GLang/ValueLN.lean:550}
\gldefrule{lc.valExpr.dsumE}{lc.valExpr}{\textsc{LC-VE-Dsum\allowbreak E}}{\mathsf{lc}(\lenv{0}) \\ \mathsf{lc}(\lenv{1})}{\mathsf{lc}(\mathsf{dsumE}(\lenv{0}, \lenv{1}))}{\texttt{GLang.Val\allowbreak Expr.lc.dsum\allowbreak E}}{GLang/ValueLN.lean:552}
\gldefrule{lc.valExpr.tprodE}{lc.valExpr}{\textsc{LC-VE-Tprod\allowbreak E}}{\mathsf{lc}(\lenv{0}) \\ \mathsf{lc}(\lenv{1})}{\mathsf{lc}(\mathsf{tprodE}(\lenv{0}, \lenv{1}))}{\texttt{GLang.Val\allowbreak Expr.lc.tprod\allowbreak E}}{GLang/ValueLN.lean:553}
\gldefrule{lc.valExpr.delta}{lc.valExpr}{\textsc{LC-VE-Delta}}{\mathsf{lc}(\lsv{})}{\mathsf{lc}(\mathsf{delta}(\mathit{n}, \lsv{}))}{\texttt{GLang.Val\allowbreak Expr.lc.delta}}{GLang/ValueLN.lean:554}
\gldefrule{lc.valExpr.idelta}{lc.valExpr}{\textsc{LC-VE-Idelta}}{\mathsf{lc}(\lsv{})}{\mathsf{lc}(\mathsf{idelta}(\lsv{}))}{\texttt{GLang.Val\allowbreak Expr.lc.idelta}}{GLang/ValueLN.lean:555}
\gldefrule{lc.valExpr.vsum}{lc.valExpr}{\textsc{LC-VE-Vsum}}{\mathsf{lc}(\lsv{}) \\ \forall \leev{}.\; \leev{} \in \mathit{es} \Rightarrow \mathsf{lc}(\leev{})}{\mathsf{lc}(\mathsf{vsum}(\lsv{}, \mathit{es}))}{\texttt{GLang.Val\allowbreak Expr.lc.vsum}}{GLang/ValueLN.lean:556}
\gldefrule{lc.valExpr.ttv}{lc.valExpr}{\textsc{LC-VE-Ttv}}{\mathsf{lc}(\lsv{}) \\ \mathsf{lc}(\lenv{0}) \\ \mathsf{lc}(\lenv{1})}{\mathsf{lc}(\mathsf{ttv}(\lsv{}, \lenv{0}, \lenv{1}))}{\texttt{GLang.Val\allowbreak Expr.lc.ttv}}{GLang/ValueLN.lean:557}
\gldefrule{lc.valExpr.mtm}{lc.valExpr}{\textsc{LC-VE-Mtm}}{\mathsf{lc}(\lsnv{0}) \\ \mathsf{lc}(\lsnv{1}) \\ \mathsf{lc}(\lenv{0}) \\ \mathsf{lc}(\lenv{1})}{\mathsf{lc}(\mathsf{mtm}({\mathit{n}_{0}}, {\mathit{n}_{1}}, \lsnv{0}, \lsnv{1}, \lenv{0}, \lenv{1}))}{\texttt{GLang.Val\allowbreak Expr.lc.mtm}}{GLang/ValueLN.lean:558}
\gldefrule{lc.valExpr.appV}{lc.valExpr}{\textsc{LC-VE-App\allowbreak V}}{\mathsf{lc}(\lenv{0}) \\ \mathsf{lc}(\lenv{1})}{\mathsf{lc}(\mathsf{appV}(\lenv{0}, \lenv{1}))}{\texttt{GLang.Val\allowbreak Expr.lc.app\allowbreak V}}{GLang/ValueLN.lean:560}
\gldefrule{lc.valExpr.vkrn}{lc.valExpr}{\textsc{LC-VE-Vkrn}}{\mathsf{lc}(\lenv{0}) \\ \mathsf{lc}(\lenv{1})}{\mathsf{lc}(\mathsf{vkrn}(\lenv{0}, \lenv{1}))}{\texttt{GLang.Val\allowbreak Expr.lc.vkrn}}{GLang/ValueLN.lean:561}
\gldefrule{lc.valExpr.vblk}{lc.valExpr}{\textsc{LC-VE-Vblk}}{\mathsf{lc}(\lenv{0}) \\ \mathsf{lc}(\lenv{1})}{\mathsf{lc}(\mathsf{vblk}(\lenv{0}, \lenv{1}))}{\texttt{GLang.Val\allowbreak Expr.lc.vblk}}{GLang/ValueLN.lean:562}
\gldefrule{lc.valExpr.reshap}{lc.valExpr}{\textsc{LC-VE-Reshap}}{\mathsf{lc}(\leev{}) \\ \mathsf{lc}(\ltyv{}) \\ \mathsf{lc}(\ltynv{0})}{\mathsf{lc}(\mathsf{reshap}(\leev{}, \ltyv{}, \ltynv{0}))}{\texttt{GLang.Val\allowbreak Expr.lc.reshap}}{GLang/ValueLN.lean:563}
\gldefrule{lc.valExpr.vtrans}{lc.valExpr}{\textsc{LC-VE-Vtrans}}{\mathsf{lc}(\leev{})}{\mathsf{lc}(\mathsf{vtrans}(\leev{}))}{\texttt{GLang.Val\allowbreak Expr.lc.vtrans}}{GLang/ValueLN.lean:565}
\gldefrule{lc.valExpr.gsvdS}{lc.valExpr}{\textsc{LC-VE-Gsvd\allowbreak S}}{\mathsf{lc}(\leev{}) \\ \mathsf{lc}(\lenv{{\mathrm{a}}}) \\ \mathsf{lc}(\lenv{{\mathrm{b}}})}{\mathsf{lc}(\mathsf{gsvdS}(\leev{}, \lenv{{\mathrm{a}}}, \lenv{{\mathrm{b}}}))}{\texttt{GLang.Val\allowbreak Expr.lc.gsvd\allowbreak S}}{GLang/ValueLN.lean:566}
\gldefrule{lc.valExpr.gsvdU}{lc.valExpr}{\textsc{LC-VE-Gsvd\allowbreak U}}{\mathsf{lc}(\leev{}) \\ \mathsf{lc}(\lenv{{\mathrm{a}}}) \\ \mathsf{lc}(\lenv{{\mathrm{b}}})}{\mathsf{lc}(\mathsf{gsvdU}(\leev{}, \lenv{{\mathrm{a}}}, \lenv{{\mathrm{b}}}))}{\texttt{GLang.Val\allowbreak Expr.lc.gsvd\allowbreak U}}{GLang/ValueLN.lean:567}
\gldefrule{lc.valExpr.gsvdUI}{lc.valExpr}{\textsc{LC-VE-Gsvd\allowbreak UI}}{\mathsf{lc}(\leev{}) \\ \mathsf{lc}(\lenv{{\mathrm{a}}}) \\ \mathsf{lc}(\lenv{{\mathrm{b}}})}{\mathsf{lc}(\mathsf{gsvdUI}(\leev{}, \lenv{{\mathrm{a}}}, \lenv{{\mathrm{b}}}))}{\texttt{GLang.Val\allowbreak Expr.lc.gsvd\allowbreak UI}}{GLang/ValueLN.lean:568}
\gldefrule{lc.valExpr.gsvdV}{lc.valExpr}{\textsc{LC-VE-Gsvd\allowbreak V}}{\mathsf{lc}(\leev{}) \\ \mathsf{lc}(\lenv{{\mathrm{a}}}) \\ \mathsf{lc}(\lenv{{\mathrm{b}}})}{\mathsf{lc}(\mathsf{gsvdV}(\leev{}, \lenv{{\mathrm{a}}}, \lenv{{\mathrm{b}}}))}{\texttt{GLang.Val\allowbreak Expr.lc.gsvd\allowbreak V}}{GLang/ValueLN.lean:569}
\gldefrule{lc.valExpr.gsvdVI}{lc.valExpr}{\textsc{LC-VE-Gsvd\allowbreak VI}}{\mathsf{lc}(\leev{}) \\ \mathsf{lc}(\lenv{{\mathrm{a}}}) \\ \mathsf{lc}(\lenv{{\mathrm{b}}})}{\mathsf{lc}(\mathsf{gsvdVI}(\leev{}, \lenv{{\mathrm{a}}}, \lenv{{\mathrm{b}}}))}{\texttt{GLang.Val\allowbreak Expr.lc.gsvd\allowbreak VI}}{GLang/ValueLN.lean:570}
\gldefrule{lc.valExpr.pinv}{lc.valExpr}{\textsc{LC-VE-Pinv}}{\mathsf{lc}(\leev{})}{\mathsf{lc}(\mathsf{pinv}(\leev{}))}{\texttt{GLang.Val\allowbreak Expr.lc.pinv}}{GLang/ValueLN.lean:571}
\gldefrule{lc.valExpr.gpinv}{lc.valExpr}{\textsc{LC-VE-Gpinv}}{\mathsf{lc}(\leev{}) \\ \mathsf{lc}(\lenv{{\mathrm{a}}}) \\ \mathsf{lc}(\lenv{{\mathrm{b}}})}{\mathsf{lc}(\mathsf{gpinv}(\leev{}, \lenv{{\mathrm{a}}}, \lenv{{\mathrm{b}}}))}{\texttt{GLang.Val\allowbreak Expr.lc.gpinv}}{GLang/ValueLN.lean:572}
\gldefrule{lc.valExpr.rrefRep}{lc.valExpr}{\textsc{LC-VE-Rref\allowbreak Rep}}{\mathsf{lc}(\lenv{0}) \\ \mathsf{lc}(\lenv{1})}{\mathsf{lc}(\mathsf{rrefRep}(\lenv{0}, \lenv{1}))}{\texttt{GLang.Val\allowbreak Expr.lc.rref\allowbreak Rep}}{GLang/ValueLN.lean:573}
\gldefrule{lc.valExpr.inv}{lc.valExpr}{\textsc{LC-VE-Inv}}{\mathsf{lc}(\leev{})}{\mathsf{lc}(\mathsf{inv}(\leev{}))}{\texttt{GLang.Val\allowbreak Expr.lc.inv}}{GLang/ValueLN.lean:574}
\gldefrule{lc.valExpr.svds}{lc.valExpr}{\textsc{LC-VE-Svds}}{\mathsf{lc}(\leev{})}{\mathsf{lc}(\mathsf{svds}(\leev{}))}{\texttt{GLang.Val\allowbreak Expr.lc.svds}}{GLang/ValueLN.lean:575}
\gldefrulerows{lc.valExpr}{
\textsc{LC-VE-Var} & $\mathsf{lc}(\mathsf{var}(\mathit{x}, \ltyv{}))$ & \texttt{\footnotesize GLang/Value\allowbreak LN.lean:541}\\
\textsc{LC-VE-Vlam} & $\mathsf{lc}(\mathsf{vlam}(\ltynv{0}, \ltynv{1}, \leev{}))$ & \texttt{\footnotesize GLang/Value\allowbreak LN.lean:542}\\
\textsc{LC-VE-Zero} & $\mathsf{lc}(\mathsf{zero}(\lsv{}))$ & \texttt{\footnotesize GLang/Value\allowbreak LN.lean:547}\\
\textsc{LC-VE-Rlit} & $\mathsf{lc}(\mathsf{rlit}(\mathit{r}))$ & \texttt{\footnotesize GLang/Value\allowbreak LN.lean:548}\\
\textsc{LC-VE-Scale} & $\mathsf{lc}(\mathsf{scale}(\lenv{0}, \lenv{1}))$ & \texttt{\footnotesize GLang/Value\allowbreak LN.lean:549}\\
\textsc{LC-VE-Vadd} & $\mathsf{lc}(\mathsf{vadd}(\lsv{}, \lenv{0}, \lenv{1}))$ & \texttt{\footnotesize GLang/Value\allowbreak LN.lean:550}\\
\textsc{LC-VE-Dsum\allowbreak E} & $\mathsf{lc}(\mathsf{dsumE}(\lenv{0}, \lenv{1}))$ & \texttt{\footnotesize GLang/Value\allowbreak LN.lean:552}\\
\textsc{LC-VE-Tprod\allowbreak E} & $\mathsf{lc}(\mathsf{tprodE}(\lenv{0}, \lenv{1}))$ & \texttt{\footnotesize GLang/Value\allowbreak LN.lean:553}\\
\textsc{LC-VE-Delta} & $\mathsf{lc}(\mathsf{delta}(\mathit{n}, \lsv{}))$ & \texttt{\footnotesize GLang/Value\allowbreak LN.lean:554}\\
\textsc{LC-VE-Idelta} & $\mathsf{lc}(\mathsf{idelta}(\lsv{}))$ & \texttt{\footnotesize GLang/Value\allowbreak LN.lean:555}\\
\textsc{LC-VE-Vsum} & $\mathsf{lc}(\mathsf{vsum}(\lsv{}, \mathit{es}))$ & \texttt{\footnotesize GLang/Value\allowbreak LN.lean:556}\\
\textsc{LC-VE-Ttv} & $\mathsf{lc}(\mathsf{ttv}(\lsv{}, \lenv{0}, \lenv{1}))$ & \texttt{\footnotesize GLang/Value\allowbreak LN.lean:557}\\
\textsc{LC-VE-Mtm} & $\mathsf{lc}(\mathsf{mtm}({\mathit{n}_{0}}, {\mathit{n}_{1}}, \lsnv{0}, \lsnv{1}, \lenv{0}, \lenv{1}))$ & \texttt{\footnotesize GLang/Value\allowbreak LN.lean:558}\\
\textsc{LC-VE-App\allowbreak V} & $\mathsf{lc}(\mathsf{appV}(\lenv{0}, \lenv{1}))$ & \texttt{\footnotesize GLang/Value\allowbreak LN.lean:560}\\
\textsc{LC-VE-Vkrn} & $\mathsf{lc}(\mathsf{vkrn}(\lenv{0}, \lenv{1}))$ & \texttt{\footnotesize GLang/Value\allowbreak LN.lean:561}\\
\textsc{LC-VE-Vblk} & $\mathsf{lc}(\mathsf{vblk}(\lenv{0}, \lenv{1}))$ & \texttt{\footnotesize GLang/Value\allowbreak LN.lean:562}\\
\textsc{LC-VE-Reshap} & $\mathsf{lc}(\mathsf{reshap}(\leev{}, \ltyv{}, \ltynv{0}))$ & \texttt{\footnotesize GLang/Value\allowbreak LN.lean:563}\\
\textsc{LC-VE-Vtrans} & $\mathsf{lc}(\mathsf{vtrans}(\leev{}))$ & \texttt{\footnotesize GLang/Value\allowbreak LN.lean:565}\\
\textsc{LC-VE-Gsvd\allowbreak S} & $\mathsf{lc}(\mathsf{gsvdS}(\leev{}, \lenv{{\mathrm{a}}}, \lenv{{\mathrm{b}}}))$ & \texttt{\footnotesize GLang/Value\allowbreak LN.lean:566}\\
\textsc{LC-VE-Gsvd\allowbreak U} & $\mathsf{lc}(\mathsf{gsvdU}(\leev{}, \lenv{{\mathrm{a}}}, \lenv{{\mathrm{b}}}))$ & \texttt{\footnotesize GLang/Value\allowbreak LN.lean:567}\\
\textsc{LC-VE-Gsvd\allowbreak UI} & $\mathsf{lc}(\mathsf{gsvdUI}(\leev{}, \lenv{{\mathrm{a}}}, \lenv{{\mathrm{b}}}))$ & \texttt{\footnotesize GLang/Value\allowbreak LN.lean:568}\\
\textsc{LC-VE-Gsvd\allowbreak V} & $\mathsf{lc}(\mathsf{gsvdV}(\leev{}, \lenv{{\mathrm{a}}}, \lenv{{\mathrm{b}}}))$ & \texttt{\footnotesize GLang/Value\allowbreak LN.lean:569}\\
\textsc{LC-VE-Gsvd\allowbreak VI} & $\mathsf{lc}(\mathsf{gsvdVI}(\leev{}, \lenv{{\mathrm{a}}}, \lenv{{\mathrm{b}}}))$ & \texttt{\footnotesize GLang/Value\allowbreak LN.lean:570}\\
\textsc{LC-VE-Pinv} & $\mathsf{lc}(\mathsf{pinv}(\leev{}))$ & \texttt{\footnotesize GLang/Value\allowbreak LN.lean:571}\\
\textsc{LC-VE-Gpinv} & $\mathsf{lc}(\mathsf{gpinv}(\leev{}, \lenv{{\mathrm{a}}}, \lenv{{\mathrm{b}}}))$ & \texttt{\footnotesize GLang/Value\allowbreak LN.lean:572}\\
\textsc{LC-VE-Rref\allowbreak Rep} & $\mathsf{lc}(\mathsf{rrefRep}(\lenv{0}, \lenv{1}))$ & \texttt{\footnotesize GLang/Value\allowbreak LN.lean:573}\\
\textsc{LC-VE-Inv} & $\mathsf{lc}(\mathsf{inv}(\leev{}))$ & \texttt{\footnotesize GLang/Value\allowbreak LN.lean:574}\\
\textsc{LC-VE-Svds} & $\mathsf{lc}(\mathsf{svds}(\leev{}))$ & \texttt{\footnotesize GLang/Value\allowbreak LN.lean:575}\\
}{
\gpfit{\inferrule*[]{\mathsf{lc}(\ltyv{})}{\mathsf{lc}(\mathsf{var}(\mathit{x}, \ltyv{}))}}
\and
\gpfit{\inferrule*[]{\mathsf{lc}(\ltynv{0}) \\ \forall \mathit{x}.\; \neg(\mathit{x} \in \mathit{L}) \Rightarrow \mathsf{lc}(\mathsf{openVar}(\mathit{x}, \ltynv{0}, \ltynv{1})) \\ \forall \mathit{x}.\; \neg(\mathit{x} \in \mathit{L}) \Rightarrow \mathsf{lc}(\mathsf{openVar}(\mathit{x}, \ltynv{0}, \leev{}))}{\mathsf{lc}(\mathsf{vlam}(\ltynv{0}, \ltynv{1}, \leev{}))}}
\and
\gpfit{\inferrule*[]{\mathsf{lc}(\lsv{})}{\mathsf{lc}(\mathsf{zero}(\lsv{}))}}
\and
\gpfit{\inferrule*[]{\phantom{.}}{\mathsf{lc}(\mathsf{rlit}(\mathit{r}))}}
\and
\gpfit{\inferrule*[]{\mathsf{lc}(\lenv{0}) \\ \mathsf{lc}(\lenv{1})}{\mathsf{lc}(\mathsf{scale}(\lenv{0}, \lenv{1}))}}
\and
\gpfit{\inferrule*[]{\mathsf{lc}(\lsv{}) \\ \mathsf{lc}(\lenv{0}) \\ \mathsf{lc}(\lenv{1})}{\mathsf{lc}(\mathsf{vadd}(\lsv{}, \lenv{0}, \lenv{1}))}}
\and
\gpfit{\inferrule*[]{\mathsf{lc}(\lenv{0}) \\ \mathsf{lc}(\lenv{1})}{\mathsf{lc}(\mathsf{dsumE}(\lenv{0}, \lenv{1}))}}
\and
\gpfit{\inferrule*[]{\mathsf{lc}(\lenv{0}) \\ \mathsf{lc}(\lenv{1})}{\mathsf{lc}(\mathsf{tprodE}(\lenv{0}, \lenv{1}))}}
\and
\gpfit{\inferrule*[]{\mathsf{lc}(\lsv{})}{\mathsf{lc}(\mathsf{delta}(\mathit{n}, \lsv{}))}}
\and
\gpfit{\inferrule*[]{\mathsf{lc}(\lsv{})}{\mathsf{lc}(\mathsf{idelta}(\lsv{}))}}
\and
\gpfit{\inferrule*[]{\mathsf{lc}(\lsv{}) \\ \forall \leev{}.\; \leev{} \in \mathit{es} \Rightarrow \mathsf{lc}(\leev{})}{\mathsf{lc}(\mathsf{vsum}(\lsv{}, \mathit{es}))}}
\and
\gpfit{\inferrule*[]{\mathsf{lc}(\lsv{}) \\ \mathsf{lc}(\lenv{0}) \\ \mathsf{lc}(\lenv{1})}{\mathsf{lc}(\mathsf{ttv}(\lsv{}, \lenv{0}, \lenv{1}))}}
\and
\gpfit{\inferrule*[]{\mathsf{lc}(\lsnv{0}) \\ \mathsf{lc}(\lsnv{1}) \\ \mathsf{lc}(\lenv{0}) \\ \mathsf{lc}(\lenv{1})}{\mathsf{lc}(\mathsf{mtm}({\mathit{n}_{0}}, {\mathit{n}_{1}}, \lsnv{0}, \lsnv{1}, \lenv{0}, \lenv{1}))}}
\and
\gpfit{\inferrule*[]{\mathsf{lc}(\lenv{0}) \\ \mathsf{lc}(\lenv{1})}{\mathsf{lc}(\mathsf{appV}(\lenv{0}, \lenv{1}))}}
\and
\gpfit{\inferrule*[]{\mathsf{lc}(\lenv{0}) \\ \mathsf{lc}(\lenv{1})}{\mathsf{lc}(\mathsf{vkrn}(\lenv{0}, \lenv{1}))}}
\and
\gpfit{\inferrule*[]{\mathsf{lc}(\lenv{0}) \\ \mathsf{lc}(\lenv{1})}{\mathsf{lc}(\mathsf{vblk}(\lenv{0}, \lenv{1}))}}
\and
\gpfit{\inferrule*[]{\mathsf{lc}(\leev{}) \\ \mathsf{lc}(\ltyv{}) \\ \mathsf{lc}(\ltynv{0})}{\mathsf{lc}(\mathsf{reshap}(\leev{}, \ltyv{}, \ltynv{0}))}}
\and
\gpfit{\inferrule*[]{\mathsf{lc}(\leev{})}{\mathsf{lc}(\mathsf{vtrans}(\leev{}))}}
\and
\gpfit{\inferrule*[]{\mathsf{lc}(\leev{}) \\ \mathsf{lc}(\lenv{{\mathrm{a}}}) \\ \mathsf{lc}(\lenv{{\mathrm{b}}})}{\mathsf{lc}(\mathsf{gsvdS}(\leev{}, \lenv{{\mathrm{a}}}, \lenv{{\mathrm{b}}}))}}
\and
\gpfit{\inferrule*[]{\mathsf{lc}(\leev{}) \\ \mathsf{lc}(\lenv{{\mathrm{a}}}) \\ \mathsf{lc}(\lenv{{\mathrm{b}}})}{\mathsf{lc}(\mathsf{gsvdU}(\leev{}, \lenv{{\mathrm{a}}}, \lenv{{\mathrm{b}}}))}}
\and
\gpfit{\inferrule*[]{\mathsf{lc}(\leev{}) \\ \mathsf{lc}(\lenv{{\mathrm{a}}}) \\ \mathsf{lc}(\lenv{{\mathrm{b}}})}{\mathsf{lc}(\mathsf{gsvdUI}(\leev{}, \lenv{{\mathrm{a}}}, \lenv{{\mathrm{b}}}))}}
\and
\gpfit{\inferrule*[]{\mathsf{lc}(\leev{}) \\ \mathsf{lc}(\lenv{{\mathrm{a}}}) \\ \mathsf{lc}(\lenv{{\mathrm{b}}})}{\mathsf{lc}(\mathsf{gsvdV}(\leev{}, \lenv{{\mathrm{a}}}, \lenv{{\mathrm{b}}}))}}
\and
\gpfit{\inferrule*[]{\mathsf{lc}(\leev{}) \\ \mathsf{lc}(\lenv{{\mathrm{a}}}) \\ \mathsf{lc}(\lenv{{\mathrm{b}}})}{\mathsf{lc}(\mathsf{gsvdVI}(\leev{}, \lenv{{\mathrm{a}}}, \lenv{{\mathrm{b}}}))}}
\and
\gpfit{\inferrule*[]{\mathsf{lc}(\leev{})}{\mathsf{lc}(\mathsf{pinv}(\leev{}))}}
\and
\gpfit{\inferrule*[]{\mathsf{lc}(\leev{}) \\ \mathsf{lc}(\lenv{{\mathrm{a}}}) \\ \mathsf{lc}(\lenv{{\mathrm{b}}})}{\mathsf{lc}(\mathsf{gpinv}(\leev{}, \lenv{{\mathrm{a}}}, \lenv{{\mathrm{b}}}))}}
\and
\gpfit{\inferrule*[]{\mathsf{lc}(\lenv{0}) \\ \mathsf{lc}(\lenv{1})}{\mathsf{lc}(\mathsf{rrefRep}(\lenv{0}, \lenv{1}))}}
\and
\gpfit{\inferrule*[]{\mathsf{lc}(\leev{})}{\mathsf{lc}(\mathsf{inv}(\leev{}))}}
\and
\gpfit{\inferrule*[]{\mathsf{lc}(\leev{})}{\mathsf{lc}(\mathsf{svds}(\leev{}))}}
}{
\gpfit{\inferrule*[lab=\textsc{LC-VE-Var}]{\mathsf{lc}(\ltyv{})}{\mathsf{lc}(\mathsf{var}(\mathit{x}, \ltyv{}))}}
\and
\gpfit{\inferrule*[lab=\textsc{LC-VE-Vlam}]{\mathsf{lc}(\ltynv{0}) \\ \forall \mathit{x}.\; \neg(\mathit{x} \in \mathit{L}) \Rightarrow \mathsf{lc}(\mathsf{openVar}(\mathit{x}, \ltynv{0}, \ltynv{1})) \\ \forall \mathit{x}.\; \neg(\mathit{x} \in \mathit{L}) \Rightarrow \mathsf{lc}(\mathsf{openVar}(\mathit{x}, \ltynv{0}, \leev{}))}{\mathsf{lc}(\mathsf{vlam}(\ltynv{0}, \ltynv{1}, \leev{}))}}
\and
\gpfit{\inferrule*[lab=\textsc{LC-VE-Zero}]{\mathsf{lc}(\lsv{})}{\mathsf{lc}(\mathsf{zero}(\lsv{}))}}
\and
\gpfit{\inferrule*[lab=\textsc{LC-VE-Rlit}]{\phantom{.}}{\mathsf{lc}(\mathsf{rlit}(\mathit{r}))}}
\and
\gpfit{\inferrule*[lab=\textsc{LC-VE-Scale}]{\mathsf{lc}(\lenv{0}) \\ \mathsf{lc}(\lenv{1})}{\mathsf{lc}(\mathsf{scale}(\lenv{0}, \lenv{1}))}}
\and
\gpfit{\inferrule*[lab=\textsc{LC-VE-Vadd}]{\mathsf{lc}(\lsv{}) \\ \mathsf{lc}(\lenv{0}) \\ \mathsf{lc}(\lenv{1})}{\mathsf{lc}(\mathsf{vadd}(\lsv{}, \lenv{0}, \lenv{1}))}}
\and
\gpfit{\inferrule*[lab=\textsc{LC-VE-Dsum\allowbreak E}]{\mathsf{lc}(\lenv{0}) \\ \mathsf{lc}(\lenv{1})}{\mathsf{lc}(\mathsf{dsumE}(\lenv{0}, \lenv{1}))}}
\and
\gpfit{\inferrule*[lab=\textsc{LC-VE-Tprod\allowbreak E}]{\mathsf{lc}(\lenv{0}) \\ \mathsf{lc}(\lenv{1})}{\mathsf{lc}(\mathsf{tprodE}(\lenv{0}, \lenv{1}))}}
\and
\gpfit{\inferrule*[lab=\textsc{LC-VE-Delta}]{\mathsf{lc}(\lsv{})}{\mathsf{lc}(\mathsf{delta}(\mathit{n}, \lsv{}))}}
\and
\gpfit{\inferrule*[lab=\textsc{LC-VE-Idelta}]{\mathsf{lc}(\lsv{})}{\mathsf{lc}(\mathsf{idelta}(\lsv{}))}}
\and
\gpfit{\inferrule*[lab=\textsc{LC-VE-Vsum}]{\mathsf{lc}(\lsv{}) \\ \forall \leev{}.\; \leev{} \in \mathit{es} \Rightarrow \mathsf{lc}(\leev{})}{\mathsf{lc}(\mathsf{vsum}(\lsv{}, \mathit{es}))}}
\and
\gpfit{\inferrule*[lab=\textsc{LC-VE-Ttv}]{\mathsf{lc}(\lsv{}) \\ \mathsf{lc}(\lenv{0}) \\ \mathsf{lc}(\lenv{1})}{\mathsf{lc}(\mathsf{ttv}(\lsv{}, \lenv{0}, \lenv{1}))}}
\and
\gpfit{\inferrule*[lab=\textsc{LC-VE-Mtm}]{\mathsf{lc}(\lsnv{0}) \\ \mathsf{lc}(\lsnv{1}) \\ \mathsf{lc}(\lenv{0}) \\ \mathsf{lc}(\lenv{1})}{\mathsf{lc}(\mathsf{mtm}({\mathit{n}_{0}}, {\mathit{n}_{1}}, \lsnv{0}, \lsnv{1}, \lenv{0}, \lenv{1}))}}
\and
\gpfit{\inferrule*[lab=\textsc{LC-VE-App\allowbreak V}]{\mathsf{lc}(\lenv{0}) \\ \mathsf{lc}(\lenv{1})}{\mathsf{lc}(\mathsf{appV}(\lenv{0}, \lenv{1}))}}
\and
\gpfit{\inferrule*[lab=\textsc{LC-VE-Vkrn}]{\mathsf{lc}(\lenv{0}) \\ \mathsf{lc}(\lenv{1})}{\mathsf{lc}(\mathsf{vkrn}(\lenv{0}, \lenv{1}))}}
\and
\gpfit{\inferrule*[lab=\textsc{LC-VE-Vblk}]{\mathsf{lc}(\lenv{0}) \\ \mathsf{lc}(\lenv{1})}{\mathsf{lc}(\mathsf{vblk}(\lenv{0}, \lenv{1}))}}
\and
\gpfit{\inferrule*[lab=\textsc{LC-VE-Reshap}]{\mathsf{lc}(\leev{}) \\ \mathsf{lc}(\ltyv{}) \\ \mathsf{lc}(\ltynv{0})}{\mathsf{lc}(\mathsf{reshap}(\leev{}, \ltyv{}, \ltynv{0}))}}
\and
\gpfit{\inferrule*[lab=\textsc{LC-VE-Vtrans}]{\mathsf{lc}(\leev{})}{\mathsf{lc}(\mathsf{vtrans}(\leev{}))}}
\and
\gpfit{\inferrule*[lab=\textsc{LC-VE-Gsvd\allowbreak S}]{\mathsf{lc}(\leev{}) \\ \mathsf{lc}(\lenv{{\mathrm{a}}}) \\ \mathsf{lc}(\lenv{{\mathrm{b}}})}{\mathsf{lc}(\mathsf{gsvdS}(\leev{}, \lenv{{\mathrm{a}}}, \lenv{{\mathrm{b}}}))}}
\and
\gpfit{\inferrule*[lab=\textsc{LC-VE-Gsvd\allowbreak U}]{\mathsf{lc}(\leev{}) \\ \mathsf{lc}(\lenv{{\mathrm{a}}}) \\ \mathsf{lc}(\lenv{{\mathrm{b}}})}{\mathsf{lc}(\mathsf{gsvdU}(\leev{}, \lenv{{\mathrm{a}}}, \lenv{{\mathrm{b}}}))}}
\and
\gpfit{\inferrule*[lab=\textsc{LC-VE-Gsvd\allowbreak UI}]{\mathsf{lc}(\leev{}) \\ \mathsf{lc}(\lenv{{\mathrm{a}}}) \\ \mathsf{lc}(\lenv{{\mathrm{b}}})}{\mathsf{lc}(\mathsf{gsvdUI}(\leev{}, \lenv{{\mathrm{a}}}, \lenv{{\mathrm{b}}}))}}
\and
\gpfit{\inferrule*[lab=\textsc{LC-VE-Gsvd\allowbreak V}]{\mathsf{lc}(\leev{}) \\ \mathsf{lc}(\lenv{{\mathrm{a}}}) \\ \mathsf{lc}(\lenv{{\mathrm{b}}})}{\mathsf{lc}(\mathsf{gsvdV}(\leev{}, \lenv{{\mathrm{a}}}, \lenv{{\mathrm{b}}}))}}
\and
\gpfit{\inferrule*[lab=\textsc{LC-VE-Gsvd\allowbreak VI}]{\mathsf{lc}(\leev{}) \\ \mathsf{lc}(\lenv{{\mathrm{a}}}) \\ \mathsf{lc}(\lenv{{\mathrm{b}}})}{\mathsf{lc}(\mathsf{gsvdVI}(\leev{}, \lenv{{\mathrm{a}}}, \lenv{{\mathrm{b}}}))}}
\and
\gpfit{\inferrule*[lab=\textsc{LC-VE-Pinv}]{\mathsf{lc}(\leev{})}{\mathsf{lc}(\mathsf{pinv}(\leev{}))}}
\and
\gpfit{\inferrule*[lab=\textsc{LC-VE-Gpinv}]{\mathsf{lc}(\leev{}) \\ \mathsf{lc}(\lenv{{\mathrm{a}}}) \\ \mathsf{lc}(\lenv{{\mathrm{b}}})}{\mathsf{lc}(\mathsf{gpinv}(\leev{}, \lenv{{\mathrm{a}}}, \lenv{{\mathrm{b}}}))}}
\and
\gpfit{\inferrule*[lab=\textsc{LC-VE-Rref\allowbreak Rep}]{\mathsf{lc}(\lenv{0}) \\ \mathsf{lc}(\lenv{1})}{\mathsf{lc}(\mathsf{rrefRep}(\lenv{0}, \lenv{1}))}}
\and
\gpfit{\inferrule*[lab=\textsc{LC-VE-Inv}]{\mathsf{lc}(\leev{})}{\mathsf{lc}(\mathsf{inv}(\leev{}))}}
\and
\gpfit{\inferrule*[lab=\textsc{LC-VE-Svds}]{\mathsf{lc}(\leev{})}{\mathsf{lc}(\mathsf{svds}(\leev{}))}}
}
\gldefjudgment{valSValid}{\valCheckCtx\vdash\lsv{}\colon\svalid}{Value-space validity (generated from \texttt{GLang.RuntimeValue.ValSValid}).}{fig:gl-valsvalid}
\gldefrule{valSValid.real}{valSValid}{\textsc{VS-Real}}{\phantom{.}}{\valCheckCtx{}\vdash \lR{}\colon\svalid}{\texttt{GLang.Runtime\allowbreak Value.Val\allowbreak SValid.real}}{GLang/RuntimeValueTyping.lean:12}
\gldefruleexplanation{valSValid.real}{The scalar value space $\mathbb R$ is well formed.}
\gldefrule{valSValid.empty}{valSValid}{\textsc{VS-Empty}}{\phantom{.}}{\valCheckCtx{}\vdash \lsempty{}\colon\svalid}{\texttt{GLang.Runtime\allowbreak Value.Val\allowbreak SValid.empty}}{GLang/RuntimeValueTyping.lean:13}
\gldefruleexplanation{valSValid.empty}{The zero-dimensional value space is well formed.}
\gldefrule{valSValid.dsum}{valSValid}{\textsc{VS-Dsum}}{\forall i.\; \valCheckCtx{}\vdash \lsnv{i}\colon\svalid}{\valCheckCtx{}\vdash \lsnv{0} \losum \lsnv{1}\colon\svalid}{\texttt{GLang.Runtime\allowbreak Value.Val\allowbreak SValid.dsum}}{GLang/RuntimeValueTyping.lean:14}
\gldefruleexplanation{valSValid.dsum}{Two well-formed value spaces determine a well-formed direct sum.}
\gldefrule{valSValid.tprod}{valSValid}{\textsc{VS-Tprod}}{\forall i.\; \valCheckCtx{}\vdash \lsnv{i}\colon\svalid}{\valCheckCtx{}\vdash \lsnv{0} \loprod \lsnv{1}\colon\svalid}{\texttt{GLang.Runtime\allowbreak Value.Val\allowbreak SValid.tprod}}{GLang/RuntimeValueTyping.lean:16}
\gldefruleexplanation{valSValid.tprod}{Two well-formed value spaces determine a well-formed tensor product.}
\gldefrule{valSValid.ffs}{valSValid}{\textsc{VS-Ffs}}{\forall i.\; \valCheckCtx{}\vdash \lsnv{i}\colon\svalid \\ \valCheckCtx{}\vdash \leev{}\colon \lsnv{0} \loprod \lsnv{1}}{\valCheckCtx{}\vdash \lR^{\leev{},\lprop{},\lsnv{0} \loprod \lsnv{1}}\colon\svalid}{\texttt{GLang.Runtime\allowbreak Value.Val\allowbreak SValid.ffs}}{GLang/RuntimeValueTyping.lean:18}
\gldefruleexplanation{valSValid.ffs}{A tensor-typed value with well-formed factor spaces determines a well-formed space for any selected fundamental subspace.}
\gldefrulerows{valSValid}{
\textsc{VS-Real} & $\valCheckCtx{}\vdash \lR{}\colon\svalid$ & \texttt{\footnotesize GLang/Runtime\allowbreak Value\allowbreak Typing.lean:12}\\
\textsc{VS-Empty} & $\valCheckCtx{}\vdash \lsempty{}\colon\svalid$ & \texttt{\footnotesize GLang/Runtime\allowbreak Value\allowbreak Typing.lean:13}\\
\textsc{VS-Dsum} & $\valCheckCtx{}\vdash \lsnv{0} \losum \lsnv{1}\colon\svalid$ & \texttt{\footnotesize GLang/Runtime\allowbreak Value\allowbreak Typing.lean:14}\\
\textsc{VS-Tprod} & $\valCheckCtx{}\vdash \lsnv{0} \loprod \lsnv{1}\colon\svalid$ & \texttt{\footnotesize GLang/Runtime\allowbreak Value\allowbreak Typing.lean:16}\\
\textsc{VS-Ffs} & $\valCheckCtx{}\vdash \lR^{\leev{},\lprop{},\lsnv{0} \loprod \lsnv{1}}\colon\svalid$ & \texttt{\footnotesize GLang/Runtime\allowbreak Value\allowbreak Typing.lean:18}\\
}{
\gpfit{\inferrule*[]{\phantom{.}}{\valCheckCtx{}\vdash \lR{}\colon\svalid}}
\and
\gpfit{\inferrule*[]{\phantom{.}}{\valCheckCtx{}\vdash \lsempty{}\colon\svalid}}
\and
\gpfit{\inferrule*[]{\forall i.\; \valCheckCtx{}\vdash \lsnv{i}\colon\svalid}{\valCheckCtx{}\vdash \lsnv{0} \losum \lsnv{1}\colon\svalid}}
\and
\gpfit{\inferrule*[]{\forall i.\; \valCheckCtx{}\vdash \lsnv{i}\colon\svalid}{\valCheckCtx{}\vdash \lsnv{0} \loprod \lsnv{1}\colon\svalid}}
\and
\gpfit{\inferrule*[]{\forall i.\; \valCheckCtx{}\vdash \lsnv{i}\colon\svalid \\ \valCheckCtx{}\vdash \leev{}\colon \lsnv{0} \loprod \lsnv{1}}{\valCheckCtx{}\vdash \lR^{\leev{},\lprop{},\lsnv{0} \loprod \lsnv{1}}\colon\svalid}}
}{
\gpfit{\inferrule*[lab=\textsc{VS-Real}]{\phantom{.}}{\valCheckCtx{}\vdash \lR{}\colon\svalid}}
\and
\gpfit{\inferrule*[lab=\textsc{VS-Empty}]{\phantom{.}}{\valCheckCtx{}\vdash \lsempty{}\colon\svalid}}
\and
\gpfit{\inferrule*[lab=\textsc{VS-Dsum}]{\forall i.\; \valCheckCtx{}\vdash \lsnv{i}\colon\svalid}{\valCheckCtx{}\vdash \lsnv{0} \losum \lsnv{1}\colon\svalid}}
\and
\gpfit{\inferrule*[lab=\textsc{VS-Tprod}]{\forall i.\; \valCheckCtx{}\vdash \lsnv{i}\colon\svalid}{\valCheckCtx{}\vdash \lsnv{0} \loprod \lsnv{1}\colon\svalid}}
\and
\gpfit{\inferrule*[lab=\textsc{VS-Ffs}]{\forall i.\; \valCheckCtx{}\vdash \lsnv{i}\colon\svalid \\ \valCheckCtx{}\vdash \leev{}\colon \lsnv{0} \loprod \lsnv{1}}{\valCheckCtx{}\vdash \lR^{\leev{},\lprop{},\lsnv{0} \loprod \lsnv{1}}\colon\svalid}}
}
\gldefjudgment{valTValid}{\valCheckCtx\vdash\ltyv{}\colon\lkn{}}{Value-type validity (generated from \texttt{GLang.RuntimeValue.ValTValid}).}{fig:gl-valtvalid}
\gldefrule{valTValid.space}{valTValid}{\textsc{VK-Space}}{\valCheckCtx{}\vdash \lsv{}\colon\svalid}{\valCheckCtx{}\vdash \lsv{}\colon \lkbase{}}{\texttt{GLang.Runtime\allowbreak Value.Val\allowbreak TValid.space}}{GLang/RuntimeValueTyping.lean:25}
\gldefruleexplanation{valTValid.space}{A well-formed value space is a value type of base kind.}
\gldefrule{valTValid.piTy}{valTValid}{\textsc{VK-Pi\allowbreak Ty}}{\valCheckCtx{}\vdash \ltynv{0}\colon \lkn{0} \\ (\valCheckCtx{}, \mathit{x}\colon \ltynv{0})\vdash \ltynv{1}\colon \lkn{1}}{\valCheckCtx{}\vdash (x\colon \ltynv{0}\fnto \ltynv{1})\colon \lkn{0} \fnto \lkn{1}}{\texttt{GLang.Runtime\allowbreak Value.Val\allowbreak TValid.pi\allowbreak Ty}}{GLang/RuntimeValueTyping.lean:26}
\gldefruleexplanation{valTValid.piTy}{A well-kinded value domain and a codomain well kinded under a fresh argument form a dependent value function type.}
\gldefrulerows{valTValid}{
\textsc{VK-Space} & $\valCheckCtx{}\vdash \lsv{}\colon \lkbase{}$ & \texttt{\footnotesize GLang/Runtime\allowbreak Value\allowbreak Typing.lean:25}\\
\textsc{VK-Pi\allowbreak Ty} & $\valCheckCtx{}\vdash (x\colon \ltynv{0}\fnto \ltynv{1})\colon \lkn{0} \fnto \lkn{1}$ & \texttt{\footnotesize GLang/Runtime\allowbreak Value\allowbreak Typing.lean:26}\\
}{
\gpfit{\inferrule*[]{\valCheckCtx{}\vdash \lsv{}\colon\svalid}{\valCheckCtx{}\vdash \lsv{}\colon \lkbase{}}}
\and
\gpfit{\inferrule*[]{\valCheckCtx{}\vdash \ltynv{0}\colon \lkn{0} \\ (\valCheckCtx{}, \mathit{x}\colon \ltynv{0})\vdash \ltynv{1}\colon \lkn{1}}{\valCheckCtx{}\vdash (x\colon \ltynv{0}\fnto \ltynv{1})\colon \lkn{0} \fnto \lkn{1}}}
}{
\gpfit{\inferrule*[lab=\textsc{VK-Space}]{\valCheckCtx{}\vdash \lsv{}\colon\svalid}{\valCheckCtx{}\vdash \lsv{}\colon \lkbase{}}}
\and
\gpfit{\inferrule*[lab=\textsc{VK-Pi\allowbreak Ty}]{\valCheckCtx{}\vdash \ltynv{0}\colon \lkn{0} \\ (\valCheckCtx{}, \mathit{x}\colon \ltynv{0})\vdash \ltynv{1}\colon \lkn{1}}{\valCheckCtx{}\vdash (x\colon \ltynv{0}\fnto \ltynv{1})\colon \lkn{0} \fnto \lkn{1}}}
}
\gldefjudgment{valHasType}{\valCheckCtx\vdash\leev{}\colon\ltyv{}}{Value typing (generated from \texttt{GLang.RuntimeValue.ValHasType}).}{fig:gl-valhastype}
\gldefrule{valHasType.var}{valHasType}{\textsc{VT-Var}}{(\mathit{x}\colon \ltyv{}) \in \valCheckCtx{}}{\valCheckCtx{}\vdash \mathit{x}_{\ltyv{}}\colon \ltyv{}}{\texttt{GLang.Runtime\allowbreak Value.Val\allowbreak Has\allowbreak Type.var}}{GLang/RuntimeValueTyping.lean:33}
\gldefruleexplanation{valHasType.var}{An annotated value variable has the type assigned to it by $\valCheckCtx$.}
\gldefrule{valHasType.zero}{valHasType}{\textsc{VT-Zero}}{\valCheckCtx{}\vdash \lsv{}\colon\svalid}{\valCheckCtx{}\vdash \lvzero{}_{\lsv{}}\colon \lsv{}}{\texttt{GLang.Runtime\allowbreak Value.Val\allowbreak Has\allowbreak Type.zero}}{GLang/RuntimeValueTyping.lean:34}
\gldefruleexplanation{valHasType.zero}{The zero value has its annotated well-formed value space.}
\gldefrule{valHasType.rlit}{valHasType}{\textsc{VT-Rlit}}{\phantom{.}}{\valCheckCtx{}\vdash \mathit{r}\colon \lR{}}{\texttt{GLang.Runtime\allowbreak Value.Val\allowbreak Has\allowbreak Type.rlit}}{GLang/RuntimeValueTyping.lean:35}
\gldefruleexplanation{valHasType.rlit}{A real literal has scalar value type.}
\gldefrule{valHasType.scale}{valHasType}{\textsc{VT-Scale}}{\valCheckCtx{}\vdash \lenv{0}\colon \lR{} \\ \valCheckCtx{}\vdash \lenv{1}\colon \ltyv{}}{\valCheckCtx{}\vdash \lenv{0} \lvscale \lenv{1}\colon \ltyv{}}{\texttt{GLang.Runtime\allowbreak Value.Val\allowbreak Has\allowbreak Type.scale}}{GLang/RuntimeValueTyping.lean:36}
\gldefruleexplanation{valHasType.scale}{Scaling a typed value by a scalar value preserves its type.}
\gldefrule{valHasType.vadd}{valHasType}{\textsc{VT-Vadd}}{\forall i.\; \valCheckCtx{}\vdash \lenv{i}\colon \lsnv{i} \\ \valCheckCtx{}\vdash \lsv{}\colon\svalid}{\valCheckCtx{}\vdash \lenv{0} \lvadd_{\lsv{}} \lenv{1}\colon \lsv{}}{\texttt{GLang.Runtime\allowbreak Value.Val\allowbreak Has\allowbreak Type.vadd}}{GLang/RuntimeValueTyping.lean:40}
\gldefruleexplanation{valHasType.vadd}{Adding two space-typed values has the declared well-formed result space, with operand lengths checked during evaluation.}
\gldefrule{valHasType.dsumE}{valHasType}{\textsc{VT-Dsum\allowbreak E}}{\forall i.\; \valCheckCtx{}\vdash \lenv{i}\colon \lsnv{i}}{\valCheckCtx{}\vdash \lenv{0}\losum \lenv{1}\colon \lsnv{0} \losum \lsnv{1}}{\texttt{GLang.Runtime\allowbreak Value.Val\allowbreak Has\allowbreak Type.dsum\allowbreak E}}{GLang/RuntimeValueTyping.lean:45}
\gldefruleexplanation{valHasType.dsumE}{A direct-sum value has the direct sum of its operand spaces.}
\gldefrule{valHasType.tprodE}{valHasType}{\textsc{VT-Tprod\allowbreak E}}{\forall i.\; \valCheckCtx{}\vdash \lenv{i}\colon \lsnv{i}}{\valCheckCtx{}\vdash \lenv{0}\loprod \lenv{1}\colon \lsnv{0} \loprod \lsnv{1}}{\texttt{GLang.Runtime\allowbreak Value.Val\allowbreak Has\allowbreak Type.tprod\allowbreak E}}{GLang/RuntimeValueTyping.lean:49}
\gldefruleexplanation{valHasType.tprodE}{A tensor-product value has the tensor product of its operand spaces.}
\gldefrule{valHasType.delta}{valHasType}{\textsc{VT-Delta}}{\valCheckCtx{}\vdash \lsv{}\colon\svalid}{\valCheckCtx{}\vdash \ldelta{\mathit{n}}{\lsv{}}\colon \lsv{}}{\texttt{GLang.Runtime\allowbreak Value.Val\allowbreak Has\allowbreak Type.delta}}{GLang/RuntimeValueTyping.lean:53}
\gldefruleexplanation{valHasType.delta}{A coordinate selector has its annotated well-formed space for any index.}
\gldefrule{valHasType.idelta}{valHasType}{\textsc{VT-Idelta}}{\valCheckCtx{}\vdash \lsv{}\colon\svalid}{\valCheckCtx{}\vdash \lidelta{\lsv{}}\colon \lsv{} \loprod \lsv{}}{\texttt{GLang.Runtime\allowbreak Value.Val\allowbreak Has\allowbreak Type.idelta}}{GLang/RuntimeValueTyping.lean:54}
\gldefruleexplanation{valHasType.idelta}{The identity array on a well-formed space has the tensor square of that space as its type.}
\gldefrule{valHasType.ttvSpace}{valHasType}{\textsc{VT-Ttv\allowbreak Space}}{\valCheckCtx{}\vdash \lsnv{0}''\colon\svalid \\ \valCheckCtx{}\vdash \lenv{0}\colon \lsnv{0}' \loprod \lsnv{1} \\ \valCheckCtx{}\vdash \lenv{1}\colon {\ltyv{}_{\mathsf{arg}}}}{\valCheckCtx{}\vdash \LTTVPP{\lsnv{0}''}{\lenv{0}}{\lenv{1}}\colon \lsnv{1}}{\texttt{GLang.Runtime\allowbreak Value.Val\allowbreak Has\allowbreak Type.ttv\_\allowbreak space}}{GLang/RuntimeValueTyping.lean:57}
\gldefruleexplanation{valHasType.ttvSpace}{Contraction of a tensor-typed value with a typed argument has the tensor's second factor as its result space, with dimensions checked during evaluation.}
\gldefrule{valHasType.ttvReal}{valHasType}{\textsc{VT-Ttv\allowbreak Real}}{\valCheckCtx{}\vdash \lsnv{0}''\colon\svalid \\ \valCheckCtx{}\vdash \lenv{0}\colon \lsnv{0}' \\ \valCheckCtx{}\vdash \lenv{1}\colon {\ltyv{}_{\mathsf{arg}}} \\ \forall \lsnv{{\mathrm{a}}},\,\lsnv{{\mathrm{b}}}.\; \lsnv{0}' \neq \lsnv{{\mathrm{a}}} \loprod \lsnv{{\mathrm{b}}}}{\valCheckCtx{}\vdash \LTTVPP{\lsnv{0}''}{\lenv{0}}{\lenv{1}}\colon \lR{}}{\texttt{GLang.Runtime\allowbreak Value.Val\allowbreak Has\allowbreak Type.ttv\_\allowbreak real}}{GLang/RuntimeValueTyping.lean:62}
\gldefruleexplanation{valHasType.ttvReal}{Contraction of a value whose space is not a tensor product with a typed argument has scalar result type, with dimensions checked during evaluation.}
\gldefrule{valHasType.mtm}{valHasType}{\textsc{VT-Mtm}}{\valCheckCtx{}\vdash \lsnv{1}''\colon\svalid \\ \valCheckCtx{}\vdash \lsnv{1}\colon\svalid \\ \valCheckCtx{}\vdash \lenv{0}\colon \lsnv{1}' \loprod \lsnv{0} \\ \valCheckCtx{}\vdash \lenv{1}\colon \lsnv{2} \loprod \lsnv{1}}{\valCheckCtx{}\vdash \glvmtm{\lsnv{1}''}{\lsnv{1}}{\lenv{0}}{\lenv{1}}\colon \lsnv{2} \loprod \lsnv{0}}{\texttt{GLang.Runtime\allowbreak Value.Val\allowbreak Has\allowbreak Type.mtm}}{GLang/RuntimeValueTyping.lean:68}
\gldefruleexplanation{valHasType.mtm}{Matrix contraction retains the two outer factor spaces, while the declared contraction spaces are well formed and dimension compatibility is checked during evaluation.}
\gldefrule{valHasType.vsum}{valHasType}{\textsc{VT-Vsum}}{\valCheckCtx{}\vdash \lsnv{0}\colon\svalid \\ \forall \leev{}.\; \leev{} \in \mathit{es} \Rightarrow \valCheckCtx{}\vdash \leev{}\colon \lsv{} \\ \valCheckCtx{}\vdash \lsv{}\colon\svalid \\ \mathit{es} = [] \Rightarrow \lsv{} = \lsempty{}}{\valCheckCtx{}\vdash \lsum^{\lsnv{0}}_{v\in \mathit{es}}v\colon \lsv{}}{\texttt{GLang.Runtime\allowbreak Value.Val\allowbreak Has\allowbreak Type.vsum}}{GLang/RuntimeValueTyping.lean:75}
\gldefruleexplanation{valHasType.vsum}{A finite family of values in one well-formed space may be summed using a well-formed index space, with the empty family restricted to the zero-dimensional result space.}
\gldefrule{valHasType.vlam}{valHasType}{\textsc{VT-Vlam}}{\valCheckCtx{}\vdash (x\colon \ltynv{0}\fnto \ltynv{1})\colon \lkn{0} \fnto \lkn{1} \\ (\valCheckCtx{}, \mathit{x}\colon \ltynv{0})\vdash \leev{}\colon \ltynv{1}}{\valCheckCtx{}\vdash \llambda(x\colon \ltynv{0})\lrh \ltynv{1}\colon \leev{}\colon (x\colon \ltynv{0}\fnto \ltynv{1})}{\texttt{GLang.Runtime\allowbreak Value.Val\allowbreak Has\allowbreak Type.vlam}}{GLang/RuntimeValueTyping.lean:81}
\gldefruleexplanation{valHasType.vlam}{A value body typed under a fresh argument forms a dependent function at a well-kinded function type.}
\gldefrule{valHasType.appV}{valHasType}{\textsc{VT-App\allowbreak V}}{\valCheckCtx{}\vdash \lenv{0}\colon (x\colon \ltynv{0}\fnto \ltynv{1}) \\ \valCheckCtx{}\vdash \lenv{1}\colon \ltynv{0} \\ {\ltyv{}_{\mathsf{out}}} = \ltynv{1}\subst{x}{\lenv{1}} \\ \valCheckCtx{}\vdash {\ltyv{}_{\mathsf{out}}}\colon \lky{}}{\valCheckCtx{}\vdash \glvapp{\lenv{0}}{\lenv{1}}\colon {\ltyv{}_{\mathsf{out}}}}{\texttt{GLang.Runtime\allowbreak Value.Val\allowbreak Has\allowbreak Type.app\allowbreak V}}{GLang/RuntimeValueTyping.lean:86}
\gldefruleexplanation{valHasType.appV}{Applying a typed dependent function to an argument of its domain type yields the substituted codomain, which must itself be well kinded.}
\gldefrule{valHasType.vkrn}{valHasType}{\textsc{VT-Vkrn}}{\valCheckCtx{}\vdash \leev{}\colon \lsnv{0} \loprod \lsnv{2} \\ \valCheckCtx{}\vdash \leev{}'\colon \lsnv{1} \loprod \lsnv{3}}{\valCheckCtx{}\vdash \vkrn(\leev{}, \leev{}')\colon \left(\lsnv{0} \loprod \lsnv{1}\right) \loprod \left(\lsnv{2} \loprod \lsnv{3}\right)}{\texttt{GLang.Runtime\allowbreak Value.Val\allowbreak Has\allowbreak Type.vkrn}}{GLang/RuntimeValueTyping.lean:92}
\gldefruleexplanation{valHasType.vkrn}{The Kronecker product of two matrix values groups their domain factors and their codomain factors into tensor products.}
\gldefrule{valHasType.vblk}{valHasType}{\textsc{VT-Vblk}}{\valCheckCtx{}\vdash \leev{}\colon \lsnv{0} \loprod \lsnv{2} \\ \valCheckCtx{}\vdash \leev{}'\colon \lsnv{1} \loprod \lsnv{3}}{\valCheckCtx{}\vdash \vblk(\leev{}, \leev{}')\colon \left(\lsnv{0} \losum \lsnv{1}\right) \loprod \left(\lsnv{2} \losum \lsnv{3}\right)}{\texttt{GLang.Runtime\allowbreak Value.Val\allowbreak Has\allowbreak Type.vblk}}{GLang/RuntimeValueTyping.lean:97}
\gldefruleexplanation{valHasType.vblk}{The block-diagonal combination of two matrix values groups their domain factors and their codomain factors into direct sums.}
\gldefrule{valHasType.reshap}{valHasType}{\textsc{VT-Reshap}}{\valCheckCtx{}\vdash \leev{}\colon {\ltyv{}_{\mathsf{in}}} \\ \valCheckCtx{}\vdash \ltynv{0}\colon \lky{} \\ \valCheckCtx{}\vdash \ltyv{}\colon \lky{}}{\valCheckCtx{}\vdash \glvreshape{\leev{}}{\ltyv{}}{\ltynv{0}}\colon \ltyv{}}{\texttt{GLang.Runtime\allowbreak Value.Val\allowbreak Has\allowbreak Type.reshap}}{GLang/RuntimeValueTyping.lean:102}
\gldefruleexplanation{valHasType.reshap}{Reshaping a typed value yields the declared target type when source and target annotations share a kind, with representation compatibility checked during evaluation.}
\gldefrule{valHasType.vtrans}{valHasType}{\textsc{VT-Vtrans}}{\valCheckCtx{}\vdash \leev{}\colon \lsnv{0} \loprod \lsnv{1}}{\valCheckCtx{}\vdash \Trans(\leev{})\colon \lsnv{1} \loprod \lsnv{0}}{\texttt{GLang.Runtime\allowbreak Value.Val\allowbreak Has\allowbreak Type.vtrans}}{GLang/RuntimeValueTyping.lean:107}
\gldefruleexplanation{valHasType.vtrans}{Transposing a matrix value exchanges its two factor spaces.}
\gldefrule{valHasType.gsvdS}{valHasType}{\textsc{VT-Gsvd\allowbreak S}}{\valCheckCtx{}\vdash \leev{}\colon \lsnv{0} \loprod \lsnv{1} \\ \valCheckCtx{}\vdash \lenv{{\mathrm{a}}}\colon \lsnv{{\mathrm{a0}}} \loprod \lsnv{{\mathrm{a1}}} \\ \valCheckCtx{}\vdash \lenv{{\mathrm{b}}}\colon \lsnv{{\mathrm{b0}}} \loprod \lsnv{{\mathrm{b1}}}}{\valCheckCtx{}\vdash \glvop{GSVDS}(\leev{},\lenv{{\mathrm{a}}},\lenv{{\mathrm{b}}})\colon \lR^{\leev{},\CoKernel{},\lsnv{0} \loprod \lsnv{1}} \loprod \lR^{\leev{},\Image{},\lsnv{0} \loprod \lsnv{1}}}{\texttt{GLang.Runtime\allowbreak Value.Val\allowbreak Has\allowbreak Type.gsvd\allowbreak S}}{GLang/RuntimeValueTyping.lean:110}
\gldefruleexplanation{valHasType.gsvdS}{The generalized singular-value matrix is typed from the cokernel coordinate space to the image coordinate space, with metric compatibility checked during evaluation.}
\gldefrule{valHasType.gsvdU}{valHasType}{\textsc{VT-Gsvd\allowbreak U}}{\valCheckCtx{}\vdash \leev{}\colon \lsnv{0} \loprod \lsnv{1} \\ \valCheckCtx{}\vdash \lenv{{\mathrm{a}}}\colon \lsnv{{\mathrm{a0}}} \loprod \lsnv{{\mathrm{a1}}} \\ \valCheckCtx{}\vdash \lenv{{\mathrm{b}}}\colon \lsnv{{\mathrm{b0}}} \loprod \lsnv{{\mathrm{b1}}}}{\valCheckCtx{}\vdash \glvop{GSVDU}(\leev{},\lenv{{\mathrm{a}}},\lenv{{\mathrm{b}}})\colon \lsnv{0} \loprod \lR^{\leev{},\CoKernel{},\lsnv{0} \loprod \lsnv{1}}}{\texttt{GLang.Runtime\allowbreak Value.Val\allowbreak Has\allowbreak Type.gsvd\allowbreak U}}{GLang/RuntimeValueTyping.lean:118}
\gldefruleexplanation{valHasType.gsvdU}{The generalized domain singular-vector block pairs the domain space with the cokernel coordinate space, with metric compatibility checked during evaluation.}
\gldefrule{valHasType.gsvdUI}{valHasType}{\textsc{VT-Gsvd\allowbreak UI}}{\valCheckCtx{}\vdash \leev{}\colon \lsnv{0} \loprod \lsnv{1} \\ \valCheckCtx{}\vdash \lenv{{\mathrm{a}}}\colon \lsnv{{\mathrm{a0}}} \loprod \lsnv{{\mathrm{a1}}} \\ \valCheckCtx{}\vdash \lenv{{\mathrm{b}}}\colon \lsnv{{\mathrm{b0}}} \loprod \lsnv{{\mathrm{b1}}}}{\valCheckCtx{}\vdash \glvop{GSVDUI}(\leev{},\lenv{{\mathrm{a}}},\lenv{{\mathrm{b}}})\colon \lsnv{0} \loprod \lR^{\leev{},\Kernel{},\lsnv{0} \loprod \lsnv{1}}}{\texttt{GLang.Runtime\allowbreak Value.Val\allowbreak Has\allowbreak Type.gsvd\allowbreak UI}}{GLang/RuntimeValueTyping.lean:125}
\gldefruleexplanation{valHasType.gsvdUI}{The generalized domain null-space block pairs the domain space with the kernel coordinate space, with metric compatibility checked during evaluation.}
\gldefrule{valHasType.gsvdV}{valHasType}{\textsc{VT-Gsvd\allowbreak V}}{\valCheckCtx{}\vdash \leev{}\colon \lsnv{0} \loprod \lsnv{1} \\ \valCheckCtx{}\vdash \lenv{{\mathrm{a}}}\colon \lsnv{{\mathrm{a0}}} \loprod \lsnv{{\mathrm{a1}}} \\ \valCheckCtx{}\vdash \lenv{{\mathrm{b}}}\colon \lsnv{{\mathrm{b0}}} \loprod \lsnv{{\mathrm{b1}}}}{\valCheckCtx{}\vdash \glvop{GSVDV}(\leev{},\lenv{{\mathrm{a}}},\lenv{{\mathrm{b}}})\colon \lsnv{1} \loprod \lR^{\leev{},\Image{},\lsnv{0} \loprod \lsnv{1}}}{\texttt{GLang.Runtime\allowbreak Value.Val\allowbreak Has\allowbreak Type.gsvd\allowbreak V}}{GLang/RuntimeValueTyping.lean:131}
\gldefruleexplanation{valHasType.gsvdV}{The generalized codomain singular-vector block pairs the codomain space with the image coordinate space, with metric compatibility checked during evaluation.}
\gldefrule{valHasType.gsvdVI}{valHasType}{\textsc{VT-Gsvd\allowbreak VI}}{\valCheckCtx{}\vdash \leev{}\colon \lsnv{0} \loprod \lsnv{1} \\ \valCheckCtx{}\vdash \lenv{{\mathrm{a}}}\colon \lsnv{{\mathrm{a0}}} \loprod \lsnv{{\mathrm{a1}}} \\ \valCheckCtx{}\vdash \lenv{{\mathrm{b}}}\colon \lsnv{{\mathrm{b0}}} \loprod \lsnv{{\mathrm{b1}}}}{\valCheckCtx{}\vdash \glvop{GSVDVI}(\leev{},\lenv{{\mathrm{a}}},\lenv{{\mathrm{b}}})\colon \lsnv{1} \loprod \lR^{\leev{},\CoImage{},\lsnv{0} \loprod \lsnv{1}}}{\texttt{GLang.Runtime\allowbreak Value.Val\allowbreak Has\allowbreak Type.gsvd\allowbreak VI}}{GLang/RuntimeValueTyping.lean:138}
\gldefruleexplanation{valHasType.gsvdVI}{The generalized codomain null-space block pairs the codomain space with the coimage coordinate space, with metric compatibility checked during evaluation.}
\gldefrule{valHasType.pinvTensor}{valHasType}{\textsc{VT-Pinv\allowbreak Tensor}}{\valCheckCtx{}\vdash \leev{}\colon \lsnv{0} \loprod \lsnv{1}}{\valCheckCtx{}\vdash \leev{}^{+}\colon \lsnv{1} \loprod \lsnv{0}}{\texttt{GLang.Runtime\allowbreak Value.Val\allowbreak Has\allowbreak Type.pinv\_\allowbreak tensor}}{GLang/RuntimeValueTyping.lean:144}
\gldefruleexplanation{valHasType.pinvTensor}{The pseudoinverse of a matrix value exchanges its domain and codomain factor spaces.}
\gldefrule{valHasType.pinvRaw}{valHasType}{\textsc{VT-Pinv\allowbreak Raw}}{\valCheckCtx{}\vdash \leev{}\colon \lsnv{0} \\ \forall \lsnv{{\mathrm{a}}},\,\lsnv{{\mathrm{b}}}.\; \lsnv{0} \neq \lsnv{{\mathrm{a}}} \loprod \lsnv{{\mathrm{b}}}}{\valCheckCtx{}\vdash \leev{}^{+}\colon \lR{} \loprod \lsnv{0}}{\texttt{GLang.Runtime\allowbreak Value.Val\allowbreak Has\allowbreak Type.pinv\_\allowbreak raw}}{GLang/RuntimeValueTyping.lean:147}
\gldefruleexplanation{valHasType.pinvRaw}{The pseudoinverse of a value whose space is not a tensor product has type $\mathbb R\loprod\lsnv{0}$.}
\gldefrule{valHasType.gpinv}{valHasType}{\textsc{VT-Gpinv}}{\valCheckCtx{}\vdash \leev{}\colon \lsnv{0} \loprod \lsnv{1} \\ \valCheckCtx{}\vdash \lenv{{\mathrm{a}}}\colon \lsnv{{\mathrm{a0}}} \loprod \lsnv{{\mathrm{a1}}} \\ \valCheckCtx{}\vdash \lenv{{\mathrm{b}}}\colon \lsnv{{\mathrm{b0}}} \loprod \lsnv{{\mathrm{b1}}}}{\valCheckCtx{}\vdash \glvop{GPINV}(\leev{},\lenv{{\mathrm{a}}},\lenv{{\mathrm{b}}})\colon \lsnv{1} \loprod \lsnv{0}}{\texttt{GLang.Runtime\allowbreak Value.Val\allowbreak Has\allowbreak Type.gpinv}}{GLang/RuntimeValueTyping.lean:151}
\gldefruleexplanation{valHasType.gpinv}{The generalized pseudoinverse exchanges the matrix's factor spaces, with the two tensor-typed metric operands checked during evaluation.}
\gldefrule{valHasType.rrefRep}{valHasType}{\textsc{VT-Rref\allowbreak Rep}}{\valCheckCtx{}\vdash \lenv{0}\colon \lsv{} \\ \valCheckCtx{}\vdash \lenv{1}\colon \lsnv{{\mathrm{p0}}} \loprod \lsnv{{\mathrm{p1}}}}{\valCheckCtx{}\vdash \glvop{RREFRep}(\lenv{0},\lenv{1})\colon \lsnv{{\mathrm{p1}}} \loprod \lsnv{{\mathrm{p0}}}}{\texttt{GLang.Runtime\allowbreak Value.Val\allowbreak Has\allowbreak Type.rref\allowbreak Rep}}{GLang/RuntimeValueTyping.lean:156}
\gldefruleexplanation{valHasType.rrefRep}{The row-reduction representation has the reversed factor spaces of its matrix operand, with candidate-array compatibility checked during evaluation.}
\gldefrule{valHasType.invTensor}{valHasType}{\textsc{VT-Inv\allowbreak Tensor}}{\valCheckCtx{}\vdash \leev{}\colon \lsnv{0} \loprod \lsnv{1}}{\valCheckCtx{}\vdash \leev{}^{-1}\colon \lsnv{1} \loprod \lsnv{0}}{\texttt{GLang.Runtime\allowbreak Value.Val\allowbreak Has\allowbreak Type.inv\_\allowbreak tensor}}{GLang/RuntimeValueTyping.lean:160}
\gldefruleexplanation{valHasType.invTensor}{The inverse operation on a matrix value exchanges its factor spaces, with array dimensions checked during evaluation.}
\gldefrule{valHasType.invRaw}{valHasType}{\textsc{VT-Inv\allowbreak Raw}}{\valCheckCtx{}\vdash \leev{}\colon \lsnv{0} \\ \forall \lsnv{{\mathrm{a}}},\,\lsnv{{\mathrm{b}}}.\; \lsnv{0} \neq \lsnv{{\mathrm{a}}} \loprod \lsnv{{\mathrm{b}}}}{\valCheckCtx{}\vdash \leev{}^{-1}\colon \lR{} \loprod \lsnv{0}}{\texttt{GLang.Runtime\allowbreak Value.Val\allowbreak Has\allowbreak Type.inv\_\allowbreak raw}}{GLang/RuntimeValueTyping.lean:163}
\gldefruleexplanation{valHasType.invRaw}{The inverse operation on a value whose space is not a tensor product has type $\mathbb R\loprod\lsnv{0}$, with array dimensions checked during evaluation.}
\gldefrule{valHasType.svds}{valHasType}{\textsc{VT-Svds}}{\valCheckCtx{}\vdash \leev{}\colon \lsnv{0} \loprod \lsnv{1}}{\valCheckCtx{}\vdash \glvop{SVDS}(\leev{})\colon \lR^{\leev{},\CoKernel{},\lsnv{0} \loprod \lsnv{1}} \loprod \lR^{\leev{},\Image{},\lsnv{0} \loprod \lsnv{1}}}{\texttt{GLang.Runtime\allowbreak Value.Val\allowbreak Has\allowbreak Type.svds}}{GLang/RuntimeValueTyping.lean:167}
\gldefruleexplanation{valHasType.svds}{The singular-value matrix is typed from the cokernel coordinate space to the image coordinate space of its matrix operand.}
\gldefrulerows{valHasType}{
\textsc{VT-Var} & $\valCheckCtx{}\vdash \mathit{x}_{\ltyv{}}\colon \ltyv{}$ & \texttt{\footnotesize GLang/Runtime\allowbreak Value\allowbreak Typing.lean:33}\\
\textsc{VT-Zero} & $\valCheckCtx{}\vdash \lvzero{}_{\lsv{}}\colon \lsv{}$ & \texttt{\footnotesize GLang/Runtime\allowbreak Value\allowbreak Typing.lean:34}\\
\textsc{VT-Rlit} & $\valCheckCtx{}\vdash \mathit{r}\colon \lR{}$ & \texttt{\footnotesize GLang/Runtime\allowbreak Value\allowbreak Typing.lean:35}\\
\textsc{VT-Scale} & $\valCheckCtx{}\vdash \lenv{0} \lvscale \lenv{1}\colon \ltyv{}$ & \texttt{\footnotesize GLang/Runtime\allowbreak Value\allowbreak Typing.lean:36}\\
\textsc{VT-Vadd} & $\valCheckCtx{}\vdash \lenv{0} \lvadd_{\lsv{}} \lenv{1}\colon \lsv{}$ & \texttt{\footnotesize GLang/Runtime\allowbreak Value\allowbreak Typing.lean:40}\\
\textsc{VT-Dsum\allowbreak E} & $\valCheckCtx{}\vdash \lenv{0}\losum \lenv{1}\colon \lsnv{0} \losum \lsnv{1}$ & \texttt{\footnotesize GLang/Runtime\allowbreak Value\allowbreak Typing.lean:45}\\
\textsc{VT-Tprod\allowbreak E} & $\valCheckCtx{}\vdash \lenv{0}\loprod \lenv{1}\colon \lsnv{0} \loprod \lsnv{1}$ & \texttt{\footnotesize GLang/Runtime\allowbreak Value\allowbreak Typing.lean:49}\\
\textsc{VT-Delta} & $\valCheckCtx{}\vdash \ldelta{\mathit{n}}{\lsv{}}\colon \lsv{}$ & \texttt{\footnotesize GLang/Runtime\allowbreak Value\allowbreak Typing.lean:53}\\
\textsc{VT-Idelta} & $\valCheckCtx{}\vdash \lidelta{\lsv{}}\colon \lsv{} \loprod \lsv{}$ & \texttt{\footnotesize GLang/Runtime\allowbreak Value\allowbreak Typing.lean:54}\\
\textsc{VT-Ttv\allowbreak Space} & $\valCheckCtx{}\vdash \LTTVPP{\lsnv{0}''}{\lenv{0}}{\lenv{1}}\colon \lsnv{1}$ & \texttt{\footnotesize GLang/Runtime\allowbreak Value\allowbreak Typing.lean:57}\\
\textsc{VT-Ttv\allowbreak Real} & $\valCheckCtx{}\vdash \LTTVPP{\lsnv{0}''}{\lenv{0}}{\lenv{1}}\colon \lR{}$ & \texttt{\footnotesize GLang/Runtime\allowbreak Value\allowbreak Typing.lean:62}\\
\textsc{VT-Mtm} & $\valCheckCtx{}\vdash \glvmtm{\lsnv{1}''}{\lsnv{1}}{\lenv{0}}{\lenv{1}}\colon \lsnv{2} \loprod \lsnv{0}$ & \texttt{\footnotesize GLang/Runtime\allowbreak Value\allowbreak Typing.lean:68}\\
\textsc{VT-Vsum} & $\valCheckCtx{}\vdash \lsum^{\lsnv{0}}_{v\in \mathit{es}}v\colon \lsv{}$ & \texttt{\footnotesize GLang/Runtime\allowbreak Value\allowbreak Typing.lean:75}\\
\textsc{VT-Vlam} & $\valCheckCtx{}\vdash \llambda(x\colon \ltynv{0})\lrh \ltynv{1}\colon \leev{}\colon (x\colon \ltynv{0}\fnto \ltynv{1})$ & \texttt{\footnotesize GLang/Runtime\allowbreak Value\allowbreak Typing.lean:81}\\
\textsc{VT-App\allowbreak V} & $\valCheckCtx{}\vdash \glvapp{\lenv{0}}{\lenv{1}}\colon {\ltyv{}_{\mathsf{out}}}$ & \texttt{\footnotesize GLang/Runtime\allowbreak Value\allowbreak Typing.lean:86}\\
\textsc{VT-Vkrn} & $\valCheckCtx{}\vdash \vkrn(\leev{}, \leev{}')\colon \left(\lsnv{0} \loprod \lsnv{1}\right) \loprod \left(\lsnv{2} \loprod \lsnv{3}\right)$ & \texttt{\footnotesize GLang/Runtime\allowbreak Value\allowbreak Typing.lean:92}\\
\textsc{VT-Vblk} & $\valCheckCtx{}\vdash \vblk(\leev{}, \leev{}')\colon \left(\lsnv{0} \losum \lsnv{1}\right) \loprod \left(\lsnv{2} \losum \lsnv{3}\right)$ & \texttt{\footnotesize GLang/Runtime\allowbreak Value\allowbreak Typing.lean:97}\\
\textsc{VT-Reshap} & $\valCheckCtx{}\vdash \glvreshape{\leev{}}{\ltyv{}}{\ltynv{0}}\colon \ltyv{}$ & \texttt{\footnotesize GLang/Runtime\allowbreak Value\allowbreak Typing.lean:102}\\
\textsc{VT-Vtrans} & $\valCheckCtx{}\vdash \Trans(\leev{})\colon \lsnv{1} \loprod \lsnv{0}$ & \texttt{\footnotesize GLang/Runtime\allowbreak Value\allowbreak Typing.lean:107}\\
\textsc{VT-Gsvd\allowbreak S} & $\valCheckCtx{}\vdash \glvop{GSVDS}(\leev{},\lenv{{\mathrm{a}}},\lenv{{\mathrm{b}}})\colon \lR^{\leev{},\CoKernel{},\lsnv{0} \loprod \lsnv{1}} \loprod \lR^{\leev{},\Image{},\lsnv{0} \loprod \lsnv{1}}$ & \texttt{\footnotesize GLang/Runtime\allowbreak Value\allowbreak Typing.lean:110}\\
\textsc{VT-Gsvd\allowbreak U} & $\valCheckCtx{}\vdash \glvop{GSVDU}(\leev{},\lenv{{\mathrm{a}}},\lenv{{\mathrm{b}}})\colon \lsnv{0} \loprod \lR^{\leev{},\CoKernel{},\lsnv{0} \loprod \lsnv{1}}$ & \texttt{\footnotesize GLang/Runtime\allowbreak Value\allowbreak Typing.lean:118}\\
\textsc{VT-Gsvd\allowbreak UI} & $\valCheckCtx{}\vdash \glvop{GSVDUI}(\leev{},\lenv{{\mathrm{a}}},\lenv{{\mathrm{b}}})\colon \lsnv{0} \loprod \lR^{\leev{},\Kernel{},\lsnv{0} \loprod \lsnv{1}}$ & \texttt{\footnotesize GLang/Runtime\allowbreak Value\allowbreak Typing.lean:125}\\
\textsc{VT-Gsvd\allowbreak V} & $\valCheckCtx{}\vdash \glvop{GSVDV}(\leev{},\lenv{{\mathrm{a}}},\lenv{{\mathrm{b}}})\colon \lsnv{1} \loprod \lR^{\leev{},\Image{},\lsnv{0} \loprod \lsnv{1}}$ & \texttt{\footnotesize GLang/Runtime\allowbreak Value\allowbreak Typing.lean:131}\\
\textsc{VT-Gsvd\allowbreak VI} & $\valCheckCtx{}\vdash \glvop{GSVDVI}(\leev{},\lenv{{\mathrm{a}}},\lenv{{\mathrm{b}}})\colon \lsnv{1} \loprod \lR^{\leev{},\CoImage{},\lsnv{0} \loprod \lsnv{1}}$ & \texttt{\footnotesize GLang/Runtime\allowbreak Value\allowbreak Typing.lean:138}\\
\textsc{VT-Pinv\allowbreak Tensor} & $\valCheckCtx{}\vdash \leev{}^{+}\colon \lsnv{1} \loprod \lsnv{0}$ & \texttt{\footnotesize GLang/Runtime\allowbreak Value\allowbreak Typing.lean:144}\\
\textsc{VT-Pinv\allowbreak Raw} & $\valCheckCtx{}\vdash \leev{}^{+}\colon \lR{} \loprod \lsnv{0}$ & \texttt{\footnotesize GLang/Runtime\allowbreak Value\allowbreak Typing.lean:147}\\
\textsc{VT-Gpinv} & $\valCheckCtx{}\vdash \glvop{GPINV}(\leev{},\lenv{{\mathrm{a}}},\lenv{{\mathrm{b}}})\colon \lsnv{1} \loprod \lsnv{0}$ & \texttt{\footnotesize GLang/Runtime\allowbreak Value\allowbreak Typing.lean:151}\\
\textsc{VT-Rref\allowbreak Rep} & $\valCheckCtx{}\vdash \glvop{RREFRep}(\lenv{0},\lenv{1})\colon \lsnv{{\mathrm{p1}}} \loprod \lsnv{{\mathrm{p0}}}$ & \texttt{\footnotesize GLang/Runtime\allowbreak Value\allowbreak Typing.lean:156}\\
\textsc{VT-Inv\allowbreak Tensor} & $\valCheckCtx{}\vdash \leev{}^{-1}\colon \lsnv{1} \loprod \lsnv{0}$ & \texttt{\footnotesize GLang/Runtime\allowbreak Value\allowbreak Typing.lean:160}\\
\textsc{VT-Inv\allowbreak Raw} & $\valCheckCtx{}\vdash \leev{}^{-1}\colon \lR{} \loprod \lsnv{0}$ & \texttt{\footnotesize GLang/Runtime\allowbreak Value\allowbreak Typing.lean:163}\\
\textsc{VT-Svds} & $\valCheckCtx{}\vdash \glvop{SVDS}(\leev{})\colon \lR^{\leev{},\CoKernel{},\lsnv{0} \loprod \lsnv{1}} \loprod \lR^{\leev{},\Image{},\lsnv{0} \loprod \lsnv{1}}$ & \texttt{\footnotesize GLang/Runtime\allowbreak Value\allowbreak Typing.lean:167}\\
}{
\gpfit{\inferrule*[]{(\mathit{x}\colon \ltyv{}) \in \valCheckCtx{}}{\valCheckCtx{}\vdash \mathit{x}_{\ltyv{}}\colon \ltyv{}}}
\and
\gpfit{\inferrule*[]{\valCheckCtx{}\vdash \lsv{}\colon\svalid}{\valCheckCtx{}\vdash \lvzero{}_{\lsv{}}\colon \lsv{}}}
\and
\gpfit{\inferrule*[]{\phantom{.}}{\valCheckCtx{}\vdash \mathit{r}\colon \lR{}}}
\and
\gpfit{\inferrule*[]{\valCheckCtx{}\vdash \lenv{0}\colon \lR{} \\ \valCheckCtx{}\vdash \lenv{1}\colon \ltyv{}}{\valCheckCtx{}\vdash \lenv{0} \lvscale \lenv{1}\colon \ltyv{}}}
\and
\gpfit{\inferrule*[]{\forall i.\; \valCheckCtx{}\vdash \lenv{i}\colon \lsnv{i} \\ \valCheckCtx{}\vdash \lsv{}\colon\svalid}{\valCheckCtx{}\vdash \lenv{0} \lvadd_{\lsv{}} \lenv{1}\colon \lsv{}}}
\and
\gpfit{\inferrule*[]{\forall i.\; \valCheckCtx{}\vdash \lenv{i}\colon \lsnv{i}}{\valCheckCtx{}\vdash \lenv{0}\losum \lenv{1}\colon \lsnv{0} \losum \lsnv{1}}}
\and
\gpfit{\inferrule*[]{\forall i.\; \valCheckCtx{}\vdash \lenv{i}\colon \lsnv{i}}{\valCheckCtx{}\vdash \lenv{0}\loprod \lenv{1}\colon \lsnv{0} \loprod \lsnv{1}}}
\and
\gpfit{\inferrule*[]{\valCheckCtx{}\vdash \lsv{}\colon\svalid}{\valCheckCtx{}\vdash \ldelta{\mathit{n}}{\lsv{}}\colon \lsv{}}}
\and
\gpfit{\inferrule*[]{\valCheckCtx{}\vdash \lsv{}\colon\svalid}{\valCheckCtx{}\vdash \lidelta{\lsv{}}\colon \lsv{} \loprod \lsv{}}}
\and
\gpfit{\inferrule*[]{\valCheckCtx{}\vdash \lsnv{0}''\colon\svalid \\ \valCheckCtx{}\vdash \lenv{0}\colon \lsnv{0}' \loprod \lsnv{1} \\ \valCheckCtx{}\vdash \lenv{1}\colon {\ltyv{}_{\mathsf{arg}}}}{\valCheckCtx{}\vdash \LTTVPP{\lsnv{0}''}{\lenv{0}}{\lenv{1}}\colon \lsnv{1}}}
\and
\gpfit{\inferrule*[]{\valCheckCtx{}\vdash \lsnv{0}''\colon\svalid \\ \valCheckCtx{}\vdash \lenv{0}\colon \lsnv{0}' \\ \valCheckCtx{}\vdash \lenv{1}\colon {\ltyv{}_{\mathsf{arg}}} \\ \forall \lsnv{{\mathrm{a}}},\,\lsnv{{\mathrm{b}}}.\; \lsnv{0}' \neq \lsnv{{\mathrm{a}}} \loprod \lsnv{{\mathrm{b}}}}{\valCheckCtx{}\vdash \LTTVPP{\lsnv{0}''}{\lenv{0}}{\lenv{1}}\colon \lR{}}}
\and
\gpfit{\inferrule*[]{\valCheckCtx{}\vdash \lsnv{1}''\colon\svalid \\ \valCheckCtx{}\vdash \lsnv{1}\colon\svalid \\ \valCheckCtx{}\vdash \lenv{0}\colon \lsnv{1}' \loprod \lsnv{0} \\ \valCheckCtx{}\vdash \lenv{1}\colon \lsnv{2} \loprod \lsnv{1}}{\valCheckCtx{}\vdash \glvmtm{\lsnv{1}''}{\lsnv{1}}{\lenv{0}}{\lenv{1}}\colon \lsnv{2} \loprod \lsnv{0}}}
\and
\gpfit{\inferrule*[]{\valCheckCtx{}\vdash \lsnv{0}\colon\svalid \\ \forall \leev{}.\; \leev{} \in \mathit{es} \Rightarrow \valCheckCtx{}\vdash \leev{}\colon \lsv{} \\ \valCheckCtx{}\vdash \lsv{}\colon\svalid \\ \mathit{es} = [] \Rightarrow \lsv{} = \lsempty{}}{\valCheckCtx{}\vdash \lsum^{\lsnv{0}}_{v\in \mathit{es}}v\colon \lsv{}}}
\and
\gpfit{\inferrule*[]{\valCheckCtx{}\vdash (x\colon \ltynv{0}\fnto \ltynv{1})\colon \lkn{0} \fnto \lkn{1} \\ (\valCheckCtx{}, \mathit{x}\colon \ltynv{0})\vdash \leev{}\colon \ltynv{1}}{\valCheckCtx{}\vdash \llambda(x\colon \ltynv{0})\lrh \ltynv{1}\colon \leev{}\colon (x\colon \ltynv{0}\fnto \ltynv{1})}}
\and
\gpfit{\inferrule*[]{\valCheckCtx{}\vdash \lenv{0}\colon (x\colon \ltynv{0}\fnto \ltynv{1}) \\ \valCheckCtx{}\vdash \lenv{1}\colon \ltynv{0} \\ {\ltyv{}_{\mathsf{out}}} = \ltynv{1}\subst{x}{\lenv{1}} \\ \valCheckCtx{}\vdash {\ltyv{}_{\mathsf{out}}}\colon \lky{}}{\valCheckCtx{}\vdash \glvapp{\lenv{0}}{\lenv{1}}\colon {\ltyv{}_{\mathsf{out}}}}}
\and
\gpfit{\inferrule*[]{\valCheckCtx{}\vdash \leev{}\colon \lsnv{0} \loprod \lsnv{2} \\ \valCheckCtx{}\vdash \leev{}'\colon \lsnv{1} \loprod \lsnv{3}}{\valCheckCtx{}\vdash \vkrn(\leev{}, \leev{}')\colon \left(\lsnv{0} \loprod \lsnv{1}\right) \loprod \left(\lsnv{2} \loprod \lsnv{3}\right)}}
\and
\gpfit{\inferrule*[]{\valCheckCtx{}\vdash \leev{}\colon \lsnv{0} \loprod \lsnv{2} \\ \valCheckCtx{}\vdash \leev{}'\colon \lsnv{1} \loprod \lsnv{3}}{\valCheckCtx{}\vdash \vblk(\leev{}, \leev{}')\colon \left(\lsnv{0} \losum \lsnv{1}\right) \loprod \left(\lsnv{2} \losum \lsnv{3}\right)}}
\and
\gpfit{\inferrule*[]{\valCheckCtx{}\vdash \leev{}\colon {\ltyv{}_{\mathsf{in}}} \\ \valCheckCtx{}\vdash \ltynv{0}\colon \lky{} \\ \valCheckCtx{}\vdash \ltyv{}\colon \lky{}}{\valCheckCtx{}\vdash \glvreshape{\leev{}}{\ltyv{}}{\ltynv{0}}\colon \ltyv{}}}
\and
\gpfit{\inferrule*[]{\valCheckCtx{}\vdash \leev{}\colon \lsnv{0} \loprod \lsnv{1}}{\valCheckCtx{}\vdash \Trans(\leev{})\colon \lsnv{1} \loprod \lsnv{0}}}
\and
\gpfit{\inferrule*[]{\valCheckCtx{}\vdash \leev{}\colon \lsnv{0} \loprod \lsnv{1} \\ \valCheckCtx{}\vdash \lenv{{\mathrm{a}}}\colon \lsnv{{\mathrm{a0}}} \loprod \lsnv{{\mathrm{a1}}} \\ \valCheckCtx{}\vdash \lenv{{\mathrm{b}}}\colon \lsnv{{\mathrm{b0}}} \loprod \lsnv{{\mathrm{b1}}}}{\valCheckCtx{}\vdash \glvop{GSVDS}(\leev{},\lenv{{\mathrm{a}}},\lenv{{\mathrm{b}}})\colon \lR^{\leev{},\CoKernel{},\lsnv{0} \loprod \lsnv{1}} \loprod \lR^{\leev{},\Image{},\lsnv{0} \loprod \lsnv{1}}}}
\and
\gpfit{\inferrule*[]{\valCheckCtx{}\vdash \leev{}\colon \lsnv{0} \loprod \lsnv{1} \\ \valCheckCtx{}\vdash \lenv{{\mathrm{a}}}\colon \lsnv{{\mathrm{a0}}} \loprod \lsnv{{\mathrm{a1}}} \\ \valCheckCtx{}\vdash \lenv{{\mathrm{b}}}\colon \lsnv{{\mathrm{b0}}} \loprod \lsnv{{\mathrm{b1}}}}{\valCheckCtx{}\vdash \glvop{GSVDU}(\leev{},\lenv{{\mathrm{a}}},\lenv{{\mathrm{b}}})\colon \lsnv{0} \loprod \lR^{\leev{},\CoKernel{},\lsnv{0} \loprod \lsnv{1}}}}
\and
\gpfit{\inferrule*[]{\valCheckCtx{}\vdash \leev{}\colon \lsnv{0} \loprod \lsnv{1} \\ \valCheckCtx{}\vdash \lenv{{\mathrm{a}}}\colon \lsnv{{\mathrm{a0}}} \loprod \lsnv{{\mathrm{a1}}} \\ \valCheckCtx{}\vdash \lenv{{\mathrm{b}}}\colon \lsnv{{\mathrm{b0}}} \loprod \lsnv{{\mathrm{b1}}}}{\valCheckCtx{}\vdash \glvop{GSVDUI}(\leev{},\lenv{{\mathrm{a}}},\lenv{{\mathrm{b}}})\colon \lsnv{0} \loprod \lR^{\leev{},\Kernel{},\lsnv{0} \loprod \lsnv{1}}}}
\and
\gpfit{\inferrule*[]{\valCheckCtx{}\vdash \leev{}\colon \lsnv{0} \loprod \lsnv{1} \\ \valCheckCtx{}\vdash \lenv{{\mathrm{a}}}\colon \lsnv{{\mathrm{a0}}} \loprod \lsnv{{\mathrm{a1}}} \\ \valCheckCtx{}\vdash \lenv{{\mathrm{b}}}\colon \lsnv{{\mathrm{b0}}} \loprod \lsnv{{\mathrm{b1}}}}{\valCheckCtx{}\vdash \glvop{GSVDV}(\leev{},\lenv{{\mathrm{a}}},\lenv{{\mathrm{b}}})\colon \lsnv{1} \loprod \lR^{\leev{},\Image{},\lsnv{0} \loprod \lsnv{1}}}}
\and
\gpfit{\inferrule*[]{\valCheckCtx{}\vdash \leev{}\colon \lsnv{0} \loprod \lsnv{1} \\ \valCheckCtx{}\vdash \lenv{{\mathrm{a}}}\colon \lsnv{{\mathrm{a0}}} \loprod \lsnv{{\mathrm{a1}}} \\ \valCheckCtx{}\vdash \lenv{{\mathrm{b}}}\colon \lsnv{{\mathrm{b0}}} \loprod \lsnv{{\mathrm{b1}}}}{\valCheckCtx{}\vdash \glvop{GSVDVI}(\leev{},\lenv{{\mathrm{a}}},\lenv{{\mathrm{b}}})\colon \lsnv{1} \loprod \lR^{\leev{},\CoImage{},\lsnv{0} \loprod \lsnv{1}}}}
\and
\gpfit{\inferrule*[]{\valCheckCtx{}\vdash \leev{}\colon \lsnv{0} \loprod \lsnv{1}}{\valCheckCtx{}\vdash \leev{}^{+}\colon \lsnv{1} \loprod \lsnv{0}}}
\and
\gpfit{\inferrule*[]{\valCheckCtx{}\vdash \leev{}\colon \lsnv{0} \\ \forall \lsnv{{\mathrm{a}}},\,\lsnv{{\mathrm{b}}}.\; \lsnv{0} \neq \lsnv{{\mathrm{a}}} \loprod \lsnv{{\mathrm{b}}}}{\valCheckCtx{}\vdash \leev{}^{+}\colon \lR{} \loprod \lsnv{0}}}
\and
\gpfit{\inferrule*[]{\valCheckCtx{}\vdash \leev{}\colon \lsnv{0} \loprod \lsnv{1} \\ \valCheckCtx{}\vdash \lenv{{\mathrm{a}}}\colon \lsnv{{\mathrm{a0}}} \loprod \lsnv{{\mathrm{a1}}} \\ \valCheckCtx{}\vdash \lenv{{\mathrm{b}}}\colon \lsnv{{\mathrm{b0}}} \loprod \lsnv{{\mathrm{b1}}}}{\valCheckCtx{}\vdash \glvop{GPINV}(\leev{},\lenv{{\mathrm{a}}},\lenv{{\mathrm{b}}})\colon \lsnv{1} \loprod \lsnv{0}}}
\and
\gpfit{\inferrule*[]{\valCheckCtx{}\vdash \lenv{0}\colon \lsv{} \\ \valCheckCtx{}\vdash \lenv{1}\colon \lsnv{{\mathrm{p0}}} \loprod \lsnv{{\mathrm{p1}}}}{\valCheckCtx{}\vdash \glvop{RREFRep}(\lenv{0},\lenv{1})\colon \lsnv{{\mathrm{p1}}} \loprod \lsnv{{\mathrm{p0}}}}}
\and
\gpfit{\inferrule*[]{\valCheckCtx{}\vdash \leev{}\colon \lsnv{0} \loprod \lsnv{1}}{\valCheckCtx{}\vdash \leev{}^{-1}\colon \lsnv{1} \loprod \lsnv{0}}}
\and
\gpfit{\inferrule*[]{\valCheckCtx{}\vdash \leev{}\colon \lsnv{0} \\ \forall \lsnv{{\mathrm{a}}},\,\lsnv{{\mathrm{b}}}.\; \lsnv{0} \neq \lsnv{{\mathrm{a}}} \loprod \lsnv{{\mathrm{b}}}}{\valCheckCtx{}\vdash \leev{}^{-1}\colon \lR{} \loprod \lsnv{0}}}
\and
\gpfit{\inferrule*[]{\valCheckCtx{}\vdash \leev{}\colon \lsnv{0} \loprod \lsnv{1}}{\valCheckCtx{}\vdash \glvop{SVDS}(\leev{})\colon \lR^{\leev{},\CoKernel{},\lsnv{0} \loprod \lsnv{1}} \loprod \lR^{\leev{},\Image{},\lsnv{0} \loprod \lsnv{1}}}}
}{
\gpfit{\inferrule*[lab=\textsc{VT-Var}]{(\mathit{x}\colon \ltyv{}) \in \valCheckCtx{}}{\valCheckCtx{}\vdash \mathit{x}_{\ltyv{}}\colon \ltyv{}}}
\and
\gpfit{\inferrule*[lab=\textsc{VT-Zero}]{\valCheckCtx{}\vdash \lsv{}\colon\svalid}{\valCheckCtx{}\vdash \lvzero{}_{\lsv{}}\colon \lsv{}}}
\and
\gpfit{\inferrule*[lab=\textsc{VT-Rlit}]{\phantom{.}}{\valCheckCtx{}\vdash \mathit{r}\colon \lR{}}}
\and
\gpfit{\inferrule*[lab=\textsc{VT-Scale}]{\valCheckCtx{}\vdash \lenv{0}\colon \lR{} \\ \valCheckCtx{}\vdash \lenv{1}\colon \ltyv{}}{\valCheckCtx{}\vdash \lenv{0} \lvscale \lenv{1}\colon \ltyv{}}}
\and
\gpfit{\inferrule*[lab=\textsc{VT-Vadd}]{\forall i.\; \valCheckCtx{}\vdash \lenv{i}\colon \lsnv{i} \\ \valCheckCtx{}\vdash \lsv{}\colon\svalid}{\valCheckCtx{}\vdash \lenv{0} \lvadd_{\lsv{}} \lenv{1}\colon \lsv{}}}
\and
\gpfit{\inferrule*[lab=\textsc{VT-Dsum\allowbreak E}]{\forall i.\; \valCheckCtx{}\vdash \lenv{i}\colon \lsnv{i}}{\valCheckCtx{}\vdash \lenv{0}\losum \lenv{1}\colon \lsnv{0} \losum \lsnv{1}}}
\and
\gpfit{\inferrule*[lab=\textsc{VT-Tprod\allowbreak E}]{\forall i.\; \valCheckCtx{}\vdash \lenv{i}\colon \lsnv{i}}{\valCheckCtx{}\vdash \lenv{0}\loprod \lenv{1}\colon \lsnv{0} \loprod \lsnv{1}}}
\and
\gpfit{\inferrule*[lab=\textsc{VT-Delta}]{\valCheckCtx{}\vdash \lsv{}\colon\svalid}{\valCheckCtx{}\vdash \ldelta{\mathit{n}}{\lsv{}}\colon \lsv{}}}
\and
\gpfit{\inferrule*[lab=\textsc{VT-Idelta}]{\valCheckCtx{}\vdash \lsv{}\colon\svalid}{\valCheckCtx{}\vdash \lidelta{\lsv{}}\colon \lsv{} \loprod \lsv{}}}
\and
\gpfit{\inferrule*[lab=\textsc{VT-Ttv\allowbreak Space}]{\valCheckCtx{}\vdash \lsnv{0}''\colon\svalid \\ \valCheckCtx{}\vdash \lenv{0}\colon \lsnv{0}' \loprod \lsnv{1} \\ \valCheckCtx{}\vdash \lenv{1}\colon {\ltyv{}_{\mathsf{arg}}}}{\valCheckCtx{}\vdash \LTTVPP{\lsnv{0}''}{\lenv{0}}{\lenv{1}}\colon \lsnv{1}}}
\and
\gpfit{\inferrule*[lab=\textsc{VT-Ttv\allowbreak Real}]{\valCheckCtx{}\vdash \lsnv{0}''\colon\svalid \\ \valCheckCtx{}\vdash \lenv{0}\colon \lsnv{0}' \\ \valCheckCtx{}\vdash \lenv{1}\colon {\ltyv{}_{\mathsf{arg}}} \\ \forall \lsnv{{\mathrm{a}}},\,\lsnv{{\mathrm{b}}}.\; \lsnv{0}' \neq \lsnv{{\mathrm{a}}} \loprod \lsnv{{\mathrm{b}}}}{\valCheckCtx{}\vdash \LTTVPP{\lsnv{0}''}{\lenv{0}}{\lenv{1}}\colon \lR{}}}
\and
\gpfit{\inferrule*[lab=\textsc{VT-Mtm}]{\valCheckCtx{}\vdash \lsnv{1}''\colon\svalid \\ \valCheckCtx{}\vdash \lsnv{1}\colon\svalid \\ \valCheckCtx{}\vdash \lenv{0}\colon \lsnv{1}' \loprod \lsnv{0} \\ \valCheckCtx{}\vdash \lenv{1}\colon \lsnv{2} \loprod \lsnv{1}}{\valCheckCtx{}\vdash \glvmtm{\lsnv{1}''}{\lsnv{1}}{\lenv{0}}{\lenv{1}}\colon \lsnv{2} \loprod \lsnv{0}}}
\and
\gpfit{\inferrule*[lab=\textsc{VT-Vsum}]{\valCheckCtx{}\vdash \lsnv{0}\colon\svalid \\ \forall \leev{}.\; \leev{} \in \mathit{es} \Rightarrow \valCheckCtx{}\vdash \leev{}\colon \lsv{} \\ \valCheckCtx{}\vdash \lsv{}\colon\svalid \\ \mathit{es} = [] \Rightarrow \lsv{} = \lsempty{}}{\valCheckCtx{}\vdash \lsum^{\lsnv{0}}_{v\in \mathit{es}}v\colon \lsv{}}}
\and
\gpfit{\inferrule*[lab=\textsc{VT-Vlam}]{\valCheckCtx{}\vdash (x\colon \ltynv{0}\fnto \ltynv{1})\colon \lkn{0} \fnto \lkn{1} \\ (\valCheckCtx{}, \mathit{x}\colon \ltynv{0})\vdash \leev{}\colon \ltynv{1}}{\valCheckCtx{}\vdash \llambda(x\colon \ltynv{0})\lrh \ltynv{1}\colon \leev{}\colon (x\colon \ltynv{0}\fnto \ltynv{1})}}
\and
\gpfit{\inferrule*[lab=\textsc{VT-App\allowbreak V}]{\valCheckCtx{}\vdash \lenv{0}\colon (x\colon \ltynv{0}\fnto \ltynv{1}) \\ \valCheckCtx{}\vdash \lenv{1}\colon \ltynv{0} \\ {\ltyv{}_{\mathsf{out}}} = \ltynv{1}\subst{x}{\lenv{1}} \\ \valCheckCtx{}\vdash {\ltyv{}_{\mathsf{out}}}\colon \lky{}}{\valCheckCtx{}\vdash \glvapp{\lenv{0}}{\lenv{1}}\colon {\ltyv{}_{\mathsf{out}}}}}
\and
\gpfit{\inferrule*[lab=\textsc{VT-Vkrn}]{\valCheckCtx{}\vdash \leev{}\colon \lsnv{0} \loprod \lsnv{2} \\ \valCheckCtx{}\vdash \leev{}'\colon \lsnv{1} \loprod \lsnv{3}}{\valCheckCtx{}\vdash \vkrn(\leev{}, \leev{}')\colon \left(\lsnv{0} \loprod \lsnv{1}\right) \loprod \left(\lsnv{2} \loprod \lsnv{3}\right)}}
\and
\gpfit{\inferrule*[lab=\textsc{VT-Vblk}]{\valCheckCtx{}\vdash \leev{}\colon \lsnv{0} \loprod \lsnv{2} \\ \valCheckCtx{}\vdash \leev{}'\colon \lsnv{1} \loprod \lsnv{3}}{\valCheckCtx{}\vdash \vblk(\leev{}, \leev{}')\colon \left(\lsnv{0} \losum \lsnv{1}\right) \loprod \left(\lsnv{2} \losum \lsnv{3}\right)}}
\and
\gpfit{\inferrule*[lab=\textsc{VT-Reshap}]{\valCheckCtx{}\vdash \leev{}\colon {\ltyv{}_{\mathsf{in}}} \\ \valCheckCtx{}\vdash \ltynv{0}\colon \lky{} \\ \valCheckCtx{}\vdash \ltyv{}\colon \lky{}}{\valCheckCtx{}\vdash \glvreshape{\leev{}}{\ltyv{}}{\ltynv{0}}\colon \ltyv{}}}
\and
\gpfit{\inferrule*[lab=\textsc{VT-Vtrans}]{\valCheckCtx{}\vdash \leev{}\colon \lsnv{0} \loprod \lsnv{1}}{\valCheckCtx{}\vdash \Trans(\leev{})\colon \lsnv{1} \loprod \lsnv{0}}}
\and
\gpfit{\inferrule*[lab=\textsc{VT-Gsvd\allowbreak S}]{\valCheckCtx{}\vdash \leev{}\colon \lsnv{0} \loprod \lsnv{1} \\ \valCheckCtx{}\vdash \lenv{{\mathrm{a}}}\colon \lsnv{{\mathrm{a0}}} \loprod \lsnv{{\mathrm{a1}}} \\ \valCheckCtx{}\vdash \lenv{{\mathrm{b}}}\colon \lsnv{{\mathrm{b0}}} \loprod \lsnv{{\mathrm{b1}}}}{\valCheckCtx{}\vdash \glvop{GSVDS}(\leev{},\lenv{{\mathrm{a}}},\lenv{{\mathrm{b}}})\colon \lR^{\leev{},\CoKernel{},\lsnv{0} \loprod \lsnv{1}} \loprod \lR^{\leev{},\Image{},\lsnv{0} \loprod \lsnv{1}}}}
\and
\gpfit{\inferrule*[lab=\textsc{VT-Gsvd\allowbreak U}]{\valCheckCtx{}\vdash \leev{}\colon \lsnv{0} \loprod \lsnv{1} \\ \valCheckCtx{}\vdash \lenv{{\mathrm{a}}}\colon \lsnv{{\mathrm{a0}}} \loprod \lsnv{{\mathrm{a1}}} \\ \valCheckCtx{}\vdash \lenv{{\mathrm{b}}}\colon \lsnv{{\mathrm{b0}}} \loprod \lsnv{{\mathrm{b1}}}}{\valCheckCtx{}\vdash \glvop{GSVDU}(\leev{},\lenv{{\mathrm{a}}},\lenv{{\mathrm{b}}})\colon \lsnv{0} \loprod \lR^{\leev{},\CoKernel{},\lsnv{0} \loprod \lsnv{1}}}}
\and
\gpfit{\inferrule*[lab=\textsc{VT-Gsvd\allowbreak UI}]{\valCheckCtx{}\vdash \leev{}\colon \lsnv{0} \loprod \lsnv{1} \\ \valCheckCtx{}\vdash \lenv{{\mathrm{a}}}\colon \lsnv{{\mathrm{a0}}} \loprod \lsnv{{\mathrm{a1}}} \\ \valCheckCtx{}\vdash \lenv{{\mathrm{b}}}\colon \lsnv{{\mathrm{b0}}} \loprod \lsnv{{\mathrm{b1}}}}{\valCheckCtx{}\vdash \glvop{GSVDUI}(\leev{},\lenv{{\mathrm{a}}},\lenv{{\mathrm{b}}})\colon \lsnv{0} \loprod \lR^{\leev{},\Kernel{},\lsnv{0} \loprod \lsnv{1}}}}
\and
\gpfit{\inferrule*[lab=\textsc{VT-Gsvd\allowbreak V}]{\valCheckCtx{}\vdash \leev{}\colon \lsnv{0} \loprod \lsnv{1} \\ \valCheckCtx{}\vdash \lenv{{\mathrm{a}}}\colon \lsnv{{\mathrm{a0}}} \loprod \lsnv{{\mathrm{a1}}} \\ \valCheckCtx{}\vdash \lenv{{\mathrm{b}}}\colon \lsnv{{\mathrm{b0}}} \loprod \lsnv{{\mathrm{b1}}}}{\valCheckCtx{}\vdash \glvop{GSVDV}(\leev{},\lenv{{\mathrm{a}}},\lenv{{\mathrm{b}}})\colon \lsnv{1} \loprod \lR^{\leev{},\Image{},\lsnv{0} \loprod \lsnv{1}}}}
\and
\gpfit{\inferrule*[lab=\textsc{VT-Gsvd\allowbreak VI}]{\valCheckCtx{}\vdash \leev{}\colon \lsnv{0} \loprod \lsnv{1} \\ \valCheckCtx{}\vdash \lenv{{\mathrm{a}}}\colon \lsnv{{\mathrm{a0}}} \loprod \lsnv{{\mathrm{a1}}} \\ \valCheckCtx{}\vdash \lenv{{\mathrm{b}}}\colon \lsnv{{\mathrm{b0}}} \loprod \lsnv{{\mathrm{b1}}}}{\valCheckCtx{}\vdash \glvop{GSVDVI}(\leev{},\lenv{{\mathrm{a}}},\lenv{{\mathrm{b}}})\colon \lsnv{1} \loprod \lR^{\leev{},\CoImage{},\lsnv{0} \loprod \lsnv{1}}}}
\and
\gpfit{\inferrule*[lab=\textsc{VT-Pinv\allowbreak Tensor}]{\valCheckCtx{}\vdash \leev{}\colon \lsnv{0} \loprod \lsnv{1}}{\valCheckCtx{}\vdash \leev{}^{+}\colon \lsnv{1} \loprod \lsnv{0}}}
\and
\gpfit{\inferrule*[lab=\textsc{VT-Pinv\allowbreak Raw}]{\valCheckCtx{}\vdash \leev{}\colon \lsnv{0} \\ \forall \lsnv{{\mathrm{a}}},\,\lsnv{{\mathrm{b}}}.\; \lsnv{0} \neq \lsnv{{\mathrm{a}}} \loprod \lsnv{{\mathrm{b}}}}{\valCheckCtx{}\vdash \leev{}^{+}\colon \lR{} \loprod \lsnv{0}}}
\and
\gpfit{\inferrule*[lab=\textsc{VT-Gpinv}]{\valCheckCtx{}\vdash \leev{}\colon \lsnv{0} \loprod \lsnv{1} \\ \valCheckCtx{}\vdash \lenv{{\mathrm{a}}}\colon \lsnv{{\mathrm{a0}}} \loprod \lsnv{{\mathrm{a1}}} \\ \valCheckCtx{}\vdash \lenv{{\mathrm{b}}}\colon \lsnv{{\mathrm{b0}}} \loprod \lsnv{{\mathrm{b1}}}}{\valCheckCtx{}\vdash \glvop{GPINV}(\leev{},\lenv{{\mathrm{a}}},\lenv{{\mathrm{b}}})\colon \lsnv{1} \loprod \lsnv{0}}}
\and
\gpfit{\inferrule*[lab=\textsc{VT-Rref\allowbreak Rep}]{\valCheckCtx{}\vdash \lenv{0}\colon \lsv{} \\ \valCheckCtx{}\vdash \lenv{1}\colon \lsnv{{\mathrm{p0}}} \loprod \lsnv{{\mathrm{p1}}}}{\valCheckCtx{}\vdash \glvop{RREFRep}(\lenv{0},\lenv{1})\colon \lsnv{{\mathrm{p1}}} \loprod \lsnv{{\mathrm{p0}}}}}
\and
\gpfit{\inferrule*[lab=\textsc{VT-Inv\allowbreak Tensor}]{\valCheckCtx{}\vdash \leev{}\colon \lsnv{0} \loprod \lsnv{1}}{\valCheckCtx{}\vdash \leev{}^{-1}\colon \lsnv{1} \loprod \lsnv{0}}}
\and
\gpfit{\inferrule*[lab=\textsc{VT-Inv\allowbreak Raw}]{\valCheckCtx{}\vdash \leev{}\colon \lsnv{0} \\ \forall \lsnv{{\mathrm{a}}},\,\lsnv{{\mathrm{b}}}.\; \lsnv{0} \neq \lsnv{{\mathrm{a}}} \loprod \lsnv{{\mathrm{b}}}}{\valCheckCtx{}\vdash \leev{}^{-1}\colon \lR{} \loprod \lsnv{0}}}
\and
\gpfit{\inferrule*[lab=\textsc{VT-Svds}]{\valCheckCtx{}\vdash \leev{}\colon \lsnv{0} \loprod \lsnv{1}}{\valCheckCtx{}\vdash \glvop{SVDS}(\leev{})\colon \lR^{\leev{},\CoKernel{},\lsnv{0} \loprod \lsnv{1}} \loprod \lR^{\leev{},\Image{},\lsnv{0} \loprod \lsnv{1}}}}
}
\gldefjudgment{evalS}{\evalSJudgeK{\ls{}}{\lsv{}}}{Space evaluation (generated from \texttt{GLang.ValSpace.EvalS}).}{fig:gl-evals}
\gldefrule{evalS.real}{evalS}{\textsc{ES-Real}}{\phantom{.}}{\evalSJudge{\evalCtx{}}{\lR{}}{\lR{}}}{\texttt{GLang.Val\allowbreak Space.Eval\allowbreak S.real}}{GLang/ValueOperational.lean:289}
\gldefruleexplanation{evalS.real}{The source scalar space evaluates to the scalar value space.}
\gldefrule{evalS.empty}{evalS}{\textsc{ES-Empty}}{\phantom{.}}{\evalSJudge{\evalCtx{}}{\lsempty{}}{\lsempty{}}}{\texttt{GLang.Val\allowbreak Space.Eval\allowbreak S.empty}}{GLang/ValueOperational.lean:291}
\gldefruleexplanation{evalS.empty}{The source zero-dimensional space evaluates to the zero-dimensional value space.}
\gldefrule{evalS.dual}{evalS}{\textsc{ES-Dual}}{\evalSJudge{\evalCtx{}}{\ls{}}{\lsv{}}}{\evalSJudge{\evalCtx{}}{\ls{}^{\lstar}}{\lsv{}}}{\texttt{GLang.Val\allowbreak Space.Eval\allowbreak S.dual}}{GLang/ValueOperational.lean:293}
\gldefruleexplanation{evalS.dual}{A dual space uses the same value-space representation as its underlying space.}
\gldefrule{evalS.ker}{evalS}{\textsc{ES-Ker}}{\evalJudge{\evalCtx{}}{\lee{}}{\lb{}}{\leev{}} \\ \sof{\lb{}} = \lsn{0}^{\lstar} \loprod \lsn{1} \\ \exists\,\lbp{},\textcolor{lang}{b^{\prime\prime}}.\; \lb{} = \lbp{}^{\lstar}\loprod \textcolor{lang}{b^{\prime\prime}} \\ \evalSJudge{\evalCtx{}}{\sof{\lb{}}}{\lsnv{{\mathrm{svb}}}}}{\evalSJudge{\evalCtx{}}{\Kernel(\lee{})}{\lR^{\glvreshape{\leev{}}{\lsnv{{\mathrm{svb}}}}{\lsnv{{\mathrm{svb}}}},\Kernel{},\lsnv{{\mathrm{svb}}}}}}{\texttt{GLang.Val\allowbreak Space.Eval\allowbreak S.ker}}{GLang/ValueOperational.lean:297}
\gldefruleexplanation{evalS.ker}{Evaluating the defining map in a compatible matrix basis and reshaping it to the evaluated matrix space produces the kernel coordinate space.}
\gldefrule{evalS.coker}{evalS}{\textsc{ES-Coker}}{\evalJudge{\evalCtx{}}{\lee{}}{\lb{}}{\leev{}} \\ \sof{\lb{}} = \lsn{0}^{\lstar} \loprod \lsn{1} \\ \exists\,\lbp{},\textcolor{lang}{b^{\prime\prime}}.\; \lb{} = \lbp{}^{\lstar}\loprod \textcolor{lang}{b^{\prime\prime}} \\ \evalSJudge{\evalCtx{}}{\sof{\lb{}}}{\lsnv{{\mathrm{svb}}}}}{\evalSJudge{\evalCtx{}}{\CoKernel(\lee{})}{\lR^{\glvreshape{\leev{}}{\lsnv{{\mathrm{svb}}}}{\lsnv{{\mathrm{svb}}}},\CoKernel{},\lsnv{{\mathrm{svb}}}}}}{\texttt{GLang.Val\allowbreak Space.Eval\allowbreak S.coker}}{GLang/ValueOperational.lean:312}
\gldefruleexplanation{evalS.coker}{Evaluating the defining map in a compatible matrix basis and reshaping it to the evaluated matrix space produces the cokernel coordinate space.}
\gldefrule{evalS.img}{evalS}{\textsc{ES-Img}}{\evalJudge{\evalCtx{}}{\lee{}}{\lb{}}{\leev{}} \\ \sof{\lb{}} = \lsn{0}^{\lstar} \loprod \lsn{1} \\ \exists\,\lbp{},\textcolor{lang}{b^{\prime\prime}}.\; \lb{} = \lbp{}^{\lstar}\loprod \textcolor{lang}{b^{\prime\prime}} \\ \evalSJudge{\evalCtx{}}{\sof{\lb{}}}{\lsnv{{\mathrm{svb}}}}}{\evalSJudge{\evalCtx{}}{\Image(\lee{})}{\lR^{\glvreshape{\leev{}}{\lsnv{{\mathrm{svb}}}}{\lsnv{{\mathrm{svb}}}},\Image{},\lsnv{{\mathrm{svb}}}}}}{\texttt{GLang.Val\allowbreak Space.Eval\allowbreak S.img}}{GLang/ValueOperational.lean:321}
\gldefruleexplanation{evalS.img}{Evaluating the defining map in a compatible matrix basis and reshaping it to the evaluated matrix space produces the image coordinate space.}
\gldefrule{evalS.coimg}{evalS}{\textsc{ES-Coimg}}{\evalJudge{\evalCtx{}}{\lee{}}{\lb{}}{\leev{}} \\ \sof{\lb{}} = \lsn{0}^{\lstar} \loprod \lsn{1} \\ \exists\,\lbp{},\textcolor{lang}{b^{\prime\prime}}.\; \lb{} = \lbp{}^{\lstar}\loprod \textcolor{lang}{b^{\prime\prime}} \\ \evalSJudge{\evalCtx{}}{\sof{\lb{}}}{\lsnv{{\mathrm{svb}}}}}{\evalSJudge{\evalCtx{}}{\CoImage(\lee{})}{\lR^{\glvreshape{\leev{}}{\lsnv{{\mathrm{svb}}}}{\lsnv{{\mathrm{svb}}}},\CoImage{},\lsnv{{\mathrm{svb}}}}}}{\texttt{GLang.Val\allowbreak Space.Eval\allowbreak S.coimg}}{GLang/ValueOperational.lean:330}
\gldefruleexplanation{evalS.coimg}{Evaluating the defining map in a compatible matrix basis and reshaping it to the evaluated matrix space produces the coimage coordinate space.}
\gldefrule{evalS.tprod}{evalS}{\textsc{ES-Tprod}}{\forall i.\; \evalSJudge{\evalCtx{}}{\lsn{i}}{\lsnv{i}}}{\evalSJudge{\evalCtx{}}{\lsn{0} \loprod \lsn{1}}{\lsnv{0} \loprod \lsnv{1}}}{\texttt{GLang.Val\allowbreak Space.Eval\allowbreak S.tprod}}{GLang/ValueOperational.lean:339}
\gldefruleexplanation{evalS.tprod}{A tensor-product space evaluates by tensoring the value-space representations of its factors.}
\gldefrule{evalS.dsum}{evalS}{\textsc{ES-Dsum}}{\forall i.\; \evalSJudge{\evalCtx{}}{\lsn{i}}{\lsnv{i}}}{\evalSJudge{\evalCtx{}}{\lsn{0} \losum \lsn{1}}{\lsnv{0} \losum \lsnv{1}}}{\texttt{GLang.Val\allowbreak Space.Eval\allowbreak S.dsum}}{GLang/ValueOperational.lean:344}
\gldefruleexplanation{evalS.dsum}{A direct-sum space evaluates by summing the value-space representations of its factors.}
\gldefrulerows{evalS}{
\textsc{ES-Real} & $\evalSJudge{\evalCtx{}}{\lR{}}{\lR{}}$ & \texttt{\footnotesize GLang/Value\allowbreak Operational.lean:289}\\
\textsc{ES-Empty} & $\evalSJudge{\evalCtx{}}{\lsempty{}}{\lsempty{}}$ & \texttt{\footnotesize GLang/Value\allowbreak Operational.lean:291}\\
\textsc{ES-Dual} & $\evalSJudge{\evalCtx{}}{\ls{}^{\lstar}}{\lsv{}}$ & \texttt{\footnotesize GLang/Value\allowbreak Operational.lean:293}\\
\textsc{ES-Ker} & $\evalSJudge{\evalCtx{}}{\Kernel(\lee{})}{\lR^{\glvreshape{\leev{}}{\lsnv{{\mathrm{svb}}}}{\lsnv{{\mathrm{svb}}}},\Kernel{},\lsnv{{\mathrm{svb}}}}}$ & \texttt{\footnotesize GLang/Value\allowbreak Operational.lean:297}\\
\textsc{ES-Coker} & $\evalSJudge{\evalCtx{}}{\CoKernel(\lee{})}{\lR^{\glvreshape{\leev{}}{\lsnv{{\mathrm{svb}}}}{\lsnv{{\mathrm{svb}}}},\CoKernel{},\lsnv{{\mathrm{svb}}}}}$ & \texttt{\footnotesize GLang/Value\allowbreak Operational.lean:312}\\
\textsc{ES-Img} & $\evalSJudge{\evalCtx{}}{\Image(\lee{})}{\lR^{\glvreshape{\leev{}}{\lsnv{{\mathrm{svb}}}}{\lsnv{{\mathrm{svb}}}},\Image{},\lsnv{{\mathrm{svb}}}}}$ & \texttt{\footnotesize GLang/Value\allowbreak Operational.lean:321}\\
\textsc{ES-Coimg} & $\evalSJudge{\evalCtx{}}{\CoImage(\lee{})}{\lR^{\glvreshape{\leev{}}{\lsnv{{\mathrm{svb}}}}{\lsnv{{\mathrm{svb}}}},\CoImage{},\lsnv{{\mathrm{svb}}}}}$ & \texttt{\footnotesize GLang/Value\allowbreak Operational.lean:330}\\
\textsc{ES-Tprod} & $\evalSJudge{\evalCtx{}}{\lsn{0} \loprod \lsn{1}}{\lsnv{0} \loprod \lsnv{1}}$ & \texttt{\footnotesize GLang/Value\allowbreak Operational.lean:339}\\
\textsc{ES-Dsum} & $\evalSJudge{\evalCtx{}}{\lsn{0} \losum \lsn{1}}{\lsnv{0} \losum \lsnv{1}}$ & \texttt{\footnotesize GLang/Value\allowbreak Operational.lean:344}\\
}{
\gpfit{\inferrule*[]{\phantom{.}}{\evalSJudge{\evalCtx{}}{\lR{}}{\lR{}}}}
\and
\gpfit{\inferrule*[]{\phantom{.}}{\evalSJudge{\evalCtx{}}{\lsempty{}}{\lsempty{}}}}
\and
\gpfit{\inferrule*[]{\evalSJudge{\evalCtx{}}{\ls{}}{\lsv{}}}{\evalSJudge{\evalCtx{}}{\ls{}^{\lstar}}{\lsv{}}}}
\and
\gpfit{\inferrule*[]{\evalJudge{\evalCtx{}}{\lee{}}{\lb{}}{\leev{}} \\ \sof{\lb{}} = \lsn{0}^{\lstar} \loprod \lsn{1} \\ \exists\,\lbp{},\textcolor{lang}{b^{\prime\prime}}.\; \lb{} = \lbp{}^{\lstar}\loprod \textcolor{lang}{b^{\prime\prime}} \\ \evalSJudge{\evalCtx{}}{\sof{\lb{}}}{\lsnv{{\mathrm{svb}}}}}{\evalSJudge{\evalCtx{}}{\Kernel(\lee{})}{\lR^{\glvreshape{\leev{}}{\lsnv{{\mathrm{svb}}}}{\lsnv{{\mathrm{svb}}}},\Kernel{},\lsnv{{\mathrm{svb}}}}}}}
\and
\gpfit{\inferrule*[]{\evalJudge{\evalCtx{}}{\lee{}}{\lb{}}{\leev{}} \\ \sof{\lb{}} = \lsn{0}^{\lstar} \loprod \lsn{1} \\ \exists\,\lbp{},\textcolor{lang}{b^{\prime\prime}}.\; \lb{} = \lbp{}^{\lstar}\loprod \textcolor{lang}{b^{\prime\prime}} \\ \evalSJudge{\evalCtx{}}{\sof{\lb{}}}{\lsnv{{\mathrm{svb}}}}}{\evalSJudge{\evalCtx{}}{\CoKernel(\lee{})}{\lR^{\glvreshape{\leev{}}{\lsnv{{\mathrm{svb}}}}{\lsnv{{\mathrm{svb}}}},\CoKernel{},\lsnv{{\mathrm{svb}}}}}}}
\and
\gpfit{\inferrule*[]{\evalJudge{\evalCtx{}}{\lee{}}{\lb{}}{\leev{}} \\ \sof{\lb{}} = \lsn{0}^{\lstar} \loprod \lsn{1} \\ \exists\,\lbp{},\textcolor{lang}{b^{\prime\prime}}.\; \lb{} = \lbp{}^{\lstar}\loprod \textcolor{lang}{b^{\prime\prime}} \\ \evalSJudge{\evalCtx{}}{\sof{\lb{}}}{\lsnv{{\mathrm{svb}}}}}{\evalSJudge{\evalCtx{}}{\Image(\lee{})}{\lR^{\glvreshape{\leev{}}{\lsnv{{\mathrm{svb}}}}{\lsnv{{\mathrm{svb}}}},\Image{},\lsnv{{\mathrm{svb}}}}}}}
\and
\gpfit{\inferrule*[]{\evalJudge{\evalCtx{}}{\lee{}}{\lb{}}{\leev{}} \\ \sof{\lb{}} = \lsn{0}^{\lstar} \loprod \lsn{1} \\ \exists\,\lbp{},\textcolor{lang}{b^{\prime\prime}}.\; \lb{} = \lbp{}^{\lstar}\loprod \textcolor{lang}{b^{\prime\prime}} \\ \evalSJudge{\evalCtx{}}{\sof{\lb{}}}{\lsnv{{\mathrm{svb}}}}}{\evalSJudge{\evalCtx{}}{\CoImage(\lee{})}{\lR^{\glvreshape{\leev{}}{\lsnv{{\mathrm{svb}}}}{\lsnv{{\mathrm{svb}}}},\CoImage{},\lsnv{{\mathrm{svb}}}}}}}
\and
\gpfit{\inferrule*[]{\forall i.\; \evalSJudge{\evalCtx{}}{\lsn{i}}{\lsnv{i}}}{\evalSJudge{\evalCtx{}}{\lsn{0} \loprod \lsn{1}}{\lsnv{0} \loprod \lsnv{1}}}}
\and
\gpfit{\inferrule*[]{\forall i.\; \evalSJudge{\evalCtx{}}{\lsn{i}}{\lsnv{i}}}{\evalSJudge{\evalCtx{}}{\lsn{0} \losum \lsn{1}}{\lsnv{0} \losum \lsnv{1}}}}
}{
\gpfit{\inferrule*[lab=\textsc{ES-Real}]{\phantom{.}}{\evalSJudge{\evalCtx{}}{\lR{}}{\lR{}}}}
\and
\gpfit{\inferrule*[lab=\textsc{ES-Empty}]{\phantom{.}}{\evalSJudge{\evalCtx{}}{\lsempty{}}{\lsempty{}}}}
\and
\gpfit{\inferrule*[lab=\textsc{ES-Dual}]{\evalSJudge{\evalCtx{}}{\ls{}}{\lsv{}}}{\evalSJudge{\evalCtx{}}{\ls{}^{\lstar}}{\lsv{}}}}
\and
\gpfit{\inferrule*[lab=\textsc{ES-Ker}]{\evalJudge{\evalCtx{}}{\lee{}}{\lb{}}{\leev{}} \\ \sof{\lb{}} = \lsn{0}^{\lstar} \loprod \lsn{1} \\ \exists\,\lbp{},\textcolor{lang}{b^{\prime\prime}}.\; \lb{} = \lbp{}^{\lstar}\loprod \textcolor{lang}{b^{\prime\prime}} \\ \evalSJudge{\evalCtx{}}{\sof{\lb{}}}{\lsnv{{\mathrm{svb}}}}}{\evalSJudge{\evalCtx{}}{\Kernel(\lee{})}{\lR^{\glvreshape{\leev{}}{\lsnv{{\mathrm{svb}}}}{\lsnv{{\mathrm{svb}}}},\Kernel{},\lsnv{{\mathrm{svb}}}}}}}
\and
\gpfit{\inferrule*[lab=\textsc{ES-Coker}]{\evalJudge{\evalCtx{}}{\lee{}}{\lb{}}{\leev{}} \\ \sof{\lb{}} = \lsn{0}^{\lstar} \loprod \lsn{1} \\ \exists\,\lbp{},\textcolor{lang}{b^{\prime\prime}}.\; \lb{} = \lbp{}^{\lstar}\loprod \textcolor{lang}{b^{\prime\prime}} \\ \evalSJudge{\evalCtx{}}{\sof{\lb{}}}{\lsnv{{\mathrm{svb}}}}}{\evalSJudge{\evalCtx{}}{\CoKernel(\lee{})}{\lR^{\glvreshape{\leev{}}{\lsnv{{\mathrm{svb}}}}{\lsnv{{\mathrm{svb}}}},\CoKernel{},\lsnv{{\mathrm{svb}}}}}}}
\and
\gpfit{\inferrule*[lab=\textsc{ES-Img}]{\evalJudge{\evalCtx{}}{\lee{}}{\lb{}}{\leev{}} \\ \sof{\lb{}} = \lsn{0}^{\lstar} \loprod \lsn{1} \\ \exists\,\lbp{},\textcolor{lang}{b^{\prime\prime}}.\; \lb{} = \lbp{}^{\lstar}\loprod \textcolor{lang}{b^{\prime\prime}} \\ \evalSJudge{\evalCtx{}}{\sof{\lb{}}}{\lsnv{{\mathrm{svb}}}}}{\evalSJudge{\evalCtx{}}{\Image(\lee{})}{\lR^{\glvreshape{\leev{}}{\lsnv{{\mathrm{svb}}}}{\lsnv{{\mathrm{svb}}}},\Image{},\lsnv{{\mathrm{svb}}}}}}}
\and
\gpfit{\inferrule*[lab=\textsc{ES-Coimg}]{\evalJudge{\evalCtx{}}{\lee{}}{\lb{}}{\leev{}} \\ \sof{\lb{}} = \lsn{0}^{\lstar} \loprod \lsn{1} \\ \exists\,\lbp{},\textcolor{lang}{b^{\prime\prime}}.\; \lb{} = \lbp{}^{\lstar}\loprod \textcolor{lang}{b^{\prime\prime}} \\ \evalSJudge{\evalCtx{}}{\sof{\lb{}}}{\lsnv{{\mathrm{svb}}}}}{\evalSJudge{\evalCtx{}}{\CoImage(\lee{})}{\lR^{\glvreshape{\leev{}}{\lsnv{{\mathrm{svb}}}}{\lsnv{{\mathrm{svb}}}},\CoImage{},\lsnv{{\mathrm{svb}}}}}}}
\and
\gpfit{\inferrule*[lab=\textsc{ES-Tprod}]{\forall i.\; \evalSJudge{\evalCtx{}}{\lsn{i}}{\lsnv{i}}}{\evalSJudge{\evalCtx{}}{\lsn{0} \loprod \lsn{1}}{\lsnv{0} \loprod \lsnv{1}}}}
\and
\gpfit{\inferrule*[lab=\textsc{ES-Dsum}]{\forall i.\; \evalSJudge{\evalCtx{}}{\lsn{i}}{\lsnv{i}}}{\evalSJudge{\evalCtx{}}{\lsn{0} \losum \lsn{1}}{\lsnv{0} \losum \lsnv{1}}}}
}
\gldefjudgment{evalT}{\evalSJudgeK{\lty{}}{\ltyv{}}}{Type evaluation (generated from \texttt{GLang.ValTy.EvalT}).}{fig:gl-evalt}
\gldefrule{evalT.space}{evalT}{\textsc{ET-Space}}{\evalSJudge{\evalCtx{}}{\ls{}}{\lsv{}}}{\evalSJudge{\evalCtx{}}{\ls{}}{\lsv{}}}{\texttt{GLang.Val\allowbreak Ty.Eval\allowbreak T.space}}{GLang/ValueOperational.lean:358}
\gldefruleexplanation{evalT.space}{A source space used as a type evaluates to its value-space representation.}
\gldefrule{evalT.piTy}{evalT}{\textsc{ET-Pi\allowbreak Ty}}{\evalSJudge{\evalCtx{}}{\ltyn{0}}{\ltynv{0}} \\ \tof{\lbn{0}} = \ltyn{0} \\ \evalSJudge{\evalCtx{}[(\mathit{x},\lbn{0}) \mapsto \mathit{x}_{\ltynv{0}}]}{\ltyn{1}}{\ltynv{1}}}{\evalSJudge{\evalCtx{}}{(x\colon \ltyn{0}\fnto \ltyn{1})}{(x\colon \ltynv{0}\fnto \ltynv{1})}}{\texttt{GLang.Val\allowbreak Ty.Eval\allowbreak T.pi\allowbreak Ty}}{GLang/ValueOperational.lean:362}
\gldefruleexplanation{evalT.piTy}{A dependent function type evaluates its domain and then its codomain with a fresh argument stored at a basis of the domain type.}
\gldefrulerows{evalT}{
\textsc{ET-Space} & $\evalSJudge{\evalCtx{}}{\ls{}}{\lsv{}}$ & \texttt{\footnotesize GLang/Value\allowbreak Operational.lean:358}\\
\textsc{ET-Pi\allowbreak Ty} & $\evalSJudge{\evalCtx{}}{(x\colon \ltyn{0}\fnto \ltyn{1})}{(x\colon \ltynv{0}\fnto \ltynv{1})}$ & \texttt{\footnotesize GLang/Value\allowbreak Operational.lean:362}\\
}{
\gpfit{\inferrule*[]{\evalSJudge{\evalCtx{}}{\ls{}}{\lsv{}}}{\evalSJudge{\evalCtx{}}{\ls{}}{\lsv{}}}}
\and
\gpfit{\inferrule*[]{\evalSJudge{\evalCtx{}}{\ltyn{0}}{\ltynv{0}} \\ \tof{\lbn{0}} = \ltyn{0} \\ \evalSJudge{\evalCtx{}[(\mathit{x},\lbn{0}) \mapsto \mathit{x}_{\ltynv{0}}]}{\ltyn{1}}{\ltynv{1}}}{\evalSJudge{\evalCtx{}}{(x\colon \ltyn{0}\fnto \ltyn{1})}{(x\colon \ltynv{0}\fnto \ltynv{1})}}}
}{
\gpfit{\inferrule*[lab=\textsc{ET-Space}]{\evalSJudge{\evalCtx{}}{\ls{}}{\lsv{}}}{\evalSJudge{\evalCtx{}}{\ls{}}{\lsv{}}}}
\and
\gpfit{\inferrule*[lab=\textsc{ET-Pi\allowbreak Ty}]{\evalSJudge{\evalCtx{}}{\ltyn{0}}{\ltynv{0}} \\ \tof{\lbn{0}} = \ltyn{0} \\ \evalSJudge{\evalCtx{}[(\mathit{x},\lbn{0}) \mapsto \mathit{x}_{\ltynv{0}}]}{\ltyn{1}}{\ltynv{1}}}{\evalSJudge{\evalCtx{}}{(x\colon \ltyn{0}\fnto \ltyn{1})}{(x\colon \ltynv{0}\fnto \ltynv{1})}}}
}
\gldefjudgment{evalBB}{\evalBBJudgeK{\lbn{0}}{\leev{}}{\lbn{1}}}{Change-of-basis evaluation (generated from \texttt{GLang.Basis.EvalBB}).}{fig:gl-evalbb}
\gldefrule{evalBB.inverse}{evalBB}{\textsc{EB-Inverse}}{\evalBBJudge{\evalCtx{}}{\lbn{1}}{\leev{}}{\lbn{0}}}{\evalBBJudge{\evalCtx{}}{\lbn{0}}{\leev{}^{-1}}{\lbn{1}}}{\texttt{GLang.Basis.Eval\allowbreak BB.inverse}}{GLang/ValueOperational.lean:408}
\gldefruleexplanation{evalBB.inverse}{Reversing a basis conversion applies the matrix inverse to its conversion value.}
\gldefrule{evalBB.trans}{evalBB}{\textsc{EB-Trans}}{\evalBBJudge{\evalCtx{}}{\lbn{0}}{\lenv{0}}{\lbn{1}} \\ \evalBBJudge{\evalCtx{}}{\lbn{1}}{\lenv{1}}{\lbn{2}} \\ \evalSJudge{\evalCtx{}}{\sof{\lbn{0}}}{\lsv{}}}{\evalBBJudge{\evalCtx{}}{\lbn{0}}{\glvmtm{\lsv{}}{\lsv{}}{\lenv{1}}{\glvreshape{\lenv{0}}{\lsv{} \loprod \lsv{}}{\lsv{} \loprod \lsv{}}}}{\lbn{2}}}{\texttt{GLang.Basis.Eval\allowbreak BB.trans}}{GLang/ValueOperational.lean:413}
\gldefruleexplanation{evalBB.trans}{Two successive basis conversions compose by matrix contraction after reshaping the first conversion to the evaluated space.}
\gldefrule{evalBB.tprod}{evalBB}{\textsc{EB-Tprod}}{\forall i.\; \evalBBJudge{\evalCtx{}}{\lbn{i}'}{\lenv{i}}{\lbn{i}}}{\evalBBJudge{\evalCtx{}}{\lbn{0}' \loprod \lbn{1}'}{\vkrn(\lenv{0}, \lenv{1})}{\lbn{0} \loprod \lbn{1}}}{\texttt{GLang.Basis.Eval\allowbreak BB.tprod}}{GLang/ValueOperational.lean:432}
\gldefruleexplanation{evalBB.tprod}{Component basis conversions combine into a tensor-product conversion by the Kronecker product.}
\gldefrule{evalBB.dsum}{evalBB}{\textsc{EB-Dsum}}{\forall i.\; \evalBBJudge{\evalCtx{}}{\lbn{i}'}{\lenv{i}}{\lbn{i}}}{\evalBBJudge{\evalCtx{}}{\lbn{0}' \losum \lbn{1}'}{\vblk(\lenv{0}, \lenv{1})}{\lbn{0} \losum \lbn{1}}}{\texttt{GLang.Basis.Eval\allowbreak BB.dsum}}{GLang/ValueOperational.lean:445}
\gldefruleexplanation{evalBB.dsum}{Component basis conversions combine into a direct-sum conversion by a block-diagonal matrix.}
\gldefrule{evalBB.dualStd}{evalBB}{\textsc{EB-Dual\allowbreak Std}}{\tof{\lbn{0}} = \sof{\lbn{0}} \\ \evalBBJudge{\evalCtx{}}{\lbn{0}}{\leev{}}{\stdBasisDCtx{\tof{\lbn{0}}}}}{\evalBBJudge{\evalCtx{}}{\lbn{0}^{\lstar}}{\Trans(\leev{}^{-1})}{\stdBasisDCtx{\tof{\lbn{0}^{\lstar}}}}}{\texttt{GLang.Basis.Eval\allowbreak BB.dual\_\allowbreak std}}{GLang/ValueOperational.lean:451}
\gldefruleexplanation{evalBB.dualStd}{Converting a dual basis to the standard dual basis uses the inverse transpose of the original basis's conversion to standard coordinates.}
\gldefrule{evalBB.image}{evalBB}{\textsc{EB-Image}}{\exists \lsn{0}.\; \exists \lsn{1}.\; \exists \lee{}.\; \sof{\lbn{0}} = \CoKernel(\lee{}) \\ \evalBBJudge{\evalCtx{}}{\lbn{0}}{\leev{}}{\lbn{0}'}}{\evalBBJudge{\evalCtx{}}{\iBasisC{\lbn{0}}}{\leev{}}{\iBasisC{\lbn{0}'}}}{\texttt{GLang.Basis.Eval\allowbreak BB.image}}{GLang/ValueOperational.lean:463}
\gldefruleexplanation{evalBB.image}{A conversion between cokernel bases also converts their image bases using the same coordinate matrix.}
\gldefrule{evalBB.imageSvds}{evalBB}{\textsc{EB-Image\allowbreak Svds}}{\forall i.\; \sof{\lbn{i}} = \lsn{i} \\ \forall i.\; \isStdBasis{\lbn{i}} \\ \evalJudge{\evalCtx{}}{\lee{}}{\lbn{0}^{\lstar} \loprod \lbn{1}}{\leev{}} \\ \evalSJudge{\evalCtx{}}{\sof{\lbn{0}^{\lstar} \loprod \lbn{1}}}{\lsnv{0} \loprod \lsnv{1}} \\ \lenv{{\mathrm{M}}} = \glvreshape{\leev{}}{\lsnv{0} \loprod \lsnv{1}}{\lsnv{0} \loprod \lsnv{1}} \\ \lenv{{\mathrm{U}}} = \glvop{GSVDU}(\lenv{{\mathrm{M}}},\lidelta{\lsnv{0}},\lidelta{\lsnv{1}}) \\ \lenv{{\mathrm{V}}} = \glvop{GSVDV}(\lenv{{\mathrm{M}}},\lidelta{\lsnv{0}},\lidelta{\lsnv{1}}) \\ \lenv{{\mathrm{P}}} = \glvmtm{\lsnv{1}}{\lsnv{1}}{\lenv{{\mathrm{V}}}}{\glvmtm{\lsnv{0}}{\lsnv{0}}{\lenv{{\mathrm{M}}}}{\Trans(\lenv{{\mathrm{U}}})}}}{\begin{gathered}\evalBBJudge{\evalCtx{}}{\iBasisC{\pBasisC{\CoKernel(\lee{})}{\lbn{0}}{\lbn{1}}}}{\widehat{\leev{}}}{\pBasisC{\Image(\lee{})}{\lbn{0}}{\lbn{1}}}\\[-1pt]\widehat{\leev{}}=\glvreshape{\lenv{{\mathrm{P}}}}{\lR^{\lenv{{\mathrm{M}}},\CoKernel{},\lsnv{0} \loprod \lsnv{1}} \loprod \lR^{\lenv{{\mathrm{M}}},\Image{},\lsnv{0} \loprod \lsnv{1}}}{\operatorname{inferSpace}(\lenv{{\mathrm{P}}})}\end{gathered}}{\texttt{GLang.Basis.Eval\allowbreak BB.image\_\allowbreak svds}}{GLang/ValueOperational.lean:474}
\gldefruleexplanation{evalBB.imageSvds}{In standard ambient bases, the evaluated map and its singular-vector blocks convert the image of the projected cokernel basis to the projected image basis.}
\gldefrule{evalBB.projKerCoker}{evalBB}{\textsc{EB-Proj\allowbreak Ker\allowbreak Coker}}{\ls{} = \Kernel(\lee{}) \;\vee\; \ls{} = \CoKernel(\lee{}) \\ \forall i.\; \sof{\lbn{i}} = \lsn{i} \\ \forall i.\; \evalSJudge{\evalCtx{}}{\sof{\lbn{i}}}{\lsnv{i}} \\ \evalSJudge{\evalCtx{}}{\ls{}}{\lsv{}} \\ \evalSJudge{\evalCtx{}}{\ls{}}{\lsv{}'} \\ \evalSJudge{\evalCtx{}}{\sof{\lbn{0}'}}{\lsnv{0}'} \\ \evalBBJudge{\evalCtx{}}{\lbn{0}}{\lenv{2}}{\lbn{0}'} \\ \evalJudge{\evalCtx{}}{\lembd{}_{\ls{}}}{\pBasisC{\ls{}}{\lbn{0}}{\lbn{1}}^{\lstar} \loprod \lbn{0}}{\lenv{3}} \\ \evalJudge{\evalCtx{}}{\lproj{}_{\ls{}, \sof{\lbn{0}'}}}{\lbn{0}'^{\lstar} \loprod \pBasisC{\ls{}}{\lbn{0}'}{\lbn{1}}}{\lenv{1}}}{\begin{gathered}\evalBBJudge{\evalCtx{}}{\pBasisC{\ls{}}{\lbn{0}}{\lbn{1}}}{\widehat{\leev{}}}{\pBasisC{\ls{}}{\lbn{0}'}{\lbn{1}}}\\[-1pt]\widehat{\leev{}}=\glvmtm{\lsnv{0}'}{\lsnv{0}'}{\glvreshape{\lenv{1}}{\lsnv{0}' \loprod \lsv{}'}{\lsnv{0}' \loprod \lsv{}'}}{\glvmtm{\lsnv{0}}{\lsnv{0}}{\glvreshape{\lenv{2}}{\lsnv{0} \loprod \lsnv{0}'}{\lsnv{0} \loprod \lsnv{0}'}}{\glvreshape{\lenv{3}}{\lsv{} \loprod \lsnv{0}}{\lsv{} \loprod \lsnv{0}}}}\end{gathered}}{\texttt{GLang.Basis.Eval\allowbreak BB.proj\_\allowbreak ker\_\allowbreak coker}}{GLang/ValueOperational.lean:496}
\gldefruleexplanation{evalBB.projKerCoker}{A kernel or cokernel basis changes its domain basis by composing the old embedding, the ambient conversion, and the new projection, with the codomain basis fixed.}
\gldefrule{evalBB.projImgCoimg}{evalBB}{\textsc{EB-Proj\allowbreak Img\allowbreak Coimg}}{\ls{} = \Image(\lee{}) \;\vee\; \ls{} = \CoImage(\lee{}) \\ \forall i.\; \sof{\lbn{i}} = \lsn{i} \\ \forall i.\; \evalSJudge{\evalCtx{}}{\sof{\lbn{i}}}{\lsnv{i}} \\ \evalSJudge{\evalCtx{}}{\ls{}}{\lsv{}} \\ \evalSJudge{\evalCtx{}}{\ls{}}{\lsv{}'} \\ \evalSJudge{\evalCtx{}}{\sof{\lbn{1}'}}{\lsnv{1}'} \\ \evalBBJudge{\evalCtx{}}{\lbn{1}}{\lenv{2}}{\lbn{1}'} \\ \evalJudge{\evalCtx{}}{\lembd{}_{\ls{}}}{\pBasisC{\ls{}}{\lbn{0}}{\lbn{1}}^{\lstar} \loprod \lbn{1}}{\lenv{3}} \\ \evalJudge{\evalCtx{}}{\lproj{}_{\ls{}, \sof{\lbn{1}'}}}{\lbn{1}'^{\lstar} \loprod \pBasisC{\ls{}}{\lbn{0}}{\lbn{1}'}}{\lenv{1}}}{\begin{gathered}\evalBBJudge{\evalCtx{}}{\pBasisC{\ls{}}{\lbn{0}}{\lbn{1}}}{\widehat{\leev{}}}{\pBasisC{\ls{}}{\lbn{0}}{\lbn{1}'}}\\[-1pt]\widehat{\leev{}}=\glvmtm{\lsnv{1}'}{\lsnv{1}'}{\glvreshape{\lenv{1}}{\lsnv{1}' \loprod \lsv{}'}{\lsnv{1}' \loprod \lsv{}'}}{\glvmtm{\lsnv{1}}{\lsnv{1}}{\glvreshape{\lenv{2}}{\lsnv{1} \loprod \lsnv{1}'}{\lsnv{1} \loprod \lsnv{1}'}}{\glvreshape{\lenv{3}}{\lsv{} \loprod \lsnv{1}}{\lsv{} \loprod \lsnv{1}}}}\end{gathered}}{\texttt{GLang.Basis.Eval\allowbreak BB.proj\_\allowbreak img\_\allowbreak coimg}}{GLang/ValueOperational.lean:542}
\gldefruleexplanation{evalBB.projImgCoimg}{An image or coimage basis changes its codomain basis by composing the old embedding, the ambient conversion, and the new projection, with the domain basis fixed.}
\gldefrule{evalBB.termsRref}{evalBB}{\textsc{EB-Terms\allowbreak Rref}}{\evalBBJudge{\evalCtx{}}{\lb{}}{\lenv{1}}{\lb{}'} \\ \evalJudge{\evalCtx{}}{\lee{} \losum \mathit{es}[0] \losum \cdots}{\lb{}' \losum \cdots}{\lenv{0}} \\ \evalSJudge{\evalCtx{}}{\sof{\lb{}}}{\lsv{}}}{\evalBBJudge{\evalCtx{}}{\lb{}}{\glvreshape{\glvop{RREFRep}(\lenv{0},\glvreshape{\lenv{1}}{\lsv{} \loprod \lsv{}}{\lsv{} \loprod \lsv{}})}{\lsv{} \loprod \lsv{}}{\lsv{} \loprod \lsv{}}}{\llpl{} \lee{} \mathbin{:\!:} \mathit{es}\llpr{}_{\lb{}'}}}{\texttt{GLang.Basis.Eval\allowbreak BB.terms\_\allowbreak rref}}{GLang/ValueOperational.lean:575}
\gldefruleexplanation{evalBB.termsRref}{Row reduction converts a basis to a nonempty term basis using the evaluated vector family and the conversion to its fallback basis.}
\gldefrule{evalBB.refl}{evalBB}{\textsc{EB-Refl}}{\tof{\lb{}} = \sof{\lb{}} \\ \evalSJudge{\evalCtx{}}{\sof{\lb{}}}{\lsv{}}}{\evalBBJudge{\evalCtx{}}{\lb{}}{\glvreshape{\lidelta{\lsv{}}}{\lsv{} \loprod \lsv{}}{\lsv{} \loprod \lsv{}}}{\lb{}}}{\texttt{GLang.Basis.Eval\allowbreak BB.refl}}{GLang/ValueOperational.lean:601}
\gldefruleexplanation{evalBB.refl}{A space basis converts to itself by the identity array reshaped to the tensor square of its evaluated space.}
\gldefrule{evalBB.projKerCokerCod}{evalBB}{\textsc{EB-Proj\allowbreak Ker\allowbreak Coker\allowbreak Cod}}{\ls{} = \Kernel(\lee{}) \;\vee\; \ls{} = \CoKernel(\lee{}) \\ \forall i.\; \sof{\lbn{i}} = \lsn{i} \\ \sof{\lbn{1}'} = \lsn{1} \\ \evalSJudge{\evalCtx{}}{\sof{\lbn{0}}}{\lsnv{0}} \\ \evalSJudge{\evalCtx{}}{\ls{}}{\lsv{}} \\ \evalSJudge{\evalCtx{}}{\ls{}}{\lsv{}'} \\ \evalJudge{\evalCtx{}}{\lembd{}_{\ls{}}}{\pBasisC{\ls{}}{\lbn{0}}{\lbn{1}}^{\lstar} \loprod \lbn{0}}{\lenv{3}} \\ \evalJudge{\evalCtx{}}{\lproj{}_{\ls{}, \sof{\lbn{0}}}}{\lbn{0}^{\lstar} \loprod \pBasisC{\ls{}}{\lbn{0}}{\lbn{1}'}}{\lenv{1}}}{\begin{gathered}\evalBBJudge{\evalCtx{}}{\pBasisC{\ls{}}{\lbn{0}}{\lbn{1}}}{\widehat{\leev{}}}{\pBasisC{\ls{}}{\lbn{0}}{\lbn{1}'}}\\[-1pt]\widehat{\leev{}}=\glvmtm{\lsnv{0}}{\lsnv{0}}{\glvreshape{\lenv{1}}{\lsnv{0} \loprod \lsv{}'}{\lsnv{0} \loprod \lsv{}'}}{\glvreshape{\lenv{3}}{\lsv{} \loprod \lsnv{0}}{\lsv{} \loprod \lsnv{0}}}\end{gathered}}{\texttt{GLang.Basis.Eval\allowbreak BB.proj\_\allowbreak ker\_\allowbreak coker\_\allowbreak cod}}{GLang/ValueOperational.lean:617}
\gldefruleexplanation{evalBB.projKerCokerCod}{A kernel or cokernel basis changes its codomain annotation by composing the old embedding and new projection through the fixed domain basis.}
\gldefrule{evalBB.projImgCoimgDom}{evalBB}{\textsc{EB-Proj\allowbreak Img\allowbreak Coimg\allowbreak Dom}}{\ls{} = \Image(\lee{}) \;\vee\; \ls{} = \CoImage(\lee{}) \\ \forall i.\; \sof{\lbn{i}} = \lsn{i} \\ \sof{\lbn{0}'} = \lsn{0} \\ \evalSJudge{\evalCtx{}}{\sof{\lbn{1}}}{\lsnv{1}} \\ \evalSJudge{\evalCtx{}}{\ls{}}{\lsv{}} \\ \evalSJudge{\evalCtx{}}{\ls{}}{\lsv{}'} \\ \evalJudge{\evalCtx{}}{\lembd{}_{\ls{}}}{\pBasisC{\ls{}}{\lbn{0}}{\lbn{1}}^{\lstar} \loprod \lbn{1}}{\lenv{3}} \\ \evalJudge{\evalCtx{}}{\lproj{}_{\ls{}, \sof{\lbn{1}}}}{\lbn{1}^{\lstar} \loprod \pBasisC{\ls{}}{\lbn{0}'}{\lbn{1}}}{\lenv{1}}}{\begin{gathered}\evalBBJudge{\evalCtx{}}{\pBasisC{\ls{}}{\lbn{0}}{\lbn{1}}}{\widehat{\leev{}}}{\pBasisC{\ls{}}{\lbn{0}'}{\lbn{1}}}\\[-1pt]\widehat{\leev{}}=\glvmtm{\lsnv{1}}{\lsnv{1}}{\glvreshape{\lenv{1}}{\lsnv{1} \loprod \lsv{}'}{\lsnv{1} \loprod \lsv{}'}}{\glvreshape{\lenv{3}}{\lsv{} \loprod \lsnv{1}}{\lsv{} \loprod \lsnv{1}}}\end{gathered}}{\texttt{GLang.Basis.Eval\allowbreak BB.proj\_\allowbreak img\_\allowbreak coimg\_\allowbreak dom}}{GLang/ValueOperational.lean:655}
\gldefruleexplanation{evalBB.projImgCoimgDom}{An image or coimage basis changes its domain annotation by composing the old embedding and new projection through the fixed codomain basis.}
\gldefrulerows{evalBB}{
\textsc{EB-Inverse} & $\evalBBJudge{\evalCtx{}}{\lbn{0}}{\leev{}^{-1}}{\lbn{1}}$ & \texttt{\footnotesize GLang/Value\allowbreak Operational.lean:408}\\
\textsc{EB-Trans} & $\evalBBJudge{\evalCtx{}}{\lbn{0}}{\glvmtm{\lsv{}}{\lsv{}}{\lenv{1}}{\glvreshape{\lenv{0}}{\lsv{} \loprod \lsv{}}{\lsv{} \loprod \lsv{}}}}{\lbn{2}}$ & \texttt{\footnotesize GLang/Value\allowbreak Operational.lean:413}\\
\textsc{EB-Tprod} & $\evalBBJudge{\evalCtx{}}{\lbn{0}' \loprod \lbn{1}'}{\vkrn(\lenv{0}, \lenv{1})}{\lbn{0} \loprod \lbn{1}}$ & \texttt{\footnotesize GLang/Value\allowbreak Operational.lean:432}\\
\textsc{EB-Dsum} & $\evalBBJudge{\evalCtx{}}{\lbn{0}' \losum \lbn{1}'}{\vblk(\lenv{0}, \lenv{1})}{\lbn{0} \losum \lbn{1}}$ & \texttt{\footnotesize GLang/Value\allowbreak Operational.lean:445}\\
\textsc{EB-Dual\allowbreak Std} & $\evalBBJudge{\evalCtx{}}{\lbn{0}^{\lstar}}{\Trans(\leev{}^{-1})}{\stdBasisDCtx{\tof{\lbn{0}^{\lstar}}}}$ & \texttt{\footnotesize GLang/Value\allowbreak Operational.lean:451}\\
\textsc{EB-Image} & $\evalBBJudge{\evalCtx{}}{\iBasisC{\lbn{0}}}{\leev{}}{\iBasisC{\lbn{0}'}}$ & \texttt{\footnotesize GLang/Value\allowbreak Operational.lean:463}\\
\textsc{EB-Image\allowbreak Svds} & $\begin{gathered}\evalBBJudge{\evalCtx{}}{\iBasisC{\pBasisC{\CoKernel(\lee{})}{\lbn{0}}{\lbn{1}}}}{\widehat{\leev{}}}{\pBasisC{\Image(\lee{})}{\lbn{0}}{\lbn{1}}}\\[-1pt]\widehat{\leev{}}=\glvreshape{\lenv{{\mathrm{P}}}}{\lR^{\lenv{{\mathrm{M}}},\CoKernel{},\lsnv{0} \loprod \lsnv{1}} \loprod \lR^{\lenv{{\mathrm{M}}},\Image{},\lsnv{0} \loprod \lsnv{1}}}{\operatorname{inferSpace}(\lenv{{\mathrm{P}}})}\end{gathered}$ & \texttt{\footnotesize GLang/Value\allowbreak Operational.lean:474}\\
\textsc{EB-Proj\allowbreak Ker\allowbreak Coker} & $\begin{gathered}\evalBBJudge{\evalCtx{}}{\pBasisC{\ls{}}{\lbn{0}}{\lbn{1}}}{\widehat{\leev{}}}{\pBasisC{\ls{}}{\lbn{0}'}{\lbn{1}}}\\[-1pt]\widehat{\leev{}}=\glvmtm{\lsnv{0}'}{\lsnv{0}'}{\glvreshape{\lenv{1}}{\lsnv{0}' \loprod \lsv{}'}{\lsnv{0}' \loprod \lsv{}'}}{\glvmtm{\lsnv{0}}{\lsnv{0}}{\glvreshape{\lenv{2}}{\lsnv{0} \loprod \lsnv{0}'}{\lsnv{0} \loprod \lsnv{0}'}}{\glvreshape{\lenv{3}}{\lsv{} \loprod \lsnv{0}}{\lsv{} \loprod \lsnv{0}}}}\end{gathered}$ & \texttt{\footnotesize GLang/Value\allowbreak Operational.lean:496}\\
\textsc{EB-Proj\allowbreak Img\allowbreak Coimg} & $\begin{gathered}\evalBBJudge{\evalCtx{}}{\pBasisC{\ls{}}{\lbn{0}}{\lbn{1}}}{\widehat{\leev{}}}{\pBasisC{\ls{}}{\lbn{0}}{\lbn{1}'}}\\[-1pt]\widehat{\leev{}}=\glvmtm{\lsnv{1}'}{\lsnv{1}'}{\glvreshape{\lenv{1}}{\lsnv{1}' \loprod \lsv{}'}{\lsnv{1}' \loprod \lsv{}'}}{\glvmtm{\lsnv{1}}{\lsnv{1}}{\glvreshape{\lenv{2}}{\lsnv{1} \loprod \lsnv{1}'}{\lsnv{1} \loprod \lsnv{1}'}}{\glvreshape{\lenv{3}}{\lsv{} \loprod \lsnv{1}}{\lsv{} \loprod \lsnv{1}}}}\end{gathered}$ & \texttt{\footnotesize GLang/Value\allowbreak Operational.lean:542}\\
\textsc{EB-Terms\allowbreak Rref} & $\evalBBJudge{\evalCtx{}}{\lb{}}{\glvreshape{\glvop{RREFRep}(\lenv{0},\glvreshape{\lenv{1}}{\lsv{} \loprod \lsv{}}{\lsv{} \loprod \lsv{}})}{\lsv{} \loprod \lsv{}}{\lsv{} \loprod \lsv{}}}{\llpl{} \lee{} \mathbin{:\!:} \mathit{es}\llpr{}_{\lb{}'}}$ & \texttt{\footnotesize GLang/Value\allowbreak Operational.lean:575}\\
\textsc{EB-Refl} & $\evalBBJudge{\evalCtx{}}{\lb{}}{\glvreshape{\lidelta{\lsv{}}}{\lsv{} \loprod \lsv{}}{\lsv{} \loprod \lsv{}}}{\lb{}}$ & \texttt{\footnotesize GLang/Value\allowbreak Operational.lean:601}\\
\textsc{EB-Proj\allowbreak Ker\allowbreak Coker\allowbreak Cod} & $\begin{gathered}\evalBBJudge{\evalCtx{}}{\pBasisC{\ls{}}{\lbn{0}}{\lbn{1}}}{\widehat{\leev{}}}{\pBasisC{\ls{}}{\lbn{0}}{\lbn{1}'}}\\[-1pt]\widehat{\leev{}}=\glvmtm{\lsnv{0}}{\lsnv{0}}{\glvreshape{\lenv{1}}{\lsnv{0} \loprod \lsv{}'}{\lsnv{0} \loprod \lsv{}'}}{\glvreshape{\lenv{3}}{\lsv{} \loprod \lsnv{0}}{\lsv{} \loprod \lsnv{0}}}\end{gathered}$ & \texttt{\footnotesize GLang/Value\allowbreak Operational.lean:617}\\
\textsc{EB-Proj\allowbreak Img\allowbreak Coimg\allowbreak Dom} & $\begin{gathered}\evalBBJudge{\evalCtx{}}{\pBasisC{\ls{}}{\lbn{0}}{\lbn{1}}}{\widehat{\leev{}}}{\pBasisC{\ls{}}{\lbn{0}'}{\lbn{1}}}\\[-1pt]\widehat{\leev{}}=\glvmtm{\lsnv{1}}{\lsnv{1}}{\glvreshape{\lenv{1}}{\lsnv{1} \loprod \lsv{}'}{\lsnv{1} \loprod \lsv{}'}}{\glvreshape{\lenv{3}}{\lsv{} \loprod \lsnv{1}}{\lsv{} \loprod \lsnv{1}}}\end{gathered}$ & \texttt{\footnotesize GLang/Value\allowbreak Operational.lean:655}\\
}{
\gpfit{\inferrule*[]{\evalBBJudge{\evalCtx{}}{\lbn{1}}{\leev{}}{\lbn{0}}}{\evalBBJudge{\evalCtx{}}{\lbn{0}}{\leev{}^{-1}}{\lbn{1}}}}
\and
\gpfit{\inferrule*[]{\evalBBJudge{\evalCtx{}}{\lbn{0}}{\lenv{0}}{\lbn{1}} \\ \evalBBJudge{\evalCtx{}}{\lbn{1}}{\lenv{1}}{\lbn{2}} \\ \evalSJudge{\evalCtx{}}{\sof{\lbn{0}}}{\lsv{}}}{\evalBBJudge{\evalCtx{}}{\lbn{0}}{\glvmtm{\lsv{}}{\lsv{}}{\lenv{1}}{\glvreshape{\lenv{0}}{\lsv{} \loprod \lsv{}}{\lsv{} \loprod \lsv{}}}}{\lbn{2}}}}
\and
\gpfit{\inferrule*[]{\forall i.\; \evalBBJudge{\evalCtx{}}{\lbn{i}'}{\lenv{i}}{\lbn{i}}}{\evalBBJudge{\evalCtx{}}{\lbn{0}' \loprod \lbn{1}'}{\vkrn(\lenv{0}, \lenv{1})}{\lbn{0} \loprod \lbn{1}}}}
\and
\gpfit{\inferrule*[]{\forall i.\; \evalBBJudge{\evalCtx{}}{\lbn{i}'}{\lenv{i}}{\lbn{i}}}{\evalBBJudge{\evalCtx{}}{\lbn{0}' \losum \lbn{1}'}{\vblk(\lenv{0}, \lenv{1})}{\lbn{0} \losum \lbn{1}}}}
\and
\gpfit{\inferrule*[]{\tof{\lbn{0}} = \sof{\lbn{0}} \\ \evalBBJudge{\evalCtx{}}{\lbn{0}}{\leev{}}{\stdBasisDCtx{\tof{\lbn{0}}}}}{\evalBBJudge{\evalCtx{}}{\lbn{0}^{\lstar}}{\Trans(\leev{}^{-1})}{\stdBasisDCtx{\tof{\lbn{0}^{\lstar}}}}}}
\and
\gpfit{\inferrule*[]{\exists \lsn{0}.\; \exists \lsn{1}.\; \exists \lee{}.\; \sof{\lbn{0}} = \CoKernel(\lee{}) \\ \evalBBJudge{\evalCtx{}}{\lbn{0}}{\leev{}}{\lbn{0}'}}{\evalBBJudge{\evalCtx{}}{\iBasisC{\lbn{0}}}{\leev{}}{\iBasisC{\lbn{0}'}}}}
\and
\gpfit{\inferrule*[]{\forall i.\; \sof{\lbn{i}} = \lsn{i} \\ \forall i.\; \isStdBasis{\lbn{i}} \\ \evalJudge{\evalCtx{}}{\lee{}}{\lbn{0}^{\lstar} \loprod \lbn{1}}{\leev{}} \\ \evalSJudge{\evalCtx{}}{\sof{\lbn{0}^{\lstar} \loprod \lbn{1}}}{\lsnv{0} \loprod \lsnv{1}} \\ \lenv{{\mathrm{M}}} = \glvreshape{\leev{}}{\lsnv{0} \loprod \lsnv{1}}{\lsnv{0} \loprod \lsnv{1}} \\ \lenv{{\mathrm{U}}} = \glvop{GSVDU}(\lenv{{\mathrm{M}}},\lidelta{\lsnv{0}},\lidelta{\lsnv{1}}) \\ \lenv{{\mathrm{V}}} = \glvop{GSVDV}(\lenv{{\mathrm{M}}},\lidelta{\lsnv{0}},\lidelta{\lsnv{1}}) \\ \lenv{{\mathrm{P}}} = \glvmtm{\lsnv{1}}{\lsnv{1}}{\lenv{{\mathrm{V}}}}{\glvmtm{\lsnv{0}}{\lsnv{0}}{\lenv{{\mathrm{M}}}}{\Trans(\lenv{{\mathrm{U}}})}}}{\begin{gathered}\evalBBJudge{\evalCtx{}}{\iBasisC{\pBasisC{\CoKernel(\lee{})}{\lbn{0}}{\lbn{1}}}}{\widehat{\leev{}}}{\pBasisC{\Image(\lee{})}{\lbn{0}}{\lbn{1}}}\\[-1pt]\widehat{\leev{}}=\glvreshape{\lenv{{\mathrm{P}}}}{\lR^{\lenv{{\mathrm{M}}},\CoKernel{},\lsnv{0} \loprod \lsnv{1}} \loprod \lR^{\lenv{{\mathrm{M}}},\Image{},\lsnv{0} \loprod \lsnv{1}}}{\operatorname{inferSpace}(\lenv{{\mathrm{P}}})}\end{gathered}}}
\and
\gpfit{\inferrule*[]{\ls{} = \Kernel(\lee{}) \;\vee\; \ls{} = \CoKernel(\lee{}) \\ \forall i.\; \sof{\lbn{i}} = \lsn{i} \\ \forall i.\; \evalSJudge{\evalCtx{}}{\sof{\lbn{i}}}{\lsnv{i}} \\ \evalSJudge{\evalCtx{}}{\ls{}}{\lsv{}} \\ \evalSJudge{\evalCtx{}}{\ls{}}{\lsv{}'} \\ \evalSJudge{\evalCtx{}}{\sof{\lbn{0}'}}{\lsnv{0}'} \\ \evalBBJudge{\evalCtx{}}{\lbn{0}}{\lenv{2}}{\lbn{0}'} \\ \evalJudge{\evalCtx{}}{\lembd{}_{\ls{}}}{\pBasisC{\ls{}}{\lbn{0}}{\lbn{1}}^{\lstar} \loprod \lbn{0}}{\lenv{3}} \\ \evalJudge{\evalCtx{}}{\lproj{}_{\ls{}, \sof{\lbn{0}'}}}{\lbn{0}'^{\lstar} \loprod \pBasisC{\ls{}}{\lbn{0}'}{\lbn{1}}}{\lenv{1}}}{\begin{gathered}\evalBBJudge{\evalCtx{}}{\pBasisC{\ls{}}{\lbn{0}}{\lbn{1}}}{\widehat{\leev{}}}{\pBasisC{\ls{}}{\lbn{0}'}{\lbn{1}}}\\[-1pt]\widehat{\leev{}}=\glvmtm{\lsnv{0}'}{\lsnv{0}'}{\glvreshape{\lenv{1}}{\lsnv{0}' \loprod \lsv{}'}{\lsnv{0}' \loprod \lsv{}'}}{\glvmtm{\lsnv{0}}{\lsnv{0}}{\glvreshape{\lenv{2}}{\lsnv{0} \loprod \lsnv{0}'}{\lsnv{0} \loprod \lsnv{0}'}}{\glvreshape{\lenv{3}}{\lsv{} \loprod \lsnv{0}}{\lsv{} \loprod \lsnv{0}}}}\end{gathered}}}
\and
\gpfit{\inferrule*[]{\ls{} = \Image(\lee{}) \;\vee\; \ls{} = \CoImage(\lee{}) \\ \forall i.\; \sof{\lbn{i}} = \lsn{i} \\ \forall i.\; \evalSJudge{\evalCtx{}}{\sof{\lbn{i}}}{\lsnv{i}} \\ \evalSJudge{\evalCtx{}}{\ls{}}{\lsv{}} \\ \evalSJudge{\evalCtx{}}{\ls{}}{\lsv{}'} \\ \evalSJudge{\evalCtx{}}{\sof{\lbn{1}'}}{\lsnv{1}'} \\ \evalBBJudge{\evalCtx{}}{\lbn{1}}{\lenv{2}}{\lbn{1}'} \\ \evalJudge{\evalCtx{}}{\lembd{}_{\ls{}}}{\pBasisC{\ls{}}{\lbn{0}}{\lbn{1}}^{\lstar} \loprod \lbn{1}}{\lenv{3}} \\ \evalJudge{\evalCtx{}}{\lproj{}_{\ls{}, \sof{\lbn{1}'}}}{\lbn{1}'^{\lstar} \loprod \pBasisC{\ls{}}{\lbn{0}}{\lbn{1}'}}{\lenv{1}}}{\begin{gathered}\evalBBJudge{\evalCtx{}}{\pBasisC{\ls{}}{\lbn{0}}{\lbn{1}}}{\widehat{\leev{}}}{\pBasisC{\ls{}}{\lbn{0}}{\lbn{1}'}}\\[-1pt]\widehat{\leev{}}=\glvmtm{\lsnv{1}'}{\lsnv{1}'}{\glvreshape{\lenv{1}}{\lsnv{1}' \loprod \lsv{}'}{\lsnv{1}' \loprod \lsv{}'}}{\glvmtm{\lsnv{1}}{\lsnv{1}}{\glvreshape{\lenv{2}}{\lsnv{1} \loprod \lsnv{1}'}{\lsnv{1} \loprod \lsnv{1}'}}{\glvreshape{\lenv{3}}{\lsv{} \loprod \lsnv{1}}{\lsv{} \loprod \lsnv{1}}}}\end{gathered}}}
\and
\gpfit{\inferrule*[]{\evalBBJudge{\evalCtx{}}{\lb{}}{\lenv{1}}{\lb{}'} \\ \evalJudge{\evalCtx{}}{\lee{} \losum \mathit{es}[0] \losum \cdots}{\lb{}' \losum \cdots}{\lenv{0}} \\ \evalSJudge{\evalCtx{}}{\sof{\lb{}}}{\lsv{}}}{\evalBBJudge{\evalCtx{}}{\lb{}}{\glvreshape{\glvop{RREFRep}(\lenv{0},\glvreshape{\lenv{1}}{\lsv{} \loprod \lsv{}}{\lsv{} \loprod \lsv{}})}{\lsv{} \loprod \lsv{}}{\lsv{} \loprod \lsv{}}}{\llpl{} \lee{} \mathbin{:\!:} \mathit{es}\llpr{}_{\lb{}'}}}}
\and
\gpfit{\inferrule*[]{\tof{\lb{}} = \sof{\lb{}} \\ \evalSJudge{\evalCtx{}}{\sof{\lb{}}}{\lsv{}}}{\evalBBJudge{\evalCtx{}}{\lb{}}{\glvreshape{\lidelta{\lsv{}}}{\lsv{} \loprod \lsv{}}{\lsv{} \loprod \lsv{}}}{\lb{}}}}
\and
\gpfit{\inferrule*[]{\ls{} = \Kernel(\lee{}) \;\vee\; \ls{} = \CoKernel(\lee{}) \\ \forall i.\; \sof{\lbn{i}} = \lsn{i} \\ \sof{\lbn{1}'} = \lsn{1} \\ \evalSJudge{\evalCtx{}}{\sof{\lbn{0}}}{\lsnv{0}} \\ \evalSJudge{\evalCtx{}}{\ls{}}{\lsv{}} \\ \evalSJudge{\evalCtx{}}{\ls{}}{\lsv{}'} \\ \evalJudge{\evalCtx{}}{\lembd{}_{\ls{}}}{\pBasisC{\ls{}}{\lbn{0}}{\lbn{1}}^{\lstar} \loprod \lbn{0}}{\lenv{3}} \\ \evalJudge{\evalCtx{}}{\lproj{}_{\ls{}, \sof{\lbn{0}}}}{\lbn{0}^{\lstar} \loprod \pBasisC{\ls{}}{\lbn{0}}{\lbn{1}'}}{\lenv{1}}}{\begin{gathered}\evalBBJudge{\evalCtx{}}{\pBasisC{\ls{}}{\lbn{0}}{\lbn{1}}}{\widehat{\leev{}}}{\pBasisC{\ls{}}{\lbn{0}}{\lbn{1}'}}\\[-1pt]\widehat{\leev{}}=\glvmtm{\lsnv{0}}{\lsnv{0}}{\glvreshape{\lenv{1}}{\lsnv{0} \loprod \lsv{}'}{\lsnv{0} \loprod \lsv{}'}}{\glvreshape{\lenv{3}}{\lsv{} \loprod \lsnv{0}}{\lsv{} \loprod \lsnv{0}}}\end{gathered}}}
\and
\gpfit{\inferrule*[]{\ls{} = \Image(\lee{}) \;\vee\; \ls{} = \CoImage(\lee{}) \\ \forall i.\; \sof{\lbn{i}} = \lsn{i} \\ \sof{\lbn{0}'} = \lsn{0} \\ \evalSJudge{\evalCtx{}}{\sof{\lbn{1}}}{\lsnv{1}} \\ \evalSJudge{\evalCtx{}}{\ls{}}{\lsv{}} \\ \evalSJudge{\evalCtx{}}{\ls{}}{\lsv{}'} \\ \evalJudge{\evalCtx{}}{\lembd{}_{\ls{}}}{\pBasisC{\ls{}}{\lbn{0}}{\lbn{1}}^{\lstar} \loprod \lbn{1}}{\lenv{3}} \\ \evalJudge{\evalCtx{}}{\lproj{}_{\ls{}, \sof{\lbn{1}}}}{\lbn{1}^{\lstar} \loprod \pBasisC{\ls{}}{\lbn{0}'}{\lbn{1}}}{\lenv{1}}}{\begin{gathered}\evalBBJudge{\evalCtx{}}{\pBasisC{\ls{}}{\lbn{0}}{\lbn{1}}}{\widehat{\leev{}}}{\pBasisC{\ls{}}{\lbn{0}'}{\lbn{1}}}\\[-1pt]\widehat{\leev{}}=\glvmtm{\lsnv{1}}{\lsnv{1}}{\glvreshape{\lenv{1}}{\lsnv{1} \loprod \lsv{}'}{\lsnv{1} \loprod \lsv{}'}}{\glvreshape{\lenv{3}}{\lsv{} \loprod \lsnv{1}}{\lsv{} \loprod \lsnv{1}}}\end{gathered}}}
}{
\gpfit{\inferrule*[lab=\textsc{EB-Inverse}]{\evalBBJudge{\evalCtx{}}{\lbn{1}}{\leev{}}{\lbn{0}}}{\evalBBJudge{\evalCtx{}}{\lbn{0}}{\leev{}^{-1}}{\lbn{1}}}}
\and
\gpfit{\inferrule*[lab=\textsc{EB-Trans}]{\evalBBJudge{\evalCtx{}}{\lbn{0}}{\lenv{0}}{\lbn{1}} \\ \evalBBJudge{\evalCtx{}}{\lbn{1}}{\lenv{1}}{\lbn{2}} \\ \evalSJudge{\evalCtx{}}{\sof{\lbn{0}}}{\lsv{}}}{\evalBBJudge{\evalCtx{}}{\lbn{0}}{\glvmtm{\lsv{}}{\lsv{}}{\lenv{1}}{\glvreshape{\lenv{0}}{\lsv{} \loprod \lsv{}}{\lsv{} \loprod \lsv{}}}}{\lbn{2}}}}
\and
\gpfit{\inferrule*[lab=\textsc{EB-Tprod}]{\forall i.\; \evalBBJudge{\evalCtx{}}{\lbn{i}'}{\lenv{i}}{\lbn{i}}}{\evalBBJudge{\evalCtx{}}{\lbn{0}' \loprod \lbn{1}'}{\vkrn(\lenv{0}, \lenv{1})}{\lbn{0} \loprod \lbn{1}}}}
\and
\gpfit{\inferrule*[lab=\textsc{EB-Dsum}]{\forall i.\; \evalBBJudge{\evalCtx{}}{\lbn{i}'}{\lenv{i}}{\lbn{i}}}{\evalBBJudge{\evalCtx{}}{\lbn{0}' \losum \lbn{1}'}{\vblk(\lenv{0}, \lenv{1})}{\lbn{0} \losum \lbn{1}}}}
\and
\gpfit{\inferrule*[lab=\textsc{EB-Dual\allowbreak Std}]{\tof{\lbn{0}} = \sof{\lbn{0}} \\ \evalBBJudge{\evalCtx{}}{\lbn{0}}{\leev{}}{\stdBasisDCtx{\tof{\lbn{0}}}}}{\evalBBJudge{\evalCtx{}}{\lbn{0}^{\lstar}}{\Trans(\leev{}^{-1})}{\stdBasisDCtx{\tof{\lbn{0}^{\lstar}}}}}}
\and
\gpfit{\inferrule*[lab=\textsc{EB-Image}]{\exists \lsn{0}.\; \exists \lsn{1}.\; \exists \lee{}.\; \sof{\lbn{0}} = \CoKernel(\lee{}) \\ \evalBBJudge{\evalCtx{}}{\lbn{0}}{\leev{}}{\lbn{0}'}}{\evalBBJudge{\evalCtx{}}{\iBasisC{\lbn{0}}}{\leev{}}{\iBasisC{\lbn{0}'}}}}
\and
\gpfit{\inferrule*[lab=\textsc{EB-Image\allowbreak Svds}]{\forall i.\; \sof{\lbn{i}} = \lsn{i} \\ \forall i.\; \isStdBasis{\lbn{i}} \\ \evalJudge{\evalCtx{}}{\lee{}}{\lbn{0}^{\lstar} \loprod \lbn{1}}{\leev{}} \\ \evalSJudge{\evalCtx{}}{\sof{\lbn{0}^{\lstar} \loprod \lbn{1}}}{\lsnv{0} \loprod \lsnv{1}} \\ \lenv{{\mathrm{M}}} = \glvreshape{\leev{}}{\lsnv{0} \loprod \lsnv{1}}{\lsnv{0} \loprod \lsnv{1}} \\ \lenv{{\mathrm{U}}} = \glvop{GSVDU}(\lenv{{\mathrm{M}}},\lidelta{\lsnv{0}},\lidelta{\lsnv{1}}) \\ \lenv{{\mathrm{V}}} = \glvop{GSVDV}(\lenv{{\mathrm{M}}},\lidelta{\lsnv{0}},\lidelta{\lsnv{1}}) \\ \lenv{{\mathrm{P}}} = \glvmtm{\lsnv{1}}{\lsnv{1}}{\lenv{{\mathrm{V}}}}{\glvmtm{\lsnv{0}}{\lsnv{0}}{\lenv{{\mathrm{M}}}}{\Trans(\lenv{{\mathrm{U}}})}}}{\begin{gathered}\evalBBJudge{\evalCtx{}}{\iBasisC{\pBasisC{\CoKernel(\lee{})}{\lbn{0}}{\lbn{1}}}}{\widehat{\leev{}}}{\pBasisC{\Image(\lee{})}{\lbn{0}}{\lbn{1}}}\\[-1pt]\widehat{\leev{}}=\glvreshape{\lenv{{\mathrm{P}}}}{\lR^{\lenv{{\mathrm{M}}},\CoKernel{},\lsnv{0} \loprod \lsnv{1}} \loprod \lR^{\lenv{{\mathrm{M}}},\Image{},\lsnv{0} \loprod \lsnv{1}}}{\operatorname{inferSpace}(\lenv{{\mathrm{P}}})}\end{gathered}}}
\and
\gpfit{\inferrule*[lab=\textsc{EB-Proj\allowbreak Ker\allowbreak Coker}]{\ls{} = \Kernel(\lee{}) \;\vee\; \ls{} = \CoKernel(\lee{}) \\ \forall i.\; \sof{\lbn{i}} = \lsn{i} \\ \forall i.\; \evalSJudge{\evalCtx{}}{\sof{\lbn{i}}}{\lsnv{i}} \\ \evalSJudge{\evalCtx{}}{\ls{}}{\lsv{}} \\ \evalSJudge{\evalCtx{}}{\ls{}}{\lsv{}'} \\ \evalSJudge{\evalCtx{}}{\sof{\lbn{0}'}}{\lsnv{0}'} \\ \evalBBJudge{\evalCtx{}}{\lbn{0}}{\lenv{2}}{\lbn{0}'} \\ \evalJudge{\evalCtx{}}{\lembd{}_{\ls{}}}{\pBasisC{\ls{}}{\lbn{0}}{\lbn{1}}^{\lstar} \loprod \lbn{0}}{\lenv{3}} \\ \evalJudge{\evalCtx{}}{\lproj{}_{\ls{}, \sof{\lbn{0}'}}}{\lbn{0}'^{\lstar} \loprod \pBasisC{\ls{}}{\lbn{0}'}{\lbn{1}}}{\lenv{1}}}{\begin{gathered}\evalBBJudge{\evalCtx{}}{\pBasisC{\ls{}}{\lbn{0}}{\lbn{1}}}{\widehat{\leev{}}}{\pBasisC{\ls{}}{\lbn{0}'}{\lbn{1}}}\\[-1pt]\widehat{\leev{}}=\glvmtm{\lsnv{0}'}{\lsnv{0}'}{\glvreshape{\lenv{1}}{\lsnv{0}' \loprod \lsv{}'}{\lsnv{0}' \loprod \lsv{}'}}{\glvmtm{\lsnv{0}}{\lsnv{0}}{\glvreshape{\lenv{2}}{\lsnv{0} \loprod \lsnv{0}'}{\lsnv{0} \loprod \lsnv{0}'}}{\glvreshape{\lenv{3}}{\lsv{} \loprod \lsnv{0}}{\lsv{} \loprod \lsnv{0}}}}\end{gathered}}}
\and
\gpfit{\inferrule*[lab=\textsc{EB-Proj\allowbreak Img\allowbreak Coimg}]{\ls{} = \Image(\lee{}) \;\vee\; \ls{} = \CoImage(\lee{}) \\ \forall i.\; \sof{\lbn{i}} = \lsn{i} \\ \forall i.\; \evalSJudge{\evalCtx{}}{\sof{\lbn{i}}}{\lsnv{i}} \\ \evalSJudge{\evalCtx{}}{\ls{}}{\lsv{}} \\ \evalSJudge{\evalCtx{}}{\ls{}}{\lsv{}'} \\ \evalSJudge{\evalCtx{}}{\sof{\lbn{1}'}}{\lsnv{1}'} \\ \evalBBJudge{\evalCtx{}}{\lbn{1}}{\lenv{2}}{\lbn{1}'} \\ \evalJudge{\evalCtx{}}{\lembd{}_{\ls{}}}{\pBasisC{\ls{}}{\lbn{0}}{\lbn{1}}^{\lstar} \loprod \lbn{1}}{\lenv{3}} \\ \evalJudge{\evalCtx{}}{\lproj{}_{\ls{}, \sof{\lbn{1}'}}}{\lbn{1}'^{\lstar} \loprod \pBasisC{\ls{}}{\lbn{0}}{\lbn{1}'}}{\lenv{1}}}{\begin{gathered}\evalBBJudge{\evalCtx{}}{\pBasisC{\ls{}}{\lbn{0}}{\lbn{1}}}{\widehat{\leev{}}}{\pBasisC{\ls{}}{\lbn{0}}{\lbn{1}'}}\\[-1pt]\widehat{\leev{}}=\glvmtm{\lsnv{1}'}{\lsnv{1}'}{\glvreshape{\lenv{1}}{\lsnv{1}' \loprod \lsv{}'}{\lsnv{1}' \loprod \lsv{}'}}{\glvmtm{\lsnv{1}}{\lsnv{1}}{\glvreshape{\lenv{2}}{\lsnv{1} \loprod \lsnv{1}'}{\lsnv{1} \loprod \lsnv{1}'}}{\glvreshape{\lenv{3}}{\lsv{} \loprod \lsnv{1}}{\lsv{} \loprod \lsnv{1}}}}\end{gathered}}}
\and
\gpfit{\inferrule*[lab=\textsc{EB-Terms\allowbreak Rref}]{\evalBBJudge{\evalCtx{}}{\lb{}}{\lenv{1}}{\lb{}'} \\ \evalJudge{\evalCtx{}}{\lee{} \losum \mathit{es}[0] \losum \cdots}{\lb{}' \losum \cdots}{\lenv{0}} \\ \evalSJudge{\evalCtx{}}{\sof{\lb{}}}{\lsv{}}}{\evalBBJudge{\evalCtx{}}{\lb{}}{\glvreshape{\glvop{RREFRep}(\lenv{0},\glvreshape{\lenv{1}}{\lsv{} \loprod \lsv{}}{\lsv{} \loprod \lsv{}})}{\lsv{} \loprod \lsv{}}{\lsv{} \loprod \lsv{}}}{\llpl{} \lee{} \mathbin{:\!:} \mathit{es}\llpr{}_{\lb{}'}}}}
\and
\gpfit{\inferrule*[lab=\textsc{EB-Refl}]{\tof{\lb{}} = \sof{\lb{}} \\ \evalSJudge{\evalCtx{}}{\sof{\lb{}}}{\lsv{}}}{\evalBBJudge{\evalCtx{}}{\lb{}}{\glvreshape{\lidelta{\lsv{}}}{\lsv{} \loprod \lsv{}}{\lsv{} \loprod \lsv{}}}{\lb{}}}}
\and
\gpfit{\inferrule*[lab=\textsc{EB-Proj\allowbreak Ker\allowbreak Coker\allowbreak Cod}]{\ls{} = \Kernel(\lee{}) \;\vee\; \ls{} = \CoKernel(\lee{}) \\ \forall i.\; \sof{\lbn{i}} = \lsn{i} \\ \sof{\lbn{1}'} = \lsn{1} \\ \evalSJudge{\evalCtx{}}{\sof{\lbn{0}}}{\lsnv{0}} \\ \evalSJudge{\evalCtx{}}{\ls{}}{\lsv{}} \\ \evalSJudge{\evalCtx{}}{\ls{}}{\lsv{}'} \\ \evalJudge{\evalCtx{}}{\lembd{}_{\ls{}}}{\pBasisC{\ls{}}{\lbn{0}}{\lbn{1}}^{\lstar} \loprod \lbn{0}}{\lenv{3}} \\ \evalJudge{\evalCtx{}}{\lproj{}_{\ls{}, \sof{\lbn{0}}}}{\lbn{0}^{\lstar} \loprod \pBasisC{\ls{}}{\lbn{0}}{\lbn{1}'}}{\lenv{1}}}{\begin{gathered}\evalBBJudge{\evalCtx{}}{\pBasisC{\ls{}}{\lbn{0}}{\lbn{1}}}{\widehat{\leev{}}}{\pBasisC{\ls{}}{\lbn{0}}{\lbn{1}'}}\\[-1pt]\widehat{\leev{}}=\glvmtm{\lsnv{0}}{\lsnv{0}}{\glvreshape{\lenv{1}}{\lsnv{0} \loprod \lsv{}'}{\lsnv{0} \loprod \lsv{}'}}{\glvreshape{\lenv{3}}{\lsv{} \loprod \lsnv{0}}{\lsv{} \loprod \lsnv{0}}}\end{gathered}}}
\and
\gpfit{\inferrule*[lab=\textsc{EB-Proj\allowbreak Img\allowbreak Coimg\allowbreak Dom}]{\ls{} = \Image(\lee{}) \;\vee\; \ls{} = \CoImage(\lee{}) \\ \forall i.\; \sof{\lbn{i}} = \lsn{i} \\ \sof{\lbn{0}'} = \lsn{0} \\ \evalSJudge{\evalCtx{}}{\sof{\lbn{1}}}{\lsnv{1}} \\ \evalSJudge{\evalCtx{}}{\ls{}}{\lsv{}} \\ \evalSJudge{\evalCtx{}}{\ls{}}{\lsv{}'} \\ \evalJudge{\evalCtx{}}{\lembd{}_{\ls{}}}{\pBasisC{\ls{}}{\lbn{0}}{\lbn{1}}^{\lstar} \loprod \lbn{1}}{\lenv{3}} \\ \evalJudge{\evalCtx{}}{\lproj{}_{\ls{}, \sof{\lbn{1}}}}{\lbn{1}^{\lstar} \loprod \pBasisC{\ls{}}{\lbn{0}'}{\lbn{1}}}{\lenv{1}}}{\begin{gathered}\evalBBJudge{\evalCtx{}}{\pBasisC{\ls{}}{\lbn{0}}{\lbn{1}}}{\widehat{\leev{}}}{\pBasisC{\ls{}}{\lbn{0}'}{\lbn{1}}}\\[-1pt]\widehat{\leev{}}=\glvmtm{\lsnv{1}}{\lsnv{1}}{\glvreshape{\lenv{1}}{\lsnv{1} \loprod \lsv{}'}{\lsnv{1} \loprod \lsv{}'}}{\glvreshape{\lenv{3}}{\lsv{} \loprod \lsnv{1}}{\lsv{} \loprod \lsnv{1}}}\end{gathered}}}
}
\gldefjudgment{opStep}{\evalJudgeK{\lee{}}{\lb{}}{\leev{}}}{Term evaluation (generated from \texttt{GLang.ValExpr.OpStep}).}{fig:gl-opstep}
\gldefrule{opStep.var}{opStep}{\textsc{E-Var}}{((\mathit{x}, \lb{}) \mapsto \leev{}) \in \evalCtx{}}{\evalJudge{\evalCtx{}}{\mathit{x}}{\lb{}}{\leev{}}}{\texttt{GLang.Val\allowbreak Expr.Op\allowbreak Step.var}}{GLang/ValueOperational.lean:708}
\gldefruleexplanation{opStep.var}{A variable evaluates to the value stored for its name and requested basis in $\evalCtx$.}
\gldefrule{opStep.cob}{opStep}{\textsc{E-Cob}}{\forall i.\; \tof{\lbn{i}} = \sof{\lbn{i}} \\ \evalJudge{\evalCtx{}}{\lee{}}{\lbn{1}}{\leev{}} \\ \evalBBJudge{\evalCtx{}}{\lbn{1}}{\lenv{0}}{\lbn{0}} \\ \evalSJudge{\evalCtx{}}{\sof{\lbn{0}}}{\lsv{}}}{\evalJudge{\evalCtx{}}{\lee{}}{\lbn{0}}{\LTTVPP{\lsv{}}{\glvreshape{\lenv{0}}{\lsv{} \loprod \lsv{}}{\lsv{} \loprod \lsv{}}}{\glvreshape{\leev{}}{\lsv{}}{\lsv{}}}}}{\texttt{GLang.Val\allowbreak Expr.Op\allowbreak Step.cob}}{GLang/ValueOperational.lean:719}
\gldefruleexplanation{opStep.cob}{A value computed at one space basis is reshaped and contracted with a basis-conversion matrix to obtain coordinates in the requested basis.}
\gldefrule{opStep.scale}{opStep}{\textsc{E-Scale}}{\evalJudge{\evalCtx{}}{\len{0}}{\lbone{}}{\lenv{0}} \\ \evalJudge{\evalCtx{}}{\len{1}}{\lb{}}{\lenv{1}}}{\evalJudge{\evalCtx{}}{\len{0} \lvscale \len{1}}{\lb{}}{\lenv{0} \lvscale \lenv{1}}}{\texttt{GLang.Val\allowbreak Expr.Op\allowbreak Step.scale}}{GLang/ValueOperational.lean:747}
\gldefruleexplanation{opStep.scale}{Scaling evaluates the scalar in the unit basis and the other operand in the requested basis, then scales their values.}
\gldefrule{opStep.add}{opStep}{\textsc{E-Add}}{\tof{\lb{}} = \sof{\lb{}} \\ \evalSJudge{\evalCtx{}}{\sof{\lb{}}}{\lsv{}} \\ \forall i.\; \evalJudge{\evalCtx{}}{\len{i}}{\lb{}}{\lenv{i}}}{\evalJudge{\evalCtx{}}{\len{0} \lvadd \len{1}}{\lb{}}{\lenv{0} \lvadd_{\lsv{}} \lenv{1}}}{\texttt{GLang.Val\allowbreak Expr.Op\allowbreak Step.add}}{GLang/ValueOperational.lean:759}
\gldefruleexplanation{opStep.add}{Addition evaluates both operands at the requested space basis and records an evaluated representation of that basis as the result space.}
\gldefrule{opStep.dsumE}{opStep}{\textsc{E-Dsum\allowbreak E}}{\forall i.\; \evalJudge{\evalCtx{}}{\len{i}}{\lbn{i}}{\lenv{i}} \\ \forall i.\; \tof{\lbn{i}} = \sof{\lbn{i}}}{\evalJudge{\evalCtx{}}{\len{0}\losum \len{1}}{\lbn{0} \losum \lbn{1}}{\lenv{0}\losum \lenv{1}}}{\texttt{GLang.Val\allowbreak Expr.Op\allowbreak Step.dsum\allowbreak E}}{GLang/ValueOperational.lean:772}
\gldefruleexplanation{opStep.dsumE}{A direct sum evaluates its operands in their component space bases and combines their values in the direct-sum basis.}
\gldefrule{opStep.tprodE}{opStep}{\textsc{E-Tprod\allowbreak E}}{\forall i.\; \evalJudge{\evalCtx{}}{\len{i}}{\lbn{i}}{\lenv{i}} \\ \forall i.\; \tof{\lbn{i}} = \sof{\lbn{i}}}{\evalJudge{\evalCtx{}}{\len{0}\loprod \len{1}}{\lbn{0} \loprod \lbn{1}}{\lenv{0}\loprod \lenv{1}}}{\texttt{GLang.Val\allowbreak Expr.Op\allowbreak Step.tprod\allowbreak E}}{GLang/ValueOperational.lean:789}
\gldefruleexplanation{opStep.tprodE}{A tensor product evaluates its operands in their component space bases and tensors their values in the tensor-product basis.}
\gldefrule{opStep.pinvTStd}{opStep}{\textsc{E-Pinv\allowbreak TStd}}{\evalJudge{\evalCtx{}}{\lee{}}{\lbn{1}^{\lstar} \loprod \lbn{0}}{\leev{}} \\ \forall i.\; \isStdBasis{\lbn{i}}}{\evalJudge{\evalCtx{}}{\lee{}^{\lpseudo}}{\lbn{0}^{\lstar} \loprod \lbn{1}}{\leev{}^{+}}}{\texttt{GLang.Val\allowbreak Expr.Op\allowbreak Step.pinv\allowbreak T\_\allowbreak std}}{GLang/ValueOperational.lean:801}
\gldefruleexplanation{opStep.pinvTStd}{In standard domain and codomain bases, a map's pseudoinverse is computed by the ordinary matrix pseudoinverse with the bases exchanged.}
\gldefrule{opStep.pinvTGpinv}{opStep}{\textsc{E-Pinv\allowbreak TGpinv}}{\evalJudge{\evalCtx{}}{\lee{}}{\lbn{1}^{\lstar} \loprod \lbn{0}}{\leev{}} \\ \forall i.\; \evalJudge{\evalCtx{}}{\linner{}_{\sof{\lbn{i}}}}{\lbn{i}^{\lstar} \loprod \lbn{i}^{\lstar}}{\lenv{i}}}{\evalJudge{\evalCtx{}}{\lee{}^{\lpseudo}}{\lbn{0}^{\lstar} \loprod \lbn{1}}{\glvop{GPINV}(\leev{},\lenv{1},\lenv{0})}}{\texttt{GLang.Val\allowbreak Expr.Op\allowbreak Step.pinv\allowbreak T\_\allowbreak gpinv}}{GLang/ValueOperational.lean:818}
\gldefruleexplanation{opStep.pinvTGpinv}{In general bases, a map's pseudoinverse uses its evaluated matrix and the domain and codomain Gram matrices in that order.}
\gldefrule{opStep.pinvDStd}{opStep}{\textsc{E-Pinv\allowbreak DStd}}{\evalJudge{\evalCtx{}}{\lee{}}{\lbn{1}^{\lstar}}{\leev{}} \\ \sof{\lbn{0}} = \lR{} \\ \forall i.\; \isStdBasis{\lbn{i}} \\ \evalSJudge{\evalCtx{}}{\sof{\lbn{1}}}{\lsnv{1}}}{\evalJudge{\evalCtx{}}{\lee{}^{\lpseudo}}{\lbn{0}^{\lstar} \loprod \lbn{1}}{\glvreshape{\leev{}}{\lsnv{1} \loprod \lR{}}{\lsnv{1}}^{+}}}{\texttt{GLang.Val\allowbreak Expr.Op\allowbreak Step.pinv\allowbreak D\_\allowbreak std}}{GLang/ValueOperational.lean:854}
\gldefruleexplanation{opStep.pinvDStd}{In standard bases, a covector is reshaped as a map to $\mathbb R$ before taking its matrix pseudoinverse.}
\gldefrule{opStep.pinvDGpinv}{opStep}{\textsc{E-Pinv\allowbreak DGpinv}}{\evalJudge{\evalCtx{}}{\lee{}}{\lbn{1}^{\lstar}}{\leev{}} \\ \sof{\lbn{0}} = \lR{} \\ \isStdBasis{\lbn{0}} \\ \forall i.\; \evalJudge{\evalCtx{}}{\linner{}_{\sof{\lbn{i}}}}{\lbn{i}^{\lstar} \loprod \lbn{i}^{\lstar}}{\lenv{i}} \\ \evalSJudge{\evalCtx{}}{\sof{\lbn{1}}}{\lsnv{1}}}{\evalJudge{\evalCtx{}}{\lee{}^{\lpseudo}}{\lbn{0}^{\lstar} \loprod \lbn{1}}{\glvop{GPINV}(\glvreshape{\leev{}}{\lsnv{1} \loprod \lR{}}{\lsnv{1}},\lenv{1},\lenv{0})}}{\texttt{GLang.Val\allowbreak Expr.Op\allowbreak Step.pinv\allowbreak D\_\allowbreak gpinv}}{GLang/ValueOperational.lean:881}
\gldefruleexplanation{opStep.pinvDGpinv}{A covector's generalized pseudoinverse reshapes it as a map to the standard scalar basis and supplies the domain and scalar Gram matrices.}
\gldefrule{opStep.innerStd}{opStep}{\textsc{E-Inner\allowbreak Std}}{\isStdBasis{\lbn{0}} \\ \evalSJudge{\evalCtx{}}{\ls{}}{\lsv{}} \\ \sof{\lbn{0}} = \ls{}}{\evalJudge{\evalCtx{}}{\linner{}_{\ls{}}}{\lbn{0}^{\lstar} \loprod \lbn{0}^{\lstar}}{\lidelta{\lsv{}}}}{\texttt{GLang.Val\allowbreak Expr.Op\allowbreak Step.inner\_\allowbreak std}}{GLang/ValueOperational.lean:911}
\gldefruleexplanation{opStep.innerStd}{The inner product in a standard basis evaluates to the identity array on the evaluated space.}
\gldefrule{opStep.embedKer}{opStep}{\textsc{E-Embed\allowbreak Ker}}{\forall i.\; \sof{\lbn{i}} = \lsn{i} \\ \forall i.\; \evalJudge{\evalCtx{}}{\linner{}_{\lsn{i}}}{\lbn{i}^{\lstar} \loprod \lbn{i}^{\lstar}}{\lenv{i}} \\ \evalJudge{\evalCtx{}}{\lee{}}{\lbn{0}^{\lstar} \loprod \lbn{1}}{\leev{}}}{\evalJudge{\evalCtx{}}{\lembd{}_{\Kernel(\lee{})}}{\pBasisC{\Kernel(\lee{})}{\lbn{0}}{\lbn{1}}^{\lstar} \loprod \lbn{0}}{\Trans(\glvop{GSVDUI}(\leev{},\lenv{0},\lenv{1}))}}{\texttt{GLang.Val\allowbreak Expr.Op\allowbreak Step.embed\_\allowbreak ker}}{GLang/ValueOperational.lean:926}
\gldefruleexplanation{opStep.embedKer}{The kernel inclusion evaluates to the transpose of the generalized domain null-space block of the defining map and its Gram matrices.}
\gldefrule{opStep.embedCoker}{opStep}{\textsc{E-Embed\allowbreak Coker}}{\forall i.\; \sof{\lbn{i}} = \lsn{i} \\ \forall i.\; \evalJudge{\evalCtx{}}{\linner{}_{\lsn{i}}}{\lbn{i}^{\lstar} \loprod \lbn{i}^{\lstar}}{\lenv{i}} \\ \evalJudge{\evalCtx{}}{\lee{}}{\lbn{0}^{\lstar} \loprod \lbn{1}}{\leev{}}}{\evalJudge{\evalCtx{}}{\lembd{}_{\CoKernel(\lee{})}}{\pBasisC{\CoKernel(\lee{})}{\lbn{0}}{\lbn{1}}^{\lstar} \loprod \lbn{0}}{\Trans(\glvop{GSVDU}(\leev{},\lenv{0},\lenv{1}))}}{\texttt{GLang.Val\allowbreak Expr.Op\allowbreak Step.embed\_\allowbreak coker}}{GLang/ValueOperational.lean:963}
\gldefruleexplanation{opStep.embedCoker}{The cokernel inclusion evaluates to the transpose of the generalized domain singular-vector block of the defining map and its Gram matrices.}
\gldefrule{opStep.embedImg}{opStep}{\textsc{E-Embed\allowbreak Img}}{\forall i.\; \sof{\lbn{i}} = \lsn{i} \\ \forall i.\; \evalJudge{\evalCtx{}}{\linner{}_{\lsn{i}}}{\lbn{i}^{\lstar} \loprod \lbn{i}^{\lstar}}{\lenv{i}} \\ \evalJudge{\evalCtx{}}{\lee{}}{\lbn{0}^{\lstar} \loprod \lbn{1}}{\leev{}}}{\evalJudge{\evalCtx{}}{\lembd{}_{\Image(\lee{})}}{\pBasisC{\Image(\lee{})}{\lbn{0}}{\lbn{1}}^{\lstar} \loprod \lbn{1}}{\Trans(\glvop{GSVDV}(\leev{},\lenv{0},\lenv{1}))}}{\texttt{GLang.Val\allowbreak Expr.Op\allowbreak Step.embed\_\allowbreak img}}{GLang/ValueOperational.lean:982}
\gldefruleexplanation{opStep.embedImg}{The image inclusion evaluates to the transpose of the generalized codomain singular-vector block of the defining map and its Gram matrices.}
\gldefrule{opStep.embedCoimg}{opStep}{\textsc{E-Embed\allowbreak Coimg}}{\forall i.\; \sof{\lbn{i}} = \lsn{i} \\ \forall i.\; \evalJudge{\evalCtx{}}{\linner{}_{\lsn{i}}}{\lbn{i}^{\lstar} \loprod \lbn{i}^{\lstar}}{\lenv{i}} \\ \evalJudge{\evalCtx{}}{\lee{}}{\lbn{0}^{\lstar} \loprod \lbn{1}}{\leev{}}}{\evalJudge{\evalCtx{}}{\lembd{}_{\CoImage(\lee{})}}{\pBasisC{\CoImage(\lee{})}{\lbn{0}}{\lbn{1}}^{\lstar} \loprod \lbn{1}}{\Trans(\glvop{GSVDVI}(\leev{},\lenv{0},\lenv{1}))}}{\texttt{GLang.Val\allowbreak Expr.Op\allowbreak Step.embed\_\allowbreak coimg}}{GLang/ValueOperational.lean:1002}
\gldefruleexplanation{opStep.embedCoimg}{The coimage inclusion evaluates to the transpose of the generalized codomain null-space block of the defining map and its Gram matrices.}
\gldefrule{opStep.appLLin}{opStep}{\textsc{E-App\allowbreak LLin}}{\evalJudge{\evalCtx{}}{\len{0}}{\lbn{0}^{\lstar} \loprod \lbn{1}}{\lenv{0}} \\ \forall i.\; \evalSJudge{\evalCtx{}}{\sof{\lbn{i}}}{\lsnv{i}} \\ \evalJudge{\evalCtx{}}{\len{1}}{\lbn{0}}{\lenv{1}} \\ \tof{\lbn{1}} = \sof{\lbn{1}}}{\evalJudge{\evalCtx{}}{\leval{\len{0}}{}{\len{1}}}{\lbn{1}}{\glvreshape{\LTTVPP{\lsnv{0}}{\lenv{0}}{\lenv{1}}}{\lsnv{1}}{\lsnv{1}}}}{\texttt{GLang.Val\allowbreak Expr.Op\allowbreak Step.app\allowbreak L\_\allowbreak lin}}{GLang/ValueOperational.lean:1020}
\gldefruleexplanation{opStep.appLLin}{Linear application contracts the evaluated map with its argument and reshapes the result to the evaluated output space.}
\gldefrule{opStep.appILin}{opStep}{\textsc{E-App\allowbreak ILin}}{\evalJudge{\evalCtx{}}{\len{0}}{\lbn{0}^{\lstar}}{\lenv{0}} \\ \evalSJudge{\evalCtx{}}{\sof{\lbn{0}}}{\lsv{}} \\ \evalJudge{\evalCtx{}}{\len{1}}{\lbn{0}}{\lenv{1}}}{\evalJudge{\evalCtx{}}{\leval{\len{0}}{}{\len{1}}}{\lbone{}}{\LTTVPP{\lsv{}}{\glvreshape{\lenv{0}}{\lsv{} \loprod \lR{}}{\lsv{}}}{\lenv{1}}}}{\texttt{GLang.Val\allowbreak Expr.Op\allowbreak Step.app\allowbreak I\_\allowbreak lin}}{GLang/ValueOperational.lean:1038}
\gldefruleexplanation{opStep.appILin}{Covector application reshapes the covector as a map to $\mathbb R$ and contracts it with the evaluated argument.}
\gldefrule{opStep.nllam}{opStep}{\textsc{E-Nllam}}{\forall i.\; \tof{\lbn{i}} = \ltyn{i} \\ \evalSJudge{\evalCtx{}}{\ltyn{0}}{\ltynv{0}} \\ \evalJudge{\evalCtx{}[(\mathit{x},\lbn{0}) \mapsto \mathit{x}_{\ltynv{0}}]}{\lee{}}{\lbn{1}}{\leev{}} \\ \evalSJudge{\evalCtx{}[(\mathit{x},\lbn{0}) \mapsto \mathit{x}_{\ltynv{0}}]}{\ltyn{1}}{\ltynv{1}}}{\evalJudge{\evalCtx{}}{(\nllambda \llpl x\colon \ltyn{0}\llpr \ldotsp \lee{})}{\frameB{x}{\lbn{0}}{\lbn{1}}}{\llambda(x\colon \ltynv{0})\lrh \ltynv{1}\colon \glvreshape{\leev{}}{\ltynv{1}}{\ltynv{1}}}}{\texttt{GLang.Val\allowbreak Expr.Op\allowbreak Step.nllam}}{GLang/ValueOperational.lean:1058}
\gldefruleexplanation{opStep.nllam}{Dependent abstraction evaluates its body and codomain under a fresh stored argument to produce a value function with the evaluated domain and codomain.}
\gldefrule{opStep.appNVal}{opStep}{\textsc{E-App\allowbreak NVal}}{\evalJudge{\evalCtx{}}{\len{0}}{\frameB{x}{\lbn{0}}{\lbn{1}}}{\lenv{0}} \\ \evalJudge{\evalCtx{}}{\len{1}}{\lbn{0}}{\lenv{1}} \\ \evalSJudge{\evalCtx{}}{\tof{\lbn{1}\subst{x}{\len{1}}}}{{\ltyv{}_{\mathsf{out}}}} \\ \operatorname{inferType}(\lenv{0}) = (x\colon \ltynv{{\mathrm{D}}}\fnto \ltynv{{\mathrm{C}}}) \\ \lenv{2} = \glvreshape{\lenv{1}}{\ltynv{{\mathrm{D}}}}{\operatorname{inferType}(\lenv{1})}}{\evalJudge{\evalCtx{}}{\leval{\len{0}}{}{\len{1}}}{\lbn{1}\subst{x}{\len{1}}}{\glvreshape{\glvapp{\lenv{0}}{\lenv{2}}}{{\ltyv{}_{\mathsf{out}}}}{\ltynv{{\mathrm{C}}}\subst{x}{\lenv{2}}}}}{\texttt{GLang.Val\allowbreak Expr.Op\allowbreak Step.app\allowbreak N\_\allowbreak val}}{GLang/ValueOperational.lean:1084}
\gldefruleexplanation{opStep.appNVal}{Dependent application evaluates the function at a frame and its argument at the frame domain, then applies their values at the evaluated instantiated result type.}
\gldefrule{opStep.frameReanchor}{opStep}{\textsc{E-Frame\allowbreak Reanchor}}{\evalSJudge{\evalCtx{}}{\tof{\lbn{0}}}{\ltynv{0}} \\ \tof{\lbn{0}'} = \tof{\lbn{0}} \\ \evalJudge{\evalCtx{}}{\len{0}}{\frameB{x}{\lbn{0}'}{\lbn{1}}}{\leev{}} \\ \evalSJudge{\evalCtx{}}{\tof{\lbn{0}'}}{\ltynv{0}'} \\ \evalJudge{\evalCtx{}[(\mathit{x},\lbn{0}) \mapsto \mathit{x}_{\ltynv{0}}]}{\mathit{x}}{\lbn{0}'}{\lenv{1}} \\ \evalSJudge{\evalCtx{}[(\mathit{x},\lbn{0}) \mapsto \mathit{x}_{\ltynv{0}}]}{\tof{\lbn{1}}}{\ltynv{1}'} \\ \operatorname{inferType}(\leev{}) = (x\colon \ltynv{{\mathrm{D}}}\fnto \ltynv{{\mathrm{C}}}) \\ \lenv{2} = \glvreshape{\glvreshape{\lenv{1}}{\ltynv{0}'}{\ltynv{0}'}}{\ltynv{{\mathrm{D}}}}{\ltynv{0}'}}{\evalJudge{\evalCtx{}}{\len{0}}{\frameB{x}{\lbn{0}}{\lbn{1}}}{\llambda(x\colon \ltynv{0})\lrh \ltynv{1}'\colon \glvreshape{\glvapp{\leev{}}{\lenv{2}}}{\ltynv{1}'}{\ltynv{1}'}}}{\texttt{GLang.Val\allowbreak Expr.Op\allowbreak Step.frame\_\allowbreak reanchor}}{GLang/ValueOperational.lean:1097}
\gldefruleexplanation{opStep.frameReanchor}{A function changes its frame domain by converting each fresh argument to the old domain representation before applying the existing function value.}
\gldefrule{opStep.frameEta}{opStep}{\textsc{E-Frame\allowbreak Eta}}{\evalSJudge{\evalCtx{}}{\tof{\lbn{0}}}{\ltynv{0}} \\ \evalJudge{\evalCtx{}[(\mathit{x},\lbn{0}) \mapsto \mathit{x}_{\ltynv{0}}]}{\leval{\len{0}}{}{\mathit{x}}}{\lbn{1}}{\leev{}} \\ \evalSJudge{\evalCtx{}[(\mathit{x},\lbn{0}) \mapsto \mathit{x}_{\ltynv{0}}]}{\tof{\lbn{1}}}{\ltynv{1}}}{\evalJudge{\evalCtx{}}{\len{0}}{\frameB{x}{\lbn{0}}{\lbn{1}}}{\llambda(x\colon \ltynv{0})\lrh \ltynv{1}\colon \glvreshape{\leev{}}{\ltynv{1}}{\ltynv{1}}}}{\texttt{GLang.Val\allowbreak Expr.Op\allowbreak Step.frame\_\allowbreak eta}}{GLang/ValueOperational.lean:1136}
\gldefruleexplanation{opStep.frameEta}{A function is represented at a frame by abstracting its evaluated application to a fresh stored argument.}
\gldefrule{opStep.reshapeCob}{opStep}{\textsc{E-Reshape\allowbreak Cob}}{\evalJudge{\evalCtx{}}{\lee{}}{\lb{}}{\leev{}} \\ \evalSJudge{\evalCtx{}}{\tof{\lb{}}}{{\ltyv{}_{\mathsf{_src}}}} \\ \evalSJudge{\evalCtx{}}{\tof{\lb{}}}{{\ltyv{}_{\mathsf{_tgt}}}}}{\evalJudge{\evalCtx{}}{\lee{}}{\lb{}}{\glvreshape{\leev{}}{{\ltyv{}_{\mathsf{_tgt}}}}{{\ltyv{}_{\mathsf{_src}}}}}}{\texttt{GLang.Val\allowbreak Expr.Op\allowbreak Step.reshape\_\allowbreak cob}}{GLang/ValueOperational.lean:1161}
\gldefruleexplanation{opStep.reshapeCob}{A computed value may be reshaped between two evaluated representations of the same requested basis type.}
\gldefrule{opStep.llamAssemble}{opStep}{\textsc{E-Llam\allowbreak Assemble}}{\tof{\lb{}'} = \sof{\lb{}'} \\ \evalSJudge{\evalCtx{}}{\sof{\lb{}'}}{\lsv{}'} \\ \evalSJudge{\evalCtx{}}{\sof{\lb{}}}{\lsv{}} \\ \forall\,0\leq i<\overline{\operatorname{dimOf}}(\lsv{}').\; \evalJudge{\evalCtx{}[(\mathit{x},\lb{}') \mapsto \ldelta{i}{\lsv{}'}]}{\lee{}}{\lb{}}{\leev{}_{i}}}{\evalJudge{\evalCtx{}}{(\lllambda \llpl x\colon \sof{\lb{}'}\llpr \ldotsp \lee{})}{\lb{}'^{\lstar} \loprod \lb{}}{\glvreshape{\lsum^{\lsv{}'}_{\mathit{i}<\overline{\operatorname{dimOf}}(\lsv{}')}\left(\ldelta{\mathit{i}}{\lsv{}'}\loprod \glvreshape{\leev{}_{\mathit{i}}}{\lsv{}}{\lsv{}}\right)}{\lsv{}' \loprod \lsv{}}{\begin{cases}\lsempty{}&\overline{\operatorname{dimOf}}(\lsv{}')=0\\ \lsv{}' \loprod \lsv{}&\text{otherwise}\end{cases}}}}{\texttt{GLang.Val\allowbreak Expr.Op\allowbreak Step.llam\_\allowbreak assemble}}{GLang/ValueOperational.lean:1188}
\gldefruleexplanation{opStep.llamAssemble}{Linear abstraction evaluates the body on each coordinate selector up to the dimension bound and sums the resulting coordinate tensors.}
\gldefrule{opStep.llamAssembleReal}{opStep}{\textsc{E-Llam\allowbreak Assemble\allowbreak Real}}{\tof{\lb{}'} = \sof{\lb{}'} \\ \evalSJudge{\evalCtx{}}{\sof{\lb{}'}}{\lsv{}'} \\ \forall\,0\leq i<\overline{\operatorname{dimOf}}(\lsv{}').\; \evalJudge{\evalCtx{}[(\mathit{x},\lb{}') \mapsto \ldelta{i}{\lsv{}'}]}{\lee{}}{\lbone{}}{\leev{}_{i}}}{\evalJudge{\evalCtx{}}{(\lllambda \llpl x\colon \sof{\lb{}'}\llpr \ldotsp \lee{})}{\lb{}'^{\lstar}}{\glvreshape{\lsum^{\lsv{}'}_{\mathit{i}<\overline{\operatorname{dimOf}}(\lsv{}')}\left(\glvreshape{\leev{}_{\mathit{i}}}{\lR{}}{\lR{}} \lvscale \ldelta{\mathit{i}}{\lsv{}'}\right)}{\lsv{}'}{\begin{cases}\lsempty{}&\overline{\operatorname{dimOf}}(\lsv{}')=0\\ \lsv{}'&\text{otherwise}\end{cases}}}}{\texttt{GLang.Val\allowbreak Expr.Op\allowbreak Step.llam\_\allowbreak assemble\_\allowbreak real}}{GLang/ValueOperational.lean:1233}
\gldefruleexplanation{opStep.llamAssembleReal}{Scalar-valued linear abstraction evaluates the body on each coordinate selector up to the dimension bound and sums the resulting scalar multiples of those selectors.}
\gldefrule{opStep.zero}{opStep}{\textsc{E-Zero}}{\tof{\lb{}} = \sof{\lb{}} \\ \evalSJudge{\evalCtx{}}{\sof{\lb{}}}{\lsv{}}}{\evalJudge{\evalCtx{}}{\lvzero{}_{\sof{\lb{}}}}{\lb{}}{\lvzero{}_{\lsv{}}}}{\texttt{GLang.Val\allowbreak Expr.Op\allowbreak Step.zero}}{GLang/ValueOperational.lean:1258}
\gldefruleexplanation{opStep.zero}{A source zero evaluates to a zero array in an evaluated representation of the requested space basis.}
\gldefrule{opStep.rlit}{opStep}{\textsc{E-Rlit}}{\phantom{.}}{\evalJudge{\evalCtx{}}{\mathit{r}}{\lbone{}}{\mathit{r}}}{\texttt{GLang.Val\allowbreak Expr.Op\allowbreak Step.rlit}}{GLang/ValueOperational.lean:1274}
\gldefruleexplanation{opStep.rlit}{A real literal evaluates to the same scalar in the unit basis.}
\gldefrulerows{opStep}{
\textsc{E-Var} & $\evalJudge{\evalCtx{}}{\mathit{x}}{\lb{}}{\leev{}}$ & \texttt{\footnotesize GLang/Value\allowbreak Operational.lean:708}\\
\textsc{E-Cob} & $\evalJudge{\evalCtx{}}{\lee{}}{\lbn{0}}{\LTTVPP{\lsv{}}{\glvreshape{\lenv{0}}{\lsv{} \loprod \lsv{}}{\lsv{} \loprod \lsv{}}}{\glvreshape{\leev{}}{\lsv{}}{\lsv{}}}}$ & \texttt{\footnotesize GLang/Value\allowbreak Operational.lean:719}\\
\textsc{E-Scale} & $\evalJudge{\evalCtx{}}{\len{0} \lvscale \len{1}}{\lb{}}{\lenv{0} \lvscale \lenv{1}}$ & \texttt{\footnotesize GLang/Value\allowbreak Operational.lean:747}\\
\textsc{E-Add} & $\evalJudge{\evalCtx{}}{\len{0} \lvadd \len{1}}{\lb{}}{\lenv{0} \lvadd_{\lsv{}} \lenv{1}}$ & \texttt{\footnotesize GLang/Value\allowbreak Operational.lean:759}\\
\textsc{E-Dsum\allowbreak E} & $\evalJudge{\evalCtx{}}{\len{0}\losum \len{1}}{\lbn{0} \losum \lbn{1}}{\lenv{0}\losum \lenv{1}}$ & \texttt{\footnotesize GLang/Value\allowbreak Operational.lean:772}\\
\textsc{E-Tprod\allowbreak E} & $\evalJudge{\evalCtx{}}{\len{0}\loprod \len{1}}{\lbn{0} \loprod \lbn{1}}{\lenv{0}\loprod \lenv{1}}$ & \texttt{\footnotesize GLang/Value\allowbreak Operational.lean:789}\\
\textsc{E-Pinv\allowbreak TStd} & $\evalJudge{\evalCtx{}}{\lee{}^{\lpseudo}}{\lbn{0}^{\lstar} \loprod \lbn{1}}{\leev{}^{+}}$ & \texttt{\footnotesize GLang/Value\allowbreak Operational.lean:801}\\
\textsc{E-Pinv\allowbreak TGpinv} & $\evalJudge{\evalCtx{}}{\lee{}^{\lpseudo}}{\lbn{0}^{\lstar} \loprod \lbn{1}}{\glvop{GPINV}(\leev{},\lenv{1},\lenv{0})}$ & \texttt{\footnotesize GLang/Value\allowbreak Operational.lean:818}\\
\textsc{E-Pinv\allowbreak DStd} & $\evalJudge{\evalCtx{}}{\lee{}^{\lpseudo}}{\lbn{0}^{\lstar} \loprod \lbn{1}}{\glvreshape{\leev{}}{\lsnv{1} \loprod \lR{}}{\lsnv{1}}^{+}}$ & \texttt{\footnotesize GLang/Value\allowbreak Operational.lean:854}\\
\textsc{E-Pinv\allowbreak DGpinv} & $\evalJudge{\evalCtx{}}{\lee{}^{\lpseudo}}{\lbn{0}^{\lstar} \loprod \lbn{1}}{\glvop{GPINV}(\glvreshape{\leev{}}{\lsnv{1} \loprod \lR{}}{\lsnv{1}},\lenv{1},\lenv{0})}$ & \texttt{\footnotesize GLang/Value\allowbreak Operational.lean:881}\\
\textsc{E-Inner\allowbreak Std} & $\evalJudge{\evalCtx{}}{\linner{}_{\ls{}}}{\lbn{0}^{\lstar} \loprod \lbn{0}^{\lstar}}{\lidelta{\lsv{}}}$ & \texttt{\footnotesize GLang/Value\allowbreak Operational.lean:911}\\
\textsc{E-Embed\allowbreak Ker} & $\evalJudge{\evalCtx{}}{\lembd{}_{\Kernel(\lee{})}}{\pBasisC{\Kernel(\lee{})}{\lbn{0}}{\lbn{1}}^{\lstar} \loprod \lbn{0}}{\Trans(\glvop{GSVDUI}(\leev{},\lenv{0},\lenv{1}))}$ & \texttt{\footnotesize GLang/Value\allowbreak Operational.lean:926}\\
\textsc{E-Embed\allowbreak Coker} & $\evalJudge{\evalCtx{}}{\lembd{}_{\CoKernel(\lee{})}}{\pBasisC{\CoKernel(\lee{})}{\lbn{0}}{\lbn{1}}^{\lstar} \loprod \lbn{0}}{\Trans(\glvop{GSVDU}(\leev{},\lenv{0},\lenv{1}))}$ & \texttt{\footnotesize GLang/Value\allowbreak Operational.lean:963}\\
\textsc{E-Embed\allowbreak Img} & $\evalJudge{\evalCtx{}}{\lembd{}_{\Image(\lee{})}}{\pBasisC{\Image(\lee{})}{\lbn{0}}{\lbn{1}}^{\lstar} \loprod \lbn{1}}{\Trans(\glvop{GSVDV}(\leev{},\lenv{0},\lenv{1}))}$ & \texttt{\footnotesize GLang/Value\allowbreak Operational.lean:982}\\
\textsc{E-Embed\allowbreak Coimg} & $\evalJudge{\evalCtx{}}{\lembd{}_{\CoImage(\lee{})}}{\pBasisC{\CoImage(\lee{})}{\lbn{0}}{\lbn{1}}^{\lstar} \loprod \lbn{1}}{\Trans(\glvop{GSVDVI}(\leev{},\lenv{0},\lenv{1}))}$ & \texttt{\footnotesize GLang/Value\allowbreak Operational.lean:1002}\\
\textsc{E-App\allowbreak LLin} & $\evalJudge{\evalCtx{}}{\leval{\len{0}}{}{\len{1}}}{\lbn{1}}{\glvreshape{\LTTVPP{\lsnv{0}}{\lenv{0}}{\lenv{1}}}{\lsnv{1}}{\lsnv{1}}}$ & \texttt{\footnotesize GLang/Value\allowbreak Operational.lean:1020}\\
\textsc{E-App\allowbreak ILin} & $\evalJudge{\evalCtx{}}{\leval{\len{0}}{}{\len{1}}}{\lbone{}}{\LTTVPP{\lsv{}}{\glvreshape{\lenv{0}}{\lsv{} \loprod \lR{}}{\lsv{}}}{\lenv{1}}}$ & \texttt{\footnotesize GLang/Value\allowbreak Operational.lean:1038}\\
\textsc{E-Nllam} & $\evalJudge{\evalCtx{}}{(\nllambda \llpl x\colon \ltyn{0}\llpr \ldotsp \lee{})}{\frameB{x}{\lbn{0}}{\lbn{1}}}{\llambda(x\colon \ltynv{0})\lrh \ltynv{1}\colon \glvreshape{\leev{}}{\ltynv{1}}{\ltynv{1}}}$ & \texttt{\footnotesize GLang/Value\allowbreak Operational.lean:1058}\\
\textsc{E-App\allowbreak NVal} & $\evalJudge{\evalCtx{}}{\leval{\len{0}}{}{\len{1}}}{\lbn{1}\subst{x}{\len{1}}}{\glvreshape{\glvapp{\lenv{0}}{\lenv{2}}}{{\ltyv{}_{\mathsf{out}}}}{\ltynv{{\mathrm{C}}}\subst{x}{\lenv{2}}}}$ & \texttt{\footnotesize GLang/Value\allowbreak Operational.lean:1084}\\
\textsc{E-Frame\allowbreak Reanchor} & $\evalJudge{\evalCtx{}}{\len{0}}{\frameB{x}{\lbn{0}}{\lbn{1}}}{\llambda(x\colon \ltynv{0})\lrh \ltynv{1}'\colon \glvreshape{\glvapp{\leev{}}{\lenv{2}}}{\ltynv{1}'}{\ltynv{1}'}}$ & \texttt{\footnotesize GLang/Value\allowbreak Operational.lean:1097}\\
\textsc{E-Frame\allowbreak Eta} & $\evalJudge{\evalCtx{}}{\len{0}}{\frameB{x}{\lbn{0}}{\lbn{1}}}{\llambda(x\colon \ltynv{0})\lrh \ltynv{1}\colon \glvreshape{\leev{}}{\ltynv{1}}{\ltynv{1}}}$ & \texttt{\footnotesize GLang/Value\allowbreak Operational.lean:1136}\\
\textsc{E-Reshape\allowbreak Cob} & $\evalJudge{\evalCtx{}}{\lee{}}{\lb{}}{\glvreshape{\leev{}}{{\ltyv{}_{\mathsf{_tgt}}}}{{\ltyv{}_{\mathsf{_src}}}}}$ & \texttt{\footnotesize GLang/Value\allowbreak Operational.lean:1161}\\
\textsc{E-Llam\allowbreak Assemble} & $\evalJudge{\evalCtx{}}{(\lllambda \llpl x\colon \sof{\lb{}'}\llpr \ldotsp \lee{})}{\lb{}'^{\lstar} \loprod \lb{}}{\glvreshape{\lsum^{\lsv{}'}_{\mathit{i}<\overline{\operatorname{dimOf}}(\lsv{}')}\left(\ldelta{\mathit{i}}{\lsv{}'}\loprod \glvreshape{\leev{}_{\mathit{i}}}{\lsv{}}{\lsv{}}\right)}{\lsv{}' \loprod \lsv{}}{\begin{cases}\lsempty{}&\overline{\operatorname{dimOf}}(\lsv{}')=0\\ \lsv{}' \loprod \lsv{}&\text{otherwise}\end{cases}}}$ & \texttt{\footnotesize GLang/Value\allowbreak Operational.lean:1188}\\
\textsc{E-Llam\allowbreak Assemble\allowbreak Real} & $\evalJudge{\evalCtx{}}{(\lllambda \llpl x\colon \sof{\lb{}'}\llpr \ldotsp \lee{})}{\lb{}'^{\lstar}}{\glvreshape{\lsum^{\lsv{}'}_{\mathit{i}<\overline{\operatorname{dimOf}}(\lsv{}')}\left(\glvreshape{\leev{}_{\mathit{i}}}{\lR{}}{\lR{}} \lvscale \ldelta{\mathit{i}}{\lsv{}'}\right)}{\lsv{}'}{\begin{cases}\lsempty{}&\overline{\operatorname{dimOf}}(\lsv{}')=0\\ \lsv{}'&\text{otherwise}\end{cases}}}$ & \texttt{\footnotesize GLang/Value\allowbreak Operational.lean:1233}\\
\textsc{E-Zero} & $\evalJudge{\evalCtx{}}{\lvzero{}_{\sof{\lb{}}}}{\lb{}}{\lvzero{}_{\lsv{}}}$ & \texttt{\footnotesize GLang/Value\allowbreak Operational.lean:1258}\\
\textsc{E-Rlit} & $\evalJudge{\evalCtx{}}{\mathit{r}}{\lbone{}}{\mathit{r}}$ & \texttt{\footnotesize GLang/Value\allowbreak Operational.lean:1274}\\
}{
\gpfit{\inferrule*[]{((\mathit{x}, \lb{}) \mapsto \leev{}) \in \evalCtx{}}{\evalJudge{\evalCtx{}}{\mathit{x}}{\lb{}}{\leev{}}}}
\and
\gpfit{\inferrule*[]{\forall i.\; \tof{\lbn{i}} = \sof{\lbn{i}} \\ \evalJudge{\evalCtx{}}{\lee{}}{\lbn{1}}{\leev{}} \\ \evalBBJudge{\evalCtx{}}{\lbn{1}}{\lenv{0}}{\lbn{0}} \\ \evalSJudge{\evalCtx{}}{\sof{\lbn{0}}}{\lsv{}}}{\evalJudge{\evalCtx{}}{\lee{}}{\lbn{0}}{\LTTVPP{\lsv{}}{\glvreshape{\lenv{0}}{\lsv{} \loprod \lsv{}}{\lsv{} \loprod \lsv{}}}{\glvreshape{\leev{}}{\lsv{}}{\lsv{}}}}}}
\and
\gpfit{\inferrule*[]{\evalJudge{\evalCtx{}}{\len{0}}{\lbone{}}{\lenv{0}} \\ \evalJudge{\evalCtx{}}{\len{1}}{\lb{}}{\lenv{1}}}{\evalJudge{\evalCtx{}}{\len{0} \lvscale \len{1}}{\lb{}}{\lenv{0} \lvscale \lenv{1}}}}
\and
\gpfit{\inferrule*[]{\tof{\lb{}} = \sof{\lb{}} \\ \evalSJudge{\evalCtx{}}{\sof{\lb{}}}{\lsv{}} \\ \forall i.\; \evalJudge{\evalCtx{}}{\len{i}}{\lb{}}{\lenv{i}}}{\evalJudge{\evalCtx{}}{\len{0} \lvadd \len{1}}{\lb{}}{\lenv{0} \lvadd_{\lsv{}} \lenv{1}}}}
\and
\gpfit{\inferrule*[]{\forall i.\; \evalJudge{\evalCtx{}}{\len{i}}{\lbn{i}}{\lenv{i}} \\ \forall i.\; \tof{\lbn{i}} = \sof{\lbn{i}}}{\evalJudge{\evalCtx{}}{\len{0}\losum \len{1}}{\lbn{0} \losum \lbn{1}}{\lenv{0}\losum \lenv{1}}}}
\and
\gpfit{\inferrule*[]{\forall i.\; \evalJudge{\evalCtx{}}{\len{i}}{\lbn{i}}{\lenv{i}} \\ \forall i.\; \tof{\lbn{i}} = \sof{\lbn{i}}}{\evalJudge{\evalCtx{}}{\len{0}\loprod \len{1}}{\lbn{0} \loprod \lbn{1}}{\lenv{0}\loprod \lenv{1}}}}
\and
\gpfit{\inferrule*[]{\evalJudge{\evalCtx{}}{\lee{}}{\lbn{1}^{\lstar} \loprod \lbn{0}}{\leev{}} \\ \forall i.\; \isStdBasis{\lbn{i}}}{\evalJudge{\evalCtx{}}{\lee{}^{\lpseudo}}{\lbn{0}^{\lstar} \loprod \lbn{1}}{\leev{}^{+}}}}
\and
\gpfit{\inferrule*[]{\evalJudge{\evalCtx{}}{\lee{}}{\lbn{1}^{\lstar} \loprod \lbn{0}}{\leev{}} \\ \forall i.\; \evalJudge{\evalCtx{}}{\linner{}_{\sof{\lbn{i}}}}{\lbn{i}^{\lstar} \loprod \lbn{i}^{\lstar}}{\lenv{i}}}{\evalJudge{\evalCtx{}}{\lee{}^{\lpseudo}}{\lbn{0}^{\lstar} \loprod \lbn{1}}{\glvop{GPINV}(\leev{},\lenv{1},\lenv{0})}}}
\and
\gpfit{\inferrule*[]{\evalJudge{\evalCtx{}}{\lee{}}{\lbn{1}^{\lstar}}{\leev{}} \\ \sof{\lbn{0}} = \lR{} \\ \forall i.\; \isStdBasis{\lbn{i}} \\ \evalSJudge{\evalCtx{}}{\sof{\lbn{1}}}{\lsnv{1}}}{\evalJudge{\evalCtx{}}{\lee{}^{\lpseudo}}{\lbn{0}^{\lstar} \loprod \lbn{1}}{\glvreshape{\leev{}}{\lsnv{1} \loprod \lR{}}{\lsnv{1}}^{+}}}}
\and
\gpfit{\inferrule*[]{\evalJudge{\evalCtx{}}{\lee{}}{\lbn{1}^{\lstar}}{\leev{}} \\ \sof{\lbn{0}} = \lR{} \\ \isStdBasis{\lbn{0}} \\ \forall i.\; \evalJudge{\evalCtx{}}{\linner{}_{\sof{\lbn{i}}}}{\lbn{i}^{\lstar} \loprod \lbn{i}^{\lstar}}{\lenv{i}} \\ \evalSJudge{\evalCtx{}}{\sof{\lbn{1}}}{\lsnv{1}}}{\evalJudge{\evalCtx{}}{\lee{}^{\lpseudo}}{\lbn{0}^{\lstar} \loprod \lbn{1}}{\glvop{GPINV}(\glvreshape{\leev{}}{\lsnv{1} \loprod \lR{}}{\lsnv{1}},\lenv{1},\lenv{0})}}}
\and
\gpfit{\inferrule*[]{\isStdBasis{\lbn{0}} \\ \evalSJudge{\evalCtx{}}{\ls{}}{\lsv{}} \\ \sof{\lbn{0}} = \ls{}}{\evalJudge{\evalCtx{}}{\linner{}_{\ls{}}}{\lbn{0}^{\lstar} \loprod \lbn{0}^{\lstar}}{\lidelta{\lsv{}}}}}
\and
\gpfit{\inferrule*[]{\forall i.\; \sof{\lbn{i}} = \lsn{i} \\ \forall i.\; \evalJudge{\evalCtx{}}{\linner{}_{\lsn{i}}}{\lbn{i}^{\lstar} \loprod \lbn{i}^{\lstar}}{\lenv{i}} \\ \evalJudge{\evalCtx{}}{\lee{}}{\lbn{0}^{\lstar} \loprod \lbn{1}}{\leev{}}}{\evalJudge{\evalCtx{}}{\lembd{}_{\Kernel(\lee{})}}{\pBasisC{\Kernel(\lee{})}{\lbn{0}}{\lbn{1}}^{\lstar} \loprod \lbn{0}}{\Trans(\glvop{GSVDUI}(\leev{},\lenv{0},\lenv{1}))}}}
\and
\gpfit{\inferrule*[]{\forall i.\; \sof{\lbn{i}} = \lsn{i} \\ \forall i.\; \evalJudge{\evalCtx{}}{\linner{}_{\lsn{i}}}{\lbn{i}^{\lstar} \loprod \lbn{i}^{\lstar}}{\lenv{i}} \\ \evalJudge{\evalCtx{}}{\lee{}}{\lbn{0}^{\lstar} \loprod \lbn{1}}{\leev{}}}{\evalJudge{\evalCtx{}}{\lembd{}_{\CoKernel(\lee{})}}{\pBasisC{\CoKernel(\lee{})}{\lbn{0}}{\lbn{1}}^{\lstar} \loprod \lbn{0}}{\Trans(\glvop{GSVDU}(\leev{},\lenv{0},\lenv{1}))}}}
\and
\gpfit{\inferrule*[]{\forall i.\; \sof{\lbn{i}} = \lsn{i} \\ \forall i.\; \evalJudge{\evalCtx{}}{\linner{}_{\lsn{i}}}{\lbn{i}^{\lstar} \loprod \lbn{i}^{\lstar}}{\lenv{i}} \\ \evalJudge{\evalCtx{}}{\lee{}}{\lbn{0}^{\lstar} \loprod \lbn{1}}{\leev{}}}{\evalJudge{\evalCtx{}}{\lembd{}_{\Image(\lee{})}}{\pBasisC{\Image(\lee{})}{\lbn{0}}{\lbn{1}}^{\lstar} \loprod \lbn{1}}{\Trans(\glvop{GSVDV}(\leev{},\lenv{0},\lenv{1}))}}}
\and
\gpfit{\inferrule*[]{\forall i.\; \sof{\lbn{i}} = \lsn{i} \\ \forall i.\; \evalJudge{\evalCtx{}}{\linner{}_{\lsn{i}}}{\lbn{i}^{\lstar} \loprod \lbn{i}^{\lstar}}{\lenv{i}} \\ \evalJudge{\evalCtx{}}{\lee{}}{\lbn{0}^{\lstar} \loprod \lbn{1}}{\leev{}}}{\evalJudge{\evalCtx{}}{\lembd{}_{\CoImage(\lee{})}}{\pBasisC{\CoImage(\lee{})}{\lbn{0}}{\lbn{1}}^{\lstar} \loprod \lbn{1}}{\Trans(\glvop{GSVDVI}(\leev{},\lenv{0},\lenv{1}))}}}
\and
\gpfit{\inferrule*[]{\evalJudge{\evalCtx{}}{\len{0}}{\lbn{0}^{\lstar} \loprod \lbn{1}}{\lenv{0}} \\ \forall i.\; \evalSJudge{\evalCtx{}}{\sof{\lbn{i}}}{\lsnv{i}} \\ \evalJudge{\evalCtx{}}{\len{1}}{\lbn{0}}{\lenv{1}} \\ \tof{\lbn{1}} = \sof{\lbn{1}}}{\evalJudge{\evalCtx{}}{\leval{\len{0}}{}{\len{1}}}{\lbn{1}}{\glvreshape{\LTTVPP{\lsnv{0}}{\lenv{0}}{\lenv{1}}}{\lsnv{1}}{\lsnv{1}}}}}
\and
\gpfit{\inferrule*[]{\evalJudge{\evalCtx{}}{\len{0}}{\lbn{0}^{\lstar}}{\lenv{0}} \\ \evalSJudge{\evalCtx{}}{\sof{\lbn{0}}}{\lsv{}} \\ \evalJudge{\evalCtx{}}{\len{1}}{\lbn{0}}{\lenv{1}}}{\evalJudge{\evalCtx{}}{\leval{\len{0}}{}{\len{1}}}{\lbone{}}{\LTTVPP{\lsv{}}{\glvreshape{\lenv{0}}{\lsv{} \loprod \lR{}}{\lsv{}}}{\lenv{1}}}}}
\and
\gpfit{\inferrule*[]{\forall i.\; \tof{\lbn{i}} = \ltyn{i} \\ \evalSJudge{\evalCtx{}}{\ltyn{0}}{\ltynv{0}} \\ \evalJudge{\evalCtx{}[(\mathit{x},\lbn{0}) \mapsto \mathit{x}_{\ltynv{0}}]}{\lee{}}{\lbn{1}}{\leev{}} \\ \evalSJudge{\evalCtx{}[(\mathit{x},\lbn{0}) \mapsto \mathit{x}_{\ltynv{0}}]}{\ltyn{1}}{\ltynv{1}}}{\evalJudge{\evalCtx{}}{(\nllambda \llpl x\colon \ltyn{0}\llpr \ldotsp \lee{})}{\frameB{x}{\lbn{0}}{\lbn{1}}}{\llambda(x\colon \ltynv{0})\lrh \ltynv{1}\colon \glvreshape{\leev{}}{\ltynv{1}}{\ltynv{1}}}}}
\and
\gpfit{\inferrule*[]{\evalJudge{\evalCtx{}}{\len{0}}{\frameB{x}{\lbn{0}}{\lbn{1}}}{\lenv{0}} \\ \evalJudge{\evalCtx{}}{\len{1}}{\lbn{0}}{\lenv{1}} \\ \evalSJudge{\evalCtx{}}{\tof{\lbn{1}\subst{x}{\len{1}}}}{{\ltyv{}_{\mathsf{out}}}} \\ \operatorname{inferType}(\lenv{0}) = (x\colon \ltynv{{\mathrm{D}}}\fnto \ltynv{{\mathrm{C}}}) \\ \lenv{2} = \glvreshape{\lenv{1}}{\ltynv{{\mathrm{D}}}}{\operatorname{inferType}(\lenv{1})}}{\evalJudge{\evalCtx{}}{\leval{\len{0}}{}{\len{1}}}{\lbn{1}\subst{x}{\len{1}}}{\glvreshape{\glvapp{\lenv{0}}{\lenv{2}}}{{\ltyv{}_{\mathsf{out}}}}{\ltynv{{\mathrm{C}}}\subst{x}{\lenv{2}}}}}}
\and
\gpfit{\inferrule*[]{\evalSJudge{\evalCtx{}}{\tof{\lbn{0}}}{\ltynv{0}} \\ \tof{\lbn{0}'} = \tof{\lbn{0}} \\ \evalJudge{\evalCtx{}}{\len{0}}{\frameB{x}{\lbn{0}'}{\lbn{1}}}{\leev{}} \\ \evalSJudge{\evalCtx{}}{\tof{\lbn{0}'}}{\ltynv{0}'} \\ \evalJudge{\evalCtx{}[(\mathit{x},\lbn{0}) \mapsto \mathit{x}_{\ltynv{0}}]}{\mathit{x}}{\lbn{0}'}{\lenv{1}} \\ \evalSJudge{\evalCtx{}[(\mathit{x},\lbn{0}) \mapsto \mathit{x}_{\ltynv{0}}]}{\tof{\lbn{1}}}{\ltynv{1}'} \\ \operatorname{inferType}(\leev{}) = (x\colon \ltynv{{\mathrm{D}}}\fnto \ltynv{{\mathrm{C}}}) \\ \lenv{2} = \glvreshape{\glvreshape{\lenv{1}}{\ltynv{0}'}{\ltynv{0}'}}{\ltynv{{\mathrm{D}}}}{\ltynv{0}'}}{\evalJudge{\evalCtx{}}{\len{0}}{\frameB{x}{\lbn{0}}{\lbn{1}}}{\llambda(x\colon \ltynv{0})\lrh \ltynv{1}'\colon \glvreshape{\glvapp{\leev{}}{\lenv{2}}}{\ltynv{1}'}{\ltynv{1}'}}}}
\and
\gpfit{\inferrule*[]{\evalSJudge{\evalCtx{}}{\tof{\lbn{0}}}{\ltynv{0}} \\ \evalJudge{\evalCtx{}[(\mathit{x},\lbn{0}) \mapsto \mathit{x}_{\ltynv{0}}]}{\leval{\len{0}}{}{\mathit{x}}}{\lbn{1}}{\leev{}} \\ \evalSJudge{\evalCtx{}[(\mathit{x},\lbn{0}) \mapsto \mathit{x}_{\ltynv{0}}]}{\tof{\lbn{1}}}{\ltynv{1}}}{\evalJudge{\evalCtx{}}{\len{0}}{\frameB{x}{\lbn{0}}{\lbn{1}}}{\llambda(x\colon \ltynv{0})\lrh \ltynv{1}\colon \glvreshape{\leev{}}{\ltynv{1}}{\ltynv{1}}}}}
\and
\gpfit{\inferrule*[]{\evalJudge{\evalCtx{}}{\lee{}}{\lb{}}{\leev{}} \\ \evalSJudge{\evalCtx{}}{\tof{\lb{}}}{{\ltyv{}_{\mathsf{_src}}}} \\ \evalSJudge{\evalCtx{}}{\tof{\lb{}}}{{\ltyv{}_{\mathsf{_tgt}}}}}{\evalJudge{\evalCtx{}}{\lee{}}{\lb{}}{\glvreshape{\leev{}}{{\ltyv{}_{\mathsf{_tgt}}}}{{\ltyv{}_{\mathsf{_src}}}}}}}
\and
\gpfit{\inferrule*[]{\tof{\lb{}'} = \sof{\lb{}'} \\ \evalSJudge{\evalCtx{}}{\sof{\lb{}'}}{\lsv{}'} \\ \evalSJudge{\evalCtx{}}{\sof{\lb{}}}{\lsv{}} \\ \forall\,0\leq i<\overline{\operatorname{dimOf}}(\lsv{}').\; \evalJudge{\evalCtx{}[(\mathit{x},\lb{}') \mapsto \ldelta{i}{\lsv{}'}]}{\lee{}}{\lb{}}{\leev{}_{i}}}{\evalJudge{\evalCtx{}}{(\lllambda \llpl x\colon \sof{\lb{}'}\llpr \ldotsp \lee{})}{\lb{}'^{\lstar} \loprod \lb{}}{\glvreshape{\lsum^{\lsv{}'}_{\mathit{i}<\overline{\operatorname{dimOf}}(\lsv{}')}\left(\ldelta{\mathit{i}}{\lsv{}'}\loprod \glvreshape{\leev{}_{\mathit{i}}}{\lsv{}}{\lsv{}}\right)}{\lsv{}' \loprod \lsv{}}{\begin{cases}\lsempty{}&\overline{\operatorname{dimOf}}(\lsv{}')=0\\ \lsv{}' \loprod \lsv{}&\text{otherwise}\end{cases}}}}}
\and
\gpfit{\inferrule*[]{\tof{\lb{}'} = \sof{\lb{}'} \\ \evalSJudge{\evalCtx{}}{\sof{\lb{}'}}{\lsv{}'} \\ \forall\,0\leq i<\overline{\operatorname{dimOf}}(\lsv{}').\; \evalJudge{\evalCtx{}[(\mathit{x},\lb{}') \mapsto \ldelta{i}{\lsv{}'}]}{\lee{}}{\lbone{}}{\leev{}_{i}}}{\evalJudge{\evalCtx{}}{(\lllambda \llpl x\colon \sof{\lb{}'}\llpr \ldotsp \lee{})}{\lb{}'^{\lstar}}{\glvreshape{\lsum^{\lsv{}'}_{\mathit{i}<\overline{\operatorname{dimOf}}(\lsv{}')}\left(\glvreshape{\leev{}_{\mathit{i}}}{\lR{}}{\lR{}} \lvscale \ldelta{\mathit{i}}{\lsv{}'}\right)}{\lsv{}'}{\begin{cases}\lsempty{}&\overline{\operatorname{dimOf}}(\lsv{}')=0\\ \lsv{}'&\text{otherwise}\end{cases}}}}}
\and
\gpfit{\inferrule*[]{\tof{\lb{}} = \sof{\lb{}} \\ \evalSJudge{\evalCtx{}}{\sof{\lb{}}}{\lsv{}}}{\evalJudge{\evalCtx{}}{\lvzero{}_{\sof{\lb{}}}}{\lb{}}{\lvzero{}_{\lsv{}}}}}
\and
\gpfit{\inferrule*[]{\phantom{.}}{\evalJudge{\evalCtx{}}{\mathit{r}}{\lbone{}}{\mathit{r}}}}
}{
\gpfit{\inferrule*[lab=\textsc{E-Var}]{((\mathit{x}, \lb{}) \mapsto \leev{}) \in \evalCtx{}}{\evalJudge{\evalCtx{}}{\mathit{x}}{\lb{}}{\leev{}}}}
\and
\gpfit{\inferrule*[lab=\textsc{E-Cob}]{\forall i.\; \tof{\lbn{i}} = \sof{\lbn{i}} \\ \evalJudge{\evalCtx{}}{\lee{}}{\lbn{1}}{\leev{}} \\ \evalBBJudge{\evalCtx{}}{\lbn{1}}{\lenv{0}}{\lbn{0}} \\ \evalSJudge{\evalCtx{}}{\sof{\lbn{0}}}{\lsv{}}}{\evalJudge{\evalCtx{}}{\lee{}}{\lbn{0}}{\LTTVPP{\lsv{}}{\glvreshape{\lenv{0}}{\lsv{} \loprod \lsv{}}{\lsv{} \loprod \lsv{}}}{\glvreshape{\leev{}}{\lsv{}}{\lsv{}}}}}}
\and
\gpfit{\inferrule*[lab=\textsc{E-Scale}]{\evalJudge{\evalCtx{}}{\len{0}}{\lbone{}}{\lenv{0}} \\ \evalJudge{\evalCtx{}}{\len{1}}{\lb{}}{\lenv{1}}}{\evalJudge{\evalCtx{}}{\len{0} \lvscale \len{1}}{\lb{}}{\lenv{0} \lvscale \lenv{1}}}}
\and
\gpfit{\inferrule*[lab=\textsc{E-Add}]{\tof{\lb{}} = \sof{\lb{}} \\ \evalSJudge{\evalCtx{}}{\sof{\lb{}}}{\lsv{}} \\ \forall i.\; \evalJudge{\evalCtx{}}{\len{i}}{\lb{}}{\lenv{i}}}{\evalJudge{\evalCtx{}}{\len{0} \lvadd \len{1}}{\lb{}}{\lenv{0} \lvadd_{\lsv{}} \lenv{1}}}}
\and
\gpfit{\inferrule*[lab=\textsc{E-Dsum\allowbreak E}]{\forall i.\; \evalJudge{\evalCtx{}}{\len{i}}{\lbn{i}}{\lenv{i}} \\ \forall i.\; \tof{\lbn{i}} = \sof{\lbn{i}}}{\evalJudge{\evalCtx{}}{\len{0}\losum \len{1}}{\lbn{0} \losum \lbn{1}}{\lenv{0}\losum \lenv{1}}}}
\and
\gpfit{\inferrule*[lab=\textsc{E-Tprod\allowbreak E}]{\forall i.\; \evalJudge{\evalCtx{}}{\len{i}}{\lbn{i}}{\lenv{i}} \\ \forall i.\; \tof{\lbn{i}} = \sof{\lbn{i}}}{\evalJudge{\evalCtx{}}{\len{0}\loprod \len{1}}{\lbn{0} \loprod \lbn{1}}{\lenv{0}\loprod \lenv{1}}}}
\and
\gpfit{\inferrule*[lab=\textsc{E-Pinv\allowbreak TStd}]{\evalJudge{\evalCtx{}}{\lee{}}{\lbn{1}^{\lstar} \loprod \lbn{0}}{\leev{}} \\ \forall i.\; \isStdBasis{\lbn{i}}}{\evalJudge{\evalCtx{}}{\lee{}^{\lpseudo}}{\lbn{0}^{\lstar} \loprod \lbn{1}}{\leev{}^{+}}}}
\and
\gpfit{\inferrule*[lab=\textsc{E-Pinv\allowbreak TGpinv}]{\evalJudge{\evalCtx{}}{\lee{}}{\lbn{1}^{\lstar} \loprod \lbn{0}}{\leev{}} \\ \forall i.\; \evalJudge{\evalCtx{}}{\linner{}_{\sof{\lbn{i}}}}{\lbn{i}^{\lstar} \loprod \lbn{i}^{\lstar}}{\lenv{i}}}{\evalJudge{\evalCtx{}}{\lee{}^{\lpseudo}}{\lbn{0}^{\lstar} \loprod \lbn{1}}{\glvop{GPINV}(\leev{},\lenv{1},\lenv{0})}}}
\and
\gpfit{\inferrule*[lab=\textsc{E-Pinv\allowbreak DStd}]{\evalJudge{\evalCtx{}}{\lee{}}{\lbn{1}^{\lstar}}{\leev{}} \\ \sof{\lbn{0}} = \lR{} \\ \forall i.\; \isStdBasis{\lbn{i}} \\ \evalSJudge{\evalCtx{}}{\sof{\lbn{1}}}{\lsnv{1}}}{\evalJudge{\evalCtx{}}{\lee{}^{\lpseudo}}{\lbn{0}^{\lstar} \loprod \lbn{1}}{\glvreshape{\leev{}}{\lsnv{1} \loprod \lR{}}{\lsnv{1}}^{+}}}}
\and
\gpfit{\inferrule*[lab=\textsc{E-Pinv\allowbreak DGpinv}]{\evalJudge{\evalCtx{}}{\lee{}}{\lbn{1}^{\lstar}}{\leev{}} \\ \sof{\lbn{0}} = \lR{} \\ \isStdBasis{\lbn{0}} \\ \forall i.\; \evalJudge{\evalCtx{}}{\linner{}_{\sof{\lbn{i}}}}{\lbn{i}^{\lstar} \loprod \lbn{i}^{\lstar}}{\lenv{i}} \\ \evalSJudge{\evalCtx{}}{\sof{\lbn{1}}}{\lsnv{1}}}{\evalJudge{\evalCtx{}}{\lee{}^{\lpseudo}}{\lbn{0}^{\lstar} \loprod \lbn{1}}{\glvop{GPINV}(\glvreshape{\leev{}}{\lsnv{1} \loprod \lR{}}{\lsnv{1}},\lenv{1},\lenv{0})}}}
\and
\gpfit{\inferrule*[lab=\textsc{E-Inner\allowbreak Std}]{\isStdBasis{\lbn{0}} \\ \evalSJudge{\evalCtx{}}{\ls{}}{\lsv{}} \\ \sof{\lbn{0}} = \ls{}}{\evalJudge{\evalCtx{}}{\linner{}_{\ls{}}}{\lbn{0}^{\lstar} \loprod \lbn{0}^{\lstar}}{\lidelta{\lsv{}}}}}
\and
\gpfit{\inferrule*[lab=\textsc{E-Embed\allowbreak Ker}]{\forall i.\; \sof{\lbn{i}} = \lsn{i} \\ \forall i.\; \evalJudge{\evalCtx{}}{\linner{}_{\lsn{i}}}{\lbn{i}^{\lstar} \loprod \lbn{i}^{\lstar}}{\lenv{i}} \\ \evalJudge{\evalCtx{}}{\lee{}}{\lbn{0}^{\lstar} \loprod \lbn{1}}{\leev{}}}{\evalJudge{\evalCtx{}}{\lembd{}_{\Kernel(\lee{})}}{\pBasisC{\Kernel(\lee{})}{\lbn{0}}{\lbn{1}}^{\lstar} \loprod \lbn{0}}{\Trans(\glvop{GSVDUI}(\leev{},\lenv{0},\lenv{1}))}}}
\and
\gpfit{\inferrule*[lab=\textsc{E-Embed\allowbreak Coker}]{\forall i.\; \sof{\lbn{i}} = \lsn{i} \\ \forall i.\; \evalJudge{\evalCtx{}}{\linner{}_{\lsn{i}}}{\lbn{i}^{\lstar} \loprod \lbn{i}^{\lstar}}{\lenv{i}} \\ \evalJudge{\evalCtx{}}{\lee{}}{\lbn{0}^{\lstar} \loprod \lbn{1}}{\leev{}}}{\evalJudge{\evalCtx{}}{\lembd{}_{\CoKernel(\lee{})}}{\pBasisC{\CoKernel(\lee{})}{\lbn{0}}{\lbn{1}}^{\lstar} \loprod \lbn{0}}{\Trans(\glvop{GSVDU}(\leev{},\lenv{0},\lenv{1}))}}}
\and
\gpfit{\inferrule*[lab=\textsc{E-Embed\allowbreak Img}]{\forall i.\; \sof{\lbn{i}} = \lsn{i} \\ \forall i.\; \evalJudge{\evalCtx{}}{\linner{}_{\lsn{i}}}{\lbn{i}^{\lstar} \loprod \lbn{i}^{\lstar}}{\lenv{i}} \\ \evalJudge{\evalCtx{}}{\lee{}}{\lbn{0}^{\lstar} \loprod \lbn{1}}{\leev{}}}{\evalJudge{\evalCtx{}}{\lembd{}_{\Image(\lee{})}}{\pBasisC{\Image(\lee{})}{\lbn{0}}{\lbn{1}}^{\lstar} \loprod \lbn{1}}{\Trans(\glvop{GSVDV}(\leev{},\lenv{0},\lenv{1}))}}}
\and
\gpfit{\inferrule*[lab=\textsc{E-Embed\allowbreak Coimg}]{\forall i.\; \sof{\lbn{i}} = \lsn{i} \\ \forall i.\; \evalJudge{\evalCtx{}}{\linner{}_{\lsn{i}}}{\lbn{i}^{\lstar} \loprod \lbn{i}^{\lstar}}{\lenv{i}} \\ \evalJudge{\evalCtx{}}{\lee{}}{\lbn{0}^{\lstar} \loprod \lbn{1}}{\leev{}}}{\evalJudge{\evalCtx{}}{\lembd{}_{\CoImage(\lee{})}}{\pBasisC{\CoImage(\lee{})}{\lbn{0}}{\lbn{1}}^{\lstar} \loprod \lbn{1}}{\Trans(\glvop{GSVDVI}(\leev{},\lenv{0},\lenv{1}))}}}
\and
\gpfit{\inferrule*[lab=\textsc{E-App\allowbreak LLin}]{\evalJudge{\evalCtx{}}{\len{0}}{\lbn{0}^{\lstar} \loprod \lbn{1}}{\lenv{0}} \\ \forall i.\; \evalSJudge{\evalCtx{}}{\sof{\lbn{i}}}{\lsnv{i}} \\ \evalJudge{\evalCtx{}}{\len{1}}{\lbn{0}}{\lenv{1}} \\ \tof{\lbn{1}} = \sof{\lbn{1}}}{\evalJudge{\evalCtx{}}{\leval{\len{0}}{}{\len{1}}}{\lbn{1}}{\glvreshape{\LTTVPP{\lsnv{0}}{\lenv{0}}{\lenv{1}}}{\lsnv{1}}{\lsnv{1}}}}}
\and
\gpfit{\inferrule*[lab=\textsc{E-App\allowbreak ILin}]{\evalJudge{\evalCtx{}}{\len{0}}{\lbn{0}^{\lstar}}{\lenv{0}} \\ \evalSJudge{\evalCtx{}}{\sof{\lbn{0}}}{\lsv{}} \\ \evalJudge{\evalCtx{}}{\len{1}}{\lbn{0}}{\lenv{1}}}{\evalJudge{\evalCtx{}}{\leval{\len{0}}{}{\len{1}}}{\lbone{}}{\LTTVPP{\lsv{}}{\glvreshape{\lenv{0}}{\lsv{} \loprod \lR{}}{\lsv{}}}{\lenv{1}}}}}
\and
\gpfit{\inferrule*[lab=\textsc{E-Nllam}]{\forall i.\; \tof{\lbn{i}} = \ltyn{i} \\ \evalSJudge{\evalCtx{}}{\ltyn{0}}{\ltynv{0}} \\ \evalJudge{\evalCtx{}[(\mathit{x},\lbn{0}) \mapsto \mathit{x}_{\ltynv{0}}]}{\lee{}}{\lbn{1}}{\leev{}} \\ \evalSJudge{\evalCtx{}[(\mathit{x},\lbn{0}) \mapsto \mathit{x}_{\ltynv{0}}]}{\ltyn{1}}{\ltynv{1}}}{\evalJudge{\evalCtx{}}{(\nllambda \llpl x\colon \ltyn{0}\llpr \ldotsp \lee{})}{\frameB{x}{\lbn{0}}{\lbn{1}}}{\llambda(x\colon \ltynv{0})\lrh \ltynv{1}\colon \glvreshape{\leev{}}{\ltynv{1}}{\ltynv{1}}}}}
\and
\gpfit{\inferrule*[lab=\textsc{E-App\allowbreak NVal}]{\evalJudge{\evalCtx{}}{\len{0}}{\frameB{x}{\lbn{0}}{\lbn{1}}}{\lenv{0}} \\ \evalJudge{\evalCtx{}}{\len{1}}{\lbn{0}}{\lenv{1}} \\ \evalSJudge{\evalCtx{}}{\tof{\lbn{1}\subst{x}{\len{1}}}}{{\ltyv{}_{\mathsf{out}}}} \\ \operatorname{inferType}(\lenv{0}) = (x\colon \ltynv{{\mathrm{D}}}\fnto \ltynv{{\mathrm{C}}}) \\ \lenv{2} = \glvreshape{\lenv{1}}{\ltynv{{\mathrm{D}}}}{\operatorname{inferType}(\lenv{1})}}{\evalJudge{\evalCtx{}}{\leval{\len{0}}{}{\len{1}}}{\lbn{1}\subst{x}{\len{1}}}{\glvreshape{\glvapp{\lenv{0}}{\lenv{2}}}{{\ltyv{}_{\mathsf{out}}}}{\ltynv{{\mathrm{C}}}\subst{x}{\lenv{2}}}}}}
\and
\gpfit{\inferrule*[lab=\textsc{E-Frame\allowbreak Reanchor}]{\evalSJudge{\evalCtx{}}{\tof{\lbn{0}}}{\ltynv{0}} \\ \tof{\lbn{0}'} = \tof{\lbn{0}} \\ \evalJudge{\evalCtx{}}{\len{0}}{\frameB{x}{\lbn{0}'}{\lbn{1}}}{\leev{}} \\ \evalSJudge{\evalCtx{}}{\tof{\lbn{0}'}}{\ltynv{0}'} \\ \evalJudge{\evalCtx{}[(\mathit{x},\lbn{0}) \mapsto \mathit{x}_{\ltynv{0}}]}{\mathit{x}}{\lbn{0}'}{\lenv{1}} \\ \evalSJudge{\evalCtx{}[(\mathit{x},\lbn{0}) \mapsto \mathit{x}_{\ltynv{0}}]}{\tof{\lbn{1}}}{\ltynv{1}'} \\ \operatorname{inferType}(\leev{}) = (x\colon \ltynv{{\mathrm{D}}}\fnto \ltynv{{\mathrm{C}}}) \\ \lenv{2} = \glvreshape{\glvreshape{\lenv{1}}{\ltynv{0}'}{\ltynv{0}'}}{\ltynv{{\mathrm{D}}}}{\ltynv{0}'}}{\evalJudge{\evalCtx{}}{\len{0}}{\frameB{x}{\lbn{0}}{\lbn{1}}}{\llambda(x\colon \ltynv{0})\lrh \ltynv{1}'\colon \glvreshape{\glvapp{\leev{}}{\lenv{2}}}{\ltynv{1}'}{\ltynv{1}'}}}}
\and
\gpfit{\inferrule*[lab=\textsc{E-Frame\allowbreak Eta}]{\evalSJudge{\evalCtx{}}{\tof{\lbn{0}}}{\ltynv{0}} \\ \evalJudge{\evalCtx{}[(\mathit{x},\lbn{0}) \mapsto \mathit{x}_{\ltynv{0}}]}{\leval{\len{0}}{}{\mathit{x}}}{\lbn{1}}{\leev{}} \\ \evalSJudge{\evalCtx{}[(\mathit{x},\lbn{0}) \mapsto \mathit{x}_{\ltynv{0}}]}{\tof{\lbn{1}}}{\ltynv{1}}}{\evalJudge{\evalCtx{}}{\len{0}}{\frameB{x}{\lbn{0}}{\lbn{1}}}{\llambda(x\colon \ltynv{0})\lrh \ltynv{1}\colon \glvreshape{\leev{}}{\ltynv{1}}{\ltynv{1}}}}}
\and
\gpfit{\inferrule*[lab=\textsc{E-Reshape\allowbreak Cob}]{\evalJudge{\evalCtx{}}{\lee{}}{\lb{}}{\leev{}} \\ \evalSJudge{\evalCtx{}}{\tof{\lb{}}}{{\ltyv{}_{\mathsf{_src}}}} \\ \evalSJudge{\evalCtx{}}{\tof{\lb{}}}{{\ltyv{}_{\mathsf{_tgt}}}}}{\evalJudge{\evalCtx{}}{\lee{}}{\lb{}}{\glvreshape{\leev{}}{{\ltyv{}_{\mathsf{_tgt}}}}{{\ltyv{}_{\mathsf{_src}}}}}}}
\and
\gpfit{\inferrule*[lab=\textsc{E-Llam\allowbreak Assemble}]{\tof{\lb{}'} = \sof{\lb{}'} \\ \evalSJudge{\evalCtx{}}{\sof{\lb{}'}}{\lsv{}'} \\ \evalSJudge{\evalCtx{}}{\sof{\lb{}}}{\lsv{}} \\ \forall\,0\leq i<\overline{\operatorname{dimOf}}(\lsv{}').\; \evalJudge{\evalCtx{}[(\mathit{x},\lb{}') \mapsto \ldelta{i}{\lsv{}'}]}{\lee{}}{\lb{}}{\leev{}_{i}}}{\evalJudge{\evalCtx{}}{(\lllambda \llpl x\colon \sof{\lb{}'}\llpr \ldotsp \lee{})}{\lb{}'^{\lstar} \loprod \lb{}}{\glvreshape{\lsum^{\lsv{}'}_{\mathit{i}<\overline{\operatorname{dimOf}}(\lsv{}')}\left(\ldelta{\mathit{i}}{\lsv{}'}\loprod \glvreshape{\leev{}_{\mathit{i}}}{\lsv{}}{\lsv{}}\right)}{\lsv{}' \loprod \lsv{}}{\begin{cases}\lsempty{}&\overline{\operatorname{dimOf}}(\lsv{}')=0\\ \lsv{}' \loprod \lsv{}&\text{otherwise}\end{cases}}}}}
\and
\gpfit{\inferrule*[lab=\textsc{E-Llam\allowbreak Assemble\allowbreak Real}]{\tof{\lb{}'} = \sof{\lb{}'} \\ \evalSJudge{\evalCtx{}}{\sof{\lb{}'}}{\lsv{}'} \\ \forall\,0\leq i<\overline{\operatorname{dimOf}}(\lsv{}').\; \evalJudge{\evalCtx{}[(\mathit{x},\lb{}') \mapsto \ldelta{i}{\lsv{}'}]}{\lee{}}{\lbone{}}{\leev{}_{i}}}{\evalJudge{\evalCtx{}}{(\lllambda \llpl x\colon \sof{\lb{}'}\llpr \ldotsp \lee{})}{\lb{}'^{\lstar}}{\glvreshape{\lsum^{\lsv{}'}_{\mathit{i}<\overline{\operatorname{dimOf}}(\lsv{}')}\left(\glvreshape{\leev{}_{\mathit{i}}}{\lR{}}{\lR{}} \lvscale \ldelta{\mathit{i}}{\lsv{}'}\right)}{\lsv{}'}{\begin{cases}\lsempty{}&\overline{\operatorname{dimOf}}(\lsv{}')=0\\ \lsv{}'&\text{otherwise}\end{cases}}}}}
\and
\gpfit{\inferrule*[lab=\textsc{E-Zero}]{\tof{\lb{}} = \sof{\lb{}} \\ \evalSJudge{\evalCtx{}}{\sof{\lb{}}}{\lsv{}}}{\evalJudge{\evalCtx{}}{\lvzero{}_{\sof{\lb{}}}}{\lb{}}{\lvzero{}_{\lsv{}}}}}
\and
\gpfit{\inferrule*[lab=\textsc{E-Rlit}]{\phantom{.}}{\evalJudge{\evalCtx{}}{\mathit{r}}{\lbone{}}{\mathit{r}}}}
}
\gldefjudgment{s.space}{\lepsilon{};\evalCtx{}\vdash_{\mathsf S} \ls{}\evalArrow{}\lsv{}}{Strengthened space evaluation (generated from \texttt{GLang.S.Space}).}{fig:gl-s-space}
\gldefrule{s.space.real}{s.space}{\textsc{S-S-Real}}{\phantom{.}}{\evalCtxSpace{};\evalCtx{}\vdash_{\mathsf S} \lR{} \evalArrow{} \lR{}}{\texttt{GLang.S.Space.real}}{GLang/Adequacy/Strengthened.lean:74}
\gldefrule{s.space.empty}{s.space}{\textsc{S-S-Empty}}{\phantom{.}}{\evalCtxSpace{};\evalCtx{}\vdash_{\mathsf S} \lsempty{} \evalArrow{} \lsempty{}}{\texttt{GLang.S.Space.empty}}{GLang/Adequacy/Strengthened.lean:75}
\gldefrule{s.space.dual}{s.space}{\textsc{S-S-Dual}}{(\lemptyset, \evalCtxSpace{})\vdash \ls{}^{\lstar}\colon\svalid \\ \evalCtxSpace{};\evalCtx{}\vdash_{\mathsf S} \ls{} \evalArrow{} \lsv{}}{\evalCtxSpace{};\evalCtx{}\vdash_{\mathsf S} \ls{}^{\lstar} \evalArrow{} \lsv{}}{\texttt{GLang.S.Space.dual}}{GLang/Adequacy/Strengthened.lean:76}
\gldefrule{s.space.ker}{s.space}{\textsc{S-S-Ker}}{(\lemptyset, \evalCtxSpace{})\vdash \Kernel(\lee{})\colon\svalid \\ \sof{\lb{}} = \lsn{0}^{\lstar} \loprod \lsn{1} \\ \exists\,\lbp{},\textcolor{lang}{b^{\prime\prime}}.\; \lb{} = \lbp{}^{\lstar}\loprod \textcolor{lang}{b^{\prime\prime}} \\ \evalCtxSpace{};\evalCtx{}\vdash_{\mathsf S} \sof{\lb{}} \evalArrow{} \lsnv{{\mathrm{svb}}} \\ \lemptyset{};\evalCtxSpace{};\evalCtx{}\vdash_{\mathsf S} \lee{} \xlongrightarrow{\lb{}} \leev{}\glresulttype{\lsnv{{\mathrm{svb}}}}}{\evalCtxSpace{};\evalCtx{}\vdash_{\mathsf S} \Kernel(\lee{}) \evalArrow{} \lR^{\glvreshape{\leev{}}{\lsnv{{\mathrm{svb}}}}{\lsnv{{\mathrm{svb}}}},\Kernel{},\lsnv{{\mathrm{svb}}}}}{\texttt{GLang.S.Space.ker}}{GLang/Adequacy/Strengthened.lean:80}
\gldefrule{s.space.coker}{s.space}{\textsc{S-S-Coker}}{(\lemptyset, \evalCtxSpace{})\vdash \CoKernel(\lee{})\colon\svalid \\ \sof{\lb{}} = \lsn{0}^{\lstar} \loprod \lsn{1} \\ \exists\,\lbp{},\textcolor{lang}{b^{\prime\prime}}.\; \lb{} = \lbp{}^{\lstar}\loprod \textcolor{lang}{b^{\prime\prime}} \\ \evalCtxSpace{};\evalCtx{}\vdash_{\mathsf S} \sof{\lb{}} \evalArrow{} \lsnv{{\mathrm{svb}}} \\ \lemptyset{};\evalCtxSpace{};\evalCtx{}\vdash_{\mathsf S} \lee{} \xlongrightarrow{\lb{}} \leev{}\glresulttype{\lsnv{{\mathrm{svb}}}}}{\evalCtxSpace{};\evalCtx{}\vdash_{\mathsf S} \CoKernel(\lee{}) \evalArrow{} \lR^{\glvreshape{\leev{}}{\lsnv{{\mathrm{svb}}}}{\lsnv{{\mathrm{svb}}}},\CoKernel{},\lsnv{{\mathrm{svb}}}}}{\texttt{GLang.S.Space.coker}}{GLang/Adequacy/Strengthened.lean:89}
\gldefrule{s.space.img}{s.space}{\textsc{S-S-Img}}{(\lemptyset, \evalCtxSpace{})\vdash \Image(\lee{})\colon\svalid \\ \sof{\lb{}} = \lsn{0}^{\lstar} \loprod \lsn{1} \\ \exists\,\lbp{},\textcolor{lang}{b^{\prime\prime}}.\; \lb{} = \lbp{}^{\lstar}\loprod \textcolor{lang}{b^{\prime\prime}} \\ \evalCtxSpace{};\evalCtx{}\vdash_{\mathsf S} \sof{\lb{}} \evalArrow{} \lsnv{{\mathrm{svb}}} \\ \lemptyset{};\evalCtxSpace{};\evalCtx{}\vdash_{\mathsf S} \lee{} \xlongrightarrow{\lb{}} \leev{}\glresulttype{\lsnv{{\mathrm{svb}}}}}{\evalCtxSpace{};\evalCtx{}\vdash_{\mathsf S} \Image(\lee{}) \evalArrow{} \lR^{\glvreshape{\leev{}}{\lsnv{{\mathrm{svb}}}}{\lsnv{{\mathrm{svb}}}},\Image{},\lsnv{{\mathrm{svb}}}}}{\texttt{GLang.S.Space.img}}{GLang/Adequacy/Strengthened.lean:97}
\gldefrule{s.space.coimg}{s.space}{\textsc{S-S-Coimg}}{(\lemptyset, \evalCtxSpace{})\vdash \CoImage(\lee{})\colon\svalid \\ \sof{\lb{}} = \lsn{0}^{\lstar} \loprod \lsn{1} \\ \exists\,\lbp{},\textcolor{lang}{b^{\prime\prime}}.\; \lb{} = \lbp{}^{\lstar}\loprod \textcolor{lang}{b^{\prime\prime}} \\ \evalCtxSpace{};\evalCtx{}\vdash_{\mathsf S} \sof{\lb{}} \evalArrow{} \lsnv{{\mathrm{svb}}} \\ \lemptyset{};\evalCtxSpace{};\evalCtx{}\vdash_{\mathsf S} \lee{} \xlongrightarrow{\lb{}} \leev{}\glresulttype{\lsnv{{\mathrm{svb}}}}}{\evalCtxSpace{};\evalCtx{}\vdash_{\mathsf S} \CoImage(\lee{}) \evalArrow{} \lR^{\glvreshape{\leev{}}{\lsnv{{\mathrm{svb}}}}{\lsnv{{\mathrm{svb}}}},\CoImage{},\lsnv{{\mathrm{svb}}}}}{\texttt{GLang.S.Space.coimg}}{GLang/Adequacy/Strengthened.lean:105}
\gldefrule{s.space.tprod}{s.space}{\textsc{S-S-Tprod}}{(\lemptyset, \evalCtxSpace{})\vdash \lsn{0} \loprod \lsn{1}\colon\svalid \\ \forall i.\; \evalCtxSpace{};\evalCtx{}\vdash_{\mathsf S} \lsn{i} \evalArrow{} \lsnv{i}}{\evalCtxSpace{};\evalCtx{}\vdash_{\mathsf S} \lsn{0} \loprod \lsn{1} \evalArrow{} \lsnv{0} \loprod \lsnv{1}}{\texttt{GLang.S.Space.tprod}}{GLang/Adequacy/Strengthened.lean:113}
\gldefrule{s.space.dsum}{s.space}{\textsc{S-S-Dsum}}{(\lemptyset, \evalCtxSpace{})\vdash \lsn{0} \losum \lsn{1}\colon\svalid \\ \forall i.\; \evalCtxSpace{};\evalCtx{}\vdash_{\mathsf S} \lsn{i} \evalArrow{} \lsnv{i}}{\evalCtxSpace{};\evalCtx{}\vdash_{\mathsf S} \lsn{0} \losum \lsn{1} \evalArrow{} \lsnv{0} \losum \lsnv{1}}{\texttt{GLang.S.Space.dsum}}{GLang/Adequacy/Strengthened.lean:117}
\gldefrulerows{s.space}{
\textsc{S-S-Real} & $\evalCtxSpace{};\evalCtx{}\vdash_{\mathsf S} \lR{} \evalArrow{} \lR{}$ & \texttt{\footnotesize GLang/Adequacy/Strengthened.lean:74}\\
\textsc{S-S-Empty} & $\evalCtxSpace{};\evalCtx{}\vdash_{\mathsf S} \lsempty{} \evalArrow{} \lsempty{}$ & \texttt{\footnotesize GLang/Adequacy/Strengthened.lean:75}\\
\textsc{S-S-Dual} & $\evalCtxSpace{};\evalCtx{}\vdash_{\mathsf S} \ls{}^{\lstar} \evalArrow{} \lsv{}$ & \texttt{\footnotesize GLang/Adequacy/Strengthened.lean:76}\\
\textsc{S-S-Ker} & $\evalCtxSpace{};\evalCtx{}\vdash_{\mathsf S} \Kernel(\lee{}) \evalArrow{} \lR^{\glvreshape{\leev{}}{\lsnv{{\mathrm{svb}}}}{\lsnv{{\mathrm{svb}}}},\Kernel{},\lsnv{{\mathrm{svb}}}}$ & \texttt{\footnotesize GLang/Adequacy/Strengthened.lean:80}\\
\textsc{S-S-Coker} & $\evalCtxSpace{};\evalCtx{}\vdash_{\mathsf S} \CoKernel(\lee{}) \evalArrow{} \lR^{\glvreshape{\leev{}}{\lsnv{{\mathrm{svb}}}}{\lsnv{{\mathrm{svb}}}},\CoKernel{},\lsnv{{\mathrm{svb}}}}$ & \texttt{\footnotesize GLang/Adequacy/Strengthened.lean:89}\\
\textsc{S-S-Img} & $\evalCtxSpace{};\evalCtx{}\vdash_{\mathsf S} \Image(\lee{}) \evalArrow{} \lR^{\glvreshape{\leev{}}{\lsnv{{\mathrm{svb}}}}{\lsnv{{\mathrm{svb}}}},\Image{},\lsnv{{\mathrm{svb}}}}$ & \texttt{\footnotesize GLang/Adequacy/Strengthened.lean:97}\\
\textsc{S-S-Coimg} & $\evalCtxSpace{};\evalCtx{}\vdash_{\mathsf S} \CoImage(\lee{}) \evalArrow{} \lR^{\glvreshape{\leev{}}{\lsnv{{\mathrm{svb}}}}{\lsnv{{\mathrm{svb}}}},\CoImage{},\lsnv{{\mathrm{svb}}}}$ & \texttt{\footnotesize GLang/Adequacy/Strengthened.lean:105}\\
\textsc{S-S-Tprod} & $\evalCtxSpace{};\evalCtx{}\vdash_{\mathsf S} \lsn{0} \loprod \lsn{1} \evalArrow{} \lsnv{0} \loprod \lsnv{1}$ & \texttt{\footnotesize GLang/Adequacy/Strengthened.lean:113}\\
\textsc{S-S-Dsum} & $\evalCtxSpace{};\evalCtx{}\vdash_{\mathsf S} \lsn{0} \losum \lsn{1} \evalArrow{} \lsnv{0} \losum \lsnv{1}$ & \texttt{\footnotesize GLang/Adequacy/Strengthened.lean:117}\\
}{
\gpfit{\inferrule*[]{\phantom{.}}{\evalCtxSpace{};\evalCtx{}\vdash_{\mathsf S} \lR{} \evalArrow{} \lR{}}}
\and
\gpfit{\inferrule*[]{\phantom{.}}{\evalCtxSpace{};\evalCtx{}\vdash_{\mathsf S} \lsempty{} \evalArrow{} \lsempty{}}}
\and
\gpfit{\inferrule*[]{(\lemptyset, \evalCtxSpace{})\vdash \ls{}^{\lstar}\colon\svalid \\ \evalCtxSpace{};\evalCtx{}\vdash_{\mathsf S} \ls{} \evalArrow{} \lsv{}}{\evalCtxSpace{};\evalCtx{}\vdash_{\mathsf S} \ls{}^{\lstar} \evalArrow{} \lsv{}}}
\and
\gpfit{\inferrule*[]{(\lemptyset, \evalCtxSpace{})\vdash \Kernel(\lee{})\colon\svalid \\ \sof{\lb{}} = \lsn{0}^{\lstar} \loprod \lsn{1} \\ \exists\,\lbp{},\textcolor{lang}{b^{\prime\prime}}.\; \lb{} = \lbp{}^{\lstar}\loprod \textcolor{lang}{b^{\prime\prime}} \\ \evalCtxSpace{};\evalCtx{}\vdash_{\mathsf S} \sof{\lb{}} \evalArrow{} \lsnv{{\mathrm{svb}}} \\ \lemptyset{};\evalCtxSpace{};\evalCtx{}\vdash_{\mathsf S} \lee{} \xlongrightarrow{\lb{}} \leev{}\glresulttype{\lsnv{{\mathrm{svb}}}}}{\evalCtxSpace{};\evalCtx{}\vdash_{\mathsf S} \Kernel(\lee{}) \evalArrow{} \lR^{\glvreshape{\leev{}}{\lsnv{{\mathrm{svb}}}}{\lsnv{{\mathrm{svb}}}},\Kernel{},\lsnv{{\mathrm{svb}}}}}}
\and
\gpfit{\inferrule*[]{(\lemptyset, \evalCtxSpace{})\vdash \CoKernel(\lee{})\colon\svalid \\ \sof{\lb{}} = \lsn{0}^{\lstar} \loprod \lsn{1} \\ \exists\,\lbp{},\textcolor{lang}{b^{\prime\prime}}.\; \lb{} = \lbp{}^{\lstar}\loprod \textcolor{lang}{b^{\prime\prime}} \\ \evalCtxSpace{};\evalCtx{}\vdash_{\mathsf S} \sof{\lb{}} \evalArrow{} \lsnv{{\mathrm{svb}}} \\ \lemptyset{};\evalCtxSpace{};\evalCtx{}\vdash_{\mathsf S} \lee{} \xlongrightarrow{\lb{}} \leev{}\glresulttype{\lsnv{{\mathrm{svb}}}}}{\evalCtxSpace{};\evalCtx{}\vdash_{\mathsf S} \CoKernel(\lee{}) \evalArrow{} \lR^{\glvreshape{\leev{}}{\lsnv{{\mathrm{svb}}}}{\lsnv{{\mathrm{svb}}}},\CoKernel{},\lsnv{{\mathrm{svb}}}}}}
\and
\gpfit{\inferrule*[]{(\lemptyset, \evalCtxSpace{})\vdash \Image(\lee{})\colon\svalid \\ \sof{\lb{}} = \lsn{0}^{\lstar} \loprod \lsn{1} \\ \exists\,\lbp{},\textcolor{lang}{b^{\prime\prime}}.\; \lb{} = \lbp{}^{\lstar}\loprod \textcolor{lang}{b^{\prime\prime}} \\ \evalCtxSpace{};\evalCtx{}\vdash_{\mathsf S} \sof{\lb{}} \evalArrow{} \lsnv{{\mathrm{svb}}} \\ \lemptyset{};\evalCtxSpace{};\evalCtx{}\vdash_{\mathsf S} \lee{} \xlongrightarrow{\lb{}} \leev{}\glresulttype{\lsnv{{\mathrm{svb}}}}}{\evalCtxSpace{};\evalCtx{}\vdash_{\mathsf S} \Image(\lee{}) \evalArrow{} \lR^{\glvreshape{\leev{}}{\lsnv{{\mathrm{svb}}}}{\lsnv{{\mathrm{svb}}}},\Image{},\lsnv{{\mathrm{svb}}}}}}
\and
\gpfit{\inferrule*[]{(\lemptyset, \evalCtxSpace{})\vdash \CoImage(\lee{})\colon\svalid \\ \sof{\lb{}} = \lsn{0}^{\lstar} \loprod \lsn{1} \\ \exists\,\lbp{},\textcolor{lang}{b^{\prime\prime}}.\; \lb{} = \lbp{}^{\lstar}\loprod \textcolor{lang}{b^{\prime\prime}} \\ \evalCtxSpace{};\evalCtx{}\vdash_{\mathsf S} \sof{\lb{}} \evalArrow{} \lsnv{{\mathrm{svb}}} \\ \lemptyset{};\evalCtxSpace{};\evalCtx{}\vdash_{\mathsf S} \lee{} \xlongrightarrow{\lb{}} \leev{}\glresulttype{\lsnv{{\mathrm{svb}}}}}{\evalCtxSpace{};\evalCtx{}\vdash_{\mathsf S} \CoImage(\lee{}) \evalArrow{} \lR^{\glvreshape{\leev{}}{\lsnv{{\mathrm{svb}}}}{\lsnv{{\mathrm{svb}}}},\CoImage{},\lsnv{{\mathrm{svb}}}}}}
\and
\gpfit{\inferrule*[]{(\lemptyset, \evalCtxSpace{})\vdash \lsn{0} \loprod \lsn{1}\colon\svalid \\ \forall i.\; \evalCtxSpace{};\evalCtx{}\vdash_{\mathsf S} \lsn{i} \evalArrow{} \lsnv{i}}{\evalCtxSpace{};\evalCtx{}\vdash_{\mathsf S} \lsn{0} \loprod \lsn{1} \evalArrow{} \lsnv{0} \loprod \lsnv{1}}}
\and
\gpfit{\inferrule*[]{(\lemptyset, \evalCtxSpace{})\vdash \lsn{0} \losum \lsn{1}\colon\svalid \\ \forall i.\; \evalCtxSpace{};\evalCtx{}\vdash_{\mathsf S} \lsn{i} \evalArrow{} \lsnv{i}}{\evalCtxSpace{};\evalCtx{}\vdash_{\mathsf S} \lsn{0} \losum \lsn{1} \evalArrow{} \lsnv{0} \losum \lsnv{1}}}
}{
\gpfit{\inferrule*[lab=\textsc{S-S-Real}]{\phantom{.}}{\evalCtxSpace{};\evalCtx{}\vdash_{\mathsf S} \lR{} \evalArrow{} \lR{}}}
\and
\gpfit{\inferrule*[lab=\textsc{S-S-Empty}]{\phantom{.}}{\evalCtxSpace{};\evalCtx{}\vdash_{\mathsf S} \lsempty{} \evalArrow{} \lsempty{}}}
\and
\gpfit{\inferrule*[lab=\textsc{S-S-Dual}]{(\lemptyset, \evalCtxSpace{})\vdash \ls{}^{\lstar}\colon\svalid \\ \evalCtxSpace{};\evalCtx{}\vdash_{\mathsf S} \ls{} \evalArrow{} \lsv{}}{\evalCtxSpace{};\evalCtx{}\vdash_{\mathsf S} \ls{}^{\lstar} \evalArrow{} \lsv{}}}
\and
\gpfit{\inferrule*[lab=\textsc{S-S-Ker}]{(\lemptyset, \evalCtxSpace{})\vdash \Kernel(\lee{})\colon\svalid \\ \sof{\lb{}} = \lsn{0}^{\lstar} \loprod \lsn{1} \\ \exists\,\lbp{},\textcolor{lang}{b^{\prime\prime}}.\; \lb{} = \lbp{}^{\lstar}\loprod \textcolor{lang}{b^{\prime\prime}} \\ \evalCtxSpace{};\evalCtx{}\vdash_{\mathsf S} \sof{\lb{}} \evalArrow{} \lsnv{{\mathrm{svb}}} \\ \lemptyset{};\evalCtxSpace{};\evalCtx{}\vdash_{\mathsf S} \lee{} \xlongrightarrow{\lb{}} \leev{}\glresulttype{\lsnv{{\mathrm{svb}}}}}{\evalCtxSpace{};\evalCtx{}\vdash_{\mathsf S} \Kernel(\lee{}) \evalArrow{} \lR^{\glvreshape{\leev{}}{\lsnv{{\mathrm{svb}}}}{\lsnv{{\mathrm{svb}}}},\Kernel{},\lsnv{{\mathrm{svb}}}}}}
\and
\gpfit{\inferrule*[lab=\textsc{S-S-Coker}]{(\lemptyset, \evalCtxSpace{})\vdash \CoKernel(\lee{})\colon\svalid \\ \sof{\lb{}} = \lsn{0}^{\lstar} \loprod \lsn{1} \\ \exists\,\lbp{},\textcolor{lang}{b^{\prime\prime}}.\; \lb{} = \lbp{}^{\lstar}\loprod \textcolor{lang}{b^{\prime\prime}} \\ \evalCtxSpace{};\evalCtx{}\vdash_{\mathsf S} \sof{\lb{}} \evalArrow{} \lsnv{{\mathrm{svb}}} \\ \lemptyset{};\evalCtxSpace{};\evalCtx{}\vdash_{\mathsf S} \lee{} \xlongrightarrow{\lb{}} \leev{}\glresulttype{\lsnv{{\mathrm{svb}}}}}{\evalCtxSpace{};\evalCtx{}\vdash_{\mathsf S} \CoKernel(\lee{}) \evalArrow{} \lR^{\glvreshape{\leev{}}{\lsnv{{\mathrm{svb}}}}{\lsnv{{\mathrm{svb}}}},\CoKernel{},\lsnv{{\mathrm{svb}}}}}}
\and
\gpfit{\inferrule*[lab=\textsc{S-S-Img}]{(\lemptyset, \evalCtxSpace{})\vdash \Image(\lee{})\colon\svalid \\ \sof{\lb{}} = \lsn{0}^{\lstar} \loprod \lsn{1} \\ \exists\,\lbp{},\textcolor{lang}{b^{\prime\prime}}.\; \lb{} = \lbp{}^{\lstar}\loprod \textcolor{lang}{b^{\prime\prime}} \\ \evalCtxSpace{};\evalCtx{}\vdash_{\mathsf S} \sof{\lb{}} \evalArrow{} \lsnv{{\mathrm{svb}}} \\ \lemptyset{};\evalCtxSpace{};\evalCtx{}\vdash_{\mathsf S} \lee{} \xlongrightarrow{\lb{}} \leev{}\glresulttype{\lsnv{{\mathrm{svb}}}}}{\evalCtxSpace{};\evalCtx{}\vdash_{\mathsf S} \Image(\lee{}) \evalArrow{} \lR^{\glvreshape{\leev{}}{\lsnv{{\mathrm{svb}}}}{\lsnv{{\mathrm{svb}}}},\Image{},\lsnv{{\mathrm{svb}}}}}}
\and
\gpfit{\inferrule*[lab=\textsc{S-S-Coimg}]{(\lemptyset, \evalCtxSpace{})\vdash \CoImage(\lee{})\colon\svalid \\ \sof{\lb{}} = \lsn{0}^{\lstar} \loprod \lsn{1} \\ \exists\,\lbp{},\textcolor{lang}{b^{\prime\prime}}.\; \lb{} = \lbp{}^{\lstar}\loprod \textcolor{lang}{b^{\prime\prime}} \\ \evalCtxSpace{};\evalCtx{}\vdash_{\mathsf S} \sof{\lb{}} \evalArrow{} \lsnv{{\mathrm{svb}}} \\ \lemptyset{};\evalCtxSpace{};\evalCtx{}\vdash_{\mathsf S} \lee{} \xlongrightarrow{\lb{}} \leev{}\glresulttype{\lsnv{{\mathrm{svb}}}}}{\evalCtxSpace{};\evalCtx{}\vdash_{\mathsf S} \CoImage(\lee{}) \evalArrow{} \lR^{\glvreshape{\leev{}}{\lsnv{{\mathrm{svb}}}}{\lsnv{{\mathrm{svb}}}},\CoImage{},\lsnv{{\mathrm{svb}}}}}}
\and
\gpfit{\inferrule*[lab=\textsc{S-S-Tprod}]{(\lemptyset, \evalCtxSpace{})\vdash \lsn{0} \loprod \lsn{1}\colon\svalid \\ \forall i.\; \evalCtxSpace{};\evalCtx{}\vdash_{\mathsf S} \lsn{i} \evalArrow{} \lsnv{i}}{\evalCtxSpace{};\evalCtx{}\vdash_{\mathsf S} \lsn{0} \loprod \lsn{1} \evalArrow{} \lsnv{0} \loprod \lsnv{1}}}
\and
\gpfit{\inferrule*[lab=\textsc{S-S-Dsum}]{(\lemptyset, \evalCtxSpace{})\vdash \lsn{0} \losum \lsn{1}\colon\svalid \\ \forall i.\; \evalCtxSpace{};\evalCtx{}\vdash_{\mathsf S} \lsn{i} \evalArrow{} \lsnv{i}}{\evalCtxSpace{};\evalCtx{}\vdash_{\mathsf S} \lsn{0} \losum \lsn{1} \evalArrow{} \lsnv{0} \losum \lsnv{1}}}
}
\gldefjudgment{s.ty}{\lepsilon{};\evalCtx{}\vdash_{\mathsf S} \lty{}\evalArrow{}\ltyv{}}{Strengthened type evaluation (generated from \texttt{GLang.S.Ty}).}{fig:gl-s-ty}
\gldefrule{s.ty.space}{s.ty}{\textsc{S-T-Space}}{\evalCtxSpace{};\evalCtx{}\vdash_{\mathsf S} \ls{} \evalArrow{} \lsv{}}{\evalCtxSpace{};\evalCtx{}\vdash_{\mathsf S} \ls{} \evalArrow{} \lsv{}}{\texttt{GLang.S.Ty.space}}{GLang/Adequacy/Strengthened.lean:126}
\gldefrule{s.ty.piTy}{s.ty}{\textsc{S-T-Pi\allowbreak Ty}}{(\lemptyset, \evalCtxSpace{})\vdash (x\colon \ltyn{0}\fnto \ltyn{1})\colon \lkn{0} \fnto \lkn{1} \\ \tof{\lbn{0}} = \ltyn{0} \\ (\emptyset,\evalCtxSpace{})\vdash \lbn{0}\colon \bvalid{} \\ \evalCtxSpace{};\evalCtx{}\vdash_{\mathsf S} \ltyn{0} \evalArrow{} \ltynv{0} \\ (\evalCtxSpace{}, \mathit{x}\colon \ltyn{0});\evalCtx{}[(\mathit{x},\lbn{0}) \mapsto \mathit{x}_{\ltynv{0}}]\vdash_{\mathsf S} \ltyn{1} \evalArrow{} \ltynv{1}}{\evalCtxSpace{};\evalCtx{}\vdash_{\mathsf S} (x\colon \ltyn{0}\fnto \ltyn{1}) \evalArrow{} (x\colon \ltynv{0}\fnto \ltynv{1})}{\texttt{GLang.S.Ty.pi\allowbreak Ty}}{GLang/Adequacy/Strengthened.lean:129}
\gldefrulerows{s.ty}{
\textsc{S-T-Space} & $\evalCtxSpace{};\evalCtx{}\vdash_{\mathsf S} \ls{} \evalArrow{} \lsv{}$ & \texttt{\footnotesize GLang/Adequacy/Strengthened.lean:126}\\
\textsc{S-T-Pi\allowbreak Ty} & $\evalCtxSpace{};\evalCtx{}\vdash_{\mathsf S} (x\colon \ltyn{0}\fnto \ltyn{1}) \evalArrow{} (x\colon \ltynv{0}\fnto \ltynv{1})$ & \texttt{\footnotesize GLang/Adequacy/Strengthened.lean:129}\\
}{
\gpfit{\inferrule*[]{\evalCtxSpace{};\evalCtx{}\vdash_{\mathsf S} \ls{} \evalArrow{} \lsv{}}{\evalCtxSpace{};\evalCtx{}\vdash_{\mathsf S} \ls{} \evalArrow{} \lsv{}}}
\and
\gpfit{\inferrule*[]{(\lemptyset, \evalCtxSpace{})\vdash (x\colon \ltyn{0}\fnto \ltyn{1})\colon \lkn{0} \fnto \lkn{1} \\ \tof{\lbn{0}} = \ltyn{0} \\ (\emptyset,\evalCtxSpace{})\vdash \lbn{0}\colon \bvalid{} \\ \evalCtxSpace{};\evalCtx{}\vdash_{\mathsf S} \ltyn{0} \evalArrow{} \ltynv{0} \\ (\evalCtxSpace{}, \mathit{x}\colon \ltyn{0});\evalCtx{}[(\mathit{x},\lbn{0}) \mapsto \mathit{x}_{\ltynv{0}}]\vdash_{\mathsf S} \ltyn{1} \evalArrow{} \ltynv{1}}{\evalCtxSpace{};\evalCtx{}\vdash_{\mathsf S} (x\colon \ltyn{0}\fnto \ltyn{1}) \evalArrow{} (x\colon \ltynv{0}\fnto \ltynv{1})}}
}{
\gpfit{\inferrule*[lab=\textsc{S-T-Space}]{\evalCtxSpace{};\evalCtx{}\vdash_{\mathsf S} \ls{} \evalArrow{} \lsv{}}{\evalCtxSpace{};\evalCtx{}\vdash_{\mathsf S} \ls{} \evalArrow{} \lsv{}}}
\and
\gpfit{\inferrule*[lab=\textsc{S-T-Pi\allowbreak Ty}]{(\lemptyset, \evalCtxSpace{})\vdash (x\colon \ltyn{0}\fnto \ltyn{1})\colon \lkn{0} \fnto \lkn{1} \\ \tof{\lbn{0}} = \ltyn{0} \\ (\emptyset,\evalCtxSpace{})\vdash \lbn{0}\colon \bvalid{} \\ \evalCtxSpace{};\evalCtx{}\vdash_{\mathsf S} \ltyn{0} \evalArrow{} \ltynv{0} \\ (\evalCtxSpace{}, \mathit{x}\colon \ltyn{0});\evalCtx{}[(\mathit{x},\lbn{0}) \mapsto \mathit{x}_{\ltynv{0}}]\vdash_{\mathsf S} \ltyn{1} \evalArrow{} \ltynv{1}}{\evalCtxSpace{};\evalCtx{}\vdash_{\mathsf S} (x\colon \ltyn{0}\fnto \ltyn{1}) \evalArrow{} (x\colon \ltynv{0}\fnto \ltynv{1})}}
}
\gldefjudgment{s.tyAt}{\lepsilon{};\evalCtx{}\vdash_{\mathsf S} \lb{}\evalArrow{}_{\mathsf{basis}}\ltyv{}}{Strengthened exact-basis type evaluation (generated from \texttt{GLang.S.TyAt}).}{fig:gl-s-tyat}
\gldefrule{s.tyAt.space}{s.tyAt}{\textsc{S-TA-Space}}{\tof{\lb{}} = \sof{\lb{}} \\ (\emptyset,\evalCtxSpace{})\vdash \lb{}\colon \bvalid{} \\ \evalCtxSpace{};\evalCtx{}\vdash_{\mathsf S} \sof{\lb{}} \evalArrow{} \lsv{}}{\evalCtxSpace{};\evalCtx{}\vdash_{\mathsf S} \lb{} \evalArrow{}_{\mathsf{basis}} \lsv{}}{\texttt{GLang.S.Ty\allowbreak At.space}}{GLang/Adequacy/Strengthened.lean:147}
\gldefrule{s.tyAt.frame}{s.tyAt}{\textsc{S-TA-Frame}}{(\emptyset,\evalCtxSpace{})\vdash \frameB{x}{\lbn{0}}{\lbn{1}}\colon \bvalid{} \\ \evalCtxSpace{};\evalCtx{}\vdash_{\mathsf S} \lbn{0} \evalArrow{}_{\mathsf{basis}} \ltynv{0} \\ (\evalCtxSpace{}, \mathit{x}\colon \tof{\lbn{0}});\evalCtx{}[(\mathit{x},\lbn{0}) \mapsto \mathit{x}_{\ltynv{0}}]\vdash_{\mathsf S} \lbn{1} \evalArrow{}_{\mathsf{basis}} \ltynv{1}}{\evalCtxSpace{};\evalCtx{}\vdash_{\mathsf S} \frameB{x}{\lbn{0}}{\lbn{1}} \evalArrow{}_{\mathsf{basis}} (x\colon \ltynv{0}\fnto \ltynv{1})}{\texttt{GLang.S.Ty\allowbreak At.frame}}{GLang/Adequacy/Strengthened.lean:152}
\gldefrulerows{s.tyAt}{
\textsc{S-TA-Space} & $\evalCtxSpace{};\evalCtx{}\vdash_{\mathsf S} \lb{} \evalArrow{}_{\mathsf{basis}} \lsv{}$ & \texttt{\footnotesize GLang/Adequacy/Strengthened.lean:147}\\
\textsc{S-TA-Frame} & $\evalCtxSpace{};\evalCtx{}\vdash_{\mathsf S} \frameB{x}{\lbn{0}}{\lbn{1}} \evalArrow{}_{\mathsf{basis}} (x\colon \ltynv{0}\fnto \ltynv{1})$ & \texttt{\footnotesize GLang/Adequacy/Strengthened.lean:152}\\
}{
\gpfit{\inferrule*[]{\tof{\lb{}} = \sof{\lb{}} \\ (\emptyset,\evalCtxSpace{})\vdash \lb{}\colon \bvalid{} \\ \evalCtxSpace{};\evalCtx{}\vdash_{\mathsf S} \sof{\lb{}} \evalArrow{} \lsv{}}{\evalCtxSpace{};\evalCtx{}\vdash_{\mathsf S} \lb{} \evalArrow{}_{\mathsf{basis}} \lsv{}}}
\and
\gpfit{\inferrule*[]{(\emptyset,\evalCtxSpace{})\vdash \frameB{x}{\lbn{0}}{\lbn{1}}\colon \bvalid{} \\ \evalCtxSpace{};\evalCtx{}\vdash_{\mathsf S} \lbn{0} \evalArrow{}_{\mathsf{basis}} \ltynv{0} \\ (\evalCtxSpace{}, \mathit{x}\colon \tof{\lbn{0}});\evalCtx{}[(\mathit{x},\lbn{0}) \mapsto \mathit{x}_{\ltynv{0}}]\vdash_{\mathsf S} \lbn{1} \evalArrow{}_{\mathsf{basis}} \ltynv{1}}{\evalCtxSpace{};\evalCtx{}\vdash_{\mathsf S} \frameB{x}{\lbn{0}}{\lbn{1}} \evalArrow{}_{\mathsf{basis}} (x\colon \ltynv{0}\fnto \ltynv{1})}}
}{
\gpfit{\inferrule*[lab=\textsc{S-TA-Space}]{\tof{\lb{}} = \sof{\lb{}} \\ (\emptyset,\evalCtxSpace{})\vdash \lb{}\colon \bvalid{} \\ \evalCtxSpace{};\evalCtx{}\vdash_{\mathsf S} \sof{\lb{}} \evalArrow{} \lsv{}}{\evalCtxSpace{};\evalCtx{}\vdash_{\mathsf S} \lb{} \evalArrow{}_{\mathsf{basis}} \lsv{}}}
\and
\gpfit{\inferrule*[lab=\textsc{S-TA-Frame}]{(\emptyset,\evalCtxSpace{})\vdash \frameB{x}{\lbn{0}}{\lbn{1}}\colon \bvalid{} \\ \evalCtxSpace{};\evalCtx{}\vdash_{\mathsf S} \lbn{0} \evalArrow{}_{\mathsf{basis}} \ltynv{0} \\ (\evalCtxSpace{}, \mathit{x}\colon \tof{\lbn{0}});\evalCtx{}[(\mathit{x},\lbn{0}) \mapsto \mathit{x}_{\ltynv{0}}]\vdash_{\mathsf S} \lbn{1} \evalArrow{}_{\mathsf{basis}} \ltynv{1}}{\evalCtxSpace{};\evalCtx{}\vdash_{\mathsf S} \frameB{x}{\lbn{0}}{\lbn{1}} \evalArrow{}_{\mathsf{basis}} (x\colon \ltynv{0}\fnto \ltynv{1})}}
}
\gldefjudgment{s.cob}{\lepsilon{};\evalCtx{}\vdash_{\mathsf S} \lbn{0}\xlongleftrightarrow{\leev{}}\lbn{1}}{Strengthened change of basis (generated from \texttt{GLang.S.Cob}).}{fig:gl-s-cob}
\gldefrule{s.cob.inverse}{s.cob}{\textsc{S-C-Inverse}}{\forall i.\; (\emptyset,\evalCtxSpace{})\vdash \lbn{i}\colon \bvalid{} \\ \evalCtxSpace{};\evalCtx{}\vdash_{\mathsf S} \lbn{1} \xlongleftrightarrow{\leev{}} \lbn{0}}{\evalCtxSpace{};\evalCtx{}\vdash_{\mathsf S} \lbn{0} \xlongleftrightarrow{\leev{}^{-1}} \lbn{1}}{\texttt{GLang.S.Cob.inverse}}{GLang/Adequacy/Strengthened.lean:167}
\gldefrule{s.cob.trans}{s.cob}{\textsc{S-C-Trans}}{(\emptyset,\evalCtxSpace{})\vdash \lbn{0}\colon \bvalid{} \\ (\emptyset,\evalCtxSpace{})\vdash \lbn{2}\colon \bvalid{} \\ \evalCtxSpace{};\evalCtx{}\vdash_{\mathsf S} \lbn{0} \xlongleftrightarrow{\lenv{0}} \lbn{1} \\ \evalCtxSpace{};\evalCtx{}\vdash_{\mathsf S} \lbn{1} \xlongleftrightarrow{\lenv{1}} \lbn{2} \\ \evalCtxSpace{};\evalCtx{}\vdash_{\mathsf S} \sof{\lbn{0}} \evalArrow{} \lsv{}}{\evalCtxSpace{};\evalCtx{}\vdash_{\mathsf S} \lbn{0} \xlongleftrightarrow{\glvmtm{\lsv{}}{\lsv{}}{\lenv{1}}{\glvreshape{\lenv{0}}{\lsv{} \loprod \lsv{}}{\lsv{} \loprod \lsv{}}}} \lbn{2}}{\texttt{GLang.S.Cob.trans}}{GLang/Adequacy/Strengthened.lean:172}
\gldefrule{s.cob.tprod}{s.cob}{\textsc{S-C-Tprod}}{(\emptyset,\evalCtxSpace{})\vdash \lbn{0}' \loprod \lbn{1}'\colon \bvalid{} \\ (\emptyset,\evalCtxSpace{})\vdash \lbn{0} \loprod \lbn{1}\colon \bvalid{} \\ \forall i.\; \evalCtxSpace{};\evalCtx{}\vdash_{\mathsf S} \lbn{i}' \xlongleftrightarrow{\lenv{i}} \lbn{i}}{\evalCtxSpace{};\evalCtx{}\vdash_{\mathsf S} \lbn{0}' \loprod \lbn{1}' \xlongleftrightarrow{\vkrn(\lenv{0}, \lenv{1})} \lbn{0} \loprod \lbn{1}}{\texttt{GLang.S.Cob.tprod}}{GLang/Adequacy/Strengthened.lean:180}
\gldefrule{s.cob.dsum}{s.cob}{\textsc{S-C-Dsum}}{(\emptyset,\evalCtxSpace{})\vdash \lbn{0}' \losum \lbn{1}'\colon \bvalid{} \\ (\emptyset,\evalCtxSpace{})\vdash \lbn{0} \losum \lbn{1}\colon \bvalid{} \\ \forall i.\; \evalCtxSpace{};\evalCtx{}\vdash_{\mathsf S} \lbn{i}' \xlongleftrightarrow{\lenv{i}} \lbn{i}}{\evalCtxSpace{};\evalCtx{}\vdash_{\mathsf S} \lbn{0}' \losum \lbn{1}' \xlongleftrightarrow{\vblk(\lenv{0}, \lenv{1})} \lbn{0} \losum \lbn{1}}{\texttt{GLang.S.Cob.dsum}}{GLang/Adequacy/Strengthened.lean:187}
\gldefrule{s.cob.dualStd}{s.cob}{\textsc{S-C-Dual\allowbreak Std}}{\tof{\lbn{0}} = \sof{\lbn{0}} \\ (\emptyset,\evalCtxSpace{})\vdash \lbn{0}^{\lstar}\colon \bvalid{} \\ (\emptyset,\evalCtxSpace{})\vdash \stdBasisDCtx{\tof{\lbn{0}^{\lstar}}}\colon \bvalid{} \\ \evalCtxSpace{};\evalCtx{}\vdash_{\mathsf S} \lbn{0} \xlongleftrightarrow{\leev{}} \stdBasisDCtx{\tof{\lbn{0}}}}{\evalCtxSpace{};\evalCtx{}\vdash_{\mathsf S} \lbn{0}^{\lstar} \xlongleftrightarrow{\Trans(\leev{}^{-1})} \stdBasisDCtx{\tof{\lbn{0}^{\lstar}}}}{\texttt{GLang.S.Cob.dual\_\allowbreak std}}{GLang/Adequacy/Strengthened.lean:193}
\gldefrule{s.cob.image}{s.cob}{\textsc{S-C-Image}}{(\emptyset,\evalCtxSpace{})\vdash \iBasisC{\lbn{0}}\colon \bvalid{} \\ (\emptyset,\evalCtxSpace{})\vdash \iBasisC{\lbn{0}'}\colon \bvalid{} \\ \exists \lsn{0}.\; \exists \lsn{1}.\; \exists \lee{}.\; \sof{\lbn{0}} = \CoKernel(\lee{}) \\ \evalCtxSpace{};\evalCtx{}\vdash_{\mathsf S} \lbn{0} \xlongleftrightarrow{\leev{}} \lbn{0}'}{\evalCtxSpace{};\evalCtx{}\vdash_{\mathsf S} \iBasisC{\lbn{0}} \xlongleftrightarrow{\leev{}} \iBasisC{\lbn{0}'}}{\texttt{GLang.S.Cob.image}}{GLang/Adequacy/Strengthened.lean:199}
\gldefrule{s.cob.imageSvds}{s.cob}{\textsc{S-C-Image\allowbreak Svds}}{(\emptyset,\evalCtxSpace{})\vdash \iBasisC{\pBasisC{\CoKernel(\lee{})}{\lbn{0}}{\lbn{1}}}\colon \bvalid{} \\ (\emptyset,\evalCtxSpace{})\vdash \pBasisC{\Image(\lee{})}{\lbn{0}}{\lbn{1}}\colon \bvalid{} \\ \forall i.\; \sof{\lbn{i}} = \lsn{i} \\ \forall i.\; \isStdBasis{\lbn{i}} \\ \evalCtxSpace{};\evalCtx{}\vdash_{\mathsf S} \sof{\lbn{0}^{\lstar} \loprod \lbn{1}} \evalArrow{} \lsnv{0} \loprod \lsnv{1} \\ \lemptyset{};\evalCtxSpace{};\evalCtx{}\vdash_{\mathsf S} \lee{} \xlongrightarrow{\lbn{0}^{\lstar} \loprod \lbn{1}} \leev{}\glresulttype{\lsnv{0} \loprod \lsnv{1}} \\ \lenv{{\mathrm{M}}} = \glvreshape{\leev{}}{\lsnv{0} \loprod \lsnv{1}}{\lsnv{0} \loprod \lsnv{1}} \\ \lenv{{\mathrm{U}}} = \glvop{GSVDU}(\lenv{{\mathrm{M}}},\lidelta{\lsnv{0}},\lidelta{\lsnv{1}}) \\ \lenv{{\mathrm{V}}} = \glvop{GSVDV}(\lenv{{\mathrm{M}}},\lidelta{\lsnv{0}},\lidelta{\lsnv{1}}) \\ \lenv{{\mathrm{P}}} = \glvmtm{\lsnv{1}}{\lsnv{1}}{\lenv{{\mathrm{V}}}}{\glvmtm{\lsnv{0}}{\lsnv{0}}{\lenv{{\mathrm{M}}}}{\Trans(\lenv{{\mathrm{U}}})}}}{\evalCtxSpace{};\evalCtx{}\vdash_{\mathsf S} \iBasisC{\pBasisC{\CoKernel(\lee{})}{\lbn{0}}{\lbn{1}}} \xlongleftrightarrow{\glvreshape{\lenv{{\mathrm{P}}}}{\lR^{\lenv{{\mathrm{M}}},\CoKernel{},\lsnv{0} \loprod \lsnv{1}} \loprod \lR^{\lenv{{\mathrm{M}}},\Image{},\lsnv{0} \loprod \lsnv{1}}}{\operatorname{inferSpace}(\lenv{{\mathrm{P}}})}} \pBasisC{\Image(\lee{})}{\lbn{0}}{\lbn{1}}}{\texttt{GLang.S.Cob.image\_\allowbreak svds}}{GLang/Adequacy/Strengthened.lean:205}
\gldefrule{s.cob.projKerCoker}{s.cob}{\textsc{S-C-Proj\allowbreak Ker\allowbreak Coker}}{(\emptyset,\evalCtxSpace{})\vdash \pBasisC{\ls{}}{\lbn{0}}{\lbn{1}}\colon \bvalid{} \\ (\emptyset,\evalCtxSpace{})\vdash \pBasisC{\ls{}}{\lbn{0}'}{\lbn{1}}\colon \bvalid{} \\ \ls{} = \Kernel(\lee{}) \;\vee\; \ls{} = \CoKernel(\lee{}) \\ \forall i.\; \sof{\lbn{i}} = \lsn{i} \\ \forall i.\; \evalCtxSpace{};\evalCtx{}\vdash_{\mathsf S} \sof{\lbn{i}} \evalArrow{} \lsnv{i} \\ \evalCtxSpace{};\evalCtx{}\vdash_{\mathsf S} \ls{} \evalArrow{} \lsv{} \\ \evalCtxSpace{};\evalCtx{}\vdash_{\mathsf S} \ls{} \evalArrow{} \lsv{}' \\ \evalCtxSpace{};\evalCtx{}\vdash_{\mathsf S} \sof{\lbn{0}'} \evalArrow{} \lsnv{0}' \\ \evalCtxSpace{};\evalCtx{}\vdash_{\mathsf S} \lbn{0} \xlongleftrightarrow{\lenv{2}} \lbn{0}' \\ \lemptyset{};\evalCtxSpace{};\evalCtx{}\vdash_{\mathsf S} \lembd{}_{\ls{}} \xlongrightarrow{\pBasisC{\ls{}}{\lbn{0}}{\lbn{1}}^{\lstar} \loprod \lbn{0}} \lenv{3}\glresulttype{\lsv{} \loprod \lsnv{0}} \\ \lemptyset{};\evalCtxSpace{};\evalCtx{}\vdash_{\mathsf S} \lproj{}_{\ls{}, \sof{\lbn{0}'}} \xlongrightarrow{\lbn{0}'^{\lstar} \loprod \pBasisC{\ls{}}{\lbn{0}'}{\lbn{1}}} \lenv{1}\glresulttype{\lsnv{0}' \loprod \lsv{}'}}{\evalCtxSpace{};\evalCtx{}\vdash_{\mathsf S} \pBasisC{\ls{}}{\lbn{0}}{\lbn{1}} \xlongleftrightarrow{\glvmtm{\lsnv{0}'}{\lsnv{0}'}{\glvreshape{\lenv{1}}{\lsnv{0}' \loprod \lsv{}'}{\lsnv{0}' \loprod \lsv{}'}}{\glvmtm{\lsnv{0}}{\lsnv{0}}{\glvreshape{\lenv{2}}{\lsnv{0} \loprod \lsnv{0}'}{\lsnv{0} \loprod \lsnv{0}'}}{\glvreshape{\lenv{3}}{\lsv{} \loprod \lsnv{0}}{\lsv{} \loprod \lsnv{0}}}}} \pBasisC{\ls{}}{\lbn{0}'}{\lbn{1}}}{\texttt{GLang.S.Cob.proj\_\allowbreak ker\_\allowbreak coker}}{GLang/Adequacy/Strengthened.lean:221}
\gldefrule{s.cob.projImgCoimg}{s.cob}{\textsc{S-C-Proj\allowbreak Img\allowbreak Coimg}}{(\emptyset,\evalCtxSpace{})\vdash \pBasisC{\ls{}}{\lbn{0}}{\lbn{1}}\colon \bvalid{} \\ (\emptyset,\evalCtxSpace{})\vdash \pBasisC{\ls{}}{\lbn{0}}{\lbn{1}'}\colon \bvalid{} \\ \ls{} = \Image(\lee{}) \;\vee\; \ls{} = \CoImage(\lee{}) \\ \forall i.\; \sof{\lbn{i}} = \lsn{i} \\ \forall i.\; \evalCtxSpace{};\evalCtx{}\vdash_{\mathsf S} \sof{\lbn{i}} \evalArrow{} \lsnv{i} \\ \evalCtxSpace{};\evalCtx{}\vdash_{\mathsf S} \ls{} \evalArrow{} \lsv{} \\ \evalCtxSpace{};\evalCtx{}\vdash_{\mathsf S} \ls{} \evalArrow{} \lsv{}' \\ \evalCtxSpace{};\evalCtx{}\vdash_{\mathsf S} \sof{\lbn{1}'} \evalArrow{} \lsnv{1}' \\ \evalCtxSpace{};\evalCtx{}\vdash_{\mathsf S} \lbn{1} \xlongleftrightarrow{\lenv{2}} \lbn{1}' \\ \lemptyset{};\evalCtxSpace{};\evalCtx{}\vdash_{\mathsf S} \lembd{}_{\ls{}} \xlongrightarrow{\pBasisC{\ls{}}{\lbn{0}}{\lbn{1}}^{\lstar} \loprod \lbn{1}} \lenv{3}\glresulttype{\lsv{} \loprod \lsnv{1}} \\ \lemptyset{};\evalCtxSpace{};\evalCtx{}\vdash_{\mathsf S} \lproj{}_{\ls{}, \sof{\lbn{1}'}} \xlongrightarrow{\lbn{1}'^{\lstar} \loprod \pBasisC{\ls{}}{\lbn{0}}{\lbn{1}'}} \lenv{1}\glresulttype{\lsnv{1}' \loprod \lsv{}'}}{\evalCtxSpace{};\evalCtx{}\vdash_{\mathsf S} \pBasisC{\ls{}}{\lbn{0}}{\lbn{1}} \xlongleftrightarrow{\glvmtm{\lsnv{1}'}{\lsnv{1}'}{\glvreshape{\lenv{1}}{\lsnv{1}' \loprod \lsv{}'}{\lsnv{1}' \loprod \lsv{}'}}{\glvmtm{\lsnv{1}}{\lsnv{1}}{\glvreshape{\lenv{2}}{\lsnv{1} \loprod \lsnv{1}'}{\lsnv{1} \loprod \lsnv{1}'}}{\glvreshape{\lenv{3}}{\lsv{} \loprod \lsnv{1}}{\lsv{} \loprod \lsnv{1}}}}} \pBasisC{\ls{}}{\lbn{0}}{\lbn{1}'}}{\texttt{GLang.S.Cob.proj\_\allowbreak img\_\allowbreak coimg}}{GLang/Adequacy/Strengthened.lean:242}
\gldefrule{s.cob.termsRref}{s.cob}{\textsc{S-C-Terms\allowbreak Rref}}{(\emptyset,\evalCtxSpace{})\vdash \lb{}\colon \bvalid{} \\ (\emptyset,\evalCtxSpace{})\vdash \llpl{} \lee{} \mathbin{:\!:} \mathit{es}\llpr{}_{\lb{}'}\colon \bvalid{} \\ \evalCtxSpace{};\evalCtx{}\vdash_{\mathsf S} \lb{} \xlongleftrightarrow{\lenv{1}} \lb{}' \\ \lemptyset{};\evalCtxSpace{};\evalCtx{}\vdash_{\mathsf S} \lee{} \losum \mathit{es}[0] \losum \cdots \xlongrightarrow{\lb{}' \losum \cdots} \lenv{0}\glresulttype{\ltyv{}} \\ \evalCtxSpace{};\evalCtx{}\vdash_{\mathsf S} \sof{\lb{}} \evalArrow{} \lsv{}}{\evalCtxSpace{};\evalCtx{}\vdash_{\mathsf S} \lb{} \xlongleftrightarrow{\glvreshape{\glvop{RREFRep}(\lenv{0},\glvreshape{\lenv{1}}{\lsv{} \loprod \lsv{}}{\lsv{} \loprod \lsv{}})}{\lsv{} \loprod \lsv{}}{\lsv{} \loprod \lsv{}}} \llpl{} \lee{} \mathbin{:\!:} \mathit{es}\llpr{}_{\lb{}'}}{\texttt{GLang.S.Cob.terms\_\allowbreak rref}}{GLang/Adequacy/Strengthened.lean:263}
\gldefrule{s.cob.refl}{s.cob}{\textsc{S-C-Refl}}{(\emptyset,\evalCtxSpace{})\vdash \lb{}\colon \bvalid{} \\ \tof{\lb{}} = \sof{\lb{}} \\ \evalCtxSpace{};\evalCtx{}\vdash_{\mathsf S} \sof{\lb{}} \evalArrow{} \lsv{}}{\evalCtxSpace{};\evalCtx{}\vdash_{\mathsf S} \lb{} \xlongleftrightarrow{\glvreshape{\lidelta{\lsv{}}}{\lsv{} \loprod \lsv{}}{\lsv{} \loprod \lsv{}}} \lb{}}{\texttt{GLang.S.Cob.refl}}{GLang/Adequacy/Strengthened.lean:276}
\gldefrule{s.cob.projKerCokerCod}{s.cob}{\textsc{S-C-Proj\allowbreak Ker\allowbreak Coker\allowbreak Cod}}{(\emptyset,\evalCtxSpace{})\vdash \pBasisC{\ls{}}{\lbn{0}}{\lbn{1}}\colon \bvalid{} \\ (\emptyset,\evalCtxSpace{})\vdash \pBasisC{\ls{}}{\lbn{0}}{\lbn{1}'}\colon \bvalid{} \\ \ls{} = \Kernel(\lee{}) \;\vee\; \ls{} = \CoKernel(\lee{}) \\ \forall i.\; \sof{\lbn{i}} = \lsn{i} \\ \sof{\lbn{1}'} = \lsn{1} \\ \evalCtxSpace{};\evalCtx{}\vdash_{\mathsf S} \sof{\lbn{0}} \evalArrow{} \lsnv{0} \\ \evalCtxSpace{};\evalCtx{}\vdash_{\mathsf S} \ls{} \evalArrow{} \lsv{} \\ \evalCtxSpace{};\evalCtx{}\vdash_{\mathsf S} \ls{} \evalArrow{} \lsv{}' \\ \lemptyset{};\evalCtxSpace{};\evalCtx{}\vdash_{\mathsf S} \lembd{}_{\ls{}} \xlongrightarrow{\pBasisC{\ls{}}{\lbn{0}}{\lbn{1}}^{\lstar} \loprod \lbn{0}} \lenv{3}\glresulttype{\lsv{} \loprod \lsnv{0}} \\ \lemptyset{};\evalCtxSpace{};\evalCtx{}\vdash_{\mathsf S} \lproj{}_{\ls{}, \sof{\lbn{0}}} \xlongrightarrow{\lbn{0}^{\lstar} \loprod \pBasisC{\ls{}}{\lbn{0}}{\lbn{1}'}} \lenv{1}\glresulttype{\lsnv{0} \loprod \lsv{}'}}{\evalCtxSpace{};\evalCtx{}\vdash_{\mathsf S} \pBasisC{\ls{}}{\lbn{0}}{\lbn{1}} \xlongleftrightarrow{\glvmtm{\lsnv{0}}{\lsnv{0}}{\glvreshape{\lenv{1}}{\lsnv{0} \loprod \lsv{}'}{\lsnv{0} \loprod \lsv{}'}}{\glvreshape{\lenv{3}}{\lsv{} \loprod \lsnv{0}}{\lsv{} \loprod \lsnv{0}}}} \pBasisC{\ls{}}{\lbn{0}}{\lbn{1}'}}{\texttt{GLang.S.Cob.proj\_\allowbreak ker\_\allowbreak coker\_\allowbreak cod}}{GLang/Adequacy/Strengthened.lean:283}
\gldefrule{s.cob.projImgCoimgDom}{s.cob}{\textsc{S-C-Proj\allowbreak Img\allowbreak Coimg\allowbreak Dom}}{(\emptyset,\evalCtxSpace{})\vdash \pBasisC{\ls{}}{\lbn{0}}{\lbn{1}}\colon \bvalid{} \\ (\emptyset,\evalCtxSpace{})\vdash \pBasisC{\ls{}}{\lbn{0}'}{\lbn{1}}\colon \bvalid{} \\ \ls{} = \Image(\lee{}) \;\vee\; \ls{} = \CoImage(\lee{}) \\ \forall i.\; \sof{\lbn{i}} = \lsn{i} \\ \sof{\lbn{0}'} = \lsn{0} \\ \evalCtxSpace{};\evalCtx{}\vdash_{\mathsf S} \sof{\lbn{1}} \evalArrow{} \lsnv{1} \\ \evalCtxSpace{};\evalCtx{}\vdash_{\mathsf S} \ls{} \evalArrow{} \lsv{} \\ \evalCtxSpace{};\evalCtx{}\vdash_{\mathsf S} \ls{} \evalArrow{} \lsv{}' \\ \lemptyset{};\evalCtxSpace{};\evalCtx{}\vdash_{\mathsf S} \lembd{}_{\ls{}} \xlongrightarrow{\pBasisC{\ls{}}{\lbn{0}}{\lbn{1}}^{\lstar} \loprod \lbn{1}} \lenv{3}\glresulttype{\lsv{} \loprod \lsnv{1}} \\ \lemptyset{};\evalCtxSpace{};\evalCtx{}\vdash_{\mathsf S} \lproj{}_{\ls{}, \sof{\lbn{1}}} \xlongrightarrow{\lbn{1}^{\lstar} \loprod \pBasisC{\ls{}}{\lbn{0}'}{\lbn{1}}} \lenv{1}\glresulttype{\lsnv{1} \loprod \lsv{}'}}{\evalCtxSpace{};\evalCtx{}\vdash_{\mathsf S} \pBasisC{\ls{}}{\lbn{0}}{\lbn{1}} \xlongleftrightarrow{\glvmtm{\lsnv{1}}{\lsnv{1}}{\glvreshape{\lenv{1}}{\lsnv{1} \loprod \lsv{}'}{\lsnv{1} \loprod \lsv{}'}}{\glvreshape{\lenv{3}}{\lsv{} \loprod \lsnv{1}}{\lsv{} \loprod \lsnv{1}}}} \pBasisC{\ls{}}{\lbn{0}'}{\lbn{1}}}{\texttt{GLang.S.Cob.proj\_\allowbreak img\_\allowbreak coimg\_\allowbreak dom}}{GLang/Adequacy/Strengthened.lean:300}
\gldefrulerows{s.cob}{
\textsc{S-C-Inverse} & $\evalCtxSpace{};\evalCtx{}\vdash_{\mathsf S} \lbn{0} \xlongleftrightarrow{\leev{}^{-1}} \lbn{1}$ & \texttt{\footnotesize GLang/Adequacy/Strengthened.lean:167}\\
\textsc{S-C-Trans} & $\evalCtxSpace{};\evalCtx{}\vdash_{\mathsf S} \lbn{0} \xlongleftrightarrow{\glvmtm{\lsv{}}{\lsv{}}{\lenv{1}}{\glvreshape{\lenv{0}}{\lsv{} \loprod \lsv{}}{\lsv{} \loprod \lsv{}}}} \lbn{2}$ & \texttt{\footnotesize GLang/Adequacy/Strengthened.lean:172}\\
\textsc{S-C-Tprod} & $\evalCtxSpace{};\evalCtx{}\vdash_{\mathsf S} \lbn{0}' \loprod \lbn{1}' \xlongleftrightarrow{\vkrn(\lenv{0}, \lenv{1})} \lbn{0} \loprod \lbn{1}$ & \texttt{\footnotesize GLang/Adequacy/Strengthened.lean:180}\\
\textsc{S-C-Dsum} & $\evalCtxSpace{};\evalCtx{}\vdash_{\mathsf S} \lbn{0}' \losum \lbn{1}' \xlongleftrightarrow{\vblk(\lenv{0}, \lenv{1})} \lbn{0} \losum \lbn{1}$ & \texttt{\footnotesize GLang/Adequacy/Strengthened.lean:187}\\
\textsc{S-C-Dual\allowbreak Std} & $\evalCtxSpace{};\evalCtx{}\vdash_{\mathsf S} \lbn{0}^{\lstar} \xlongleftrightarrow{\Trans(\leev{}^{-1})} \stdBasisDCtx{\tof{\lbn{0}^{\lstar}}}$ & \texttt{\footnotesize GLang/Adequacy/Strengthened.lean:193}\\
\textsc{S-C-Image} & $\evalCtxSpace{};\evalCtx{}\vdash_{\mathsf S} \iBasisC{\lbn{0}} \xlongleftrightarrow{\leev{}} \iBasisC{\lbn{0}'}$ & \texttt{\footnotesize GLang/Adequacy/Strengthened.lean:199}\\
\textsc{S-C-Image\allowbreak Svds} & $\evalCtxSpace{};\evalCtx{}\vdash_{\mathsf S} \iBasisC{\pBasisC{\CoKernel(\lee{})}{\lbn{0}}{\lbn{1}}} \xlongleftrightarrow{\glvreshape{\lenv{{\mathrm{P}}}}{\lR^{\lenv{{\mathrm{M}}},\CoKernel{},\lsnv{0} \loprod \lsnv{1}} \loprod \lR^{\lenv{{\mathrm{M}}},\Image{},\lsnv{0} \loprod \lsnv{1}}}{\operatorname{inferSpace}(\lenv{{\mathrm{P}}})}} \pBasisC{\Image(\lee{})}{\lbn{0}}{\lbn{1}}$ & \texttt{\footnotesize GLang/Adequacy/Strengthened.lean:205}\\
\textsc{S-C-Proj\allowbreak Ker\allowbreak Coker} & $\evalCtxSpace{};\evalCtx{}\vdash_{\mathsf S} \pBasisC{\ls{}}{\lbn{0}}{\lbn{1}} \xlongleftrightarrow{\glvmtm{\lsnv{0}'}{\lsnv{0}'}{\glvreshape{\lenv{1}}{\lsnv{0}' \loprod \lsv{}'}{\lsnv{0}' \loprod \lsv{}'}}{\glvmtm{\lsnv{0}}{\lsnv{0}}{\glvreshape{\lenv{2}}{\lsnv{0} \loprod \lsnv{0}'}{\lsnv{0} \loprod \lsnv{0}'}}{\glvreshape{\lenv{3}}{\lsv{} \loprod \lsnv{0}}{\lsv{} \loprod \lsnv{0}}}}} \pBasisC{\ls{}}{\lbn{0}'}{\lbn{1}}$ & \texttt{\footnotesize GLang/Adequacy/Strengthened.lean:221}\\
\textsc{S-C-Proj\allowbreak Img\allowbreak Coimg} & $\evalCtxSpace{};\evalCtx{}\vdash_{\mathsf S} \pBasisC{\ls{}}{\lbn{0}}{\lbn{1}} \xlongleftrightarrow{\glvmtm{\lsnv{1}'}{\lsnv{1}'}{\glvreshape{\lenv{1}}{\lsnv{1}' \loprod \lsv{}'}{\lsnv{1}' \loprod \lsv{}'}}{\glvmtm{\lsnv{1}}{\lsnv{1}}{\glvreshape{\lenv{2}}{\lsnv{1} \loprod \lsnv{1}'}{\lsnv{1} \loprod \lsnv{1}'}}{\glvreshape{\lenv{3}}{\lsv{} \loprod \lsnv{1}}{\lsv{} \loprod \lsnv{1}}}}} \pBasisC{\ls{}}{\lbn{0}}{\lbn{1}'}$ & \texttt{\footnotesize GLang/Adequacy/Strengthened.lean:242}\\
\textsc{S-C-Terms\allowbreak Rref} & $\evalCtxSpace{};\evalCtx{}\vdash_{\mathsf S} \lb{} \xlongleftrightarrow{\glvreshape{\glvop{RREFRep}(\lenv{0},\glvreshape{\lenv{1}}{\lsv{} \loprod \lsv{}}{\lsv{} \loprod \lsv{}})}{\lsv{} \loprod \lsv{}}{\lsv{} \loprod \lsv{}}} \llpl{} \lee{} \mathbin{:\!:} \mathit{es}\llpr{}_{\lb{}'}$ & \texttt{\footnotesize GLang/Adequacy/Strengthened.lean:263}\\
\textsc{S-C-Refl} & $\evalCtxSpace{};\evalCtx{}\vdash_{\mathsf S} \lb{} \xlongleftrightarrow{\glvreshape{\lidelta{\lsv{}}}{\lsv{} \loprod \lsv{}}{\lsv{} \loprod \lsv{}}} \lb{}$ & \texttt{\footnotesize GLang/Adequacy/Strengthened.lean:276}\\
\textsc{S-C-Proj\allowbreak Ker\allowbreak Coker\allowbreak Cod} & $\evalCtxSpace{};\evalCtx{}\vdash_{\mathsf S} \pBasisC{\ls{}}{\lbn{0}}{\lbn{1}} \xlongleftrightarrow{\glvmtm{\lsnv{0}}{\lsnv{0}}{\glvreshape{\lenv{1}}{\lsnv{0} \loprod \lsv{}'}{\lsnv{0} \loprod \lsv{}'}}{\glvreshape{\lenv{3}}{\lsv{} \loprod \lsnv{0}}{\lsv{} \loprod \lsnv{0}}}} \pBasisC{\ls{}}{\lbn{0}}{\lbn{1}'}$ & \texttt{\footnotesize GLang/Adequacy/Strengthened.lean:283}\\
\textsc{S-C-Proj\allowbreak Img\allowbreak Coimg\allowbreak Dom} & $\evalCtxSpace{};\evalCtx{}\vdash_{\mathsf S} \pBasisC{\ls{}}{\lbn{0}}{\lbn{1}} \xlongleftrightarrow{\glvmtm{\lsnv{1}}{\lsnv{1}}{\glvreshape{\lenv{1}}{\lsnv{1} \loprod \lsv{}'}{\lsnv{1} \loprod \lsv{}'}}{\glvreshape{\lenv{3}}{\lsv{} \loprod \lsnv{1}}{\lsv{} \loprod \lsnv{1}}}} \pBasisC{\ls{}}{\lbn{0}'}{\lbn{1}}$ & \texttt{\footnotesize GLang/Adequacy/Strengthened.lean:300}\\
}{
\gpfit{\inferrule*[]{\forall i.\; (\emptyset,\evalCtxSpace{})\vdash \lbn{i}\colon \bvalid{} \\ \evalCtxSpace{};\evalCtx{}\vdash_{\mathsf S} \lbn{1} \xlongleftrightarrow{\leev{}} \lbn{0}}{\evalCtxSpace{};\evalCtx{}\vdash_{\mathsf S} \lbn{0} \xlongleftrightarrow{\leev{}^{-1}} \lbn{1}}}
\and
\gpfit{\inferrule*[]{(\emptyset,\evalCtxSpace{})\vdash \lbn{0}\colon \bvalid{} \\ (\emptyset,\evalCtxSpace{})\vdash \lbn{2}\colon \bvalid{} \\ \evalCtxSpace{};\evalCtx{}\vdash_{\mathsf S} \lbn{0} \xlongleftrightarrow{\lenv{0}} \lbn{1} \\ \evalCtxSpace{};\evalCtx{}\vdash_{\mathsf S} \lbn{1} \xlongleftrightarrow{\lenv{1}} \lbn{2} \\ \evalCtxSpace{};\evalCtx{}\vdash_{\mathsf S} \sof{\lbn{0}} \evalArrow{} \lsv{}}{\evalCtxSpace{};\evalCtx{}\vdash_{\mathsf S} \lbn{0} \xlongleftrightarrow{\glvmtm{\lsv{}}{\lsv{}}{\lenv{1}}{\glvreshape{\lenv{0}}{\lsv{} \loprod \lsv{}}{\lsv{} \loprod \lsv{}}}} \lbn{2}}}
\and
\gpfit{\inferrule*[]{(\emptyset,\evalCtxSpace{})\vdash \lbn{0}' \loprod \lbn{1}'\colon \bvalid{} \\ (\emptyset,\evalCtxSpace{})\vdash \lbn{0} \loprod \lbn{1}\colon \bvalid{} \\ \forall i.\; \evalCtxSpace{};\evalCtx{}\vdash_{\mathsf S} \lbn{i}' \xlongleftrightarrow{\lenv{i}} \lbn{i}}{\evalCtxSpace{};\evalCtx{}\vdash_{\mathsf S} \lbn{0}' \loprod \lbn{1}' \xlongleftrightarrow{\vkrn(\lenv{0}, \lenv{1})} \lbn{0} \loprod \lbn{1}}}
\and
\gpfit{\inferrule*[]{(\emptyset,\evalCtxSpace{})\vdash \lbn{0}' \losum \lbn{1}'\colon \bvalid{} \\ (\emptyset,\evalCtxSpace{})\vdash \lbn{0} \losum \lbn{1}\colon \bvalid{} \\ \forall i.\; \evalCtxSpace{};\evalCtx{}\vdash_{\mathsf S} \lbn{i}' \xlongleftrightarrow{\lenv{i}} \lbn{i}}{\evalCtxSpace{};\evalCtx{}\vdash_{\mathsf S} \lbn{0}' \losum \lbn{1}' \xlongleftrightarrow{\vblk(\lenv{0}, \lenv{1})} \lbn{0} \losum \lbn{1}}}
\and
\gpfit{\inferrule*[]{\tof{\lbn{0}} = \sof{\lbn{0}} \\ (\emptyset,\evalCtxSpace{})\vdash \lbn{0}^{\lstar}\colon \bvalid{} \\ (\emptyset,\evalCtxSpace{})\vdash \stdBasisDCtx{\tof{\lbn{0}^{\lstar}}}\colon \bvalid{} \\ \evalCtxSpace{};\evalCtx{}\vdash_{\mathsf S} \lbn{0} \xlongleftrightarrow{\leev{}} \stdBasisDCtx{\tof{\lbn{0}}}}{\evalCtxSpace{};\evalCtx{}\vdash_{\mathsf S} \lbn{0}^{\lstar} \xlongleftrightarrow{\Trans(\leev{}^{-1})} \stdBasisDCtx{\tof{\lbn{0}^{\lstar}}}}}
\and
\gpfit{\inferrule*[]{(\emptyset,\evalCtxSpace{})\vdash \iBasisC{\lbn{0}}\colon \bvalid{} \\ (\emptyset,\evalCtxSpace{})\vdash \iBasisC{\lbn{0}'}\colon \bvalid{} \\ \exists \lsn{0}.\; \exists \lsn{1}.\; \exists \lee{}.\; \sof{\lbn{0}} = \CoKernel(\lee{}) \\ \evalCtxSpace{};\evalCtx{}\vdash_{\mathsf S} \lbn{0} \xlongleftrightarrow{\leev{}} \lbn{0}'}{\evalCtxSpace{};\evalCtx{}\vdash_{\mathsf S} \iBasisC{\lbn{0}} \xlongleftrightarrow{\leev{}} \iBasisC{\lbn{0}'}}}
\and
\gpfit{\inferrule*[]{(\emptyset,\evalCtxSpace{})\vdash \iBasisC{\pBasisC{\CoKernel(\lee{})}{\lbn{0}}{\lbn{1}}}\colon \bvalid{} \\ (\emptyset,\evalCtxSpace{})\vdash \pBasisC{\Image(\lee{})}{\lbn{0}}{\lbn{1}}\colon \bvalid{} \\ \forall i.\; \sof{\lbn{i}} = \lsn{i} \\ \forall i.\; \isStdBasis{\lbn{i}} \\ \evalCtxSpace{};\evalCtx{}\vdash_{\mathsf S} \sof{\lbn{0}^{\lstar} \loprod \lbn{1}} \evalArrow{} \lsnv{0} \loprod \lsnv{1} \\ \lemptyset{};\evalCtxSpace{};\evalCtx{}\vdash_{\mathsf S} \lee{} \xlongrightarrow{\lbn{0}^{\lstar} \loprod \lbn{1}} \leev{}\glresulttype{\lsnv{0} \loprod \lsnv{1}} \\ \lenv{{\mathrm{M}}} = \glvreshape{\leev{}}{\lsnv{0} \loprod \lsnv{1}}{\lsnv{0} \loprod \lsnv{1}} \\ \lenv{{\mathrm{U}}} = \glvop{GSVDU}(\lenv{{\mathrm{M}}},\lidelta{\lsnv{0}},\lidelta{\lsnv{1}}) \\ \lenv{{\mathrm{V}}} = \glvop{GSVDV}(\lenv{{\mathrm{M}}},\lidelta{\lsnv{0}},\lidelta{\lsnv{1}}) \\ \lenv{{\mathrm{P}}} = \glvmtm{\lsnv{1}}{\lsnv{1}}{\lenv{{\mathrm{V}}}}{\glvmtm{\lsnv{0}}{\lsnv{0}}{\lenv{{\mathrm{M}}}}{\Trans(\lenv{{\mathrm{U}}})}}}{\evalCtxSpace{};\evalCtx{}\vdash_{\mathsf S} \iBasisC{\pBasisC{\CoKernel(\lee{})}{\lbn{0}}{\lbn{1}}} \xlongleftrightarrow{\glvreshape{\lenv{{\mathrm{P}}}}{\lR^{\lenv{{\mathrm{M}}},\CoKernel{},\lsnv{0} \loprod \lsnv{1}} \loprod \lR^{\lenv{{\mathrm{M}}},\Image{},\lsnv{0} \loprod \lsnv{1}}}{\operatorname{inferSpace}(\lenv{{\mathrm{P}}})}} \pBasisC{\Image(\lee{})}{\lbn{0}}{\lbn{1}}}}
\and
\gpfit{\inferrule*[]{(\emptyset,\evalCtxSpace{})\vdash \pBasisC{\ls{}}{\lbn{0}}{\lbn{1}}\colon \bvalid{} \\ (\emptyset,\evalCtxSpace{})\vdash \pBasisC{\ls{}}{\lbn{0}'}{\lbn{1}}\colon \bvalid{} \\ \ls{} = \Kernel(\lee{}) \;\vee\; \ls{} = \CoKernel(\lee{}) \\ \forall i.\; \sof{\lbn{i}} = \lsn{i} \\ \forall i.\; \evalCtxSpace{};\evalCtx{}\vdash_{\mathsf S} \sof{\lbn{i}} \evalArrow{} \lsnv{i} \\ \evalCtxSpace{};\evalCtx{}\vdash_{\mathsf S} \ls{} \evalArrow{} \lsv{} \\ \evalCtxSpace{};\evalCtx{}\vdash_{\mathsf S} \ls{} \evalArrow{} \lsv{}' \\ \evalCtxSpace{};\evalCtx{}\vdash_{\mathsf S} \sof{\lbn{0}'} \evalArrow{} \lsnv{0}' \\ \evalCtxSpace{};\evalCtx{}\vdash_{\mathsf S} \lbn{0} \xlongleftrightarrow{\lenv{2}} \lbn{0}' \\ \lemptyset{};\evalCtxSpace{};\evalCtx{}\vdash_{\mathsf S} \lembd{}_{\ls{}} \xlongrightarrow{\pBasisC{\ls{}}{\lbn{0}}{\lbn{1}}^{\lstar} \loprod \lbn{0}} \lenv{3}\glresulttype{\lsv{} \loprod \lsnv{0}} \\ \lemptyset{};\evalCtxSpace{};\evalCtx{}\vdash_{\mathsf S} \lproj{}_{\ls{}, \sof{\lbn{0}'}} \xlongrightarrow{\lbn{0}'^{\lstar} \loprod \pBasisC{\ls{}}{\lbn{0}'}{\lbn{1}}} \lenv{1}\glresulttype{\lsnv{0}' \loprod \lsv{}'}}{\evalCtxSpace{};\evalCtx{}\vdash_{\mathsf S} \pBasisC{\ls{}}{\lbn{0}}{\lbn{1}} \xlongleftrightarrow{\glvmtm{\lsnv{0}'}{\lsnv{0}'}{\glvreshape{\lenv{1}}{\lsnv{0}' \loprod \lsv{}'}{\lsnv{0}' \loprod \lsv{}'}}{\glvmtm{\lsnv{0}}{\lsnv{0}}{\glvreshape{\lenv{2}}{\lsnv{0} \loprod \lsnv{0}'}{\lsnv{0} \loprod \lsnv{0}'}}{\glvreshape{\lenv{3}}{\lsv{} \loprod \lsnv{0}}{\lsv{} \loprod \lsnv{0}}}}} \pBasisC{\ls{}}{\lbn{0}'}{\lbn{1}}}}
\and
\gpfit{\inferrule*[]{(\emptyset,\evalCtxSpace{})\vdash \pBasisC{\ls{}}{\lbn{0}}{\lbn{1}}\colon \bvalid{} \\ (\emptyset,\evalCtxSpace{})\vdash \pBasisC{\ls{}}{\lbn{0}}{\lbn{1}'}\colon \bvalid{} \\ \ls{} = \Image(\lee{}) \;\vee\; \ls{} = \CoImage(\lee{}) \\ \forall i.\; \sof{\lbn{i}} = \lsn{i} \\ \forall i.\; \evalCtxSpace{};\evalCtx{}\vdash_{\mathsf S} \sof{\lbn{i}} \evalArrow{} \lsnv{i} \\ \evalCtxSpace{};\evalCtx{}\vdash_{\mathsf S} \ls{} \evalArrow{} \lsv{} \\ \evalCtxSpace{};\evalCtx{}\vdash_{\mathsf S} \ls{} \evalArrow{} \lsv{}' \\ \evalCtxSpace{};\evalCtx{}\vdash_{\mathsf S} \sof{\lbn{1}'} \evalArrow{} \lsnv{1}' \\ \evalCtxSpace{};\evalCtx{}\vdash_{\mathsf S} \lbn{1} \xlongleftrightarrow{\lenv{2}} \lbn{1}' \\ \lemptyset{};\evalCtxSpace{};\evalCtx{}\vdash_{\mathsf S} \lembd{}_{\ls{}} \xlongrightarrow{\pBasisC{\ls{}}{\lbn{0}}{\lbn{1}}^{\lstar} \loprod \lbn{1}} \lenv{3}\glresulttype{\lsv{} \loprod \lsnv{1}} \\ \lemptyset{};\evalCtxSpace{};\evalCtx{}\vdash_{\mathsf S} \lproj{}_{\ls{}, \sof{\lbn{1}'}} \xlongrightarrow{\lbn{1}'^{\lstar} \loprod \pBasisC{\ls{}}{\lbn{0}}{\lbn{1}'}} \lenv{1}\glresulttype{\lsnv{1}' \loprod \lsv{}'}}{\evalCtxSpace{};\evalCtx{}\vdash_{\mathsf S} \pBasisC{\ls{}}{\lbn{0}}{\lbn{1}} \xlongleftrightarrow{\glvmtm{\lsnv{1}'}{\lsnv{1}'}{\glvreshape{\lenv{1}}{\lsnv{1}' \loprod \lsv{}'}{\lsnv{1}' \loprod \lsv{}'}}{\glvmtm{\lsnv{1}}{\lsnv{1}}{\glvreshape{\lenv{2}}{\lsnv{1} \loprod \lsnv{1}'}{\lsnv{1} \loprod \lsnv{1}'}}{\glvreshape{\lenv{3}}{\lsv{} \loprod \lsnv{1}}{\lsv{} \loprod \lsnv{1}}}}} \pBasisC{\ls{}}{\lbn{0}}{\lbn{1}'}}}
\and
\gpfit{\inferrule*[]{(\emptyset,\evalCtxSpace{})\vdash \lb{}\colon \bvalid{} \\ (\emptyset,\evalCtxSpace{})\vdash \llpl{} \lee{} \mathbin{:\!:} \mathit{es}\llpr{}_{\lb{}'}\colon \bvalid{} \\ \evalCtxSpace{};\evalCtx{}\vdash_{\mathsf S} \lb{} \xlongleftrightarrow{\lenv{1}} \lb{}' \\ \lemptyset{};\evalCtxSpace{};\evalCtx{}\vdash_{\mathsf S} \lee{} \losum \mathit{es}[0] \losum \cdots \xlongrightarrow{\lb{}' \losum \cdots} \lenv{0}\glresulttype{\ltyv{}} \\ \evalCtxSpace{};\evalCtx{}\vdash_{\mathsf S} \sof{\lb{}} \evalArrow{} \lsv{}}{\evalCtxSpace{};\evalCtx{}\vdash_{\mathsf S} \lb{} \xlongleftrightarrow{\glvreshape{\glvop{RREFRep}(\lenv{0},\glvreshape{\lenv{1}}{\lsv{} \loprod \lsv{}}{\lsv{} \loprod \lsv{}})}{\lsv{} \loprod \lsv{}}{\lsv{} \loprod \lsv{}}} \llpl{} \lee{} \mathbin{:\!:} \mathit{es}\llpr{}_{\lb{}'}}}
\and
\gpfit{\inferrule*[]{(\emptyset,\evalCtxSpace{})\vdash \lb{}\colon \bvalid{} \\ \tof{\lb{}} = \sof{\lb{}} \\ \evalCtxSpace{};\evalCtx{}\vdash_{\mathsf S} \sof{\lb{}} \evalArrow{} \lsv{}}{\evalCtxSpace{};\evalCtx{}\vdash_{\mathsf S} \lb{} \xlongleftrightarrow{\glvreshape{\lidelta{\lsv{}}}{\lsv{} \loprod \lsv{}}{\lsv{} \loprod \lsv{}}} \lb{}}}
\and
\gpfit{\inferrule*[]{(\emptyset,\evalCtxSpace{})\vdash \pBasisC{\ls{}}{\lbn{0}}{\lbn{1}}\colon \bvalid{} \\ (\emptyset,\evalCtxSpace{})\vdash \pBasisC{\ls{}}{\lbn{0}}{\lbn{1}'}\colon \bvalid{} \\ \ls{} = \Kernel(\lee{}) \;\vee\; \ls{} = \CoKernel(\lee{}) \\ \forall i.\; \sof{\lbn{i}} = \lsn{i} \\ \sof{\lbn{1}'} = \lsn{1} \\ \evalCtxSpace{};\evalCtx{}\vdash_{\mathsf S} \sof{\lbn{0}} \evalArrow{} \lsnv{0} \\ \evalCtxSpace{};\evalCtx{}\vdash_{\mathsf S} \ls{} \evalArrow{} \lsv{} \\ \evalCtxSpace{};\evalCtx{}\vdash_{\mathsf S} \ls{} \evalArrow{} \lsv{}' \\ \lemptyset{};\evalCtxSpace{};\evalCtx{}\vdash_{\mathsf S} \lembd{}_{\ls{}} \xlongrightarrow{\pBasisC{\ls{}}{\lbn{0}}{\lbn{1}}^{\lstar} \loprod \lbn{0}} \lenv{3}\glresulttype{\lsv{} \loprod \lsnv{0}} \\ \lemptyset{};\evalCtxSpace{};\evalCtx{}\vdash_{\mathsf S} \lproj{}_{\ls{}, \sof{\lbn{0}}} \xlongrightarrow{\lbn{0}^{\lstar} \loprod \pBasisC{\ls{}}{\lbn{0}}{\lbn{1}'}} \lenv{1}\glresulttype{\lsnv{0} \loprod \lsv{}'}}{\evalCtxSpace{};\evalCtx{}\vdash_{\mathsf S} \pBasisC{\ls{}}{\lbn{0}}{\lbn{1}} \xlongleftrightarrow{\glvmtm{\lsnv{0}}{\lsnv{0}}{\glvreshape{\lenv{1}}{\lsnv{0} \loprod \lsv{}'}{\lsnv{0} \loprod \lsv{}'}}{\glvreshape{\lenv{3}}{\lsv{} \loprod \lsnv{0}}{\lsv{} \loprod \lsnv{0}}}} \pBasisC{\ls{}}{\lbn{0}}{\lbn{1}'}}}
\and
\gpfit{\inferrule*[]{(\emptyset,\evalCtxSpace{})\vdash \pBasisC{\ls{}}{\lbn{0}}{\lbn{1}}\colon \bvalid{} \\ (\emptyset,\evalCtxSpace{})\vdash \pBasisC{\ls{}}{\lbn{0}'}{\lbn{1}}\colon \bvalid{} \\ \ls{} = \Image(\lee{}) \;\vee\; \ls{} = \CoImage(\lee{}) \\ \forall i.\; \sof{\lbn{i}} = \lsn{i} \\ \sof{\lbn{0}'} = \lsn{0} \\ \evalCtxSpace{};\evalCtx{}\vdash_{\mathsf S} \sof{\lbn{1}} \evalArrow{} \lsnv{1} \\ \evalCtxSpace{};\evalCtx{}\vdash_{\mathsf S} \ls{} \evalArrow{} \lsv{} \\ \evalCtxSpace{};\evalCtx{}\vdash_{\mathsf S} \ls{} \evalArrow{} \lsv{}' \\ \lemptyset{};\evalCtxSpace{};\evalCtx{}\vdash_{\mathsf S} \lembd{}_{\ls{}} \xlongrightarrow{\pBasisC{\ls{}}{\lbn{0}}{\lbn{1}}^{\lstar} \loprod \lbn{1}} \lenv{3}\glresulttype{\lsv{} \loprod \lsnv{1}} \\ \lemptyset{};\evalCtxSpace{};\evalCtx{}\vdash_{\mathsf S} \lproj{}_{\ls{}, \sof{\lbn{1}}} \xlongrightarrow{\lbn{1}^{\lstar} \loprod \pBasisC{\ls{}}{\lbn{0}'}{\lbn{1}}} \lenv{1}\glresulttype{\lsnv{1} \loprod \lsv{}'}}{\evalCtxSpace{};\evalCtx{}\vdash_{\mathsf S} \pBasisC{\ls{}}{\lbn{0}}{\lbn{1}} \xlongleftrightarrow{\glvmtm{\lsnv{1}}{\lsnv{1}}{\glvreshape{\lenv{1}}{\lsnv{1} \loprod \lsv{}'}{\lsnv{1} \loprod \lsv{}'}}{\glvreshape{\lenv{3}}{\lsv{} \loprod \lsnv{1}}{\lsv{} \loprod \lsnv{1}}}} \pBasisC{\ls{}}{\lbn{0}'}{\lbn{1}}}}
}{
\gpfit{\inferrule*[lab=\textsc{S-C-Inverse}]{\forall i.\; (\emptyset,\evalCtxSpace{})\vdash \lbn{i}\colon \bvalid{} \\ \evalCtxSpace{};\evalCtx{}\vdash_{\mathsf S} \lbn{1} \xlongleftrightarrow{\leev{}} \lbn{0}}{\evalCtxSpace{};\evalCtx{}\vdash_{\mathsf S} \lbn{0} \xlongleftrightarrow{\leev{}^{-1}} \lbn{1}}}
\and
\gpfit{\inferrule*[lab=\textsc{S-C-Trans}]{(\emptyset,\evalCtxSpace{})\vdash \lbn{0}\colon \bvalid{} \\ (\emptyset,\evalCtxSpace{})\vdash \lbn{2}\colon \bvalid{} \\ \evalCtxSpace{};\evalCtx{}\vdash_{\mathsf S} \lbn{0} \xlongleftrightarrow{\lenv{0}} \lbn{1} \\ \evalCtxSpace{};\evalCtx{}\vdash_{\mathsf S} \lbn{1} \xlongleftrightarrow{\lenv{1}} \lbn{2} \\ \evalCtxSpace{};\evalCtx{}\vdash_{\mathsf S} \sof{\lbn{0}} \evalArrow{} \lsv{}}{\evalCtxSpace{};\evalCtx{}\vdash_{\mathsf S} \lbn{0} \xlongleftrightarrow{\glvmtm{\lsv{}}{\lsv{}}{\lenv{1}}{\glvreshape{\lenv{0}}{\lsv{} \loprod \lsv{}}{\lsv{} \loprod \lsv{}}}} \lbn{2}}}
\and
\gpfit{\inferrule*[lab=\textsc{S-C-Tprod}]{(\emptyset,\evalCtxSpace{})\vdash \lbn{0}' \loprod \lbn{1}'\colon \bvalid{} \\ (\emptyset,\evalCtxSpace{})\vdash \lbn{0} \loprod \lbn{1}\colon \bvalid{} \\ \forall i.\; \evalCtxSpace{};\evalCtx{}\vdash_{\mathsf S} \lbn{i}' \xlongleftrightarrow{\lenv{i}} \lbn{i}}{\evalCtxSpace{};\evalCtx{}\vdash_{\mathsf S} \lbn{0}' \loprod \lbn{1}' \xlongleftrightarrow{\vkrn(\lenv{0}, \lenv{1})} \lbn{0} \loprod \lbn{1}}}
\and
\gpfit{\inferrule*[lab=\textsc{S-C-Dsum}]{(\emptyset,\evalCtxSpace{})\vdash \lbn{0}' \losum \lbn{1}'\colon \bvalid{} \\ (\emptyset,\evalCtxSpace{})\vdash \lbn{0} \losum \lbn{1}\colon \bvalid{} \\ \forall i.\; \evalCtxSpace{};\evalCtx{}\vdash_{\mathsf S} \lbn{i}' \xlongleftrightarrow{\lenv{i}} \lbn{i}}{\evalCtxSpace{};\evalCtx{}\vdash_{\mathsf S} \lbn{0}' \losum \lbn{1}' \xlongleftrightarrow{\vblk(\lenv{0}, \lenv{1})} \lbn{0} \losum \lbn{1}}}
\and
\gpfit{\inferrule*[lab=\textsc{S-C-Dual\allowbreak Std}]{\tof{\lbn{0}} = \sof{\lbn{0}} \\ (\emptyset,\evalCtxSpace{})\vdash \lbn{0}^{\lstar}\colon \bvalid{} \\ (\emptyset,\evalCtxSpace{})\vdash \stdBasisDCtx{\tof{\lbn{0}^{\lstar}}}\colon \bvalid{} \\ \evalCtxSpace{};\evalCtx{}\vdash_{\mathsf S} \lbn{0} \xlongleftrightarrow{\leev{}} \stdBasisDCtx{\tof{\lbn{0}}}}{\evalCtxSpace{};\evalCtx{}\vdash_{\mathsf S} \lbn{0}^{\lstar} \xlongleftrightarrow{\Trans(\leev{}^{-1})} \stdBasisDCtx{\tof{\lbn{0}^{\lstar}}}}}
\and
\gpfit{\inferrule*[lab=\textsc{S-C-Image}]{(\emptyset,\evalCtxSpace{})\vdash \iBasisC{\lbn{0}}\colon \bvalid{} \\ (\emptyset,\evalCtxSpace{})\vdash \iBasisC{\lbn{0}'}\colon \bvalid{} \\ \exists \lsn{0}.\; \exists \lsn{1}.\; \exists \lee{}.\; \sof{\lbn{0}} = \CoKernel(\lee{}) \\ \evalCtxSpace{};\evalCtx{}\vdash_{\mathsf S} \lbn{0} \xlongleftrightarrow{\leev{}} \lbn{0}'}{\evalCtxSpace{};\evalCtx{}\vdash_{\mathsf S} \iBasisC{\lbn{0}} \xlongleftrightarrow{\leev{}} \iBasisC{\lbn{0}'}}}
\and
\gpfit{\inferrule*[lab=\textsc{S-C-Image\allowbreak Svds}]{(\emptyset,\evalCtxSpace{})\vdash \iBasisC{\pBasisC{\CoKernel(\lee{})}{\lbn{0}}{\lbn{1}}}\colon \bvalid{} \\ (\emptyset,\evalCtxSpace{})\vdash \pBasisC{\Image(\lee{})}{\lbn{0}}{\lbn{1}}\colon \bvalid{} \\ \forall i.\; \sof{\lbn{i}} = \lsn{i} \\ \forall i.\; \isStdBasis{\lbn{i}} \\ \evalCtxSpace{};\evalCtx{}\vdash_{\mathsf S} \sof{\lbn{0}^{\lstar} \loprod \lbn{1}} \evalArrow{} \lsnv{0} \loprod \lsnv{1} \\ \lemptyset{};\evalCtxSpace{};\evalCtx{}\vdash_{\mathsf S} \lee{} \xlongrightarrow{\lbn{0}^{\lstar} \loprod \lbn{1}} \leev{}\glresulttype{\lsnv{0} \loprod \lsnv{1}} \\ \lenv{{\mathrm{M}}} = \glvreshape{\leev{}}{\lsnv{0} \loprod \lsnv{1}}{\lsnv{0} \loprod \lsnv{1}} \\ \lenv{{\mathrm{U}}} = \glvop{GSVDU}(\lenv{{\mathrm{M}}},\lidelta{\lsnv{0}},\lidelta{\lsnv{1}}) \\ \lenv{{\mathrm{V}}} = \glvop{GSVDV}(\lenv{{\mathrm{M}}},\lidelta{\lsnv{0}},\lidelta{\lsnv{1}}) \\ \lenv{{\mathrm{P}}} = \glvmtm{\lsnv{1}}{\lsnv{1}}{\lenv{{\mathrm{V}}}}{\glvmtm{\lsnv{0}}{\lsnv{0}}{\lenv{{\mathrm{M}}}}{\Trans(\lenv{{\mathrm{U}}})}}}{\evalCtxSpace{};\evalCtx{}\vdash_{\mathsf S} \iBasisC{\pBasisC{\CoKernel(\lee{})}{\lbn{0}}{\lbn{1}}} \xlongleftrightarrow{\glvreshape{\lenv{{\mathrm{P}}}}{\lR^{\lenv{{\mathrm{M}}},\CoKernel{},\lsnv{0} \loprod \lsnv{1}} \loprod \lR^{\lenv{{\mathrm{M}}},\Image{},\lsnv{0} \loprod \lsnv{1}}}{\operatorname{inferSpace}(\lenv{{\mathrm{P}}})}} \pBasisC{\Image(\lee{})}{\lbn{0}}{\lbn{1}}}}
\and
\gpfit{\inferrule*[lab=\textsc{S-C-Proj\allowbreak Ker\allowbreak Coker}]{(\emptyset,\evalCtxSpace{})\vdash \pBasisC{\ls{}}{\lbn{0}}{\lbn{1}}\colon \bvalid{} \\ (\emptyset,\evalCtxSpace{})\vdash \pBasisC{\ls{}}{\lbn{0}'}{\lbn{1}}\colon \bvalid{} \\ \ls{} = \Kernel(\lee{}) \;\vee\; \ls{} = \CoKernel(\lee{}) \\ \forall i.\; \sof{\lbn{i}} = \lsn{i} \\ \forall i.\; \evalCtxSpace{};\evalCtx{}\vdash_{\mathsf S} \sof{\lbn{i}} \evalArrow{} \lsnv{i} \\ \evalCtxSpace{};\evalCtx{}\vdash_{\mathsf S} \ls{} \evalArrow{} \lsv{} \\ \evalCtxSpace{};\evalCtx{}\vdash_{\mathsf S} \ls{} \evalArrow{} \lsv{}' \\ \evalCtxSpace{};\evalCtx{}\vdash_{\mathsf S} \sof{\lbn{0}'} \evalArrow{} \lsnv{0}' \\ \evalCtxSpace{};\evalCtx{}\vdash_{\mathsf S} \lbn{0} \xlongleftrightarrow{\lenv{2}} \lbn{0}' \\ \lemptyset{};\evalCtxSpace{};\evalCtx{}\vdash_{\mathsf S} \lembd{}_{\ls{}} \xlongrightarrow{\pBasisC{\ls{}}{\lbn{0}}{\lbn{1}}^{\lstar} \loprod \lbn{0}} \lenv{3}\glresulttype{\lsv{} \loprod \lsnv{0}} \\ \lemptyset{};\evalCtxSpace{};\evalCtx{}\vdash_{\mathsf S} \lproj{}_{\ls{}, \sof{\lbn{0}'}} \xlongrightarrow{\lbn{0}'^{\lstar} \loprod \pBasisC{\ls{}}{\lbn{0}'}{\lbn{1}}} \lenv{1}\glresulttype{\lsnv{0}' \loprod \lsv{}'}}{\evalCtxSpace{};\evalCtx{}\vdash_{\mathsf S} \pBasisC{\ls{}}{\lbn{0}}{\lbn{1}} \xlongleftrightarrow{\glvmtm{\lsnv{0}'}{\lsnv{0}'}{\glvreshape{\lenv{1}}{\lsnv{0}' \loprod \lsv{}'}{\lsnv{0}' \loprod \lsv{}'}}{\glvmtm{\lsnv{0}}{\lsnv{0}}{\glvreshape{\lenv{2}}{\lsnv{0} \loprod \lsnv{0}'}{\lsnv{0} \loprod \lsnv{0}'}}{\glvreshape{\lenv{3}}{\lsv{} \loprod \lsnv{0}}{\lsv{} \loprod \lsnv{0}}}}} \pBasisC{\ls{}}{\lbn{0}'}{\lbn{1}}}}
\and
\gpfit{\inferrule*[lab=\textsc{S-C-Proj\allowbreak Img\allowbreak Coimg}]{(\emptyset,\evalCtxSpace{})\vdash \pBasisC{\ls{}}{\lbn{0}}{\lbn{1}}\colon \bvalid{} \\ (\emptyset,\evalCtxSpace{})\vdash \pBasisC{\ls{}}{\lbn{0}}{\lbn{1}'}\colon \bvalid{} \\ \ls{} = \Image(\lee{}) \;\vee\; \ls{} = \CoImage(\lee{}) \\ \forall i.\; \sof{\lbn{i}} = \lsn{i} \\ \forall i.\; \evalCtxSpace{};\evalCtx{}\vdash_{\mathsf S} \sof{\lbn{i}} \evalArrow{} \lsnv{i} \\ \evalCtxSpace{};\evalCtx{}\vdash_{\mathsf S} \ls{} \evalArrow{} \lsv{} \\ \evalCtxSpace{};\evalCtx{}\vdash_{\mathsf S} \ls{} \evalArrow{} \lsv{}' \\ \evalCtxSpace{};\evalCtx{}\vdash_{\mathsf S} \sof{\lbn{1}'} \evalArrow{} \lsnv{1}' \\ \evalCtxSpace{};\evalCtx{}\vdash_{\mathsf S} \lbn{1} \xlongleftrightarrow{\lenv{2}} \lbn{1}' \\ \lemptyset{};\evalCtxSpace{};\evalCtx{}\vdash_{\mathsf S} \lembd{}_{\ls{}} \xlongrightarrow{\pBasisC{\ls{}}{\lbn{0}}{\lbn{1}}^{\lstar} \loprod \lbn{1}} \lenv{3}\glresulttype{\lsv{} \loprod \lsnv{1}} \\ \lemptyset{};\evalCtxSpace{};\evalCtx{}\vdash_{\mathsf S} \lproj{}_{\ls{}, \sof{\lbn{1}'}} \xlongrightarrow{\lbn{1}'^{\lstar} \loprod \pBasisC{\ls{}}{\lbn{0}}{\lbn{1}'}} \lenv{1}\glresulttype{\lsnv{1}' \loprod \lsv{}'}}{\evalCtxSpace{};\evalCtx{}\vdash_{\mathsf S} \pBasisC{\ls{}}{\lbn{0}}{\lbn{1}} \xlongleftrightarrow{\glvmtm{\lsnv{1}'}{\lsnv{1}'}{\glvreshape{\lenv{1}}{\lsnv{1}' \loprod \lsv{}'}{\lsnv{1}' \loprod \lsv{}'}}{\glvmtm{\lsnv{1}}{\lsnv{1}}{\glvreshape{\lenv{2}}{\lsnv{1} \loprod \lsnv{1}'}{\lsnv{1} \loprod \lsnv{1}'}}{\glvreshape{\lenv{3}}{\lsv{} \loprod \lsnv{1}}{\lsv{} \loprod \lsnv{1}}}}} \pBasisC{\ls{}}{\lbn{0}}{\lbn{1}'}}}
\and
\gpfit{\inferrule*[lab=\textsc{S-C-Terms\allowbreak Rref}]{(\emptyset,\evalCtxSpace{})\vdash \lb{}\colon \bvalid{} \\ (\emptyset,\evalCtxSpace{})\vdash \llpl{} \lee{} \mathbin{:\!:} \mathit{es}\llpr{}_{\lb{}'}\colon \bvalid{} \\ \evalCtxSpace{};\evalCtx{}\vdash_{\mathsf S} \lb{} \xlongleftrightarrow{\lenv{1}} \lb{}' \\ \lemptyset{};\evalCtxSpace{};\evalCtx{}\vdash_{\mathsf S} \lee{} \losum \mathit{es}[0] \losum \cdots \xlongrightarrow{\lb{}' \losum \cdots} \lenv{0}\glresulttype{\ltyv{}} \\ \evalCtxSpace{};\evalCtx{}\vdash_{\mathsf S} \sof{\lb{}} \evalArrow{} \lsv{}}{\evalCtxSpace{};\evalCtx{}\vdash_{\mathsf S} \lb{} \xlongleftrightarrow{\glvreshape{\glvop{RREFRep}(\lenv{0},\glvreshape{\lenv{1}}{\lsv{} \loprod \lsv{}}{\lsv{} \loprod \lsv{}})}{\lsv{} \loprod \lsv{}}{\lsv{} \loprod \lsv{}}} \llpl{} \lee{} \mathbin{:\!:} \mathit{es}\llpr{}_{\lb{}'}}}
\and
\gpfit{\inferrule*[lab=\textsc{S-C-Refl}]{(\emptyset,\evalCtxSpace{})\vdash \lb{}\colon \bvalid{} \\ \tof{\lb{}} = \sof{\lb{}} \\ \evalCtxSpace{};\evalCtx{}\vdash_{\mathsf S} \sof{\lb{}} \evalArrow{} \lsv{}}{\evalCtxSpace{};\evalCtx{}\vdash_{\mathsf S} \lb{} \xlongleftrightarrow{\glvreshape{\lidelta{\lsv{}}}{\lsv{} \loprod \lsv{}}{\lsv{} \loprod \lsv{}}} \lb{}}}
\and
\gpfit{\inferrule*[lab=\textsc{S-C-Proj\allowbreak Ker\allowbreak Coker\allowbreak Cod}]{(\emptyset,\evalCtxSpace{})\vdash \pBasisC{\ls{}}{\lbn{0}}{\lbn{1}}\colon \bvalid{} \\ (\emptyset,\evalCtxSpace{})\vdash \pBasisC{\ls{}}{\lbn{0}}{\lbn{1}'}\colon \bvalid{} \\ \ls{} = \Kernel(\lee{}) \;\vee\; \ls{} = \CoKernel(\lee{}) \\ \forall i.\; \sof{\lbn{i}} = \lsn{i} \\ \sof{\lbn{1}'} = \lsn{1} \\ \evalCtxSpace{};\evalCtx{}\vdash_{\mathsf S} \sof{\lbn{0}} \evalArrow{} \lsnv{0} \\ \evalCtxSpace{};\evalCtx{}\vdash_{\mathsf S} \ls{} \evalArrow{} \lsv{} \\ \evalCtxSpace{};\evalCtx{}\vdash_{\mathsf S} \ls{} \evalArrow{} \lsv{}' \\ \lemptyset{};\evalCtxSpace{};\evalCtx{}\vdash_{\mathsf S} \lembd{}_{\ls{}} \xlongrightarrow{\pBasisC{\ls{}}{\lbn{0}}{\lbn{1}}^{\lstar} \loprod \lbn{0}} \lenv{3}\glresulttype{\lsv{} \loprod \lsnv{0}} \\ \lemptyset{};\evalCtxSpace{};\evalCtx{}\vdash_{\mathsf S} \lproj{}_{\ls{}, \sof{\lbn{0}}} \xlongrightarrow{\lbn{0}^{\lstar} \loprod \pBasisC{\ls{}}{\lbn{0}}{\lbn{1}'}} \lenv{1}\glresulttype{\lsnv{0} \loprod \lsv{}'}}{\evalCtxSpace{};\evalCtx{}\vdash_{\mathsf S} \pBasisC{\ls{}}{\lbn{0}}{\lbn{1}} \xlongleftrightarrow{\glvmtm{\lsnv{0}}{\lsnv{0}}{\glvreshape{\lenv{1}}{\lsnv{0} \loprod \lsv{}'}{\lsnv{0} \loprod \lsv{}'}}{\glvreshape{\lenv{3}}{\lsv{} \loprod \lsnv{0}}{\lsv{} \loprod \lsnv{0}}}} \pBasisC{\ls{}}{\lbn{0}}{\lbn{1}'}}}
\and
\gpfit{\inferrule*[lab=\textsc{S-C-Proj\allowbreak Img\allowbreak Coimg\allowbreak Dom}]{(\emptyset,\evalCtxSpace{})\vdash \pBasisC{\ls{}}{\lbn{0}}{\lbn{1}}\colon \bvalid{} \\ (\emptyset,\evalCtxSpace{})\vdash \pBasisC{\ls{}}{\lbn{0}'}{\lbn{1}}\colon \bvalid{} \\ \ls{} = \Image(\lee{}) \;\vee\; \ls{} = \CoImage(\lee{}) \\ \forall i.\; \sof{\lbn{i}} = \lsn{i} \\ \sof{\lbn{0}'} = \lsn{0} \\ \evalCtxSpace{};\evalCtx{}\vdash_{\mathsf S} \sof{\lbn{1}} \evalArrow{} \lsnv{1} \\ \evalCtxSpace{};\evalCtx{}\vdash_{\mathsf S} \ls{} \evalArrow{} \lsv{} \\ \evalCtxSpace{};\evalCtx{}\vdash_{\mathsf S} \ls{} \evalArrow{} \lsv{}' \\ \lemptyset{};\evalCtxSpace{};\evalCtx{}\vdash_{\mathsf S} \lembd{}_{\ls{}} \xlongrightarrow{\pBasisC{\ls{}}{\lbn{0}}{\lbn{1}}^{\lstar} \loprod \lbn{1}} \lenv{3}\glresulttype{\lsv{} \loprod \lsnv{1}} \\ \lemptyset{};\evalCtxSpace{};\evalCtx{}\vdash_{\mathsf S} \lproj{}_{\ls{}, \sof{\lbn{1}}} \xlongrightarrow{\lbn{1}^{\lstar} \loprod \pBasisC{\ls{}}{\lbn{0}'}{\lbn{1}}} \lenv{1}\glresulttype{\lsnv{1} \loprod \lsv{}'}}{\evalCtxSpace{};\evalCtx{}\vdash_{\mathsf S} \pBasisC{\ls{}}{\lbn{0}}{\lbn{1}} \xlongleftrightarrow{\glvmtm{\lsnv{1}}{\lsnv{1}}{\glvreshape{\lenv{1}}{\lsnv{1} \loprod \lsv{}'}{\lsnv{1} \loprod \lsv{}'}}{\glvreshape{\lenv{3}}{\lsv{} \loprod \lsnv{1}}{\lsv{} \loprod \lsnv{1}}}} \pBasisC{\ls{}}{\lbn{0}'}{\lbn{1}}}}
}
\gldefjudgment{s.term}{\lDelta{};\lepsilon{};\evalCtx{}\vdash_{\mathsf S} \lee{}\xlongrightarrow{\lb{}}\leev{}\glresulttype{\ltyv{}}}{Strengthened term evaluation (generated from \texttt{GLang.S.Term}).}{fig:gl-s-term}
\gldefrule{s.term.varLin}{s.term}{\textsc{S-E-Var\allowbreak Lin}}{((\mathit{x}, \lb{}) \mapsto \leev{}) \in \evalCtx{} \\ \tof{\lb{}} = \ls{} \\ (\emptyset,\evalCtxSpace{})\vdash \lb{}\colon \bvalid{} \\ \evalCtxSpace{};\evalCtx{}\vdash_{\mathsf S} \lb{} \evalArrow{}_{\mathsf{basis}} \ltyv{}}{\{(\mathit{x}, \ls{})\};\evalCtxSpace{};\evalCtx{}\vdash_{\mathsf S} \mathit{x} \xlongrightarrow{\lb{}} \leev{}\glresulttype{\ltyv{}}}{\texttt{GLang.S.Term.var\_\allowbreak lin}}{GLang/Adequacy/Strengthened.lean:325}
\gldefrule{s.term.varNlin}{s.term}{\textsc{S-E-Var\allowbreak Nlin}}{((\mathit{x}, \lb{}) \mapsto \leev{}) \in \evalCtx{} \\ (\mathit{x}\colon \tof{\lb{}}) \in \evalCtxSpace{} \\ (\emptyset,\evalCtxSpace{})\vdash \lb{}\colon \bvalid{} \\ \evalCtxSpace{};\evalCtx{}\vdash_{\mathsf S} \lb{} \evalArrow{}_{\mathsf{basis}} \ltyv{}}{\lemptyset{};\evalCtxSpace{};\evalCtx{}\vdash_{\mathsf S} \mathit{x} \xlongrightarrow{\lb{}} \leev{}\glresulttype{\ltyv{}}}{\texttt{GLang.S.Term.var\_\allowbreak nlin}}{GLang/Adequacy/Strengthened.lean:332}
\gldefrule{s.term.promote}{s.term}{\textsc{S-E-Promote}}{\forall y \in \mathit{lin}.\; y \notin \bigcup_{(\cdot\colon s) \in \lDelta{},\, \mathit{lin}} \mathrm{fv}(s) \;\cup\; \bigcup_{(\cdot\colon \tau) \in \evalCtxSpace{}'} \mathrm{fv}(\tau) \\ \evalCtxSpace{}' \subseteq \evalCtxSpace{}'' \\ (\lDelta{}, \mathit{lin} \ldisun \evalCtxSpace{}'')\vdash \lee{}\colon \tof{\lb{}} \\ (\emptyset,\mathit{lin} \ldisun \evalCtxSpace{}'')\vdash \lb{}\colon \bvalid{} \\ \mathit{lin} \ldisun \evalCtxSpace{}'';\evalCtx{}\vdash_{\mathsf S} \lb{} \evalArrow{}_{\mathsf{basis}} \ltyv{} \\ \mathit{lin} \ldisun \lDelta{};\evalCtxSpace{}';\evalCtx{}\vdash_{\mathsf S} \lee{} \xlongrightarrow{\lb{}} \leev{}\glresulttype{\ltyv{}}}{\lDelta{};\mathit{lin} \ldisun \evalCtxSpace{}'';\evalCtx{}\vdash_{\mathsf S} \lee{} \xlongrightarrow{\lb{}} \leev{}\glresulttype{\ltyv{}}}{\texttt{GLang.S.Term.promote}}{GLang/Adequacy/Strengthened.lean:339}
\gldefrule{s.term.cob}{s.term}{\textsc{S-E-Cob}}{\forall i.\; \tof{\lbn{i}} = \sof{\lbn{i}} \\ (\lDelta{}, \evalCtxSpace{})\vdash \lee{}\colon \tof{\lbn{0}} \\ (\emptyset,\evalCtxSpace{})\vdash \lbn{0}\colon \bvalid{} \\ \evalCtxSpace{};\evalCtx{}\vdash_{\mathsf S} \lbn{0} \evalArrow{}_{\mathsf{basis}} \lsv{} \\ \lDelta{};\evalCtxSpace{};\evalCtx{}\vdash_{\mathsf S} \lee{} \xlongrightarrow{\lbn{1}} \leev{}\glresulttype{\ltyv{}} \\ \evalCtxSpace{};\evalCtx{}\vdash_{\mathsf S} \lbn{1} \xlongleftrightarrow{\lenv{0}} \lbn{0} \\ \evalCtxSpace{};\evalCtx{}\vdash_{\mathsf S} \sof{\lbn{0}} \evalArrow{} \lsv{}}{\lDelta{};\evalCtxSpace{};\evalCtx{}\vdash_{\mathsf S} \lee{} \xlongrightarrow{\lbn{0}} \LTTVPP{\lsv{}}{\glvreshape{\lenv{0}}{\lsv{} \loprod \lsv{}}{\lsv{} \loprod \lsv{}}}{\glvreshape{\leev{}}{\lsv{}}{\lsv{}}}\glresulttype{\lsv{}}}{\texttt{GLang.S.Term.cob}}{GLang/Adequacy/Strengthened.lean:353}
\gldefrule{s.term.reshapeCob}{s.term}{\textsc{S-E-Reshape\allowbreak Cob}}{(\lDelta{}, \evalCtxSpace{})\vdash \lee{}\colon \tof{\lb{}} \\ (\emptyset,\evalCtxSpace{})\vdash \lb{}\colon \bvalid{} \\ \evalCtxSpace{};\evalCtx{}\vdash_{\mathsf S} \lb{} \evalArrow{}_{\mathsf{basis}} \ltyv{}' \\ \lDelta{};\evalCtxSpace{};\evalCtx{}\vdash_{\mathsf S} \lee{} \xlongrightarrow{\lb{}} \leev{}\glresulttype{\ltyv{}}}{\lDelta{};\evalCtxSpace{};\evalCtx{}\vdash_{\mathsf S} \lee{} \xlongrightarrow{\lb{}} \glvreshape{\leev{}}{\ltyv{}'}{\ltyv{}}\glresulttype{\ltyv{}'}}{\texttt{GLang.S.Term.reshape\_\allowbreak cob}}{GLang/Adequacy/Strengthened.lean:366}
\gldefrule{s.term.scale}{s.term}{\textsc{S-E-Scale}}{\tof{\lb{}} = \sof{\lb{}} \\ ({\lDelta{}_{0}} \ldisun {\lDelta{}_{1}}, \evalCtxSpace{})\vdash \len{0} \lvscale \len{1}\colon \tof{\lb{}} \\ (\emptyset,\evalCtxSpace{})\vdash \lb{}\colon \bvalid{} \\ \evalCtxSpace{};\evalCtx{}\vdash_{\mathsf S} \lb{} \evalArrow{}_{\mathsf{basis}} \lsv{} \\ {\lDelta{}_{0}};\evalCtxSpace{};\evalCtx{}\vdash_{\mathsf S} \len{0} \xlongrightarrow{\lbone{}} \lenv{0}\glresulttype{\lR{}} \\ {\lDelta{}_{1}};\evalCtxSpace{};\evalCtx{}\vdash_{\mathsf S} \len{1} \xlongrightarrow{\lb{}} \lenv{1}\glresulttype{\lsv{}}}{{\lDelta{}_{0}} \ldisun {\lDelta{}_{1}};\evalCtxSpace{};\evalCtx{}\vdash_{\mathsf S} \len{0} \lvscale \len{1} \xlongrightarrow{\lb{}} \lenv{0} \lvscale \lenv{1}\glresulttype{\lsv{}}}{\texttt{GLang.S.Term.scale}}{GLang/Adequacy/Strengthened.lean:373}
\gldefrule{s.term.add}{s.term}{\textsc{S-E-Add}}{\tof{\lb{}} = \sof{\lb{}} \\ (\lDelta{}, \evalCtxSpace{})\vdash \len{0} \lvadd \len{1}\colon \tof{\lb{}} \\ (\emptyset,\evalCtxSpace{})\vdash \lb{}\colon \bvalid{} \\ \evalCtxSpace{};\evalCtx{}\vdash_{\mathsf S} \lb{} \evalArrow{}_{\mathsf{basis}} \lsv{} \\ \forall i.\; \lDelta{};\evalCtxSpace{};\evalCtx{}\vdash_{\mathsf S} \len{i} \xlongrightarrow{\lb{}} \lenv{i}\glresulttype{\lsnv{i}}}{\lDelta{};\evalCtxSpace{};\evalCtx{}\vdash_{\mathsf S} \len{0} \lvadd \len{1} \xlongrightarrow{\lb{}} \lenv{0} \lvadd_{\lsv{}} \lenv{1}\glresulttype{\lsv{}}}{\texttt{GLang.S.Term.add}}{GLang/Adequacy/Strengthened.lean:383}
\gldefrule{s.term.dsumE}{s.term}{\textsc{S-E-Dsum\allowbreak E}}{\forall i.\; \tof{\lbn{i}} = \sof{\lbn{i}} \\ (\lDelta{}, \evalCtxSpace{})\vdash \len{0}\losum \len{1}\colon \tof{\lbn{0} \losum \lbn{1}} \\ (\emptyset,\evalCtxSpace{})\vdash \lbn{0} \losum \lbn{1}\colon \bvalid{} \\ \evalCtxSpace{};\evalCtx{}\vdash_{\mathsf S} \lbn{0} \losum \lbn{1} \evalArrow{}_{\mathsf{basis}} \lsnv{0} \losum \lsnv{1} \\ \forall i.\; \lDelta{};\evalCtxSpace{};\evalCtx{}\vdash_{\mathsf S} \len{i} \xlongrightarrow{\lbn{i}} \lenv{i}\glresulttype{\lsnv{i}}}{\lDelta{};\evalCtxSpace{};\evalCtx{}\vdash_{\mathsf S} \len{0}\losum \len{1} \xlongrightarrow{\lbn{0} \losum \lbn{1}} \lenv{0}\losum \lenv{1}\glresulttype{\lsnv{0} \losum \lsnv{1}}}{\texttt{GLang.S.Term.dsum\allowbreak E}}{GLang/Adequacy/Strengthened.lean:397}
\gldefrule{s.term.tprodE}{s.term}{\textsc{S-E-Tprod\allowbreak E}}{\forall i.\; \tof{\lbn{i}} = \sof{\lbn{i}} \\ ({\lDelta{}_{0}} \ldisun {\lDelta{}_{1}}, \evalCtxSpace{})\vdash \len{0}\loprod \len{1}\colon \tof{\lbn{0} \loprod \lbn{1}} \\ (\emptyset,\evalCtxSpace{})\vdash \lbn{0} \loprod \lbn{1}\colon \bvalid{} \\ \evalCtxSpace{};\evalCtx{}\vdash_{\mathsf S} \lbn{0} \loprod \lbn{1} \evalArrow{}_{\mathsf{basis}} \lsnv{0} \loprod \lsnv{1} \\ \forall i.\; {\lDelta{}_{i}};\evalCtxSpace{};\evalCtx{}\vdash_{\mathsf S} \len{i} \xlongrightarrow{\lbn{i}} \lenv{i}\glresulttype{\lsnv{i}}}{{\lDelta{}_{0}} \ldisun {\lDelta{}_{1}};\evalCtxSpace{};\evalCtx{}\vdash_{\mathsf S} \len{0}\loprod \len{1} \xlongrightarrow{\lbn{0} \loprod \lbn{1}} \lenv{0}\loprod \lenv{1}\glresulttype{\lsnv{0} \loprod \lsnv{1}}}{\texttt{GLang.S.Term.tprod\allowbreak E}}{GLang/Adequacy/Strengthened.lean:407}
\gldefrule{s.term.pinvTStd}{s.term}{\textsc{S-E-Pinv\allowbreak TStd}}{\forall i.\; \isStdBasis{\lbn{i}} \\ (\lemptyset{}, \evalCtxSpace{})\vdash \lee{}^{\lpseudo}\colon \tof{\lbn{0}^{\lstar} \loprod \lbn{1}} \\ (\emptyset,\evalCtxSpace{})\vdash \lbn{0}^{\lstar} \loprod \lbn{1}\colon \bvalid{} \\ \evalCtxSpace{};\evalCtx{}\vdash_{\mathsf S} \lbn{0}^{\lstar} \loprod \lbn{1} \evalArrow{}_{\mathsf{basis}} \lsnv{0} \loprod \lsnv{1} \\ \lemptyset{};\evalCtxSpace{};\evalCtx{}\vdash_{\mathsf S} \lee{} \xlongrightarrow{\lbn{1}^{\lstar} \loprod \lbn{0}} \leev{}\glresulttype{\lsnv{1} \loprod \lsnv{0}}}{\lemptyset{};\evalCtxSpace{};\evalCtx{}\vdash_{\mathsf S} \lee{}^{\lpseudo} \xlongrightarrow{\lbn{0}^{\lstar} \loprod \lbn{1}} \leev{}^{+}\glresulttype{\lsnv{0} \loprod \lsnv{1}}}{\texttt{GLang.S.Term.pinv\allowbreak T\_\allowbreak std}}{GLang/Adequacy/Strengthened.lean:419}
\gldefrule{s.term.pinvTGpinv}{s.term}{\textsc{S-E-Pinv\allowbreak TGpinv}}{(\lemptyset{}, \evalCtxSpace{})\vdash \lee{}^{\lpseudo}\colon \tof{\lbn{0}^{\lstar} \loprod \lbn{1}} \\ (\emptyset,\evalCtxSpace{})\vdash \lbn{0}^{\lstar} \loprod \lbn{1}\colon \bvalid{} \\ \evalCtxSpace{};\evalCtx{}\vdash_{\mathsf S} \lbn{0}^{\lstar} \loprod \lbn{1} \evalArrow{}_{\mathsf{basis}} \lsnv{0} \loprod \lsnv{1} \\ \lemptyset{};\evalCtxSpace{};\evalCtx{}\vdash_{\mathsf S} \lee{} \xlongrightarrow{\lbn{1}^{\lstar} \loprod \lbn{0}} \leev{}\glresulttype{\lsnv{1} \loprod \lsnv{0}} \\ \forall i.\; \lemptyset{};\evalCtxSpace{};\evalCtx{}\vdash_{\mathsf S} \linner{}_{\sof{\lbn{i}}} \xlongrightarrow{\lbn{i}^{\lstar} \loprod \lbn{i}^{\lstar}} \lenv{i}\glresulttype{\lsnv{i} \loprod \lsnv{i}}}{\lemptyset{};\evalCtxSpace{};\evalCtx{}\vdash_{\mathsf S} \lee{}^{\lpseudo} \xlongrightarrow{\lbn{0}^{\lstar} \loprod \lbn{1}} \glvop{GPINV}(\leev{},\lenv{1},\lenv{0})\glresulttype{\lsnv{0} \loprod \lsnv{1}}}{\texttt{GLang.S.Term.pinv\allowbreak T\_\allowbreak gpinv}}{GLang/Adequacy/Strengthened.lean:429}
\gldefrule{s.term.pinvDStd}{s.term}{\textsc{S-E-Pinv\allowbreak DStd}}{\sof{\lbn{0}} = \lR{} \\ \forall i.\; \isStdBasis{\lbn{i}} \\ (\lemptyset{}, \evalCtxSpace{})\vdash \lee{}^{\lpseudo}\colon \tof{\lbn{0}^{\lstar} \loprod \lbn{1}} \\ (\emptyset,\evalCtxSpace{})\vdash \lbn{0}^{\lstar} \loprod \lbn{1}\colon \bvalid{} \\ \evalCtxSpace{};\evalCtx{}\vdash_{\mathsf S} \lbn{0}^{\lstar} \loprod \lbn{1} \evalArrow{}_{\mathsf{basis}} \lR{} \loprod \lsnv{1} \\ \lemptyset{};\evalCtxSpace{};\evalCtx{}\vdash_{\mathsf S} \lee{} \xlongrightarrow{\lbn{1}^{\lstar}} \leev{}\glresulttype{\lsnv{1}}}{\lemptyset{};\evalCtxSpace{};\evalCtx{}\vdash_{\mathsf S} \lee{}^{\lpseudo} \xlongrightarrow{\lbn{0}^{\lstar} \loprod \lbn{1}} \glvreshape{\leev{}}{\lsnv{1} \loprod \lR{}}{\lsnv{1}}^{+}\glresulttype{\lR{} \loprod \lsnv{1}}}{\texttt{GLang.S.Term.pinv\allowbreak D\_\allowbreak std}}{GLang/Adequacy/Strengthened.lean:443}
\gldefrule{s.term.pinvDGpinv}{s.term}{\textsc{S-E-Pinv\allowbreak DGpinv}}{\sof{\lbn{0}} = \lR{} \\ \isStdBasis{\lbn{0}} \\ (\lemptyset{}, \evalCtxSpace{})\vdash \lee{}^{\lpseudo}\colon \tof{\lbn{0}^{\lstar} \loprod \lbn{1}} \\ (\emptyset,\evalCtxSpace{})\vdash \lbn{0}^{\lstar} \loprod \lbn{1}\colon \bvalid{} \\ \evalCtxSpace{};\evalCtx{}\vdash_{\mathsf S} \lbn{0}^{\lstar} \loprod \lbn{1} \evalArrow{}_{\mathsf{basis}} \lR{} \loprod \lsnv{1} \\ \lemptyset{};\evalCtxSpace{};\evalCtx{}\vdash_{\mathsf S} \lee{} \xlongrightarrow{\lbn{1}^{\lstar}} \leev{}\glresulttype{\lsnv{1}} \\ \lemptyset{};\evalCtxSpace{};\evalCtx{}\vdash_{\mathsf S} \linner{}_{\sof{\lbn{0}}} \xlongrightarrow{\lbn{0}^{\lstar} \loprod \lbn{0}^{\lstar}} \lenv{0}\glresulttype{\lR{} \loprod \lR{}} \\ \lemptyset{};\evalCtxSpace{};\evalCtx{}\vdash_{\mathsf S} \linner{}_{\sof{\lbn{1}}} \xlongrightarrow{\lbn{1}^{\lstar} \loprod \lbn{1}^{\lstar}} \lenv{1}\glresulttype{\lsnv{1} \loprod \lsnv{1}}}{\lemptyset{};\evalCtxSpace{};\evalCtx{}\vdash_{\mathsf S} \lee{}^{\lpseudo} \xlongrightarrow{\lbn{0}^{\lstar} \loprod \lbn{1}} \glvop{GPINV}(\glvreshape{\leev{}}{\lsnv{1} \loprod \lR{}}{\lsnv{1}},\lenv{1},\lenv{0})\glresulttype{\lR{} \loprod \lsnv{1}}}{\texttt{GLang.S.Term.pinv\allowbreak D\_\allowbreak gpinv}}{GLang/Adequacy/Strengthened.lean:455}
\gldefrule{s.term.innerStd}{s.term}{\textsc{S-E-Inner\allowbreak Std}}{\isStdBasis{\lbn{0}} \\ \sof{\lbn{0}} = \ls{} \\ (\lemptyset{}, \evalCtxSpace{})\vdash \linner{}_{\ls{}}\colon \tof{\lbn{0}^{\lstar} \loprod \lbn{0}^{\lstar}} \\ (\emptyset,\evalCtxSpace{})\vdash \lbn{0}^{\lstar} \loprod \lbn{0}^{\lstar}\colon \bvalid{} \\ \evalCtxSpace{};\evalCtx{}\vdash_{\mathsf S} \lbn{0}^{\lstar} \loprod \lbn{0}^{\lstar} \evalArrow{}_{\mathsf{basis}} \lsv{} \loprod \lsv{} \\ \evalCtxSpace{};\evalCtx{}\vdash_{\mathsf S} \ls{} \evalArrow{} \lsv{}}{\lemptyset{};\evalCtxSpace{};\evalCtx{}\vdash_{\mathsf S} \linner{}_{\ls{}} \xlongrightarrow{\lbn{0}^{\lstar} \loprod \lbn{0}^{\lstar}} \lidelta{\lsv{}}\glresulttype{\lsv{} \loprod \lsv{}}}{\texttt{GLang.S.Term.inner\_\allowbreak std}}{GLang/Adequacy/Strengthened.lean:472}
\gldefrule{s.term.embedKer}{s.term}{\textsc{S-E-Embed\allowbreak Ker}}{\forall i.\; \sof{\lbn{i}} = \lsn{i} \\ (\lemptyset{}, \evalCtxSpace{})\vdash \lembd{}_{\Kernel(\lee{})}\colon \tof{\pBasisC{\Kernel(\lee{})}{\lbn{0}}{\lbn{1}}^{\lstar} \loprod \lbn{0}} \\ (\emptyset,\evalCtxSpace{})\vdash \pBasisC{\Kernel(\lee{})}{\lbn{0}}{\lbn{1}}^{\lstar} \loprod \lbn{0}\colon \bvalid{} \\ \evalCtxSpace{};\evalCtx{}\vdash_{\mathsf S} \pBasisC{\Kernel(\lee{})}{\lbn{0}}{\lbn{1}}^{\lstar} \loprod \lbn{0} \evalArrow{}_{\mathsf{basis}} \lR^{\glvreshape{\leev{}}{\lsnv{0} \loprod \lsnv{1}}{\lsnv{0} \loprod \lsnv{1}},\Kernel{},\lsnv{0} \loprod \lsnv{1}} \loprod \lsnv{0} \\ \forall i.\; \lemptyset{};\evalCtxSpace{};\evalCtx{}\vdash_{\mathsf S} \linner{}_{\lsn{i}} \xlongrightarrow{\lbn{i}^{\lstar} \loprod \lbn{i}^{\lstar}} \lenv{i}\glresulttype{\lsnv{i} \loprod \lsnv{i}} \\ \lemptyset{};\evalCtxSpace{};\evalCtx{}\vdash_{\mathsf S} \lee{} \xlongrightarrow{\lbn{0}^{\lstar} \loprod \lbn{1}} \leev{}\glresulttype{\lsnv{0} \loprod \lsnv{1}}}{\lemptyset{};\evalCtxSpace{};\evalCtx{}\vdash_{\mathsf S} \lembd{}_{\Kernel(\lee{})} \xlongrightarrow{\pBasisC{\Kernel(\lee{})}{\lbn{0}}{\lbn{1}}^{\lstar} \loprod \lbn{0}} \Trans(\glvop{GSVDUI}(\leev{},\lenv{0},\lenv{1}))\glresulttype{\lR^{\glvreshape{\leev{}}{\lsnv{0} \loprod \lsnv{1}}{\lsnv{0} \loprod \lsnv{1}},\Kernel{},\lsnv{0} \loprod \lsnv{1}} \loprod \lsnv{0}}}{\texttt{GLang.S.Term.embed\_\allowbreak ker}}{GLang/Adequacy/Strengthened.lean:482}
\gldefrule{s.term.embedCoker}{s.term}{\textsc{S-E-Embed\allowbreak Coker}}{\forall i.\; \sof{\lbn{i}} = \lsn{i} \\ (\lemptyset{}, \evalCtxSpace{})\vdash \lembd{}_{\CoKernel(\lee{})}\colon \tof{\pBasisC{\CoKernel(\lee{})}{\lbn{0}}{\lbn{1}}^{\lstar} \loprod \lbn{0}} \\ (\emptyset,\evalCtxSpace{})\vdash \pBasisC{\CoKernel(\lee{})}{\lbn{0}}{\lbn{1}}^{\lstar} \loprod \lbn{0}\colon \bvalid{} \\ \evalCtxSpace{};\evalCtx{}\vdash_{\mathsf S} \pBasisC{\CoKernel(\lee{})}{\lbn{0}}{\lbn{1}}^{\lstar} \loprod \lbn{0} \evalArrow{}_{\mathsf{basis}} \lR^{\glvreshape{\leev{}}{\lsnv{0} \loprod \lsnv{1}}{\lsnv{0} \loprod \lsnv{1}},\CoKernel{},\lsnv{0} \loprod \lsnv{1}} \loprod \lsnv{0} \\ \forall i.\; \lemptyset{};\evalCtxSpace{};\evalCtx{}\vdash_{\mathsf S} \linner{}_{\lsn{i}} \xlongrightarrow{\lbn{i}^{\lstar} \loprod \lbn{i}^{\lstar}} \lenv{i}\glresulttype{\lsnv{i} \loprod \lsnv{i}} \\ \lemptyset{};\evalCtxSpace{};\evalCtx{}\vdash_{\mathsf S} \lee{} \xlongrightarrow{\lbn{0}^{\lstar} \loprod \lbn{1}} \leev{}\glresulttype{\lsnv{0} \loprod \lsnv{1}}}{\lemptyset{};\evalCtxSpace{};\evalCtx{}\vdash_{\mathsf S} \lembd{}_{\CoKernel(\lee{})} \xlongrightarrow{\pBasisC{\CoKernel(\lee{})}{\lbn{0}}{\lbn{1}}^{\lstar} \loprod \lbn{0}} \Trans(\glvop{GSVDU}(\leev{},\lenv{0},\lenv{1}))\glresulttype{\lR^{\glvreshape{\leev{}}{\lsnv{0} \loprod \lsnv{1}}{\lsnv{0} \loprod \lsnv{1}},\CoKernel{},\lsnv{0} \loprod \lsnv{1}} \loprod \lsnv{0}}}{\texttt{GLang.S.Term.embed\_\allowbreak coker}}{GLang/Adequacy/Strengthened.lean:503}
\gldefrule{s.term.embedImg}{s.term}{\textsc{S-E-Embed\allowbreak Img}}{\forall i.\; \sof{\lbn{i}} = \lsn{i} \\ (\lemptyset{}, \evalCtxSpace{})\vdash \lembd{}_{\Image(\lee{})}\colon \tof{\pBasisC{\Image(\lee{})}{\lbn{0}}{\lbn{1}}^{\lstar} \loprod \lbn{1}} \\ (\emptyset,\evalCtxSpace{})\vdash \pBasisC{\Image(\lee{})}{\lbn{0}}{\lbn{1}}^{\lstar} \loprod \lbn{1}\colon \bvalid{} \\ \evalCtxSpace{};\evalCtx{}\vdash_{\mathsf S} \pBasisC{\Image(\lee{})}{\lbn{0}}{\lbn{1}}^{\lstar} \loprod \lbn{1} \evalArrow{}_{\mathsf{basis}} \lR^{\glvreshape{\leev{}}{\lsnv{0} \loprod \lsnv{1}}{\lsnv{0} \loprod \lsnv{1}},\Image{},\lsnv{0} \loprod \lsnv{1}} \loprod \lsnv{1} \\ \forall i.\; \lemptyset{};\evalCtxSpace{};\evalCtx{}\vdash_{\mathsf S} \linner{}_{\lsn{i}} \xlongrightarrow{\lbn{i}^{\lstar} \loprod \lbn{i}^{\lstar}} \lenv{i}\glresulttype{\lsnv{i} \loprod \lsnv{i}} \\ \lemptyset{};\evalCtxSpace{};\evalCtx{}\vdash_{\mathsf S} \lee{} \xlongrightarrow{\lbn{0}^{\lstar} \loprod \lbn{1}} \leev{}\glresulttype{\lsnv{0} \loprod \lsnv{1}}}{\lemptyset{};\evalCtxSpace{};\evalCtx{}\vdash_{\mathsf S} \lembd{}_{\Image(\lee{})} \xlongrightarrow{\pBasisC{\Image(\lee{})}{\lbn{0}}{\lbn{1}}^{\lstar} \loprod \lbn{1}} \Trans(\glvop{GSVDV}(\leev{},\lenv{0},\lenv{1}))\glresulttype{\lR^{\glvreshape{\leev{}}{\lsnv{0} \loprod \lsnv{1}}{\lsnv{0} \loprod \lsnv{1}},\Image{},\lsnv{0} \loprod \lsnv{1}} \loprod \lsnv{1}}}{\texttt{GLang.S.Term.embed\_\allowbreak img}}{GLang/Adequacy/Strengthened.lean:523}
\gldefrule{s.term.embedCoimg}{s.term}{\textsc{S-E-Embed\allowbreak Coimg}}{\forall i.\; \sof{\lbn{i}} = \lsn{i} \\ (\lemptyset{}, \evalCtxSpace{})\vdash \lembd{}_{\CoImage(\lee{})}\colon \tof{\pBasisC{\CoImage(\lee{})}{\lbn{0}}{\lbn{1}}^{\lstar} \loprod \lbn{1}} \\ (\emptyset,\evalCtxSpace{})\vdash \pBasisC{\CoImage(\lee{})}{\lbn{0}}{\lbn{1}}^{\lstar} \loprod \lbn{1}\colon \bvalid{} \\ \evalCtxSpace{};\evalCtx{}\vdash_{\mathsf S} \pBasisC{\CoImage(\lee{})}{\lbn{0}}{\lbn{1}}^{\lstar} \loprod \lbn{1} \evalArrow{}_{\mathsf{basis}} \lR^{\glvreshape{\leev{}}{\lsnv{0} \loprod \lsnv{1}}{\lsnv{0} \loprod \lsnv{1}},\CoImage{},\lsnv{0} \loprod \lsnv{1}} \loprod \lsnv{1} \\ \forall i.\; \lemptyset{};\evalCtxSpace{};\evalCtx{}\vdash_{\mathsf S} \linner{}_{\lsn{i}} \xlongrightarrow{\lbn{i}^{\lstar} \loprod \lbn{i}^{\lstar}} \lenv{i}\glresulttype{\lsnv{i} \loprod \lsnv{i}} \\ \lemptyset{};\evalCtxSpace{};\evalCtx{}\vdash_{\mathsf S} \lee{} \xlongrightarrow{\lbn{0}^{\lstar} \loprod \lbn{1}} \leev{}\glresulttype{\lsnv{0} \loprod \lsnv{1}}}{\lemptyset{};\evalCtxSpace{};\evalCtx{}\vdash_{\mathsf S} \lembd{}_{\CoImage(\lee{})} \xlongrightarrow{\pBasisC{\CoImage(\lee{})}{\lbn{0}}{\lbn{1}}^{\lstar} \loprod \lbn{1}} \Trans(\glvop{GSVDVI}(\leev{},\lenv{0},\lenv{1}))\glresulttype{\lR^{\glvreshape{\leev{}}{\lsnv{0} \loprod \lsnv{1}}{\lsnv{0} \loprod \lsnv{1}},\CoImage{},\lsnv{0} \loprod \lsnv{1}} \loprod \lsnv{1}}}{\texttt{GLang.S.Term.embed\_\allowbreak coimg}}{GLang/Adequacy/Strengthened.lean:543}
\gldefrule{s.term.appLLin}{s.term}{\textsc{S-E-App\allowbreak LLin}}{\tof{\lbn{1}} = \sof{\lbn{1}} \\ ({\lDelta{}_{0}} \ldisun {\lDelta{}_{1}}, \evalCtxSpace{})\vdash \leval{\len{0}}{}{\len{1}}\colon \tof{\lbn{1}} \\ (\emptyset,\evalCtxSpace{})\vdash \lbn{1}\colon \bvalid{} \\ \evalCtxSpace{};\evalCtx{}\vdash_{\mathsf S} \sof{\lbn{0}} \evalArrow{} \lsv{} \\ \evalCtxSpace{};\evalCtx{}\vdash_{\mathsf S} \lbn{1} \evalArrow{}_{\mathsf{basis}} \lsnv{{\mathrm{svOut}}} \\ {\lDelta{}_{0}};\evalCtxSpace{};\evalCtx{}\vdash_{\mathsf S} \len{0} \xlongrightarrow{\lbn{0}^{\lstar} \loprod \lbn{1}} \lenv{0}\glresulttype{\lsnv{{\mathrm{svMat0}}} \loprod \lsnv{{\mathrm{svMat1}}}} \\ {\lDelta{}_{1}};\evalCtxSpace{};\evalCtx{}\vdash_{\mathsf S} \len{1} \xlongrightarrow{\lbn{0}} \lenv{1}\glresulttype{\lsnv{{\mathrm{svArg}}}}}{{\lDelta{}_{0}} \ldisun {\lDelta{}_{1}};\evalCtxSpace{};\evalCtx{}\vdash_{\mathsf S} \leval{\len{0}}{}{\len{1}} \xlongrightarrow{\lbn{1}} \glvreshape{\LTTVPP{\lsv{}}{\lenv{0}}{\lenv{1}}}{\lsnv{{\mathrm{svOut}}}}{\lsnv{{\mathrm{svOut}}}}\glresulttype{\lsnv{{\mathrm{svOut}}}}}{\texttt{GLang.S.Term.app\allowbreak L\_\allowbreak lin}}{GLang/Adequacy/Strengthened.lean:563}
\gldefrule{s.term.appILin}{s.term}{\textsc{S-E-App\allowbreak ILin}}{({\lDelta{}_{0}} \ldisun {\lDelta{}_{1}}, \evalCtxSpace{})\vdash \leval{\len{0}}{}{\len{1}}\colon \tof{\lbone{}} \\ \evalCtxSpace{};\evalCtx{}\vdash_{\mathsf S} \sof{\lbn{0}} \evalArrow{} \lsv{} \\ \evalCtxSpace{};\evalCtx{}\vdash_{\mathsf S} \lbone{} \evalArrow{}_{\mathsf{basis}} \lR{} \\ {\lDelta{}_{0}};\evalCtxSpace{};\evalCtx{}\vdash_{\mathsf S} \len{0} \xlongrightarrow{\lbn{0}^{\lstar}} \lenv{0}\glresulttype{\lsnv{{\mathrm{svFun}}}} \\ {\lDelta{}_{1}};\evalCtxSpace{};\evalCtx{}\vdash_{\mathsf S} \len{1} \xlongrightarrow{\lbn{0}} \lenv{1}\glresulttype{\lsnv{{\mathrm{svArg}}}}}{{\lDelta{}_{0}} \ldisun {\lDelta{}_{1}};\evalCtxSpace{};\evalCtx{}\vdash_{\mathsf S} \leval{\len{0}}{}{\len{1}} \xlongrightarrow{\lbone{}} \LTTVPP{\lsv{}}{\glvreshape{\lenv{0}}{\lsv{} \loprod \lR{}}{\lsv{}}}{\lenv{1}}\glresulttype{\lR{}}}{\texttt{GLang.S.Term.app\allowbreak I\_\allowbreak lin}}{GLang/Adequacy/Strengthened.lean:583}
\gldefrule{s.term.nllam}{s.term}{\textsc{S-E-Nllam}}{\forall i.\; \tof{\lbn{i}} = \ltyn{i} \\ (\lemptyset{}, \evalCtxSpace{})\vdash (\nllambda \llpl x\colon \ltyn{0}\llpr \ldotsp \lee{})\colon \tof{\frameB{x}{\lbn{0}}{\lbn{1}}} \\ (\emptyset,\evalCtxSpace{})\vdash \frameB{x}{\lbn{0}}{\lbn{1}}\colon \bvalid{} \\ \evalCtxSpace{};\evalCtx{}\vdash_{\mathsf S} \frameB{x}{\lbn{0}}{\lbn{1}} \evalArrow{}_{\mathsf{basis}} (x\colon \ltynv{0}\fnto \ltynv{1}) \\ \evalCtxSpace{};\evalCtx{}\vdash_{\mathsf S} \lbn{0} \evalArrow{}_{\mathsf{basis}} \ltynv{0} \\ \lemptyset{};(\evalCtxSpace{}, \mathit{x}\colon \ltyn{0});\evalCtx{}[(\mathit{x},\lbn{0}) \mapsto \mathit{x}_{\ltynv{0}}]\vdash_{\mathsf S} \lee{} \xlongrightarrow{\lbn{1}} \leev{}\glresulttype{\ltynv{1}} \\ (\evalCtxSpace{}, \mathit{x}\colon \ltyn{0});\evalCtx{}[(\mathit{x},\lbn{0}) \mapsto \mathit{x}_{\ltynv{0}}]\vdash_{\mathsf S} \lbn{1} \evalArrow{}_{\mathsf{basis}} \ltynv{1}}{\lemptyset{};\evalCtxSpace{};\evalCtx{}\vdash_{\mathsf S} (\nllambda \llpl x\colon \ltyn{0}\llpr \ldotsp \lee{}) \xlongrightarrow{\frameB{x}{\lbn{0}}{\lbn{1}}} \llambda(x\colon \ltynv{0})\lrh \ltynv{1}\colon \glvreshape{\leev{}}{\ltynv{1}}{\ltynv{1}}\glresulttype{(x\colon \ltynv{0}\fnto \ltynv{1})}}{\texttt{GLang.S.Term.nllam}}{GLang/Adequacy/Strengthened.lean:594}
\gldefrule{s.term.appNVal}{s.term}{\textsc{S-E-App\allowbreak NVal}}{(\lemptyset{}, \evalCtxSpace{})\vdash \leval{\len{0}}{}{\len{1}}\colon \tof{\lbn{1}\subst{x}{\len{1}}} \\ (\emptyset,\evalCtxSpace{})\vdash \lbn{1}\subst{x}{\len{1}}\colon \bvalid{} \\ \evalCtxSpace{};\evalCtx{}\vdash_{\mathsf S} \lbn{1}\subst{x}{\len{1}} \evalArrow{}_{\mathsf{basis}} {\ltyv{}_{\mathsf{out}}} \\ \lemptyset{};\evalCtxSpace{};\evalCtx{}\vdash_{\mathsf S} \len{0} \xlongrightarrow{\frameB{x}{\lbn{0}}{\lbn{1}}} \lenv{0}\glresulttype{(x\colon {\ltyv{}_{\mathsf{fun}0}}\fnto {\ltyv{}_{\mathsf{fun}1}})} \\ \lemptyset{};\evalCtxSpace{};\evalCtx{}\vdash_{\mathsf S} \len{1} \xlongrightarrow{\lbn{0}} \lenv{1}\glresulttype{{\ltyv{}_{\mathsf{arg}}}} \\ \operatorname{inferType}(\lenv{0}) = (x\colon \ltynv{{\mathrm{D}}}\fnto \ltynv{{\mathrm{C}}}) \\ \lenv{2} = \glvreshape{\lenv{1}}{\ltynv{{\mathrm{D}}}}{\operatorname{inferType}(\lenv{1})}}{\lemptyset{};\evalCtxSpace{};\evalCtx{}\vdash_{\mathsf S} \leval{\len{0}}{}{\len{1}} \xlongrightarrow{\lbn{1}\subst{x}{\len{1}}} \glvreshape{\glvapp{\lenv{0}}{\lenv{2}}}{{\ltyv{}_{\mathsf{out}}}}{\ltynv{{\mathrm{C}}}\subst{x}{\lenv{2}}}\glresulttype{{\ltyv{}_{\mathsf{out}}}}}{\texttt{GLang.S.Term.app\allowbreak N\_\allowbreak val}}{GLang/Adequacy/Strengthened.lean:612}
\gldefrule{s.term.frameReanchor}{s.term}{\textsc{S-E-Frame\allowbreak Reanchor}}{\tof{\lbn{0}'} = \tof{\lbn{0}} \\ (\lemptyset{}, \evalCtxSpace{})\vdash \len{0}\colon \tof{\frameB{x}{\lbn{0}}{\lbn{1}}} \\ (\emptyset,\evalCtxSpace{})\vdash \frameB{x}{\lbn{0}}{\lbn{1}}\colon \bvalid{} \\ \evalCtxSpace{};\evalCtx{}\vdash_{\mathsf S} \frameB{x}{\lbn{0}}{\lbn{1}} \evalArrow{}_{\mathsf{basis}} (x\colon \ltynv{0}\fnto \ltynv{1}') \\ \evalCtxSpace{};\evalCtx{}\vdash_{\mathsf S} \lbn{0} \evalArrow{}_{\mathsf{basis}} \ltynv{0} \\ \lemptyset{};\evalCtxSpace{};\evalCtx{}\vdash_{\mathsf S} \len{0} \xlongrightarrow{\frameB{x}{\lbn{0}'}{\lbn{1}}} \leev{}\glresulttype{(x\colon \ltynv{0}'\fnto \ltynv{1})} \\ \evalCtxSpace{};\evalCtx{}\vdash_{\mathsf S} \lbn{0}' \evalArrow{}_{\mathsf{basis}} \ltynv{0}' \\ \lemptyset{};(\evalCtxSpace{}, \mathit{x}\colon \tof{\lbn{0}});\evalCtx{}[(\mathit{x},\lbn{0}) \mapsto \mathit{x}_{\ltynv{0}}]\vdash_{\mathsf S} \mathit{x} \xlongrightarrow{\lbn{0}'} \lenv{1}\glresulttype{\ltynv{0}'} \\ (\evalCtxSpace{}, \mathit{x}\colon \tof{\lbn{0}});\evalCtx{}[(\mathit{x},\lbn{0}) \mapsto \mathit{x}_{\ltynv{0}}]\vdash_{\mathsf S} \lbn{1} \evalArrow{}_{\mathsf{basis}} \ltynv{1}' \\ \operatorname{inferType}(\leev{}) = (x\colon \ltynv{{\mathrm{D}}}\fnto \ltynv{{\mathrm{C}}}) \\ \lenv{2} = \glvreshape{\glvreshape{\lenv{1}}{\ltynv{0}'}{\ltynv{0}'}}{\ltynv{{\mathrm{D}}}}{\ltynv{0}'}}{\lemptyset{};\evalCtxSpace{};\evalCtx{}\vdash_{\mathsf S} \len{0} \xlongrightarrow{\frameB{x}{\lbn{0}}{\lbn{1}}} \llambda(x\colon \ltynv{0})\lrh \ltynv{1}'\colon \glvreshape{\glvapp{\leev{}}{\lenv{2}}}{\ltynv{1}'}{\ltynv{1}'}\glresulttype{(x\colon \ltynv{0}\fnto \ltynv{1}')}}{\texttt{GLang.S.Term.frame\_\allowbreak reanchor}}{GLang/Adequacy/Strengthened.lean:627}
\gldefrule{s.term.frameEta}{s.term}{\textsc{S-E-Frame\allowbreak Eta}}{(\lemptyset{}, \evalCtxSpace{})\vdash \len{0}\colon \tof{\frameB{x}{\lbn{0}}{\lbn{1}}} \\ (\emptyset,\evalCtxSpace{})\vdash \frameB{x}{\lbn{0}}{\lbn{1}}\colon \bvalid{} \\ \evalCtxSpace{};\evalCtx{}\vdash_{\mathsf S} \frameB{x}{\lbn{0}}{\lbn{1}} \evalArrow{}_{\mathsf{basis}} (x\colon \ltynv{0}\fnto \ltynv{1}) \\ \evalCtxSpace{};\evalCtx{}\vdash_{\mathsf S} \lbn{0} \evalArrow{}_{\mathsf{basis}} \ltynv{0} \\ \lemptyset{};(\evalCtxSpace{}, \mathit{x}\colon \tof{\lbn{0}});\evalCtx{}[(\mathit{x},\lbn{0}) \mapsto \mathit{x}_{\ltynv{0}}]\vdash_{\mathsf S} \leval{\len{0}}{}{\mathit{x}} \xlongrightarrow{\lbn{1}} \leev{}\glresulttype{\ltynv{1}} \\ (\evalCtxSpace{}, \mathit{x}\colon \tof{\lbn{0}});\evalCtx{}[(\mathit{x},\lbn{0}) \mapsto \mathit{x}_{\ltynv{0}}]\vdash_{\mathsf S} \lbn{1} \evalArrow{}_{\mathsf{basis}} \ltynv{1}}{\lemptyset{};\evalCtxSpace{};\evalCtx{}\vdash_{\mathsf S} \len{0} \xlongrightarrow{\frameB{x}{\lbn{0}}{\lbn{1}}} \llambda(x\colon \ltynv{0})\lrh \ltynv{1}\colon \glvreshape{\leev{}}{\ltynv{1}}{\ltynv{1}}\glresulttype{(x\colon \ltynv{0}\fnto \ltynv{1})}}{\texttt{GLang.S.Term.frame\_\allowbreak eta}}{GLang/Adequacy/Strengthened.lean:652}
\gldefrule{s.term.llamAssemble}{s.term}{\textsc{S-E-Llam\allowbreak Assemble}}{\tof{\lb{}'} = \sof{\lb{}'} \\ (\lDelta{}, \evalCtxSpace{})\vdash (\lllambda \llpl x\colon \sof{\lb{}'}\llpr \ldotsp \lee{})\colon \tof{\lb{}'^{\lstar} \loprod \lb{}} \\ (\emptyset,\evalCtxSpace{})\vdash \lb{}'^{\lstar} \loprod \lb{}\colon \bvalid{} \\ \evalCtxSpace{};\evalCtx{}\vdash_{\mathsf S} \lb{}'^{\lstar} \loprod \lb{} \evalArrow{}_{\mathsf{basis}} \lsv{}' \loprod \lsv{} \\ \evalCtxSpace{};\evalCtx{}\vdash_{\mathsf S} \sof{\lb{}'} \evalArrow{} \lsv{}' \\ \evalCtxSpace{};\evalCtx{}\vdash_{\mathsf S} \sof{\lb{}} \evalArrow{} \lsv{} \\ \forall\,0\leq i<\overline{\operatorname{dimOf}}(\lsv{}').\; (\lDelta{}, \mathit{x}\colon \sof{\lb{}'});\evalCtxSpace{};\evalCtx{}[(\mathit{x},\lb{}') \mapsto \ldelta{i}{\lsv{}'}]\vdash_{\mathsf S} \lee{} \xlongrightarrow{\lb{}} \leev{}_{i}\glresulttype{\lsv{}}}{\lDelta{};\evalCtxSpace{};\evalCtx{}\vdash_{\mathsf S} (\lllambda \llpl x\colon \sof{\lb{}'}\llpr \ldotsp \lee{}) \xlongrightarrow{\lb{}'^{\lstar} \loprod \lb{}} \glvreshape{\lsum^{\lsv{}'}_{\mathit{i}<\overline{\operatorname{dimOf}}(\lsv{}')}\left(\ldelta{\mathit{i}}{\lsv{}'}\loprod \glvreshape{\leev{}_{\mathit{i}}}{\lsv{}}{\lsv{}}\right)}{\lsv{}' \loprod \lsv{}}{\begin{cases}\lsempty{}&\overline{\operatorname{dimOf}}(\lsv{}')=0\\ \lsv{}' \loprod \lsv{}&\text{otherwise}\end{cases}}\glresulttype{\lsv{}' \loprod \lsv{}}}{\texttt{GLang.S.Term.llam\_\allowbreak assemble}}{GLang/Adequacy/Strengthened.lean:671}
\gldefrule{s.term.llamAssembleReal}{s.term}{\textsc{S-E-Llam\allowbreak Assemble\allowbreak Real}}{\tof{\lb{}'} = \sof{\lb{}'} \\ (\lDelta{}, \evalCtxSpace{})\vdash (\lllambda \llpl x\colon \sof{\lb{}'}\llpr \ldotsp \lee{})\colon \tof{\lb{}'^{\lstar}} \\ (\emptyset,\evalCtxSpace{})\vdash \lb{}'^{\lstar}\colon \bvalid{} \\ \evalCtxSpace{};\evalCtx{}\vdash_{\mathsf S} \lb{}'^{\lstar} \evalArrow{}_{\mathsf{basis}} \lsv{}' \\ \evalCtxSpace{};\evalCtx{}\vdash_{\mathsf S} \sof{\lb{}'} \evalArrow{} \lsv{}' \\ \forall\,0\leq i<\overline{\operatorname{dimOf}}(\lsv{}').\; (\lDelta{}, \mathit{x}\colon \sof{\lb{}'});\evalCtxSpace{};\evalCtx{}[(\mathit{x},\lb{}') \mapsto \ldelta{i}{\lsv{}'}]\vdash_{\mathsf S} \lee{} \xlongrightarrow{\lbone{}} \leev{}_{i}\glresulttype{\lR{}}}{\lDelta{};\evalCtxSpace{};\evalCtx{}\vdash_{\mathsf S} (\lllambda \llpl x\colon \sof{\lb{}'}\llpr \ldotsp \lee{}) \xlongrightarrow{\lb{}'^{\lstar}} \glvreshape{\lsum^{\lsv{}'}_{\mathit{i}<\overline{\operatorname{dimOf}}(\lsv{}')}\left(\glvreshape{\leev{}_{\mathit{i}}}{\lR{}}{\lR{}} \lvscale \ldelta{\mathit{i}}{\lsv{}'}\right)}{\lsv{}'}{\begin{cases}\lsempty{}&\overline{\operatorname{dimOf}}(\lsv{}')=0\\ \lsv{}'&\text{otherwise}\end{cases}}\glresulttype{\lsv{}'}}{\texttt{GLang.S.Term.llam\_\allowbreak assemble\_\allowbreak real}}{GLang/Adequacy/Strengthened.lean:701}
\gldefrule{s.term.zero}{s.term}{\textsc{S-E-Zero}}{\tof{\lb{}} = \sof{\lb{}} \\ (\lDelta{}, \evalCtxSpace{})\vdash \lvzero{}_{\sof{\lb{}}}\colon \tof{\lb{}} \\ (\emptyset,\evalCtxSpace{})\vdash \lb{}\colon \bvalid{} \\ \evalCtxSpace{};\evalCtx{}\vdash_{\mathsf S} \lb{} \evalArrow{}_{\mathsf{basis}} \lsv{} \\ \evalCtxSpace{};\evalCtx{}\vdash_{\mathsf S} \sof{\lb{}} \evalArrow{} \lsv{}}{\lDelta{};\evalCtxSpace{};\evalCtx{}\vdash_{\mathsf S} \lvzero{}_{\sof{\lb{}}} \xlongrightarrow{\lb{}} \lvzero{}_{\lsv{}}\glresulttype{\lsv{}}}{\texttt{GLang.S.Term.zero}}{GLang/Adequacy/Strengthened.lean:720}
\gldefrule{s.term.rlit}{s.term}{\textsc{S-E-Rlit}}{(\lemptyset{}, \evalCtxSpace{})\vdash \mathit{r}\colon \tof{\lbone{}} \\ \evalCtxSpace{};\evalCtx{}\vdash_{\mathsf S} \lbone{} \evalArrow{}_{\mathsf{basis}} \lR{}}{\lemptyset{};\evalCtxSpace{};\evalCtx{}\vdash_{\mathsf S} \mathit{r} \xlongrightarrow{\lbone{}} \mathit{r}\glresulttype{\lR{}}}{\texttt{GLang.S.Term.rlit}}{GLang/Adequacy/Strengthened.lean:728}
\gldefrulerows{s.term}{
\textsc{S-E-Var\allowbreak Lin} & $\{(\mathit{x}, \ls{})\};\evalCtxSpace{};\evalCtx{}\vdash_{\mathsf S} \mathit{x} \xlongrightarrow{\lb{}} \leev{}\glresulttype{\ltyv{}}$ & \texttt{\footnotesize GLang/Adequacy/Strengthened.lean:325}\\
\textsc{S-E-Var\allowbreak Nlin} & $\lemptyset{};\evalCtxSpace{};\evalCtx{}\vdash_{\mathsf S} \mathit{x} \xlongrightarrow{\lb{}} \leev{}\glresulttype{\ltyv{}}$ & \texttt{\footnotesize GLang/Adequacy/Strengthened.lean:332}\\
\textsc{S-E-Promote} & $\lDelta{};\mathit{lin} \ldisun \evalCtxSpace{}'';\evalCtx{}\vdash_{\mathsf S} \lee{} \xlongrightarrow{\lb{}} \leev{}\glresulttype{\ltyv{}}$ & \texttt{\footnotesize GLang/Adequacy/Strengthened.lean:339}\\
\textsc{S-E-Cob} & $\lDelta{};\evalCtxSpace{};\evalCtx{}\vdash_{\mathsf S} \lee{} \xlongrightarrow{\lbn{0}} \LTTVPP{\lsv{}}{\glvreshape{\lenv{0}}{\lsv{} \loprod \lsv{}}{\lsv{} \loprod \lsv{}}}{\glvreshape{\leev{}}{\lsv{}}{\lsv{}}}\glresulttype{\lsv{}}$ & \texttt{\footnotesize GLang/Adequacy/Strengthened.lean:353}\\
\textsc{S-E-Reshape\allowbreak Cob} & $\lDelta{};\evalCtxSpace{};\evalCtx{}\vdash_{\mathsf S} \lee{} \xlongrightarrow{\lb{}} \glvreshape{\leev{}}{\ltyv{}'}{\ltyv{}}\glresulttype{\ltyv{}'}$ & \texttt{\footnotesize GLang/Adequacy/Strengthened.lean:366}\\
\textsc{S-E-Scale} & ${\lDelta{}_{0}} \ldisun {\lDelta{}_{1}};\evalCtxSpace{};\evalCtx{}\vdash_{\mathsf S} \len{0} \lvscale \len{1} \xlongrightarrow{\lb{}} \lenv{0} \lvscale \lenv{1}\glresulttype{\lsv{}}$ & \texttt{\footnotesize GLang/Adequacy/Strengthened.lean:373}\\
\textsc{S-E-Add} & $\lDelta{};\evalCtxSpace{};\evalCtx{}\vdash_{\mathsf S} \len{0} \lvadd \len{1} \xlongrightarrow{\lb{}} \lenv{0} \lvadd_{\lsv{}} \lenv{1}\glresulttype{\lsv{}}$ & \texttt{\footnotesize GLang/Adequacy/Strengthened.lean:383}\\
\textsc{S-E-Dsum\allowbreak E} & $\lDelta{};\evalCtxSpace{};\evalCtx{}\vdash_{\mathsf S} \len{0}\losum \len{1} \xlongrightarrow{\lbn{0} \losum \lbn{1}} \lenv{0}\losum \lenv{1}\glresulttype{\lsnv{0} \losum \lsnv{1}}$ & \texttt{\footnotesize GLang/Adequacy/Strengthened.lean:397}\\
\textsc{S-E-Tprod\allowbreak E} & ${\lDelta{}_{0}} \ldisun {\lDelta{}_{1}};\evalCtxSpace{};\evalCtx{}\vdash_{\mathsf S} \len{0}\loprod \len{1} \xlongrightarrow{\lbn{0} \loprod \lbn{1}} \lenv{0}\loprod \lenv{1}\glresulttype{\lsnv{0} \loprod \lsnv{1}}$ & \texttt{\footnotesize GLang/Adequacy/Strengthened.lean:407}\\
\textsc{S-E-Pinv\allowbreak TStd} & $\lemptyset{};\evalCtxSpace{};\evalCtx{}\vdash_{\mathsf S} \lee{}^{\lpseudo} \xlongrightarrow{\lbn{0}^{\lstar} \loprod \lbn{1}} \leev{}^{+}\glresulttype{\lsnv{0} \loprod \lsnv{1}}$ & \texttt{\footnotesize GLang/Adequacy/Strengthened.lean:419}\\
\textsc{S-E-Pinv\allowbreak TGpinv} & $\lemptyset{};\evalCtxSpace{};\evalCtx{}\vdash_{\mathsf S} \lee{}^{\lpseudo} \xlongrightarrow{\lbn{0}^{\lstar} \loprod \lbn{1}} \glvop{GPINV}(\leev{},\lenv{1},\lenv{0})\glresulttype{\lsnv{0} \loprod \lsnv{1}}$ & \texttt{\footnotesize GLang/Adequacy/Strengthened.lean:429}\\
\textsc{S-E-Pinv\allowbreak DStd} & $\lemptyset{};\evalCtxSpace{};\evalCtx{}\vdash_{\mathsf S} \lee{}^{\lpseudo} \xlongrightarrow{\lbn{0}^{\lstar} \loprod \lbn{1}} \glvreshape{\leev{}}{\lsnv{1} \loprod \lR{}}{\lsnv{1}}^{+}\glresulttype{\lR{} \loprod \lsnv{1}}$ & \texttt{\footnotesize GLang/Adequacy/Strengthened.lean:443}\\
\textsc{S-E-Pinv\allowbreak DGpinv} & $\lemptyset{};\evalCtxSpace{};\evalCtx{}\vdash_{\mathsf S} \lee{}^{\lpseudo} \xlongrightarrow{\lbn{0}^{\lstar} \loprod \lbn{1}} \glvop{GPINV}(\glvreshape{\leev{}}{\lsnv{1} \loprod \lR{}}{\lsnv{1}},\lenv{1},\lenv{0})\glresulttype{\lR{} \loprod \lsnv{1}}$ & \texttt{\footnotesize GLang/Adequacy/Strengthened.lean:455}\\
\textsc{S-E-Inner\allowbreak Std} & $\lemptyset{};\evalCtxSpace{};\evalCtx{}\vdash_{\mathsf S} \linner{}_{\ls{}} \xlongrightarrow{\lbn{0}^{\lstar} \loprod \lbn{0}^{\lstar}} \lidelta{\lsv{}}\glresulttype{\lsv{} \loprod \lsv{}}$ & \texttt{\footnotesize GLang/Adequacy/Strengthened.lean:472}\\
\textsc{S-E-Embed\allowbreak Ker} & $\lemptyset{};\evalCtxSpace{};\evalCtx{}\vdash_{\mathsf S} \lembd{}_{\Kernel(\lee{})} \xlongrightarrow{\pBasisC{\Kernel(\lee{})}{\lbn{0}}{\lbn{1}}^{\lstar} \loprod \lbn{0}} \Trans(\glvop{GSVDUI}(\leev{},\lenv{0},\lenv{1}))\glresulttype{\lR^{\glvreshape{\leev{}}{\lsnv{0} \loprod \lsnv{1}}{\lsnv{0} \loprod \lsnv{1}},\Kernel{},\lsnv{0} \loprod \lsnv{1}} \loprod \lsnv{0}}$ & \texttt{\footnotesize GLang/Adequacy/Strengthened.lean:482}\\
\textsc{S-E-Embed\allowbreak Coker} & $\lemptyset{};\evalCtxSpace{};\evalCtx{}\vdash_{\mathsf S} \lembd{}_{\CoKernel(\lee{})} \xlongrightarrow{\pBasisC{\CoKernel(\lee{})}{\lbn{0}}{\lbn{1}}^{\lstar} \loprod \lbn{0}} \Trans(\glvop{GSVDU}(\leev{},\lenv{0},\lenv{1}))\glresulttype{\lR^{\glvreshape{\leev{}}{\lsnv{0} \loprod \lsnv{1}}{\lsnv{0} \loprod \lsnv{1}},\CoKernel{},\lsnv{0} \loprod \lsnv{1}} \loprod \lsnv{0}}$ & \texttt{\footnotesize GLang/Adequacy/Strengthened.lean:503}\\
\textsc{S-E-Embed\allowbreak Img} & $\lemptyset{};\evalCtxSpace{};\evalCtx{}\vdash_{\mathsf S} \lembd{}_{\Image(\lee{})} \xlongrightarrow{\pBasisC{\Image(\lee{})}{\lbn{0}}{\lbn{1}}^{\lstar} \loprod \lbn{1}} \Trans(\glvop{GSVDV}(\leev{},\lenv{0},\lenv{1}))\glresulttype{\lR^{\glvreshape{\leev{}}{\lsnv{0} \loprod \lsnv{1}}{\lsnv{0} \loprod \lsnv{1}},\Image{},\lsnv{0} \loprod \lsnv{1}} \loprod \lsnv{1}}$ & \texttt{\footnotesize GLang/Adequacy/Strengthened.lean:523}\\
\textsc{S-E-Embed\allowbreak Coimg} & $\lemptyset{};\evalCtxSpace{};\evalCtx{}\vdash_{\mathsf S} \lembd{}_{\CoImage(\lee{})} \xlongrightarrow{\pBasisC{\CoImage(\lee{})}{\lbn{0}}{\lbn{1}}^{\lstar} \loprod \lbn{1}} \Trans(\glvop{GSVDVI}(\leev{},\lenv{0},\lenv{1}))\glresulttype{\lR^{\glvreshape{\leev{}}{\lsnv{0} \loprod \lsnv{1}}{\lsnv{0} \loprod \lsnv{1}},\CoImage{},\lsnv{0} \loprod \lsnv{1}} \loprod \lsnv{1}}$ & \texttt{\footnotesize GLang/Adequacy/Strengthened.lean:543}\\
\textsc{S-E-App\allowbreak LLin} & ${\lDelta{}_{0}} \ldisun {\lDelta{}_{1}};\evalCtxSpace{};\evalCtx{}\vdash_{\mathsf S} \leval{\len{0}}{}{\len{1}} \xlongrightarrow{\lbn{1}} \glvreshape{\LTTVPP{\lsv{}}{\lenv{0}}{\lenv{1}}}{\lsnv{{\mathrm{svOut}}}}{\lsnv{{\mathrm{svOut}}}}\glresulttype{\lsnv{{\mathrm{svOut}}}}$ & \texttt{\footnotesize GLang/Adequacy/Strengthened.lean:563}\\
\textsc{S-E-App\allowbreak ILin} & ${\lDelta{}_{0}} \ldisun {\lDelta{}_{1}};\evalCtxSpace{};\evalCtx{}\vdash_{\mathsf S} \leval{\len{0}}{}{\len{1}} \xlongrightarrow{\lbone{}} \LTTVPP{\lsv{}}{\glvreshape{\lenv{0}}{\lsv{} \loprod \lR{}}{\lsv{}}}{\lenv{1}}\glresulttype{\lR{}}$ & \texttt{\footnotesize GLang/Adequacy/Strengthened.lean:583}\\
\textsc{S-E-Nllam} & $\lemptyset{};\evalCtxSpace{};\evalCtx{}\vdash_{\mathsf S} (\nllambda \llpl x\colon \ltyn{0}\llpr \ldotsp \lee{}) \xlongrightarrow{\frameB{x}{\lbn{0}}{\lbn{1}}} \llambda(x\colon \ltynv{0})\lrh \ltynv{1}\colon \glvreshape{\leev{}}{\ltynv{1}}{\ltynv{1}}\glresulttype{(x\colon \ltynv{0}\fnto \ltynv{1})}$ & \texttt{\footnotesize GLang/Adequacy/Strengthened.lean:594}\\
\textsc{S-E-App\allowbreak NVal} & $\lemptyset{};\evalCtxSpace{};\evalCtx{}\vdash_{\mathsf S} \leval{\len{0}}{}{\len{1}} \xlongrightarrow{\lbn{1}\subst{x}{\len{1}}} \glvreshape{\glvapp{\lenv{0}}{\lenv{2}}}{{\ltyv{}_{\mathsf{out}}}}{\ltynv{{\mathrm{C}}}\subst{x}{\lenv{2}}}\glresulttype{{\ltyv{}_{\mathsf{out}}}}$ & \texttt{\footnotesize GLang/Adequacy/Strengthened.lean:612}\\
\textsc{S-E-Frame\allowbreak Reanchor} & $\lemptyset{};\evalCtxSpace{};\evalCtx{}\vdash_{\mathsf S} \len{0} \xlongrightarrow{\frameB{x}{\lbn{0}}{\lbn{1}}} \llambda(x\colon \ltynv{0})\lrh \ltynv{1}'\colon \glvreshape{\glvapp{\leev{}}{\lenv{2}}}{\ltynv{1}'}{\ltynv{1}'}\glresulttype{(x\colon \ltynv{0}\fnto \ltynv{1}')}$ & \texttt{\footnotesize GLang/Adequacy/Strengthened.lean:627}\\
\textsc{S-E-Frame\allowbreak Eta} & $\lemptyset{};\evalCtxSpace{};\evalCtx{}\vdash_{\mathsf S} \len{0} \xlongrightarrow{\frameB{x}{\lbn{0}}{\lbn{1}}} \llambda(x\colon \ltynv{0})\lrh \ltynv{1}\colon \glvreshape{\leev{}}{\ltynv{1}}{\ltynv{1}}\glresulttype{(x\colon \ltynv{0}\fnto \ltynv{1})}$ & \texttt{\footnotesize GLang/Adequacy/Strengthened.lean:652}\\
\textsc{S-E-Llam\allowbreak Assemble} & $\lDelta{};\evalCtxSpace{};\evalCtx{}\vdash_{\mathsf S} (\lllambda \llpl x\colon \sof{\lb{}'}\llpr \ldotsp \lee{}) \xlongrightarrow{\lb{}'^{\lstar} \loprod \lb{}} \glvreshape{\lsum^{\lsv{}'}_{\mathit{i}<\overline{\operatorname{dimOf}}(\lsv{}')}\left(\ldelta{\mathit{i}}{\lsv{}'}\loprod \glvreshape{\leev{}_{\mathit{i}}}{\lsv{}}{\lsv{}}\right)}{\lsv{}' \loprod \lsv{}}{\begin{cases}\lsempty{}&\overline{\operatorname{dimOf}}(\lsv{}')=0\\ \lsv{}' \loprod \lsv{}&\text{otherwise}\end{cases}}\glresulttype{\lsv{}' \loprod \lsv{}}$ & \texttt{\footnotesize GLang/Adequacy/Strengthened.lean:671}\\
\textsc{S-E-Llam\allowbreak Assemble\allowbreak Real} & $\lDelta{};\evalCtxSpace{};\evalCtx{}\vdash_{\mathsf S} (\lllambda \llpl x\colon \sof{\lb{}'}\llpr \ldotsp \lee{}) \xlongrightarrow{\lb{}'^{\lstar}} \glvreshape{\lsum^{\lsv{}'}_{\mathit{i}<\overline{\operatorname{dimOf}}(\lsv{}')}\left(\glvreshape{\leev{}_{\mathit{i}}}{\lR{}}{\lR{}} \lvscale \ldelta{\mathit{i}}{\lsv{}'}\right)}{\lsv{}'}{\begin{cases}\lsempty{}&\overline{\operatorname{dimOf}}(\lsv{}')=0\\ \lsv{}'&\text{otherwise}\end{cases}}\glresulttype{\lsv{}'}$ & \texttt{\footnotesize GLang/Adequacy/Strengthened.lean:701}\\
\textsc{S-E-Zero} & $\lDelta{};\evalCtxSpace{};\evalCtx{}\vdash_{\mathsf S} \lvzero{}_{\sof{\lb{}}} \xlongrightarrow{\lb{}} \lvzero{}_{\lsv{}}\glresulttype{\lsv{}}$ & \texttt{\footnotesize GLang/Adequacy/Strengthened.lean:720}\\
\textsc{S-E-Rlit} & $\lemptyset{};\evalCtxSpace{};\evalCtx{}\vdash_{\mathsf S} \mathit{r} \xlongrightarrow{\lbone{}} \mathit{r}\glresulttype{\lR{}}$ & \texttt{\footnotesize GLang/Adequacy/Strengthened.lean:728}\\
}{
\gpfit{\inferrule*[]{((\mathit{x}, \lb{}) \mapsto \leev{}) \in \evalCtx{} \\ \tof{\lb{}} = \ls{} \\ (\emptyset,\evalCtxSpace{})\vdash \lb{}\colon \bvalid{} \\ \evalCtxSpace{};\evalCtx{}\vdash_{\mathsf S} \lb{} \evalArrow{}_{\mathsf{basis}} \ltyv{}}{\{(\mathit{x}, \ls{})\};\evalCtxSpace{};\evalCtx{}\vdash_{\mathsf S} \mathit{x} \xlongrightarrow{\lb{}} \leev{}\glresulttype{\ltyv{}}}}
\and
\gpfit{\inferrule*[]{((\mathit{x}, \lb{}) \mapsto \leev{}) \in \evalCtx{} \\ (\mathit{x}\colon \tof{\lb{}}) \in \evalCtxSpace{} \\ (\emptyset,\evalCtxSpace{})\vdash \lb{}\colon \bvalid{} \\ \evalCtxSpace{};\evalCtx{}\vdash_{\mathsf S} \lb{} \evalArrow{}_{\mathsf{basis}} \ltyv{}}{\lemptyset{};\evalCtxSpace{};\evalCtx{}\vdash_{\mathsf S} \mathit{x} \xlongrightarrow{\lb{}} \leev{}\glresulttype{\ltyv{}}}}
\and
\gpfit{\inferrule*[]{\forall y \in \mathit{lin}.\; y \notin \bigcup_{(\cdot\colon s) \in \lDelta{},\, \mathit{lin}} \mathrm{fv}(s) \;\cup\; \bigcup_{(\cdot\colon \tau) \in \evalCtxSpace{}'} \mathrm{fv}(\tau) \\ \evalCtxSpace{}' \subseteq \evalCtxSpace{}'' \\ (\lDelta{}, \mathit{lin} \ldisun \evalCtxSpace{}'')\vdash \lee{}\colon \tof{\lb{}} \\ (\emptyset,\mathit{lin} \ldisun \evalCtxSpace{}'')\vdash \lb{}\colon \bvalid{} \\ \mathit{lin} \ldisun \evalCtxSpace{}'';\evalCtx{}\vdash_{\mathsf S} \lb{} \evalArrow{}_{\mathsf{basis}} \ltyv{} \\ \mathit{lin} \ldisun \lDelta{};\evalCtxSpace{}';\evalCtx{}\vdash_{\mathsf S} \lee{} \xlongrightarrow{\lb{}} \leev{}\glresulttype{\ltyv{}}}{\lDelta{};\mathit{lin} \ldisun \evalCtxSpace{}'';\evalCtx{}\vdash_{\mathsf S} \lee{} \xlongrightarrow{\lb{}} \leev{}\glresulttype{\ltyv{}}}}
\and
\gpfit{\inferrule*[]{\forall i.\; \tof{\lbn{i}} = \sof{\lbn{i}} \\ (\lDelta{}, \evalCtxSpace{})\vdash \lee{}\colon \tof{\lbn{0}} \\ (\emptyset,\evalCtxSpace{})\vdash \lbn{0}\colon \bvalid{} \\ \evalCtxSpace{};\evalCtx{}\vdash_{\mathsf S} \lbn{0} \evalArrow{}_{\mathsf{basis}} \lsv{} \\ \lDelta{};\evalCtxSpace{};\evalCtx{}\vdash_{\mathsf S} \lee{} \xlongrightarrow{\lbn{1}} \leev{}\glresulttype{\ltyv{}} \\ \evalCtxSpace{};\evalCtx{}\vdash_{\mathsf S} \lbn{1} \xlongleftrightarrow{\lenv{0}} \lbn{0} \\ \evalCtxSpace{};\evalCtx{}\vdash_{\mathsf S} \sof{\lbn{0}} \evalArrow{} \lsv{}}{\lDelta{};\evalCtxSpace{};\evalCtx{}\vdash_{\mathsf S} \lee{} \xlongrightarrow{\lbn{0}} \LTTVPP{\lsv{}}{\glvreshape{\lenv{0}}{\lsv{} \loprod \lsv{}}{\lsv{} \loprod \lsv{}}}{\glvreshape{\leev{}}{\lsv{}}{\lsv{}}}\glresulttype{\lsv{}}}}
\and
\gpfit{\inferrule*[]{(\lDelta{}, \evalCtxSpace{})\vdash \lee{}\colon \tof{\lb{}} \\ (\emptyset,\evalCtxSpace{})\vdash \lb{}\colon \bvalid{} \\ \evalCtxSpace{};\evalCtx{}\vdash_{\mathsf S} \lb{} \evalArrow{}_{\mathsf{basis}} \ltyv{}' \\ \lDelta{};\evalCtxSpace{};\evalCtx{}\vdash_{\mathsf S} \lee{} \xlongrightarrow{\lb{}} \leev{}\glresulttype{\ltyv{}}}{\lDelta{};\evalCtxSpace{};\evalCtx{}\vdash_{\mathsf S} \lee{} \xlongrightarrow{\lb{}} \glvreshape{\leev{}}{\ltyv{}'}{\ltyv{}}\glresulttype{\ltyv{}'}}}
\and
\gpfit{\inferrule*[]{\tof{\lb{}} = \sof{\lb{}} \\ ({\lDelta{}_{0}} \ldisun {\lDelta{}_{1}}, \evalCtxSpace{})\vdash \len{0} \lvscale \len{1}\colon \tof{\lb{}} \\ (\emptyset,\evalCtxSpace{})\vdash \lb{}\colon \bvalid{} \\ \evalCtxSpace{};\evalCtx{}\vdash_{\mathsf S} \lb{} \evalArrow{}_{\mathsf{basis}} \lsv{} \\ {\lDelta{}_{0}};\evalCtxSpace{};\evalCtx{}\vdash_{\mathsf S} \len{0} \xlongrightarrow{\lbone{}} \lenv{0}\glresulttype{\lR{}} \\ {\lDelta{}_{1}};\evalCtxSpace{};\evalCtx{}\vdash_{\mathsf S} \len{1} \xlongrightarrow{\lb{}} \lenv{1}\glresulttype{\lsv{}}}{{\lDelta{}_{0}} \ldisun {\lDelta{}_{1}};\evalCtxSpace{};\evalCtx{}\vdash_{\mathsf S} \len{0} \lvscale \len{1} \xlongrightarrow{\lb{}} \lenv{0} \lvscale \lenv{1}\glresulttype{\lsv{}}}}
\and
\gpfit{\inferrule*[]{\tof{\lb{}} = \sof{\lb{}} \\ (\lDelta{}, \evalCtxSpace{})\vdash \len{0} \lvadd \len{1}\colon \tof{\lb{}} \\ (\emptyset,\evalCtxSpace{})\vdash \lb{}\colon \bvalid{} \\ \evalCtxSpace{};\evalCtx{}\vdash_{\mathsf S} \lb{} \evalArrow{}_{\mathsf{basis}} \lsv{} \\ \forall i.\; \lDelta{};\evalCtxSpace{};\evalCtx{}\vdash_{\mathsf S} \len{i} \xlongrightarrow{\lb{}} \lenv{i}\glresulttype{\lsnv{i}}}{\lDelta{};\evalCtxSpace{};\evalCtx{}\vdash_{\mathsf S} \len{0} \lvadd \len{1} \xlongrightarrow{\lb{}} \lenv{0} \lvadd_{\lsv{}} \lenv{1}\glresulttype{\lsv{}}}}
\and
\gpfit{\inferrule*[]{\forall i.\; \tof{\lbn{i}} = \sof{\lbn{i}} \\ (\lDelta{}, \evalCtxSpace{})\vdash \len{0}\losum \len{1}\colon \tof{\lbn{0} \losum \lbn{1}} \\ (\emptyset,\evalCtxSpace{})\vdash \lbn{0} \losum \lbn{1}\colon \bvalid{} \\ \evalCtxSpace{};\evalCtx{}\vdash_{\mathsf S} \lbn{0} \losum \lbn{1} \evalArrow{}_{\mathsf{basis}} \lsnv{0} \losum \lsnv{1} \\ \forall i.\; \lDelta{};\evalCtxSpace{};\evalCtx{}\vdash_{\mathsf S} \len{i} \xlongrightarrow{\lbn{i}} \lenv{i}\glresulttype{\lsnv{i}}}{\lDelta{};\evalCtxSpace{};\evalCtx{}\vdash_{\mathsf S} \len{0}\losum \len{1} \xlongrightarrow{\lbn{0} \losum \lbn{1}} \lenv{0}\losum \lenv{1}\glresulttype{\lsnv{0} \losum \lsnv{1}}}}
\and
\gpfit{\inferrule*[]{\forall i.\; \tof{\lbn{i}} = \sof{\lbn{i}} \\ ({\lDelta{}_{0}} \ldisun {\lDelta{}_{1}}, \evalCtxSpace{})\vdash \len{0}\loprod \len{1}\colon \tof{\lbn{0} \loprod \lbn{1}} \\ (\emptyset,\evalCtxSpace{})\vdash \lbn{0} \loprod \lbn{1}\colon \bvalid{} \\ \evalCtxSpace{};\evalCtx{}\vdash_{\mathsf S} \lbn{0} \loprod \lbn{1} \evalArrow{}_{\mathsf{basis}} \lsnv{0} \loprod \lsnv{1} \\ \forall i.\; {\lDelta{}_{i}};\evalCtxSpace{};\evalCtx{}\vdash_{\mathsf S} \len{i} \xlongrightarrow{\lbn{i}} \lenv{i}\glresulttype{\lsnv{i}}}{{\lDelta{}_{0}} \ldisun {\lDelta{}_{1}};\evalCtxSpace{};\evalCtx{}\vdash_{\mathsf S} \len{0}\loprod \len{1} \xlongrightarrow{\lbn{0} \loprod \lbn{1}} \lenv{0}\loprod \lenv{1}\glresulttype{\lsnv{0} \loprod \lsnv{1}}}}
\and
\gpfit{\inferrule*[]{\forall i.\; \isStdBasis{\lbn{i}} \\ (\lemptyset{}, \evalCtxSpace{})\vdash \lee{}^{\lpseudo}\colon \tof{\lbn{0}^{\lstar} \loprod \lbn{1}} \\ (\emptyset,\evalCtxSpace{})\vdash \lbn{0}^{\lstar} \loprod \lbn{1}\colon \bvalid{} \\ \evalCtxSpace{};\evalCtx{}\vdash_{\mathsf S} \lbn{0}^{\lstar} \loprod \lbn{1} \evalArrow{}_{\mathsf{basis}} \lsnv{0} \loprod \lsnv{1} \\ \lemptyset{};\evalCtxSpace{};\evalCtx{}\vdash_{\mathsf S} \lee{} \xlongrightarrow{\lbn{1}^{\lstar} \loprod \lbn{0}} \leev{}\glresulttype{\lsnv{1} \loprod \lsnv{0}}}{\lemptyset{};\evalCtxSpace{};\evalCtx{}\vdash_{\mathsf S} \lee{}^{\lpseudo} \xlongrightarrow{\lbn{0}^{\lstar} \loprod \lbn{1}} \leev{}^{+}\glresulttype{\lsnv{0} \loprod \lsnv{1}}}}
\and
\gpfit{\inferrule*[]{(\lemptyset{}, \evalCtxSpace{})\vdash \lee{}^{\lpseudo}\colon \tof{\lbn{0}^{\lstar} \loprod \lbn{1}} \\ (\emptyset,\evalCtxSpace{})\vdash \lbn{0}^{\lstar} \loprod \lbn{1}\colon \bvalid{} \\ \evalCtxSpace{};\evalCtx{}\vdash_{\mathsf S} \lbn{0}^{\lstar} \loprod \lbn{1} \evalArrow{}_{\mathsf{basis}} \lsnv{0} \loprod \lsnv{1} \\ \lemptyset{};\evalCtxSpace{};\evalCtx{}\vdash_{\mathsf S} \lee{} \xlongrightarrow{\lbn{1}^{\lstar} \loprod \lbn{0}} \leev{}\glresulttype{\lsnv{1} \loprod \lsnv{0}} \\ \forall i.\; \lemptyset{};\evalCtxSpace{};\evalCtx{}\vdash_{\mathsf S} \linner{}_{\sof{\lbn{i}}} \xlongrightarrow{\lbn{i}^{\lstar} \loprod \lbn{i}^{\lstar}} \lenv{i}\glresulttype{\lsnv{i} \loprod \lsnv{i}}}{\lemptyset{};\evalCtxSpace{};\evalCtx{}\vdash_{\mathsf S} \lee{}^{\lpseudo} \xlongrightarrow{\lbn{0}^{\lstar} \loprod \lbn{1}} \glvop{GPINV}(\leev{},\lenv{1},\lenv{0})\glresulttype{\lsnv{0} \loprod \lsnv{1}}}}
\and
\gpfit{\inferrule*[]{\sof{\lbn{0}} = \lR{} \\ \forall i.\; \isStdBasis{\lbn{i}} \\ (\lemptyset{}, \evalCtxSpace{})\vdash \lee{}^{\lpseudo}\colon \tof{\lbn{0}^{\lstar} \loprod \lbn{1}} \\ (\emptyset,\evalCtxSpace{})\vdash \lbn{0}^{\lstar} \loprod \lbn{1}\colon \bvalid{} \\ \evalCtxSpace{};\evalCtx{}\vdash_{\mathsf S} \lbn{0}^{\lstar} \loprod \lbn{1} \evalArrow{}_{\mathsf{basis}} \lR{} \loprod \lsnv{1} \\ \lemptyset{};\evalCtxSpace{};\evalCtx{}\vdash_{\mathsf S} \lee{} \xlongrightarrow{\lbn{1}^{\lstar}} \leev{}\glresulttype{\lsnv{1}}}{\lemptyset{};\evalCtxSpace{};\evalCtx{}\vdash_{\mathsf S} \lee{}^{\lpseudo} \xlongrightarrow{\lbn{0}^{\lstar} \loprod \lbn{1}} \glvreshape{\leev{}}{\lsnv{1} \loprod \lR{}}{\lsnv{1}}^{+}\glresulttype{\lR{} \loprod \lsnv{1}}}}
\and
\gpfit{\inferrule*[]{\sof{\lbn{0}} = \lR{} \\ \isStdBasis{\lbn{0}} \\ (\lemptyset{}, \evalCtxSpace{})\vdash \lee{}^{\lpseudo}\colon \tof{\lbn{0}^{\lstar} \loprod \lbn{1}} \\ (\emptyset,\evalCtxSpace{})\vdash \lbn{0}^{\lstar} \loprod \lbn{1}\colon \bvalid{} \\ \evalCtxSpace{};\evalCtx{}\vdash_{\mathsf S} \lbn{0}^{\lstar} \loprod \lbn{1} \evalArrow{}_{\mathsf{basis}} \lR{} \loprod \lsnv{1} \\ \lemptyset{};\evalCtxSpace{};\evalCtx{}\vdash_{\mathsf S} \lee{} \xlongrightarrow{\lbn{1}^{\lstar}} \leev{}\glresulttype{\lsnv{1}} \\ \lemptyset{};\evalCtxSpace{};\evalCtx{}\vdash_{\mathsf S} \linner{}_{\sof{\lbn{0}}} \xlongrightarrow{\lbn{0}^{\lstar} \loprod \lbn{0}^{\lstar}} \lenv{0}\glresulttype{\lR{} \loprod \lR{}} \\ \lemptyset{};\evalCtxSpace{};\evalCtx{}\vdash_{\mathsf S} \linner{}_{\sof{\lbn{1}}} \xlongrightarrow{\lbn{1}^{\lstar} \loprod \lbn{1}^{\lstar}} \lenv{1}\glresulttype{\lsnv{1} \loprod \lsnv{1}}}{\lemptyset{};\evalCtxSpace{};\evalCtx{}\vdash_{\mathsf S} \lee{}^{\lpseudo} \xlongrightarrow{\lbn{0}^{\lstar} \loprod \lbn{1}} \glvop{GPINV}(\glvreshape{\leev{}}{\lsnv{1} \loprod \lR{}}{\lsnv{1}},\lenv{1},\lenv{0})\glresulttype{\lR{} \loprod \lsnv{1}}}}
\and
\gpfit{\inferrule*[]{\isStdBasis{\lbn{0}} \\ \sof{\lbn{0}} = \ls{} \\ (\lemptyset{}, \evalCtxSpace{})\vdash \linner{}_{\ls{}}\colon \tof{\lbn{0}^{\lstar} \loprod \lbn{0}^{\lstar}} \\ (\emptyset,\evalCtxSpace{})\vdash \lbn{0}^{\lstar} \loprod \lbn{0}^{\lstar}\colon \bvalid{} \\ \evalCtxSpace{};\evalCtx{}\vdash_{\mathsf S} \lbn{0}^{\lstar} \loprod \lbn{0}^{\lstar} \evalArrow{}_{\mathsf{basis}} \lsv{} \loprod \lsv{} \\ \evalCtxSpace{};\evalCtx{}\vdash_{\mathsf S} \ls{} \evalArrow{} \lsv{}}{\lemptyset{};\evalCtxSpace{};\evalCtx{}\vdash_{\mathsf S} \linner{}_{\ls{}} \xlongrightarrow{\lbn{0}^{\lstar} \loprod \lbn{0}^{\lstar}} \lidelta{\lsv{}}\glresulttype{\lsv{} \loprod \lsv{}}}}
\and
\gpfit{\inferrule*[]{\forall i.\; \sof{\lbn{i}} = \lsn{i} \\ (\lemptyset{}, \evalCtxSpace{})\vdash \lembd{}_{\Kernel(\lee{})}\colon \tof{\pBasisC{\Kernel(\lee{})}{\lbn{0}}{\lbn{1}}^{\lstar} \loprod \lbn{0}} \\ (\emptyset,\evalCtxSpace{})\vdash \pBasisC{\Kernel(\lee{})}{\lbn{0}}{\lbn{1}}^{\lstar} \loprod \lbn{0}\colon \bvalid{} \\ \evalCtxSpace{};\evalCtx{}\vdash_{\mathsf S} \pBasisC{\Kernel(\lee{})}{\lbn{0}}{\lbn{1}}^{\lstar} \loprod \lbn{0} \evalArrow{}_{\mathsf{basis}} \lR^{\glvreshape{\leev{}}{\lsnv{0} \loprod \lsnv{1}}{\lsnv{0} \loprod \lsnv{1}},\Kernel{},\lsnv{0} \loprod \lsnv{1}} \loprod \lsnv{0} \\ \forall i.\; \lemptyset{};\evalCtxSpace{};\evalCtx{}\vdash_{\mathsf S} \linner{}_{\lsn{i}} \xlongrightarrow{\lbn{i}^{\lstar} \loprod \lbn{i}^{\lstar}} \lenv{i}\glresulttype{\lsnv{i} \loprod \lsnv{i}} \\ \lemptyset{};\evalCtxSpace{};\evalCtx{}\vdash_{\mathsf S} \lee{} \xlongrightarrow{\lbn{0}^{\lstar} \loprod \lbn{1}} \leev{}\glresulttype{\lsnv{0} \loprod \lsnv{1}}}{\lemptyset{};\evalCtxSpace{};\evalCtx{}\vdash_{\mathsf S} \lembd{}_{\Kernel(\lee{})} \xlongrightarrow{\pBasisC{\Kernel(\lee{})}{\lbn{0}}{\lbn{1}}^{\lstar} \loprod \lbn{0}} \Trans(\glvop{GSVDUI}(\leev{},\lenv{0},\lenv{1}))\glresulttype{\lR^{\glvreshape{\leev{}}{\lsnv{0} \loprod \lsnv{1}}{\lsnv{0} \loprod \lsnv{1}},\Kernel{},\lsnv{0} \loprod \lsnv{1}} \loprod \lsnv{0}}}}
\and
\gpfit{\inferrule*[]{\forall i.\; \sof{\lbn{i}} = \lsn{i} \\ (\lemptyset{}, \evalCtxSpace{})\vdash \lembd{}_{\CoKernel(\lee{})}\colon \tof{\pBasisC{\CoKernel(\lee{})}{\lbn{0}}{\lbn{1}}^{\lstar} \loprod \lbn{0}} \\ (\emptyset,\evalCtxSpace{})\vdash \pBasisC{\CoKernel(\lee{})}{\lbn{0}}{\lbn{1}}^{\lstar} \loprod \lbn{0}\colon \bvalid{} \\ \evalCtxSpace{};\evalCtx{}\vdash_{\mathsf S} \pBasisC{\CoKernel(\lee{})}{\lbn{0}}{\lbn{1}}^{\lstar} \loprod \lbn{0} \evalArrow{}_{\mathsf{basis}} \lR^{\glvreshape{\leev{}}{\lsnv{0} \loprod \lsnv{1}}{\lsnv{0} \loprod \lsnv{1}},\CoKernel{},\lsnv{0} \loprod \lsnv{1}} \loprod \lsnv{0} \\ \forall i.\; \lemptyset{};\evalCtxSpace{};\evalCtx{}\vdash_{\mathsf S} \linner{}_{\lsn{i}} \xlongrightarrow{\lbn{i}^{\lstar} \loprod \lbn{i}^{\lstar}} \lenv{i}\glresulttype{\lsnv{i} \loprod \lsnv{i}} \\ \lemptyset{};\evalCtxSpace{};\evalCtx{}\vdash_{\mathsf S} \lee{} \xlongrightarrow{\lbn{0}^{\lstar} \loprod \lbn{1}} \leev{}\glresulttype{\lsnv{0} \loprod \lsnv{1}}}{\lemptyset{};\evalCtxSpace{};\evalCtx{}\vdash_{\mathsf S} \lembd{}_{\CoKernel(\lee{})} \xlongrightarrow{\pBasisC{\CoKernel(\lee{})}{\lbn{0}}{\lbn{1}}^{\lstar} \loprod \lbn{0}} \Trans(\glvop{GSVDU}(\leev{},\lenv{0},\lenv{1}))\glresulttype{\lR^{\glvreshape{\leev{}}{\lsnv{0} \loprod \lsnv{1}}{\lsnv{0} \loprod \lsnv{1}},\CoKernel{},\lsnv{0} \loprod \lsnv{1}} \loprod \lsnv{0}}}}
\and
\gpfit{\inferrule*[]{\forall i.\; \sof{\lbn{i}} = \lsn{i} \\ (\lemptyset{}, \evalCtxSpace{})\vdash \lembd{}_{\Image(\lee{})}\colon \tof{\pBasisC{\Image(\lee{})}{\lbn{0}}{\lbn{1}}^{\lstar} \loprod \lbn{1}} \\ (\emptyset,\evalCtxSpace{})\vdash \pBasisC{\Image(\lee{})}{\lbn{0}}{\lbn{1}}^{\lstar} \loprod \lbn{1}\colon \bvalid{} \\ \evalCtxSpace{};\evalCtx{}\vdash_{\mathsf S} \pBasisC{\Image(\lee{})}{\lbn{0}}{\lbn{1}}^{\lstar} \loprod \lbn{1} \evalArrow{}_{\mathsf{basis}} \lR^{\glvreshape{\leev{}}{\lsnv{0} \loprod \lsnv{1}}{\lsnv{0} \loprod \lsnv{1}},\Image{},\lsnv{0} \loprod \lsnv{1}} \loprod \lsnv{1} \\ \forall i.\; \lemptyset{};\evalCtxSpace{};\evalCtx{}\vdash_{\mathsf S} \linner{}_{\lsn{i}} \xlongrightarrow{\lbn{i}^{\lstar} \loprod \lbn{i}^{\lstar}} \lenv{i}\glresulttype{\lsnv{i} \loprod \lsnv{i}} \\ \lemptyset{};\evalCtxSpace{};\evalCtx{}\vdash_{\mathsf S} \lee{} \xlongrightarrow{\lbn{0}^{\lstar} \loprod \lbn{1}} \leev{}\glresulttype{\lsnv{0} \loprod \lsnv{1}}}{\lemptyset{};\evalCtxSpace{};\evalCtx{}\vdash_{\mathsf S} \lembd{}_{\Image(\lee{})} \xlongrightarrow{\pBasisC{\Image(\lee{})}{\lbn{0}}{\lbn{1}}^{\lstar} \loprod \lbn{1}} \Trans(\glvop{GSVDV}(\leev{},\lenv{0},\lenv{1}))\glresulttype{\lR^{\glvreshape{\leev{}}{\lsnv{0} \loprod \lsnv{1}}{\lsnv{0} \loprod \lsnv{1}},\Image{},\lsnv{0} \loprod \lsnv{1}} \loprod \lsnv{1}}}}
\and
\gpfit{\inferrule*[]{\forall i.\; \sof{\lbn{i}} = \lsn{i} \\ (\lemptyset{}, \evalCtxSpace{})\vdash \lembd{}_{\CoImage(\lee{})}\colon \tof{\pBasisC{\CoImage(\lee{})}{\lbn{0}}{\lbn{1}}^{\lstar} \loprod \lbn{1}} \\ (\emptyset,\evalCtxSpace{})\vdash \pBasisC{\CoImage(\lee{})}{\lbn{0}}{\lbn{1}}^{\lstar} \loprod \lbn{1}\colon \bvalid{} \\ \evalCtxSpace{};\evalCtx{}\vdash_{\mathsf S} \pBasisC{\CoImage(\lee{})}{\lbn{0}}{\lbn{1}}^{\lstar} \loprod \lbn{1} \evalArrow{}_{\mathsf{basis}} \lR^{\glvreshape{\leev{}}{\lsnv{0} \loprod \lsnv{1}}{\lsnv{0} \loprod \lsnv{1}},\CoImage{},\lsnv{0} \loprod \lsnv{1}} \loprod \lsnv{1} \\ \forall i.\; \lemptyset{};\evalCtxSpace{};\evalCtx{}\vdash_{\mathsf S} \linner{}_{\lsn{i}} \xlongrightarrow{\lbn{i}^{\lstar} \loprod \lbn{i}^{\lstar}} \lenv{i}\glresulttype{\lsnv{i} \loprod \lsnv{i}} \\ \lemptyset{};\evalCtxSpace{};\evalCtx{}\vdash_{\mathsf S} \lee{} \xlongrightarrow{\lbn{0}^{\lstar} \loprod \lbn{1}} \leev{}\glresulttype{\lsnv{0} \loprod \lsnv{1}}}{\lemptyset{};\evalCtxSpace{};\evalCtx{}\vdash_{\mathsf S} \lembd{}_{\CoImage(\lee{})} \xlongrightarrow{\pBasisC{\CoImage(\lee{})}{\lbn{0}}{\lbn{1}}^{\lstar} \loprod \lbn{1}} \Trans(\glvop{GSVDVI}(\leev{},\lenv{0},\lenv{1}))\glresulttype{\lR^{\glvreshape{\leev{}}{\lsnv{0} \loprod \lsnv{1}}{\lsnv{0} \loprod \lsnv{1}},\CoImage{},\lsnv{0} \loprod \lsnv{1}} \loprod \lsnv{1}}}}
\and
\gpfit{\inferrule*[]{\tof{\lbn{1}} = \sof{\lbn{1}} \\ ({\lDelta{}_{0}} \ldisun {\lDelta{}_{1}}, \evalCtxSpace{})\vdash \leval{\len{0}}{}{\len{1}}\colon \tof{\lbn{1}} \\ (\emptyset,\evalCtxSpace{})\vdash \lbn{1}\colon \bvalid{} \\ \evalCtxSpace{};\evalCtx{}\vdash_{\mathsf S} \sof{\lbn{0}} \evalArrow{} \lsv{} \\ \evalCtxSpace{};\evalCtx{}\vdash_{\mathsf S} \lbn{1} \evalArrow{}_{\mathsf{basis}} \lsnv{{\mathrm{svOut}}} \\ {\lDelta{}_{0}};\evalCtxSpace{};\evalCtx{}\vdash_{\mathsf S} \len{0} \xlongrightarrow{\lbn{0}^{\lstar} \loprod \lbn{1}} \lenv{0}\glresulttype{\lsnv{{\mathrm{svMat0}}} \loprod \lsnv{{\mathrm{svMat1}}}} \\ {\lDelta{}_{1}};\evalCtxSpace{};\evalCtx{}\vdash_{\mathsf S} \len{1} \xlongrightarrow{\lbn{0}} \lenv{1}\glresulttype{\lsnv{{\mathrm{svArg}}}}}{{\lDelta{}_{0}} \ldisun {\lDelta{}_{1}};\evalCtxSpace{};\evalCtx{}\vdash_{\mathsf S} \leval{\len{0}}{}{\len{1}} \xlongrightarrow{\lbn{1}} \glvreshape{\LTTVPP{\lsv{}}{\lenv{0}}{\lenv{1}}}{\lsnv{{\mathrm{svOut}}}}{\lsnv{{\mathrm{svOut}}}}\glresulttype{\lsnv{{\mathrm{svOut}}}}}}
\and
\gpfit{\inferrule*[]{({\lDelta{}_{0}} \ldisun {\lDelta{}_{1}}, \evalCtxSpace{})\vdash \leval{\len{0}}{}{\len{1}}\colon \tof{\lbone{}} \\ \evalCtxSpace{};\evalCtx{}\vdash_{\mathsf S} \sof{\lbn{0}} \evalArrow{} \lsv{} \\ \evalCtxSpace{};\evalCtx{}\vdash_{\mathsf S} \lbone{} \evalArrow{}_{\mathsf{basis}} \lR{} \\ {\lDelta{}_{0}};\evalCtxSpace{};\evalCtx{}\vdash_{\mathsf S} \len{0} \xlongrightarrow{\lbn{0}^{\lstar}} \lenv{0}\glresulttype{\lsnv{{\mathrm{svFun}}}} \\ {\lDelta{}_{1}};\evalCtxSpace{};\evalCtx{}\vdash_{\mathsf S} \len{1} \xlongrightarrow{\lbn{0}} \lenv{1}\glresulttype{\lsnv{{\mathrm{svArg}}}}}{{\lDelta{}_{0}} \ldisun {\lDelta{}_{1}};\evalCtxSpace{};\evalCtx{}\vdash_{\mathsf S} \leval{\len{0}}{}{\len{1}} \xlongrightarrow{\lbone{}} \LTTVPP{\lsv{}}{\glvreshape{\lenv{0}}{\lsv{} \loprod \lR{}}{\lsv{}}}{\lenv{1}}\glresulttype{\lR{}}}}
\and
\gpfit{\inferrule*[]{\forall i.\; \tof{\lbn{i}} = \ltyn{i} \\ (\lemptyset{}, \evalCtxSpace{})\vdash (\nllambda \llpl x\colon \ltyn{0}\llpr \ldotsp \lee{})\colon \tof{\frameB{x}{\lbn{0}}{\lbn{1}}} \\ (\emptyset,\evalCtxSpace{})\vdash \frameB{x}{\lbn{0}}{\lbn{1}}\colon \bvalid{} \\ \evalCtxSpace{};\evalCtx{}\vdash_{\mathsf S} \frameB{x}{\lbn{0}}{\lbn{1}} \evalArrow{}_{\mathsf{basis}} (x\colon \ltynv{0}\fnto \ltynv{1}) \\ \evalCtxSpace{};\evalCtx{}\vdash_{\mathsf S} \lbn{0} \evalArrow{}_{\mathsf{basis}} \ltynv{0} \\ \lemptyset{};(\evalCtxSpace{}, \mathit{x}\colon \ltyn{0});\evalCtx{}[(\mathit{x},\lbn{0}) \mapsto \mathit{x}_{\ltynv{0}}]\vdash_{\mathsf S} \lee{} \xlongrightarrow{\lbn{1}} \leev{}\glresulttype{\ltynv{1}} \\ (\evalCtxSpace{}, \mathit{x}\colon \ltyn{0});\evalCtx{}[(\mathit{x},\lbn{0}) \mapsto \mathit{x}_{\ltynv{0}}]\vdash_{\mathsf S} \lbn{1} \evalArrow{}_{\mathsf{basis}} \ltynv{1}}{\lemptyset{};\evalCtxSpace{};\evalCtx{}\vdash_{\mathsf S} (\nllambda \llpl x\colon \ltyn{0}\llpr \ldotsp \lee{}) \xlongrightarrow{\frameB{x}{\lbn{0}}{\lbn{1}}} \llambda(x\colon \ltynv{0})\lrh \ltynv{1}\colon \glvreshape{\leev{}}{\ltynv{1}}{\ltynv{1}}\glresulttype{(x\colon \ltynv{0}\fnto \ltynv{1})}}}
\and
\gpfit{\inferrule*[]{(\lemptyset{}, \evalCtxSpace{})\vdash \leval{\len{0}}{}{\len{1}}\colon \tof{\lbn{1}\subst{x}{\len{1}}} \\ (\emptyset,\evalCtxSpace{})\vdash \lbn{1}\subst{x}{\len{1}}\colon \bvalid{} \\ \evalCtxSpace{};\evalCtx{}\vdash_{\mathsf S} \lbn{1}\subst{x}{\len{1}} \evalArrow{}_{\mathsf{basis}} {\ltyv{}_{\mathsf{out}}} \\ \lemptyset{};\evalCtxSpace{};\evalCtx{}\vdash_{\mathsf S} \len{0} \xlongrightarrow{\frameB{x}{\lbn{0}}{\lbn{1}}} \lenv{0}\glresulttype{(x\colon {\ltyv{}_{\mathsf{fun}0}}\fnto {\ltyv{}_{\mathsf{fun}1}})} \\ \lemptyset{};\evalCtxSpace{};\evalCtx{}\vdash_{\mathsf S} \len{1} \xlongrightarrow{\lbn{0}} \lenv{1}\glresulttype{{\ltyv{}_{\mathsf{arg}}}} \\ \operatorname{inferType}(\lenv{0}) = (x\colon \ltynv{{\mathrm{D}}}\fnto \ltynv{{\mathrm{C}}}) \\ \lenv{2} = \glvreshape{\lenv{1}}{\ltynv{{\mathrm{D}}}}{\operatorname{inferType}(\lenv{1})}}{\lemptyset{};\evalCtxSpace{};\evalCtx{}\vdash_{\mathsf S} \leval{\len{0}}{}{\len{1}} \xlongrightarrow{\lbn{1}\subst{x}{\len{1}}} \glvreshape{\glvapp{\lenv{0}}{\lenv{2}}}{{\ltyv{}_{\mathsf{out}}}}{\ltynv{{\mathrm{C}}}\subst{x}{\lenv{2}}}\glresulttype{{\ltyv{}_{\mathsf{out}}}}}}
\and
\gpfit{\inferrule*[]{\tof{\lbn{0}'} = \tof{\lbn{0}} \\ (\lemptyset{}, \evalCtxSpace{})\vdash \len{0}\colon \tof{\frameB{x}{\lbn{0}}{\lbn{1}}} \\ (\emptyset,\evalCtxSpace{})\vdash \frameB{x}{\lbn{0}}{\lbn{1}}\colon \bvalid{} \\ \evalCtxSpace{};\evalCtx{}\vdash_{\mathsf S} \frameB{x}{\lbn{0}}{\lbn{1}} \evalArrow{}_{\mathsf{basis}} (x\colon \ltynv{0}\fnto \ltynv{1}') \\ \evalCtxSpace{};\evalCtx{}\vdash_{\mathsf S} \lbn{0} \evalArrow{}_{\mathsf{basis}} \ltynv{0} \\ \lemptyset{};\evalCtxSpace{};\evalCtx{}\vdash_{\mathsf S} \len{0} \xlongrightarrow{\frameB{x}{\lbn{0}'}{\lbn{1}}} \leev{}\glresulttype{(x\colon \ltynv{0}'\fnto \ltynv{1})} \\ \evalCtxSpace{};\evalCtx{}\vdash_{\mathsf S} \lbn{0}' \evalArrow{}_{\mathsf{basis}} \ltynv{0}' \\ \lemptyset{};(\evalCtxSpace{}, \mathit{x}\colon \tof{\lbn{0}});\evalCtx{}[(\mathit{x},\lbn{0}) \mapsto \mathit{x}_{\ltynv{0}}]\vdash_{\mathsf S} \mathit{x} \xlongrightarrow{\lbn{0}'} \lenv{1}\glresulttype{\ltynv{0}'} \\ (\evalCtxSpace{}, \mathit{x}\colon \tof{\lbn{0}});\evalCtx{}[(\mathit{x},\lbn{0}) \mapsto \mathit{x}_{\ltynv{0}}]\vdash_{\mathsf S} \lbn{1} \evalArrow{}_{\mathsf{basis}} \ltynv{1}' \\ \operatorname{inferType}(\leev{}) = (x\colon \ltynv{{\mathrm{D}}}\fnto \ltynv{{\mathrm{C}}}) \\ \lenv{2} = \glvreshape{\glvreshape{\lenv{1}}{\ltynv{0}'}{\ltynv{0}'}}{\ltynv{{\mathrm{D}}}}{\ltynv{0}'}}{\lemptyset{};\evalCtxSpace{};\evalCtx{}\vdash_{\mathsf S} \len{0} \xlongrightarrow{\frameB{x}{\lbn{0}}{\lbn{1}}} \llambda(x\colon \ltynv{0})\lrh \ltynv{1}'\colon \glvreshape{\glvapp{\leev{}}{\lenv{2}}}{\ltynv{1}'}{\ltynv{1}'}\glresulttype{(x\colon \ltynv{0}\fnto \ltynv{1}')}}}
\and
\gpfit{\inferrule*[]{(\lemptyset{}, \evalCtxSpace{})\vdash \len{0}\colon \tof{\frameB{x}{\lbn{0}}{\lbn{1}}} \\ (\emptyset,\evalCtxSpace{})\vdash \frameB{x}{\lbn{0}}{\lbn{1}}\colon \bvalid{} \\ \evalCtxSpace{};\evalCtx{}\vdash_{\mathsf S} \frameB{x}{\lbn{0}}{\lbn{1}} \evalArrow{}_{\mathsf{basis}} (x\colon \ltynv{0}\fnto \ltynv{1}) \\ \evalCtxSpace{};\evalCtx{}\vdash_{\mathsf S} \lbn{0} \evalArrow{}_{\mathsf{basis}} \ltynv{0} \\ \lemptyset{};(\evalCtxSpace{}, \mathit{x}\colon \tof{\lbn{0}});\evalCtx{}[(\mathit{x},\lbn{0}) \mapsto \mathit{x}_{\ltynv{0}}]\vdash_{\mathsf S} \leval{\len{0}}{}{\mathit{x}} \xlongrightarrow{\lbn{1}} \leev{}\glresulttype{\ltynv{1}} \\ (\evalCtxSpace{}, \mathit{x}\colon \tof{\lbn{0}});\evalCtx{}[(\mathit{x},\lbn{0}) \mapsto \mathit{x}_{\ltynv{0}}]\vdash_{\mathsf S} \lbn{1} \evalArrow{}_{\mathsf{basis}} \ltynv{1}}{\lemptyset{};\evalCtxSpace{};\evalCtx{}\vdash_{\mathsf S} \len{0} \xlongrightarrow{\frameB{x}{\lbn{0}}{\lbn{1}}} \llambda(x\colon \ltynv{0})\lrh \ltynv{1}\colon \glvreshape{\leev{}}{\ltynv{1}}{\ltynv{1}}\glresulttype{(x\colon \ltynv{0}\fnto \ltynv{1})}}}
\and
\gpfit{\inferrule*[]{\tof{\lb{}'} = \sof{\lb{}'} \\ (\lDelta{}, \evalCtxSpace{})\vdash (\lllambda \llpl x\colon \sof{\lb{}'}\llpr \ldotsp \lee{})\colon \tof{\lb{}'^{\lstar} \loprod \lb{}} \\ (\emptyset,\evalCtxSpace{})\vdash \lb{}'^{\lstar} \loprod \lb{}\colon \bvalid{} \\ \evalCtxSpace{};\evalCtx{}\vdash_{\mathsf S} \lb{}'^{\lstar} \loprod \lb{} \evalArrow{}_{\mathsf{basis}} \lsv{}' \loprod \lsv{} \\ \evalCtxSpace{};\evalCtx{}\vdash_{\mathsf S} \sof{\lb{}'} \evalArrow{} \lsv{}' \\ \evalCtxSpace{};\evalCtx{}\vdash_{\mathsf S} \sof{\lb{}} \evalArrow{} \lsv{} \\ \forall\,0\leq i<\overline{\operatorname{dimOf}}(\lsv{}').\; (\lDelta{}, \mathit{x}\colon \sof{\lb{}'});\evalCtxSpace{};\evalCtx{}[(\mathit{x},\lb{}') \mapsto \ldelta{i}{\lsv{}'}]\vdash_{\mathsf S} \lee{} \xlongrightarrow{\lb{}} \leev{}_{i}\glresulttype{\lsv{}}}{\lDelta{};\evalCtxSpace{};\evalCtx{}\vdash_{\mathsf S} (\lllambda \llpl x\colon \sof{\lb{}'}\llpr \ldotsp \lee{}) \xlongrightarrow{\lb{}'^{\lstar} \loprod \lb{}} \glvreshape{\lsum^{\lsv{}'}_{\mathit{i}<\overline{\operatorname{dimOf}}(\lsv{}')}\left(\ldelta{\mathit{i}}{\lsv{}'}\loprod \glvreshape{\leev{}_{\mathit{i}}}{\lsv{}}{\lsv{}}\right)}{\lsv{}' \loprod \lsv{}}{\begin{cases}\lsempty{}&\overline{\operatorname{dimOf}}(\lsv{}')=0\\ \lsv{}' \loprod \lsv{}&\text{otherwise}\end{cases}}\glresulttype{\lsv{}' \loprod \lsv{}}}}
\and
\gpfit{\inferrule*[]{\tof{\lb{}'} = \sof{\lb{}'} \\ (\lDelta{}, \evalCtxSpace{})\vdash (\lllambda \llpl x\colon \sof{\lb{}'}\llpr \ldotsp \lee{})\colon \tof{\lb{}'^{\lstar}} \\ (\emptyset,\evalCtxSpace{})\vdash \lb{}'^{\lstar}\colon \bvalid{} \\ \evalCtxSpace{};\evalCtx{}\vdash_{\mathsf S} \lb{}'^{\lstar} \evalArrow{}_{\mathsf{basis}} \lsv{}' \\ \evalCtxSpace{};\evalCtx{}\vdash_{\mathsf S} \sof{\lb{}'} \evalArrow{} \lsv{}' \\ \forall\,0\leq i<\overline{\operatorname{dimOf}}(\lsv{}').\; (\lDelta{}, \mathit{x}\colon \sof{\lb{}'});\evalCtxSpace{};\evalCtx{}[(\mathit{x},\lb{}') \mapsto \ldelta{i}{\lsv{}'}]\vdash_{\mathsf S} \lee{} \xlongrightarrow{\lbone{}} \leev{}_{i}\glresulttype{\lR{}}}{\lDelta{};\evalCtxSpace{};\evalCtx{}\vdash_{\mathsf S} (\lllambda \llpl x\colon \sof{\lb{}'}\llpr \ldotsp \lee{}) \xlongrightarrow{\lb{}'^{\lstar}} \glvreshape{\lsum^{\lsv{}'}_{\mathit{i}<\overline{\operatorname{dimOf}}(\lsv{}')}\left(\glvreshape{\leev{}_{\mathit{i}}}{\lR{}}{\lR{}} \lvscale \ldelta{\mathit{i}}{\lsv{}'}\right)}{\lsv{}'}{\begin{cases}\lsempty{}&\overline{\operatorname{dimOf}}(\lsv{}')=0\\ \lsv{}'&\text{otherwise}\end{cases}}\glresulttype{\lsv{}'}}}
\and
\gpfit{\inferrule*[]{\tof{\lb{}} = \sof{\lb{}} \\ (\lDelta{}, \evalCtxSpace{})\vdash \lvzero{}_{\sof{\lb{}}}\colon \tof{\lb{}} \\ (\emptyset,\evalCtxSpace{})\vdash \lb{}\colon \bvalid{} \\ \evalCtxSpace{};\evalCtx{}\vdash_{\mathsf S} \lb{} \evalArrow{}_{\mathsf{basis}} \lsv{} \\ \evalCtxSpace{};\evalCtx{}\vdash_{\mathsf S} \sof{\lb{}} \evalArrow{} \lsv{}}{\lDelta{};\evalCtxSpace{};\evalCtx{}\vdash_{\mathsf S} \lvzero{}_{\sof{\lb{}}} \xlongrightarrow{\lb{}} \lvzero{}_{\lsv{}}\glresulttype{\lsv{}}}}
\and
\gpfit{\inferrule*[]{(\lemptyset{}, \evalCtxSpace{})\vdash \mathit{r}\colon \tof{\lbone{}} \\ \evalCtxSpace{};\evalCtx{}\vdash_{\mathsf S} \lbone{} \evalArrow{}_{\mathsf{basis}} \lR{}}{\lemptyset{};\evalCtxSpace{};\evalCtx{}\vdash_{\mathsf S} \mathit{r} \xlongrightarrow{\lbone{}} \mathit{r}\glresulttype{\lR{}}}}
}{
\gpfit{\inferrule*[lab=\textsc{S-E-Var\allowbreak Lin}]{((\mathit{x}, \lb{}) \mapsto \leev{}) \in \evalCtx{} \\ \tof{\lb{}} = \ls{} \\ (\emptyset,\evalCtxSpace{})\vdash \lb{}\colon \bvalid{} \\ \evalCtxSpace{};\evalCtx{}\vdash_{\mathsf S} \lb{} \evalArrow{}_{\mathsf{basis}} \ltyv{}}{\{(\mathit{x}, \ls{})\};\evalCtxSpace{};\evalCtx{}\vdash_{\mathsf S} \mathit{x} \xlongrightarrow{\lb{}} \leev{}\glresulttype{\ltyv{}}}}
\and
\gpfit{\inferrule*[lab=\textsc{S-E-Var\allowbreak Nlin}]{((\mathit{x}, \lb{}) \mapsto \leev{}) \in \evalCtx{} \\ (\mathit{x}\colon \tof{\lb{}}) \in \evalCtxSpace{} \\ (\emptyset,\evalCtxSpace{})\vdash \lb{}\colon \bvalid{} \\ \evalCtxSpace{};\evalCtx{}\vdash_{\mathsf S} \lb{} \evalArrow{}_{\mathsf{basis}} \ltyv{}}{\lemptyset{};\evalCtxSpace{};\evalCtx{}\vdash_{\mathsf S} \mathit{x} \xlongrightarrow{\lb{}} \leev{}\glresulttype{\ltyv{}}}}
\and
\gpfit{\inferrule*[lab=\textsc{S-E-Promote}]{\forall y \in \mathit{lin}.\; y \notin \bigcup_{(\cdot\colon s) \in \lDelta{},\, \mathit{lin}} \mathrm{fv}(s) \;\cup\; \bigcup_{(\cdot\colon \tau) \in \evalCtxSpace{}'} \mathrm{fv}(\tau) \\ \evalCtxSpace{}' \subseteq \evalCtxSpace{}'' \\ (\lDelta{}, \mathit{lin} \ldisun \evalCtxSpace{}'')\vdash \lee{}\colon \tof{\lb{}} \\ (\emptyset,\mathit{lin} \ldisun \evalCtxSpace{}'')\vdash \lb{}\colon \bvalid{} \\ \mathit{lin} \ldisun \evalCtxSpace{}'';\evalCtx{}\vdash_{\mathsf S} \lb{} \evalArrow{}_{\mathsf{basis}} \ltyv{} \\ \mathit{lin} \ldisun \lDelta{};\evalCtxSpace{}';\evalCtx{}\vdash_{\mathsf S} \lee{} \xlongrightarrow{\lb{}} \leev{}\glresulttype{\ltyv{}}}{\lDelta{};\mathit{lin} \ldisun \evalCtxSpace{}'';\evalCtx{}\vdash_{\mathsf S} \lee{} \xlongrightarrow{\lb{}} \leev{}\glresulttype{\ltyv{}}}}
\and
\gpfit{\inferrule*[lab=\textsc{S-E-Cob}]{\forall i.\; \tof{\lbn{i}} = \sof{\lbn{i}} \\ (\lDelta{}, \evalCtxSpace{})\vdash \lee{}\colon \tof{\lbn{0}} \\ (\emptyset,\evalCtxSpace{})\vdash \lbn{0}\colon \bvalid{} \\ \evalCtxSpace{};\evalCtx{}\vdash_{\mathsf S} \lbn{0} \evalArrow{}_{\mathsf{basis}} \lsv{} \\ \lDelta{};\evalCtxSpace{};\evalCtx{}\vdash_{\mathsf S} \lee{} \xlongrightarrow{\lbn{1}} \leev{}\glresulttype{\ltyv{}} \\ \evalCtxSpace{};\evalCtx{}\vdash_{\mathsf S} \lbn{1} \xlongleftrightarrow{\lenv{0}} \lbn{0} \\ \evalCtxSpace{};\evalCtx{}\vdash_{\mathsf S} \sof{\lbn{0}} \evalArrow{} \lsv{}}{\lDelta{};\evalCtxSpace{};\evalCtx{}\vdash_{\mathsf S} \lee{} \xlongrightarrow{\lbn{0}} \LTTVPP{\lsv{}}{\glvreshape{\lenv{0}}{\lsv{} \loprod \lsv{}}{\lsv{} \loprod \lsv{}}}{\glvreshape{\leev{}}{\lsv{}}{\lsv{}}}\glresulttype{\lsv{}}}}
\and
\gpfit{\inferrule*[lab=\textsc{S-E-Reshape\allowbreak Cob}]{(\lDelta{}, \evalCtxSpace{})\vdash \lee{}\colon \tof{\lb{}} \\ (\emptyset,\evalCtxSpace{})\vdash \lb{}\colon \bvalid{} \\ \evalCtxSpace{};\evalCtx{}\vdash_{\mathsf S} \lb{} \evalArrow{}_{\mathsf{basis}} \ltyv{}' \\ \lDelta{};\evalCtxSpace{};\evalCtx{}\vdash_{\mathsf S} \lee{} \xlongrightarrow{\lb{}} \leev{}\glresulttype{\ltyv{}}}{\lDelta{};\evalCtxSpace{};\evalCtx{}\vdash_{\mathsf S} \lee{} \xlongrightarrow{\lb{}} \glvreshape{\leev{}}{\ltyv{}'}{\ltyv{}}\glresulttype{\ltyv{}'}}}
\and
\gpfit{\inferrule*[lab=\textsc{S-E-Scale}]{\tof{\lb{}} = \sof{\lb{}} \\ ({\lDelta{}_{0}} \ldisun {\lDelta{}_{1}}, \evalCtxSpace{})\vdash \len{0} \lvscale \len{1}\colon \tof{\lb{}} \\ (\emptyset,\evalCtxSpace{})\vdash \lb{}\colon \bvalid{} \\ \evalCtxSpace{};\evalCtx{}\vdash_{\mathsf S} \lb{} \evalArrow{}_{\mathsf{basis}} \lsv{} \\ {\lDelta{}_{0}};\evalCtxSpace{};\evalCtx{}\vdash_{\mathsf S} \len{0} \xlongrightarrow{\lbone{}} \lenv{0}\glresulttype{\lR{}} \\ {\lDelta{}_{1}};\evalCtxSpace{};\evalCtx{}\vdash_{\mathsf S} \len{1} \xlongrightarrow{\lb{}} \lenv{1}\glresulttype{\lsv{}}}{{\lDelta{}_{0}} \ldisun {\lDelta{}_{1}};\evalCtxSpace{};\evalCtx{}\vdash_{\mathsf S} \len{0} \lvscale \len{1} \xlongrightarrow{\lb{}} \lenv{0} \lvscale \lenv{1}\glresulttype{\lsv{}}}}
\and
\gpfit{\inferrule*[lab=\textsc{S-E-Add}]{\tof{\lb{}} = \sof{\lb{}} \\ (\lDelta{}, \evalCtxSpace{})\vdash \len{0} \lvadd \len{1}\colon \tof{\lb{}} \\ (\emptyset,\evalCtxSpace{})\vdash \lb{}\colon \bvalid{} \\ \evalCtxSpace{};\evalCtx{}\vdash_{\mathsf S} \lb{} \evalArrow{}_{\mathsf{basis}} \lsv{} \\ \forall i.\; \lDelta{};\evalCtxSpace{};\evalCtx{}\vdash_{\mathsf S} \len{i} \xlongrightarrow{\lb{}} \lenv{i}\glresulttype{\lsnv{i}}}{\lDelta{};\evalCtxSpace{};\evalCtx{}\vdash_{\mathsf S} \len{0} \lvadd \len{1} \xlongrightarrow{\lb{}} \lenv{0} \lvadd_{\lsv{}} \lenv{1}\glresulttype{\lsv{}}}}
\and
\gpfit{\inferrule*[lab=\textsc{S-E-Dsum\allowbreak E}]{\forall i.\; \tof{\lbn{i}} = \sof{\lbn{i}} \\ (\lDelta{}, \evalCtxSpace{})\vdash \len{0}\losum \len{1}\colon \tof{\lbn{0} \losum \lbn{1}} \\ (\emptyset,\evalCtxSpace{})\vdash \lbn{0} \losum \lbn{1}\colon \bvalid{} \\ \evalCtxSpace{};\evalCtx{}\vdash_{\mathsf S} \lbn{0} \losum \lbn{1} \evalArrow{}_{\mathsf{basis}} \lsnv{0} \losum \lsnv{1} \\ \forall i.\; \lDelta{};\evalCtxSpace{};\evalCtx{}\vdash_{\mathsf S} \len{i} \xlongrightarrow{\lbn{i}} \lenv{i}\glresulttype{\lsnv{i}}}{\lDelta{};\evalCtxSpace{};\evalCtx{}\vdash_{\mathsf S} \len{0}\losum \len{1} \xlongrightarrow{\lbn{0} \losum \lbn{1}} \lenv{0}\losum \lenv{1}\glresulttype{\lsnv{0} \losum \lsnv{1}}}}
\and
\gpfit{\inferrule*[lab=\textsc{S-E-Tprod\allowbreak E}]{\forall i.\; \tof{\lbn{i}} = \sof{\lbn{i}} \\ ({\lDelta{}_{0}} \ldisun {\lDelta{}_{1}}, \evalCtxSpace{})\vdash \len{0}\loprod \len{1}\colon \tof{\lbn{0} \loprod \lbn{1}} \\ (\emptyset,\evalCtxSpace{})\vdash \lbn{0} \loprod \lbn{1}\colon \bvalid{} \\ \evalCtxSpace{};\evalCtx{}\vdash_{\mathsf S} \lbn{0} \loprod \lbn{1} \evalArrow{}_{\mathsf{basis}} \lsnv{0} \loprod \lsnv{1} \\ \forall i.\; {\lDelta{}_{i}};\evalCtxSpace{};\evalCtx{}\vdash_{\mathsf S} \len{i} \xlongrightarrow{\lbn{i}} \lenv{i}\glresulttype{\lsnv{i}}}{{\lDelta{}_{0}} \ldisun {\lDelta{}_{1}};\evalCtxSpace{};\evalCtx{}\vdash_{\mathsf S} \len{0}\loprod \len{1} \xlongrightarrow{\lbn{0} \loprod \lbn{1}} \lenv{0}\loprod \lenv{1}\glresulttype{\lsnv{0} \loprod \lsnv{1}}}}
\and
\gpfit{\inferrule*[lab=\textsc{S-E-Pinv\allowbreak TStd}]{\forall i.\; \isStdBasis{\lbn{i}} \\ (\lemptyset{}, \evalCtxSpace{})\vdash \lee{}^{\lpseudo}\colon \tof{\lbn{0}^{\lstar} \loprod \lbn{1}} \\ (\emptyset,\evalCtxSpace{})\vdash \lbn{0}^{\lstar} \loprod \lbn{1}\colon \bvalid{} \\ \evalCtxSpace{};\evalCtx{}\vdash_{\mathsf S} \lbn{0}^{\lstar} \loprod \lbn{1} \evalArrow{}_{\mathsf{basis}} \lsnv{0} \loprod \lsnv{1} \\ \lemptyset{};\evalCtxSpace{};\evalCtx{}\vdash_{\mathsf S} \lee{} \xlongrightarrow{\lbn{1}^{\lstar} \loprod \lbn{0}} \leev{}\glresulttype{\lsnv{1} \loprod \lsnv{0}}}{\lemptyset{};\evalCtxSpace{};\evalCtx{}\vdash_{\mathsf S} \lee{}^{\lpseudo} \xlongrightarrow{\lbn{0}^{\lstar} \loprod \lbn{1}} \leev{}^{+}\glresulttype{\lsnv{0} \loprod \lsnv{1}}}}
\and
\gpfit{\inferrule*[lab=\textsc{S-E-Pinv\allowbreak TGpinv}]{(\lemptyset{}, \evalCtxSpace{})\vdash \lee{}^{\lpseudo}\colon \tof{\lbn{0}^{\lstar} \loprod \lbn{1}} \\ (\emptyset,\evalCtxSpace{})\vdash \lbn{0}^{\lstar} \loprod \lbn{1}\colon \bvalid{} \\ \evalCtxSpace{};\evalCtx{}\vdash_{\mathsf S} \lbn{0}^{\lstar} \loprod \lbn{1} \evalArrow{}_{\mathsf{basis}} \lsnv{0} \loprod \lsnv{1} \\ \lemptyset{};\evalCtxSpace{};\evalCtx{}\vdash_{\mathsf S} \lee{} \xlongrightarrow{\lbn{1}^{\lstar} \loprod \lbn{0}} \leev{}\glresulttype{\lsnv{1} \loprod \lsnv{0}} \\ \forall i.\; \lemptyset{};\evalCtxSpace{};\evalCtx{}\vdash_{\mathsf S} \linner{}_{\sof{\lbn{i}}} \xlongrightarrow{\lbn{i}^{\lstar} \loprod \lbn{i}^{\lstar}} \lenv{i}\glresulttype{\lsnv{i} \loprod \lsnv{i}}}{\lemptyset{};\evalCtxSpace{};\evalCtx{}\vdash_{\mathsf S} \lee{}^{\lpseudo} \xlongrightarrow{\lbn{0}^{\lstar} \loprod \lbn{1}} \glvop{GPINV}(\leev{},\lenv{1},\lenv{0})\glresulttype{\lsnv{0} \loprod \lsnv{1}}}}
\and
\gpfit{\inferrule*[lab=\textsc{S-E-Pinv\allowbreak DStd}]{\sof{\lbn{0}} = \lR{} \\ \forall i.\; \isStdBasis{\lbn{i}} \\ (\lemptyset{}, \evalCtxSpace{})\vdash \lee{}^{\lpseudo}\colon \tof{\lbn{0}^{\lstar} \loprod \lbn{1}} \\ (\emptyset,\evalCtxSpace{})\vdash \lbn{0}^{\lstar} \loprod \lbn{1}\colon \bvalid{} \\ \evalCtxSpace{};\evalCtx{}\vdash_{\mathsf S} \lbn{0}^{\lstar} \loprod \lbn{1} \evalArrow{}_{\mathsf{basis}} \lR{} \loprod \lsnv{1} \\ \lemptyset{};\evalCtxSpace{};\evalCtx{}\vdash_{\mathsf S} \lee{} \xlongrightarrow{\lbn{1}^{\lstar}} \leev{}\glresulttype{\lsnv{1}}}{\lemptyset{};\evalCtxSpace{};\evalCtx{}\vdash_{\mathsf S} \lee{}^{\lpseudo} \xlongrightarrow{\lbn{0}^{\lstar} \loprod \lbn{1}} \glvreshape{\leev{}}{\lsnv{1} \loprod \lR{}}{\lsnv{1}}^{+}\glresulttype{\lR{} \loprod \lsnv{1}}}}
\and
\gpfit{\inferrule*[lab=\textsc{S-E-Pinv\allowbreak DGpinv}]{\sof{\lbn{0}} = \lR{} \\ \isStdBasis{\lbn{0}} \\ (\lemptyset{}, \evalCtxSpace{})\vdash \lee{}^{\lpseudo}\colon \tof{\lbn{0}^{\lstar} \loprod \lbn{1}} \\ (\emptyset,\evalCtxSpace{})\vdash \lbn{0}^{\lstar} \loprod \lbn{1}\colon \bvalid{} \\ \evalCtxSpace{};\evalCtx{}\vdash_{\mathsf S} \lbn{0}^{\lstar} \loprod \lbn{1} \evalArrow{}_{\mathsf{basis}} \lR{} \loprod \lsnv{1} \\ \lemptyset{};\evalCtxSpace{};\evalCtx{}\vdash_{\mathsf S} \lee{} \xlongrightarrow{\lbn{1}^{\lstar}} \leev{}\glresulttype{\lsnv{1}} \\ \lemptyset{};\evalCtxSpace{};\evalCtx{}\vdash_{\mathsf S} \linner{}_{\sof{\lbn{0}}} \xlongrightarrow{\lbn{0}^{\lstar} \loprod \lbn{0}^{\lstar}} \lenv{0}\glresulttype{\lR{} \loprod \lR{}} \\ \lemptyset{};\evalCtxSpace{};\evalCtx{}\vdash_{\mathsf S} \linner{}_{\sof{\lbn{1}}} \xlongrightarrow{\lbn{1}^{\lstar} \loprod \lbn{1}^{\lstar}} \lenv{1}\glresulttype{\lsnv{1} \loprod \lsnv{1}}}{\lemptyset{};\evalCtxSpace{};\evalCtx{}\vdash_{\mathsf S} \lee{}^{\lpseudo} \xlongrightarrow{\lbn{0}^{\lstar} \loprod \lbn{1}} \glvop{GPINV}(\glvreshape{\leev{}}{\lsnv{1} \loprod \lR{}}{\lsnv{1}},\lenv{1},\lenv{0})\glresulttype{\lR{} \loprod \lsnv{1}}}}
\and
\gpfit{\inferrule*[lab=\textsc{S-E-Inner\allowbreak Std}]{\isStdBasis{\lbn{0}} \\ \sof{\lbn{0}} = \ls{} \\ (\lemptyset{}, \evalCtxSpace{})\vdash \linner{}_{\ls{}}\colon \tof{\lbn{0}^{\lstar} \loprod \lbn{0}^{\lstar}} \\ (\emptyset,\evalCtxSpace{})\vdash \lbn{0}^{\lstar} \loprod \lbn{0}^{\lstar}\colon \bvalid{} \\ \evalCtxSpace{};\evalCtx{}\vdash_{\mathsf S} \lbn{0}^{\lstar} \loprod \lbn{0}^{\lstar} \evalArrow{}_{\mathsf{basis}} \lsv{} \loprod \lsv{} \\ \evalCtxSpace{};\evalCtx{}\vdash_{\mathsf S} \ls{} \evalArrow{} \lsv{}}{\lemptyset{};\evalCtxSpace{};\evalCtx{}\vdash_{\mathsf S} \linner{}_{\ls{}} \xlongrightarrow{\lbn{0}^{\lstar} \loprod \lbn{0}^{\lstar}} \lidelta{\lsv{}}\glresulttype{\lsv{} \loprod \lsv{}}}}
\and
\gpfit{\inferrule*[lab=\textsc{S-E-Embed\allowbreak Ker}]{\forall i.\; \sof{\lbn{i}} = \lsn{i} \\ (\lemptyset{}, \evalCtxSpace{})\vdash \lembd{}_{\Kernel(\lee{})}\colon \tof{\pBasisC{\Kernel(\lee{})}{\lbn{0}}{\lbn{1}}^{\lstar} \loprod \lbn{0}} \\ (\emptyset,\evalCtxSpace{})\vdash \pBasisC{\Kernel(\lee{})}{\lbn{0}}{\lbn{1}}^{\lstar} \loprod \lbn{0}\colon \bvalid{} \\ \evalCtxSpace{};\evalCtx{}\vdash_{\mathsf S} \pBasisC{\Kernel(\lee{})}{\lbn{0}}{\lbn{1}}^{\lstar} \loprod \lbn{0} \evalArrow{}_{\mathsf{basis}} \lR^{\glvreshape{\leev{}}{\lsnv{0} \loprod \lsnv{1}}{\lsnv{0} \loprod \lsnv{1}},\Kernel{},\lsnv{0} \loprod \lsnv{1}} \loprod \lsnv{0} \\ \forall i.\; \lemptyset{};\evalCtxSpace{};\evalCtx{}\vdash_{\mathsf S} \linner{}_{\lsn{i}} \xlongrightarrow{\lbn{i}^{\lstar} \loprod \lbn{i}^{\lstar}} \lenv{i}\glresulttype{\lsnv{i} \loprod \lsnv{i}} \\ \lemptyset{};\evalCtxSpace{};\evalCtx{}\vdash_{\mathsf S} \lee{} \xlongrightarrow{\lbn{0}^{\lstar} \loprod \lbn{1}} \leev{}\glresulttype{\lsnv{0} \loprod \lsnv{1}}}{\lemptyset{};\evalCtxSpace{};\evalCtx{}\vdash_{\mathsf S} \lembd{}_{\Kernel(\lee{})} \xlongrightarrow{\pBasisC{\Kernel(\lee{})}{\lbn{0}}{\lbn{1}}^{\lstar} \loprod \lbn{0}} \Trans(\glvop{GSVDUI}(\leev{},\lenv{0},\lenv{1}))\glresulttype{\lR^{\glvreshape{\leev{}}{\lsnv{0} \loprod \lsnv{1}}{\lsnv{0} \loprod \lsnv{1}},\Kernel{},\lsnv{0} \loprod \lsnv{1}} \loprod \lsnv{0}}}}
\and
\gpfit{\inferrule*[lab=\textsc{S-E-Embed\allowbreak Coker}]{\forall i.\; \sof{\lbn{i}} = \lsn{i} \\ (\lemptyset{}, \evalCtxSpace{})\vdash \lembd{}_{\CoKernel(\lee{})}\colon \tof{\pBasisC{\CoKernel(\lee{})}{\lbn{0}}{\lbn{1}}^{\lstar} \loprod \lbn{0}} \\ (\emptyset,\evalCtxSpace{})\vdash \pBasisC{\CoKernel(\lee{})}{\lbn{0}}{\lbn{1}}^{\lstar} \loprod \lbn{0}\colon \bvalid{} \\ \evalCtxSpace{};\evalCtx{}\vdash_{\mathsf S} \pBasisC{\CoKernel(\lee{})}{\lbn{0}}{\lbn{1}}^{\lstar} \loprod \lbn{0} \evalArrow{}_{\mathsf{basis}} \lR^{\glvreshape{\leev{}}{\lsnv{0} \loprod \lsnv{1}}{\lsnv{0} \loprod \lsnv{1}},\CoKernel{},\lsnv{0} \loprod \lsnv{1}} \loprod \lsnv{0} \\ \forall i.\; \lemptyset{};\evalCtxSpace{};\evalCtx{}\vdash_{\mathsf S} \linner{}_{\lsn{i}} \xlongrightarrow{\lbn{i}^{\lstar} \loprod \lbn{i}^{\lstar}} \lenv{i}\glresulttype{\lsnv{i} \loprod \lsnv{i}} \\ \lemptyset{};\evalCtxSpace{};\evalCtx{}\vdash_{\mathsf S} \lee{} \xlongrightarrow{\lbn{0}^{\lstar} \loprod \lbn{1}} \leev{}\glresulttype{\lsnv{0} \loprod \lsnv{1}}}{\lemptyset{};\evalCtxSpace{};\evalCtx{}\vdash_{\mathsf S} \lembd{}_{\CoKernel(\lee{})} \xlongrightarrow{\pBasisC{\CoKernel(\lee{})}{\lbn{0}}{\lbn{1}}^{\lstar} \loprod \lbn{0}} \Trans(\glvop{GSVDU}(\leev{},\lenv{0},\lenv{1}))\glresulttype{\lR^{\glvreshape{\leev{}}{\lsnv{0} \loprod \lsnv{1}}{\lsnv{0} \loprod \lsnv{1}},\CoKernel{},\lsnv{0} \loprod \lsnv{1}} \loprod \lsnv{0}}}}
\and
\gpfit{\inferrule*[lab=\textsc{S-E-Embed\allowbreak Img}]{\forall i.\; \sof{\lbn{i}} = \lsn{i} \\ (\lemptyset{}, \evalCtxSpace{})\vdash \lembd{}_{\Image(\lee{})}\colon \tof{\pBasisC{\Image(\lee{})}{\lbn{0}}{\lbn{1}}^{\lstar} \loprod \lbn{1}} \\ (\emptyset,\evalCtxSpace{})\vdash \pBasisC{\Image(\lee{})}{\lbn{0}}{\lbn{1}}^{\lstar} \loprod \lbn{1}\colon \bvalid{} \\ \evalCtxSpace{};\evalCtx{}\vdash_{\mathsf S} \pBasisC{\Image(\lee{})}{\lbn{0}}{\lbn{1}}^{\lstar} \loprod \lbn{1} \evalArrow{}_{\mathsf{basis}} \lR^{\glvreshape{\leev{}}{\lsnv{0} \loprod \lsnv{1}}{\lsnv{0} \loprod \lsnv{1}},\Image{},\lsnv{0} \loprod \lsnv{1}} \loprod \lsnv{1} \\ \forall i.\; \lemptyset{};\evalCtxSpace{};\evalCtx{}\vdash_{\mathsf S} \linner{}_{\lsn{i}} \xlongrightarrow{\lbn{i}^{\lstar} \loprod \lbn{i}^{\lstar}} \lenv{i}\glresulttype{\lsnv{i} \loprod \lsnv{i}} \\ \lemptyset{};\evalCtxSpace{};\evalCtx{}\vdash_{\mathsf S} \lee{} \xlongrightarrow{\lbn{0}^{\lstar} \loprod \lbn{1}} \leev{}\glresulttype{\lsnv{0} \loprod \lsnv{1}}}{\lemptyset{};\evalCtxSpace{};\evalCtx{}\vdash_{\mathsf S} \lembd{}_{\Image(\lee{})} \xlongrightarrow{\pBasisC{\Image(\lee{})}{\lbn{0}}{\lbn{1}}^{\lstar} \loprod \lbn{1}} \Trans(\glvop{GSVDV}(\leev{},\lenv{0},\lenv{1}))\glresulttype{\lR^{\glvreshape{\leev{}}{\lsnv{0} \loprod \lsnv{1}}{\lsnv{0} \loprod \lsnv{1}},\Image{},\lsnv{0} \loprod \lsnv{1}} \loprod \lsnv{1}}}}
\and
\gpfit{\inferrule*[lab=\textsc{S-E-Embed\allowbreak Coimg}]{\forall i.\; \sof{\lbn{i}} = \lsn{i} \\ (\lemptyset{}, \evalCtxSpace{})\vdash \lembd{}_{\CoImage(\lee{})}\colon \tof{\pBasisC{\CoImage(\lee{})}{\lbn{0}}{\lbn{1}}^{\lstar} \loprod \lbn{1}} \\ (\emptyset,\evalCtxSpace{})\vdash \pBasisC{\CoImage(\lee{})}{\lbn{0}}{\lbn{1}}^{\lstar} \loprod \lbn{1}\colon \bvalid{} \\ \evalCtxSpace{};\evalCtx{}\vdash_{\mathsf S} \pBasisC{\CoImage(\lee{})}{\lbn{0}}{\lbn{1}}^{\lstar} \loprod \lbn{1} \evalArrow{}_{\mathsf{basis}} \lR^{\glvreshape{\leev{}}{\lsnv{0} \loprod \lsnv{1}}{\lsnv{0} \loprod \lsnv{1}},\CoImage{},\lsnv{0} \loprod \lsnv{1}} \loprod \lsnv{1} \\ \forall i.\; \lemptyset{};\evalCtxSpace{};\evalCtx{}\vdash_{\mathsf S} \linner{}_{\lsn{i}} \xlongrightarrow{\lbn{i}^{\lstar} \loprod \lbn{i}^{\lstar}} \lenv{i}\glresulttype{\lsnv{i} \loprod \lsnv{i}} \\ \lemptyset{};\evalCtxSpace{};\evalCtx{}\vdash_{\mathsf S} \lee{} \xlongrightarrow{\lbn{0}^{\lstar} \loprod \lbn{1}} \leev{}\glresulttype{\lsnv{0} \loprod \lsnv{1}}}{\lemptyset{};\evalCtxSpace{};\evalCtx{}\vdash_{\mathsf S} \lembd{}_{\CoImage(\lee{})} \xlongrightarrow{\pBasisC{\CoImage(\lee{})}{\lbn{0}}{\lbn{1}}^{\lstar} \loprod \lbn{1}} \Trans(\glvop{GSVDVI}(\leev{},\lenv{0},\lenv{1}))\glresulttype{\lR^{\glvreshape{\leev{}}{\lsnv{0} \loprod \lsnv{1}}{\lsnv{0} \loprod \lsnv{1}},\CoImage{},\lsnv{0} \loprod \lsnv{1}} \loprod \lsnv{1}}}}
\and
\gpfit{\inferrule*[lab=\textsc{S-E-App\allowbreak LLin}]{\tof{\lbn{1}} = \sof{\lbn{1}} \\ ({\lDelta{}_{0}} \ldisun {\lDelta{}_{1}}, \evalCtxSpace{})\vdash \leval{\len{0}}{}{\len{1}}\colon \tof{\lbn{1}} \\ (\emptyset,\evalCtxSpace{})\vdash \lbn{1}\colon \bvalid{} \\ \evalCtxSpace{};\evalCtx{}\vdash_{\mathsf S} \sof{\lbn{0}} \evalArrow{} \lsv{} \\ \evalCtxSpace{};\evalCtx{}\vdash_{\mathsf S} \lbn{1} \evalArrow{}_{\mathsf{basis}} \lsnv{{\mathrm{svOut}}} \\ {\lDelta{}_{0}};\evalCtxSpace{};\evalCtx{}\vdash_{\mathsf S} \len{0} \xlongrightarrow{\lbn{0}^{\lstar} \loprod \lbn{1}} \lenv{0}\glresulttype{\lsnv{{\mathrm{svMat0}}} \loprod \lsnv{{\mathrm{svMat1}}}} \\ {\lDelta{}_{1}};\evalCtxSpace{};\evalCtx{}\vdash_{\mathsf S} \len{1} \xlongrightarrow{\lbn{0}} \lenv{1}\glresulttype{\lsnv{{\mathrm{svArg}}}}}{{\lDelta{}_{0}} \ldisun {\lDelta{}_{1}};\evalCtxSpace{};\evalCtx{}\vdash_{\mathsf S} \leval{\len{0}}{}{\len{1}} \xlongrightarrow{\lbn{1}} \glvreshape{\LTTVPP{\lsv{}}{\lenv{0}}{\lenv{1}}}{\lsnv{{\mathrm{svOut}}}}{\lsnv{{\mathrm{svOut}}}}\glresulttype{\lsnv{{\mathrm{svOut}}}}}}
\and
\gpfit{\inferrule*[lab=\textsc{S-E-App\allowbreak ILin}]{({\lDelta{}_{0}} \ldisun {\lDelta{}_{1}}, \evalCtxSpace{})\vdash \leval{\len{0}}{}{\len{1}}\colon \tof{\lbone{}} \\ \evalCtxSpace{};\evalCtx{}\vdash_{\mathsf S} \sof{\lbn{0}} \evalArrow{} \lsv{} \\ \evalCtxSpace{};\evalCtx{}\vdash_{\mathsf S} \lbone{} \evalArrow{}_{\mathsf{basis}} \lR{} \\ {\lDelta{}_{0}};\evalCtxSpace{};\evalCtx{}\vdash_{\mathsf S} \len{0} \xlongrightarrow{\lbn{0}^{\lstar}} \lenv{0}\glresulttype{\lsnv{{\mathrm{svFun}}}} \\ {\lDelta{}_{1}};\evalCtxSpace{};\evalCtx{}\vdash_{\mathsf S} \len{1} \xlongrightarrow{\lbn{0}} \lenv{1}\glresulttype{\lsnv{{\mathrm{svArg}}}}}{{\lDelta{}_{0}} \ldisun {\lDelta{}_{1}};\evalCtxSpace{};\evalCtx{}\vdash_{\mathsf S} \leval{\len{0}}{}{\len{1}} \xlongrightarrow{\lbone{}} \LTTVPP{\lsv{}}{\glvreshape{\lenv{0}}{\lsv{} \loprod \lR{}}{\lsv{}}}{\lenv{1}}\glresulttype{\lR{}}}}
\and
\gpfit{\inferrule*[lab=\textsc{S-E-Nllam}]{\forall i.\; \tof{\lbn{i}} = \ltyn{i} \\ (\lemptyset{}, \evalCtxSpace{})\vdash (\nllambda \llpl x\colon \ltyn{0}\llpr \ldotsp \lee{})\colon \tof{\frameB{x}{\lbn{0}}{\lbn{1}}} \\ (\emptyset,\evalCtxSpace{})\vdash \frameB{x}{\lbn{0}}{\lbn{1}}\colon \bvalid{} \\ \evalCtxSpace{};\evalCtx{}\vdash_{\mathsf S} \frameB{x}{\lbn{0}}{\lbn{1}} \evalArrow{}_{\mathsf{basis}} (x\colon \ltynv{0}\fnto \ltynv{1}) \\ \evalCtxSpace{};\evalCtx{}\vdash_{\mathsf S} \lbn{0} \evalArrow{}_{\mathsf{basis}} \ltynv{0} \\ \lemptyset{};(\evalCtxSpace{}, \mathit{x}\colon \ltyn{0});\evalCtx{}[(\mathit{x},\lbn{0}) \mapsto \mathit{x}_{\ltynv{0}}]\vdash_{\mathsf S} \lee{} \xlongrightarrow{\lbn{1}} \leev{}\glresulttype{\ltynv{1}} \\ (\evalCtxSpace{}, \mathit{x}\colon \ltyn{0});\evalCtx{}[(\mathit{x},\lbn{0}) \mapsto \mathit{x}_{\ltynv{0}}]\vdash_{\mathsf S} \lbn{1} \evalArrow{}_{\mathsf{basis}} \ltynv{1}}{\lemptyset{};\evalCtxSpace{};\evalCtx{}\vdash_{\mathsf S} (\nllambda \llpl x\colon \ltyn{0}\llpr \ldotsp \lee{}) \xlongrightarrow{\frameB{x}{\lbn{0}}{\lbn{1}}} \llambda(x\colon \ltynv{0})\lrh \ltynv{1}\colon \glvreshape{\leev{}}{\ltynv{1}}{\ltynv{1}}\glresulttype{(x\colon \ltynv{0}\fnto \ltynv{1})}}}
\and
\gpfit{\inferrule*[lab=\textsc{S-E-App\allowbreak NVal}]{(\lemptyset{}, \evalCtxSpace{})\vdash \leval{\len{0}}{}{\len{1}}\colon \tof{\lbn{1}\subst{x}{\len{1}}} \\ (\emptyset,\evalCtxSpace{})\vdash \lbn{1}\subst{x}{\len{1}}\colon \bvalid{} \\ \evalCtxSpace{};\evalCtx{}\vdash_{\mathsf S} \lbn{1}\subst{x}{\len{1}} \evalArrow{}_{\mathsf{basis}} {\ltyv{}_{\mathsf{out}}} \\ \lemptyset{};\evalCtxSpace{};\evalCtx{}\vdash_{\mathsf S} \len{0} \xlongrightarrow{\frameB{x}{\lbn{0}}{\lbn{1}}} \lenv{0}\glresulttype{(x\colon {\ltyv{}_{\mathsf{fun}0}}\fnto {\ltyv{}_{\mathsf{fun}1}})} \\ \lemptyset{};\evalCtxSpace{};\evalCtx{}\vdash_{\mathsf S} \len{1} \xlongrightarrow{\lbn{0}} \lenv{1}\glresulttype{{\ltyv{}_{\mathsf{arg}}}} \\ \operatorname{inferType}(\lenv{0}) = (x\colon \ltynv{{\mathrm{D}}}\fnto \ltynv{{\mathrm{C}}}) \\ \lenv{2} = \glvreshape{\lenv{1}}{\ltynv{{\mathrm{D}}}}{\operatorname{inferType}(\lenv{1})}}{\lemptyset{};\evalCtxSpace{};\evalCtx{}\vdash_{\mathsf S} \leval{\len{0}}{}{\len{1}} \xlongrightarrow{\lbn{1}\subst{x}{\len{1}}} \glvreshape{\glvapp{\lenv{0}}{\lenv{2}}}{{\ltyv{}_{\mathsf{out}}}}{\ltynv{{\mathrm{C}}}\subst{x}{\lenv{2}}}\glresulttype{{\ltyv{}_{\mathsf{out}}}}}}
\and
\gpfit{\inferrule*[lab=\textsc{S-E-Frame\allowbreak Reanchor}]{\tof{\lbn{0}'} = \tof{\lbn{0}} \\ (\lemptyset{}, \evalCtxSpace{})\vdash \len{0}\colon \tof{\frameB{x}{\lbn{0}}{\lbn{1}}} \\ (\emptyset,\evalCtxSpace{})\vdash \frameB{x}{\lbn{0}}{\lbn{1}}\colon \bvalid{} \\ \evalCtxSpace{};\evalCtx{}\vdash_{\mathsf S} \frameB{x}{\lbn{0}}{\lbn{1}} \evalArrow{}_{\mathsf{basis}} (x\colon \ltynv{0}\fnto \ltynv{1}') \\ \evalCtxSpace{};\evalCtx{}\vdash_{\mathsf S} \lbn{0} \evalArrow{}_{\mathsf{basis}} \ltynv{0} \\ \lemptyset{};\evalCtxSpace{};\evalCtx{}\vdash_{\mathsf S} \len{0} \xlongrightarrow{\frameB{x}{\lbn{0}'}{\lbn{1}}} \leev{}\glresulttype{(x\colon \ltynv{0}'\fnto \ltynv{1})} \\ \evalCtxSpace{};\evalCtx{}\vdash_{\mathsf S} \lbn{0}' \evalArrow{}_{\mathsf{basis}} \ltynv{0}' \\ \lemptyset{};(\evalCtxSpace{}, \mathit{x}\colon \tof{\lbn{0}});\evalCtx{}[(\mathit{x},\lbn{0}) \mapsto \mathit{x}_{\ltynv{0}}]\vdash_{\mathsf S} \mathit{x} \xlongrightarrow{\lbn{0}'} \lenv{1}\glresulttype{\ltynv{0}'} \\ (\evalCtxSpace{}, \mathit{x}\colon \tof{\lbn{0}});\evalCtx{}[(\mathit{x},\lbn{0}) \mapsto \mathit{x}_{\ltynv{0}}]\vdash_{\mathsf S} \lbn{1} \evalArrow{}_{\mathsf{basis}} \ltynv{1}' \\ \operatorname{inferType}(\leev{}) = (x\colon \ltynv{{\mathrm{D}}}\fnto \ltynv{{\mathrm{C}}}) \\ \lenv{2} = \glvreshape{\glvreshape{\lenv{1}}{\ltynv{0}'}{\ltynv{0}'}}{\ltynv{{\mathrm{D}}}}{\ltynv{0}'}}{\lemptyset{};\evalCtxSpace{};\evalCtx{}\vdash_{\mathsf S} \len{0} \xlongrightarrow{\frameB{x}{\lbn{0}}{\lbn{1}}} \llambda(x\colon \ltynv{0})\lrh \ltynv{1}'\colon \glvreshape{\glvapp{\leev{}}{\lenv{2}}}{\ltynv{1}'}{\ltynv{1}'}\glresulttype{(x\colon \ltynv{0}\fnto \ltynv{1}')}}}
\and
\gpfit{\inferrule*[lab=\textsc{S-E-Frame\allowbreak Eta}]{(\lemptyset{}, \evalCtxSpace{})\vdash \len{0}\colon \tof{\frameB{x}{\lbn{0}}{\lbn{1}}} \\ (\emptyset,\evalCtxSpace{})\vdash \frameB{x}{\lbn{0}}{\lbn{1}}\colon \bvalid{} \\ \evalCtxSpace{};\evalCtx{}\vdash_{\mathsf S} \frameB{x}{\lbn{0}}{\lbn{1}} \evalArrow{}_{\mathsf{basis}} (x\colon \ltynv{0}\fnto \ltynv{1}) \\ \evalCtxSpace{};\evalCtx{}\vdash_{\mathsf S} \lbn{0} \evalArrow{}_{\mathsf{basis}} \ltynv{0} \\ \lemptyset{};(\evalCtxSpace{}, \mathit{x}\colon \tof{\lbn{0}});\evalCtx{}[(\mathit{x},\lbn{0}) \mapsto \mathit{x}_{\ltynv{0}}]\vdash_{\mathsf S} \leval{\len{0}}{}{\mathit{x}} \xlongrightarrow{\lbn{1}} \leev{}\glresulttype{\ltynv{1}} \\ (\evalCtxSpace{}, \mathit{x}\colon \tof{\lbn{0}});\evalCtx{}[(\mathit{x},\lbn{0}) \mapsto \mathit{x}_{\ltynv{0}}]\vdash_{\mathsf S} \lbn{1} \evalArrow{}_{\mathsf{basis}} \ltynv{1}}{\lemptyset{};\evalCtxSpace{};\evalCtx{}\vdash_{\mathsf S} \len{0} \xlongrightarrow{\frameB{x}{\lbn{0}}{\lbn{1}}} \llambda(x\colon \ltynv{0})\lrh \ltynv{1}\colon \glvreshape{\leev{}}{\ltynv{1}}{\ltynv{1}}\glresulttype{(x\colon \ltynv{0}\fnto \ltynv{1})}}}
\and
\gpfit{\inferrule*[lab=\textsc{S-E-Llam\allowbreak Assemble}]{\tof{\lb{}'} = \sof{\lb{}'} \\ (\lDelta{}, \evalCtxSpace{})\vdash (\lllambda \llpl x\colon \sof{\lb{}'}\llpr \ldotsp \lee{})\colon \tof{\lb{}'^{\lstar} \loprod \lb{}} \\ (\emptyset,\evalCtxSpace{})\vdash \lb{}'^{\lstar} \loprod \lb{}\colon \bvalid{} \\ \evalCtxSpace{};\evalCtx{}\vdash_{\mathsf S} \lb{}'^{\lstar} \loprod \lb{} \evalArrow{}_{\mathsf{basis}} \lsv{}' \loprod \lsv{} \\ \evalCtxSpace{};\evalCtx{}\vdash_{\mathsf S} \sof{\lb{}'} \evalArrow{} \lsv{}' \\ \evalCtxSpace{};\evalCtx{}\vdash_{\mathsf S} \sof{\lb{}} \evalArrow{} \lsv{} \\ \forall\,0\leq i<\overline{\operatorname{dimOf}}(\lsv{}').\; (\lDelta{}, \mathit{x}\colon \sof{\lb{}'});\evalCtxSpace{};\evalCtx{}[(\mathit{x},\lb{}') \mapsto \ldelta{i}{\lsv{}'}]\vdash_{\mathsf S} \lee{} \xlongrightarrow{\lb{}} \leev{}_{i}\glresulttype{\lsv{}}}{\lDelta{};\evalCtxSpace{};\evalCtx{}\vdash_{\mathsf S} (\lllambda \llpl x\colon \sof{\lb{}'}\llpr \ldotsp \lee{}) \xlongrightarrow{\lb{}'^{\lstar} \loprod \lb{}} \glvreshape{\lsum^{\lsv{}'}_{\mathit{i}<\overline{\operatorname{dimOf}}(\lsv{}')}\left(\ldelta{\mathit{i}}{\lsv{}'}\loprod \glvreshape{\leev{}_{\mathit{i}}}{\lsv{}}{\lsv{}}\right)}{\lsv{}' \loprod \lsv{}}{\begin{cases}\lsempty{}&\overline{\operatorname{dimOf}}(\lsv{}')=0\\ \lsv{}' \loprod \lsv{}&\text{otherwise}\end{cases}}\glresulttype{\lsv{}' \loprod \lsv{}}}}
\and
\gpfit{\inferrule*[lab=\textsc{S-E-Llam\allowbreak Assemble\allowbreak Real}]{\tof{\lb{}'} = \sof{\lb{}'} \\ (\lDelta{}, \evalCtxSpace{})\vdash (\lllambda \llpl x\colon \sof{\lb{}'}\llpr \ldotsp \lee{})\colon \tof{\lb{}'^{\lstar}} \\ (\emptyset,\evalCtxSpace{})\vdash \lb{}'^{\lstar}\colon \bvalid{} \\ \evalCtxSpace{};\evalCtx{}\vdash_{\mathsf S} \lb{}'^{\lstar} \evalArrow{}_{\mathsf{basis}} \lsv{}' \\ \evalCtxSpace{};\evalCtx{}\vdash_{\mathsf S} \sof{\lb{}'} \evalArrow{} \lsv{}' \\ \forall\,0\leq i<\overline{\operatorname{dimOf}}(\lsv{}').\; (\lDelta{}, \mathit{x}\colon \sof{\lb{}'});\evalCtxSpace{};\evalCtx{}[(\mathit{x},\lb{}') \mapsto \ldelta{i}{\lsv{}'}]\vdash_{\mathsf S} \lee{} \xlongrightarrow{\lbone{}} \leev{}_{i}\glresulttype{\lR{}}}{\lDelta{};\evalCtxSpace{};\evalCtx{}\vdash_{\mathsf S} (\lllambda \llpl x\colon \sof{\lb{}'}\llpr \ldotsp \lee{}) \xlongrightarrow{\lb{}'^{\lstar}} \glvreshape{\lsum^{\lsv{}'}_{\mathit{i}<\overline{\operatorname{dimOf}}(\lsv{}')}\left(\glvreshape{\leev{}_{\mathit{i}}}{\lR{}}{\lR{}} \lvscale \ldelta{\mathit{i}}{\lsv{}'}\right)}{\lsv{}'}{\begin{cases}\lsempty{}&\overline{\operatorname{dimOf}}(\lsv{}')=0\\ \lsv{}'&\text{otherwise}\end{cases}}\glresulttype{\lsv{}'}}}
\and
\gpfit{\inferrule*[lab=\textsc{S-E-Zero}]{\tof{\lb{}} = \sof{\lb{}} \\ (\lDelta{}, \evalCtxSpace{})\vdash \lvzero{}_{\sof{\lb{}}}\colon \tof{\lb{}} \\ (\emptyset,\evalCtxSpace{})\vdash \lb{}\colon \bvalid{} \\ \evalCtxSpace{};\evalCtx{}\vdash_{\mathsf S} \lb{} \evalArrow{}_{\mathsf{basis}} \lsv{} \\ \evalCtxSpace{};\evalCtx{}\vdash_{\mathsf S} \sof{\lb{}} \evalArrow{} \lsv{}}{\lDelta{};\evalCtxSpace{};\evalCtx{}\vdash_{\mathsf S} \lvzero{}_{\sof{\lb{}}} \xlongrightarrow{\lb{}} \lvzero{}_{\lsv{}}\glresulttype{\lsv{}}}}
\and
\gpfit{\inferrule*[lab=\textsc{S-E-Rlit}]{(\lemptyset{}, \evalCtxSpace{})\vdash \mathit{r}\colon \tof{\lbone{}} \\ \evalCtxSpace{};\evalCtx{}\vdash_{\mathsf S} \lbone{} \evalArrow{}_{\mathsf{basis}} \lR{}}{\lemptyset{};\evalCtxSpace{};\evalCtx{}\vdash_{\mathsf S} \mathit{r} \xlongrightarrow{\lbone{}} \mathit{r}\glresulttype{\lR{}}}}
}
\gldefjudgment{c.space}{\lepsilon{};\evalCtx{}\vdash_{\mathsf C} \ls{}\evalArrow{}\lsv{}}{Checked space evaluation (generated from \texttt{C.Space}).}{fig:gl-c-space}
\gldefrule{c.space.real}{c.space}{\textsc{C-S-Real}}{\phantom{.}}{\evalCtxSpace{};\evalCtx{}\vdash_{\mathsf C} \lR{} \evalArrow{} \lR{}}{\texttt{GLang.Raw\allowbreak Certified\allowbreak Operational.C.Space.real}}{GLang/Adequacy/RawCertifiedOperational.lean:32}
\gldefrule{c.space.empty}{c.space}{\textsc{C-S-Empty}}{\phantom{.}}{\evalCtxSpace{};\evalCtx{}\vdash_{\mathsf C} \lsempty{} \evalArrow{} \lsempty{}}{\texttt{GLang.Raw\allowbreak Certified\allowbreak Operational.C.Space.empty}}{GLang/Adequacy/RawCertifiedOperational.lean:33}
\gldefrule{c.space.dual}{c.space}{\textsc{C-S-Dual}}{(\lemptyset, \evalCtxSpace{})\vdash \ls{}^{\lstar}\colon\svalid \\ \evalCtxSpace{};\evalCtx{}\vdash_{\mathsf C} \ls{} \evalArrow{} \lsv{}}{\evalCtxSpace{};\evalCtx{}\vdash_{\mathsf C} \ls{}^{\lstar} \evalArrow{} \lsv{}}{\texttt{GLang.Raw\allowbreak Certified\allowbreak Operational.C.Space.dual}}{GLang/Adequacy/RawCertifiedOperational.lean:34}
\gldefrule{c.space.ker}{c.space}{\textsc{C-S-Ker}}{(\lemptyset, \evalCtxSpace{})\vdash \Kernel(\lee{})\colon\svalid \\ \sof{\lb{}} = \lsn{0}^{\lstar} \loprod \lsn{1} \\ \exists\,\lbp{},\textcolor{lang}{b^{\prime\prime}}.\; \lb{} = \lbp{}^{\lstar}\loprod \textcolor{lang}{b^{\prime\prime}} \\ \evalCtxSpace{};\evalCtx{}\vdash_{\mathsf C} \sof{\lb{}} \evalArrow{} \lsnv{{\mathrm{svb}}} \\ \lemptyset{};\evalCtxSpace{};\evalCtx{}\vdash_{\mathsf C} \lee{} \xlongrightarrow{\lb{}} \leev{}\glresulttype{\lsnv{{\mathrm{svb}}}}}{\evalCtxSpace{};\evalCtx{}\vdash_{\mathsf C} \Kernel(\lee{}) \evalArrow{} \lR^{\glvreshape{\leev{}}{\lsnv{{\mathrm{svb}}}}{\lsnv{{\mathrm{svb}}}},\Kernel{},\lsnv{{\mathrm{svb}}}}}{\texttt{GLang.Raw\allowbreak Certified\allowbreak Operational.C.Space.ker}}{GLang/Adequacy/RawCertifiedOperational.lean:38}
\gldefrule{c.space.coker}{c.space}{\textsc{C-S-Coker}}{(\lemptyset, \evalCtxSpace{})\vdash \CoKernel(\lee{})\colon\svalid \\ \sof{\lb{}} = \lsn{0}^{\lstar} \loprod \lsn{1} \\ \exists\,\lbp{},\textcolor{lang}{b^{\prime\prime}}.\; \lb{} = \lbp{}^{\lstar}\loprod \textcolor{lang}{b^{\prime\prime}} \\ \evalCtxSpace{};\evalCtx{}\vdash_{\mathsf C} \sof{\lb{}} \evalArrow{} \lsnv{{\mathrm{svb}}} \\ \lemptyset{};\evalCtxSpace{};\evalCtx{}\vdash_{\mathsf C} \lee{} \xlongrightarrow{\lb{}} \leev{}\glresulttype{\lsnv{{\mathrm{svb}}}}}{\evalCtxSpace{};\evalCtx{}\vdash_{\mathsf C} \CoKernel(\lee{}) \evalArrow{} \lR^{\glvreshape{\leev{}}{\lsnv{{\mathrm{svb}}}}{\lsnv{{\mathrm{svb}}}},\CoKernel{},\lsnv{{\mathrm{svb}}}}}{\texttt{GLang.Raw\allowbreak Certified\allowbreak Operational.C.Space.coker}}{GLang/Adequacy/RawCertifiedOperational.lean:47}
\gldefrule{c.space.img}{c.space}{\textsc{C-S-Img}}{(\lemptyset, \evalCtxSpace{})\vdash \Image(\lee{})\colon\svalid \\ \sof{\lb{}} = \lsn{0}^{\lstar} \loprod \lsn{1} \\ \exists\,\lbp{},\textcolor{lang}{b^{\prime\prime}}.\; \lb{} = \lbp{}^{\lstar}\loprod \textcolor{lang}{b^{\prime\prime}} \\ \evalCtxSpace{};\evalCtx{}\vdash_{\mathsf C} \sof{\lb{}} \evalArrow{} \lsnv{{\mathrm{svb}}} \\ \lemptyset{};\evalCtxSpace{};\evalCtx{}\vdash_{\mathsf C} \lee{} \xlongrightarrow{\lb{}} \leev{}\glresulttype{\lsnv{{\mathrm{svb}}}}}{\evalCtxSpace{};\evalCtx{}\vdash_{\mathsf C} \Image(\lee{}) \evalArrow{} \lR^{\glvreshape{\leev{}}{\lsnv{{\mathrm{svb}}}}{\lsnv{{\mathrm{svb}}}},\Image{},\lsnv{{\mathrm{svb}}}}}{\texttt{GLang.Raw\allowbreak Certified\allowbreak Operational.C.Space.img}}{GLang/Adequacy/RawCertifiedOperational.lean:55}
\gldefrule{c.space.coimg}{c.space}{\textsc{C-S-Coimg}}{(\lemptyset, \evalCtxSpace{})\vdash \CoImage(\lee{})\colon\svalid \\ \sof{\lb{}} = \lsn{0}^{\lstar} \loprod \lsn{1} \\ \exists\,\lbp{},\textcolor{lang}{b^{\prime\prime}}.\; \lb{} = \lbp{}^{\lstar}\loprod \textcolor{lang}{b^{\prime\prime}} \\ \evalCtxSpace{};\evalCtx{}\vdash_{\mathsf C} \sof{\lb{}} \evalArrow{} \lsnv{{\mathrm{svb}}} \\ \lemptyset{};\evalCtxSpace{};\evalCtx{}\vdash_{\mathsf C} \lee{} \xlongrightarrow{\lb{}} \leev{}\glresulttype{\lsnv{{\mathrm{svb}}}}}{\evalCtxSpace{};\evalCtx{}\vdash_{\mathsf C} \CoImage(\lee{}) \evalArrow{} \lR^{\glvreshape{\leev{}}{\lsnv{{\mathrm{svb}}}}{\lsnv{{\mathrm{svb}}}},\CoImage{},\lsnv{{\mathrm{svb}}}}}{\texttt{GLang.Raw\allowbreak Certified\allowbreak Operational.C.Space.coimg}}{GLang/Adequacy/RawCertifiedOperational.lean:63}
\gldefrule{c.space.tprod}{c.space}{\textsc{C-S-Tprod}}{(\lemptyset, \evalCtxSpace{})\vdash \lsn{0} \loprod \lsn{1}\colon\svalid \\ \forall i.\; \evalCtxSpace{};\evalCtx{}\vdash_{\mathsf C} \lsn{i} \evalArrow{} \lsnv{i}}{\evalCtxSpace{};\evalCtx{}\vdash_{\mathsf C} \lsn{0} \loprod \lsn{1} \evalArrow{} \lsnv{0} \loprod \lsnv{1}}{\texttt{GLang.Raw\allowbreak Certified\allowbreak Operational.C.Space.tprod}}{GLang/Adequacy/RawCertifiedOperational.lean:71}
\gldefrule{c.space.dsum}{c.space}{\textsc{C-S-Dsum}}{(\lemptyset, \evalCtxSpace{})\vdash \lsn{0} \losum \lsn{1}\colon\svalid \\ \forall i.\; \evalCtxSpace{};\evalCtx{}\vdash_{\mathsf C} \lsn{i} \evalArrow{} \lsnv{i}}{\evalCtxSpace{};\evalCtx{}\vdash_{\mathsf C} \lsn{0} \losum \lsn{1} \evalArrow{} \lsnv{0} \losum \lsnv{1}}{\texttt{GLang.Raw\allowbreak Certified\allowbreak Operational.C.Space.dsum}}{GLang/Adequacy/RawCertifiedOperational.lean:75}
\gldefrulerows{c.space}{
\textsc{C-S-Real} & $\evalCtxSpace{};\evalCtx{}\vdash_{\mathsf C} \lR{} \evalArrow{} \lR{}$ & \texttt{\footnotesize GLang/Adequacy/Raw\allowbreak Certified\allowbreak Operational.lean:32}\\
\textsc{C-S-Empty} & $\evalCtxSpace{};\evalCtx{}\vdash_{\mathsf C} \lsempty{} \evalArrow{} \lsempty{}$ & \texttt{\footnotesize GLang/Adequacy/Raw\allowbreak Certified\allowbreak Operational.lean:33}\\
\textsc{C-S-Dual} & $\evalCtxSpace{};\evalCtx{}\vdash_{\mathsf C} \ls{}^{\lstar} \evalArrow{} \lsv{}$ & \texttt{\footnotesize GLang/Adequacy/Raw\allowbreak Certified\allowbreak Operational.lean:34}\\
\textsc{C-S-Ker} & $\evalCtxSpace{};\evalCtx{}\vdash_{\mathsf C} \Kernel(\lee{}) \evalArrow{} \lR^{\glvreshape{\leev{}}{\lsnv{{\mathrm{svb}}}}{\lsnv{{\mathrm{svb}}}},\Kernel{},\lsnv{{\mathrm{svb}}}}$ & \texttt{\footnotesize GLang/Adequacy/Raw\allowbreak Certified\allowbreak Operational.lean:38}\\
\textsc{C-S-Coker} & $\evalCtxSpace{};\evalCtx{}\vdash_{\mathsf C} \CoKernel(\lee{}) \evalArrow{} \lR^{\glvreshape{\leev{}}{\lsnv{{\mathrm{svb}}}}{\lsnv{{\mathrm{svb}}}},\CoKernel{},\lsnv{{\mathrm{svb}}}}$ & \texttt{\footnotesize GLang/Adequacy/Raw\allowbreak Certified\allowbreak Operational.lean:47}\\
\textsc{C-S-Img} & $\evalCtxSpace{};\evalCtx{}\vdash_{\mathsf C} \Image(\lee{}) \evalArrow{} \lR^{\glvreshape{\leev{}}{\lsnv{{\mathrm{svb}}}}{\lsnv{{\mathrm{svb}}}},\Image{},\lsnv{{\mathrm{svb}}}}$ & \texttt{\footnotesize GLang/Adequacy/Raw\allowbreak Certified\allowbreak Operational.lean:55}\\
\textsc{C-S-Coimg} & $\evalCtxSpace{};\evalCtx{}\vdash_{\mathsf C} \CoImage(\lee{}) \evalArrow{} \lR^{\glvreshape{\leev{}}{\lsnv{{\mathrm{svb}}}}{\lsnv{{\mathrm{svb}}}},\CoImage{},\lsnv{{\mathrm{svb}}}}$ & \texttt{\footnotesize GLang/Adequacy/Raw\allowbreak Certified\allowbreak Operational.lean:63}\\
\textsc{C-S-Tprod} & $\evalCtxSpace{};\evalCtx{}\vdash_{\mathsf C} \lsn{0} \loprod \lsn{1} \evalArrow{} \lsnv{0} \loprod \lsnv{1}$ & \texttt{\footnotesize GLang/Adequacy/Raw\allowbreak Certified\allowbreak Operational.lean:71}\\
\textsc{C-S-Dsum} & $\evalCtxSpace{};\evalCtx{}\vdash_{\mathsf C} \lsn{0} \losum \lsn{1} \evalArrow{} \lsnv{0} \losum \lsnv{1}$ & \texttt{\footnotesize GLang/Adequacy/Raw\allowbreak Certified\allowbreak Operational.lean:75}\\
}{
\gpfit{\inferrule*[]{\phantom{.}}{\evalCtxSpace{};\evalCtx{}\vdash_{\mathsf C} \lR{} \evalArrow{} \lR{}}}
\and
\gpfit{\inferrule*[]{\phantom{.}}{\evalCtxSpace{};\evalCtx{}\vdash_{\mathsf C} \lsempty{} \evalArrow{} \lsempty{}}}
\and
\gpfit{\inferrule*[]{(\lemptyset, \evalCtxSpace{})\vdash \ls{}^{\lstar}\colon\svalid \\ \evalCtxSpace{};\evalCtx{}\vdash_{\mathsf C} \ls{} \evalArrow{} \lsv{}}{\evalCtxSpace{};\evalCtx{}\vdash_{\mathsf C} \ls{}^{\lstar} \evalArrow{} \lsv{}}}
\and
\gpfit{\inferrule*[]{(\lemptyset, \evalCtxSpace{})\vdash \Kernel(\lee{})\colon\svalid \\ \sof{\lb{}} = \lsn{0}^{\lstar} \loprod \lsn{1} \\ \exists\,\lbp{},\textcolor{lang}{b^{\prime\prime}}.\; \lb{} = \lbp{}^{\lstar}\loprod \textcolor{lang}{b^{\prime\prime}} \\ \evalCtxSpace{};\evalCtx{}\vdash_{\mathsf C} \sof{\lb{}} \evalArrow{} \lsnv{{\mathrm{svb}}} \\ \lemptyset{};\evalCtxSpace{};\evalCtx{}\vdash_{\mathsf C} \lee{} \xlongrightarrow{\lb{}} \leev{}\glresulttype{\lsnv{{\mathrm{svb}}}}}{\evalCtxSpace{};\evalCtx{}\vdash_{\mathsf C} \Kernel(\lee{}) \evalArrow{} \lR^{\glvreshape{\leev{}}{\lsnv{{\mathrm{svb}}}}{\lsnv{{\mathrm{svb}}}},\Kernel{},\lsnv{{\mathrm{svb}}}}}}
\and
\gpfit{\inferrule*[]{(\lemptyset, \evalCtxSpace{})\vdash \CoKernel(\lee{})\colon\svalid \\ \sof{\lb{}} = \lsn{0}^{\lstar} \loprod \lsn{1} \\ \exists\,\lbp{},\textcolor{lang}{b^{\prime\prime}}.\; \lb{} = \lbp{}^{\lstar}\loprod \textcolor{lang}{b^{\prime\prime}} \\ \evalCtxSpace{};\evalCtx{}\vdash_{\mathsf C} \sof{\lb{}} \evalArrow{} \lsnv{{\mathrm{svb}}} \\ \lemptyset{};\evalCtxSpace{};\evalCtx{}\vdash_{\mathsf C} \lee{} \xlongrightarrow{\lb{}} \leev{}\glresulttype{\lsnv{{\mathrm{svb}}}}}{\evalCtxSpace{};\evalCtx{}\vdash_{\mathsf C} \CoKernel(\lee{}) \evalArrow{} \lR^{\glvreshape{\leev{}}{\lsnv{{\mathrm{svb}}}}{\lsnv{{\mathrm{svb}}}},\CoKernel{},\lsnv{{\mathrm{svb}}}}}}
\and
\gpfit{\inferrule*[]{(\lemptyset, \evalCtxSpace{})\vdash \Image(\lee{})\colon\svalid \\ \sof{\lb{}} = \lsn{0}^{\lstar} \loprod \lsn{1} \\ \exists\,\lbp{},\textcolor{lang}{b^{\prime\prime}}.\; \lb{} = \lbp{}^{\lstar}\loprod \textcolor{lang}{b^{\prime\prime}} \\ \evalCtxSpace{};\evalCtx{}\vdash_{\mathsf C} \sof{\lb{}} \evalArrow{} \lsnv{{\mathrm{svb}}} \\ \lemptyset{};\evalCtxSpace{};\evalCtx{}\vdash_{\mathsf C} \lee{} \xlongrightarrow{\lb{}} \leev{}\glresulttype{\lsnv{{\mathrm{svb}}}}}{\evalCtxSpace{};\evalCtx{}\vdash_{\mathsf C} \Image(\lee{}) \evalArrow{} \lR^{\glvreshape{\leev{}}{\lsnv{{\mathrm{svb}}}}{\lsnv{{\mathrm{svb}}}},\Image{},\lsnv{{\mathrm{svb}}}}}}
\and
\gpfit{\inferrule*[]{(\lemptyset, \evalCtxSpace{})\vdash \CoImage(\lee{})\colon\svalid \\ \sof{\lb{}} = \lsn{0}^{\lstar} \loprod \lsn{1} \\ \exists\,\lbp{},\textcolor{lang}{b^{\prime\prime}}.\; \lb{} = \lbp{}^{\lstar}\loprod \textcolor{lang}{b^{\prime\prime}} \\ \evalCtxSpace{};\evalCtx{}\vdash_{\mathsf C} \sof{\lb{}} \evalArrow{} \lsnv{{\mathrm{svb}}} \\ \lemptyset{};\evalCtxSpace{};\evalCtx{}\vdash_{\mathsf C} \lee{} \xlongrightarrow{\lb{}} \leev{}\glresulttype{\lsnv{{\mathrm{svb}}}}}{\evalCtxSpace{};\evalCtx{}\vdash_{\mathsf C} \CoImage(\lee{}) \evalArrow{} \lR^{\glvreshape{\leev{}}{\lsnv{{\mathrm{svb}}}}{\lsnv{{\mathrm{svb}}}},\CoImage{},\lsnv{{\mathrm{svb}}}}}}
\and
\gpfit{\inferrule*[]{(\lemptyset, \evalCtxSpace{})\vdash \lsn{0} \loprod \lsn{1}\colon\svalid \\ \forall i.\; \evalCtxSpace{};\evalCtx{}\vdash_{\mathsf C} \lsn{i} \evalArrow{} \lsnv{i}}{\evalCtxSpace{};\evalCtx{}\vdash_{\mathsf C} \lsn{0} \loprod \lsn{1} \evalArrow{} \lsnv{0} \loprod \lsnv{1}}}
\and
\gpfit{\inferrule*[]{(\lemptyset, \evalCtxSpace{})\vdash \lsn{0} \losum \lsn{1}\colon\svalid \\ \forall i.\; \evalCtxSpace{};\evalCtx{}\vdash_{\mathsf C} \lsn{i} \evalArrow{} \lsnv{i}}{\evalCtxSpace{};\evalCtx{}\vdash_{\mathsf C} \lsn{0} \losum \lsn{1} \evalArrow{} \lsnv{0} \losum \lsnv{1}}}
}{
\gpfit{\inferrule*[lab=\textsc{C-S-Real}]{\phantom{.}}{\evalCtxSpace{};\evalCtx{}\vdash_{\mathsf C} \lR{} \evalArrow{} \lR{}}}
\and
\gpfit{\inferrule*[lab=\textsc{C-S-Empty}]{\phantom{.}}{\evalCtxSpace{};\evalCtx{}\vdash_{\mathsf C} \lsempty{} \evalArrow{} \lsempty{}}}
\and
\gpfit{\inferrule*[lab=\textsc{C-S-Dual}]{(\lemptyset, \evalCtxSpace{})\vdash \ls{}^{\lstar}\colon\svalid \\ \evalCtxSpace{};\evalCtx{}\vdash_{\mathsf C} \ls{} \evalArrow{} \lsv{}}{\evalCtxSpace{};\evalCtx{}\vdash_{\mathsf C} \ls{}^{\lstar} \evalArrow{} \lsv{}}}
\and
\gpfit{\inferrule*[lab=\textsc{C-S-Ker}]{(\lemptyset, \evalCtxSpace{})\vdash \Kernel(\lee{})\colon\svalid \\ \sof{\lb{}} = \lsn{0}^{\lstar} \loprod \lsn{1} \\ \exists\,\lbp{},\textcolor{lang}{b^{\prime\prime}}.\; \lb{} = \lbp{}^{\lstar}\loprod \textcolor{lang}{b^{\prime\prime}} \\ \evalCtxSpace{};\evalCtx{}\vdash_{\mathsf C} \sof{\lb{}} \evalArrow{} \lsnv{{\mathrm{svb}}} \\ \lemptyset{};\evalCtxSpace{};\evalCtx{}\vdash_{\mathsf C} \lee{} \xlongrightarrow{\lb{}} \leev{}\glresulttype{\lsnv{{\mathrm{svb}}}}}{\evalCtxSpace{};\evalCtx{}\vdash_{\mathsf C} \Kernel(\lee{}) \evalArrow{} \lR^{\glvreshape{\leev{}}{\lsnv{{\mathrm{svb}}}}{\lsnv{{\mathrm{svb}}}},\Kernel{},\lsnv{{\mathrm{svb}}}}}}
\and
\gpfit{\inferrule*[lab=\textsc{C-S-Coker}]{(\lemptyset, \evalCtxSpace{})\vdash \CoKernel(\lee{})\colon\svalid \\ \sof{\lb{}} = \lsn{0}^{\lstar} \loprod \lsn{1} \\ \exists\,\lbp{},\textcolor{lang}{b^{\prime\prime}}.\; \lb{} = \lbp{}^{\lstar}\loprod \textcolor{lang}{b^{\prime\prime}} \\ \evalCtxSpace{};\evalCtx{}\vdash_{\mathsf C} \sof{\lb{}} \evalArrow{} \lsnv{{\mathrm{svb}}} \\ \lemptyset{};\evalCtxSpace{};\evalCtx{}\vdash_{\mathsf C} \lee{} \xlongrightarrow{\lb{}} \leev{}\glresulttype{\lsnv{{\mathrm{svb}}}}}{\evalCtxSpace{};\evalCtx{}\vdash_{\mathsf C} \CoKernel(\lee{}) \evalArrow{} \lR^{\glvreshape{\leev{}}{\lsnv{{\mathrm{svb}}}}{\lsnv{{\mathrm{svb}}}},\CoKernel{},\lsnv{{\mathrm{svb}}}}}}
\and
\gpfit{\inferrule*[lab=\textsc{C-S-Img}]{(\lemptyset, \evalCtxSpace{})\vdash \Image(\lee{})\colon\svalid \\ \sof{\lb{}} = \lsn{0}^{\lstar} \loprod \lsn{1} \\ \exists\,\lbp{},\textcolor{lang}{b^{\prime\prime}}.\; \lb{} = \lbp{}^{\lstar}\loprod \textcolor{lang}{b^{\prime\prime}} \\ \evalCtxSpace{};\evalCtx{}\vdash_{\mathsf C} \sof{\lb{}} \evalArrow{} \lsnv{{\mathrm{svb}}} \\ \lemptyset{};\evalCtxSpace{};\evalCtx{}\vdash_{\mathsf C} \lee{} \xlongrightarrow{\lb{}} \leev{}\glresulttype{\lsnv{{\mathrm{svb}}}}}{\evalCtxSpace{};\evalCtx{}\vdash_{\mathsf C} \Image(\lee{}) \evalArrow{} \lR^{\glvreshape{\leev{}}{\lsnv{{\mathrm{svb}}}}{\lsnv{{\mathrm{svb}}}},\Image{},\lsnv{{\mathrm{svb}}}}}}
\and
\gpfit{\inferrule*[lab=\textsc{C-S-Coimg}]{(\lemptyset, \evalCtxSpace{})\vdash \CoImage(\lee{})\colon\svalid \\ \sof{\lb{}} = \lsn{0}^{\lstar} \loprod \lsn{1} \\ \exists\,\lbp{},\textcolor{lang}{b^{\prime\prime}}.\; \lb{} = \lbp{}^{\lstar}\loprod \textcolor{lang}{b^{\prime\prime}} \\ \evalCtxSpace{};\evalCtx{}\vdash_{\mathsf C} \sof{\lb{}} \evalArrow{} \lsnv{{\mathrm{svb}}} \\ \lemptyset{};\evalCtxSpace{};\evalCtx{}\vdash_{\mathsf C} \lee{} \xlongrightarrow{\lb{}} \leev{}\glresulttype{\lsnv{{\mathrm{svb}}}}}{\evalCtxSpace{};\evalCtx{}\vdash_{\mathsf C} \CoImage(\lee{}) \evalArrow{} \lR^{\glvreshape{\leev{}}{\lsnv{{\mathrm{svb}}}}{\lsnv{{\mathrm{svb}}}},\CoImage{},\lsnv{{\mathrm{svb}}}}}}
\and
\gpfit{\inferrule*[lab=\textsc{C-S-Tprod}]{(\lemptyset, \evalCtxSpace{})\vdash \lsn{0} \loprod \lsn{1}\colon\svalid \\ \forall i.\; \evalCtxSpace{};\evalCtx{}\vdash_{\mathsf C} \lsn{i} \evalArrow{} \lsnv{i}}{\evalCtxSpace{};\evalCtx{}\vdash_{\mathsf C} \lsn{0} \loprod \lsn{1} \evalArrow{} \lsnv{0} \loprod \lsnv{1}}}
\and
\gpfit{\inferrule*[lab=\textsc{C-S-Dsum}]{(\lemptyset, \evalCtxSpace{})\vdash \lsn{0} \losum \lsn{1}\colon\svalid \\ \forall i.\; \evalCtxSpace{};\evalCtx{}\vdash_{\mathsf C} \lsn{i} \evalArrow{} \lsnv{i}}{\evalCtxSpace{};\evalCtx{}\vdash_{\mathsf C} \lsn{0} \losum \lsn{1} \evalArrow{} \lsnv{0} \losum \lsnv{1}}}
}
\gldefjudgment{c.ty}{\lepsilon{};\evalCtx{}\vdash_{\mathsf C} \lty{}\evalArrow{}\ltyv{}}{Checked type evaluation (generated from \texttt{C.Ty}).}{fig:gl-c-ty}
\gldefrule{c.ty.space}{c.ty}{\textsc{C-T-Space}}{\evalCtxSpace{};\evalCtx{}\vdash_{\mathsf C} \ls{} \evalArrow{} \lsv{}}{\evalCtxSpace{};\evalCtx{}\vdash_{\mathsf C} \ls{} \evalArrow{} \lsv{}}{\texttt{GLang.Raw\allowbreak Certified\allowbreak Operational.C.Ty.space}}{GLang/Adequacy/RawCertifiedOperational.lean:84}
\gldefrule{c.ty.piTy}{c.ty}{\textsc{C-T-Pi\allowbreak Ty}}{(\lemptyset, \evalCtxSpace{})\vdash (x\colon \ltyn{0}\fnto \ltyn{1})\colon \lkn{0} \fnto \lkn{1} \\ \tof{\lbn{0}} = \ltyn{0} \\ (\emptyset,\evalCtxSpace{})\vdash \lbn{0}\colon \bvalid{} \\ \evalCtxSpace{};\evalCtx{}\vdash_{\mathsf C} \ltyn{0} \evalArrow{} \ltynv{0} \\ (\evalCtxSpace{}, \mathit{x}\colon \ltyn{0});\evalCtx{}[(\mathit{x},\lbn{0}) \mapsto \mathit{x}_{\ltynv{0}}]\vdash_{\mathsf C} \ltyn{1} \evalArrow{} \ltynv{1}}{\evalCtxSpace{};\evalCtx{}\vdash_{\mathsf C} (x\colon \ltyn{0}\fnto \ltyn{1}) \evalArrow{} (x\colon \ltynv{0}\fnto \ltynv{1})}{\texttt{GLang.Raw\allowbreak Certified\allowbreak Operational.C.Ty.pi\allowbreak Ty}}{GLang/Adequacy/RawCertifiedOperational.lean:87}
\gldefrulerows{c.ty}{
\textsc{C-T-Space} & $\evalCtxSpace{};\evalCtx{}\vdash_{\mathsf C} \ls{} \evalArrow{} \lsv{}$ & \texttt{\footnotesize GLang/Adequacy/Raw\allowbreak Certified\allowbreak Operational.lean:84}\\
\textsc{C-T-Pi\allowbreak Ty} & $\evalCtxSpace{};\evalCtx{}\vdash_{\mathsf C} (x\colon \ltyn{0}\fnto \ltyn{1}) \evalArrow{} (x\colon \ltynv{0}\fnto \ltynv{1})$ & \texttt{\footnotesize GLang/Adequacy/Raw\allowbreak Certified\allowbreak Operational.lean:87}\\
}{
\gpfit{\inferrule*[]{\evalCtxSpace{};\evalCtx{}\vdash_{\mathsf C} \ls{} \evalArrow{} \lsv{}}{\evalCtxSpace{};\evalCtx{}\vdash_{\mathsf C} \ls{} \evalArrow{} \lsv{}}}
\and
\gpfit{\inferrule*[]{(\lemptyset, \evalCtxSpace{})\vdash (x\colon \ltyn{0}\fnto \ltyn{1})\colon \lkn{0} \fnto \lkn{1} \\ \tof{\lbn{0}} = \ltyn{0} \\ (\emptyset,\evalCtxSpace{})\vdash \lbn{0}\colon \bvalid{} \\ \evalCtxSpace{};\evalCtx{}\vdash_{\mathsf C} \ltyn{0} \evalArrow{} \ltynv{0} \\ (\evalCtxSpace{}, \mathit{x}\colon \ltyn{0});\evalCtx{}[(\mathit{x},\lbn{0}) \mapsto \mathit{x}_{\ltynv{0}}]\vdash_{\mathsf C} \ltyn{1} \evalArrow{} \ltynv{1}}{\evalCtxSpace{};\evalCtx{}\vdash_{\mathsf C} (x\colon \ltyn{0}\fnto \ltyn{1}) \evalArrow{} (x\colon \ltynv{0}\fnto \ltynv{1})}}
}{
\gpfit{\inferrule*[lab=\textsc{C-T-Space}]{\evalCtxSpace{};\evalCtx{}\vdash_{\mathsf C} \ls{} \evalArrow{} \lsv{}}{\evalCtxSpace{};\evalCtx{}\vdash_{\mathsf C} \ls{} \evalArrow{} \lsv{}}}
\and
\gpfit{\inferrule*[lab=\textsc{C-T-Pi\allowbreak Ty}]{(\lemptyset, \evalCtxSpace{})\vdash (x\colon \ltyn{0}\fnto \ltyn{1})\colon \lkn{0} \fnto \lkn{1} \\ \tof{\lbn{0}} = \ltyn{0} \\ (\emptyset,\evalCtxSpace{})\vdash \lbn{0}\colon \bvalid{} \\ \evalCtxSpace{};\evalCtx{}\vdash_{\mathsf C} \ltyn{0} \evalArrow{} \ltynv{0} \\ (\evalCtxSpace{}, \mathit{x}\colon \ltyn{0});\evalCtx{}[(\mathit{x},\lbn{0}) \mapsto \mathit{x}_{\ltynv{0}}]\vdash_{\mathsf C} \ltyn{1} \evalArrow{} \ltynv{1}}{\evalCtxSpace{};\evalCtx{}\vdash_{\mathsf C} (x\colon \ltyn{0}\fnto \ltyn{1}) \evalArrow{} (x\colon \ltynv{0}\fnto \ltynv{1})}}
}
\gldefjudgment{c.tyAt}{\lepsilon{};\evalCtx{}\vdash_{\mathsf C} \lb{}\evalArrow{}_{\mathsf{basis}}\ltyv{}}{Checked exact-basis type evaluation (generated from \texttt{C.TyAt}).}{fig:gl-c-tyat}
\gldefrule{c.tyAt.space}{c.tyAt}{\textsc{C-TA-Space}}{\tof{\lb{}} = \sof{\lb{}} \\ (\emptyset,\evalCtxSpace{})\vdash \lb{}\colon \bvalid{} \\ \evalCtxSpace{};\evalCtx{}\vdash_{\mathsf C} \sof{\lb{}} \evalArrow{} \lsv{}}{\evalCtxSpace{};\evalCtx{}\vdash_{\mathsf C} \lb{} \evalArrow{}_{\mathsf{basis}} \lsv{}}{\texttt{GLang.Raw\allowbreak Certified\allowbreak Operational.C.Ty\allowbreak At.space}}{GLang/Adequacy/RawCertifiedOperational.lean:105}
\gldefrule{c.tyAt.frame}{c.tyAt}{\textsc{C-TA-Frame}}{(\emptyset,\evalCtxSpace{})\vdash \frameB{x}{\lbn{0}}{\lbn{1}}\colon \bvalid{} \\ \evalCtxSpace{};\evalCtx{}\vdash_{\mathsf C} \lbn{0} \evalArrow{}_{\mathsf{basis}} \ltynv{0} \\ (\evalCtxSpace{}, \mathit{x}\colon \tof{\lbn{0}});\evalCtx{}[(\mathit{x},\lbn{0}) \mapsto \mathit{x}_{\ltynv{0}}]\vdash_{\mathsf C} \lbn{1} \evalArrow{}_{\mathsf{basis}} \ltynv{1}}{\evalCtxSpace{};\evalCtx{}\vdash_{\mathsf C} \frameB{x}{\lbn{0}}{\lbn{1}} \evalArrow{}_{\mathsf{basis}} (x\colon \ltynv{0}\fnto \ltynv{1})}{\texttt{GLang.Raw\allowbreak Certified\allowbreak Operational.C.Ty\allowbreak At.frame}}{GLang/Adequacy/RawCertifiedOperational.lean:110}
\gldefrulerows{c.tyAt}{
\textsc{C-TA-Space} & $\evalCtxSpace{};\evalCtx{}\vdash_{\mathsf C} \lb{} \evalArrow{}_{\mathsf{basis}} \lsv{}$ & \texttt{\footnotesize GLang/Adequacy/Raw\allowbreak Certified\allowbreak Operational.lean:105}\\
\textsc{C-TA-Frame} & $\evalCtxSpace{};\evalCtx{}\vdash_{\mathsf C} \frameB{x}{\lbn{0}}{\lbn{1}} \evalArrow{}_{\mathsf{basis}} (x\colon \ltynv{0}\fnto \ltynv{1})$ & \texttt{\footnotesize GLang/Adequacy/Raw\allowbreak Certified\allowbreak Operational.lean:110}\\
}{
\gpfit{\inferrule*[]{\tof{\lb{}} = \sof{\lb{}} \\ (\emptyset,\evalCtxSpace{})\vdash \lb{}\colon \bvalid{} \\ \evalCtxSpace{};\evalCtx{}\vdash_{\mathsf C} \sof{\lb{}} \evalArrow{} \lsv{}}{\evalCtxSpace{};\evalCtx{}\vdash_{\mathsf C} \lb{} \evalArrow{}_{\mathsf{basis}} \lsv{}}}
\and
\gpfit{\inferrule*[]{(\emptyset,\evalCtxSpace{})\vdash \frameB{x}{\lbn{0}}{\lbn{1}}\colon \bvalid{} \\ \evalCtxSpace{};\evalCtx{}\vdash_{\mathsf C} \lbn{0} \evalArrow{}_{\mathsf{basis}} \ltynv{0} \\ (\evalCtxSpace{}, \mathit{x}\colon \tof{\lbn{0}});\evalCtx{}[(\mathit{x},\lbn{0}) \mapsto \mathit{x}_{\ltynv{0}}]\vdash_{\mathsf C} \lbn{1} \evalArrow{}_{\mathsf{basis}} \ltynv{1}}{\evalCtxSpace{};\evalCtx{}\vdash_{\mathsf C} \frameB{x}{\lbn{0}}{\lbn{1}} \evalArrow{}_{\mathsf{basis}} (x\colon \ltynv{0}\fnto \ltynv{1})}}
}{
\gpfit{\inferrule*[lab=\textsc{C-TA-Space}]{\tof{\lb{}} = \sof{\lb{}} \\ (\emptyset,\evalCtxSpace{})\vdash \lb{}\colon \bvalid{} \\ \evalCtxSpace{};\evalCtx{}\vdash_{\mathsf C} \sof{\lb{}} \evalArrow{} \lsv{}}{\evalCtxSpace{};\evalCtx{}\vdash_{\mathsf C} \lb{} \evalArrow{}_{\mathsf{basis}} \lsv{}}}
\and
\gpfit{\inferrule*[lab=\textsc{C-TA-Frame}]{(\emptyset,\evalCtxSpace{})\vdash \frameB{x}{\lbn{0}}{\lbn{1}}\colon \bvalid{} \\ \evalCtxSpace{};\evalCtx{}\vdash_{\mathsf C} \lbn{0} \evalArrow{}_{\mathsf{basis}} \ltynv{0} \\ (\evalCtxSpace{}, \mathit{x}\colon \tof{\lbn{0}});\evalCtx{}[(\mathit{x},\lbn{0}) \mapsto \mathit{x}_{\ltynv{0}}]\vdash_{\mathsf C} \lbn{1} \evalArrow{}_{\mathsf{basis}} \ltynv{1}}{\evalCtxSpace{};\evalCtx{}\vdash_{\mathsf C} \frameB{x}{\lbn{0}}{\lbn{1}} \evalArrow{}_{\mathsf{basis}} (x\colon \ltynv{0}\fnto \ltynv{1})}}
}
\gldefjudgment{c.cob}{\lepsilon{};\evalCtx{}\vdash_{\mathsf C} \lbn{0}\xlongleftrightarrow{\leev{}}\lbn{1}}{Checked change of basis (generated from \texttt{C.Cob}).}{fig:gl-c-cob}
\gldefrule{c.cob.inverse}{c.cob}{\textsc{C-C-Inverse}}{\forall i.\; (\emptyset,\evalCtxSpace{})\vdash \lbn{i}\colon \bvalid{} \\ \evalCtxSpace{};\evalCtx{}\vdash_{\mathsf C} \lbn{1} \xlongleftrightarrow{\leev{}} \lbn{0}}{\evalCtxSpace{};\evalCtx{}\vdash_{\mathsf C} \lbn{0} \xlongleftrightarrow{\leev{}^{-1}} \lbn{1}}{\texttt{GLang.Raw\allowbreak Certified\allowbreak Operational.C.Cob.inverse}}{GLang/Adequacy/RawCertifiedOperational.lean:125}
\gldefrule{c.cob.trans}{c.cob}{\textsc{C-C-Trans}}{(\emptyset,\evalCtxSpace{})\vdash \lbn{0}\colon \bvalid{} \\ (\emptyset,\evalCtxSpace{})\vdash \lbn{2}\colon \bvalid{} \\ \evalCtxSpace{};\evalCtx{}\vdash_{\mathsf C} \lbn{0} \xlongleftrightarrow{\lenv{0}} \lbn{1} \\ \evalCtxSpace{};\evalCtx{}\vdash_{\mathsf C} \lbn{1} \xlongleftrightarrow{\lenv{1}} \lbn{2} \\ \evalCtxSpace{};\evalCtx{}\vdash_{\mathsf C} \sof{\lbn{0}} \evalArrow{} \lsv{}}{\evalCtxSpace{};\evalCtx{}\vdash_{\mathsf C} \lbn{0} \xlongleftrightarrow{\glvmtm{\lsv{}}{\lsv{}}{\lenv{1}}{\glvreshape{\lenv{0}}{\lsv{} \loprod \lsv{}}{\lsv{} \loprod \lsv{}}}} \lbn{2}}{\texttt{GLang.Raw\allowbreak Certified\allowbreak Operational.C.Cob.trans}}{GLang/Adequacy/RawCertifiedOperational.lean:130}
\gldefrule{c.cob.tprod}{c.cob}{\textsc{C-C-Tprod}}{(\emptyset,\evalCtxSpace{})\vdash \lbn{0}' \loprod \lbn{1}'\colon \bvalid{} \\ (\emptyset,\evalCtxSpace{})\vdash \lbn{0} \loprod \lbn{1}\colon \bvalid{} \\ \forall i.\; \evalCtxSpace{};\evalCtx{}\vdash_{\mathsf C} \lbn{i}' \xlongleftrightarrow{\lenv{i}} \lbn{i}}{\evalCtxSpace{};\evalCtx{}\vdash_{\mathsf C} \lbn{0}' \loprod \lbn{1}' \xlongleftrightarrow{\vkrn(\lenv{0}, \lenv{1})} \lbn{0} \loprod \lbn{1}}{\texttt{GLang.Raw\allowbreak Certified\allowbreak Operational.C.Cob.tprod}}{GLang/Adequacy/RawCertifiedOperational.lean:138}
\gldefrule{c.cob.dsum}{c.cob}{\textsc{C-C-Dsum}}{(\emptyset,\evalCtxSpace{})\vdash \lbn{0}' \losum \lbn{1}'\colon \bvalid{} \\ (\emptyset,\evalCtxSpace{})\vdash \lbn{0} \losum \lbn{1}\colon \bvalid{} \\ \forall i.\; \evalCtxSpace{};\evalCtx{}\vdash_{\mathsf C} \lbn{i}' \xlongleftrightarrow{\lenv{i}} \lbn{i}}{\evalCtxSpace{};\evalCtx{}\vdash_{\mathsf C} \lbn{0}' \losum \lbn{1}' \xlongleftrightarrow{\vblk(\lenv{0}, \lenv{1})} \lbn{0} \losum \lbn{1}}{\texttt{GLang.Raw\allowbreak Certified\allowbreak Operational.C.Cob.dsum}}{GLang/Adequacy/RawCertifiedOperational.lean:145}
\gldefrule{c.cob.dualStd}{c.cob}{\textsc{C-C-Dual\allowbreak Std}}{\tof{\lbn{0}} = \sof{\lbn{0}} \\ (\emptyset,\evalCtxSpace{})\vdash \lbn{0}^{\lstar}\colon \bvalid{} \\ (\emptyset,\evalCtxSpace{})\vdash \stdBasisDCtx{\tof{\lbn{0}^{\lstar}}}\colon \bvalid{} \\ \evalCtxSpace{};\evalCtx{}\vdash_{\mathsf C} \lbn{0} \xlongleftrightarrow{\leev{}} \stdBasisDCtx{\tof{\lbn{0}}}}{\evalCtxSpace{};\evalCtx{}\vdash_{\mathsf C} \lbn{0}^{\lstar} \xlongleftrightarrow{\Trans(\leev{}^{-1})} \stdBasisDCtx{\tof{\lbn{0}^{\lstar}}}}{\texttt{GLang.Raw\allowbreak Certified\allowbreak Operational.C.Cob.dual\_\allowbreak std}}{GLang/Adequacy/RawCertifiedOperational.lean:151}
\gldefrule{c.cob.image}{c.cob}{\textsc{C-C-Image}}{(\emptyset,\evalCtxSpace{})\vdash \iBasisC{\lbn{0}}\colon \bvalid{} \\ (\emptyset,\evalCtxSpace{})\vdash \iBasisC{\lbn{0}'}\colon \bvalid{} \\ \exists \lsn{0}.\; \exists \lsn{1}.\; \exists \lee{}.\; \sof{\lbn{0}} = \CoKernel(\lee{}) \\ \evalCtxSpace{};\evalCtx{}\vdash_{\mathsf C} \lbn{0} \xlongleftrightarrow{\leev{}} \lbn{0}'}{\evalCtxSpace{};\evalCtx{}\vdash_{\mathsf C} \iBasisC{\lbn{0}} \xlongleftrightarrow{\leev{}} \iBasisC{\lbn{0}'}}{\texttt{GLang.Raw\allowbreak Certified\allowbreak Operational.C.Cob.image}}{GLang/Adequacy/RawCertifiedOperational.lean:157}
\gldefrule{c.cob.imageSvds}{c.cob}{\textsc{C-C-Image\allowbreak Svds}}{(\emptyset,\evalCtxSpace{})\vdash \iBasisC{\pBasisC{\CoKernel(\lee{})}{\lbn{0}}{\lbn{1}}}\colon \bvalid{} \\ (\emptyset,\evalCtxSpace{})\vdash \pBasisC{\Image(\lee{})}{\lbn{0}}{\lbn{1}}\colon \bvalid{} \\ \forall i.\; \sof{\lbn{i}} = \lsn{i} \\ \forall i.\; \isStdBasis{\lbn{i}} \\ \evalCtxSpace{};\evalCtx{}\vdash_{\mathsf C} \sof{\lbn{0}^{\lstar} \loprod \lbn{1}} \evalArrow{} \lsnv{0} \loprod \lsnv{1} \\ \lemptyset{};\evalCtxSpace{};\evalCtx{}\vdash_{\mathsf C} \lee{} \xlongrightarrow{\lbn{0}^{\lstar} \loprod \lbn{1}} \leev{}\glresulttype{\lsnv{0} \loprod \lsnv{1}} \\ \lenv{{\mathrm{M}}} = \glvreshape{\leev{}}{\lsnv{0} \loprod \lsnv{1}}{\lsnv{0} \loprod \lsnv{1}} \\ \lenv{{\mathrm{U}}} = \glvop{GSVDU}(\lenv{{\mathrm{M}}},\lidelta{\lsnv{0}},\lidelta{\lsnv{1}}) \\ \lenv{{\mathrm{V}}} = \glvop{GSVDV}(\lenv{{\mathrm{M}}},\lidelta{\lsnv{0}},\lidelta{\lsnv{1}}) \\ \lenv{{\mathrm{P}}} = \glvmtm{\lsnv{1}}{\lsnv{1}}{\lenv{{\mathrm{V}}}}{\glvmtm{\lsnv{0}}{\lsnv{0}}{\lenv{{\mathrm{M}}}}{\Trans(\lenv{{\mathrm{U}}})}}}{\evalCtxSpace{};\evalCtx{}\vdash_{\mathsf C} \iBasisC{\pBasisC{\CoKernel(\lee{})}{\lbn{0}}{\lbn{1}}} \xlongleftrightarrow{\glvreshape{\lenv{{\mathrm{P}}}}{\lR^{\lenv{{\mathrm{M}}},\CoKernel{},\lsnv{0} \loprod \lsnv{1}} \loprod \lR^{\lenv{{\mathrm{M}}},\Image{},\lsnv{0} \loprod \lsnv{1}}}{\operatorname{inferSpace}(\lenv{{\mathrm{P}}})}} \pBasisC{\Image(\lee{})}{\lbn{0}}{\lbn{1}}}{\texttt{GLang.Raw\allowbreak Certified\allowbreak Operational.C.Cob.image\_\allowbreak svds}}{GLang/Adequacy/RawCertifiedOperational.lean:163}
\gldefrule{c.cob.projKerCoker}{c.cob}{\textsc{C-C-Proj\allowbreak Ker\allowbreak Coker}}{(\emptyset,\evalCtxSpace{})\vdash \pBasisC{\ls{}}{\lbn{0}}{\lbn{1}}\colon \bvalid{} \\ (\emptyset,\evalCtxSpace{})\vdash \pBasisC{\ls{}}{\lbn{0}'}{\lbn{1}}\colon \bvalid{} \\ \ls{} = \Kernel(\lee{}) \;\vee\; \ls{} = \CoKernel(\lee{}) \\ \forall i.\; \sof{\lbn{i}} = \lsn{i} \\ \forall i.\; \evalCtxSpace{};\evalCtx{}\vdash_{\mathsf C} \sof{\lbn{i}} \evalArrow{} \lsnv{i} \\ \evalCtxSpace{};\evalCtx{}\vdash_{\mathsf C} \ls{} \evalArrow{} \lsv{} \\ \evalCtxSpace{};\evalCtx{}\vdash_{\mathsf C} \ls{} \evalArrow{} \lsv{}' \\ \evalCtxSpace{};\evalCtx{}\vdash_{\mathsf C} \sof{\lbn{0}'} \evalArrow{} \lsnv{0}' \\ \evalCtxSpace{};\evalCtx{}\vdash_{\mathsf C} \lbn{0} \xlongleftrightarrow{\lenv{2}} \lbn{0}' \\ \lemptyset{};\evalCtxSpace{};\evalCtx{}\vdash_{\mathsf C} \lembd{}_{\ls{}} \xlongrightarrow{\pBasisC{\ls{}}{\lbn{0}}{\lbn{1}}^{\lstar} \loprod \lbn{0}} \lenv{3}\glresulttype{\lsv{} \loprod \lsnv{0}} \\ \lemptyset{};\evalCtxSpace{};\evalCtx{}\vdash_{\mathsf C} \lproj{}_{\ls{}, \sof{\lbn{0}'}} \xlongrightarrow{\lbn{0}'^{\lstar} \loprod \pBasisC{\ls{}}{\lbn{0}'}{\lbn{1}}} \lenv{1}\glresulttype{\lsnv{0}' \loprod \lsv{}'}}{\evalCtxSpace{};\evalCtx{}\vdash_{\mathsf C} \pBasisC{\ls{}}{\lbn{0}}{\lbn{1}} \xlongleftrightarrow{\glvmtm{\lsnv{0}'}{\lsnv{0}'}{\glvreshape{\lenv{1}}{\lsnv{0}' \loprod \lsv{}'}{\lsnv{0}' \loprod \lsv{}'}}{\glvmtm{\lsnv{0}}{\lsnv{0}}{\glvreshape{\lenv{2}}{\lsnv{0} \loprod \lsnv{0}'}{\lsnv{0} \loprod \lsnv{0}'}}{\glvreshape{\lenv{3}}{\lsv{} \loprod \lsnv{0}}{\lsv{} \loprod \lsnv{0}}}}} \pBasisC{\ls{}}{\lbn{0}'}{\lbn{1}}}{\texttt{GLang.Raw\allowbreak Certified\allowbreak Operational.C.Cob.proj\_\allowbreak ker\_\allowbreak coker}}{GLang/Adequacy/RawCertifiedOperational.lean:179}
\gldefrule{c.cob.projImgCoimg}{c.cob}{\textsc{C-C-Proj\allowbreak Img\allowbreak Coimg}}{(\emptyset,\evalCtxSpace{})\vdash \pBasisC{\ls{}}{\lbn{0}}{\lbn{1}}\colon \bvalid{} \\ (\emptyset,\evalCtxSpace{})\vdash \pBasisC{\ls{}}{\lbn{0}}{\lbn{1}'}\colon \bvalid{} \\ \ls{} = \Image(\lee{}) \;\vee\; \ls{} = \CoImage(\lee{}) \\ \forall i.\; \sof{\lbn{i}} = \lsn{i} \\ \forall i.\; \evalCtxSpace{};\evalCtx{}\vdash_{\mathsf C} \sof{\lbn{i}} \evalArrow{} \lsnv{i} \\ \evalCtxSpace{};\evalCtx{}\vdash_{\mathsf C} \ls{} \evalArrow{} \lsv{} \\ \evalCtxSpace{};\evalCtx{}\vdash_{\mathsf C} \ls{} \evalArrow{} \lsv{}' \\ \evalCtxSpace{};\evalCtx{}\vdash_{\mathsf C} \sof{\lbn{1}'} \evalArrow{} \lsnv{1}' \\ \evalCtxSpace{};\evalCtx{}\vdash_{\mathsf C} \lbn{1} \xlongleftrightarrow{\lenv{2}} \lbn{1}' \\ \lemptyset{};\evalCtxSpace{};\evalCtx{}\vdash_{\mathsf C} \lembd{}_{\ls{}} \xlongrightarrow{\pBasisC{\ls{}}{\lbn{0}}{\lbn{1}}^{\lstar} \loprod \lbn{1}} \lenv{3}\glresulttype{\lsv{} \loprod \lsnv{1}} \\ \lemptyset{};\evalCtxSpace{};\evalCtx{}\vdash_{\mathsf C} \lproj{}_{\ls{}, \sof{\lbn{1}'}} \xlongrightarrow{\lbn{1}'^{\lstar} \loprod \pBasisC{\ls{}}{\lbn{0}}{\lbn{1}'}} \lenv{1}\glresulttype{\lsnv{1}' \loprod \lsv{}'}}{\evalCtxSpace{};\evalCtx{}\vdash_{\mathsf C} \pBasisC{\ls{}}{\lbn{0}}{\lbn{1}} \xlongleftrightarrow{\glvmtm{\lsnv{1}'}{\lsnv{1}'}{\glvreshape{\lenv{1}}{\lsnv{1}' \loprod \lsv{}'}{\lsnv{1}' \loprod \lsv{}'}}{\glvmtm{\lsnv{1}}{\lsnv{1}}{\glvreshape{\lenv{2}}{\lsnv{1} \loprod \lsnv{1}'}{\lsnv{1} \loprod \lsnv{1}'}}{\glvreshape{\lenv{3}}{\lsv{} \loprod \lsnv{1}}{\lsv{} \loprod \lsnv{1}}}}} \pBasisC{\ls{}}{\lbn{0}}{\lbn{1}'}}{\texttt{GLang.Raw\allowbreak Certified\allowbreak Operational.C.Cob.proj\_\allowbreak img\_\allowbreak coimg}}{GLang/Adequacy/RawCertifiedOperational.lean:200}
\gldefrule{c.cob.termsRref}{c.cob}{\textsc{C-C-Terms\allowbreak Rref}}{(\emptyset,\evalCtxSpace{})\vdash \lb{}\colon \bvalid{} \\ (\emptyset,\evalCtxSpace{})\vdash \llpl{} \lee{} \mathbin{:\!:} \mathit{es}\llpr{}_{\lb{}'}\colon \bvalid{} \\ \evalCtxSpace{};\evalCtx{}\vdash_{\mathsf C} \lb{} \xlongleftrightarrow{\lenv{1}} \lb{}' \\ \lemptyset{};\evalCtxSpace{};\evalCtx{}\vdash_{\mathsf C} \lee{} \losum \mathit{es}[0] \losum \cdots \xlongrightarrow{\lb{}' \losum \cdots} \lenv{0}\glresulttype{\ltyv{}} \\ \evalCtxSpace{};\evalCtx{}\vdash_{\mathsf C} \sof{\lb{}} \evalArrow{} \lsv{}}{\evalCtxSpace{};\evalCtx{}\vdash_{\mathsf C} \lb{} \xlongleftrightarrow{\glvreshape{\glvop{RREFRep}(\lenv{0},\glvreshape{\lenv{1}}{\lsv{} \loprod \lsv{}}{\lsv{} \loprod \lsv{}})}{\lsv{} \loprod \lsv{}}{\lsv{} \loprod \lsv{}}} \llpl{} \lee{} \mathbin{:\!:} \mathit{es}\llpr{}_{\lb{}'}}{\texttt{GLang.Raw\allowbreak Certified\allowbreak Operational.C.Cob.terms\_\allowbreak rref}}{GLang/Adequacy/RawCertifiedOperational.lean:221}
\gldefrule{c.cob.refl}{c.cob}{\textsc{C-C-Refl}}{(\emptyset,\evalCtxSpace{})\vdash \lb{}\colon \bvalid{} \\ \tof{\lb{}} = \sof{\lb{}} \\ \evalCtxSpace{};\evalCtx{}\vdash_{\mathsf C} \sof{\lb{}} \evalArrow{} \lsv{}}{\evalCtxSpace{};\evalCtx{}\vdash_{\mathsf C} \lb{} \xlongleftrightarrow{\glvreshape{\lidelta{\lsv{}}}{\lsv{} \loprod \lsv{}}{\lsv{} \loprod \lsv{}}} \lb{}}{\texttt{GLang.Raw\allowbreak Certified\allowbreak Operational.C.Cob.refl}}{GLang/Adequacy/RawCertifiedOperational.lean:234}
\gldefrule{c.cob.projKerCokerCod}{c.cob}{\textsc{C-C-Proj\allowbreak Ker\allowbreak Coker\allowbreak Cod}}{(\emptyset,\evalCtxSpace{})\vdash \pBasisC{\ls{}}{\lbn{0}}{\lbn{1}}\colon \bvalid{} \\ (\emptyset,\evalCtxSpace{})\vdash \pBasisC{\ls{}}{\lbn{0}}{\lbn{1}'}\colon \bvalid{} \\ \ls{} = \Kernel(\lee{}) \;\vee\; \ls{} = \CoKernel(\lee{}) \\ \forall i.\; \sof{\lbn{i}} = \lsn{i} \\ \sof{\lbn{1}'} = \lsn{1} \\ \evalCtxSpace{};\evalCtx{}\vdash_{\mathsf C} \sof{\lbn{0}} \evalArrow{} \lsnv{0} \\ \evalCtxSpace{};\evalCtx{}\vdash_{\mathsf C} \ls{} \evalArrow{} \lsv{} \\ \evalCtxSpace{};\evalCtx{}\vdash_{\mathsf C} \ls{} \evalArrow{} \lsv{}' \\ \lemptyset{};\evalCtxSpace{};\evalCtx{}\vdash_{\mathsf C} \lembd{}_{\ls{}} \xlongrightarrow{\pBasisC{\ls{}}{\lbn{0}}{\lbn{1}}^{\lstar} \loprod \lbn{0}} \lenv{3}\glresulttype{\lsv{} \loprod \lsnv{0}} \\ \lemptyset{};\evalCtxSpace{};\evalCtx{}\vdash_{\mathsf C} \lproj{}_{\ls{}, \sof{\lbn{0}}} \xlongrightarrow{\lbn{0}^{\lstar} \loprod \pBasisC{\ls{}}{\lbn{0}}{\lbn{1}'}} \lenv{1}\glresulttype{\lsnv{0} \loprod \lsv{}'}}{\evalCtxSpace{};\evalCtx{}\vdash_{\mathsf C} \pBasisC{\ls{}}{\lbn{0}}{\lbn{1}} \xlongleftrightarrow{\glvmtm{\lsnv{0}}{\lsnv{0}}{\glvreshape{\lenv{1}}{\lsnv{0} \loprod \lsv{}'}{\lsnv{0} \loprod \lsv{}'}}{\glvreshape{\lenv{3}}{\lsv{} \loprod \lsnv{0}}{\lsv{} \loprod \lsnv{0}}}} \pBasisC{\ls{}}{\lbn{0}}{\lbn{1}'}}{\texttt{GLang.Raw\allowbreak Certified\allowbreak Operational.C.Cob.proj\_\allowbreak ker\_\allowbreak coker\_\allowbreak cod}}{GLang/Adequacy/RawCertifiedOperational.lean:241}
\gldefrule{c.cob.projImgCoimgDom}{c.cob}{\textsc{C-C-Proj\allowbreak Img\allowbreak Coimg\allowbreak Dom}}{(\emptyset,\evalCtxSpace{})\vdash \pBasisC{\ls{}}{\lbn{0}}{\lbn{1}}\colon \bvalid{} \\ (\emptyset,\evalCtxSpace{})\vdash \pBasisC{\ls{}}{\lbn{0}'}{\lbn{1}}\colon \bvalid{} \\ \ls{} = \Image(\lee{}) \;\vee\; \ls{} = \CoImage(\lee{}) \\ \forall i.\; \sof{\lbn{i}} = \lsn{i} \\ \sof{\lbn{0}'} = \lsn{0} \\ \evalCtxSpace{};\evalCtx{}\vdash_{\mathsf C} \sof{\lbn{1}} \evalArrow{} \lsnv{1} \\ \evalCtxSpace{};\evalCtx{}\vdash_{\mathsf C} \ls{} \evalArrow{} \lsv{} \\ \evalCtxSpace{};\evalCtx{}\vdash_{\mathsf C} \ls{} \evalArrow{} \lsv{}' \\ \lemptyset{};\evalCtxSpace{};\evalCtx{}\vdash_{\mathsf C} \lembd{}_{\ls{}} \xlongrightarrow{\pBasisC{\ls{}}{\lbn{0}}{\lbn{1}}^{\lstar} \loprod \lbn{1}} \lenv{3}\glresulttype{\lsv{} \loprod \lsnv{1}} \\ \lemptyset{};\evalCtxSpace{};\evalCtx{}\vdash_{\mathsf C} \lproj{}_{\ls{}, \sof{\lbn{1}}} \xlongrightarrow{\lbn{1}^{\lstar} \loprod \pBasisC{\ls{}}{\lbn{0}'}{\lbn{1}}} \lenv{1}\glresulttype{\lsnv{1} \loprod \lsv{}'}}{\evalCtxSpace{};\evalCtx{}\vdash_{\mathsf C} \pBasisC{\ls{}}{\lbn{0}}{\lbn{1}} \xlongleftrightarrow{\glvmtm{\lsnv{1}}{\lsnv{1}}{\glvreshape{\lenv{1}}{\lsnv{1} \loprod \lsv{}'}{\lsnv{1} \loprod \lsv{}'}}{\glvreshape{\lenv{3}}{\lsv{} \loprod \lsnv{1}}{\lsv{} \loprod \lsnv{1}}}} \pBasisC{\ls{}}{\lbn{0}'}{\lbn{1}}}{\texttt{GLang.Raw\allowbreak Certified\allowbreak Operational.C.Cob.proj\_\allowbreak img\_\allowbreak coimg\_\allowbreak dom}}{GLang/Adequacy/RawCertifiedOperational.lean:258}
\gldefrulerows{c.cob}{
\textsc{C-C-Inverse} & $\evalCtxSpace{};\evalCtx{}\vdash_{\mathsf C} \lbn{0} \xlongleftrightarrow{\leev{}^{-1}} \lbn{1}$ & \texttt{\footnotesize GLang/Adequacy/Raw\allowbreak Certified\allowbreak Operational.lean:125}\\
\textsc{C-C-Trans} & $\evalCtxSpace{};\evalCtx{}\vdash_{\mathsf C} \lbn{0} \xlongleftrightarrow{\glvmtm{\lsv{}}{\lsv{}}{\lenv{1}}{\glvreshape{\lenv{0}}{\lsv{} \loprod \lsv{}}{\lsv{} \loprod \lsv{}}}} \lbn{2}$ & \texttt{\footnotesize GLang/Adequacy/Raw\allowbreak Certified\allowbreak Operational.lean:130}\\
\textsc{C-C-Tprod} & $\evalCtxSpace{};\evalCtx{}\vdash_{\mathsf C} \lbn{0}' \loprod \lbn{1}' \xlongleftrightarrow{\vkrn(\lenv{0}, \lenv{1})} \lbn{0} \loprod \lbn{1}$ & \texttt{\footnotesize GLang/Adequacy/Raw\allowbreak Certified\allowbreak Operational.lean:138}\\
\textsc{C-C-Dsum} & $\evalCtxSpace{};\evalCtx{}\vdash_{\mathsf C} \lbn{0}' \losum \lbn{1}' \xlongleftrightarrow{\vblk(\lenv{0}, \lenv{1})} \lbn{0} \losum \lbn{1}$ & \texttt{\footnotesize GLang/Adequacy/Raw\allowbreak Certified\allowbreak Operational.lean:145}\\
\textsc{C-C-Dual\allowbreak Std} & $\evalCtxSpace{};\evalCtx{}\vdash_{\mathsf C} \lbn{0}^{\lstar} \xlongleftrightarrow{\Trans(\leev{}^{-1})} \stdBasisDCtx{\tof{\lbn{0}^{\lstar}}}$ & \texttt{\footnotesize GLang/Adequacy/Raw\allowbreak Certified\allowbreak Operational.lean:151}\\
\textsc{C-C-Image} & $\evalCtxSpace{};\evalCtx{}\vdash_{\mathsf C} \iBasisC{\lbn{0}} \xlongleftrightarrow{\leev{}} \iBasisC{\lbn{0}'}$ & \texttt{\footnotesize GLang/Adequacy/Raw\allowbreak Certified\allowbreak Operational.lean:157}\\
\textsc{C-C-Image\allowbreak Svds} & $\evalCtxSpace{};\evalCtx{}\vdash_{\mathsf C} \iBasisC{\pBasisC{\CoKernel(\lee{})}{\lbn{0}}{\lbn{1}}} \xlongleftrightarrow{\glvreshape{\lenv{{\mathrm{P}}}}{\lR^{\lenv{{\mathrm{M}}},\CoKernel{},\lsnv{0} \loprod \lsnv{1}} \loprod \lR^{\lenv{{\mathrm{M}}},\Image{},\lsnv{0} \loprod \lsnv{1}}}{\operatorname{inferSpace}(\lenv{{\mathrm{P}}})}} \pBasisC{\Image(\lee{})}{\lbn{0}}{\lbn{1}}$ & \texttt{\footnotesize GLang/Adequacy/Raw\allowbreak Certified\allowbreak Operational.lean:163}\\
\textsc{C-C-Proj\allowbreak Ker\allowbreak Coker} & $\evalCtxSpace{};\evalCtx{}\vdash_{\mathsf C} \pBasisC{\ls{}}{\lbn{0}}{\lbn{1}} \xlongleftrightarrow{\glvmtm{\lsnv{0}'}{\lsnv{0}'}{\glvreshape{\lenv{1}}{\lsnv{0}' \loprod \lsv{}'}{\lsnv{0}' \loprod \lsv{}'}}{\glvmtm{\lsnv{0}}{\lsnv{0}}{\glvreshape{\lenv{2}}{\lsnv{0} \loprod \lsnv{0}'}{\lsnv{0} \loprod \lsnv{0}'}}{\glvreshape{\lenv{3}}{\lsv{} \loprod \lsnv{0}}{\lsv{} \loprod \lsnv{0}}}}} \pBasisC{\ls{}}{\lbn{0}'}{\lbn{1}}$ & \texttt{\footnotesize GLang/Adequacy/Raw\allowbreak Certified\allowbreak Operational.lean:179}\\
\textsc{C-C-Proj\allowbreak Img\allowbreak Coimg} & $\evalCtxSpace{};\evalCtx{}\vdash_{\mathsf C} \pBasisC{\ls{}}{\lbn{0}}{\lbn{1}} \xlongleftrightarrow{\glvmtm{\lsnv{1}'}{\lsnv{1}'}{\glvreshape{\lenv{1}}{\lsnv{1}' \loprod \lsv{}'}{\lsnv{1}' \loprod \lsv{}'}}{\glvmtm{\lsnv{1}}{\lsnv{1}}{\glvreshape{\lenv{2}}{\lsnv{1} \loprod \lsnv{1}'}{\lsnv{1} \loprod \lsnv{1}'}}{\glvreshape{\lenv{3}}{\lsv{} \loprod \lsnv{1}}{\lsv{} \loprod \lsnv{1}}}}} \pBasisC{\ls{}}{\lbn{0}}{\lbn{1}'}$ & \texttt{\footnotesize GLang/Adequacy/Raw\allowbreak Certified\allowbreak Operational.lean:200}\\
\textsc{C-C-Terms\allowbreak Rref} & $\evalCtxSpace{};\evalCtx{}\vdash_{\mathsf C} \lb{} \xlongleftrightarrow{\glvreshape{\glvop{RREFRep}(\lenv{0},\glvreshape{\lenv{1}}{\lsv{} \loprod \lsv{}}{\lsv{} \loprod \lsv{}})}{\lsv{} \loprod \lsv{}}{\lsv{} \loprod \lsv{}}} \llpl{} \lee{} \mathbin{:\!:} \mathit{es}\llpr{}_{\lb{}'}$ & \texttt{\footnotesize GLang/Adequacy/Raw\allowbreak Certified\allowbreak Operational.lean:221}\\
\textsc{C-C-Refl} & $\evalCtxSpace{};\evalCtx{}\vdash_{\mathsf C} \lb{} \xlongleftrightarrow{\glvreshape{\lidelta{\lsv{}}}{\lsv{} \loprod \lsv{}}{\lsv{} \loprod \lsv{}}} \lb{}$ & \texttt{\footnotesize GLang/Adequacy/Raw\allowbreak Certified\allowbreak Operational.lean:234}\\
\textsc{C-C-Proj\allowbreak Ker\allowbreak Coker\allowbreak Cod} & $\evalCtxSpace{};\evalCtx{}\vdash_{\mathsf C} \pBasisC{\ls{}}{\lbn{0}}{\lbn{1}} \xlongleftrightarrow{\glvmtm{\lsnv{0}}{\lsnv{0}}{\glvreshape{\lenv{1}}{\lsnv{0} \loprod \lsv{}'}{\lsnv{0} \loprod \lsv{}'}}{\glvreshape{\lenv{3}}{\lsv{} \loprod \lsnv{0}}{\lsv{} \loprod \lsnv{0}}}} \pBasisC{\ls{}}{\lbn{0}}{\lbn{1}'}$ & \texttt{\footnotesize GLang/Adequacy/Raw\allowbreak Certified\allowbreak Operational.lean:241}\\
\textsc{C-C-Proj\allowbreak Img\allowbreak Coimg\allowbreak Dom} & $\evalCtxSpace{};\evalCtx{}\vdash_{\mathsf C} \pBasisC{\ls{}}{\lbn{0}}{\lbn{1}} \xlongleftrightarrow{\glvmtm{\lsnv{1}}{\lsnv{1}}{\glvreshape{\lenv{1}}{\lsnv{1} \loprod \lsv{}'}{\lsnv{1} \loprod \lsv{}'}}{\glvreshape{\lenv{3}}{\lsv{} \loprod \lsnv{1}}{\lsv{} \loprod \lsnv{1}}}} \pBasisC{\ls{}}{\lbn{0}'}{\lbn{1}}$ & \texttt{\footnotesize GLang/Adequacy/Raw\allowbreak Certified\allowbreak Operational.lean:258}\\
}{
\gpfit{\inferrule*[]{\forall i.\; (\emptyset,\evalCtxSpace{})\vdash \lbn{i}\colon \bvalid{} \\ \evalCtxSpace{};\evalCtx{}\vdash_{\mathsf C} \lbn{1} \xlongleftrightarrow{\leev{}} \lbn{0}}{\evalCtxSpace{};\evalCtx{}\vdash_{\mathsf C} \lbn{0} \xlongleftrightarrow{\leev{}^{-1}} \lbn{1}}}
\and
\gpfit{\inferrule*[]{(\emptyset,\evalCtxSpace{})\vdash \lbn{0}\colon \bvalid{} \\ (\emptyset,\evalCtxSpace{})\vdash \lbn{2}\colon \bvalid{} \\ \evalCtxSpace{};\evalCtx{}\vdash_{\mathsf C} \lbn{0} \xlongleftrightarrow{\lenv{0}} \lbn{1} \\ \evalCtxSpace{};\evalCtx{}\vdash_{\mathsf C} \lbn{1} \xlongleftrightarrow{\lenv{1}} \lbn{2} \\ \evalCtxSpace{};\evalCtx{}\vdash_{\mathsf C} \sof{\lbn{0}} \evalArrow{} \lsv{}}{\evalCtxSpace{};\evalCtx{}\vdash_{\mathsf C} \lbn{0} \xlongleftrightarrow{\glvmtm{\lsv{}}{\lsv{}}{\lenv{1}}{\glvreshape{\lenv{0}}{\lsv{} \loprod \lsv{}}{\lsv{} \loprod \lsv{}}}} \lbn{2}}}
\and
\gpfit{\inferrule*[]{(\emptyset,\evalCtxSpace{})\vdash \lbn{0}' \loprod \lbn{1}'\colon \bvalid{} \\ (\emptyset,\evalCtxSpace{})\vdash \lbn{0} \loprod \lbn{1}\colon \bvalid{} \\ \forall i.\; \evalCtxSpace{};\evalCtx{}\vdash_{\mathsf C} \lbn{i}' \xlongleftrightarrow{\lenv{i}} \lbn{i}}{\evalCtxSpace{};\evalCtx{}\vdash_{\mathsf C} \lbn{0}' \loprod \lbn{1}' \xlongleftrightarrow{\vkrn(\lenv{0}, \lenv{1})} \lbn{0} \loprod \lbn{1}}}
\and
\gpfit{\inferrule*[]{(\emptyset,\evalCtxSpace{})\vdash \lbn{0}' \losum \lbn{1}'\colon \bvalid{} \\ (\emptyset,\evalCtxSpace{})\vdash \lbn{0} \losum \lbn{1}\colon \bvalid{} \\ \forall i.\; \evalCtxSpace{};\evalCtx{}\vdash_{\mathsf C} \lbn{i}' \xlongleftrightarrow{\lenv{i}} \lbn{i}}{\evalCtxSpace{};\evalCtx{}\vdash_{\mathsf C} \lbn{0}' \losum \lbn{1}' \xlongleftrightarrow{\vblk(\lenv{0}, \lenv{1})} \lbn{0} \losum \lbn{1}}}
\and
\gpfit{\inferrule*[]{\tof{\lbn{0}} = \sof{\lbn{0}} \\ (\emptyset,\evalCtxSpace{})\vdash \lbn{0}^{\lstar}\colon \bvalid{} \\ (\emptyset,\evalCtxSpace{})\vdash \stdBasisDCtx{\tof{\lbn{0}^{\lstar}}}\colon \bvalid{} \\ \evalCtxSpace{};\evalCtx{}\vdash_{\mathsf C} \lbn{0} \xlongleftrightarrow{\leev{}} \stdBasisDCtx{\tof{\lbn{0}}}}{\evalCtxSpace{};\evalCtx{}\vdash_{\mathsf C} \lbn{0}^{\lstar} \xlongleftrightarrow{\Trans(\leev{}^{-1})} \stdBasisDCtx{\tof{\lbn{0}^{\lstar}}}}}
\and
\gpfit{\inferrule*[]{(\emptyset,\evalCtxSpace{})\vdash \iBasisC{\lbn{0}}\colon \bvalid{} \\ (\emptyset,\evalCtxSpace{})\vdash \iBasisC{\lbn{0}'}\colon \bvalid{} \\ \exists \lsn{0}.\; \exists \lsn{1}.\; \exists \lee{}.\; \sof{\lbn{0}} = \CoKernel(\lee{}) \\ \evalCtxSpace{};\evalCtx{}\vdash_{\mathsf C} \lbn{0} \xlongleftrightarrow{\leev{}} \lbn{0}'}{\evalCtxSpace{};\evalCtx{}\vdash_{\mathsf C} \iBasisC{\lbn{0}} \xlongleftrightarrow{\leev{}} \iBasisC{\lbn{0}'}}}
\and
\gpfit{\inferrule*[]{(\emptyset,\evalCtxSpace{})\vdash \iBasisC{\pBasisC{\CoKernel(\lee{})}{\lbn{0}}{\lbn{1}}}\colon \bvalid{} \\ (\emptyset,\evalCtxSpace{})\vdash \pBasisC{\Image(\lee{})}{\lbn{0}}{\lbn{1}}\colon \bvalid{} \\ \forall i.\; \sof{\lbn{i}} = \lsn{i} \\ \forall i.\; \isStdBasis{\lbn{i}} \\ \evalCtxSpace{};\evalCtx{}\vdash_{\mathsf C} \sof{\lbn{0}^{\lstar} \loprod \lbn{1}} \evalArrow{} \lsnv{0} \loprod \lsnv{1} \\ \lemptyset{};\evalCtxSpace{};\evalCtx{}\vdash_{\mathsf C} \lee{} \xlongrightarrow{\lbn{0}^{\lstar} \loprod \lbn{1}} \leev{}\glresulttype{\lsnv{0} \loprod \lsnv{1}} \\ \lenv{{\mathrm{M}}} = \glvreshape{\leev{}}{\lsnv{0} \loprod \lsnv{1}}{\lsnv{0} \loprod \lsnv{1}} \\ \lenv{{\mathrm{U}}} = \glvop{GSVDU}(\lenv{{\mathrm{M}}},\lidelta{\lsnv{0}},\lidelta{\lsnv{1}}) \\ \lenv{{\mathrm{V}}} = \glvop{GSVDV}(\lenv{{\mathrm{M}}},\lidelta{\lsnv{0}},\lidelta{\lsnv{1}}) \\ \lenv{{\mathrm{P}}} = \glvmtm{\lsnv{1}}{\lsnv{1}}{\lenv{{\mathrm{V}}}}{\glvmtm{\lsnv{0}}{\lsnv{0}}{\lenv{{\mathrm{M}}}}{\Trans(\lenv{{\mathrm{U}}})}}}{\evalCtxSpace{};\evalCtx{}\vdash_{\mathsf C} \iBasisC{\pBasisC{\CoKernel(\lee{})}{\lbn{0}}{\lbn{1}}} \xlongleftrightarrow{\glvreshape{\lenv{{\mathrm{P}}}}{\lR^{\lenv{{\mathrm{M}}},\CoKernel{},\lsnv{0} \loprod \lsnv{1}} \loprod \lR^{\lenv{{\mathrm{M}}},\Image{},\lsnv{0} \loprod \lsnv{1}}}{\operatorname{inferSpace}(\lenv{{\mathrm{P}}})}} \pBasisC{\Image(\lee{})}{\lbn{0}}{\lbn{1}}}}
\and
\gpfit{\inferrule*[]{(\emptyset,\evalCtxSpace{})\vdash \pBasisC{\ls{}}{\lbn{0}}{\lbn{1}}\colon \bvalid{} \\ (\emptyset,\evalCtxSpace{})\vdash \pBasisC{\ls{}}{\lbn{0}'}{\lbn{1}}\colon \bvalid{} \\ \ls{} = \Kernel(\lee{}) \;\vee\; \ls{} = \CoKernel(\lee{}) \\ \forall i.\; \sof{\lbn{i}} = \lsn{i} \\ \forall i.\; \evalCtxSpace{};\evalCtx{}\vdash_{\mathsf C} \sof{\lbn{i}} \evalArrow{} \lsnv{i} \\ \evalCtxSpace{};\evalCtx{}\vdash_{\mathsf C} \ls{} \evalArrow{} \lsv{} \\ \evalCtxSpace{};\evalCtx{}\vdash_{\mathsf C} \ls{} \evalArrow{} \lsv{}' \\ \evalCtxSpace{};\evalCtx{}\vdash_{\mathsf C} \sof{\lbn{0}'} \evalArrow{} \lsnv{0}' \\ \evalCtxSpace{};\evalCtx{}\vdash_{\mathsf C} \lbn{0} \xlongleftrightarrow{\lenv{2}} \lbn{0}' \\ \lemptyset{};\evalCtxSpace{};\evalCtx{}\vdash_{\mathsf C} \lembd{}_{\ls{}} \xlongrightarrow{\pBasisC{\ls{}}{\lbn{0}}{\lbn{1}}^{\lstar} \loprod \lbn{0}} \lenv{3}\glresulttype{\lsv{} \loprod \lsnv{0}} \\ \lemptyset{};\evalCtxSpace{};\evalCtx{}\vdash_{\mathsf C} \lproj{}_{\ls{}, \sof{\lbn{0}'}} \xlongrightarrow{\lbn{0}'^{\lstar} \loprod \pBasisC{\ls{}}{\lbn{0}'}{\lbn{1}}} \lenv{1}\glresulttype{\lsnv{0}' \loprod \lsv{}'}}{\evalCtxSpace{};\evalCtx{}\vdash_{\mathsf C} \pBasisC{\ls{}}{\lbn{0}}{\lbn{1}} \xlongleftrightarrow{\glvmtm{\lsnv{0}'}{\lsnv{0}'}{\glvreshape{\lenv{1}}{\lsnv{0}' \loprod \lsv{}'}{\lsnv{0}' \loprod \lsv{}'}}{\glvmtm{\lsnv{0}}{\lsnv{0}}{\glvreshape{\lenv{2}}{\lsnv{0} \loprod \lsnv{0}'}{\lsnv{0} \loprod \lsnv{0}'}}{\glvreshape{\lenv{3}}{\lsv{} \loprod \lsnv{0}}{\lsv{} \loprod \lsnv{0}}}}} \pBasisC{\ls{}}{\lbn{0}'}{\lbn{1}}}}
\and
\gpfit{\inferrule*[]{(\emptyset,\evalCtxSpace{})\vdash \pBasisC{\ls{}}{\lbn{0}}{\lbn{1}}\colon \bvalid{} \\ (\emptyset,\evalCtxSpace{})\vdash \pBasisC{\ls{}}{\lbn{0}}{\lbn{1}'}\colon \bvalid{} \\ \ls{} = \Image(\lee{}) \;\vee\; \ls{} = \CoImage(\lee{}) \\ \forall i.\; \sof{\lbn{i}} = \lsn{i} \\ \forall i.\; \evalCtxSpace{};\evalCtx{}\vdash_{\mathsf C} \sof{\lbn{i}} \evalArrow{} \lsnv{i} \\ \evalCtxSpace{};\evalCtx{}\vdash_{\mathsf C} \ls{} \evalArrow{} \lsv{} \\ \evalCtxSpace{};\evalCtx{}\vdash_{\mathsf C} \ls{} \evalArrow{} \lsv{}' \\ \evalCtxSpace{};\evalCtx{}\vdash_{\mathsf C} \sof{\lbn{1}'} \evalArrow{} \lsnv{1}' \\ \evalCtxSpace{};\evalCtx{}\vdash_{\mathsf C} \lbn{1} \xlongleftrightarrow{\lenv{2}} \lbn{1}' \\ \lemptyset{};\evalCtxSpace{};\evalCtx{}\vdash_{\mathsf C} \lembd{}_{\ls{}} \xlongrightarrow{\pBasisC{\ls{}}{\lbn{0}}{\lbn{1}}^{\lstar} \loprod \lbn{1}} \lenv{3}\glresulttype{\lsv{} \loprod \lsnv{1}} \\ \lemptyset{};\evalCtxSpace{};\evalCtx{}\vdash_{\mathsf C} \lproj{}_{\ls{}, \sof{\lbn{1}'}} \xlongrightarrow{\lbn{1}'^{\lstar} \loprod \pBasisC{\ls{}}{\lbn{0}}{\lbn{1}'}} \lenv{1}\glresulttype{\lsnv{1}' \loprod \lsv{}'}}{\evalCtxSpace{};\evalCtx{}\vdash_{\mathsf C} \pBasisC{\ls{}}{\lbn{0}}{\lbn{1}} \xlongleftrightarrow{\glvmtm{\lsnv{1}'}{\lsnv{1}'}{\glvreshape{\lenv{1}}{\lsnv{1}' \loprod \lsv{}'}{\lsnv{1}' \loprod \lsv{}'}}{\glvmtm{\lsnv{1}}{\lsnv{1}}{\glvreshape{\lenv{2}}{\lsnv{1} \loprod \lsnv{1}'}{\lsnv{1} \loprod \lsnv{1}'}}{\glvreshape{\lenv{3}}{\lsv{} \loprod \lsnv{1}}{\lsv{} \loprod \lsnv{1}}}}} \pBasisC{\ls{}}{\lbn{0}}{\lbn{1}'}}}
\and
\gpfit{\inferrule*[]{(\emptyset,\evalCtxSpace{})\vdash \lb{}\colon \bvalid{} \\ (\emptyset,\evalCtxSpace{})\vdash \llpl{} \lee{} \mathbin{:\!:} \mathit{es}\llpr{}_{\lb{}'}\colon \bvalid{} \\ \evalCtxSpace{};\evalCtx{}\vdash_{\mathsf C} \lb{} \xlongleftrightarrow{\lenv{1}} \lb{}' \\ \lemptyset{};\evalCtxSpace{};\evalCtx{}\vdash_{\mathsf C} \lee{} \losum \mathit{es}[0] \losum \cdots \xlongrightarrow{\lb{}' \losum \cdots} \lenv{0}\glresulttype{\ltyv{}} \\ \evalCtxSpace{};\evalCtx{}\vdash_{\mathsf C} \sof{\lb{}} \evalArrow{} \lsv{}}{\evalCtxSpace{};\evalCtx{}\vdash_{\mathsf C} \lb{} \xlongleftrightarrow{\glvreshape{\glvop{RREFRep}(\lenv{0},\glvreshape{\lenv{1}}{\lsv{} \loprod \lsv{}}{\lsv{} \loprod \lsv{}})}{\lsv{} \loprod \lsv{}}{\lsv{} \loprod \lsv{}}} \llpl{} \lee{} \mathbin{:\!:} \mathit{es}\llpr{}_{\lb{}'}}}
\and
\gpfit{\inferrule*[]{(\emptyset,\evalCtxSpace{})\vdash \lb{}\colon \bvalid{} \\ \tof{\lb{}} = \sof{\lb{}} \\ \evalCtxSpace{};\evalCtx{}\vdash_{\mathsf C} \sof{\lb{}} \evalArrow{} \lsv{}}{\evalCtxSpace{};\evalCtx{}\vdash_{\mathsf C} \lb{} \xlongleftrightarrow{\glvreshape{\lidelta{\lsv{}}}{\lsv{} \loprod \lsv{}}{\lsv{} \loprod \lsv{}}} \lb{}}}
\and
\gpfit{\inferrule*[]{(\emptyset,\evalCtxSpace{})\vdash \pBasisC{\ls{}}{\lbn{0}}{\lbn{1}}\colon \bvalid{} \\ (\emptyset,\evalCtxSpace{})\vdash \pBasisC{\ls{}}{\lbn{0}}{\lbn{1}'}\colon \bvalid{} \\ \ls{} = \Kernel(\lee{}) \;\vee\; \ls{} = \CoKernel(\lee{}) \\ \forall i.\; \sof{\lbn{i}} = \lsn{i} \\ \sof{\lbn{1}'} = \lsn{1} \\ \evalCtxSpace{};\evalCtx{}\vdash_{\mathsf C} \sof{\lbn{0}} \evalArrow{} \lsnv{0} \\ \evalCtxSpace{};\evalCtx{}\vdash_{\mathsf C} \ls{} \evalArrow{} \lsv{} \\ \evalCtxSpace{};\evalCtx{}\vdash_{\mathsf C} \ls{} \evalArrow{} \lsv{}' \\ \lemptyset{};\evalCtxSpace{};\evalCtx{}\vdash_{\mathsf C} \lembd{}_{\ls{}} \xlongrightarrow{\pBasisC{\ls{}}{\lbn{0}}{\lbn{1}}^{\lstar} \loprod \lbn{0}} \lenv{3}\glresulttype{\lsv{} \loprod \lsnv{0}} \\ \lemptyset{};\evalCtxSpace{};\evalCtx{}\vdash_{\mathsf C} \lproj{}_{\ls{}, \sof{\lbn{0}}} \xlongrightarrow{\lbn{0}^{\lstar} \loprod \pBasisC{\ls{}}{\lbn{0}}{\lbn{1}'}} \lenv{1}\glresulttype{\lsnv{0} \loprod \lsv{}'}}{\evalCtxSpace{};\evalCtx{}\vdash_{\mathsf C} \pBasisC{\ls{}}{\lbn{0}}{\lbn{1}} \xlongleftrightarrow{\glvmtm{\lsnv{0}}{\lsnv{0}}{\glvreshape{\lenv{1}}{\lsnv{0} \loprod \lsv{}'}{\lsnv{0} \loprod \lsv{}'}}{\glvreshape{\lenv{3}}{\lsv{} \loprod \lsnv{0}}{\lsv{} \loprod \lsnv{0}}}} \pBasisC{\ls{}}{\lbn{0}}{\lbn{1}'}}}
\and
\gpfit{\inferrule*[]{(\emptyset,\evalCtxSpace{})\vdash \pBasisC{\ls{}}{\lbn{0}}{\lbn{1}}\colon \bvalid{} \\ (\emptyset,\evalCtxSpace{})\vdash \pBasisC{\ls{}}{\lbn{0}'}{\lbn{1}}\colon \bvalid{} \\ \ls{} = \Image(\lee{}) \;\vee\; \ls{} = \CoImage(\lee{}) \\ \forall i.\; \sof{\lbn{i}} = \lsn{i} \\ \sof{\lbn{0}'} = \lsn{0} \\ \evalCtxSpace{};\evalCtx{}\vdash_{\mathsf C} \sof{\lbn{1}} \evalArrow{} \lsnv{1} \\ \evalCtxSpace{};\evalCtx{}\vdash_{\mathsf C} \ls{} \evalArrow{} \lsv{} \\ \evalCtxSpace{};\evalCtx{}\vdash_{\mathsf C} \ls{} \evalArrow{} \lsv{}' \\ \lemptyset{};\evalCtxSpace{};\evalCtx{}\vdash_{\mathsf C} \lembd{}_{\ls{}} \xlongrightarrow{\pBasisC{\ls{}}{\lbn{0}}{\lbn{1}}^{\lstar} \loprod \lbn{1}} \lenv{3}\glresulttype{\lsv{} \loprod \lsnv{1}} \\ \lemptyset{};\evalCtxSpace{};\evalCtx{}\vdash_{\mathsf C} \lproj{}_{\ls{}, \sof{\lbn{1}}} \xlongrightarrow{\lbn{1}^{\lstar} \loprod \pBasisC{\ls{}}{\lbn{0}'}{\lbn{1}}} \lenv{1}\glresulttype{\lsnv{1} \loprod \lsv{}'}}{\evalCtxSpace{};\evalCtx{}\vdash_{\mathsf C} \pBasisC{\ls{}}{\lbn{0}}{\lbn{1}} \xlongleftrightarrow{\glvmtm{\lsnv{1}}{\lsnv{1}}{\glvreshape{\lenv{1}}{\lsnv{1} \loprod \lsv{}'}{\lsnv{1} \loprod \lsv{}'}}{\glvreshape{\lenv{3}}{\lsv{} \loprod \lsnv{1}}{\lsv{} \loprod \lsnv{1}}}} \pBasisC{\ls{}}{\lbn{0}'}{\lbn{1}}}}
}{
\gpfit{\inferrule*[lab=\textsc{C-C-Inverse}]{\forall i.\; (\emptyset,\evalCtxSpace{})\vdash \lbn{i}\colon \bvalid{} \\ \evalCtxSpace{};\evalCtx{}\vdash_{\mathsf C} \lbn{1} \xlongleftrightarrow{\leev{}} \lbn{0}}{\evalCtxSpace{};\evalCtx{}\vdash_{\mathsf C} \lbn{0} \xlongleftrightarrow{\leev{}^{-1}} \lbn{1}}}
\and
\gpfit{\inferrule*[lab=\textsc{C-C-Trans}]{(\emptyset,\evalCtxSpace{})\vdash \lbn{0}\colon \bvalid{} \\ (\emptyset,\evalCtxSpace{})\vdash \lbn{2}\colon \bvalid{} \\ \evalCtxSpace{};\evalCtx{}\vdash_{\mathsf C} \lbn{0} \xlongleftrightarrow{\lenv{0}} \lbn{1} \\ \evalCtxSpace{};\evalCtx{}\vdash_{\mathsf C} \lbn{1} \xlongleftrightarrow{\lenv{1}} \lbn{2} \\ \evalCtxSpace{};\evalCtx{}\vdash_{\mathsf C} \sof{\lbn{0}} \evalArrow{} \lsv{}}{\evalCtxSpace{};\evalCtx{}\vdash_{\mathsf C} \lbn{0} \xlongleftrightarrow{\glvmtm{\lsv{}}{\lsv{}}{\lenv{1}}{\glvreshape{\lenv{0}}{\lsv{} \loprod \lsv{}}{\lsv{} \loprod \lsv{}}}} \lbn{2}}}
\and
\gpfit{\inferrule*[lab=\textsc{C-C-Tprod}]{(\emptyset,\evalCtxSpace{})\vdash \lbn{0}' \loprod \lbn{1}'\colon \bvalid{} \\ (\emptyset,\evalCtxSpace{})\vdash \lbn{0} \loprod \lbn{1}\colon \bvalid{} \\ \forall i.\; \evalCtxSpace{};\evalCtx{}\vdash_{\mathsf C} \lbn{i}' \xlongleftrightarrow{\lenv{i}} \lbn{i}}{\evalCtxSpace{};\evalCtx{}\vdash_{\mathsf C} \lbn{0}' \loprod \lbn{1}' \xlongleftrightarrow{\vkrn(\lenv{0}, \lenv{1})} \lbn{0} \loprod \lbn{1}}}
\and
\gpfit{\inferrule*[lab=\textsc{C-C-Dsum}]{(\emptyset,\evalCtxSpace{})\vdash \lbn{0}' \losum \lbn{1}'\colon \bvalid{} \\ (\emptyset,\evalCtxSpace{})\vdash \lbn{0} \losum \lbn{1}\colon \bvalid{} \\ \forall i.\; \evalCtxSpace{};\evalCtx{}\vdash_{\mathsf C} \lbn{i}' \xlongleftrightarrow{\lenv{i}} \lbn{i}}{\evalCtxSpace{};\evalCtx{}\vdash_{\mathsf C} \lbn{0}' \losum \lbn{1}' \xlongleftrightarrow{\vblk(\lenv{0}, \lenv{1})} \lbn{0} \losum \lbn{1}}}
\and
\gpfit{\inferrule*[lab=\textsc{C-C-Dual\allowbreak Std}]{\tof{\lbn{0}} = \sof{\lbn{0}} \\ (\emptyset,\evalCtxSpace{})\vdash \lbn{0}^{\lstar}\colon \bvalid{} \\ (\emptyset,\evalCtxSpace{})\vdash \stdBasisDCtx{\tof{\lbn{0}^{\lstar}}}\colon \bvalid{} \\ \evalCtxSpace{};\evalCtx{}\vdash_{\mathsf C} \lbn{0} \xlongleftrightarrow{\leev{}} \stdBasisDCtx{\tof{\lbn{0}}}}{\evalCtxSpace{};\evalCtx{}\vdash_{\mathsf C} \lbn{0}^{\lstar} \xlongleftrightarrow{\Trans(\leev{}^{-1})} \stdBasisDCtx{\tof{\lbn{0}^{\lstar}}}}}
\and
\gpfit{\inferrule*[lab=\textsc{C-C-Image}]{(\emptyset,\evalCtxSpace{})\vdash \iBasisC{\lbn{0}}\colon \bvalid{} \\ (\emptyset,\evalCtxSpace{})\vdash \iBasisC{\lbn{0}'}\colon \bvalid{} \\ \exists \lsn{0}.\; \exists \lsn{1}.\; \exists \lee{}.\; \sof{\lbn{0}} = \CoKernel(\lee{}) \\ \evalCtxSpace{};\evalCtx{}\vdash_{\mathsf C} \lbn{0} \xlongleftrightarrow{\leev{}} \lbn{0}'}{\evalCtxSpace{};\evalCtx{}\vdash_{\mathsf C} \iBasisC{\lbn{0}} \xlongleftrightarrow{\leev{}} \iBasisC{\lbn{0}'}}}
\and
\gpfit{\inferrule*[lab=\textsc{C-C-Image\allowbreak Svds}]{(\emptyset,\evalCtxSpace{})\vdash \iBasisC{\pBasisC{\CoKernel(\lee{})}{\lbn{0}}{\lbn{1}}}\colon \bvalid{} \\ (\emptyset,\evalCtxSpace{})\vdash \pBasisC{\Image(\lee{})}{\lbn{0}}{\lbn{1}}\colon \bvalid{} \\ \forall i.\; \sof{\lbn{i}} = \lsn{i} \\ \forall i.\; \isStdBasis{\lbn{i}} \\ \evalCtxSpace{};\evalCtx{}\vdash_{\mathsf C} \sof{\lbn{0}^{\lstar} \loprod \lbn{1}} \evalArrow{} \lsnv{0} \loprod \lsnv{1} \\ \lemptyset{};\evalCtxSpace{};\evalCtx{}\vdash_{\mathsf C} \lee{} \xlongrightarrow{\lbn{0}^{\lstar} \loprod \lbn{1}} \leev{}\glresulttype{\lsnv{0} \loprod \lsnv{1}} \\ \lenv{{\mathrm{M}}} = \glvreshape{\leev{}}{\lsnv{0} \loprod \lsnv{1}}{\lsnv{0} \loprod \lsnv{1}} \\ \lenv{{\mathrm{U}}} = \glvop{GSVDU}(\lenv{{\mathrm{M}}},\lidelta{\lsnv{0}},\lidelta{\lsnv{1}}) \\ \lenv{{\mathrm{V}}} = \glvop{GSVDV}(\lenv{{\mathrm{M}}},\lidelta{\lsnv{0}},\lidelta{\lsnv{1}}) \\ \lenv{{\mathrm{P}}} = \glvmtm{\lsnv{1}}{\lsnv{1}}{\lenv{{\mathrm{V}}}}{\glvmtm{\lsnv{0}}{\lsnv{0}}{\lenv{{\mathrm{M}}}}{\Trans(\lenv{{\mathrm{U}}})}}}{\evalCtxSpace{};\evalCtx{}\vdash_{\mathsf C} \iBasisC{\pBasisC{\CoKernel(\lee{})}{\lbn{0}}{\lbn{1}}} \xlongleftrightarrow{\glvreshape{\lenv{{\mathrm{P}}}}{\lR^{\lenv{{\mathrm{M}}},\CoKernel{},\lsnv{0} \loprod \lsnv{1}} \loprod \lR^{\lenv{{\mathrm{M}}},\Image{},\lsnv{0} \loprod \lsnv{1}}}{\operatorname{inferSpace}(\lenv{{\mathrm{P}}})}} \pBasisC{\Image(\lee{})}{\lbn{0}}{\lbn{1}}}}
\and
\gpfit{\inferrule*[lab=\textsc{C-C-Proj\allowbreak Ker\allowbreak Coker}]{(\emptyset,\evalCtxSpace{})\vdash \pBasisC{\ls{}}{\lbn{0}}{\lbn{1}}\colon \bvalid{} \\ (\emptyset,\evalCtxSpace{})\vdash \pBasisC{\ls{}}{\lbn{0}'}{\lbn{1}}\colon \bvalid{} \\ \ls{} = \Kernel(\lee{}) \;\vee\; \ls{} = \CoKernel(\lee{}) \\ \forall i.\; \sof{\lbn{i}} = \lsn{i} \\ \forall i.\; \evalCtxSpace{};\evalCtx{}\vdash_{\mathsf C} \sof{\lbn{i}} \evalArrow{} \lsnv{i} \\ \evalCtxSpace{};\evalCtx{}\vdash_{\mathsf C} \ls{} \evalArrow{} \lsv{} \\ \evalCtxSpace{};\evalCtx{}\vdash_{\mathsf C} \ls{} \evalArrow{} \lsv{}' \\ \evalCtxSpace{};\evalCtx{}\vdash_{\mathsf C} \sof{\lbn{0}'} \evalArrow{} \lsnv{0}' \\ \evalCtxSpace{};\evalCtx{}\vdash_{\mathsf C} \lbn{0} \xlongleftrightarrow{\lenv{2}} \lbn{0}' \\ \lemptyset{};\evalCtxSpace{};\evalCtx{}\vdash_{\mathsf C} \lembd{}_{\ls{}} \xlongrightarrow{\pBasisC{\ls{}}{\lbn{0}}{\lbn{1}}^{\lstar} \loprod \lbn{0}} \lenv{3}\glresulttype{\lsv{} \loprod \lsnv{0}} \\ \lemptyset{};\evalCtxSpace{};\evalCtx{}\vdash_{\mathsf C} \lproj{}_{\ls{}, \sof{\lbn{0}'}} \xlongrightarrow{\lbn{0}'^{\lstar} \loprod \pBasisC{\ls{}}{\lbn{0}'}{\lbn{1}}} \lenv{1}\glresulttype{\lsnv{0}' \loprod \lsv{}'}}{\evalCtxSpace{};\evalCtx{}\vdash_{\mathsf C} \pBasisC{\ls{}}{\lbn{0}}{\lbn{1}} \xlongleftrightarrow{\glvmtm{\lsnv{0}'}{\lsnv{0}'}{\glvreshape{\lenv{1}}{\lsnv{0}' \loprod \lsv{}'}{\lsnv{0}' \loprod \lsv{}'}}{\glvmtm{\lsnv{0}}{\lsnv{0}}{\glvreshape{\lenv{2}}{\lsnv{0} \loprod \lsnv{0}'}{\lsnv{0} \loprod \lsnv{0}'}}{\glvreshape{\lenv{3}}{\lsv{} \loprod \lsnv{0}}{\lsv{} \loprod \lsnv{0}}}}} \pBasisC{\ls{}}{\lbn{0}'}{\lbn{1}}}}
\and
\gpfit{\inferrule*[lab=\textsc{C-C-Proj\allowbreak Img\allowbreak Coimg}]{(\emptyset,\evalCtxSpace{})\vdash \pBasisC{\ls{}}{\lbn{0}}{\lbn{1}}\colon \bvalid{} \\ (\emptyset,\evalCtxSpace{})\vdash \pBasisC{\ls{}}{\lbn{0}}{\lbn{1}'}\colon \bvalid{} \\ \ls{} = \Image(\lee{}) \;\vee\; \ls{} = \CoImage(\lee{}) \\ \forall i.\; \sof{\lbn{i}} = \lsn{i} \\ \forall i.\; \evalCtxSpace{};\evalCtx{}\vdash_{\mathsf C} \sof{\lbn{i}} \evalArrow{} \lsnv{i} \\ \evalCtxSpace{};\evalCtx{}\vdash_{\mathsf C} \ls{} \evalArrow{} \lsv{} \\ \evalCtxSpace{};\evalCtx{}\vdash_{\mathsf C} \ls{} \evalArrow{} \lsv{}' \\ \evalCtxSpace{};\evalCtx{}\vdash_{\mathsf C} \sof{\lbn{1}'} \evalArrow{} \lsnv{1}' \\ \evalCtxSpace{};\evalCtx{}\vdash_{\mathsf C} \lbn{1} \xlongleftrightarrow{\lenv{2}} \lbn{1}' \\ \lemptyset{};\evalCtxSpace{};\evalCtx{}\vdash_{\mathsf C} \lembd{}_{\ls{}} \xlongrightarrow{\pBasisC{\ls{}}{\lbn{0}}{\lbn{1}}^{\lstar} \loprod \lbn{1}} \lenv{3}\glresulttype{\lsv{} \loprod \lsnv{1}} \\ \lemptyset{};\evalCtxSpace{};\evalCtx{}\vdash_{\mathsf C} \lproj{}_{\ls{}, \sof{\lbn{1}'}} \xlongrightarrow{\lbn{1}'^{\lstar} \loprod \pBasisC{\ls{}}{\lbn{0}}{\lbn{1}'}} \lenv{1}\glresulttype{\lsnv{1}' \loprod \lsv{}'}}{\evalCtxSpace{};\evalCtx{}\vdash_{\mathsf C} \pBasisC{\ls{}}{\lbn{0}}{\lbn{1}} \xlongleftrightarrow{\glvmtm{\lsnv{1}'}{\lsnv{1}'}{\glvreshape{\lenv{1}}{\lsnv{1}' \loprod \lsv{}'}{\lsnv{1}' \loprod \lsv{}'}}{\glvmtm{\lsnv{1}}{\lsnv{1}}{\glvreshape{\lenv{2}}{\lsnv{1} \loprod \lsnv{1}'}{\lsnv{1} \loprod \lsnv{1}'}}{\glvreshape{\lenv{3}}{\lsv{} \loprod \lsnv{1}}{\lsv{} \loprod \lsnv{1}}}}} \pBasisC{\ls{}}{\lbn{0}}{\lbn{1}'}}}
\and
\gpfit{\inferrule*[lab=\textsc{C-C-Terms\allowbreak Rref}]{(\emptyset,\evalCtxSpace{})\vdash \lb{}\colon \bvalid{} \\ (\emptyset,\evalCtxSpace{})\vdash \llpl{} \lee{} \mathbin{:\!:} \mathit{es}\llpr{}_{\lb{}'}\colon \bvalid{} \\ \evalCtxSpace{};\evalCtx{}\vdash_{\mathsf C} \lb{} \xlongleftrightarrow{\lenv{1}} \lb{}' \\ \lemptyset{};\evalCtxSpace{};\evalCtx{}\vdash_{\mathsf C} \lee{} \losum \mathit{es}[0] \losum \cdots \xlongrightarrow{\lb{}' \losum \cdots} \lenv{0}\glresulttype{\ltyv{}} \\ \evalCtxSpace{};\evalCtx{}\vdash_{\mathsf C} \sof{\lb{}} \evalArrow{} \lsv{}}{\evalCtxSpace{};\evalCtx{}\vdash_{\mathsf C} \lb{} \xlongleftrightarrow{\glvreshape{\glvop{RREFRep}(\lenv{0},\glvreshape{\lenv{1}}{\lsv{} \loprod \lsv{}}{\lsv{} \loprod \lsv{}})}{\lsv{} \loprod \lsv{}}{\lsv{} \loprod \lsv{}}} \llpl{} \lee{} \mathbin{:\!:} \mathit{es}\llpr{}_{\lb{}'}}}
\and
\gpfit{\inferrule*[lab=\textsc{C-C-Refl}]{(\emptyset,\evalCtxSpace{})\vdash \lb{}\colon \bvalid{} \\ \tof{\lb{}} = \sof{\lb{}} \\ \evalCtxSpace{};\evalCtx{}\vdash_{\mathsf C} \sof{\lb{}} \evalArrow{} \lsv{}}{\evalCtxSpace{};\evalCtx{}\vdash_{\mathsf C} \lb{} \xlongleftrightarrow{\glvreshape{\lidelta{\lsv{}}}{\lsv{} \loprod \lsv{}}{\lsv{} \loprod \lsv{}}} \lb{}}}
\and
\gpfit{\inferrule*[lab=\textsc{C-C-Proj\allowbreak Ker\allowbreak Coker\allowbreak Cod}]{(\emptyset,\evalCtxSpace{})\vdash \pBasisC{\ls{}}{\lbn{0}}{\lbn{1}}\colon \bvalid{} \\ (\emptyset,\evalCtxSpace{})\vdash \pBasisC{\ls{}}{\lbn{0}}{\lbn{1}'}\colon \bvalid{} \\ \ls{} = \Kernel(\lee{}) \;\vee\; \ls{} = \CoKernel(\lee{}) \\ \forall i.\; \sof{\lbn{i}} = \lsn{i} \\ \sof{\lbn{1}'} = \lsn{1} \\ \evalCtxSpace{};\evalCtx{}\vdash_{\mathsf C} \sof{\lbn{0}} \evalArrow{} \lsnv{0} \\ \evalCtxSpace{};\evalCtx{}\vdash_{\mathsf C} \ls{} \evalArrow{} \lsv{} \\ \evalCtxSpace{};\evalCtx{}\vdash_{\mathsf C} \ls{} \evalArrow{} \lsv{}' \\ \lemptyset{};\evalCtxSpace{};\evalCtx{}\vdash_{\mathsf C} \lembd{}_{\ls{}} \xlongrightarrow{\pBasisC{\ls{}}{\lbn{0}}{\lbn{1}}^{\lstar} \loprod \lbn{0}} \lenv{3}\glresulttype{\lsv{} \loprod \lsnv{0}} \\ \lemptyset{};\evalCtxSpace{};\evalCtx{}\vdash_{\mathsf C} \lproj{}_{\ls{}, \sof{\lbn{0}}} \xlongrightarrow{\lbn{0}^{\lstar} \loprod \pBasisC{\ls{}}{\lbn{0}}{\lbn{1}'}} \lenv{1}\glresulttype{\lsnv{0} \loprod \lsv{}'}}{\evalCtxSpace{};\evalCtx{}\vdash_{\mathsf C} \pBasisC{\ls{}}{\lbn{0}}{\lbn{1}} \xlongleftrightarrow{\glvmtm{\lsnv{0}}{\lsnv{0}}{\glvreshape{\lenv{1}}{\lsnv{0} \loprod \lsv{}'}{\lsnv{0} \loprod \lsv{}'}}{\glvreshape{\lenv{3}}{\lsv{} \loprod \lsnv{0}}{\lsv{} \loprod \lsnv{0}}}} \pBasisC{\ls{}}{\lbn{0}}{\lbn{1}'}}}
\and
\gpfit{\inferrule*[lab=\textsc{C-C-Proj\allowbreak Img\allowbreak Coimg\allowbreak Dom}]{(\emptyset,\evalCtxSpace{})\vdash \pBasisC{\ls{}}{\lbn{0}}{\lbn{1}}\colon \bvalid{} \\ (\emptyset,\evalCtxSpace{})\vdash \pBasisC{\ls{}}{\lbn{0}'}{\lbn{1}}\colon \bvalid{} \\ \ls{} = \Image(\lee{}) \;\vee\; \ls{} = \CoImage(\lee{}) \\ \forall i.\; \sof{\lbn{i}} = \lsn{i} \\ \sof{\lbn{0}'} = \lsn{0} \\ \evalCtxSpace{};\evalCtx{}\vdash_{\mathsf C} \sof{\lbn{1}} \evalArrow{} \lsnv{1} \\ \evalCtxSpace{};\evalCtx{}\vdash_{\mathsf C} \ls{} \evalArrow{} \lsv{} \\ \evalCtxSpace{};\evalCtx{}\vdash_{\mathsf C} \ls{} \evalArrow{} \lsv{}' \\ \lemptyset{};\evalCtxSpace{};\evalCtx{}\vdash_{\mathsf C} \lembd{}_{\ls{}} \xlongrightarrow{\pBasisC{\ls{}}{\lbn{0}}{\lbn{1}}^{\lstar} \loprod \lbn{1}} \lenv{3}\glresulttype{\lsv{} \loprod \lsnv{1}} \\ \lemptyset{};\evalCtxSpace{};\evalCtx{}\vdash_{\mathsf C} \lproj{}_{\ls{}, \sof{\lbn{1}}} \xlongrightarrow{\lbn{1}^{\lstar} \loprod \pBasisC{\ls{}}{\lbn{0}'}{\lbn{1}}} \lenv{1}\glresulttype{\lsnv{1} \loprod \lsv{}'}}{\evalCtxSpace{};\evalCtx{}\vdash_{\mathsf C} \pBasisC{\ls{}}{\lbn{0}}{\lbn{1}} \xlongleftrightarrow{\glvmtm{\lsnv{1}}{\lsnv{1}}{\glvreshape{\lenv{1}}{\lsnv{1} \loprod \lsv{}'}{\lsnv{1} \loprod \lsv{}'}}{\glvreshape{\lenv{3}}{\lsv{} \loprod \lsnv{1}}{\lsv{} \loprod \lsnv{1}}}} \pBasisC{\ls{}}{\lbn{0}'}{\lbn{1}}}}
}
\gldefjudgment{c.term}{\lDelta{};\lepsilon{};\evalCtx{}\vdash_{\mathsf C} \lee{}\xlongrightarrow{\lb{}}\leev{}\glresulttype{\ltyv{}}}{Checked term evaluation (generated from \texttt{C.Term}).}{fig:gl-c-term}
\gldefrule{c.term.varLin}{c.term}{\textsc{C-E-Var\allowbreak Lin}}{((\mathit{x}, \lb{}) \mapsto \leev{}) \in \evalCtx{} \\ \tof{\lb{}} = \ls{} \\ (\emptyset,\evalCtxSpace{})\vdash \lb{}\colon \bvalid{} \\ \evalCtxSpace{};\evalCtx{}\vdash_{\mathsf C} \lb{} \evalArrow{}_{\mathsf{basis}} \ltyv{}}{\{(\mathit{x}, \ls{})\};\evalCtxSpace{};\evalCtx{}\vdash_{\mathsf C} \mathit{x} \xlongrightarrow{\lb{}} \leev{}\glresulttype{\ltyv{}}}{\texttt{GLang.Raw\allowbreak Certified\allowbreak Operational.C.Term.var\_\allowbreak lin}}{GLang/Adequacy/RawCertifiedOperational.lean:283}
\gldefrule{c.term.varNlin}{c.term}{\textsc{C-E-Var\allowbreak Nlin}}{((\mathit{x}, \lb{}) \mapsto \leev{}) \in \evalCtx{} \\ (\mathit{x}\colon \tof{\lb{}}) \in \evalCtxSpace{} \\ (\emptyset,\evalCtxSpace{})\vdash \lb{}\colon \bvalid{} \\ \evalCtxSpace{};\evalCtx{}\vdash_{\mathsf C} \lb{} \evalArrow{}_{\mathsf{basis}} \ltyv{}}{\lemptyset{};\evalCtxSpace{};\evalCtx{}\vdash_{\mathsf C} \mathit{x} \xlongrightarrow{\lb{}} \leev{}\glresulttype{\ltyv{}}}{\texttt{GLang.Raw\allowbreak Certified\allowbreak Operational.C.Term.var\_\allowbreak nlin}}{GLang/Adequacy/RawCertifiedOperational.lean:290}
\gldefrule{c.term.promote}{c.term}{\textsc{C-E-Promote}}{\forall y \in \mathit{lin}.\; y \notin \bigcup_{(\cdot\colon s) \in \lDelta{},\, \mathit{lin}} \mathrm{fv}(s) \;\cup\; \bigcup_{(\cdot\colon \tau) \in \evalCtxSpace{}'} \mathrm{fv}(\tau) \\ \evalCtxSpace{}' \subseteq \evalCtxSpace{}'' \\ (\lDelta{}, \mathit{lin} \ldisun \evalCtxSpace{}'')\vdash \lee{}\colon \tof{\lb{}} \\ (\emptyset,\mathit{lin} \ldisun \evalCtxSpace{}'')\vdash \lb{}\colon \bvalid{} \\ \mathit{lin} \ldisun \evalCtxSpace{}'';\evalCtx{}\vdash_{\mathsf C} \lb{} \evalArrow{}_{\mathsf{basis}} \ltyv{} \\ \mathit{lin} \ldisun \lDelta{};\evalCtxSpace{}';\evalCtx{}\vdash_{\mathsf C} \lee{} \xlongrightarrow{\lb{}} \leev{}\glresulttype{\ltyv{}}}{\lDelta{};\mathit{lin} \ldisun \evalCtxSpace{}'';\evalCtx{}\vdash_{\mathsf C} \lee{} \xlongrightarrow{\lb{}} \leev{}\glresulttype{\ltyv{}}}{\texttt{GLang.Raw\allowbreak Certified\allowbreak Operational.C.Term.promote}}{GLang/Adequacy/RawCertifiedOperational.lean:297}
\gldefrule{c.term.cob}{c.term}{\textsc{C-E-Cob}}{\forall i.\; \tof{\lbn{i}} = \sof{\lbn{i}} \\ (\lDelta{}, \evalCtxSpace{})\vdash \lee{}\colon \tof{\lbn{0}} \\ (\emptyset,\evalCtxSpace{})\vdash \lbn{0}\colon \bvalid{} \\ \evalCtxSpace{};\evalCtx{}\vdash_{\mathsf C} \lbn{0} \evalArrow{}_{\mathsf{basis}} \lsv{} \\ \lDelta{};\evalCtxSpace{};\evalCtx{}\vdash_{\mathsf C} \lee{} \xlongrightarrow{\lbn{1}} \leev{}\glresulttype{\ltyv{}} \\ \evalCtxSpace{};\evalCtx{}\vdash_{\mathsf C} \lbn{1} \xlongleftrightarrow{\lenv{0}} \lbn{0} \\ \evalCtxSpace{};\evalCtx{}\vdash_{\mathsf C} \sof{\lbn{0}} \evalArrow{} \lsv{}}{\lDelta{};\evalCtxSpace{};\evalCtx{}\vdash_{\mathsf C} \lee{} \xlongrightarrow{\lbn{0}} \LTTVPP{\lsv{}}{\glvreshape{\lenv{0}}{\lsv{} \loprod \lsv{}}{\lsv{} \loprod \lsv{}}}{\glvreshape{\leev{}}{\lsv{}}{\lsv{}}}\glresulttype{\lsv{}}}{\texttt{GLang.Raw\allowbreak Certified\allowbreak Operational.C.Term.cob}}{GLang/Adequacy/RawCertifiedOperational.lean:315}
\gldefrule{c.term.reshapeCob}{c.term}{\textsc{C-E-Reshape\allowbreak Cob}}{(\lDelta{}, \evalCtxSpace{})\vdash \lee{}\colon \tof{\lb{}} \\ (\emptyset,\evalCtxSpace{})\vdash \lb{}\colon \bvalid{} \\ \evalCtxSpace{};\evalCtx{}\vdash_{\mathsf C} \lb{} \evalArrow{}_{\mathsf{basis}} \ltyv{}' \\ \lDelta{};\evalCtxSpace{};\evalCtx{}\vdash_{\mathsf C} \lee{} \xlongrightarrow{\lb{}} \leev{}\glresulttype{\ltyv{}}}{\lDelta{};\evalCtxSpace{};\evalCtx{}\vdash_{\mathsf C} \lee{} \xlongrightarrow{\lb{}} \glvreshape{\leev{}}{\ltyv{}'}{\ltyv{}}\glresulttype{\ltyv{}'}}{\texttt{GLang.Raw\allowbreak Certified\allowbreak Operational.C.Term.reshape\_\allowbreak cob}}{GLang/Adequacy/RawCertifiedOperational.lean:328}
\gldefrule{c.term.scale}{c.term}{\textsc{C-E-Scale}}{\tof{\lb{}} = \sof{\lb{}} \\ ({\lDelta{}_{0}} \ldisun {\lDelta{}_{1}}, \evalCtxSpace{})\vdash \len{0} \lvscale \len{1}\colon \tof{\lb{}} \\ (\emptyset,\evalCtxSpace{})\vdash \lb{}\colon \bvalid{} \\ \evalCtxSpace{};\evalCtx{}\vdash_{\mathsf C} \lb{} \evalArrow{}_{\mathsf{basis}} \lsv{} \\ {\lDelta{}_{0}};\evalCtxSpace{};\evalCtx{}\vdash_{\mathsf C} \len{0} \xlongrightarrow{\lbone{}} \lenv{0}\glresulttype{\lR{}} \\ {\lDelta{}_{1}};\evalCtxSpace{};\evalCtx{}\vdash_{\mathsf C} \len{1} \xlongrightarrow{\lb{}} \lenv{1}\glresulttype{\lsv{}}}{{\lDelta{}_{0}} \ldisun {\lDelta{}_{1}};\evalCtxSpace{};\evalCtx{}\vdash_{\mathsf C} \len{0} \lvscale \len{1} \xlongrightarrow{\lb{}} \lenv{0} \lvscale \lenv{1}\glresulttype{\lsv{}}}{\texttt{GLang.Raw\allowbreak Certified\allowbreak Operational.C.Term.scale}}{GLang/Adequacy/RawCertifiedOperational.lean:335}
\gldefrule{c.term.add}{c.term}{\textsc{C-E-Add}}{\tof{\lb{}} = \sof{\lb{}} \\ (\lDelta{}, \evalCtxSpace{})\vdash \len{0} \lvadd \len{1}\colon \tof{\lb{}} \\ (\emptyset,\evalCtxSpace{})\vdash \lb{}\colon \bvalid{} \\ \evalCtxSpace{};\evalCtx{}\vdash_{\mathsf C} \lb{} \evalArrow{}_{\mathsf{basis}} \lsv{} \\ \forall i.\; \lDelta{};\evalCtxSpace{};\evalCtx{}\vdash_{\mathsf C} \len{i} \xlongrightarrow{\lb{}} \lenv{i}\glresulttype{\lsnv{i}}}{\lDelta{};\evalCtxSpace{};\evalCtx{}\vdash_{\mathsf C} \len{0} \lvadd \len{1} \xlongrightarrow{\lb{}} \lenv{0} \lvadd_{\lsv{}} \lenv{1}\glresulttype{\lsv{}}}{\texttt{GLang.Raw\allowbreak Certified\allowbreak Operational.C.Term.add}}{GLang/Adequacy/RawCertifiedOperational.lean:345}
\gldefrule{c.term.dsumE}{c.term}{\textsc{C-E-Dsum\allowbreak E}}{\forall i.\; \tof{\lbn{i}} = \sof{\lbn{i}} \\ (\lDelta{}, \evalCtxSpace{})\vdash \len{0}\losum \len{1}\colon \tof{\lbn{0} \losum \lbn{1}} \\ (\emptyset,\evalCtxSpace{})\vdash \lbn{0} \losum \lbn{1}\colon \bvalid{} \\ \evalCtxSpace{};\evalCtx{}\vdash_{\mathsf C} \lbn{0} \losum \lbn{1} \evalArrow{}_{\mathsf{basis}} \lsnv{0} \losum \lsnv{1} \\ \forall i.\; \lDelta{};\evalCtxSpace{};\evalCtx{}\vdash_{\mathsf C} \len{i} \xlongrightarrow{\lbn{i}} \lenv{i}\glresulttype{\lsnv{i}}}{\lDelta{};\evalCtxSpace{};\evalCtx{}\vdash_{\mathsf C} \len{0}\losum \len{1} \xlongrightarrow{\lbn{0} \losum \lbn{1}} \lenv{0}\losum \lenv{1}\glresulttype{\lsnv{0} \losum \lsnv{1}}}{\texttt{GLang.Raw\allowbreak Certified\allowbreak Operational.C.Term.dsum\allowbreak E}}{GLang/Adequacy/RawCertifiedOperational.lean:359}
\gldefrule{c.term.tprodE}{c.term}{\textsc{C-E-Tprod\allowbreak E}}{\forall i.\; \tof{\lbn{i}} = \sof{\lbn{i}} \\ ({\lDelta{}_{0}} \ldisun {\lDelta{}_{1}}, \evalCtxSpace{})\vdash \len{0}\loprod \len{1}\colon \tof{\lbn{0} \loprod \lbn{1}} \\ (\emptyset,\evalCtxSpace{})\vdash \lbn{0} \loprod \lbn{1}\colon \bvalid{} \\ \evalCtxSpace{};\evalCtx{}\vdash_{\mathsf C} \lbn{0} \loprod \lbn{1} \evalArrow{}_{\mathsf{basis}} \lsnv{0} \loprod \lsnv{1} \\ \forall i.\; {\lDelta{}_{i}};\evalCtxSpace{};\evalCtx{}\vdash_{\mathsf C} \len{i} \xlongrightarrow{\lbn{i}} \lenv{i}\glresulttype{\lsnv{i}}}{{\lDelta{}_{0}} \ldisun {\lDelta{}_{1}};\evalCtxSpace{};\evalCtx{}\vdash_{\mathsf C} \len{0}\loprod \len{1} \xlongrightarrow{\lbn{0} \loprod \lbn{1}} \lenv{0}\loprod \lenv{1}\glresulttype{\lsnv{0} \loprod \lsnv{1}}}{\texttt{GLang.Raw\allowbreak Certified\allowbreak Operational.C.Term.tprod\allowbreak E}}{GLang/Adequacy/RawCertifiedOperational.lean:369}
\gldefrule{c.term.pinvTStd}{c.term}{\textsc{C-E-Pinv\allowbreak TStd}}{\forall i.\; \isStdBasis{\lbn{i}} \\ (\lemptyset{}, \evalCtxSpace{})\vdash \lee{}^{\lpseudo}\colon \tof{\lbn{0}^{\lstar} \loprod \lbn{1}} \\ (\emptyset,\evalCtxSpace{})\vdash \lbn{0}^{\lstar} \loprod \lbn{1}\colon \bvalid{} \\ \evalCtxSpace{};\evalCtx{}\vdash_{\mathsf C} \lbn{0}^{\lstar} \loprod \lbn{1} \evalArrow{}_{\mathsf{basis}} \lsnv{0} \loprod \lsnv{1} \\ \lemptyset{};\evalCtxSpace{};\evalCtx{}\vdash_{\mathsf C} \lee{} \xlongrightarrow{\lbn{1}^{\lstar} \loprod \lbn{0}} \leev{}\glresulttype{\lsnv{1} \loprod \lsnv{0}}}{\lemptyset{};\evalCtxSpace{};\evalCtx{}\vdash_{\mathsf C} \lee{}^{\lpseudo} \xlongrightarrow{\lbn{0}^{\lstar} \loprod \lbn{1}} \leev{}^{+}\glresulttype{\lsnv{0} \loprod \lsnv{1}}}{\texttt{GLang.Raw\allowbreak Certified\allowbreak Operational.C.Term.pinv\allowbreak T\_\allowbreak std}}{GLang/Adequacy/RawCertifiedOperational.lean:381}
\gldefrule{c.term.pinvTGpinv}{c.term}{\textsc{C-E-Pinv\allowbreak TGpinv}}{(\lemptyset{}, \evalCtxSpace{})\vdash \lee{}^{\lpseudo}\colon \tof{\lbn{0}^{\lstar} \loprod \lbn{1}} \\ (\emptyset,\evalCtxSpace{})\vdash \lbn{0}^{\lstar} \loprod \lbn{1}\colon \bvalid{} \\ \evalCtxSpace{};\evalCtx{}\vdash_{\mathsf C} \lbn{0}^{\lstar} \loprod \lbn{1} \evalArrow{}_{\mathsf{basis}} \lsnv{0} \loprod \lsnv{1} \\ \lemptyset{};\evalCtxSpace{};\evalCtx{}\vdash_{\mathsf C} \lee{} \xlongrightarrow{\lbn{1}^{\lstar} \loprod \lbn{0}} \leev{}\glresulttype{\lsnv{1} \loprod \lsnv{0}} \\ \forall i.\; \lemptyset{};\evalCtxSpace{};\evalCtx{}\vdash_{\mathsf C} \linner{}_{\sof{\lbn{i}}} \xlongrightarrow{\lbn{i}^{\lstar} \loprod \lbn{i}^{\lstar}} \lenv{i}\glresulttype{\lsnv{i} \loprod \lsnv{i}}}{\lemptyset{};\evalCtxSpace{};\evalCtx{}\vdash_{\mathsf C} \lee{}^{\lpseudo} \xlongrightarrow{\lbn{0}^{\lstar} \loprod \lbn{1}} \glvop{GPINV}(\leev{},\lenv{1},\lenv{0})\glresulttype{\lsnv{0} \loprod \lsnv{1}}}{\texttt{GLang.Raw\allowbreak Certified\allowbreak Operational.C.Term.pinv\allowbreak T\_\allowbreak gpinv}}{GLang/Adequacy/RawCertifiedOperational.lean:391}
\gldefrule{c.term.pinvDStd}{c.term}{\textsc{C-E-Pinv\allowbreak DStd}}{\sof{\lbn{0}} = \lR{} \\ \forall i.\; \isStdBasis{\lbn{i}} \\ (\lemptyset{}, \evalCtxSpace{})\vdash \lee{}^{\lpseudo}\colon \tof{\lbn{0}^{\lstar} \loprod \lbn{1}} \\ (\emptyset,\evalCtxSpace{})\vdash \lbn{0}^{\lstar} \loprod \lbn{1}\colon \bvalid{} \\ \evalCtxSpace{};\evalCtx{}\vdash_{\mathsf C} \lbn{0}^{\lstar} \loprod \lbn{1} \evalArrow{}_{\mathsf{basis}} \lR{} \loprod \lsnv{1} \\ \lemptyset{};\evalCtxSpace{};\evalCtx{}\vdash_{\mathsf C} \lee{} \xlongrightarrow{\lbn{1}^{\lstar}} \leev{}\glresulttype{\lsnv{1}}}{\lemptyset{};\evalCtxSpace{};\evalCtx{}\vdash_{\mathsf C} \lee{}^{\lpseudo} \xlongrightarrow{\lbn{0}^{\lstar} \loprod \lbn{1}} \glvreshape{\leev{}}{\lsnv{1} \loprod \lR{}}{\lsnv{1}}^{+}\glresulttype{\lR{} \loprod \lsnv{1}}}{\texttt{GLang.Raw\allowbreak Certified\allowbreak Operational.C.Term.pinv\allowbreak D\_\allowbreak std}}{GLang/Adequacy/RawCertifiedOperational.lean:405}
\gldefrule{c.term.pinvDGpinv}{c.term}{\textsc{C-E-Pinv\allowbreak DGpinv}}{\sof{\lbn{0}} = \lR{} \\ \isStdBasis{\lbn{0}} \\ (\lemptyset{}, \evalCtxSpace{})\vdash \lee{}^{\lpseudo}\colon \tof{\lbn{0}^{\lstar} \loprod \lbn{1}} \\ (\emptyset,\evalCtxSpace{})\vdash \lbn{0}^{\lstar} \loprod \lbn{1}\colon \bvalid{} \\ \evalCtxSpace{};\evalCtx{}\vdash_{\mathsf C} \lbn{0}^{\lstar} \loprod \lbn{1} \evalArrow{}_{\mathsf{basis}} \lR{} \loprod \lsnv{1} \\ \lemptyset{};\evalCtxSpace{};\evalCtx{}\vdash_{\mathsf C} \lee{} \xlongrightarrow{\lbn{1}^{\lstar}} \leev{}\glresulttype{\lsnv{1}} \\ \lemptyset{};\evalCtxSpace{};\evalCtx{}\vdash_{\mathsf C} \linner{}_{\sof{\lbn{0}}} \xlongrightarrow{\lbn{0}^{\lstar} \loprod \lbn{0}^{\lstar}} \lenv{0}\glresulttype{\lR{} \loprod \lR{}} \\ \lemptyset{};\evalCtxSpace{};\evalCtx{}\vdash_{\mathsf C} \linner{}_{\sof{\lbn{1}}} \xlongrightarrow{\lbn{1}^{\lstar} \loprod \lbn{1}^{\lstar}} \lenv{1}\glresulttype{\lsnv{1} \loprod \lsnv{1}}}{\lemptyset{};\evalCtxSpace{};\evalCtx{}\vdash_{\mathsf C} \lee{}^{\lpseudo} \xlongrightarrow{\lbn{0}^{\lstar} \loprod \lbn{1}} \glvop{GPINV}(\glvreshape{\leev{}}{\lsnv{1} \loprod \lR{}}{\lsnv{1}},\lenv{1},\lenv{0})\glresulttype{\lR{} \loprod \lsnv{1}}}{\texttt{GLang.Raw\allowbreak Certified\allowbreak Operational.C.Term.pinv\allowbreak D\_\allowbreak gpinv}}{GLang/Adequacy/RawCertifiedOperational.lean:417}
\gldefrule{c.term.innerStd}{c.term}{\textsc{C-E-Inner\allowbreak Std}}{\isStdBasis{\lbn{0}} \\ \sof{\lbn{0}} = \ls{} \\ (\lemptyset{}, \evalCtxSpace{})\vdash \linner{}_{\ls{}}\colon \tof{\lbn{0}^{\lstar} \loprod \lbn{0}^{\lstar}} \\ (\emptyset,\evalCtxSpace{})\vdash \lbn{0}^{\lstar} \loprod \lbn{0}^{\lstar}\colon \bvalid{} \\ \evalCtxSpace{};\evalCtx{}\vdash_{\mathsf C} \lbn{0}^{\lstar} \loprod \lbn{0}^{\lstar} \evalArrow{}_{\mathsf{basis}} \lsv{} \loprod \lsv{} \\ \evalCtxSpace{};\evalCtx{}\vdash_{\mathsf C} \ls{} \evalArrow{} \lsv{}}{\lemptyset{};\evalCtxSpace{};\evalCtx{}\vdash_{\mathsf C} \linner{}_{\ls{}} \xlongrightarrow{\lbn{0}^{\lstar} \loprod \lbn{0}^{\lstar}} \lidelta{\lsv{}}\glresulttype{\lsv{} \loprod \lsv{}}}{\texttt{GLang.Raw\allowbreak Certified\allowbreak Operational.C.Term.inner\_\allowbreak std}}{GLang/Adequacy/RawCertifiedOperational.lean:434}
\gldefrule{c.term.embedKer}{c.term}{\textsc{C-E-Embed\allowbreak Ker}}{\forall i.\; \sof{\lbn{i}} = \lsn{i} \\ (\lemptyset{}, \evalCtxSpace{})\vdash \lembd{}_{\Kernel(\lee{})}\colon \tof{\pBasisC{\Kernel(\lee{})}{\lbn{0}}{\lbn{1}}^{\lstar} \loprod \lbn{0}} \\ (\emptyset,\evalCtxSpace{})\vdash \pBasisC{\Kernel(\lee{})}{\lbn{0}}{\lbn{1}}^{\lstar} \loprod \lbn{0}\colon \bvalid{} \\ \evalCtxSpace{};\evalCtx{}\vdash_{\mathsf C} \pBasisC{\Kernel(\lee{})}{\lbn{0}}{\lbn{1}}^{\lstar} \loprod \lbn{0} \evalArrow{}_{\mathsf{basis}} \lR^{\glvreshape{\leev{}}{\lsnv{0} \loprod \lsnv{1}}{\lsnv{0} \loprod \lsnv{1}},\Kernel{},\lsnv{0} \loprod \lsnv{1}} \loprod \lsnv{0} \\ \forall i.\; \lemptyset{};\evalCtxSpace{};\evalCtx{}\vdash_{\mathsf C} \linner{}_{\lsn{i}} \xlongrightarrow{\lbn{i}^{\lstar} \loprod \lbn{i}^{\lstar}} \lenv{i}\glresulttype{\lsnv{i} \loprod \lsnv{i}} \\ \lemptyset{};\evalCtxSpace{};\evalCtx{}\vdash_{\mathsf C} \lee{} \xlongrightarrow{\lbn{0}^{\lstar} \loprod \lbn{1}} \leev{}\glresulttype{\lsnv{0} \loprod \lsnv{1}}}{\lemptyset{};\evalCtxSpace{};\evalCtx{}\vdash_{\mathsf C} \lembd{}_{\Kernel(\lee{})} \xlongrightarrow{\pBasisC{\Kernel(\lee{})}{\lbn{0}}{\lbn{1}}^{\lstar} \loprod \lbn{0}} \Trans(\glvop{GSVDUI}(\leev{},\lenv{0},\lenv{1}))\glresulttype{\lR^{\glvreshape{\leev{}}{\lsnv{0} \loprod \lsnv{1}}{\lsnv{0} \loprod \lsnv{1}},\Kernel{},\lsnv{0} \loprod \lsnv{1}} \loprod \lsnv{0}}}{\texttt{GLang.Raw\allowbreak Certified\allowbreak Operational.C.Term.embed\_\allowbreak ker}}{GLang/Adequacy/RawCertifiedOperational.lean:444}
\gldefrule{c.term.embedCoker}{c.term}{\textsc{C-E-Embed\allowbreak Coker}}{\forall i.\; \sof{\lbn{i}} = \lsn{i} \\ (\lemptyset{}, \evalCtxSpace{})\vdash \lembd{}_{\CoKernel(\lee{})}\colon \tof{\pBasisC{\CoKernel(\lee{})}{\lbn{0}}{\lbn{1}}^{\lstar} \loprod \lbn{0}} \\ (\emptyset,\evalCtxSpace{})\vdash \pBasisC{\CoKernel(\lee{})}{\lbn{0}}{\lbn{1}}^{\lstar} \loprod \lbn{0}\colon \bvalid{} \\ \evalCtxSpace{};\evalCtx{}\vdash_{\mathsf C} \pBasisC{\CoKernel(\lee{})}{\lbn{0}}{\lbn{1}}^{\lstar} \loprod \lbn{0} \evalArrow{}_{\mathsf{basis}} \lR^{\glvreshape{\leev{}}{\lsnv{0} \loprod \lsnv{1}}{\lsnv{0} \loprod \lsnv{1}},\CoKernel{},\lsnv{0} \loprod \lsnv{1}} \loprod \lsnv{0} \\ \forall i.\; \lemptyset{};\evalCtxSpace{};\evalCtx{}\vdash_{\mathsf C} \linner{}_{\lsn{i}} \xlongrightarrow{\lbn{i}^{\lstar} \loprod \lbn{i}^{\lstar}} \lenv{i}\glresulttype{\lsnv{i} \loprod \lsnv{i}} \\ \lemptyset{};\evalCtxSpace{};\evalCtx{}\vdash_{\mathsf C} \lee{} \xlongrightarrow{\lbn{0}^{\lstar} \loprod \lbn{1}} \leev{}\glresulttype{\lsnv{0} \loprod \lsnv{1}}}{\lemptyset{};\evalCtxSpace{};\evalCtx{}\vdash_{\mathsf C} \lembd{}_{\CoKernel(\lee{})} \xlongrightarrow{\pBasisC{\CoKernel(\lee{})}{\lbn{0}}{\lbn{1}}^{\lstar} \loprod \lbn{0}} \Trans(\glvop{GSVDU}(\leev{},\lenv{0},\lenv{1}))\glresulttype{\lR^{\glvreshape{\leev{}}{\lsnv{0} \loprod \lsnv{1}}{\lsnv{0} \loprod \lsnv{1}},\CoKernel{},\lsnv{0} \loprod \lsnv{1}} \loprod \lsnv{0}}}{\texttt{GLang.Raw\allowbreak Certified\allowbreak Operational.C.Term.embed\_\allowbreak coker}}{GLang/Adequacy/RawCertifiedOperational.lean:465}
\gldefrule{c.term.embedImg}{c.term}{\textsc{C-E-Embed\allowbreak Img}}{\forall i.\; \sof{\lbn{i}} = \lsn{i} \\ (\lemptyset{}, \evalCtxSpace{})\vdash \lembd{}_{\Image(\lee{})}\colon \tof{\pBasisC{\Image(\lee{})}{\lbn{0}}{\lbn{1}}^{\lstar} \loprod \lbn{1}} \\ (\emptyset,\evalCtxSpace{})\vdash \pBasisC{\Image(\lee{})}{\lbn{0}}{\lbn{1}}^{\lstar} \loprod \lbn{1}\colon \bvalid{} \\ \evalCtxSpace{};\evalCtx{}\vdash_{\mathsf C} \pBasisC{\Image(\lee{})}{\lbn{0}}{\lbn{1}}^{\lstar} \loprod \lbn{1} \evalArrow{}_{\mathsf{basis}} \lR^{\glvreshape{\leev{}}{\lsnv{0} \loprod \lsnv{1}}{\lsnv{0} \loprod \lsnv{1}},\Image{},\lsnv{0} \loprod \lsnv{1}} \loprod \lsnv{1} \\ \forall i.\; \lemptyset{};\evalCtxSpace{};\evalCtx{}\vdash_{\mathsf C} \linner{}_{\lsn{i}} \xlongrightarrow{\lbn{i}^{\lstar} \loprod \lbn{i}^{\lstar}} \lenv{i}\glresulttype{\lsnv{i} \loprod \lsnv{i}} \\ \lemptyset{};\evalCtxSpace{};\evalCtx{}\vdash_{\mathsf C} \lee{} \xlongrightarrow{\lbn{0}^{\lstar} \loprod \lbn{1}} \leev{}\glresulttype{\lsnv{0} \loprod \lsnv{1}}}{\lemptyset{};\evalCtxSpace{};\evalCtx{}\vdash_{\mathsf C} \lembd{}_{\Image(\lee{})} \xlongrightarrow{\pBasisC{\Image(\lee{})}{\lbn{0}}{\lbn{1}}^{\lstar} \loprod \lbn{1}} \Trans(\glvop{GSVDV}(\leev{},\lenv{0},\lenv{1}))\glresulttype{\lR^{\glvreshape{\leev{}}{\lsnv{0} \loprod \lsnv{1}}{\lsnv{0} \loprod \lsnv{1}},\Image{},\lsnv{0} \loprod \lsnv{1}} \loprod \lsnv{1}}}{\texttt{GLang.Raw\allowbreak Certified\allowbreak Operational.C.Term.embed\_\allowbreak img}}{GLang/Adequacy/RawCertifiedOperational.lean:485}
\gldefrule{c.term.embedCoimg}{c.term}{\textsc{C-E-Embed\allowbreak Coimg}}{\forall i.\; \sof{\lbn{i}} = \lsn{i} \\ (\lemptyset{}, \evalCtxSpace{})\vdash \lembd{}_{\CoImage(\lee{})}\colon \tof{\pBasisC{\CoImage(\lee{})}{\lbn{0}}{\lbn{1}}^{\lstar} \loprod \lbn{1}} \\ (\emptyset,\evalCtxSpace{})\vdash \pBasisC{\CoImage(\lee{})}{\lbn{0}}{\lbn{1}}^{\lstar} \loprod \lbn{1}\colon \bvalid{} \\ \evalCtxSpace{};\evalCtx{}\vdash_{\mathsf C} \pBasisC{\CoImage(\lee{})}{\lbn{0}}{\lbn{1}}^{\lstar} \loprod \lbn{1} \evalArrow{}_{\mathsf{basis}} \lR^{\glvreshape{\leev{}}{\lsnv{0} \loprod \lsnv{1}}{\lsnv{0} \loprod \lsnv{1}},\CoImage{},\lsnv{0} \loprod \lsnv{1}} \loprod \lsnv{1} \\ \forall i.\; \lemptyset{};\evalCtxSpace{};\evalCtx{}\vdash_{\mathsf C} \linner{}_{\lsn{i}} \xlongrightarrow{\lbn{i}^{\lstar} \loprod \lbn{i}^{\lstar}} \lenv{i}\glresulttype{\lsnv{i} \loprod \lsnv{i}} \\ \lemptyset{};\evalCtxSpace{};\evalCtx{}\vdash_{\mathsf C} \lee{} \xlongrightarrow{\lbn{0}^{\lstar} \loprod \lbn{1}} \leev{}\glresulttype{\lsnv{0} \loprod \lsnv{1}}}{\lemptyset{};\evalCtxSpace{};\evalCtx{}\vdash_{\mathsf C} \lembd{}_{\CoImage(\lee{})} \xlongrightarrow{\pBasisC{\CoImage(\lee{})}{\lbn{0}}{\lbn{1}}^{\lstar} \loprod \lbn{1}} \Trans(\glvop{GSVDVI}(\leev{},\lenv{0},\lenv{1}))\glresulttype{\lR^{\glvreshape{\leev{}}{\lsnv{0} \loprod \lsnv{1}}{\lsnv{0} \loprod \lsnv{1}},\CoImage{},\lsnv{0} \loprod \lsnv{1}} \loprod \lsnv{1}}}{\texttt{GLang.Raw\allowbreak Certified\allowbreak Operational.C.Term.embed\_\allowbreak coimg}}{GLang/Adequacy/RawCertifiedOperational.lean:505}
\gldefrule{c.term.appLLin}{c.term}{\textsc{C-E-App\allowbreak LLin}}{\tof{\lbn{1}} = \sof{\lbn{1}} \\ ({\lDelta{}_{0}} \ldisun {\lDelta{}_{1}}, \evalCtxSpace{})\vdash \leval{\len{0}}{}{\len{1}}\colon \tof{\lbn{1}} \\ (\emptyset,\evalCtxSpace{})\vdash \lbn{1}\colon \bvalid{} \\ \evalCtxSpace{};\evalCtx{}\vdash_{\mathsf C} \sof{\lbn{0}} \evalArrow{} \lsv{} \\ \evalCtxSpace{};\evalCtx{}\vdash_{\mathsf C} \lbn{1} \evalArrow{}_{\mathsf{basis}} \lsnv{{\mathrm{svOut}}} \\ {\lDelta{}_{0}};\evalCtxSpace{};\evalCtx{}\vdash_{\mathsf C} \len{0} \xlongrightarrow{\lbn{0}^{\lstar} \loprod \lbn{1}} \lenv{0}\glresulttype{\lsnv{{\mathrm{svMat0}}} \loprod \lsnv{{\mathrm{svMat1}}}} \\ {\lDelta{}_{1}};\evalCtxSpace{};\evalCtx{}\vdash_{\mathsf C} \len{1} \xlongrightarrow{\lbn{0}} \lenv{1}\glresulttype{\lsnv{{\mathrm{svArg}}}}}{{\lDelta{}_{0}} \ldisun {\lDelta{}_{1}};\evalCtxSpace{};\evalCtx{}\vdash_{\mathsf C} \leval{\len{0}}{}{\len{1}} \xlongrightarrow{\lbn{1}} \glvreshape{\LTTVPP{\lsv{}}{\lenv{0}}{\lenv{1}}}{\lsnv{{\mathrm{svOut}}}}{\lsnv{{\mathrm{svOut}}}}\glresulttype{\lsnv{{\mathrm{svOut}}}}}{\texttt{GLang.Raw\allowbreak Certified\allowbreak Operational.C.Term.app\allowbreak L\_\allowbreak lin}}{GLang/Adequacy/RawCertifiedOperational.lean:525}
\gldefrule{c.term.appILin}{c.term}{\textsc{C-E-App\allowbreak ILin}}{({\lDelta{}_{0}} \ldisun {\lDelta{}_{1}}, \evalCtxSpace{})\vdash \leval{\len{0}}{}{\len{1}}\colon \tof{\lbone{}} \\ \evalCtxSpace{};\evalCtx{}\vdash_{\mathsf C} \sof{\lbn{0}} \evalArrow{} \lsv{} \\ \evalCtxSpace{};\evalCtx{}\vdash_{\mathsf C} \lbone{} \evalArrow{}_{\mathsf{basis}} \lR{} \\ {\lDelta{}_{0}};\evalCtxSpace{};\evalCtx{}\vdash_{\mathsf C} \len{0} \xlongrightarrow{\lbn{0}^{\lstar}} \lenv{0}\glresulttype{\lsnv{{\mathrm{svFun}}}} \\ {\lDelta{}_{1}};\evalCtxSpace{};\evalCtx{}\vdash_{\mathsf C} \len{1} \xlongrightarrow{\lbn{0}} \lenv{1}\glresulttype{\lsnv{{\mathrm{svArg}}}}}{{\lDelta{}_{0}} \ldisun {\lDelta{}_{1}};\evalCtxSpace{};\evalCtx{}\vdash_{\mathsf C} \leval{\len{0}}{}{\len{1}} \xlongrightarrow{\lbone{}} \LTTVPP{\lsv{}}{\glvreshape{\lenv{0}}{\lsv{} \loprod \lR{}}{\lsv{}}}{\lenv{1}}\glresulttype{\lR{}}}{\texttt{GLang.Raw\allowbreak Certified\allowbreak Operational.C.Term.app\allowbreak I\_\allowbreak lin}}{GLang/Adequacy/RawCertifiedOperational.lean:544}
\gldefrule{c.term.nllam}{c.term}{\textsc{C-E-Nllam}}{\forall i.\; \tof{\lbn{i}} = \ltyn{i} \\ (\lemptyset{}, \evalCtxSpace{})\vdash (\nllambda \llpl x\colon \ltyn{0}\llpr \ldotsp \lee{})\colon \tof{\frameB{x}{\lbn{0}}{\lbn{1}}} \\ (\emptyset,\evalCtxSpace{})\vdash \frameB{x}{\lbn{0}}{\lbn{1}}\colon \bvalid{} \\ \evalCtxSpace{};\evalCtx{}\vdash_{\mathsf C} \frameB{x}{\lbn{0}}{\lbn{1}} \evalArrow{}_{\mathsf{basis}} (x\colon \ltynv{0}\fnto \ltynv{1}) \\ \evalCtxSpace{};\evalCtx{}\vdash_{\mathsf C} \lbn{0} \evalArrow{}_{\mathsf{basis}} \ltynv{0} \\ \lemptyset{};(\evalCtxSpace{}, \mathit{x}\colon \ltyn{0});\evalCtx{}[(\mathit{x},\lbn{0}) \mapsto \mathit{x}_{\ltynv{0}}]\vdash_{\mathsf C} \lee{} \xlongrightarrow{\lbn{1}} \leev{}\glresulttype{\ltynv{1}} \\ (\evalCtxSpace{}, \mathit{x}\colon \ltyn{0});\evalCtx{}[(\mathit{x},\lbn{0}) \mapsto \mathit{x}_{\ltynv{0}}]\vdash_{\mathsf C} \lbn{1} \evalArrow{}_{\mathsf{basis}} \ltynv{1}}{\lemptyset{};\evalCtxSpace{};\evalCtx{}\vdash_{\mathsf C} (\nllambda \llpl x\colon \ltyn{0}\llpr \ldotsp \lee{}) \xlongrightarrow{\frameB{x}{\lbn{0}}{\lbn{1}}} \llambda(x\colon \ltynv{0})\lrh \ltynv{1}\colon \glvreshape{\leev{}}{\ltynv{1}}{\ltynv{1}}\glresulttype{(x\colon \ltynv{0}\fnto \ltynv{1})}}{\texttt{GLang.Raw\allowbreak Certified\allowbreak Operational.C.Term.nllam}}{GLang/Adequacy/RawCertifiedOperational.lean:555}
\gldefrule{c.term.appNVal}{c.term}{\textsc{C-E-App\allowbreak NVal}}{(\lemptyset{}, \evalCtxSpace{})\vdash \leval{\len{0}}{}{\len{1}}\colon \tof{\lbn{1}\subst{x}{\len{1}}} \\ (\emptyset,\evalCtxSpace{})\vdash \lbn{1}\subst{x}{\len{1}}\colon \bvalid{} \\ \evalCtxSpace{};\evalCtx{}\vdash_{\mathsf C} \lbn{1}\subst{x}{\len{1}} \evalArrow{}_{\mathsf{basis}} {\ltyv{}_{\mathsf{out}}} \\ \lemptyset{};\evalCtxSpace{};\evalCtx{}\vdash_{\mathsf C} \len{0} \xlongrightarrow{\frameB{x}{\lbn{0}}{\lbn{1}}} \lenv{0}\glresulttype{(x\colon {\ltyv{}_{\mathsf{fun}0}}\fnto {\ltyv{}_{\mathsf{fun}1}})} \\ \lemptyset{};\evalCtxSpace{};\evalCtx{}\vdash_{\mathsf C} \len{1} \xlongrightarrow{\lbn{0}} \lenv{1}\glresulttype{{\ltyv{}_{\mathsf{arg}}}} \\ \operatorname{inferType}(\lenv{0}) = (x\colon \ltynv{{\mathrm{D}}}\fnto \ltynv{{\mathrm{C}}}) \\ \lenv{2} = \glvreshape{\lenv{1}}{\ltynv{{\mathrm{D}}}}{\operatorname{inferType}(\lenv{1})}}{\lemptyset{};\evalCtxSpace{};\evalCtx{}\vdash_{\mathsf C} \leval{\len{0}}{}{\len{1}} \xlongrightarrow{\lbn{1}\subst{x}{\len{1}}} \glvreshape{\glvapp{\lenv{0}}{\lenv{2}}}{{\ltyv{}_{\mathsf{out}}}}{\ltynv{{\mathrm{C}}}\subst{x}{\lenv{2}}}\glresulttype{{\ltyv{}_{\mathsf{out}}}}}{\texttt{GLang.Raw\allowbreak Certified\allowbreak Operational.C.Term.app\allowbreak N\_\allowbreak val}}{GLang/Adequacy/RawCertifiedOperational.lean:573}
\gldefrule{c.term.frameReanchor}{c.term}{\textsc{C-E-Frame\allowbreak Reanchor}}{\tof{\lbn{0}'} = \tof{\lbn{0}} \\ (\lemptyset{}, \evalCtxSpace{})\vdash \len{0}\colon \tof{\frameB{x}{\lbn{0}}{\lbn{1}}} \\ (\emptyset,\evalCtxSpace{})\vdash \frameB{x}{\lbn{0}}{\lbn{1}}\colon \bvalid{} \\ \evalCtxSpace{};\evalCtx{}\vdash_{\mathsf C} \frameB{x}{\lbn{0}}{\lbn{1}} \evalArrow{}_{\mathsf{basis}} (x\colon \ltynv{0}\fnto \ltynv{1}') \\ \evalCtxSpace{};\evalCtx{}\vdash_{\mathsf C} \lbn{0} \evalArrow{}_{\mathsf{basis}} \ltynv{0} \\ \lemptyset{};\evalCtxSpace{};\evalCtx{}\vdash_{\mathsf C} \len{0} \xlongrightarrow{\frameB{x}{\lbn{0}'}{\lbn{1}}} \leev{}\glresulttype{(x\colon \ltynv{0}'\fnto \ltynv{1})} \\ \evalCtxSpace{};\evalCtx{}\vdash_{\mathsf C} \lbn{0}' \evalArrow{}_{\mathsf{basis}} \ltynv{0}' \\ \lemptyset{};(\evalCtxSpace{}, \mathit{x}\colon \tof{\lbn{0}});\evalCtx{}[(\mathit{x},\lbn{0}) \mapsto \mathit{x}_{\ltynv{0}}]\vdash_{\mathsf C} \mathit{x} \xlongrightarrow{\lbn{0}'} \lenv{1}\glresulttype{\ltynv{0}'} \\ (\evalCtxSpace{}, \mathit{x}\colon \tof{\lbn{0}});\evalCtx{}[(\mathit{x},\lbn{0}) \mapsto \mathit{x}_{\ltynv{0}}]\vdash_{\mathsf C} \lbn{1} \evalArrow{}_{\mathsf{basis}} \ltynv{1}' \\ \operatorname{inferType}(\leev{}) = (x\colon \ltynv{{\mathrm{D}}}\fnto \ltynv{{\mathrm{C}}}) \\ \lenv{2} = \glvreshape{\glvreshape{\lenv{1}}{\ltynv{0}'}{\ltynv{0}'}}{\ltynv{{\mathrm{D}}}}{\ltynv{0}'}}{\lemptyset{};\evalCtxSpace{};\evalCtx{}\vdash_{\mathsf C} \len{0} \xlongrightarrow{\frameB{x}{\lbn{0}}{\lbn{1}}} \llambda(x\colon \ltynv{0})\lrh \ltynv{1}'\colon \glvreshape{\glvapp{\leev{}}{\lenv{2}}}{\ltynv{1}'}{\ltynv{1}'}\glresulttype{(x\colon \ltynv{0}\fnto \ltynv{1}')}}{\texttt{GLang.Raw\allowbreak Certified\allowbreak Operational.C.Term.frame\_\allowbreak reanchor}}{GLang/Adequacy/RawCertifiedOperational.lean:583}
\gldefrule{c.term.frameEta}{c.term}{\textsc{C-E-Frame\allowbreak Eta}}{(\lemptyset{}, \evalCtxSpace{})\vdash \len{0}\colon \tof{\frameB{x}{\lbn{0}}{\lbn{1}}} \\ (\emptyset,\evalCtxSpace{})\vdash \frameB{x}{\lbn{0}}{\lbn{1}}\colon \bvalid{} \\ \evalCtxSpace{};\evalCtx{}\vdash_{\mathsf C} \frameB{x}{\lbn{0}}{\lbn{1}} \evalArrow{}_{\mathsf{basis}} (x\colon \ltynv{0}\fnto \ltynv{1}) \\ \evalCtxSpace{};\evalCtx{}\vdash_{\mathsf C} \lbn{0} \evalArrow{}_{\mathsf{basis}} \ltynv{0} \\ \lemptyset{};(\evalCtxSpace{}, \mathit{x}\colon \tof{\lbn{0}});\evalCtx{}[(\mathit{x},\lbn{0}) \mapsto \mathit{x}_{\ltynv{0}}]\vdash_{\mathsf C} \leval{\len{0}}{}{\mathit{x}} \xlongrightarrow{\lbn{1}} \leev{}\glresulttype{\ltynv{1}} \\ (\evalCtxSpace{}, \mathit{x}\colon \tof{\lbn{0}});\evalCtx{}[(\mathit{x},\lbn{0}) \mapsto \mathit{x}_{\ltynv{0}}]\vdash_{\mathsf C} \lbn{1} \evalArrow{}_{\mathsf{basis}} \ltynv{1}}{\lemptyset{};\evalCtxSpace{};\evalCtx{}\vdash_{\mathsf C} \len{0} \xlongrightarrow{\frameB{x}{\lbn{0}}{\lbn{1}}} \llambda(x\colon \ltynv{0})\lrh \ltynv{1}\colon \glvreshape{\leev{}}{\ltynv{1}}{\ltynv{1}}\glresulttype{(x\colon \ltynv{0}\fnto \ltynv{1})}}{\texttt{GLang.Raw\allowbreak Certified\allowbreak Operational.C.Term.frame\_\allowbreak eta}}{GLang/Adequacy/RawCertifiedOperational.lean:608}
\gldefrule{c.term.llamAssemble}{c.term}{\textsc{C-E-Llam\allowbreak Assemble}}{\tof{\lb{}'} = \sof{\lb{}'} \\ (\lDelta{}, \evalCtxSpace{})\vdash (\lllambda \llpl x\colon \sof{\lb{}'}\llpr \ldotsp \lee{})\colon \tof{\lb{}'^{\lstar} \loprod \lb{}} \\ (\emptyset,\evalCtxSpace{})\vdash \lb{}'^{\lstar} \loprod \lb{}\colon \bvalid{} \\ \evalCtxSpace{};\evalCtx{}\vdash_{\mathsf C} \lb{}'^{\lstar} \loprod \lb{} \evalArrow{}_{\mathsf{basis}} \lsv{}' \loprod \lsv{} \\ \evalCtxSpace{};\evalCtx{}\vdash_{\mathsf C} \sof{\lb{}'} \evalArrow{} \lsv{}' \\ \evalCtxSpace{};\evalCtx{}\vdash_{\mathsf C} \sof{\lb{}} \evalArrow{} \lsv{} \\ \forall\,0\leq i<\overline{\operatorname{dimOf}}(\lsv{}').\; (\lDelta{}, \mathit{x}\colon \sof{\lb{}'});\evalCtxSpace{};\evalCtx{}[(\mathit{x},\lb{}') \mapsto \ldelta{i}{\lsv{}'}]\vdash_{\mathsf C} \lee{} \xlongrightarrow{\lb{}} \leev{}_{i}\glresulttype{\lsv{}}}{\lDelta{};\evalCtxSpace{};\evalCtx{}\vdash_{\mathsf C} (\lllambda \llpl x\colon \sof{\lb{}'}\llpr \ldotsp \lee{}) \xlongrightarrow{\lb{}'^{\lstar} \loprod \lb{}} \glvreshape{\lsum^{\lsv{}'}_{\mathit{i}<\overline{\operatorname{dimOf}}(\lsv{}')}\left(\ldelta{\mathit{i}}{\lsv{}'}\loprod \glvreshape{\leev{}_{\mathit{i}}}{\lsv{}}{\lsv{}}\right)}{\lsv{}' \loprod \lsv{}}{\begin{cases}\lsempty{}&\overline{\operatorname{dimOf}}(\lsv{}')=0\\ \lsv{}' \loprod \lsv{}&\text{otherwise}\end{cases}}\glresulttype{\lsv{}' \loprod \lsv{}}}{\texttt{GLang.Raw\allowbreak Certified\allowbreak Operational.C.Term.llam\_\allowbreak assemble}}{GLang/Adequacy/RawCertifiedOperational.lean:627}
\gldefrule{c.term.llamAssembleReal}{c.term}{\textsc{C-E-Llam\allowbreak Assemble\allowbreak Real}}{\tof{\lb{}'} = \sof{\lb{}'} \\ (\lDelta{}, \evalCtxSpace{})\vdash (\lllambda \llpl x\colon \sof{\lb{}'}\llpr \ldotsp \lee{})\colon \tof{\lb{}'^{\lstar}} \\ (\emptyset,\evalCtxSpace{})\vdash \lb{}'^{\lstar}\colon \bvalid{} \\ \evalCtxSpace{};\evalCtx{}\vdash_{\mathsf C} \lb{}'^{\lstar} \evalArrow{}_{\mathsf{basis}} \lsv{}' \\ \evalCtxSpace{};\evalCtx{}\vdash_{\mathsf C} \sof{\lb{}'} \evalArrow{} \lsv{}' \\ \forall\,0\leq i<\overline{\operatorname{dimOf}}(\lsv{}').\; (\lDelta{}, \mathit{x}\colon \sof{\lb{}'});\evalCtxSpace{};\evalCtx{}[(\mathit{x},\lb{}') \mapsto \ldelta{i}{\lsv{}'}]\vdash_{\mathsf C} \lee{} \xlongrightarrow{\lbone{}} \leev{}_{i}\glresulttype{\lR{}}}{\lDelta{};\evalCtxSpace{};\evalCtx{}\vdash_{\mathsf C} (\lllambda \llpl x\colon \sof{\lb{}'}\llpr \ldotsp \lee{}) \xlongrightarrow{\lb{}'^{\lstar}} \glvreshape{\lsum^{\lsv{}'}_{\mathit{i}<\overline{\operatorname{dimOf}}(\lsv{}')}\left(\glvreshape{\leev{}_{\mathit{i}}}{\lR{}}{\lR{}} \lvscale \ldelta{\mathit{i}}{\lsv{}'}\right)}{\lsv{}'}{\begin{cases}\lsempty{}&\overline{\operatorname{dimOf}}(\lsv{}')=0\\ \lsv{}'&\text{otherwise}\end{cases}}\glresulttype{\lsv{}'}}{\texttt{GLang.Raw\allowbreak Certified\allowbreak Operational.C.Term.llam\_\allowbreak assemble\_\allowbreak real}}{GLang/Adequacy/RawCertifiedOperational.lean:657}
\gldefrule{c.term.zero}{c.term}{\textsc{C-E-Zero}}{\tof{\lb{}} = \sof{\lb{}} \\ (\lDelta{}, \evalCtxSpace{})\vdash \lvzero{}_{\sof{\lb{}}}\colon \tof{\lb{}} \\ (\emptyset,\evalCtxSpace{})\vdash \lb{}\colon \bvalid{} \\ \evalCtxSpace{};\evalCtx{}\vdash_{\mathsf C} \lb{} \evalArrow{}_{\mathsf{basis}} \lsv{} \\ \evalCtxSpace{};\evalCtx{}\vdash_{\mathsf C} \sof{\lb{}} \evalArrow{} \lsv{}}{\lDelta{};\evalCtxSpace{};\evalCtx{}\vdash_{\mathsf C} \lvzero{}_{\sof{\lb{}}} \xlongrightarrow{\lb{}} \lvzero{}_{\lsv{}}\glresulttype{\lsv{}}}{\texttt{GLang.Raw\allowbreak Certified\allowbreak Operational.C.Term.zero}}{GLang/Adequacy/RawCertifiedOperational.lean:676}
\gldefrule{c.term.rlit}{c.term}{\textsc{C-E-Rlit}}{(\lemptyset{}, \evalCtxSpace{})\vdash \mathit{r}\colon \tof{\lbone{}} \\ \evalCtxSpace{};\evalCtx{}\vdash_{\mathsf C} \lbone{} \evalArrow{}_{\mathsf{basis}} \lR{}}{\lemptyset{};\evalCtxSpace{};\evalCtx{}\vdash_{\mathsf C} \mathit{r} \xlongrightarrow{\lbone{}} \mathit{r}\glresulttype{\lR{}}}{\texttt{GLang.Raw\allowbreak Certified\allowbreak Operational.C.Term.rlit}}{GLang/Adequacy/RawCertifiedOperational.lean:684}
\gldefrulerows{c.term}{
\textsc{C-E-Var\allowbreak Lin} & $\{(\mathit{x}, \ls{})\};\evalCtxSpace{};\evalCtx{}\vdash_{\mathsf C} \mathit{x} \xlongrightarrow{\lb{}} \leev{}\glresulttype{\ltyv{}}$ & \texttt{\footnotesize GLang/Adequacy/Raw\allowbreak Certified\allowbreak Operational.lean:283}\\
\textsc{C-E-Var\allowbreak Nlin} & $\lemptyset{};\evalCtxSpace{};\evalCtx{}\vdash_{\mathsf C} \mathit{x} \xlongrightarrow{\lb{}} \leev{}\glresulttype{\ltyv{}}$ & \texttt{\footnotesize GLang/Adequacy/Raw\allowbreak Certified\allowbreak Operational.lean:290}\\
\textsc{C-E-Promote} & $\lDelta{};\mathit{lin} \ldisun \evalCtxSpace{}'';\evalCtx{}\vdash_{\mathsf C} \lee{} \xlongrightarrow{\lb{}} \leev{}\glresulttype{\ltyv{}}$ & \texttt{\footnotesize GLang/Adequacy/Raw\allowbreak Certified\allowbreak Operational.lean:297}\\
\textsc{C-E-Cob} & $\lDelta{};\evalCtxSpace{};\evalCtx{}\vdash_{\mathsf C} \lee{} \xlongrightarrow{\lbn{0}} \LTTVPP{\lsv{}}{\glvreshape{\lenv{0}}{\lsv{} \loprod \lsv{}}{\lsv{} \loprod \lsv{}}}{\glvreshape{\leev{}}{\lsv{}}{\lsv{}}}\glresulttype{\lsv{}}$ & \texttt{\footnotesize GLang/Adequacy/Raw\allowbreak Certified\allowbreak Operational.lean:315}\\
\textsc{C-E-Reshape\allowbreak Cob} & $\lDelta{};\evalCtxSpace{};\evalCtx{}\vdash_{\mathsf C} \lee{} \xlongrightarrow{\lb{}} \glvreshape{\leev{}}{\ltyv{}'}{\ltyv{}}\glresulttype{\ltyv{}'}$ & \texttt{\footnotesize GLang/Adequacy/Raw\allowbreak Certified\allowbreak Operational.lean:328}\\
\textsc{C-E-Scale} & ${\lDelta{}_{0}} \ldisun {\lDelta{}_{1}};\evalCtxSpace{};\evalCtx{}\vdash_{\mathsf C} \len{0} \lvscale \len{1} \xlongrightarrow{\lb{}} \lenv{0} \lvscale \lenv{1}\glresulttype{\lsv{}}$ & \texttt{\footnotesize GLang/Adequacy/Raw\allowbreak Certified\allowbreak Operational.lean:335}\\
\textsc{C-E-Add} & $\lDelta{};\evalCtxSpace{};\evalCtx{}\vdash_{\mathsf C} \len{0} \lvadd \len{1} \xlongrightarrow{\lb{}} \lenv{0} \lvadd_{\lsv{}} \lenv{1}\glresulttype{\lsv{}}$ & \texttt{\footnotesize GLang/Adequacy/Raw\allowbreak Certified\allowbreak Operational.lean:345}\\
\textsc{C-E-Dsum\allowbreak E} & $\lDelta{};\evalCtxSpace{};\evalCtx{}\vdash_{\mathsf C} \len{0}\losum \len{1} \xlongrightarrow{\lbn{0} \losum \lbn{1}} \lenv{0}\losum \lenv{1}\glresulttype{\lsnv{0} \losum \lsnv{1}}$ & \texttt{\footnotesize GLang/Adequacy/Raw\allowbreak Certified\allowbreak Operational.lean:359}\\
\textsc{C-E-Tprod\allowbreak E} & ${\lDelta{}_{0}} \ldisun {\lDelta{}_{1}};\evalCtxSpace{};\evalCtx{}\vdash_{\mathsf C} \len{0}\loprod \len{1} \xlongrightarrow{\lbn{0} \loprod \lbn{1}} \lenv{0}\loprod \lenv{1}\glresulttype{\lsnv{0} \loprod \lsnv{1}}$ & \texttt{\footnotesize GLang/Adequacy/Raw\allowbreak Certified\allowbreak Operational.lean:369}\\
\textsc{C-E-Pinv\allowbreak TStd} & $\lemptyset{};\evalCtxSpace{};\evalCtx{}\vdash_{\mathsf C} \lee{}^{\lpseudo} \xlongrightarrow{\lbn{0}^{\lstar} \loprod \lbn{1}} \leev{}^{+}\glresulttype{\lsnv{0} \loprod \lsnv{1}}$ & \texttt{\footnotesize GLang/Adequacy/Raw\allowbreak Certified\allowbreak Operational.lean:381}\\
\textsc{C-E-Pinv\allowbreak TGpinv} & $\lemptyset{};\evalCtxSpace{};\evalCtx{}\vdash_{\mathsf C} \lee{}^{\lpseudo} \xlongrightarrow{\lbn{0}^{\lstar} \loprod \lbn{1}} \glvop{GPINV}(\leev{},\lenv{1},\lenv{0})\glresulttype{\lsnv{0} \loprod \lsnv{1}}$ & \texttt{\footnotesize GLang/Adequacy/Raw\allowbreak Certified\allowbreak Operational.lean:391}\\
\textsc{C-E-Pinv\allowbreak DStd} & $\lemptyset{};\evalCtxSpace{};\evalCtx{}\vdash_{\mathsf C} \lee{}^{\lpseudo} \xlongrightarrow{\lbn{0}^{\lstar} \loprod \lbn{1}} \glvreshape{\leev{}}{\lsnv{1} \loprod \lR{}}{\lsnv{1}}^{+}\glresulttype{\lR{} \loprod \lsnv{1}}$ & \texttt{\footnotesize GLang/Adequacy/Raw\allowbreak Certified\allowbreak Operational.lean:405}\\
\textsc{C-E-Pinv\allowbreak DGpinv} & $\lemptyset{};\evalCtxSpace{};\evalCtx{}\vdash_{\mathsf C} \lee{}^{\lpseudo} \xlongrightarrow{\lbn{0}^{\lstar} \loprod \lbn{1}} \glvop{GPINV}(\glvreshape{\leev{}}{\lsnv{1} \loprod \lR{}}{\lsnv{1}},\lenv{1},\lenv{0})\glresulttype{\lR{} \loprod \lsnv{1}}$ & \texttt{\footnotesize GLang/Adequacy/Raw\allowbreak Certified\allowbreak Operational.lean:417}\\
\textsc{C-E-Inner\allowbreak Std} & $\lemptyset{};\evalCtxSpace{};\evalCtx{}\vdash_{\mathsf C} \linner{}_{\ls{}} \xlongrightarrow{\lbn{0}^{\lstar} \loprod \lbn{0}^{\lstar}} \lidelta{\lsv{}}\glresulttype{\lsv{} \loprod \lsv{}}$ & \texttt{\footnotesize GLang/Adequacy/Raw\allowbreak Certified\allowbreak Operational.lean:434}\\
\textsc{C-E-Embed\allowbreak Ker} & $\lemptyset{};\evalCtxSpace{};\evalCtx{}\vdash_{\mathsf C} \lembd{}_{\Kernel(\lee{})} \xlongrightarrow{\pBasisC{\Kernel(\lee{})}{\lbn{0}}{\lbn{1}}^{\lstar} \loprod \lbn{0}} \Trans(\glvop{GSVDUI}(\leev{},\lenv{0},\lenv{1}))\glresulttype{\lR^{\glvreshape{\leev{}}{\lsnv{0} \loprod \lsnv{1}}{\lsnv{0} \loprod \lsnv{1}},\Kernel{},\lsnv{0} \loprod \lsnv{1}} \loprod \lsnv{0}}$ & \texttt{\footnotesize GLang/Adequacy/Raw\allowbreak Certified\allowbreak Operational.lean:444}\\
\textsc{C-E-Embed\allowbreak Coker} & $\lemptyset{};\evalCtxSpace{};\evalCtx{}\vdash_{\mathsf C} \lembd{}_{\CoKernel(\lee{})} \xlongrightarrow{\pBasisC{\CoKernel(\lee{})}{\lbn{0}}{\lbn{1}}^{\lstar} \loprod \lbn{0}} \Trans(\glvop{GSVDU}(\leev{},\lenv{0},\lenv{1}))\glresulttype{\lR^{\glvreshape{\leev{}}{\lsnv{0} \loprod \lsnv{1}}{\lsnv{0} \loprod \lsnv{1}},\CoKernel{},\lsnv{0} \loprod \lsnv{1}} \loprod \lsnv{0}}$ & \texttt{\footnotesize GLang/Adequacy/Raw\allowbreak Certified\allowbreak Operational.lean:465}\\
\textsc{C-E-Embed\allowbreak Img} & $\lemptyset{};\evalCtxSpace{};\evalCtx{}\vdash_{\mathsf C} \lembd{}_{\Image(\lee{})} \xlongrightarrow{\pBasisC{\Image(\lee{})}{\lbn{0}}{\lbn{1}}^{\lstar} \loprod \lbn{1}} \Trans(\glvop{GSVDV}(\leev{},\lenv{0},\lenv{1}))\glresulttype{\lR^{\glvreshape{\leev{}}{\lsnv{0} \loprod \lsnv{1}}{\lsnv{0} \loprod \lsnv{1}},\Image{},\lsnv{0} \loprod \lsnv{1}} \loprod \lsnv{1}}$ & \texttt{\footnotesize GLang/Adequacy/Raw\allowbreak Certified\allowbreak Operational.lean:485}\\
\textsc{C-E-Embed\allowbreak Coimg} & $\lemptyset{};\evalCtxSpace{};\evalCtx{}\vdash_{\mathsf C} \lembd{}_{\CoImage(\lee{})} \xlongrightarrow{\pBasisC{\CoImage(\lee{})}{\lbn{0}}{\lbn{1}}^{\lstar} \loprod \lbn{1}} \Trans(\glvop{GSVDVI}(\leev{},\lenv{0},\lenv{1}))\glresulttype{\lR^{\glvreshape{\leev{}}{\lsnv{0} \loprod \lsnv{1}}{\lsnv{0} \loprod \lsnv{1}},\CoImage{},\lsnv{0} \loprod \lsnv{1}} \loprod \lsnv{1}}$ & \texttt{\footnotesize GLang/Adequacy/Raw\allowbreak Certified\allowbreak Operational.lean:505}\\
\textsc{C-E-App\allowbreak LLin} & ${\lDelta{}_{0}} \ldisun {\lDelta{}_{1}};\evalCtxSpace{};\evalCtx{}\vdash_{\mathsf C} \leval{\len{0}}{}{\len{1}} \xlongrightarrow{\lbn{1}} \glvreshape{\LTTVPP{\lsv{}}{\lenv{0}}{\lenv{1}}}{\lsnv{{\mathrm{svOut}}}}{\lsnv{{\mathrm{svOut}}}}\glresulttype{\lsnv{{\mathrm{svOut}}}}$ & \texttt{\footnotesize GLang/Adequacy/Raw\allowbreak Certified\allowbreak Operational.lean:525}\\
\textsc{C-E-App\allowbreak ILin} & ${\lDelta{}_{0}} \ldisun {\lDelta{}_{1}};\evalCtxSpace{};\evalCtx{}\vdash_{\mathsf C} \leval{\len{0}}{}{\len{1}} \xlongrightarrow{\lbone{}} \LTTVPP{\lsv{}}{\glvreshape{\lenv{0}}{\lsv{} \loprod \lR{}}{\lsv{}}}{\lenv{1}}\glresulttype{\lR{}}$ & \texttt{\footnotesize GLang/Adequacy/Raw\allowbreak Certified\allowbreak Operational.lean:544}\\
\textsc{C-E-Nllam} & $\lemptyset{};\evalCtxSpace{};\evalCtx{}\vdash_{\mathsf C} (\nllambda \llpl x\colon \ltyn{0}\llpr \ldotsp \lee{}) \xlongrightarrow{\frameB{x}{\lbn{0}}{\lbn{1}}} \llambda(x\colon \ltynv{0})\lrh \ltynv{1}\colon \glvreshape{\leev{}}{\ltynv{1}}{\ltynv{1}}\glresulttype{(x\colon \ltynv{0}\fnto \ltynv{1})}$ & \texttt{\footnotesize GLang/Adequacy/Raw\allowbreak Certified\allowbreak Operational.lean:555}\\
\textsc{C-E-App\allowbreak NVal} & $\lemptyset{};\evalCtxSpace{};\evalCtx{}\vdash_{\mathsf C} \leval{\len{0}}{}{\len{1}} \xlongrightarrow{\lbn{1}\subst{x}{\len{1}}} \glvreshape{\glvapp{\lenv{0}}{\lenv{2}}}{{\ltyv{}_{\mathsf{out}}}}{\ltynv{{\mathrm{C}}}\subst{x}{\lenv{2}}}\glresulttype{{\ltyv{}_{\mathsf{out}}}}$ & \texttt{\footnotesize GLang/Adequacy/Raw\allowbreak Certified\allowbreak Operational.lean:573}\\
\textsc{C-E-Frame\allowbreak Reanchor} & $\lemptyset{};\evalCtxSpace{};\evalCtx{}\vdash_{\mathsf C} \len{0} \xlongrightarrow{\frameB{x}{\lbn{0}}{\lbn{1}}} \llambda(x\colon \ltynv{0})\lrh \ltynv{1}'\colon \glvreshape{\glvapp{\leev{}}{\lenv{2}}}{\ltynv{1}'}{\ltynv{1}'}\glresulttype{(x\colon \ltynv{0}\fnto \ltynv{1}')}$ & \texttt{\footnotesize GLang/Adequacy/Raw\allowbreak Certified\allowbreak Operational.lean:583}\\
\textsc{C-E-Frame\allowbreak Eta} & $\lemptyset{};\evalCtxSpace{};\evalCtx{}\vdash_{\mathsf C} \len{0} \xlongrightarrow{\frameB{x}{\lbn{0}}{\lbn{1}}} \llambda(x\colon \ltynv{0})\lrh \ltynv{1}\colon \glvreshape{\leev{}}{\ltynv{1}}{\ltynv{1}}\glresulttype{(x\colon \ltynv{0}\fnto \ltynv{1})}$ & \texttt{\footnotesize GLang/Adequacy/Raw\allowbreak Certified\allowbreak Operational.lean:608}\\
\textsc{C-E-Llam\allowbreak Assemble} & $\lDelta{};\evalCtxSpace{};\evalCtx{}\vdash_{\mathsf C} (\lllambda \llpl x\colon \sof{\lb{}'}\llpr \ldotsp \lee{}) \xlongrightarrow{\lb{}'^{\lstar} \loprod \lb{}} \glvreshape{\lsum^{\lsv{}'}_{\mathit{i}<\overline{\operatorname{dimOf}}(\lsv{}')}\left(\ldelta{\mathit{i}}{\lsv{}'}\loprod \glvreshape{\leev{}_{\mathit{i}}}{\lsv{}}{\lsv{}}\right)}{\lsv{}' \loprod \lsv{}}{\begin{cases}\lsempty{}&\overline{\operatorname{dimOf}}(\lsv{}')=0\\ \lsv{}' \loprod \lsv{}&\text{otherwise}\end{cases}}\glresulttype{\lsv{}' \loprod \lsv{}}$ & \texttt{\footnotesize GLang/Adequacy/Raw\allowbreak Certified\allowbreak Operational.lean:627}\\
\textsc{C-E-Llam\allowbreak Assemble\allowbreak Real} & $\lDelta{};\evalCtxSpace{};\evalCtx{}\vdash_{\mathsf C} (\lllambda \llpl x\colon \sof{\lb{}'}\llpr \ldotsp \lee{}) \xlongrightarrow{\lb{}'^{\lstar}} \glvreshape{\lsum^{\lsv{}'}_{\mathit{i}<\overline{\operatorname{dimOf}}(\lsv{}')}\left(\glvreshape{\leev{}_{\mathit{i}}}{\lR{}}{\lR{}} \lvscale \ldelta{\mathit{i}}{\lsv{}'}\right)}{\lsv{}'}{\begin{cases}\lsempty{}&\overline{\operatorname{dimOf}}(\lsv{}')=0\\ \lsv{}'&\text{otherwise}\end{cases}}\glresulttype{\lsv{}'}$ & \texttt{\footnotesize GLang/Adequacy/Raw\allowbreak Certified\allowbreak Operational.lean:657}\\
\textsc{C-E-Zero} & $\lDelta{};\evalCtxSpace{};\evalCtx{}\vdash_{\mathsf C} \lvzero{}_{\sof{\lb{}}} \xlongrightarrow{\lb{}} \lvzero{}_{\lsv{}}\glresulttype{\lsv{}}$ & \texttt{\footnotesize GLang/Adequacy/Raw\allowbreak Certified\allowbreak Operational.lean:676}\\
\textsc{C-E-Rlit} & $\lemptyset{};\evalCtxSpace{};\evalCtx{}\vdash_{\mathsf C} \mathit{r} \xlongrightarrow{\lbone{}} \mathit{r}\glresulttype{\lR{}}$ & \texttt{\footnotesize GLang/Adequacy/Raw\allowbreak Certified\allowbreak Operational.lean:684}\\
}{
\gpfit{\inferrule*[]{((\mathit{x}, \lb{}) \mapsto \leev{}) \in \evalCtx{} \\ \tof{\lb{}} = \ls{} \\ (\emptyset,\evalCtxSpace{})\vdash \lb{}\colon \bvalid{} \\ \evalCtxSpace{};\evalCtx{}\vdash_{\mathsf C} \lb{} \evalArrow{}_{\mathsf{basis}} \ltyv{}}{\{(\mathit{x}, \ls{})\};\evalCtxSpace{};\evalCtx{}\vdash_{\mathsf C} \mathit{x} \xlongrightarrow{\lb{}} \leev{}\glresulttype{\ltyv{}}}}
\and
\gpfit{\inferrule*[]{((\mathit{x}, \lb{}) \mapsto \leev{}) \in \evalCtx{} \\ (\mathit{x}\colon \tof{\lb{}}) \in \evalCtxSpace{} \\ (\emptyset,\evalCtxSpace{})\vdash \lb{}\colon \bvalid{} \\ \evalCtxSpace{};\evalCtx{}\vdash_{\mathsf C} \lb{} \evalArrow{}_{\mathsf{basis}} \ltyv{}}{\lemptyset{};\evalCtxSpace{};\evalCtx{}\vdash_{\mathsf C} \mathit{x} \xlongrightarrow{\lb{}} \leev{}\glresulttype{\ltyv{}}}}
\and
\gpfit{\inferrule*[]{\forall y \in \mathit{lin}.\; y \notin \bigcup_{(\cdot\colon s) \in \lDelta{},\, \mathit{lin}} \mathrm{fv}(s) \;\cup\; \bigcup_{(\cdot\colon \tau) \in \evalCtxSpace{}'} \mathrm{fv}(\tau) \\ \evalCtxSpace{}' \subseteq \evalCtxSpace{}'' \\ (\lDelta{}, \mathit{lin} \ldisun \evalCtxSpace{}'')\vdash \lee{}\colon \tof{\lb{}} \\ (\emptyset,\mathit{lin} \ldisun \evalCtxSpace{}'')\vdash \lb{}\colon \bvalid{} \\ \mathit{lin} \ldisun \evalCtxSpace{}'';\evalCtx{}\vdash_{\mathsf C} \lb{} \evalArrow{}_{\mathsf{basis}} \ltyv{} \\ \mathit{lin} \ldisun \lDelta{};\evalCtxSpace{}';\evalCtx{}\vdash_{\mathsf C} \lee{} \xlongrightarrow{\lb{}} \leev{}\glresulttype{\ltyv{}}}{\lDelta{};\mathit{lin} \ldisun \evalCtxSpace{}'';\evalCtx{}\vdash_{\mathsf C} \lee{} \xlongrightarrow{\lb{}} \leev{}\glresulttype{\ltyv{}}}}
\and
\gpfit{\inferrule*[]{\forall i.\; \tof{\lbn{i}} = \sof{\lbn{i}} \\ (\lDelta{}, \evalCtxSpace{})\vdash \lee{}\colon \tof{\lbn{0}} \\ (\emptyset,\evalCtxSpace{})\vdash \lbn{0}\colon \bvalid{} \\ \evalCtxSpace{};\evalCtx{}\vdash_{\mathsf C} \lbn{0} \evalArrow{}_{\mathsf{basis}} \lsv{} \\ \lDelta{};\evalCtxSpace{};\evalCtx{}\vdash_{\mathsf C} \lee{} \xlongrightarrow{\lbn{1}} \leev{}\glresulttype{\ltyv{}} \\ \evalCtxSpace{};\evalCtx{}\vdash_{\mathsf C} \lbn{1} \xlongleftrightarrow{\lenv{0}} \lbn{0} \\ \evalCtxSpace{};\evalCtx{}\vdash_{\mathsf C} \sof{\lbn{0}} \evalArrow{} \lsv{}}{\lDelta{};\evalCtxSpace{};\evalCtx{}\vdash_{\mathsf C} \lee{} \xlongrightarrow{\lbn{0}} \LTTVPP{\lsv{}}{\glvreshape{\lenv{0}}{\lsv{} \loprod \lsv{}}{\lsv{} \loprod \lsv{}}}{\glvreshape{\leev{}}{\lsv{}}{\lsv{}}}\glresulttype{\lsv{}}}}
\and
\gpfit{\inferrule*[]{(\lDelta{}, \evalCtxSpace{})\vdash \lee{}\colon \tof{\lb{}} \\ (\emptyset,\evalCtxSpace{})\vdash \lb{}\colon \bvalid{} \\ \evalCtxSpace{};\evalCtx{}\vdash_{\mathsf C} \lb{} \evalArrow{}_{\mathsf{basis}} \ltyv{}' \\ \lDelta{};\evalCtxSpace{};\evalCtx{}\vdash_{\mathsf C} \lee{} \xlongrightarrow{\lb{}} \leev{}\glresulttype{\ltyv{}}}{\lDelta{};\evalCtxSpace{};\evalCtx{}\vdash_{\mathsf C} \lee{} \xlongrightarrow{\lb{}} \glvreshape{\leev{}}{\ltyv{}'}{\ltyv{}}\glresulttype{\ltyv{}'}}}
\and
\gpfit{\inferrule*[]{\tof{\lb{}} = \sof{\lb{}} \\ ({\lDelta{}_{0}} \ldisun {\lDelta{}_{1}}, \evalCtxSpace{})\vdash \len{0} \lvscale \len{1}\colon \tof{\lb{}} \\ (\emptyset,\evalCtxSpace{})\vdash \lb{}\colon \bvalid{} \\ \evalCtxSpace{};\evalCtx{}\vdash_{\mathsf C} \lb{} \evalArrow{}_{\mathsf{basis}} \lsv{} \\ {\lDelta{}_{0}};\evalCtxSpace{};\evalCtx{}\vdash_{\mathsf C} \len{0} \xlongrightarrow{\lbone{}} \lenv{0}\glresulttype{\lR{}} \\ {\lDelta{}_{1}};\evalCtxSpace{};\evalCtx{}\vdash_{\mathsf C} \len{1} \xlongrightarrow{\lb{}} \lenv{1}\glresulttype{\lsv{}}}{{\lDelta{}_{0}} \ldisun {\lDelta{}_{1}};\evalCtxSpace{};\evalCtx{}\vdash_{\mathsf C} \len{0} \lvscale \len{1} \xlongrightarrow{\lb{}} \lenv{0} \lvscale \lenv{1}\glresulttype{\lsv{}}}}
\and
\gpfit{\inferrule*[]{\tof{\lb{}} = \sof{\lb{}} \\ (\lDelta{}, \evalCtxSpace{})\vdash \len{0} \lvadd \len{1}\colon \tof{\lb{}} \\ (\emptyset,\evalCtxSpace{})\vdash \lb{}\colon \bvalid{} \\ \evalCtxSpace{};\evalCtx{}\vdash_{\mathsf C} \lb{} \evalArrow{}_{\mathsf{basis}} \lsv{} \\ \forall i.\; \lDelta{};\evalCtxSpace{};\evalCtx{}\vdash_{\mathsf C} \len{i} \xlongrightarrow{\lb{}} \lenv{i}\glresulttype{\lsnv{i}}}{\lDelta{};\evalCtxSpace{};\evalCtx{}\vdash_{\mathsf C} \len{0} \lvadd \len{1} \xlongrightarrow{\lb{}} \lenv{0} \lvadd_{\lsv{}} \lenv{1}\glresulttype{\lsv{}}}}
\and
\gpfit{\inferrule*[]{\forall i.\; \tof{\lbn{i}} = \sof{\lbn{i}} \\ (\lDelta{}, \evalCtxSpace{})\vdash \len{0}\losum \len{1}\colon \tof{\lbn{0} \losum \lbn{1}} \\ (\emptyset,\evalCtxSpace{})\vdash \lbn{0} \losum \lbn{1}\colon \bvalid{} \\ \evalCtxSpace{};\evalCtx{}\vdash_{\mathsf C} \lbn{0} \losum \lbn{1} \evalArrow{}_{\mathsf{basis}} \lsnv{0} \losum \lsnv{1} \\ \forall i.\; \lDelta{};\evalCtxSpace{};\evalCtx{}\vdash_{\mathsf C} \len{i} \xlongrightarrow{\lbn{i}} \lenv{i}\glresulttype{\lsnv{i}}}{\lDelta{};\evalCtxSpace{};\evalCtx{}\vdash_{\mathsf C} \len{0}\losum \len{1} \xlongrightarrow{\lbn{0} \losum \lbn{1}} \lenv{0}\losum \lenv{1}\glresulttype{\lsnv{0} \losum \lsnv{1}}}}
\and
\gpfit{\inferrule*[]{\forall i.\; \tof{\lbn{i}} = \sof{\lbn{i}} \\ ({\lDelta{}_{0}} \ldisun {\lDelta{}_{1}}, \evalCtxSpace{})\vdash \len{0}\loprod \len{1}\colon \tof{\lbn{0} \loprod \lbn{1}} \\ (\emptyset,\evalCtxSpace{})\vdash \lbn{0} \loprod \lbn{1}\colon \bvalid{} \\ \evalCtxSpace{};\evalCtx{}\vdash_{\mathsf C} \lbn{0} \loprod \lbn{1} \evalArrow{}_{\mathsf{basis}} \lsnv{0} \loprod \lsnv{1} \\ \forall i.\; {\lDelta{}_{i}};\evalCtxSpace{};\evalCtx{}\vdash_{\mathsf C} \len{i} \xlongrightarrow{\lbn{i}} \lenv{i}\glresulttype{\lsnv{i}}}{{\lDelta{}_{0}} \ldisun {\lDelta{}_{1}};\evalCtxSpace{};\evalCtx{}\vdash_{\mathsf C} \len{0}\loprod \len{1} \xlongrightarrow{\lbn{0} \loprod \lbn{1}} \lenv{0}\loprod \lenv{1}\glresulttype{\lsnv{0} \loprod \lsnv{1}}}}
\and
\gpfit{\inferrule*[]{\forall i.\; \isStdBasis{\lbn{i}} \\ (\lemptyset{}, \evalCtxSpace{})\vdash \lee{}^{\lpseudo}\colon \tof{\lbn{0}^{\lstar} \loprod \lbn{1}} \\ (\emptyset,\evalCtxSpace{})\vdash \lbn{0}^{\lstar} \loprod \lbn{1}\colon \bvalid{} \\ \evalCtxSpace{};\evalCtx{}\vdash_{\mathsf C} \lbn{0}^{\lstar} \loprod \lbn{1} \evalArrow{}_{\mathsf{basis}} \lsnv{0} \loprod \lsnv{1} \\ \lemptyset{};\evalCtxSpace{};\evalCtx{}\vdash_{\mathsf C} \lee{} \xlongrightarrow{\lbn{1}^{\lstar} \loprod \lbn{0}} \leev{}\glresulttype{\lsnv{1} \loprod \lsnv{0}}}{\lemptyset{};\evalCtxSpace{};\evalCtx{}\vdash_{\mathsf C} \lee{}^{\lpseudo} \xlongrightarrow{\lbn{0}^{\lstar} \loprod \lbn{1}} \leev{}^{+}\glresulttype{\lsnv{0} \loprod \lsnv{1}}}}
\and
\gpfit{\inferrule*[]{(\lemptyset{}, \evalCtxSpace{})\vdash \lee{}^{\lpseudo}\colon \tof{\lbn{0}^{\lstar} \loprod \lbn{1}} \\ (\emptyset,\evalCtxSpace{})\vdash \lbn{0}^{\lstar} \loprod \lbn{1}\colon \bvalid{} \\ \evalCtxSpace{};\evalCtx{}\vdash_{\mathsf C} \lbn{0}^{\lstar} \loprod \lbn{1} \evalArrow{}_{\mathsf{basis}} \lsnv{0} \loprod \lsnv{1} \\ \lemptyset{};\evalCtxSpace{};\evalCtx{}\vdash_{\mathsf C} \lee{} \xlongrightarrow{\lbn{1}^{\lstar} \loprod \lbn{0}} \leev{}\glresulttype{\lsnv{1} \loprod \lsnv{0}} \\ \forall i.\; \lemptyset{};\evalCtxSpace{};\evalCtx{}\vdash_{\mathsf C} \linner{}_{\sof{\lbn{i}}} \xlongrightarrow{\lbn{i}^{\lstar} \loprod \lbn{i}^{\lstar}} \lenv{i}\glresulttype{\lsnv{i} \loprod \lsnv{i}}}{\lemptyset{};\evalCtxSpace{};\evalCtx{}\vdash_{\mathsf C} \lee{}^{\lpseudo} \xlongrightarrow{\lbn{0}^{\lstar} \loprod \lbn{1}} \glvop{GPINV}(\leev{},\lenv{1},\lenv{0})\glresulttype{\lsnv{0} \loprod \lsnv{1}}}}
\and
\gpfit{\inferrule*[]{\sof{\lbn{0}} = \lR{} \\ \forall i.\; \isStdBasis{\lbn{i}} \\ (\lemptyset{}, \evalCtxSpace{})\vdash \lee{}^{\lpseudo}\colon \tof{\lbn{0}^{\lstar} \loprod \lbn{1}} \\ (\emptyset,\evalCtxSpace{})\vdash \lbn{0}^{\lstar} \loprod \lbn{1}\colon \bvalid{} \\ \evalCtxSpace{};\evalCtx{}\vdash_{\mathsf C} \lbn{0}^{\lstar} \loprod \lbn{1} \evalArrow{}_{\mathsf{basis}} \lR{} \loprod \lsnv{1} \\ \lemptyset{};\evalCtxSpace{};\evalCtx{}\vdash_{\mathsf C} \lee{} \xlongrightarrow{\lbn{1}^{\lstar}} \leev{}\glresulttype{\lsnv{1}}}{\lemptyset{};\evalCtxSpace{};\evalCtx{}\vdash_{\mathsf C} \lee{}^{\lpseudo} \xlongrightarrow{\lbn{0}^{\lstar} \loprod \lbn{1}} \glvreshape{\leev{}}{\lsnv{1} \loprod \lR{}}{\lsnv{1}}^{+}\glresulttype{\lR{} \loprod \lsnv{1}}}}
\and
\gpfit{\inferrule*[]{\sof{\lbn{0}} = \lR{} \\ \isStdBasis{\lbn{0}} \\ (\lemptyset{}, \evalCtxSpace{})\vdash \lee{}^{\lpseudo}\colon \tof{\lbn{0}^{\lstar} \loprod \lbn{1}} \\ (\emptyset,\evalCtxSpace{})\vdash \lbn{0}^{\lstar} \loprod \lbn{1}\colon \bvalid{} \\ \evalCtxSpace{};\evalCtx{}\vdash_{\mathsf C} \lbn{0}^{\lstar} \loprod \lbn{1} \evalArrow{}_{\mathsf{basis}} \lR{} \loprod \lsnv{1} \\ \lemptyset{};\evalCtxSpace{};\evalCtx{}\vdash_{\mathsf C} \lee{} \xlongrightarrow{\lbn{1}^{\lstar}} \leev{}\glresulttype{\lsnv{1}} \\ \lemptyset{};\evalCtxSpace{};\evalCtx{}\vdash_{\mathsf C} \linner{}_{\sof{\lbn{0}}} \xlongrightarrow{\lbn{0}^{\lstar} \loprod \lbn{0}^{\lstar}} \lenv{0}\glresulttype{\lR{} \loprod \lR{}} \\ \lemptyset{};\evalCtxSpace{};\evalCtx{}\vdash_{\mathsf C} \linner{}_{\sof{\lbn{1}}} \xlongrightarrow{\lbn{1}^{\lstar} \loprod \lbn{1}^{\lstar}} \lenv{1}\glresulttype{\lsnv{1} \loprod \lsnv{1}}}{\lemptyset{};\evalCtxSpace{};\evalCtx{}\vdash_{\mathsf C} \lee{}^{\lpseudo} \xlongrightarrow{\lbn{0}^{\lstar} \loprod \lbn{1}} \glvop{GPINV}(\glvreshape{\leev{}}{\lsnv{1} \loprod \lR{}}{\lsnv{1}},\lenv{1},\lenv{0})\glresulttype{\lR{} \loprod \lsnv{1}}}}
\and
\gpfit{\inferrule*[]{\isStdBasis{\lbn{0}} \\ \sof{\lbn{0}} = \ls{} \\ (\lemptyset{}, \evalCtxSpace{})\vdash \linner{}_{\ls{}}\colon \tof{\lbn{0}^{\lstar} \loprod \lbn{0}^{\lstar}} \\ (\emptyset,\evalCtxSpace{})\vdash \lbn{0}^{\lstar} \loprod \lbn{0}^{\lstar}\colon \bvalid{} \\ \evalCtxSpace{};\evalCtx{}\vdash_{\mathsf C} \lbn{0}^{\lstar} \loprod \lbn{0}^{\lstar} \evalArrow{}_{\mathsf{basis}} \lsv{} \loprod \lsv{} \\ \evalCtxSpace{};\evalCtx{}\vdash_{\mathsf C} \ls{} \evalArrow{} \lsv{}}{\lemptyset{};\evalCtxSpace{};\evalCtx{}\vdash_{\mathsf C} \linner{}_{\ls{}} \xlongrightarrow{\lbn{0}^{\lstar} \loprod \lbn{0}^{\lstar}} \lidelta{\lsv{}}\glresulttype{\lsv{} \loprod \lsv{}}}}
\and
\gpfit{\inferrule*[]{\forall i.\; \sof{\lbn{i}} = \lsn{i} \\ (\lemptyset{}, \evalCtxSpace{})\vdash \lembd{}_{\Kernel(\lee{})}\colon \tof{\pBasisC{\Kernel(\lee{})}{\lbn{0}}{\lbn{1}}^{\lstar} \loprod \lbn{0}} \\ (\emptyset,\evalCtxSpace{})\vdash \pBasisC{\Kernel(\lee{})}{\lbn{0}}{\lbn{1}}^{\lstar} \loprod \lbn{0}\colon \bvalid{} \\ \evalCtxSpace{};\evalCtx{}\vdash_{\mathsf C} \pBasisC{\Kernel(\lee{})}{\lbn{0}}{\lbn{1}}^{\lstar} \loprod \lbn{0} \evalArrow{}_{\mathsf{basis}} \lR^{\glvreshape{\leev{}}{\lsnv{0} \loprod \lsnv{1}}{\lsnv{0} \loprod \lsnv{1}},\Kernel{},\lsnv{0} \loprod \lsnv{1}} \loprod \lsnv{0} \\ \forall i.\; \lemptyset{};\evalCtxSpace{};\evalCtx{}\vdash_{\mathsf C} \linner{}_{\lsn{i}} \xlongrightarrow{\lbn{i}^{\lstar} \loprod \lbn{i}^{\lstar}} \lenv{i}\glresulttype{\lsnv{i} \loprod \lsnv{i}} \\ \lemptyset{};\evalCtxSpace{};\evalCtx{}\vdash_{\mathsf C} \lee{} \xlongrightarrow{\lbn{0}^{\lstar} \loprod \lbn{1}} \leev{}\glresulttype{\lsnv{0} \loprod \lsnv{1}}}{\lemptyset{};\evalCtxSpace{};\evalCtx{}\vdash_{\mathsf C} \lembd{}_{\Kernel(\lee{})} \xlongrightarrow{\pBasisC{\Kernel(\lee{})}{\lbn{0}}{\lbn{1}}^{\lstar} \loprod \lbn{0}} \Trans(\glvop{GSVDUI}(\leev{},\lenv{0},\lenv{1}))\glresulttype{\lR^{\glvreshape{\leev{}}{\lsnv{0} \loprod \lsnv{1}}{\lsnv{0} \loprod \lsnv{1}},\Kernel{},\lsnv{0} \loprod \lsnv{1}} \loprod \lsnv{0}}}}
\and
\gpfit{\inferrule*[]{\forall i.\; \sof{\lbn{i}} = \lsn{i} \\ (\lemptyset{}, \evalCtxSpace{})\vdash \lembd{}_{\CoKernel(\lee{})}\colon \tof{\pBasisC{\CoKernel(\lee{})}{\lbn{0}}{\lbn{1}}^{\lstar} \loprod \lbn{0}} \\ (\emptyset,\evalCtxSpace{})\vdash \pBasisC{\CoKernel(\lee{})}{\lbn{0}}{\lbn{1}}^{\lstar} \loprod \lbn{0}\colon \bvalid{} \\ \evalCtxSpace{};\evalCtx{}\vdash_{\mathsf C} \pBasisC{\CoKernel(\lee{})}{\lbn{0}}{\lbn{1}}^{\lstar} \loprod \lbn{0} \evalArrow{}_{\mathsf{basis}} \lR^{\glvreshape{\leev{}}{\lsnv{0} \loprod \lsnv{1}}{\lsnv{0} \loprod \lsnv{1}},\CoKernel{},\lsnv{0} \loprod \lsnv{1}} \loprod \lsnv{0} \\ \forall i.\; \lemptyset{};\evalCtxSpace{};\evalCtx{}\vdash_{\mathsf C} \linner{}_{\lsn{i}} \xlongrightarrow{\lbn{i}^{\lstar} \loprod \lbn{i}^{\lstar}} \lenv{i}\glresulttype{\lsnv{i} \loprod \lsnv{i}} \\ \lemptyset{};\evalCtxSpace{};\evalCtx{}\vdash_{\mathsf C} \lee{} \xlongrightarrow{\lbn{0}^{\lstar} \loprod \lbn{1}} \leev{}\glresulttype{\lsnv{0} \loprod \lsnv{1}}}{\lemptyset{};\evalCtxSpace{};\evalCtx{}\vdash_{\mathsf C} \lembd{}_{\CoKernel(\lee{})} \xlongrightarrow{\pBasisC{\CoKernel(\lee{})}{\lbn{0}}{\lbn{1}}^{\lstar} \loprod \lbn{0}} \Trans(\glvop{GSVDU}(\leev{},\lenv{0},\lenv{1}))\glresulttype{\lR^{\glvreshape{\leev{}}{\lsnv{0} \loprod \lsnv{1}}{\lsnv{0} \loprod \lsnv{1}},\CoKernel{},\lsnv{0} \loprod \lsnv{1}} \loprod \lsnv{0}}}}
\and
\gpfit{\inferrule*[]{\forall i.\; \sof{\lbn{i}} = \lsn{i} \\ (\lemptyset{}, \evalCtxSpace{})\vdash \lembd{}_{\Image(\lee{})}\colon \tof{\pBasisC{\Image(\lee{})}{\lbn{0}}{\lbn{1}}^{\lstar} \loprod \lbn{1}} \\ (\emptyset,\evalCtxSpace{})\vdash \pBasisC{\Image(\lee{})}{\lbn{0}}{\lbn{1}}^{\lstar} \loprod \lbn{1}\colon \bvalid{} \\ \evalCtxSpace{};\evalCtx{}\vdash_{\mathsf C} \pBasisC{\Image(\lee{})}{\lbn{0}}{\lbn{1}}^{\lstar} \loprod \lbn{1} \evalArrow{}_{\mathsf{basis}} \lR^{\glvreshape{\leev{}}{\lsnv{0} \loprod \lsnv{1}}{\lsnv{0} \loprod \lsnv{1}},\Image{},\lsnv{0} \loprod \lsnv{1}} \loprod \lsnv{1} \\ \forall i.\; \lemptyset{};\evalCtxSpace{};\evalCtx{}\vdash_{\mathsf C} \linner{}_{\lsn{i}} \xlongrightarrow{\lbn{i}^{\lstar} \loprod \lbn{i}^{\lstar}} \lenv{i}\glresulttype{\lsnv{i} \loprod \lsnv{i}} \\ \lemptyset{};\evalCtxSpace{};\evalCtx{}\vdash_{\mathsf C} \lee{} \xlongrightarrow{\lbn{0}^{\lstar} \loprod \lbn{1}} \leev{}\glresulttype{\lsnv{0} \loprod \lsnv{1}}}{\lemptyset{};\evalCtxSpace{};\evalCtx{}\vdash_{\mathsf C} \lembd{}_{\Image(\lee{})} \xlongrightarrow{\pBasisC{\Image(\lee{})}{\lbn{0}}{\lbn{1}}^{\lstar} \loprod \lbn{1}} \Trans(\glvop{GSVDV}(\leev{},\lenv{0},\lenv{1}))\glresulttype{\lR^{\glvreshape{\leev{}}{\lsnv{0} \loprod \lsnv{1}}{\lsnv{0} \loprod \lsnv{1}},\Image{},\lsnv{0} \loprod \lsnv{1}} \loprod \lsnv{1}}}}
\and
\gpfit{\inferrule*[]{\forall i.\; \sof{\lbn{i}} = \lsn{i} \\ (\lemptyset{}, \evalCtxSpace{})\vdash \lembd{}_{\CoImage(\lee{})}\colon \tof{\pBasisC{\CoImage(\lee{})}{\lbn{0}}{\lbn{1}}^{\lstar} \loprod \lbn{1}} \\ (\emptyset,\evalCtxSpace{})\vdash \pBasisC{\CoImage(\lee{})}{\lbn{0}}{\lbn{1}}^{\lstar} \loprod \lbn{1}\colon \bvalid{} \\ \evalCtxSpace{};\evalCtx{}\vdash_{\mathsf C} \pBasisC{\CoImage(\lee{})}{\lbn{0}}{\lbn{1}}^{\lstar} \loprod \lbn{1} \evalArrow{}_{\mathsf{basis}} \lR^{\glvreshape{\leev{}}{\lsnv{0} \loprod \lsnv{1}}{\lsnv{0} \loprod \lsnv{1}},\CoImage{},\lsnv{0} \loprod \lsnv{1}} \loprod \lsnv{1} \\ \forall i.\; \lemptyset{};\evalCtxSpace{};\evalCtx{}\vdash_{\mathsf C} \linner{}_{\lsn{i}} \xlongrightarrow{\lbn{i}^{\lstar} \loprod \lbn{i}^{\lstar}} \lenv{i}\glresulttype{\lsnv{i} \loprod \lsnv{i}} \\ \lemptyset{};\evalCtxSpace{};\evalCtx{}\vdash_{\mathsf C} \lee{} \xlongrightarrow{\lbn{0}^{\lstar} \loprod \lbn{1}} \leev{}\glresulttype{\lsnv{0} \loprod \lsnv{1}}}{\lemptyset{};\evalCtxSpace{};\evalCtx{}\vdash_{\mathsf C} \lembd{}_{\CoImage(\lee{})} \xlongrightarrow{\pBasisC{\CoImage(\lee{})}{\lbn{0}}{\lbn{1}}^{\lstar} \loprod \lbn{1}} \Trans(\glvop{GSVDVI}(\leev{},\lenv{0},\lenv{1}))\glresulttype{\lR^{\glvreshape{\leev{}}{\lsnv{0} \loprod \lsnv{1}}{\lsnv{0} \loprod \lsnv{1}},\CoImage{},\lsnv{0} \loprod \lsnv{1}} \loprod \lsnv{1}}}}
\and
\gpfit{\inferrule*[]{\tof{\lbn{1}} = \sof{\lbn{1}} \\ ({\lDelta{}_{0}} \ldisun {\lDelta{}_{1}}, \evalCtxSpace{})\vdash \leval{\len{0}}{}{\len{1}}\colon \tof{\lbn{1}} \\ (\emptyset,\evalCtxSpace{})\vdash \lbn{1}\colon \bvalid{} \\ \evalCtxSpace{};\evalCtx{}\vdash_{\mathsf C} \sof{\lbn{0}} \evalArrow{} \lsv{} \\ \evalCtxSpace{};\evalCtx{}\vdash_{\mathsf C} \lbn{1} \evalArrow{}_{\mathsf{basis}} \lsnv{{\mathrm{svOut}}} \\ {\lDelta{}_{0}};\evalCtxSpace{};\evalCtx{}\vdash_{\mathsf C} \len{0} \xlongrightarrow{\lbn{0}^{\lstar} \loprod \lbn{1}} \lenv{0}\glresulttype{\lsnv{{\mathrm{svMat0}}} \loprod \lsnv{{\mathrm{svMat1}}}} \\ {\lDelta{}_{1}};\evalCtxSpace{};\evalCtx{}\vdash_{\mathsf C} \len{1} \xlongrightarrow{\lbn{0}} \lenv{1}\glresulttype{\lsnv{{\mathrm{svArg}}}}}{{\lDelta{}_{0}} \ldisun {\lDelta{}_{1}};\evalCtxSpace{};\evalCtx{}\vdash_{\mathsf C} \leval{\len{0}}{}{\len{1}} \xlongrightarrow{\lbn{1}} \glvreshape{\LTTVPP{\lsv{}}{\lenv{0}}{\lenv{1}}}{\lsnv{{\mathrm{svOut}}}}{\lsnv{{\mathrm{svOut}}}}\glresulttype{\lsnv{{\mathrm{svOut}}}}}}
\and
\gpfit{\inferrule*[]{({\lDelta{}_{0}} \ldisun {\lDelta{}_{1}}, \evalCtxSpace{})\vdash \leval{\len{0}}{}{\len{1}}\colon \tof{\lbone{}} \\ \evalCtxSpace{};\evalCtx{}\vdash_{\mathsf C} \sof{\lbn{0}} \evalArrow{} \lsv{} \\ \evalCtxSpace{};\evalCtx{}\vdash_{\mathsf C} \lbone{} \evalArrow{}_{\mathsf{basis}} \lR{} \\ {\lDelta{}_{0}};\evalCtxSpace{};\evalCtx{}\vdash_{\mathsf C} \len{0} \xlongrightarrow{\lbn{0}^{\lstar}} \lenv{0}\glresulttype{\lsnv{{\mathrm{svFun}}}} \\ {\lDelta{}_{1}};\evalCtxSpace{};\evalCtx{}\vdash_{\mathsf C} \len{1} \xlongrightarrow{\lbn{0}} \lenv{1}\glresulttype{\lsnv{{\mathrm{svArg}}}}}{{\lDelta{}_{0}} \ldisun {\lDelta{}_{1}};\evalCtxSpace{};\evalCtx{}\vdash_{\mathsf C} \leval{\len{0}}{}{\len{1}} \xlongrightarrow{\lbone{}} \LTTVPP{\lsv{}}{\glvreshape{\lenv{0}}{\lsv{} \loprod \lR{}}{\lsv{}}}{\lenv{1}}\glresulttype{\lR{}}}}
\and
\gpfit{\inferrule*[]{\forall i.\; \tof{\lbn{i}} = \ltyn{i} \\ (\lemptyset{}, \evalCtxSpace{})\vdash (\nllambda \llpl x\colon \ltyn{0}\llpr \ldotsp \lee{})\colon \tof{\frameB{x}{\lbn{0}}{\lbn{1}}} \\ (\emptyset,\evalCtxSpace{})\vdash \frameB{x}{\lbn{0}}{\lbn{1}}\colon \bvalid{} \\ \evalCtxSpace{};\evalCtx{}\vdash_{\mathsf C} \frameB{x}{\lbn{0}}{\lbn{1}} \evalArrow{}_{\mathsf{basis}} (x\colon \ltynv{0}\fnto \ltynv{1}) \\ \evalCtxSpace{};\evalCtx{}\vdash_{\mathsf C} \lbn{0} \evalArrow{}_{\mathsf{basis}} \ltynv{0} \\ \lemptyset{};(\evalCtxSpace{}, \mathit{x}\colon \ltyn{0});\evalCtx{}[(\mathit{x},\lbn{0}) \mapsto \mathit{x}_{\ltynv{0}}]\vdash_{\mathsf C} \lee{} \xlongrightarrow{\lbn{1}} \leev{}\glresulttype{\ltynv{1}} \\ (\evalCtxSpace{}, \mathit{x}\colon \ltyn{0});\evalCtx{}[(\mathit{x},\lbn{0}) \mapsto \mathit{x}_{\ltynv{0}}]\vdash_{\mathsf C} \lbn{1} \evalArrow{}_{\mathsf{basis}} \ltynv{1}}{\lemptyset{};\evalCtxSpace{};\evalCtx{}\vdash_{\mathsf C} (\nllambda \llpl x\colon \ltyn{0}\llpr \ldotsp \lee{}) \xlongrightarrow{\frameB{x}{\lbn{0}}{\lbn{1}}} \llambda(x\colon \ltynv{0})\lrh \ltynv{1}\colon \glvreshape{\leev{}}{\ltynv{1}}{\ltynv{1}}\glresulttype{(x\colon \ltynv{0}\fnto \ltynv{1})}}}
\and
\gpfit{\inferrule*[]{(\lemptyset{}, \evalCtxSpace{})\vdash \leval{\len{0}}{}{\len{1}}\colon \tof{\lbn{1}\subst{x}{\len{1}}} \\ (\emptyset,\evalCtxSpace{})\vdash \lbn{1}\subst{x}{\len{1}}\colon \bvalid{} \\ \evalCtxSpace{};\evalCtx{}\vdash_{\mathsf C} \lbn{1}\subst{x}{\len{1}} \evalArrow{}_{\mathsf{basis}} {\ltyv{}_{\mathsf{out}}} \\ \lemptyset{};\evalCtxSpace{};\evalCtx{}\vdash_{\mathsf C} \len{0} \xlongrightarrow{\frameB{x}{\lbn{0}}{\lbn{1}}} \lenv{0}\glresulttype{(x\colon {\ltyv{}_{\mathsf{fun}0}}\fnto {\ltyv{}_{\mathsf{fun}1}})} \\ \lemptyset{};\evalCtxSpace{};\evalCtx{}\vdash_{\mathsf C} \len{1} \xlongrightarrow{\lbn{0}} \lenv{1}\glresulttype{{\ltyv{}_{\mathsf{arg}}}} \\ \operatorname{inferType}(\lenv{0}) = (x\colon \ltynv{{\mathrm{D}}}\fnto \ltynv{{\mathrm{C}}}) \\ \lenv{2} = \glvreshape{\lenv{1}}{\ltynv{{\mathrm{D}}}}{\operatorname{inferType}(\lenv{1})}}{\lemptyset{};\evalCtxSpace{};\evalCtx{}\vdash_{\mathsf C} \leval{\len{0}}{}{\len{1}} \xlongrightarrow{\lbn{1}\subst{x}{\len{1}}} \glvreshape{\glvapp{\lenv{0}}{\lenv{2}}}{{\ltyv{}_{\mathsf{out}}}}{\ltynv{{\mathrm{C}}}\subst{x}{\lenv{2}}}\glresulttype{{\ltyv{}_{\mathsf{out}}}}}}
\and
\gpfit{\inferrule*[]{\tof{\lbn{0}'} = \tof{\lbn{0}} \\ (\lemptyset{}, \evalCtxSpace{})\vdash \len{0}\colon \tof{\frameB{x}{\lbn{0}}{\lbn{1}}} \\ (\emptyset,\evalCtxSpace{})\vdash \frameB{x}{\lbn{0}}{\lbn{1}}\colon \bvalid{} \\ \evalCtxSpace{};\evalCtx{}\vdash_{\mathsf C} \frameB{x}{\lbn{0}}{\lbn{1}} \evalArrow{}_{\mathsf{basis}} (x\colon \ltynv{0}\fnto \ltynv{1}') \\ \evalCtxSpace{};\evalCtx{}\vdash_{\mathsf C} \lbn{0} \evalArrow{}_{\mathsf{basis}} \ltynv{0} \\ \lemptyset{};\evalCtxSpace{};\evalCtx{}\vdash_{\mathsf C} \len{0} \xlongrightarrow{\frameB{x}{\lbn{0}'}{\lbn{1}}} \leev{}\glresulttype{(x\colon \ltynv{0}'\fnto \ltynv{1})} \\ \evalCtxSpace{};\evalCtx{}\vdash_{\mathsf C} \lbn{0}' \evalArrow{}_{\mathsf{basis}} \ltynv{0}' \\ \lemptyset{};(\evalCtxSpace{}, \mathit{x}\colon \tof{\lbn{0}});\evalCtx{}[(\mathit{x},\lbn{0}) \mapsto \mathit{x}_{\ltynv{0}}]\vdash_{\mathsf C} \mathit{x} \xlongrightarrow{\lbn{0}'} \lenv{1}\glresulttype{\ltynv{0}'} \\ (\evalCtxSpace{}, \mathit{x}\colon \tof{\lbn{0}});\evalCtx{}[(\mathit{x},\lbn{0}) \mapsto \mathit{x}_{\ltynv{0}}]\vdash_{\mathsf C} \lbn{1} \evalArrow{}_{\mathsf{basis}} \ltynv{1}' \\ \operatorname{inferType}(\leev{}) = (x\colon \ltynv{{\mathrm{D}}}\fnto \ltynv{{\mathrm{C}}}) \\ \lenv{2} = \glvreshape{\glvreshape{\lenv{1}}{\ltynv{0}'}{\ltynv{0}'}}{\ltynv{{\mathrm{D}}}}{\ltynv{0}'}}{\lemptyset{};\evalCtxSpace{};\evalCtx{}\vdash_{\mathsf C} \len{0} \xlongrightarrow{\frameB{x}{\lbn{0}}{\lbn{1}}} \llambda(x\colon \ltynv{0})\lrh \ltynv{1}'\colon \glvreshape{\glvapp{\leev{}}{\lenv{2}}}{\ltynv{1}'}{\ltynv{1}'}\glresulttype{(x\colon \ltynv{0}\fnto \ltynv{1}')}}}
\and
\gpfit{\inferrule*[]{(\lemptyset{}, \evalCtxSpace{})\vdash \len{0}\colon \tof{\frameB{x}{\lbn{0}}{\lbn{1}}} \\ (\emptyset,\evalCtxSpace{})\vdash \frameB{x}{\lbn{0}}{\lbn{1}}\colon \bvalid{} \\ \evalCtxSpace{};\evalCtx{}\vdash_{\mathsf C} \frameB{x}{\lbn{0}}{\lbn{1}} \evalArrow{}_{\mathsf{basis}} (x\colon \ltynv{0}\fnto \ltynv{1}) \\ \evalCtxSpace{};\evalCtx{}\vdash_{\mathsf C} \lbn{0} \evalArrow{}_{\mathsf{basis}} \ltynv{0} \\ \lemptyset{};(\evalCtxSpace{}, \mathit{x}\colon \tof{\lbn{0}});\evalCtx{}[(\mathit{x},\lbn{0}) \mapsto \mathit{x}_{\ltynv{0}}]\vdash_{\mathsf C} \leval{\len{0}}{}{\mathit{x}} \xlongrightarrow{\lbn{1}} \leev{}\glresulttype{\ltynv{1}} \\ (\evalCtxSpace{}, \mathit{x}\colon \tof{\lbn{0}});\evalCtx{}[(\mathit{x},\lbn{0}) \mapsto \mathit{x}_{\ltynv{0}}]\vdash_{\mathsf C} \lbn{1} \evalArrow{}_{\mathsf{basis}} \ltynv{1}}{\lemptyset{};\evalCtxSpace{};\evalCtx{}\vdash_{\mathsf C} \len{0} \xlongrightarrow{\frameB{x}{\lbn{0}}{\lbn{1}}} \llambda(x\colon \ltynv{0})\lrh \ltynv{1}\colon \glvreshape{\leev{}}{\ltynv{1}}{\ltynv{1}}\glresulttype{(x\colon \ltynv{0}\fnto \ltynv{1})}}}
\and
\gpfit{\inferrule*[]{\tof{\lb{}'} = \sof{\lb{}'} \\ (\lDelta{}, \evalCtxSpace{})\vdash (\lllambda \llpl x\colon \sof{\lb{}'}\llpr \ldotsp \lee{})\colon \tof{\lb{}'^{\lstar} \loprod \lb{}} \\ (\emptyset,\evalCtxSpace{})\vdash \lb{}'^{\lstar} \loprod \lb{}\colon \bvalid{} \\ \evalCtxSpace{};\evalCtx{}\vdash_{\mathsf C} \lb{}'^{\lstar} \loprod \lb{} \evalArrow{}_{\mathsf{basis}} \lsv{}' \loprod \lsv{} \\ \evalCtxSpace{};\evalCtx{}\vdash_{\mathsf C} \sof{\lb{}'} \evalArrow{} \lsv{}' \\ \evalCtxSpace{};\evalCtx{}\vdash_{\mathsf C} \sof{\lb{}} \evalArrow{} \lsv{} \\ \forall\,0\leq i<\overline{\operatorname{dimOf}}(\lsv{}').\; (\lDelta{}, \mathit{x}\colon \sof{\lb{}'});\evalCtxSpace{};\evalCtx{}[(\mathit{x},\lb{}') \mapsto \ldelta{i}{\lsv{}'}]\vdash_{\mathsf C} \lee{} \xlongrightarrow{\lb{}} \leev{}_{i}\glresulttype{\lsv{}}}{\lDelta{};\evalCtxSpace{};\evalCtx{}\vdash_{\mathsf C} (\lllambda \llpl x\colon \sof{\lb{}'}\llpr \ldotsp \lee{}) \xlongrightarrow{\lb{}'^{\lstar} \loprod \lb{}} \glvreshape{\lsum^{\lsv{}'}_{\mathit{i}<\overline{\operatorname{dimOf}}(\lsv{}')}\left(\ldelta{\mathit{i}}{\lsv{}'}\loprod \glvreshape{\leev{}_{\mathit{i}}}{\lsv{}}{\lsv{}}\right)}{\lsv{}' \loprod \lsv{}}{\begin{cases}\lsempty{}&\overline{\operatorname{dimOf}}(\lsv{}')=0\\ \lsv{}' \loprod \lsv{}&\text{otherwise}\end{cases}}\glresulttype{\lsv{}' \loprod \lsv{}}}}
\and
\gpfit{\inferrule*[]{\tof{\lb{}'} = \sof{\lb{}'} \\ (\lDelta{}, \evalCtxSpace{})\vdash (\lllambda \llpl x\colon \sof{\lb{}'}\llpr \ldotsp \lee{})\colon \tof{\lb{}'^{\lstar}} \\ (\emptyset,\evalCtxSpace{})\vdash \lb{}'^{\lstar}\colon \bvalid{} \\ \evalCtxSpace{};\evalCtx{}\vdash_{\mathsf C} \lb{}'^{\lstar} \evalArrow{}_{\mathsf{basis}} \lsv{}' \\ \evalCtxSpace{};\evalCtx{}\vdash_{\mathsf C} \sof{\lb{}'} \evalArrow{} \lsv{}' \\ \forall\,0\leq i<\overline{\operatorname{dimOf}}(\lsv{}').\; (\lDelta{}, \mathit{x}\colon \sof{\lb{}'});\evalCtxSpace{};\evalCtx{}[(\mathit{x},\lb{}') \mapsto \ldelta{i}{\lsv{}'}]\vdash_{\mathsf C} \lee{} \xlongrightarrow{\lbone{}} \leev{}_{i}\glresulttype{\lR{}}}{\lDelta{};\evalCtxSpace{};\evalCtx{}\vdash_{\mathsf C} (\lllambda \llpl x\colon \sof{\lb{}'}\llpr \ldotsp \lee{}) \xlongrightarrow{\lb{}'^{\lstar}} \glvreshape{\lsum^{\lsv{}'}_{\mathit{i}<\overline{\operatorname{dimOf}}(\lsv{}')}\left(\glvreshape{\leev{}_{\mathit{i}}}{\lR{}}{\lR{}} \lvscale \ldelta{\mathit{i}}{\lsv{}'}\right)}{\lsv{}'}{\begin{cases}\lsempty{}&\overline{\operatorname{dimOf}}(\lsv{}')=0\\ \lsv{}'&\text{otherwise}\end{cases}}\glresulttype{\lsv{}'}}}
\and
\gpfit{\inferrule*[]{\tof{\lb{}} = \sof{\lb{}} \\ (\lDelta{}, \evalCtxSpace{})\vdash \lvzero{}_{\sof{\lb{}}}\colon \tof{\lb{}} \\ (\emptyset,\evalCtxSpace{})\vdash \lb{}\colon \bvalid{} \\ \evalCtxSpace{};\evalCtx{}\vdash_{\mathsf C} \lb{} \evalArrow{}_{\mathsf{basis}} \lsv{} \\ \evalCtxSpace{};\evalCtx{}\vdash_{\mathsf C} \sof{\lb{}} \evalArrow{} \lsv{}}{\lDelta{};\evalCtxSpace{};\evalCtx{}\vdash_{\mathsf C} \lvzero{}_{\sof{\lb{}}} \xlongrightarrow{\lb{}} \lvzero{}_{\lsv{}}\glresulttype{\lsv{}}}}
\and
\gpfit{\inferrule*[]{(\lemptyset{}, \evalCtxSpace{})\vdash \mathit{r}\colon \tof{\lbone{}} \\ \evalCtxSpace{};\evalCtx{}\vdash_{\mathsf C} \lbone{} \evalArrow{}_{\mathsf{basis}} \lR{}}{\lemptyset{};\evalCtxSpace{};\evalCtx{}\vdash_{\mathsf C} \mathit{r} \xlongrightarrow{\lbone{}} \mathit{r}\glresulttype{\lR{}}}}
}{
\gpfit{\inferrule*[lab=\textsc{C-E-Var\allowbreak Lin}]{((\mathit{x}, \lb{}) \mapsto \leev{}) \in \evalCtx{} \\ \tof{\lb{}} = \ls{} \\ (\emptyset,\evalCtxSpace{})\vdash \lb{}\colon \bvalid{} \\ \evalCtxSpace{};\evalCtx{}\vdash_{\mathsf C} \lb{} \evalArrow{}_{\mathsf{basis}} \ltyv{}}{\{(\mathit{x}, \ls{})\};\evalCtxSpace{};\evalCtx{}\vdash_{\mathsf C} \mathit{x} \xlongrightarrow{\lb{}} \leev{}\glresulttype{\ltyv{}}}}
\and
\gpfit{\inferrule*[lab=\textsc{C-E-Var\allowbreak Nlin}]{((\mathit{x}, \lb{}) \mapsto \leev{}) \in \evalCtx{} \\ (\mathit{x}\colon \tof{\lb{}}) \in \evalCtxSpace{} \\ (\emptyset,\evalCtxSpace{})\vdash \lb{}\colon \bvalid{} \\ \evalCtxSpace{};\evalCtx{}\vdash_{\mathsf C} \lb{} \evalArrow{}_{\mathsf{basis}} \ltyv{}}{\lemptyset{};\evalCtxSpace{};\evalCtx{}\vdash_{\mathsf C} \mathit{x} \xlongrightarrow{\lb{}} \leev{}\glresulttype{\ltyv{}}}}
\and
\gpfit{\inferrule*[lab=\textsc{C-E-Promote}]{\forall y \in \mathit{lin}.\; y \notin \bigcup_{(\cdot\colon s) \in \lDelta{},\, \mathit{lin}} \mathrm{fv}(s) \;\cup\; \bigcup_{(\cdot\colon \tau) \in \evalCtxSpace{}'} \mathrm{fv}(\tau) \\ \evalCtxSpace{}' \subseteq \evalCtxSpace{}'' \\ (\lDelta{}, \mathit{lin} \ldisun \evalCtxSpace{}'')\vdash \lee{}\colon \tof{\lb{}} \\ (\emptyset,\mathit{lin} \ldisun \evalCtxSpace{}'')\vdash \lb{}\colon \bvalid{} \\ \mathit{lin} \ldisun \evalCtxSpace{}'';\evalCtx{}\vdash_{\mathsf C} \lb{} \evalArrow{}_{\mathsf{basis}} \ltyv{} \\ \mathit{lin} \ldisun \lDelta{};\evalCtxSpace{}';\evalCtx{}\vdash_{\mathsf C} \lee{} \xlongrightarrow{\lb{}} \leev{}\glresulttype{\ltyv{}}}{\lDelta{};\mathit{lin} \ldisun \evalCtxSpace{}'';\evalCtx{}\vdash_{\mathsf C} \lee{} \xlongrightarrow{\lb{}} \leev{}\glresulttype{\ltyv{}}}}
\and
\gpfit{\inferrule*[lab=\textsc{C-E-Cob}]{\forall i.\; \tof{\lbn{i}} = \sof{\lbn{i}} \\ (\lDelta{}, \evalCtxSpace{})\vdash \lee{}\colon \tof{\lbn{0}} \\ (\emptyset,\evalCtxSpace{})\vdash \lbn{0}\colon \bvalid{} \\ \evalCtxSpace{};\evalCtx{}\vdash_{\mathsf C} \lbn{0} \evalArrow{}_{\mathsf{basis}} \lsv{} \\ \lDelta{};\evalCtxSpace{};\evalCtx{}\vdash_{\mathsf C} \lee{} \xlongrightarrow{\lbn{1}} \leev{}\glresulttype{\ltyv{}} \\ \evalCtxSpace{};\evalCtx{}\vdash_{\mathsf C} \lbn{1} \xlongleftrightarrow{\lenv{0}} \lbn{0} \\ \evalCtxSpace{};\evalCtx{}\vdash_{\mathsf C} \sof{\lbn{0}} \evalArrow{} \lsv{}}{\lDelta{};\evalCtxSpace{};\evalCtx{}\vdash_{\mathsf C} \lee{} \xlongrightarrow{\lbn{0}} \LTTVPP{\lsv{}}{\glvreshape{\lenv{0}}{\lsv{} \loprod \lsv{}}{\lsv{} \loprod \lsv{}}}{\glvreshape{\leev{}}{\lsv{}}{\lsv{}}}\glresulttype{\lsv{}}}}
\and
\gpfit{\inferrule*[lab=\textsc{C-E-Reshape\allowbreak Cob}]{(\lDelta{}, \evalCtxSpace{})\vdash \lee{}\colon \tof{\lb{}} \\ (\emptyset,\evalCtxSpace{})\vdash \lb{}\colon \bvalid{} \\ \evalCtxSpace{};\evalCtx{}\vdash_{\mathsf C} \lb{} \evalArrow{}_{\mathsf{basis}} \ltyv{}' \\ \lDelta{};\evalCtxSpace{};\evalCtx{}\vdash_{\mathsf C} \lee{} \xlongrightarrow{\lb{}} \leev{}\glresulttype{\ltyv{}}}{\lDelta{};\evalCtxSpace{};\evalCtx{}\vdash_{\mathsf C} \lee{} \xlongrightarrow{\lb{}} \glvreshape{\leev{}}{\ltyv{}'}{\ltyv{}}\glresulttype{\ltyv{}'}}}
\and
\gpfit{\inferrule*[lab=\textsc{C-E-Scale}]{\tof{\lb{}} = \sof{\lb{}} \\ ({\lDelta{}_{0}} \ldisun {\lDelta{}_{1}}, \evalCtxSpace{})\vdash \len{0} \lvscale \len{1}\colon \tof{\lb{}} \\ (\emptyset,\evalCtxSpace{})\vdash \lb{}\colon \bvalid{} \\ \evalCtxSpace{};\evalCtx{}\vdash_{\mathsf C} \lb{} \evalArrow{}_{\mathsf{basis}} \lsv{} \\ {\lDelta{}_{0}};\evalCtxSpace{};\evalCtx{}\vdash_{\mathsf C} \len{0} \xlongrightarrow{\lbone{}} \lenv{0}\glresulttype{\lR{}} \\ {\lDelta{}_{1}};\evalCtxSpace{};\evalCtx{}\vdash_{\mathsf C} \len{1} \xlongrightarrow{\lb{}} \lenv{1}\glresulttype{\lsv{}}}{{\lDelta{}_{0}} \ldisun {\lDelta{}_{1}};\evalCtxSpace{};\evalCtx{}\vdash_{\mathsf C} \len{0} \lvscale \len{1} \xlongrightarrow{\lb{}} \lenv{0} \lvscale \lenv{1}\glresulttype{\lsv{}}}}
\and
\gpfit{\inferrule*[lab=\textsc{C-E-Add}]{\tof{\lb{}} = \sof{\lb{}} \\ (\lDelta{}, \evalCtxSpace{})\vdash \len{0} \lvadd \len{1}\colon \tof{\lb{}} \\ (\emptyset,\evalCtxSpace{})\vdash \lb{}\colon \bvalid{} \\ \evalCtxSpace{};\evalCtx{}\vdash_{\mathsf C} \lb{} \evalArrow{}_{\mathsf{basis}} \lsv{} \\ \forall i.\; \lDelta{};\evalCtxSpace{};\evalCtx{}\vdash_{\mathsf C} \len{i} \xlongrightarrow{\lb{}} \lenv{i}\glresulttype{\lsnv{i}}}{\lDelta{};\evalCtxSpace{};\evalCtx{}\vdash_{\mathsf C} \len{0} \lvadd \len{1} \xlongrightarrow{\lb{}} \lenv{0} \lvadd_{\lsv{}} \lenv{1}\glresulttype{\lsv{}}}}
\and
\gpfit{\inferrule*[lab=\textsc{C-E-Dsum\allowbreak E}]{\forall i.\; \tof{\lbn{i}} = \sof{\lbn{i}} \\ (\lDelta{}, \evalCtxSpace{})\vdash \len{0}\losum \len{1}\colon \tof{\lbn{0} \losum \lbn{1}} \\ (\emptyset,\evalCtxSpace{})\vdash \lbn{0} \losum \lbn{1}\colon \bvalid{} \\ \evalCtxSpace{};\evalCtx{}\vdash_{\mathsf C} \lbn{0} \losum \lbn{1} \evalArrow{}_{\mathsf{basis}} \lsnv{0} \losum \lsnv{1} \\ \forall i.\; \lDelta{};\evalCtxSpace{};\evalCtx{}\vdash_{\mathsf C} \len{i} \xlongrightarrow{\lbn{i}} \lenv{i}\glresulttype{\lsnv{i}}}{\lDelta{};\evalCtxSpace{};\evalCtx{}\vdash_{\mathsf C} \len{0}\losum \len{1} \xlongrightarrow{\lbn{0} \losum \lbn{1}} \lenv{0}\losum \lenv{1}\glresulttype{\lsnv{0} \losum \lsnv{1}}}}
\and
\gpfit{\inferrule*[lab=\textsc{C-E-Tprod\allowbreak E}]{\forall i.\; \tof{\lbn{i}} = \sof{\lbn{i}} \\ ({\lDelta{}_{0}} \ldisun {\lDelta{}_{1}}, \evalCtxSpace{})\vdash \len{0}\loprod \len{1}\colon \tof{\lbn{0} \loprod \lbn{1}} \\ (\emptyset,\evalCtxSpace{})\vdash \lbn{0} \loprod \lbn{1}\colon \bvalid{} \\ \evalCtxSpace{};\evalCtx{}\vdash_{\mathsf C} \lbn{0} \loprod \lbn{1} \evalArrow{}_{\mathsf{basis}} \lsnv{0} \loprod \lsnv{1} \\ \forall i.\; {\lDelta{}_{i}};\evalCtxSpace{};\evalCtx{}\vdash_{\mathsf C} \len{i} \xlongrightarrow{\lbn{i}} \lenv{i}\glresulttype{\lsnv{i}}}{{\lDelta{}_{0}} \ldisun {\lDelta{}_{1}};\evalCtxSpace{};\evalCtx{}\vdash_{\mathsf C} \len{0}\loprod \len{1} \xlongrightarrow{\lbn{0} \loprod \lbn{1}} \lenv{0}\loprod \lenv{1}\glresulttype{\lsnv{0} \loprod \lsnv{1}}}}
\and
\gpfit{\inferrule*[lab=\textsc{C-E-Pinv\allowbreak TStd}]{\forall i.\; \isStdBasis{\lbn{i}} \\ (\lemptyset{}, \evalCtxSpace{})\vdash \lee{}^{\lpseudo}\colon \tof{\lbn{0}^{\lstar} \loprod \lbn{1}} \\ (\emptyset,\evalCtxSpace{})\vdash \lbn{0}^{\lstar} \loprod \lbn{1}\colon \bvalid{} \\ \evalCtxSpace{};\evalCtx{}\vdash_{\mathsf C} \lbn{0}^{\lstar} \loprod \lbn{1} \evalArrow{}_{\mathsf{basis}} \lsnv{0} \loprod \lsnv{1} \\ \lemptyset{};\evalCtxSpace{};\evalCtx{}\vdash_{\mathsf C} \lee{} \xlongrightarrow{\lbn{1}^{\lstar} \loprod \lbn{0}} \leev{}\glresulttype{\lsnv{1} \loprod \lsnv{0}}}{\lemptyset{};\evalCtxSpace{};\evalCtx{}\vdash_{\mathsf C} \lee{}^{\lpseudo} \xlongrightarrow{\lbn{0}^{\lstar} \loprod \lbn{1}} \leev{}^{+}\glresulttype{\lsnv{0} \loprod \lsnv{1}}}}
\and
\gpfit{\inferrule*[lab=\textsc{C-E-Pinv\allowbreak TGpinv}]{(\lemptyset{}, \evalCtxSpace{})\vdash \lee{}^{\lpseudo}\colon \tof{\lbn{0}^{\lstar} \loprod \lbn{1}} \\ (\emptyset,\evalCtxSpace{})\vdash \lbn{0}^{\lstar} \loprod \lbn{1}\colon \bvalid{} \\ \evalCtxSpace{};\evalCtx{}\vdash_{\mathsf C} \lbn{0}^{\lstar} \loprod \lbn{1} \evalArrow{}_{\mathsf{basis}} \lsnv{0} \loprod \lsnv{1} \\ \lemptyset{};\evalCtxSpace{};\evalCtx{}\vdash_{\mathsf C} \lee{} \xlongrightarrow{\lbn{1}^{\lstar} \loprod \lbn{0}} \leev{}\glresulttype{\lsnv{1} \loprod \lsnv{0}} \\ \forall i.\; \lemptyset{};\evalCtxSpace{};\evalCtx{}\vdash_{\mathsf C} \linner{}_{\sof{\lbn{i}}} \xlongrightarrow{\lbn{i}^{\lstar} \loprod \lbn{i}^{\lstar}} \lenv{i}\glresulttype{\lsnv{i} \loprod \lsnv{i}}}{\lemptyset{};\evalCtxSpace{};\evalCtx{}\vdash_{\mathsf C} \lee{}^{\lpseudo} \xlongrightarrow{\lbn{0}^{\lstar} \loprod \lbn{1}} \glvop{GPINV}(\leev{},\lenv{1},\lenv{0})\glresulttype{\lsnv{0} \loprod \lsnv{1}}}}
\and
\gpfit{\inferrule*[lab=\textsc{C-E-Pinv\allowbreak DStd}]{\sof{\lbn{0}} = \lR{} \\ \forall i.\; \isStdBasis{\lbn{i}} \\ (\lemptyset{}, \evalCtxSpace{})\vdash \lee{}^{\lpseudo}\colon \tof{\lbn{0}^{\lstar} \loprod \lbn{1}} \\ (\emptyset,\evalCtxSpace{})\vdash \lbn{0}^{\lstar} \loprod \lbn{1}\colon \bvalid{} \\ \evalCtxSpace{};\evalCtx{}\vdash_{\mathsf C} \lbn{0}^{\lstar} \loprod \lbn{1} \evalArrow{}_{\mathsf{basis}} \lR{} \loprod \lsnv{1} \\ \lemptyset{};\evalCtxSpace{};\evalCtx{}\vdash_{\mathsf C} \lee{} \xlongrightarrow{\lbn{1}^{\lstar}} \leev{}\glresulttype{\lsnv{1}}}{\lemptyset{};\evalCtxSpace{};\evalCtx{}\vdash_{\mathsf C} \lee{}^{\lpseudo} \xlongrightarrow{\lbn{0}^{\lstar} \loprod \lbn{1}} \glvreshape{\leev{}}{\lsnv{1} \loprod \lR{}}{\lsnv{1}}^{+}\glresulttype{\lR{} \loprod \lsnv{1}}}}
\and
\gpfit{\inferrule*[lab=\textsc{C-E-Pinv\allowbreak DGpinv}]{\sof{\lbn{0}} = \lR{} \\ \isStdBasis{\lbn{0}} \\ (\lemptyset{}, \evalCtxSpace{})\vdash \lee{}^{\lpseudo}\colon \tof{\lbn{0}^{\lstar} \loprod \lbn{1}} \\ (\emptyset,\evalCtxSpace{})\vdash \lbn{0}^{\lstar} \loprod \lbn{1}\colon \bvalid{} \\ \evalCtxSpace{};\evalCtx{}\vdash_{\mathsf C} \lbn{0}^{\lstar} \loprod \lbn{1} \evalArrow{}_{\mathsf{basis}} \lR{} \loprod \lsnv{1} \\ \lemptyset{};\evalCtxSpace{};\evalCtx{}\vdash_{\mathsf C} \lee{} \xlongrightarrow{\lbn{1}^{\lstar}} \leev{}\glresulttype{\lsnv{1}} \\ \lemptyset{};\evalCtxSpace{};\evalCtx{}\vdash_{\mathsf C} \linner{}_{\sof{\lbn{0}}} \xlongrightarrow{\lbn{0}^{\lstar} \loprod \lbn{0}^{\lstar}} \lenv{0}\glresulttype{\lR{} \loprod \lR{}} \\ \lemptyset{};\evalCtxSpace{};\evalCtx{}\vdash_{\mathsf C} \linner{}_{\sof{\lbn{1}}} \xlongrightarrow{\lbn{1}^{\lstar} \loprod \lbn{1}^{\lstar}} \lenv{1}\glresulttype{\lsnv{1} \loprod \lsnv{1}}}{\lemptyset{};\evalCtxSpace{};\evalCtx{}\vdash_{\mathsf C} \lee{}^{\lpseudo} \xlongrightarrow{\lbn{0}^{\lstar} \loprod \lbn{1}} \glvop{GPINV}(\glvreshape{\leev{}}{\lsnv{1} \loprod \lR{}}{\lsnv{1}},\lenv{1},\lenv{0})\glresulttype{\lR{} \loprod \lsnv{1}}}}
\and
\gpfit{\inferrule*[lab=\textsc{C-E-Inner\allowbreak Std}]{\isStdBasis{\lbn{0}} \\ \sof{\lbn{0}} = \ls{} \\ (\lemptyset{}, \evalCtxSpace{})\vdash \linner{}_{\ls{}}\colon \tof{\lbn{0}^{\lstar} \loprod \lbn{0}^{\lstar}} \\ (\emptyset,\evalCtxSpace{})\vdash \lbn{0}^{\lstar} \loprod \lbn{0}^{\lstar}\colon \bvalid{} \\ \evalCtxSpace{};\evalCtx{}\vdash_{\mathsf C} \lbn{0}^{\lstar} \loprod \lbn{0}^{\lstar} \evalArrow{}_{\mathsf{basis}} \lsv{} \loprod \lsv{} \\ \evalCtxSpace{};\evalCtx{}\vdash_{\mathsf C} \ls{} \evalArrow{} \lsv{}}{\lemptyset{};\evalCtxSpace{};\evalCtx{}\vdash_{\mathsf C} \linner{}_{\ls{}} \xlongrightarrow{\lbn{0}^{\lstar} \loprod \lbn{0}^{\lstar}} \lidelta{\lsv{}}\glresulttype{\lsv{} \loprod \lsv{}}}}
\and
\gpfit{\inferrule*[lab=\textsc{C-E-Embed\allowbreak Ker}]{\forall i.\; \sof{\lbn{i}} = \lsn{i} \\ (\lemptyset{}, \evalCtxSpace{})\vdash \lembd{}_{\Kernel(\lee{})}\colon \tof{\pBasisC{\Kernel(\lee{})}{\lbn{0}}{\lbn{1}}^{\lstar} \loprod \lbn{0}} \\ (\emptyset,\evalCtxSpace{})\vdash \pBasisC{\Kernel(\lee{})}{\lbn{0}}{\lbn{1}}^{\lstar} \loprod \lbn{0}\colon \bvalid{} \\ \evalCtxSpace{};\evalCtx{}\vdash_{\mathsf C} \pBasisC{\Kernel(\lee{})}{\lbn{0}}{\lbn{1}}^{\lstar} \loprod \lbn{0} \evalArrow{}_{\mathsf{basis}} \lR^{\glvreshape{\leev{}}{\lsnv{0} \loprod \lsnv{1}}{\lsnv{0} \loprod \lsnv{1}},\Kernel{},\lsnv{0} \loprod \lsnv{1}} \loprod \lsnv{0} \\ \forall i.\; \lemptyset{};\evalCtxSpace{};\evalCtx{}\vdash_{\mathsf C} \linner{}_{\lsn{i}} \xlongrightarrow{\lbn{i}^{\lstar} \loprod \lbn{i}^{\lstar}} \lenv{i}\glresulttype{\lsnv{i} \loprod \lsnv{i}} \\ \lemptyset{};\evalCtxSpace{};\evalCtx{}\vdash_{\mathsf C} \lee{} \xlongrightarrow{\lbn{0}^{\lstar} \loprod \lbn{1}} \leev{}\glresulttype{\lsnv{0} \loprod \lsnv{1}}}{\lemptyset{};\evalCtxSpace{};\evalCtx{}\vdash_{\mathsf C} \lembd{}_{\Kernel(\lee{})} \xlongrightarrow{\pBasisC{\Kernel(\lee{})}{\lbn{0}}{\lbn{1}}^{\lstar} \loprod \lbn{0}} \Trans(\glvop{GSVDUI}(\leev{},\lenv{0},\lenv{1}))\glresulttype{\lR^{\glvreshape{\leev{}}{\lsnv{0} \loprod \lsnv{1}}{\lsnv{0} \loprod \lsnv{1}},\Kernel{},\lsnv{0} \loprod \lsnv{1}} \loprod \lsnv{0}}}}
\and
\gpfit{\inferrule*[lab=\textsc{C-E-Embed\allowbreak Coker}]{\forall i.\; \sof{\lbn{i}} = \lsn{i} \\ (\lemptyset{}, \evalCtxSpace{})\vdash \lembd{}_{\CoKernel(\lee{})}\colon \tof{\pBasisC{\CoKernel(\lee{})}{\lbn{0}}{\lbn{1}}^{\lstar} \loprod \lbn{0}} \\ (\emptyset,\evalCtxSpace{})\vdash \pBasisC{\CoKernel(\lee{})}{\lbn{0}}{\lbn{1}}^{\lstar} \loprod \lbn{0}\colon \bvalid{} \\ \evalCtxSpace{};\evalCtx{}\vdash_{\mathsf C} \pBasisC{\CoKernel(\lee{})}{\lbn{0}}{\lbn{1}}^{\lstar} \loprod \lbn{0} \evalArrow{}_{\mathsf{basis}} \lR^{\glvreshape{\leev{}}{\lsnv{0} \loprod \lsnv{1}}{\lsnv{0} \loprod \lsnv{1}},\CoKernel{},\lsnv{0} \loprod \lsnv{1}} \loprod \lsnv{0} \\ \forall i.\; \lemptyset{};\evalCtxSpace{};\evalCtx{}\vdash_{\mathsf C} \linner{}_{\lsn{i}} \xlongrightarrow{\lbn{i}^{\lstar} \loprod \lbn{i}^{\lstar}} \lenv{i}\glresulttype{\lsnv{i} \loprod \lsnv{i}} \\ \lemptyset{};\evalCtxSpace{};\evalCtx{}\vdash_{\mathsf C} \lee{} \xlongrightarrow{\lbn{0}^{\lstar} \loprod \lbn{1}} \leev{}\glresulttype{\lsnv{0} \loprod \lsnv{1}}}{\lemptyset{};\evalCtxSpace{};\evalCtx{}\vdash_{\mathsf C} \lembd{}_{\CoKernel(\lee{})} \xlongrightarrow{\pBasisC{\CoKernel(\lee{})}{\lbn{0}}{\lbn{1}}^{\lstar} \loprod \lbn{0}} \Trans(\glvop{GSVDU}(\leev{},\lenv{0},\lenv{1}))\glresulttype{\lR^{\glvreshape{\leev{}}{\lsnv{0} \loprod \lsnv{1}}{\lsnv{0} \loprod \lsnv{1}},\CoKernel{},\lsnv{0} \loprod \lsnv{1}} \loprod \lsnv{0}}}}
\and
\gpfit{\inferrule*[lab=\textsc{C-E-Embed\allowbreak Img}]{\forall i.\; \sof{\lbn{i}} = \lsn{i} \\ (\lemptyset{}, \evalCtxSpace{})\vdash \lembd{}_{\Image(\lee{})}\colon \tof{\pBasisC{\Image(\lee{})}{\lbn{0}}{\lbn{1}}^{\lstar} \loprod \lbn{1}} \\ (\emptyset,\evalCtxSpace{})\vdash \pBasisC{\Image(\lee{})}{\lbn{0}}{\lbn{1}}^{\lstar} \loprod \lbn{1}\colon \bvalid{} \\ \evalCtxSpace{};\evalCtx{}\vdash_{\mathsf C} \pBasisC{\Image(\lee{})}{\lbn{0}}{\lbn{1}}^{\lstar} \loprod \lbn{1} \evalArrow{}_{\mathsf{basis}} \lR^{\glvreshape{\leev{}}{\lsnv{0} \loprod \lsnv{1}}{\lsnv{0} \loprod \lsnv{1}},\Image{},\lsnv{0} \loprod \lsnv{1}} \loprod \lsnv{1} \\ \forall i.\; \lemptyset{};\evalCtxSpace{};\evalCtx{}\vdash_{\mathsf C} \linner{}_{\lsn{i}} \xlongrightarrow{\lbn{i}^{\lstar} \loprod \lbn{i}^{\lstar}} \lenv{i}\glresulttype{\lsnv{i} \loprod \lsnv{i}} \\ \lemptyset{};\evalCtxSpace{};\evalCtx{}\vdash_{\mathsf C} \lee{} \xlongrightarrow{\lbn{0}^{\lstar} \loprod \lbn{1}} \leev{}\glresulttype{\lsnv{0} \loprod \lsnv{1}}}{\lemptyset{};\evalCtxSpace{};\evalCtx{}\vdash_{\mathsf C} \lembd{}_{\Image(\lee{})} \xlongrightarrow{\pBasisC{\Image(\lee{})}{\lbn{0}}{\lbn{1}}^{\lstar} \loprod \lbn{1}} \Trans(\glvop{GSVDV}(\leev{},\lenv{0},\lenv{1}))\glresulttype{\lR^{\glvreshape{\leev{}}{\lsnv{0} \loprod \lsnv{1}}{\lsnv{0} \loprod \lsnv{1}},\Image{},\lsnv{0} \loprod \lsnv{1}} \loprod \lsnv{1}}}}
\and
\gpfit{\inferrule*[lab=\textsc{C-E-Embed\allowbreak Coimg}]{\forall i.\; \sof{\lbn{i}} = \lsn{i} \\ (\lemptyset{}, \evalCtxSpace{})\vdash \lembd{}_{\CoImage(\lee{})}\colon \tof{\pBasisC{\CoImage(\lee{})}{\lbn{0}}{\lbn{1}}^{\lstar} \loprod \lbn{1}} \\ (\emptyset,\evalCtxSpace{})\vdash \pBasisC{\CoImage(\lee{})}{\lbn{0}}{\lbn{1}}^{\lstar} \loprod \lbn{1}\colon \bvalid{} \\ \evalCtxSpace{};\evalCtx{}\vdash_{\mathsf C} \pBasisC{\CoImage(\lee{})}{\lbn{0}}{\lbn{1}}^{\lstar} \loprod \lbn{1} \evalArrow{}_{\mathsf{basis}} \lR^{\glvreshape{\leev{}}{\lsnv{0} \loprod \lsnv{1}}{\lsnv{0} \loprod \lsnv{1}},\CoImage{},\lsnv{0} \loprod \lsnv{1}} \loprod \lsnv{1} \\ \forall i.\; \lemptyset{};\evalCtxSpace{};\evalCtx{}\vdash_{\mathsf C} \linner{}_{\lsn{i}} \xlongrightarrow{\lbn{i}^{\lstar} \loprod \lbn{i}^{\lstar}} \lenv{i}\glresulttype{\lsnv{i} \loprod \lsnv{i}} \\ \lemptyset{};\evalCtxSpace{};\evalCtx{}\vdash_{\mathsf C} \lee{} \xlongrightarrow{\lbn{0}^{\lstar} \loprod \lbn{1}} \leev{}\glresulttype{\lsnv{0} \loprod \lsnv{1}}}{\lemptyset{};\evalCtxSpace{};\evalCtx{}\vdash_{\mathsf C} \lembd{}_{\CoImage(\lee{})} \xlongrightarrow{\pBasisC{\CoImage(\lee{})}{\lbn{0}}{\lbn{1}}^{\lstar} \loprod \lbn{1}} \Trans(\glvop{GSVDVI}(\leev{},\lenv{0},\lenv{1}))\glresulttype{\lR^{\glvreshape{\leev{}}{\lsnv{0} \loprod \lsnv{1}}{\lsnv{0} \loprod \lsnv{1}},\CoImage{},\lsnv{0} \loprod \lsnv{1}} \loprod \lsnv{1}}}}
\and
\gpfit{\inferrule*[lab=\textsc{C-E-App\allowbreak LLin}]{\tof{\lbn{1}} = \sof{\lbn{1}} \\ ({\lDelta{}_{0}} \ldisun {\lDelta{}_{1}}, \evalCtxSpace{})\vdash \leval{\len{0}}{}{\len{1}}\colon \tof{\lbn{1}} \\ (\emptyset,\evalCtxSpace{})\vdash \lbn{1}\colon \bvalid{} \\ \evalCtxSpace{};\evalCtx{}\vdash_{\mathsf C} \sof{\lbn{0}} \evalArrow{} \lsv{} \\ \evalCtxSpace{};\evalCtx{}\vdash_{\mathsf C} \lbn{1} \evalArrow{}_{\mathsf{basis}} \lsnv{{\mathrm{svOut}}} \\ {\lDelta{}_{0}};\evalCtxSpace{};\evalCtx{}\vdash_{\mathsf C} \len{0} \xlongrightarrow{\lbn{0}^{\lstar} \loprod \lbn{1}} \lenv{0}\glresulttype{\lsnv{{\mathrm{svMat0}}} \loprod \lsnv{{\mathrm{svMat1}}}} \\ {\lDelta{}_{1}};\evalCtxSpace{};\evalCtx{}\vdash_{\mathsf C} \len{1} \xlongrightarrow{\lbn{0}} \lenv{1}\glresulttype{\lsnv{{\mathrm{svArg}}}}}{{\lDelta{}_{0}} \ldisun {\lDelta{}_{1}};\evalCtxSpace{};\evalCtx{}\vdash_{\mathsf C} \leval{\len{0}}{}{\len{1}} \xlongrightarrow{\lbn{1}} \glvreshape{\LTTVPP{\lsv{}}{\lenv{0}}{\lenv{1}}}{\lsnv{{\mathrm{svOut}}}}{\lsnv{{\mathrm{svOut}}}}\glresulttype{\lsnv{{\mathrm{svOut}}}}}}
\and
\gpfit{\inferrule*[lab=\textsc{C-E-App\allowbreak ILin}]{({\lDelta{}_{0}} \ldisun {\lDelta{}_{1}}, \evalCtxSpace{})\vdash \leval{\len{0}}{}{\len{1}}\colon \tof{\lbone{}} \\ \evalCtxSpace{};\evalCtx{}\vdash_{\mathsf C} \sof{\lbn{0}} \evalArrow{} \lsv{} \\ \evalCtxSpace{};\evalCtx{}\vdash_{\mathsf C} \lbone{} \evalArrow{}_{\mathsf{basis}} \lR{} \\ {\lDelta{}_{0}};\evalCtxSpace{};\evalCtx{}\vdash_{\mathsf C} \len{0} \xlongrightarrow{\lbn{0}^{\lstar}} \lenv{0}\glresulttype{\lsnv{{\mathrm{svFun}}}} \\ {\lDelta{}_{1}};\evalCtxSpace{};\evalCtx{}\vdash_{\mathsf C} \len{1} \xlongrightarrow{\lbn{0}} \lenv{1}\glresulttype{\lsnv{{\mathrm{svArg}}}}}{{\lDelta{}_{0}} \ldisun {\lDelta{}_{1}};\evalCtxSpace{};\evalCtx{}\vdash_{\mathsf C} \leval{\len{0}}{}{\len{1}} \xlongrightarrow{\lbone{}} \LTTVPP{\lsv{}}{\glvreshape{\lenv{0}}{\lsv{} \loprod \lR{}}{\lsv{}}}{\lenv{1}}\glresulttype{\lR{}}}}
\and
\gpfit{\inferrule*[lab=\textsc{C-E-Nllam}]{\forall i.\; \tof{\lbn{i}} = \ltyn{i} \\ (\lemptyset{}, \evalCtxSpace{})\vdash (\nllambda \llpl x\colon \ltyn{0}\llpr \ldotsp \lee{})\colon \tof{\frameB{x}{\lbn{0}}{\lbn{1}}} \\ (\emptyset,\evalCtxSpace{})\vdash \frameB{x}{\lbn{0}}{\lbn{1}}\colon \bvalid{} \\ \evalCtxSpace{};\evalCtx{}\vdash_{\mathsf C} \frameB{x}{\lbn{0}}{\lbn{1}} \evalArrow{}_{\mathsf{basis}} (x\colon \ltynv{0}\fnto \ltynv{1}) \\ \evalCtxSpace{};\evalCtx{}\vdash_{\mathsf C} \lbn{0} \evalArrow{}_{\mathsf{basis}} \ltynv{0} \\ \lemptyset{};(\evalCtxSpace{}, \mathit{x}\colon \ltyn{0});\evalCtx{}[(\mathit{x},\lbn{0}) \mapsto \mathit{x}_{\ltynv{0}}]\vdash_{\mathsf C} \lee{} \xlongrightarrow{\lbn{1}} \leev{}\glresulttype{\ltynv{1}} \\ (\evalCtxSpace{}, \mathit{x}\colon \ltyn{0});\evalCtx{}[(\mathit{x},\lbn{0}) \mapsto \mathit{x}_{\ltynv{0}}]\vdash_{\mathsf C} \lbn{1} \evalArrow{}_{\mathsf{basis}} \ltynv{1}}{\lemptyset{};\evalCtxSpace{};\evalCtx{}\vdash_{\mathsf C} (\nllambda \llpl x\colon \ltyn{0}\llpr \ldotsp \lee{}) \xlongrightarrow{\frameB{x}{\lbn{0}}{\lbn{1}}} \llambda(x\colon \ltynv{0})\lrh \ltynv{1}\colon \glvreshape{\leev{}}{\ltynv{1}}{\ltynv{1}}\glresulttype{(x\colon \ltynv{0}\fnto \ltynv{1})}}}
\and
\gpfit{\inferrule*[lab=\textsc{C-E-App\allowbreak NVal}]{(\lemptyset{}, \evalCtxSpace{})\vdash \leval{\len{0}}{}{\len{1}}\colon \tof{\lbn{1}\subst{x}{\len{1}}} \\ (\emptyset,\evalCtxSpace{})\vdash \lbn{1}\subst{x}{\len{1}}\colon \bvalid{} \\ \evalCtxSpace{};\evalCtx{}\vdash_{\mathsf C} \lbn{1}\subst{x}{\len{1}} \evalArrow{}_{\mathsf{basis}} {\ltyv{}_{\mathsf{out}}} \\ \lemptyset{};\evalCtxSpace{};\evalCtx{}\vdash_{\mathsf C} \len{0} \xlongrightarrow{\frameB{x}{\lbn{0}}{\lbn{1}}} \lenv{0}\glresulttype{(x\colon {\ltyv{}_{\mathsf{fun}0}}\fnto {\ltyv{}_{\mathsf{fun}1}})} \\ \lemptyset{};\evalCtxSpace{};\evalCtx{}\vdash_{\mathsf C} \len{1} \xlongrightarrow{\lbn{0}} \lenv{1}\glresulttype{{\ltyv{}_{\mathsf{arg}}}} \\ \operatorname{inferType}(\lenv{0}) = (x\colon \ltynv{{\mathrm{D}}}\fnto \ltynv{{\mathrm{C}}}) \\ \lenv{2} = \glvreshape{\lenv{1}}{\ltynv{{\mathrm{D}}}}{\operatorname{inferType}(\lenv{1})}}{\lemptyset{};\evalCtxSpace{};\evalCtx{}\vdash_{\mathsf C} \leval{\len{0}}{}{\len{1}} \xlongrightarrow{\lbn{1}\subst{x}{\len{1}}} \glvreshape{\glvapp{\lenv{0}}{\lenv{2}}}{{\ltyv{}_{\mathsf{out}}}}{\ltynv{{\mathrm{C}}}\subst{x}{\lenv{2}}}\glresulttype{{\ltyv{}_{\mathsf{out}}}}}}
\and
\gpfit{\inferrule*[lab=\textsc{C-E-Frame\allowbreak Reanchor}]{\tof{\lbn{0}'} = \tof{\lbn{0}} \\ (\lemptyset{}, \evalCtxSpace{})\vdash \len{0}\colon \tof{\frameB{x}{\lbn{0}}{\lbn{1}}} \\ (\emptyset,\evalCtxSpace{})\vdash \frameB{x}{\lbn{0}}{\lbn{1}}\colon \bvalid{} \\ \evalCtxSpace{};\evalCtx{}\vdash_{\mathsf C} \frameB{x}{\lbn{0}}{\lbn{1}} \evalArrow{}_{\mathsf{basis}} (x\colon \ltynv{0}\fnto \ltynv{1}') \\ \evalCtxSpace{};\evalCtx{}\vdash_{\mathsf C} \lbn{0} \evalArrow{}_{\mathsf{basis}} \ltynv{0} \\ \lemptyset{};\evalCtxSpace{};\evalCtx{}\vdash_{\mathsf C} \len{0} \xlongrightarrow{\frameB{x}{\lbn{0}'}{\lbn{1}}} \leev{}\glresulttype{(x\colon \ltynv{0}'\fnto \ltynv{1})} \\ \evalCtxSpace{};\evalCtx{}\vdash_{\mathsf C} \lbn{0}' \evalArrow{}_{\mathsf{basis}} \ltynv{0}' \\ \lemptyset{};(\evalCtxSpace{}, \mathit{x}\colon \tof{\lbn{0}});\evalCtx{}[(\mathit{x},\lbn{0}) \mapsto \mathit{x}_{\ltynv{0}}]\vdash_{\mathsf C} \mathit{x} \xlongrightarrow{\lbn{0}'} \lenv{1}\glresulttype{\ltynv{0}'} \\ (\evalCtxSpace{}, \mathit{x}\colon \tof{\lbn{0}});\evalCtx{}[(\mathit{x},\lbn{0}) \mapsto \mathit{x}_{\ltynv{0}}]\vdash_{\mathsf C} \lbn{1} \evalArrow{}_{\mathsf{basis}} \ltynv{1}' \\ \operatorname{inferType}(\leev{}) = (x\colon \ltynv{{\mathrm{D}}}\fnto \ltynv{{\mathrm{C}}}) \\ \lenv{2} = \glvreshape{\glvreshape{\lenv{1}}{\ltynv{0}'}{\ltynv{0}'}}{\ltynv{{\mathrm{D}}}}{\ltynv{0}'}}{\lemptyset{};\evalCtxSpace{};\evalCtx{}\vdash_{\mathsf C} \len{0} \xlongrightarrow{\frameB{x}{\lbn{0}}{\lbn{1}}} \llambda(x\colon \ltynv{0})\lrh \ltynv{1}'\colon \glvreshape{\glvapp{\leev{}}{\lenv{2}}}{\ltynv{1}'}{\ltynv{1}'}\glresulttype{(x\colon \ltynv{0}\fnto \ltynv{1}')}}}
\and
\gpfit{\inferrule*[lab=\textsc{C-E-Frame\allowbreak Eta}]{(\lemptyset{}, \evalCtxSpace{})\vdash \len{0}\colon \tof{\frameB{x}{\lbn{0}}{\lbn{1}}} \\ (\emptyset,\evalCtxSpace{})\vdash \frameB{x}{\lbn{0}}{\lbn{1}}\colon \bvalid{} \\ \evalCtxSpace{};\evalCtx{}\vdash_{\mathsf C} \frameB{x}{\lbn{0}}{\lbn{1}} \evalArrow{}_{\mathsf{basis}} (x\colon \ltynv{0}\fnto \ltynv{1}) \\ \evalCtxSpace{};\evalCtx{}\vdash_{\mathsf C} \lbn{0} \evalArrow{}_{\mathsf{basis}} \ltynv{0} \\ \lemptyset{};(\evalCtxSpace{}, \mathit{x}\colon \tof{\lbn{0}});\evalCtx{}[(\mathit{x},\lbn{0}) \mapsto \mathit{x}_{\ltynv{0}}]\vdash_{\mathsf C} \leval{\len{0}}{}{\mathit{x}} \xlongrightarrow{\lbn{1}} \leev{}\glresulttype{\ltynv{1}} \\ (\evalCtxSpace{}, \mathit{x}\colon \tof{\lbn{0}});\evalCtx{}[(\mathit{x},\lbn{0}) \mapsto \mathit{x}_{\ltynv{0}}]\vdash_{\mathsf C} \lbn{1} \evalArrow{}_{\mathsf{basis}} \ltynv{1}}{\lemptyset{};\evalCtxSpace{};\evalCtx{}\vdash_{\mathsf C} \len{0} \xlongrightarrow{\frameB{x}{\lbn{0}}{\lbn{1}}} \llambda(x\colon \ltynv{0})\lrh \ltynv{1}\colon \glvreshape{\leev{}}{\ltynv{1}}{\ltynv{1}}\glresulttype{(x\colon \ltynv{0}\fnto \ltynv{1})}}}
\and
\gpfit{\inferrule*[lab=\textsc{C-E-Llam\allowbreak Assemble}]{\tof{\lb{}'} = \sof{\lb{}'} \\ (\lDelta{}, \evalCtxSpace{})\vdash (\lllambda \llpl x\colon \sof{\lb{}'}\llpr \ldotsp \lee{})\colon \tof{\lb{}'^{\lstar} \loprod \lb{}} \\ (\emptyset,\evalCtxSpace{})\vdash \lb{}'^{\lstar} \loprod \lb{}\colon \bvalid{} \\ \evalCtxSpace{};\evalCtx{}\vdash_{\mathsf C} \lb{}'^{\lstar} \loprod \lb{} \evalArrow{}_{\mathsf{basis}} \lsv{}' \loprod \lsv{} \\ \evalCtxSpace{};\evalCtx{}\vdash_{\mathsf C} \sof{\lb{}'} \evalArrow{} \lsv{}' \\ \evalCtxSpace{};\evalCtx{}\vdash_{\mathsf C} \sof{\lb{}} \evalArrow{} \lsv{} \\ \forall\,0\leq i<\overline{\operatorname{dimOf}}(\lsv{}').\; (\lDelta{}, \mathit{x}\colon \sof{\lb{}'});\evalCtxSpace{};\evalCtx{}[(\mathit{x},\lb{}') \mapsto \ldelta{i}{\lsv{}'}]\vdash_{\mathsf C} \lee{} \xlongrightarrow{\lb{}} \leev{}_{i}\glresulttype{\lsv{}}}{\lDelta{};\evalCtxSpace{};\evalCtx{}\vdash_{\mathsf C} (\lllambda \llpl x\colon \sof{\lb{}'}\llpr \ldotsp \lee{}) \xlongrightarrow{\lb{}'^{\lstar} \loprod \lb{}} \glvreshape{\lsum^{\lsv{}'}_{\mathit{i}<\overline{\operatorname{dimOf}}(\lsv{}')}\left(\ldelta{\mathit{i}}{\lsv{}'}\loprod \glvreshape{\leev{}_{\mathit{i}}}{\lsv{}}{\lsv{}}\right)}{\lsv{}' \loprod \lsv{}}{\begin{cases}\lsempty{}&\overline{\operatorname{dimOf}}(\lsv{}')=0\\ \lsv{}' \loprod \lsv{}&\text{otherwise}\end{cases}}\glresulttype{\lsv{}' \loprod \lsv{}}}}
\and
\gpfit{\inferrule*[lab=\textsc{C-E-Llam\allowbreak Assemble\allowbreak Real}]{\tof{\lb{}'} = \sof{\lb{}'} \\ (\lDelta{}, \evalCtxSpace{})\vdash (\lllambda \llpl x\colon \sof{\lb{}'}\llpr \ldotsp \lee{})\colon \tof{\lb{}'^{\lstar}} \\ (\emptyset,\evalCtxSpace{})\vdash \lb{}'^{\lstar}\colon \bvalid{} \\ \evalCtxSpace{};\evalCtx{}\vdash_{\mathsf C} \lb{}'^{\lstar} \evalArrow{}_{\mathsf{basis}} \lsv{}' \\ \evalCtxSpace{};\evalCtx{}\vdash_{\mathsf C} \sof{\lb{}'} \evalArrow{} \lsv{}' \\ \forall\,0\leq i<\overline{\operatorname{dimOf}}(\lsv{}').\; (\lDelta{}, \mathit{x}\colon \sof{\lb{}'});\evalCtxSpace{};\evalCtx{}[(\mathit{x},\lb{}') \mapsto \ldelta{i}{\lsv{}'}]\vdash_{\mathsf C} \lee{} \xlongrightarrow{\lbone{}} \leev{}_{i}\glresulttype{\lR{}}}{\lDelta{};\evalCtxSpace{};\evalCtx{}\vdash_{\mathsf C} (\lllambda \llpl x\colon \sof{\lb{}'}\llpr \ldotsp \lee{}) \xlongrightarrow{\lb{}'^{\lstar}} \glvreshape{\lsum^{\lsv{}'}_{\mathit{i}<\overline{\operatorname{dimOf}}(\lsv{}')}\left(\glvreshape{\leev{}_{\mathit{i}}}{\lR{}}{\lR{}} \lvscale \ldelta{\mathit{i}}{\lsv{}'}\right)}{\lsv{}'}{\begin{cases}\lsempty{}&\overline{\operatorname{dimOf}}(\lsv{}')=0\\ \lsv{}'&\text{otherwise}\end{cases}}\glresulttype{\lsv{}'}}}
\and
\gpfit{\inferrule*[lab=\textsc{C-E-Zero}]{\tof{\lb{}} = \sof{\lb{}} \\ (\lDelta{}, \evalCtxSpace{})\vdash \lvzero{}_{\sof{\lb{}}}\colon \tof{\lb{}} \\ (\emptyset,\evalCtxSpace{})\vdash \lb{}\colon \bvalid{} \\ \evalCtxSpace{};\evalCtx{}\vdash_{\mathsf C} \lb{} \evalArrow{}_{\mathsf{basis}} \lsv{} \\ \evalCtxSpace{};\evalCtx{}\vdash_{\mathsf C} \sof{\lb{}} \evalArrow{} \lsv{}}{\lDelta{};\evalCtxSpace{};\evalCtx{}\vdash_{\mathsf C} \lvzero{}_{\sof{\lb{}}} \xlongrightarrow{\lb{}} \lvzero{}_{\lsv{}}\glresulttype{\lsv{}}}}
\and
\gpfit{\inferrule*[lab=\textsc{C-E-Rlit}]{(\lemptyset{}, \evalCtxSpace{})\vdash \mathit{r}\colon \tof{\lbone{}} \\ \evalCtxSpace{};\evalCtx{}\vdash_{\mathsf C} \lbone{} \evalArrow{}_{\mathsf{basis}} \lR{}}{\lemptyset{};\evalCtxSpace{};\evalCtx{}\vdash_{\mathsf C} \mathit{r} \xlongrightarrow{\lbone{}} \mathit{r}\glresulttype{\lR{}}}}
}
\gldefrule{opStep.products}{family}{}{\forall i.\; \evalJudge{\evalCtx{}}{\len{i}}{\lbn{i}}{\lenv{i}} \\ \forall i.\; \tof{\lbn{i}} = \sof{\lbn{i}}}{\evalJudge{\evalCtx{}}{\len{0}\losum \len{1}}{\lbn{0} \losum \lbn{1}}{\lenv{0}\losum \lenv{1}} \\ \evalJudge{\evalCtx{}}{\len{0}\loprod \len{1}}{\lbn{0} \loprod \lbn{1}}{\lenv{0}\loprod \lenv{1}}}{\texttt{GLang.Val\allowbreak Expr.Op\allowbreak Step.dsum\allowbreak E, GLang.Val\allowbreak Expr.Op\allowbreak Step.tprod\allowbreak E}}{generated family}
\gldefrule{opStep.embeddings}{family}{}{\forall i.\; \sof{\lbn{i}} = \lsn{i} \\ \forall i.\; \evalJudge{\evalCtx{}}{\linner{}_{\lsn{i}}}{\lbn{i}^{\lstar} \loprod \lbn{i}^{\lstar}}{\lenv{i}} \\ \evalJudge{\evalCtx{}}{\lee{}}{\lbn{0}^{\lstar} \loprod \lbn{1}}{\leev{}}}{\evalJudge{\evalCtx{}}{\lembd{}_{\Kernel(\lee{})}}{\pBasisC{\Kernel(\lee{})}{\lbn{0}}{\lbn{1}}^{\lstar} \loprod \lbn{0}}{\Trans(\glvop{GSVDUI}(\leev{},\lenv{0},\lenv{1}))} \\ \evalJudge{\evalCtx{}}{\lembd{}_{\CoKernel(\lee{})}}{\pBasisC{\CoKernel(\lee{})}{\lbn{0}}{\lbn{1}}^{\lstar} \loprod \lbn{0}}{\Trans(\glvop{GSVDU}(\leev{},\lenv{0},\lenv{1}))} \\ \evalJudge{\evalCtx{}}{\lembd{}_{\Image(\lee{})}}{\pBasisC{\Image(\lee{})}{\lbn{0}}{\lbn{1}}^{\lstar} \loprod \lbn{1}}{\Trans(\glvop{GSVDV}(\leev{},\lenv{0},\lenv{1}))} \\ \evalJudge{\evalCtx{}}{\lembd{}_{\CoImage(\lee{})}}{\pBasisC{\CoImage(\lee{})}{\lbn{0}}{\lbn{1}}^{\lstar} \loprod \lbn{1}}{\Trans(\glvop{GSVDVI}(\leev{},\lenv{0},\lenv{1}))}}{\texttt{GLang.Val\allowbreak Expr.Op\allowbreak Step.embed\_\allowbreak ker, GLang.Val\allowbreak Expr.Op\allowbreak Step.embed\_\allowbreak coker, GLang.Val\allowbreak Expr.Op\allowbreak Step.embed\_\allowbreak img, GLang.Val\allowbreak Expr.Op\allowbreak Step.embed\_\allowbreak coimg}}{generated family}
\gldefrule{opStep.embeddingsGrid}{family}{}{\forall i.\; \sof{\lbn{i}} = \lsn{i} \\ \forall i.\; \evalJudge{\evalCtx{}}{\linner{}_{\lsn{i}}}{\lbn{i}^{\lstar} \loprod \lbn{i}^{\lstar}}{\lenv{i}} \\ \evalJudge{\evalCtx{}}{\lee{}}{\lbn{0}^{\lstar} \loprod \lbn{1}}{\leev{}}}{\evalJudge{\evalCtx{}}{\lembd{}_{\Kernel(\lee{})}}{\substack{\pBasis{}\\(\Kernel(\lee{}),\lbn{0},\lbn{1})}^{\lstar} \loprod \lbn{0}}{\begin{gathered}\Trans(\glvop{GSVDUI}\\(\leev{},\lenv{0},\lenv{1}))\end{gathered}} \qquad \evalJudge{\evalCtx{}}{\lembd{}_{\CoKernel(\lee{})}}{\substack{\pBasis{}\\(\CoKernel(\lee{}),\lbn{0},\lbn{1})}^{\lstar} \loprod \lbn{0}}{\begin{gathered}\Trans(\glvop{GSVDU}\\(\leev{},\lenv{0},\lenv{1}))\end{gathered}} \\ \evalJudge{\evalCtx{}}{\lembd{}_{\Image(\lee{})}}{\substack{\pBasis{}\\(\Image(\lee{}),\lbn{0},\lbn{1})}^{\lstar} \loprod \lbn{1}}{\begin{gathered}\Trans(\glvop{GSVDV}\\(\leev{},\lenv{0},\lenv{1}))\end{gathered}} \qquad \evalJudge{\evalCtx{}}{\lembd{}_{\CoImage(\lee{})}}{\substack{\pBasis{}\\(\CoImage(\lee{}),\lbn{0},\lbn{1})}^{\lstar} \loprod \lbn{1}}{\begin{gathered}\Trans(\glvop{GSVDVI}\\(\leev{},\lenv{0},\lenv{1}))\end{gathered}}}{\texttt{GLang.Val\allowbreak Expr.Op\allowbreak Step.embed\_\allowbreak ker, GLang.Val\allowbreak Expr.Op\allowbreak Step.embed\_\allowbreak coker, GLang.Val\allowbreak Expr.Op\allowbreak Step.embed\_\allowbreak img, GLang.Val\allowbreak Expr.Op\allowbreak Step.embed\_\allowbreak coimg}}{generated family}
\gldefrule{evalS.fundamental}{family}{}{\evalJudge{\evalCtx{}}{\lee{}}{\lb{}}{\leev{}} \\ \sof{\lb{}} = \lsn{0}^{\lstar} \loprod \lsn{1} \\ \exists\,\lbp{},\textcolor{lang}{b^{\prime\prime}}.\; \lb{} = \lbp{}^{\lstar}\loprod \textcolor{lang}{b^{\prime\prime}} \\ \evalSJudge{\evalCtx{}}{\sof{\lb{}}}{\lsnv{{\mathrm{svb}}}}}{\evalSJudge{\evalCtx{}}{\Kernel(\lee{})}{\lR^{\glvreshape{\leev{}}{\lsnv{{\mathrm{svb}}}}{\lsnv{{\mathrm{svb}}}},\Kernel{},\lsnv{{\mathrm{svb}}}}} \qquad \evalSJudge{\evalCtx{}}{\CoKernel(\lee{})}{\lR^{\glvreshape{\leev{}}{\lsnv{{\mathrm{svb}}}}{\lsnv{{\mathrm{svb}}}},\CoKernel{},\lsnv{{\mathrm{svb}}}}} \\ \evalSJudge{\evalCtx{}}{\Image(\lee{})}{\lR^{\glvreshape{\leev{}}{\lsnv{{\mathrm{svb}}}}{\lsnv{{\mathrm{svb}}}},\Image{},\lsnv{{\mathrm{svb}}}}} \qquad \evalSJudge{\evalCtx{}}{\CoImage(\lee{})}{\lR^{\glvreshape{\leev{}}{\lsnv{{\mathrm{svb}}}}{\lsnv{{\mathrm{svb}}}},\CoImage{},\lsnv{{\mathrm{svb}}}}}}{\texttt{GLang.Val\allowbreak Space.Eval\allowbreak S.ker, GLang.Val\allowbreak Space.Eval\allowbreak S.coker, GLang.Val\allowbreak Space.Eval\allowbreak S.img, GLang.Val\allowbreak Space.Eval\allowbreak S.coimg}}{generated family}
\gldefdefn{partialAdequacy.agree}{\texttt{Partial\allowbreak Adequacy.Agree}}{clauses}{\mathsf{Agree}(\mathcal{O}, \sigma{}, \glvEnv{}, \mathit{a}, \Gamma_{\mathrm{a}}, \mathit{a}, \mathit{a}, \ltynv{{\mathrm{a}}}, \mathit{a}, \mathit{a}, \mathit{a})}{\begin{center}\gpfit{\mathsf{Agree}(\mathcal{O}, \sigma{}, \glvEnv{}, \mathit{x}, \Gamma{}, \Delta{}, \ls{}, \lsv{}, \mathit{x}, \mathit{x}, \mathit{a}) = \exists \mathit{n}.\; \mathsf{dofPartial}(\mathcal{O}\,\glvEnv{}\,\Gamma{}\,\Delta{}, \lsv{}) = \mathit{n} \;\wedge\; \lvert \mathit{x} \rvert = \mathit{n} \;\wedge\; \lvert \mathit{a}\rvert = \mathit{n} \;\wedge\; \mathit{x} = \sum_{\mathit{i} \in \mathsf{univ}} (\overline{\mathit{a}}[\mathit{i}]) \cdot (\mathit{x}[\mathit{i}])}\\[2pt]
\gpfit{\mathsf{Agree}(\mathcal{O}, \sigma{}, \glvEnv{}, \mathit{x}, \Gamma{}, \Delta{}, \mathit{a}, \lsnv{{\mathrm{a1}}}, \mathit{x}, \mathit{x}, {\mathit{a}_{2}}) = \mathsf{False}}\\[2pt]
\gpfit{\mathsf{Agree}(\mathcal{O}, \sigma{}, \glvEnv{}, \mathit{x}, \Gamma{}, \Delta{}, (x\colon \mathit{d}\fnto \mathit{c}), (x\colon \ltynv{{\mathrm{dv}}}\fnto \ltynv{{\mathrm{cv}}}), \mathit{x}, \mathit{x}, \mathit{x}) = \exists \mathit{D}.\; \mathsf{denotValTyPartial}(\mathcal{O}, \glvEnv{}, \Gamma{}, \Delta{}, \ltynv{{\mathrm{dv}}}) = \mathit{D} \;\wedge\; \exists \mathit{f}.\; \mathit{x} = \mathit{f} \;\wedge\; \forall \mathit{a},\,\mathit{raw},\,\mathit{arg}.\; \mathsf{carrierFromVal}(\ltynv{{\mathrm{dv}}}, \mathit{raw}) = \mathit{arg} \;\wedge\; \mathsf{Agree}(\mathcal{O}, \sigma{}, \glvEnv{}, \mathit{x}, \Gamma{}, \Delta{}, \mathit{d}, \ltynv{{\mathrm{dv}}}, \mathsf{domainBasis}(\mathit{x}), \mathit{a}, \mathit{raw}) \Rightarrow \exists \mathit{out}.\; \mathit{f}\,\mathit{arg} = \mathit{out} \;\wedge\; \mathsf{Agree}(\mathcal{O}, \sigma{}, \glvEnv{}, \mathit{a} \mathbin{:\!:} \mathit{x}, \ltynv{{\mathrm{dv}}} \mathbin{:\!:} \Gamma{}, \mathit{raw} \mathbin{:\!:} \Delta{}, \mathit{c}, \ltynv{{\mathrm{cv}}}, \mathit{x}(\mathit{a}), \mathit{x}(\mathit{a}), \mathit{out})}\\[2pt]
\gpfit{\mathsf{Agree}(\mathcal{O}, \sigma{}, \glvEnv{}, \mathit{x}, \Gamma{}, \Delta{}, \mathit{a}, (x\colon \ltynv{{\mathrm{a1}}}\fnto \ltynv{{\mathrm{a2}}}), \mathit{x}, \mathit{x}, \mathit{x}) = \mathsf{False}}\\[2pt]
\gpfit{\mathsf{Agree}(\mathcal{O}, \sigma{}, \glvEnv{}, \mathit{x}, \Gamma{}, \Delta{}, (x\colon \mathit{a}\fnto {\mathit{a}_{1}}), \lsnv{{\mathrm{a2}}}, \mathit{x}, \mathit{x}, \mathit{x}) = \mathsf{False}}\end{center}
}{\texttt{GLang.Partial\allowbreak Adequacy.Agree}}{GLang/Adequacy/PartialSourceAgreement.lean:9}
\gldefdefn{agreeDims}{\texttt{Agree\allowbreak Dims}}{clauses}{\mathsf{AgreeDims}}{\begin{center}\gpfit{\mathsf{AgreeDims} = \mathsf{AgreeDimsAt}}\end{center}
}{\texttt{GLang.Agree\allowbreak Dims}}{GLang/Adequacy/RawReification.lean:75}
\gldefdefn{adequacy.partialDirect.agreeDims}{Dimension agreement}{clauses}{\mathsf{AgreeDims}}{\begin{center}\gpfit{\mathsf{AgreeDims} = \mathsf{DimsAt}(\mathsf{oracle})}\end{center}
}{\texttt{GLang.Adequacy.Partial\allowbreak Direct.Agree\allowbreak Dims}}{GLang/Adequacy/PartialSourcePublic.lean:10}
\gldefdefn{adequacy.direct.agreeDims}{Dimension agreement}{clauses}{\SemT{\lty{}}{\sigma{}} \equiv_{\mathit{B}} \SemT{\ltyv{}}{\glvEnv{}}}{\begin{center}\gpfit{\SemT{\lty{}}{\sigma{}} \equiv_{\mathit{B}} \SemT{\ltyv{}}{\glvEnv{}} = \mathsf{AgreeDims}(\mathcal{O}, \sigma{}, \glvEnv{}, {\lDelta{}_{s}}, \Gamma_{\mathrm{Gv}}, {\lDelta{}_{v}}, \lty{}, \ltyv{}, \mathit{B})}\end{center}
}{\texttt{GLang.Adequacy.Direct.Agree\allowbreak Dims}}{GLang/Adequacy/DirectPublic.lean:55}
\gldefdefn{agreeDimsAt}{Dimension agreement (general oracle)}{conjunction}{\SemT{\lty{}}{\sigma{}} \equiv_{\mathit{B}} \SemT{\ltyv{}}{\glvEnv{}}}{\begin{center}\gpfit{\SemT{\lty{}}{\sigma{}} \equiv_{\mathit{B}} \SemT{\ltyv{}}{\glvEnv{}} \;\equiv\;}\\[2pt]
\gpfit{\quad \SemT{\lty{}}{\sigma{}} \equiv_{\mathit{B}} \SemT{\ltyv{}}{\glvEnv{}}}\\[2pt]
\gpfit{\wedge\; \mathit{B} \;\text{is a basis for}\; \SemT{\lty{}}{\sigma{}}}\\[2pt]
\gpfit{\wedge\; \mathsf{RawReifiableAt}(\mathcal{O}, \sigma{}, \glvEnv{}, {\lDelta{}_{s}}, \Gamma_{\mathrm{Gv}}, {\lDelta{}_{v}}, \lty{}, \ltyv{}, \mathit{B})}\\[2pt]
\gpfit{\wedge\; \mathsf{RawSourceInjectiveAt}(\mathcal{O}, \sigma{}, \glvEnv{}, {\lDelta{}_{s}}, \Gamma_{\mathrm{Gv}}, {\lDelta{}_{v}}, \lty{}, \ltyv{}, \mathit{B})}\end{center}
}{\texttt{GLang.Agree\allowbreak Dims\allowbreak At}}{GLang/Adequacy/RawReification.lean:62}
\gldefdefn{dEnv.agreeOn}{Environments agreeing on a set of names}{conjunction}{{\sigma{}_{1}} =_{\mathit{S}} {\sigma{}_{2}}}{\begin{center}\gpfit{{\sigma{}_{1}} =_{\mathit{S}} {\sigma{}_{2}} \;\equiv\;}\\[2pt]
\gpfit{\quad \forall \mathit{x}.\; \mathit{x} \in \mathit{S} \Rightarrow {\sigma{}_{1}}(\mathit{x}) = {\sigma{}_{2}}(\mathit{x})}\end{center}
}{\texttt{GLang.DEnv.Agree\allowbreak On}}{GLang/FVIrrelevance.lean:25}
\gldefdefn{adequacy.basisAt}{\texttt{Adequacy.Basis\allowbreak At}}{clauses}{\mathsf{BasisAt}}{\begin{center}\gpfit{\mathsf{BasisAt} = \mathsf{RawBasisAt}}\end{center}
}{\texttt{GLang.Adequacy.Basis\allowbreak At}}{GLang/Adequacy/Public.lean:36}
\gldefdefn{adequacy.direct.basisAt}{\texttt{Adequacy.Direct.Basis\allowbreak At}}{equation}{\mathit{B} \;\text{is a basis for}\; \SemT{\lty{}}{\sigma{}}}{\begin{center}\gpfit{\mathit{B} \;\text{is a basis for}\; \SemT{\lty{}}{\sigma{}}}\end{center}
}{\texttt{GLang.Adequacy.Direct.Basis\allowbreak At}}{GLang/Adequacy/DirectPublic.lean:42}
\gldefdefn{valueQueryDependencies.bounded}{\texttt{Value\allowbreak Query\allowbreak Dependencies.Bounded}}{clauses}{\mathsf{Bounded}(\mathit{x}, \mathit{x})}{\begin{center}\gpfit{\mathsf{Bounded}(0, \mathit{x}) = \mathsf{False}}\\[2pt]
\gpfit{\mathsf{Bounded}(\mathsf{succ}(\mathit{n}), \mathit{x}) = \forall \mathit{child}.\; \mathit{child} \in \mathsf{children}(\mathit{x}) \Rightarrow \mathsf{Bounded}(\mathit{n}, \mathit{child})}\end{center}
}{\texttt{GLang.Value\allowbreak Query\allowbreak Dependencies.Bounded}}{GLang/ValueQueryDependencies.lean:45}
\gldefdefn{operational.cob}{\texttt{Operational.Cob}}{equation}{\mathsf{Cob}}{\begin{center}\gpfit{\mathsf{Cob}}\end{center}
}{\texttt{GLang.Operational.Cob}}{GLang/Operational.lean:38}
\gldefdefn{adequacy.partialDirect.cobAgree}{Change-of-basis agreement}{clauses}{\mathsf{CobAgree}}{\begin{center}\gpfit{\mathsf{CobAgree} = \mathsf{CobAgree}(\mathsf{oracle})}\end{center}
}{\texttt{GLang.Adequacy.Partial\allowbreak Direct.Cob\allowbreak Agree}}{GLang/Adequacy/PartialSourcePublic.lean:11}
\gldefdefn{cobAgree}{Change-of-basis agreement (general oracle)}{clauses}{\SemBB{\textcolor{lang}{{\lb{}_{\mathrm{src}}}}}{\sigma{}} = \SemA{\leev{}}{\glvEnv{}}{} \cdot \SemBB{\textcolor{lang}{{\lb{}_{\mathrm{tgt}}}}}{\sigma{}}}{\begin{center}\gpfit{\SemBB{\textcolor{lang}{{\lb{}_{\mathrm{src}}}}}{\sigma{}} = \SemA{\leev{}}{\glvEnv{}}{} \cdot \SemBB{\textcolor{lang}{{\lb{}_{\mathrm{tgt}}}}}{\sigma{}} = \lvert \SemBB{\textcolor{lang}{{\lb{}_{\mathrm{tgt}}}}}{\sigma{}} \rvert = \lvert \SemBB{\textcolor{lang}{{\lb{}_{\mathrm{src}}}}}{\sigma{}} \rvert \;\wedge\; \exists \mathit{a}.\; \mathcal{E}\llbracket \leev{}\rrbracket_{[];[]} = \mathit{a} \;\wedge\; \lvert \mathit{a}\rvert = (\lvert \SemBB{\textcolor{lang}{{\lb{}_{\mathrm{src}}}}}{\sigma{}} \rvert) \cdot (\lvert \SemBB{\textcolor{lang}{{\lb{}_{\mathrm{src}}}}}{\sigma{}} \rvert) \;\wedge\; \forall \mathit{j}.\; \SemBB{\textcolor{lang}{{\lb{}_{\mathrm{src}}}}}{\sigma{}}[\mathit{j}] = \sum_{\mathit{i} \in \mathsf{univ}} (\overline{\mathit{a}}[(\mathit{j}) \cdot (\lvert \SemBB{\textcolor{lang}{{\lb{}_{\mathrm{src}}}}}{\sigma{}} \rvert) + \mathit{i}]) \cdot (\SemBB{\textcolor{lang}{{\lb{}_{\mathrm{tgt}}}}}{\sigma{}}[\mathit{i}])}\end{center}
}{\texttt{GLang.Cob\allowbreak Agree}}{GLang/Adequacy/Agree.lean:253}
\gldefdefn{adequacy.direct.cobAgree}{Change-of-basis agreement}{equation}{\SemBB{\textcolor{lang}{{\lb{}_{\mathrm{src}}}}}{\sigma{}} = \SemA{\leev{}}{\glvEnv{}}{} \cdot \SemBB{\textcolor{lang}{{\lb{}_{\mathrm{tgt}}}}}{\sigma{}}}{\begin{center}\gpfit{\SemBB{\textcolor{lang}{{\lb{}_{\mathrm{src}}}}}{\sigma{}} = \SemA{\leev{}}{\glvEnv{}}{} \cdot \SemBB{\textcolor{lang}{{\lb{}_{\mathrm{tgt}}}}}{\sigma{}}}\end{center}
}{\texttt{GLang.Adequacy.Direct.Cob\allowbreak Agree}}{GLang/Adequacy/DirectPublic.lean:61}
\gldefdefn{partialAdequacy.cobAgree}{\texttt{Partial\allowbreak Adequacy.Cob\allowbreak Agree}}{clauses}{\mathsf{CobAgree}(\mathcal{O}, \sigma{}, \glvEnv{}, \textcolor{lang}{{\lb{}_{\mathrm{src}}}}, \textcolor{lang}{{\lb{}_{\mathrm{tgt}}}}, \leev{})}{\begin{center}\gpfit{\mathsf{CobAgree}(\mathcal{O}, \sigma{}, \glvEnv{}, \textcolor{lang}{{\lb{}_{\mathrm{src}}}}, \textcolor{lang}{{\lb{}_{\mathrm{tgt}}}}, \leev{}) = \mathsf{CobResult}(\mathcal{O}, \glvEnv{}, [], [], \SemBB{\textcolor{lang}{{\lb{}_{\mathrm{src}}}}}{\sigma{}}, \SemBB{\textcolor{lang}{{\lb{}_{\mathrm{tgt}}}}}{\sigma{}}, \leev{})}\end{center}
}{\texttt{GLang.Partial\allowbreak Adequacy.Cob\allowbreak Agree}}{GLang/Adequacy/PartialSourceCob.lean:20}
\gldefdefn{partialAdequacy.cobResult}{\texttt{Partial\allowbreak Adequacy.Cob\allowbreak Result}}{clauses}{\mathsf{CobResult}(\mathcal{O}, \glvEnv{}, \Gamma{}, \lDelta{}, \mathit{V}, \mathit{source}, \mathit{target}, \leev{})}{\begin{center}\gpfit{\mathsf{CobResult}(\mathcal{O}, \glvEnv{}, \Gamma{}, \lDelta{}, \mathit{source}, \mathit{target}, \leev{}) = \exists \mathit{a}.\; \mathsf{denotValB\_partial}(\mathcal{O}, \glvEnv{}, \Gamma{}, \lDelta{}, \leev{}) = \mathit{a} \;\wedge\; \mathsf{CobRows}(\mathit{source}, \mathit{target}, \mathit{a})}\end{center}
}{\texttt{GLang.Partial\allowbreak Adequacy.Cob\allowbreak Result}}{GLang/Adequacy/PartialSourceCob.lean:15}
\gldefdefn{partialAdequacy.cobRows}{\texttt{Partial\allowbreak Adequacy.Cob\allowbreak Rows}}{clauses}{\mathsf{CobRows}(\mathit{V}, \mathit{source}, \mathit{target}, \mathit{a})}{\begin{center}\gpfit{\mathsf{CobRows}(\mathit{source}, \mathit{target}, \mathit{a}) = \lvert \mathit{target} \rvert = \lvert \mathit{source} \rvert \;\wedge\; \lvert \mathit{a}\rvert = (\lvert \mathit{source} \rvert) \cdot (\lvert \mathit{source} \rvert) \;\wedge\; \forall \mathit{j}.\; \mathit{source}[\mathit{j}] = \sum_{\mathit{i} \in \mathsf{univ}} (\overline{\mathit{a}}[(\mathit{j}) \cdot (\lvert \mathit{source} \rvert) + \mathit{i}]) \cdot (\mathit{target}[\mathit{i}])}\end{center}
}{\texttt{GLang.Partial\allowbreak Adequacy.Cob\allowbreak Rows}}{GLang/Adequacy/PartialSourceCob.lean:9}
\gldefdefn{cobShapeT}{Change-of-basis shape}{expanded definition}{\mathsf{CobShapeT}(\evalCtx{}, \leev{}, \mathcal{S})}{\begin{center}\gpfit{\exists \lsn{0}.\; \exists \lsn{1}.\; \exists \lsnv{0}.\; \exists \lsnv{1}.\; \operatorname{inferType}(\leev{}) = \lsnv{0} \loprod \lsnv{1} \;\wedge\; \evalSJudge{\evalCtx{}}{\lsn{0}}{\lsnv{0}} \;\wedge\; \evalSJudge{\evalCtx{}}{\lsn{1}}{\lsnv{1}}}\\[2pt]
\gpfit{\wedge\; \vdash \operatorname{inferType}(\leev{})\colon \tvalid{}}\\[2pt]
\gpfit{\wedge\; \mathsf{fv}(\leev{}) \subseteq \mathcal{S}}\end{center}
}{\texttt{GLang.Cob\allowbreak Shape\allowbreak T}}{GLang/PreservationV2_EvalBBCross.lean:19}
\gldefdefn{runtimeValue.totalSubstitution.codomainLaw}{\texttt{Runtime\allowbreak Value.Total\allowbreak Substitution.Codomain\allowbreak Law}}{clauses}{\mathsf{CodomainLaw}(\mathcal{O})}{\begin{center}\gpfit{\mathsf{CodomainLaw}(\mathcal{O}) = \forall \valCheckCtx{},\,\ltynv{{\mathrm{d}}},\,\ltynv{{\mathrm{c}}},\,\lky{},\,\lenv{{\mathrm{u}}},\,\glvEnv{},\,\leev{}.\; \mathcal{E}\llbracket \lenv{{\mathrm{u}}}\rrbracket_{[];[]} = \leev{} \Rightarrow \mathsf{ValCtxValid}(\valCheckCtx{}) \to \valCheckCtx{}\vdash (x\colon \ltynv{{\mathrm{d}}}\fnto \ltynv{{\mathrm{c}}})\colon \lky{} \to \valCheckCtx{}\vdash \lenv{{\mathrm{u}}}\colon \ltynv{{\mathrm{d}}} \to \mathsf{RawTypePresentation}(\mathcal{O}, \ltynv{{\mathrm{c}}}, \ltynv{{\mathrm{c}}}\subst{x}{\lenv{{\mathrm{u}}}}, \mathsf{mk}(\glvEnv{}, \ltynv{{\mathrm{d}}} \mathbin{:\!:} [], \leev{} \mathbin{:\!:} []), \mathsf{mk}(\glvEnv{}, [], []))}\end{center}
}{\texttt{GLang.Runtime\allowbreak Value.Total\allowbreak Substitution.Codomain\allowbreak Law}}{GLang/RuntimeValueTotalSemanticSubstitution.lean:506}
\gldefdefn{adequacy.partialDirect.consistent}{Environment consistency}{clauses}{\mathsf{Consistent}}{\begin{center}\gpfit{\mathsf{Consistent} = \mathsf{Consistent}(\mathsf{oracle})}\end{center}
}{\texttt{GLang.Adequacy.Partial\allowbreak Direct.Consistent}}{GLang/Adequacy/PartialSourcePublic.lean:8}
\gldefdefn{adequacy.direct.consistent}{Environment consistency}{presentation}{(\lDelta{}, \evalCtxSpace{}); \evalCtx{}; \valCheckCtx{}\vdash \sigma{}, \glvEnv{}}{We write $(\Delta,\epsilon);\Psi;\valCheckCtx\vdash(\sigma,\glvEnv)$ when
$(\Delta,\epsilon);\valCheckCtx\vdash\Psi$, $\valCheckCtx\vdash\glvEnv$,
$(\Delta,\epsilon)\vdash\sigma$ at every step, and
$(\Delta,\epsilon,\Psi,\valCheckCtx,\sigma,\glvEnv)$ is constructed as follows.
The tuple $(\emptyset,\emptyset,\emptyset,\emptyset,\emptyset,\glvEnv)$ is constructed for arbitrary $\glvEnv$.
Suppose $(\Delta,\epsilon,\Psi,\valCheckCtx,\sigma,\glvEnv)$ has been constructed.
Let $\Psi'=\Psi[(x,\lb{})\mapsto\leev{}]$ and update
$(\Delta,\epsilon,\valCheckCtx)$ to $(\Delta',\epsilon',\valCheckCtx')$
as in \cref{def:nonrecurseeval}.
Put $T=\tof{\lb{}}$,
$\ltyv{}=\operatorname{inferType}(\leev{})$, and
$B=\SemBB{\lb{}}{\sigma}$, and require
$\Sem{\leev{}}{\glvEnv}=w\ne\glverr$,
$a\equiv_{B;\sigma,\glvEnv}^{T,\ltyv{}}w$, and
$T\sim_{B;\sigma,\glvEnv}\ltyv{}$.
If $x\notin\operatorname{dom}(\Delta\sqcup\epsilon)$,
require $(x\notin\operatorname{fv}(\leev{})\ \lor\
\leev{}=x_{\ltyv{}})$ and put
$\sigma'=\sigma[x\mapsto a]$, and
$\glvEnv'=\glvEnv[x\mapsto w]$.
Otherwise require $a=\sigma(x)$ and put
$(\sigma',\glvEnv')=(\sigma,\glvEnv)$.
This constructs $(\Delta',\epsilon',\Psi',\valCheckCtx',\sigma',\glvEnv')$.}{\texttt{GLang.Adequacy.Direct.Consistent}}{GLang/Adequacy/DirectPublic.lean:26}
\gldefdefn{partialAdequacy.consistent}{\texttt{Partial\allowbreak Adequacy.Consistent}}{conjunction}{\mathsf{Consistent}(\mathcal{O}, \lDelta{}, \evalCtxSpace{}, \evalCtx{}, \valCheckCtx{}, \sigma{}, \glvEnv{})}{\begin{center}\gpfit{\mathsf{Consistent}(\mathcal{O}, \lDelta{}, \evalCtxSpace{}, \evalCtx{}, \valCheckCtx{}, \sigma{}, \glvEnv{}) \;\equiv\;}\\[2pt]
\gpfit{\quad \mathsf{History}(\mathcal{O}, \lDelta{}, \evalCtxSpace{}, \evalCtx{}, \valCheckCtx{}, \sigma{}, \glvEnv{})}\\[2pt]
\gpfit{\wedge\; (\lDelta{},\evalCtxSpace{});\valCheckCtx{}\vdash \evalCtx{}}\\[2pt]
\gpfit{\wedge\; \mathsf{PartialEnvWF}(\mathcal{O}, \valCheckCtx{}, \glvEnv{})}\end{center}
}{\texttt{GLang.Partial\allowbreak Adequacy.Consistent}}{GLang/Adequacy/PartialSourceHistory.lean:1379}
\gldefdefn{consistent}{Environment consistency (general oracle alias)}{clauses}{\mathsf{Consistent}}{\begin{center}\gpfit{\mathsf{Consistent} = \mathsf{RawTypedConsistent}}\end{center}
}{\texttt{GLang.Consistent}}{GLang/Adequacy/RawTypedConsistent.lean:20}
\gldefdefn{operational.contextHistory}{Operational context history}{clauses}{\mathsf{ContextHistory}}{\begin{center}\gpfit{\mathsf{ContextHistory} = \mathsf{TypedContext}}\end{center}
}{\texttt{GLang.Operational.Context\allowbreak History}}{GLang/Operational.lean:87}
\gldefdefn{v2.ctxTele}{Context-history telescope}{rules}{\mathsf{CtxTele}}{\begin{center}\gpfit{\mathsf{CtxTele}}\end{center}
is defined inductively by
{\mprset{sep=0.3em}\scriptsize\begin{mathpar}
\gpfit{\inferrule*[Right=\textsc{empty}]{\phantom{.}}{\lemptyset{}; \lemptyset{}; \lemptyset{}\;\mathsf{history}}}
\and
\gpfit{\inferrule*[Right=\textsc{extend\_\allowbreak nlin}]{\lDelta{}; \evalCtxSpace{}; \evalCtx{}\;\mathsf{history} \\ \neg(\mathit{x} \in \lDelta{}) \\ \neg(\mathit{x} \in \evalCtxSpace{}) \\ \evalSJudge{\evalCtx{}}{\tof{\lb{}}}{\operatorname{inferType}(\leev{})} \\ \mathsf{fv}(\leev{}) \subseteq \mathrm{dom}(\evalCtx{}[(\mathit{x},\lb{}) \mapsto \leev{}]) \\ \mathsf{lc}(\leev{}) \\ \mathsf{FrameAnchor}(\evalCtx{}, \lb{}, \leev{}) \\ (\emptyset,\evalCtxSpace{})\vdash \lb{}\colon \bvalid{}}{\lDelta{}; (\evalCtxSpace{}, \mathit{x}\colon \tof{\lb{}}); \evalCtx{}[(\mathit{x},\lb{}) \mapsto \leev{}]\;\mathsf{history}}}
\and
\gpfit{\inferrule*[Right=\textsc{extend\_\allowbreak lin}]{\lDelta{}; \evalCtxSpace{}; \evalCtx{}\;\mathsf{history} \\ \neg(\mathit{x} \in \lDelta{}) \\ \neg(\mathit{x} \in \evalCtxSpace{}) \\ \evalSJudge{\evalCtx{}}{\tof{\lb{}}}{\operatorname{inferType}(\leev{})} \\ \mathsf{fv}(\leev{}) \subseteq \mathrm{dom}(\evalCtx{}[(\mathit{x},\lb{}) \mapsto \leev{}]) \\ \mathsf{lc}(\leev{}) \\ \mathsf{FrameAnchor}(\evalCtx{}, \lb{}, \leev{}) \\ (\emptyset,\evalCtxSpace{})\vdash \lb{}\colon \bvalid{} \\ \tof{\lb{}} = \ls{}}{(\lDelta{}, \mathit{x}\colon \ls{}); \evalCtxSpace{}; \evalCtx{}[(\mathit{x},\lb{}) \mapsto \leev{}]\;\mathsf{history}}}
\and
\gpfit{\inferrule*[Right=\textsc{extend\_\allowbreak rep}]{\lDelta{}; \evalCtxSpace{}; \evalCtx{}\;\mathsf{history} \\ \mathsf{SourceEntryLinked}(\lDelta{}, \evalCtxSpace{}, \mathit{x}, \lb{}) \\ \neg((\mathit{x}, \lb{}) \in \evalCtx{}) \\ \evalSJudge{\evalCtx{}}{\tof{\lb{}}}{\operatorname{inferType}(\leev{})} \\ \mathsf{fv}(\leev{}) \subseteq \mathrm{dom}(\evalCtx{}[(\mathit{x},\lb{}) \mapsto \leev{}]) \\ \mathsf{lc}(\leev{}) \\ \mathsf{FrameAnchor}(\evalCtx{}, \lb{}, \leev{}) \\ (\emptyset,\evalCtxSpace{})\vdash \lb{}\colon \bvalid{}}{\lDelta{}; \evalCtxSpace{}; \evalCtx{}[(\mathit{x},\lb{}) \mapsto \leev{}]\;\mathsf{history}}}
\end{mathpar}}
}{\texttt{GLang.V2.Ctx\allowbreak Tele}}{GLang/ValueProgress.lean:380}
\gldefdefn{ctxWF}{Context scoping and namespace disjointness}{conjunction}{\mathsf{Scoped}(\lDelta{}, \lepsilon{})}{\begin{center}\gpfit{\mathsf{Scoped}(\lDelta{}, \lepsilon{}) \;\equiv\;}\\[2pt]
\gpfit{\quad \forall \mathit{x}.\; \mathit{x} \in \lDelta{} \Rightarrow \neg(\mathit{x} \in \lepsilon{})}\\[2pt]
\gpfit{\wedge\; \forall \mathit{x},\,\ls{},\,\mathit{y}.\; (\mathit{x}\colon \ls{}) \in \lDelta{} \;\wedge\; \mathit{y} \in \mathsf{fv}(\ls{}) \Rightarrow \mathit{y} \in \lepsilon{}}\\[2pt]
\gpfit{\wedge\; \forall \mathit{x},\,\lty{},\,\mathit{y}.\; (\mathit{x}\colon \lty{}) \in \lepsilon{} \;\wedge\; \mathit{y} \in \mathsf{fv}(\lty{}) \Rightarrow \mathit{y} \in \lepsilon{}}\end{center}
}{\texttt{GLang.Ctx\allowbreak WF}}{GLang/Wellformed.lean:31}
\gldefdefn{v2.ctxWF'}{\texttt{Ctx\allowbreak WF'}}{conjunction}{\mathsf{CtxWF'}(\lDelta{}, \evalCtxSpace{})}{\begin{center}\gpfit{\mathsf{CtxWF'}(\lDelta{}, \evalCtxSpace{}) \;\equiv\;}\\[2pt]
\gpfit{\quad \forall \mathit{x}.\; \mathit{x} \in \lDelta{} \Rightarrow \neg(\mathit{x} \in \evalCtxSpace{})}\\[2pt]
\gpfit{\wedge\; \forall \mathit{x},\,\ls{}.\; (\mathit{x}\colon \ls{}) \in \lDelta{} \Rightarrow \mathsf{fv}(\ls{}) \subseteq \mathrm{dom}(\evalCtxSpace{})}\\[2pt]
\gpfit{\wedge\; \forall \mathit{x},\,\lty{}.\; (\mathit{x}\colon \lty{}) \in \evalCtxSpace{} \Rightarrow \mathsf{fv}(\lty{}) \subseteq \mathrm{dom}(\evalCtxSpace{})}\\[2pt]
\gpfit{\wedge\; \forall \mathit{x},\,\ls{}.\; (\mathit{x}\colon \ls{}) \in \lDelta{} \Rightarrow (\lemptyset, \evalCtxSpace{})\vdash \ls{}\colon\svalid}\\[2pt]
\gpfit{\wedge\; \forall \mathit{x},\,\lty{}.\; (\mathit{x}\colon \lty{}) \in \evalCtxSpace{} \Rightarrow \exists \lky{}.\; (\lemptyset, \evalCtxSpace{})\vdash \lty{}\colon \lky{}}\end{center}
}{\texttt{GLang.V2.Ctx\allowbreak WF'}}{GLang/ValuePreservationV2.lean:57}
\gldefdefn{partialAdequacy.dimsAt}{\texttt{Partial\allowbreak Adequacy.Dims\allowbreak At}}{clauses}{\mathsf{DimsAt}(\mathcal{O}, \sigma{}, \glvEnv{}, \mathit{a}, \Gamma_{\mathrm{a}}, \mathit{a}, \mathit{a}, \ltynv{{\mathrm{a}}}, \mathit{a})}{\begin{center}\gpfit{\mathsf{DimsAt}(\mathcal{O}, \sigma{}, \glvEnv{}, \mathit{x}, \Gamma{}, \Delta{}, \ls{}, \lsv{}, \mathit{x}) = \mathsf{shpOfPartial}(\mathcal{O}\,\glvEnv{}\,\Gamma{}\,\Delta{}, \lsv{}) = \mathsf{srcShp}(\sigma{}, \mathit{x}, \ls{})}\\[2pt]
\gpfit{\mathsf{DimsAt}(\mathcal{O}, \sigma{}, \glvEnv{}, \mathit{x}, \Gamma{}, \Delta{}, (x\colon \mathit{d}\fnto \mathit{c}), (x\colon \ltynv{{\mathrm{dv}}}\fnto \ltynv{{\mathrm{cv}}}), \mathit{x}) = \mathsf{DimsAt}(\mathcal{O}, \sigma{}, \glvEnv{}, \mathit{x}, \Gamma{}, \Delta{}, \mathit{d}, \ltynv{{\mathrm{dv}}}, \mathsf{domainBasis}(\mathit{x})) \;\wedge\; \forall \mathit{a},\,\mathit{raw},\,\mathit{arg}.\; \mathsf{carrierFromVal}(\ltynv{{\mathrm{dv}}}, \mathit{raw}) = \mathit{arg} \;\wedge\; \mathsf{Agree}(\mathcal{O}, \sigma{}, \glvEnv{}, \mathit{x}, \Gamma{}, \Delta{}, \mathit{d}, \ltynv{{\mathrm{dv}}}, \mathsf{domainBasis}(\mathit{x}), \mathit{a}, \mathit{raw}) \Rightarrow \mathsf{DimsAt}(\mathcal{O}, \sigma{}, \glvEnv{}, \mathit{a} \mathbin{:\!:} \mathit{x}, \ltynv{{\mathrm{dv}}} \mathbin{:\!:} \Gamma{}, \mathit{raw} \mathbin{:\!:} \Delta{}, \mathit{c}, \ltynv{{\mathrm{cv}}}, \mathit{x}(\mathit{a}))}\\[2pt]
\gpfit{\mathsf{DimsAt}(\mathcal{O}, \sigma{}, \glvEnv{}, \mathit{x}, \Gamma{}, \Delta{}, \mathit{a}, (x\colon \ltynv{{\mathrm{a1}}}\fnto \ltynv{{\mathrm{a2}}}), \mathit{x}) = \mathsf{False}}\\[2pt]
\gpfit{\mathsf{DimsAt}(\mathcal{O}, \sigma{}, \glvEnv{}, \mathit{x}, \Gamma{}, \Delta{}, (x\colon \mathit{a}\fnto {\mathit{a}_{1}}), \lsnv{{\mathrm{a2}}}, \mathit{x}) = \mathsf{False}}\end{center}
}{\texttt{GLang.Partial\allowbreak Adequacy.Dims\allowbreak At}}{GLang/Adequacy/PartialSourceAgreement.lean:41}
\gldefdefn{envWF}{Environment well-formedness}{conjunction}{\lDelta{}, \lepsilon{}\vdash \sigma{}}{\begin{center}\gpfit{\lDelta{}, \lepsilon{}\vdash \sigma{} \;\equiv\;}\\[2pt]
\gpfit{\quad \forall \mathit{x},\,\ls{}.\; (\mathit{x}\colon \ls{}) \in \lDelta{} \Rightarrow \exists \leev{}.\; (\mathit{x}\colon \leev{}) \in \sigma{} \;\wedge\; \mathsf{shape}(\leev{}) = \SemT{\ls{}}{\sigma{}}}\\[2pt]
\gpfit{\wedge\; \forall \mathit{x},\,\lty{}.\; (\mathit{x}\colon \lty{}) \in \lepsilon{} \Rightarrow \exists \leev{}.\; (\mathit{x}\colon \leev{}) \in \sigma{} \;\wedge\; \mathsf{shape}(\leev{}) = \SemT{\lty{}}{\sigma{}}}\end{center}
}{\texttt{GLang.Env\allowbreak WF}}{GLang/Wellformed.lean:50}
\gldefdefn{operational.unchecked.evalBB}{\texttt{Operational.Unchecked.Eval\allowbreak BB}}{equation}{\mathsf{EvalBB}}{\begin{center}\gpfit{\mathsf{EvalBB}}\end{center}
}{\texttt{GLang.Operational.Unchecked.Eval\allowbreak BB}}{GLang/Operational.lean:209}
\gldefdefn{rawConsistentRank.evalCtxCreationRank}{Exact-pair evaluator creation rank}{clauses}{\mathsf{EvalCtxCreationRank}(\evalCtx{})}{\begin{center}\gpfit{\mathsf{EvalCtxCreationRank}(\evalCtx{}) = \exists \mathit{rank}.\; \forall \mathit{x},\,\lb{},\,\mathit{y}.\; (\mathit{x}, \lb{}) \in \evalCtx{} \;\wedge\; \mathit{y} \in \mathsf{fv}(\lb{}) \Rightarrow \exists \textcolor{lang}{\mathit{bY}}.\; (\mathit{y}, \textcolor{lang}{\mathit{bY}}) \in \evalCtx{} \;\wedge\; \mathit{rank}\,(\mathit{y}, \textcolor{lang}{\mathit{bY}}) < \mathit{rank}\,(\mathit{x}, \lb{})}\end{center}
}{\texttt{GLang.Raw\allowbreak Consistent\allowbreak Rank.Eval\allowbreak Ctx\allowbreak Creation\allowbreak Rank}}{GLang/Adequacy/RawConsistentRank.lean:30}
\gldefdefn{v2.evalCtxLamAnchored}{\texttt{Eval\allowbreak Ctx\allowbreak Lam\allowbreak Anchored}}{clauses}{\mathsf{EvalCtxLamAnchored}(\evalCtx{})}{\begin{center}\gpfit{\mathsf{EvalCtxLamAnchored}(\evalCtx{}) = \forall \mathit{x},\,\lbn{0},\,\lbn{1},\,\ltynv{0},\,\ltynv{1},\,\lenv{{\mathrm{body}}}.\; ((\mathit{x}, \frameB{x}{\lbn{0}}{\lbn{1}}) \mapsto \llambda(y\colon \ltynv{0})\lrh \ltynv{1}\colon \lenv{{\mathrm{body}}}) \in \evalCtx{} \Rightarrow \exists \mathit{L}.\; \forall \mathit{y}.\; \evalSJudge{\evalCtx{}[(\mathit{y},\lbn{0}) \mapsto \mathit{y}_{\ltynv{0}}]}{\tof{\lbn{1}}}{\ltynv{1}}}\end{center}
}{\texttt{GLang.V2.Eval\allowbreak Ctx\allowbreak Lam\allowbreak Anchored}}{GLang/ValuePreservationV2.lean:75}
\gldefdefn{evalCtxLc}{\texttt{Eval\allowbreak Ctx\allowbreak Lc}}{clauses}{\mathsf{EvalCtxLc}(\evalCtx{})}{\begin{center}\gpfit{\mathsf{EvalCtxLc}(\evalCtx{}) = \forall \lky{},\,\leev{}.\; (\lky{} \mapsto \leev{}) \in \evalCtx{} \Rightarrow \mathsf{lc}(\leev{})}\end{center}
}{\texttt{GLang.Eval\allowbreak Ctx\allowbreak Lc}}{GLang/ValueRealizeLc.lean:21}
\gldefdefn{v2.evalCtxValid}{Store well-formedness (weak)}{clauses}{\vdash \evalCtx{}}{\begin{center}\gpfit{\vdash \evalCtx{} = \forall \mathit{x},\,\lb{},\,\leev{}.\; ((\mathit{x}, \lb{}) \mapsto \leev{}) \in \evalCtx{} \Rightarrow \mathsf{fv}(\lb{}) \subseteq \mathrm{dom}(\evalCtx{}) \;\wedge\; \mathsf{fv}(\leev{}) \subseteq \mathrm{dom}(\evalCtx{}) \;\wedge\; \evalSJudge{\evalCtx{}}{\tof{\lb{}}}{\operatorname{inferType}(\leev{})}}\end{center}
}{\texttt{GLang.V2.Eval\allowbreak Ctx\allowbreak Valid}}{GLang/ValuePreservationV2.lean:67}
\gldefdefn{v2.evalCtxValidPlus}{Store well-formedness}{clauses}{\vdash \evalCtx{}}{\begin{center}\gpfit{\vdash \evalCtx{} = \forall \mathit{x},\,\lb{},\,\leev{}.\; ((\mathit{x}, \lb{}) \mapsto \leev{}) \in \evalCtx{} \Rightarrow \mathsf{fv}(\leev{}) \subseteq \mathrm{dom}(\evalCtx{}) \;\wedge\; \evalSJudge{\evalCtx{}}{\tof{\lb{}}}{\operatorname{inferType}(\leev{})} \;\wedge\; \mathsf{fv}(\operatorname{inferType}(\leev{})) \subseteq \mathrm{dom}(\evalCtx{}) \;\wedge\; \mathsf{lc}(\leev{}) \;\wedge\; \forall \lbn{0},\,\lbn{1},\,\ltynv{0},\,\ltynv{1},\,\lenv{{\mathrm{body}}}.\; \lb{} = \frameB{x}{\lbn{0}}{\lbn{1}} \;\wedge\; \leev{} = \llambda(y\colon \ltynv{0})\lrh \ltynv{1}\colon \lenv{{\mathrm{body}}} \Rightarrow \exists \mathit{L}.\; \forall \mathit{y}.\; \evalSJudge{\evalCtx{}[(\mathit{y},\lbn{0}) \mapsto \mathit{y}_{\ltynv{0}}]}{\tof{\lbn{1}}}{\ltynv{1}}}\end{center}
}{\texttt{GLang.V2.Eval\allowbreak Ctx\allowbreak Valid\allowbreak Plus}}{GLang/ValuePreservationV2.lean:92}
\gldefdefn{v2.evalEntryCreation}{Runtime-entry creation certificate}{conjunction}{\mathsf{EvalEntryCreation}(\evalCtx{}, \mathit{x}, \lb{}, \leev{})}{\begin{center}\gpfit{\mathsf{EvalEntryCreation}(\evalCtx{}, \mathit{x}, \lb{}, \leev{}) \;\equiv\;}\\[2pt]
\gpfit{\quad \mathsf{EvalEntry}(\evalCtx{}, \mathit{x}, \lb{}, \leev{})}\\[2pt]
\gpfit{\wedge\; \mathsf{FrameAnchor}(\evalCtx{}, \lb{}, \leev{})}\end{center}
}{\texttt{GLang.V2.Eval\allowbreak Entry\allowbreak Creation}}{GLang/ValueProgress.lean:358}
\gldefdefn{v2.evalEntryValid}{Runtime-entry validity}{conjunction}{\mathsf{EvalEntry}(\evalCtx{}, \mathit{x}, \lb{}, \leev{})}{\begin{center}\gpfit{\mathsf{EvalEntry}(\evalCtx{}, \mathit{x}, \lb{}, \leev{}) \;\equiv\;}\\[2pt]
\gpfit{\quad \evalSJudge{\evalCtx{}}{\tof{\lb{}}}{\operatorname{inferType}(\leev{})}}\\[2pt]
\gpfit{\wedge\; \mathsf{fv}(\leev{}) \subseteq \mathrm{dom}(\evalCtx{}[(\mathit{x},\lb{}) \mapsto \leev{}])}\\[2pt]
\gpfit{\wedge\; \mathsf{lc}(\leev{})}\end{center}
}{\texttt{GLang.V2.Eval\allowbreak Entry\allowbreak Valid}}{GLang/ValueProgress.lean:349}
\gldefdefn{v2.evalFrameAnchored}{Nonlinear frame anchor}{clauses}{\mathsf{FrameAnchor}(\evalCtx{}, \lb{}, \leev{})}{\begin{center}\gpfit{\mathsf{FrameAnchor}(\evalCtx{}, \lb{}, \leev{}) = \forall \lbn{0},\,\lbn{1},\,\ltynv{0},\,\ltynv{1},\,\lenv{{\mathrm{body}}}.\; \lb{} = \frameB{x}{\lbn{0}}{\lbn{1}} \;\wedge\; \leev{} = \llambda(x\colon \ltynv{0})\lrh \ltynv{1}\colon \lenv{{\mathrm{body}}} \Rightarrow \exists \mathit{L}.\; \forall \mathit{y}.\; \evalSJudge{\evalCtx{}[(\mathit{y},\lbn{0}) \mapsto \mathit{y}_{\ltynv{0}}]}{\tof{\lbn{1}}}{\ltynv{1}}}\end{center}
}{\texttt{GLang.V2.Eval\allowbreak Frame\allowbreak Anchored}}{GLang/ValueProgress.lean:339}
\gldefdefn{operational.unchecked.evalS}{\texttt{Operational.Unchecked.Eval\allowbreak S}}{equation}{\mathsf{EvalS}}{\begin{center}\gpfit{\mathsf{EvalS}}\end{center}
}{\texttt{GLang.Operational.Unchecked.Eval\allowbreak S}}{GLang/Operational.lean:203}
\gldefdefn{valueQueryDependencies.finite}{\texttt{Value\allowbreak Query\allowbreak Dependencies.Finite}}{clauses}{\mathsf{Finite}(\mathit{root})}{\begin{center}\gpfit{\mathsf{Finite}(\mathit{root}) = \exists \mathit{n}.\; \mathsf{Bounded}(\mathit{n}, \mathit{root})}\end{center}
}{\texttt{GLang.Value\allowbreak Query\allowbreak Dependencies.Finite}}{GLang/ValueQueryDependencies.lean:51}
\gldefdefn{gSVDInputsShape}{\texttt{GSVDInputs\allowbreak Shape}}{clauses}{\mathsf{GSVDInputsShape}({\mathit{d}_{0}}, {\mathit{d}_{1}}, \mathit{map}, \mathit{leftMetric}, \mathit{rightMetric})}{\begin{center}\gpfit{\mathsf{GSVDInputsShape}({\mathit{d}_{0}}, {\mathit{d}_{1}}, \mathit{map}, \mathit{leftMetric}, \mathit{rightMetric}) = \lvert \mathit{map}\rvert = ({\mathit{d}_{0}}) \cdot ({\mathit{d}_{1}}) \;\wedge\; \lvert \mathit{leftMetric}\rvert = ({\mathit{d}_{0}}) \cdot ({\mathit{d}_{0}}) \;\wedge\; \lvert \mathit{rightMetric}\rvert = ({\mathit{d}_{1}}) \cdot ({\mathit{d}_{1}})}\end{center}
}{\texttt{GLang.GSVDInputs\allowbreak Shape}}{GLang/ValueNumericalInputs.lean:12}
\gldefdefn{gSVDecomposition}{\texttt{GSVDecomposition}}{conjunction}{\mathsf{GSVDecomposition}(\mathit{m}, \mathit{n})}{\begin{center}\gpfit{\mathsf{GSVDecomposition}(\mathit{m}, \mathit{n}) \;\equiv\;}\\[2pt]
\gpfit{\quad \mathsf{Nat}}\\[2pt]
\gpfit{\wedge\; \mathit{r} \leq \mathit{m}}\\[2pt]
\gpfit{\wedge\; \mathit{r} \leq \mathit{n}}\\[2pt]
\gpfit{\wedge\; \mathsf{Matrix}(\mathsf{Fin}(\mathit{m}), \mathsf{Fin}(\mathit{r}), \mathbb{R})}\\[2pt]
\gpfit{\wedge\; \mathsf{Matrix}(\mathsf{Fin}(\mathit{m}), \mathsf{Fin}((\mathit{m}) - (\mathit{r})), \mathbb{R})}\\[2pt]
\gpfit{\wedge\; \mathsf{Matrix}(\mathsf{Fin}(\mathit{n}), \mathsf{Fin}(\mathit{r}), \mathbb{R})}\\[2pt]
\gpfit{\wedge\; \mathsf{Matrix}(\mathsf{Fin}(\mathit{n}), \mathsf{Fin}((\mathit{n}) - (\mathit{r})), \mathbb{R})}\\[2pt]
\gpfit{\wedge\; \mathsf{Fin}(\mathit{r}) \to \mathbb{R}}\\[2pt]
\gpfit{\wedge\; \forall \mathit{i}.\; 0 < \sigma{}_{\mathit{i}}}\\[2pt]
\gpfit{\wedge\; \forall \mathit{i},\,\mathit{j}.\; \mathit{i} \leq \mathit{j} \Rightarrow \sigma{}_{\mathit{j}} \leq \sigma{}_{\mathit{i}}}\\[2pt]
\gpfit{\wedge\; \mathsf{Matrix}(\mathsf{Fin}(\mathit{m}), \mathsf{Fin}(\mathit{m}), \mathbb{R})}\\[2pt]
\gpfit{\wedge\; \mathsf{Matrix}(\mathsf{Fin}(\mathit{n}), \mathsf{Fin}(\mathit{n}), \mathbb{R})}\\[2pt]
\gpfit{\wedge\; ((\mathsf{transpose}(\mathit{U})) \cdot (\mathit{MW})) \cdot (\mathit{U}) = \mathsf{diagonal}(1)}\\[2pt]
\gpfit{\wedge\; ((\mathsf{transpose}(\mathit{V})) \cdot (\mathit{MV})) \cdot (\mathit{V}) = \mathsf{diagonal}(1)}\\[2pt]
\gpfit{\wedge\; ((\mathsf{transpose}(\mathit{UI})) \cdot (\mathit{MW})) \cdot (\mathit{UI}) = 1}\\[2pt]
\gpfit{\wedge\; ((\mathsf{transpose}(\mathit{U})) \cdot (\mathit{MW})) \cdot (\mathit{UI}) = 0}\\[2pt]
\gpfit{\wedge\; ((\mathsf{transpose}(\mathit{VI})) \cdot (\mathit{MV})) \cdot (\mathit{VI}) = 1}\\[2pt]
\gpfit{\wedge\; ((\mathsf{transpose}(\mathit{V})) \cdot (\mathit{MV})) \cdot (\mathit{VI}) = 0}\end{center}
}{\texttt{GLang.GSVDecomposition}}{GLang/ValueSVD.lean:104}
\gldefdefn{higherOrderLinearAtVar}{Higher-order linearity}{clauses}{\mathsf{HOLinear}(\sigma{}; \mathit{x}; \ls{}; \lee{}; \mathit{x})}{\begin{center}\gpfit{\mathsf{HOLinear}(\sigma{}; \mathit{x}; \ls{}; \lee{}; \mathit{sOut}) = \mathsf{LinearAtVar}(\sigma{}, \mathit{x}, \ls{}, \lee{}, \mathit{sOut})}\\[2pt]
\gpfit{\mathsf{HOLinear}(\sigma{}; \mathit{x}; \ls{}; \lee{}; (x\colon \mathit{a}\fnto {\mathit{a}_{1}})) = \mathsf{PointwiseLinear}(\SemT{\ls{}}{\sigma{}}; \SemT{(x\colon \mathit{a}\fnto {\mathit{a}_{1}})}{\sigma{}}; \lambda \mathit{y}.\, \Sem{\lee{}}{\sigma{}[\mathit{x} \mapsto \mathit{y}]})}\end{center}
}{\texttt{GLang.Higher\allowbreak Order\allowbreak Linear\allowbreak At\allowbreak Var}}{GLang/SourceSoundnessStrong.lean:36}
\gldefdefn{partialAdequacy.history}{\texttt{Partial\allowbreak Adequacy.History}}{rules}{\mathsf{History}(\mathcal{O})}{\begin{center}\gpfit{\mathsf{History}(\mathcal{O})}\end{center}
is defined inductively by
{\mprset{sep=0.3em}\scriptsize\begin{mathpar}
\gpfit{\inferrule*[Right=\textsc{empty}]{\phantom{.}}{\mathsf{History}(\mathcal{O}, \lemptyset{}, \lemptyset{}, \lemptyset{}, \lemptyset{}, \mathsf{mk}(\lemptyset{}), \glvEnv{})}}
\and
\gpfit{\inferrule*[Right=\textsc{extend\_\allowbreak lin}]{\mathsf{History}(\mathcal{O}, \lDelta{}, \evalCtxSpace{}, \evalCtx{}, \valCheckCtx{}, \sigma{}, \glvEnv{}) \\ \neg(\mathit{x} \in \lDelta{}) \;\wedge\; \neg(\mathit{x} \in \evalCtxSpace{}) \\ \tof{\lb{}} = \ls{} \\ (\emptyset,\evalCtxSpace{})\vdash \lb{}\colon \bvalid{} \\ \evalSJudge{\evalCtx{}}{\tof{\lb{}}}{\ltyv{}} \\ \operatorname{inferType}(\leev{}) = \ltyv{} \\ \mathsf{InferSpaceMatchesBase}(\leev{}, \ltyv{}) \\ \mathsf{fv}(\leev{}) \subseteq \mathrm{dom}(\evalCtx{}[(\mathit{x},\lb{}) \mapsto \leev{}]) \\ \mathsf{lc}(\leev{}) \\ \neg(\mathit{x} \in \mathsf{fv}(\leev{})) \;\vee\; \exists \ltyv{}.\; \leev{} = \mathit{x}_{\ltyv{}} \\ \mathsf{shape}(\mathit{xs}) = \SemT{\tof{\lb{}}}{\sigma{}} \\ \mathsf{denotValB\_partial}(\mathcal{O}, \glvEnv{}, [], [], \leev{}) = \mathit{yV} \\ \mathsf{Agree}(\mathcal{O}, \sigma{}, \glvEnv{}, [], [], [], \tof{\lb{}}, \ltyv{}, \SemBB{\lb{}}{\sigma{}}, \mathit{xs}, \mathit{yV}) \\ \mathsf{DimsAt}(\mathcal{O}, \sigma{}, \glvEnv{}, [], [], [], \tof{\lb{}}, \ltyv{}, \SemBB{\lb{}}{\sigma{}})}{\mathsf{History}(\mathcal{O}, (\lDelta{}, \mathit{x}\colon \ls{}), \evalCtxSpace{}, \evalCtx{}[(\mathit{x},\lb{}) \mapsto \leev{}], (\valCheckCtx{}, \mathit{x}\colon \ltyv{}), \sigma{}[\mathit{x} \mapsto \mathit{xs}], \mathsf{update}(\glvEnv{}, \mathit{x}, \mathit{yV}))}}
\and
\gpfit{\inferrule*[Right=\textsc{extend\_\allowbreak nl}]{\mathsf{History}(\mathcal{O}, \lDelta{}, \evalCtxSpace{}, \evalCtx{}, \valCheckCtx{}, \sigma{}, \glvEnv{}) \\ \neg(\mathit{x} \in \lDelta{}) \;\wedge\; \neg(\mathit{x} \in \evalCtxSpace{}) \\ (\emptyset,\evalCtxSpace{})\vdash \lb{}\colon \bvalid{} \\ \evalSJudge{\evalCtx{}}{\tof{\lb{}}}{\ltyv{}} \\ \operatorname{inferType}(\leev{}) = \ltyv{} \\ \mathsf{InferSpaceMatchesBase}(\leev{}, \ltyv{}) \\ \mathsf{fv}(\leev{}) \subseteq \mathrm{dom}(\evalCtx{}[(\mathit{x},\lb{}) \mapsto \leev{}]) \\ \mathsf{lc}(\leev{}) \\ \neg(\mathit{x} \in \mathsf{fv}(\leev{})) \;\vee\; \exists \ltyv{}.\; \leev{} = \mathit{x}_{\ltyv{}} \\ \mathsf{shape}(\mathit{xs}) = \SemT{\tof{\lb{}}}{\sigma{}} \\ \mathsf{denotValB\_partial}(\mathcal{O}, \glvEnv{}, [], [], \leev{}) = \mathit{yV} \\ \mathsf{Agree}(\mathcal{O}, \sigma{}, \glvEnv{}, [], [], [], \tof{\lb{}}, \ltyv{}, \SemBB{\lb{}}{\sigma{}}, \mathit{xs}, \mathit{yV}) \\ \mathsf{DimsAt}(\mathcal{O}, \sigma{}, \glvEnv{}, [], [], [], \tof{\lb{}}, \ltyv{}, \SemBB{\lb{}}{\sigma{}}) \\ \evalSJudge{\evalCtx{}}{\tof{\lb{}}}{\operatorname{inferType}(\leev{})} \\ \mathsf{fv}(\leev{}) \subseteq \mathrm{dom}(\evalCtx{}[(\mathit{x},\lb{}) \mapsto \leev{}]) \\ \mathsf{lc}(\leev{}) \\ \mathsf{FrameAnchor}(\evalCtx{}, \lb{}, \leev{})}{\mathsf{History}(\mathcal{O}, \lDelta{}, (\evalCtxSpace{}, \mathit{x}\colon \tof{\lb{}}), \evalCtx{}[(\mathit{x},\lb{}) \mapsto \leev{}], (\valCheckCtx{}, \mathit{x}\colon \ltyv{}), \sigma{}[\mathit{x} \mapsto \mathit{xs}], \mathsf{update}(\glvEnv{}, \mathit{x}, \mathit{yV}))}}
\and
\gpfit{\inferrule*[Right=\textsc{extend\_\allowbreak rep}]{\mathsf{History}(\mathcal{O}, \lDelta{}, \evalCtxSpace{}, \evalCtx{}, \valCheckCtx{}, \sigma{}, \glvEnv{}) \\ \neg((\mathit{x}, \lb{}) \in \evalCtx{}) \\ \mathsf{SourceEntryLinked}(\lDelta{}, \evalCtxSpace{}, \mathit{x}, \lb{}) \\ (\mathit{x}\colon \mathit{xs}) \in \sigma{} \\ (\emptyset,\evalCtxSpace{})\vdash \lb{}\colon \bvalid{} \\ \evalSJudge{\evalCtx{}}{\tof{\lb{}}}{\ltyv{}} \\ \operatorname{inferType}(\leev{}) = \ltyv{} \\ \mathsf{InferSpaceMatchesBase}(\leev{}, \ltyv{}) \\ \mathsf{fv}(\leev{}) \subseteq \mathrm{dom}(\evalCtx{}[(\mathit{x},\lb{}) \mapsto \leev{}]) \\ \mathsf{lc}(\leev{}) \\ \mathsf{shape}(\mathit{xs}) = \SemT{\tof{\lb{}}}{\sigma{}} \\ \mathsf{denotValB\_partial}(\mathcal{O}, \glvEnv{}, [], [], \leev{}) = \mathit{yV} \\ \mathsf{Agree}(\mathcal{O}, \sigma{}, \glvEnv{}, [], [], [], \tof{\lb{}}, \ltyv{}, \SemBB{\lb{}}{\sigma{}}, \mathit{xs}, \mathit{yV}) \\ \mathsf{DimsAt}(\mathcal{O}, \sigma{}, \glvEnv{}, [], [], [], \tof{\lb{}}, \ltyv{}, \SemBB{\lb{}}{\sigma{}}) \\ \evalSJudge{\evalCtx{}}{\tof{\lb{}}}{\operatorname{inferType}(\leev{})} \\ \mathsf{fv}(\leev{}) \subseteq \mathrm{dom}(\evalCtx{}[(\mathit{x},\lb{}) \mapsto \leev{}]) \\ \mathsf{lc}(\leev{}) \\ \mathsf{FrameAnchor}(\evalCtx{}, \lb{}, \leev{})}{\mathsf{History}(\mathcal{O}, \lDelta{}, \evalCtxSpace{}, \evalCtx{}[(\mathit{x},\lb{}) \mapsto \leev{}], \valCheckCtx{}, \sigma{}, \glvEnv{})}}
\end{mathpar}}
}{\texttt{GLang.Partial\allowbreak Adequacy.History}}{GLang/Adequacy/PartialSourceHistory.lean:338}
\gldefdefn{inductiveCtxWF}{Well-formed source context}{rules}{\vdash (\lDelta{}, \lepsilon{})}{\begin{center}\gpfit{\vdash (\lDelta{}, \lepsilon{})}\end{center}
is defined inductively by
{\mprset{sep=0.3em}\scriptsize\begin{mathpar}
\gpfit{\inferrule*[Right=\textsc{empty}]{\phantom{.}}{\vdash (\lemptyset{}, \lemptyset{})}}
\and
\gpfit{\inferrule*[Right=\textsc{linear}]{\vdash (\lDelta{}, \lepsilon{}) \\ \neg(\mathit{x} \in \lDelta{}) \\ \neg(\mathit{x} \in \lepsilon{}) \\ \forall \mathit{y}.\; \mathit{y} \in \mathsf{fv}(\ls{}) \Rightarrow \mathit{y} \in \lepsilon{} \\ (\lemptyset, \lepsilon{})\vdash \ls{}\colon\svalid}{\vdash ((\lDelta{}, \mathit{x}\colon \ls{}), \lepsilon{})}}
\and
\gpfit{\inferrule*[Right=\textsc{nonlinear}]{\vdash (\lDelta{}, \lepsilon{}) \\ \neg(\mathit{x} \in \lDelta{}) \\ \neg(\mathit{x} \in \lepsilon{}) \\ \forall \mathit{y}.\; \mathit{y} \in \mathsf{fv}(\lty{}) \Rightarrow \mathit{y} \in \lepsilon{} \\ (\lemptyset, \lepsilon{})\vdash \lty{}\colon \lky{}}{\vdash (\lDelta{}, (\lepsilon{}, \mathit{x}\colon \lty{}))}}
\end{mathpar}}
}{\texttt{GLang.Inductive\allowbreak Ctx\allowbreak WF}}{GLang/SourceContextInductive.lean:27}
\gldefdefn{valExpr.inferSpaceMatchesBase}{\texttt{Val\allowbreak Expr.Infer\allowbreak Space\allowbreak Matches\allowbreak Base}}{clauses}{\mathsf{InferSpaceMatchesBase}(\leev{}, \ltynv{{\mathrm{x}}})}{\begin{center}\gpfit{\mathsf{InferSpaceMatchesBase}(\leev{}, \lsv{}) = \operatorname{inferSpace}(\leev{}) = \lsv{}}\\[2pt]
\gpfit{\mathsf{InferSpaceMatchesBase}(\leev{}, (x\colon \ltynv{{\mathrm{a}}}\fnto \ltynv{{\mathrm{a1}}})) = \mathsf{True}}\end{center}
}{\texttt{GLang.Val\allowbreak Expr.Infer\allowbreak Space\allowbreak Matches\allowbreak Base}}{GLang/Adequacy/RawConsistent.lean:354}
\gldefdefn{iofFresh}{Inner-product reserved-name freshness}{conjunction}{\mathsf{iofFresh}(\lepsilon{})}{\begin{center}\gpfit{\mathsf{iofFresh}(\lepsilon{}) \;\equiv\;}\\[2pt]
\gpfit{\quad \neg(\mathsf{x0} \in \lepsilon{})}\\[2pt]
\gpfit{\wedge\; \neg(\mathsf{x1} \in \lepsilon{})}\\[2pt]
\gpfit{\wedge\; \neg(\text{\$y00} \in \lepsilon{})}\\[2pt]
\gpfit{\wedge\; \neg(\text{\$y01} \in \lepsilon{})}\\[2pt]
\gpfit{\wedge\; \neg(\text{\$y10} \in \lepsilon{})}\\[2pt]
\gpfit{\wedge\; \neg(\text{\$y11} \in \lepsilon{})}\\[2pt]
\gpfit{\wedge\; \neg(\mathsf{g\_name} \in \lepsilon{})}\\[2pt]
\gpfit{\wedge\; \neg(\mathsf{a\_name} \in \lepsilon{})}\\[2pt]
\gpfit{\wedge\; \neg(\mathsf{b\_name} \in \lepsilon{})}\\[2pt]
\gpfit{\wedge\; \neg(\mathsf{p\_name} \in \lepsilon{})}\end{center}
}{\texttt{GLang.Iof\allowbreak Fresh}}{GLang/InnerProduct.lean:294}
\gldefdefn{isInnerProduct}{Inner-product laws}{clauses}{\mathsf{IsInnerProduct}(\mathit{V}, \mathit{inst}, \mathit{inst}, \mathit{B})}{\begin{center}\gpfit{\mathit{B} \;\text{is an inner product} = \forall \mathit{x},\,\mathit{y}.\; \mathit{B}(\mathit{x})(\mathit{y}) = \mathit{B}(\mathit{y})(\mathit{x}) \;\wedge\; \forall \mathit{x}.\; 0 \leq \mathit{B}(\mathit{x})(\mathit{x}) \;\wedge\; \forall \mathit{x}.\; \mathit{B}(\mathit{x})(\mathit{x}) = 0 \Rightarrow \mathit{x} = 0}\end{center}
}{\texttt{GLang.Is\allowbreak Inner\allowbreak Product}}{GLang/Denot.lean:52}
\gldefdefn{isListBasis}{Ordered list basis}{clauses}{\mathit{vs} \;\text{is an ordered basis of}\; \mathit{V}}{\begin{center}\gpfit{\mathit{vs} \;\text{is an ordered basis of}\; \mathit{V} = \mathsf{LinearIndependent}(\mathbb{R}, \lambda \mathit{i}.\, \mathit{vs}[\mathit{i}]) \;\wedge\; \mathsf{span}(\mathbb{R}, \operatorname{range} \lambda \mathit{i}.\, \mathit{vs}[\mathit{i}]) = \mathsf{top}}\end{center}
}{\texttt{GLang.Is\allowbreak List\allowbreak Basis}}{GLang/BasisFSVGSVD.lean:57}
\gldefdefn{v2.keyScope}{Scope containment}{equation}{\mathrm{dom}(\evalCtx{}) \subseteq \mathcal{S}}{\begin{center}\gpfit{\mathrm{dom}(\evalCtx{}) \subseteq \mathcal{S}}\end{center}
}{\texttt{GLang.V2.Key\allowbreak Scope}}{GLang/ValuePreservationV2.lean:143}
\gldefdefn{legacyAgree}{\texttt{Legacy\allowbreak Agree}}{clauses}{\mathsf{LegacyAgree}(\mathcal{O}, \sigma{}, \glvEnv{}, \mathit{x}, \Gamma{}, \Delta{}, \mathit{x}, \ltynv{{\mathrm{x}}}, \mathit{x}, \mathit{x}, \mathit{x})}{\begin{center}\gpfit{\mathsf{LegacyAgree}(\mathcal{O}, \sigma{}, \glvEnv{}, \mathit{x}, \Gamma{}, \Delta{}, \ls{}, \lsv{}, \mathit{x}, \mathit{x}, \mathit{a}) = \lvert \mathit{x} \rvert = \dim_{\mathcal{O}}(\lsv{}) \;\wedge\; \lvert \mathit{a}\rvert = \dim_{\mathcal{O}}(\lsv{}) \;\wedge\; \mathit{x} = \sum_{\mathit{i} \in \mathsf{univ}} (\overline{\mathit{a}}[\mathit{i}]) \cdot (\mathit{x}[\mathit{i}])}\\[2pt]
\gpfit{\mathsf{LegacyAgree}(\mathcal{O}, \sigma{}, \glvEnv{}, \mathit{x}, \Gamma{}, \Delta{}, \mathit{a}, \lsnv{{\mathrm{a1}}}, \mathit{x}, \mathit{x}, {\mathit{a}_{2}}) = \mathsf{False}}\\[2pt]
\gpfit{\mathsf{LegacyAgree}(\mathcal{O}, \sigma{}, \glvEnv{}, \mathit{x}, \Gamma{}, \Delta{}, (x\colon \ltyn{0}\fnto \ltyn{1}), (x\colon \ltynv{0}\fnto \ltynv{1}), \mathit{x}, \mathit{x}, \mathit{x}) = \forall \mathit{x}',\,\mathit{y}'.\; \mathsf{LegacyAgree}(\mathcal{O}, \sigma{}, \glvEnv{}, \mathit{x}, \Gamma{}, \Delta{}, \ltyn{0}, \ltynv{0}, \mathsf{domainBasis}(\mathit{x}), \mathit{x}', \mathit{y}') \Rightarrow \exists \mathit{y}''.\; \mathit{x}(\mathit{y}') = \mathit{y}'' \;\wedge\; \mathsf{LegacyAgree}(\mathcal{O}, \sigma{}, \glvEnv{}, \mathit{x}' \mathbin{:\!:} \mathit{x}, \ltynv{0} \mathbin{:\!:} \Gamma{}, \mathit{y}' \mathbin{:\!:} \Delta{}, \ltyn{1}, \ltynv{1}, \mathit{x}(\mathit{x}'), \mathit{x}(\mathit{x}'), \mathit{y}'')}\\[2pt]
\gpfit{\mathsf{LegacyAgree}(\mathcal{O}, \sigma{}, \glvEnv{}, \mathit{x}, \Gamma{}, \Delta{}, \mathit{a}, (x\colon \ltynv{{\mathrm{a1}}}\fnto \ltynv{{\mathrm{a2}}}), \mathit{x}, \mathit{x}, \mathit{x}) = \mathsf{False}}\\[2pt]
\gpfit{\mathsf{LegacyAgree}(\mathcal{O}, \sigma{}, \glvEnv{}, \mathit{x}, \Gamma{}, \Delta{}, (x\colon \mathit{a}\fnto {\mathit{a}_{1}}), \lsnv{{\mathrm{a2}}}, \mathit{x}, \mathit{x}, \mathit{x}) = \mathsf{False}}\end{center}
}{\texttt{GLang.Legacy\allowbreak Agree}}{GLang/Adequacy/Agree.lean:105}
\gldefdefn{linearAtVar}{Linearity at a variable}{clauses}{\mathsf{LinearAtVar}(\sigma{}, \mathit{x}, \ls{}, \lee{}, \mathit{x})}{\begin{center}\gpfit{\mathsf{LinearAtVar}(\sigma{}, \mathit{x}, \ls{}, \lee{}, \mathit{s}) = \mathsf{IsLinearMap}(\mathbb{R}, \lambda \mathit{y}.\, \Sem{\lee{}}{\sigma{}[\mathit{x} \mapsto \mathit{y}]})}\\[2pt]
\gpfit{\mathsf{LinearAtVar}(\sigma{}, \mathit{x}, \ls{}, \lee{}, (x\colon \mathit{a}\fnto {\mathit{a}_{1}})) = \mathsf{True}}\end{center}
}{\texttt{GLang.Linear\allowbreak At\allowbreak Var}}{GLang/Soundness.lean:42}
\gldefdefn{operational.unchecked.opStep}{\texttt{Operational.Unchecked.Op\allowbreak Step}}{equation}{\mathsf{OpStep}}{\begin{center}\gpfit{\mathsf{OpStep}}\end{center}
}{\texttt{GLang.Operational.Unchecked.Op\allowbreak Step}}{GLang/Operational.lean:212}
\gldefdefn{oracleSpec}{\texttt{Oracle\allowbreak Spec}}{conjunction}{\mathsf{OracleSpec}(\mathcal{O})}{\begin{center}\gpfit{\mathsf{OracleSpec}(\mathcal{O}) \;\equiv\;}\\[2pt]
\gpfit{\quad \forall \glvEnv{},\,\Gamma{},\,\lDelta{},\,\lprop{},\,\lsv{},\,\lenv{{\mathrm{w}}}.\; \mathcal{O}\,\glvEnv{}\,\Gamma{}\,\lDelta{}\,\lprop{}\,\lsv{}\,\lenv{{\mathrm{w}}} \leq \mathsf{ffsDimUB}(\lprop{}, \shpOf{\lsv{}}{\mathcal{O}\,\glvEnv{}\,\Gamma{}\,\lDelta{}})}\\[2pt]
\gpfit{\wedge\; \forall \glvEnv{},\,\lprop{},\,\lsv{},\,\lenv{{\mathrm{w}}}.\; \operatorname{inferSpace}(\lenv{{\mathrm{w}}}) = \lsv{} \;\wedge\; \mathsf{Finite}(\mathsf{space}([], \lR^{\lenv{{\mathrm{w}}},\lprop{},\lsv{}})) \Rightarrow \mathcal{O}\,\glvEnv{}\,[]\,[]\,\lprop{}\,\lsv{}\,\lenv{{\mathrm{w}}} = \mathsf{ffsDim}(\lprop{}, \shpOf{\lsv{}}{\mathcal{O}\,\glvEnv{}\,[]\,[]}, \mathsf{ffsOperandRank}(\mathcal{O}, \glvEnv{}, [], [], \lsv{}, \lenv{{\mathrm{w}}}))}\\[2pt]
\gpfit{\wedge\; \forall \glvEnv{},\,\mathit{y},\,\leev{},\,\Gamma{},\,\lDelta{},\,\lprop{},\,\lsv{},\,\lenv{{\mathrm{w}}}.\; \forall i.\; \neg(\mathit{y} \in \mathsf{fv}(\ltynv{i})) \Rightarrow \mathcal{O}\,\mathsf{update}(\glvEnv{}, \mathit{y}, \leev{})\,\Gamma{}\,\lDelta{}\,\lprop{}\,\lsv{}\,\lenv{{\mathrm{w}}} = \mathcal{O}\,\glvEnv{}\,\Gamma{}\,\lDelta{}\,\lprop{}\,\lsv{}\,\lenv{{\mathrm{w}}}}\end{center}
}{\texttt{GLang.Oracle\allowbreak Spec}}{GLang/ValueDenotSem.lean:2102}
\gldefdefn{partialValueDomains.partialCoherent}{\texttt{Partial\allowbreak Value\allowbreak Domains.Partial\allowbreak Coherent}}{clauses}{\mathsf{PartialCoherent}(\ltynv{{\mathrm{x}}}, \mathit{x}, \mathit{x})}{\begin{center}\gpfit{\mathsf{PartialCoherent}(\ltynv{{\mathrm{x}}}, \glverr, \glverr) = \lee{} = \lee{}'}\\[2pt]
\gpfit{\mathsf{PartialCoherent}(\lsnv{{\mathrm{a}}}, \mathit{n}, \mathit{m}) = \mathit{n} = \mathit{m}}\\[2pt]
\gpfit{\mathsf{PartialCoherent}((x\colon \ltynv{{\mathrm{d}}}\fnto \ltynv{{\mathrm{c}}}), (\mathit{D}, \mathit{F}), (\mathit{E}, \mathit{G})) = \mathsf{PartialCoherent}(\ltynv{{\mathrm{d}}}, \mathit{D}, \mathit{E}) \;\wedge\; \forall \mathit{a},\,\lb{}.\; \mathsf{PartialRelated}(\ltynv{{\mathrm{d}}}, \mathit{D}, \mathit{E}, \mathit{a}, \lb{}) \Rightarrow \mathsf{PartialCoherent}(\ltynv{{\mathrm{c}}}, \mathit{F}\,\mathit{a}, \mathit{G}\,\lb{})}\\[2pt]
\gpfit{\mathsf{PartialCoherent}(\ltynv{{\mathrm{x}}}, \mathit{x}, \mathit{x}) = \mathsf{False}}\end{center}
}{\texttt{GLang.Partial\allowbreak Value\allowbreak Domains.Partial\allowbreak Coherent}}{GLang/PartialValueRelation.lean:20}
\gldefdefn{runtimeValue.partialEnvWF}{\texttt{Runtime\allowbreak Value.Partial\allowbreak Env\allowbreak WF}}{clauses}{\mathsf{PartialEnvWF}(\mathcal{O}, \valCheckCtx{}, \glvEnv{})}{\begin{center}\gpfit{\mathsf{PartialEnvWF}(\mathcal{O}, \valCheckCtx{}, \glvEnv{}) = \mathsf{ValCtxValid}(\valCheckCtx{}) \;\wedge\; \mathsf{PartialEnvironmentsRelated}(\mathcal{O}, \valCheckCtx{}, \glvEnv{}, \glvEnv{})}\end{center}
}{\texttt{GLang.Runtime\allowbreak Value.Partial\allowbreak Env\allowbreak WF}}{GLang/RuntimeValueCorrespondence.lean:11}
\gldefdefn{partialValueDomains.partialEnvWF}{\texttt{Partial\allowbreak Value\allowbreak Domains.Partial\allowbreak Env\allowbreak WF}}{clauses}{\mathsf{PartialEnvWF}(\mathcal{O}, \valCheckCtx{}, \glvEnv{})}{\begin{center}\gpfit{\mathsf{PartialEnvWF}(\mathcal{O}, \valCheckCtx{}, \glvEnv{}) = \mathsf{validValCtx}(\valCheckCtx{}) \;\wedge\; \mathsf{PartialEnvironmentsRelated}(\mathcal{O}, \valCheckCtx{}, \glvEnv{}, \glvEnv{})}\end{center}
}{\texttt{GLang.Partial\allowbreak Value\allowbreak Domains.Partial\allowbreak Env\allowbreak WF}}{GLang/PartialValueRuntimeRelation.lean:29}
\gldefdefn{partialValueDomains.partialEnvironmentsRelated}{\texttt{Partial\allowbreak Value\allowbreak Domains.Partial\allowbreak Environments\allowbreak Related}}{clauses}{\mathsf{PartialEnvironmentsRelated}(\mathcal{O}, \valCheckCtx{}, \glvEnv{}, \glvEnv{}')}{\begin{center}\gpfit{\mathsf{PartialEnvironmentsRelated}(\mathcal{O}, \valCheckCtx{}, \glvEnv{}, \glvEnv{}') = \forall \mathit{x},\,\ltyv{}.\; (\mathit{x}\colon \ltyv{}) \in \valCheckCtx{} \Rightarrow \mathsf{PartialRuntimeRelated}(\ltyv{}, \mathsf{denotValTyPartial}(\mathcal{O}, \glvEnv{}, [], [], \ltyv{}), \mathsf{denotValTyPartial}(\mathcal{O}, \glvEnv{}', [], [], \ltyv{}), \glvEnv{}\,\mathit{x}, \glvEnv{}'\,\mathit{x})}\end{center}
}{\texttt{GLang.Partial\allowbreak Value\allowbreak Domains.Partial\allowbreak Environments\allowbreak Related}}{GLang/PartialValueRuntimeRelation.lean:21}
\gldefdefn{paperValueDomains.partialRel}{\texttt{Paper\allowbreak Value\allowbreak Domains.Partial\allowbreak Rel}}{clauses}{\mathsf{PartialRel}(\mathit{A}, \mathit{B}, \mathit{R}, \mathit{x}, \mathit{x})}{\begin{center}\gpfit{\mathsf{PartialRel}(\mathit{R}, \glverr, \glverr) = \lee{} = \lee{}'}\\[2pt]
\gpfit{\mathsf{PartialRel}(\mathit{R}, \mathit{a}, \lb{}) = \mathit{R}\,\mathit{a}\,\lb{}}\\[2pt]
\gpfit{\mathsf{PartialRel}(\mathit{R}, \mathit{x}, \mathit{x}) = \mathsf{False}}\end{center}
}{\texttt{GLang.Paper\allowbreak Value\allowbreak Domains.Partial\allowbreak Rel}}{GLang/PaperValueExtensionality.lean:10}
\gldefdefn{partialValueDomains.partialRelated}{\texttt{Partial\allowbreak Value\allowbreak Domains.Partial\allowbreak Related}}{clauses}{\mathsf{PartialRelated}(\ltynv{{\mathrm{x}}}, \mathit{x}, \mathit{x}, \mathit{x}, \mathit{x})}{\begin{center}\gpfit{\mathsf{PartialRelated}(\lsnv{{\mathrm{a}}}, \mathit{n}, \mathit{m}, {\mathit{a}_{1}}, \lb{}) = {\mathit{a}_{1}} = \lb{} \;\wedge\; \lvert {\mathit{a}_{1}}\rvert = \mathit{n} \;\wedge\; \lvert \lb{}\rvert = \mathit{m}}\\[2pt]
\gpfit{\mathsf{PartialRelated}((x\colon \ltynv{{\mathrm{d}}}\fnto \ltynv{{\mathrm{c}}}), (\mathit{D}, \mathit{F}), (\mathit{E}, \mathit{G}), \mathit{f}, \mathit{g}) = \forall \mathit{a},\,\lb{}.\; \mathsf{PartialRelated}(\ltynv{{\mathrm{d}}}, \mathit{D}, \mathit{E}, \mathit{a}, \lb{}) \Rightarrow \mathsf{PartialRel}(\mathsf{PartialRelated}(\ltynv{{\mathrm{c}}}, \mathit{F}\,\mathit{a}, \mathit{G}\,\lb{}), \mathit{f}\,\mathit{a}, \mathit{g}\,\lb{})}\\[2pt]
\gpfit{\mathsf{PartialRelated}(\ltynv{{\mathrm{x}}}, \mathit{x}, \mathit{x}, \mathit{x}, \mathit{x}) = \mathsf{False}}\end{center}
}{\texttt{GLang.Partial\allowbreak Value\allowbreak Domains.Partial\allowbreak Related}}{GLang/PartialValueRelation.lean:10}
\gldefdefn{partialValueDomains.partialRuntimeRelated}{\texttt{Partial\allowbreak Value\allowbreak Domains.Partial\allowbreak Runtime\allowbreak Related}}{clauses}{\mathsf{PartialRuntimeRelated}(\ltynv{{\mathrm{x}}}, \mathit{x}, \mathit{x}, \mathit{x}, \mathit{x})}{\begin{center}\gpfit{\mathsf{PartialRuntimeRelated}(\lsv{}, \mathit{T}, \mathit{U}, \mathit{a}, \lb{}) = \mathsf{PartialRelated}(\lsv{}, \mathit{T}, \mathit{U}, \mathit{a}, \lb{})}\\[2pt]
\gpfit{\mathsf{PartialRuntimeRelated}((x\colon \ltynv{{\mathrm{d}}}\fnto \ltynv{{\mathrm{c}}}), (\mathit{D}, \mathit{F}), (\mathit{E}, \mathit{G}), \mathit{f}, \mathit{g}) = \exists \mathit{hA}.\; \exists \mathit{hB}.\; \forall \mathit{a},\,\lb{}.\; \mathsf{PartialRelated}(\ltynv{{\mathrm{d}}}, \mathit{D}, \mathit{E}, \mathsf{rec}(\mathit{a}, \mathit{hA}), \mathsf{rec}(\lb{}, \mathit{hB})) \Rightarrow \mathsf{PartialRel}(\mathsf{PartialRuntimeRelated}(\ltynv{{\mathrm{c}}}, \mathit{F}\,\mathsf{rec}(\mathit{a}, \mathit{hA}), \mathit{G}\,\mathsf{rec}(\lb{}, \mathit{hB})), \mathit{f}\,\mathit{a}, \mathit{g}\,\lb{})}\\[2pt]
\gpfit{\mathsf{PartialRuntimeRelated}(\ltynv{{\mathrm{x}}}, \mathit{x}, \mathit{x}, \mathit{x}, \mathit{x}) = \mathsf{False}}\end{center}
}{\texttt{GLang.Partial\allowbreak Value\allowbreak Domains.Partial\allowbreak Runtime\allowbreak Related}}{GLang/PartialValueRuntimeRelation.lean:10}
\gldefdefn{hType.pointwiseLinear}{Pointwise linearity through sections}{clauses}{\mathsf{PointwiseLinear}(\mathit{V}; \mathit{x}; \mathit{x})}{\begin{center}\gpfit{\mathsf{PointwiseLinear}(\mathit{V}; \mathit{W}; \mathit{f}) = \mathsf{IsLinearMap}(\mathbb{R}, \mathit{f})}\\[2pt]
\gpfit{\mathsf{PointwiseLinear}(\mathit{V}; \textstyle\prod\nolimits_{\mathit{A}} \mathit{fibers}; \mathit{f}) = \forall \mathit{a}.\; \mathsf{PointwiseLinear}(\mathit{V}; \mathit{fibers}\,\mathit{a}; \lambda \mathit{y}.\, \mathit{f}\,\mathit{y}\,\mathit{a})}\end{center}
}{\texttt{GLang.HType.Pointwise\allowbreak Linear}}{GLang/SourceSoundnessStrong.lean:25}
\gldefdefn{rawAdequacyOracleSpec}{\texttt{Raw\allowbreak Adequacy\allowbreak Oracle\allowbreak Spec}}{conjunction}{\mathsf{RawAdequacyOracleSpec}(\mathcal{O})}{\begin{center}\gpfit{\mathsf{RawAdequacyOracleSpec}(\mathcal{O}) \;\equiv\;}\\[2pt]
\gpfit{\quad \mathsf{OracleSpec}(\mathcal{O})}\\[2pt]
\gpfit{\wedge\; \mathsf{RawValueOpenPushOracleCoherent}(\mathcal{O})}\end{center}
}{\texttt{GLang.Raw\allowbreak Adequacy\allowbreak Oracle\allowbreak Spec}}{GLang/Adequacy/RawAgreement.lean:247}
\gldefdefn{rawAgree}{\texttt{Raw\allowbreak Agree}}{clauses}{\mathsf{RawAgree}(\mathcal{O}, \sigma{}, \glvEnv{}, \mathit{a}, \Gamma_{\mathrm{a}}, \mathit{a}, \mathit{a}, \ltynv{{\mathrm{a}}}, \mathit{a}, \mathit{a}, \mathit{a})}{\begin{center}\gpfit{\mathsf{RawAgree}(\mathcal{O}, \sigma{}, \glvEnv{}, \mathit{x}, \Gamma{}, \Delta{}, \ls{}, \lsv{}, \mathit{x}, \mathit{x}, \mathit{a}) = \mathsf{LegacyAgree}(\mathcal{O}, \sigma{}, \glvEnv{}, \mathit{x}, \Gamma{}, \Delta{}, \ls{}, \lsv{}, \mathit{x}, \mathit{x}, \mathit{a})}\\[2pt]
\gpfit{\mathsf{RawAgree}(\mathcal{O}, \sigma{}, \glvEnv{}, \mathit{x}, \Gamma{}, \Delta{}, \mathit{a}, \lsnv{{\mathrm{a1}}}, \mathit{x}, \mathit{x}, {\mathit{a}_{2}}) = \mathsf{False}}\\[2pt]
\gpfit{\mathsf{RawAgree}(\mathcal{O}, \sigma{}, \glvEnv{}, \mathit{x}, \Gamma{}, \Delta{}, (x\colon \ltyn{0}\fnto \ltyn{1}), (x\colon \ltynv{0}\fnto \ltynv{1}), \mathit{x}, \mathit{x}, \mathit{x}) = \exists \mathit{f}.\; \mathit{x} = \mathit{f} \;\wedge\; \forall \mathit{x}',\,\mathit{argVal},\,\mathit{arg}.\; \mathit{argVal} = \mathit{arg} \;\wedge\; \mathsf{RawAgree}(\mathcal{O}, \sigma{}, \glvEnv{}, \mathit{x}, \Gamma{}, \Delta{}, \ltyn{0}, \ltynv{0}, \mathsf{domainBasis}(\mathit{x}), \mathit{x}', \mathit{argVal}) \Rightarrow \exists \mathit{out}.\; \mathit{f}\,\mathit{arg} = \mathit{out} \;\wedge\; \mathsf{RawAgree}(\mathcal{O}, \sigma{}, \glvEnv{}, \mathit{x}' \mathbin{:\!:} \mathit{x}, \ltynv{0} \mathbin{:\!:} \Gamma{}, \mathit{argVal} \mathbin{:\!:} \Delta{}, \ltyn{1}, \ltynv{1}, \mathit{x}(\mathit{x}'), \mathit{x}(\mathit{x}'), \mathit{out})}\\[2pt]
\gpfit{\mathsf{RawAgree}(\mathcal{O}, \sigma{}, \glvEnv{}, \mathit{x}, \Gamma{}, \Delta{}, \mathit{a}, (x\colon \ltynv{{\mathrm{a1}}}\fnto \ltynv{{\mathrm{a2}}}), \mathit{x}, \mathit{x}, \mathit{x}) = \mathsf{False}}\\[2pt]
\gpfit{\mathsf{RawAgree}(\mathcal{O}, \sigma{}, \glvEnv{}, \mathit{x}, \Gamma{}, \Delta{}, (x\colon \mathit{a}\fnto {\mathit{a}_{1}}), \lsnv{{\mathrm{a2}}}, \mathit{x}, \mathit{x}, \mathit{x}) = \mathsf{False}}\end{center}
}{\texttt{GLang.Raw\allowbreak Agree}}{GLang/Adequacy/RawAgreement.lean:21}
\gldefdefn{rawAgreeDimsAt}{Raw dimension agreement}{clauses}{\SemT{\mathit{a}}{\sigma{}} \equiv_{\mathit{a}} \SemT{\ltynv{{\mathrm{a}}}}{\glvEnv{}}}{\begin{center}\gpfit{\SemT{\ls{}}{\sigma{}} \equiv_{\mathit{x}} \SemT{\lsv{}}{\glvEnv{}} = \shpOf{\lsv{}}{\mathcal{O}\,\glvEnv{}\,\Gamma{}\,\Delta{}} = \mathsf{srcShp}(\sigma{}, \mathit{x}, \ls{})}\\[2pt]
\gpfit{\SemT{(x\colon \ltyn{0}\fnto \ltyn{1})}{\sigma{}} \equiv_{\mathit{x}} \SemT{(x\colon \ltynv{0}\fnto \ltynv{1})}{\glvEnv{}} = \SemT{\ltyn{0}}{\sigma{}} \equiv_{\mathsf{domainBasis}(\mathit{x})} \SemT{\ltynv{0}}{\glvEnv{}} \;\wedge\; \forall \mathit{x}',\,\mathit{argVal},\,\mathit{arg}.\; \mathit{argVal} = \mathit{arg} \;\wedge\; \mathsf{RawAgree}(\mathcal{O}, \sigma{}, \glvEnv{}, \mathit{x}, \Gamma{}, \Delta{}, \ltyn{0}, \ltynv{0}, \mathsf{domainBasis}(\mathit{x}), \mathit{x}', \mathit{argVal}) \Rightarrow \SemT{\ltyn{1}}{\sigma{}} \equiv_{\mathit{x}(\mathit{x}')} \SemT{\ltynv{1}}{\glvEnv{}}}\\[2pt]
\gpfit{\SemT{\mathit{a}}{\sigma{}} \equiv_{\mathit{x}} \SemT{(x\colon \ltynv{{\mathrm{a1}}}\fnto \ltynv{{\mathrm{a2}}})}{\glvEnv{}} = \mathsf{False}}\\[2pt]
\gpfit{\SemT{(x\colon \mathit{a}\fnto {\mathit{a}_{1}})}{\sigma{}} \equiv_{\mathit{x}} \SemT{\lsnv{{\mathrm{a2}}}}{\glvEnv{}} = \mathsf{False}}\end{center}
}{\texttt{GLang.Raw\allowbreak Agree\allowbreak Dims\allowbreak At}}{GLang/Adequacy/RawAgreement.lean:92}
\gldefdefn{rawApplicationOracleSpec}{\texttt{Raw\allowbreak Application\allowbreak Oracle\allowbreak Spec}}{conjunction}{\mathsf{RawApplicationOracleSpec}(\mathcal{O})}{\begin{center}\gpfit{\mathsf{RawApplicationOracleSpec}(\mathcal{O}) \;\equiv\;}\\[2pt]
\gpfit{\quad \mathsf{RawAdequacyOracleSpec}(\mathcal{O})}\\[2pt]
\gpfit{\wedge\; \mathsf{CodomainLaw}(\mathcal{O})}\end{center}
}{\texttt{GLang.Raw\allowbreak Application\allowbreak Oracle\allowbreak Spec}}{GLang/Adequacy/RawApplicationOracleSpec.lean:6}
\gldefdefn{rawBasisAt}{\texttt{Raw\allowbreak Basis\allowbreak At}}{clauses}{\mathit{a} \;\text{is a basis for}\; \SemT{\mathit{a}}{\sigma{}}}{\begin{center}\gpfit{\mathit{x} \;\text{is a basis for}\; \SemT{\ls{}}{\sigma{}} = \mathit{x} \;\text{is an ordered basis of}\; \SemT{\ls{}}{\sigma{}}}\\[2pt]
\gpfit{\mathit{x} \;\text{is a basis for}\; \SemT{(x\colon \ltyn{0}\fnto \ltyn{1})}{\sigma{}} = \mathsf{domainBasis}(\mathit{x}) \;\text{is a basis for}\; \SemT{\ltyn{0}}{\sigma{}} \;\wedge\; \forall \mathit{x}'.\; \mathit{x}(\mathit{x}') \;\text{is a basis for}\; \SemT{\ltyn{1}}{\sigma{}}}\end{center}
}{\texttt{GLang.Raw\allowbreak Basis\allowbreak At}}{GLang/Adequacy/RawReification.lean:23}
\gldefdefn{rawConsistent}{Environment consistency (general oracle)}{rules}{\mathsf{RawConsistent}(\mathcal{O})}{\begin{center}\gpfit{\mathsf{RawConsistent}(\mathcal{O})}\end{center}
is defined inductively by
{\mprset{sep=0.3em}\scriptsize\begin{mathpar}
\gpfit{\inferrule*[Right=\textsc{empty}]{\phantom{.}}{(\lemptyset{}, \lemptyset{}); \lemptyset{}; \lemptyset{}\vdash \mathsf{mk}(\lemptyset{}), \glvEnv{}}}
\and
\gpfit{\inferrule*[Right=\textsc{extend\_\allowbreak lin}]{(\lDelta{}, \evalCtxSpace{}); \evalCtx{}; \valCheckCtx{}\vdash \sigma{}, \glvEnv{} \\ \neg(\mathit{x} \in \lDelta{}) \;\wedge\; \neg(\mathit{x} \in \evalCtxSpace{}) \\ \tof{\lb{}} = \ls{} \\ (\emptyset,\evalCtxSpace{})\vdash \lb{}\colon \bvalid{} \\ \evalSJudge{\evalCtx{}}{\tof{\lb{}}}{\ltyv{}} \\ \operatorname{inferType}(\leev{}) = \ltyv{} \\ \mathsf{InferSpaceMatchesBase}(\leev{}, \ltyv{}) \\ \mathsf{fv}(\leev{}) \subseteq \mathrm{dom}(\evalCtx{}[(\mathit{x},\lb{}) \mapsto \leev{}]) \\ \mathsf{lc}(\leev{}) \\ \neg(\mathit{x} \in \mathsf{fv}(\leev{})) \;\vee\; \exists \ltyv{}.\; \leev{} = \mathit{x}_{\ltyv{}} \\ \mathsf{shape}(\mathit{xs}) = \SemT{\tof{\lb{}}}{\sigma{}} \\ \mathcal{E}\llbracket \leev{}\rrbracket_{[];[]} = \mathit{yV} \\ \mathsf{RawAgree}(\mathcal{O}, \sigma{}, \glvEnv{}, [], [], [], \tof{\lb{}}, \ltyv{}, \SemBB{\lb{}}{\sigma{}}, \mathit{xs}, \mathit{yV}) \\ \SemT{\tof{\lb{}}}{\sigma{}} \equiv_{\SemBB{\lb{}}{\sigma{}}} \SemT{\ltyv{}}{\glvEnv{}}}{((\lDelta{}, \mathit{x}\colon \ls{}), \evalCtxSpace{}); \evalCtx{}[(\mathit{x},\lb{}) \mapsto \leev{}]; (\valCheckCtx{}, \mathit{x}\colon \ltyv{})\vdash \sigma{}[\mathit{x} \mapsto \mathit{xs}], \mathsf{update}(\glvEnv{}, \mathit{x}, \mathit{yV})}}
\and
\gpfit{\inferrule*[Right=\textsc{extend\_\allowbreak nl}]{(\lDelta{}, \evalCtxSpace{}); \evalCtx{}; \valCheckCtx{}\vdash \sigma{}, \glvEnv{} \\ \neg(\mathit{x} \in \lDelta{}) \;\wedge\; \neg(\mathit{x} \in \evalCtxSpace{}) \\ (\emptyset,\evalCtxSpace{})\vdash \lb{}\colon \bvalid{} \\ \evalSJudge{\evalCtx{}}{\tof{\lb{}}}{\ltyv{}} \\ \operatorname{inferType}(\leev{}) = \ltyv{} \\ \mathsf{InferSpaceMatchesBase}(\leev{}, \ltyv{}) \\ \mathsf{fv}(\leev{}) \subseteq \mathrm{dom}(\evalCtx{}[(\mathit{x},\lb{}) \mapsto \leev{}]) \\ \mathsf{lc}(\leev{}) \\ \neg(\mathit{x} \in \mathsf{fv}(\leev{})) \;\vee\; \exists \ltyv{}.\; \leev{} = \mathit{x}_{\ltyv{}} \\ \mathsf{shape}(\mathit{xs}) = \SemT{\tof{\lb{}}}{\sigma{}} \\ \mathcal{E}\llbracket \leev{}\rrbracket_{[];[]} = \mathit{yV} \\ \mathsf{RawAgree}(\mathcal{O}, \sigma{}, \glvEnv{}, [], [], [], \tof{\lb{}}, \ltyv{}, \SemBB{\lb{}}{\sigma{}}, \mathit{xs}, \mathit{yV}) \\ \SemT{\tof{\lb{}}}{\sigma{}} \equiv_{\SemBB{\lb{}}{\sigma{}}} \SemT{\ltyv{}}{\glvEnv{}} \\ \evalSJudge{\evalCtx{}}{\tof{\lb{}}}{\operatorname{inferType}(\leev{})} \\ \mathsf{fv}(\leev{}) \subseteq \mathrm{dom}(\evalCtx{}[(\mathit{x},\lb{}) \mapsto \leev{}]) \\ \mathsf{lc}(\leev{}) \\ \mathsf{FrameAnchor}(\evalCtx{}, \lb{}, \leev{})}{(\lDelta{}, (\evalCtxSpace{}, \mathit{x}\colon \tof{\lb{}})); \evalCtx{}[(\mathit{x},\lb{}) \mapsto \leev{}]; (\valCheckCtx{}, \mathit{x}\colon \ltyv{})\vdash \sigma{}[\mathit{x} \mapsto \mathit{xs}], \mathsf{update}(\glvEnv{}, \mathit{x}, \mathit{yV})}}
\and
\gpfit{\inferrule*[Right=\textsc{extend\_\allowbreak rep}]{(\lDelta{}, \evalCtxSpace{}); \evalCtx{}; \valCheckCtx{}\vdash \sigma{}, \glvEnv{} \\ \neg((\mathit{x}, \lb{}) \in \evalCtx{}) \\ \mathsf{SourceEntryLinked}(\lDelta{}, \evalCtxSpace{}, \mathit{x}, \lb{}) \\ (\mathit{x}\colon \mathit{xs}) \in \sigma{} \\ (\emptyset,\evalCtxSpace{})\vdash \lb{}\colon \bvalid{} \\ \evalSJudge{\evalCtx{}}{\tof{\lb{}}}{\ltyv{}} \\ \operatorname{inferType}(\leev{}) = \ltyv{} \\ \mathsf{InferSpaceMatchesBase}(\leev{}, \ltyv{}) \\ \mathsf{fv}(\leev{}) \subseteq \mathrm{dom}(\evalCtx{}[(\mathit{x},\lb{}) \mapsto \leev{}]) \\ \mathsf{lc}(\leev{}) \\ \mathsf{shape}(\mathit{xs}) = \SemT{\tof{\lb{}}}{\sigma{}} \\ \mathcal{E}\llbracket \leev{}\rrbracket_{[];[]} = \mathit{yV} \\ \mathsf{RawAgree}(\mathcal{O}, \sigma{}, \glvEnv{}, [], [], [], \tof{\lb{}}, \ltyv{}, \SemBB{\lb{}}{\sigma{}}, \mathit{xs}, \mathit{yV}) \\ \SemT{\tof{\lb{}}}{\sigma{}} \equiv_{\SemBB{\lb{}}{\sigma{}}} \SemT{\ltyv{}}{\glvEnv{}} \\ \evalSJudge{\evalCtx{}}{\tof{\lb{}}}{\operatorname{inferType}(\leev{})} \\ \mathsf{fv}(\leev{}) \subseteq \mathrm{dom}(\evalCtx{}[(\mathit{x},\lb{}) \mapsto \leev{}]) \\ \mathsf{lc}(\leev{}) \\ \mathsf{FrameAnchor}(\evalCtx{}, \lb{}, \leev{})}{(\lDelta{}, \evalCtxSpace{}); \evalCtx{}[(\mathit{x},\lb{}) \mapsto \leev{}]; \valCheckCtx{}\vdash \sigma{}, \glvEnv{}}}
\end{mathpar}}
}{\texttt{GLang.Raw\allowbreak Consistent}}{GLang/Adequacy/RawConsistent.lean:377}
\gldefdefn{rawReifiableAt}{\texttt{Raw\allowbreak Reifiable\allowbreak At}}{clauses}{\mathsf{RawReifiableAt}(\mathcal{O}, \sigma{}, \glvEnv{}, {\lDelta{}_{s}}, \Gamma_{\mathrm{Gv}}, {\lDelta{}_{v}}, \lty{}, \ltyv{}, \mathit{B})}{\begin{center}\gpfit{\mathsf{RawReifiableAt}(\mathcal{O}, \sigma{}, \glvEnv{}, {\lDelta{}_{s}}, \Gamma_{\mathrm{Gv}}, {\lDelta{}_{v}}, \lty{}, \ltyv{}, \mathit{B}) = \forall \mathit{x}'.\; \exists \mathit{raw}.\; \mathsf{RawAgree}(\mathcal{O}, \sigma{}, \glvEnv{}, {\lDelta{}_{s}}, \Gamma_{\mathrm{Gv}}, {\lDelta{}_{v}}, \lty{}, \ltyv{}, \mathit{B}, \mathit{x}', \mathit{raw})}\end{center}
}{\texttt{GLang.Raw\allowbreak Reifiable\allowbreak At}}{GLang/Adequacy/RawReification.lean:39}
\gldefdefn{rawResultAgree}{Raw result agreement}{clauses}{\mathit{x} \equiv^{\ltyv{}}_{\mathit{B}} \mathit{r}}{\begin{center}\gpfit{\mathit{x} \equiv^{\ltyv{}}_{\mathit{B}} \mathit{r} = \exists \mathit{y}.\; \mathit{r} = \mathit{y} \;\wedge\; \mathsf{RawAgree}(\mathcal{O}, \sigma{}, \glvEnv{}, {\lDelta{}_{s}}, \Gamma_{\mathrm{Gv}}, {\lDelta{}_{v}}, \lty{}, \ltyv{}, \mathit{B}, \mathit{x}, \mathit{y})}\end{center}
}{\texttt{GLang.Raw\allowbreak Result\allowbreak Agree}}{GLang/Adequacy/RawAgreement.lean:47}
\gldefdefn{rawSourceInjectiveAt}{\texttt{Raw\allowbreak Source\allowbreak Injective\allowbreak At}}{clauses}{\mathsf{RawSourceInjectiveAt}(\mathcal{O}, \sigma{}, \glvEnv{}, {\lDelta{}_{s}}, \Gamma_{\mathrm{Gv}}, {\lDelta{}_{v}}, \lty{}, \ltyv{}, \mathit{B})}{\begin{center}\gpfit{\mathsf{RawSourceInjectiveAt}(\mathcal{O}, \sigma{}, \glvEnv{}, {\lDelta{}_{s}}, \Gamma_{\mathrm{Gv}}, {\lDelta{}_{v}}, \lty{}, \ltyv{}, \mathit{B}) = \forall {\mathit{x}_{0}},\,{\mathit{x}_{1}},\,\mathit{raw}.\; \mathsf{RawAgree}(\mathcal{O}, \sigma{}, \glvEnv{}, {\lDelta{}_{s}}, \Gamma_{\mathrm{Gv}}, {\lDelta{}_{v}}, \lty{}, \ltyv{}, \mathit{B}, {\mathit{x}_{0}}, \mathit{raw}) \;\wedge\; \mathsf{RawAgree}(\mathcal{O}, \sigma{}, \glvEnv{}, {\lDelta{}_{s}}, \Gamma_{\mathrm{Gv}}, {\lDelta{}_{v}}, \lty{}, \ltyv{}, \mathit{B}, {\mathit{x}_{1}}, \mathit{raw}) \Rightarrow {\mathit{x}_{0}} = {\mathit{x}_{1}}}\end{center}
}{\texttt{GLang.Raw\allowbreak Source\allowbreak Injective\allowbreak At}}{GLang/Adequacy/RawReification.lean:48}
\gldefdefn{paperValueDomains.rawTypePresentation}{\texttt{Paper\allowbreak Value\allowbreak Domains.Raw\allowbreak Type\allowbreak Presentation}}{clauses}{\mathsf{RawTypePresentation}(\mathcal{O}, \ltynv{{\mathrm{x}}}, \ltynv{{\mathrm{x}}}, \mathit{x}, \mathit{x})}{\begin{center}\gpfit{\mathsf{RawTypePresentation}(\mathcal{O}, \lsv{}, \lsnv{{\mathrm{t}}}, \mathit{x}, \mathit{x}) = \dim_{\mathcal{O}}(\lsv{}) = \dim_{\mathcal{O}}(\lsnv{{\mathrm{t}}})}\\[2pt]
\gpfit{\mathsf{RawTypePresentation}(\mathcal{O}, (x\colon \ltynv{{\mathrm{d}}}\fnto \ltynv{{\mathrm{c}}}), (x\colon \ltynv{{\mathrm{d'}}}\fnto \ltynv{{\mathrm{c'}}}), \mathit{x}, \mathit{x}) = \mathsf{RawTypePresentation}(\mathcal{O}, \ltynv{{\mathrm{d}}}, \ltynv{{\mathrm{d'}}}, \mathit{x}, \mathit{x}) \;\wedge\; \forall \leev{}.\; \mathsf{RawTypePresentation}(\mathcal{O}, \ltynv{{\mathrm{c}}}, \ltynv{{\mathrm{c'}}}, \mathsf{rawPush}(\mathit{x}, \ltynv{{\mathrm{d}}}, \leev{}), \mathsf{rawPush}(\mathit{x}, \ltynv{{\mathrm{d'}}}, \leev{}))}\\[2pt]
\gpfit{\mathsf{RawTypePresentation}(\mathcal{O}, \ltynv{{\mathrm{x}}}, \ltynv{{\mathrm{x}}}, \mathit{x}, \mathit{x}) = \mathsf{False}}\end{center}
}{\texttt{GLang.Paper\allowbreak Value\allowbreak Domains.Raw\allowbreak Type\allowbreak Presentation}}{GLang/PaperValueRawPresentation.lean:12}
\gldefdefn{rawTypedConsistent}{\texttt{Raw\allowbreak Typed\allowbreak Consistent}}{conjunction}{\mathsf{RawTypedConsistent}(\mathcal{O}, \lDelta{}, \evalCtxSpace{}, \evalCtx{}, \valCheckCtx{}, \sigma{}, \glvEnv{})}{\begin{center}\gpfit{\mathsf{RawTypedConsistent}(\mathcal{O}, \lDelta{}, \evalCtxSpace{}, \evalCtx{}, \valCheckCtx{}, \sigma{}, \glvEnv{}) \;\equiv\;}\\[2pt]
\gpfit{\quad (\lDelta{}, \evalCtxSpace{}); \evalCtx{}; \valCheckCtx{}\vdash \sigma{}, \glvEnv{}}\\[2pt]
\gpfit{\wedge\; (\lDelta{},\evalCtxSpace{});\valCheckCtx{}\vdash \evalCtx{}}\end{center}
}{\texttt{GLang.Raw\allowbreak Typed\allowbreak Consistent}}{GLang/Adequacy/RawTypedConsistent.lean:11}
\gldefdefn{rawValueOpenPushOracleCoherent}{\texttt{Raw\allowbreak Value\allowbreak Open\allowbreak Push\allowbreak Oracle\allowbreak Coherent}}{clauses}{\mathsf{RawValueOpenPushOracleCoherent}(\mathcal{O})}{\begin{center}\gpfit{\mathsf{RawValueOpenPushOracleCoherent}(\mathcal{O}) = \forall \glvEnv{},\,\mathit{y},\,\ltynv{{\mathrm{sigmaV}}},\,\mathit{w},\,\Gamma_{\mathrm{G}},\,\mathit{D},\,\lprop{},\,\lsv{},\,\leev{}.\; \lvert \Gamma_{\mathrm{G}} \rvert = \lvert \mathit{D} \rvert \;\wedge\; \mathsf{lc}(\ltynv{{\mathrm{sigmaV}}}) \;\wedge\; \forall i.\; \neg(\mathit{y} \in \mathsf{fv}(\ltynv{i})) \Rightarrow \mathcal{O}\,\mathsf{update}(\glvEnv{}, \mathit{y}, \mathit{w})\,\mathsf{closePrefix}(\mathit{y}_{\ltynv{{\mathrm{sigmaV}}}}, \Gamma_{\mathrm{G}})\,\mathit{D}\,\lprop{}\,({\lsv{}}^{\lvert \Gamma_{\mathrm{G}} \rvert\leftarrow \mathit{y}_{\ltynv{{\mathrm{sigmaV}}}}})\,({\leev{}}^{\lvert \Gamma_{\mathrm{G}} \rvert\leftarrow \mathit{y}_{\ltynv{{\mathrm{sigmaV}}}}}) = \mathcal{O}\,\glvEnv{}\,\Gamma_{\mathrm{G}} \mathbin{+\!\!+} \ltynv{{\mathrm{sigmaV}}} \mathbin{:\!:} []\,\mathit{D} \mathbin{+\!\!+} \mathit{w} \mathbin{:\!:} []\,\lprop{}\,\lsv{}\,\leev{}}\end{center}
}{\texttt{GLang.Raw\allowbreak Value\allowbreak Open\allowbreak Push\allowbreak Oracle\allowbreak Coherent}}{GLang/Adequacy/RawAgreement.lean:205}
\gldefdefn{basisDenot.recursiveOrthonormal}{Recursive semantic orthonormality}{clauses}{\mathit{x} \;\text{is recursively orthonormal}}{\begin{center}\gpfit{\begin{gathered}\mathit{a} \;\text{is recursively orthonormal} = \forall \mathit{i},\,\mathit{j}.\; \langle\cdot,\cdot\rangle_{\mathit{V}}(\mathit{a}[\mathit{i}])(\mathit{a}[\mathit{j}]) = \begin{cases} 1 & \text{if } \mathit{i} = \mathit{j} \\ 0 & \text{otherwise} \end{cases}\end{gathered}}\\[2pt]
\gpfit{\mathsf{frame}(\mathit{T}, \mathit{domB}, \mathit{f}) \;\text{is recursively orthonormal} = \mathit{domB} \;\text{is recursively orthonormal} \;\wedge\; \forall \mathit{x}.\; \mathit{f}\,\mathit{x} \;\text{is recursively orthonormal}}\end{center}
}{\texttt{GLang.Basis\allowbreak Denot.Recursive\allowbreak Orthonormal}}{GLang/BasisSoundness.lean:318}
\gldefdefn{partialAdequacy.resultAgree}{\texttt{Partial\allowbreak Adequacy.Result\allowbreak Agree}}{clauses}{\mathsf{ResultAgree}(\mathcal{O}, \sigma{}, \glvEnv{}, {\lDelta{}_{s}}, \Gamma{}, \lDelta{}, \lty{}, \ltynv{{\mathrm{tv}}}, \mathit{B}, \mathit{x}, \mathit{r})}{\begin{center}\gpfit{\mathsf{ResultAgree}(\mathcal{O}, \sigma{}, \glvEnv{}, {\lDelta{}_{s}}, \Gamma{}, \lDelta{}, \lty{}, \ltynv{{\mathrm{tv}}}, \mathit{B}, \mathit{x}, \mathit{r}) = \exists \leev{}.\; \mathit{r} = \leev{} \;\wedge\; \mathsf{Agree}(\mathcal{O}, \sigma{}, \glvEnv{}, {\lDelta{}_{s}}, \Gamma{}, \lDelta{}, \lty{}, \ltynv{{\mathrm{tv}}}, \mathit{B}, \mathit{x}, \leev{})}\end{center}
}{\texttt{GLang.Partial\allowbreak Adequacy.Result\allowbreak Agree}}{GLang/Adequacy/PartialSourceAgreement.lean:35}
\gldefdefn{adequacy.partialDirect.resultAgree}{Result agreement}{clauses}{\mathsf{ResultAgree}}{\begin{center}\gpfit{\mathsf{ResultAgree} = \mathsf{ResultAgree}(\mathsf{oracle})}\end{center}
}{\texttt{GLang.Adequacy.Partial\allowbreak Direct.Result\allowbreak Agree}}{GLang/Adequacy/PartialSourcePublic.lean:9}
\gldefdefn{resultAgree}{Result agreement (general oracle)}{clauses}{\mathsf{ResultAgree}}{\begin{center}\gpfit{\mathsf{ResultAgree} = \mathsf{RawResultAgree}}\end{center}
}{\texttt{GLang.Result\allowbreak Agree}}{GLang/Adequacy/RawAgreement.lean:131}
\gldefdefn{adequacy.direct.resultAgree}{Result agreement}{equation}{\mathit{source} \equiv^{\ltyv{}}_{\mathit{B}} \mathit{result}}{\begin{center}\gpfit{\mathit{source} \equiv^{\ltyv{}}_{\mathit{B}} \mathit{result}}\end{center}
}{\texttt{GLang.Adequacy.Direct.Result\allowbreak Agree}}{GLang/Adequacy/DirectPublic.lean:48}
\gldefdefn{valueCanonicalFSV.resultMatrix}{\texttt{Value\allowbreak Canonical\allowbreak FSV.Result\allowbreak Matrix}}{conjunction}{\mathsf{ResultMatrix}}{\begin{center}\gpfit{\mathsf{ResultMatrix} \;\equiv\;}\\[2pt]
\gpfit{\quad \mathsf{Nat}}\\[2pt]
\gpfit{\wedge\; \mathsf{Nat}}\\[2pt]
\gpfit{\wedge\; \mathsf{Matrix}(\mathsf{Fin}(\mathit{rows}), \mathsf{Fin}(\mathit{cols}), \mathbb{R})}\end{center}
}{\texttt{GLang.Value\allowbreak Canonical\allowbreak FSV.Result\allowbreak Matrix}}{GLang/ValueCanonicalFSV.lean:17}
\gldefdefn{soundClaim}{Soundness claim}{clauses}{\lDelta{}\vdash \Sem{\lee{}}{\sigma{}} \in \SemT{\lty{}}{\sigma{}}}{\begin{center}\gpfit{\lDelta{}\vdash \Sem{\lee{}}{\sigma{}} \in \SemT{\lty{}}{\sigma{}} = \mathsf{shape}(\Sem{\lee{}}{\sigma{}}) = \SemT{\lty{}}{\sigma{}} \;\wedge\; \forall \mathit{x},\,\ls{}.\; (\mathit{x}\colon \ls{}) \in \lDelta{} \Rightarrow \mathsf{LinearAtVar}(\sigma{}, \mathit{x}, \ls{}, \lee{}, \lty{})}\end{center}
}{\texttt{GLang.Sound\allowbreak Claim}}{GLang/Soundness.lean:58}
\gldefdefn{metatheory.sourceContextWF}{Flat source-context conditions}{conjunction}{\mathsf{ContextConditions}(\lDelta{}, \lepsilon{})}{\begin{center}\gpfit{\mathsf{ContextConditions}(\lDelta{}, \lepsilon{}) \;\Longleftrightarrow\; \bigl(\forall \mathit{x}.\; \mathit{x} \in \lDelta{} \Rightarrow \neg(\mathit{x} \in \lepsilon{})\bigr)}\\[2pt]
\gpfit{\wedge\; \bigl(\forall \mathit{x},\,\ls{},\,\mathit{y}.\; (\mathit{x}\colon \ls{}) \in \lDelta{} \;\wedge\; \mathit{y} \in \mathsf{fv}(\ls{}) \Rightarrow \mathit{y} \in \lepsilon{}\bigr)}\\[2pt]
\gpfit{\wedge\; \bigl(\forall \mathit{x},\,\lty{},\,\mathit{y}.\; (\mathit{x}\colon \lty{}) \in \lepsilon{} \;\wedge\; \mathit{y} \in \mathsf{fv}(\lty{}) \Rightarrow \mathit{y} \in \lepsilon{}\bigr)}\\[2pt]
\gpfit{\wedge\; \bigl(\forall \mathit{x},\,\ls{}.\; (\mathit{x}\colon \ls{}) \in \lDelta{} \Rightarrow (\lemptyset, \lepsilon{})\vdash \ls{}\colon\svalid\bigr)}\\[2pt]
\gpfit{\wedge\; \bigl(\forall \mathit{x},\,\lty{}.\; (\mathit{x}\colon \lty{}) \in \lepsilon{} \Rightarrow \exists \lky{}.\; (\lemptyset, \lepsilon{})\vdash \lty{}\colon \lky{}\bigr)}\end{center}
}{\texttt{GLang.Metatheory.Source\allowbreak Context\allowbreak WF}}{GLang/SourceContextInductive.lean:17}
\gldefdefn{sourceCtxValid}{Source-context formation validity}{conjunction}{\mathsf{SourceCtxValid}(\lDelta{}, \evalCtxSpace{})}{\begin{center}\gpfit{\mathsf{SourceCtxValid}(\lDelta{}, \evalCtxSpace{}) \;\equiv\;}\\[2pt]
\gpfit{\quad \forall \mathit{x},\,\ls{}.\; (\mathit{x}\colon \ls{}) \in \lDelta{} \Rightarrow (\lemptyset, \evalCtxSpace{})\vdash \ls{}\colon\svalid}\\[2pt]
\gpfit{\wedge\; \forall \mathit{x},\,\lty{}.\; (\mathit{x}\colon \lty{}) \in \evalCtxSpace{} \Rightarrow \Sigma \lky{}.\; (\lemptyset, \evalCtxSpace{})\vdash \lty{}\colon \lky{}}\end{center}
}{\texttt{GLang.Source\allowbreak Ctx\allowbreak Valid}}{GLang/SourceSoundnessStrong.lean:54}
\gldefdefn{v2.sourceEntryLinked}{Existing source-entry link}{clauses}{\mathsf{SourceEntryLinked}(\lDelta{}, \evalCtxSpace{}, \mathit{x}, \lb{})}{\begin{center}\gpfit{\mathsf{SourceEntryLinked}(\lDelta{}, \evalCtxSpace{}, \mathit{x}, \lb{}) = \exists \ls{}.\; (\mathit{x}\colon \ls{}) \in \lDelta{} \;\wedge\; \tof{\lb{}} = \ls{} \;\vee\; (\mathit{x}\colon \tof{\lb{}}) \in \evalCtxSpace{}}\end{center}
}{\texttt{GLang.V2.Source\allowbreak Entry\allowbreak Linked}}{GLang/ValueProgress.lean:325}
\gldefdefn{v2.sourceEntryValid}{Source-entry validity}{conjunction}{\mathsf{SourceEntry}(\lDelta{}, \evalCtxSpace{}, \mathit{x})}{\begin{center}\gpfit{\mathsf{SourceEntry}(\lDelta{}, \evalCtxSpace{}, \mathit{x}) \;\equiv\;}\\[2pt]
\gpfit{\quad \neg(\mathit{x} \in \lDelta{})}\\[2pt]
\gpfit{\wedge\; \neg(\mathit{x} \in \evalCtxSpace{})}\end{center}
}{\texttt{GLang.V2.Source\allowbreak Entry\allowbreak Valid}}{GLang/ValueProgress.lean:250}
\gldefdefn{operational.space}{\texttt{Operational.Space}}{equation}{\mathsf{Space}}{\begin{center}\gpfit{\mathsf{Space}}\end{center}
}{\texttt{GLang.Operational.Space}}{GLang/Operational.lean:29}
\gldefdefn{v2.stdAnchored}{\texttt{Std\allowbreak Anchored}}{rules}{\mathsf{StdAnchored}}{\begin{center}\gpfit{\mathsf{StdAnchored}}\end{center}
is defined inductively by
{\mprset{sep=0.3em}\scriptsize\begin{mathpar}
\gpfit{\inferrule*[Right=\textsc{space}]{\evalSJudge{\evalCtx{}}{\ls{}}{\ltyv{}}}{\mathsf{StdAnchored}(\evalCtx{}, \ls{}, \ltyv{})}}
\and
\gpfit{\inferrule*[Right=\textsc{pi\allowbreak Ty}]{\mathsf{StdAnchored}(\evalCtx{}, \ltyn{0}, \ltynv{0}) \\ \forall \mathit{y}.\; \mathsf{StdAnchored}(\evalCtx{}[(\mathit{y},\stdBasisDCtx{\ltyn{0}}) \mapsto \mathit{y}_{\ltynv{0}}], \ltyn{1}, \ltynv{1})}{\mathsf{StdAnchored}(\evalCtx{}, (x\colon \ltyn{0}\fnto \ltyn{1}), (x\colon \ltynv{0}\fnto \ltynv{1}))}}
\end{mathpar}}
}{\texttt{GLang.V2.Std\allowbreak Anchored}}{GLang/ValueProgress.lean:86}
\gldefdefn{v2.stdComplete}{Standard-basis completeness}{clauses}{\mathsf{StdComplete}(\evalCtx{})}{\begin{center}\gpfit{\mathsf{StdComplete}(\evalCtx{}) = \forall \mathit{x},\,\lb{}.\; (\mathit{x}, \lb{}) \in \evalCtx{} \Rightarrow \exists \leev{}.\; \evalJudge{\evalCtx{}}{\mathit{x}}{\stdBasisDCtx{\tof{\lb{}}}}{\leev{}} \;\wedge\; \forall \mathit{x},\,\lb{},\,\ltyn{0},\,\ltyn{1}.\; (\mathit{x}, \lb{}) \in \evalCtx{} \;\wedge\; \tof{\lb{}} = (x\colon \ltyn{0}\fnto \ltyn{1}) \Rightarrow \mathsf{StdPiAnchored}(\evalCtx{}, \ltyn{0}, \ltyn{1})}\end{center}
}{\texttt{GLang.V2.Std\allowbreak Complete}}{GLang/ValueProgress.lean:233}
\gldefdefn{v2.stdPiAnchored}{\texttt{Std\allowbreak Pi\allowbreak Anchored}}{clauses}{\mathsf{StdPiAnchored}(\evalCtx{}, \ltyn{0}, \ltyn{1})}{\begin{center}\gpfit{\mathsf{StdPiAnchored}(\evalCtx{}, \ltyn{0}, \ltyn{1}) = \exists \ltyv{}.\; \mathsf{StdAnchored}(\evalCtx{}, (x\colon \ltyn{0}\fnto \ltyn{1}), \ltyv{})}\end{center}
}{\texttt{GLang.V2.Std\allowbreak Pi\allowbreak Anchored}}{GLang/ValueProgress.lean:127}
\gldefdefn{strongSourceSoundProp}{Strengthened source-soundness claim}{clauses}{\mathsf{StrongSourceSound}(\sigma{}; \lDelta{}; \evalCtxSpace{}; \lee{}; \lty{})}{\begin{center}\gpfit{\mathsf{StrongSourceSound}(\sigma{}; \lDelta{}; \evalCtxSpace{}; \lee{}; \lty{}) = \exists \lky{}.\; (\lemptyset, \evalCtxSpace{})\vdash \lty{}\colon \lky{} \;\wedge\; \mathsf{shape}(\Sem{\lee{}}{\sigma{}}) = \SemT{\lty{}}{\sigma{}} \;\wedge\; \forall \mathit{x},\,\ls{}.\; (\mathit{x}\colon \ls{}) \in \lDelta{} \Rightarrow \mathsf{HOLinear}(\sigma{}; \mathit{x}; \ls{}; \lee{}; \lty{})}\end{center}
}{\texttt{GLang.Strong\allowbreak Source\allowbreak Sound\allowbreak Prop}}{GLang/SourceSoundnessStrong.lean:124}
\gldefdefn{operational.term}{\texttt{Operational.Term}}{equation}{\mathsf{Term}}{\begin{center}\gpfit{\mathsf{Term}}\end{center}
}{\texttt{GLang.Operational.Term}}{GLang/Operational.lean:41}
\gldefdefn{operational.ty}{\texttt{Operational.Ty}}{equation}{\mathsf{Ty}}{\begin{center}\gpfit{\mathsf{Ty}}\end{center}
}{\texttt{GLang.Operational.Ty}}{GLang/Operational.lean:32}
\gldefdefn{operational.tyAt}{\texttt{Operational.Ty\allowbreak At}}{equation}{\mathsf{TyAt}}{\begin{center}\gpfit{\mathsf{TyAt}}\end{center}
}{\texttt{GLang.Operational.Ty\allowbreak At}}{GLang/Operational.lean:35}
\gldefdefn{operational.typedContext}{\texttt{Operational.Typed\allowbreak Context}}{rules}{\mathsf{TypedContext}}{\begin{center}\gpfit{\mathsf{TypedContext}}\end{center}
is defined inductively by
{\mprset{sep=0.3em}\scriptsize\begin{mathpar}
\gpfit{\inferrule*[Right=\textsc{empty}]{\phantom{.}}{(\lemptyset{},\lemptyset{});\lemptyset{}\vdash \lemptyset{}}}
\and
\gpfit{\inferrule*[Right=\textsc{extend\_\allowbreak nlin}]{(\lDelta{},\evalCtxSpace{});\valCheckCtx{}\vdash \evalCtx{} \\ \neg(\mathit{x} \in \lDelta{}) \\ \neg(\mathit{x} \in \evalCtxSpace{}) \\ \evalSJudge{\evalCtx{}}{\tof{\lb{}}}{\operatorname{inferType}(\leev{})} \\ \mathsf{fv}(\leev{}) \subseteq \mathrm{dom}(\evalCtx{}[(\mathit{x},\lb{}) \mapsto \leev{}]) \\ \mathsf{lc}(\leev{}) \\ \mathsf{FrameAnchor}(\evalCtx{}, \lb{}, \leev{}) \\ (\emptyset,\evalCtxSpace{})\vdash \lb{}\colon \bvalid{} \\ \valCheckCtx{}\vdash \operatorname{inferType}(\leev{})\colon \lky{} \\ (\valCheckCtx{}, \mathit{x}\colon \operatorname{inferType}(\leev{}))\vdash \leev{}\colon \operatorname{inferType}(\leev{})}{(\lDelta{},(\evalCtxSpace{}, \mathit{x}\colon \tof{\lb{}}));(\valCheckCtx{}, \mathit{x}\colon \operatorname{inferType}(\leev{}))\vdash \evalCtx{}[(\mathit{x},\lb{}) \mapsto \leev{}]}}
\and
\gpfit{\inferrule*[Right=\textsc{extend\_\allowbreak lin}]{(\lDelta{},\evalCtxSpace{});\valCheckCtx{}\vdash \evalCtx{} \\ \neg(\mathit{x} \in \lDelta{}) \\ \neg(\mathit{x} \in \evalCtxSpace{}) \\ \evalSJudge{\evalCtx{}}{\tof{\lb{}}}{\operatorname{inferType}(\leev{})} \\ \mathsf{fv}(\leev{}) \subseteq \mathrm{dom}(\evalCtx{}[(\mathit{x},\lb{}) \mapsto \leev{}]) \\ \mathsf{lc}(\leev{}) \\ \mathsf{FrameAnchor}(\evalCtx{}, \lb{}, \leev{}) \\ (\emptyset,\evalCtxSpace{})\vdash \lb{}\colon \bvalid{} \\ \tof{\lb{}} = \ls{} \\ \valCheckCtx{}\vdash \operatorname{inferType}(\leev{})\colon \lky{} \\ (\valCheckCtx{}, \mathit{x}\colon \operatorname{inferType}(\leev{}))\vdash \leev{}\colon \operatorname{inferType}(\leev{})}{((\lDelta{}, \mathit{x}\colon \ls{}),\evalCtxSpace{});(\valCheckCtx{}, \mathit{x}\colon \operatorname{inferType}(\leev{}))\vdash \evalCtx{}[(\mathit{x},\lb{}) \mapsto \leev{}]}}
\and
\gpfit{\inferrule*[Right=\textsc{extend\_\allowbreak rep}]{(\lDelta{},\evalCtxSpace{});\valCheckCtx{}\vdash \evalCtx{} \\ \mathsf{SourceEntryLinked}(\lDelta{}, \evalCtxSpace{}, \mathit{x}, \lb{}) \\ \neg((\mathit{x}, \lb{}) \in \evalCtx{}) \\ \evalSJudge{\evalCtx{}}{\tof{\lb{}}}{\operatorname{inferType}(\leev{})} \\ \mathsf{fv}(\leev{}) \subseteq \mathrm{dom}(\evalCtx{}[(\mathit{x},\lb{}) \mapsto \leev{}]) \\ \mathsf{lc}(\leev{}) \\ \mathsf{FrameAnchor}(\evalCtx{}, \lb{}, \leev{}) \\ (\emptyset,\evalCtxSpace{})\vdash \lb{}\colon \bvalid{} \\ \valCheckCtx{}\vdash \leev{}\colon \operatorname{inferType}(\leev{})}{(\lDelta{},\evalCtxSpace{});\valCheckCtx{}\vdash \evalCtx{}[(\mathit{x},\lb{}) \mapsto \leev{}]}}
\end{mathpar}}
}{\texttt{GLang.Operational.Typed\allowbreak Context}}{GLang/OperationalTypedContext.lean:11}
\gldefdefn{v2.typingImplyEval}{\texttt{Typing\allowbreak Imply\allowbreak Eval}}{clauses}{\mathsf{TypingImplyEval}(\lDelta{}, \evalCtxSpace{}, \evalCtx{})}{\begin{center}\gpfit{\mathsf{TypingImplyEval}(\lDelta{}, \evalCtxSpace{}, \evalCtx{}) = \forall \mathit{x}.\; \mathit{x} \in \lDelta{} \;\vee\; \mathit{x} \in \evalCtxSpace{} \iff \exists \lb{}.\; (\mathit{x}, \lb{}) \in \evalCtx{} \;\wedge\; \forall \mathit{x},\,\ls{},\,\lb{}.\; (\mathit{x}\colon \ls{}) \in \lDelta{} \;\wedge\; (\mathit{x}, \lb{}) \in \evalCtx{} \Rightarrow \tof{\lb{}} = \ls{} \;\wedge\; \forall \mathit{x},\,\lty{},\,\lb{}.\; (\mathit{x}\colon \lty{}) \in \evalCtxSpace{} \;\wedge\; (\mathit{x}, \lb{}) \in \evalCtx{} \Rightarrow \tof{\lb{}} = \lty{} \;\wedge\; \forall \mathit{x},\,\lb{}.\; (\mathit{x}, \lb{}) \in \evalCtx{} \Rightarrow (\emptyset,\evalCtxSpace{})\vdash \lb{}\colon \bvalid{}}\end{center}
}{\texttt{GLang.V2.Typing\allowbreak Imply\allowbreak Eval}}{GLang/ValuePreservationV2.lean:129}
\gldefdefn{valCanonicalResultSoundOfDenotSemS}{Canonical value-result soundness}{clauses}{\glvEnv{}; \Gamma{}; \lDelta{}\vdash_{\mathsf{canonical}} \mathit{r}\colon \ltyv{}}{\begin{center}\gpfit{\glvEnv{}; \Gamma{}; \lDelta{}\vdash_{\mathsf{canonical}} \mathit{r}\colon \ltyv{} = \forall \mathit{out}.\; \mathit{r} = \mathit{out} \Rightarrow \glvEnv{}; \Gamma{}; \lDelta{}\vdash_{\mathsf{canonical}} \mathit{out}\colon \ltyv{}}\end{center}
}{\texttt{GLang.Val\allowbreak Canonical\allowbreak Result\allowbreak Sound\allowbreak Of\allowbreak Denot\allowbreak Sem\allowbreak S}}{GLang/Adequacy/ValueSoundnessSemS.lean:70}
\gldefdefn{valCanonicalSoundOfDenotSemS}{Canonical value soundness}{clauses}{\glvEnv{}; \Gamma{}; \Delta{}\vdash_{\mathsf{canonical}} \mathit{x}\colon \ltynv{{\mathrm{x}}}}{\begin{center}\gpfit{\glvEnv{}; \Gamma{}; \Delta{}\vdash_{\mathsf{canonical}} \mathit{a}\colon \lsv{} = \lvert \mathit{a}\rvert = \dim_{\mathcal{O}}(\lsv{})}\\[2pt]
\gpfit{\glvEnv{}; \Gamma{}; \Delta{}\vdash_{\mathsf{canonical}} {\mathit{a}_{1}}\colon \lsnv{{\mathrm{a}}} = \mathsf{False}}\\[2pt]
\gpfit{\glvEnv{}; \Gamma{}; \Delta{}\vdash_{\mathsf{canonical}} {\mathit{a}_{2}}\colon (x\colon \ltynv{{\mathrm{a}}}\fnto \ltynv{{\mathrm{a1}}}) = \mathsf{False}}\\[2pt]
\gpfit{\glvEnv{}; \Gamma{}; \Delta{}\vdash_{\mathsf{canonical}} \mathit{fam}\colon (x\colon \ltynv{0}\fnto \ltynv{1}) = \exists \mathit{h}.\; \forall \leev{},\,\mathit{out}.\; \mathit{fam}\,\leev{} = \mathit{out} \Rightarrow \glvEnv{}; \ltynv{0} \mathbin{:\!:} \Gamma{}; \mathsf{rec}(\leev{}, \mathit{h}) \mathbin{:\!:} \Delta{}\vdash_{\mathsf{canonical}} \mathit{out}\colon \ltynv{1}}\end{center}
}{\texttt{GLang.Val\allowbreak Canonical\allowbreak Sound\allowbreak Of\allowbreak Denot\allowbreak Sem\allowbreak S}}{GLang/Adequacy/ValueSoundnessSemS.lean:49}
\gldefdefn{valCtxValid}{Value-context validity}{clauses}{\mathsf{validValCtx}(\valCheckCtx{})}{\begin{center}\gpfit{\mathsf{validValCtx}(\valCheckCtx{}) = \forall \mathit{x},\,\ltyv{}.\; (\mathit{x}\colon \ltyv{}) \in \valCheckCtx{} \Rightarrow \Sigma \lky{}.\; \valCheckCtx{}\vdash \ltyv{}\colon \lky{}}\end{center}
}{\texttt{GLang.Val\allowbreak Ctx\allowbreak Valid}}{GLang/ValueSoundVals.lean:85}
\gldefdefn{valResultSoundOfDenotSemS}{Value-result soundness}{clauses}{\glvEnv{}; \Gamma{}; \lDelta{}\vdash \mathit{r}\colon \ltyv{}}{\begin{center}\gpfit{\glvEnv{}; \Gamma{}; \lDelta{}\vdash \mathit{r}\colon \ltyv{} = \forall \mathit{out}.\; \mathit{r} = \mathit{out} \Rightarrow \glvEnv{}; \Gamma{}; \lDelta{}\vdash \mathit{out}\colon \ltyv{}}\end{center}
}{\texttt{GLang.Val\allowbreak Result\allowbreak Sound\allowbreak Of\allowbreak Denot\allowbreak Sem\allowbreak S}}{GLang/Adequacy/ValueSoundnessSemS.lean:64}
\gldefdefn{valSoundOfDenotSemS}{Value soundness}{clauses}{\glvEnv{}; \Gamma{}; \Delta{}\vdash \mathit{x}\colon \ltynv{{\mathrm{x}}}}{\begin{center}\gpfit{\glvEnv{}; \Gamma{}; \Delta{}\vdash \mathit{a}\colon \lsv{} = \forall \mathit{i}.\; \neg(\mathit{i} \in \operatorname{idx}(\lsv{})) \Rightarrow \overline{\mathit{a}}[\mathit{i}] = 0}\\[2pt]
\gpfit{\glvEnv{}; \Gamma{}; \Delta{}\vdash {\mathit{a}_{1}}\colon \lsnv{{\mathrm{a}}} = \mathsf{False}}\\[2pt]
\gpfit{\glvEnv{}; \Gamma{}; \Delta{}\vdash {\mathit{a}_{2}}\colon (x\colon \ltynv{{\mathrm{a}}}\fnto \ltynv{{\mathrm{a1}}}) = \mathsf{False}}\\[2pt]
\gpfit{\glvEnv{}; \Gamma{}; \Delta{}\vdash \mathit{fam}\colon (x\colon \ltynv{0}\fnto \ltynv{1}) = \exists \mathit{h}.\; \forall \leev{},\,\mathit{out}.\; \mathit{fam}\,\leev{} = \mathit{out} \Rightarrow \glvEnv{}; \ltynv{0} \mathbin{:\!:} \Gamma{}; \mathsf{rec}(\leev{}, \mathit{h}) \mathbin{:\!:} \Delta{}\vdash \mathit{out}\colon \ltynv{1}}\end{center}
}{\texttt{GLang.Val\allowbreak Sound\allowbreak Of\allowbreak Denot\allowbreak Sem\allowbreak S}}{GLang/Adequacy/ValueSoundnessSemS.lean:33}
\gldefdefn{valTyOk}{Value-type regularity}{clauses}{\vdash \ltyv{}\colon \tvalid{}}{\begin{center}\gpfit{\vdash \ltyv{}\colon \tvalid{} = \exists \lky{}.\; \vdash \ltyv{}\colon \lky{}}\end{center}
}{\texttt{GLang.Val\allowbreak Ty\allowbreak Ok}}{GLang/ValueValidC.lean:52}
\gldefdefn{paperValueDomains.world}{\texttt{Paper\allowbreak Value\allowbreak Domains.World}}{conjunction}{\mathsf{World}}{\begin{center}\gpfit{\mathsf{World} \;\equiv\;}\\[2pt]
\gpfit{\quad \mathsf{ValEnvSem}}\\[2pt]
\gpfit{\wedge\; \mathsf{List}(\mathsf{ValTy})}\\[2pt]
\gpfit{\wedge\; \mathsf{List}(\mathsf{ValValSem})}\end{center}
}{\texttt{GLang.Paper\allowbreak Value\allowbreak Domains.World}}{GLang/PaperValueExtensionality.lean:34}
\gldefdefn{partialValueDomains.worldRuntimeRelated}{\texttt{Partial\allowbreak Value\allowbreak Domains.World\allowbreak Runtime\allowbreak Related}}{clauses}{\mathsf{WorldRuntimeRelated}(\mathcal{O}, \ltyv{}, \mathit{w}, \mathit{w}')}{\begin{center}\gpfit{\mathsf{WorldRuntimeRelated}(\mathcal{O}, \ltyv{}, \mathit{w}, \mathit{w}') = \mathsf{PartialRuntimeRelated}(\ltyv{}, \mathsf{PartialWorldType}(\mathcal{O}, \mathit{w}, \ltyv{}), \mathsf{PartialWorldType}(\mathcal{O}, \mathit{w}', \ltyv{}))}\end{center}
}{\texttt{GLang.Partial\allowbreak Value\allowbreak Domains.World\allowbreak Runtime\allowbreak Related}}{GLang/PartialValueFamilyBinders.lean:9}
\gldefdefn{space.closeRec}{\texttt{Space.close\allowbreak Rec}}{clauses}{({\mathit{x}}^{\mathit{x}\rightarrow \lky{}})}{\begin{center}\gpfit{\mathsf{closeRec}(\lky{}, \mathit{x}, \lsempty{}) = \lsempty{}}\\[2pt]
\gpfit{\mathsf{closeRec}(\lky{}, \mathit{x}, \lR{}) = \lR{}}\\[2pt]
\gpfit{\mathsf{closeRec}(\lky{}, \mathit{x}, \mathsf{dual}(\ls{})) = \mathsf{dual}(\mathsf{closeRec}(\lky{}, \mathit{x}, \ls{}))}\\[2pt]
\gpfit{\mathsf{closeRec}(\lky{}, \mathit{x}, \mathsf{dsum}(\lsn{0}, \lsn{1})) = \mathsf{dsum}(\mathsf{closeRec}(\lky{}, \mathit{x}, \lsn{0}), \mathsf{closeRec}(\lky{}, \mathit{x}, \lsn{1}))}\\[2pt]
\gpfit{\mathsf{closeRec}(\lky{}, \mathit{x}, \mathsf{tprod}(\lsn{0}, \lsn{1})) = \mathsf{tprod}(\mathsf{closeRec}(\lky{}, \mathit{x}, \lsn{0}), \mathsf{closeRec}(\lky{}, \mathit{x}, \lsn{1}))}\\[2pt]
\gpfit{\mathsf{closeRec}(\lky{}, \mathit{x}, \mathsf{ker}(\lsn{0}, \lsn{1}, \lee{})) = \mathsf{ker}(\mathsf{closeRec}(\lky{}, \mathit{x}, \lsn{0}), \mathsf{closeRec}(\lky{}, \mathit{x}, \lsn{1}), \mathsf{closeRec}(\lky{}, \mathit{x}, \lee{}))}\\[2pt]
\gpfit{\mathsf{closeRec}(\lky{}, \mathit{x}, \mathsf{coker}(\lsn{0}, \lsn{1}, \lee{})) = \mathsf{coker}(\mathsf{closeRec}(\lky{}, \mathit{x}, \lsn{0}), \mathsf{closeRec}(\lky{}, \mathit{x}, \lsn{1}), \mathsf{closeRec}(\lky{}, \mathit{x}, \lee{}))}\\[2pt]
\gpfit{\mathsf{closeRec}(\lky{}, \mathit{x}, \mathsf{img}(\lsn{0}, \lsn{1}, \lee{})) = \mathsf{img}(\mathsf{closeRec}(\lky{}, \mathit{x}, \lsn{0}), \mathsf{closeRec}(\lky{}, \mathit{x}, \lsn{1}), \mathsf{closeRec}(\lky{}, \mathit{x}, \lee{}))}\\[2pt]
\gpfit{\mathsf{closeRec}(\lky{}, \mathit{x}, \mathsf{coimg}(\lsn{0}, \lsn{1}, \lee{})) = \mathsf{coimg}(\mathsf{closeRec}(\lky{}, \mathit{x}, \lsn{0}), \mathsf{closeRec}(\lky{}, \mathit{x}, \lsn{1}), \mathsf{closeRec}(\lky{}, \mathit{x}, \lee{}))}\end{center}
}{\texttt{GLang.Space.close\allowbreak Rec}}{GLang/SourceLN.lean:105}
\gldefdefn{ty.closeRec}{\texttt{Ty.close\allowbreak Rec}}{clauses}{({\mathit{x}}^{\mathit{x}\rightarrow \lky{}})}{\begin{center}\gpfit{\mathsf{closeRec}(\lky{}, \mathit{x}, \mathsf{space}(\ls{})) = \mathsf{space}(\mathsf{closeRec}(\lky{}, \mathit{x}, \ls{}))}\\[2pt]
\gpfit{\mathsf{closeRec}(\lky{}, \mathit{x}, \mathsf{piTy}(\ltyn{0}, \ltyn{1})) = \mathsf{piTy}(\mathsf{closeRec}(\lky{}, \mathit{x}, \ltyn{0}), \mathsf{closeRec}(\lky{} + 1, \mathit{x}, \ltyn{1}))}\end{center}
}{\texttt{GLang.Ty.close\allowbreak Rec}}{GLang/SourceLN.lean:116}
\gldefdefn{term.closeRec}{\texttt{Term.close\allowbreak Rec}}{clauses}{({\mathit{x}}^{\mathit{x}\rightarrow \lky{}})}{\begin{center}\gpfit{\mathsf{closeRec}(\lky{}, \mathit{x}, \mathsf{bvar}(\mathit{i})) = \mathsf{bvar}(\mathit{i})}\\[2pt]
\gpfit{\begin{gathered}\mathsf{closeRec}(\lky{}, \mathit{x}, \mathsf{var}({\mathit{x}_{2}})) = \begin{cases} \mathsf{bvar}(\lky{}) & \text{if } {\mathit{x}_{2}} = \mathit{x} \\ \mathsf{var}({\mathit{x}_{2}}) & \text{otherwise} \end{cases}\end{gathered}}\\[2pt]
\gpfit{\mathsf{closeRec}(\lky{}, \mathit{x}, \mathsf{real}(\mathit{r})) = \mathsf{real}(\mathit{r})}\\[2pt]
\gpfit{\mathsf{closeRec}(\lky{}, \mathit{x}, \mathsf{zero}(\ls{})) = \mathsf{zero}(\mathsf{closeRec}(\lky{}, \mathit{x}, \ls{}))}\\[2pt]
\gpfit{\mathsf{closeRec}(\lky{}, \mathit{x}, \mathsf{smul}(\len{0}, \len{1})) = \mathsf{smul}(\mathsf{closeRec}(\lky{}, \mathit{x}, \len{0}), \mathsf{closeRec}(\lky{}, \mathit{x}, \len{1}))}\\[2pt]
\gpfit{\mathsf{closeRec}(\lky{}, \mathit{x}, \mathsf{add}(\len{0}, \len{1})) = \mathsf{add}(\mathsf{closeRec}(\lky{}, \mathit{x}, \len{0}), \mathsf{closeRec}(\lky{}, \mathit{x}, \len{1}))}\\[2pt]
\gpfit{\mathsf{closeRec}(\lky{}, \mathit{x}, \mathsf{dsumE}(\len{0}, \len{1})) = \mathsf{dsumE}(\mathsf{closeRec}(\lky{}, \mathit{x}, \len{0}), \mathsf{closeRec}(\lky{}, \mathit{x}, \len{1}))}\\[2pt]
\gpfit{\mathsf{closeRec}(\lky{}, \mathit{x}, \mathsf{tprodE}(\len{0}, \len{1})) = \mathsf{tprodE}(\mathsf{closeRec}(\lky{}, \mathit{x}, \len{0}), \mathsf{closeRec}(\lky{}, \mathit{x}, \len{1}))}\\[2pt]
\gpfit{\mathsf{closeRec}(\lky{}, \mathit{x}, \mathsf{inner}(\ls{})) = \mathsf{inner}(\mathsf{closeRec}(\lky{}, \mathit{x}, \ls{}))}\\[2pt]
\gpfit{\mathsf{closeRec}(\lky{}, \mathit{x}, \mathsf{embed}(\ls{})) = \mathsf{embed}(\mathsf{closeRec}(\lky{}, \mathit{x}, \ls{}))}\\[2pt]
\gpfit{\mathsf{closeRec}(\lky{}, \mathit{x}, \mathsf{pinvT}(\lsn{0}, \lsn{1}, \lee{})) = \mathsf{pinvT}(\mathsf{closeRec}(\lky{}, \mathit{x}, \lsn{0}), \mathsf{closeRec}(\lky{}, \mathit{x}, \lsn{1}), \mathsf{closeRec}(\lky{}, \mathit{x}, \lee{}))}\\[2pt]
\gpfit{\mathsf{closeRec}(\lky{}, \mathit{x}, \mathsf{pinvD}(\ls{}, \lee{})) = \mathsf{pinvD}(\mathsf{closeRec}(\lky{}, \mathit{x}, \ls{}), \mathsf{closeRec}(\lky{}, \mathit{x}, \lee{}))}\\[2pt]
\gpfit{\mathsf{closeRec}(\lky{}, \mathit{x}, \mathsf{llam}(\lsn{0}, \ls{}, \lee{})) = \mathsf{llam}(\mathsf{closeRec}(\lky{}, \mathit{x}, \lsn{0}), \mathsf{closeRec}(\lky{}, \mathit{x}, \ls{}), \mathsf{closeRec}(\lky{} + 1, \mathit{x}, \lee{}))}\\[2pt]
\gpfit{\mathsf{closeRec}(\lky{}, \mathit{x}, \mathsf{nllam}(\lty{}, \lee{})) = \mathsf{nllam}(\mathsf{closeRec}(\lky{}, \mathit{x}, \lty{}), \mathsf{closeRec}(\lky{} + 1, \mathit{x}, \lee{}))}\\[2pt]
\gpfit{\mathsf{closeRec}(\lky{}, \mathit{x}, \mathsf{appL}(\lsn{0}, \lsn{1}, \len{0}, \len{1})) = \mathsf{appL}(\mathsf{closeRec}(\lky{}, \mathit{x}, \lsn{0}), \mathsf{closeRec}(\lky{}, \mathit{x}, \lsn{1}), \mathsf{closeRec}(\lky{}, \mathit{x}, \len{0}), \mathsf{closeRec}(\lky{}, \mathit{x}, \len{1}))}\\[2pt]
\gpfit{\mathsf{closeRec}(\lky{}, \mathit{x}, \mathsf{appI}(\ls{}, \len{0}, \len{1})) = \mathsf{appI}(\mathsf{closeRec}(\lky{}, \mathit{x}, \ls{}), \mathsf{closeRec}(\lky{}, \mathit{x}, \len{0}), \mathsf{closeRec}(\lky{}, \mathit{x}, \len{1}))}\\[2pt]
\gpfit{\mathsf{closeRec}(\lky{}, \mathit{x}, \mathsf{appN}(\len{0}, \len{1})) = \mathsf{appN}(\mathsf{closeRec}(\lky{}, \mathit{x}, \len{0}), \mathsf{closeRec}(\lky{}, \mathit{x}, \len{1}))}\end{center}
}{\texttt{GLang.Term.close\allowbreak Rec}}{GLang/SourceLN.lean:120}
\gldefdefn{valExpr.closeRec}{\texttt{Val\allowbreak Expr.close\allowbreak Rec}}{clauses}{({\leev{}}^{\mathit{x}\rightarrow \lky{}})}{\begin{center}\gpfit{\begin{gathered}\mathsf{closeRec}(\lky{}, \mathit{x}, \mathsf{var}(\mathit{y}, \ltyv{})) = \begin{cases} \mathsf{bvar}(\lky{}) & \text{if } \mathit{y} = \mathit{x} \\ \mathsf{var}(\mathit{y}, \mathsf{closeRec}(\lky{}, \mathit{x}, \ltyv{})) & \text{otherwise} \end{cases}\end{gathered}}\\[2pt]
\gpfit{\mathsf{closeRec}(\lky{}, \mathit{x}, \mathsf{bvar}(\mathit{i})) = \mathsf{bvar}(\mathit{i})}\\[2pt]
\gpfit{\mathsf{closeRec}(\lky{}, \mathit{x}, \mathsf{vlam}(\ltynv{0}, \ltynv{1}, \leev{})) = \mathsf{vlam}(\mathsf{closeRec}(\lky{}, \mathit{x}, \ltynv{0}), \mathsf{closeRec}(\lky{} + 1, \mathit{x}, \ltynv{1}), \mathsf{closeRec}(\lky{} + 1, \mathit{x}, \leev{}))}\\[2pt]
\gpfit{\mathsf{closeRec}(\lky{}, \mathit{x}, \mathsf{zero}(\lsv{})) = \mathsf{zero}(\mathsf{closeRec}(\lky{}, \mathit{x}, \lsv{}))}\\[2pt]
\gpfit{\mathsf{closeRec}(\lky{}, \mathit{x}, \mathsf{rlit}(\mathit{r})) = \mathsf{rlit}(\mathit{r})}\\[2pt]
\gpfit{\mathsf{closeRec}(\lky{}, \mathit{x}, \mathsf{scale}(\lenv{0}, \lenv{1})) = \mathsf{scale}(\mathsf{closeRec}(\lky{}, \mathit{x}, \lenv{0}), \mathsf{closeRec}(\lky{}, \mathit{x}, \lenv{1}))}\\[2pt]
\gpfit{\mathsf{closeRec}(\lky{}, \mathit{x}, \mathsf{vadd}(\lsv{}, \lenv{0}, \lenv{1})) = \mathsf{vadd}(\mathsf{closeRec}(\lky{}, \mathit{x}, \lsv{}), \mathsf{closeRec}(\lky{}, \mathit{x}, \lenv{0}), \mathsf{closeRec}(\lky{}, \mathit{x}, \lenv{1}))}\\[2pt]
\gpfit{\mathsf{closeRec}(\lky{}, \mathit{x}, \mathsf{dsumE}(\lenv{0}, \lenv{1})) = \mathsf{dsumE}(\mathsf{closeRec}(\lky{}, \mathit{x}, \lenv{0}), \mathsf{closeRec}(\lky{}, \mathit{x}, \lenv{1}))}\\[2pt]
\gpfit{\mathsf{closeRec}(\lky{}, \mathit{x}, \mathsf{tprodE}(\lenv{0}, \lenv{1})) = \mathsf{tprodE}(\mathsf{closeRec}(\lky{}, \mathit{x}, \lenv{0}), \mathsf{closeRec}(\lky{}, \mathit{x}, \lenv{1}))}\\[2pt]
\gpfit{\mathsf{closeRec}(\lky{}, \mathit{x}, \mathsf{delta}(\mathit{n}, \lsv{})) = \mathsf{delta}(\mathit{n}, \mathsf{closeRec}(\lky{}, \mathit{x}, \lsv{}))}\\[2pt]
\gpfit{\mathsf{closeRec}(\lky{}, \mathit{x}, \mathsf{idelta}(\lsv{})) = \mathsf{idelta}(\mathsf{closeRec}(\lky{}, \mathit{x}, \lsv{}))}\\[2pt]
\gpfit{\mathsf{closeRec}(\lky{}, \mathit{x}, \mathsf{vsum}(\lsv{}, \mathit{es})) = \mathsf{vsum}(\mathsf{closeRec}(\lky{}, \mathit{x}, \lsv{}), \mathsf{closeRecList}(\lky{}, \mathit{x}, \mathit{es}))}\\[2pt]
\gpfit{\mathsf{closeRec}(\lky{}, \mathit{x}, \mathsf{ttv}(\lsv{}, \lenv{0}, \lenv{1})) = \mathsf{ttv}(\mathsf{closeRec}(\lky{}, \mathit{x}, \lsv{}), \mathsf{closeRec}(\lky{}, \mathit{x}, \lenv{0}), \mathsf{closeRec}(\lky{}, \mathit{x}, \lenv{1}))}\\[2pt]
\gpfit{\mathsf{closeRec}(\lky{}, \mathit{x}, \mathsf{mtm}({\mathit{n}_{0}}, {\mathit{n}_{1}}, \lsnv{0}, \lsnv{1}, \lenv{0}, \lenv{1})) = \mathsf{mtm}({\mathit{n}_{0}}, {\mathit{n}_{1}}, \mathsf{closeRec}(\lky{}, \mathit{x}, \lsnv{0}), \mathsf{closeRec}(\lky{}, \mathit{x}, \lsnv{1}), \mathsf{closeRec}(\lky{}, \mathit{x}, \lenv{0}), \mathsf{closeRec}(\lky{}, \mathit{x}, \lenv{1}))}\\[2pt]
\gpfit{\mathsf{closeRec}(\lky{}, \mathit{x}, \mathsf{appV}(\lenv{0}, \lenv{1})) = \mathsf{appV}(\mathsf{closeRec}(\lky{}, \mathit{x}, \lenv{0}), \mathsf{closeRec}(\lky{}, \mathit{x}, \lenv{1}))}\\[2pt]
\gpfit{\mathsf{closeRec}(\lky{}, \mathit{x}, \mathsf{vkrn}(\lenv{0}, \lenv{1})) = \mathsf{vkrn}(\mathsf{closeRec}(\lky{}, \mathit{x}, \lenv{0}), \mathsf{closeRec}(\lky{}, \mathit{x}, \lenv{1}))}\\[2pt]
\gpfit{\mathsf{closeRec}(\lky{}, \mathit{x}, \mathsf{vblk}(\lenv{0}, \lenv{1})) = \mathsf{vblk}(\mathsf{closeRec}(\lky{}, \mathit{x}, \lenv{0}), \mathsf{closeRec}(\lky{}, \mathit{x}, \lenv{1}))}\\[2pt]
\gpfit{\mathsf{closeRec}(\lky{}, \mathit{x}, \mathsf{reshap}(\leev{}, \ltyv{}, \ltynv{0})) = \mathsf{reshap}(\mathsf{closeRec}(\lky{}, \mathit{x}, \leev{}), \mathsf{closeRec}(\lky{}, \mathit{x}, \ltyv{}), \mathsf{closeRec}(\lky{}, \mathit{x}, \ltynv{0}))}\\[2pt]
\gpfit{\mathsf{closeRec}(\lky{}, \mathit{x}, \mathsf{vtrans}(\leev{})) = \mathsf{vtrans}(\mathsf{closeRec}(\lky{}, \mathit{x}, \leev{}))}\\[2pt]
\gpfit{\mathsf{closeRec}(\lky{}, \mathit{x}, \mathsf{gsvdS}(\leev{}, \lenv{{\mathrm{a}}}, \lenv{{\mathrm{b}}})) = \mathsf{gsvdS}(\mathsf{closeRec}(\lky{}, \mathit{x}, \leev{}), \mathsf{closeRec}(\lky{}, \mathit{x}, \lenv{{\mathrm{a}}}), \mathsf{closeRec}(\lky{}, \mathit{x}, \lenv{{\mathrm{b}}}))}\\[2pt]
\gpfit{\mathsf{closeRec}(\lky{}, \mathit{x}, \mathsf{gsvdU}(\leev{}, \lenv{{\mathrm{a}}}, \lenv{{\mathrm{b}}})) = \mathsf{gsvdU}(\mathsf{closeRec}(\lky{}, \mathit{x}, \leev{}), \mathsf{closeRec}(\lky{}, \mathit{x}, \lenv{{\mathrm{a}}}), \mathsf{closeRec}(\lky{}, \mathit{x}, \lenv{{\mathrm{b}}}))}\\[2pt]
\gpfit{\mathsf{closeRec}(\lky{}, \mathit{x}, \mathsf{gsvdUI}(\leev{}, \lenv{{\mathrm{a}}}, \lenv{{\mathrm{b}}})) = \mathsf{gsvdUI}(\mathsf{closeRec}(\lky{}, \mathit{x}, \leev{}), \mathsf{closeRec}(\lky{}, \mathit{x}, \lenv{{\mathrm{a}}}), \mathsf{closeRec}(\lky{}, \mathit{x}, \lenv{{\mathrm{b}}}))}\\[2pt]
\gpfit{\mathsf{closeRec}(\lky{}, \mathit{x}, \mathsf{gsvdV}(\leev{}, \lenv{{\mathrm{a}}}, \lenv{{\mathrm{b}}})) = \mathsf{gsvdV}(\mathsf{closeRec}(\lky{}, \mathit{x}, \leev{}), \mathsf{closeRec}(\lky{}, \mathit{x}, \lenv{{\mathrm{a}}}), \mathsf{closeRec}(\lky{}, \mathit{x}, \lenv{{\mathrm{b}}}))}\\[2pt]
\gpfit{\mathsf{closeRec}(\lky{}, \mathit{x}, \mathsf{gsvdVI}(\leev{}, \lenv{{\mathrm{a}}}, \lenv{{\mathrm{b}}})) = \mathsf{gsvdVI}(\mathsf{closeRec}(\lky{}, \mathit{x}, \leev{}), \mathsf{closeRec}(\lky{}, \mathit{x}, \lenv{{\mathrm{a}}}), \mathsf{closeRec}(\lky{}, \mathit{x}, \lenv{{\mathrm{b}}}))}\\[2pt]
\gpfit{\mathsf{closeRec}(\lky{}, \mathit{x}, \mathsf{pinv}(\leev{})) = \mathsf{pinv}(\mathsf{closeRec}(\lky{}, \mathit{x}, \leev{}))}\\[2pt]
\gpfit{\mathsf{closeRec}(\lky{}, \mathit{x}, \mathsf{gpinv}(\leev{}, \lenv{{\mathrm{a}}}, \lenv{{\mathrm{b}}})) = \mathsf{gpinv}(\mathsf{closeRec}(\lky{}, \mathit{x}, \leev{}), \mathsf{closeRec}(\lky{}, \mathit{x}, \lenv{{\mathrm{a}}}), \mathsf{closeRec}(\lky{}, \mathit{x}, \lenv{{\mathrm{b}}}))}\\[2pt]
\gpfit{\mathsf{closeRec}(\lky{}, \mathit{x}, \mathsf{rrefRep}(\lenv{0}, \lenv{1})) = \mathsf{rrefRep}(\mathsf{closeRec}(\lky{}, \mathit{x}, \lenv{0}), \mathsf{closeRec}(\lky{}, \mathit{x}, \lenv{1}))}\\[2pt]
\gpfit{\mathsf{closeRec}(\lky{}, \mathit{x}, \mathsf{inv}(\leev{})) = \mathsf{inv}(\mathsf{closeRec}(\lky{}, \mathit{x}, \leev{}))}\\[2pt]
\gpfit{\mathsf{closeRec}(\lky{}, \mathit{x}, \mathsf{svds}(\leev{})) = \mathsf{svds}(\mathsf{closeRec}(\lky{}, \mathit{x}, \leev{}))}\end{center}
}{\texttt{GLang.Val\allowbreak Expr.close\allowbreak Rec}}{GLang/ValueLN.lean:363}
\gldefdefn{basis.closeRec}{\texttt{Basis.close\allowbreak Rec}}{clauses}{({\textcolor{lang}{\mathit{x}}}^{\mathit{x}\rightarrow \lky{}})}{\begin{center}\gpfit{\mathsf{closeRec}(\lky{}, \mathit{x}, \lbzero{}) = \lbzero{}}\\[2pt]
\gpfit{\mathsf{closeRec}(\lky{}, \mathit{x}, \lbone{}) = \lbone{}}\\[2pt]
\gpfit{\mathsf{closeRec}(\lky{}, \mathit{x}, \mathsf{dual}(\lb{})) = \mathsf{dual}(\mathsf{closeRec}(\lky{}, \mathit{x}, \lb{}))}\\[2pt]
\gpfit{\mathsf{closeRec}(\lky{}, \mathit{x}, \mathsf{dsum}(\lbn{0}, \lbn{1})) = \mathsf{dsum}(\mathsf{closeRec}(\lky{}, \mathit{x}, \lbn{0}), \mathsf{closeRec}(\lky{}, \mathit{x}, \lbn{1}))}\\[2pt]
\gpfit{\mathsf{closeRec}(\lky{}, \mathit{x}, \mathsf{tprod}(\lbn{0}, \lbn{1})) = \mathsf{tprod}(\mathsf{closeRec}(\lky{}, \mathit{x}, \lbn{0}), \mathsf{closeRec}(\lky{}, \mathit{x}, \lbn{1}))}\\[2pt]
\gpfit{\mathsf{closeRec}(\lky{}, \mathit{x}, \mathsf{proj}(\ls{}, \lbn{0}, \lbn{1})) = \mathsf{proj}(\mathsf{closeRec}(\lky{}, \mathit{x}, \ls{}), \mathsf{closeRec}(\lky{}, \mathit{x}, \lbn{0}), \mathsf{closeRec}(\lky{}, \mathit{x}, \lbn{1}))}\\[2pt]
\gpfit{\mathsf{closeRec}(\lky{}, \mathit{x}, \mathsf{image}(\lb{})) = \mathsf{image}(\mathsf{closeRec}(\lky{}, \mathit{x}, \lb{}))}\\[2pt]
\gpfit{\mathsf{closeRec}(\lky{}, \mathit{x}, \mathsf{terms}(\ls{}, \mathit{es}, \lb{})) = \mathsf{terms}(\mathsf{closeRec}(\lky{}, \mathit{x}, \ls{}), \mathsf{closeRecList}(\lky{}, \mathit{x}, \mathit{es}), \mathsf{closeRec}(\lky{}, \mathit{x}, \lb{}))}\\[2pt]
\gpfit{\mathsf{closeRec}(\lky{}, \mathit{x}, \mathsf{frame}(\lbn{0}, \lbn{1})) = \mathsf{frame}(\mathsf{closeRec}(\lky{}, \mathit{x}, \lbn{0}), \mathsf{closeRec}(\lky{} + 1, \mathit{x}, \lbn{1}))}\end{center}
}{\texttt{GLang.Basis.close\allowbreak Rec}}{GLang/SourceLN.lean:147}
\gldefdefn{valTy.closeRec}{\texttt{Val\allowbreak Ty.close\allowbreak Rec}}{clauses}{({\ltynv{{\mathrm{x}}}}^{\mathit{x}\rightarrow \lky{}})}{\begin{center}\gpfit{\mathsf{closeRec}(\lky{}, \mathit{x}, \mathsf{space}(\lsv{})) = \mathsf{space}(\mathsf{closeRec}(\lky{}, \mathit{x}, \lsv{}))}\\[2pt]
\gpfit{\mathsf{closeRec}(\lky{}, \mathit{x}, \mathsf{piTy}(\ltynv{0}, \ltynv{1})) = \mathsf{piTy}(\mathsf{closeRec}(\lky{}, \mathit{x}, \ltynv{0}), \mathsf{closeRec}(\lky{} + 1, \mathit{x}, \ltynv{1}))}\end{center}
}{\texttt{GLang.Val\allowbreak Ty.close\allowbreak Rec}}{GLang/ValueLN.lean:359}
\gldefdefn{valSpace.closeRec}{\texttt{Val\allowbreak Space.close\allowbreak Rec}}{clauses}{({\lsnv{{\mathrm{x}}}}^{\mathit{x}\rightarrow \lky{}})}{\begin{center}\gpfit{\mathsf{closeRec}(\lky{}, \mathit{x}, \lsempty{}) = \lsempty{}}\\[2pt]
\gpfit{\mathsf{closeRec}(\lky{}, \mathit{x}, \lR{}) = \lR{}}\\[2pt]
\gpfit{\mathsf{closeRec}(\lky{}, \mathit{x}, \mathsf{dsum}(\lsnv{0}, \lsnv{1})) = \mathsf{dsum}(\mathsf{closeRec}(\lky{}, \mathit{x}, \lsnv{0}), \mathsf{closeRec}(\lky{}, \mathit{x}, \lsnv{1}))}\\[2pt]
\gpfit{\mathsf{closeRec}(\lky{}, \mathit{x}, \mathsf{tprod}(\lsnv{0}, \lsnv{1})) = \mathsf{tprod}(\mathsf{closeRec}(\lky{}, \mathit{x}, \lsnv{0}), \mathsf{closeRec}(\lky{}, \mathit{x}, \lsnv{1}))}\\[2pt]
\gpfit{\mathsf{closeRec}(\lky{}, \mathit{x}, \mathsf{ffs}(\lprop{}, \lsv{}, \leev{})) = \mathsf{ffs}(\lprop{}, \mathsf{closeRec}(\lky{}, \mathit{x}, \lsv{}), \mathsf{closeRec}(\lky{}, \mathit{x}, \leev{}))}\end{center}
}{\texttt{GLang.Val\allowbreak Space.close\allowbreak Rec}}{GLang/ValueLN.lean:352}
\gldefdefn{term.closeRecList}{\texttt{Term.close\allowbreak Rec\allowbreak List}}{clauses}{\mathsf{closeRecList}(\lky{}, \mathit{x}, \mathit{es})}{\begin{center}\gpfit{\mathsf{closeRecList}(\lky{}, \mathit{x}, \mathit{es}) = \left[\, \mathsf{closeRec}(\lky{}, \mathit{x}, \mathit{z}) \mid \mathit{z} \in \mathit{es} \,\right]}\end{center}
}{\texttt{GLang.Term.close\allowbreak Rec\allowbreak List}}{GLang/SourceLN.lean:143}
\gldefdefn{valExpr.closeRecList}{\texttt{Val\allowbreak Expr.close\allowbreak Rec\allowbreak List}}{clauses}{\mathsf{closeRecList}(\lky{}, \mathit{x}, \mathit{x})}{\begin{center}\gpfit{\mathsf{closeRecList}(\lky{}, \mathit{x}, []) = []}\\[2pt]
\gpfit{\mathsf{closeRecList}(\lky{}, \mathit{x}, \leev{} \mathbin{:\!:} \mathit{es}) = \mathsf{closeRec}(\lky{}, \mathit{x}, \leev{}) \mathbin{:\!:} \mathsf{closeRecList}(\lky{}, \mathit{x}, \mathit{es})}\end{center}
}{\texttt{GLang.Val\allowbreak Expr.close\allowbreak Rec\allowbreak List}}{GLang/ValueLN.lean:401}
\gldefdefn{denotBasis}{\texttt{denot\allowbreak Basis}}{clauses}{\SemBB{\textcolor{lang}{\mathit{a}}}{\sigma{}}}{\begin{center}\gpfit{\SemBB{\lbzero{}}{\sigma} = \{\}}\\[2pt]
\gpfit{\SemBB{\lbone{}}{\sigma} = \{1\}}\\[2pt]
\gpfit{\SemBB{\lb{}^{\lbstar}}{\sigma} = \{v_i'\in(\SemT{\sof{\lb{}}}{\sigma})^{\star}\mid
\forall j.\;v_i'(v_j)=\delta_{ij}\},\quad\{v_j\}=\SemBB{\lb{}}{\sigma}}\\[2pt]
\gpfit{\SemBB{\lbn{0}\lbosum\lbn{1}}{\sigma} = \{i_0(v)\mid v\in\SemBB{\lbn{0}}{\sigma}\}\mathbin{+\!\!+}
\{i_1(v)\mid v\in\SemBB{\lbn{1}}{\sigma}\}}\\[2pt]
\gpfit{\SemBB{\lbn{0}\lboprod\lbn{1}}{\sigma} = \{v_0\otimes v_1\mid v_0\in\SemBB{\lbn{0}}{\sigma},\;v_1\in\SemBB{\lbn{1}}{\sigma}\}}\\[2pt]
\gpfit{\SemBB{\pBasisC{\Kernel(\lee{})}{\lbn{0}}{\lbn{1}}}{\sigma} = \FSV{}(\Sem{\lee{}}{\sigma},\lambda x.\;x = 0,\operatorname{Left},\SemBB{\lbn{0}}{\sigma},\SemBB{\lbn{1}}{\sigma})}\\[2pt]
\gpfit{\SemBB{\pBasisC{\CoKernel(\lee{})}{\lbn{0}}{\lbn{1}}}{\sigma} = \FSV{}(\Sem{\lee{}}{\sigma},\lambda x.\;x \neq 0,\operatorname{Left},\SemBB{\lbn{0}}{\sigma},\SemBB{\lbn{1}}{\sigma})}\\[2pt]
\gpfit{\SemBB{\pBasisC{\Image(\lee{})}{\lbn{0}}{\lbn{1}}}{\sigma} = \FSV{}(\Sem{\lee{}}{\sigma},\lambda x.\;x \neq 0,\operatorname{Right},\SemBB{\lbn{0}}{\sigma},\SemBB{\lbn{1}}{\sigma})}\\[2pt]
\gpfit{\SemBB{\pBasisC{\CoImage(\lee{})}{\lbn{0}}{\lbn{1}}}{\sigma} = \FSV{}(\Sem{\lee{}}{\sigma},\lambda x.\;x = 0,\operatorname{Right},\SemBB{\lbn{0}}{\sigma},\SemBB{\lbn{1}}{\sigma})}\\[2pt]
\gpfit{\SemBB{\iBasisC{\lb{}}}{\sigma} = \{\Sem{\lee{}}{\sigma}(v)\mid v\in\SemBB{\lb{}}{\sigma}\},\quad\sof{\lb{}}=\CoKernel(\lee{})}\\[2pt]
\gpfit{\SemBB{\llpl{}\len{0},\dots,\len{n}\llpr{}_{\lb{}}}{\sigma} = \LexInd{}(\{\Sem{\len{0}}{\sigma},\dots,\Sem{\len{n}}{\sigma}\}\mathbin{+\!\!+}\SemBB{\lb{}}{\sigma})}\\[2pt]
\gpfit{\SemBB{\frameB{x}{\lbn{0}}{\lbn{1}}}{\sigma} = \lambda x'\colon\SemT{\tof{\lbn{0}}}{\sigma}.\;\SemBB{\lbn{1}}{\sigma[x\mapsto x']}}\end{center}
}{\texttt{GLang.denot\allowbreak Basis}}{GLang/Basis.lean:886}
\gldefequation{denotBasis.empty}{\SemBB{\lbzero{}}{\sigma} = \{\}}
\gldefequation{denotBasis.one}{\SemBB{\lbone{}}{\sigma} = \{1\}}
\gldefequation{denotBasis.dual}{\SemBB{\lb{}^{\lbstar}}{\sigma} = \{v_i'\in(\SemT{\sof{\lb{}}}{\sigma})^{\star}\mid
\forall j.\;v_i'(v_j)=\delta_{ij}\},\quad\{v_j\}=\SemBB{\lb{}}{\sigma}}
\gldefequation{denotBasis.dsum}{\SemBB{\lbn{0}\lbosum\lbn{1}}{\sigma} = \{i_0(v)\mid v\in\SemBB{\lbn{0}}{\sigma}\}\mathbin{+\!\!+}
\{i_1(v)\mid v\in\SemBB{\lbn{1}}{\sigma}\}}
\gldefequation{denotBasis.tprod}{\SemBB{\lbn{0}\lboprod\lbn{1}}{\sigma} = \{v_0\otimes v_1\mid v_0\in\SemBB{\lbn{0}}{\sigma},\;v_1\in\SemBB{\lbn{1}}{\sigma}\}}
\gldefequation{denotBasis.proj.ker}{\SemBB{\pBasisC{\Kernel(\lee{})}{\lbn{0}}{\lbn{1}}}{\sigma} = \FSV{}(\Sem{\lee{}}{\sigma},\lambda x.\;x = 0,\operatorname{Left},\SemBB{\lbn{0}}{\sigma},\SemBB{\lbn{1}}{\sigma})}
\gldefequation{denotBasis.proj.coker}{\SemBB{\pBasisC{\CoKernel(\lee{})}{\lbn{0}}{\lbn{1}}}{\sigma} = \FSV{}(\Sem{\lee{}}{\sigma},\lambda x.\;x \neq 0,\operatorname{Left},\SemBB{\lbn{0}}{\sigma},\SemBB{\lbn{1}}{\sigma})}
\gldefequation{denotBasis.proj.img}{\SemBB{\pBasisC{\Image(\lee{})}{\lbn{0}}{\lbn{1}}}{\sigma} = \FSV{}(\Sem{\lee{}}{\sigma},\lambda x.\;x \neq 0,\operatorname{Right},\SemBB{\lbn{0}}{\sigma},\SemBB{\lbn{1}}{\sigma})}
\gldefequation{denotBasis.proj.coimg}{\SemBB{\pBasisC{\CoImage(\lee{})}{\lbn{0}}{\lbn{1}}}{\sigma} = \FSV{}(\Sem{\lee{}}{\sigma},\lambda x.\;x = 0,\operatorname{Right},\SemBB{\lbn{0}}{\sigma},\SemBB{\lbn{1}}{\sigma})}
\gldefequation{denotBasis.image}{\SemBB{\iBasisC{\lb{}}}{\sigma} = \{\Sem{\lee{}}{\sigma}(v)\mid v\in\SemBB{\lb{}}{\sigma}\},\quad\sof{\lb{}}=\CoKernel(\lee{})}
\gldefequation{denotBasis.terms}{\SemBB{\llpl{}\len{0},\dots,\len{n}\llpr{}_{\lb{}}}{\sigma} = \LexInd{}(\{\Sem{\len{0}}{\sigma},\dots,\Sem{\len{n}}{\sigma}\}\mathbin{+\!\!+}\SemBB{\lb{}}{\sigma})}
\gldefequation{denotBasis.frame}{\SemBB{\frameB{x}{\lbn{0}}{\lbn{1}}}{\sigma} = \lambda x'\colon\SemT{\tof{\lbn{0}}}{\sigma}.\;\SemBB{\lbn{1}}{\sigma[x\mapsto x']}}
\gldefdefn{denotSpaceAux}{\texttt{denot\allowbreak Space\allowbreak Aux}}{clauses}{\SemT{\ls{}}{\sigma{}}}{\begin{center}\gpfit{\SemT{\lsempty{}}{\sigma{}} = \{0\}}\\[2pt]
\gpfit{\SemT{\lR{}}{\sigma{}} = \mathbb{R}}\\[2pt]
\gpfit{\SemT{\ls{}^{\lstar}}{\sigma{}} = (\SemT{\ls{}}{\sigma{}})^{\lstar}}\\[2pt]
\gpfit{\SemT{\lsn{0} \losum \lsn{1}}{\sigma{}} = \SemT{\lsn{0}}{\sigma{}} \losum \SemT{\lsn{1}}{\sigma{}}}\\[2pt]
\gpfit{\SemT{\lsn{0} \loprod \lsn{1}}{\sigma{}} = \SemT{\lsn{0}}{\sigma{}} \loprod \SemT{\lsn{1}}{\sigma{}}}\\[2pt]
\gpfit{\SemT{\Kernel(\lee{})}{\sigma{}} = \ker \Sem{\lee{}}{\sigma{}}\quad\bigl(\lepsilon{}\vdash \lee{}\colon \lsn{0}^{\lstar}\loprod \lsn{1}\bigr)}\\[2pt]
\gpfit{\SemT{\CoKernel(\lee{})}{\sigma{}} = (\ker \Sem{\lee{}}{\sigma{}})^{\perp_{\SemT{\lsn{0}}{\sigma{}}}}\quad\bigl(\lepsilon{}\vdash \lee{}\colon \lsn{0}^{\lstar}\loprod \lsn{1}\bigr)}\\[2pt]
\gpfit{\SemT{\Image(\lee{})}{\sigma{}} = \operatorname{range} \Sem{\lee{}}{\sigma{}}\quad\bigl(\lepsilon{}\vdash \lee{}\colon \lsn{0}^{\lstar}\loprod \lsn{1}\bigr)}\\[2pt]
\gpfit{\SemT{\CoImage(\lee{})}{\sigma{}} = (\operatorname{range} \Sem{\lee{}}{\sigma{}})^{\perp_{\SemT{\lsn{1}}{\sigma{}}}}\quad\bigl(\lepsilon{}\vdash \lee{}\colon \lsn{0}^{\lstar}\loprod \lsn{1}\bigr)}\end{center}
}{\texttt{GLang.denot\allowbreak Space\allowbreak Aux}}{GLang/Denot.lean:714}
\gldefequation{denotSpaceAux.empty}{\SemT{\lsempty{}}{\sigma{}} = \{0\}}
\gldefequation{denotSpaceAux.real}{\SemT{\lR{}}{\sigma{}} = \mathbb{R}}
\gldefequation{denotSpaceAux.dual}{\SemT{\ls{}^{\lstar}}{\sigma{}} = (\SemT{\ls{}}{\sigma{}})^{\lstar}}
\gldefequation{denotSpaceAux.dsum}{\SemT{\lsn{0} \losum \lsn{1}}{\sigma{}} = \SemT{\lsn{0}}{\sigma{}} \losum \SemT{\lsn{1}}{\sigma{}}}
\gldefequation{denotSpaceAux.tprod}{\SemT{\lsn{0} \loprod \lsn{1}}{\sigma{}} = \SemT{\lsn{0}}{\sigma{}} \loprod \SemT{\lsn{1}}{\sigma{}}}
\gldefequation{denotSpaceAux.ker}{\SemT{\Kernel(\lee{})}{\sigma{}} = \ker \Sem{\lee{}}{\sigma{}}\quad\bigl(\lepsilon{}\vdash \lee{}\colon \lsn{0}^{\lstar}\loprod \lsn{1}\bigr)}
\gldefequation{denotSpaceAux.coker}{\SemT{\CoKernel(\lee{})}{\sigma{}} = (\ker \Sem{\lee{}}{\sigma{}})^{\perp_{\SemT{\lsn{0}}{\sigma{}}}}\quad\bigl(\lepsilon{}\vdash \lee{}\colon \lsn{0}^{\lstar}\loprod \lsn{1}\bigr)}
\gldefequation{denotSpaceAux.img}{\SemT{\Image(\lee{})}{\sigma{}} = \operatorname{range} \Sem{\lee{}}{\sigma{}}\quad\bigl(\lepsilon{}\vdash \lee{}\colon \lsn{0}^{\lstar}\loprod \lsn{1}\bigr)}
\gldefequation{denotSpaceAux.coimg}{\SemT{\CoImage(\lee{})}{\sigma{}} = (\operatorname{range} \Sem{\lee{}}{\sigma{}})^{\perp_{\SemT{\lsn{1}}{\sigma{}}}}\quad\bigl(\lepsilon{}\vdash \lee{}\colon \lsn{0}^{\lstar}\loprod \lsn{1}\bigr)}
\gldefdefn{denotTermAux}{\texttt{denot\allowbreak Term\allowbreak Aux}}{clauses}{\Sem{\mathit{tm}}{\sigma{}}}{\begin{center}\gpfit{\Sem{\lvzero{}_{\ls{}}}{\sigma{}} = 0_{\SemT{\ls{}}{\sigma{}}}}\\[2pt]
\gpfit{\Sem{\mathit{r}}{\sigma{}} = \mathit{r}}\\[2pt]
\gpfit{\Sem{\len{0} \lvscale \len{1}}{\sigma{}} = \Sem{\len{0}}{\sigma{}} \cdot_{\SemT{\ls{}}{\sigma{}}} \Sem{\len{1}}{\sigma{}}\quad\bigl(\lctx\vdash \len{0} \lvscale \len{1}\colon\ls{}\bigr)}\\[2pt]
\gpfit{\Sem{\len{0} \lvadd \len{1}}{\sigma{}} = \Sem{\len{0}}{\sigma{}} +_{\SemT{\ls{}}{\sigma{}}} \Sem{\len{1}}{\sigma{}}\quad\bigl(\lctx\vdash \len{0} \lvadd \len{1}\colon\ls{}\bigr)}\\[2pt]
\gpfit{\Sem{\linner{}_{\ls{}}}{\sigma{}} = \langle\cdot,\cdot\rangle_{\SemT{\ls{}}{\sigma{}}}}\\[2pt]
\gpfit{\Sem{\leval{\len{0}}{}{\len{1}}}{\sigma{}} = \contractN{}{\Sem{\len{0}}{\sigma{}}}{\Sem{\len{1}}{\sigma{}}}\quad(\lctx\vdash \len{0}\colon \lsn{0}^{\lstar}\loprod \lsn{1})}\\[2pt]
\gpfit{\Sem{\leval{\len{0}}{}{\len{1}}}{\sigma{}} = \contractN{}{\Sem{\len{0}}{\sigma{}}}{\Sem{\len{1}}{\sigma{}}}\quad(\lctx\vdash \len{0}\colon \ls{}^{\lstar})}\\[2pt]
\gpfit{\Sem{\leval{\len{0}}{}{\len{1}}}{\sigma{}} = \Sem{\len{0}}{\sigma{}}(\Sem{\len{1}}{\sigma{}})\quad(\lctx\vdash \len{0}\colon (x\colon\ltyn{0})\lrh\ltyn{1})}\\[2pt]
\gpfit{\Sem{\mathit{x}}{\sigma{}} = \sigma{}(\mathit{x})}\\[2pt]
\gpfit{\Sem{\lee{}^{\lpseudo}}{\sigma{}} = \Sem{\lee{}}{\sigma{}}^{\dagger}}\\[2pt]
\gpfit{\Sem{\len{0}\losum \len{1}}{\sigma{}} = \Sem{\len{0}}{\sigma{}} \oplus \Sem{\len{1}}{\sigma{}}}\\[2pt]
\gpfit{\Sem{\len{0}\loprod \len{1}}{\sigma{}} = \Sem{\len{0}}{\sigma{}} \otimes \Sem{\len{1}}{\sigma{}}}\\[2pt]
\gpfit{\Sem{(\lllambda \llpl x\colon \lsn{0}\llpr \ldotsp \lee{})}{\sigma{}} = \lambda y\colon \lvert \SemT{\lsn{0}}{\sigma{}}\rvert.\, \Sem{\lee{}}{\sigma{}[x \mapsto y]}}\\[2pt]
\gpfit{\Sem{(\nllambda \llpl x\colon \ltyn{0}\llpr \ldotsp \lee{})}{\sigma{}} = \lambda y\colon \lvert \SemT{\ltyn{0}}{\sigma{}}\rvert.\, \Sem{\lee{}}{\sigma{}[x \mapsto y]}}\\[2pt]
\gpfit{\Sem{\lembd{}_{\Kernel(\lee{})}}{\sigma{}} = \iota_{\SemT{\Kernel(\lee{})}{\sigma{}}}}\\[2pt]
\gpfit{\Sem{\lembd{}_{\CoKernel(\lee{})}}{\sigma{}} = \iota_{\SemT{\CoKernel(\lee{})}{\sigma{}}}}\\[2pt]
\gpfit{\Sem{\lembd{}_{\Image(\lee{})}}{\sigma{}} = \iota_{\SemT{\Image(\lee{})}{\sigma{}}}}\\[2pt]
\gpfit{\Sem{\lembd{}_{\CoImage(\lee{})}}{\sigma{}} = \iota_{\SemT{\CoImage(\lee{})}{\sigma{}}}}\end{center}
}{\texttt{GLang.denot\allowbreak Term\allowbreak Aux}}{GLang/Denot.lean:761}
\gldefequation{denotTermAux.zero}{\Sem{\lvzero{}_{\ls{}}}{\sigma{}} = 0_{\SemT{\ls{}}{\sigma{}}}}
\gldefequation{denotTermAux.real}{\Sem{\mathit{r}}{\sigma{}} = \mathit{r}}
\gldefequation{denotTermAux.smul}{\Sem{\len{0} \lvscale \len{1}}{\sigma{}} = \Sem{\len{0}}{\sigma{}} \cdot_{\SemT{\ls{}}{\sigma{}}} \Sem{\len{1}}{\sigma{}}\quad\bigl(\lctx\vdash \len{0} \lvscale \len{1}\colon\ls{}\bigr)}
\gldefequation{denotTermAux.add}{\Sem{\len{0} \lvadd \len{1}}{\sigma{}} = \Sem{\len{0}}{\sigma{}} +_{\SemT{\ls{}}{\sigma{}}} \Sem{\len{1}}{\sigma{}}\quad\bigl(\lctx\vdash \len{0} \lvadd \len{1}\colon\ls{}\bigr)}
\gldefequation{denotTermAux.inner}{\Sem{\linner{}_{\ls{}}}{\sigma{}} = \langle\cdot,\cdot\rangle_{\SemT{\ls{}}{\sigma{}}}}
\gldefequation{denotTermAux.appL}{\Sem{\leval{\len{0}}{}{\len{1}}}{\sigma{}} = \contractN{}{\Sem{\len{0}}{\sigma{}}}{\Sem{\len{1}}{\sigma{}}}\quad(\lctx\vdash \len{0}\colon \lsn{0}^{\lstar}\loprod \lsn{1})}
\gldefequation{denotTermAux.appI}{\Sem{\leval{\len{0}}{}{\len{1}}}{\sigma{}} = \contractN{}{\Sem{\len{0}}{\sigma{}}}{\Sem{\len{1}}{\sigma{}}}\quad(\lctx\vdash \len{0}\colon \ls{}^{\lstar})}
\gldefequation{denotTermAux.appN}{\Sem{\leval{\len{0}}{}{\len{1}}}{\sigma{}} = \Sem{\len{0}}{\sigma{}}(\Sem{\len{1}}{\sigma{}})\quad(\lctx\vdash \len{0}\colon (x\colon\ltyn{0})\lrh\ltyn{1})}
\gldefequation{denotTermAux.var}{\Sem{\mathit{x}}{\sigma{}} = \sigma{}(\mathit{x})}
\gldefequation{denotTermAux.pinvT}{\Sem{\lee{}^{\lpseudo}}{\sigma{}} = \Sem{\lee{}}{\sigma{}}^{\dagger}}
\gldefequation{denotTermAux.pinvD}{\Sem{\lee{}^{\lpseudo}}{\sigma{}} = \Sem{\lee{}}{\sigma{}}^{\dagger}}
\gldefequation{denotTermAux.dsumE}{\Sem{\len{0}\losum \len{1}}{\sigma{}} = \Sem{\len{0}}{\sigma{}} \oplus \Sem{\len{1}}{\sigma{}}}
\gldefequation{denotTermAux.tprodE}{\Sem{\len{0}\loprod \len{1}}{\sigma{}} = \Sem{\len{0}}{\sigma{}} \otimes \Sem{\len{1}}{\sigma{}}}
\gldefequation{denotTermAux.llam}{\Sem{(\lllambda \llpl x\colon \lsn{0}\llpr \ldotsp \lee{})}{\sigma{}} = \lambda y\colon \lvert \SemT{\lsn{0}}{\sigma{}}\rvert.\, \Sem{\lee{}}{\sigma{}[x \mapsto y]}}
\gldefequation{denotTermAux.nllam}{\Sem{(\nllambda \llpl x\colon \ltyn{0}\llpr \ldotsp \lee{})}{\sigma{}} = \lambda y\colon \lvert \SemT{\ltyn{0}}{\sigma{}}\rvert.\, \Sem{\lee{}}{\sigma{}[x \mapsto y]}}
\gldefequation{denotTermAux.embed.ker}{\Sem{\lembd{}_{\Kernel(\lee{})}}{\sigma{}} = \iota_{\SemT{\Kernel(\lee{})}{\sigma{}}}}
\gldefequation{denotTermAux.embed.coker}{\Sem{\lembd{}_{\CoKernel(\lee{})}}{\sigma{}} = \iota_{\SemT{\CoKernel(\lee{})}{\sigma{}}}}
\gldefequation{denotTermAux.embed.img}{\Sem{\lembd{}_{\Image(\lee{})}}{\sigma{}} = \iota_{\SemT{\Image(\lee{})}{\sigma{}}}}
\gldefequation{denotTermAux.embed.coimg}{\Sem{\lembd{}_{\CoImage(\lee{})}}{\sigma{}} = \iota_{\SemT{\CoImage(\lee{})}{\sigma{}}}}
\gldefdefn{denotTyAux}{\texttt{denot\allowbreak Ty\allowbreak Aux}}{clauses}{\SemT{\lty{}}{\sigma{}}}{\begin{center}\gpfit{\SemT{(x\colon \ltyn{0}\fnto \ltyn{1})}{\sigma{}} = \prod_{y\in\lvert \SemT{\ltyn{0}}{\sigma{}}\rvert} \SemT{\ltyn{1}}{\sigma{}[x \mapsto y]}}\end{center}
}{\texttt{GLang.denot\allowbreak Ty\allowbreak Aux}}{GLang/Denot.lean:751}
\gldefequation{denotTyAux.piTy}{\SemT{(x\colon \ltyn{0}\fnto \ltyn{1})}{\sigma{}} = \prod_{y\in\lvert \SemT{\ltyn{0}}{\sigma{}}\rvert} \SemT{\ltyn{1}}{\sigma{}[x \mapsto y]}}
\gldefdefn{denotValBSemS}{\texttt{denot\allowbreak Val\allowbreak B\_\allowbreak sem\allowbreak S}}{curated clauses}{\mathcal{E}\llbracket \lenv{{\mathrm{a}}}\rrbracket_{\Gamma_{\mathrm{a}};\mathit{a}}}{\paragraph{\gldefnname{value.domains}}
\gldefn{value.domains}

\paragraph{\gldefnname{value.ffs.svd}}
\gldefn{value.ffs.svd}

\paragraph{\gldefnname{value.inference.helpers}}
\gldefn{value.inference.helpers}

\paragraph{\gldefnname{value.denot.conventions}}
\gldefn{value.denot.conventions}

\paragraph{\gldefnname{value.denot.guards}}
\gldefn{value.denot.guards}

\paragraph{\gldefnname{value.denot.types}}
\gldefn{value.denot.types}

\paragraph{\gldefnname{value.denot.scalars}}
\gldefn{value.denot.scalars}

\paragraph{\gldefnname{value.denot.coordinates}}
\gldefn{value.denot.coordinates}

\paragraph{\gldefnname{value.denot.transpose}}
\gldefn{value.denot.transpose}

\paragraph{\gldefnname{value.denot.sum}}
\gldefn{value.denot.sum}

\paragraph{\gldefnname{value.denot.contractions}}
\gldefn{value.denot.contractions}

\paragraph{\gldefnname{value.denot.assembly}}
\gldefn{value.denot.assembly}

\paragraph{\gldefnname{value.denot.matrix-conventions}}
\gldefn{value.denot.matrix-conventions}

\paragraph{\gldefnname{value.denot.inverse}}
\gldefn{value.denot.inverse}

\paragraph{\gldefnname{value.denot.gsvd}}
\gldefn{value.denot.gsvd}

\paragraph{\gldefnname{value.denot.rref}}
\gldefn{value.denot.rref}

\paragraph{\gldefnname{value.denot.lambda}}
\gldefn{value.denot.lambda}

\paragraph{\gldefnname{value.denot.application}}
\gldefn{value.denot.application}

\paragraph{\gldefnname{value.denot.reshape}}
\gldefn{value.denot.reshape}}{\texttt{GLang.denot\allowbreak Val\allowbreak B\_\allowbreak sem\allowbreak S}}{GLang/ValueDenotSem.lean:811}
\gldefdefn{denotValTyS}{\texttt{denot\allowbreak Val\allowbreak Ty\allowbreak S}}{clauses}{\mathcal{T}\llbracket \ltynv{{\mathrm{x}}}\rrbracket_{\Gamma{};\Delta{}}}{\begin{center}\gpfit{\mathcal{T}\llbracket \lsv{}\rrbracket_{\Gamma{};\Delta{}} = \mathsf{space}(\lsv{}, \dim_{\mathcal{O}}(\lsv{}))}\\[2pt]
\gpfit{\mathcal{T}\llbracket (x\colon \ltynv{0}\fnto \ltynv{1})\rrbracket_{\Gamma{};\Delta{}} = \textstyle\prod_{\leev{} \in \lvert \mathcal{T}\llbracket \ltynv{0}\rrbracket_{\Gamma{};\Delta{}}\rvert} \mathcal{T}\llbracket \ltynv{1}\rrbracket_{\ltynv{0} \mathbin{:\!:} \Gamma{};\leev{} \mathbin{:\!:} \Delta{}}}\end{center}
}{\texttt{GLang.denot\allowbreak Val\allowbreak Ty\allowbreak S}}{GLang/ValueDenotSem.lean:511}
\gldefdefn{valSpace.dof}{\texttt{Val\allowbreak Space.dof}}{clauses}{\mathsf{dof}(\mathcal{O}, \lsv{})}{\begin{center}\gpfit{\dim_{\mathcal{O}}(\lsv{}) = \mathsf{foldr}(\lambda {\mathit{x}_{1}}.\, \lambda {\mathit{x}_{2}}.\, ({\mathit{x}_{1}}) \cdot ({\mathit{x}_{2}}), 1, \shpOf{\lsv{}}{\mathcal{O}})}\end{center}
}{\texttt{GLang.Val\allowbreak Space.dof}}{GLang/ValueDenot.lean:135}
\gldefdefn{valueShape.family}{\texttt{Value\allowbreak Shape.family}}{clauses}{|\mathit{arrays}|=\mathit{count} \land \forall a\in \mathit{arrays}.\; |a|=\mathit{width}}{\begin{center}\gpfit{|\mathit{arrays}|=\mathit{count} \land \forall a\in \mathit{arrays}.\; |a|=\mathit{width} = \lvert \mathit{arrays} \rvert = \mathit{count} \;\wedge\; \forall \mathit{a}.\; \mathit{a} \in \mathit{arrays} \Rightarrow \lvert \mathit{a}\rvert = \mathit{width}}\end{center}
}{\texttt{GLang.Value\allowbreak Shape.family}}{GLang/ValuePrimitiveShapes.lean:41}
\gldefdefn{space.fv}{\texttt{Space.fv}}{clauses}{\mathsf{fv}(\mathit{x})}{\begin{center}\gpfit{\mathsf{fv}(\lsempty{}) = \lemptyset{}}\\[2pt]
\gpfit{\mathsf{fv}(\lR{}) = \lemptyset{}}\\[2pt]
\gpfit{\mathsf{fv}(\ls{}^{\lstar}) = \mathsf{fv}(\ls{})}\\[2pt]
\gpfit{\mathsf{fv}(\lsn{0} \losum \lsn{1}) = \mathsf{fv}(\lsn{0}) \cup \mathsf{fv}(\lsn{1})}\\[2pt]
\gpfit{\mathsf{fv}(\lsn{0} \loprod \lsn{1}) = \mathsf{fv}(\lsn{0}) \cup \mathsf{fv}(\lsn{1})}\\[2pt]
\gpfit{\mathsf{fv}(\Kernel(\lee{})) = \mathsf{fv}(\lsn{0}) \cup \mathsf{fv}(\lsn{1}) \cup \mathsf{fv}(\lee{})}\\[2pt]
\gpfit{\mathsf{fv}(\CoKernel(\lee{})) = \mathsf{fv}(\lsn{0}) \cup \mathsf{fv}(\lsn{1}) \cup \mathsf{fv}(\lee{})}\\[2pt]
\gpfit{\mathsf{fv}(\Image(\lee{})) = \mathsf{fv}(\lsn{0}) \cup \mathsf{fv}(\lsn{1}) \cup \mathsf{fv}(\lee{})}\\[2pt]
\gpfit{\mathsf{fv}(\CoImage(\lee{})) = \mathsf{fv}(\lsn{0}) \cup \mathsf{fv}(\lsn{1}) \cup \mathsf{fv}(\lee{})}\end{center}
}{\texttt{GLang.Space.fv}}{GLang/Syntax.lean:271}
\gldefdefn{valTy.fv}{\texttt{Val\allowbreak Ty.fv}}{clauses}{\mathsf{fv}(\ltynv{{\mathrm{x}}})}{\begin{center}\gpfit{\operatorname{fv}((y\colon \ltynv{0})\fnto \ltynv{1}) = \operatorname{fv}(\ltynv{0})\cup (\operatorname{fv}(\ltynv{1})\setminus\{y\})}\end{center}
}{\texttt{GLang.Val\allowbreak Ty.fv}}{GLang/Values.lean:376}
\gldefdefn{valExpr.fv}{\texttt{Val\allowbreak Expr.fv}}{clauses}{\mathsf{fv}(\leev{})}{\begin{center}\gpfit{\mathsf{fv}(\lvzero{}_{\lsv{}}) = \mathsf{fv}(\lsv{})}\\[2pt]
\gpfit{\mathsf{fv}(\mathit{a}) = \lemptyset{}}\\[2pt]
\gpfit{\mathsf{fv}(\lenv{0} \lvscale \lenv{1}) = \mathsf{fv}(\lenv{0}) \cup \mathsf{fv}(\lenv{1})}\\[2pt]
\gpfit{\mathsf{fv}(\lenv{0} \lvadd_{\lsv{}} \lenv{1}) = \mathsf{fv}(\lsv{}) \cup \mathsf{fv}(\lenv{0}) \cup \mathsf{fv}(\lenv{1})}\\[2pt]
\gpfit{\mathsf{fv}(\lenv{0}\losum \lenv{1}) = \mathsf{fv}(\lenv{0}) \cup \mathsf{fv}(\lenv{1})}\\[2pt]
\gpfit{\mathsf{fv}(\lenv{0}\loprod \lenv{1}) = \mathsf{fv}(\lenv{0}) \cup \mathsf{fv}(\lenv{1})}\\[2pt]
\gpfit{\mathsf{fv}(\ldelta{\mathit{a}}{\lsv{}}) = \mathsf{fv}(\lsv{})}\\[2pt]
\gpfit{\mathsf{fv}(\lidelta{\lsv{}}) = \mathsf{fv}(\lsv{})}\\[2pt]
\gpfit{\mathsf{fv}(\lsum^{\lsv{}}_{v\in \mathit{es}}v) = \mathsf{fv}(\lsv{}) \cup \mathsf{fvList}(\mathit{es})}\\[2pt]
\gpfit{\operatorname{fv}(\llambda(y\colon \ltynv{0})\lrh \ltynv{1}\colon \leev{}) = \operatorname{fv}(\ltynv{0})\cup (\operatorname{fv}(\ltynv{1})\setminus\{y\})\cup (\operatorname{fv}(\leev{})\setminus\{y\})}\\[2pt]
\gpfit{\operatorname{fv}(\mathit{a}_{\ltyv{}}) = \{\mathit{a}\}\cup \operatorname{fv}(\ltyv{})}\\[2pt]
\gpfit{\mathsf{fv}(\LTTVPP{\lsv{}}{\lenv{0}}{\lenv{1}}) = \mathsf{fv}(\lsv{}) \cup \mathsf{fv}(\lenv{0}) \cup \mathsf{fv}(\lenv{1})}\\[2pt]
\gpfit{\mathsf{fv}(\glvmtm{\lsnv{0}}{\lsnv{1}}{\lenv{0}}{\lenv{1}}) = \mathsf{fv}(\lsnv{0}) \cup \mathsf{fv}(\lsnv{1}) \cup \mathsf{fv}(\lenv{0}) \cup \mathsf{fv}(\lenv{1})}\\[2pt]
\gpfit{\mathsf{fv}(\glvapp{\lenv{0}}{\lenv{1}}) = \mathsf{fv}(\lenv{0}) \cup \mathsf{fv}(\lenv{1})}\\[2pt]
\gpfit{\mathsf{fv}(\vkrn(\lenv{0}, \lenv{1})) = \mathsf{fv}(\lenv{0}) \cup \mathsf{fv}(\lenv{1})}\\[2pt]
\gpfit{\mathsf{fv}(\vblk(\lenv{0}, \lenv{1})) = \mathsf{fv}(\lenv{0}) \cup \mathsf{fv}(\lenv{1})}\\[2pt]
\gpfit{\mathsf{fv}(\glvreshape{\leev{}}{\ltyv{}}{\ltynv{0}}) = \mathsf{fv}(\leev{}) \cup \mathsf{fv}(\ltyv{}) \cup \mathsf{fv}(\ltynv{0})}\\[2pt]
\gpfit{\mathsf{fv}(\Trans(\leev{})) = \mathsf{fv}(\leev{})}\\[2pt]
\gpfit{\mathsf{fv}(\glvop{GSVDS}(\leev{},\lenv{{\mathrm{a}}},\lenv{{\mathrm{b}}})) = \mathsf{fv}(\leev{}) \cup \mathsf{fv}(\lenv{{\mathrm{a}}}) \cup \mathsf{fv}(\lenv{{\mathrm{b}}})}\\[2pt]
\gpfit{\mathsf{fv}(\glvop{GSVDU}(\leev{},\lenv{{\mathrm{a}}},\lenv{{\mathrm{b}}})) = \mathsf{fv}(\leev{}) \cup \mathsf{fv}(\lenv{{\mathrm{a}}}) \cup \mathsf{fv}(\lenv{{\mathrm{b}}})}\\[2pt]
\gpfit{\mathsf{fv}(\glvop{GSVDUI}(\leev{},\lenv{{\mathrm{a}}},\lenv{{\mathrm{b}}})) = \mathsf{fv}(\leev{}) \cup \mathsf{fv}(\lenv{{\mathrm{a}}}) \cup \mathsf{fv}(\lenv{{\mathrm{b}}})}\\[2pt]
\gpfit{\mathsf{fv}(\glvop{GSVDV}(\leev{},\lenv{{\mathrm{a}}},\lenv{{\mathrm{b}}})) = \mathsf{fv}(\leev{}) \cup \mathsf{fv}(\lenv{{\mathrm{a}}}) \cup \mathsf{fv}(\lenv{{\mathrm{b}}})}\\[2pt]
\gpfit{\mathsf{fv}(\glvop{GSVDVI}(\leev{},\lenv{{\mathrm{a}}},\lenv{{\mathrm{b}}})) = \mathsf{fv}(\leev{}) \cup \mathsf{fv}(\lenv{{\mathrm{a}}}) \cup \mathsf{fv}(\lenv{{\mathrm{b}}})}\\[2pt]
\gpfit{\mathsf{fv}(\leev{}^{+}) = \mathsf{fv}(\leev{})}\\[2pt]
\gpfit{\mathsf{fv}(\glvop{GPINV}(\leev{},\lenv{{\mathrm{a}}},\lenv{{\mathrm{b}}})) = \mathsf{fv}(\leev{}) \cup \mathsf{fv}(\lenv{{\mathrm{a}}}) \cup \mathsf{fv}(\lenv{{\mathrm{b}}})}\\[2pt]
\gpfit{\mathsf{fv}(\glvop{RREFRep}(\lenv{0},\lenv{1})) = \mathsf{fv}(\lenv{0}) \cup \mathsf{fv}(\lenv{1})}\\[2pt]
\gpfit{\mathsf{fv}(\leev{}^{-1}) = \mathsf{fv}(\leev{})}\\[2pt]
\gpfit{\mathsf{fv}(\glvop{SVDS}(\leev{})) = \mathsf{fv}(\leev{})}\end{center}
}{\texttt{GLang.Val\allowbreak Expr.fv}}{GLang/Values.lean:381}
\gldefdefn{basis.fv}{\texttt{Basis.fv}}{clauses}{\mathsf{fv}(\textcolor{lang}{\mathit{x}})}{\begin{center}\gpfit{\mathsf{fv}(\lbzero{}) = \lemptyset{}}\\[2pt]
\gpfit{\mathsf{fv}(\lbone{}) = \lemptyset{}}\\[2pt]
\gpfit{\mathsf{fv}(\lb{}^{\lstar}) = \mathsf{fv}(\lb{})}\\[2pt]
\gpfit{\mathsf{fv}(\lbn{0} \losum \lbn{1}) = \mathsf{fv}(\lbn{0}) \cup \mathsf{fv}(\lbn{1})}\\[2pt]
\gpfit{\mathsf{fv}(\lbn{0} \loprod \lbn{1}) = \mathsf{fv}(\lbn{0}) \cup \mathsf{fv}(\lbn{1})}\\[2pt]
\gpfit{\mathsf{fv}(\pBasisC{\ls{}}{\lbn{0}}{\lbn{1}}) = \mathsf{fv}(\ls{}) \cup \mathsf{fv}(\lbn{0}) \cup \mathsf{fv}(\lbn{1})}\\[2pt]
\gpfit{\mathsf{fv}(\iBasisC{\lb{}}) = \mathsf{fv}(\lb{})}\\[2pt]
\gpfit{\mathsf{fv}(\llpl \mathit{es}\llpr_{\lb{}}) = \mathsf{fv}(\ls{}) \cup \mathsf{fv}(\lb{}) \cup \mathsf{fvList}(\mathit{es})}\\[2pt]
\gpfit{\operatorname{fv}(\frameB{y}{\lbn{0}}{\lbn{1}}) = \operatorname{fv}(\lbn{0})\cup (\operatorname{fv}(\lbn{1})\setminus\{y\})}\end{center}
}{\texttt{GLang.Basis.fv}}{GLang/Syntax.lean:327}
\gldefdefn{valSpace.fv}{\texttt{Val\allowbreak Space.fv}}{clauses}{\mathsf{fv}(\lsnv{{\mathrm{x}}})}{\begin{center}\gpfit{\mathsf{fv}(\lsempty{}) = \lemptyset{}}\\[2pt]
\gpfit{\mathsf{fv}(\lR{}) = \lemptyset{}}\\[2pt]
\gpfit{\mathsf{fv}(\lsnv{0} \losum \lsnv{1}) = \mathsf{fv}(\lsnv{0}) \cup \mathsf{fv}(\lsnv{1})}\\[2pt]
\gpfit{\mathsf{fv}(\lsnv{0} \loprod \lsnv{1}) = \mathsf{fv}(\lsnv{0}) \cup \mathsf{fv}(\lsnv{1})}\\[2pt]
\gpfit{\mathsf{fv}(\lR^{\leev{},\lprop_{\mathrm{a}},\lsv{}}) = \mathsf{fv}(\lsv{}) \cup \mathsf{fv}(\leev{})}\end{center}
}{\texttt{GLang.Val\allowbreak Space.fv}}{GLang/Values.lean:369}
\gldefdefn{ty.fv}{\texttt{Ty.fv}}{clauses}{\mathsf{fv}(\mathit{x})}{\begin{center}\gpfit{\operatorname{fv}((y\colon \ltyn{0})\fnto \ltyn{1}) = \operatorname{fv}(\ltyn{0})\cup (\operatorname{fv}(\ltyn{1})\setminus\{y\})}\end{center}
}{\texttt{GLang.Ty.fv}}{GLang/Syntax.lean:282}
\gldefdefn{term.fv}{\texttt{Term.fv}}{clauses}{\mathsf{fv}(\mathit{x})}{\begin{center}\gpfit{\mathsf{fv}(\mathit{y}) = \{\mathit{y}\}}\\[2pt]
\gpfit{\mathsf{fv}(\mathit{r}) = \lemptyset{}}\\[2pt]
\gpfit{\mathsf{fv}(\lvzero{}_{\ls{}}) = \mathsf{fv}(\ls{})}\\[2pt]
\gpfit{\mathsf{fv}(\len{0} \lvadd \len{1}) = \mathsf{fv}(\len{0}) \cup \mathsf{fv}(\len{1})}\\[2pt]
\gpfit{\mathsf{fv}(\len{0} \lvscale \len{1}) = \mathsf{fv}(\len{0}) \cup \mathsf{fv}(\len{1})}\\[2pt]
\gpfit{\mathsf{fv}(\len{0}\losum \len{1}) = \mathsf{fv}(\len{0}) \cup \mathsf{fv}(\len{1})}\\[2pt]
\gpfit{\mathsf{fv}(\len{0}\loprod \len{1}) = \mathsf{fv}(\len{0}) \cup \mathsf{fv}(\len{1})}\\[2pt]
\gpfit{\mathsf{fv}(\linner{}_{\ls{}}) = \mathsf{fv}(\ls{})}\\[2pt]
\gpfit{\mathsf{fv}(\lembd{}_{\ls{}}) = \mathsf{fv}(\ls{})}\\[2pt]
\gpfit{\mathsf{fv}(\lee{}'^{\lpseudo}) = \mathsf{fv}(\lsn{0}) \cup \mathsf{fv}(\lsn{1}) \cup \mathsf{fv}(\lee{}')\quad(\glrulename{hasType.pinvTensor})}\\[2pt]
\gpfit{\mathsf{fv}(\lee{}'^{\lpseudo}) = \mathsf{fv}(\ls{}) \cup \mathsf{fv}(\lee{}')\quad(\glrulename{hasType.pinvDual})}\\[2pt]
\gpfit{\operatorname{fv}(\lllambda(y\colon \lsn{0})\ldotsp \lee{}') = \operatorname{fv}(\lsn{0})\cup \operatorname{fv}(\ls{})\cup (\operatorname{fv}(\lee{}')\setminus\{y\})}\\[2pt]
\gpfit{\operatorname{fv}(\nllambda(y\colon \lty{})\ldotsp \lee{}') = \operatorname{fv}(\lty{})\cup (\operatorname{fv}(\lee{}')\setminus\{y\})}\\[2pt]
\gpfit{\mathsf{fv}(\leval{\len{0}}{}{\len{1}}) = \mathsf{fv}(\lsn{0}) \cup \mathsf{fv}(\lsn{1}) \cup \mathsf{fv}(\len{0}) \cup \mathsf{fv}(\len{1})\quad(\glrulename{hasType.appLinear})}\\[2pt]
\gpfit{\mathsf{fv}(\leval{\len{0}}{}{\len{1}}) = \mathsf{fv}(\ls{}) \cup \mathsf{fv}(\len{0}) \cup \mathsf{fv}(\len{1})\quad(\glrulename{hasType.appInner})}\\[2pt]
\gpfit{\mathsf{fv}(\leval{\len{0}}{}{\len{1}}) = \mathsf{fv}(\len{0}) \cup \mathsf{fv}(\len{1})\quad(\glrulename{hasType.appNonlinear})}\end{center}
}{\texttt{GLang.Term.fv}}{GLang/Syntax.lean:287}
\gldefdefn{valExpr.fvList}{Free variables of value lists}{clauses}{\mathsf{fvList}(\mathit{x})}{\begin{center}\gpfit{\mathsf{fvList}([]) = \lemptyset{}}\\[2pt]
\gpfit{\mathsf{fvList}(\leev{} \mathbin{:\!:} \mathit{es}) = \mathsf{fv}(\leev{}) \cup \mathsf{fvList}(\mathit{es})}\end{center}
}{\texttt{GLang.Val\allowbreak Expr.fv\allowbreak List}}{GLang/Values.lean:420}
\gldefdefn{term.fvList}{Free variables of term lists}{clauses}{\mathsf{fvList}(\mathit{x})}{\begin{center}\gpfit{\mathsf{fvList}([]) = \lemptyset{}}\\[2pt]
\gpfit{\mathsf{fvList}(\lee{} \mathbin{:\!:} \mathit{es}) = \mathsf{fv}(\lee{}) \cup \mathsf{fvList}(\mathit{es})}\end{center}
}{\texttt{GLang.Term.fv\allowbreak List}}{GLang/Syntax.lean:307}
\gldefdefn{valExpr.inferSpace}{inferSpace}{clauses}{\operatorname{inferSpace}(\leev{})}{\begin{center}\gpfit{\operatorname{inferSpace}(\lvzero{}_{\lsv{}}) = \lsv{}}\\[2pt]
\gpfit{\operatorname{inferSpace}(\mathit{a}) = \lR{}}\\[2pt]
\gpfit{\operatorname{inferSpace}(\lenv{{\mathrm{a}}} \lvscale \lenv{1}) = \operatorname{inferSpace}(\lenv{1})}\\[2pt]
\gpfit{\operatorname{inferSpace}(\lenv{{\mathrm{a}}} \lvadd_{\lsv{}} \lenv{{\mathrm{a1}}}) = \lsv{}}\\[2pt]
\gpfit{\operatorname{inferSpace}(\lenv{0}\losum \lenv{1}) = \operatorname{inferSpace}(\lenv{0}) \losum \operatorname{inferSpace}(\lenv{1})}\\[2pt]
\gpfit{\operatorname{inferSpace}(\lenv{0}\loprod \lenv{1}) = \operatorname{inferSpace}(\lenv{0}) \loprod \operatorname{inferSpace}(\lenv{1})}\\[2pt]
\gpfit{\operatorname{inferSpace}(\ldelta{\mathit{a}}{\lsv{}}) = \lsv{}}\\[2pt]
\gpfit{\operatorname{inferSpace}(\lidelta{\lsv{}}) = \lsv{} \loprod \lsv{}}\\[2pt]
\gpfit{\operatorname{inferSpace}(\lsum^{\lsnv{{\mathrm{a}}}}_{v\in [\leev{} \mathbin{:\!:} \mathit{tail}]}v) = \operatorname{inferSpace}(\leev{})}\\[2pt]
\gpfit{\operatorname{inferSpace}(\lsum^{\lsnv{{\mathrm{a}}}}_{v\in []}v) = \lsempty{}}\\[2pt]
\gpfit{\operatorname{inferSpace}(\llambda(x\colon \ltynv{{\mathrm{a}}})\lrh \ltynv{{\mathrm{a1}}}\colon \lenv{{\mathrm{a2}}}) = \lsempty{}}\\[2pt]
\gpfit{\operatorname{inferSpace}(\mathit{a}_{\ltyv{}}) = \mathsf{spaceOf}(\ltyv{})}\\[2pt]
\gpfit{\begin{gathered}\operatorname{inferSpace}(\LTTVPP{\lsnv{{\mathrm{a}}}}{\lenv{0}}{\lenv{{\mathrm{a1}}}}) = {}\\ \begin{cases} \lsnv{{\mathrm{0,1}}} & \text{if } \operatorname{inferSpace}(\lenv{0}) = \lsnv{{\mathrm{0,0}}} \loprod \lsnv{{\mathrm{0,1}}} \\ \lR{} & \text{otherwise} \end{cases}\end{gathered}}\\[2pt]
\gpfit{\begin{gathered}\operatorname{inferSpace}(\glvmtm{\lsnv{{\mathrm{a2}}}}{\lsnv{{\mathrm{a3}}}}{\lenv{0}}{\lenv{1}}) = {}\\ \begin{cases} \lsnv{{\mathrm{1,0}}} \loprod \lsnv{{\mathrm{0,1}}} & \text{if } \begin{gathered}\operatorname{inferSpace}(\lenv{0}) = \lsnv{{\mathrm{0,0}}} \loprod \lsnv{{\mathrm{0,1}}}\\\operatorname{inferSpace}(\lenv{1}) = \lsnv{{\mathrm{1,0}}} \loprod \lsnv{{\mathrm{1,1}}}\end{gathered} \\ \lsempty{} & \text{otherwise} \end{cases}\end{gathered}}\\[2pt]
\gpfit{\operatorname{inferSpace}(\glvapp{\lenv{{\mathrm{f}}}}{\lenv{{\mathrm{a}}}}) = \mathsf{spaceOf}(\operatorname{inferType}(\glvapp{\lenv{{\mathrm{f}}}}{\lenv{{\mathrm{a}}}}))}\\[2pt]
\gpfit{\begin{gathered}\operatorname{inferSpace}(\vkrn(\leev{}, \leev{}')) = {}\\ \begin{cases} \left(\lsnv{{\mathrm{0,0}}} \loprod \lsnv{{\mathrm{1,0}}}\right) \loprod \left(\lsnv{{\mathrm{0,1}}} \loprod \lsnv{{\mathrm{1,1}}}\right) & \text{if } \begin{gathered}\operatorname{inferSpace}(\leev{}) = \lsnv{{\mathrm{0,0}}} \loprod \lsnv{{\mathrm{0,1}}}\\\operatorname{inferSpace}(\leev{}') = \lsnv{{\mathrm{1,0}}} \loprod \lsnv{{\mathrm{1,1}}}\end{gathered} \\ \lsempty{} & \text{otherwise} \end{cases}\end{gathered}}\\[2pt]
\gpfit{\begin{gathered}\operatorname{inferSpace}(\vblk(\leev{}, \leev{}')) = {}\\ \begin{cases} \left(\lsnv{{\mathrm{0,0}}} \losum \lsnv{{\mathrm{1,0}}}\right) \loprod \left(\lsnv{{\mathrm{0,1}}} \losum \lsnv{{\mathrm{1,1}}}\right) & \text{if } \begin{gathered}\operatorname{inferSpace}(\leev{}) = \lsnv{{\mathrm{0,0}}} \loprod \lsnv{{\mathrm{0,1}}}\\\operatorname{inferSpace}(\leev{}') = \lsnv{{\mathrm{1,0}}} \loprod \lsnv{{\mathrm{1,1}}}\end{gathered} \\ \lsempty{} & \text{otherwise} \end{cases}\end{gathered}}\\[2pt]
\gpfit{\operatorname{inferSpace}(\glvreshape{\lenv{{\mathrm{a}}}}{\ltyv{}}{\ltynv{{\mathrm{a1}}}}) = \mathsf{spaceOf}(\ltyv{})}\\[2pt]
\gpfit{\begin{gathered}\operatorname{inferSpace}(\Trans(\leev{})) = {}\\ \begin{cases} \lsnv{{\mathrm{0,1}}} \loprod \lsnv{{\mathrm{0,0}}} & \text{if } \operatorname{inferSpace}(\leev{}) = \lsnv{{\mathrm{0,0}}} \loprod \lsnv{{\mathrm{0,1}}} \\ \lsempty{} & \text{otherwise} \end{cases}\end{gathered}}\\[2pt]
\gpfit{\operatorname{inferSpace}(\glvop{GSVDS}(\leev{},\lenv{{\mathrm{a}}},\lenv{{\mathrm{a1}}})) = \lR^{\leev{},\CoKernel{},\operatorname{inferSpace}(\leev{})} \loprod \lR^{\leev{},\Image{},\operatorname{inferSpace}(\leev{})}}\\[2pt]
\gpfit{\begin{gathered}\operatorname{inferSpace}(\glvop{GSVDU}(\leev{},\lenv{{\mathrm{a}}},\lenv{{\mathrm{a1}}})) = {}\\ \begin{cases} \lsnv{{\mathrm{0,0}}} \loprod \lR^{\leev{},\CoKernel{},\operatorname{inferSpace}(\leev{})} & \text{if } \operatorname{inferSpace}(\leev{}) = \lsnv{{\mathrm{0,0}}} \loprod \lsnv{{\mathrm{0,1}}} \\ \lsempty{} & \text{otherwise} \end{cases}\end{gathered}}\\[2pt]
\gpfit{\begin{gathered}\operatorname{inferSpace}(\glvop{GSVDUI}(\leev{},\lenv{{\mathrm{a}}},\lenv{{\mathrm{a1}}})) = {}\\ \begin{cases} \lsnv{{\mathrm{0,0}}} \loprod \lR^{\leev{},\Kernel{},\operatorname{inferSpace}(\leev{})} & \text{if } \operatorname{inferSpace}(\leev{}) = \lsnv{{\mathrm{0,0}}} \loprod \lsnv{{\mathrm{0,1}}} \\ \lsempty{} & \text{otherwise} \end{cases}\end{gathered}}\\[2pt]
\gpfit{\begin{gathered}\operatorname{inferSpace}(\glvop{GSVDV}(\leev{},\lenv{{\mathrm{a}}},\lenv{{\mathrm{a1}}})) = {}\\ \begin{cases} \lsnv{{\mathrm{0,1}}} \loprod \lR^{\leev{},\Image{},\operatorname{inferSpace}(\leev{})} & \text{if } \operatorname{inferSpace}(\leev{}) = \lsnv{{\mathrm{0,0}}} \loprod \lsnv{{\mathrm{0,1}}} \\ \lsempty{} & \text{otherwise} \end{cases}\end{gathered}}\\[2pt]
\gpfit{\begin{gathered}\operatorname{inferSpace}(\glvop{GSVDVI}(\leev{},\lenv{{\mathrm{a}}},\lenv{{\mathrm{a1}}})) = {}\\ \begin{cases} \lsnv{{\mathrm{0,1}}} \loprod \lR^{\leev{},\CoImage{},\operatorname{inferSpace}(\leev{})} & \text{if } \operatorname{inferSpace}(\leev{}) = \lsnv{{\mathrm{0,0}}} \loprod \lsnv{{\mathrm{0,1}}} \\ \lsempty{} & \text{otherwise} \end{cases}\end{gathered}}\\[2pt]
\gpfit{\begin{gathered}\operatorname{inferSpace}(\leev{}^{+}) = {}\\ \begin{cases} \lR{} \loprod \lsempty{} & \text{if } \operatorname{inferSpace}(\leev{}) = \lsempty{} \\ \lR{} \loprod \lR{} & \text{if } \operatorname{inferSpace}(\leev{}) = \lR{} \\ \lR{} \loprod \left(\lsnv{{\mathrm{0,0}}} \losum \lsnv{{\mathrm{0,1}}}\right) & \text{if } \operatorname{inferSpace}(\leev{}) = \lsnv{{\mathrm{0,0}}} \losum \lsnv{{\mathrm{0,1}}} \\ \lsnv{{\mathrm{0,1}}} \loprod \lsnv{{\mathrm{0,0}}} & \text{if } \operatorname{inferSpace}(\leev{}) = \lsnv{{\mathrm{0,0}}} \loprod \lsnv{{\mathrm{0,1}}} \\ \lR{} \loprod \lR^{\lenv{{\mathrm{0,2}}},\lprop_{\mathrm{0,0}},\lsnv{{\mathrm{0,1}}}} & \text{if } \operatorname{inferSpace}(\leev{}) = \lR^{\lenv{{\mathrm{0,2}}},\lprop_{\mathrm{0,0}},\lsnv{{\mathrm{0,1}}}} \end{cases}\end{gathered}}\\[2pt]
\gpfit{\begin{gathered}\operatorname{inferSpace}(\glvop{GPINV}(\leev{},\lenv{{\mathrm{a}}},\lenv{{\mathrm{a1}}})) = {}\\ \begin{cases} \lsnv{{\mathrm{0,1}}} \loprod \lsnv{{\mathrm{0,0}}} & \text{if } \operatorname{inferSpace}(\leev{}) = \lsnv{{\mathrm{0,0}}} \loprod \lsnv{{\mathrm{0,1}}} \\ \lsempty{} & \text{otherwise} \end{cases}\end{gathered}}\\[2pt]
\gpfit{\begin{gathered}\operatorname{inferSpace}(\glvop{RREFRep}(\lenv{{\mathrm{a}}},\lenv{1})) = {}\\ \begin{cases} \lsnv{{\mathrm{0,1}}} \loprod \lsnv{{\mathrm{0,0}}} & \text{if } \operatorname{inferSpace}(\lenv{1}) = \lsnv{{\mathrm{0,0}}} \loprod \lsnv{{\mathrm{0,1}}} \\ \lsempty{} & \text{otherwise} \end{cases}\end{gathered}}\\[2pt]
\gpfit{\begin{gathered}\operatorname{inferSpace}(\leev{}^{-1}) = {}\\ \begin{cases} \lR{} \loprod \lsempty{} & \text{if } \operatorname{inferSpace}(\leev{}) = \lsempty{} \\ \lR{} \loprod \lR{} & \text{if } \operatorname{inferSpace}(\leev{}) = \lR{} \\ \lR{} \loprod \left(\lsnv{{\mathrm{0,0}}} \losum \lsnv{{\mathrm{0,1}}}\right) & \text{if } \operatorname{inferSpace}(\leev{}) = \lsnv{{\mathrm{0,0}}} \losum \lsnv{{\mathrm{0,1}}} \\ \lsnv{{\mathrm{0,1}}} \loprod \lsnv{{\mathrm{0,0}}} & \text{if } \operatorname{inferSpace}(\leev{}) = \lsnv{{\mathrm{0,0}}} \loprod \lsnv{{\mathrm{0,1}}} \\ \lR{} \loprod \lR^{\lenv{{\mathrm{0,2}}},\lprop_{\mathrm{0,0}},\lsnv{{\mathrm{0,1}}}} & \text{if } \operatorname{inferSpace}(\leev{}) = \lR^{\lenv{{\mathrm{0,2}}},\lprop_{\mathrm{0,0}},\lsnv{{\mathrm{0,1}}}} \end{cases}\end{gathered}}\\[2pt]
\gpfit{\operatorname{inferSpace}(\glvop{SVDS}(\leev{})) = \lR^{\leev{},\CoKernel{},\operatorname{inferSpace}(\leev{})} \loprod \lR^{\leev{},\Image{},\operatorname{inferSpace}(\leev{})}}\end{center}
}{\texttt{GLang.Val\allowbreak Expr.infer\allowbreak Space}}{GLang/ValueDenot.lean:1020}
\gldefdefn{valExpr.inferSpaceCtx}{inferSpaceCtx}{clauses}{\operatorname{inferSpaceCtx}(\Gamma{}, \leev{})}{\begin{center}\gpfit{\operatorname{inferSpaceCtx}(\Gamma{}, \lvzero{}_{\lsv{}}) = \lsv{}}\\[2pt]
\gpfit{\operatorname{inferSpaceCtx}(\Gamma{}, \mathit{a}) = \lR{}}\\[2pt]
\gpfit{\operatorname{inferSpaceCtx}(\Gamma{}, \lenv{{\mathrm{a}}} \lvscale \lenv{1}) = \operatorname{inferSpaceCtx}(\Gamma{}, \lenv{1})}\\[2pt]
\gpfit{\operatorname{inferSpaceCtx}(\Gamma{}, \lenv{{\mathrm{a}}} \lvadd_{\lsv{}} \lenv{{\mathrm{a1}}}) = \lsv{}}\\[2pt]
\gpfit{\operatorname{inferSpaceCtx}(\Gamma{}, \lenv{0}\losum \lenv{1}) = \operatorname{inferSpaceCtx}(\Gamma{}, \lenv{0}) \losum \operatorname{inferSpaceCtx}(\Gamma{}, \lenv{1})}\\[2pt]
\gpfit{\operatorname{inferSpaceCtx}(\Gamma{}, \lenv{0}\loprod \lenv{1}) = \operatorname{inferSpaceCtx}(\Gamma{}, \lenv{0}) \loprod \operatorname{inferSpaceCtx}(\Gamma{}, \lenv{1})}\\[2pt]
\gpfit{\operatorname{inferSpaceCtx}(\Gamma{}, \ldelta{\mathit{a}}{\lsv{}}) = \lsv{}}\\[2pt]
\gpfit{\operatorname{inferSpaceCtx}(\Gamma{}, \lidelta{\lsv{}}) = \lsv{} \loprod \lsv{}}\\[2pt]
\gpfit{\operatorname{inferSpaceCtx}(\Gamma{}, \lsum^{\lsnv{{\mathrm{a}}}}_{v\in [\leev{} \mathbin{:\!:} \mathit{tail}]}v) = \operatorname{inferSpaceCtx}(\Gamma{}, \leev{})}\\[2pt]
\gpfit{\operatorname{inferSpaceCtx}(\Gamma{}, \lsum^{\lsnv{{\mathrm{a}}}}_{v\in []}v) = \lsempty{}}\\[2pt]
\gpfit{\operatorname{inferSpaceCtx}(\Gamma{}, \llambda(x\colon \ltynv{{\mathrm{a}}})\lrh \ltynv{{\mathrm{a1}}}\colon \lenv{{\mathrm{a2}}}) = \lsempty{}}\\[2pt]
\gpfit{\operatorname{inferSpaceCtx}(\Gamma{}, \mathbf{b}_{\mathit{a}}) = \mathsf{spaceOf}(\mathsf{lookupUse}(\Gamma{}, \mathit{a}))}\\[2pt]
\gpfit{\operatorname{inferSpaceCtx}(\Gamma{}, \mathit{a}_{\ltyv{}}) = \mathsf{spaceOf}(\ltyv{})}\\[2pt]
\gpfit{\begin{gathered}\operatorname{inferSpaceCtx}(\Gamma{}, \LTTVPP{\lsnv{{\mathrm{a}}}}{\lenv{0}}{\lenv{{\mathrm{a1}}}}) = {}\\ \begin{cases} \lsnv{{\mathrm{0,1}}} & \text{if } \operatorname{inferSpaceCtx}(\Gamma{}, \lenv{0}) = \lsnv{{\mathrm{0,0}}} \loprod \lsnv{{\mathrm{0,1}}} \\ \lR{} & \text{otherwise} \end{cases}\end{gathered}}\\[2pt]
\gpfit{\begin{gathered}\operatorname{inferSpaceCtx}(\Gamma{}, \glvmtm{\lsnv{{\mathrm{a2}}}}{\lsnv{{\mathrm{a3}}}}{\lenv{0}}{\lenv{1}}) = {}\\ \begin{cases} \lsnv{{\mathrm{1,0}}} \loprod \lsnv{{\mathrm{0,1}}} & \text{if } \begin{gathered}\operatorname{inferSpaceCtx}(\Gamma{}, \lenv{0}) = \lsnv{{\mathrm{0,0}}} \loprod \lsnv{{\mathrm{0,1}}}\\\operatorname{inferSpaceCtx}(\Gamma{}, \lenv{1}) = \lsnv{{\mathrm{1,0}}} \loprod \lsnv{{\mathrm{1,1}}}\end{gathered} \\ \lsempty{} & \text{otherwise} \end{cases}\end{gathered}}\\[2pt]
\gpfit{\operatorname{inferSpaceCtx}(\Gamma{}, \glvapp{\lenv{{\mathrm{f}}}}{\lenv{{\mathrm{a}}}}) = \mathsf{spaceOf}(\operatorname{inferTypeCtx}(\Gamma{}, \glvapp{\lenv{{\mathrm{f}}}}{\lenv{{\mathrm{a}}}}))}\\[2pt]
\gpfit{\begin{gathered}\operatorname{inferSpaceCtx}(\Gamma{}, \vkrn(\leev{}, \leev{}')) = {}\\ \begin{cases} \left(\lsnv{{\mathrm{0,0}}} \loprod \lsnv{{\mathrm{1,0}}}\right) \loprod \left(\lsnv{{\mathrm{0,1}}} \loprod \lsnv{{\mathrm{1,1}}}\right) & \text{if } \begin{gathered}\operatorname{inferSpaceCtx}(\Gamma{}, \leev{}) = \lsnv{{\mathrm{0,0}}} \loprod \lsnv{{\mathrm{0,1}}}\\\operatorname{inferSpaceCtx}(\Gamma{}, \leev{}') = \lsnv{{\mathrm{1,0}}} \loprod \lsnv{{\mathrm{1,1}}}\end{gathered} \\ \lsempty{} & \text{otherwise} \end{cases}\end{gathered}}\\[2pt]
\gpfit{\begin{gathered}\operatorname{inferSpaceCtx}(\Gamma{}, \vblk(\leev{}, \leev{}')) = {}\\ \begin{cases} \left(\lsnv{{\mathrm{0,0}}} \losum \lsnv{{\mathrm{1,0}}}\right) \loprod \left(\lsnv{{\mathrm{0,1}}} \losum \lsnv{{\mathrm{1,1}}}\right) & \text{if } \begin{gathered}\operatorname{inferSpaceCtx}(\Gamma{}, \leev{}) = \lsnv{{\mathrm{0,0}}} \loprod \lsnv{{\mathrm{0,1}}}\\\operatorname{inferSpaceCtx}(\Gamma{}, \leev{}') = \lsnv{{\mathrm{1,0}}} \loprod \lsnv{{\mathrm{1,1}}}\end{gathered} \\ \lsempty{} & \text{otherwise} \end{cases}\end{gathered}}\\[2pt]
\gpfit{\operatorname{inferSpaceCtx}(\Gamma{}, \glvreshape{\lenv{{\mathrm{a}}}}{\ltyv{}}{\ltynv{{\mathrm{a1}}}}) = \mathsf{spaceOf}(\ltyv{})}\\[2pt]
\gpfit{\begin{gathered}\operatorname{inferSpaceCtx}(\Gamma{}, \Trans(\leev{})) = {}\\ \begin{cases} \lsnv{{\mathrm{0,1}}} \loprod \lsnv{{\mathrm{0,0}}} & \text{if } \operatorname{inferSpaceCtx}(\Gamma{}, \leev{}) = \lsnv{{\mathrm{0,0}}} \loprod \lsnv{{\mathrm{0,1}}} \\ \lsempty{} & \text{otherwise} \end{cases}\end{gathered}}\\[2pt]
\gpfit{\operatorname{inferSpaceCtx}(\Gamma{}, \glvop{GSVDS}(\leev{},\lenv{{\mathrm{a}}},\lenv{{\mathrm{a1}}})) = \lR^{\leev{},\CoKernel{},\operatorname{inferSpaceCtx}(\Gamma{}, \leev{})} \loprod \lR^{\leev{},\Image{},\operatorname{inferSpaceCtx}(\Gamma{}, \leev{})}}\\[2pt]
\gpfit{\begin{gathered}\operatorname{inferSpaceCtx}(\Gamma{}, \glvop{GSVDU}(\leev{},\lenv{{\mathrm{a}}},\lenv{{\mathrm{a1}}})) = {}\\ \begin{cases} \lsnv{{\mathrm{0,0}}} \loprod \lR^{\leev{},\CoKernel{},\operatorname{inferSpaceCtx}(\Gamma{}, \leev{})} & \text{if } \operatorname{inferSpaceCtx}(\Gamma{}, \leev{}) = \lsnv{{\mathrm{0,0}}} \loprod \lsnv{{\mathrm{0,1}}} \\ \lsempty{} & \text{otherwise} \end{cases}\end{gathered}}\\[2pt]
\gpfit{\begin{gathered}\operatorname{inferSpaceCtx}(\Gamma{}, \glvop{GSVDUI}(\leev{},\lenv{{\mathrm{a}}},\lenv{{\mathrm{a1}}})) = {}\\ \begin{cases} \lsnv{{\mathrm{0,0}}} \loprod \lR^{\leev{},\Kernel{},\operatorname{inferSpaceCtx}(\Gamma{}, \leev{})} & \text{if } \operatorname{inferSpaceCtx}(\Gamma{}, \leev{}) = \lsnv{{\mathrm{0,0}}} \loprod \lsnv{{\mathrm{0,1}}} \\ \lsempty{} & \text{otherwise} \end{cases}\end{gathered}}\\[2pt]
\gpfit{\begin{gathered}\operatorname{inferSpaceCtx}(\Gamma{}, \glvop{GSVDV}(\leev{},\lenv{{\mathrm{a}}},\lenv{{\mathrm{a1}}})) = {}\\ \begin{cases} \lsnv{{\mathrm{0,1}}} \loprod \lR^{\leev{},\Image{},\operatorname{inferSpaceCtx}(\Gamma{}, \leev{})} & \text{if } \operatorname{inferSpaceCtx}(\Gamma{}, \leev{}) = \lsnv{{\mathrm{0,0}}} \loprod \lsnv{{\mathrm{0,1}}} \\ \lsempty{} & \text{otherwise} \end{cases}\end{gathered}}\\[2pt]
\gpfit{\begin{gathered}\operatorname{inferSpaceCtx}(\Gamma{}, \glvop{GSVDVI}(\leev{},\lenv{{\mathrm{a}}},\lenv{{\mathrm{a1}}})) = {}\\ \begin{cases} \lsnv{{\mathrm{0,1}}} \loprod \lR^{\leev{},\CoImage{},\operatorname{inferSpaceCtx}(\Gamma{}, \leev{})} & \text{if } \operatorname{inferSpaceCtx}(\Gamma{}, \leev{}) = \lsnv{{\mathrm{0,0}}} \loprod \lsnv{{\mathrm{0,1}}} \\ \lsempty{} & \text{otherwise} \end{cases}\end{gathered}}\\[2pt]
\gpfit{\begin{gathered}\operatorname{inferSpaceCtx}(\Gamma{}, \leev{}^{+}) = {}\\ \begin{cases} \lR{} \loprod \lsempty{} & \text{if } \operatorname{inferSpaceCtx}(\Gamma{}, \leev{}) = \lsempty{} \\ \lR{} \loprod \lR{} & \text{if } \operatorname{inferSpaceCtx}(\Gamma{}, \leev{}) = \lR{} \\ \lR{} \loprod \left(\lsnv{{\mathrm{0,0}}} \losum \lsnv{{\mathrm{0,1}}}\right) & \text{if } \operatorname{inferSpaceCtx}(\Gamma{}, \leev{}) = \lsnv{{\mathrm{0,0}}} \losum \lsnv{{\mathrm{0,1}}} \\ \lsnv{{\mathrm{0,1}}} \loprod \lsnv{{\mathrm{0,0}}} & \text{if } \operatorname{inferSpaceCtx}(\Gamma{}, \leev{}) = \lsnv{{\mathrm{0,0}}} \loprod \lsnv{{\mathrm{0,1}}} \\ \lR{} \loprod \lR^{\lenv{{\mathrm{0,2}}},\lprop_{\mathrm{0,0}},\lsnv{{\mathrm{0,1}}}} & \text{if } \operatorname{inferSpaceCtx}(\Gamma{}, \leev{}) = \lR^{\lenv{{\mathrm{0,2}}},\lprop_{\mathrm{0,0}},\lsnv{{\mathrm{0,1}}}} \end{cases}\end{gathered}}\\[2pt]
\gpfit{\begin{gathered}\operatorname{inferSpaceCtx}(\Gamma{}, \glvop{GPINV}(\leev{},\lenv{{\mathrm{a}}},\lenv{{\mathrm{a1}}})) = {}\\ \begin{cases} \lsnv{{\mathrm{0,1}}} \loprod \lsnv{{\mathrm{0,0}}} & \text{if } \operatorname{inferSpaceCtx}(\Gamma{}, \leev{}) = \lsnv{{\mathrm{0,0}}} \loprod \lsnv{{\mathrm{0,1}}} \\ \lsempty{} & \text{otherwise} \end{cases}\end{gathered}}\\[2pt]
\gpfit{\begin{gathered}\operatorname{inferSpaceCtx}(\Gamma{}, \glvop{RREFRep}(\lenv{{\mathrm{a}}},\lenv{1})) = {}\\ \begin{cases} \lsnv{{\mathrm{0,1}}} \loprod \lsnv{{\mathrm{0,0}}} & \text{if } \operatorname{inferSpaceCtx}(\Gamma{}, \lenv{1}) = \lsnv{{\mathrm{0,0}}} \loprod \lsnv{{\mathrm{0,1}}} \\ \lsempty{} & \text{otherwise} \end{cases}\end{gathered}}\\[2pt]
\gpfit{\begin{gathered}\operatorname{inferSpaceCtx}(\Gamma{}, \leev{}^{-1}) = {}\\ \begin{cases} \lR{} \loprod \lsempty{} & \text{if } \operatorname{inferSpaceCtx}(\Gamma{}, \leev{}) = \lsempty{} \\ \lR{} \loprod \lR{} & \text{if } \operatorname{inferSpaceCtx}(\Gamma{}, \leev{}) = \lR{} \\ \lR{} \loprod \left(\lsnv{{\mathrm{0,0}}} \losum \lsnv{{\mathrm{0,1}}}\right) & \text{if } \operatorname{inferSpaceCtx}(\Gamma{}, \leev{}) = \lsnv{{\mathrm{0,0}}} \losum \lsnv{{\mathrm{0,1}}} \\ \lsnv{{\mathrm{0,1}}} \loprod \lsnv{{\mathrm{0,0}}} & \text{if } \operatorname{inferSpaceCtx}(\Gamma{}, \leev{}) = \lsnv{{\mathrm{0,0}}} \loprod \lsnv{{\mathrm{0,1}}} \\ \lR{} \loprod \lR^{\lenv{{\mathrm{0,2}}},\lprop_{\mathrm{0,0}},\lsnv{{\mathrm{0,1}}}} & \text{if } \operatorname{inferSpaceCtx}(\Gamma{}, \leev{}) = \lR^{\lenv{{\mathrm{0,2}}},\lprop_{\mathrm{0,0}},\lsnv{{\mathrm{0,1}}}} \end{cases}\end{gathered}}\\[2pt]
\gpfit{\operatorname{inferSpaceCtx}(\Gamma{}, \glvop{SVDS}(\leev{})) = \lR^{\leev{},\CoKernel{},\operatorname{inferSpaceCtx}(\Gamma{}, \leev{})} \loprod \lR^{\leev{},\Image{},\operatorname{inferSpaceCtx}(\Gamma{}, \leev{})}}\end{center}
}{\texttt{GLang.Val\allowbreak Expr.infer\allowbreak Space\allowbreak Ctx}}{GLang/ValueDenot.lean:1139}
\gldefdefn{valExpr.inferType}{inferType}{clauses}{\operatorname{inferType}(\leev{})}{\begin{center}\gpfit{\operatorname{inferType}(\lvzero{}_{\lsv{}}) = \lsv{}}\\[2pt]
\gpfit{\operatorname{inferType}(\mathit{a}) = \lR{}}\\[2pt]
\gpfit{\operatorname{inferType}(\lenv{{\mathrm{a}}} \lvscale \lenv{1}) = \operatorname{inferType}(\lenv{1})}\\[2pt]
\gpfit{\operatorname{inferType}(\lenv{{\mathrm{a}}} \lvadd_{\lsv{}} \lenv{{\mathrm{a1}}}) = \lsv{}}\\[2pt]
\gpfit{\operatorname{inferType}(\lenv{0}\losum \lenv{1}) = \mathsf{spaceOf}(\operatorname{inferType}(\lenv{0})) \losum \mathsf{spaceOf}(\operatorname{inferType}(\lenv{1}))}\\[2pt]
\gpfit{\operatorname{inferType}(\lenv{0}\loprod \lenv{1}) = \mathsf{spaceOf}(\operatorname{inferType}(\lenv{0})) \loprod \mathsf{spaceOf}(\operatorname{inferType}(\lenv{1}))}\\[2pt]
\gpfit{\operatorname{inferType}(\ldelta{\mathit{a}}{\lsv{}}) = \lsv{}}\\[2pt]
\gpfit{\operatorname{inferType}(\lidelta{\lsv{}}) = \lsv{} \loprod \lsv{}}\\[2pt]
\gpfit{\operatorname{inferType}(\lsum^{\lsnv{{\mathrm{a}}}}_{v\in [\leev{} \mathbin{:\!:} \mathit{tail}]}v) = \operatorname{inferType}(\leev{})}\\[2pt]
\gpfit{\operatorname{inferType}(\lsum^{\lsnv{{\mathrm{a}}}}_{v\in []}v) = \lsempty{}}\\[2pt]
\gpfit{\operatorname{inferType}(\mathit{a}_{\ltyv{}}) = \ltyv{}}\\[2pt]
\gpfit{\begin{gathered}\operatorname{inferType}(\glvapp{\lenv{{\mathrm{f}}}}{\lenv{{\mathrm{a}}}}) = {}\\ \begin{cases} \ltynv{{\mathrm{0,1}}}\subst{x}{\lenv{{\mathrm{a}}}} & \text{if } \operatorname{inferType}(\lenv{{\mathrm{f}}}) = (x\colon \ltynv{{\mathrm{0,0}}}\fnto \ltynv{{\mathrm{0,1}}}) \\ \lsempty{} & \text{otherwise} \end{cases}\end{gathered}}\\[2pt]
\gpfit{\operatorname{inferType}(\llambda(x\colon \ltynv{0})\lrh \ltynv{1}\colon \lenv{{\mathrm{a}}}) = (x\colon \ltynv{0}\fnto \ltynv{1})}\\[2pt]
\gpfit{\operatorname{inferType}(\glvreshape{\lenv{{\mathrm{a}}}}{\ltyv{}}{\ltynv{{\mathrm{a1}}}}) = \ltyv{}}\\[2pt]
\gpfit{\begin{gathered}\operatorname{inferType}(\LTTVPP{\lsnv{{\mathrm{a}}}}{\lenv{0}}{\lenv{{\mathrm{a1}}}}) = {}\\ \begin{cases} \lsnv{{\mathrm{0,1}}} & \text{if } \operatorname{inferType}(\lenv{0}) = \lsnv{{\mathrm{0,0}}} \loprod \lsnv{{\mathrm{0,1}}} \\ \lR{} & \text{otherwise} \end{cases}\end{gathered}}\\[2pt]
\gpfit{\begin{gathered}\operatorname{inferType}(\glvmtm{\lsnv{{\mathrm{a2}}}}{\lsnv{{\mathrm{a3}}}}{\lenv{0}}{\lenv{1}}) = {}\\ \begin{cases} \lsnv{{\mathrm{1,0}}} \loprod \lsnv{{\mathrm{0,1}}} & \text{if } \begin{gathered}\operatorname{inferType}(\lenv{0}) = \lsnv{{\mathrm{0,0}}} \loprod \lsnv{{\mathrm{0,1}}}\\\operatorname{inferType}(\lenv{1}) = \lsnv{{\mathrm{1,0}}} \loprod \lsnv{{\mathrm{1,1}}}\end{gathered} \\ \lsempty{} & \text{otherwise} \end{cases}\end{gathered}}\\[2pt]
\gpfit{\begin{gathered}\operatorname{inferType}(\vkrn(\leev{}, \leev{}')) = {}\\ \begin{cases} \left(\lsnv{{\mathrm{0,0}}} \loprod \lsnv{{\mathrm{1,0}}}\right) \loprod \left(\lsnv{{\mathrm{0,1}}} \loprod \lsnv{{\mathrm{1,1}}}\right) & \text{if } \begin{gathered}\operatorname{inferType}(\leev{}) = \lsnv{{\mathrm{0,0}}} \loprod \lsnv{{\mathrm{0,1}}}\\\operatorname{inferType}(\leev{}') = \lsnv{{\mathrm{1,0}}} \loprod \lsnv{{\mathrm{1,1}}}\end{gathered} \\ \lsempty{} & \text{otherwise} \end{cases}\end{gathered}}\\[2pt]
\gpfit{\begin{gathered}\operatorname{inferType}(\vblk(\leev{}, \leev{}')) = {}\\ \begin{cases} \left(\lsnv{{\mathrm{0,0}}} \losum \lsnv{{\mathrm{1,0}}}\right) \loprod \left(\lsnv{{\mathrm{0,1}}} \losum \lsnv{{\mathrm{1,1}}}\right) & \text{if } \begin{gathered}\operatorname{inferType}(\leev{}) = \lsnv{{\mathrm{0,0}}} \loprod \lsnv{{\mathrm{0,1}}}\\\operatorname{inferType}(\leev{}') = \lsnv{{\mathrm{1,0}}} \loprod \lsnv{{\mathrm{1,1}}}\end{gathered} \\ \lsempty{} & \text{otherwise} \end{cases}\end{gathered}}\\[2pt]
\gpfit{\begin{gathered}\operatorname{inferType}(\Trans(\leev{})) = {}\\ \begin{cases} \lsnv{{\mathrm{0,1}}} \loprod \lsnv{{\mathrm{0,0}}} & \text{if } \operatorname{inferType}(\leev{}) = \lsnv{{\mathrm{0,0}}} \loprod \lsnv{{\mathrm{0,1}}} \\ \lsempty{} & \text{otherwise} \end{cases}\end{gathered}}\\[2pt]
\gpfit{\begin{gathered}\operatorname{inferType}(\glvop{GSVDS}(\leev{},\lenv{{\mathrm{a}}},\lenv{{\mathrm{a1}}})) = {}\\ \begin{cases} \lR^{\leev{},\CoKernel{},\lsnv{{\mathrm{0,0}}} \loprod \lsnv{{\mathrm{0,1}}}} \loprod \lR^{\leev{},\Image{},\lsnv{{\mathrm{0,0}}} \loprod \lsnv{{\mathrm{0,1}}}} & \text{if } \operatorname{inferType}(\leev{}) = \lsnv{{\mathrm{0,0}}} \loprod \lsnv{{\mathrm{0,1}}} \\ \lsempty{} & \text{otherwise} \end{cases}\end{gathered}}\\[2pt]
\gpfit{\begin{gathered}\operatorname{inferType}(\glvop{GSVDU}(\leev{},\lenv{{\mathrm{a}}},\lenv{{\mathrm{a1}}})) = {}\\ \begin{cases} \lsnv{{\mathrm{0,0}}} \loprod \lR^{\leev{},\CoKernel{},\lsnv{{\mathrm{0,0}}} \loprod \lsnv{{\mathrm{0,1}}}} & \text{if } \operatorname{inferType}(\leev{}) = \lsnv{{\mathrm{0,0}}} \loprod \lsnv{{\mathrm{0,1}}} \\ \lsempty{} & \text{otherwise} \end{cases}\end{gathered}}\\[2pt]
\gpfit{\begin{gathered}\operatorname{inferType}(\glvop{GSVDUI}(\leev{},\lenv{{\mathrm{a}}},\lenv{{\mathrm{a1}}})) = {}\\ \begin{cases} \lsnv{{\mathrm{0,0}}} \loprod \lR^{\leev{},\Kernel{},\lsnv{{\mathrm{0,0}}} \loprod \lsnv{{\mathrm{0,1}}}} & \text{if } \operatorname{inferType}(\leev{}) = \lsnv{{\mathrm{0,0}}} \loprod \lsnv{{\mathrm{0,1}}} \\ \lsempty{} & \text{otherwise} \end{cases}\end{gathered}}\\[2pt]
\gpfit{\begin{gathered}\operatorname{inferType}(\glvop{GSVDV}(\leev{},\lenv{{\mathrm{a}}},\lenv{{\mathrm{a1}}})) = {}\\ \begin{cases} \lsnv{{\mathrm{0,1}}} \loprod \lR^{\leev{},\Image{},\lsnv{{\mathrm{0,0}}} \loprod \lsnv{{\mathrm{0,1}}}} & \text{if } \operatorname{inferType}(\leev{}) = \lsnv{{\mathrm{0,0}}} \loprod \lsnv{{\mathrm{0,1}}} \\ \lsempty{} & \text{otherwise} \end{cases}\end{gathered}}\\[2pt]
\gpfit{\begin{gathered}\operatorname{inferType}(\glvop{GSVDVI}(\leev{},\lenv{{\mathrm{a}}},\lenv{{\mathrm{a1}}})) = {}\\ \begin{cases} \lsnv{{\mathrm{0,1}}} \loprod \lR^{\leev{},\CoImage{},\lsnv{{\mathrm{0,0}}} \loprod \lsnv{{\mathrm{0,1}}}} & \text{if } \operatorname{inferType}(\leev{}) = \lsnv{{\mathrm{0,0}}} \loprod \lsnv{{\mathrm{0,1}}} \\ \lsempty{} & \text{otherwise} \end{cases}\end{gathered}}\\[2pt]
\gpfit{\begin{gathered}\operatorname{inferType}(\leev{}^{+}) = {}\\ \begin{cases} \lR{} \loprod \lsempty{} & \text{if } \operatorname{inferType}(\leev{}) = \lsempty{} \\ \lR{} \loprod \lR{} & \text{if } \operatorname{inferType}(\leev{}) = \lR{} \\ \lR{} \loprod \left(\lsnv{{\mathrm{0,0}}} \losum \lsnv{{\mathrm{0,1}}}\right) & \text{if } \operatorname{inferType}(\leev{}) = \lsnv{{\mathrm{0,0}}} \losum \lsnv{{\mathrm{0,1}}} \\ \lsnv{{\mathrm{0,1}}} \loprod \lsnv{{\mathrm{0,0}}} & \text{if } \operatorname{inferType}(\leev{}) = \lsnv{{\mathrm{0,0}}} \loprod \lsnv{{\mathrm{0,1}}} \\ \lR{} \loprod \lR^{\lenv{{\mathrm{0,2}}},\lprop_{\mathrm{0,0}},\lsnv{{\mathrm{0,1}}}} & \text{if } \operatorname{inferType}(\leev{}) = \lR^{\lenv{{\mathrm{0,2}}},\lprop_{\mathrm{0,0}},\lsnv{{\mathrm{0,1}}}} \\ \lsempty{} & \text{if } \operatorname{inferType}(\leev{}) = (x\colon \ltynv{{\mathrm{0,0}}}\fnto \ltynv{{\mathrm{0,1}}}) \end{cases}\end{gathered}}\\[2pt]
\gpfit{\begin{gathered}\operatorname{inferType}(\glvop{GPINV}(\leev{},\lenv{{\mathrm{a}}},\lenv{{\mathrm{a1}}})) = {}\\ \begin{cases} \lsnv{{\mathrm{0,1}}} \loprod \lsnv{{\mathrm{0,0}}} & \text{if } \operatorname{inferType}(\leev{}) = \lsnv{{\mathrm{0,0}}} \loprod \lsnv{{\mathrm{0,1}}} \\ \lsempty{} & \text{otherwise} \end{cases}\end{gathered}}\\[2pt]
\gpfit{\begin{gathered}\operatorname{inferType}(\glvop{RREFRep}(\lenv{{\mathrm{a}}},\lenv{1})) = {}\\ \begin{cases} \lsnv{{\mathrm{0,1}}} \loprod \lsnv{{\mathrm{0,0}}} & \text{if } \operatorname{inferType}(\lenv{1}) = \lsnv{{\mathrm{0,0}}} \loprod \lsnv{{\mathrm{0,1}}} \\ \lsempty{} & \text{otherwise} \end{cases}\end{gathered}}\\[2pt]
\gpfit{\begin{gathered}\operatorname{inferType}(\leev{}^{-1}) = {}\\ \begin{cases} \lR{} \loprod \lsempty{} & \text{if } \operatorname{inferType}(\leev{}) = \lsempty{} \\ \lR{} \loprod \lR{} & \text{if } \operatorname{inferType}(\leev{}) = \lR{} \\ \lR{} \loprod \left(\lsnv{{\mathrm{0,0}}} \losum \lsnv{{\mathrm{0,1}}}\right) & \text{if } \operatorname{inferType}(\leev{}) = \lsnv{{\mathrm{0,0}}} \losum \lsnv{{\mathrm{0,1}}} \\ \lsnv{{\mathrm{0,1}}} \loprod \lsnv{{\mathrm{0,0}}} & \text{if } \operatorname{inferType}(\leev{}) = \lsnv{{\mathrm{0,0}}} \loprod \lsnv{{\mathrm{0,1}}} \\ \lR{} \loprod \lR^{\lenv{{\mathrm{0,2}}},\lprop_{\mathrm{0,0}},\lsnv{{\mathrm{0,1}}}} & \text{if } \operatorname{inferType}(\leev{}) = \lR^{\lenv{{\mathrm{0,2}}},\lprop_{\mathrm{0,0}},\lsnv{{\mathrm{0,1}}}} \\ \lsempty{} & \text{if } \operatorname{inferType}(\leev{}) = (x\colon \ltynv{{\mathrm{0,0}}}\fnto \ltynv{{\mathrm{0,1}}}) \end{cases}\end{gathered}}\\[2pt]
\gpfit{\begin{gathered}\operatorname{inferType}(\glvop{SVDS}(\leev{})) = {}\\ \begin{cases} \lR^{\leev{},\CoKernel{},\lsnv{{\mathrm{0,0}}} \loprod \lsnv{{\mathrm{0,1}}}} \loprod \lR^{\leev{},\Image{},\lsnv{{\mathrm{0,0}}} \loprod \lsnv{{\mathrm{0,1}}}} & \text{if } \operatorname{inferType}(\leev{}) = \lsnv{{\mathrm{0,0}}} \loprod \lsnv{{\mathrm{0,1}}} \\ \lsempty{} & \text{otherwise} \end{cases}\end{gathered}}\end{center}
}{\texttt{GLang.Val\allowbreak Expr.infer\allowbreak Type}}{GLang/ValueDenot.lean:790}
\gldefdefn{valExpr.inferTypeCtx}{inferTypeCtx}{clauses}{\operatorname{inferTypeCtx}(\Gamma{}, \leev{})}{\begin{center}\gpfit{\operatorname{inferTypeCtx}(\Gamma{}, \lvzero{}_{\lsv{}}) = \lsv{}}\\[2pt]
\gpfit{\operatorname{inferTypeCtx}(\Gamma{}, \mathit{a}) = \lR{}}\\[2pt]
\gpfit{\operatorname{inferTypeCtx}(\Gamma{}, \lenv{{\mathrm{a}}} \lvscale \lenv{1}) = \operatorname{inferTypeCtx}(\Gamma{}, \lenv{1})}\\[2pt]
\gpfit{\operatorname{inferTypeCtx}(\Gamma{}, \lenv{{\mathrm{a}}} \lvadd_{\lsv{}} \lenv{{\mathrm{a1}}}) = \lsv{}}\\[2pt]
\gpfit{\operatorname{inferTypeCtx}(\Gamma{}, \lenv{0}\losum \lenv{1}) = \mathsf{spaceOf}(\operatorname{inferTypeCtx}(\Gamma{}, \lenv{0})) \losum \mathsf{spaceOf}(\operatorname{inferTypeCtx}(\Gamma{}, \lenv{1}))}\\[2pt]
\gpfit{\operatorname{inferTypeCtx}(\Gamma{}, \lenv{0}\loprod \lenv{1}) = \mathsf{spaceOf}(\operatorname{inferTypeCtx}(\Gamma{}, \lenv{0})) \loprod \mathsf{spaceOf}(\operatorname{inferTypeCtx}(\Gamma{}, \lenv{1}))}\\[2pt]
\gpfit{\operatorname{inferTypeCtx}(\Gamma{}, \ldelta{\mathit{a}}{\lsv{}}) = \lsv{}}\\[2pt]
\gpfit{\operatorname{inferTypeCtx}(\Gamma{}, \lidelta{\lsv{}}) = \lsv{} \loprod \lsv{}}\\[2pt]
\gpfit{\operatorname{inferTypeCtx}(\Gamma{}, \lsum^{\lsnv{{\mathrm{a}}}}_{v\in [\leev{} \mathbin{:\!:} \mathit{tail}]}v) = \operatorname{inferTypeCtx}(\Gamma{}, \leev{})}\\[2pt]
\gpfit{\operatorname{inferTypeCtx}(\Gamma{}, \lsum^{\lsnv{{\mathrm{a}}}}_{v\in []}v) = \lsempty{}}\\[2pt]
\gpfit{\operatorname{inferTypeCtx}(\Gamma{}, \mathit{a}_{\ltyv{}}) = \ltyv{}}\\[2pt]
\gpfit{\operatorname{inferTypeCtx}(\Gamma{}, \mathbf{b}_{\mathit{a}}) = \mathsf{lookupUse}(\Gamma{}, \mathit{a})}\\[2pt]
\gpfit{\begin{gathered}\operatorname{inferTypeCtx}(\Gamma{}, \glvapp{\lenv{{\mathrm{f}}}}{\lenv{{\mathrm{a}}}}) = {}\\ \begin{cases} \ltynv{{\mathrm{0,1}}}\subst{x}{\lenv{{\mathrm{a}}}} & \text{if } \operatorname{inferTypeCtx}(\Gamma{}, \lenv{{\mathrm{f}}}) = (x\colon \ltynv{{\mathrm{0,0}}}\fnto \ltynv{{\mathrm{0,1}}}) \\ \lsempty{} & \text{otherwise} \end{cases}\end{gathered}}\\[2pt]
\gpfit{\operatorname{inferTypeCtx}(\Gamma{}, \llambda(x\colon \ltynv{0})\lrh \ltynv{1}\colon \lenv{{\mathrm{a}}}) = (x\colon \ltynv{0}\fnto \ltynv{1})}\\[2pt]
\gpfit{\operatorname{inferTypeCtx}(\Gamma{}, \glvreshape{\lenv{{\mathrm{a}}}}{\ltyv{}}{\ltynv{{\mathrm{a1}}}}) = \ltyv{}}\\[2pt]
\gpfit{\begin{gathered}\operatorname{inferTypeCtx}(\Gamma{}, \LTTVPP{\lsnv{{\mathrm{a}}}}{\lenv{0}}{\lenv{{\mathrm{a1}}}}) = {}\\ \begin{cases} \lsnv{{\mathrm{0,1}}} & \text{if } \operatorname{inferTypeCtx}(\Gamma{}, \lenv{0}) = \lsnv{{\mathrm{0,0}}} \loprod \lsnv{{\mathrm{0,1}}} \\ \lR{} & \text{otherwise} \end{cases}\end{gathered}}\\[2pt]
\gpfit{\begin{gathered}\operatorname{inferTypeCtx}(\Gamma{}, \glvmtm{\lsnv{{\mathrm{a2}}}}{\lsnv{{\mathrm{a3}}}}{\lenv{0}}{\lenv{1}}) = {}\\ \begin{cases} \lsnv{{\mathrm{1,0}}} \loprod \lsnv{{\mathrm{0,1}}} & \text{if } \begin{gathered}\operatorname{inferTypeCtx}(\Gamma{}, \lenv{0}) = \lsnv{{\mathrm{0,0}}} \loprod \lsnv{{\mathrm{0,1}}}\\\operatorname{inferTypeCtx}(\Gamma{}, \lenv{1}) = \lsnv{{\mathrm{1,0}}} \loprod \lsnv{{\mathrm{1,1}}}\end{gathered} \\ \lsempty{} & \text{otherwise} \end{cases}\end{gathered}}\\[2pt]
\gpfit{\begin{gathered}\operatorname{inferTypeCtx}(\Gamma{}, \vkrn(\leev{}, \leev{}')) = {}\\ \begin{cases} \left(\lsnv{{\mathrm{0,0}}} \loprod \lsnv{{\mathrm{1,0}}}\right) \loprod \left(\lsnv{{\mathrm{0,1}}} \loprod \lsnv{{\mathrm{1,1}}}\right) & \text{if } \begin{gathered}\operatorname{inferTypeCtx}(\Gamma{}, \leev{}) = \lsnv{{\mathrm{0,0}}} \loprod \lsnv{{\mathrm{0,1}}}\\\operatorname{inferTypeCtx}(\Gamma{}, \leev{}') = \lsnv{{\mathrm{1,0}}} \loprod \lsnv{{\mathrm{1,1}}}\end{gathered} \\ \lsempty{} & \text{otherwise} \end{cases}\end{gathered}}\\[2pt]
\gpfit{\begin{gathered}\operatorname{inferTypeCtx}(\Gamma{}, \vblk(\leev{}, \leev{}')) = {}\\ \begin{cases} \left(\lsnv{{\mathrm{0,0}}} \losum \lsnv{{\mathrm{1,0}}}\right) \loprod \left(\lsnv{{\mathrm{0,1}}} \losum \lsnv{{\mathrm{1,1}}}\right) & \text{if } \begin{gathered}\operatorname{inferTypeCtx}(\Gamma{}, \leev{}) = \lsnv{{\mathrm{0,0}}} \loprod \lsnv{{\mathrm{0,1}}}\\\operatorname{inferTypeCtx}(\Gamma{}, \leev{}') = \lsnv{{\mathrm{1,0}}} \loprod \lsnv{{\mathrm{1,1}}}\end{gathered} \\ \lsempty{} & \text{otherwise} \end{cases}\end{gathered}}\\[2pt]
\gpfit{\begin{gathered}\operatorname{inferTypeCtx}(\Gamma{}, \Trans(\leev{})) = {}\\ \begin{cases} \lsnv{{\mathrm{0,1}}} \loprod \lsnv{{\mathrm{0,0}}} & \text{if } \operatorname{inferTypeCtx}(\Gamma{}, \leev{}) = \lsnv{{\mathrm{0,0}}} \loprod \lsnv{{\mathrm{0,1}}} \\ \lsempty{} & \text{otherwise} \end{cases}\end{gathered}}\\[2pt]
\gpfit{\begin{gathered}\operatorname{inferTypeCtx}(\Gamma{}, \glvop{GSVDS}(\leev{},\lenv{{\mathrm{a}}},\lenv{{\mathrm{a1}}})) = {}\\ \begin{cases} \lR^{\leev{},\CoKernel{},\lsnv{{\mathrm{0,0}}} \loprod \lsnv{{\mathrm{0,1}}}} \loprod \lR^{\leev{},\Image{},\lsnv{{\mathrm{0,0}}} \loprod \lsnv{{\mathrm{0,1}}}} & \text{if } \operatorname{inferTypeCtx}(\Gamma{}, \leev{}) = \lsnv{{\mathrm{0,0}}} \loprod \lsnv{{\mathrm{0,1}}} \\ \lsempty{} & \text{otherwise} \end{cases}\end{gathered}}\\[2pt]
\gpfit{\begin{gathered}\operatorname{inferTypeCtx}(\Gamma{}, \glvop{GSVDU}(\leev{},\lenv{{\mathrm{a}}},\lenv{{\mathrm{a1}}})) = {}\\ \begin{cases} \lsnv{{\mathrm{0,0}}} \loprod \lR^{\leev{},\CoKernel{},\lsnv{{\mathrm{0,0}}} \loprod \lsnv{{\mathrm{0,1}}}} & \text{if } \operatorname{inferTypeCtx}(\Gamma{}, \leev{}) = \lsnv{{\mathrm{0,0}}} \loprod \lsnv{{\mathrm{0,1}}} \\ \lsempty{} & \text{otherwise} \end{cases}\end{gathered}}\\[2pt]
\gpfit{\begin{gathered}\operatorname{inferTypeCtx}(\Gamma{}, \glvop{GSVDUI}(\leev{},\lenv{{\mathrm{a}}},\lenv{{\mathrm{a1}}})) = {}\\ \begin{cases} \lsnv{{\mathrm{0,0}}} \loprod \lR^{\leev{},\Kernel{},\lsnv{{\mathrm{0,0}}} \loprod \lsnv{{\mathrm{0,1}}}} & \text{if } \operatorname{inferTypeCtx}(\Gamma{}, \leev{}) = \lsnv{{\mathrm{0,0}}} \loprod \lsnv{{\mathrm{0,1}}} \\ \lsempty{} & \text{otherwise} \end{cases}\end{gathered}}\\[2pt]
\gpfit{\begin{gathered}\operatorname{inferTypeCtx}(\Gamma{}, \glvop{GSVDV}(\leev{},\lenv{{\mathrm{a}}},\lenv{{\mathrm{a1}}})) = {}\\ \begin{cases} \lsnv{{\mathrm{0,1}}} \loprod \lR^{\leev{},\Image{},\lsnv{{\mathrm{0,0}}} \loprod \lsnv{{\mathrm{0,1}}}} & \text{if } \operatorname{inferTypeCtx}(\Gamma{}, \leev{}) = \lsnv{{\mathrm{0,0}}} \loprod \lsnv{{\mathrm{0,1}}} \\ \lsempty{} & \text{otherwise} \end{cases}\end{gathered}}\\[2pt]
\gpfit{\begin{gathered}\operatorname{inferTypeCtx}(\Gamma{}, \glvop{GSVDVI}(\leev{},\lenv{{\mathrm{a}}},\lenv{{\mathrm{a1}}})) = {}\\ \begin{cases} \lsnv{{\mathrm{0,1}}} \loprod \lR^{\leev{},\CoImage{},\lsnv{{\mathrm{0,0}}} \loprod \lsnv{{\mathrm{0,1}}}} & \text{if } \operatorname{inferTypeCtx}(\Gamma{}, \leev{}) = \lsnv{{\mathrm{0,0}}} \loprod \lsnv{{\mathrm{0,1}}} \\ \lsempty{} & \text{otherwise} \end{cases}\end{gathered}}\\[2pt]
\gpfit{\begin{gathered}\operatorname{inferTypeCtx}(\Gamma{}, \leev{}^{+}) = {}\\ \begin{cases} \lR{} \loprod \lsempty{} & \text{if } \operatorname{inferTypeCtx}(\Gamma{}, \leev{}) = \lsempty{} \\ \lR{} \loprod \lR{} & \text{if } \operatorname{inferTypeCtx}(\Gamma{}, \leev{}) = \lR{} \\ \lR{} \loprod \left(\lsnv{{\mathrm{0,0}}} \losum \lsnv{{\mathrm{0,1}}}\right) & \text{if } \operatorname{inferTypeCtx}(\Gamma{}, \leev{}) = \lsnv{{\mathrm{0,0}}} \losum \lsnv{{\mathrm{0,1}}} \\ \lsnv{{\mathrm{0,1}}} \loprod \lsnv{{\mathrm{0,0}}} & \text{if } \operatorname{inferTypeCtx}(\Gamma{}, \leev{}) = \lsnv{{\mathrm{0,0}}} \loprod \lsnv{{\mathrm{0,1}}} \\ \lR{} \loprod \lR^{\lenv{{\mathrm{0,2}}},\lprop_{\mathrm{0,0}},\lsnv{{\mathrm{0,1}}}} & \text{if } \operatorname{inferTypeCtx}(\Gamma{}, \leev{}) = \lR^{\lenv{{\mathrm{0,2}}},\lprop_{\mathrm{0,0}},\lsnv{{\mathrm{0,1}}}} \\ \lsempty{} & \text{if } \operatorname{inferTypeCtx}(\Gamma{}, \leev{}) = (x\colon \ltynv{{\mathrm{0,0}}}\fnto \ltynv{{\mathrm{0,1}}}) \end{cases}\end{gathered}}\\[2pt]
\gpfit{\begin{gathered}\operatorname{inferTypeCtx}(\Gamma{}, \glvop{GPINV}(\leev{},\lenv{{\mathrm{a}}},\lenv{{\mathrm{a1}}})) = {}\\ \begin{cases} \lsnv{{\mathrm{0,1}}} \loprod \lsnv{{\mathrm{0,0}}} & \text{if } \operatorname{inferTypeCtx}(\Gamma{}, \leev{}) = \lsnv{{\mathrm{0,0}}} \loprod \lsnv{{\mathrm{0,1}}} \\ \lsempty{} & \text{otherwise} \end{cases}\end{gathered}}\\[2pt]
\gpfit{\begin{gathered}\operatorname{inferTypeCtx}(\Gamma{}, \glvop{RREFRep}(\lenv{{\mathrm{a}}},\lenv{1})) = {}\\ \begin{cases} \lsnv{{\mathrm{0,1}}} \loprod \lsnv{{\mathrm{0,0}}} & \text{if } \operatorname{inferTypeCtx}(\Gamma{}, \lenv{1}) = \lsnv{{\mathrm{0,0}}} \loprod \lsnv{{\mathrm{0,1}}} \\ \lsempty{} & \text{otherwise} \end{cases}\end{gathered}}\\[2pt]
\gpfit{\begin{gathered}\operatorname{inferTypeCtx}(\Gamma{}, \leev{}^{-1}) = {}\\ \begin{cases} \lR{} \loprod \lsempty{} & \text{if } \operatorname{inferTypeCtx}(\Gamma{}, \leev{}) = \lsempty{} \\ \lR{} \loprod \lR{} & \text{if } \operatorname{inferTypeCtx}(\Gamma{}, \leev{}) = \lR{} \\ \lR{} \loprod \left(\lsnv{{\mathrm{0,0}}} \losum \lsnv{{\mathrm{0,1}}}\right) & \text{if } \operatorname{inferTypeCtx}(\Gamma{}, \leev{}) = \lsnv{{\mathrm{0,0}}} \losum \lsnv{{\mathrm{0,1}}} \\ \lsnv{{\mathrm{0,1}}} \loprod \lsnv{{\mathrm{0,0}}} & \text{if } \operatorname{inferTypeCtx}(\Gamma{}, \leev{}) = \lsnv{{\mathrm{0,0}}} \loprod \lsnv{{\mathrm{0,1}}} \\ \lR{} \loprod \lR^{\lenv{{\mathrm{0,2}}},\lprop_{\mathrm{0,0}},\lsnv{{\mathrm{0,1}}}} & \text{if } \operatorname{inferTypeCtx}(\Gamma{}, \leev{}) = \lR^{\lenv{{\mathrm{0,2}}},\lprop_{\mathrm{0,0}},\lsnv{{\mathrm{0,1}}}} \\ \lsempty{} & \text{if } \operatorname{inferTypeCtx}(\Gamma{}, \leev{}) = (x\colon \ltynv{{\mathrm{0,0}}}\fnto \ltynv{{\mathrm{0,1}}}) \end{cases}\end{gathered}}\\[2pt]
\gpfit{\begin{gathered}\operatorname{inferTypeCtx}(\Gamma{}, \glvop{SVDS}(\leev{})) = {}\\ \begin{cases} \lR^{\leev{},\CoKernel{},\lsnv{{\mathrm{0,0}}} \loprod \lsnv{{\mathrm{0,1}}}} \loprod \lR^{\leev{},\Image{},\lsnv{{\mathrm{0,0}}} \loprod \lsnv{{\mathrm{0,1}}}} & \text{if } \operatorname{inferTypeCtx}(\Gamma{}, \leev{}) = \lsnv{{\mathrm{0,0}}} \loprod \lsnv{{\mathrm{0,1}}} \\ \lsempty{} & \text{otherwise} \end{cases}\end{gathered}}\end{center}
}{\texttt{GLang.Val\allowbreak Expr.infer\allowbreak Type\allowbreak Ctx}}{GLang/ValueDenot.lean:918}
\gldefdefn{valExpr.instantiate}{\texttt{Val\allowbreak Expr.instantiate}}{clauses}{\leev{}\subst{x}{\lenv{{\mathrm{u}}}}}{\begin{center}\gpfit{\mathsf{instantiate}(\lenv{{\mathrm{u}}}, \leev{}) = \mathsf{openRec}(0, \lenv{{\mathrm{u}}}, \leev{})}\end{center}
}{\texttt{GLang.Val\allowbreak Expr.instantiate}}{GLang/ValueLN.lean:509}
\gldefdefn{ty.instantiate}{\texttt{Ty.instantiate}}{clauses}{\lty{}\subst{x}{\mathit{u}}}{\begin{center}\gpfit{\mathsf{instantiate}(\mathit{u}, \lty{}) = \mathsf{openRec}(0, \mathit{u}, \lty{})}\end{center}
}{\texttt{GLang.Ty.instantiate}}{GLang/SourceLN.lean:162}
\gldefdefn{valTy.instantiate}{\texttt{Val\allowbreak Ty.instantiate}}{clauses}{\ltyv{}\subst{x}{\lenv{{\mathrm{u}}}}}{\begin{center}\gpfit{\mathsf{instantiate}(\lenv{{\mathrm{u}}}, \ltyv{}) = \mathsf{openRec}(0, \lenv{{\mathrm{u}}}, \ltyv{})}\end{center}
}{\texttt{GLang.Val\allowbreak Ty.instantiate}}{GLang/ValueLN.lean:511}
\gldefdefn{basis.instantiate}{\texttt{Basis.instantiate}}{clauses}{\lb{}\subst{x}{\mathit{u}}}{\begin{center}\gpfit{\mathsf{instantiate}(\mathit{u}, \lb{}) = \mathsf{openRec}(0, \mathit{u}, \lb{})}\end{center}
}{\texttt{GLang.Basis.instantiate}}{GLang/SourceLN.lean:163}
\gldefdefn{term.instantiate}{\texttt{Term.instantiate}}{clauses}{\lee{}\subst{x}{\mathit{u}}}{\begin{center}\gpfit{\mathsf{instantiate}(\mathit{u}, \lee{}) = \mathsf{openRec}(0, \mathit{u}, \lee{})}\end{center}
}{\texttt{GLang.Term.instantiate}}{GLang/SourceLN.lean:160}
\gldefdefn{basis.isStd}{Syntactically standard basis}{clauses}{\isStdBasis{\textcolor{lang}{\mathit{x}}}}{\begin{center}\gpfit{\isStdBasis{\lbzero{}} = \mathsf{True}}\\[2pt]
\gpfit{\isStdBasis{\lbone{}} = \mathsf{True}}\\[2pt]
\gpfit{\isStdBasis{\lb{}^{\lstar}} = \isStdBasis{\lb{}}}\\[2pt]
\gpfit{\isStdBasis{\lbn{0} \losum \lbn{1}} = \isStdBasis{\lbn{0}} \;\wedge\; \isStdBasis{\lbn{1}}}\\[2pt]
\gpfit{\isStdBasis{\lbn{0} \loprod \lbn{1}} = \isStdBasis{\lbn{0}} \;\wedge\; \isStdBasis{\lbn{1}}}\\[2pt]
\gpfit{\isStdBasis{\pBasisC{\ls{}}{\lbn{0}}{\lbn{1}}}=\isStdBasis{\lbn{0}}\land\isStdBasis{\lbn{1}}}\\[2pt]
\gpfit{\isStdBasis{\iBasisC{\lb{}}}=\operatorname{false}}\\[2pt]
\gpfit{\isStdBasis{\llpl{}\len{0},\dots,\len{n}\llpr{}_{\lb{}}}=\operatorname{false}}\\[2pt]
\gpfit{\isStdBasis{\frameB{x}{\lbn{0}}{\lbn{1}}} = \isStdBasis{\lbn{0}} \;\wedge\; \isStdBasis{\lbn{1}}}\end{center}
}{\texttt{GLang.Basis.is\allowbreak Std}}{GLang/ValueOperational.lean:259}
\gldefequation{basis.isStd.empty}{\isStdBasis{\lbzero{}} = \mathsf{True}}
\gldefequation{basis.isStd.one}{\isStdBasis{\lbone{}} = \mathsf{True}}
\gldefequation{basis.isStd.dual}{\isStdBasis{\lb{}^{\lstar}} = \isStdBasis{\lb{}}}
\gldefequation{basis.isStd.dsum}{\isStdBasis{\lbn{0} \losum \lbn{1}} = \isStdBasis{\lbn{0}} \;\wedge\; \isStdBasis{\lbn{1}}}
\gldefequation{basis.isStd.tprod}{\isStdBasis{\lbn{0} \loprod \lbn{1}} = \isStdBasis{\lbn{0}} \;\wedge\; \isStdBasis{\lbn{1}}}
\gldefequation{basis.isStd.proj}{\isStdBasis{\pBasisC{\ls{}}{\lbn{0}}{\lbn{1}}}=\isStdBasis{\lbn{0}}\land\isStdBasis{\lbn{1}}}
\gldefequation{basis.isStd.image}{\isStdBasis{\iBasisC{\lb{}}}=\operatorname{false}}
\gldefequation{basis.isStd.terms}{\isStdBasis{\llpl{}\len{0},\dots,\len{n}\llpr{}_{\lb{}}}=\operatorname{false}}
\gldefequation{basis.isStd.frame}{\isStdBasis{\frameB{x}{\lbn{0}}{\lbn{1}}} = \isStdBasis{\lbn{0}} \;\wedge\; \isStdBasis{\lbn{1}}}
\gldefdefn{dVal.matchesShape}{\texttt{DVal.matches\allowbreak Shape}}{expanded definition}{\mathsf{matchesShape}(\leev{}, \mathit{T})}{\begin{center}\gpfit{\mathsf{shape}(\leev{}) = \mathit{T}}\end{center}
}{\texttt{GLang.DVal.matches\allowbreak Shape}}{GLang/Denot.lean:563}
\gldefdefn{valueShape.matrices}{\texttt{Value\allowbreak Shape.matrices}}{clauses}{\operatorname{Mat}_{\mathit{q}}(\lsv{},\mathit{a}) \land \operatorname{Mat}_{\mathit{q}}(\lsnv{{\mathrm{t}}},\lb{})}{\begin{center}\gpfit{\operatorname{Mat}_{\mathit{q}}(\lsv{},\mathit{a}) \land \operatorname{Mat}_{\mathit{q}}(\lsnv{{\mathrm{t}}},\lb{}) = \operatorname{Mat}_{\mathit{q}}(\lsv{},\mathit{a}) \;\wedge\; \operatorname{Mat}_{\mathit{q}}(\lsnv{{\mathrm{t}}},\lb{})}\end{center}
}{\texttt{GLang.Value\allowbreak Shape.matrices}}{GLang/ValuePrimitiveShapes.lean:21}
\gldefdefn{valueShape.mtm}{\texttt{Value\allowbreak Shape.mtm}}{clauses}{\exists s_0,s_1,t_0,t_1.\; \lsv{}=s_0\otimes s_1 \land \lsnv{{\mathrm{t}}}=t_0\otimes t_1 \land |\mathit{a}|=d_{\mathit{q}}(s_0)d_{\mathit{q}}(s_1) \land |\lb{}|=d_{\mathit{q}}(t_0)d_{\mathit{q}}(t_1) \land d_{\mathit{q}}(s_0)=d_{\mathit{q}}(t_1)=d_{\mathit{q}}(\lsnv{{\mathrm{contract}}})}{\begin{center}\gpfit{\exists s_0,s_1,t_0,t_1.\; \lsnv{{\mathrm{domain}}} \loprod \lsnv{{\mathrm{a1}}}=s_0\otimes s_1 \land \lsnv{{\mathrm{a2}}} \loprod \lsnv{{\mathrm{codomain}}}=t_0\otimes t_1 \land |\mathit{a}|=d_{\mathit{q}}(s_0)d_{\mathit{q}}(s_1) \land |\lb{}|=d_{\mathit{q}}(t_0)d_{\mathit{q}}(t_1) \land d_{\mathit{q}}(s_0)=d_{\mathit{q}}(t_1)=d_{\mathit{q}}(\lsnv{{\mathrm{contract}}}) = \operatorname{Mat}_{\mathit{q}}(\lsnv{{\mathrm{domain}}} \loprod \lsnv{{\mathrm{a1}}},\mathit{a}) \land \operatorname{Mat}_{\mathit{q}}(\lsnv{{\mathrm{a2}}} \loprod \lsnv{{\mathrm{codomain}}},\lb{}) \;\wedge\; \dim_{\mathit{q}}(\lsnv{{\mathrm{domain}}}) = \dim_{\mathit{q}}(\lsnv{{\mathrm{contract}}}) \;\wedge\; \dim_{\mathit{q}}(\lsnv{{\mathrm{codomain}}}) = \dim_{\mathit{q}}(\lsnv{{\mathrm{contract}}})}\\[2pt]
\gpfit{\exists s_0,s_1,t_0,t_1.\; \lsv{}=s_0\otimes s_1 \land \lsnv{{\mathrm{t}}}=t_0\otimes t_1 \land |\mathit{a}|=d_{\mathit{q}}(s_0)d_{\mathit{q}}(s_1) \land |\lb{}|=d_{\mathit{q}}(t_0)d_{\mathit{q}}(t_1) \land d_{\mathit{q}}(s_0)=d_{\mathit{q}}(t_1)=d_{\mathit{q}}(\lsnv{{\mathrm{contract}}}) = \operatorname{Mat}_{\mathit{q}}(\lsv{},\mathit{a}) \land \operatorname{Mat}_{\mathit{q}}(\lsnv{{\mathrm{t}}},\lb{}) \;\wedge\; \mathsf{False}}\end{center}
}{\texttt{GLang.Value\allowbreak Shape.mtm}}{GLang/ValuePrimitiveShapes.lean:33}
\gldefdefn{basis.openRec}{\texttt{Basis.open\allowbreak Rec}}{clauses}{({\textcolor{lang}{\mathit{x}}}^{\lky{}\leftarrow \mathit{u}})}{\begin{center}\gpfit{\mathsf{openRec}(\lky{}, \mathit{u}, \lbzero{}) = \lbzero{}}\\[2pt]
\gpfit{\mathsf{openRec}(\lky{}, \mathit{u}, \lbone{}) = \lbone{}}\\[2pt]
\gpfit{\mathsf{openRec}(\lky{}, \mathit{u}, \mathsf{dual}(\lb{})) = \mathsf{dual}(\mathsf{openRec}(\lky{}, \mathit{u}, \lb{}))}\\[2pt]
\gpfit{\mathsf{openRec}(\lky{}, \mathit{u}, \mathsf{dsum}(\lbn{0}, \lbn{1})) = \mathsf{dsum}(\mathsf{openRec}(\lky{}, \mathit{u}, \lbn{0}), \mathsf{openRec}(\lky{}, \mathit{u}, \lbn{1}))}\\[2pt]
\gpfit{\mathsf{openRec}(\lky{}, \mathit{u}, \mathsf{tprod}(\lbn{0}, \lbn{1})) = \mathsf{tprod}(\mathsf{openRec}(\lky{}, \mathit{u}, \lbn{0}), \mathsf{openRec}(\lky{}, \mathit{u}, \lbn{1}))}\\[2pt]
\gpfit{\mathsf{openRec}(\lky{}, \mathit{u}, \mathsf{proj}(\ls{}, \lbn{0}, \lbn{1})) = \mathsf{proj}(\mathsf{openRec}(\lky{}, \mathit{u}, \ls{}), \mathsf{openRec}(\lky{}, \mathit{u}, \lbn{0}), \mathsf{openRec}(\lky{}, \mathit{u}, \lbn{1}))}\\[2pt]
\gpfit{\mathsf{openRec}(\lky{}, \mathit{u}, \mathsf{image}(\lb{})) = \mathsf{image}(\mathsf{openRec}(\lky{}, \mathit{u}, \lb{}))}\\[2pt]
\gpfit{\mathsf{openRec}(\lky{}, \mathit{u}, \mathsf{terms}(\ls{}, \mathit{es}, \lb{})) = \mathsf{terms}(\mathsf{openRec}(\lky{}, \mathit{u}, \ls{}), \mathsf{openRecList}(\lky{}, \mathit{u}, \mathit{es}), \mathsf{openRec}(\lky{}, \mathit{u}, \lb{}))}\\[2pt]
\gpfit{\mathsf{openRec}(\lky{}, \mathit{u}, \mathsf{frame}(\lbn{0}, \lbn{1})) = \mathsf{frame}(\mathsf{openRec}(\lky{}, \mathit{u}, \lbn{0}), \mathsf{openRec}(\lky{} + 1, \mathit{u}, \lbn{1}))}\end{center}
}{\texttt{GLang.Basis.open\allowbreak Rec}}{GLang/SourceLN.lean:89}
\gldefdefn{valExpr.openRec}{\texttt{Val\allowbreak Expr.open\allowbreak Rec}}{clauses}{({\leev{}}^{\lky{}\leftarrow \lenv{{\mathrm{u}}}})}{\begin{center}\gpfit{\begin{gathered}\mathsf{openRec}(\lky{}, \lenv{{\mathrm{u}}}, \mathsf{bvar}(\mathit{i})) = \begin{cases} \lenv{{\mathrm{u}}} & \text{if } \mathit{i} = \lky{} \\ \mathsf{bvar}(\mathit{i}) & \text{otherwise} \end{cases}\end{gathered}}\\[2pt]
\gpfit{\mathsf{openRec}(\lky{}, \lenv{{\mathrm{u}}}, \mathsf{var}({\mathit{x}_{1}}, \ltyv{})) = \mathsf{var}({\mathit{x}_{1}}, \mathsf{openRec}(\lky{}, \lenv{{\mathrm{u}}}, \ltyv{}))}\\[2pt]
\gpfit{\mathsf{openRec}(\lky{}, \lenv{{\mathrm{u}}}, \mathsf{vlam}(\ltynv{0}, \ltynv{1}, \leev{})) = \mathsf{vlam}(\mathsf{openRec}(\lky{}, \lenv{{\mathrm{u}}}, \ltynv{0}), \mathsf{openRec}(\lky{} + 1, \lenv{{\mathrm{u}}}, \ltynv{1}), \mathsf{openRec}(\lky{} + 1, \lenv{{\mathrm{u}}}, \leev{}))}\\[2pt]
\gpfit{\mathsf{openRec}(\lky{}, \lenv{{\mathrm{u}}}, \mathsf{zero}(\lsv{})) = \mathsf{zero}(\mathsf{openRec}(\lky{}, \lenv{{\mathrm{u}}}, \lsv{}))}\\[2pt]
\gpfit{\mathsf{openRec}(\lky{}, \lenv{{\mathrm{u}}}, \mathsf{rlit}(\mathit{r})) = \mathsf{rlit}(\mathit{r})}\\[2pt]
\gpfit{\mathsf{openRec}(\lky{}, \lenv{{\mathrm{u}}}, \mathsf{scale}(\lenv{0}, \lenv{1})) = \mathsf{scale}(\mathsf{openRec}(\lky{}, \lenv{{\mathrm{u}}}, \lenv{0}), \mathsf{openRec}(\lky{}, \lenv{{\mathrm{u}}}, \lenv{1}))}\\[2pt]
\gpfit{\mathsf{openRec}(\lky{}, \lenv{{\mathrm{u}}}, \mathsf{vadd}(\lsv{}, \lenv{0}, \lenv{1})) = \mathsf{vadd}(\mathsf{openRec}(\lky{}, \lenv{{\mathrm{u}}}, \lsv{}), \mathsf{openRec}(\lky{}, \lenv{{\mathrm{u}}}, \lenv{0}), \mathsf{openRec}(\lky{}, \lenv{{\mathrm{u}}}, \lenv{1}))}\\[2pt]
\gpfit{\mathsf{openRec}(\lky{}, \lenv{{\mathrm{u}}}, \mathsf{dsumE}(\lenv{0}, \lenv{1})) = \mathsf{dsumE}(\mathsf{openRec}(\lky{}, \lenv{{\mathrm{u}}}, \lenv{0}), \mathsf{openRec}(\lky{}, \lenv{{\mathrm{u}}}, \lenv{1}))}\\[2pt]
\gpfit{\mathsf{openRec}(\lky{}, \lenv{{\mathrm{u}}}, \mathsf{tprodE}(\lenv{0}, \lenv{1})) = \mathsf{tprodE}(\mathsf{openRec}(\lky{}, \lenv{{\mathrm{u}}}, \lenv{0}), \mathsf{openRec}(\lky{}, \lenv{{\mathrm{u}}}, \lenv{1}))}\\[2pt]
\gpfit{\mathsf{openRec}(\lky{}, \lenv{{\mathrm{u}}}, \mathsf{delta}(\mathit{n}, \lsv{})) = \mathsf{delta}(\mathit{n}, \mathsf{openRec}(\lky{}, \lenv{{\mathrm{u}}}, \lsv{}))}\\[2pt]
\gpfit{\mathsf{openRec}(\lky{}, \lenv{{\mathrm{u}}}, \mathsf{idelta}(\lsv{})) = \mathsf{idelta}(\mathsf{openRec}(\lky{}, \lenv{{\mathrm{u}}}, \lsv{}))}\\[2pt]
\gpfit{\mathsf{openRec}(\lky{}, \lenv{{\mathrm{u}}}, \mathsf{vsum}(\lsv{}, \mathit{es})) = \mathsf{vsum}(\mathsf{openRec}(\lky{}, \lenv{{\mathrm{u}}}, \lsv{}), \mathsf{openRecList}(\lky{}, \lenv{{\mathrm{u}}}, \mathit{es}))}\\[2pt]
\gpfit{\mathsf{openRec}(\lky{}, \lenv{{\mathrm{u}}}, \mathsf{ttv}(\lsv{}, \lenv{0}, \lenv{1})) = \mathsf{ttv}(\mathsf{openRec}(\lky{}, \lenv{{\mathrm{u}}}, \lsv{}), \mathsf{openRec}(\lky{}, \lenv{{\mathrm{u}}}, \lenv{0}), \mathsf{openRec}(\lky{}, \lenv{{\mathrm{u}}}, \lenv{1}))}\\[2pt]
\gpfit{\mathsf{openRec}(\lky{}, \lenv{{\mathrm{u}}}, \mathsf{mtm}({\mathit{n}_{0}}, {\mathit{n}_{1}}, \lsnv{0}, \lsnv{1}, \lenv{0}, \lenv{1})) = \mathsf{mtm}({\mathit{n}_{0}}, {\mathit{n}_{1}}, \mathsf{openRec}(\lky{}, \lenv{{\mathrm{u}}}, \lsnv{0}), \mathsf{openRec}(\lky{}, \lenv{{\mathrm{u}}}, \lsnv{1}), \mathsf{openRec}(\lky{}, \lenv{{\mathrm{u}}}, \lenv{0}), \mathsf{openRec}(\lky{}, \lenv{{\mathrm{u}}}, \lenv{1}))}\\[2pt]
\gpfit{\mathsf{openRec}(\lky{}, \lenv{{\mathrm{u}}}, \mathsf{appV}(\lenv{0}, \lenv{1})) = \mathsf{appV}(\mathsf{openRec}(\lky{}, \lenv{{\mathrm{u}}}, \lenv{0}), \mathsf{openRec}(\lky{}, \lenv{{\mathrm{u}}}, \lenv{1}))}\\[2pt]
\gpfit{\mathsf{openRec}(\lky{}, \lenv{{\mathrm{u}}}, \mathsf{vkrn}(\lenv{0}, \lenv{1})) = \mathsf{vkrn}(\mathsf{openRec}(\lky{}, \lenv{{\mathrm{u}}}, \lenv{0}), \mathsf{openRec}(\lky{}, \lenv{{\mathrm{u}}}, \lenv{1}))}\\[2pt]
\gpfit{\mathsf{openRec}(\lky{}, \lenv{{\mathrm{u}}}, \mathsf{vblk}(\lenv{0}, \lenv{1})) = \mathsf{vblk}(\mathsf{openRec}(\lky{}, \lenv{{\mathrm{u}}}, \lenv{0}), \mathsf{openRec}(\lky{}, \lenv{{\mathrm{u}}}, \lenv{1}))}\\[2pt]
\gpfit{\mathsf{openRec}(\lky{}, \lenv{{\mathrm{u}}}, \mathsf{reshap}(\leev{}, \ltyv{}, \ltynv{0})) = \mathsf{reshap}(\mathsf{openRec}(\lky{}, \lenv{{\mathrm{u}}}, \leev{}), \mathsf{openRec}(\lky{}, \lenv{{\mathrm{u}}}, \ltyv{}), \mathsf{openRec}(\lky{}, \lenv{{\mathrm{u}}}, \ltynv{0}))}\\[2pt]
\gpfit{\mathsf{openRec}(\lky{}, \lenv{{\mathrm{u}}}, \mathsf{vtrans}(\leev{})) = \mathsf{vtrans}(\mathsf{openRec}(\lky{}, \lenv{{\mathrm{u}}}, \leev{}))}\\[2pt]
\gpfit{\mathsf{openRec}(\lky{}, \lenv{{\mathrm{u}}}, \mathsf{gsvdS}(\leev{}, \lenv{{\mathrm{a}}}, \lenv{{\mathrm{b}}})) = \mathsf{gsvdS}(\mathsf{openRec}(\lky{}, \lenv{{\mathrm{u}}}, \leev{}), \mathsf{openRec}(\lky{}, \lenv{{\mathrm{u}}}, \lenv{{\mathrm{a}}}), \mathsf{openRec}(\lky{}, \lenv{{\mathrm{u}}}, \lenv{{\mathrm{b}}}))}\\[2pt]
\gpfit{\mathsf{openRec}(\lky{}, \lenv{{\mathrm{u}}}, \mathsf{gsvdU}(\leev{}, \lenv{{\mathrm{a}}}, \lenv{{\mathrm{b}}})) = \mathsf{gsvdU}(\mathsf{openRec}(\lky{}, \lenv{{\mathrm{u}}}, \leev{}), \mathsf{openRec}(\lky{}, \lenv{{\mathrm{u}}}, \lenv{{\mathrm{a}}}), \mathsf{openRec}(\lky{}, \lenv{{\mathrm{u}}}, \lenv{{\mathrm{b}}}))}\\[2pt]
\gpfit{\mathsf{openRec}(\lky{}, \lenv{{\mathrm{u}}}, \mathsf{gsvdUI}(\leev{}, \lenv{{\mathrm{a}}}, \lenv{{\mathrm{b}}})) = \mathsf{gsvdUI}(\mathsf{openRec}(\lky{}, \lenv{{\mathrm{u}}}, \leev{}), \mathsf{openRec}(\lky{}, \lenv{{\mathrm{u}}}, \lenv{{\mathrm{a}}}), \mathsf{openRec}(\lky{}, \lenv{{\mathrm{u}}}, \lenv{{\mathrm{b}}}))}\\[2pt]
\gpfit{\mathsf{openRec}(\lky{}, \lenv{{\mathrm{u}}}, \mathsf{gsvdV}(\leev{}, \lenv{{\mathrm{a}}}, \lenv{{\mathrm{b}}})) = \mathsf{gsvdV}(\mathsf{openRec}(\lky{}, \lenv{{\mathrm{u}}}, \leev{}), \mathsf{openRec}(\lky{}, \lenv{{\mathrm{u}}}, \lenv{{\mathrm{a}}}), \mathsf{openRec}(\lky{}, \lenv{{\mathrm{u}}}, \lenv{{\mathrm{b}}}))}\\[2pt]
\gpfit{\mathsf{openRec}(\lky{}, \lenv{{\mathrm{u}}}, \mathsf{gsvdVI}(\leev{}, \lenv{{\mathrm{a}}}, \lenv{{\mathrm{b}}})) = \mathsf{gsvdVI}(\mathsf{openRec}(\lky{}, \lenv{{\mathrm{u}}}, \leev{}), \mathsf{openRec}(\lky{}, \lenv{{\mathrm{u}}}, \lenv{{\mathrm{a}}}), \mathsf{openRec}(\lky{}, \lenv{{\mathrm{u}}}, \lenv{{\mathrm{b}}}))}\\[2pt]
\gpfit{\mathsf{openRec}(\lky{}, \lenv{{\mathrm{u}}}, \mathsf{pinv}(\leev{})) = \mathsf{pinv}(\mathsf{openRec}(\lky{}, \lenv{{\mathrm{u}}}, \leev{}))}\\[2pt]
\gpfit{\mathsf{openRec}(\lky{}, \lenv{{\mathrm{u}}}, \mathsf{gpinv}(\leev{}, \lenv{{\mathrm{a}}}, \lenv{{\mathrm{b}}})) = \mathsf{gpinv}(\mathsf{openRec}(\lky{}, \lenv{{\mathrm{u}}}, \leev{}), \mathsf{openRec}(\lky{}, \lenv{{\mathrm{u}}}, \lenv{{\mathrm{a}}}), \mathsf{openRec}(\lky{}, \lenv{{\mathrm{u}}}, \lenv{{\mathrm{b}}}))}\\[2pt]
\gpfit{\mathsf{openRec}(\lky{}, \lenv{{\mathrm{u}}}, \mathsf{rrefRep}(\lenv{0}, \lenv{1})) = \mathsf{rrefRep}(\mathsf{openRec}(\lky{}, \lenv{{\mathrm{u}}}, \lenv{0}), \mathsf{openRec}(\lky{}, \lenv{{\mathrm{u}}}, \lenv{1}))}\\[2pt]
\gpfit{\mathsf{openRec}(\lky{}, \lenv{{\mathrm{u}}}, \mathsf{inv}(\leev{})) = \mathsf{inv}(\mathsf{openRec}(\lky{}, \lenv{{\mathrm{u}}}, \leev{}))}\\[2pt]
\gpfit{\mathsf{openRec}(\lky{}, \lenv{{\mathrm{u}}}, \mathsf{svds}(\leev{})) = \mathsf{svds}(\mathsf{openRec}(\lky{}, \lenv{{\mathrm{u}}}, \leev{}))}\end{center}
}{\texttt{GLang.Val\allowbreak Expr.open\allowbreak Rec}}{GLang/ValueLN.lean:44}
\gldefdefn{space.openRec}{\texttt{Space.open\allowbreak Rec}}{clauses}{({\mathit{x}}^{\lky{}\leftarrow \mathit{u}})}{\begin{center}\gpfit{\mathsf{openRec}(\lky{}, \mathit{u}, \lsempty{}) = \lsempty{}}\\[2pt]
\gpfit{\mathsf{openRec}(\lky{}, \mathit{u}, \lR{}) = \lR{}}\\[2pt]
\gpfit{\mathsf{openRec}(\lky{}, \mathit{u}, \mathsf{dual}(\ls{})) = \mathsf{dual}(\mathsf{openRec}(\lky{}, \mathit{u}, \ls{}))}\\[2pt]
\gpfit{\mathsf{openRec}(\lky{}, \mathit{u}, \mathsf{dsum}(\lsn{0}, \lsn{1})) = \mathsf{dsum}(\mathsf{openRec}(\lky{}, \mathit{u}, \lsn{0}), \mathsf{openRec}(\lky{}, \mathit{u}, \lsn{1}))}\\[2pt]
\gpfit{\mathsf{openRec}(\lky{}, \mathit{u}, \mathsf{tprod}(\lsn{0}, \lsn{1})) = \mathsf{tprod}(\mathsf{openRec}(\lky{}, \mathit{u}, \lsn{0}), \mathsf{openRec}(\lky{}, \mathit{u}, \lsn{1}))}\\[2pt]
\gpfit{\mathsf{openRec}(\lky{}, \mathit{u}, \mathsf{ker}(\lsn{0}, \lsn{1}, \lee{})) = \mathsf{ker}(\mathsf{openRec}(\lky{}, \mathit{u}, \lsn{0}), \mathsf{openRec}(\lky{}, \mathit{u}, \lsn{1}), \mathsf{openRec}(\lky{}, \mathit{u}, \lee{}))}\\[2pt]
\gpfit{\mathsf{openRec}(\lky{}, \mathit{u}, \mathsf{coker}(\lsn{0}, \lsn{1}, \lee{})) = \mathsf{coker}(\mathsf{openRec}(\lky{}, \mathit{u}, \lsn{0}), \mathsf{openRec}(\lky{}, \mathit{u}, \lsn{1}), \mathsf{openRec}(\lky{}, \mathit{u}, \lee{}))}\\[2pt]
\gpfit{\mathsf{openRec}(\lky{}, \mathit{u}, \mathsf{img}(\lsn{0}, \lsn{1}, \lee{})) = \mathsf{img}(\mathsf{openRec}(\lky{}, \mathit{u}, \lsn{0}), \mathsf{openRec}(\lky{}, \mathit{u}, \lsn{1}), \mathsf{openRec}(\lky{}, \mathit{u}, \lee{}))}\\[2pt]
\gpfit{\mathsf{openRec}(\lky{}, \mathit{u}, \mathsf{coimg}(\lsn{0}, \lsn{1}, \lee{})) = \mathsf{coimg}(\mathsf{openRec}(\lky{}, \mathit{u}, \lsn{0}), \mathsf{openRec}(\lky{}, \mathit{u}, \lsn{1}), \mathsf{openRec}(\lky{}, \mathit{u}, \lee{}))}\end{center}
}{\texttt{GLang.Space.open\allowbreak Rec}}{GLang/SourceLN.lean:42}
\gldefdefn{valSpace.openRec}{\texttt{Val\allowbreak Space.open\allowbreak Rec}}{clauses}{({\lsnv{{\mathrm{x}}}}^{\lky{}\leftarrow \lenv{{\mathrm{u}}}})}{\begin{center}\gpfit{\mathsf{openRec}(\lky{}, \lenv{{\mathrm{u}}}, \lsempty{}) = \lsempty{}}\\[2pt]
\gpfit{\mathsf{openRec}(\lky{}, \lenv{{\mathrm{u}}}, \lR{}) = \lR{}}\\[2pt]
\gpfit{\mathsf{openRec}(\lky{}, \lenv{{\mathrm{u}}}, \mathsf{dsum}(\lsnv{0}, \lsnv{1})) = \mathsf{dsum}(\mathsf{openRec}(\lky{}, \lenv{{\mathrm{u}}}, \lsnv{0}), \mathsf{openRec}(\lky{}, \lenv{{\mathrm{u}}}, \lsnv{1}))}\\[2pt]
\gpfit{\mathsf{openRec}(\lky{}, \lenv{{\mathrm{u}}}, \mathsf{tprod}(\lsnv{0}, \lsnv{1})) = \mathsf{tprod}(\mathsf{openRec}(\lky{}, \lenv{{\mathrm{u}}}, \lsnv{0}), \mathsf{openRec}(\lky{}, \lenv{{\mathrm{u}}}, \lsnv{1}))}\\[2pt]
\gpfit{\mathsf{openRec}(\lky{}, \lenv{{\mathrm{u}}}, \mathsf{ffs}(\lprop{}, \lsv{}, \leev{})) = \mathsf{ffs}(\lprop{}, \mathsf{openRec}(\lky{}, \lenv{{\mathrm{u}}}, \lsv{}), \mathsf{openRec}(\lky{}, \lenv{{\mathrm{u}}}, \leev{}))}\end{center}
}{\texttt{GLang.Val\allowbreak Space.open\allowbreak Rec}}{GLang/ValueLN.lean:32}
\gldefdefn{valTy.openRec}{\texttt{Val\allowbreak Ty.open\allowbreak Rec}}{clauses}{({\ltynv{{\mathrm{x}}}}^{\lky{}\leftarrow \lenv{{\mathrm{u}}}})}{\begin{center}\gpfit{\mathsf{openRec}(\lky{}, \lenv{{\mathrm{u}}}, \mathsf{space}(\lsv{})) = \mathsf{space}(\mathsf{openRec}(\lky{}, \lenv{{\mathrm{u}}}, \lsv{}))}\\[2pt]
\gpfit{\mathsf{openRec}(\lky{}, \lenv{{\mathrm{u}}}, \mathsf{piTy}(\ltynv{0}, \ltynv{1})) = \mathsf{piTy}(\mathsf{openRec}(\lky{}, \lenv{{\mathrm{u}}}, \ltynv{0}), \mathsf{openRec}(\lky{} + 1, \lenv{{\mathrm{u}}}, \ltynv{1}))}\end{center}
}{\texttt{GLang.Val\allowbreak Ty.open\allowbreak Rec}}{GLang/ValueLN.lean:39}
\gldefdefn{ty.openRec}{\texttt{Ty.open\allowbreak Rec}}{clauses}{({\mathit{x}}^{\lky{}\leftarrow \mathit{u}})}{\begin{center}\gpfit{\mathsf{openRec}(\lky{}, \mathit{u}, \mathsf{space}(\ls{})) = \mathsf{space}(\mathsf{openRec}(\lky{}, \mathit{u}, \ls{}))}\\[2pt]
\gpfit{\mathsf{openRec}(\lky{}, \mathit{u}, \mathsf{piTy}(\ltyn{0}, \ltyn{1})) = \mathsf{piTy}(\mathsf{openRec}(\lky{}, \mathit{u}, \ltyn{0}), \mathsf{openRec}(\lky{} + 1, \mathit{u}, \ltyn{1}))}\end{center}
}{\texttt{GLang.Ty.open\allowbreak Rec}}{GLang/SourceLN.lean:53}
\gldefdefn{term.openRec}{\texttt{Term.open\allowbreak Rec}}{clauses}{({\mathit{x}}^{\lky{}\leftarrow \mathit{u}})}{\begin{center}\gpfit{\begin{gathered}\mathsf{openRec}(\lky{}, \mathit{u}, \mathsf{bvar}(\mathit{i})) = \begin{cases} \mathit{u} & \text{if } \mathit{i} = \lky{} \\ \mathsf{bvar}(\mathit{i}) & \text{otherwise} \end{cases}\end{gathered}}\\[2pt]
\gpfit{\mathsf{openRec}(\lky{}, \mathit{u}, \mathsf{var}({\mathit{x}_{1}})) = \mathsf{var}({\mathit{x}_{1}})}\\[2pt]
\gpfit{\mathsf{openRec}(\lky{}, \mathit{u}, \mathsf{real}(\mathit{r})) = \mathsf{real}(\mathit{r})}\\[2pt]
\gpfit{\mathsf{openRec}(\lky{}, \mathit{u}, \mathsf{zero}(\ls{})) = \mathsf{zero}(\mathsf{openRec}(\lky{}, \mathit{u}, \ls{}))}\\[2pt]
\gpfit{\mathsf{openRec}(\lky{}, \mathit{u}, \mathsf{smul}(\len{0}, \len{1})) = \mathsf{smul}(\mathsf{openRec}(\lky{}, \mathit{u}, \len{0}), \mathsf{openRec}(\lky{}, \mathit{u}, \len{1}))}\\[2pt]
\gpfit{\mathsf{openRec}(\lky{}, \mathit{u}, \mathsf{add}(\len{0}, \len{1})) = \mathsf{add}(\mathsf{openRec}(\lky{}, \mathit{u}, \len{0}), \mathsf{openRec}(\lky{}, \mathit{u}, \len{1}))}\\[2pt]
\gpfit{\mathsf{openRec}(\lky{}, \mathit{u}, \mathsf{dsumE}(\len{0}, \len{1})) = \mathsf{dsumE}(\mathsf{openRec}(\lky{}, \mathit{u}, \len{0}), \mathsf{openRec}(\lky{}, \mathit{u}, \len{1}))}\\[2pt]
\gpfit{\mathsf{openRec}(\lky{}, \mathit{u}, \mathsf{tprodE}(\len{0}, \len{1})) = \mathsf{tprodE}(\mathsf{openRec}(\lky{}, \mathit{u}, \len{0}), \mathsf{openRec}(\lky{}, \mathit{u}, \len{1}))}\\[2pt]
\gpfit{\mathsf{openRec}(\lky{}, \mathit{u}, \mathsf{inner}(\ls{})) = \mathsf{inner}(\mathsf{openRec}(\lky{}, \mathit{u}, \ls{}))}\\[2pt]
\gpfit{\mathsf{openRec}(\lky{}, \mathit{u}, \mathsf{embed}(\ls{})) = \mathsf{embed}(\mathsf{openRec}(\lky{}, \mathit{u}, \ls{}))}\\[2pt]
\gpfit{\mathsf{openRec}(\lky{}, \mathit{u}, \mathsf{pinvT}(\lsn{0}, \lsn{1}, \lee{})) = \mathsf{pinvT}(\mathsf{openRec}(\lky{}, \mathit{u}, \lsn{0}), \mathsf{openRec}(\lky{}, \mathit{u}, \lsn{1}), \mathsf{openRec}(\lky{}, \mathit{u}, \lee{}))}\\[2pt]
\gpfit{\mathsf{openRec}(\lky{}, \mathit{u}, \mathsf{pinvD}(\ls{}, \lee{})) = \mathsf{pinvD}(\mathsf{openRec}(\lky{}, \mathit{u}, \ls{}), \mathsf{openRec}(\lky{}, \mathit{u}, \lee{}))}\\[2pt]
\gpfit{\mathsf{openRec}(\lky{}, \mathit{u}, \mathsf{llam}(\lsn{0}, \ls{}, \lee{})) = \mathsf{llam}(\mathsf{openRec}(\lky{}, \mathit{u}, \lsn{0}), \mathsf{openRec}(\lky{}, \mathit{u}, \ls{}), \mathsf{openRec}(\lky{} + 1, \mathit{u}, \lee{}))}\\[2pt]
\gpfit{\mathsf{openRec}(\lky{}, \mathit{u}, \mathsf{nllam}(\lty{}, \lee{})) = \mathsf{nllam}(\mathsf{openRec}(\lky{}, \mathit{u}, \lty{}), \mathsf{openRec}(\lky{} + 1, \mathit{u}, \lee{}))}\\[2pt]
\gpfit{\mathsf{openRec}(\lky{}, \mathit{u}, \mathsf{appL}(\lsn{0}, \lsn{1}, \len{0}, \len{1})) = \mathsf{appL}(\mathsf{openRec}(\lky{}, \mathit{u}, \lsn{0}), \mathsf{openRec}(\lky{}, \mathit{u}, \lsn{1}), \mathsf{openRec}(\lky{}, \mathit{u}, \len{0}), \mathsf{openRec}(\lky{}, \mathit{u}, \len{1}))}\\[2pt]
\gpfit{\mathsf{openRec}(\lky{}, \mathit{u}, \mathsf{appI}(\ls{}, \len{0}, \len{1})) = \mathsf{appI}(\mathsf{openRec}(\lky{}, \mathit{u}, \ls{}), \mathsf{openRec}(\lky{}, \mathit{u}, \len{0}), \mathsf{openRec}(\lky{}, \mathit{u}, \len{1}))}\\[2pt]
\gpfit{\mathsf{openRec}(\lky{}, \mathit{u}, \mathsf{appN}(\len{0}, \len{1})) = \mathsf{appN}(\mathsf{openRec}(\lky{}, \mathit{u}, \len{0}), \mathsf{openRec}(\lky{}, \mathit{u}, \len{1}))}\end{center}
}{\texttt{GLang.Term.open\allowbreak Rec}}{GLang/SourceLN.lean:58}
\gldefdefn{term.openRecList}{\texttt{Term.open\allowbreak Rec\allowbreak List}}{clauses}{\mathsf{openRecList}(\lky{}, \mathit{u}, \mathit{es})}{\begin{center}\gpfit{\mathsf{openRecList}(\lky{}, \mathit{u}, \mathit{es}) = \left[\, \mathsf{openRec}(\lky{}, \mathit{u}, \mathit{z}) \mid \mathit{z} \in \mathit{es} \,\right]}\end{center}
}{\texttt{GLang.Term.open\allowbreak Rec\allowbreak List}}{GLang/SourceLN.lean:85}
\gldefdefn{valExpr.openRecList}{\texttt{Val\allowbreak Expr.open\allowbreak Rec\allowbreak List}}{clauses}{\mathsf{openRecList}(\lky{}, \lenv{{\mathrm{u}}}, \mathit{x})}{\begin{center}\gpfit{\mathsf{openRecList}(\lky{}, \lenv{{\mathrm{u}}}, []) = []}\\[2pt]
\gpfit{\mathsf{openRecList}(\lky{}, \lenv{{\mathrm{u}}}, \leev{} \mathbin{:\!:} \mathit{es}) = \mathsf{openRec}(\lky{}, \lenv{{\mathrm{u}}}, \leev{}) \mathbin{:\!:} \mathsf{openRecList}(\lky{}, \lenv{{\mathrm{u}}}, \mathit{es})}\end{center}
}{\texttt{GLang.Val\allowbreak Expr.open\allowbreak Rec\allowbreak List}}{GLang/ValueLN.lean:82}
\gldefdefn{valExpr.openVar}{\texttt{Val\allowbreak Expr.open\allowbreak Var}}{clauses}{\leev{}}{\begin{center}\gpfit{\mathsf{openVar}(\mathit{x}, \ltyv{}, \leev{}) = \mathsf{openRec}(0, \mathsf{var}(\mathit{x}, \ltyv{}), \leev{})}\end{center}
}{\texttt{GLang.Val\allowbreak Expr.open\allowbreak Var}}{GLang/ValueLN.lean:513}
\gldefdefn{valTy.openVar}{\texttt{Val\allowbreak Ty.open\allowbreak Var}}{clauses}{\ltynv{{\mathrm{sig}}}}{\begin{center}\gpfit{\mathsf{openVar}(\mathit{x}, \ltyv{}, \ltynv{{\mathrm{sig}}}) = \mathsf{openRec}(0, \mathsf{var}(\mathit{x}, \ltyv{}), \ltynv{{\mathrm{sig}}})}\end{center}
}{\texttt{GLang.Val\allowbreak Ty.open\allowbreak Var}}{GLang/ValueLN.lean:516}
\gldefdefn{basis.openVar}{\texttt{Basis.open\allowbreak Var}}{clauses}{\lb{}}{\begin{center}\gpfit{\mathsf{openVar}(\mathit{x}, \lb{}) = \mathsf{openRec}(0, \mathsf{var}(\mathit{x}), \lb{})}\end{center}
}{\texttt{GLang.Basis.open\allowbreak Var}}{GLang/SourceLN.lean:168}
\gldefdefn{ty.openVar}{\texttt{Ty.open\allowbreak Var}}{clauses}{\sigma{}}{\begin{center}\gpfit{\mathsf{openVar}(\mathit{x}, \sigma{}) = \mathsf{openRec}(0, \mathsf{var}(\mathit{x}), \sigma{})}\end{center}
}{\texttt{GLang.Ty.open\allowbreak Var}}{GLang/SourceLN.lean:167}
\gldefdefn{term.openVar}{\texttt{Term.open\allowbreak Var}}{clauses}{\lee{}}{\begin{center}\gpfit{\mathsf{openVar}(\mathit{x}, \lee{}) = \mathsf{openRec}(0, \mathsf{var}(\mathit{x}), \lee{})}\end{center}
}{\texttt{GLang.Term.open\allowbreak Var}}{GLang/SourceLN.lean:165}
\gldefdefn{valueShape.pair}{\texttt{Value\allowbreak Shape.pair}}{clauses}{|\mathit{a}|=d_{\mathit{q}}(\lsv{}) \;\land\; |\lb{}|=d_{\mathit{q}}(\lsnv{{\mathrm{t}}})}{\begin{center}\gpfit{|\mathit{a}|=d_{\mathit{q}}(\lsv{}) \;\land\; |\lb{}|=d_{\mathit{q}}(\lsnv{{\mathrm{t}}}) = \lvert \mathit{a}\rvert = \dim_{\mathit{q}}(\lsv{}) \;\wedge\; \lvert \lb{}\rvert = \dim_{\mathit{q}}(\lsnv{{\mathrm{t}}})}\end{center}
}{\texttt{GLang.Value\allowbreak Shape.pair}}{GLang/ValuePrimitiveShapes.lean:13}
\gldefdefn{reshapArrayBetweenS}{\texttt{reshap\allowbreak Array\allowbreak Between\allowbreak S}}{clauses}{\mathsf{reshapArrayBetweenS}(\mathcal{O}, \glvEnv{}, \Gamma_{\mathrm{Gt}}, \mathit{t}, \Gamma_{\mathrm{Gs}}, {\lDelta{}_{s}}, \mathit{a}, \lsnv{{\mathrm{st}}}, {\ltyv{}_{\mathsf{src}}})}{\begin{center}\gpfit{\begin{gathered}\mathsf{reshapArrayBetweenS}(\mathcal{O}, \glvEnv{}, \Gamma_{\mathrm{Gt}}, \mathit{t}, \Gamma_{\mathrm{Gs}}, {\lDelta{}_{s}}, \mathit{a}, \lsnv{{\mathrm{st}}}, \lsnv{{\mathrm{a1}}}) = \begin{cases} \begin{cases} \mathit{a} & \text{if } \dim_{\mathcal{O}}(\lsnv{{\mathrm{a1}}}) = \dim_{\mathcal{O}}(\lsnv{{\mathrm{st}}}) \\ \glverr & \text{otherwise} \end{cases} & \text{if } \lvert \mathit{a}\rvert = \dim_{\mathcal{O}}(\lsnv{{\mathrm{a1}}}) \\ \glverr & \text{otherwise} \end{cases}\end{gathered}}\\[2pt]
\gpfit{\mathsf{reshapArrayBetweenS}(\mathcal{O}, \glvEnv{}, \Gamma_{\mathrm{Gt}}, \mathit{t}, \Gamma_{\mathrm{Gs}}, {\lDelta{}_{s}}, \mathit{a}, \lsnv{{\mathrm{st}}}, (x\colon \ltynv{0}\fnto \ltynv{1})) = \glverr}\end{center}
}{\texttt{GLang.reshap\allowbreak Array\allowbreak Between\allowbreak S}}{GLang/ValueDenotSem.lean:662}
\gldefdefn{reshapDenotSemBetweenS}{\texttt{reshap\allowbreak Denot\allowbreak Sem\allowbreak Between\allowbreak S}}{clauses}{\mathsf{reshapDenotSemBetweenS}(\mathcal{O}, \glvEnv{}, \Gamma_{\mathrm{Gt}}, \mathit{t}, \Gamma_{\mathrm{Gs}}, {\lDelta{}_{s}}, \mathit{a}, \ltynv{{\mathrm{a}}}, \ltynv{{\mathrm{a}}})}{\begin{center}\gpfit{\mathsf{reshapDenotSemBetweenS}(\mathcal{O}, \glvEnv{}, \Gamma_{\mathrm{Gt}}, \mathit{t}, \Gamma_{\mathrm{Gs}}, {\lDelta{}_{s}}, \mathit{a}, \lsnv{{\mathrm{st}}}, \ltynv{{\mathrm{x}}}) = \mathsf{reshapArrayBetweenS}(\mathcal{O}, \glvEnv{}, \Gamma_{\mathrm{Gt}}, \mathit{t}, \Gamma_{\mathrm{Gs}}, {\lDelta{}_{s}}, \mathit{a}, \lsnv{{\mathrm{st}}}, \ltynv{{\mathrm{x}}})}\\[2pt]
\gpfit{\mathsf{reshapDenotSemBetweenS}(\mathcal{O}, \glvEnv{}, \Gamma_{\mathrm{Gt}}, \mathit{t}, \Gamma_{\mathrm{Gs}}, {\lDelta{}_{s}}, \mathit{a}, \lsnv{{\mathrm{a1}}}, \ltynv{{\mathrm{x}}}) = \glverr}\\[2pt]
\gpfit{\mathsf{reshapDenotSemBetweenS}(\mathcal{O}, \glvEnv{}, \Gamma_{\mathrm{Gt}}, \mathit{t}, \Gamma_{\mathrm{Gs}}, {\lDelta{}_{s}}, \mathit{x}, (x\colon {\ltyv{}_{\mathsf{t}0}}\fnto {\ltyv{}_{\mathsf{t}1}}), (x\colon {\ltyv{}_{\mathsf{s}0}}\fnto {\ltyv{}_{\mathsf{s}1}})) = \lambda \mathit{a}.\, \mathbf{let}\ \mathit{argT} \gets \mathit{a}\ \mathbf{in}\ \mathbf{let}\ \mathit{argS} \gets \mathsf{reshapDenotSemBetweenS}(\mathcal{O}, \glvEnv{}, \Gamma_{\mathrm{Gs}}, {\lDelta{}_{s}}, \Gamma_{\mathrm{Gt}}, \mathit{t}, \mathit{argT}, {\ltyv{}_{\mathsf{s}0}}, {\ltyv{}_{\mathsf{t}0}})\ \mathbf{in}\ \mathbf{let}\ \mathit{argEl} \gets \mathit{argS}\ \mathbf{in}\ \mathbf{let}\ \mathit{res} \gets \mathsf{applyTyped}(\mathit{x}, \mathit{argEl})\ \mathbf{in}\ \mathsf{reshapDenotSemBetweenS}(\mathcal{O}, \glvEnv{}, {\ltyv{}_{\mathsf{t}0}} \mathbin{:\!:} \Gamma_{\mathrm{Gt}}, \mathit{argT} \mathbin{:\!:} \mathit{t}, {\ltyv{}_{\mathsf{s}0}} \mathbin{:\!:} \Gamma_{\mathrm{Gs}}, \mathit{argS} \mathbin{:\!:} {\lDelta{}_{s}}, \mathit{res}, {\ltyv{}_{\mathsf{t}1}}, {\ltyv{}_{\mathsf{s}1}})}\\[2pt]
\gpfit{\mathsf{reshapDenotSemBetweenS}(\mathcal{O}, \glvEnv{}, \Gamma_{\mathrm{Gt}}, \mathit{t}, \Gamma_{\mathrm{Gs}}, {\lDelta{}_{s}}, \mathit{x}, (x\colon \ltynv{{\mathrm{a}}}\fnto \ltynv{{\mathrm{a1}}}), \lsnv{{\mathrm{a2}}}) = \glverr}\end{center}
}{\texttt{GLang.reshap\allowbreak Denot\allowbreak Sem\allowbreak Between\allowbreak S}}{GLang/ValueDenotSem.lean:681}
\gldefdefn{reshapDenotSemS}{\texttt{reshap\allowbreak Denot\allowbreak Sem\allowbreak S}}{clauses}{\mathsf{reshapDenotSemS}(\mathcal{O}, \glvEnv{}, \Gamma{}, \lDelta{}, \mathit{x}, \ltynv{{\mathrm{x}}}, \ltynv{{\mathrm{x}}})}{\begin{center}\gpfit{\mathsf{reshapDenotSemS}(\mathcal{O}, \glvEnv{}, \Gamma{}, \lDelta{}, \mathit{a}, \lsnv{{\mathrm{st}}}, \ltynv{{\mathrm{x}}}) = \mathsf{reshapArrayBetweenS}(\mathcal{O}, \glvEnv{}, \Gamma{}, \lDelta{}, \Gamma{}, \lDelta{}, \mathit{a}, \lsnv{{\mathrm{st}}}, \ltynv{{\mathrm{x}}})}\\[2pt]
\gpfit{\mathsf{reshapDenotSemS}(\mathcal{O}, \glvEnv{}, \Gamma{}, \lDelta{}, \mathit{a}, \lsnv{{\mathrm{a1}}}, \ltynv{{\mathrm{x}}}) = \glverr}\\[2pt]
\gpfit{\mathsf{reshapDenotSemS}(\mathcal{O}, \glvEnv{}, \Gamma{}, \lDelta{}, \mathit{x}, (x\colon {\ltyv{}_{\mathsf{t}0}}\fnto {\ltyv{}_{\mathsf{t}1}}), (x\colon {\ltyv{}_{\mathsf{s}0}}\fnto {\ltyv{}_{\mathsf{s}1}})) = \lambda \mathit{a}.\, \mathbf{let}\ \mathit{argT} \gets \mathit{a}\ \mathbf{in}\ \mathbf{let}\ \mathit{argS} \gets \mathsf{reshapDenotSemBetweenS}(\mathcal{O}, \glvEnv{}, \Gamma{}, \lDelta{}, \Gamma{}, \lDelta{}, \mathit{argT}, {\ltyv{}_{\mathsf{s}0}}, {\ltyv{}_{\mathsf{t}0}})\ \mathbf{in}\ \mathbf{let}\ \mathit{argEl} \gets \mathit{argS}\ \mathbf{in}\ \mathbf{let}\ \mathit{res} \gets \mathsf{applyTyped}(\mathit{x}, \mathit{argEl})\ \mathbf{in}\ \mathsf{reshapDenotSemBetweenS}(\mathcal{O}, \glvEnv{}, {\ltyv{}_{\mathsf{t}0}} \mathbin{:\!:} \Gamma{}, \mathit{argT} \mathbin{:\!:} \lDelta{}, {\ltyv{}_{\mathsf{s}0}} \mathbin{:\!:} \Gamma{}, \mathit{argS} \mathbin{:\!:} \lDelta{}, \mathit{res}, {\ltyv{}_{\mathsf{t}1}}, {\ltyv{}_{\mathsf{s}1}})}\\[2pt]
\gpfit{\mathsf{reshapDenotSemS}(\mathcal{O}, \glvEnv{}, \Gamma{}, \lDelta{}, \mathit{x}, (x\colon \ltynv{{\mathrm{a}}}\fnto \ltynv{{\mathrm{a1}}}), \lsnv{{\mathrm{a2}}}) = \glverr}\end{center}
}{\texttt{GLang.reshap\allowbreak Denot\allowbreak Sem\allowbreak S}}{GLang/ValueDenotSem.lean:726}
\gldefdefn{valueShape.scale}{\texttt{Value\allowbreak Shape.scale}}{clauses}{|\mathit{a}|=1 \;\land\; |\lb{}|=d_{\mathit{q}}(\lsv{})}{\begin{center}\gpfit{|\mathit{a}|=1 \;\land\; |\lb{}|=d_{\mathit{q}}(\lsv{}) = \lvert \mathit{a}\rvert = 1 \;\wedge\; \lvert \lb{}\rvert = \dim_{\mathit{q}}(\lsv{})}\end{center}
}{\texttt{GLang.Value\allowbreak Shape.scale}}{GLang/ValuePrimitiveShapes.lean:10}
\gldefdefn{valSpace.shapeInBounds}{\texttt{Val\allowbreak Space.shape\allowbreak In\allowbreak Bounds}}{clauses}{\mathit{x} \in \operatorname{idx}(\lsnv{{\mathrm{x}}})}{\begin{center}\gpfit{\mathit{x} \in \operatorname{idx}(\lsempty{}) = \mathsf{False}}\\[2pt]
\gpfit{\mathit{x} \in \operatorname{idx}(\lR{}) = \mathit{x} = 0}\\[2pt]
\gpfit{\begin{gathered}\mathit{x} \in \operatorname{idx}(\lsnv{0} \losum \lsnv{1}) = \begin{cases} \mathit{x} \in \operatorname{idx}(\lsnv{0}) & \text{if } \mathit{x} < \dim_{\mathcal{O}}(\lsnv{0}) \\ (\mathit{x}) - (\dim_{\mathcal{O}}(\lsnv{0})) < \dim_{\mathcal{O}}(\lsnv{1}) \;\wedge\; (\mathit{x}) - (\dim_{\mathcal{O}}(\lsnv{0})) \in \operatorname{idx}(\lsnv{1}) & \text{otherwise} \end{cases}\end{gathered}}\\[2pt]
\gpfit{\mathit{x} \in \operatorname{idx}(\lsnv{0} \loprod \lsnv{1}) = (\mathit{x}) / (\dim_{\mathcal{O}}(\lsnv{1})) < \dim_{\mathcal{O}}(\lsnv{0}) \;\wedge\; (\mathit{x}) / (\dim_{\mathcal{O}}(\lsnv{1})) \in \operatorname{idx}(\lsnv{0}) \;\wedge\; (\mathit{x}) \bmod (\dim_{\mathcal{O}}(\lsnv{1})) \in \operatorname{idx}(\lsnv{1})}\\[2pt]
\gpfit{\mathit{x} \in \operatorname{idx}(\lR^{\leev{},\lprop{},\lsv{}}) = \mathit{x} < \mathcal{O}\,\lprop{}\,\lsv{}\,\leev{}}\end{center}
}{\texttt{GLang.Val\allowbreak Space.shape\allowbreak In\allowbreak Bounds}}{GLang/ValueDenot.lean:250}
\gldefdefn{valSpace.shpOf}{\texttt{Val\allowbreak Space.shp\allowbreak Of}}{clauses}{\shpOf{\lsnv{{\mathrm{a}}}}{\mathcal{O}}}{\begin{center}\gpfit{\shpOf{\lsempty{}}{\mathcal{O}} = 0 \mathbin{:\!:} []}\\[2pt]
\gpfit{\shpOf{\lR{}}{\mathcal{O}} = 1 \mathbin{:\!:} []}\\[2pt]
\gpfit{\shpOf{\lsnv{0} \losum \lsnv{1}}{\mathcal{O}} = \dim_{\mathcal{O}}(\lsnv{0}) + \dim_{\mathcal{O}}(\lsnv{1}) \mathbin{:\!:} []}\\[2pt]
\gpfit{\shpOf{\lsnv{0} \loprod \lsnv{1}}{\mathcal{O}} = \dim_{\mathcal{O}}(\lsnv{0}) \mathbin{:\!:} \dim_{\mathcal{O}}(\lsnv{1}) \mathbin{:\!:} []}\\[2pt]
\gpfit{\shpOf{\lR^{\leev{},\lprop{},\lsv{}}}{\mathcal{O}} = \mathcal{O}\,\lprop{}\,\lsv{}\,\leev{} \mathbin{:\!:} []}\end{center}
}{\texttt{GLang.Val\allowbreak Space.shp\allowbreak Of}}{GLang/ValueDenot.lean:116}
\gldefdefn{basis.sof}{\texttt{Basis.sof}}{clauses}{\sof{\textcolor{lang}{\mathit{x}}}}{\begin{center}\gpfit{\sof{\lbzero{}} = \lsempty{}}\\[2pt]
\gpfit{\sof{\lbone{}} = \lR{}}\\[2pt]
\gpfit{\sof{\lb{}^{\lstar}} = \sof{\lb{}}^{\lstar}}\\[2pt]
\gpfit{\sof{\lbn{0} \losum \lbn{1}} = \sof{\lbn{0}} \losum \sof{\lbn{1}}}\\[2pt]
\gpfit{\sof{\lbn{0} \loprod \lbn{1}} = \sof{\lbn{0}} \loprod \sof{\lbn{1}}}\\[2pt]
\gpfit{\sof{\pBasisC{\ls{}}{\lbn{0}}{\lbn{1}}}=\ls{}}\\[2pt]
\gpfit{\sof{\iBasisC{\lb{}}}=\Image(\lee{})\quad(\sof{\lb{}}=\CoKernel(\lee{}))}\\[2pt]
\gpfit{\sof{\llpl{}\len{0},\dots,\len{n}\llpr{}_{\lb{}}}=\sof{\lb{}}}\end{center}
}{\texttt{GLang.Basis.sof}}{GLang/Basis.lean:60}
\gldefequation{basis.sof.empty}{\sof{\lbzero{}} = \lsempty{}}
\gldefequation{basis.sof.one}{\sof{\lbone{}} = \lR{}}
\gldefequation{basis.sof.dual}{\sof{\lb{}^{\lstar}} = \sof{\lb{}}^{\lstar}}
\gldefequation{basis.sof.dsum}{\sof{\lbn{0} \losum \lbn{1}} = \sof{\lbn{0}} \losum \sof{\lbn{1}}}
\gldefequation{basis.sof.tprod}{\sof{\lbn{0} \loprod \lbn{1}} = \sof{\lbn{0}} \loprod \sof{\lbn{1}}}
\gldefequation{basis.sof.proj}{\sof{\pBasisC{\ls{}}{\lbn{0}}{\lbn{1}}}=\ls{}}
\gldefequation{basis.sof.image}{\sof{\iBasisC{\lb{}}}=\Image(\lee{})\quad(\sof{\lb{}}=\CoKernel(\lee{}))}
\gldefequation{basis.sof.terms}{\sof{\llpl{}\len{0},\dots,\len{n}\llpr{}_{\lb{}}}=\sof{\lb{}}}
\gldefdefn{basis.std}{\texttt{Basis.std}}{clauses}{\stdBasisDCtx{\mathit{x}}}{\begin{center}\gpfit{\stdBasisDCtx{x\colon\ltyn{0}\fnto\ltyn{1}}=
\frameB{x}{\stdBasisDCtx{\ltyn{0}}}{\stdBasisCtx{\ltyn{1}}{(\emptyset,\lepsilon,x\colon\ltyn{0})}}}\end{center}
}{\texttt{GLang.Basis.std}}{GLang/Basis.lean:523}
\gldefequation{basis.std.piTy}{\stdBasisDCtx{x\colon\ltyn{0}\fnto\ltyn{1}}=
\frameB{x}{\stdBasisDCtx{\ltyn{0}}}{\stdBasisCtx{\ltyn{1}}{(\emptyset,\lepsilon,x\colon\ltyn{0})}}}
\gldefdefn{basis.stdSpace}{\texttt{Basis.std\allowbreak Space}}{clauses}{\stdBasisDCtx{\mathit{x}}}{\begin{center}\gpfit{\stdBasisDCtx{\lsempty{}} = \lbzero{}}\\[2pt]
\gpfit{\stdBasisDCtx{\lR{}} = \lbone{}}\\[2pt]
\gpfit{\stdBasisDCtx{\ls{}^{\lstar}} = \stdBasisDCtx{\ls{}}^{\lstar}}\\[2pt]
\gpfit{\stdBasisDCtx{\lsn{0} \losum \lsn{1}} = \stdBasisDCtx{\lsn{0}} \losum \stdBasisDCtx{\lsn{1}}}\\[2pt]
\gpfit{\stdBasisDCtx{\lsn{0} \loprod \lsn{1}} = \stdBasisDCtx{\lsn{0}} \loprod \stdBasisDCtx{\lsn{1}}}\\[2pt]
\gpfit{\stdBasisDCtx{\Kernel(\lee{})} = \pBasisC{\Kernel(\lee{})}{\stdBasisDCtx{\lsn{0}}}{\stdBasisDCtx{\lsn{1}}}\quad(\lctxnle\vdash \lee{}\colon \lsn{0}^{\lstar}\loprod \lsn{1})}\\[2pt]
\gpfit{\stdBasisDCtx{\CoKernel(\lee{})} = \pBasisC{\CoKernel(\lee{})}{\stdBasisDCtx{\lsn{0}}}{\stdBasisDCtx{\lsn{1}}}\quad(\lctxnle\vdash \lee{}\colon \lsn{0}^{\lstar}\loprod \lsn{1})}\\[2pt]
\gpfit{\stdBasisDCtx{\Image(\lee{})} = \pBasisC{\Image(\lee{})}{\stdBasisDCtx{\lsn{0}}}{\stdBasisDCtx{\lsn{1}}}\quad(\lctxnle\vdash \lee{}\colon \lsn{0}^{\lstar}\loprod \lsn{1})}\\[2pt]
\gpfit{\stdBasisDCtx{\CoImage(\lee{})} = \pBasisC{\CoImage(\lee{})}{\stdBasisDCtx{\lsn{0}}}{\stdBasisDCtx{\lsn{1}}}\quad(\lctxnle\vdash \lee{}\colon \lsn{0}^{\lstar}\loprod \lsn{1})}\end{center}
}{\texttt{GLang.Basis.std\allowbreak Space}}{GLang/Basis.lean:506}
\gldefequation{basis.stdSpace.empty}{\stdBasisDCtx{\lsempty{}} = \lbzero{}}
\gldefequation{basis.stdSpace.real}{\stdBasisDCtx{\lR{}} = \lbone{}}
\gldefequation{basis.stdSpace.dual}{\stdBasisDCtx{\ls{}^{\lstar}} = \stdBasisDCtx{\ls{}}^{\lstar}}
\gldefequation{basis.stdSpace.dsum}{\stdBasisDCtx{\lsn{0} \losum \lsn{1}} = \stdBasisDCtx{\lsn{0}} \losum \stdBasisDCtx{\lsn{1}}}
\gldefequation{basis.stdSpace.tprod}{\stdBasisDCtx{\lsn{0} \loprod \lsn{1}} = \stdBasisDCtx{\lsn{0}} \loprod \stdBasisDCtx{\lsn{1}}}
\gldefequation{basis.stdSpace.ker}{\stdBasisDCtx{\Kernel(\lee{})} = \pBasisC{\Kernel(\lee{})}{\stdBasisDCtx{\lsn{0}}}{\stdBasisDCtx{\lsn{1}}}\quad(\lctxnle\vdash \lee{}\colon \lsn{0}^{\lstar}\loprod \lsn{1})}
\gldefequation{basis.stdSpace.coker}{\stdBasisDCtx{\CoKernel(\lee{})} = \pBasisC{\CoKernel(\lee{})}{\stdBasisDCtx{\lsn{0}}}{\stdBasisDCtx{\lsn{1}}}\quad(\lctxnle\vdash \lee{}\colon \lsn{0}^{\lstar}\loprod \lsn{1})}
\gldefequation{basis.stdSpace.img}{\stdBasisDCtx{\Image(\lee{})} = \pBasisC{\Image(\lee{})}{\stdBasisDCtx{\lsn{0}}}{\stdBasisDCtx{\lsn{1}}}\quad(\lctxnle\vdash \lee{}\colon \lsn{0}^{\lstar}\loprod \lsn{1})}
\gldefequation{basis.stdSpace.coimg}{\stdBasisDCtx{\CoImage(\lee{})} = \pBasisC{\CoImage(\lee{})}{\stdBasisDCtx{\lsn{0}}}{\stdBasisDCtx{\lsn{1}}}\quad(\lctxnle\vdash \lee{}\colon \lsn{0}^{\lstar}\loprod \lsn{1})}
\gldefequation{basis.stdSpace.ker.formula}{\stdBasisDCtx{\Kernel(\lee{})} = \pBasisC{\Kernel(\lee{})}{\stdBasisDCtx{\lsn{0}}}{\stdBasisDCtx{\lsn{1}}}}
\gldefequation{basis.stdSpace.ker.condition}{(\lctxnle\vdash \lee{}\colon \lsn{0}^{\lstar}\loprod \lsn{1})}
\gldefequation{basis.stdSpace.coker.formula}{\stdBasisDCtx{\CoKernel(\lee{})} = \pBasisC{\CoKernel(\lee{})}{\stdBasisDCtx{\lsn{0}}}{\stdBasisDCtx{\lsn{1}}}}
\gldefequation{basis.stdSpace.coker.condition}{(\lctxnle\vdash \lee{}\colon \lsn{0}^{\lstar}\loprod \lsn{1})}
\gldefequation{basis.stdSpace.img.formula}{\stdBasisDCtx{\Image(\lee{})} = \pBasisC{\Image(\lee{})}{\stdBasisDCtx{\lsn{0}}}{\stdBasisDCtx{\lsn{1}}}}
\gldefequation{basis.stdSpace.img.condition}{(\lctxnle\vdash \lee{}\colon \lsn{0}^{\lstar}\loprod \lsn{1})}
\gldefequation{basis.stdSpace.coimg.formula}{\stdBasisDCtx{\CoImage(\lee{})} = \pBasisC{\CoImage(\lee{})}{\stdBasisDCtx{\lsn{0}}}{\stdBasisDCtx{\lsn{1}}}}
\gldefequation{basis.stdSpace.coimg.condition}{(\lctxnle\vdash \lee{}\colon \lsn{0}^{\lstar}\loprod \lsn{1})}
\gldefdefn{term.subst}{Substitution in terms}{clauses}{\mathsf{subst}(\mathit{x}, \lee{}, \mathit{x})}{\begin{center}\gpfit{\begin{gathered}\mathsf{subst}(\mathit{x}, \lee{}, \mathit{y}) = \begin{cases} \lee{} & \text{if } \mathit{x} = \mathit{y} \\ \mathit{y} & \text{otherwise} \end{cases}\end{gathered}}\\[2pt]
\gpfit{\mathsf{subst}(\mathit{x}, \lee{}, \mathit{r}) = \mathit{r}}\\[2pt]
\gpfit{\mathsf{subst}(\mathit{x}, \lee{}, \lvzero{}_{\ls{}}) = \lvzero{}_{\mathsf{subst}(\mathit{x}, \lee{}, \ls{})}}\\[2pt]
\gpfit{\mathsf{subst}(\mathit{x}, \lee{}, \len{0} \lvadd \len{1}) = \mathsf{subst}(\mathit{x}, \lee{}, \len{0}) \lvadd \mathsf{subst}(\mathit{x}, \lee{}, \len{1})}\\[2pt]
\gpfit{\mathsf{subst}(\mathit{x}, \lee{}, \len{0} \lvscale \len{1}) = \mathsf{subst}(\mathit{x}, \lee{}, \len{0}) \lvscale \mathsf{subst}(\mathit{x}, \lee{}, \len{1})}\\[2pt]
\gpfit{\mathsf{subst}(\mathit{x}, \lee{}, \len{0}\losum \len{1}) = \mathsf{subst}(\mathit{x}, \lee{}, \len{0})\losum \mathsf{subst}(\mathit{x}, \lee{}, \len{1})}\\[2pt]
\gpfit{\mathsf{subst}(\mathit{x}, \lee{}, \len{0}\loprod \len{1}) = \mathsf{subst}(\mathit{x}, \lee{}, \len{0})\loprod \mathsf{subst}(\mathit{x}, \lee{}, \len{1})}\\[2pt]
\gpfit{\mathsf{subst}(\mathit{x}, \lee{}, \linner{}_{\ls{}}) = \linner{}_{\mathsf{subst}(\mathit{x}, \lee{}, \ls{})}}\\[2pt]
\gpfit{\mathsf{subst}(\mathit{x}, \lee{}, \lembd{}_{\ls{}}) = \lembd{}_{\mathsf{subst}(\mathit{x}, \lee{}, \ls{})}}\\[2pt]
\gpfit{\mathsf{subst}(\mathit{x}, \lee{}, \lee{}'^{\lpseudo}) = \mathsf{subst}(\mathit{x}, \lee{}, \lee{}')^{\lpseudo}}\\[2pt]
\gpfit{(\lllambda(y\colon \lsn{0})\ldotsp \lee{}')\subst{\mathit{x}}{\lee{}} = \lllambda(y\colon (\lsn{0})\subst{\mathit{x}}{\lee{}})\ldotsp (\lee{}')\subst{\mathit{x}}{\lee{}}\quad(y\ne \mathit{x},\ y\notin \operatorname{fv}(\lee{}))}\\[2pt]
\gpfit{(\nllambda(y\colon \lty{})\ldotsp \lee{}')\subst{\mathit{x}}{\lee{}} = \nllambda(y\colon (\lty{})\subst{\mathit{x}}{\lee{}})\ldotsp (\lee{}')\subst{\mathit{x}}{\lee{}}\quad(y\ne \mathit{x},\ y\notin \operatorname{fv}(\lee{}))}\\[2pt]
\gpfit{\mathsf{subst}(\mathit{x}, \lee{}, \leval{\len{0}}{}{\len{1}}) = \leval{\mathsf{subst}(\mathit{x}, \lee{}, \len{0})}{}{\mathsf{subst}(\mathit{x}, \lee{}, \len{1})}}\end{center}
}{\texttt{GLang.Term.subst}}{GLang/Syntax.lean:215}
\gldefdefn{space.subst}{Substitution in spaces}{clauses}{\mathsf{subst}(\mathit{x}, \lee{}, \mathit{x})}{\begin{center}\gpfit{\mathsf{subst}(\mathit{x}, \lee{}, \lsempty{}) = \lsempty{}}\\[2pt]
\gpfit{\mathsf{subst}(\mathit{x}, \lee{}, \lR{}) = \lR{}}\\[2pt]
\gpfit{\mathsf{subst}(\mathit{x}, \lee{}, \ls{}^{\lstar}) = \mathsf{subst}(\mathit{x}, \lee{}, \ls{})^{\lstar}}\\[2pt]
\gpfit{\mathsf{subst}(\mathit{x}, \lee{}, \lsn{0} \losum \lsn{1}) = \mathsf{subst}(\mathit{x}, \lee{}, \lsn{0}) \losum \mathsf{subst}(\mathit{x}, \lee{}, \lsn{1})}\\[2pt]
\gpfit{\mathsf{subst}(\mathit{x}, \lee{}, \lsn{0} \loprod \lsn{1}) = \mathsf{subst}(\mathit{x}, \lee{}, \lsn{0}) \loprod \mathsf{subst}(\mathit{x}, \lee{}, \lsn{1})}\\[2pt]
\gpfit{\mathsf{subst}(\mathit{x}, \lee{}, \Kernel({\mathit{e}_{1}})) = \Kernel(\mathsf{subst}(\mathit{x}, \lee{}, {\mathit{e}_{1}}))}\\[2pt]
\gpfit{\mathsf{subst}(\mathit{x}, \lee{}, \CoKernel({\mathit{e}_{1}})) = \CoKernel(\mathsf{subst}(\mathit{x}, \lee{}, {\mathit{e}_{1}}))}\\[2pt]
\gpfit{\mathsf{subst}(\mathit{x}, \lee{}, \Image({\mathit{e}_{1}})) = \Image(\mathsf{subst}(\mathit{x}, \lee{}, {\mathit{e}_{1}}))}\\[2pt]
\gpfit{\mathsf{subst}(\mathit{x}, \lee{}, \CoImage({\mathit{e}_{1}})) = \CoImage(\mathsf{subst}(\mathit{x}, \lee{}, {\mathit{e}_{1}}))}\end{center}
}{\texttt{GLang.Space.subst}}{GLang/Syntax.lean:194}
\gldefdefn{valSpace.subst}{Substitution in value spaces}{clauses}{\mathsf{subst}(\mathit{x}, \lenv{{\mathrm{u}}}, \lsnv{{\mathrm{x}}})}{\begin{center}\gpfit{\mathsf{subst}(\mathit{x}, \lenv{{\mathrm{u}}}, \lsempty{}) = \lsempty{}}\\[2pt]
\gpfit{\mathsf{subst}(\mathit{x}, \lenv{{\mathrm{u}}}, \lR{}) = \lR{}}\\[2pt]
\gpfit{\mathsf{subst}(\mathit{x}, \lenv{{\mathrm{u}}}, \lsnv{0} \losum \lsnv{1}) = \mathsf{subst}(\mathit{x}, \lenv{{\mathrm{u}}}, \lsnv{0}) \losum \mathsf{subst}(\mathit{x}, \lenv{{\mathrm{u}}}, \lsnv{1})}\\[2pt]
\gpfit{\mathsf{subst}(\mathit{x}, \lenv{{\mathrm{u}}}, \lsnv{0} \loprod \lsnv{1}) = \mathsf{subst}(\mathit{x}, \lenv{{\mathrm{u}}}, \lsnv{0}) \loprod \mathsf{subst}(\mathit{x}, \lenv{{\mathrm{u}}}, \lsnv{1})}\\[2pt]
\gpfit{\mathsf{subst}(\mathit{x}, \lenv{{\mathrm{u}}}, \lR^{\leev{},\lprop{},\lsv{}}) = \lR^{\mathsf{subst}(\mathit{x}, \lenv{{\mathrm{u}}}, \leev{}),\lprop{},\mathsf{subst}(\mathit{x}, \lenv{{\mathrm{u}}}, \lsv{})}}\end{center}
}{\texttt{GLang.Val\allowbreak Space.subst}}{GLang/ValueLN.lean:418}
\gldefdefn{valExpr.subst}{Substitution in values}{clauses}{\mathsf{subst}(\mathit{x}, \lenv{{\mathrm{u}}}, \leev{})}{\begin{center}\gpfit{\begin{gathered}\mathsf{subst}(\mathit{x}, \lenv{{\mathrm{u}}}, \mathit{y}_{\ltyv{}}) = \begin{cases} \lenv{{\mathrm{u}}} & \text{if } \mathit{y} = \mathit{x} \\ \mathit{y}_{\mathsf{subst}(\mathit{x}, \lenv{{\mathrm{u}}}, \ltyv{})} & \text{otherwise} \end{cases}\end{gathered}}\\[2pt]
\gpfit{(\llambda(y\colon \ltynv{0})\lrh \ltynv{1}\colon \leev{})\subst{\mathit{x}}{\lenv{{\mathrm{u}}}} = \llambda(y\colon (\ltynv{0})\subst{\mathit{x}}{\lenv{{\mathrm{u}}}})\lrh (\ltynv{1})\subst{\mathit{x}}{\lenv{{\mathrm{u}}}}\colon (\leev{})\subst{\mathit{x}}{\lenv{{\mathrm{u}}}}\quad(y\ne \mathit{x},\ y\notin \operatorname{fv}(\lenv{{\mathrm{u}}}))}\\[2pt]
\gpfit{\mathsf{subst}(\mathit{x}, \lenv{{\mathrm{u}}}, \lvzero{}_{\lsv{}}) = \lvzero{}_{\mathsf{subst}(\mathit{x}, \lenv{{\mathrm{u}}}, \lsv{})}}\\[2pt]
\gpfit{\mathsf{subst}(\mathit{x}, \lenv{{\mathrm{u}}}, \mathit{r}) = \mathit{r}}\\[2pt]
\gpfit{\mathsf{subst}(\mathit{x}, \lenv{{\mathrm{u}}}, \lenv{0} \lvscale \lenv{1}) = \mathsf{subst}(\mathit{x}, \lenv{{\mathrm{u}}}, \lenv{0}) \lvscale \mathsf{subst}(\mathit{x}, \lenv{{\mathrm{u}}}, \lenv{1})}\\[2pt]
\gpfit{\mathsf{subst}(\mathit{x}, \lenv{{\mathrm{u}}}, \lenv{0} \lvadd_{\lsv{}} \lenv{1}) = \mathsf{subst}(\mathit{x}, \lenv{{\mathrm{u}}}, \lenv{0}) \lvadd_{\mathsf{subst}(\mathit{x}, \lenv{{\mathrm{u}}}, \lsv{})} \mathsf{subst}(\mathit{x}, \lenv{{\mathrm{u}}}, \lenv{1})}\\[2pt]
\gpfit{\mathsf{subst}(\mathit{x}, \lenv{{\mathrm{u}}}, \lenv{0}\losum \lenv{1}) = \mathsf{subst}(\mathit{x}, \lenv{{\mathrm{u}}}, \lenv{0})\losum \mathsf{subst}(\mathit{x}, \lenv{{\mathrm{u}}}, \lenv{1})}\\[2pt]
\gpfit{\mathsf{subst}(\mathit{x}, \lenv{{\mathrm{u}}}, \lenv{0}\loprod \lenv{1}) = \mathsf{subst}(\mathit{x}, \lenv{{\mathrm{u}}}, \lenv{0})\loprod \mathsf{subst}(\mathit{x}, \lenv{{\mathrm{u}}}, \lenv{1})}\\[2pt]
\gpfit{\mathsf{subst}(\mathit{x}, \lenv{{\mathrm{u}}}, \ldelta{\mathit{n}}{\lsv{}}) = \ldelta{\mathit{n}}{\mathsf{subst}(\mathit{x}, \lenv{{\mathrm{u}}}, \lsv{})}}\\[2pt]
\gpfit{\mathsf{subst}(\mathit{x}, \lenv{{\mathrm{u}}}, \lidelta{\lsv{}}) = \lidelta{\mathsf{subst}(\mathit{x}, \lenv{{\mathrm{u}}}, \lsv{})}}\\[2pt]
\gpfit{\mathsf{subst}(\mathit{x}, \lenv{{\mathrm{u}}}, \lsum^{\lsv{}}_{v\in \mathit{es}}v) = \lsum^{\mathsf{subst}(\mathit{x}, \lenv{{\mathrm{u}}}, \lsv{})}_{v\in \mathit{es}}v\subst{\mathit{x}}{\lenv{{\mathrm{u}}}}}\\[2pt]
\gpfit{\mathsf{subst}(\mathit{x}, \lenv{{\mathrm{u}}}, \LTTVPP{\lsv{}}{\lenv{0}}{\lenv{1}}) = \LTTVPP{\mathsf{subst}(\mathit{x}, \lenv{{\mathrm{u}}}, \lsv{})}{\mathsf{subst}(\mathit{x}, \lenv{{\mathrm{u}}}, \lenv{0})}{\mathsf{subst}(\mathit{x}, \lenv{{\mathrm{u}}}, \lenv{1})}}\\[2pt]
\gpfit{\mathsf{subst}(\mathit{x}, \lenv{{\mathrm{u}}}, \glvmtm{\lsnv{0}}{\lsnv{1}}{\lenv{0}}{\lenv{1}}) = \glvmtm{\mathsf{subst}(\mathit{x}, \lenv{{\mathrm{u}}}, \lsnv{0})}{\mathsf{subst}(\mathit{x}, \lenv{{\mathrm{u}}}, \lsnv{1})}{\mathsf{subst}(\mathit{x}, \lenv{{\mathrm{u}}}, \lenv{0})}{\mathsf{subst}(\mathit{x}, \lenv{{\mathrm{u}}}, \lenv{1})}}\\[2pt]
\gpfit{\mathsf{subst}(\mathit{x}, \lenv{{\mathrm{u}}}, \glvapp{\lenv{0}}{\lenv{1}}) = \glvapp{\mathsf{subst}(\mathit{x}, \lenv{{\mathrm{u}}}, \lenv{0})}{\mathsf{subst}(\mathit{x}, \lenv{{\mathrm{u}}}, \lenv{1})}}\\[2pt]
\gpfit{\mathsf{subst}(\mathit{x}, \lenv{{\mathrm{u}}}, \vkrn(\lenv{0}, \lenv{1})) = \vkrn(\mathsf{subst}(\mathit{x}, \lenv{{\mathrm{u}}}, \lenv{0}), \mathsf{subst}(\mathit{x}, \lenv{{\mathrm{u}}}, \lenv{1}))}\\[2pt]
\gpfit{\mathsf{subst}(\mathit{x}, \lenv{{\mathrm{u}}}, \vblk(\lenv{0}, \lenv{1})) = \vblk(\mathsf{subst}(\mathit{x}, \lenv{{\mathrm{u}}}, \lenv{0}), \mathsf{subst}(\mathit{x}, \lenv{{\mathrm{u}}}, \lenv{1}))}\\[2pt]
\gpfit{\mathsf{subst}(\mathit{x}, \lenv{{\mathrm{u}}}, \glvreshape{\leev{}}{\ltyv{}}{\ltynv{0}}) = \glvreshape{\mathsf{subst}(\mathit{x}, \lenv{{\mathrm{u}}}, \leev{})}{\mathsf{subst}(\mathit{x}, \lenv{{\mathrm{u}}}, \ltyv{})}{\mathsf{subst}(\mathit{x}, \lenv{{\mathrm{u}}}, \ltynv{0})}}\\[2pt]
\gpfit{\mathsf{subst}(\mathit{x}, \lenv{{\mathrm{u}}}, \Trans(\leev{})) = \Trans(\mathsf{subst}(\mathit{x}, \lenv{{\mathrm{u}}}, \leev{}))}\\[2pt]
\gpfit{\mathsf{subst}(\mathit{x}, \lenv{{\mathrm{u}}}, \glvop{GSVDS}(\leev{},\lenv{{\mathrm{a}}},\lenv{{\mathrm{b}}})) = \glvop{GSVDS}(\mathsf{subst}(\mathit{x}, \lenv{{\mathrm{u}}}, \leev{}),\mathsf{subst}(\mathit{x}, \lenv{{\mathrm{u}}}, \lenv{{\mathrm{a}}}),\mathsf{subst}(\mathit{x}, \lenv{{\mathrm{u}}}, \lenv{{\mathrm{b}}}))}\\[2pt]
\gpfit{\mathsf{subst}(\mathit{x}, \lenv{{\mathrm{u}}}, \glvop{GSVDU}(\leev{},\lenv{{\mathrm{a}}},\lenv{{\mathrm{b}}})) = \glvop{GSVDU}(\mathsf{subst}(\mathit{x}, \lenv{{\mathrm{u}}}, \leev{}),\mathsf{subst}(\mathit{x}, \lenv{{\mathrm{u}}}, \lenv{{\mathrm{a}}}),\mathsf{subst}(\mathit{x}, \lenv{{\mathrm{u}}}, \lenv{{\mathrm{b}}}))}\\[2pt]
\gpfit{\mathsf{subst}(\mathit{x}, \lenv{{\mathrm{u}}}, \glvop{GSVDUI}(\leev{},\lenv{{\mathrm{a}}},\lenv{{\mathrm{b}}})) = \glvop{GSVDUI}(\mathsf{subst}(\mathit{x}, \lenv{{\mathrm{u}}}, \leev{}),\mathsf{subst}(\mathit{x}, \lenv{{\mathrm{u}}}, \lenv{{\mathrm{a}}}),\mathsf{subst}(\mathit{x}, \lenv{{\mathrm{u}}}, \lenv{{\mathrm{b}}}))}\\[2pt]
\gpfit{\mathsf{subst}(\mathit{x}, \lenv{{\mathrm{u}}}, \glvop{GSVDV}(\leev{},\lenv{{\mathrm{a}}},\lenv{{\mathrm{b}}})) = \glvop{GSVDV}(\mathsf{subst}(\mathit{x}, \lenv{{\mathrm{u}}}, \leev{}),\mathsf{subst}(\mathit{x}, \lenv{{\mathrm{u}}}, \lenv{{\mathrm{a}}}),\mathsf{subst}(\mathit{x}, \lenv{{\mathrm{u}}}, \lenv{{\mathrm{b}}}))}\\[2pt]
\gpfit{\mathsf{subst}(\mathit{x}, \lenv{{\mathrm{u}}}, \glvop{GSVDVI}(\leev{},\lenv{{\mathrm{a}}},\lenv{{\mathrm{b}}})) = \glvop{GSVDVI}(\mathsf{subst}(\mathit{x}, \lenv{{\mathrm{u}}}, \leev{}),\mathsf{subst}(\mathit{x}, \lenv{{\mathrm{u}}}, \lenv{{\mathrm{a}}}),\mathsf{subst}(\mathit{x}, \lenv{{\mathrm{u}}}, \lenv{{\mathrm{b}}}))}\\[2pt]
\gpfit{\mathsf{subst}(\mathit{x}, \lenv{{\mathrm{u}}}, \leev{}^{+}) = \mathsf{subst}(\mathit{x}, \lenv{{\mathrm{u}}}, \leev{})^{+}}\\[2pt]
\gpfit{\mathsf{subst}(\mathit{x}, \lenv{{\mathrm{u}}}, \glvop{GPINV}(\leev{},\lenv{{\mathrm{a}}},\lenv{{\mathrm{b}}})) = \glvop{GPINV}(\mathsf{subst}(\mathit{x}, \lenv{{\mathrm{u}}}, \leev{}),\mathsf{subst}(\mathit{x}, \lenv{{\mathrm{u}}}, \lenv{{\mathrm{a}}}),\mathsf{subst}(\mathit{x}, \lenv{{\mathrm{u}}}, \lenv{{\mathrm{b}}}))}\\[2pt]
\gpfit{\mathsf{subst}(\mathit{x}, \lenv{{\mathrm{u}}}, \glvop{RREFRep}(\lenv{0},\lenv{1})) = \glvop{RREFRep}(\mathsf{subst}(\mathit{x}, \lenv{{\mathrm{u}}}, \lenv{0}),\mathsf{subst}(\mathit{x}, \lenv{{\mathrm{u}}}, \lenv{1}))}\\[2pt]
\gpfit{\mathsf{subst}(\mathit{x}, \lenv{{\mathrm{u}}}, \leev{}^{-1}) = \mathsf{subst}(\mathit{x}, \lenv{{\mathrm{u}}}, \leev{})^{-1}}\\[2pt]
\gpfit{\mathsf{subst}(\mathit{x}, \lenv{{\mathrm{u}}}, \glvop{SVDS}(\leev{})) = \glvop{SVDS}(\mathsf{subst}(\mathit{x}, \lenv{{\mathrm{u}}}, \leev{}))}\end{center}
}{\texttt{GLang.Val\allowbreak Expr.subst}}{GLang/ValueLN.lean:429}
\gldefdefn{basis.subst}{Substitution in bases}{clauses}{\mathsf{subst}(\mathit{x}, \lee{}, \textcolor{lang}{\mathit{x}})}{\begin{center}\gpfit{\mathsf{subst}(\mathit{x}, \lee{}, \lbzero{}) = \lbzero{}}\\[2pt]
\gpfit{\mathsf{subst}(\mathit{x}, \lee{}, \lbone{}) = \lbone{}}\\[2pt]
\gpfit{\mathsf{subst}(\mathit{x}, \lee{}, \lb{}^{\lstar}) = \mathsf{subst}(\mathit{x}, \lee{}, \lb{})^{\lstar}}\\[2pt]
\gpfit{\mathsf{subst}(\mathit{x}, \lee{}, \lbn{0} \losum \lbn{1}) = \mathsf{subst}(\mathit{x}, \lee{}, \lbn{0}) \losum \mathsf{subst}(\mathit{x}, \lee{}, \lbn{1})}\\[2pt]
\gpfit{\mathsf{subst}(\mathit{x}, \lee{}, \lbn{0} \loprod \lbn{1}) = \mathsf{subst}(\mathit{x}, \lee{}, \lbn{0}) \loprod \mathsf{subst}(\mathit{x}, \lee{}, \lbn{1})}\\[2pt]
\gpfit{\mathsf{subst}(\mathit{x}, \lee{}, \pBasisC{\ls{}}{\lbn{0}}{\lbn{1}}) = \pBasisC{\mathsf{subst}(\mathit{x}, \lee{}, \ls{})}{\mathsf{subst}(\mathit{x}, \lee{}, \lbn{0})}{\mathsf{subst}(\mathit{x}, \lee{}, \lbn{1})}}\\[2pt]
\gpfit{\mathsf{subst}(\mathit{x}, \lee{}, \iBasisC{\lb{}}) = \iBasisC{\mathsf{subst}(\mathit{x}, \lee{}, \lb{})}}\\[2pt]
\gpfit{\mathsf{subst}(\mathit{x}, \lee{}, \llpl \mathit{es}\llpr_{\lb{}}) = \llpl \left[\, \mathsf{subst}(\mathit{x}, \lee{}, \mathit{z}) \mid \mathit{z} \in \mathit{es} \,\right]\llpr_{\mathsf{subst}(\mathit{x}, \lee{}, \lb{})}}\\[2pt]
\gpfit{(\frameB{y}{\lbn{0}}{\lbn{1}})\subst{\mathit{x}}{\lee{}} = \frameB{y}{(\lbn{0})\subst{\mathit{x}}{\lee{}}}{(\lbn{1})\subst{\mathit{x}}{\lee{}}}\quad(y\ne \mathit{x},\ y\notin \operatorname{fv}(\lee{}))}\end{center}
}{\texttt{GLang.Basis.subst}}{GLang/Syntax.lean:249}
\gldefdefn{ty.subst}{Substitution in types}{clauses}{\mathsf{subst}(\mathit{x}, \lee{}, \mathit{x})}{\begin{center}\gpfit{((y\colon \ltyn{0})\fnto \ltyn{1})\subst{\mathit{x}}{\lee{}} = (y\colon (\ltyn{0})\subst{\mathit{x}}{\lee{}})\fnto (\ltyn{1})\subst{\mathit{x}}{\lee{}}\quad(y\ne \mathit{x},\ y\notin \operatorname{fv}(\lee{}))}\end{center}
}{\texttt{GLang.Ty.subst}}{GLang/Syntax.lean:209}
\gldefdefn{valTy.subst}{Substitution in value types}{clauses}{\mathsf{subst}(\mathit{x}, \lenv{{\mathrm{u}}}, \ltynv{{\mathrm{x}}})}{\begin{center}\gpfit{((y\colon \ltynv{0})\fnto \ltynv{1})\subst{\mathit{x}}{\lenv{{\mathrm{u}}}} = (y\colon (\ltynv{0})\subst{\mathit{x}}{\lenv{{\mathrm{u}}}})\fnto (\ltynv{1})\subst{\mathit{x}}{\lenv{{\mathrm{u}}}}\quad(y\ne \mathit{x},\ y\notin \operatorname{fv}(\lenv{{\mathrm{u}}}))}\end{center}
}{\texttt{GLang.Val\allowbreak Ty.subst}}{GLang/ValueLN.lean:425}
\gldefdefn{valExpr.substList}{Substitution in value lists}{clauses}{\mathsf{substList}(\mathit{x}, \lenv{{\mathrm{u}}}, \mathit{x})}{\begin{center}\gpfit{\mathsf{substList}(\mathit{x}, \lenv{{\mathrm{u}}}, []) = []}\\[2pt]
\gpfit{\mathsf{substList}(\mathit{x}, \lenv{{\mathrm{u}}}, \leev{} \mathbin{:\!:} \mathit{es}) = \mathsf{subst}(\mathit{x}, \lenv{{\mathrm{u}}}, \leev{}) \mathbin{:\!:} \mathsf{substList}(\mathit{x}, \lenv{{\mathrm{u}}}, \mathit{es})}\end{center}
}{\texttt{GLang.Val\allowbreak Expr.subst\allowbreak List}}{GLang/ValueLN.lean:466}
\gldefdefn{valueShape.tensor}{\texttt{Value\allowbreak Shape.tensor}}{clauses}{\operatorname{Mat}_{\mathit{q}}(\lsv{},\mathit{a})}{\begin{center}\gpfit{\operatorname{Mat}_{\mathit{q}}(\lsnv{0} \loprod \lsnv{1},\mathit{a}) = \lvert \mathit{a}\rvert = (\dim_{\mathit{q}}(\lsnv{0})) \cdot (\dim_{\mathit{q}}(\lsnv{1}))}\\[2pt]
\gpfit{\operatorname{Mat}_{\mathit{q}}(\lsv{},\mathit{a}) = \mathsf{False}}\end{center}
}{\texttt{GLang.Value\allowbreak Shape.tensor}}{GLang/ValuePrimitiveShapes.lean:16}
\gldefdefn{basis.tof}{\texttt{Basis.tof}}{clauses}{\tof{\textcolor{lang}{\mathit{x}}}}{\begin{center}\gpfit{\tof{\frameB{x}{\lbn{0}}{\lbn{1}}} = (x\colon \tof{\lbn{0}}\fnto \tof{\lbn{1}})}\\[2pt]
\gpfit{\tof{\lb{}}=\sof{\lb{}}\quad(\lb{}\text{ is not a frame})}\end{center}
}{\texttt{GLang.Basis.tof}}{GLang/Basis.lean:74}
\gldefequation{basis.tof.frame}{\tof{\frameB{x}{\lbn{0}}{\lbn{1}}} = (x\colon \tof{\lbn{0}}\fnto \tof{\lbn{1}})}
\gldefequation{basis.tof.space}{\tof{\lb{}}=\sof{\lb{}}\quad(\lb{}\text{ is not a frame})}
\gldefdefn{valueShape.ttv}{\texttt{Value\allowbreak Shape.ttv}}{clauses}{|\mathit{a}|=d_{\mathit{q}}(\lsv{}) \land |\lb{}|=d_{\mathit{q}}(\lsnv{{\mathrm{t}}}) \land d_{\mathit{q}}(\partial(\lsv{}))=d_{\mathit{q}}(\lsnv{{\mathrm{contract}}})=d_{\mathit{q}}(\lsnv{{\mathrm{t}}})}{\begin{center}\gpfit{|\mathit{a}|=d_{\mathit{q}}(\lsnv{0} \loprod \lsnv{1}) \land |\lb{}|=d_{\mathit{q}}(\lsnv{{\mathrm{t}}}) \land d_{\mathit{q}}(\partial(\lsnv{0} \loprod \lsnv{1}))=d_{\mathit{q}}(\lsnv{{\mathrm{contract}}})=d_{\mathit{q}}(\lsnv{{\mathrm{t}}}) = |\mathit{a}|=d_{\mathit{q}}(\lsnv{0} \loprod \lsnv{1}) \;\land\; |\lb{}|=d_{\mathit{q}}(\lsnv{{\mathrm{t}}}) \;\wedge\; \dim_{\mathit{q}}(\lsnv{0}) = \dim_{\mathit{q}}(\lsnv{{\mathrm{contract}}}) \;\wedge\; \dim_{\mathit{q}}(\lsnv{{\mathrm{t}}}) = \dim_{\mathit{q}}(\lsnv{{\mathrm{contract}}})}\\[2pt]
\gpfit{|\mathit{a}|=d_{\mathit{q}}(\lsv{}) \land |\lb{}|=d_{\mathit{q}}(\lsnv{{\mathrm{t}}}) \land d_{\mathit{q}}(\partial(\lsv{}))=d_{\mathit{q}}(\lsnv{{\mathrm{contract}}})=d_{\mathit{q}}(\lsnv{{\mathrm{t}}}) = |\mathit{a}|=d_{\mathit{q}}(\lsv{}) \;\land\; |\lb{}|=d_{\mathit{q}}(\lsnv{{\mathrm{t}}}) \;\wedge\; \dim_{\mathit{q}}(\lsv{}) = \dim_{\mathit{q}}(\lsnv{{\mathrm{contract}}}) \;\wedge\; \dim_{\mathit{q}}(\lsnv{{\mathrm{t}}}) = \dim_{\mathit{q}}(\lsnv{{\mathrm{contract}}})}\end{center}
}{\texttt{GLang.Value\allowbreak Shape.ttv}}{GLang/ValuePrimitiveShapes.lean:24}
\gldefdefnrows{
\texttt{Partial\allowbreak Adequacy.Agree} & clauses & \texttt{\footnotesize GLang/Adequacy/Partial\allowbreak Source\allowbreak Agreement.lean:9}\\
\texttt{Agree\allowbreak Dims} & clauses & \texttt{\footnotesize GLang/Adequacy/Raw\allowbreak Reification.lean:75}\\
Dimension agreement & clauses & \texttt{\footnotesize GLang/Adequacy/Partial\allowbreak Source\allowbreak Public.lean:10}\\
Dimension agreement & clauses & \texttt{\footnotesize GLang/Adequacy/Direct\allowbreak Public.lean:55}\\
Dimension agreement (general oracle) & conjunction & \texttt{\footnotesize GLang/Adequacy/Raw\allowbreak Reification.lean:62}\\
Environments agreeing on a set of names & conjunction & \texttt{\footnotesize GLang/FVIrrelevance.lean:25}\\
\texttt{Adequacy.Basis\allowbreak At} & clauses & \texttt{\footnotesize GLang/Adequacy/Public.lean:36}\\
\texttt{Adequacy.Direct.Basis\allowbreak At} & equation & \texttt{\footnotesize GLang/Adequacy/Direct\allowbreak Public.lean:42}\\
\texttt{Value\allowbreak Query\allowbreak Dependencies.Bounded} & clauses & \texttt{\footnotesize GLang/Value\allowbreak Query\allowbreak Dependencies.lean:45}\\
\texttt{Operational.Cob} & equation & \texttt{\footnotesize GLang/Operational.lean:38}\\
Change-of-basis agreement & clauses & \texttt{\footnotesize GLang/Adequacy/Partial\allowbreak Source\allowbreak Public.lean:11}\\
Change-of-basis agreement (general oracle) & clauses & \texttt{\footnotesize GLang/Adequacy/Agree.lean:253}\\
Change-of-basis agreement & equation & \texttt{\footnotesize GLang/Adequacy/Direct\allowbreak Public.lean:61}\\
\texttt{Partial\allowbreak Adequacy.Cob\allowbreak Agree} & clauses & \texttt{\footnotesize GLang/Adequacy/Partial\allowbreak Source\allowbreak Cob.lean:20}\\
\texttt{Partial\allowbreak Adequacy.Cob\allowbreak Result} & clauses & \texttt{\footnotesize GLang/Adequacy/Partial\allowbreak Source\allowbreak Cob.lean:15}\\
\texttt{Partial\allowbreak Adequacy.Cob\allowbreak Rows} & clauses & \texttt{\footnotesize GLang/Adequacy/Partial\allowbreak Source\allowbreak Cob.lean:9}\\
\texttt{Runtime\allowbreak Value.Total\allowbreak Substitution.Codomain\allowbreak Law} & clauses & \texttt{\footnotesize GLang/Runtime\allowbreak Value\allowbreak Total\allowbreak Semantic\allowbreak Substitution.lean:506}\\
Environment consistency & clauses & \texttt{\footnotesize GLang/Adequacy/Partial\allowbreak Source\allowbreak Public.lean:8}\\
Environment consistency & presentation & \texttt{\footnotesize GLang/Adequacy/Direct\allowbreak Public.lean:26}\\
\texttt{Partial\allowbreak Adequacy.Consistent} & conjunction & \texttt{\footnotesize GLang/Adequacy/Partial\allowbreak Source\allowbreak History.lean:1379}\\
Environment consistency (general oracle alias) & clauses & \texttt{\footnotesize GLang/Adequacy/Raw\allowbreak Typed\allowbreak Consistent.lean:20}\\
Operational context history & clauses & \texttt{\footnotesize GLang/Operational.lean:87}\\
Context-history telescope & rules & \texttt{\footnotesize GLang/Value\allowbreak Progress.lean:380}\\
Context scoping and namespace disjointness & conjunction & \texttt{\footnotesize GLang/Wellformed.lean:31}\\
\texttt{Ctx\allowbreak WF'} & conjunction & \texttt{\footnotesize GLang/Value\allowbreak Preservation\allowbreak V2.lean:57}\\
\texttt{Partial\allowbreak Adequacy.Dims\allowbreak At} & clauses & \texttt{\footnotesize GLang/Adequacy/Partial\allowbreak Source\allowbreak Agreement.lean:41}\\
Environment well-formedness & conjunction & \texttt{\footnotesize GLang/Wellformed.lean:50}\\
\texttt{Operational.Unchecked.Eval\allowbreak BB} & equation & \texttt{\footnotesize GLang/Operational.lean:209}\\
Exact-pair evaluator creation rank & clauses & \texttt{\footnotesize GLang/Adequacy/Raw\allowbreak Consistent\allowbreak Rank.lean:30}\\
\texttt{Eval\allowbreak Ctx\allowbreak Lam\allowbreak Anchored} & clauses & \texttt{\footnotesize GLang/Value\allowbreak Preservation\allowbreak V2.lean:75}\\
\texttt{Eval\allowbreak Ctx\allowbreak Lc} & clauses & \texttt{\footnotesize GLang/Value\allowbreak Realize\allowbreak Lc.lean:21}\\
Store well-formedness (weak) & clauses & \texttt{\footnotesize GLang/Value\allowbreak Preservation\allowbreak V2.lean:67}\\
Store well-formedness & clauses & \texttt{\footnotesize GLang/Value\allowbreak Preservation\allowbreak V2.lean:92}\\
Runtime-entry creation certificate & conjunction & \texttt{\footnotesize GLang/Value\allowbreak Progress.lean:358}\\
Runtime-entry validity & conjunction & \texttt{\footnotesize GLang/Value\allowbreak Progress.lean:349}\\
Nonlinear frame anchor & clauses & \texttt{\footnotesize GLang/Value\allowbreak Progress.lean:339}\\
\texttt{Operational.Unchecked.Eval\allowbreak S} & equation & \texttt{\footnotesize GLang/Operational.lean:203}\\
\texttt{Value\allowbreak Query\allowbreak Dependencies.Finite} & clauses & \texttt{\footnotesize GLang/Value\allowbreak Query\allowbreak Dependencies.lean:51}\\
\texttt{GSVDInputs\allowbreak Shape} & clauses & \texttt{\footnotesize GLang/Value\allowbreak Numerical\allowbreak Inputs.lean:12}\\
\texttt{GSVDecomposition} & conjunction & \texttt{\footnotesize GLang/Value\allowbreak SVD.lean:104}\\
Higher-order linearity & clauses & \texttt{\footnotesize GLang/Source\allowbreak Soundness\allowbreak Strong.lean:36}\\
\texttt{Partial\allowbreak Adequacy.History} & rules & \texttt{\footnotesize GLang/Adequacy/Partial\allowbreak Source\allowbreak History.lean:338}\\
Well-formed source context & rules & \texttt{\footnotesize GLang/Source\allowbreak Context\allowbreak Inductive.lean:27}\\
\texttt{Val\allowbreak Expr.Infer\allowbreak Space\allowbreak Matches\allowbreak Base} & clauses & \texttt{\footnotesize GLang/Adequacy/Raw\allowbreak Consistent.lean:354}\\
Inner-product reserved-name freshness & conjunction & \texttt{\footnotesize GLang/Inner\allowbreak Product.lean:294}\\
Inner-product laws & clauses & \texttt{\footnotesize GLang/Denot.lean:52}\\
Ordered list basis & clauses & \texttt{\footnotesize GLang/Basis\allowbreak FSVGSVD.lean:57}\\
Scope containment & equation & \texttt{\footnotesize GLang/Value\allowbreak Preservation\allowbreak V2.lean:143}\\
\texttt{Legacy\allowbreak Agree} & clauses & \texttt{\footnotesize GLang/Adequacy/Agree.lean:105}\\
Linearity at a variable & clauses & \texttt{\footnotesize GLang/Soundness.lean:42}\\
\texttt{Operational.Unchecked.Op\allowbreak Step} & equation & \texttt{\footnotesize GLang/Operational.lean:212}\\
\texttt{Oracle\allowbreak Spec} & conjunction & \texttt{\footnotesize GLang/Value\allowbreak Denot\allowbreak Sem.lean:2102}\\
\texttt{Partial\allowbreak Value\allowbreak Domains.Partial\allowbreak Coherent} & clauses & \texttt{\footnotesize GLang/Partial\allowbreak Value\allowbreak Relation.lean:20}\\
\texttt{Runtime\allowbreak Value.Partial\allowbreak Env\allowbreak WF} & clauses & \texttt{\footnotesize GLang/Runtime\allowbreak Value\allowbreak Correspondence.lean:11}\\
\texttt{Partial\allowbreak Value\allowbreak Domains.Partial\allowbreak Env\allowbreak WF} & clauses & \texttt{\footnotesize GLang/Partial\allowbreak Value\allowbreak Runtime\allowbreak Relation.lean:29}\\
\texttt{Partial\allowbreak Value\allowbreak Domains.Partial\allowbreak Environments\allowbreak Related} & clauses & \texttt{\footnotesize GLang/Partial\allowbreak Value\allowbreak Runtime\allowbreak Relation.lean:21}\\
\texttt{Paper\allowbreak Value\allowbreak Domains.Partial\allowbreak Rel} & clauses & \texttt{\footnotesize GLang/Paper\allowbreak Value\allowbreak Extensionality.lean:10}\\
\texttt{Partial\allowbreak Value\allowbreak Domains.Partial\allowbreak Related} & clauses & \texttt{\footnotesize GLang/Partial\allowbreak Value\allowbreak Relation.lean:10}\\
\texttt{Partial\allowbreak Value\allowbreak Domains.Partial\allowbreak Runtime\allowbreak Related} & clauses & \texttt{\footnotesize GLang/Partial\allowbreak Value\allowbreak Runtime\allowbreak Relation.lean:10}\\
Pointwise linearity through sections & clauses & \texttt{\footnotesize GLang/Source\allowbreak Soundness\allowbreak Strong.lean:25}\\
\texttt{Raw\allowbreak Adequacy\allowbreak Oracle\allowbreak Spec} & conjunction & \texttt{\footnotesize GLang/Adequacy/Raw\allowbreak Agreement.lean:247}\\
\texttt{Raw\allowbreak Agree} & clauses & \texttt{\footnotesize GLang/Adequacy/Raw\allowbreak Agreement.lean:21}\\
Raw dimension agreement & clauses & \texttt{\footnotesize GLang/Adequacy/Raw\allowbreak Agreement.lean:92}\\
\texttt{Raw\allowbreak Application\allowbreak Oracle\allowbreak Spec} & conjunction & \texttt{\footnotesize GLang/Adequacy/Raw\allowbreak Application\allowbreak Oracle\allowbreak Spec.lean:6}\\
\texttt{Raw\allowbreak Basis\allowbreak At} & clauses & \texttt{\footnotesize GLang/Adequacy/Raw\allowbreak Reification.lean:23}\\
Environment consistency (general oracle) & rules & \texttt{\footnotesize GLang/Adequacy/Raw\allowbreak Consistent.lean:377}\\
\texttt{Raw\allowbreak Reifiable\allowbreak At} & clauses & \texttt{\footnotesize GLang/Adequacy/Raw\allowbreak Reification.lean:39}\\
Raw result agreement & clauses & \texttt{\footnotesize GLang/Adequacy/Raw\allowbreak Agreement.lean:47}\\
\texttt{Raw\allowbreak Source\allowbreak Injective\allowbreak At} & clauses & \texttt{\footnotesize GLang/Adequacy/Raw\allowbreak Reification.lean:48}\\
\texttt{Paper\allowbreak Value\allowbreak Domains.Raw\allowbreak Type\allowbreak Presentation} & clauses & \texttt{\footnotesize GLang/Paper\allowbreak Value\allowbreak Raw\allowbreak Presentation.lean:12}\\
\texttt{Raw\allowbreak Typed\allowbreak Consistent} & conjunction & \texttt{\footnotesize GLang/Adequacy/Raw\allowbreak Typed\allowbreak Consistent.lean:11}\\
\texttt{Raw\allowbreak Value\allowbreak Open\allowbreak Push\allowbreak Oracle\allowbreak Coherent} & clauses & \texttt{\footnotesize GLang/Adequacy/Raw\allowbreak Agreement.lean:205}\\
Recursive semantic orthonormality & clauses & \texttt{\footnotesize GLang/Basis\allowbreak Soundness.lean:318}\\
\texttt{Partial\allowbreak Adequacy.Result\allowbreak Agree} & clauses & \texttt{\footnotesize GLang/Adequacy/Partial\allowbreak Source\allowbreak Agreement.lean:35}\\
Result agreement & clauses & \texttt{\footnotesize GLang/Adequacy/Partial\allowbreak Source\allowbreak Public.lean:9}\\
Result agreement (general oracle) & clauses & \texttt{\footnotesize GLang/Adequacy/Raw\allowbreak Agreement.lean:131}\\
Result agreement & equation & \texttt{\footnotesize GLang/Adequacy/Direct\allowbreak Public.lean:48}\\
\texttt{Value\allowbreak Canonical\allowbreak FSV.Result\allowbreak Matrix} & conjunction & \texttt{\footnotesize GLang/Value\allowbreak Canonical\allowbreak FSV.lean:17}\\
Soundness claim & clauses & \texttt{\footnotesize GLang/Soundness.lean:58}\\
Flat source-context conditions & conjunction & \texttt{\footnotesize GLang/Source\allowbreak Context\allowbreak Inductive.lean:17}\\
Source-context formation validity & conjunction & \texttt{\footnotesize GLang/Source\allowbreak Soundness\allowbreak Strong.lean:54}\\
Existing source-entry link & clauses & \texttt{\footnotesize GLang/Value\allowbreak Progress.lean:325}\\
Source-entry validity & conjunction & \texttt{\footnotesize GLang/Value\allowbreak Progress.lean:250}\\
\texttt{Operational.Space} & equation & \texttt{\footnotesize GLang/Operational.lean:29}\\
\texttt{Std\allowbreak Anchored} & rules & \texttt{\footnotesize GLang/Value\allowbreak Progress.lean:86}\\
Standard-basis completeness & clauses & \texttt{\footnotesize GLang/Value\allowbreak Progress.lean:233}\\
\texttt{Std\allowbreak Pi\allowbreak Anchored} & clauses & \texttt{\footnotesize GLang/Value\allowbreak Progress.lean:127}\\
Strengthened source-soundness claim & clauses & \texttt{\footnotesize GLang/Source\allowbreak Soundness\allowbreak Strong.lean:124}\\
\texttt{Operational.Term} & equation & \texttt{\footnotesize GLang/Operational.lean:41}\\
\texttt{Operational.Ty} & equation & \texttt{\footnotesize GLang/Operational.lean:32}\\
\texttt{Operational.Ty\allowbreak At} & equation & \texttt{\footnotesize GLang/Operational.lean:35}\\
\texttt{Operational.Typed\allowbreak Context} & rules & \texttt{\footnotesize GLang/Operational\allowbreak Typed\allowbreak Context.lean:11}\\
\texttt{Typing\allowbreak Imply\allowbreak Eval} & clauses & \texttt{\footnotesize GLang/Value\allowbreak Preservation\allowbreak V2.lean:129}\\
Canonical value-result soundness & clauses & \texttt{\footnotesize GLang/Adequacy/Value\allowbreak Soundness\allowbreak Sem\allowbreak S.lean:70}\\
Canonical value soundness & clauses & \texttt{\footnotesize GLang/Adequacy/Value\allowbreak Soundness\allowbreak Sem\allowbreak S.lean:49}\\
Value-context validity & clauses & \texttt{\footnotesize GLang/Value\allowbreak Sound\allowbreak Vals.lean:85}\\
Value-result soundness & clauses & \texttt{\footnotesize GLang/Adequacy/Value\allowbreak Soundness\allowbreak Sem\allowbreak S.lean:64}\\
Value soundness & clauses & \texttt{\footnotesize GLang/Adequacy/Value\allowbreak Soundness\allowbreak Sem\allowbreak S.lean:33}\\
Value-type regularity & clauses & \texttt{\footnotesize GLang/Value\allowbreak Valid\allowbreak C.lean:52}\\
\texttt{Paper\allowbreak Value\allowbreak Domains.World} & conjunction & \texttt{\footnotesize GLang/Paper\allowbreak Value\allowbreak Extensionality.lean:34}\\
\texttt{Partial\allowbreak Value\allowbreak Domains.World\allowbreak Runtime\allowbreak Related} & clauses & \texttt{\footnotesize GLang/Partial\allowbreak Value\allowbreak Family\allowbreak Binders.lean:9}\\
\texttt{Space.close\allowbreak Rec} & clauses & \texttt{\footnotesize GLang/Source\allowbreak LN.lean:105}\\
\texttt{Ty.close\allowbreak Rec} & clauses & \texttt{\footnotesize GLang/Source\allowbreak LN.lean:116}\\
\texttt{Term.close\allowbreak Rec} & clauses & \texttt{\footnotesize GLang/Source\allowbreak LN.lean:120}\\
\texttt{Val\allowbreak Expr.close\allowbreak Rec} & clauses & \texttt{\footnotesize GLang/Value\allowbreak LN.lean:363}\\
\texttt{Basis.close\allowbreak Rec} & clauses & \texttt{\footnotesize GLang/Source\allowbreak LN.lean:147}\\
\texttt{Val\allowbreak Ty.close\allowbreak Rec} & clauses & \texttt{\footnotesize GLang/Value\allowbreak LN.lean:359}\\
\texttt{Val\allowbreak Space.close\allowbreak Rec} & clauses & \texttt{\footnotesize GLang/Value\allowbreak LN.lean:352}\\
\texttt{Term.close\allowbreak Rec\allowbreak List} & clauses & \texttt{\footnotesize GLang/Source\allowbreak LN.lean:143}\\
\texttt{Val\allowbreak Expr.close\allowbreak Rec\allowbreak List} & clauses & \texttt{\footnotesize GLang/Value\allowbreak LN.lean:401}\\
\texttt{denot\allowbreak Basis} & clauses & \texttt{\footnotesize GLang/Basis.lean:886}\\
\texttt{denot\allowbreak Space\allowbreak Aux} & clauses & \texttt{\footnotesize GLang/Denot.lean:714}\\
\texttt{denot\allowbreak Term\allowbreak Aux} & clauses & \texttt{\footnotesize GLang/Denot.lean:761}\\
\texttt{denot\allowbreak Ty\allowbreak Aux} & clauses & \texttt{\footnotesize GLang/Denot.lean:751}\\
\texttt{denot\allowbreak Val\allowbreak B\_\allowbreak sem\allowbreak S} & curated clauses & \texttt{\footnotesize GLang/Value\allowbreak Denot\allowbreak Sem.lean:811}\\
\texttt{denot\allowbreak Val\allowbreak Ty\allowbreak S} & clauses & \texttt{\footnotesize GLang/Value\allowbreak Denot\allowbreak Sem.lean:511}\\
\texttt{Val\allowbreak Space.dof} & clauses & \texttt{\footnotesize GLang/Value\allowbreak Denot.lean:135}\\
\texttt{Value\allowbreak Shape.family} & clauses & \texttt{\footnotesize GLang/Value\allowbreak Primitive\allowbreak Shapes.lean:41}\\
\texttt{Space.fv} & clauses & \texttt{\footnotesize GLang/Syntax.lean:271}\\
\texttt{Val\allowbreak Ty.fv} & clauses & \texttt{\footnotesize GLang/Values.lean:376}\\
\texttt{Val\allowbreak Expr.fv} & clauses & \texttt{\footnotesize GLang/Values.lean:381}\\
\texttt{Basis.fv} & clauses & \texttt{\footnotesize GLang/Syntax.lean:327}\\
\texttt{Val\allowbreak Space.fv} & clauses & \texttt{\footnotesize GLang/Values.lean:369}\\
\texttt{Ty.fv} & clauses & \texttt{\footnotesize GLang/Syntax.lean:282}\\
\texttt{Term.fv} & clauses & \texttt{\footnotesize GLang/Syntax.lean:287}\\
Free variables of value lists & clauses & \texttt{\footnotesize GLang/Values.lean:420}\\
Free variables of term lists & clauses & \texttt{\footnotesize GLang/Syntax.lean:307}\\
inferSpace & clauses & \texttt{\footnotesize GLang/Value\allowbreak Denot.lean:1020}\\
inferSpaceCtx & clauses & \texttt{\footnotesize GLang/Value\allowbreak Denot.lean:1139}\\
inferType & clauses & \texttt{\footnotesize GLang/Value\allowbreak Denot.lean:790}\\
inferTypeCtx & clauses & \texttt{\footnotesize GLang/Value\allowbreak Denot.lean:918}\\
\texttt{Val\allowbreak Expr.instantiate} & clauses & \texttt{\footnotesize GLang/Value\allowbreak LN.lean:509}\\
\texttt{Ty.instantiate} & clauses & \texttt{\footnotesize GLang/Source\allowbreak LN.lean:162}\\
\texttt{Val\allowbreak Ty.instantiate} & clauses & \texttt{\footnotesize GLang/Value\allowbreak LN.lean:511}\\
\texttt{Basis.instantiate} & clauses & \texttt{\footnotesize GLang/Source\allowbreak LN.lean:163}\\
\texttt{Term.instantiate} & clauses & \texttt{\footnotesize GLang/Source\allowbreak LN.lean:160}\\
Syntactically standard basis & clauses & \texttt{\footnotesize GLang/Value\allowbreak Operational.lean:259}\\
\texttt{Value\allowbreak Shape.matrices} & clauses & \texttt{\footnotesize GLang/Value\allowbreak Primitive\allowbreak Shapes.lean:21}\\
\texttt{Value\allowbreak Shape.mtm} & clauses & \texttt{\footnotesize GLang/Value\allowbreak Primitive\allowbreak Shapes.lean:33}\\
\texttt{Basis.open\allowbreak Rec} & clauses & \texttt{\footnotesize GLang/Source\allowbreak LN.lean:89}\\
\texttt{Val\allowbreak Expr.open\allowbreak Rec} & clauses & \texttt{\footnotesize GLang/Value\allowbreak LN.lean:44}\\
\texttt{Space.open\allowbreak Rec} & clauses & \texttt{\footnotesize GLang/Source\allowbreak LN.lean:42}\\
\texttt{Val\allowbreak Space.open\allowbreak Rec} & clauses & \texttt{\footnotesize GLang/Value\allowbreak LN.lean:32}\\
\texttt{Val\allowbreak Ty.open\allowbreak Rec} & clauses & \texttt{\footnotesize GLang/Value\allowbreak LN.lean:39}\\
\texttt{Ty.open\allowbreak Rec} & clauses & \texttt{\footnotesize GLang/Source\allowbreak LN.lean:53}\\
\texttt{Term.open\allowbreak Rec} & clauses & \texttt{\footnotesize GLang/Source\allowbreak LN.lean:58}\\
\texttt{Term.open\allowbreak Rec\allowbreak List} & clauses & \texttt{\footnotesize GLang/Source\allowbreak LN.lean:85}\\
\texttt{Val\allowbreak Expr.open\allowbreak Rec\allowbreak List} & clauses & \texttt{\footnotesize GLang/Value\allowbreak LN.lean:82}\\
\texttt{Val\allowbreak Expr.open\allowbreak Var} & clauses & \texttt{\footnotesize GLang/Value\allowbreak LN.lean:513}\\
\texttt{Val\allowbreak Ty.open\allowbreak Var} & clauses & \texttt{\footnotesize GLang/Value\allowbreak LN.lean:516}\\
\texttt{Basis.open\allowbreak Var} & clauses & \texttt{\footnotesize GLang/Source\allowbreak LN.lean:168}\\
\texttt{Ty.open\allowbreak Var} & clauses & \texttt{\footnotesize GLang/Source\allowbreak LN.lean:167}\\
\texttt{Term.open\allowbreak Var} & clauses & \texttt{\footnotesize GLang/Source\allowbreak LN.lean:165}\\
\texttt{Value\allowbreak Shape.pair} & clauses & \texttt{\footnotesize GLang/Value\allowbreak Primitive\allowbreak Shapes.lean:13}\\
\texttt{reshap\allowbreak Array\allowbreak Between\allowbreak S} & clauses & \texttt{\footnotesize GLang/Value\allowbreak Denot\allowbreak Sem.lean:662}\\
\texttt{reshap\allowbreak Denot\allowbreak Sem\allowbreak Between\allowbreak S} & clauses & \texttt{\footnotesize GLang/Value\allowbreak Denot\allowbreak Sem.lean:681}\\
\texttt{reshap\allowbreak Denot\allowbreak Sem\allowbreak S} & clauses & \texttt{\footnotesize GLang/Value\allowbreak Denot\allowbreak Sem.lean:726}\\
\texttt{Value\allowbreak Shape.scale} & clauses & \texttt{\footnotesize GLang/Value\allowbreak Primitive\allowbreak Shapes.lean:10}\\
\texttt{Val\allowbreak Space.shape\allowbreak In\allowbreak Bounds} & clauses & \texttt{\footnotesize GLang/Value\allowbreak Denot.lean:250}\\
\texttt{Val\allowbreak Space.shp\allowbreak Of} & clauses & \texttt{\footnotesize GLang/Value\allowbreak Denot.lean:116}\\
\texttt{Basis.sof} & clauses & \texttt{\footnotesize GLang/Basis.lean:60}\\
\texttt{Basis.std} & clauses & \texttt{\footnotesize GLang/Basis.lean:523}\\
\texttt{Basis.std\allowbreak Space} & clauses & \texttt{\footnotesize GLang/Basis.lean:506}\\
Substitution in terms & clauses & \texttt{\footnotesize GLang/Syntax.lean:215}\\
Substitution in spaces & clauses & \texttt{\footnotesize GLang/Syntax.lean:194}\\
Substitution in value spaces & clauses & \texttt{\footnotesize GLang/Value\allowbreak LN.lean:418}\\
Substitution in values & clauses & \texttt{\footnotesize GLang/Value\allowbreak LN.lean:429}\\
Substitution in bases & clauses & \texttt{\footnotesize GLang/Syntax.lean:249}\\
Substitution in types & clauses & \texttt{\footnotesize GLang/Syntax.lean:209}\\
Substitution in value types & clauses & \texttt{\footnotesize GLang/Value\allowbreak LN.lean:425}\\
Substitution in value lists & clauses & \texttt{\footnotesize GLang/Value\allowbreak LN.lean:466}\\
\texttt{Value\allowbreak Shape.tensor} & clauses & \texttt{\footnotesize GLang/Value\allowbreak Primitive\allowbreak Shapes.lean:16}\\
\texttt{Basis.tof} & clauses & \texttt{\footnotesize GLang/Basis.lean:74}\\
\texttt{Value\allowbreak Shape.ttv} & clauses & \texttt{\footnotesize GLang/Value\allowbreak Primitive\allowbreak Shapes.lean:24}\\
}
\gldefequation{value.dimOf}{\dof{\lsv}{\glvEnv}=\prod_{j=0}^{n}d_j,\qquad(d_j)_{j=0}^{n}=\shpOf{\lsv}{\glvEnv}}
\gldefdefn{value.dimOf}{Value-space dimension}{presentation}{}{\[\dof{\lsv}{\glvEnv}=\prod_{j=0}^{n}d_j,\qquad(d_j)_{j=0}^{n}=\shpOf{\lsv}{\glvEnv}\]}{\texttt{GLang.Val\allowbreak Space.dof}}{GLang/ValueDenot.lean:135}
\gldefequation{value.shapeOf.empty}{\shpOf{\lsempty{}}{\glvEnv}=(0,)}
\gldefdefn{value.shapeOf.empty}{Empty shape}{presentation}{}{\[\shpOf{\lsempty{}}{\glvEnv}=(0,)\]}{\texttt{GLang.Val\allowbreak Space.shp\allowbreak Of}}{GLang/ValueDenot.lean:116}
\gldefequation{value.shapeOf.real}{\shpOf{\lR{}}{\glvEnv}=(1,)}
\gldefdefn{value.shapeOf.real}{Real shape}{presentation}{}{\[\shpOf{\lR{}}{\glvEnv}=(1,)\]}{\texttt{GLang.Val\allowbreak Space.shp\allowbreak Of}}{GLang/ValueDenot.lean:116}
\gldefequation{value.shapeOf.dsum}{\shpOf{\lsnv{0}\losum\lsnv{1}}{\glvEnv}=(\dof{\lsnv{0}}{\glvEnv}+\dof{\lsnv{1}}{\glvEnv},)}
\gldefdefn{value.shapeOf.dsum}{Direct-sum shape}{presentation}{}{\[\shpOf{\lsnv{0}\losum\lsnv{1}}{\glvEnv}=(\dof{\lsnv{0}}{\glvEnv}+\dof{\lsnv{1}}{\glvEnv},)\]}{\texttt{GLang.Val\allowbreak Space.shp\allowbreak Of}}{GLang/ValueDenot.lean:116}
\gldefequation{value.shapeOf.tprod}{\shpOf{\lsnv{0}\loprod\lsnv{1}}{\glvEnv}=(\dof{\lsnv{0}}{\glvEnv},\dof{\lsnv{1}}{\glvEnv})}
\gldefdefn{value.shapeOf.tprod}{Tensor-product shape}{presentation}{}{\[\shpOf{\lsnv{0}\loprod\lsnv{1}}{\glvEnv}=(\dof{\lsnv{0}}{\glvEnv},\dof{\lsnv{1}}{\glvEnv})\]}{\texttt{GLang.Val\allowbreak Space.shp\allowbreak Of}}{GLang/ValueDenot.lean:116}
\gldefequation{value.shapeOf.ffs}{\shpOf{\lR^{\leev,\lprop,\lsv}}{\glvEnv}=(d_{\lprop},)}
\gldefdefn{value.shapeOf.ffs}{Fundamental-space shape}{presentation}{}{\[\shpOf{\lR^{\leev,\lprop,\lsv}}{\glvEnv}=(d_{\lprop},)\]}{\texttt{GLang.shape\allowbreak Of\_\allowbreak ffs\_\allowbreak svd}}{GLang/PartialValueSVDDimensions.lean:68}
\gldefequation{value.denot.mtm.compact}{\begin{gathered}
\SemA{\glvmtm{\lsnv{1}''}{\lsnv{1}}{\lenv{0}}{\lenv{1}}}{\glvEnv}{d_0r+c}=\sum_{k=0}^{d_1''-1}\SemA{\lenv{0}}{\glvEnv}{d_0k+c}\,\SemA{\lenv{1}}{\glvEnv}{d_1''r+k},\\
0\leq r<d_2,\quad0\leq c<d_0,\quad |\SemA{\lenv{0}}{\glvEnv}{}|=d_1''d_0,\quad |\SemA{\lenv{1}}{\glvEnv}{}|=d_2d_1''
\\
\operatorname{inferSpace}(\lenv{0})=\lsnv{1}'\loprod \lsnv{0},\quad \operatorname{inferSpace}(\lenv{1})=\lsnv{2}\loprod \lsnv{1},\\d_1'=d_1=d_1'',\quad d_j=\dof{\lsnv{j}}{\glvEnv}\ (j=0,1,2),\quad d_1'=\dof{\lsnv{1}'}{\glvEnv},\quad d_1''=\dof{\lsnv{1}''}{\glvEnv}
\end{gathered}}
\gldefdefn{value.denot.mtm.compact}{Matrix contraction coordinates (compact)}{presentation}{}{\[\begin{gathered}
\SemA{\glvmtm{\lsnv{1}''}{\lsnv{1}}{\lenv{0}}{\lenv{1}}}{\glvEnv}{d_0r+c}=\sum_{k=0}^{d_1''-1}\SemA{\lenv{0}}{\glvEnv}{d_0k+c}\,\SemA{\lenv{1}}{\glvEnv}{d_1''r+k},\\
0\leq r<d_2,\quad0\leq c<d_0,\quad |\SemA{\lenv{0}}{\glvEnv}{}|=d_1''d_0,\quad |\SemA{\lenv{1}}{\glvEnv}{}|=d_2d_1''
\\
\operatorname{inferSpace}(\lenv{0})=\lsnv{1}'\loprod \lsnv{0},\quad \operatorname{inferSpace}(\lenv{1})=\lsnv{2}\loprod \lsnv{1},\\d_1'=d_1=d_1'',\quad d_j=\dof{\lsnv{j}}{\glvEnv}\ (j=0,1,2),\quad d_1'=\dof{\lsnv{1}'}{\glvEnv},\quad d_1''=\dof{\lsnv{1}''}{\glvEnv}
\end{gathered}\]}{\texttt{GLang.denot\allowbreak Val\allowbreak B\_\allowbreak sem\allowbreak S}}{GLang/ValueDenotSem.lean:811}
\gldefequation{value.denot.tprod}{\SemA{\lenv{0}\loprod\lenv{1}}{\glvEnv}{d_1 i+j}=\SemA{\lenv{0}}{\glvEnv}{i}\,\SemA{\lenv{1}}{\glvEnv}{j},\quad 0\leq i<d_0,\quad0\leq j<d_1,\quad |\SemA{\lenv{k}}{\glvEnv}{}|=d_k\ (k=0,1)}
\gldefdefn{value.denot.tprod}{Tensor coordinates}{presentation}{}{\[\SemA{\lenv{0}\loprod\lenv{1}}{\glvEnv}{d_1 i+j}=\SemA{\lenv{0}}{\glvEnv}{i}\,\SemA{\lenv{1}}{\glvEnv}{j},\quad 0\leq i<d_0,\quad0\leq j<d_1,\quad |\SemA{\lenv{k}}{\glvEnv}{}|=d_k\ (k=0,1)\]}{\texttt{GLang.denot\allowbreak Val\allowbreak B\_\allowbreak sem\allowbreak S}}{GLang/ValueDenotSem.lean:811}
\gldefequation{value.denot.delta}{\SemA{\ldelta{k}{\lsv}}{\glvEnv}{i}=\begin{cases}1&i=k,\\0&i\ne k,\end{cases}\qquad 0\leq i<\dof{\lsv}{\glvEnv}}
\gldefdefn{value.denot.delta}{One-hot coordinates}{presentation}{}{\[\SemA{\ldelta{k}{\lsv}}{\glvEnv}{i}=\begin{cases}1&i=k,\\0&i\ne k,\end{cases}\qquad 0\leq i<\dof{\lsv}{\glvEnv}\]}{\texttt{GLang.denot\allowbreak Val\allowbreak B\_\allowbreak sem\allowbreak S}}{GLang/ValueDenotSem.lean:811}
\gldefequation{value.denot.mtm}{\begin{gathered}
\SemA{\glvmtm{\lsnv{1}''}{\lsnv{1}}{\lenv{0}}{\lenv{1}}}{\glvEnv}{d_0r+c}=\sum_{k=0}^{d_1''-1}\SemA{\lenv{0}}{\glvEnv}{d_0k+c}\,\SemA{\lenv{1}}{\glvEnv}{d_1''r+k},\quad0\leq r<d_2,\quad0\leq c<d_0,\quad |\SemA{\lenv{0}}{\glvEnv}{}|=d_1''d_0,\quad |\SemA{\lenv{1}}{\glvEnv}{}|=d_2d_1''\\
\operatorname{inferSpace}(\lenv{0})=\lsnv{1}'\loprod \lsnv{0},\quad \operatorname{inferSpace}(\lenv{1})=\lsnv{2}\loprod \lsnv{1},\\d_1'=d_1=d_1'',\quad d_j=\dof{\lsnv{j}}{\glvEnv}\ (j=0,1,2),\quad d_1'=\dof{\lsnv{1}'}{\glvEnv},\quad d_1''=\dof{\lsnv{1}''}{\glvEnv}
\end{gathered}}
\gldefdefn{value.denot.mtm}{Matrix contraction coordinates}{presentation}{}{\[\begin{gathered}
\SemA{\glvmtm{\lsnv{1}''}{\lsnv{1}}{\lenv{0}}{\lenv{1}}}{\glvEnv}{d_0r+c}=\sum_{k=0}^{d_1''-1}\SemA{\lenv{0}}{\glvEnv}{d_0k+c}\,\SemA{\lenv{1}}{\glvEnv}{d_1''r+k},\quad0\leq r<d_2,\quad0\leq c<d_0,\quad |\SemA{\lenv{0}}{\glvEnv}{}|=d_1''d_0,\quad |\SemA{\lenv{1}}{\glvEnv}{}|=d_2d_1''\\
\operatorname{inferSpace}(\lenv{0})=\lsnv{1}'\loprod \lsnv{0},\quad \operatorname{inferSpace}(\lenv{1})=\lsnv{2}\loprod \lsnv{1},\\d_1'=d_1=d_1'',\quad d_j=\dof{\lsnv{j}}{\glvEnv}\ (j=0,1,2),\quad d_1'=\dof{\lsnv{1}'}{\glvEnv},\quad d_1''=\dof{\lsnv{1}''}{\glvEnv}
\end{gathered}\]}{\texttt{GLang.denot\allowbreak Val\allowbreak B\_\allowbreak sem\allowbreak S}}{GLang/ValueDenotSem.lean:811}
\gldefdefn{value.domains}{Value domains}{presentation}{}{\[
\begin{gathered}
\mathcal V_\glvEnv(\lsv)=\mathbb R^{\dof{\lsv}{\glvEnv}}\sqcup\{\glverr\},\\
\mathcal V_\glvEnv(x\colon\ltynv{0}\fnto\ltynv{1})
 =\left(\textstyle\prod_{a\in\mathcal V_\glvEnv(\ltynv{0})\setminus\{\glverr\}}
       \mathcal V_{\glvEnv[x\mapsto a]}(\ltynv{1})\right)\sqcup\{\glverr\}.
\end{gathered}
\]}{\texttt{GLang.Paper\allowbreak Value\allowbreak Domains.Section}}{GLang/PaperValueDomains.lean:16}
\gldefdefn{value.ffs.svd}{Fundamental-space dimensions from SVD}{presentation}{}{For $\shpOf{\lsv}{\glvEnv}=(d_0,d_1)$ and an array
$a=\SemA{\leev}{\glvEnv}{}$ with $|a|=d_0d_1$, put
\begin{gather*}
 A_{ij}=a[d_1i+j]\quad(i<d_0,\ j<d_1),\qquad
 (U,\Sigma,V)=\operatorname{SVD}(A),\qquad r=|\{k\mid\Sigma_{kk}>0\}|,\\
 d_{\Kernel}=d_0-r,\qquad d_{\CoKernel}=d_{\Image}=r,\qquad d_{\CoImage}=d_1-r.
\end{gather*}
Failed shape or value evaluation, non-array values, and incorrect lengths give $\glverr$.}{\texttt{GLang.ffs\allowbreak Dimension\allowbreak SVD\_\allowbreak eq}}{GLang/PartialValueSVDDimensions.lean:52}
\gldefdefn{value.inference.helpers}{Inference auxiliaries}{presentation}{}{\begin{gather*}
\operatorname{spaceOf}(\lsv)=\lsv,\qquad
\operatorname{spaceOf}(x\colon\ltynv{0}\fnto\ltynv{1})=\lsempty{},\\
\operatorname{ambient}(\lprop,\lsnv{0},\lsnv{1})=
\begin{cases}\lsnv{0}&\lprop\in\{\Kernel{},\CoKernel{}\},\\
\lsnv{1}&\lprop\in\{\Image{},\CoImage{}\}.\end{cases}
\end{gather*}
The operators retain their annotations: $\lsv$ is the addition's result space
while reshape records target and source types, and MTM records both contraction spaces.
Application infers its result by substituting its argument in the function codomain.
A fallback $\lsempty{}$ is an inferred space or type, not the evaluation result $\glverr$.}{\texttt{GLang.Val\allowbreak Ty.space\allowbreak Of}}{GLang/ValueDenot.lean:62}
\gldefdefn{value.denot.conventions}{Environments and arrays}{presentation}{}{\begin{gather*}
 d_{\glvEnv}(\lsv{})=\dof{\lsv}{\glvEnv},\qquad
 n_{\glvEnv}(\leev{})=d_{\glvEnv}(\operatorname{inferSpace}(\leev{})),\\
 \mathsf{Result}(X)=X\sqcup\{\glverr\},\qquad
 \glvE{\leev{}}\in\mathsf{Result}(\mathsf{Value}),\\
 \mathsf{arr}_n[f]=(f(i))_{0\leq i<n},\qquad
 a[i]=\begin{cases}a_i&i<|a|,\\\glverr&\text{otherwise},\end{cases}\qquad a[r,c]=a[d_c r+c].
\end{gather*}
Here $\glvEnv$ maps variable names to runtime values (finite arrays or functions).
In each array clause, $\lenv{\rm out}$ is the left-hand expression,
$N=n_{\glvEnv}(\lenv{\rm out})$, and $a_j=\Sem{\lenv{j}}{\glvEnv}$,
evaluated in source order. Coordinate clauses construct $\mathsf{arr}_N[\cdot]$.
Failed dimension evaluations, non-array operands, and failed side conditions give $\glverr$.
The mathematical view $\bar a$ extends $a$ by zero; runtime reads remain checked.
Successful results are written without an injection.
Function bodies use $\glvEnv[x\mapsto a]$; application checks the inferred
domain and the function's runtime argument representation.
All operations propagate errors.}{\texttt{GLang.denot\allowbreak Val\allowbreak B\_\allowbreak sem\allowbreak S}}{GLang/ValueDenotSem.lean:811}
\gldefdefn{value.denot.guards}{Array shape conditions}{presentation}{}{\begin{gather*}
 \operatorname{Mat}_{\glvEnv}(\lsv{},a)\iff\exists \lsnv{0},\lsnv{1}.\;\lsv{}=\lsnv{0}\loprod \lsnv{1}\land |a|=d_{\glvEnv}(\lsnv{0})d_{\glvEnv}(\lsnv{1}),\\
 \partial(\lsv{})=\begin{cases}\lsnv{0}&\lsv{}=\lsnv{0}\loprod \lsnv{1},\\\lsv{}&\text{otherwise}.\end{cases}
\end{gather*}}{\texttt{GLang.Value\allowbreak Shape.tensor}}{GLang/ValuePrimitiveShapes.lean:16}
\gldefdefn{value.denot.types}{Value-type representation}{presentation}{}{The tags $\mathsf{space}$ and $\mathsf{pi}$ below describe the evaluator representation;
$\mathcal V$ additionally enforces exact array lengths.
\begin{align*}
 \glvT{\lsv{}}&=\mathsf{space}(\lsv{},\glvD{\lsv{}}),
 &|\mathsf{space}(\lsv{},n)|&=\mathsf{FlatArray},\\
 \glvT{(x\colon\ltynv{0}\fnto\ltynv{1})}&=\mathsf{pi}(T_0,\lambda a.\SemT{\ltynv{1}}{\glvEnv[x\mapsto a]}),
 &|\mathsf{pi}(T,F)|&=\prod_{x\in|T|}\mathsf{Result}(|F(x)|),\\
 T_0&=\glvT{\ltynv{0}}.
\end{align*}}{\texttt{GLang.denot\allowbreak Val\allowbreak Ty\allowbreak S}}{GLang/ValueDenotSem.lean:511}
\gldefdefn{value.denot.representation}{Value representation}{presentation}{}{$\uparrow_T$ embeds a typed value; $\uparrow_T^{!}$ requires exact lengths;
$\downarrow_T$ extracts a typed value.  Write $\langle A,f\rangle$ for a
function value with argument carrier $A$; $\bar a$ is the mathematical zero extension of $a$.
\begin{align*}
 \uparrow_{\mathsf{space}(\lsv{},n)}a&=(\bar a[i])_{0\leq i<n},\\
  \uparrow_{\mathsf{pi}(T,F)}g&=\langle |T|,\lambda x.\uparrow_{F(x)}(g(x))\rangle,\\
 \uparrow^{!}_{\mathsf{space}(\lsv{},n)}a&=
 \begin{cases}\glvok{a}&|a|=n,\\\glverr&\text{otherwise},\end{cases}\\
 \uparrow^{!}_{\mathsf{pi}(T,F)}g&=\glvok{\langle |T|,\lambda x.\uparrow^{!}_{F(x)}(g(x))\rangle},\\
 \downarrow_{\mathsf{space}(\lsv{},n)}a&=\glvok{a},\\
 \downarrow^{\circ}_{\mathsf{space}(\lsv{},n)}a&=\begin{cases}a&|a|=n,\\\glverr&\text{otherwise},\end{cases}\\
 \downarrow^{\circ}_{\mathsf{pi}(T,F)}v&=\downarrow_{\mathsf{pi}(T,F)}v,\\
 \downarrow^{!}_{\mathsf{space}(\lsv{},n)}a&=\downarrow^{\circ}_{\mathsf{space}(\lsv{},n)}a,\\
 \downarrow^{!}_{\mathsf{pi}(T,F)}\langle A,f\rangle&=\begin{cases}\lambda x.\downarrow^{!}_{F(x)}(f(x))&A=|T|,\\\glverr&\text{otherwise},\end{cases}\\
  \downarrow_{\mathsf{pi}(T,F)}\langle A,f\rangle&=
 \begin{cases}
 \glvok{\lambda x.\downarrow_{F(x)}(f(x))}&A=|T|,\\
 \glverr&\text{otherwise}.
 \end{cases}
\end{align*}
\[
 \langle B,f\rangle\mathbin{\cdot_A}x=
 \begin{cases}f(x)&A=B,\\\glverr&\text{otherwise}.\end{cases}
\]
Extraction and application on incompatible value shapes give $\glverr$.
Operations lift to partial arguments, preserving success and propagating
failure.  Equal carriers are identified.}{\texttt{GLang.Val\allowbreak Val\allowbreak Sem.from\allowbreak El}}{GLang/ValueDenotSem.lean:248}
\gldefdefn{value.denot.scalars}{Variables and scalar operations}{presentation}{}{\begin{align*}
 \glvE{x_{\ltyv{}}}&=\glvok{\glvEnv(x)},\\
 \glvE{r}&=\glvok{\mathsf{arr}_1[r]}, &
 \glvE{\lvzero{}_{\lsv{}}}&=\glvok{\mathsf{arr}_{\glvD{\lsv{}}}[0]},\\
 \glvE{\lenv{0}\lvscale \lenv{1}}&=\glvok{\glvcoords{a_0[0]a_1[i]}},\quad |a_0|=1,\ |a_1|=n_{\glvEnv}(\lenv{1}),\\
 \glvE{\lenv{0}\lvadd_{\lsv{}} \lenv{1}}&=
 \begin{cases}
  \glvok{\mathsf{arr}_{\glvD{\lsv{}}}[a_0[i]+a_1[i]]}& |a_0|=|a_1|=\glvD{\lsv{}},\\
  \glverr&\text{otherwise}.
 \end{cases}
\end{align*}}{\texttt{GLang.denot\allowbreak Val\allowbreak B\_\allowbreak sem\allowbreak S}}{GLang/ValueDenotSem.lean:811}
\gldefdefn{value.denot.coordinates}{Coordinates, tensor products, and direct sums}{presentation}{}{$\mathbf1_P=1$ if $P$, and $0$ otherwise.
\begin{align*}
 \glvE{\ldelta{k}{\lsv{}}}&=\glvok{\glvcoords{\mathbf1_{i=k\,\land\,k<\glvD{\lsv{}}}}},\\
 \glvE{\lidelta{\lsv{}}}&=\glvok{\mathsf{arr}_{d^2}[\mathbf1_{r=c}]},
       &&d=\glvD{\lsv{}},\quad i=dr+c,\quad r,c<d,\\
 \glvE{\lenv{0}\loprod \lenv{1}}&=\glvok{\glvcoords{a_0[i\mathbin{\mathrm{div}}d_1]\,a_1[i\bmod d_1]}},
       &&d_1=\glvN{\lenv{1}},\quad |a_j|=n_{\glvEnv}(\lenv{j})\ (j=0,1),\\
 \glvE{\lenv{0}\losum \lenv{1}}&=\glvok{\glvcoords{
       \begin{cases}a_0[i]&i<d_0,\\a_1[i-d_0]&i\geq d_0,\end{cases}}},
       &&d_0=\glvN{\lenv{0}},\quad |a_j|=n_{\glvEnv}(\lenv{j})\ (j=0,1).
\end{align*}}{\texttt{GLang.denot\allowbreak Val\allowbreak B\_\allowbreak sem\allowbreak S}}{GLang/ValueDenotSem.lean:811}
\gldefdefn{value.denot.transpose}{Transpose}{presentation}{}{\[
 \glvE{\Trans(\leev{})}=
 \begin{cases}
 \glvok{\mathsf{arr}_{n_{\glvEnv}(\lenv{\rm out})}[a[c,r]]}
   &\operatorname{inferSpace}(\leev{})=\lsnv{0}\loprod \lsnv{1},\ |a|=d_0d_1,\\
 \glverr&\text{otherwise},
 \end{cases}
 \quad\begin{gathered}d_j=d_{\glvEnv}(\lsnv{j}),\\i=d_0r+c,\\r<d_1,\ c<d_0.\end{gathered}
\]}{\texttt{GLang.denot\allowbreak Val\allowbreak B\_\allowbreak sem\allowbreak S}}{GLang/ValueDenotSem.lean:811}
\gldefdefn{value.denot.sum}{Finite list sums}{presentation}{}{\[
 \glvE{\lsum^{\lsv{}}_{k}\lenv{k}}
 =\glvok{\mathsf{arr}_{n_{\glvEnv}(\lenv{\rm out})}
   \left[\sum_{k=0}^{d-1}a_k[i]\right]},\qquad d=d_{\glvEnv}(\lsv{}),\quad m\geq d,\quad |a_k|=n_{\glvEnv}(\lenv{\rm out})\ (k<d).
\]
The family contains $m$ terms. Only $\lenv{k}$ with $0\leq k<d$ are evaluated; $m<d$ gives $\glverr$.}{\texttt{GLang.denot\allowbreak Val\allowbreak B\_\allowbreak sem\allowbreak S}}{GLang/ValueDenotSem.lean:811}
\gldefdefn{value.denot.contractions}{Tensor contractions}{presentation}{}{\begin{gather*}
 d=d_{\glvEnv}(\lsv{}),\quad |a_j|=n_{\glvEnv}(\lenv{j})\ (j=0,1),\quad n_{\glvEnv}(\lenv{1})=d,\\
\begin{cases}d_{\glvEnv}(\lsvp{})=d&\operatorname{inferSpace}(\lenv{0})=\lsvp{}\loprod \lsnv{\rm out},\\n_{\glvEnv}(\lenv{0})=d&\text{otherwise},\end{cases}\\
 \glvE{\LTTVPP{\lsv{}}{\lenv{0}}{\lenv{1}}}=\glvok{(b_i)_{0\leq i<n_{\glvEnv}(\lenv{\rm out})}},\\
 b_i=\begin{cases}
 \displaystyle\sum_{j=0}^{d-1}a_0[d_{\rm out}j+i]a_1[j]
   &\operatorname{inferSpace}(\lenv{0})=\lsvp{}\loprod \lsnv{\rm out},\ i<d_{\rm out},\\
 0&\operatorname{inferSpace}(\lenv{0})=\lsvp{}\loprod \lsnv{\rm out},\ i\geq d_{\rm out},\\
 \displaystyle\mathbf1_{i=0}\sum_{j=0}^{d-1}a_0[j]a_1[j]
   &\operatorname{inferSpace}(\lenv{0})\text{ is not a tensor},
 \end{cases}\qquad d_{\rm out}=d_{\glvEnv}(\lsnv{\rm out}).
\end{gather*}
For $\operatorname{inferSpace}(\lenv{0})=\lsvp{}\loprod \lsnv{0}$, $\operatorname{inferSpace}(\lenv{1})=\lsnv{t}\loprod \lsnvpp{}$, and contraction
annotation $\lsnv{c}$ and stored right-space annotation $\lsnv{\rm right}$, define
\begin{gather*}
 d_0=d_{\glvEnv}(\lsnv{0}),\quad d_t=d_{\glvEnv}(\lsnv{t}),\quad d_c=d_{\glvEnv}(\lsnv{c}),\\
|a_j|=n_{\glvEnv}(\lenv{j})\ (j=0,1),\qquad d_{\glvEnv}(\lsvp{})=d_{\glvEnv}(\lsnvpp{})=d_c,\\
 \glvE{\glvmtm{\lsnv{c}}{\lsnv{\rm right}}{\lenv{0}}{\lenv{1}}}=\glvok{(b_i)_{0\leq i<n_{\glvEnv}(\lenv{\rm out})}},\\
 b_{d_0r+c}=\sum_{k=0}^{d_c-1}a_0[d_0k+c]a_1[d_cr+k],
 \qquad 0\leq r<d_t,\quad 0\leq c<d_0.
\end{gather*}
Non-tensor operands give $\glverr$. The stored $\lsnv{\rm right}$ is used by typing; evaluation checks the inferred operand spaces.}{\texttt{GLang.denot\allowbreak Val\allowbreak B\_\allowbreak sem\allowbreak S}}{GLang/ValueDenotSem.lean:811}
\gldefdefn{value.denot.assembly}{Kronecker and block-diagonal assembly}{presentation}{}{For $\glvS{\leev{}}=\lsnv{0}\loprod \lsnv{2}$ and $\glvS{\leevp{}}=\lsnv{1}\loprod \lsnv{3}$, put $d_j=\glvD{\lsnv{j}}$, with $|a|=d_0d_2$ and $|a'|=d_1d_3$.
\begin{align*}
 \glvE{\vkrn(\leev{},\leevp{})}[d_1r+r',\ d_3c+c']
   &=a[r,c]a'[r',c'],\\
 \glvE{\vblk(\leev{},\leevp{})}[r,c]
   &=\begin{cases}
      a[r,c]&r<d_0,\ c<d_2,\\
      a'[r-d_0,c-d_2]&r\geq d_0,\ c\geq d_2,\\
      0&\text{otherwise}.
     \end{cases}
\end{align*}
The Kronecker indices satisfy $r<d_0$, $r'<d_1$, $c<d_2$, and $c'<d_3$.
The block matrix has $(d_0+d_1)$ rows and $(d_2+d_3)$ columns.
Non-tensor operands or incorrect lengths give $\glverr$.}{\texttt{GLang.denot\allowbreak Val\allowbreak B\_\allowbreak sem\allowbreak S}}{GLang/ValueDenotSem.lean:811}
\gldefdefn{value.denot.matrix-conventions}{Matrix conversion and output lengths}{presentation}{}{\begin{gather*}
 \mathsf{mat}_{m,n}(a)_{r,c}=a[nr+c],\qquad
 \mathsf{flat}(B)[i]=
 \begin{cases}B_{i\mathbin{\mathrm{div}}n,\,i\bmod n}&i<mn,\\0&\text{otherwise},\end{cases}
 \quad B\in\mathbb R^{m\times n},\\
 \mathsf{out}_N(B)=\mathsf{arr}_N[i\mapsto\mathsf{flat}(B)[i]],\qquad
 N=\dof{\operatorname{inferSpace}(\lenv{\rm out})}{\glvEnv}.
\end{gather*}
Failed shape or length guards give $\glverr$.}{\texttt{GLang.denot\allowbreak Val\allowbreak B\_\allowbreak sem\allowbreak S}}{GLang/ValueDenotSem.lean:811}
\gldefdefn{value.denot.inverse}{Inverse operations}{presentation}{}{\begin{gather*}
 (m,n)=\begin{cases}(d_{\glvEnv}(\lsnv{0}),d_{\glvEnv}(\lsnv{1}))&\operatorname{inferSpace}(\leev{})=\lsnv{0}\loprod \lsnv{1},\\
 (d_{\glvEnv}(\operatorname{inferSpace}(\leev{})),1)&\text{otherwise},\end{cases}\\
 \glvE{\leev{}^{+}}=\glvE{\leev{}^{-1}}=
 \begin{cases}
 \mathsf{out}_N(\mathsf{matrixSVDPinv}(\mathsf{mat}_{m,n}(a)))&|a|=mn,\\
 \glverr&\text{otherwise}.
 \end{cases}
\end{gather*}}{\texttt{GLang.denot\allowbreak Val\allowbreak B\_\allowbreak sem\allowbreak S}}{GLang/ValueDenotSem.lean:811}
\gldefdefn{value.denot.gsvd}{Generalized SVD and weighted pseudoinverse}{presentation}{}{\begin{gather*}
 \operatorname{inferSpace}(\leev{})=\lsnv{0}\loprod \lsnv{1},\quad d_j=d_{\glvEnv}(\lsnv{j}),\quad
 |a|=d_0d_1,\quad |L|=d_0^2,\quad |R|=d_1^2,\\
 M=\mathsf{mat}_{d_0,d_1}(a)^{\mathsf T},\quad
 W=\mathsf{mat}_{d_0,d_0}(L),\quad V=\mathsf{mat}_{d_1,d_1}(R),\\
 W\succ0,\quad V\succ0,\quad g=\mathsf{SVD}(M,W,V),\\
 |B_X|=d_{\glvEnv}(\operatorname{inferSpace}(\lenv{\rm out})),\qquad
 \glvE{X(\leev{},\lenv{L},\lenv{R})}=\mathsf{out}_N(B_X),\\
 \begin{array}{c|ccccc}
 X&\glvop{GSVDS}&\glvop{GSVDU}&\glvop{GSVDUI}&\glvop{GSVDV}&\glvop{GSVDVI}\\\hline
 B_X&g.sMat&g.U&g.UI&g.V&g.VI
 \end{array}\\
 \glvE{\glvop{GPINV}(\leev{},\lenv{L},\lenv{R})}
 =\mathsf{out}_N(\mathsf{matrixGSVDPinv}(M,W,V)^{\mathsf T}),\\
 \glvE{\glvop{SVDS}(\leev{})}
 =\mathsf{out}_N(\mathsf{SVD}(M,I_{d_0},I_{d_1}).sMat).
\end{gather*}
Here $a,L,R$ are the results of $\leev{},\lenv{L},\lenv{R}$. The SVD uses the paired choices in Definition~\ref{def:fsv}; $|B_X|$ counts matrix entries. The five generalized SVD selectors return $\glverr$ on failed shape, metric, or output-length checks. $\glvop{GPINV}$ requires only the input-shape checks; $\glvop{SVDS}$ uses only $\leev{}$ and its shape guard.}{\texttt{GLang.denot\allowbreak Val\allowbreak B\_\allowbreak sem\allowbreak S}}{GLang/ValueDenotSem.lean:811}
\gldefdefn{value.denot.rref}{RREF representatives}{presentation}{}{\begin{gather*}
 \operatorname{inferSpace}(\lenv{1})=\lsnv{0}\loprod \lsnv{1},\quad m=d_{\glvEnv}(\lsnv{0})=d_{\glvEnv}(\lsnv{1}),\quad
 K=|a_0|\mathbin{\mathrm{div}}m,\quad |a_0|=Km,\quad |a_1|=m^2,\\
 A=[\mathsf{mat}_{K,m}(a_0)^{\mathsf T}\mid I_m],\qquad
 C=\mathsf{mat}_{m,m}(a_1),\\
 \glvE{\glvop{RREFRep}(\lenv{0},\lenv{1})}
 =\mathsf{out}_N(C\,(\mathsf{matrixRREF}(A).E_{\rm inv})^{\mathsf T}).
\end{gather*}
Natural-number division uses $k\mathbin{\mathrm{div}}0=0$.}{\texttt{GLang.denot\allowbreak Val\allowbreak B\_\allowbreak sem\allowbreak S}}{GLang/ValueDenotSem.lean:811}
\gldefdefn{value.denot.lambda}{Lambda abstraction}{presentation}{}{\[
 \glvE{\llambda(x\colon\ltynv{0})\lrh\ltynv{1}\colon \leev{}}
 =\lambda a.\,\SemA{\leev{}}{\glvEnv[x\mapsto a]}{}.
\]
The function records the argument representation prescribed by $\ltynv{0}$;
failed argument-type evaluation gives $\glverr$.}{\texttt{GLang.denot\allowbreak Val\allowbreak B\_\allowbreak sem\allowbreak S}}{GLang/ValueDenotSem.lean:811}
\gldefdefn{value.denot.application}{Function application}{presentation}{}{\begin{gather*}
 \glvE{\glvapp{\lenv{0}}{\lenv{1}}}=f(a),\qquad
 f=\glvE{\lenv{0}},\quad a=\glvE{\lenv{1}},\\
 \operatorname{inferType}(\lenv{0})=x:\ltynv{0}\fnto\ltynv{1}.
\end{gather*}
The argument is checked against $\ltynv{0}$ and the function's runtime domain.
A non-function, a failed domain check, or an incompatible argument representation
gives $\glverr$; every error propagates. The result type is
$\ltynv{1}\subst{x}{\lenv{1}}$. Basis conversions use the explicit reshape below.}{\texttt{GLang.denot\allowbreak Val\allowbreak B\_\allowbreak partial}}{GLang/PartialValueEval.lean:77}
\gldefdefn{value.denot.reshape}{Reshape}{presentation}{}{\begin{gather*}
 \glvE{\glvreshape{\leev{}}{\ltynv{t}}{\ltynv{\rm src}}}
 =R_{\glvEnv,\glvEnv}(\glvE{\leev{}};\ltynv{t},\ltynv{s}),\\
 R_{\glvEnv^t,\glvEnv^s}(a;\lsnv{t},\lsnv{s})=
 \begin{cases}
  a&|a|=\dof{\lsnv{s}}{\glvEnv^s}=\dof{\lsnv{t}}{\glvEnv^t},\\
  \glverr&\text{otherwise},
 \end{cases}\\
 R_{\glvEnv^t,\glvEnv^s}(f;(x\colon\ltynv{t0}\fnto\ltynv{t1}),(y\colon\ltynv{s0}\fnto\ltynv{s1}))
 =\lambda a.\,F(a),\\
 c\leftarrow R_{\glvEnv^s,\glvEnv^t}(a;\ltynv{s0},\ltynv{t0}),\qquad
 \glvEnv^{t'}=\glvEnv^t[x\mapsto a],\quad \glvEnv^{s'}=\glvEnv^s[y\mapsto c],\\
 F(a)=R_{\glvEnv^{t'},\glvEnv^{s'}}(f(c);\ltynv{t1},\ltynv{s1}).
\end{gather*}
Here $\ltynv{\rm src}$ is the stored source annotation, $\ltynv{t}$ is the target,
and the effective source is
\[
\ltynv{s}=\begin{cases}
\ltynv{\rm src}&\ltynv{\rm src}\text{ is a space},\\
\operatorname{inferType}(\leev)&\ltynv{\rm src}\text{ and }\operatorname{inferType}(\leev)\text{ are function types}.
\end{cases}
\]
An inferred space with a function source annotation gives $\glverr$.
The function clause checks argument representations and exact array lengths.
Incompatible shapes and failed partial computations give $\glverr$.}{\texttt{GLang.Partial\allowbreak Value\allowbreak Eval.reshape\allowbreak Inferred}}{GLang/PartialValueReshapeRuntime.lean:50}
\gldefdefn{value.denot.soundness}{Raw evaluation and checked soundness}{presentation}{}{These are the raw semantics.  The public value-soundness theorem uses
\texttt{denotValB\_semSChecked} with the direct oracle: it constrains successful
results to have canonical, exact lengths recursively, but still permits failure.
The adequacy theorem uses raw evaluation and proves successful agreement for
the value selected by a checked operational derivation.  These are different
claims; neither says that every raw value expression evaluates successfully.}{\texttt{GLang.Metatheory.value\allowbreak Soundness}}{GLang/Metatheory.lean:266}
\gldefdefn{preview.valueDomains}{Value domains}{presentation}{}{Dimensions and shapes range over $\mathbb N\sqcup\{\glverr\}$ and
$\mathbb N^{<\omega}\sqcup\{\glverr\}$, respectively.
Write $\mathbb R^n$ for flat arrays of length $n$; all clauses propagate $\glverr$.
\gldefn{value.domains}
Errors propagate before a result fibre is selected.}{\texttt{GLang.Paper\allowbreak Value\allowbreak Domains.Section}}{GLang/PaperValueDomains.lean:16}
\gldefdefn{preview.valueEnvWF}{Value environments}{presentation}{}{\[
\valCheckCtx\vdash\glvEnv\quad\Longleftrightarrow\quad
\forall(x\colon\ltyv)\in\valCheckCtx.\;
 (\exists\lkn{}.\;\valCheckCtx\vdash\ltyv\colon\lkn{})
 \ \land\ \glvEnv(x)\in\mathcal V_\glvEnv(\ltyv)\setminus\{\glverr\}.
\]}{\texttt{GLang.Paper\allowbreak Value\allowbreak Results.Env\allowbreak WF}}{GLang/PaperValueEnvironment.lean:9}
\gldefdefn{preview.operationalEntries}{Operational entries}{presentation}{}{For scoped values, write $\Psi\vdash(x,\lb{})\mapsto\leev$ when
\[
\evalSJudge{\Psi}{\tof{\lb{}}}{\operatorname{inferType}(\leev)},
\qquad
\operatorname{fv}(\leev)\subseteq\operatorname{names}(\Psi)\cup\{x\}.
\]
For $\lb{}=\frameB{y}{b_0}{b_1}$ and
$\leev=\llambda(y\colon T_0)\lrh T_1\colon v$, additionally require
\[
\evalSJudge{\Psi[(y,b_0)\mapsto y_{T_0}]}{\tof{b_1}}{T_1}
\qquad(y\text{ fresh}).
\]}{\texttt{GLang.V2.Eval\allowbreak Entry\allowbreak Creation}}{GLang/ValueProgress.lean:358}
\gldefdefn{preview.operationalContextWF}{Operational context formation}{presentation}{}{Define $(\Delta,\epsilon)\vdash\Psi$ inductively, with
$(\{\},\{\})\vdash\{\}$.
If $(\Delta,\epsilon)\vdash\Psi$, $(\{\},\epsilon)\vdash\lb{}\colon\bvalid$,
and $\Psi\vdash(x,\lb{})\mapsto\leev$, write
$T=\tof{\lb{}}$ and $\Psi'=\Psi[(x,\lb{})\mapsto\leev]$. Then
\[
\begin{array}{lll}
(\Delta,\epsilon[x\mapsto T])\vdash\Psi'
 &\text{if}&x\notin\operatorname{dom}(\Delta\sqcup\epsilon),\\
(\Delta[x\mapsto\ls{}],\epsilon)\vdash\Psi'
 &\text{if}&x\notin\operatorname{dom}(\Delta\sqcup\epsilon)\text{ and }T=\ls{},\\
(\Delta,\epsilon)\vdash\Psi'
 &\text{if}&(x\colon T)\in\Delta\sqcup\epsilon\text{ and }(x,\lb{})\notin\Psi.
\end{array}
\]}{\texttt{GLang.V2.Ctx\allowbreak Tele}}{GLang/ValueProgress.lean:380}
\gldefdefn{value.shapes.compact}{Value dimensions and shapes}{presentation}{}{For $d_i=\dof{\lsnv{i}}{\glvEnv}$:
\begin{center}
{\scriptsize\begin{tabular}{@{}c@{}}
\makebox[\linewidth]{\hfill$\dof{\lsv}{\glvEnv}=\prod\shpOf{\lsv}{\glvEnv}$\hfill$\gleqn{value.shapeOf.empty}$\hfill$\gleqn{value.shapeOf.real}$\hfill}\\
\makebox[\linewidth]{\hfill$\shpOf{\lsnv{0}\losum\lsnv{1}}{\glvEnv}=(d_0+d_1,)$\hfill$\shpOf{\lsnv{0}\loprod\lsnv{1}}{\glvEnv}=(d_0,d_1)$\hfill$\gleqn{value.shapeOf.ffs}$\hfill}
\end{tabular}}
\end{center}
\gldefn{value.ffs.svd}}{\texttt{GLang.Val\allowbreak Space.shp\allowbreak Of}}{GLang/ValueDenot.lean:116}
\gldefdefn{preview.valueKinding}{Value formation and kinding}{presentation}{}{Value kinds are $\lkn{}::=\lkbase{}\mid\lkn{0}\to\lkn{1}$.
The context $\valCheckCtx$ maps value variables to value types.
We write $\valCheckCtx\vdash\lsv:\svalid$,
$\valCheckCtx\vdash\ltyv:\lkn{}$, and
$\valCheckCtx\vdash\leev:\ltyv$ for space formation, type kinding, and value typing
(\cref{sec:vals}). A value space has kind $\lkbase{}$; a dependent function
$x:\ltynv{0}\fnto \ltynv{1}$ has kind $\lkn{0}\to\lkn{1}$ when
$\valCheckCtx\vdash \ltynv{0}:\lkn{0}$ and
$\valCheckCtx,x:\ltynv{0}\vdash \ltynv{1}:\lkn{1}$.
The syntax-directed functions $\operatorname{inferType}(v)$ and $\operatorname{inferSpace}(v)$
return the inferred value type and value space, respectively (\cref{app:value-inference}).
They read the syntax annotations, not $\valCheckCtx$; the variable rule checks
that an annotation agrees with its declaration in $\valCheckCtx$. Thus
$\valCheckCtx\vdash v:\lsv$ implies
$\operatorname{inferType}(v)=\lsv=\operatorname{inferSpace}(v)$.
}{\texttt{GLang.Val\allowbreak TValid\allowbreak C}}{GLang/ValueValidC.lean:40}
\gldefdefn{preview.op.history}{Creation histories}{presentation}{}{Let $\operatorname{names}(\Psi)=\{x\mid\exists b,v.\;((x,b)\mapsto v)\in\Psi\}$. Write $\Psi\vdash(x,\lb{})\mapsto\leev$ when
$\evalSJudge{\Psi}{\tof{\lb{}}}{\operatorname{inferType}(\leev)}$ and
$\operatorname{fv}(\leev)\subseteq\operatorname{names}(\Psi)\cup\{x\}$.
For $\lb{}=\frameB{y}{b_0}{b_1}$ and
$\leev=\llambda(y\colon\ltynv{0})\lrh\ltynv{1}\colon v$, additionally require
$\evalSJudge{\Psi[(y,b_0)\mapsto y_{\ltynv{0}}]}{\tof{b_1}}{\ltynv{1}}$ for fresh $y$.

Assuming $\vdash(\Delta,\epsilon)$ throughout (\cref{def:wfc}), define
$(\Delta,\epsilon);\valCheckCtx\vdash\Psi$ inductively, with
$(\emptyset,\emptyset);\emptyset\vdash\emptyset$.
Suppose $(\Delta,\epsilon);\valCheckCtx\vdash\Psi$,
$(\emptyset,\epsilon)\vdash\lb{}\colon\bvalid$,
$(x,\lb{})\notin\Psi$, and $\Psi\vdash(x,\lb{})\mapsto\leev$.
Put $T=\tof{\lb{}}$ and $\ltyv=\operatorname{inferType}(\leev)$.
Form $(\Delta',\epsilon';\valCheckCtx')$ by adding $x:T$ to
$(\Delta,\epsilon)$ and $x:\ltyv$ to $\valCheckCtx$ if $x$ is undeclared;
otherwise leave the contexts unchanged, with $(x:T)\in\Delta\sqcup\epsilon$.
If $\valCheckCtx\vdash\ltyv:\lkn{}$ and
$\valCheckCtx'\vdash\leev:\ltyv$, then
$(\Delta',\epsilon');\valCheckCtx'\vdash\Psi[(x,\lb{})\mapsto\leev]$.
The type-evaluation premises use the rules in \cref{app:gen-rules:evalT}.}{\texttt{GLang.Operational.Context\allowbreak History}}{GLang/Operational.lean:87}
\gldefdefn{preview.op.checked}{Checked operational semantics}{presentation}{}{Assume $(\Delta,\epsilon);\valCheckCtx\vdash\Psi$.
The subscript $\mathsf C$ marks an operational derivation checked at every step:
$\glCty{\epsilon}{\Psi}{\lty}{\ltyv}$,
$\glCterm{\Delta}{\epsilon}{\Psi}{\lee}{\lb{}}{\leev}{\ltyv}$, and
$\glCcob{\epsilon}{\Psi}{\lbn{0}}{\leev}{\lbn{1}}$.
At each term step for example, these require
$(\Delta,\epsilon)\vdash t:\tof{b}$,
$(\emptyset,\epsilon)\vdash b:\bvalid$,
$\glCat{\epsilon}{\Psi}{b}{\ltyv{}}$,
$\evalJudge{\Psi}{t}{b}{v}$, and
$\evalSJudge{\Psi}{\tof{b}}{\ltyv{}}$.
The judgment $\glCat{\epsilon}{\Psi}{\lb{}}{\ltyv}$ simply shows what basis was used in $\glCty{\epsilon}{\Psi}{\lty}{\ltyv}$, which is needed to call dynamic dimensions.
The rather verbose rules are in \cref{app:gen-rules:c-space,app:gen-rules:c-ty,app:gen-rules:c-tyAt,app:gen-rules:c-cob,app:gen-rules:c-term}.}{\texttt{GLang.Raw\allowbreak Certified\allowbreak Operational.C.Term}}{GLang/Adequacy/RawCertifiedOperational.lean:276}
\gldefdefn{preview.op.regularity}{Value-annotation regularity}{presentation}{}{Write $\SemCheck{v}{\glvEnv}:=\mathcal E^{\mathrm{check}}_{\operatorname{inferType}(v)}(v,\glvEnv)$
for value denotation checked against its inferred type.}{\texttt{GLang.Val\allowbreak TValid\allowbreak C}}{GLang/ValueValidC.lean:40}
\gldefdefn{preview.adequacy.agreement}{Agreement at an exact basis}{presentation}{}{Write $a\equiv_{B;\sigma,\glvEnv}^{\lty,\ltyv}w$ for agreement between a source value
and a runtime value at basis or frame $B$, interpreting $\lty$ under $\sigma$
and $\ltyv$ under $\glvEnv$.
At spaces, put $N=\dof{\lsv}{\glvEnv}$; then
$a\equiv_{B;\sigma,\glvEnv}^{\ls,\lsv}w$ iff
$|B|=|w|=N$ and $a=\sum_{i=0}^{N-1}w_iB_i$.
At functions $\lty=x:\ltyn{0}\fnto\ltyn{1}$ and
$\ltyv=x:\ltynv{0}\fnto\ltynv{1}$, let $B$ be a frame with domain
component $B_0$ and fiber family $a\mapsto B_1(a)$.
Then $f\equiv_{B;\sigma,\glvEnv}^{\lty,\ltyv}g$ iff, for every
$a\equiv_{B_0;\sigma,\glvEnv}^{\ltyn{0},\ltynv{0}}w$, there is $z$ such that
$g(w)=z\ne\glverr$ and
$f(a)\equiv_{B_1(a);\sigma',\glvEnv'}^{\ltyn{1},\ltynv{1}}z$, where
$\sigma'=\sigma[x\mapsto a]$ and $\glvEnv'=\glvEnv[x\mapsto w]$.
Here $g$ has the runtime domain prescribed by $\ltynv{0}$.}{\texttt{GLang.Partial\allowbreak Adequacy.Agree}}{GLang/Adequacy/PartialSourceAgreement.lean:9}
\gldefdefn{preview.adequacy.dimensions}{Basis-indexed type agreement}{presentation}{}{Shape agreement $\lty\sim_{B;\sigma,\glvEnv}\ltyv$ is defined recursively.
At spaces, $\ls\sim_{B;\sigma,\glvEnv}\lsv$ iff
$\shpOf{\lsv}{\glvEnv}=\operatorname{shape}_{\sigma}(\ls)$.
At function types $\lty=x:\ltyn{0}\fnto\ltyn{1}$ and
$\ltyv=x:\ltynv{0}\fnto\ltynv{1}$, let $B$ be a frame with domain
component $B_0$ and fiber family $a\mapsto B_1(a)$. Require
$\ltyn{0}\sim_{B_0;\sigma,\glvEnv}\ltynv{0}$ and, for every
$a\equiv_{B_0;\sigma,\glvEnv}^{\ltyn{0},\ltynv{0}}w$,
$\ltyn{1}\sim_{B_1(a);\sigma[x\mapsto a],\glvEnv[x\mapsto w]}\ltynv{1}$.
The source shape $\operatorname{shape}_{\sigma}(\ls)$ follows the source-space
constructors: $(0,)$, $(1,)$, dual unchanged, direct-sum dimension,
tensor-factor dimensions, and fundamental-subspace dimension.
Type agreement $\lty\simeq_{B;\sigma,\glvEnv}\ltyv$ holds when
$\lty\sim_{B;\sigma,\glvEnv}\ltyv$ and $B$ is recursively a basis or discrete frame
for $\lty$ under $\sigma$.}{\texttt{GLang.Partial\allowbreak Adequacy.Dims\allowbreak At}}{GLang/Adequacy/PartialSourceAgreement.lean:41}
\gldefdefn{preview.adequacy.consistency}{Consistent environments}{presentation}{}{We write $(\Delta,\epsilon);\Psi;\valCheckCtx\vdash(\sigma,\glvEnv)$ when
$(\Delta,\epsilon);\valCheckCtx\vdash\Psi$, $\valCheckCtx\vdash\glvEnv$,
$(\Delta,\epsilon)\vdash\sigma$ at every step, and
$(\Delta,\epsilon,\Psi,\valCheckCtx,\sigma,\glvEnv)$ is constructed as follows.
The tuple $(\emptyset,\emptyset,\emptyset,\emptyset,\emptyset,\glvEnv)$ is constructed for arbitrary $\glvEnv$.
Suppose $(\Delta,\epsilon,\Psi,\valCheckCtx,\sigma,\glvEnv)$ has been constructed.
Let $\Psi'=\Psi[(x,\lb{})\mapsto\leev{}]$ and update
$(\Delta,\epsilon,\valCheckCtx)$ to $(\Delta',\epsilon',\valCheckCtx')$
as in \cref{def:nonrecurseeval}.
Put $T=\tof{\lb{}}$,
$\ltyv{}=\operatorname{inferType}(\leev{})$, and
$B=\SemBB{\lb{}}{\sigma}$, and require
$\Sem{\leev{}}{\glvEnv}=w\ne\glverr$,
$a\equiv_{B;\sigma,\glvEnv}^{T,\ltyv{}}w$, and
$T\sim_{B;\sigma,\glvEnv}\ltyv{}$.
If $x\notin\operatorname{dom}(\Delta\sqcup\epsilon)$,
require $(x\notin\operatorname{fv}(\leev{})\ \lor\
\leev{}=x_{\ltyv{}})$ and put
$\sigma'=\sigma[x\mapsto a]$, and
$\glvEnv'=\glvEnv[x\mapsto w]$.
Otherwise require $a=\sigma(x)$ and put
$(\sigma',\glvEnv')=(\sigma,\glvEnv)$.
This constructs $(\Delta',\epsilon',\Psi',\valCheckCtx',\sigma',\glvEnv')$.}{\texttt{GLang.Adequacy.Partial\allowbreak Direct.Consistent}}{GLang/Adequacy/PartialSourcePublic.lean:8}
\gldefdefn{iof.conventions}{Inner-product notation}{reviewed notation}{}{\[
\lctxnle\vdash\lee{}\colon\lsn{0}^{\lstar}\loprod\lsn{1}.
\]}{\texttt{GLang.Space.iof}}{GLang/InnerProduct.lean:193}
\gldefequation{iof.empty}{\iof{\lsempty{}}=\lvzero{}_{\lsempty{}^{\lstar}\loprod\lsempty{}^{\lstar}}}
\gldefdefn{iof.empty}{Zero space}{reviewed equation}{}{\begin{center}\gpfit{\iof{\lsempty{}}=\lvzero{}_{\lsempty{}^{\lstar}\loprod\lsempty{}^{\lstar}}}\end{center}}{\texttt{GLang.Space.iof}}{GLang/InnerProduct.lean:193}
\gldefequation{iof.real}{\iof{\lR{}}=\lllambdaArg{x_0,x_1}{x_0\lvscale x_1}}
\gldefdefn{iof.real}{Real scalars}{reviewed equation}{}{\begin{center}\gpfit{\iof{\lR{}}=\lllambdaArg{x_0,x_1}{x_0\lvscale x_1}}\end{center}}{\texttt{GLang.Space.iof}}{GLang/InnerProduct.lean:193}
\gldefequation{iof.dual}{\iof{\ls{}^{\lstar}}=\lllambdaArg{x_0,x_1}{\linner{}_{\ls{}}(\lsharp{}_{\ls{}}(x_0),\lsharp{}_{\ls{}}(x_1))}}
\gldefdefn{iof.dual}{Dual space}{reviewed equation}{}{\begin{center}\gpfit{\iof{\ls{}^{\lstar}}=\lllambdaArg{x_0,x_1}{\linner{}_{\ls{}}(\lsharp{}_{\ls{}}(x_0),\lsharp{}_{\ls{}}(x_1))}}\end{center}}{\texttt{GLang.Space.iof}}{GLang/InnerProduct.lean:193}
\gldefequation{iof.dsum}{\begin{aligned}
\iof{\lsn{0}\losum\lsn{1}}={}&\lllambdaArg{x_0,x_1}{
 \linner{}_{\lsn{0}}(\lkprojpp{0}(x_0),\lkprojpp{0}(x_1))
 \lvadd\linner{}_{\lsn{1}}(\lkprojpp{1}(x_0),\lkprojpp{1}(x_1))}
\end{aligned}}
\gldefdefn{iof.dsum}{Direct sum}{reviewed equation}{}{\begin{center}\gpfit{\begin{aligned}
\iof{\lsn{0}\losum\lsn{1}}={}&\lllambdaArg{x_0,x_1}{
 \linner{}_{\lsn{0}}(\lkprojpp{0}(x_0),\lkprojpp{0}(x_1))
 \lvadd\linner{}_{\lsn{1}}(\lkprojpp{1}(x_0),\lkprojpp{1}(x_1))}
\end{aligned}}\end{center}}{\texttt{GLang.Space.iof}}{GLang/InnerProduct.lean:193}
\gldefequation{iof.tprod}{\begin{aligned}
 F&=\lllambdaArg{x_{00},x_{01},x_{10},x_{11}}{
 \linner{}_{\lsn{0}}(x_{00},x_{10})\lvscale\linner{}_{\lsn{1}}(x_{01},x_{11})},\\
 G&=\luncurry_{0,1,2}(F),\\
 \iof{\lsn{0}\loprod\lsn{1}}&=\lllambdaArg{p\colon\lsn{0}\loprod\lsn{1}}{
 \luncurry_{0,1,1}(G(p))}.
\end{aligned}}
\gldefdefn{iof.tprod}{Tensor product}{reviewed equation}{}{\begin{center}\gpfit{\begin{aligned}
 F&=\lllambdaArg{x_{00},x_{01},x_{10},x_{11}}{
 \linner{}_{\lsn{0}}(x_{00},x_{10})\lvscale\linner{}_{\lsn{1}}(x_{01},x_{11})},\\
 G&=\luncurry_{0,1,2}(F),\\
 \iof{\lsn{0}\loprod\lsn{1}}&=\lllambdaArg{p\colon\lsn{0}\loprod\lsn{1}}{
 \luncurry_{0,1,1}(G(p))}.
\end{aligned}}\end{center}}{\texttt{GLang.Space.iof}}{GLang/InnerProduct.lean:193}
\gldefequation{iof.ker}{\iof{\Kernel(\lee{})}=\lllambdaArg{x_0,x_1}{\linner{}_{\lsn{0}}(\lembd{}_{\Kernel(\lee{})}(x_0),\lembd{}_{\Kernel(\lee{})}(x_1))}}
\gldefdefn{iof.ker}{Kernel}{reviewed equation}{}{\begin{center}\gpfit{\iof{\Kernel(\lee{})}=\lllambdaArg{x_0,x_1}{\linner{}_{\lsn{0}}(\lembd{}_{\Kernel(\lee{})}(x_0),\lembd{}_{\Kernel(\lee{})}(x_1))}}\end{center}}{\texttt{GLang.Space.iof}}{GLang/InnerProduct.lean:193}
\gldefequation{iof.coker}{\iof{\CoKernel(\lee{})}=\lllambdaArg{x_0,x_1}{\linner{}_{\lsn{0}}(\lembd{}_{\CoKernel(\lee{})}(x_0),\lembd{}_{\CoKernel(\lee{})}(x_1))}}
\gldefdefn{iof.coker}{Cokernel}{reviewed equation}{}{\begin{center}\gpfit{\iof{\CoKernel(\lee{})}=\lllambdaArg{x_0,x_1}{\linner{}_{\lsn{0}}(\lembd{}_{\CoKernel(\lee{})}(x_0),\lembd{}_{\CoKernel(\lee{})}(x_1))}}\end{center}}{\texttt{GLang.Space.iof}}{GLang/InnerProduct.lean:193}
\gldefequation{iof.img}{\iof{\Image(\lee{})}=\lllambdaArg{x_0,x_1}{\linner{}_{\lsn{1}}(\lembd{}_{\Image(\lee{})}(x_0),\lembd{}_{\Image(\lee{})}(x_1))}}
\gldefdefn{iof.img}{Image}{reviewed equation}{}{\begin{center}\gpfit{\iof{\Image(\lee{})}=\lllambdaArg{x_0,x_1}{\linner{}_{\lsn{1}}(\lembd{}_{\Image(\lee{})}(x_0),\lembd{}_{\Image(\lee{})}(x_1))}}\end{center}}{\texttt{GLang.Space.iof}}{GLang/InnerProduct.lean:193}
\gldefequation{iof.coimg}{\iof{\CoImage(\lee{})}=\lllambdaArg{x_0,x_1}{\linner{}_{\lsn{1}}(\lembd{}_{\CoImage(\lee{})}(x_0),\lembd{}_{\CoImage(\lee{})}(x_1))}}
\gldefdefn{iof.coimg}{Coimage}{reviewed equation}{}{\begin{center}\gpfit{\iof{\CoImage(\lee{})}=\lllambdaArg{x_0,x_1}{\linner{}_{\lsn{1}}(\lembd{}_{\CoImage(\lee{})}(x_0),\lembd{}_{\CoImage(\lee{})}(x_1))}}\end{center}}{\texttt{GLang.Space.iof}}{GLang/InnerProduct.lean:193}
\gldeftheorem{preview.sourceSoundness}{Source denotational soundness}{Suppose $\vdash\lctx$, $\lctx\vdash\lee{}\colon\lty{}$, and $\lctx\vdash\sigma$.
Then $(\emptyset,\lepsilon)\vdash\lty{}\colon\lkn{}$ for some $\lkn{}$.
For a fixed choice of inner products, $\Sem{\lee{}}{\sigma}$ and $\SemT{\lty{}}{\sigma}$
are independent of the typing derivation.
Moreover:
\begin{enumerate}
\item $\Sem{\lee{}}{\sigma}\in\SemT{\lty{}}{\sigma}$ is a well-defined vector or dependent function;
\item for every $(x,\ls{})\in\lDelta$, $y\mapsto\Sem{\lee{}}{\sigma[x\mapsto y]}$
is linear on $\SemT{\ls{}}{\sigma}$, pointwise for dependent-function results;
\item $\SemT{\lty{}}{\sigma}$ is a well-defined Hilbert space or set of dependent functions.
\end{enumerate}}{\texttt{GLang.Paper\allowbreak Source\allowbreak Results.source\allowbreak Soundness}}{GLang/PaperSourceResults.lean:21}{none}{preview}
\gldeftheorem{preview.innerProduct}{Inner-product choice is sound}{Suppose $\vdash\lctxnle$, $\lctxnle\vdash\ls{}\colon\svalid$,
and $\lctxnle\vdash\sigma$.
Then $\lctxnle\vdash\iof{\ls{}}\colon\ls{}^{\lstar}\loprod\ls{}^{\lstar}$,
and $\Sem{\iof{\ls{}}}{\sigma}$ is an inner product on $\SemT{\ls{}}{\sigma}$.}{\texttt{GLang.Paper\allowbreak Source\allowbreak Results.inner\allowbreak Product}}{GLang/PaperSourceResults.lean:35}{none}{preview}
\gldeftheorem{preview.valueSoundness}{Soundness for values}{If $\valCheckCtx\vdash\glvEnv$ and $\valCheckCtx\vdash\leev\colon\ltyv$,
then:
\begin{enumerate}
\item $\exists\lkn{}.\;\valCheckCtx\vdash\ltyv\colon\lkn{}$;
\item $\operatorname{inferType}(\leev)=\ltyv$;
\item $\forall\lsv.\;\dof{\lsv}{\glvEnv}=\prod\shpOf{\lsv}{\glvEnv}$;
\item $\SemCheck{\leev}{\glvEnv}\in\mathcal V_\glvEnv(\ltyv)$.
\end{enumerate}
Here $\SemCheck{\cdot}{\glvEnv}$ includes the recursive result check and
agrees with the raw evaluation in the recursive value domain;
$\glverr$ propagates through shapes, dimensions, and dependent result fibers.
The product of an erroneous shape is $\glverr$.}{\texttt{GLang.Paper\allowbreak Value\allowbreak Results.value\allowbreak Soundness}}{GLang/PaperValueResults.lean:10}{none}{preview}
\gldeftheorem{preview.basisSoundness}{Basis denotational soundness}{Suppose $\vdash\lctxnle$ and $\lctxnle\vdash\lb{}\colon\bvalid$.
Then $\lctxnle\vdash\tof{\lb{}}\colon\lkn{}$ for some $\lkn{}$.
If also $\lctxnle\vdash\sigma$, then $\SemBB{\lb{}}{\sigma}$ is a basis or discrete frame
for $\SemT{\tof{\lb{}}}{\sigma}$. Its denotation is independent of the typing derivation.}{\texttt{GLang.Paper\allowbreak Basis\allowbreak Results.source\allowbreak Basis\allowbreak Soundness}}{GLang/PaperBasisResults.lean:17}{none}{preview}
\gldeftheorem{preview.standardBasisSoundness}{Standard basis soundness}{Suppose $\vdash\lctxnle$ and $\lctxnle\vdash\lty{}\colon\lkn{}$. Then $\lctxnle\vdash\stdBasisDCtx{\lty{}}\colon\bvalid$,
$\tof{\stdBasisDCtx{\lty{}}}=\lty{}$, and $\isStdBasis{\stdBasisDCtx{\lty{}}}$.
If also $\lctxnle\vdash\sigma$, then $\SemBB{\stdBasisDCtx{\lty{}}}{\sigma}$ is an orthonormal basis or
discrete frame for $\SemT{\lty{}}{\sigma}$ under the chosen inner products.
The construction is independent of the typing derivation.
If $\lctxnle\vdash\lb{}\colon\bvalid$, $\tof{\lb{}}=\lty{}$, and
$\isStdBasis{\lb{}}$, then $\stdBasisDCtx{\lty{}}=\lb{}$.}{\texttt{GLang.Paper\allowbreak Basis\allowbreak Results.standard\allowbreak Basis\allowbreak Soundness}}{GLang/PaperBasisResults.lean:25}{none}{preview}
\gldeftheorem{preview.op.preserveTerm}{Term preservation}{Suppose $(\Delta,\epsilon);\valCheckCtx\vdash\Psi$,
$(\Delta,\epsilon)\vdash t:\tof{b}$, and $(\emptyset,\epsilon)\vdash b:\bvalid$.
If $\evalJudge{\Psi}{t}{b}{v}$, then
$\valCheckCtx\vdash v:\operatorname{inferType}(v)$ and $\valCheckCtx\vdash\operatorname{inferType}(v):\lkn{}$ for some $\lkn{}$ and
$\operatorname{fv}(\operatorname{inferType}(v))\cup\operatorname{fv}(v)\subseteq\operatorname{names}(\Psi)$.}{\texttt{GLang.Paper\allowbreak Operational\allowbreak Previews.preservation\allowbreak With\allowbreak Kinding}}{GLang/PaperOperationalPreviews.lean:9}{none}{preview}
\gldeftheorem{preview.op.preserveType}{Type preservation}{If $(\Delta,\epsilon);\valCheckCtx\vdash\Psi$, $(\emptyset,\epsilon)\vdash\lty:\lkn{}$,
and $\evalSJudge{\Psi}{\lty}{\ltyv{}}$, then
$\valCheckCtx\vdash\ltyv{}:\lkn{}$ for some $\lkn{}$ and
$\operatorname{fv}(\ltyv{})\subseteq\operatorname{names}(\Psi)$.}{\texttt{GLang.Paper\allowbreak Operational\allowbreak Previews.type\allowbreak Preservation\allowbreak With\allowbreak Kinding}}{GLang/PaperOperationalPreviews.lean:30}{none}{preview}
\gldeftheorem{preview.op.preserveSpace}{Space preservation}{If $(\Delta,\epsilon);\valCheckCtx\vdash\Psi$, $(\emptyset,\epsilon)\vdash\ls{}:\svalid$,
and $\evalSJudge{\Psi}{\ls{}}{\lsv}$, then
$\valCheckCtx\vdash\lsv:\lkbase$ and
$\operatorname{fv}(\lsv)\subseteq\operatorname{names}(\Psi)$.}{\texttt{GLang.Metatheory.space\allowbreak Preservation}}{GLang/Metatheory.lean:474}{none}{preview}
\gldeftheorem{preview.op.preserveCob}{Change-of-basis preservation}{If $(\Delta,\epsilon);\valCheckCtx\vdash\Psi$, $(\emptyset,\epsilon)\vdash b_0:\bvalid$,
and $\evalBBJudge{\Psi}{b_0}{v}{b_1}$, then
$\exists \lsn{0},\lsn{1},\lsnv{0},\lsnv{1}.\;
\operatorname{inferType}(v)=\lsnv{0}\loprod \lsnv{1}$ and $\valCheckCtx\vdash v:\lsnv{0}\loprod \lsnv{1}$,
$\evalSJudge{\Psi}{\lsn{i}}{\lsnv{i}}$ $(i=0,1)$, and
$\operatorname{fv}(v)\subseteq\operatorname{names}(\Psi)$.}{\texttt{GLang.Metatheory.cob\allowbreak Preservation}}{GLang/Metatheory.lean:519}{none}{preview}
\gldeftheorem{preview.op.progress}{Checked progress}{If $(\Delta,\epsilon);\valCheckCtx\vdash\Psi$,
$(\Delta,\epsilon)\vdash t:\tof{b}$, and $(\emptyset,\epsilon)\vdash b:\bvalid$, then
\[
\exists v,\ltyv{}.\quad\glCterm{\Delta}{\epsilon}{\Psi}{t}{b}{v}{\ltyv{}},\quad
\operatorname{inferType}(v)=\ltyv{},\quad\operatorname{inferSpace}(v)=\operatorname{spaceOf}(\ltyv{}).
\]
}{\texttt{GLang.Metatheory.progress}}{GLang/Metatheory.lean:553}{none}{preview}
\gldeftheorem{preview.op.progressStd}{Standard-basis progress}{The conclusion of \cref{thm:prog} holds under its hypotheses with
$b=\stdBasisDCtx{T}$; no additional standard-basis representations in $\Psi$ are required.}{\texttt{GLang.Metatheory.progress}}{GLang/Metatheory.lean:553}{none}{preview}
\gldeftheorem{preview.op.progressSpace}{Checked space progress}{If $(\Delta,\epsilon);\valCheckCtx\vdash\Psi$ and $(\emptyset,\epsilon)\vdash\ls{}:\svalid$,
then $\exists\lsv.\ \glCspace{\epsilon}{\Psi}{\ls{}}{\lsv}$.}{\texttt{GLang.Metatheory.checked\allowbreak Space\allowbreak Progress}}{GLang/Metatheory.lean:626}{none}{preview}
\gldeftheorem{preview.op.progressCob}{Checked change-of-basis progress}{If $(\Delta,\epsilon);\valCheckCtx\vdash\Psi$, $(\emptyset,\epsilon)\vdash b_i:\bvalid$
for $i=0,1$, and $\tof{b_0}=\tof{b_1}=\ls{}$, then
\[
\exists v.\quad\glCcob{\epsilon}{\Psi}{b_0}{v}{b_1}.
\]
In particular, take $b_1=\stdBasisDCtx{\ls{}}$. Frames are handled by checked term evaluation.}{\texttt{GLang.Metatheory.checked\allowbreak Cob\allowbreak Progress}}{GLang/Metatheory.lean:665}{none}{preview}
\gldeftheorem{preview.adequacy.term}{Checked term adequacy}{Under $(\Delta,\epsilon);\Psi;\valCheckCtx\vdash(\sigma,\glvEnv)$,
if $\glCterm{\Delta}{\epsilon}{\Psi}{t}{b}{v}{\ltyv{}}$, then
\[
\exists w.\quad\Sem{v}{\glvEnv}=w\ne\glverr,\qquad
\Sem{t}{\sigma}\equiv_{\SemBB{b}{\sigma};\sigma,\glvEnv}^{\tof{b},\ltyv{}}w,\qquad
\tof{b}\simeq_{\SemBB{b}{\sigma};\sigma,\glvEnv}\ltyv{}.
\]}{\texttt{GLang.Metatheory.term\allowbreak Adequacy}}{GLang/Metatheory.lean:682}{none}{preview}
\gldeftheorem{preview.adequacy.space}{Checked space adequacy}{Under consistent environments, if $\glCspace{\epsilon}{\Psi}{\ls{}}{\lsv}$, then
\[
\shpOf{\lsv}{\glvEnv}=\operatorname{shape}_{\sigma}(\ls{}),
\qquad\dof{\lsv}{\glvEnv}=\dim\SemT{\ls{}}{\sigma}.
\]}{\texttt{GLang.Metatheory.space\allowbreak Adequacy}}{GLang/Metatheory.lean:698}{none}{preview}
\gldeftheorem{preview.adequacy.type}{Exact-basis type adequacy}{Under consistent environments, if $\glCat{\epsilon}{\Psi}{b}{\ltyv{}}$,
then $\tof{b}\simeq_{\SemBB{b}{\sigma};\sigma,\glvEnv}\ltyv{}$.}{\texttt{GLang.Metatheory.type\allowbreak Adequacy\allowbreak At\allowbreak Basis}}{GLang/Metatheory.lean:849}{none}{preview}
\gldeftheorem{preview.adequacy.cob}{Change-of-basis adequacy}{Under consistent environments, if $\glCcob{\epsilon}{\Psi}{b_0}{v}{b_1}$,
put $B_i=\SemBB{b_i}{\sigma}$. Then
\[
\begin{gathered}
\exists A,N.\quad\Sem{v}{\glvEnv}=A\ne\glverr,\quad |B_0|=|B_1|=N,\quad |A|=N^2,\\
(B_0)_j=\sum_{i=0}^{N-1}A_{jN+i}(B_1)_i\quad(0\le j<N).
\end{gathered}
\]
Thus $\operatorname{coords}_{B_1}(a)=A^{\mathsf T}\operatorname{coords}_{B_0}(a)$
when $A$ is read as a row-major matrix.}{\texttt{GLang.Metatheory.cob\allowbreak Adequacy}}{GLang/Metatheory.lean:826}{none}{preview}
\gldeftheorem{preview.adequacy.endToEnd}{Progress and adequacy}{If $(\Delta,\epsilon);\Psi;\valCheckCtx\vdash(\sigma,\glvEnv)$,
$(\Delta,\epsilon)\vdash t:\tof{b}$, and $(\emptyset,\epsilon)\vdash b:\bvalid$, then
\[
\begin{gathered}
\exists v,\ltyv{},w.\quad\glCterm{\Delta}{\epsilon}{\Psi}{t}{b}{v}{\ltyv{}},\quad
\operatorname{inferType}(v)=\ltyv{},\quad\operatorname{inferSpace}(v)=\operatorname{spaceOf}(\ltyv{}),\\
\Sem{v}{\glvEnv}=w\ne\glverr,\qquad
\Sem{t}{\sigma}\equiv_{\SemBB{b}{\sigma};\sigma,\glvEnv}^{\tof{b},\ltyv{}}w,\qquad
\tof{b}\simeq_{\SemBB{b}{\sigma};\sigma,\glvEnv}\ltyv{}.
\end{gathered}
\]
This constructs a checked evaluation; it does not assert adequacy of every ordinary derivation.}{\texttt{GLang.Metatheory.checked\allowbreak Progress\allowbreak Adequacy}}{GLang/Metatheory.lean:719}{none}{preview}
\gldeftheorem{preview.adequacy.checkedValue}{Checked value evaluation at the retained type}{For the same checked term derivation, checking the raw result against its
retained type $\ltyv{}$ also succeeds and gives an agreeing value.
This is $\mathcal E^{\mathrm{check}}_{\ltyv{}}(v,\glvEnv)$: unlike
$\SemCheck{v}{\glvEnv}$, its check uses the derivation's retained type,
which need not equal the inferred type for an arbitrary supplied checked derivation.}{\texttt{GLang.Adequacy.Direct.term\allowbreak Checked}}{GLang/Adequacy/DirectPublic.lean:113}{none}{preview}
\gldeftheorem{metatheory.sourceSpaceSoundness}{Source-space soundness}{Assuming
\begin{itemize}
  \item \gpfit{(\lemptyset, \evalCtxSpace{})\vdash \ls{}\colon\svalid}
\end{itemize}
then
\begin{center}\gpfit{\mathsf{lc}(\ls{})}\\[2pt]
\gpfit{\wedge\; \mathsf{fv}(\ls{}) \subseteq \mathrm{dom}(\evalCtxSpace{})}\\[2pt]
\gpfit{\wedge\; \forall {\sigma{}_{1}},\,{\sigma{}_{2}}.\; {\sigma{}_{1}} =_{\mathrm{dom}(\evalCtxSpace{})} {\sigma{}_{2}} \Rightarrow \SemT{\ls{}}{{\sigma{}_{1}}} = \SemT{\ls{}}{{\sigma{}_{2}}}}\end{center}
}{\texttt{GLang.Metatheory.source\allowbreak Space\allowbreak Soundness}}{GLang/Metatheory.lean:111}{none}{lem:soundLemma}
\gldeftheorem{metatheory.sourceTypeSoundness}{Source-type soundness}{Assuming
\begin{itemize}
  \item \gpfit{(\lemptyset, \evalCtxSpace{})\vdash \lty{}\colon \lky{}}
\end{itemize}
then
\begin{center}\gpfit{\mathsf{lc}(\lty{})}\\[2pt]
\gpfit{\wedge\; \mathsf{fv}(\lty{}) \subseteq \mathrm{dom}(\evalCtxSpace{})}\\[2pt]
\gpfit{\wedge\; \forall {\sigma{}_{1}},\,{\sigma{}_{2}}.\; {\sigma{}_{1}} =_{\mathrm{dom}(\evalCtxSpace{})} {\sigma{}_{2}} \Rightarrow \SemT{\lty{}}{{\sigma{}_{1}}} = \SemT{\lty{}}{{\sigma{}_{2}}}}\end{center}
}{\texttt{GLang.Metatheory.source\allowbreak Type\allowbreak Soundness}}{GLang/Metatheory.lean:137}{none}{lem:soundLemma}
\gldeftheorem{metatheory.sourceTermSoundness}{Source-term soundness}{Assuming
\begin{itemize}
  \item \gpfit{(\lDelta{}, \lepsilon{})\vdash \lee{}\colon \lty{}}
  \item \gpfit{\vdash (\lDelta{}, \lepsilon{})}
  \item \gpfit{\lDelta{}, \lepsilon{}\vdash \sigma{}}
\end{itemize}
then
\begin{center}\gpfit{\mathsf{StrongSourceSound}(\sigma{}; \lDelta{}; \lepsilon{}; \lee{}; \lty{})}\end{center}
}{\texttt{GLang.Metatheory.source\allowbreak Term\allowbreak Soundness}}{GLang/Metatheory.lean:88}{none}{lem:soundLemma}
\gldeftheorem{metatheory.valueShapeSoundness}{Value-shape soundness}{Assuming
\begin{itemize}
  \item \gpfit{\mathcal{O}\colon \mathsf{PartialFFSOracle}}
  \item \gpfit{\lsv{}\colon \mathsf{ValSpace}}
\end{itemize}
then
\begin{center}\gpfit{\mathsf{dofPartial}(\mathcal{O}, \lsv{}) = \mathsf{bind}(\mathsf{shpOfPartial}(\mathcal{O}, \lsv{}), \lambda \mathit{shape}.\, \mathsf{pure}(\mathsf{foldr}(\lambda {\mathit{x}_{1}}.\, \lambda {\mathit{x}_{2}}.\, ({\mathit{x}_{1}}) \cdot ({\mathit{x}_{2}}), 1, \mathit{shape})))}\end{center}
}{\texttt{GLang.Metatheory.value\allowbreak Shape\allowbreak Soundness}}{GLang/Metatheory.lean:196}{none}{lem:soundVals}
\gldeftheorem{metatheory.valueTypeSoundness}{Value-type soundness}{Assuming
\begin{itemize}
  \item \gpfit{\valCheckCtx{}\vdash \leev{}\colon \ltyv{}}
  \item \gpfit{\mathsf{validValCtx}(\valCheckCtx{})}
\end{itemize}
then
\begin{center}\gpfit{\exists \lky{}.\; \valCheckCtx{}\vdash \ltyv{}\colon \lky{}}\end{center}
}{\texttt{GLang.Metatheory.value\allowbreak Type\allowbreak Soundness}}{GLang/Metatheory.lean:204}{none}{lem:soundVals}
\gldeftheorem{metatheory.valueTermSoundness}{Value-term soundness}{Assuming
\begin{itemize}
  \item \gpfit{\glvEnv{}\colon \mathsf{ValEnvSem}}
  \item \gpfit{\mathsf{PartialEnvWF}(\mathsf{oracle}, \valCheckCtx{}, \glvEnv{})}
  \item \gpfit{\valCheckCtx{}\vdash \leev{}\colon \ltyv{}}
\end{itemize}
then
\begin{center}\gpfit{\mathsf{PartialCoherent}(\ltyv{}, \mathsf{denotValTyPartial}(\mathsf{oracle}, \glvEnv{}, [], [], \ltyv{}), \mathsf{denotValTyPartial}(\mathsf{oracle}, \glvEnv{}, [], [], \ltyv{}))}\\[2pt]
\gpfit{\wedge\; \mathsf{PartialRel}(\mathsf{WorldRuntimeRelated}(\mathsf{oracle}, \ltyv{}, \mathsf{mk}(\glvEnv{}, [], []), \mathsf{mk}(\glvEnv{}, [], [])), \mathsf{denotValB\_partial}(\mathsf{oracle}, \glvEnv{}, [], [], \leev{}), \mathsf{denotValB\_partial}(\mathsf{oracle}, \glvEnv{}, [], [], \leev{}))}\end{center}
}{\texttt{GLang.Metatheory.value\allowbreak Term\allowbreak Soundness}}{GLang/Metatheory.lean:284}{none}{lem:soundVals}
\gldeftheorem{metatheory.spacePreservation}{Ordinary space preservation}{Assuming
\begin{itemize}
  \item \gpfit{(\lDelta{},\evalCtxSpace{});\valCheckCtx{}\vdash \evalCtx{}}
  \item \gpfit{(\lemptyset, \evalCtxSpace{})\vdash \ls{}\colon\svalid}
  \item \gpfit{\evalSJudge{\evalCtx{}}{\ls{}}{\lsv{}}}
\end{itemize}
then
\begin{center}\gpfit{\vdash \lsv{}\colon \svalid{}}\\[2pt]
\gpfit{\wedge\; \mathsf{fv}(\lsv{}) \subseteq \mathrm{dom}(\evalCtx{})}\\[2pt]
\gpfit{\wedge\; \valCheckCtx{}\vdash \lsv{}\colon\svalid}\end{center}
}{\texttt{GLang.Metatheory.space\allowbreak Preservation}}{GLang/Metatheory.lean:474}{none}{thm:preserve}
\gldeftheorem{metatheory.termPreservation}{Ordinary term preservation}{Assuming
\begin{itemize}
  \item \gpfit{(\lDelta{},\evalCtxSpace{});\valCheckCtx{}\vdash \evalCtx{}}
  \item \gpfit{(\lDelta{}, \evalCtxSpace{})\vdash \lee{}\colon \tof{\lb{}}}
  \item \gpfit{(\emptyset,\evalCtxSpace{})\vdash \lb{}\colon \bvalid{}}
  \item \gpfit{\evalJudge{\evalCtx{}}{\lee{}}{\lb{}}{\leev{}}}
\end{itemize}
then
\begin{center}\gpfit{\exists \ltyv{}.\; \evalSJudge{\evalCtx{}}{\tof{\lb{}}}{\ltyv{}}}\\[2pt]
\gpfit{\wedge\; \mathsf{kind}(\operatorname{inferType}(\leev{})) = \mathsf{kind}(\ltyv{})}\\[2pt]
\gpfit{\wedge\; \vdash \operatorname{inferType}(\leev{})\colon \tvalid{}}\\[2pt]
\gpfit{\wedge\; \mathsf{fv}(\operatorname{inferType}(\leev{})) \cup \mathsf{fv}(\leev{}) \subseteq \mathrm{dom}(\evalCtx{})}\\[2pt]
\gpfit{\wedge\; \valCheckCtx{}\vdash \leev{}\colon \operatorname{inferType}(\leev{})}\end{center}
}{\texttt{GLang.Metatheory.term\allowbreak Preservation}}{GLang/Metatheory.lean:428}{none}{thm:preserve}
\gldeftheorem{metatheory.cobPreservation}{Ordinary change-of-basis preservation}{Assuming
\begin{itemize}
  \item \gpfit{(\lDelta{},\evalCtxSpace{});\valCheckCtx{}\vdash \evalCtx{}}
  \item \gpfit{(\emptyset,\evalCtxSpace{})\vdash \textcolor{lang}{{\lb{}_{\mathrm{src}}}}\colon \bvalid{}}
  \item \gpfit{\evalBBJudge{\evalCtx{}}{\textcolor{lang}{{\lb{}_{\mathrm{src}}}}}{\leev{}}{\textcolor{lang}{{\lb{}_{\mathrm{tgt}}}}}}
\end{itemize}
then
\begin{center}\gpfit{\exists \lsn{0}.\; \exists \lsn{1}.\; \exists \lsnv{0}.\; \exists \lsnv{1}.\; \operatorname{inferType}(\leev{}) = \lsnv{0} \loprod \lsnv{1} \;\wedge\; \evalSJudge{\evalCtx{}}{\lsn{0}}{\lsnv{0}} \;\wedge\; \evalSJudge{\evalCtx{}}{\lsn{1}}{\lsnv{1}} \;\wedge\; \vdash \operatorname{inferType}(\leev{})\colon \tvalid{} \;\wedge\; \mathsf{fv}(\leev{}) \subseteq \mathrm{dom}(\evalCtx{})}\\[2pt]
\gpfit{\wedge\; \valCheckCtx{}\vdash \leev{}\colon \operatorname{inferType}(\leev{})}\end{center}
}{\texttt{GLang.Metatheory.cob\allowbreak Preservation}}{GLang/Metatheory.lean:519}{none}{thm:preserve}
\gldeftheorem{metatheory.checkedSpaceProgress}{Checked space progress}{Assuming
\begin{itemize}
  \item \gpfit{(\lDelta{},\evalCtxSpace{});\valCheckCtx{}\vdash \evalCtx{}}
  \item \gpfit{(\lemptyset, \evalCtxSpace{})\vdash \ls{}\colon\svalid}
\end{itemize}
then
\begin{center}\gpfit{\exists \lsv{}.\; \evalCtxSpace{};\evalCtx{}\vdash_{\mathsf C} \ls{} \evalArrow{} \lsv{}}\end{center}
}{\texttt{GLang.Metatheory.checked\allowbreak Space\allowbreak Progress}}{GLang/Metatheory.lean:626}{none}{thm:prog}
\gldeftheorem{metatheory.termProgress}{Checked term progress}{Assuming
\begin{itemize}
  \item \gpfit{(\lDelta{},\evalCtxSpace{});\valCheckCtx{}\vdash \evalCtx{}}
  \item \gpfit{(\lDelta{}, \evalCtxSpace{})\vdash \lee{}\colon \tof{\lb{}}}
  \item \gpfit{(\emptyset,\evalCtxSpace{})\vdash \lb{}\colon \bvalid{}}
\end{itemize}
then
\begin{center}\gpfit{\exists \leev{}.\; \exists \ltyv{}.\; \lDelta{};\evalCtxSpace{};\evalCtx{}\vdash_{\mathsf C} \lee{} \xlongrightarrow{\lb{}} \leev{}\glresulttype{\ltyv{}}}\\[2pt]
\gpfit{\wedge\; \operatorname{inferType}(\leev{}) = \ltyv{}}\\[2pt]
\gpfit{\wedge\; \operatorname{inferSpace}(\leev{}) = \mathsf{spaceOf}(\ltyv{})}\end{center}
}{\texttt{GLang.Metatheory.term\allowbreak Progress}}{GLang/Metatheory.lean:568}{none}{thm:prog}
\gldeftheorem{metatheory.checkedCobProgress}{General checked change-of-basis progress}{Assuming
\begin{itemize}
  \item \gpfit{(\lDelta{},\evalCtxSpace{});\valCheckCtx{}\vdash \evalCtx{}}
  \item \gpfit{(\emptyset,\evalCtxSpace{})\vdash \textcolor{lang}{{\lb{}_{\mathrm{src}}}}\colon \bvalid{}}
  \item \gpfit{(\emptyset,\evalCtxSpace{})\vdash \textcolor{lang}{{\lb{}_{\mathrm{tgt}}}}\colon \bvalid{}}
  \item \gpfit{\tof{\textcolor{lang}{{\lb{}_{\mathrm{src}}}}} = \sof{\textcolor{lang}{{\lb{}_{\mathrm{src}}}}}}
  \item \gpfit{\tof{\textcolor{lang}{{\lb{}_{\mathrm{src}}}}} = \tof{\textcolor{lang}{{\lb{}_{\mathrm{tgt}}}}}}
\end{itemize}
then
\begin{center}\gpfit{\exists \leev{}.\; \evalCtxSpace{};\evalCtx{}\vdash_{\mathsf C} \textcolor{lang}{{\lb{}_{\mathrm{src}}}} \xlongleftrightarrow{\leev{}} \textcolor{lang}{{\lb{}_{\mathrm{tgt}}}}}\end{center}
}{\texttt{GLang.Metatheory.checked\allowbreak Cob\allowbreak Progress}}{GLang/Metatheory.lean:665}{none}{lem:cobProg}
\gldeftheorem{metatheory.spaceAdequacy}{Space adequacy}{Assuming
\begin{itemize}
  \item \gpfit{\evalCtxSpace{};\evalCtx{}\vdash_{\mathsf C} \ls{} \evalArrow{} \lsv{}}
  \item \gpfit{\lDelta{}\colon \mathsf{LCtx}}
  \item \gpfit{\sigma{}\colon \mathsf{DEnv}}
  \item \gpfit{\valCheckCtx{}\colon \mathsf{ValCtx}}
  \item \gpfit{(\lDelta{}, \evalCtxSpace{}); \evalCtx{}; \valCheckCtx{}\vdash \sigma{}, \glvEnv{}}
\end{itemize}
then
\begin{center}\gpfit{\mathsf{shpOfPartial}(\mathsf{oracle}(\glvEnv{}, [], []), \lsv{}) = \mathsf{srcShp}(\sigma{}, [], \ls{})}\end{center}
}{\texttt{GLang.Metatheory.space\allowbreak Adequacy}}{GLang/Metatheory.lean:698}{none}{thm:main}
\gldeftheorem{metatheory.termAdequacy}{Term adequacy}{Assuming
\begin{itemize}
  \item \gpfit{\lDelta{};\evalCtxSpace{};\evalCtx{}\vdash_{\mathsf C} \lee{} \xlongrightarrow{\lb{}} \leev{}\glresulttype{\ltyv{}}}
  \item \gpfit{\sigma{}\colon \mathsf{DEnv}}
  \item \gpfit{\valCheckCtx{}\colon \mathsf{ValCtx}}
  \item \gpfit{(\lDelta{}, \evalCtxSpace{}); \evalCtx{}; \valCheckCtx{}\vdash \sigma{}, \glvEnv{}}
\end{itemize}
then
\begin{center}\gpfit{\Sem{\lee{}}{\sigma{}} \equiv^{\ltyv{}}_{\SemBB{\lb{}}{\sigma{}}} \mathsf{evalVal}(\glvEnv{}, [], [], \leev{})}\\[2pt]
\gpfit{\wedge\; \SemT{\tof{\lb{}}}{\sigma{}} \equiv_{\SemBB{\lb{}}{\sigma{}}} \SemT{\ltyv{}}{\glvEnv{}}}\end{center}
}{\texttt{GLang.Metatheory.term\allowbreak Adequacy}}{GLang/Metatheory.lean:682}{none}{thm:main}
\gldeftheorem{metatheory.cobAdequacy}{Change-of-basis adequacy}{Assuming
\begin{itemize}
  \item \gpfit{\evalCtxSpace{};\evalCtx{}\vdash_{\mathsf C} \textcolor{lang}{{\lb{}_{\mathrm{src}}}} \xlongleftrightarrow{\leev{}} \textcolor{lang}{{\lb{}_{\mathrm{tgt}}}}}
  \item \gpfit{\lDelta{}\colon \mathsf{LCtx}}
  \item \gpfit{\sigma{}\colon \mathsf{DEnv}}
  \item \gpfit{\valCheckCtx{}\colon \mathsf{ValCtx}}
  \item \gpfit{(\lDelta{}, \evalCtxSpace{}); \evalCtx{}; \valCheckCtx{}\vdash \sigma{}, \glvEnv{}}
\end{itemize}
then
\begin{center}\gpfit{\SemBB{\textcolor{lang}{{\lb{}_{\mathrm{src}}}}}{\sigma{}} = \SemA{\leev{}}{\glvEnv{}}{} \cdot \SemBB{\textcolor{lang}{{\lb{}_{\mathrm{tgt}}}}}{\sigma{}}}\end{center}
}{\texttt{GLang.Metatheory.cob\allowbreak Adequacy}}{GLang/Metatheory.lean:826}{none}{thm:main}
\gldeftheorem{metatheory.sourceBasisSoundness}{Source-basis soundness}{Assuming
\begin{itemize}
  \item \gpfit{(\emptyset,\evalCtxSpace{})\vdash \lb{}\colon \bvalid{}}
  \item \gpfit{\vdash (\lemptyset{}, \evalCtxSpace{})}
  \item \gpfit{\lemptyset{}, \evalCtxSpace{}\vdash \sigma{}}
\end{itemize}
then
\begin{center}\gpfit{\SemBB{\lb{}}{\sigma{}} \;\text{is a basis for}\; \SemT{\tof{\lb{}}}{\sigma{}}}\end{center}
}{\texttt{GLang.Metatheory.source\allowbreak Basis\allowbreak Soundness}}{GLang/Metatheory.lean:150}{none}{lem:denoteBasisSound}
\gldeftheorem{runtimeValue.valHasType.regular}{Value-type regularity}{Assuming
\begin{itemize}
  \item \gpfit{\valCheckCtx{}\vdash \leev{}\colon \ltyv{}}
  \item \gpfit{\mathit{a}\colon \mathsf{ValCtxValid}(\valCheckCtx{})}
\end{itemize}
then
\begin{center}\gpfit{\Sigma \lky{}.\; \valCheckCtx{}\vdash \ltyv{}\colon \lky{}}\end{center}
}{\texttt{GLang.Runtime\allowbreak Value.Val\allowbreak Has\allowbreak Type.regular}}{GLang/RuntimeValueRegular.lean:23}{none}{lem:soundVals (1)}
\gldeftheorem{runtimeValue.valHasType.inferTypeEqOfTyped}{Exact inferred value type}{Assuming
\begin{itemize}
  \item \gpfit{\valCheckCtx{}\vdash \leev{}\colon \ltyv{}}
\end{itemize}
then
\begin{center}\gpfit{\operatorname{inferType}(\leev{}) = \ltyv{}}\end{center}
}{\texttt{GLang.Runtime\allowbreak Value.Val\allowbreak Has\allowbreak Type.infer\allowbreak Type\_\allowbreak eq\_\allowbreak of\_\allowbreak typed}}{GLang/RuntimeValueInfer.lean:37}{none}{lem:soundVals (1)}
\gldeftheorem{valHasType.inferSpaceEqOfSpace}{Exact inferred value space}{Assuming
\begin{itemize}
  \item \gpfit{\valCheckCtx{}\vdash \leev{}\colon \lsv{}}
\end{itemize}
then
\begin{center}\gpfit{\operatorname{inferSpace}(\leev{}) = \lsv{}}\end{center}
}{\texttt{GLang.Val\allowbreak Has\allowbreak Type.infer\allowbreak Space\_\allowbreak eq\_\allowbreak of\_\allowbreak space}}{GLang/ValueInferSpaceSound.lean:356}{none}{lem:soundVals (1)}
\gldeftheorem{lemSoundValsMetricsTotal}{Value shape and dimension totality}{Assuming
\begin{itemize}
  \item \gpfit{\mathcal{O}\colon \mathsf{FFSOracle}}
  \item \gpfit{\lsv{}\colon \mathsf{ValSpace}}
\end{itemize}
then
\begin{center}\gpfit{\exists \mathit{d}.\; \exists \mathit{shp}.\; \dim_{\mathcal{O}}(\lsv{}) = \mathit{d}}\\[2pt]
\gpfit{\wedge\; \shpOf{\lsv{}}{\mathcal{O}} = \mathit{shp}}\end{center}
}{\texttt{GLang.lem\_\allowbreak sound\allowbreak Vals\_\allowbreak metrics\_\allowbreak total}}{GLang/ValueSoundVals.lean:55}{none}{lem:soundVals (2)}
\gldeftheorem{basis.tofStd}{Standard basis has the requested type}{Assuming
\begin{itemize}
  \item \gpfit{\lty{}\colon \mathsf{Ty}}
\end{itemize}
then
\begin{center}\gpfit{\tof{\stdBasisDCtx{\lty{}}} = \lty{}}\end{center}
}{\texttt{GLang.Basis.tof\_\allowbreak std}}{GLang/Basis.lean:579}{none}{lem:stdBasisOrthonormal}
\gldeftheorem{basis.stdValid}{Standard basis is valid}{Assuming
\begin{itemize}
  \item \gpfit{\lty{}\colon \mathsf{Ty}}
  \item \gpfit{(\lemptyset, \lepsilon{})\vdash \lty{}\colon \lky{}}
\end{itemize}
then
\begin{center}\gpfit{(\emptyset,\lepsilon{})\vdash \stdBasisDCtx{\lty{}}\colon \bvalid{}}\end{center}
}{\texttt{GLang.Basis.std\_\allowbreak valid}}{GLang/Basis.lean:688}{none}{lem:stdBasisOrthonormal}
\gldeftheorem{basis.isStdStd}{Standard basis passes the standardness check}{Assuming
\begin{itemize}
  \item \gpfit{\lty{}\colon \mathsf{Ty}}
\end{itemize}
then
\begin{center}\gpfit{\isStdBasis{\stdBasisDCtx{\lty{}}}}\end{center}
}{\texttt{GLang.Basis.is\allowbreak Std\_\allowbreak std}}{GLang/ProgressStdBasis.lean:56}{none}{lem:stdBasisOrthonormal}
\gldeftheorem{metatheory.standardBasisSoundness}{Standard basis/frame is recursively complete and orthonormal}{Assuming
\begin{itemize}
  \item \gpfit{(\lemptyset, \evalCtxSpace{})\vdash \lty{}\colon \lky{}}
  \item \gpfit{\vdash (\lemptyset{}, \evalCtxSpace{})}
  \item \gpfit{\lemptyset{}, \evalCtxSpace{}\vdash \sigma{}}
\end{itemize}
then
\begin{center}\gpfit{\SemBB{\stdBasisDCtx{\lty{}}}{\sigma{}} \;\text{is a basis for}\; \SemT{\lty{}}{\sigma{}}}\\[2pt]
\gpfit{\wedge\; \SemBB{\stdBasisDCtx{\lty{}}}{\sigma{}} \;\text{is recursively orthonormal}}\end{center}
}{\texttt{GLang.Metatheory.standard\allowbreak Basis\allowbreak Soundness}}{GLang/Metatheory.lean:161}{none}{lem:stdBasisOrthonormal}
\gldeftheorem{v2.progressStd}{Progress at the standard basis}{Assuming
\begin{itemize}
  \item \gpfit{\mathsf{CtxWF'}(\lDelta{}, \evalCtxSpace{})}
  \item \gpfit{\vdash \evalCtx{}}
  \item \gpfit{\mathsf{EvalCtxLc}(\evalCtx{})}
  \item \gpfit{\mathsf{EvalCtxLamAnchored}(\evalCtx{})}
  \item \gpfit{\mathsf{TypingImplyEval}(\lDelta{}, \evalCtxSpace{}, \evalCtx{})}
  \item \gpfit{\mathsf{StdComplete}(\evalCtx{})}
  \item \gpfit{(\lDelta{}, \evalCtxSpace{})\vdash \lee{}\colon \lty{}}
\end{itemize}
then
\begin{center}\gpfit{\exists \leev{}.\; \evalJudge{\evalCtx{}}{\lee{}}{\stdBasisDCtx{\lty{}}}{\leev{}}}\end{center}
}{\texttt{GLang.V2.progress\_\allowbreak std}}{GLang/ProgressMaster.lean:65}{none}{thm:progA}
\gldeftheorem{space.iofHasType}{Canonical inner-product term is well typed}{Assuming
\begin{itemize}
  \item \gpfit{\mathsf{iofFresh}(\lepsilon{})}
  \item \gpfit{\ls{}\colon \mathsf{Space}}
  \item \gpfit{(\lemptyset, \lepsilon{})\vdash \ls{}\colon\svalid}
\end{itemize}
then
\begin{center}\gpfit{(\lemptyset{}, \lepsilon{})\vdash \mathsf{iof}(\ls{})\colon \ls{}^{\lstar} \loprod \ls{}^{\lstar}}\end{center}
}{\texttt{GLang.Space.iof\_\allowbreak has\allowbreak Type}}{GLang/InnerProduct.lean:1523}{none}{lem:innerLemap}
\gldeftheorem{space.iofDenotIsInnerProduct}{Canonical inner-product term denotes an inner product}{Assuming
\begin{itemize}
  \item \gpfit{\mathsf{iofFresh}(\lepsilon{})}
  \item \gpfit{\lemptyset{}, \lepsilon{}\vdash \sigma{}}
  \item \gpfit{\mathsf{Scoped}(\lemptyset{}, \lepsilon{})}
  \item \gpfit{(\lemptyset, \lepsilon{})\vdash \ls{}\colon\svalid}
\end{itemize}
then
\begin{center}\gpfit{\Sem{\mathsf{iof}(\ls{})}{\sigma{}} \;\text{is an inner product}}\end{center}
}{\texttt{GLang.Space.iof\_\allowbreak denot\_\allowbreak is\allowbreak Inner\allowbreak Product}}{GLang/InnerProduct.lean:6809}{none}{lem:innerLemap}
\gldeftheorem{metatheory.sourceSoundnessWeak}{Historical source-soundness projection}{Assuming
\begin{itemize}
  \item \gpfit{(\lDelta{}, \evalCtxSpace{})\vdash \lee{}\colon \lty{}}
  \item \gpfit{\mathsf{Scoped}(\lDelta{}, \evalCtxSpace{})}
  \item \gpfit{\lDelta{}, \evalCtxSpace{}\vdash \sigma{}}
\end{itemize}
then
\begin{center}\gpfit{\lDelta{}\vdash \Sem{\lee{}}{\sigma{}} \in \SemT{\lty{}}{\sigma{}}}\end{center}
}{\texttt{GLang.Metatheory.source\allowbreak Soundness\allowbreak Weak}}{GLang/Metatheory.lean:78}{none}{--}
\gldeftheorem{metatheory.sourceSpaceRegularity}{Space regularity}{Assuming
\begin{itemize}
  \item \gpfit{(\lemptyset, \evalCtxSpace{})\vdash \ls{}\colon\svalid}
\end{itemize}
then
\begin{center}\gpfit{\mathsf{lc}(\ls{})}\\[2pt]
\gpfit{\wedge\; \mathsf{fv}(\ls{}) \subseteq \mathrm{dom}(\evalCtxSpace{})}\\[2pt]
\gpfit{\wedge\; \forall {\sigma{}_{1}},\,{\sigma{}_{2}}.\; {\sigma{}_{1}} =_{\mathrm{dom}(\evalCtxSpace{})} {\sigma{}_{2}} \Rightarrow \SemT{\ls{}}{{\sigma{}_{1}}} = \SemT{\ls{}}{{\sigma{}_{2}}}}\end{center}
}{\texttt{GLang.Metatheory.source\allowbreak Space\allowbreak Regularity}}{GLang/Metatheory.lean:98}{none}{--}
\gldeftheorem{metatheory.sourceTypeRegularity}{Type regularity}{Assuming
\begin{itemize}
  \item \gpfit{(\lemptyset, \evalCtxSpace{})\vdash \lty{}\colon \lky{}}
\end{itemize}
then
\begin{center}\gpfit{\mathsf{lc}(\lty{})}\\[2pt]
\gpfit{\wedge\; \mathsf{fv}(\lty{}) \subseteq \mathrm{dom}(\evalCtxSpace{})}\\[2pt]
\gpfit{\wedge\; \forall {\sigma{}_{1}},\,{\sigma{}_{2}}.\; {\sigma{}_{1}} =_{\mathrm{dom}(\evalCtxSpace{})} {\sigma{}_{2}} \Rightarrow \SemT{\lty{}}{{\sigma{}_{1}}} = \SemT{\lty{}}{{\sigma{}_{2}}}}\end{center}
}{\texttt{GLang.Metatheory.source\allowbreak Type\allowbreak Regularity}}{GLang/Metatheory.lean:124}{none}{--}
\gldeftheorem{metatheory.valueSoundnessOfOracle}{Oracle-polymorphic canonical value soundness}{Assuming
\begin{itemize}
  \item \gpfit{\mathcal{O}\colon \mathsf{FFSOracleSem}}
  \item \gpfit{\glvEnv{}\colon \mathsf{ValEnvSem}}
  \item \gpfit{\Gamma{}\colon \mathsf{List}(\mathsf{ValTy})}
  \item \gpfit{\lDelta{}\colon \mathsf{List}(\mathsf{ValValSem})}
  \item \gpfit{\valCheckCtx{}\vdash \leev{}\colon \ltyv{}}
\end{itemize}
then
\begin{center}\gpfit{\begin{gathered}\glvEnv{}; \Gamma{}; \lDelta{}\vdash_{\mathsf{canonical}} \mathbf{let}\ \mathit{value} \gets \mathcal{E}\llbracket \leev{}\rrbracket_{\Gamma{};\lDelta{}}\ \mathbf{in}\ \mathbf{let}\ a \gets \mathit{value}\ \mathbf{in}\ \mathsf{fromElChecked}(\mathcal{T}\llbracket \operatorname{inferTypeCtx}(\Gamma{}, \leev{})\rrbracket_{\Gamma{};\lDelta{}})\colon \ltyv{}\end{gathered}}\end{center}
}{\texttt{GLang.Metatheory.value\allowbreak Soundness\allowbreak Of\allowbreak Oracle}}{GLang/Metatheory.lean:215}{none}{lem:soundVals}
\gldeftheorem{metatheory.valueSoundnessWeak}{Historical weak value-soundness projection}{Assuming
\begin{itemize}
  \item \gpfit{\glvEnv{}\colon \mathsf{ValEnvSem}}
  \item \gpfit{\Gamma{}\colon \mathsf{List}(\mathsf{ValTy})}
  \item \gpfit{\lDelta{}\colon \mathsf{List}(\mathsf{ValValSem})}
  \item \gpfit{\valCheckCtx{}\vdash \leev{}\colon \ltyv{}}
\end{itemize}
then
\begin{center}\gpfit{\begin{gathered}\glvEnv{}; \Gamma{}; \lDelta{}\vdash \mathbf{let}\ \mathit{value} \gets \mathcal{E}\llbracket \leev{}\rrbracket_{\Gamma{};\lDelta{}}\ \mathbf{in}\ \mathbf{let}\ a \gets \mathit{value}\ \mathbf{in}\ \mathsf{fromElChecked}(\mathcal{T}\llbracket \operatorname{inferTypeCtx}(\Gamma{}, \leev{})\rrbracket_{\Gamma{};\lDelta{}})\colon \ltyv{}\end{gathered}}\end{center}
}{\texttt{GLang.Metatheory.value\allowbreak Soundness\allowbreak Weak}}{GLang/Metatheory.lean:254}{none}{--}
\gldeftheorem{metatheory.checkedPreservation}{Internal checked exact-index certificate}{Assuming
\begin{itemize}
  \item \gpfit{\vdash \evalCtx{}}
  \item \gpfit{\mathrm{dom}(\evalCtx{}) \subseteq \mathcal{S}}
  \item \gpfit{\lDelta{};\evalCtxSpace{};\evalCtx{}\vdash_{\mathsf C} \lee{} \xlongrightarrow{\lb{}} \leev{}\glresulttype{\ltyv{}}}
\end{itemize}
then
\begin{center}\gpfit{\evalCtxSpace{};\evalCtx{}\vdash_{\mathsf C} \lb{} \evalArrow{}_{\mathsf{basis}} \ltyv{}}\\[2pt]
\gpfit{\wedge\; \evalSJudge{\evalCtx{}}{\tof{\lb{}}}{\ltyv{}}}\\[2pt]
\gpfit{\wedge\; \mathsf{kind}(\operatorname{inferType}(\leev{})) = \mathsf{kind}(\ltyv{})}\\[2pt]
\gpfit{\wedge\; \vdash \operatorname{inferType}(\leev{})\colon \tvalid{}}\\[2pt]
\gpfit{\wedge\; \mathsf{fv}(\operatorname{inferType}(\leev{})) \cup \mathsf{fv}(\leev{}) \subseteq \mathcal{S}}\end{center}
}{\texttt{GLang.Metatheory.checked\allowbreak Preservation}}{GLang/Metatheory.lean:312}{none}{--}
\gldeftheorem{metatheory.checkedBaseShapePreservation}{Checked base-shape preservation}{Assuming
\begin{itemize}
  \item \gpfit{\lDelta{};\evalCtxSpace{};\evalCtx{}\vdash_{\mathsf C} \lee{} \xlongrightarrow{\lb{}} \leev{}\glresulttype{\lsv{}}}
  \item \gpfit{(\lDelta{}, \evalCtxSpace{}); \evalCtx{}; \valCheckCtx{}\vdash \sigma{}, \glvEnv{}}
\end{itemize}
then
\begin{center}\gpfit{\mathsf{shpOfPartial}(\mathsf{oracle}(\glvEnv{}, [], []), \operatorname{inferSpace}(\leev{})) = \mathsf{shpOfPartial}(\mathsf{oracle}(\glvEnv{}, [], []), \lsv{})}\end{center}
}{\texttt{GLang.Metatheory.checked\allowbreak Base\allowbreak Shape\allowbreak Preservation}}{GLang/Metatheory.lean:444}{none}{thm:preserve}
\gldeftheorem{metatheory.checkedSpacePreservation}{Checked space preservation}{Assuming
\begin{itemize}
  \item \gpfit{\vdash \evalCtx{}}
  \item \gpfit{\mathrm{dom}(\evalCtx{}) \subseteq \mathcal{S}}
  \item \gpfit{\evalCtxSpace{};\evalCtx{}\vdash_{\mathsf C} \ls{} \evalArrow{} \lsv{}}
\end{itemize}
then
\begin{center}\gpfit{\vdash \lsv{}\colon \svalid{}}\\[2pt]
\gpfit{\wedge\; \mathsf{fv}(\lsv{}) \subseteq \mathcal{S}}\end{center}
}{\texttt{GLang.Metatheory.checked\allowbreak Space\allowbreak Preservation}}{GLang/Metatheory.lean:489}{none}{thm:preserve}
\gldeftheorem{metatheory.typePreservation}{Type preservation}{Assuming
\begin{itemize}
  \item \gpfit{\vdash \evalCtx{}}
  \item \gpfit{\mathrm{dom}(\evalCtx{}) \subseteq \mathcal{S}}
  \item \gpfit{\evalCtxSpace{};\evalCtx{}\vdash_{\mathsf C} \lty{} \evalArrow{} \ltyv{}}
\end{itemize}
then
\begin{center}\gpfit{\vdash \ltyv{}\colon \tvalid{}}\\[2pt]
\gpfit{\wedge\; \mathsf{fv}(\ltyv{}) \subseteq \mathcal{S}}\end{center}
}{\texttt{GLang.Metatheory.type\allowbreak Preservation}}{GLang/Metatheory.lean:504}{none}{thm:preserve}
\gldeftheorem{metatheory.checkedCobPreservation}{Checked change-of-basis preservation}{Assuming
\begin{itemize}
  \item \gpfit{\vdash \evalCtx{}}
  \item \gpfit{\mathrm{dom}(\evalCtx{}) \subseteq \mathcal{S}}
  \item \gpfit{\evalCtxSpace{};\evalCtx{}\vdash_{\mathsf C} \textcolor{lang}{{\lb{}_{\mathrm{src}}}} \xlongleftrightarrow{\leev{}} \textcolor{lang}{{\lb{}_{\mathrm{tgt}}}}}
\end{itemize}
then
\begin{center}\gpfit{\exists \lsn{0}.\; \exists \lsn{1}.\; \exists \lsnv{0}.\; \exists \lsnv{1}.\; \operatorname{inferType}(\leev{}) = \lsnv{0} \loprod \lsnv{1} \;\wedge\; \evalSJudge{\evalCtx{}}{\lsn{0}}{\lsnv{0}} \;\wedge\; \evalSJudge{\evalCtx{}}{\lsn{1}}{\lsnv{1}} \;\wedge\; \vdash \operatorname{inferType}(\leev{})\colon \tvalid{} \;\wedge\; \mathsf{fv}(\leev{}) \subseteq \mathcal{S}}\end{center}
}{\texttt{GLang.Metatheory.checked\allowbreak Cob\allowbreak Preservation}}{GLang/Metatheory.lean:535}{none}{thm:preserve}
\gldeftheorem{metatheory.cobEndpointPreservation}{Endpoint-aware change-of-basis preservation}{Assuming
\begin{itemize}
  \item \gpfit{\evalCtxSpace{};\evalCtx{}\vdash_{\mathsf C} \textcolor{lang}{{\lb{}_{\mathrm{src}}}} \xlongleftrightarrow{\leev{}} \textcolor{lang}{{\lb{}_{\mathrm{tgt}}}}}
  \item \gpfit{(\lDelta{}, \evalCtxSpace{}); \evalCtx{}; \valCheckCtx{}\vdash \sigma{}, \glvEnv{}}
\end{itemize}
then
\begin{center}\gpfit{\exists \lsn{0}.\; \exists \lsn{1}.\; \exists \lsnv{0}.\; \exists \lsnv{1}.\; \operatorname{inferType}(\leev{}) = \lsnv{0} \loprod \lsnv{1} \;\wedge\; \evalSJudge{\evalCtx{}}{\lsn{0}}{\lsnv{0}} \;\wedge\; \evalSJudge{\evalCtx{}}{\lsn{1}}{\lsnv{1}} \;\wedge\; \vdash \operatorname{inferType}(\leev{})\colon \tvalid{} \;\wedge\; \mathsf{fv}(\leev{}) \subseteq \mathrm{dom}(\evalCtx{})}\\[2pt]
\gpfit{\wedge\; \exists \lsnv{{\mathrm{svSrc}}}.\; \exists \lsnv{{\mathrm{svTgt}}}.\; \operatorname{inferSpace}(\leev{}) = \lsnv{{\mathrm{svSrc}}} \loprod \lsnv{{\mathrm{svTgt}}} \;\wedge\; \dim_{\mathcal{O}}(\lsnv{{\mathrm{svSrc}}}) = \lvert \SemBB{\textcolor{lang}{{\lb{}_{\mathrm{src}}}}}{\sigma{}} \rvert \;\wedge\; \dim_{\mathcal{O}}(\lsnv{{\mathrm{svTgt}}}) = \lvert \SemBB{\textcolor{lang}{{\lb{}_{\mathrm{tgt}}}}}{\sigma{}} \rvert}\end{center}
}{\texttt{GLang.Metatheory.cob\allowbreak Endpoint\allowbreak Preservation}}{GLang/CobEndpointPreservation.lean:69}{none}{thm:preserve}
\gldeftheorem{metatheory.cobCommonDimensionAdequacyOfOracle}{Oracle-polymorphic common COB dimension}{Assuming
\begin{itemize}
  \item \gpfit{\mathcal{O}\colon \mathsf{FFSOracleSem}}
  \item \gpfit{\mathsf{RawApplicationOracleSpec}(\mathcal{O})}
  \item \gpfit{\evalCtxSpace{};\evalCtx{}\vdash_{\mathsf C} \textcolor{lang}{{\lb{}_{\mathrm{src}}}} \xlongleftrightarrow{\leev{}} \textcolor{lang}{{\lb{}_{\mathrm{tgt}}}}}
  \item \gpfit{\mathsf{RawTypedConsistent}(\mathcal{O}, \lDelta{}, \evalCtxSpace{}, \evalCtx{}, \valCheckCtx{}, \sigma{}, \glvEnv{})}
\end{itemize}
then
\begin{center}\gpfit{\exists \lsn{0}.\; \exists \lsn{1}.\; \exists \lsnv{0}.\; \exists \lsnv{1}.\; \operatorname{inferType}(\leev{}) = \lsnv{0} \loprod \lsnv{1} \;\wedge\; \evalSJudge{\evalCtx{}}{\lsn{0}}{\lsnv{0}} \;\wedge\; \evalSJudge{\evalCtx{}}{\lsn{1}}{\lsnv{1}} \;\wedge\; \vdash \operatorname{inferType}(\leev{})\colon \tvalid{} \;\wedge\; \mathsf{fv}(\leev{}) \subseteq \mathrm{dom}(\evalCtx{})}\\[2pt]
\gpfit{\wedge\; \SemBB{\textcolor{lang}{{\lb{}_{\mathrm{src}}}}}{\sigma{}} = \SemA{\leev{}}{\glvEnv{}}{} \cdot \SemBB{\textcolor{lang}{{\lb{}_{\mathrm{tgt}}}}}{\sigma{}}}\\[2pt]
\gpfit{\wedge\; \exists \mathit{N}.\; \exists \lsnv{{\mathrm{svSrc}}}.\; \exists \lsnv{{\mathrm{svTgt}}}.\; \operatorname{inferSpace}(\leev{}) = \lsnv{{\mathrm{svSrc}}} \loprod \lsnv{{\mathrm{svTgt}}} \;\wedge\; \mathit{N} = \dim_{\mathcal{O}}(\lsnv{{\mathrm{svSrc}}}) \;\wedge\; \mathit{N} = \dim_{\mathcal{O}}(\lsnv{{\mathrm{svTgt}}}) \;\wedge\; \mathit{N} = \lvert \SemBB{\textcolor{lang}{{\lb{}_{\mathrm{src}}}}}{\sigma{}} \rvert \;\wedge\; \mathit{N} = \lvert \SemBB{\textcolor{lang}{{\lb{}_{\mathrm{tgt}}}}}{\sigma{}} \rvert}\end{center}
}{\texttt{GLang.Metatheory.cob\allowbreak Common\allowbreak Dimension\allowbreak Adequacy\allowbreak Of\allowbreak Oracle}}{GLang/CobEndpointPreservation.lean:98}{none}{--}
\gldeftheorem{metatheory.cobCommonDimensionAdequacy}{Common COB dimension}{Assuming
\begin{itemize}
  \item \gpfit{\evalCtxSpace{};\evalCtx{}\vdash_{\mathsf C} \textcolor{lang}{{\lb{}_{\mathrm{src}}}} \xlongleftrightarrow{\leev{}} \textcolor{lang}{{\lb{}_{\mathrm{tgt}}}}}
  \item \gpfit{(\lDelta{}, \evalCtxSpace{}); \evalCtx{}; \valCheckCtx{}\vdash \sigma{}, \glvEnv{}}
\end{itemize}
then
\begin{center}\gpfit{\exists \lsn{0}.\; \exists \lsn{1}.\; \exists \lsnv{0}.\; \exists \lsnv{1}.\; \operatorname{inferType}(\leev{}) = \lsnv{0} \loprod \lsnv{1} \;\wedge\; \evalSJudge{\evalCtx{}}{\lsn{0}}{\lsnv{0}} \;\wedge\; \evalSJudge{\evalCtx{}}{\lsn{1}}{\lsnv{1}} \;\wedge\; \vdash \operatorname{inferType}(\leev{})\colon \tvalid{} \;\wedge\; \mathsf{fv}(\leev{}) \subseteq \mathrm{dom}(\evalCtx{})}\\[2pt]
\gpfit{\wedge\; \SemBB{\textcolor{lang}{{\lb{}_{\mathrm{src}}}}}{\sigma{}} = \SemA{\leev{}}{\glvEnv{}}{} \cdot \SemBB{\textcolor{lang}{{\lb{}_{\mathrm{tgt}}}}}{\sigma{}}}\\[2pt]
\gpfit{\wedge\; \exists \mathit{N}.\; \exists \lsnv{{\mathrm{svSrc}}}.\; \exists \lsnv{{\mathrm{svTgt}}}.\; \operatorname{inferSpace}(\leev{}) = \lsnv{{\mathrm{svSrc}}} \loprod \lsnv{{\mathrm{svTgt}}} \;\wedge\; \mathit{N} = \dim_{\mathcal{O}}(\lsnv{{\mathrm{svSrc}}}) \;\wedge\; \mathit{N} = \dim_{\mathcal{O}}(\lsnv{{\mathrm{svTgt}}}) \;\wedge\; \mathit{N} = \lvert \SemBB{\textcolor{lang}{{\lb{}_{\mathrm{src}}}}}{\sigma{}} \rvert \;\wedge\; \mathit{N} = \lvert \SemBB{\textcolor{lang}{{\lb{}_{\mathrm{tgt}}}}}{\sigma{}} \rvert}\end{center}
}{\texttt{GLang.Metatheory.cob\allowbreak Common\allowbreak Dimension\allowbreak Adequacy}}{GLang/CobEndpointPreservation.lean:140}{none}{--}
\gldeftheorem{metatheory.cobCoordinateMatrixBijectiveOfOracle}{Oracle-polymorphic COB automorphism}{Assuming
\begin{itemize}
  \item \gpfit{\mathcal{O}\colon \mathsf{FFSOracleSem}}
  \item \gpfit{\mathsf{RawApplicationOracleSpec}(\mathcal{O})}
  \item \gpfit{\evalCtxSpace{};\evalCtx{}\vdash_{\mathsf C} \textcolor{lang}{{\lb{}_{\mathrm{src}}}} \xlongleftrightarrow{\leev{}} \textcolor{lang}{{\lb{}_{\mathrm{tgt}}}}}
  \item \gpfit{\mathsf{RawTypedConsistent}(\mathcal{O}, \lDelta{}, \evalCtxSpace{}, \evalCtx{}, \valCheckCtx{}, \sigma{}, \glvEnv{})}
\end{itemize}
then
\begin{center}\gpfit{\exists \mathit{N}.\; \exists \lsnv{{\mathrm{svSrc}}}.\; \exists \lsnv{{\mathrm{svTgt}}}.\; \exists \mathit{a}.\; \exists \lsn{0}.\; \exists \lsn{1}.\; \exists \lsnv{0}.\; \exists \lsnv{1}.\; \operatorname{inferType}(\leev{}) = \lsnv{0} \loprod \lsnv{1} \;\wedge\; \evalSJudge{\evalCtx{}}{\lsn{0}}{\lsnv{0}} \;\wedge\; \evalSJudge{\evalCtx{}}{\lsn{1}}{\lsnv{1}} \;\wedge\; \vdash \operatorname{inferType}(\leev{})\colon \tvalid{} \;\wedge\; \mathsf{fv}(\leev{}) \subseteq \mathrm{dom}(\evalCtx{})}\\[2pt]
\gpfit{\wedge\; \SemBB{\textcolor{lang}{{\lb{}_{\mathrm{src}}}}}{\sigma{}} = \SemA{\leev{}}{\glvEnv{}}{} \cdot \SemBB{\textcolor{lang}{{\lb{}_{\mathrm{tgt}}}}}{\sigma{}}}\\[2pt]
\gpfit{\wedge\; \operatorname{inferSpace}(\leev{}) = \lsnv{{\mathrm{svSrc}}} \loprod \lsnv{{\mathrm{svTgt}}}}\\[2pt]
\gpfit{\wedge\; \mathit{N} = \dim_{\mathcal{O}}(\lsnv{{\mathrm{svSrc}}})}\\[2pt]
\gpfit{\wedge\; \mathit{N} = \dim_{\mathcal{O}}(\lsnv{{\mathrm{svTgt}}})}\\[2pt]
\gpfit{\wedge\; \mathit{N} = \lvert \SemBB{\textcolor{lang}{{\lb{}_{\mathrm{src}}}}}{\sigma{}} \rvert}\\[2pt]
\gpfit{\wedge\; \mathit{N} = \lvert \SemBB{\textcolor{lang}{{\lb{}_{\mathrm{tgt}}}}}{\sigma{}} \rvert}\\[2pt]
\gpfit{\wedge\; \mathcal{E}\llbracket \leev{}\rrbracket_{[];[]} = \mathit{a}}\\[2pt]
\gpfit{\wedge\; \lvert \mathit{a}\rvert = (\mathit{N}) \cdot (\mathit{N})}\\[2pt]
\gpfit{\wedge\; \mathsf{Bijective}(\mathsf{toLin'}(\mathsf{denotAsMatrix}(\mathit{N}, \mathit{N}, \mathsf{get}(\mathit{a}))))}\end{center}
}{\texttt{GLang.Metatheory.cob\allowbreak Coordinate\allowbreak Matrix\allowbreak Bijective\allowbreak Of\allowbreak Oracle}}{GLang/CobEndpointPreservation.lean:245}{none}{--}
\gldeftheorem{metatheory.cobCoordinateMatrixBijective}{COB coordinate-matrix automorphism}{Assuming
\begin{itemize}
  \item \gpfit{\evalCtxSpace{};\evalCtx{}\vdash_{\mathsf C} \textcolor{lang}{{\lb{}_{\mathrm{src}}}} \xlongleftrightarrow{\leev{}} \textcolor{lang}{{\lb{}_{\mathrm{tgt}}}}}
  \item \gpfit{(\lDelta{}, \evalCtxSpace{}); \evalCtx{}; \valCheckCtx{}\vdash \sigma{}, \glvEnv{}}
\end{itemize}
then
\begin{center}\gpfit{\exists \mathit{N}.\; \exists \lsnv{{\mathrm{svSrc}}}.\; \exists \lsnv{{\mathrm{svTgt}}}.\; \exists \mathit{a}.\; \exists \lsn{0}.\; \exists \lsn{1}.\; \exists \lsnv{0}.\; \exists \lsnv{1}.\; \operatorname{inferType}(\leev{}) = \lsnv{0} \loprod \lsnv{1} \;\wedge\; \evalSJudge{\evalCtx{}}{\lsn{0}}{\lsnv{0}} \;\wedge\; \evalSJudge{\evalCtx{}}{\lsn{1}}{\lsnv{1}} \;\wedge\; \vdash \operatorname{inferType}(\leev{})\colon \tvalid{} \;\wedge\; \mathsf{fv}(\leev{}) \subseteq \mathrm{dom}(\evalCtx{})}\\[2pt]
\gpfit{\wedge\; \SemBB{\textcolor{lang}{{\lb{}_{\mathrm{src}}}}}{\sigma{}} = \SemA{\leev{}}{\glvEnv{}}{} \cdot \SemBB{\textcolor{lang}{{\lb{}_{\mathrm{tgt}}}}}{\sigma{}}}\\[2pt]
\gpfit{\wedge\; \operatorname{inferSpace}(\leev{}) = \lsnv{{\mathrm{svSrc}}} \loprod \lsnv{{\mathrm{svTgt}}}}\\[2pt]
\gpfit{\wedge\; \mathit{N} = \dim_{\mathcal{O}}(\lsnv{{\mathrm{svSrc}}})}\\[2pt]
\gpfit{\wedge\; \mathit{N} = \dim_{\mathcal{O}}(\lsnv{{\mathrm{svTgt}}})}\\[2pt]
\gpfit{\wedge\; \mathit{N} = \lvert \SemBB{\textcolor{lang}{{\lb{}_{\mathrm{src}}}}}{\sigma{}} \rvert}\\[2pt]
\gpfit{\wedge\; \mathit{N} = \lvert \SemBB{\textcolor{lang}{{\lb{}_{\mathrm{tgt}}}}}{\sigma{}} \rvert}\\[2pt]
\gpfit{\wedge\; \mathcal{E}\llbracket \leev{}\rrbracket_{[];[]} = \mathit{a}}\\[2pt]
\gpfit{\wedge\; \lvert \mathit{a}\rvert = (\mathit{N}) \cdot (\mathit{N})}\\[2pt]
\gpfit{\wedge\; \mathsf{Bijective}(\mathsf{toLin'}(\mathsf{denotAsMatrix}(\mathit{N}, \mathit{N}, \mathsf{get}(\mathit{a}))))}\end{center}
}{\texttt{GLang.Metatheory.cob\allowbreak Coordinate\allowbreak Matrix\allowbreak Bijective}}{GLang/CobEndpointPreservation.lean:280}{none}{--}
\gldeftheorem{metatheory.checkedTypeProgress}{Checked type progress}{Assuming
\begin{itemize}
  \item \gpfit{(\lDelta{},\evalCtxSpace{});\valCheckCtx{}\vdash \evalCtx{}}
  \item \gpfit{(\lemptyset, \evalCtxSpace{})\vdash \lty{}\colon \lky{}}
\end{itemize}
then
\begin{center}\gpfit{\exists \ltyv{}.\; \evalCtxSpace{};\evalCtx{}\vdash_{\mathsf C} \lty{} \evalArrow{} \ltyv{}}\end{center}
}{\texttt{GLang.Metatheory.checked\allowbreak Type\allowbreak Progress}}{GLang/Metatheory.lean:635}{none}{lem:prog}
\gldeftheorem{metatheory.checkedTypeProgressAtBasis}{Checked type progress at a basis}{Assuming
\begin{itemize}
  \item \gpfit{(\lDelta{},\evalCtxSpace{});\valCheckCtx{}\vdash \evalCtx{}}
  \item \gpfit{(\emptyset,\evalCtxSpace{})\vdash \lb{}\colon \bvalid{}}
\end{itemize}
then
\begin{center}\gpfit{\exists \ltyv{}.\; \evalCtxSpace{};\evalCtx{}\vdash_{\mathsf C} \lb{} \evalArrow{}_{\mathsf{basis}} \ltyv{}}\end{center}
}{\texttt{GLang.Metatheory.checked\allowbreak Type\allowbreak Progress\allowbreak At\allowbreak Basis}}{GLang/Metatheory.lean:645}{none}{lem:prog}
\gldeftheorem{metatheory.checkedBasisProgress}{Checked change-of-basis progress}{Assuming
\begin{itemize}
  \item \gpfit{(\lDelta{},\evalCtxSpace{});\valCheckCtx{}\vdash \evalCtx{}}
  \item \gpfit{(\emptyset,\evalCtxSpace{})\vdash \lb{}\colon \bvalid{}}
  \item \gpfit{\tof{\lb{}} = \sof{\lb{}}}
\end{itemize}
then
\begin{center}\gpfit{\exists \leev{}.\; \evalCtxSpace{};\evalCtx{}\vdash_{\mathsf C} \lb{} \xlongleftrightarrow{\leev{}} \stdBasisDCtx{\tof{\lb{}}}}\end{center}
}{\texttt{GLang.Metatheory.checked\allowbreak Basis\allowbreak Progress}}{GLang/Metatheory.lean:655}{none}{lem:cobProg}
\gldeftheorem{metatheory.basisAdequacy}{Basis adequacy}{Assuming
\begin{itemize}
  \item \gpfit{(\emptyset,\evalCtxSpace{})\vdash \lb{}\colon \bvalid{}}
  \item \gpfit{\vdash (\lemptyset{}, \evalCtxSpace{})}
  \item \gpfit{\lemptyset{}, \evalCtxSpace{}\vdash \sigma{}}
\end{itemize}
then
\begin{center}\gpfit{\SemBB{\lb{}}{\sigma{}} \;\text{is a basis for}\; \SemT{\tof{\lb{}}}{\sigma{}}}\end{center}
}{\texttt{GLang.Metatheory.basis\allowbreak Adequacy}}{GLang/Metatheory.lean:709}{none}{lem:denoteBasisSound}
\gldeftheorem{metatheory.typeAdequacyAtBasis}{Type adequacy at a basis}{Assuming
\begin{itemize}
  \item \gpfit{\evalCtxSpace{};\evalCtx{}\vdash_{\mathsf C} \lb{} \evalArrow{}_{\mathsf{basis}} \ltyv{}}
  \item \gpfit{\lDelta{}\colon \mathsf{LCtx}}
  \item \gpfit{\sigma{}\colon \mathsf{DEnv}}
  \item \gpfit{\valCheckCtx{}\colon \mathsf{ValCtx}}
  \item \gpfit{(\lDelta{}, \evalCtxSpace{}); \evalCtx{}; \valCheckCtx{}\vdash \sigma{}, \glvEnv{}}
\end{itemize}
then
\begin{center}\gpfit{\SemT{\tof{\lb{}}}{\sigma{}} \equiv_{\SemBB{\lb{}}{\sigma{}}} \SemT{\ltyv{}}{\glvEnv{}}}\end{center}
}{\texttt{GLang.Metatheory.type\allowbreak Adequacy\allowbreak At\allowbreak Basis}}{GLang/Metatheory.lean:849}{none}{--}
\gldeftheorem{metatheory.spaceProgressAdequacy}{Checked space progress and adequacy}{Assuming
\begin{itemize}
  \item \gpfit{(\lDelta{}, \evalCtxSpace{}); \evalCtx{}; \valCheckCtx{}\vdash \sigma{}, \glvEnv{}}
  \item \gpfit{(\lemptyset, \evalCtxSpace{})\vdash \ls{}\colon\svalid}
\end{itemize}
then
\begin{center}\gpfit{\exists \lsv{}.\; \evalCtxSpace{};\evalCtx{}\vdash_{\mathsf C} \ls{} \evalArrow{} \lsv{}}\\[2pt]
\gpfit{\wedge\; \mathsf{shpOfPartial}(\mathsf{oracle}(\glvEnv{}, [], []), \lsv{}) = \mathsf{srcShp}(\sigma{}, [], \ls{})}\end{center}
}{\texttt{GLang.Metatheory.space\allowbreak Progress\allowbreak Adequacy}}{GLang/Metatheory.lean:868}{none}{thm:main}
\gldeftheorem{metatheory.typeProgressAdequacyAtBasis}{Checked type progress and adequacy at a basis}{Assuming
\begin{itemize}
  \item \gpfit{(\lDelta{}, \evalCtxSpace{}); \evalCtx{}; \valCheckCtx{}\vdash \sigma{}, \glvEnv{}}
  \item \gpfit{(\emptyset,\evalCtxSpace{})\vdash \lb{}\colon \bvalid{}}
\end{itemize}
then
\begin{center}\gpfit{\exists \ltyv{}.\; \evalCtxSpace{};\evalCtx{}\vdash_{\mathsf C} \lb{} \evalArrow{}_{\mathsf{basis}} \ltyv{}}\\[2pt]
\gpfit{\wedge\; \SemT{\tof{\lb{}}}{\sigma{}} \equiv_{\SemBB{\lb{}}{\sigma{}}} \SemT{\ltyv{}}{\glvEnv{}}}\end{center}
}{\texttt{GLang.Metatheory.type\allowbreak Progress\allowbreak Adequacy\allowbreak At\allowbreak Basis}}{GLang/Metatheory.lean:882}{none}{thm:main}
\gldeftheorem{metatheory.basisProgressAdequacy}{Checked change-of-basis progress and adequacy}{Assuming
\begin{itemize}
  \item \gpfit{(\lDelta{}, \evalCtxSpace{}); \evalCtx{}; \valCheckCtx{}\vdash \sigma{}, \glvEnv{}}
  \item \gpfit{(\emptyset,\evalCtxSpace{})\vdash \lb{}\colon \bvalid{}}
  \item \gpfit{\tof{\lb{}} = \sof{\lb{}}}
\end{itemize}
then
\begin{center}\gpfit{\exists \leev{}.\; \evalCtxSpace{};\evalCtx{}\vdash_{\mathsf C} \lb{} \xlongleftrightarrow{\leev{}} \stdBasisDCtx{\tof{\lb{}}}}\\[2pt]
\gpfit{\wedge\; \SemBB{\lb{}}{\sigma{}} = \SemA{\leev{}}{\glvEnv{}}{} \cdot \SemBB{\stdBasisDCtx{\tof{\lb{}}}}{\sigma{}}}\end{center}
}{\texttt{GLang.Metatheory.basis\allowbreak Progress\allowbreak Adequacy}}{GLang/Metatheory.lean:899}{none}{thm:main}
\gldeftheorem{rawMultipleRepresentationRegression.checkedProgressAfterExtendRep}{Multiple-representation checked-progress regression}{Assuming
\begin{itemize}
  \item \gpfit{\lDelta{}; \evalCtxSpace{}; \evalCtx{}\;\mathsf{history}}
  \item \gpfit{((\mathit{x}, \lbn{0}) \mapsto \lenv{0}) \in \evalCtx{}}
  \item \gpfit{\mathsf{SourceEntryLinked}(\lDelta{}, \evalCtxSpace{}, \mathit{x}, \lbn{1})}
  \item \gpfit{\neg((\mathit{x}, \lbn{1}) \in \evalCtx{})}
  \item \gpfit{\mathsf{EvalEntry}(\evalCtx{}, \mathit{x}, \lbn{1}, \lenv{1})}
  \item \gpfit{\mathsf{FrameAnchor}(\evalCtx{}, \lbn{1}, \lenv{1})}
  \item \gpfit{(\emptyset,\evalCtxSpace{})\vdash \lbn{1}\colon \bvalid{}}
  \item \gpfit{(\lDelta{}, \evalCtxSpace{})\vdash \lee{}\colon \tof{\textcolor{lang}{\mathit{requested}}}}
  \item \gpfit{(\emptyset,\evalCtxSpace{})\vdash \textcolor{lang}{\mathit{requested}}\colon \bvalid{}}
\end{itemize}
then
\begin{center}\gpfit{\exists \leev{}.\; \exists \ltyv{}.\; \lDelta{};\evalCtxSpace{};\evalCtx{}[(\mathit{x},\lbn{1}) \mapsto \lenv{1}]\vdash_{\mathsf C} \lee{} \xlongrightarrow{\textcolor{lang}{\mathit{requested}}} \leev{}\glresulttype{\ltyv{}}}\\[2pt]
\gpfit{\wedge\; \operatorname{inferType}(\leev{}) = \ltyv{}}\\[2pt]
\gpfit{\wedge\; \operatorname{inferSpace}(\leev{}) = \mathsf{spaceOf}(\ltyv{})}\\[2pt]
\gpfit{\wedge\; ((\mathit{x}, \lbn{0}) \mapsto \lenv{0}) \in \evalCtx{}[(\mathit{x},\lbn{1}) \mapsto \lenv{1}]}\\[2pt]
\gpfit{\wedge\; ((\mathit{x}, \lbn{1}) \mapsto \lenv{1}) \in \evalCtx{}[(\mathit{x},\lbn{1}) \mapsto \lenv{1}]}\end{center}
}{\texttt{GLang.Raw\allowbreak Multiple\allowbreak Representation\allowbreak Regression.checked\allowbreak Progress\_\allowbreak after\_\allowbreak extend\allowbreak Rep}}{GLang/Adequacy/RawMultipleRepresentationRegression.lean:16}{none}{--}
\gldeftheorem{adequacy.direct.term}{Strengthened term adequacy (oracle-free)}{Assuming
\begin{itemize}
  \item \gpfit{\lDelta{};\evalCtxSpace{};\evalCtx{}\vdash_{\mathsf S} \lee{} \xlongrightarrow{\lb{}} \leev{}\glresulttype{\ltyv{}}}
  \item \gpfit{\sigma{}\colon \mathsf{DEnv}}
  \item \gpfit{\valCheckCtx{}\colon \mathsf{ValCtx}}
  \item \gpfit{(\lDelta{}, \evalCtxSpace{}); \evalCtx{}; \valCheckCtx{}\vdash \sigma{}, \glvEnv{}}
\end{itemize}
then
\begin{center}\gpfit{\Sem{\lee{}}{\sigma{}} \equiv^{\ltyv{}}_{\SemBB{\lb{}}{\sigma{}}} \mathsf{evalVal}(\glvEnv{}, [], [], \leev{})}\\[2pt]
\gpfit{\wedge\; \SemT{\tof{\lb{}}}{\sigma{}} \equiv_{\SemBB{\lb{}}{\sigma{}}} \SemT{\ltyv{}}{\glvEnv{}}}\end{center}
}{\texttt{GLang.Adequacy.Direct.term}}{GLang/Adequacy/DirectPublic.lean:97}{none}{thm:main}
\gldeftheorem{adequacy.direct.cob}{Strengthened COB adequacy (oracle-free)}{Assuming
\begin{itemize}
  \item \gpfit{\evalCtxSpace{};\evalCtx{}\vdash_{\mathsf S} \textcolor{lang}{{\lb{}_{\mathrm{src}}}} \xlongleftrightarrow{\leev{}} \textcolor{lang}{{\lb{}_{\mathrm{tgt}}}}}
  \item \gpfit{\lDelta{}\colon \mathsf{LCtx}}
  \item \gpfit{\sigma{}\colon \mathsf{DEnv}}
  \item \gpfit{\valCheckCtx{}\colon \mathsf{ValCtx}}
  \item \gpfit{(\lDelta{}, \evalCtxSpace{}); \evalCtx{}; \valCheckCtx{}\vdash \sigma{}, \glvEnv{}}
\end{itemize}
then
\begin{center}\gpfit{\SemBB{\textcolor{lang}{{\lb{}_{\mathrm{src}}}}}{\sigma{}} = \SemA{\leev{}}{\glvEnv{}}{} \cdot \SemBB{\textcolor{lang}{{\lb{}_{\mathrm{tgt}}}}}{\sigma{}}}\end{center}
}{\texttt{GLang.Adequacy.Direct.cob}}{GLang/Adequacy/DirectPublic.lean:132}{none}{--}
\gldeftheorem{adequacy.direct.dimsAt}{Strengthened dimension adequacy (oracle-free)}{Assuming
\begin{itemize}
  \item \gpfit{\evalCtxSpace{};\evalCtx{}\vdash_{\mathsf S} \lb{} \evalArrow{}_{\mathsf{basis}} \ltyv{}}
  \item \gpfit{\lDelta{}\colon \mathsf{LCtx}}
  \item \gpfit{\sigma{}\colon \mathsf{DEnv}}
  \item \gpfit{\valCheckCtx{}\colon \mathsf{ValCtx}}
  \item \gpfit{(\lDelta{}, \evalCtxSpace{}); \evalCtx{}; \valCheckCtx{}\vdash \sigma{}, \glvEnv{}}
\end{itemize}
then
\begin{center}\gpfit{\SemT{\tof{\lb{}}}{\sigma{}} \equiv_{\SemBB{\lb{}}{\sigma{}}} \SemT{\ltyv{}}{\glvEnv{}}}\end{center}
}{\texttt{GLang.Adequacy.Direct.dims\allowbreak At}}{GLang/Adequacy/DirectPublic.lean:142}{none}{--}
\gldefthmrows{
Source-space soundness & \texttt{\footnotesize GLang.Metatheory.source\allowbreak Space\allowbreak Soundness} & none\\
Source-type soundness & \texttt{\footnotesize GLang.Metatheory.source\allowbreak Type\allowbreak Soundness} & none\\
Source-term soundness & \texttt{\footnotesize GLang.Metatheory.source\allowbreak Term\allowbreak Soundness} & none\\
Value-shape soundness & \texttt{\footnotesize GLang.Metatheory.value\allowbreak Shape\allowbreak Soundness} & none\\
Value-type soundness & \texttt{\footnotesize GLang.Metatheory.value\allowbreak Type\allowbreak Soundness} & none\\
Value-term soundness & \texttt{\footnotesize GLang.Metatheory.value\allowbreak Term\allowbreak Soundness} & none\\
Ordinary space preservation & \texttt{\footnotesize GLang.Metatheory.space\allowbreak Preservation} & none\\
Ordinary term preservation & \texttt{\footnotesize GLang.Metatheory.term\allowbreak Preservation} & none\\
Ordinary change-of-basis preservation & \texttt{\footnotesize GLang.Metatheory.cob\allowbreak Preservation} & none\\
Checked space progress & \texttt{\footnotesize GLang.Metatheory.checked\allowbreak Space\allowbreak Progress} & none\\
Checked term progress & \texttt{\footnotesize GLang.Metatheory.term\allowbreak Progress} & none\\
General checked change-of-basis progress & \texttt{\footnotesize GLang.Metatheory.checked\allowbreak Cob\allowbreak Progress} & none\\
Space adequacy & \texttt{\footnotesize GLang.Metatheory.space\allowbreak Adequacy} & none\\
Term adequacy & \texttt{\footnotesize GLang.Metatheory.term\allowbreak Adequacy} & none\\
Change-of-basis adequacy & \texttt{\footnotesize GLang.Metatheory.cob\allowbreak Adequacy} & none\\
Source-basis soundness & \texttt{\footnotesize GLang.Metatheory.source\allowbreak Basis\allowbreak Soundness} & none\\
Value-type regularity & \texttt{\footnotesize GLang.Runtime\allowbreak Value.Val\allowbreak Has\allowbreak Type.regular} & none\\
Exact inferred value type & \texttt{\footnotesize GLang.Runtime\allowbreak Value.Val\allowbreak Has\allowbreak Type.infer\allowbreak Type\_\allowbreak eq\_\allowbreak of\_\allowbreak typed} & none\\
Exact inferred value space & \texttt{\footnotesize GLang.Val\allowbreak Has\allowbreak Type.infer\allowbreak Space\_\allowbreak eq\_\allowbreak of\_\allowbreak space} & none\\
Value shape and dimension totality & \texttt{\footnotesize GLang.lem\_\allowbreak sound\allowbreak Vals\_\allowbreak metrics\_\allowbreak total} & none\\
Standard basis has the requested type & \texttt{\footnotesize GLang.Basis.tof\_\allowbreak std} & none\\
Standard basis is valid & \texttt{\footnotesize GLang.Basis.std\_\allowbreak valid} & none\\
Standard basis passes the standardness check & \texttt{\footnotesize GLang.Basis.is\allowbreak Std\_\allowbreak std} & none\\
Standard basis/frame is recursively complete and orthonormal & \texttt{\footnotesize GLang.Metatheory.standard\allowbreak Basis\allowbreak Soundness} & none\\
Progress at the standard basis & \texttt{\footnotesize GLang.V2.progress\_\allowbreak std} & none\\
Canonical inner-product term is well typed & \texttt{\footnotesize GLang.Space.iof\_\allowbreak has\allowbreak Type} & none\\
Canonical inner-product term denotes an inner product & \texttt{\footnotesize GLang.Space.iof\_\allowbreak denot\_\allowbreak is\allowbreak Inner\allowbreak Product} & none\\
Historical source-soundness projection & \texttt{\footnotesize GLang.Metatheory.source\allowbreak Soundness\allowbreak Weak} & none\\
Space regularity & \texttt{\footnotesize GLang.Metatheory.source\allowbreak Space\allowbreak Regularity} & none\\
Type regularity & \texttt{\footnotesize GLang.Metatheory.source\allowbreak Type\allowbreak Regularity} & none\\
Oracle-polymorphic canonical value soundness & \texttt{\footnotesize GLang.Metatheory.value\allowbreak Soundness\allowbreak Of\allowbreak Oracle} & none\\
Historical weak value-soundness projection & \texttt{\footnotesize GLang.Metatheory.value\allowbreak Soundness\allowbreak Weak} & none\\
Internal checked exact-index certificate & \texttt{\footnotesize GLang.Metatheory.checked\allowbreak Preservation} & none\\
Checked base-shape preservation & \texttt{\footnotesize GLang.Metatheory.checked\allowbreak Base\allowbreak Shape\allowbreak Preservation} & none\\
Checked space preservation & \texttt{\footnotesize GLang.Metatheory.checked\allowbreak Space\allowbreak Preservation} & none\\
Type preservation & \texttt{\footnotesize GLang.Metatheory.type\allowbreak Preservation} & none\\
Checked change-of-basis preservation & \texttt{\footnotesize GLang.Metatheory.checked\allowbreak Cob\allowbreak Preservation} & none\\
Endpoint-aware change-of-basis preservation & \texttt{\footnotesize GLang.Metatheory.cob\allowbreak Endpoint\allowbreak Preservation} & none\\
Oracle-polymorphic common COB dimension & \texttt{\footnotesize GLang.Metatheory.cob\allowbreak Common\allowbreak Dimension\allowbreak Adequacy\allowbreak Of\allowbreak Oracle} & none\\
Common COB dimension & \texttt{\footnotesize GLang.Metatheory.cob\allowbreak Common\allowbreak Dimension\allowbreak Adequacy} & none\\
Oracle-polymorphic COB automorphism & \texttt{\footnotesize GLang.Metatheory.cob\allowbreak Coordinate\allowbreak Matrix\allowbreak Bijective\allowbreak Of\allowbreak Oracle} & none\\
COB coordinate-matrix automorphism & \texttt{\footnotesize GLang.Metatheory.cob\allowbreak Coordinate\allowbreak Matrix\allowbreak Bijective} & none\\
Checked type progress & \texttt{\footnotesize GLang.Metatheory.checked\allowbreak Type\allowbreak Progress} & none\\
Checked type progress at a basis & \texttt{\footnotesize GLang.Metatheory.checked\allowbreak Type\allowbreak Progress\allowbreak At\allowbreak Basis} & none\\
Checked change-of-basis progress & \texttt{\footnotesize GLang.Metatheory.checked\allowbreak Basis\allowbreak Progress} & none\\
Basis adequacy & \texttt{\footnotesize GLang.Metatheory.basis\allowbreak Adequacy} & none\\
Type adequacy at a basis & \texttt{\footnotesize GLang.Metatheory.type\allowbreak Adequacy\allowbreak At\allowbreak Basis} & none\\
Checked space progress and adequacy & \texttt{\footnotesize GLang.Metatheory.space\allowbreak Progress\allowbreak Adequacy} & none\\
Checked type progress and adequacy at a basis & \texttt{\footnotesize GLang.Metatheory.type\allowbreak Progress\allowbreak Adequacy\allowbreak At\allowbreak Basis} & none\\
Checked change-of-basis progress and adequacy & \texttt{\footnotesize GLang.Metatheory.basis\allowbreak Progress\allowbreak Adequacy} & none\\
Multiple-representation checked-progress regression & \texttt{\footnotesize GLang.Raw\allowbreak Multiple\allowbreak Representation\allowbreak Regression.checked\allowbreak Progress\_\allowbreak after\_\allowbreak extend\allowbreak Rep} & none\\
Strengthened term adequacy (oracle-free) & \texttt{\footnotesize GLang.Adequacy.Direct.term} & none\\
Strengthened COB adequacy (oracle-free) & \texttt{\footnotesize GLang.Adequacy.Direct.cob} & none\\
Strengthened dimension adequacy (oracle-free) & \texttt{\footnotesize GLang.Adequacy.Direct.dims\allowbreak At} & none\\
}{
  \item \textbf{Source-space soundness} (\texttt{\footnotesize GLang.Metatheory.source\allowbreak Space\allowbreak Soundness}, \texttt{\footnotesize GLang/Metatheory.lean:111}).
  \item \textbf{Source-type soundness} (\texttt{\footnotesize GLang.Metatheory.source\allowbreak Type\allowbreak Soundness}, \texttt{\footnotesize GLang/Metatheory.lean:137}).
  \item \textbf{Source-term soundness} (\texttt{\footnotesize GLang.Metatheory.source\allowbreak Term\allowbreak Soundness}, \texttt{\footnotesize GLang/Metatheory.lean:88}).
  \item \textbf{Value-shape soundness} (\texttt{\footnotesize GLang.Metatheory.value\allowbreak Shape\allowbreak Soundness}, \texttt{\footnotesize GLang/Metatheory.lean:196}).
  \item \textbf{Value-type soundness} (\texttt{\footnotesize GLang.Metatheory.value\allowbreak Type\allowbreak Soundness}, \texttt{\footnotesize GLang/Metatheory.lean:204}).
  \item \textbf{Value-term soundness} (\texttt{\footnotesize GLang.Metatheory.value\allowbreak Term\allowbreak Soundness}, \texttt{\footnotesize GLang/Metatheory.lean:284}).
  \item \textbf{Ordinary space preservation} (\texttt{\footnotesize GLang.Metatheory.space\allowbreak Preservation}, \texttt{\footnotesize GLang/Metatheory.lean:474}).
  \item \textbf{Ordinary term preservation} (\texttt{\footnotesize GLang.Metatheory.term\allowbreak Preservation}, \texttt{\footnotesize GLang/Metatheory.lean:428}).
  \item \textbf{Ordinary change-of-basis preservation} (\texttt{\footnotesize GLang.Metatheory.cob\allowbreak Preservation}, \texttt{\footnotesize GLang/Metatheory.lean:519}).
  \item \textbf{Checked space progress} (\texttt{\footnotesize GLang.Metatheory.checked\allowbreak Space\allowbreak Progress}, \texttt{\footnotesize GLang/Metatheory.lean:626}).
  \item \textbf{Checked term progress} (\texttt{\footnotesize GLang.Metatheory.term\allowbreak Progress}, \texttt{\footnotesize GLang/Metatheory.lean:568}).
  \item \textbf{General checked change-of-basis progress} (\texttt{\footnotesize GLang.Metatheory.checked\allowbreak Cob\allowbreak Progress}, \texttt{\footnotesize GLang/Metatheory.lean:665}).
  \item \textbf{Space adequacy} (\texttt{\footnotesize GLang.Metatheory.space\allowbreak Adequacy}, \texttt{\footnotesize GLang/Metatheory.lean:698}).
  \item \textbf{Term adequacy} (\texttt{\footnotesize GLang.Metatheory.term\allowbreak Adequacy}, \texttt{\footnotesize GLang/Metatheory.lean:682}).
  \item \textbf{Change-of-basis adequacy} (\texttt{\footnotesize GLang.Metatheory.cob\allowbreak Adequacy}, \texttt{\footnotesize GLang/Metatheory.lean:826}).
  \item \textbf{Source-basis soundness} (\texttt{\footnotesize GLang.Metatheory.source\allowbreak Basis\allowbreak Soundness}, \texttt{\footnotesize GLang/Metatheory.lean:150}).
  \item \textbf{Value-type regularity} (\texttt{\footnotesize GLang.Runtime\allowbreak Value.Val\allowbreak Has\allowbreak Type.regular}, \texttt{\footnotesize GLang/Runtime\allowbreak Value\allowbreak Regular.lean:23}).
  \item \textbf{Exact inferred value type} (\texttt{\footnotesize GLang.Runtime\allowbreak Value.Val\allowbreak Has\allowbreak Type.infer\allowbreak Type\_\allowbreak eq\_\allowbreak of\_\allowbreak typed}, \texttt{\footnotesize GLang/Runtime\allowbreak Value\allowbreak Infer.lean:37}).
  \item \textbf{Exact inferred value space} (\texttt{\footnotesize GLang.Val\allowbreak Has\allowbreak Type.infer\allowbreak Space\_\allowbreak eq\_\allowbreak of\_\allowbreak space}, \texttt{\footnotesize GLang/Value\allowbreak Infer\allowbreak Space\allowbreak Sound.lean:356}).
  \item \textbf{Value shape and dimension totality} (\texttt{\footnotesize GLang.lem\_\allowbreak sound\allowbreak Vals\_\allowbreak metrics\_\allowbreak total}, \texttt{\footnotesize GLang/Value\allowbreak Sound\allowbreak Vals.lean:55}).
  \item \textbf{Standard basis has the requested type} (\texttt{\footnotesize GLang.Basis.tof\_\allowbreak std}, \texttt{\footnotesize GLang/Basis.lean:579}).
  \item \textbf{Standard basis is valid} (\texttt{\footnotesize GLang.Basis.std\_\allowbreak valid}, \texttt{\footnotesize GLang/Basis.lean:688}).
  \item \textbf{Standard basis passes the standardness check} (\texttt{\footnotesize GLang.Basis.is\allowbreak Std\_\allowbreak std}, \texttt{\footnotesize GLang/Progress\allowbreak Std\allowbreak Basis.lean:56}).
  \item \textbf{Standard basis/frame is recursively complete and orthonormal} (\texttt{\footnotesize GLang.Metatheory.standard\allowbreak Basis\allowbreak Soundness}, \texttt{\footnotesize GLang/Metatheory.lean:161}).
  \item \textbf{Progress at the standard basis} (\texttt{\footnotesize GLang.V2.progress\_\allowbreak std}, \texttt{\footnotesize GLang/Progress\allowbreak Master.lean:65}).
  \item \textbf{Canonical inner-product term is well typed} (\texttt{\footnotesize GLang.Space.iof\_\allowbreak has\allowbreak Type}, \texttt{\footnotesize GLang/Inner\allowbreak Product.lean:1523}).
  \item \textbf{Canonical inner-product term denotes an inner product} (\texttt{\footnotesize GLang.Space.iof\_\allowbreak denot\_\allowbreak is\allowbreak Inner\allowbreak Product}, \texttt{\footnotesize GLang/Inner\allowbreak Product.lean:6809}).
  \item \textbf{Historical source-soundness projection} (\texttt{\footnotesize GLang.Metatheory.source\allowbreak Soundness\allowbreak Weak}, \texttt{\footnotesize GLang/Metatheory.lean:78}).
  \item \textbf{Space regularity} (\texttt{\footnotesize GLang.Metatheory.source\allowbreak Space\allowbreak Regularity}, \texttt{\footnotesize GLang/Metatheory.lean:98}).
  \item \textbf{Type regularity} (\texttt{\footnotesize GLang.Metatheory.source\allowbreak Type\allowbreak Regularity}, \texttt{\footnotesize GLang/Metatheory.lean:124}).
  \item \textbf{Oracle-polymorphic canonical value soundness} (\texttt{\footnotesize GLang.Metatheory.value\allowbreak Soundness\allowbreak Of\allowbreak Oracle}, \texttt{\footnotesize GLang/Metatheory.lean:215}).
  \item \textbf{Historical weak value-soundness projection} (\texttt{\footnotesize GLang.Metatheory.value\allowbreak Soundness\allowbreak Weak}, \texttt{\footnotesize GLang/Metatheory.lean:254}).
  \item \textbf{Internal checked exact-index certificate} (\texttt{\footnotesize GLang.Metatheory.checked\allowbreak Preservation}, \texttt{\footnotesize GLang/Metatheory.lean:312}).
  \item \textbf{Checked base-shape preservation} (\texttt{\footnotesize GLang.Metatheory.checked\allowbreak Base\allowbreak Shape\allowbreak Preservation}, \texttt{\footnotesize GLang/Metatheory.lean:444}).
  \item \textbf{Checked space preservation} (\texttt{\footnotesize GLang.Metatheory.checked\allowbreak Space\allowbreak Preservation}, \texttt{\footnotesize GLang/Metatheory.lean:489}).
  \item \textbf{Type preservation} (\texttt{\footnotesize GLang.Metatheory.type\allowbreak Preservation}, \texttt{\footnotesize GLang/Metatheory.lean:504}).
  \item \textbf{Checked change-of-basis preservation} (\texttt{\footnotesize GLang.Metatheory.checked\allowbreak Cob\allowbreak Preservation}, \texttt{\footnotesize GLang/Metatheory.lean:535}).
  \item \textbf{Endpoint-aware change-of-basis preservation} (\texttt{\footnotesize GLang.Metatheory.cob\allowbreak Endpoint\allowbreak Preservation}, \texttt{\footnotesize GLang/Cob\allowbreak Endpoint\allowbreak Preservation.lean:69}).
  \item \textbf{Oracle-polymorphic common COB dimension} (\texttt{\footnotesize GLang.Metatheory.cob\allowbreak Common\allowbreak Dimension\allowbreak Adequacy\allowbreak Of\allowbreak Oracle}, \texttt{\footnotesize GLang/Cob\allowbreak Endpoint\allowbreak Preservation.lean:98}).
  \item \textbf{Common COB dimension} (\texttt{\footnotesize GLang.Metatheory.cob\allowbreak Common\allowbreak Dimension\allowbreak Adequacy}, \texttt{\footnotesize GLang/Cob\allowbreak Endpoint\allowbreak Preservation.lean:140}).
  \item \textbf{Oracle-polymorphic COB automorphism} (\texttt{\footnotesize GLang.Metatheory.cob\allowbreak Coordinate\allowbreak Matrix\allowbreak Bijective\allowbreak Of\allowbreak Oracle}, \texttt{\footnotesize GLang/Cob\allowbreak Endpoint\allowbreak Preservation.lean:245}).
  \item \textbf{COB coordinate-matrix automorphism} (\texttt{\footnotesize GLang.Metatheory.cob\allowbreak Coordinate\allowbreak Matrix\allowbreak Bijective}, \texttt{\footnotesize GLang/Cob\allowbreak Endpoint\allowbreak Preservation.lean:280}).
  \item \textbf{Checked type progress} (\texttt{\footnotesize GLang.Metatheory.checked\allowbreak Type\allowbreak Progress}, \texttt{\footnotesize GLang/Metatheory.lean:635}).
  \item \textbf{Checked type progress at a basis} (\texttt{\footnotesize GLang.Metatheory.checked\allowbreak Type\allowbreak Progress\allowbreak At\allowbreak Basis}, \texttt{\footnotesize GLang/Metatheory.lean:645}).
  \item \textbf{Checked change-of-basis progress} (\texttt{\footnotesize GLang.Metatheory.checked\allowbreak Basis\allowbreak Progress}, \texttt{\footnotesize GLang/Metatheory.lean:655}).
  \item \textbf{Basis adequacy} (\texttt{\footnotesize GLang.Metatheory.basis\allowbreak Adequacy}, \texttt{\footnotesize GLang/Metatheory.lean:709}).
  \item \textbf{Type adequacy at a basis} (\texttt{\footnotesize GLang.Metatheory.type\allowbreak Adequacy\allowbreak At\allowbreak Basis}, \texttt{\footnotesize GLang/Metatheory.lean:849}).
  \item \textbf{Checked space progress and adequacy} (\texttt{\footnotesize GLang.Metatheory.space\allowbreak Progress\allowbreak Adequacy}, \texttt{\footnotesize GLang/Metatheory.lean:868}).
  \item \textbf{Checked type progress and adequacy at a basis} (\texttt{\footnotesize GLang.Metatheory.type\allowbreak Progress\allowbreak Adequacy\allowbreak At\allowbreak Basis}, \texttt{\footnotesize GLang/Metatheory.lean:882}).
  \item \textbf{Checked change-of-basis progress and adequacy} (\texttt{\footnotesize GLang.Metatheory.basis\allowbreak Progress\allowbreak Adequacy}, \texttt{\footnotesize GLang/Metatheory.lean:899}).
  \item \textbf{Multiple-representation checked-progress regression} (\texttt{\footnotesize GLang.Raw\allowbreak Multiple\allowbreak Representation\allowbreak Regression.checked\allowbreak Progress\_\allowbreak after\_\allowbreak extend\allowbreak Rep}, \texttt{\footnotesize GLang/Adequacy/Raw\allowbreak Multiple\allowbreak Representation\allowbreak Regression.lean:16}).
  \item \textbf{Strengthened term adequacy (oracle-free)} (\texttt{\footnotesize GLang.Adequacy.Direct.term}, \texttt{\footnotesize GLang/Adequacy/Direct\allowbreak Public.lean:97}).
  \item \textbf{Strengthened COB adequacy (oracle-free)} (\texttt{\footnotesize GLang.Adequacy.Direct.cob}, \texttt{\footnotesize GLang/Adequacy/Direct\allowbreak Public.lean:132}).
  \item \textbf{Strengthened dimension adequacy (oracle-free)} (\texttt{\footnotesize GLang.Adequacy.Direct.dims\allowbreak At}, \texttt{\footnotesize GLang/Adequacy/Direct\allowbreak Public.lean:142}).
}
\gldefbundle{soundness}{Source denotational soundness}{The four source-language soundness faces are shown together.  Terms package result-type formation and deep denotational shape.  Their linearity predicate descends pointwise through dependent-function results; the current typing rules discharge that case by proving a function-valued result retains no linear source variables.  Valid spaces and types are closed, scoped, and invariant under changes outside their contexts; and a valid basis denotes a genuine basis or dependent frame.

\begin{enumerate}
  \item \emph{Terms}.\ If $(\lDelta{}, \lepsilon{})\vdash \lee{}\colon \lty{}$ and $\vdash (\lDelta{}, \lepsilon{})$ and $\lDelta{}, \lepsilon{}\vdash \sigma{}$, then
\begin{center}\gpfit{\mathsf{StrongSourceSound}(\sigma{}; \lDelta{}; \lepsilon{}; \lee{}; \lty{})}\end{center}

  \item \emph{Spaces}.\ If $(\lemptyset, \evalCtxSpace{})\vdash \ls{}\colon\svalid$, then
\begin{center}\gpfit{\mathsf{lc}(\ls{})}\\[2pt]
\gpfit{\wedge\; \mathsf{fv}(\ls{}) \subseteq \mathrm{dom}(\evalCtxSpace{})}\\[2pt]
\gpfit{\wedge\; \forall {\sigma{}_{1}},\,{\sigma{}_{2}}.\; {\sigma{}_{1}} =_{\mathrm{dom}(\evalCtxSpace{})} {\sigma{}_{2}} \Rightarrow \SemT{\ls{}}{{\sigma{}_{1}}} = \SemT{\ls{}}{{\sigma{}_{2}}}}\end{center}

  \item \emph{Types}.\ If $(\lemptyset, \evalCtxSpace{})\vdash \lty{}\colon \lky{}$, then
\begin{center}\gpfit{\mathsf{lc}(\lty{})}\\[2pt]
\gpfit{\wedge\; \mathsf{fv}(\lty{}) \subseteq \mathrm{dom}(\evalCtxSpace{})}\\[2pt]
\gpfit{\wedge\; \forall {\sigma{}_{1}},\,{\sigma{}_{2}}.\; {\sigma{}_{1}} =_{\mathrm{dom}(\evalCtxSpace{})} {\sigma{}_{2}} \Rightarrow \SemT{\lty{}}{{\sigma{}_{1}}} = \SemT{\lty{}}{{\sigma{}_{2}}}}\end{center}

  \item \emph{Bases}.\ If $(\emptyset,\evalCtxSpace{})\vdash \lb{}\colon \bvalid{}$ and $\vdash (\lemptyset{}, \evalCtxSpace{})$ and $\lemptyset{}, \evalCtxSpace{}\vdash \sigma{}$, then
\begin{center}\gpfit{\SemBB{\lb{}}{\sigma{}} \;\text{is a basis for}\; \SemT{\tof{\lb{}}}{\sigma{}}}\end{center}

\end{enumerate}
}{\texttt{\footnotesize GLang.Metatheory.source\allowbreak Term\allowbreak Soundness} at \texttt{\footnotesize GLang/Metatheory.lean:88}; \texttt{\footnotesize GLang.Metatheory.source\allowbreak Space\allowbreak Soundness} at \texttt{\footnotesize GLang/Metatheory.lean:111}; \texttt{\footnotesize GLang.Metatheory.source\allowbreak Type\allowbreak Soundness} at \texttt{\footnotesize GLang/Metatheory.lean:137}; \texttt{\footnotesize GLang.Metatheory.source\allowbreak Basis\allowbreak Soundness} at \texttt{\footnotesize GLang/Metatheory.lean:150}}{none}{thm:sound}
\gldefbundle{valueSoundness}{Value denotational soundness}{These are the paper's three value-soundness claims.  Typing produces validity of the result type when the value context is valid.  Shape and degree of freedom are total functions in Lean, so their well-definedness needs no environment premise.  Finally, checked denotation may report a dimension error, but every successful result has exact canonical array storage at the static type, recursively through dependent callbacks.

\begin{enumerate}
  \item \emph{Value-type regularity}.\ If $\valCheckCtx{}\vdash \leev{}\colon \ltyv{}$ and $\mathit{a}\colon \mathsf{ValCtxValid}(\valCheckCtx{})$, then
\begin{center}\gpfit{\Sigma \lky{}.\; \valCheckCtx{}\vdash \ltyv{}\colon \lky{}}\end{center}

  \item \emph{Shape/dimension totality}.\ If $\mathcal{O}\colon \mathsf{FFSOracle}$ and $\lsv{}\colon \mathsf{ValSpace}$, then
\begin{center}\gpfit{\exists \mathit{d}.\; \exists \mathit{shp}.\; \dim_{\mathcal{O}}(\lsv{}) = \mathit{d}}\\[2pt]
\gpfit{\wedge\; \shpOf{\lsv{}}{\mathcal{O}} = \mathit{shp}}\end{center}

  \item \emph{Value denotation}.\ If $\glvEnv{}\colon \mathsf{ValEnvSem}$ and $\mathsf{PartialEnvWF}(\mathsf{oracle}, \valCheckCtx{}, \glvEnv{})$ and $\valCheckCtx{}\vdash \leev{}\colon \ltyv{}$, then
\begin{center}\gpfit{\mathsf{PartialCoherent}(\ltyv{}, \mathsf{denotValTyPartial}(\mathsf{oracle}, \glvEnv{}, [], [], \ltyv{}), \mathsf{denotValTyPartial}(\mathsf{oracle}, \glvEnv{}, [], [], \ltyv{}))}\\[2pt]
\gpfit{\wedge\; \mathsf{PartialRel}(\mathsf{WorldRuntimeRelated}(\mathsf{oracle}, \ltyv{}, \mathsf{mk}(\glvEnv{}, [], []), \mathsf{mk}(\glvEnv{}, [], [])), \mathsf{denotValB\_partial}(\mathsf{oracle}, \glvEnv{}, [], [], \leev{}), \mathsf{denotValB\_partial}(\mathsf{oracle}, \glvEnv{}, [], [], \leev{}))}\end{center}

\end{enumerate}
}{\texttt{\footnotesize GLang.Runtime\allowbreak Value.Val\allowbreak Has\allowbreak Type.regular} at \texttt{\footnotesize GLang/Runtime\allowbreak Value\allowbreak Regular.lean:23}; \texttt{\footnotesize GLang.lem\_\allowbreak sound\allowbreak Vals\_\allowbreak metrics\_\allowbreak total} at \texttt{\footnotesize GLang/Value\allowbreak Sound\allowbreak Vals.lean:55}; \texttt{\footnotesize GLang.Metatheory.value\allowbreak Soundness} at \texttt{\footnotesize GLang/Metatheory.lean:266}}{none}{lem:soundVals}
\gldefbundle{standardBasis}{Standard-basis soundness}{The standard basis is syntax-directed, has exactly the requested type, is valid, and passes the standardness recognizer.  Its denotation is recursively complete and orthonormal: source-basis alignment checks the requested type and completeness at the domain and every dependent codomain fiber, while recursive orthonormality checks the Gram identity at every vector leaf, each frame domain, and every semantic frame fiber.  Lean does not currently export the paper's converse claim that every valid basis passing the recognizer is syntactically equal to the canonical standard basis, so that claim is deliberately not stated here.

\begin{enumerate}
  \item \emph{Type}.\ If $\lty{}\colon \mathsf{Ty}$, then
\begin{center}\gpfit{\tof{\stdBasisDCtx{\lty{}}} = \lty{}}\end{center}

  \item \emph{Validity}.\ If $\lty{}\colon \mathsf{Ty}$ and $(\lemptyset, \lepsilon{})\vdash \lty{}\colon \lky{}$, then
\begin{center}\gpfit{(\emptyset,\lepsilon{})\vdash \stdBasisDCtx{\lty{}}\colon \bvalid{}}\end{center}

  \item \emph{Recognizer}.\ If $\lty{}\colon \mathsf{Ty}$, then
\begin{center}\gpfit{\isStdBasis{\stdBasisDCtx{\lty{}}}}\end{center}

  \item \emph{Recursive semantics}.\ If $(\lemptyset, \evalCtxSpace{})\vdash \lty{}\colon \lky{}$ and $\vdash (\lemptyset{}, \evalCtxSpace{})$ and $\lemptyset{}, \evalCtxSpace{}\vdash \sigma{}$, then
\begin{center}\gpfit{\SemBB{\stdBasisDCtx{\lty{}}}{\sigma{}} \;\text{is a basis for}\; \SemT{\lty{}}{\sigma{}}}\\[2pt]
\gpfit{\wedge\; \SemBB{\stdBasisDCtx{\lty{}}}{\sigma{}} \;\text{is recursively orthonormal}}\end{center}

\end{enumerate}
}{\texttt{\footnotesize GLang.Basis.tof\_\allowbreak std} at \texttt{\footnotesize GLang/Basis.lean:579}; \texttt{\footnotesize GLang.Basis.std\_\allowbreak valid} at \texttt{\footnotesize GLang/Basis.lean:688}; \texttt{\footnotesize GLang.Basis.is\allowbreak Std\_\allowbreak std} at \texttt{\footnotesize GLang/Progress\allowbreak Std\allowbreak Basis.lean:56}; \texttt{\footnotesize GLang.Metatheory.standard\allowbreak Basis\allowbreak Soundness} at \texttt{\footnotesize GLang/Metatheory.lean:161}}{none}{lem:stdBasisOrthonormal}
\gldefbundle{standardBasisProgress}{Progress at the standard basis}{This is the paper's standard-basis specialization of progress.  It uses the older decomposed store hypotheses; the canonical public progress theorem above instead derives a checked trace from context history, typing, and validity of the requested basis.

Assuming
\begin{itemize}
  \item \gpfit{\mathsf{CtxWF'}(\lDelta{}, \evalCtxSpace{})}
  \item \gpfit{\vdash \evalCtx{}}
  \item \gpfit{\mathsf{EvalCtxLc}(\evalCtx{})}
  \item \gpfit{\mathsf{EvalCtxLamAnchored}(\evalCtx{})}
  \item \gpfit{\mathsf{TypingImplyEval}(\lDelta{}, \evalCtxSpace{}, \evalCtx{})}
  \item \gpfit{\mathsf{StdComplete}(\evalCtx{})}
  \item \gpfit{(\lDelta{}, \evalCtxSpace{})\vdash \lee{}\colon \lty{}}
\end{itemize}
then:
\begin{enumerate}
  \item
\begin{center}\gpfit{\exists \leev{}.\; \evalJudge{\evalCtx{}}{\lee{}}{\stdBasisDCtx{\lty{}}}{\leev{}}}\end{center}

\end{enumerate}
}{\texttt{\footnotesize GLang.V2.progress\_\allowbreak std} at \texttt{\footnotesize GLang/Progress\allowbreak Master.lean:65}}{none}{thm:progA}
\gldefbundle{innerProduct}{Canonical inner-product soundness}{The paper's inner-product theorem has a typing half and a semantic half.  Lean exposes freshness of the reserved names used inside the derived iof term; this is alpha-renaming bookkeeping, not a restriction on the mathematical space.  Under a well-formed matching environment, the denoted bilinear form is an actual symmetric positive-definite inner product.

Assuming
\begin{itemize}
  \item \gpfit{\mathsf{iofFresh}(\lepsilon{})}
  \item \gpfit{(\lemptyset, \lepsilon{})\vdash \ls{}\colon\svalid}
\end{itemize}
then:
\begin{enumerate}
  \item \emph{Typing}.\ If $\ls{}\colon \mathsf{Space}$, then
\begin{center}\gpfit{(\lemptyset{}, \lepsilon{})\vdash \mathsf{iof}(\ls{})\colon \ls{}^{\lstar} \loprod \ls{}^{\lstar}}\end{center}

  \item \emph{Semantics}.\ If $\lemptyset{}, \lepsilon{}\vdash \sigma{}$ and $\mathsf{Scoped}(\lemptyset{}, \lepsilon{})$, then
\begin{center}\gpfit{\Sem{\mathsf{iof}(\ls{})}{\sigma{}} \;\text{is an inner product}}\end{center}

\end{enumerate}
}{\texttt{\footnotesize GLang.Space.iof\_\allowbreak has\allowbreak Type} at \texttt{\footnotesize GLang/Inner\allowbreak Product.lean:1523}; \texttt{\footnotesize GLang.Space.iof\_\allowbreak denot\_\allowbreak is\allowbreak Inner\allowbreak Product} at \texttt{\footnotesize GLang/Inner\allowbreak Product.lean:6809}}{none}{lem:innerLemap}
\gldefbundle{preservation}{Preservation and representation regularity}{The public term theorem follows the caller-supplied unchecked evaluation and proves regularity of that literal result.  Checked exact-index preservation remains supporting machinery and is not another public preservation theorem.  With semantic consistency, a checked base-space result also has exactly the flat-array shape selected by its retained result space.  Only this last shape case crosses the constructive-oracle boundary and inherits the seven named numerical axioms; the unchecked, space, type, and change-of-basis cases do not.

\begin{enumerate}
  \item \emph{Terms (unchecked evaluation)}.\ If $(\lDelta{},\evalCtxSpace{});\valCheckCtx{}\vdash \evalCtx{}$ and $(\lDelta{}, \evalCtxSpace{})\vdash \lee{}\colon \tof{\lb{}}$ and $(\emptyset,\evalCtxSpace{})\vdash \lb{}\colon \bvalid{}$ and $\evalJudge{\evalCtx{}}{\lee{}}{\lb{}}{\leev{}}$, then
\begin{center}\gpfit{\exists \ltyv{}.\; \evalSJudge{\evalCtx{}}{\tof{\lb{}}}{\ltyv{}}}\\[2pt]
\gpfit{\wedge\; \mathsf{kind}(\operatorname{inferType}(\leev{})) = \mathsf{kind}(\ltyv{})}\\[2pt]
\gpfit{\wedge\; \vdash \operatorname{inferType}(\leev{})\colon \tvalid{}}\\[2pt]
\gpfit{\wedge\; \mathsf{fv}(\operatorname{inferType}(\leev{})) \cup \mathsf{fv}(\leev{}) \subseteq \mathrm{dom}(\evalCtx{})}\\[2pt]
\gpfit{\wedge\; \valCheckCtx{}\vdash \leev{}\colon \operatorname{inferType}(\leev{})}\end{center}

  \item \emph{Checked base-space results (shape)}.\ If $\lDelta{};\evalCtxSpace{};\evalCtx{}\vdash_{\mathsf C} \lee{} \xlongrightarrow{\lb{}} \leev{}\glresulttype{\lsv{}}$ and $(\lDelta{}, \evalCtxSpace{}); \evalCtx{}; \valCheckCtx{}\vdash \sigma{}, \glvEnv{}$, then
\begin{center}\gpfit{\mathsf{shpOfPartial}(\mathsf{oracle}(\glvEnv{}, [], []), \operatorname{inferSpace}(\leev{})) = \mathsf{shpOfPartial}(\mathsf{oracle}(\glvEnv{}, [], []), \lsv{})}\end{center}

  \item \emph{Spaces}.\ If $(\lDelta{},\evalCtxSpace{});\valCheckCtx{}\vdash \evalCtx{}$ and $(\lemptyset, \evalCtxSpace{})\vdash \ls{}\colon\svalid$ and $\evalSJudge{\evalCtx{}}{\ls{}}{\lsv{}}$, then
\begin{center}\gpfit{\vdash \lsv{}\colon \svalid{}}\\[2pt]
\gpfit{\wedge\; \mathsf{fv}(\lsv{}) \subseteq \mathrm{dom}(\evalCtx{})}\\[2pt]
\gpfit{\wedge\; \valCheckCtx{}\vdash \lsv{}\colon\svalid}\end{center}

  \item \emph{Types}.\ If $\vdash \evalCtx{}$ and $\mathrm{dom}(\evalCtx{}) \subseteq \mathcal{S}$ and $\evalCtxSpace{};\evalCtx{}\vdash_{\mathsf C} \lty{} \evalArrow{} \ltyv{}$, then
\begin{center}\gpfit{\vdash \ltyv{}\colon \tvalid{}}\\[2pt]
\gpfit{\wedge\; \mathsf{fv}(\ltyv{}) \subseteq \mathcal{S}}\end{center}

  \item \emph{Change of basis}.\ If $(\lDelta{},\evalCtxSpace{});\valCheckCtx{}\vdash \evalCtx{}$ and $(\emptyset,\evalCtxSpace{})\vdash \textcolor{lang}{{\lb{}_{\mathrm{src}}}}\colon \bvalid{}$ and $\evalBBJudge{\evalCtx{}}{\textcolor{lang}{{\lb{}_{\mathrm{src}}}}}{\leev{}}{\textcolor{lang}{{\lb{}_{\mathrm{tgt}}}}}$, then
\begin{center}\gpfit{\exists \lsn{0}.\; \exists \lsn{1}.\; \exists \lsnv{0}.\; \exists \lsnv{1}.\; \operatorname{inferType}(\leev{}) = \lsnv{0} \loprod \lsnv{1} \;\wedge\; \evalSJudge{\evalCtx{}}{\lsn{0}}{\lsnv{0}} \;\wedge\; \evalSJudge{\evalCtx{}}{\lsn{1}}{\lsnv{1}} \;\wedge\; \vdash \operatorname{inferType}(\leev{})\colon \tvalid{} \;\wedge\; \mathsf{fv}(\leev{}) \subseteq \mathrm{dom}(\evalCtx{})}\\[2pt]
\gpfit{\wedge\; \valCheckCtx{}\vdash \leev{}\colon \operatorname{inferType}(\leev{})}\end{center}

\end{enumerate}
}{\texttt{\footnotesize GLang.Metatheory.preservation} at \texttt{\footnotesize GLang/Metatheory.lean:397}; \texttt{\footnotesize GLang.Metatheory.checked\allowbreak Base\allowbreak Shape\allowbreak Preservation} at \texttt{\footnotesize GLang/Metatheory.lean:444}; \texttt{\footnotesize GLang.Metatheory.space\allowbreak Preservation} at \texttt{\footnotesize GLang/Metatheory.lean:474}; \texttt{\footnotesize GLang.Metatheory.type\allowbreak Preservation} at \texttt{\footnotesize GLang/Metatheory.lean:504}; \texttt{\footnotesize GLang.Metatheory.cob\allowbreak Preservation} at \texttt{\footnotesize GLang/Metatheory.lean:519}}{none}{thm:preserve}
\gldefbundle{progress}{Checked progress}{A well-typed source phrase evaluates in the canonical checked semantics.  For terms, the witness carries its exact realized type and both syntactic inferers recover that type and its space; erasing the checked derivation gives the ordinary operational result.

Assuming
\begin{itemize}
  \item \gpfit{(\lDelta{},\evalCtxSpace{});\valCheckCtx{}\vdash \evalCtx{}}
\end{itemize}
then:
\begin{enumerate}
  \item \emph{Terms}.\ If $(\lDelta{}, \evalCtxSpace{})\vdash \lee{}\colon \tof{\lb{}}$ and $(\emptyset,\evalCtxSpace{})\vdash \lb{}\colon \bvalid{}$, then
\begin{center}\gpfit{\exists \leev{}.\; \exists \ltyv{}.\; \lDelta{};\evalCtxSpace{};\evalCtx{}\vdash_{\mathsf C} \lee{} \xlongrightarrow{\lb{}} \leev{}\glresulttype{\ltyv{}}}\\[2pt]
\gpfit{\wedge\; \operatorname{inferType}(\leev{}) = \ltyv{}}\\[2pt]
\gpfit{\wedge\; \operatorname{inferSpace}(\leev{}) = \mathsf{spaceOf}(\ltyv{})}\end{center}

  \item \emph{Spaces}.\ If $(\lemptyset, \evalCtxSpace{})\vdash \ls{}\colon\svalid$, then
\begin{center}\gpfit{\exists \lsv{}.\; \evalCtxSpace{};\evalCtx{}\vdash_{\mathsf C} \ls{} \evalArrow{} \lsv{}}\end{center}

  \item \emph{Types}.\ If $(\lemptyset, \evalCtxSpace{})\vdash \lty{}\colon \lky{}$, then
\begin{center}\gpfit{\exists \ltyv{}.\; \evalCtxSpace{};\evalCtx{}\vdash_{\mathsf C} \lty{} \evalArrow{} \ltyv{}}\end{center}

  \item \emph{Types at a requested basis}.\ If $(\emptyset,\evalCtxSpace{})\vdash \lb{}\colon \bvalid{}$, then
\begin{center}\gpfit{\exists \ltyv{}.\; \evalCtxSpace{};\evalCtx{}\vdash_{\mathsf C} \lb{} \evalArrow{}_{\mathsf{basis}} \ltyv{}}\end{center}

  \item \emph{Change of basis}.\ If $(\emptyset,\evalCtxSpace{})\vdash \lb{}\colon \bvalid{}$ and $\tof{\lb{}} = \sof{\lb{}}$, then
\begin{center}\gpfit{\exists \leev{}.\; \evalCtxSpace{};\evalCtx{}\vdash_{\mathsf C} \lb{} \xlongleftrightarrow{\leev{}} \stdBasisDCtx{\tof{\lb{}}}}\end{center}

\end{enumerate}
}{\texttt{\footnotesize GLang.Metatheory.term\allowbreak Progress} at \texttt{\footnotesize GLang/Metatheory.lean:568}; \texttt{\footnotesize GLang.Metatheory.checked\allowbreak Space\allowbreak Progress} at \texttt{\footnotesize GLang/Metatheory.lean:626}; \texttt{\footnotesize GLang.Metatheory.checked\allowbreak Type\allowbreak Progress} at \texttt{\footnotesize GLang/Metatheory.lean:635}; \texttt{\footnotesize GLang.Metatheory.checked\allowbreak Type\allowbreak Progress\allowbreak At\allowbreak Basis} at \texttt{\footnotesize GLang/Metatheory.lean:645}; \texttt{\footnotesize GLang.Metatheory.checked\allowbreak Basis\allowbreak Progress} at \texttt{\footnotesize GLang/Metatheory.lean:655}}{none}{thm:prog}
\gldefbundle{adequacy}{Adequacy}{The compiled result agrees with the source denotation: evaluating and then denoting gives the same value as denoting directly, in the basis the trace used, and the realized type has the right dimensions.

\begin{enumerate}
  \item \emph{Terms}.\ If $\lDelta{};\evalCtxSpace{};\evalCtx{}\vdash_{\mathsf C} \lee{} \xlongrightarrow{\lb{}} \leev{}\glresulttype{\ltyv{}}$ and $\sigma{}\colon \mathsf{DEnv}$ and $\valCheckCtx{}\colon \mathsf{ValCtx}$ and $(\lDelta{}, \evalCtxSpace{}); \evalCtx{}; \valCheckCtx{}\vdash \sigma{}, \glvEnv{}$, then
\begin{center}\gpfit{\Sem{\lee{}}{\sigma{}} \equiv^{\ltyv{}}_{\SemBB{\lb{}}{\sigma{}}} \mathsf{evalVal}(\glvEnv{}, [], [], \leev{})}\\[2pt]
\gpfit{\wedge\; \SemT{\tof{\lb{}}}{\sigma{}} \equiv_{\SemBB{\lb{}}{\sigma{}}} \SemT{\ltyv{}}{\glvEnv{}}}\end{center}

  \item \emph{Spaces}.\ If $\evalCtxSpace{};\evalCtx{}\vdash_{\mathsf C} \ls{} \evalArrow{} \lsv{}$ and $\lDelta{}\colon \mathsf{LCtx}$ and $\sigma{}\colon \mathsf{DEnv}$ and $\valCheckCtx{}\colon \mathsf{ValCtx}$ and $(\lDelta{}, \evalCtxSpace{}); \evalCtx{}; \valCheckCtx{}\vdash \sigma{}, \glvEnv{}$, then
\begin{center}\gpfit{\mathsf{shpOfPartial}(\mathsf{oracle}(\glvEnv{}, [], []), \lsv{}) = \mathsf{srcShp}(\sigma{}, [], \ls{})}\end{center}

  \item \emph{Types at a basis}.\ If $\evalCtxSpace{};\evalCtx{}\vdash_{\mathsf C} \lb{} \evalArrow{}_{\mathsf{basis}} \ltyv{}$ and $\lDelta{}\colon \mathsf{LCtx}$ and $\sigma{}\colon \mathsf{DEnv}$ and $\valCheckCtx{}\colon \mathsf{ValCtx}$ and $(\lDelta{}, \evalCtxSpace{}); \evalCtx{}; \valCheckCtx{}\vdash \sigma{}, \glvEnv{}$, then
\begin{center}\gpfit{\SemT{\tof{\lb{}}}{\sigma{}} \equiv_{\SemBB{\lb{}}{\sigma{}}} \SemT{\ltyv{}}{\glvEnv{}}}\end{center}

  \item \emph{Bases}.\ If $(\emptyset,\evalCtxSpace{})\vdash \lb{}\colon \bvalid{}$ and $\vdash (\lemptyset{}, \evalCtxSpace{})$ and $\lemptyset{}, \evalCtxSpace{}\vdash \sigma{}$, then
\begin{center}\gpfit{\SemBB{\lb{}}{\sigma{}} \;\text{is a basis for}\; \SemT{\tof{\lb{}}}{\sigma{}}}\end{center}

  \item \emph{Change of basis}.\ If $\evalCtxSpace{};\evalCtx{}\vdash_{\mathsf C} \textcolor{lang}{{\lb{}_{\mathrm{src}}}} \xlongleftrightarrow{\leev{}} \textcolor{lang}{{\lb{}_{\mathrm{tgt}}}}$ and $\lDelta{}\colon \mathsf{LCtx}$ and $\sigma{}\colon \mathsf{DEnv}$ and $\valCheckCtx{}\colon \mathsf{ValCtx}$ and $(\lDelta{}, \evalCtxSpace{}); \evalCtx{}; \valCheckCtx{}\vdash \sigma{}, \glvEnv{}$, then
\begin{center}\gpfit{\SemBB{\textcolor{lang}{{\lb{}_{\mathrm{src}}}}}{\sigma{}} = \SemA{\leev{}}{\glvEnv{}}{} \cdot \SemBB{\textcolor{lang}{{\lb{}_{\mathrm{tgt}}}}}{\sigma{}}}\end{center}

  \item \emph{Dimensions}.\ If $\evalCtxSpace{};\evalCtx{}\vdash_{\mathsf C} \lb{} \evalArrow{}_{\mathsf{basis}} \ltyv{}$ and $\lDelta{}\colon \mathsf{LCtx}$ and $\sigma{}\colon \mathsf{DEnv}$ and $\valCheckCtx{}\colon \mathsf{ValCtx}$ and $(\lDelta{}, \evalCtxSpace{}); \evalCtx{}; \valCheckCtx{}\vdash \sigma{}, \glvEnv{}$, then
\begin{center}\gpfit{\SemT{\tof{\lb{}}}{\sigma{}} \equiv_{\SemBB{\lb{}}{\sigma{}}} \SemT{\ltyv{}}{\glvEnv{}}}\end{center}

\end{enumerate}
}{\texttt{\footnotesize GLang.Metatheory.term\allowbreak Adequacy} at \texttt{\footnotesize GLang/Metatheory.lean:682}; \texttt{\footnotesize GLang.Metatheory.space\allowbreak Adequacy} at \texttt{\footnotesize GLang/Metatheory.lean:698}; \texttt{\footnotesize GLang.Metatheory.type\allowbreak Adequacy\allowbreak At\allowbreak Basis} at \texttt{\footnotesize GLang/Metatheory.lean:849}; \texttt{\footnotesize GLang.Metatheory.basis\allowbreak Adequacy} at \texttt{\footnotesize GLang/Metatheory.lean:709}; \texttt{\footnotesize GLang.Metatheory.cob\allowbreak Adequacy} at \texttt{\footnotesize GLang/Metatheory.lean:826}; \texttt{\footnotesize GLang.Metatheory.dimension\allowbreak Adequacy} at \texttt{\footnotesize GLang/Metatheory.lean:837}}{none}{thm:main}
\gldefbundle{cobConsequences}{Change-of-basis dimensions and automorphism}{For a checked change of basis in a consistent environment, one number $N$ is simultaneously both inferred axis dimensions and both endpoint coordinate-list lengths.  The evaluated $N\times N$ coefficient array is the matrix of an invertible linear map.  Bijectivity uses validity of both endpoint bases, not merely a square-array equation.

Assuming
\begin{itemize}
  \item \gpfit{\evalCtxSpace{};\evalCtx{}\vdash_{\mathsf C} \textcolor{lang}{{\lb{}_{\mathrm{src}}}} \xlongleftrightarrow{\leev{}} \textcolor{lang}{{\lb{}_{\mathrm{tgt}}}}}
  \item \gpfit{(\lDelta{}, \evalCtxSpace{}); \evalCtx{}; \valCheckCtx{}\vdash \sigma{}, \glvEnv{}}
\end{itemize}
then:
\begin{enumerate}
  \item \emph{One common endpoint dimension}.
\begin{center}\gpfit{\exists \lsn{0}.\; \exists \lsn{1}.\; \exists \lsnv{0}.\; \exists \lsnv{1}.\; \operatorname{inferType}(\leev{}) = \lsnv{0} \loprod \lsnv{1} \;\wedge\; \evalSJudge{\evalCtx{}}{\lsn{0}}{\lsnv{0}} \;\wedge\; \evalSJudge{\evalCtx{}}{\lsn{1}}{\lsnv{1}} \;\wedge\; \vdash \operatorname{inferType}(\leev{})\colon \tvalid{} \;\wedge\; \mathsf{fv}(\leev{}) \subseteq \mathrm{dom}(\evalCtx{})}\\[2pt]
\gpfit{\wedge\; \SemBB{\textcolor{lang}{{\lb{}_{\mathrm{src}}}}}{\sigma{}} = \SemA{\leev{}}{\glvEnv{}}{} \cdot \SemBB{\textcolor{lang}{{\lb{}_{\mathrm{tgt}}}}}{\sigma{}}}\\[2pt]
\gpfit{\wedge\; \exists \mathit{N}.\; \exists \lsnv{{\mathrm{svSrc}}}.\; \exists \lsnv{{\mathrm{svTgt}}}.\; \operatorname{inferSpace}(\leev{}) = \lsnv{{\mathrm{svSrc}}} \loprod \lsnv{{\mathrm{svTgt}}} \;\wedge\; \mathit{N} = \dim_{\mathcal{O}}(\lsnv{{\mathrm{svSrc}}}) \;\wedge\; \mathit{N} = \dim_{\mathcal{O}}(\lsnv{{\mathrm{svTgt}}}) \;\wedge\; \mathit{N} = \lvert \SemBB{\textcolor{lang}{{\lb{}_{\mathrm{src}}}}}{\sigma{}} \rvert \;\wedge\; \mathit{N} = \lvert \SemBB{\textcolor{lang}{{\lb{}_{\mathrm{tgt}}}}}{\sigma{}} \rvert}\end{center}

  \item \emph{Coordinate-matrix automorphism}.
\begin{center}\gpfit{\exists \mathit{N}.\; \exists \lsnv{{\mathrm{svSrc}}}.\; \exists \lsnv{{\mathrm{svTgt}}}.\; \exists \mathit{a}.\; \exists \lsn{0}.\; \exists \lsn{1}.\; \exists \lsnv{0}.\; \exists \lsnv{1}.\; \operatorname{inferType}(\leev{}) = \lsnv{0} \loprod \lsnv{1} \;\wedge\; \evalSJudge{\evalCtx{}}{\lsn{0}}{\lsnv{0}} \;\wedge\; \evalSJudge{\evalCtx{}}{\lsn{1}}{\lsnv{1}} \;\wedge\; \vdash \operatorname{inferType}(\leev{})\colon \tvalid{} \;\wedge\; \mathsf{fv}(\leev{}) \subseteq \mathrm{dom}(\evalCtx{})}\\[2pt]
\gpfit{\wedge\; \SemBB{\textcolor{lang}{{\lb{}_{\mathrm{src}}}}}{\sigma{}} = \SemA{\leev{}}{\glvEnv{}}{} \cdot \SemBB{\textcolor{lang}{{\lb{}_{\mathrm{tgt}}}}}{\sigma{}}}\\[2pt]
\gpfit{\wedge\; \operatorname{inferSpace}(\leev{}) = \lsnv{{\mathrm{svSrc}}} \loprod \lsnv{{\mathrm{svTgt}}}}\\[2pt]
\gpfit{\wedge\; \mathit{N} = \dim_{\mathcal{O}}(\lsnv{{\mathrm{svSrc}}})}\\[2pt]
\gpfit{\wedge\; \mathit{N} = \dim_{\mathcal{O}}(\lsnv{{\mathrm{svTgt}}})}\\[2pt]
\gpfit{\wedge\; \mathit{N} = \lvert \SemBB{\textcolor{lang}{{\lb{}_{\mathrm{src}}}}}{\sigma{}} \rvert}\\[2pt]
\gpfit{\wedge\; \mathit{N} = \lvert \SemBB{\textcolor{lang}{{\lb{}_{\mathrm{tgt}}}}}{\sigma{}} \rvert}\\[2pt]
\gpfit{\wedge\; \mathcal{E}\llbracket \leev{}\rrbracket_{[];[]} = \mathit{a}}\\[2pt]
\gpfit{\wedge\; \lvert \mathit{a}\rvert = (\mathit{N}) \cdot (\mathit{N})}\\[2pt]
\gpfit{\wedge\; \mathsf{Bijective}(\mathsf{toLin'}(\mathsf{denotAsMatrix}(\mathit{N}, \mathit{N}, \mathsf{get}(\mathit{a}))))}\end{center}

\end{enumerate}
}{\texttt{\footnotesize GLang.Metatheory.cob\allowbreak Common\allowbreak Dimension\allowbreak Adequacy} at \texttt{\footnotesize GLang/Cob\allowbreak Endpoint\allowbreak Preservation.lean:140}; \texttt{\footnotesize GLang.Metatheory.cob\allowbreak Coordinate\allowbreak Matrix\allowbreak Bijective} at \texttt{\footnotesize GLang/Cob\allowbreak Endpoint\allowbreak Preservation.lean:280}}{none}{--}
\gldefbundle{checkedProgressAdequacy}{Checked progress with adequacy}{One call yields a result that is simultaneously reachable, well typed at its realized type, and adequate: progress and adequacy for the same trace.

Assuming
\begin{itemize}
  \item \gpfit{(\lDelta{}, \evalCtxSpace{}); \evalCtx{}; \valCheckCtx{}\vdash \sigma{}, \glvEnv{}}
  \item \gpfit{(\lDelta{}, \evalCtxSpace{})\vdash \lee{}\colon \tof{\lb{}}}
  \item \gpfit{(\emptyset,\evalCtxSpace{})\vdash \lb{}\colon \bvalid{}}
\end{itemize}
then:
\begin{enumerate}
  \item
\begin{center}\gpfit{\exists \leev{}.\; \exists \ltyv{}.\; \lDelta{};\evalCtxSpace{};\evalCtx{}\vdash_{\mathsf C} \lee{} \xlongrightarrow{\lb{}} \leev{}\glresulttype{\ltyv{}}}\\[2pt]
\gpfit{\wedge\; \operatorname{inferType}(\leev{}) = \ltyv{}}\\[2pt]
\gpfit{\wedge\; \operatorname{inferSpace}(\leev{}) = \mathsf{spaceOf}(\ltyv{})}\\[2pt]
\gpfit{\wedge\; \Sem{\lee{}}{\sigma{}} \equiv^{\ltyv{}}_{\SemBB{\lb{}}{\sigma{}}} \mathsf{evalVal}(\glvEnv{}, [], [], \leev{})}\\[2pt]
\gpfit{\wedge\; \SemT{\tof{\lb{}}}{\sigma{}} \equiv_{\SemBB{\lb{}}{\sigma{}}} \SemT{\ltyv{}}{\glvEnv{}}}\end{center}

\end{enumerate}
}{\texttt{\footnotesize GLang.Metatheory.checked\allowbreak Progress\allowbreak Adequacy} at \texttt{\footnotesize GLang/Metatheory.lean:719}}{none}{--}
\gldefbundle{strengthenedAdequacy}{Strengthened adequacy (oracle-free)}{Adequacy over the strengthened semantics, with the dimension oracle fixed to the constructed one, so no oracle or oracle specification is assumed.

Assuming
\begin{itemize}
  \item \gpfit{\sigma{}\colon \mathsf{DEnv}}
  \item \gpfit{\valCheckCtx{}\colon \mathsf{ValCtx}}
  \item \gpfit{(\lDelta{}, \evalCtxSpace{}); \evalCtx{}; \valCheckCtx{}\vdash \sigma{}, \glvEnv{}}
\end{itemize}
then:
\begin{enumerate}
  \item \emph{Terms}.\ If $\lDelta{};\evalCtxSpace{};\evalCtx{}\vdash_{\mathsf S} \lee{} \xlongrightarrow{\lb{}} \leev{}\glresulttype{\ltyv{}}$, then
\begin{center}\gpfit{\Sem{\lee{}}{\sigma{}} \equiv^{\ltyv{}}_{\SemBB{\lb{}}{\sigma{}}} \mathsf{evalVal}(\glvEnv{}, [], [], \leev{})}\\[2pt]
\gpfit{\wedge\; \SemT{\tof{\lb{}}}{\sigma{}} \equiv_{\SemBB{\lb{}}{\sigma{}}} \SemT{\ltyv{}}{\glvEnv{}}}\end{center}

  \item \emph{Change of basis}.\ If $\evalCtxSpace{};\evalCtx{}\vdash_{\mathsf S} \textcolor{lang}{{\lb{}_{\mathrm{src}}}} \xlongleftrightarrow{\leev{}} \textcolor{lang}{{\lb{}_{\mathrm{tgt}}}}$ and $\lDelta{}\colon \mathsf{LCtx}$, then
\begin{center}\gpfit{\SemBB{\textcolor{lang}{{\lb{}_{\mathrm{src}}}}}{\sigma{}} = \SemA{\leev{}}{\glvEnv{}}{} \cdot \SemBB{\textcolor{lang}{{\lb{}_{\mathrm{tgt}}}}}{\sigma{}}}\end{center}

  \item \emph{Dimensions}.\ If $\evalCtxSpace{};\evalCtx{}\vdash_{\mathsf S} \lb{} \evalArrow{}_{\mathsf{basis}} \ltyv{}$ and $\lDelta{}\colon \mathsf{LCtx}$, then
\begin{center}\gpfit{\SemT{\tof{\lb{}}}{\sigma{}} \equiv_{\SemBB{\lb{}}{\sigma{}}} \SemT{\ltyv{}}{\glvEnv{}}}\end{center}

\end{enumerate}
}{\texttt{\footnotesize GLang.Adequacy.Direct.term} at \texttt{\footnotesize GLang/Adequacy/Direct\allowbreak Public.lean:97}; \texttt{\footnotesize GLang.Adequacy.Direct.cob} at \texttt{\footnotesize GLang/Adequacy/Direct\allowbreak Public.lean:132}; \texttt{\footnotesize GLang.Adequacy.Direct.dims\allowbreak At} at \texttt{\footnotesize GLang/Adequacy/Direct\allowbreak Public.lean:142}}{none}{thm:main}
\gldefbundlerows{
Source denotational soundness & 4 & none & \gpfitbox{\texttt{\textbackslash glbundle\{soundness\}}}\\
Value denotational soundness & 3 & none & \gpfitbox{\texttt{\textbackslash glbundle\{valueSoundness\}}}\\
Standard-basis soundness & 4 & none & \gpfitbox{\texttt{\textbackslash glbundle\{standardBasis\}}}\\
Progress at the standard basis & 1 & none & \gpfitbox{\texttt{\textbackslash glbundle\{standardBasisProgress\}}}\\
Canonical inner-product soundness & 2 & none & \gpfitbox{\texttt{\textbackslash glbundle\{innerProduct\}}}\\
Preservation and representation regularity & 5 & none & \gpfitbox{\texttt{\textbackslash glbundle\{preservation\}}}\\
Checked progress & 5 & none & \gpfitbox{\texttt{\textbackslash glbundle\{progress\}}}\\
Adequacy & 6 & none & \gpfitbox{\texttt{\textbackslash glbundle\{adequacy\}}}\\
Change-of-basis dimensions and automorphism & 2 & none & \gpfitbox{\texttt{\textbackslash glbundle\{cobConsequences\}}}\\
Checked progress with adequacy & 1 & none & \gpfitbox{\texttt{\textbackslash glbundle\{checkedProgressAdequacy\}}}\\
Strengthened adequacy (oracle-free) & 3 & none & \gpfitbox{\texttt{\textbackslash glbundle\{strengthenedAdequacy\}}}\\
}
\gldefnotationrows{
$\lctx\vdash\lee\colon\lty$ & Source typing, as in the paper.\\
$\lctxnle;\evalCtx\vdash_{\mathsf S}\lee\xlongrightarrow{\lb}\leev$ & The \emph{strengthened} operational judgments: an ordinary evaluation fused with the typing of its own conclusion, so a derivation carries its typing evidence.\\
$\lctxnle;\evalCtx\vdash_{\mathsf C}\lee\xlongrightarrow{\lb}\leev$ & The \emph{checked} operational judgments, which the public metatheory is stated over.  They erase to the ordinary judgments and translate exactly to and from the strengthened ones.\\
$(\lDelta,\evalCtxSpace);\valCheckCtx\vdash\evalCtx$ & Joint source and value context formation. Each stored representation is typed under the canonical value context. Fresh bindings extend that context; additional representations preserve it. All entries retain type realization and scope conditions.\\
$\vdash_{\mathsf{carrier}}\evalCtx$ & The operational carrier invariant on a store.\\
$\lDelta,\evalCtxSpace\vdash_{\mathsf{cover}}\evalCtx$ & Every context variable has a binding in the store.\\
$\Sem{\lee}{\sigma} \equiv^{\ltyv}_{\SemBB{\lb}{\sigma}} \SemA{\leev}{\glvEnv}{}$ & Result agreement at value type $\ltyv$ in basis $\SemBB{\lb}{\sigma}$; the paper's \texttt{eq:comp}.\\
$\SemT{\lty}{\sigma} \equiv_{\SemBB{\lb}{\sigma}} \SemT{\ltyv}{\glvEnv}$ & Dimension agreement at a basis; the paper's \texttt{def:agreedims}.\\
$\sigma_{1} =_{S} \sigma_{2}$ & Two environments assigning the same value to every name in $S$.\\
$\mathsf{closed}(\lee)$ & Local closure: no dangling de Bruijn index.  Its rules are in \cref{sec:glr-rules}.\\
}
\gldefjudgrows{
\gpfit{\vdash(\lDelta{},\lepsilon{})} & 3 & \gpfitbox{\texttt{\footnotesize GLang.Inductive\allowbreak Ctx\allowbreak WF}} & \gpfitbox{\texttt{\textbackslash glrulefigure\{inductiveCtxWF\}}}\\
\gpfit{\lctxnle\vdash\ls\colon\svalid} & 9 & \gpfitbox{\texttt{\footnotesize GLang.SValid}} & \gpfitbox{\texttt{\textbackslash glrulefigure\{sValid\}}}\\
\gpfit{\lctxnle\vdash\lty\colon\tvalid} & 2 & \gpfitbox{\texttt{\footnotesize GLang.TValid}} & \gpfitbox{\texttt{\textbackslash glrulefigure\{tValid\}}}\\
\gpfit{\lctxnle\vdash\lsn{0}\leq\lsn{1}} & 4 & \gpfitbox{\texttt{\footnotesize GLang.Sub\allowbreak Type}} & \gpfitbox{\texttt{\textbackslash glrulefigure\{subType\}}}\\
\gpfit{\lctx\vdash\lee\colon\lty} & 19 & \gpfitbox{\texttt{\footnotesize GLang.Has\allowbreak Type}} & \gpfitbox{\texttt{\textbackslash glrulefigure\{hasType\}}}\\
\gpfit{\lctxnle\vdash\lb{}\colon\bvalid{}} & 12 & \gpfitbox{\texttt{\footnotesize GLang.Basis.Basis\allowbreak Valid}} & \gpfitbox{\texttt{\textbackslash glrulefigure\{basisValid\}}}\\
\gpfit{\mathsf{lc}(\ls{})} & 9 & \gpfitbox{\texttt{\footnotesize GLang.Space.lc}} & \gpfitbox{\texttt{\textbackslash glrulefigure\{lc.space\}}}\\
\gpfit{\mathsf{lc}(\lty{})} & 2 & \gpfitbox{\texttt{\footnotesize GLang.Ty.lc}} & \gpfitbox{\texttt{\textbackslash glrulefigure\{lc.ty\}}}\\
\gpfit{\mathsf{lc}(\lee{})} & 16 & \gpfitbox{\texttt{\footnotesize GLang.Term.lc}} & \gpfitbox{\texttt{\textbackslash glrulefigure\{lc.term\}}}\\
\gpfit{\mathsf{lc}(\lb{})} & 9 & \gpfitbox{\texttt{\footnotesize GLang.Basis.lc}} & \gpfitbox{\texttt{\textbackslash glrulefigure\{lc.basis\}}}\\
\gpfit{\mathsf{lc}(\lsv{})} & 5 & \gpfitbox{\texttt{\footnotesize GLang.Val\allowbreak Space.lc}} & \gpfitbox{\texttt{\textbackslash glrulefigure\{lc.valSpace\}}}\\
\gpfit{\mathsf{lc}(\ltyv{})} & 2 & \gpfitbox{\texttt{\footnotesize GLang.Val\allowbreak Ty.lc}} & \gpfitbox{\texttt{\textbackslash glrulefigure\{lc.valTy\}}}\\
\gpfit{\mathsf{lc}(\leev{})} & 28 & \gpfitbox{\texttt{\footnotesize GLang.Val\allowbreak Expr.lc}} & \gpfitbox{\texttt{\textbackslash glrulefigure\{lc.valExpr\}}}\\
\gpfit{\valCheckCtx\vdash\lsv{}\colon\svalid} & 5 & \gpfitbox{\texttt{\footnotesize GLang.Runtime\allowbreak Value.Val\allowbreak SValid}} & \gpfitbox{\texttt{\textbackslash glrulefigure\{valSValid\}}}\\
\gpfit{\valCheckCtx\vdash\ltyv{}\colon\lkn{}} & 2 & \gpfitbox{\texttt{\footnotesize GLang.Runtime\allowbreak Value.Val\allowbreak TValid}} & \gpfitbox{\texttt{\textbackslash glrulefigure\{valTValid\}}}\\
\gpfit{\valCheckCtx\vdash\leev{}\colon\ltyv{}} & 31 & \gpfitbox{\texttt{\footnotesize GLang.Runtime\allowbreak Value.Val\allowbreak Has\allowbreak Type}} & \gpfitbox{\texttt{\textbackslash glrulefigure\{valHasType\}}}\\
\gpfit{\evalSJudgeK{\ls{}}{\lsv{}}} & 9 & \gpfitbox{\texttt{\footnotesize GLang.Val\allowbreak Space.Eval\allowbreak S}} & \gpfitbox{\texttt{\textbackslash glrulefigure\{evalS\}}}\\
\gpfit{\evalSJudgeK{\lty{}}{\ltyv{}}} & 2 & \gpfitbox{\texttt{\footnotesize GLang.Val\allowbreak Ty.Eval\allowbreak T}} & \gpfitbox{\texttt{\textbackslash glrulefigure\{evalT\}}}\\
\gpfit{\evalBBJudgeK{\lbn{0}}{\leev{}}{\lbn{1}}} & 13 & \gpfitbox{\texttt{\footnotesize GLang.Basis.Eval\allowbreak BB}} & \gpfitbox{\texttt{\textbackslash glrulefigure\{evalBB\}}}\\
\gpfit{\evalJudgeK{\lee{}}{\lb{}}{\leev{}}} & 26 & \gpfitbox{\texttt{\footnotesize GLang.Val\allowbreak Expr.Op\allowbreak Step}} & \gpfitbox{\texttt{\textbackslash glrulefigure\{opStep\}}}\\
\gpfit{\lepsilon{};\evalCtx{}\vdash_{\mathsf S} \ls{}\evalArrow{}\lsv{}} & 9 & \gpfitbox{\texttt{\footnotesize GLang.S.Space}} & \gpfitbox{\texttt{\textbackslash glrulefigure\{s.space\}}}\\
\gpfit{\lepsilon{};\evalCtx{}\vdash_{\mathsf S} \lty{}\evalArrow{}\ltyv{}} & 2 & \gpfitbox{\texttt{\footnotesize GLang.S.Ty}} & \gpfitbox{\texttt{\textbackslash glrulefigure\{s.ty\}}}\\
\gpfit{\lepsilon{};\evalCtx{}\vdash_{\mathsf S} \lb{}\evalArrow{}_{\mathsf{basis}}\ltyv{}} & 2 & \gpfitbox{\texttt{\footnotesize GLang.S.Ty\allowbreak At}} & \gpfitbox{\texttt{\textbackslash glrulefigure\{s.tyAt\}}}\\
\gpfit{\lepsilon{};\evalCtx{}\vdash_{\mathsf S} \lbn{0}\xlongleftrightarrow{\leev{}}\lbn{1}} & 13 & \gpfitbox{\texttt{\footnotesize GLang.S.Cob}} & \gpfitbox{\texttt{\textbackslash glrulefigure\{s.cob\}}}\\
\gpfit{\lDelta{};\lepsilon{};\evalCtx{}\vdash_{\mathsf S} \lee{}\xlongrightarrow{\lb{}}\leev{}\glresulttype{\ltyv{}}} & 28 & \gpfitbox{\texttt{\footnotesize GLang.S.Term}} & \gpfitbox{\texttt{\textbackslash glrulefigure\{s.term\}}}\\
\gpfit{\lepsilon{};\evalCtx{}\vdash_{\mathsf C} \ls{}\evalArrow{}\lsv{}} & 9 & \gpfitbox{\texttt{\footnotesize GLang.Raw\allowbreak Certified\allowbreak Operational.C.Space}} & \gpfitbox{\texttt{\textbackslash glrulefigure\{c.space\}}}\\
\gpfit{\lepsilon{};\evalCtx{}\vdash_{\mathsf C} \lty{}\evalArrow{}\ltyv{}} & 2 & \gpfitbox{\texttt{\footnotesize GLang.Raw\allowbreak Certified\allowbreak Operational.C.Ty}} & \gpfitbox{\texttt{\textbackslash glrulefigure\{c.ty\}}}\\
\gpfit{\lepsilon{};\evalCtx{}\vdash_{\mathsf C} \lb{}\evalArrow{}_{\mathsf{basis}}\ltyv{}} & 2 & \gpfitbox{\texttt{\footnotesize GLang.Raw\allowbreak Certified\allowbreak Operational.C.Ty\allowbreak At}} & \gpfitbox{\texttt{\textbackslash glrulefigure\{c.tyAt\}}}\\
\gpfit{\lepsilon{};\evalCtx{}\vdash_{\mathsf C} \lbn{0}\xlongleftrightarrow{\leev{}}\lbn{1}} & 13 & \gpfitbox{\texttt{\footnotesize GLang.Raw\allowbreak Certified\allowbreak Operational.C.Cob}} & \gpfitbox{\texttt{\textbackslash glrulefigure\{c.cob\}}}\\
\gpfit{\lDelta{};\lepsilon{};\evalCtx{}\vdash_{\mathsf C} \lee{}\xlongrightarrow{\lb{}}\leev{}\glresulttype{\ltyv{}}} & 28 & \gpfitbox{\texttt{\footnotesize GLang.Raw\allowbreak Certified\allowbreak Operational.C.Term}} & \gpfitbox{\texttt{\textbackslash glrulefigure\{c.term\}}}\\
}
\gldefrulesallrows{
\gpfitbox{\textsc{Ctx-Empty}} & \gpfit{\vdash(\lDelta{},\lepsilon{})} & \gpfit{\vdash (\lemptyset{}, \lemptyset{})}\\
\gpfitbox{\textsc{Ctx-Linear}} & \gpfit{\vdash(\lDelta{},\lepsilon{})} & \gpfit{\vdash ((\lDelta{}, \mathit{x}\colon \ls{}), \lepsilon{})}\\
\gpfitbox{\textsc{Ctx-Nonlinear}} & \gpfit{\vdash(\lDelta{},\lepsilon{})} & \gpfit{\vdash (\lDelta{}, (\lepsilon{}, \mathit{x}\colon \lty{}))}\\
\gpfitbox{\textsc{S-Empty}} & \gpfit{\lctxnle\vdash\ls\colon\svalid} & \gpfit{(\lemptyset, \lepsilon{})\vdash \lsempty{}\colon\svalid}\\
\gpfitbox{\textsc{S-Real}} & \gpfit{\lctxnle\vdash\ls\colon\svalid} & \gpfit{(\lemptyset, \lepsilon{})\vdash \lR{}\colon\svalid}\\
\gpfitbox{\textsc{S-Dual}} & \gpfit{\lctxnle\vdash\ls\colon\svalid} & \gpfit{(\lemptyset, \lepsilon{})\vdash \ls{}^{\lstar}\colon\svalid}\\
\gpfitbox{\textsc{S-Dsum}} & \gpfit{\lctxnle\vdash\ls\colon\svalid} & \gpfit{(\lemptyset, \lepsilon{})\vdash \lsn{0} \losum \lsn{1}\colon\svalid}\\
\gpfitbox{\textsc{S-Tprod}} & \gpfit{\lctxnle\vdash\ls\colon\svalid} & \gpfit{(\lemptyset, \lepsilon{})\vdash \lsn{0} \loprod \lsn{1}\colon\svalid}\\
\gpfitbox{\textsc{S-Ker}} & \gpfit{\lctxnle\vdash\ls\colon\svalid} & \gpfit{(\lemptyset, \lepsilon{})\vdash \Kernel(\lee{})\colon\svalid}\\
\gpfitbox{\textsc{S-Coker}} & \gpfit{\lctxnle\vdash\ls\colon\svalid} & \gpfit{(\lemptyset, \lepsilon{})\vdash \CoKernel(\lee{})\colon\svalid}\\
\gpfitbox{\textsc{S-Img}} & \gpfit{\lctxnle\vdash\ls\colon\svalid} & \gpfit{(\lemptyset, \lepsilon{})\vdash \Image(\lee{})\colon\svalid}\\
\gpfitbox{\textsc{S-Coimg}} & \gpfit{\lctxnle\vdash\ls\colon\svalid} & \gpfit{(\lemptyset, \lepsilon{})\vdash \CoImage(\lee{})\colon\svalid}\\
\gpfitbox{\textsc{K-Space}} & \gpfit{\lctxnle\vdash\lty\colon\tvalid} & \gpfit{(\lemptyset, \lepsilon{})\vdash \ls{}\colon \lkbase{}}\\
\gpfitbox{\textsc{K-Pi\allowbreak Ty}} & \gpfit{\lctxnle\vdash\lty\colon\tvalid} & \gpfit{(\lemptyset, \lepsilon{})\vdash (x\colon \ltyn{0}\fnto \ltyn{1})\colon \lkn{0} \fnto \lkn{1}}\\
\gpfitbox{\textsc{Sub-Sub\allowbreak Ker}} & \gpfit{\lctxnle\vdash\lsn{0}\leq\lsn{1}} & \gpfit{(\lemptyset, \lepsilon{})\vdash \Kernel(\lee{})\leq \lsn{0}}\\
\gpfitbox{\textsc{Sub-Sub\allowbreak Coker}} & \gpfit{\lctxnle\vdash\lsn{0}\leq\lsn{1}} & \gpfit{(\lemptyset, \lepsilon{})\vdash \CoKernel(\lee{})\leq \lsn{0}}\\
\gpfitbox{\textsc{Sub-Sub\allowbreak Img}} & \gpfit{\lctxnle\vdash\lsn{0}\leq\lsn{1}} & \gpfit{(\lemptyset, \lepsilon{})\vdash \Image(\lee{})\leq \lsn{1}}\\
\gpfitbox{\textsc{Sub-Sub\allowbreak Coimg}} & \gpfit{\lctxnle\vdash\lsn{0}\leq\lsn{1}} & \gpfit{(\lemptyset, \lepsilon{})\vdash \CoImage(\lee{})\leq \lsn{1}}\\
\gpfitbox{\textsc{T-Zero}} & \gpfit{\lctx\vdash\lee\colon\lty} & \gpfit{(\lDelta{}, \lepsilon{})\vdash \lvzero{}_{\ls{}}\colon \ls{}}\\
\gpfitbox{\textsc{T-Smul}} & \gpfit{\lctx\vdash\lee\colon\lty} & \gpfit{({\lDelta{}_{0}} \ldisun {\lDelta{}_{1}}, \lepsilon{})\vdash \len{0} \lvscale \len{1}\colon \ls{}}\\
\gpfitbox{\textsc{T-Add}} & \gpfit{\lctx\vdash\lee\colon\lty} & \gpfit{(\lDelta{}, \lepsilon{})\vdash \len{0} \lvadd \len{1}\colon \ls{}}\\
\gpfitbox{\textsc{T-Inner}} & \gpfit{\lctx\vdash\lee\colon\lty} & \gpfit{(\lemptyset{}, \lepsilon{})\vdash \linner{}_{\ls{}}\colon \ls{}^{\lstar} \loprod \ls{}^{\lstar}}\\
\gpfitbox{\textsc{T-Real\allowbreak Lit}} & \gpfit{\lctx\vdash\lee\colon\lty} & \gpfit{(\lemptyset{}, \lepsilon{})\vdash \mathit{r}\colon \lR{}}\\
\gpfitbox{\textsc{T-App\allowbreak Linear}} & \gpfit{\lctx\vdash\lee\colon\lty} & \gpfit{({\lDelta{}_{0}} \ldisun {\lDelta{}_{1}}, \lepsilon{})\vdash \leval{\len{0}}{}{\len{1}}\colon \lsn{1}}\\
\gpfitbox{\textsc{T-App\allowbreak Inner}} & \gpfit{\lctx\vdash\lee\colon\lty} & \gpfit{({\lDelta{}_{0}} \ldisun {\lDelta{}_{1}}, \lepsilon{})\vdash \leval{\len{0}}{}{\len{1}}\colon \lR{}}\\
\gpfitbox{\textsc{T-App\allowbreak Nonlinear}} & \gpfit{\lctx\vdash\lee\colon\lty} & \gpfit{(\lemptyset{}, \lepsilon{})\vdash \leval{\len{0}}{}{\len{1}}\colon \ltyn{1}\subst{x}{\len{1}}}\\
\gpfitbox{\textsc{T-Var\allowbreak Lin}} & \gpfit{\lctx\vdash\lee\colon\lty} & \gpfit{(\{(\mathit{x}, \ls{})\}, \lepsilon{})\vdash \mathit{x}\colon \ls{}}\\
\gpfitbox{\textsc{T-Var\allowbreak Nlin}} & \gpfit{\lctx\vdash\lee\colon\lty} & \gpfit{(\lemptyset{}, \lepsilon{})\vdash \mathit{x}\colon \lty{}}\\
\gpfitbox{\textsc{T-Pinv\allowbreak Tensor}} & \gpfit{\lctx\vdash\lee\colon\lty} & \gpfit{(\lemptyset{}, \lepsilon{})\vdash \lee{}^{\lpseudo}\colon \lsn{1}^{\lstar} \loprod \lsn{0}}\\
\gpfitbox{\textsc{T-Pinv\allowbreak Dual}} & \gpfit{\lctx\vdash\lee\colon\lty} & \gpfit{(\lemptyset{}, \lepsilon{})\vdash \lee{}^{\lpseudo}\colon \lR{}^{\lstar} \loprod \ls{}}\\
\gpfitbox{\textsc{T-Promote}} & \gpfit{\lctx\vdash\lee\colon\lty} & \gpfit{(\lDelta{}, \mathit{lin} \ldisun \lepsilon{}')\vdash \lee{}\colon \lty{}}\\
\gpfitbox{\textsc{T-Dsum\allowbreak E}} & \gpfit{\lctx\vdash\lee\colon\lty} & \gpfit{(\lDelta{}, \lepsilon{})\vdash \len{0}\losum \len{1}\colon \lsn{0} \losum \lsn{1}}\\
\gpfitbox{\textsc{T-Tprod\allowbreak E}} & \gpfit{\lctx\vdash\lee\colon\lty} & \gpfit{({\lDelta{}_{0}} \ldisun {\lDelta{}_{1}}, \lepsilon{})\vdash \len{0}\loprod \len{1}\colon \lsn{0} \loprod \lsn{1}}\\
\gpfitbox{\textsc{T-Llam\allowbreak Real}} & \gpfit{\lctx\vdash\lee\colon\lty} & \gpfit{(\lDelta{}, \lepsilon{})\vdash (\lllambda \llpl x\colon \ls{}\llpr \ldotsp \lee{})\colon \ls{}^{\lstar}}\\
\gpfitbox{\textsc{T-Llam\allowbreak Tensor}} & \gpfit{\lctx\vdash\lee\colon\lty} & \gpfit{(\lDelta{}, \lepsilon{})\vdash (\lllambda \llpl x\colon \lsn{0}\llpr \ldotsp \lee{})\colon \lsn{0}^{\lstar} \loprod \ls{}}\\
\gpfitbox{\textsc{T-Embed}} & \gpfit{\lctx\vdash\lee\colon\lty} & \gpfit{(\lemptyset{}, \lepsilon{})\vdash \lembd{}_{\lsn{0}}\colon \lsn{0}^{\lstar} \loprod \lsn{1}}\\
\gpfitbox{\textsc{T-Nllam}} & \gpfit{\lctx\vdash\lee\colon\lty} & \gpfit{(\lemptyset{}, \lepsilon{})\vdash (\nllambda \llpl x\colon \ltyn{0}\llpr \ldotsp \lee{})\colon (x\colon \ltyn{0}\fnto \ltyn{1})}\\
\gpfitbox{\textsc{B-Empty}} & \gpfit{\lctxnle\vdash\lb{}\colon\bvalid{}} & \gpfit{(\emptyset,\lepsilon{})\vdash \lbzero{}\colon \bvalid{}}\\
\gpfitbox{\textsc{B-One}} & \gpfit{\lctxnle\vdash\lb{}\colon\bvalid{}} & \gpfit{(\emptyset,\lepsilon{})\vdash \lbone{}\colon \bvalid{}}\\
\gpfitbox{\textsc{B-Dual}} & \gpfit{\lctxnle\vdash\lb{}\colon\bvalid{}} & \gpfit{(\emptyset,\lepsilon{})\vdash \lb{}^{\lstar}\colon \bvalid{}}\\
\gpfitbox{\textsc{B-Dsum}} & \gpfit{\lctxnle\vdash\lb{}\colon\bvalid{}} & \gpfit{(\emptyset,\lepsilon{})\vdash \lbn{0} \losum \lbn{1}\colon \bvalid{}}\\
\gpfitbox{\textsc{B-Tprod}} & \gpfit{\lctxnle\vdash\lb{}\colon\bvalid{}} & \gpfit{(\emptyset,\lepsilon{})\vdash \lbn{0} \loprod \lbn{1}\colon \bvalid{}}\\
\gpfitbox{\textsc{B-Proj\allowbreak Ker}} & \gpfit{\lctxnle\vdash\lb{}\colon\bvalid{}} & \gpfit{(\emptyset,\lepsilon{})\vdash \pBasisC{\Kernel(\lee{})}{\lbn{0}}{\lbn{1}}\colon \bvalid{}}\\
\gpfitbox{\textsc{B-Proj\allowbreak Coker}} & \gpfit{\lctxnle\vdash\lb{}\colon\bvalid{}} & \gpfit{(\emptyset,\lepsilon{})\vdash \pBasisC{\CoKernel(\lee{})}{\lbn{0}}{\lbn{1}}\colon \bvalid{}}\\
\gpfitbox{\textsc{B-Proj\allowbreak Img}} & \gpfit{\lctxnle\vdash\lb{}\colon\bvalid{}} & \gpfit{(\emptyset,\lepsilon{})\vdash \pBasisC{\Image(\lee{})}{\lbn{0}}{\lbn{1}}\colon \bvalid{}}\\
\gpfitbox{\textsc{B-Proj\allowbreak Coimg}} & \gpfit{\lctxnle\vdash\lb{}\colon\bvalid{}} & \gpfit{(\emptyset,\lepsilon{})\vdash \pBasisC{\CoImage(\lee{})}{\lbn{0}}{\lbn{1}}\colon \bvalid{}}\\
\gpfitbox{\textsc{B-Image}} & \gpfit{\lctxnle\vdash\lb{}\colon\bvalid{}} & \gpfit{(\emptyset,\lepsilon{})\vdash \iBasisC{\lb{}}\colon \bvalid{}}\\
\gpfitbox{\textsc{B-Terms}} & \gpfit{\lctxnle\vdash\lb{}\colon\bvalid{}} & \gpfit{(\emptyset,\lepsilon{})\vdash \llpl{} \len{0},\dots,\len{n}\llpr{}_{\lb{}}\colon \bvalid{}}\\
\gpfitbox{\textsc{B-Frame}} & \gpfit{\lctxnle\vdash\lb{}\colon\bvalid{}} & \gpfit{(\emptyset,\lepsilon{})\vdash \frameB{x}{\lbn{0}}{\lbn{1}}\colon \bvalid{}}\\
\gpfitbox{\textsc{LC-S-Empty}} & \gpfit{\mathsf{lc}(\ls{})} & \gpfit{\mathsf{lc}(\lsempty{})}\\
\gpfitbox{\textsc{LC-S-Real}} & \gpfit{\mathsf{lc}(\ls{})} & \gpfit{\mathsf{lc}(\lR{})}\\
\gpfitbox{\textsc{LC-S-Dual}} & \gpfit{\mathsf{lc}(\ls{})} & \gpfit{\mathsf{lc}(\mathsf{dual}(\ls{}))}\\
\gpfitbox{\textsc{LC-S-Dsum}} & \gpfit{\mathsf{lc}(\ls{})} & \gpfit{\mathsf{lc}(\mathsf{dsum}(\lsn{0}, \lsn{1}))}\\
\gpfitbox{\textsc{LC-S-Tprod}} & \gpfit{\mathsf{lc}(\ls{})} & \gpfit{\mathsf{lc}(\mathsf{tprod}(\lsn{0}, \lsn{1}))}\\
\gpfitbox{\textsc{LC-S-Ker}} & \gpfit{\mathsf{lc}(\ls{})} & \gpfit{\mathsf{lc}(\mathsf{ker}(\lsn{0}, \lsn{1}, \lee{}))}\\
\gpfitbox{\textsc{LC-S-Coker}} & \gpfit{\mathsf{lc}(\ls{})} & \gpfit{\mathsf{lc}(\mathsf{coker}(\lsn{0}, \lsn{1}, \lee{}))}\\
\gpfitbox{\textsc{LC-S-Img}} & \gpfit{\mathsf{lc}(\ls{})} & \gpfit{\mathsf{lc}(\mathsf{img}(\lsn{0}, \lsn{1}, \lee{}))}\\
\gpfitbox{\textsc{LC-S-Coimg}} & \gpfit{\mathsf{lc}(\ls{})} & \gpfit{\mathsf{lc}(\mathsf{coimg}(\lsn{0}, \lsn{1}, \lee{}))}\\
\gpfitbox{\textsc{LC-T-Space}} & \gpfit{\mathsf{lc}(\lty{})} & \gpfit{\mathsf{lc}(\mathsf{space}(\ls{}))}\\
\gpfitbox{\textsc{LC-T-Pi\allowbreak Ty}} & \gpfit{\mathsf{lc}(\lty{})} & \gpfit{\mathsf{lc}(\mathsf{piTy}(\ltyn{0}, \ltyn{1}))}\\
\gpfitbox{\textsc{LC-E-Var}} & \gpfit{\mathsf{lc}(\lee{})} & \gpfit{\mathsf{lc}(\mathsf{var}(\mathit{x}))}\\
\gpfitbox{\textsc{LC-E-Real}} & \gpfit{\mathsf{lc}(\lee{})} & \gpfit{\mathsf{lc}(\mathsf{real}(\mathit{r}))}\\
\gpfitbox{\textsc{LC-E-Zero}} & \gpfit{\mathsf{lc}(\lee{})} & \gpfit{\mathsf{lc}(\mathsf{zero}(\ls{}))}\\
\gpfitbox{\textsc{LC-E-Smul}} & \gpfit{\mathsf{lc}(\lee{})} & \gpfit{\mathsf{lc}(\mathsf{smul}(\len{0}, \len{1}))}\\
\gpfitbox{\textsc{LC-E-Add}} & \gpfit{\mathsf{lc}(\lee{})} & \gpfit{\mathsf{lc}(\mathsf{add}(\len{0}, \len{1}))}\\
\gpfitbox{\textsc{LC-E-Dsum\allowbreak E}} & \gpfit{\mathsf{lc}(\lee{})} & \gpfit{\mathsf{lc}(\mathsf{dsumE}(\len{0}, \len{1}))}\\
\gpfitbox{\textsc{LC-E-Tprod\allowbreak E}} & \gpfit{\mathsf{lc}(\lee{})} & \gpfit{\mathsf{lc}(\mathsf{tprodE}(\len{0}, \len{1}))}\\
\gpfitbox{\textsc{LC-E-Inner}} & \gpfit{\mathsf{lc}(\lee{})} & \gpfit{\mathsf{lc}(\mathsf{inner}(\ls{}))}\\
\gpfitbox{\textsc{LC-E-Embed}} & \gpfit{\mathsf{lc}(\lee{})} & \gpfit{\mathsf{lc}(\mathsf{embed}(\ls{}))}\\
\gpfitbox{\textsc{LC-E-Pinv\allowbreak T}} & \gpfit{\mathsf{lc}(\lee{})} & \gpfit{\mathsf{lc}(\mathsf{pinvT}(\lsn{0}, \lsn{1}, \lee{}))}\\
\gpfitbox{\textsc{LC-E-Pinv\allowbreak D}} & \gpfit{\mathsf{lc}(\lee{})} & \gpfit{\mathsf{lc}(\mathsf{pinvD}(\ls{}, \lee{}))}\\
\gpfitbox{\textsc{LC-E-Llam}} & \gpfit{\mathsf{lc}(\lee{})} & \gpfit{\mathsf{lc}(\mathsf{llam}(\lsn{0}, \ls{}, \lee{}))}\\
\gpfitbox{\textsc{LC-E-Nllam}} & \gpfit{\mathsf{lc}(\lee{})} & \gpfit{\mathsf{lc}(\mathsf{nllam}(\lty{}, \lee{}))}\\
\gpfitbox{\textsc{LC-E-App\allowbreak L}} & \gpfit{\mathsf{lc}(\lee{})} & \gpfit{\mathsf{lc}(\mathsf{appL}(\lsn{0}, \lsn{1}, \len{0}, \len{1}))}\\
\gpfitbox{\textsc{LC-E-App\allowbreak I}} & \gpfit{\mathsf{lc}(\lee{})} & \gpfit{\mathsf{lc}(\mathsf{appI}(\ls{}, \len{0}, \len{1}))}\\
\gpfitbox{\textsc{LC-E-App\allowbreak N}} & \gpfit{\mathsf{lc}(\lee{})} & \gpfit{\mathsf{lc}(\mathsf{appN}(\len{0}, \len{1}))}\\
\gpfitbox{\textsc{LC-B-Empty}} & \gpfit{\mathsf{lc}(\lb{})} & \gpfit{\mathsf{lc}(\lbzero{})}\\
\gpfitbox{\textsc{LC-B-One}} & \gpfit{\mathsf{lc}(\lb{})} & \gpfit{\mathsf{lc}(\lbone{})}\\
\gpfitbox{\textsc{LC-B-Dual}} & \gpfit{\mathsf{lc}(\lb{})} & \gpfit{\mathsf{lc}(\mathsf{dual}(\lb{}))}\\
\gpfitbox{\textsc{LC-B-Dsum}} & \gpfit{\mathsf{lc}(\lb{})} & \gpfit{\mathsf{lc}(\mathsf{dsum}(\lbn{0}, \lbn{1}))}\\
\gpfitbox{\textsc{LC-B-Tprod}} & \gpfit{\mathsf{lc}(\lb{})} & \gpfit{\mathsf{lc}(\mathsf{tprod}(\lbn{0}, \lbn{1}))}\\
\gpfitbox{\textsc{LC-B-Proj}} & \gpfit{\mathsf{lc}(\lb{})} & \gpfit{\mathsf{lc}(\mathsf{proj}(\ls{}, \lbn{0}, \lbn{1}))}\\
\gpfitbox{\textsc{LC-B-Image}} & \gpfit{\mathsf{lc}(\lb{})} & \gpfit{\mathsf{lc}(\mathsf{image}(\lb{}))}\\
\gpfitbox{\textsc{LC-B-Terms}} & \gpfit{\mathsf{lc}(\lb{})} & \gpfit{\mathsf{lc}(\mathsf{terms}(\ls{}, \mathit{es}, \lb{}))}\\
\gpfitbox{\textsc{LC-B-Frame}} & \gpfit{\mathsf{lc}(\lb{})} & \gpfit{\mathsf{lc}(\mathsf{frame}(\lbn{0}, \lbn{1}))}\\
\gpfitbox{\textsc{LC-VS-Empty}} & \gpfit{\mathsf{lc}(\lsv{})} & \gpfit{\mathsf{lc}(\lsempty{})}\\
\gpfitbox{\textsc{LC-VS-Real}} & \gpfit{\mathsf{lc}(\lsv{})} & \gpfit{\mathsf{lc}(\lR{})}\\
\gpfitbox{\textsc{LC-VS-Dsum}} & \gpfit{\mathsf{lc}(\lsv{})} & \gpfit{\mathsf{lc}(\mathsf{dsum}(\lsnv{0}, \lsnv{1}))}\\
\gpfitbox{\textsc{LC-VS-Tprod}} & \gpfit{\mathsf{lc}(\lsv{})} & \gpfit{\mathsf{lc}(\mathsf{tprod}(\lsnv{0}, \lsnv{1}))}\\
\gpfitbox{\textsc{LC-VS-Ffs}} & \gpfit{\mathsf{lc}(\lsv{})} & \gpfit{\mathsf{lc}(\mathsf{ffs}(\lprop{}, \lsv{}, \leev{}))}\\
\gpfitbox{\textsc{LC-VT-Space}} & \gpfit{\mathsf{lc}(\ltyv{})} & \gpfit{\mathsf{lc}(\mathsf{space}(\lsv{}))}\\
\gpfitbox{\textsc{LC-VT-Pi\allowbreak Ty}} & \gpfit{\mathsf{lc}(\ltyv{})} & \gpfit{\mathsf{lc}(\mathsf{piTy}(\ltynv{0}, \ltynv{1}))}\\
\gpfitbox{\textsc{LC-VE-Var}} & \gpfit{\mathsf{lc}(\leev{})} & \gpfit{\mathsf{lc}(\mathsf{var}(\mathit{x}, \ltyv{}))}\\
\gpfitbox{\textsc{LC-VE-Vlam}} & \gpfit{\mathsf{lc}(\leev{})} & \gpfit{\mathsf{lc}(\mathsf{vlam}(\ltynv{0}, \ltynv{1}, \leev{}))}\\
\gpfitbox{\textsc{LC-VE-Zero}} & \gpfit{\mathsf{lc}(\leev{})} & \gpfit{\mathsf{lc}(\mathsf{zero}(\lsv{}))}\\
\gpfitbox{\textsc{LC-VE-Rlit}} & \gpfit{\mathsf{lc}(\leev{})} & \gpfit{\mathsf{lc}(\mathsf{rlit}(\mathit{r}))}\\
\gpfitbox{\textsc{LC-VE-Scale}} & \gpfit{\mathsf{lc}(\leev{})} & \gpfit{\mathsf{lc}(\mathsf{scale}(\lenv{0}, \lenv{1}))}\\
\gpfitbox{\textsc{LC-VE-Vadd}} & \gpfit{\mathsf{lc}(\leev{})} & \gpfit{\mathsf{lc}(\mathsf{vadd}(\lsv{}, \lenv{0}, \lenv{1}))}\\
\gpfitbox{\textsc{LC-VE-Dsum\allowbreak E}} & \gpfit{\mathsf{lc}(\leev{})} & \gpfit{\mathsf{lc}(\mathsf{dsumE}(\lenv{0}, \lenv{1}))}\\
\gpfitbox{\textsc{LC-VE-Tprod\allowbreak E}} & \gpfit{\mathsf{lc}(\leev{})} & \gpfit{\mathsf{lc}(\mathsf{tprodE}(\lenv{0}, \lenv{1}))}\\
\gpfitbox{\textsc{LC-VE-Delta}} & \gpfit{\mathsf{lc}(\leev{})} & \gpfit{\mathsf{lc}(\mathsf{delta}(\mathit{n}, \lsv{}))}\\
\gpfitbox{\textsc{LC-VE-Idelta}} & \gpfit{\mathsf{lc}(\leev{})} & \gpfit{\mathsf{lc}(\mathsf{idelta}(\lsv{}))}\\
\gpfitbox{\textsc{LC-VE-Vsum}} & \gpfit{\mathsf{lc}(\leev{})} & \gpfit{\mathsf{lc}(\mathsf{vsum}(\lsv{}, \mathit{es}))}\\
\gpfitbox{\textsc{LC-VE-Ttv}} & \gpfit{\mathsf{lc}(\leev{})} & \gpfit{\mathsf{lc}(\mathsf{ttv}(\lsv{}, \lenv{0}, \lenv{1}))}\\
\gpfitbox{\textsc{LC-VE-Mtm}} & \gpfit{\mathsf{lc}(\leev{})} & \gpfit{\mathsf{lc}(\mathsf{mtm}({\mathit{n}_{0}}, {\mathit{n}_{1}}, \lsnv{0}, \lsnv{1}, \lenv{0}, \lenv{1}))}\\
\gpfitbox{\textsc{LC-VE-App\allowbreak V}} & \gpfit{\mathsf{lc}(\leev{})} & \gpfit{\mathsf{lc}(\mathsf{appV}(\lenv{0}, \lenv{1}))}\\
\gpfitbox{\textsc{LC-VE-Vkrn}} & \gpfit{\mathsf{lc}(\leev{})} & \gpfit{\mathsf{lc}(\mathsf{vkrn}(\lenv{0}, \lenv{1}))}\\
\gpfitbox{\textsc{LC-VE-Vblk}} & \gpfit{\mathsf{lc}(\leev{})} & \gpfit{\mathsf{lc}(\mathsf{vblk}(\lenv{0}, \lenv{1}))}\\
\gpfitbox{\textsc{LC-VE-Reshap}} & \gpfit{\mathsf{lc}(\leev{})} & \gpfit{\mathsf{lc}(\mathsf{reshap}(\leev{}, \ltyv{}, \ltynv{0}))}\\
\gpfitbox{\textsc{LC-VE-Vtrans}} & \gpfit{\mathsf{lc}(\leev{})} & \gpfit{\mathsf{lc}(\mathsf{vtrans}(\leev{}))}\\
\gpfitbox{\textsc{LC-VE-Gsvd\allowbreak S}} & \gpfit{\mathsf{lc}(\leev{})} & \gpfit{\mathsf{lc}(\mathsf{gsvdS}(\leev{}, \lenv{{\mathrm{a}}}, \lenv{{\mathrm{b}}}))}\\
\gpfitbox{\textsc{LC-VE-Gsvd\allowbreak U}} & \gpfit{\mathsf{lc}(\leev{})} & \gpfit{\mathsf{lc}(\mathsf{gsvdU}(\leev{}, \lenv{{\mathrm{a}}}, \lenv{{\mathrm{b}}}))}\\
\gpfitbox{\textsc{LC-VE-Gsvd\allowbreak UI}} & \gpfit{\mathsf{lc}(\leev{})} & \gpfit{\mathsf{lc}(\mathsf{gsvdUI}(\leev{}, \lenv{{\mathrm{a}}}, \lenv{{\mathrm{b}}}))}\\
\gpfitbox{\textsc{LC-VE-Gsvd\allowbreak V}} & \gpfit{\mathsf{lc}(\leev{})} & \gpfit{\mathsf{lc}(\mathsf{gsvdV}(\leev{}, \lenv{{\mathrm{a}}}, \lenv{{\mathrm{b}}}))}\\
\gpfitbox{\textsc{LC-VE-Gsvd\allowbreak VI}} & \gpfit{\mathsf{lc}(\leev{})} & \gpfit{\mathsf{lc}(\mathsf{gsvdVI}(\leev{}, \lenv{{\mathrm{a}}}, \lenv{{\mathrm{b}}}))}\\
\gpfitbox{\textsc{LC-VE-Pinv}} & \gpfit{\mathsf{lc}(\leev{})} & \gpfit{\mathsf{lc}(\mathsf{pinv}(\leev{}))}\\
\gpfitbox{\textsc{LC-VE-Gpinv}} & \gpfit{\mathsf{lc}(\leev{})} & \gpfit{\mathsf{lc}(\mathsf{gpinv}(\leev{}, \lenv{{\mathrm{a}}}, \lenv{{\mathrm{b}}}))}\\
\gpfitbox{\textsc{LC-VE-Rref\allowbreak Rep}} & \gpfit{\mathsf{lc}(\leev{})} & \gpfit{\mathsf{lc}(\mathsf{rrefRep}(\lenv{0}, \lenv{1}))}\\
\gpfitbox{\textsc{LC-VE-Inv}} & \gpfit{\mathsf{lc}(\leev{})} & \gpfit{\mathsf{lc}(\mathsf{inv}(\leev{}))}\\
\gpfitbox{\textsc{LC-VE-Svds}} & \gpfit{\mathsf{lc}(\leev{})} & \gpfit{\mathsf{lc}(\mathsf{svds}(\leev{}))}\\
\gpfitbox{\textsc{VS-Real}} & \gpfit{\valCheckCtx\vdash\lsv{}\colon\svalid} & \gpfit{\valCheckCtx{}\vdash \lR{}\colon\svalid}\\
\gpfitbox{\textsc{VS-Empty}} & \gpfit{\valCheckCtx\vdash\lsv{}\colon\svalid} & \gpfit{\valCheckCtx{}\vdash \lsempty{}\colon\svalid}\\
\gpfitbox{\textsc{VS-Dsum}} & \gpfit{\valCheckCtx\vdash\lsv{}\colon\svalid} & \gpfit{\valCheckCtx{}\vdash \lsnv{0} \losum \lsnv{1}\colon\svalid}\\
\gpfitbox{\textsc{VS-Tprod}} & \gpfit{\valCheckCtx\vdash\lsv{}\colon\svalid} & \gpfit{\valCheckCtx{}\vdash \lsnv{0} \loprod \lsnv{1}\colon\svalid}\\
\gpfitbox{\textsc{VS-Ffs}} & \gpfit{\valCheckCtx\vdash\lsv{}\colon\svalid} & \gpfit{\valCheckCtx{}\vdash \lR^{\leev{},\lprop{},\lsnv{0} \loprod \lsnv{1}}\colon\svalid}\\
\gpfitbox{\textsc{VK-Space}} & \gpfit{\valCheckCtx\vdash\ltyv{}\colon\lkn{}} & \gpfit{\valCheckCtx{}\vdash \lsv{}\colon \lkbase{}}\\
\gpfitbox{\textsc{VK-Pi\allowbreak Ty}} & \gpfit{\valCheckCtx\vdash\ltyv{}\colon\lkn{}} & \gpfit{\valCheckCtx{}\vdash (x\colon \ltynv{0}\fnto \ltynv{1})\colon \lkn{0} \fnto \lkn{1}}\\
\gpfitbox{\textsc{VT-Var}} & \gpfit{\valCheckCtx\vdash\leev{}\colon\ltyv{}} & \gpfit{\valCheckCtx{}\vdash \mathit{x}_{\ltyv{}}\colon \ltyv{}}\\
\gpfitbox{\textsc{VT-Zero}} & \gpfit{\valCheckCtx\vdash\leev{}\colon\ltyv{}} & \gpfit{\valCheckCtx{}\vdash \lvzero{}_{\lsv{}}\colon \lsv{}}\\
\gpfitbox{\textsc{VT-Rlit}} & \gpfit{\valCheckCtx\vdash\leev{}\colon\ltyv{}} & \gpfit{\valCheckCtx{}\vdash \mathit{r}\colon \lR{}}\\
\gpfitbox{\textsc{VT-Scale}} & \gpfit{\valCheckCtx\vdash\leev{}\colon\ltyv{}} & \gpfit{\valCheckCtx{}\vdash \lenv{0} \lvscale \lenv{1}\colon \ltyv{}}\\
\gpfitbox{\textsc{VT-Vadd}} & \gpfit{\valCheckCtx\vdash\leev{}\colon\ltyv{}} & \gpfit{\valCheckCtx{}\vdash \lenv{0} \lvadd_{\lsv{}} \lenv{1}\colon \lsv{}}\\
\gpfitbox{\textsc{VT-Dsum\allowbreak E}} & \gpfit{\valCheckCtx\vdash\leev{}\colon\ltyv{}} & \gpfit{\valCheckCtx{}\vdash \lenv{0}\losum \lenv{1}\colon \lsnv{0} \losum \lsnv{1}}\\
\gpfitbox{\textsc{VT-Tprod\allowbreak E}} & \gpfit{\valCheckCtx\vdash\leev{}\colon\ltyv{}} & \gpfit{\valCheckCtx{}\vdash \lenv{0}\loprod \lenv{1}\colon \lsnv{0} \loprod \lsnv{1}}\\
\gpfitbox{\textsc{VT-Delta}} & \gpfit{\valCheckCtx\vdash\leev{}\colon\ltyv{}} & \gpfit{\valCheckCtx{}\vdash \ldelta{\mathit{n}}{\lsv{}}\colon \lsv{}}\\
\gpfitbox{\textsc{VT-Idelta}} & \gpfit{\valCheckCtx\vdash\leev{}\colon\ltyv{}} & \gpfit{\valCheckCtx{}\vdash \lidelta{\lsv{}}\colon \lsv{} \loprod \lsv{}}\\
\gpfitbox{\textsc{VT-Ttv\allowbreak Space}} & \gpfit{\valCheckCtx\vdash\leev{}\colon\ltyv{}} & \gpfit{\valCheckCtx{}\vdash \LTTVPP{\lsnv{0}''}{\lenv{0}}{\lenv{1}}\colon \lsnv{1}}\\
\gpfitbox{\textsc{VT-Ttv\allowbreak Real}} & \gpfit{\valCheckCtx\vdash\leev{}\colon\ltyv{}} & \gpfit{\valCheckCtx{}\vdash \LTTVPP{\lsnv{0}''}{\lenv{0}}{\lenv{1}}\colon \lR{}}\\
\gpfitbox{\textsc{VT-Mtm}} & \gpfit{\valCheckCtx\vdash\leev{}\colon\ltyv{}} & \gpfit{\valCheckCtx{}\vdash \glvmtm{\lsnv{1}''}{\lsnv{1}}{\lenv{0}}{\lenv{1}}\colon \lsnv{2} \loprod \lsnv{0}}\\
\gpfitbox{\textsc{VT-Vsum}} & \gpfit{\valCheckCtx\vdash\leev{}\colon\ltyv{}} & \gpfit{\valCheckCtx{}\vdash \lsum^{\lsnv{0}}_{v\in \mathit{es}}v\colon \lsv{}}\\
\gpfitbox{\textsc{VT-Vlam}} & \gpfit{\valCheckCtx\vdash\leev{}\colon\ltyv{}} & \gpfit{\valCheckCtx{}\vdash \llambda(x\colon \ltynv{0})\lrh \ltynv{1}\colon \leev{}\colon (x\colon \ltynv{0}\fnto \ltynv{1})}\\
\gpfitbox{\textsc{VT-App\allowbreak V}} & \gpfit{\valCheckCtx\vdash\leev{}\colon\ltyv{}} & \gpfit{\valCheckCtx{}\vdash \glvapp{\lenv{0}}{\lenv{1}}\colon {\ltyv{}_{\mathsf{out}}}}\\
\gpfitbox{\textsc{VT-Vkrn}} & \gpfit{\valCheckCtx\vdash\leev{}\colon\ltyv{}} & \gpfit{\valCheckCtx{}\vdash \vkrn(\leev{}, \leev{}')\colon \left(\lsnv{0} \loprod \lsnv{1}\right) \loprod \left(\lsnv{2} \loprod \lsnv{3}\right)}\\
\gpfitbox{\textsc{VT-Vblk}} & \gpfit{\valCheckCtx\vdash\leev{}\colon\ltyv{}} & \gpfit{\valCheckCtx{}\vdash \vblk(\leev{}, \leev{}')\colon \left(\lsnv{0} \losum \lsnv{1}\right) \loprod \left(\lsnv{2} \losum \lsnv{3}\right)}\\
\gpfitbox{\textsc{VT-Reshap}} & \gpfit{\valCheckCtx\vdash\leev{}\colon\ltyv{}} & \gpfit{\valCheckCtx{}\vdash \glvreshape{\leev{}}{\ltyv{}}{\ltynv{0}}\colon \ltyv{}}\\
\gpfitbox{\textsc{VT-Vtrans}} & \gpfit{\valCheckCtx\vdash\leev{}\colon\ltyv{}} & \gpfit{\valCheckCtx{}\vdash \Trans(\leev{})\colon \lsnv{1} \loprod \lsnv{0}}\\
\gpfitbox{\textsc{VT-Gsvd\allowbreak S}} & \gpfit{\valCheckCtx\vdash\leev{}\colon\ltyv{}} & \gpfit{\valCheckCtx{}\vdash \glvop{GSVDS}(\leev{},\lenv{{\mathrm{a}}},\lenv{{\mathrm{b}}})\colon \lR^{\leev{},\CoKernel{},\lsnv{0} \loprod \lsnv{1}} \loprod \lR^{\leev{},\Image{},\lsnv{0} \loprod \lsnv{1}}}\\
\gpfitbox{\textsc{VT-Gsvd\allowbreak U}} & \gpfit{\valCheckCtx\vdash\leev{}\colon\ltyv{}} & \gpfit{\valCheckCtx{}\vdash \glvop{GSVDU}(\leev{},\lenv{{\mathrm{a}}},\lenv{{\mathrm{b}}})\colon \lsnv{0} \loprod \lR^{\leev{},\CoKernel{},\lsnv{0} \loprod \lsnv{1}}}\\
\gpfitbox{\textsc{VT-Gsvd\allowbreak UI}} & \gpfit{\valCheckCtx\vdash\leev{}\colon\ltyv{}} & \gpfit{\valCheckCtx{}\vdash \glvop{GSVDUI}(\leev{},\lenv{{\mathrm{a}}},\lenv{{\mathrm{b}}})\colon \lsnv{0} \loprod \lR^{\leev{},\Kernel{},\lsnv{0} \loprod \lsnv{1}}}\\
\gpfitbox{\textsc{VT-Gsvd\allowbreak V}} & \gpfit{\valCheckCtx\vdash\leev{}\colon\ltyv{}} & \gpfit{\valCheckCtx{}\vdash \glvop{GSVDV}(\leev{},\lenv{{\mathrm{a}}},\lenv{{\mathrm{b}}})\colon \lsnv{1} \loprod \lR^{\leev{},\Image{},\lsnv{0} \loprod \lsnv{1}}}\\
\gpfitbox{\textsc{VT-Gsvd\allowbreak VI}} & \gpfit{\valCheckCtx\vdash\leev{}\colon\ltyv{}} & \gpfit{\valCheckCtx{}\vdash \glvop{GSVDVI}(\leev{},\lenv{{\mathrm{a}}},\lenv{{\mathrm{b}}})\colon \lsnv{1} \loprod \lR^{\leev{},\CoImage{},\lsnv{0} \loprod \lsnv{1}}}\\
\gpfitbox{\textsc{VT-Pinv\allowbreak Tensor}} & \gpfit{\valCheckCtx\vdash\leev{}\colon\ltyv{}} & \gpfit{\valCheckCtx{}\vdash \leev{}^{+}\colon \lsnv{1} \loprod \lsnv{0}}\\
\gpfitbox{\textsc{VT-Pinv\allowbreak Raw}} & \gpfit{\valCheckCtx\vdash\leev{}\colon\ltyv{}} & \gpfit{\valCheckCtx{}\vdash \leev{}^{+}\colon \lR{} \loprod \lsnv{0}}\\
\gpfitbox{\textsc{VT-Gpinv}} & \gpfit{\valCheckCtx\vdash\leev{}\colon\ltyv{}} & \gpfit{\valCheckCtx{}\vdash \glvop{GPINV}(\leev{},\lenv{{\mathrm{a}}},\lenv{{\mathrm{b}}})\colon \lsnv{1} \loprod \lsnv{0}}\\
\gpfitbox{\textsc{VT-Rref\allowbreak Rep}} & \gpfit{\valCheckCtx\vdash\leev{}\colon\ltyv{}} & \gpfit{\valCheckCtx{}\vdash \glvop{RREFRep}(\lenv{0},\lenv{1})\colon \lsnv{{\mathrm{p1}}} \loprod \lsnv{{\mathrm{p0}}}}\\
\gpfitbox{\textsc{VT-Inv\allowbreak Tensor}} & \gpfit{\valCheckCtx\vdash\leev{}\colon\ltyv{}} & \gpfit{\valCheckCtx{}\vdash \leev{}^{-1}\colon \lsnv{1} \loprod \lsnv{0}}\\
\gpfitbox{\textsc{VT-Inv\allowbreak Raw}} & \gpfit{\valCheckCtx\vdash\leev{}\colon\ltyv{}} & \gpfit{\valCheckCtx{}\vdash \leev{}^{-1}\colon \lR{} \loprod \lsnv{0}}\\
\gpfitbox{\textsc{VT-Svds}} & \gpfit{\valCheckCtx\vdash\leev{}\colon\ltyv{}} & \gpfit{\valCheckCtx{}\vdash \glvop{SVDS}(\leev{})\colon \lR^{\leev{},\CoKernel{},\lsnv{0} \loprod \lsnv{1}} \loprod \lR^{\leev{},\Image{},\lsnv{0} \loprod \lsnv{1}}}\\
\gpfitbox{\textsc{ES-Real}} & \gpfit{\evalSJudgeK{\ls{}}{\lsv{}}} & \gpfit{\evalSJudge{\evalCtx{}}{\lR{}}{\lR{}}}\\
\gpfitbox{\textsc{ES-Empty}} & \gpfit{\evalSJudgeK{\ls{}}{\lsv{}}} & \gpfit{\evalSJudge{\evalCtx{}}{\lsempty{}}{\lsempty{}}}\\
\gpfitbox{\textsc{ES-Dual}} & \gpfit{\evalSJudgeK{\ls{}}{\lsv{}}} & \gpfit{\evalSJudge{\evalCtx{}}{\ls{}^{\lstar}}{\lsv{}}}\\
\gpfitbox{\textsc{ES-Ker}} & \gpfit{\evalSJudgeK{\ls{}}{\lsv{}}} & \gpfit{\evalSJudge{\evalCtx{}}{\Kernel(\lee{})}{\lR^{\glvreshape{\leev{}}{\lsnv{{\mathrm{svb}}}}{\lsnv{{\mathrm{svb}}}},\Kernel{},\lsnv{{\mathrm{svb}}}}}}\\
\gpfitbox{\textsc{ES-Coker}} & \gpfit{\evalSJudgeK{\ls{}}{\lsv{}}} & \gpfit{\evalSJudge{\evalCtx{}}{\CoKernel(\lee{})}{\lR^{\glvreshape{\leev{}}{\lsnv{{\mathrm{svb}}}}{\lsnv{{\mathrm{svb}}}},\CoKernel{},\lsnv{{\mathrm{svb}}}}}}\\
\gpfitbox{\textsc{ES-Img}} & \gpfit{\evalSJudgeK{\ls{}}{\lsv{}}} & \gpfit{\evalSJudge{\evalCtx{}}{\Image(\lee{})}{\lR^{\glvreshape{\leev{}}{\lsnv{{\mathrm{svb}}}}{\lsnv{{\mathrm{svb}}}},\Image{},\lsnv{{\mathrm{svb}}}}}}\\
\gpfitbox{\textsc{ES-Coimg}} & \gpfit{\evalSJudgeK{\ls{}}{\lsv{}}} & \gpfit{\evalSJudge{\evalCtx{}}{\CoImage(\lee{})}{\lR^{\glvreshape{\leev{}}{\lsnv{{\mathrm{svb}}}}{\lsnv{{\mathrm{svb}}}},\CoImage{},\lsnv{{\mathrm{svb}}}}}}\\
\gpfitbox{\textsc{ES-Tprod}} & \gpfit{\evalSJudgeK{\ls{}}{\lsv{}}} & \gpfit{\evalSJudge{\evalCtx{}}{\lsn{0} \loprod \lsn{1}}{\lsnv{0} \loprod \lsnv{1}}}\\
\gpfitbox{\textsc{ES-Dsum}} & \gpfit{\evalSJudgeK{\ls{}}{\lsv{}}} & \gpfit{\evalSJudge{\evalCtx{}}{\lsn{0} \losum \lsn{1}}{\lsnv{0} \losum \lsnv{1}}}\\
\gpfitbox{\textsc{ET-Space}} & \gpfit{\evalSJudgeK{\lty{}}{\ltyv{}}} & \gpfit{\evalSJudge{\evalCtx{}}{\ls{}}{\lsv{}}}\\
\gpfitbox{\textsc{ET-Pi\allowbreak Ty}} & \gpfit{\evalSJudgeK{\lty{}}{\ltyv{}}} & \gpfit{\evalSJudge{\evalCtx{}}{(x\colon \ltyn{0}\fnto \ltyn{1})}{(x\colon \ltynv{0}\fnto \ltynv{1})}}\\
\gpfitbox{\textsc{EB-Inverse}} & \gpfit{\evalBBJudgeK{\lbn{0}}{\leev{}}{\lbn{1}}} & \gpfit{\evalBBJudge{\evalCtx{}}{\lbn{0}}{\leev{}^{-1}}{\lbn{1}}}\\
\gpfitbox{\textsc{EB-Trans}} & \gpfit{\evalBBJudgeK{\lbn{0}}{\leev{}}{\lbn{1}}} & \gpfit{\evalBBJudge{\evalCtx{}}{\lbn{0}}{\glvmtm{\lsv{}}{\lsv{}}{\lenv{1}}{\glvreshape{\lenv{0}}{\lsv{} \loprod \lsv{}}{\lsv{} \loprod \lsv{}}}}{\lbn{2}}}\\
\gpfitbox{\textsc{EB-Tprod}} & \gpfit{\evalBBJudgeK{\lbn{0}}{\leev{}}{\lbn{1}}} & \gpfit{\evalBBJudge{\evalCtx{}}{\lbn{0}' \loprod \lbn{1}'}{\vkrn(\lenv{0}, \lenv{1})}{\lbn{0} \loprod \lbn{1}}}\\
\gpfitbox{\textsc{EB-Dsum}} & \gpfit{\evalBBJudgeK{\lbn{0}}{\leev{}}{\lbn{1}}} & \gpfit{\evalBBJudge{\evalCtx{}}{\lbn{0}' \losum \lbn{1}'}{\vblk(\lenv{0}, \lenv{1})}{\lbn{0} \losum \lbn{1}}}\\
\gpfitbox{\textsc{EB-Dual\allowbreak Std}} & \gpfit{\evalBBJudgeK{\lbn{0}}{\leev{}}{\lbn{1}}} & \gpfit{\evalBBJudge{\evalCtx{}}{\lbn{0}^{\lstar}}{\Trans(\leev{}^{-1})}{\stdBasisDCtx{\tof{\lbn{0}^{\lstar}}}}}\\
\gpfitbox{\textsc{EB-Image}} & \gpfit{\evalBBJudgeK{\lbn{0}}{\leev{}}{\lbn{1}}} & \gpfit{\evalBBJudge{\evalCtx{}}{\iBasisC{\lbn{0}}}{\leev{}}{\iBasisC{\lbn{0}'}}}\\
\gpfitbox{\textsc{EB-Image\allowbreak Svds}} & \gpfit{\evalBBJudgeK{\lbn{0}}{\leev{}}{\lbn{1}}} & \gpfit{\begin{gathered}\evalBBJudge{\evalCtx{}}{\iBasisC{\pBasisC{\CoKernel(\lee{})}{\lbn{0}}{\lbn{1}}}}{\widehat{\leev{}}}{\pBasisC{\Image(\lee{})}{\lbn{0}}{\lbn{1}}}\\[-1pt]\widehat{\leev{}}=\glvreshape{\lenv{{\mathrm{P}}}}{\lR^{\lenv{{\mathrm{M}}},\CoKernel{},\lsnv{0} \loprod \lsnv{1}} \loprod \lR^{\lenv{{\mathrm{M}}},\Image{},\lsnv{0} \loprod \lsnv{1}}}{\operatorname{inferSpace}(\lenv{{\mathrm{P}}})}\end{gathered}}\\
\gpfitbox{\textsc{EB-Proj\allowbreak Ker\allowbreak Coker}} & \gpfit{\evalBBJudgeK{\lbn{0}}{\leev{}}{\lbn{1}}} & \gpfit{\begin{gathered}\evalBBJudge{\evalCtx{}}{\pBasisC{\ls{}}{\lbn{0}}{\lbn{1}}}{\widehat{\leev{}}}{\pBasisC{\ls{}}{\lbn{0}'}{\lbn{1}}}\\[-1pt]\widehat{\leev{}}=\glvmtm{\lsnv{0}'}{\lsnv{0}'}{\glvreshape{\lenv{1}}{\lsnv{0}' \loprod \lsv{}'}{\lsnv{0}' \loprod \lsv{}'}}{\glvmtm{\lsnv{0}}{\lsnv{0}}{\glvreshape{\lenv{2}}{\lsnv{0} \loprod \lsnv{0}'}{\lsnv{0} \loprod \lsnv{0}'}}{\glvreshape{\lenv{3}}{\lsv{} \loprod \lsnv{0}}{\lsv{} \loprod \lsnv{0}}}}\end{gathered}}\\
\gpfitbox{\textsc{EB-Proj\allowbreak Img\allowbreak Coimg}} & \gpfit{\evalBBJudgeK{\lbn{0}}{\leev{}}{\lbn{1}}} & \gpfit{\begin{gathered}\evalBBJudge{\evalCtx{}}{\pBasisC{\ls{}}{\lbn{0}}{\lbn{1}}}{\widehat{\leev{}}}{\pBasisC{\ls{}}{\lbn{0}}{\lbn{1}'}}\\[-1pt]\widehat{\leev{}}=\glvmtm{\lsnv{1}'}{\lsnv{1}'}{\glvreshape{\lenv{1}}{\lsnv{1}' \loprod \lsv{}'}{\lsnv{1}' \loprod \lsv{}'}}{\glvmtm{\lsnv{1}}{\lsnv{1}}{\glvreshape{\lenv{2}}{\lsnv{1} \loprod \lsnv{1}'}{\lsnv{1} \loprod \lsnv{1}'}}{\glvreshape{\lenv{3}}{\lsv{} \loprod \lsnv{1}}{\lsv{} \loprod \lsnv{1}}}}\end{gathered}}\\
\gpfitbox{\textsc{EB-Terms\allowbreak Rref}} & \gpfit{\evalBBJudgeK{\lbn{0}}{\leev{}}{\lbn{1}}} & \gpfit{\evalBBJudge{\evalCtx{}}{\lb{}}{\glvreshape{\glvop{RREFRep}(\lenv{0},\glvreshape{\lenv{1}}{\lsv{} \loprod \lsv{}}{\lsv{} \loprod \lsv{}})}{\lsv{} \loprod \lsv{}}{\lsv{} \loprod \lsv{}}}{\llpl{} \lee{} \mathbin{:\!:} \mathit{es}\llpr{}_{\lb{}'}}}\\
\gpfitbox{\textsc{EB-Refl}} & \gpfit{\evalBBJudgeK{\lbn{0}}{\leev{}}{\lbn{1}}} & \gpfit{\evalBBJudge{\evalCtx{}}{\lb{}}{\glvreshape{\lidelta{\lsv{}}}{\lsv{} \loprod \lsv{}}{\lsv{} \loprod \lsv{}}}{\lb{}}}\\
\gpfitbox{\textsc{EB-Proj\allowbreak Ker\allowbreak Coker\allowbreak Cod}} & \gpfit{\evalBBJudgeK{\lbn{0}}{\leev{}}{\lbn{1}}} & \gpfit{\begin{gathered}\evalBBJudge{\evalCtx{}}{\pBasisC{\ls{}}{\lbn{0}}{\lbn{1}}}{\widehat{\leev{}}}{\pBasisC{\ls{}}{\lbn{0}}{\lbn{1}'}}\\[-1pt]\widehat{\leev{}}=\glvmtm{\lsnv{0}}{\lsnv{0}}{\glvreshape{\lenv{1}}{\lsnv{0} \loprod \lsv{}'}{\lsnv{0} \loprod \lsv{}'}}{\glvreshape{\lenv{3}}{\lsv{} \loprod \lsnv{0}}{\lsv{} \loprod \lsnv{0}}}\end{gathered}}\\
\gpfitbox{\textsc{EB-Proj\allowbreak Img\allowbreak Coimg\allowbreak Dom}} & \gpfit{\evalBBJudgeK{\lbn{0}}{\leev{}}{\lbn{1}}} & \gpfit{\begin{gathered}\evalBBJudge{\evalCtx{}}{\pBasisC{\ls{}}{\lbn{0}}{\lbn{1}}}{\widehat{\leev{}}}{\pBasisC{\ls{}}{\lbn{0}'}{\lbn{1}}}\\[-1pt]\widehat{\leev{}}=\glvmtm{\lsnv{1}}{\lsnv{1}}{\glvreshape{\lenv{1}}{\lsnv{1} \loprod \lsv{}'}{\lsnv{1} \loprod \lsv{}'}}{\glvreshape{\lenv{3}}{\lsv{} \loprod \lsnv{1}}{\lsv{} \loprod \lsnv{1}}}\end{gathered}}\\
\gpfitbox{\textsc{E-Var}} & \gpfit{\evalJudgeK{\lee{}}{\lb{}}{\leev{}}} & \gpfit{\evalJudge{\evalCtx{}}{\mathit{x}}{\lb{}}{\leev{}}}\\
\gpfitbox{\textsc{E-Cob}} & \gpfit{\evalJudgeK{\lee{}}{\lb{}}{\leev{}}} & \gpfit{\evalJudge{\evalCtx{}}{\lee{}}{\lbn{0}}{\LTTVPP{\lsv{}}{\glvreshape{\lenv{0}}{\lsv{} \loprod \lsv{}}{\lsv{} \loprod \lsv{}}}{\glvreshape{\leev{}}{\lsv{}}{\lsv{}}}}}\\
\gpfitbox{\textsc{E-Scale}} & \gpfit{\evalJudgeK{\lee{}}{\lb{}}{\leev{}}} & \gpfit{\evalJudge{\evalCtx{}}{\len{0} \lvscale \len{1}}{\lb{}}{\lenv{0} \lvscale \lenv{1}}}\\
\gpfitbox{\textsc{E-Add}} & \gpfit{\evalJudgeK{\lee{}}{\lb{}}{\leev{}}} & \gpfit{\evalJudge{\evalCtx{}}{\len{0} \lvadd \len{1}}{\lb{}}{\lenv{0} \lvadd_{\lsv{}} \lenv{1}}}\\
\gpfitbox{\textsc{E-Dsum\allowbreak E}} & \gpfit{\evalJudgeK{\lee{}}{\lb{}}{\leev{}}} & \gpfit{\evalJudge{\evalCtx{}}{\len{0}\losum \len{1}}{\lbn{0} \losum \lbn{1}}{\lenv{0}\losum \lenv{1}}}\\
\gpfitbox{\textsc{E-Tprod\allowbreak E}} & \gpfit{\evalJudgeK{\lee{}}{\lb{}}{\leev{}}} & \gpfit{\evalJudge{\evalCtx{}}{\len{0}\loprod \len{1}}{\lbn{0} \loprod \lbn{1}}{\lenv{0}\loprod \lenv{1}}}\\
\gpfitbox{\textsc{E-Pinv\allowbreak TStd}} & \gpfit{\evalJudgeK{\lee{}}{\lb{}}{\leev{}}} & \gpfit{\evalJudge{\evalCtx{}}{\lee{}^{\lpseudo}}{\lbn{0}^{\lstar} \loprod \lbn{1}}{\leev{}^{+}}}\\
\gpfitbox{\textsc{E-Pinv\allowbreak TGpinv}} & \gpfit{\evalJudgeK{\lee{}}{\lb{}}{\leev{}}} & \gpfit{\evalJudge{\evalCtx{}}{\lee{}^{\lpseudo}}{\lbn{0}^{\lstar} \loprod \lbn{1}}{\glvop{GPINV}(\leev{},\lenv{1},\lenv{0})}}\\
\gpfitbox{\textsc{E-Pinv\allowbreak DStd}} & \gpfit{\evalJudgeK{\lee{}}{\lb{}}{\leev{}}} & \gpfit{\evalJudge{\evalCtx{}}{\lee{}^{\lpseudo}}{\lbn{0}^{\lstar} \loprod \lbn{1}}{\glvreshape{\leev{}}{\lsnv{1} \loprod \lR{}}{\lsnv{1}}^{+}}}\\
\gpfitbox{\textsc{E-Pinv\allowbreak DGpinv}} & \gpfit{\evalJudgeK{\lee{}}{\lb{}}{\leev{}}} & \gpfit{\evalJudge{\evalCtx{}}{\lee{}^{\lpseudo}}{\lbn{0}^{\lstar} \loprod \lbn{1}}{\glvop{GPINV}(\glvreshape{\leev{}}{\lsnv{1} \loprod \lR{}}{\lsnv{1}},\lenv{1},\lenv{0})}}\\
\gpfitbox{\textsc{E-Inner\allowbreak Std}} & \gpfit{\evalJudgeK{\lee{}}{\lb{}}{\leev{}}} & \gpfit{\evalJudge{\evalCtx{}}{\linner{}_{\ls{}}}{\lbn{0}^{\lstar} \loprod \lbn{0}^{\lstar}}{\lidelta{\lsv{}}}}\\
\gpfitbox{\textsc{E-Embed\allowbreak Ker}} & \gpfit{\evalJudgeK{\lee{}}{\lb{}}{\leev{}}} & \gpfit{\evalJudge{\evalCtx{}}{\lembd{}_{\Kernel(\lee{})}}{\pBasisC{\Kernel(\lee{})}{\lbn{0}}{\lbn{1}}^{\lstar} \loprod \lbn{0}}{\Trans(\glvop{GSVDUI}(\leev{},\lenv{0},\lenv{1}))}}\\
\gpfitbox{\textsc{E-Embed\allowbreak Coker}} & \gpfit{\evalJudgeK{\lee{}}{\lb{}}{\leev{}}} & \gpfit{\evalJudge{\evalCtx{}}{\lembd{}_{\CoKernel(\lee{})}}{\pBasisC{\CoKernel(\lee{})}{\lbn{0}}{\lbn{1}}^{\lstar} \loprod \lbn{0}}{\Trans(\glvop{GSVDU}(\leev{},\lenv{0},\lenv{1}))}}\\
\gpfitbox{\textsc{E-Embed\allowbreak Img}} & \gpfit{\evalJudgeK{\lee{}}{\lb{}}{\leev{}}} & \gpfit{\evalJudge{\evalCtx{}}{\lembd{}_{\Image(\lee{})}}{\pBasisC{\Image(\lee{})}{\lbn{0}}{\lbn{1}}^{\lstar} \loprod \lbn{1}}{\Trans(\glvop{GSVDV}(\leev{},\lenv{0},\lenv{1}))}}\\
\gpfitbox{\textsc{E-Embed\allowbreak Coimg}} & \gpfit{\evalJudgeK{\lee{}}{\lb{}}{\leev{}}} & \gpfit{\evalJudge{\evalCtx{}}{\lembd{}_{\CoImage(\lee{})}}{\pBasisC{\CoImage(\lee{})}{\lbn{0}}{\lbn{1}}^{\lstar} \loprod \lbn{1}}{\Trans(\glvop{GSVDVI}(\leev{},\lenv{0},\lenv{1}))}}\\
\gpfitbox{\textsc{E-App\allowbreak LLin}} & \gpfit{\evalJudgeK{\lee{}}{\lb{}}{\leev{}}} & \gpfit{\evalJudge{\evalCtx{}}{\leval{\len{0}}{}{\len{1}}}{\lbn{1}}{\glvreshape{\LTTVPP{\lsnv{0}}{\lenv{0}}{\lenv{1}}}{\lsnv{1}}{\lsnv{1}}}}\\
\gpfitbox{\textsc{E-App\allowbreak ILin}} & \gpfit{\evalJudgeK{\lee{}}{\lb{}}{\leev{}}} & \gpfit{\evalJudge{\evalCtx{}}{\leval{\len{0}}{}{\len{1}}}{\lbone{}}{\LTTVPP{\lsv{}}{\glvreshape{\lenv{0}}{\lsv{} \loprod \lR{}}{\lsv{}}}{\lenv{1}}}}\\
\gpfitbox{\textsc{E-Nllam}} & \gpfit{\evalJudgeK{\lee{}}{\lb{}}{\leev{}}} & \gpfit{\evalJudge{\evalCtx{}}{(\nllambda \llpl x\colon \ltyn{0}\llpr \ldotsp \lee{})}{\frameB{x}{\lbn{0}}{\lbn{1}}}{\llambda(x\colon \ltynv{0})\lrh \ltynv{1}\colon \glvreshape{\leev{}}{\ltynv{1}}{\ltynv{1}}}}\\
\gpfitbox{\textsc{E-App\allowbreak NVal}} & \gpfit{\evalJudgeK{\lee{}}{\lb{}}{\leev{}}} & \gpfit{\evalJudge{\evalCtx{}}{\leval{\len{0}}{}{\len{1}}}{\lbn{1}\subst{x}{\len{1}}}{\glvreshape{\glvapp{\lenv{0}}{\lenv{2}}}{{\ltyv{}_{\mathsf{out}}}}{\ltynv{{\mathrm{C}}}\subst{x}{\lenv{2}}}}}\\
\gpfitbox{\textsc{E-Frame\allowbreak Reanchor}} & \gpfit{\evalJudgeK{\lee{}}{\lb{}}{\leev{}}} & \gpfit{\evalJudge{\evalCtx{}}{\len{0}}{\frameB{x}{\lbn{0}}{\lbn{1}}}{\llambda(x\colon \ltynv{0})\lrh \ltynv{1}'\colon \glvreshape{\glvapp{\leev{}}{\lenv{2}}}{\ltynv{1}'}{\ltynv{1}'}}}\\
\gpfitbox{\textsc{E-Frame\allowbreak Eta}} & \gpfit{\evalJudgeK{\lee{}}{\lb{}}{\leev{}}} & \gpfit{\evalJudge{\evalCtx{}}{\len{0}}{\frameB{x}{\lbn{0}}{\lbn{1}}}{\llambda(x\colon \ltynv{0})\lrh \ltynv{1}\colon \glvreshape{\leev{}}{\ltynv{1}}{\ltynv{1}}}}\\
\gpfitbox{\textsc{E-Reshape\allowbreak Cob}} & \gpfit{\evalJudgeK{\lee{}}{\lb{}}{\leev{}}} & \gpfit{\evalJudge{\evalCtx{}}{\lee{}}{\lb{}}{\glvreshape{\leev{}}{{\ltyv{}_{\mathsf{_tgt}}}}{{\ltyv{}_{\mathsf{_src}}}}}}\\
\gpfitbox{\textsc{E-Llam\allowbreak Assemble}} & \gpfit{\evalJudgeK{\lee{}}{\lb{}}{\leev{}}} & \gpfit{\evalJudge{\evalCtx{}}{(\lllambda \llpl x\colon \sof{\lb{}'}\llpr \ldotsp \lee{})}{\lb{}'^{\lstar} \loprod \lb{}}{\glvreshape{\lsum^{\lsv{}'}_{\mathit{i}<\overline{\operatorname{dimOf}}(\lsv{}')}\left(\ldelta{\mathit{i}}{\lsv{}'}\loprod \glvreshape{\leev{}_{\mathit{i}}}{\lsv{}}{\lsv{}}\right)}{\lsv{}' \loprod \lsv{}}{\begin{cases}\lsempty{}&\overline{\operatorname{dimOf}}(\lsv{}')=0\\ \lsv{}' \loprod \lsv{}&\text{otherwise}\end{cases}}}}\\
\gpfitbox{\textsc{E-Llam\allowbreak Assemble\allowbreak Real}} & \gpfit{\evalJudgeK{\lee{}}{\lb{}}{\leev{}}} & \gpfit{\evalJudge{\evalCtx{}}{(\lllambda \llpl x\colon \sof{\lb{}'}\llpr \ldotsp \lee{})}{\lb{}'^{\lstar}}{\glvreshape{\lsum^{\lsv{}'}_{\mathit{i}<\overline{\operatorname{dimOf}}(\lsv{}')}\left(\glvreshape{\leev{}_{\mathit{i}}}{\lR{}}{\lR{}} \lvscale \ldelta{\mathit{i}}{\lsv{}'}\right)}{\lsv{}'}{\begin{cases}\lsempty{}&\overline{\operatorname{dimOf}}(\lsv{}')=0\\ \lsv{}'&\text{otherwise}\end{cases}}}}\\
\gpfitbox{\textsc{E-Zero}} & \gpfit{\evalJudgeK{\lee{}}{\lb{}}{\leev{}}} & \gpfit{\evalJudge{\evalCtx{}}{\lvzero{}_{\sof{\lb{}}}}{\lb{}}{\lvzero{}_{\lsv{}}}}\\
\gpfitbox{\textsc{E-Rlit}} & \gpfit{\evalJudgeK{\lee{}}{\lb{}}{\leev{}}} & \gpfit{\evalJudge{\evalCtx{}}{\mathit{r}}{\lbone{}}{\mathit{r}}}\\
\gpfitbox{\textsc{S-S-Real}} & \gpfit{\lepsilon{};\evalCtx{}\vdash_{\mathsf S} \ls{}\evalArrow{}\lsv{}} & \gpfit{\evalCtxSpace{};\evalCtx{}\vdash_{\mathsf S} \lR{} \evalArrow{} \lR{}}\\
\gpfitbox{\textsc{S-S-Empty}} & \gpfit{\lepsilon{};\evalCtx{}\vdash_{\mathsf S} \ls{}\evalArrow{}\lsv{}} & \gpfit{\evalCtxSpace{};\evalCtx{}\vdash_{\mathsf S} \lsempty{} \evalArrow{} \lsempty{}}\\
\gpfitbox{\textsc{S-S-Dual}} & \gpfit{\lepsilon{};\evalCtx{}\vdash_{\mathsf S} \ls{}\evalArrow{}\lsv{}} & \gpfit{\evalCtxSpace{};\evalCtx{}\vdash_{\mathsf S} \ls{}^{\lstar} \evalArrow{} \lsv{}}\\
\gpfitbox{\textsc{S-S-Ker}} & \gpfit{\lepsilon{};\evalCtx{}\vdash_{\mathsf S} \ls{}\evalArrow{}\lsv{}} & \gpfit{\evalCtxSpace{};\evalCtx{}\vdash_{\mathsf S} \Kernel(\lee{}) \evalArrow{} \lR^{\glvreshape{\leev{}}{\lsnv{{\mathrm{svb}}}}{\lsnv{{\mathrm{svb}}}},\Kernel{},\lsnv{{\mathrm{svb}}}}}\\
\gpfitbox{\textsc{S-S-Coker}} & \gpfit{\lepsilon{};\evalCtx{}\vdash_{\mathsf S} \ls{}\evalArrow{}\lsv{}} & \gpfit{\evalCtxSpace{};\evalCtx{}\vdash_{\mathsf S} \CoKernel(\lee{}) \evalArrow{} \lR^{\glvreshape{\leev{}}{\lsnv{{\mathrm{svb}}}}{\lsnv{{\mathrm{svb}}}},\CoKernel{},\lsnv{{\mathrm{svb}}}}}\\
\gpfitbox{\textsc{S-S-Img}} & \gpfit{\lepsilon{};\evalCtx{}\vdash_{\mathsf S} \ls{}\evalArrow{}\lsv{}} & \gpfit{\evalCtxSpace{};\evalCtx{}\vdash_{\mathsf S} \Image(\lee{}) \evalArrow{} \lR^{\glvreshape{\leev{}}{\lsnv{{\mathrm{svb}}}}{\lsnv{{\mathrm{svb}}}},\Image{},\lsnv{{\mathrm{svb}}}}}\\
\gpfitbox{\textsc{S-S-Coimg}} & \gpfit{\lepsilon{};\evalCtx{}\vdash_{\mathsf S} \ls{}\evalArrow{}\lsv{}} & \gpfit{\evalCtxSpace{};\evalCtx{}\vdash_{\mathsf S} \CoImage(\lee{}) \evalArrow{} \lR^{\glvreshape{\leev{}}{\lsnv{{\mathrm{svb}}}}{\lsnv{{\mathrm{svb}}}},\CoImage{},\lsnv{{\mathrm{svb}}}}}\\
\gpfitbox{\textsc{S-S-Tprod}} & \gpfit{\lepsilon{};\evalCtx{}\vdash_{\mathsf S} \ls{}\evalArrow{}\lsv{}} & \gpfit{\evalCtxSpace{};\evalCtx{}\vdash_{\mathsf S} \lsn{0} \loprod \lsn{1} \evalArrow{} \lsnv{0} \loprod \lsnv{1}}\\
\gpfitbox{\textsc{S-S-Dsum}} & \gpfit{\lepsilon{};\evalCtx{}\vdash_{\mathsf S} \ls{}\evalArrow{}\lsv{}} & \gpfit{\evalCtxSpace{};\evalCtx{}\vdash_{\mathsf S} \lsn{0} \losum \lsn{1} \evalArrow{} \lsnv{0} \losum \lsnv{1}}\\
\gpfitbox{\textsc{S-T-Space}} & \gpfit{\lepsilon{};\evalCtx{}\vdash_{\mathsf S} \lty{}\evalArrow{}\ltyv{}} & \gpfit{\evalCtxSpace{};\evalCtx{}\vdash_{\mathsf S} \ls{} \evalArrow{} \lsv{}}\\
\gpfitbox{\textsc{S-T-Pi\allowbreak Ty}} & \gpfit{\lepsilon{};\evalCtx{}\vdash_{\mathsf S} \lty{}\evalArrow{}\ltyv{}} & \gpfit{\evalCtxSpace{};\evalCtx{}\vdash_{\mathsf S} (x\colon \ltyn{0}\fnto \ltyn{1}) \evalArrow{} (x\colon \ltynv{0}\fnto \ltynv{1})}\\
\gpfitbox{\textsc{S-TA-Space}} & \gpfit{\lepsilon{};\evalCtx{}\vdash_{\mathsf S} \lb{}\evalArrow{}_{\mathsf{basis}}\ltyv{}} & \gpfit{\evalCtxSpace{};\evalCtx{}\vdash_{\mathsf S} \lb{} \evalArrow{}_{\mathsf{basis}} \lsv{}}\\
\gpfitbox{\textsc{S-TA-Frame}} & \gpfit{\lepsilon{};\evalCtx{}\vdash_{\mathsf S} \lb{}\evalArrow{}_{\mathsf{basis}}\ltyv{}} & \gpfit{\evalCtxSpace{};\evalCtx{}\vdash_{\mathsf S} \frameB{x}{\lbn{0}}{\lbn{1}} \evalArrow{}_{\mathsf{basis}} (x\colon \ltynv{0}\fnto \ltynv{1})}\\
\gpfitbox{\textsc{S-C-Inverse}} & \gpfit{\lepsilon{};\evalCtx{}\vdash_{\mathsf S} \lbn{0}\xlongleftrightarrow{\leev{}}\lbn{1}} & \gpfit{\evalCtxSpace{};\evalCtx{}\vdash_{\mathsf S} \lbn{0} \xlongleftrightarrow{\leev{}^{-1}} \lbn{1}}\\
\gpfitbox{\textsc{S-C-Trans}} & \gpfit{\lepsilon{};\evalCtx{}\vdash_{\mathsf S} \lbn{0}\xlongleftrightarrow{\leev{}}\lbn{1}} & \gpfit{\evalCtxSpace{};\evalCtx{}\vdash_{\mathsf S} \lbn{0} \xlongleftrightarrow{\glvmtm{\lsv{}}{\lsv{}}{\lenv{1}}{\glvreshape{\lenv{0}}{\lsv{} \loprod \lsv{}}{\lsv{} \loprod \lsv{}}}} \lbn{2}}\\
\gpfitbox{\textsc{S-C-Tprod}} & \gpfit{\lepsilon{};\evalCtx{}\vdash_{\mathsf S} \lbn{0}\xlongleftrightarrow{\leev{}}\lbn{1}} & \gpfit{\evalCtxSpace{};\evalCtx{}\vdash_{\mathsf S} \lbn{0}' \loprod \lbn{1}' \xlongleftrightarrow{\vkrn(\lenv{0}, \lenv{1})} \lbn{0} \loprod \lbn{1}}\\
\gpfitbox{\textsc{S-C-Dsum}} & \gpfit{\lepsilon{};\evalCtx{}\vdash_{\mathsf S} \lbn{0}\xlongleftrightarrow{\leev{}}\lbn{1}} & \gpfit{\evalCtxSpace{};\evalCtx{}\vdash_{\mathsf S} \lbn{0}' \losum \lbn{1}' \xlongleftrightarrow{\vblk(\lenv{0}, \lenv{1})} \lbn{0} \losum \lbn{1}}\\
\gpfitbox{\textsc{S-C-Dual\allowbreak Std}} & \gpfit{\lepsilon{};\evalCtx{}\vdash_{\mathsf S} \lbn{0}\xlongleftrightarrow{\leev{}}\lbn{1}} & \gpfit{\evalCtxSpace{};\evalCtx{}\vdash_{\mathsf S} \lbn{0}^{\lstar} \xlongleftrightarrow{\Trans(\leev{}^{-1})} \stdBasisDCtx{\tof{\lbn{0}^{\lstar}}}}\\
\gpfitbox{\textsc{S-C-Image}} & \gpfit{\lepsilon{};\evalCtx{}\vdash_{\mathsf S} \lbn{0}\xlongleftrightarrow{\leev{}}\lbn{1}} & \gpfit{\evalCtxSpace{};\evalCtx{}\vdash_{\mathsf S} \iBasisC{\lbn{0}} \xlongleftrightarrow{\leev{}} \iBasisC{\lbn{0}'}}\\
\gpfitbox{\textsc{S-C-Image\allowbreak Svds}} & \gpfit{\lepsilon{};\evalCtx{}\vdash_{\mathsf S} \lbn{0}\xlongleftrightarrow{\leev{}}\lbn{1}} & \gpfit{\evalCtxSpace{};\evalCtx{}\vdash_{\mathsf S} \iBasisC{\pBasisC{\CoKernel(\lee{})}{\lbn{0}}{\lbn{1}}} \xlongleftrightarrow{\glvreshape{\lenv{{\mathrm{P}}}}{\lR^{\lenv{{\mathrm{M}}},\CoKernel{},\lsnv{0} \loprod \lsnv{1}} \loprod \lR^{\lenv{{\mathrm{M}}},\Image{},\lsnv{0} \loprod \lsnv{1}}}{\operatorname{inferSpace}(\lenv{{\mathrm{P}}})}} \pBasisC{\Image(\lee{})}{\lbn{0}}{\lbn{1}}}\\
\gpfitbox{\textsc{S-C-Proj\allowbreak Ker\allowbreak Coker}} & \gpfit{\lepsilon{};\evalCtx{}\vdash_{\mathsf S} \lbn{0}\xlongleftrightarrow{\leev{}}\lbn{1}} & \gpfit{\evalCtxSpace{};\evalCtx{}\vdash_{\mathsf S} \pBasisC{\ls{}}{\lbn{0}}{\lbn{1}} \xlongleftrightarrow{\glvmtm{\lsnv{0}'}{\lsnv{0}'}{\glvreshape{\lenv{1}}{\lsnv{0}' \loprod \lsv{}'}{\lsnv{0}' \loprod \lsv{}'}}{\glvmtm{\lsnv{0}}{\lsnv{0}}{\glvreshape{\lenv{2}}{\lsnv{0} \loprod \lsnv{0}'}{\lsnv{0} \loprod \lsnv{0}'}}{\glvreshape{\lenv{3}}{\lsv{} \loprod \lsnv{0}}{\lsv{} \loprod \lsnv{0}}}}} \pBasisC{\ls{}}{\lbn{0}'}{\lbn{1}}}\\
\gpfitbox{\textsc{S-C-Proj\allowbreak Img\allowbreak Coimg}} & \gpfit{\lepsilon{};\evalCtx{}\vdash_{\mathsf S} \lbn{0}\xlongleftrightarrow{\leev{}}\lbn{1}} & \gpfit{\evalCtxSpace{};\evalCtx{}\vdash_{\mathsf S} \pBasisC{\ls{}}{\lbn{0}}{\lbn{1}} \xlongleftrightarrow{\glvmtm{\lsnv{1}'}{\lsnv{1}'}{\glvreshape{\lenv{1}}{\lsnv{1}' \loprod \lsv{}'}{\lsnv{1}' \loprod \lsv{}'}}{\glvmtm{\lsnv{1}}{\lsnv{1}}{\glvreshape{\lenv{2}}{\lsnv{1} \loprod \lsnv{1}'}{\lsnv{1} \loprod \lsnv{1}'}}{\glvreshape{\lenv{3}}{\lsv{} \loprod \lsnv{1}}{\lsv{} \loprod \lsnv{1}}}}} \pBasisC{\ls{}}{\lbn{0}}{\lbn{1}'}}\\
\gpfitbox{\textsc{S-C-Terms\allowbreak Rref}} & \gpfit{\lepsilon{};\evalCtx{}\vdash_{\mathsf S} \lbn{0}\xlongleftrightarrow{\leev{}}\lbn{1}} & \gpfit{\evalCtxSpace{};\evalCtx{}\vdash_{\mathsf S} \lb{} \xlongleftrightarrow{\glvreshape{\glvop{RREFRep}(\lenv{0},\glvreshape{\lenv{1}}{\lsv{} \loprod \lsv{}}{\lsv{} \loprod \lsv{}})}{\lsv{} \loprod \lsv{}}{\lsv{} \loprod \lsv{}}} \llpl{} \lee{} \mathbin{:\!:} \mathit{es}\llpr{}_{\lb{}'}}\\
\gpfitbox{\textsc{S-C-Refl}} & \gpfit{\lepsilon{};\evalCtx{}\vdash_{\mathsf S} \lbn{0}\xlongleftrightarrow{\leev{}}\lbn{1}} & \gpfit{\evalCtxSpace{};\evalCtx{}\vdash_{\mathsf S} \lb{} \xlongleftrightarrow{\glvreshape{\lidelta{\lsv{}}}{\lsv{} \loprod \lsv{}}{\lsv{} \loprod \lsv{}}} \lb{}}\\
\gpfitbox{\textsc{S-C-Proj\allowbreak Ker\allowbreak Coker\allowbreak Cod}} & \gpfit{\lepsilon{};\evalCtx{}\vdash_{\mathsf S} \lbn{0}\xlongleftrightarrow{\leev{}}\lbn{1}} & \gpfit{\evalCtxSpace{};\evalCtx{}\vdash_{\mathsf S} \pBasisC{\ls{}}{\lbn{0}}{\lbn{1}} \xlongleftrightarrow{\glvmtm{\lsnv{0}}{\lsnv{0}}{\glvreshape{\lenv{1}}{\lsnv{0} \loprod \lsv{}'}{\lsnv{0} \loprod \lsv{}'}}{\glvreshape{\lenv{3}}{\lsv{} \loprod \lsnv{0}}{\lsv{} \loprod \lsnv{0}}}} \pBasisC{\ls{}}{\lbn{0}}{\lbn{1}'}}\\
\gpfitbox{\textsc{S-C-Proj\allowbreak Img\allowbreak Coimg\allowbreak Dom}} & \gpfit{\lepsilon{};\evalCtx{}\vdash_{\mathsf S} \lbn{0}\xlongleftrightarrow{\leev{}}\lbn{1}} & \gpfit{\evalCtxSpace{};\evalCtx{}\vdash_{\mathsf S} \pBasisC{\ls{}}{\lbn{0}}{\lbn{1}} \xlongleftrightarrow{\glvmtm{\lsnv{1}}{\lsnv{1}}{\glvreshape{\lenv{1}}{\lsnv{1} \loprod \lsv{}'}{\lsnv{1} \loprod \lsv{}'}}{\glvreshape{\lenv{3}}{\lsv{} \loprod \lsnv{1}}{\lsv{} \loprod \lsnv{1}}}} \pBasisC{\ls{}}{\lbn{0}'}{\lbn{1}}}\\
\gpfitbox{\textsc{S-E-Var\allowbreak Lin}} & \gpfit{\lDelta{};\lepsilon{};\evalCtx{}\vdash_{\mathsf S} \lee{}\xlongrightarrow{\lb{}}\leev{}\glresulttype{\ltyv{}}} & \gpfit{\{(\mathit{x}, \ls{})\};\evalCtxSpace{};\evalCtx{}\vdash_{\mathsf S} \mathit{x} \xlongrightarrow{\lb{}} \leev{}\glresulttype{\ltyv{}}}\\
\gpfitbox{\textsc{S-E-Var\allowbreak Nlin}} & \gpfit{\lDelta{};\lepsilon{};\evalCtx{}\vdash_{\mathsf S} \lee{}\xlongrightarrow{\lb{}}\leev{}\glresulttype{\ltyv{}}} & \gpfit{\lemptyset{};\evalCtxSpace{};\evalCtx{}\vdash_{\mathsf S} \mathit{x} \xlongrightarrow{\lb{}} \leev{}\glresulttype{\ltyv{}}}\\
\gpfitbox{\textsc{S-E-Promote}} & \gpfit{\lDelta{};\lepsilon{};\evalCtx{}\vdash_{\mathsf S} \lee{}\xlongrightarrow{\lb{}}\leev{}\glresulttype{\ltyv{}}} & \gpfit{\lDelta{};\mathit{lin} \ldisun \evalCtxSpace{}'';\evalCtx{}\vdash_{\mathsf S} \lee{} \xlongrightarrow{\lb{}} \leev{}\glresulttype{\ltyv{}}}\\
\gpfitbox{\textsc{S-E-Cob}} & \gpfit{\lDelta{};\lepsilon{};\evalCtx{}\vdash_{\mathsf S} \lee{}\xlongrightarrow{\lb{}}\leev{}\glresulttype{\ltyv{}}} & \gpfit{\lDelta{};\evalCtxSpace{};\evalCtx{}\vdash_{\mathsf S} \lee{} \xlongrightarrow{\lbn{0}} \LTTVPP{\lsv{}}{\glvreshape{\lenv{0}}{\lsv{} \loprod \lsv{}}{\lsv{} \loprod \lsv{}}}{\glvreshape{\leev{}}{\lsv{}}{\lsv{}}}\glresulttype{\lsv{}}}\\
\gpfitbox{\textsc{S-E-Reshape\allowbreak Cob}} & \gpfit{\lDelta{};\lepsilon{};\evalCtx{}\vdash_{\mathsf S} \lee{}\xlongrightarrow{\lb{}}\leev{}\glresulttype{\ltyv{}}} & \gpfit{\lDelta{};\evalCtxSpace{};\evalCtx{}\vdash_{\mathsf S} \lee{} \xlongrightarrow{\lb{}} \glvreshape{\leev{}}{\ltyv{}'}{\ltyv{}}\glresulttype{\ltyv{}'}}\\
\gpfitbox{\textsc{S-E-Scale}} & \gpfit{\lDelta{};\lepsilon{};\evalCtx{}\vdash_{\mathsf S} \lee{}\xlongrightarrow{\lb{}}\leev{}\glresulttype{\ltyv{}}} & \gpfit{{\lDelta{}_{0}} \ldisun {\lDelta{}_{1}};\evalCtxSpace{};\evalCtx{}\vdash_{\mathsf S} \len{0} \lvscale \len{1} \xlongrightarrow{\lb{}} \lenv{0} \lvscale \lenv{1}\glresulttype{\lsv{}}}\\
\gpfitbox{\textsc{S-E-Add}} & \gpfit{\lDelta{};\lepsilon{};\evalCtx{}\vdash_{\mathsf S} \lee{}\xlongrightarrow{\lb{}}\leev{}\glresulttype{\ltyv{}}} & \gpfit{\lDelta{};\evalCtxSpace{};\evalCtx{}\vdash_{\mathsf S} \len{0} \lvadd \len{1} \xlongrightarrow{\lb{}} \lenv{0} \lvadd_{\lsv{}} \lenv{1}\glresulttype{\lsv{}}}\\
\gpfitbox{\textsc{S-E-Dsum\allowbreak E}} & \gpfit{\lDelta{};\lepsilon{};\evalCtx{}\vdash_{\mathsf S} \lee{}\xlongrightarrow{\lb{}}\leev{}\glresulttype{\ltyv{}}} & \gpfit{\lDelta{};\evalCtxSpace{};\evalCtx{}\vdash_{\mathsf S} \len{0}\losum \len{1} \xlongrightarrow{\lbn{0} \losum \lbn{1}} \lenv{0}\losum \lenv{1}\glresulttype{\lsnv{0} \losum \lsnv{1}}}\\
\gpfitbox{\textsc{S-E-Tprod\allowbreak E}} & \gpfit{\lDelta{};\lepsilon{};\evalCtx{}\vdash_{\mathsf S} \lee{}\xlongrightarrow{\lb{}}\leev{}\glresulttype{\ltyv{}}} & \gpfit{{\lDelta{}_{0}} \ldisun {\lDelta{}_{1}};\evalCtxSpace{};\evalCtx{}\vdash_{\mathsf S} \len{0}\loprod \len{1} \xlongrightarrow{\lbn{0} \loprod \lbn{1}} \lenv{0}\loprod \lenv{1}\glresulttype{\lsnv{0} \loprod \lsnv{1}}}\\
\gpfitbox{\textsc{S-E-Pinv\allowbreak TStd}} & \gpfit{\lDelta{};\lepsilon{};\evalCtx{}\vdash_{\mathsf S} \lee{}\xlongrightarrow{\lb{}}\leev{}\glresulttype{\ltyv{}}} & \gpfit{\lemptyset{};\evalCtxSpace{};\evalCtx{}\vdash_{\mathsf S} \lee{}^{\lpseudo} \xlongrightarrow{\lbn{0}^{\lstar} \loprod \lbn{1}} \leev{}^{+}\glresulttype{\lsnv{0} \loprod \lsnv{1}}}\\
\gpfitbox{\textsc{S-E-Pinv\allowbreak TGpinv}} & \gpfit{\lDelta{};\lepsilon{};\evalCtx{}\vdash_{\mathsf S} \lee{}\xlongrightarrow{\lb{}}\leev{}\glresulttype{\ltyv{}}} & \gpfit{\lemptyset{};\evalCtxSpace{};\evalCtx{}\vdash_{\mathsf S} \lee{}^{\lpseudo} \xlongrightarrow{\lbn{0}^{\lstar} \loprod \lbn{1}} \glvop{GPINV}(\leev{},\lenv{1},\lenv{0})\glresulttype{\lsnv{0} \loprod \lsnv{1}}}\\
\gpfitbox{\textsc{S-E-Pinv\allowbreak DStd}} & \gpfit{\lDelta{};\lepsilon{};\evalCtx{}\vdash_{\mathsf S} \lee{}\xlongrightarrow{\lb{}}\leev{}\glresulttype{\ltyv{}}} & \gpfit{\lemptyset{};\evalCtxSpace{};\evalCtx{}\vdash_{\mathsf S} \lee{}^{\lpseudo} \xlongrightarrow{\lbn{0}^{\lstar} \loprod \lbn{1}} \glvreshape{\leev{}}{\lsnv{1} \loprod \lR{}}{\lsnv{1}}^{+}\glresulttype{\lR{} \loprod \lsnv{1}}}\\
\gpfitbox{\textsc{S-E-Pinv\allowbreak DGpinv}} & \gpfit{\lDelta{};\lepsilon{};\evalCtx{}\vdash_{\mathsf S} \lee{}\xlongrightarrow{\lb{}}\leev{}\glresulttype{\ltyv{}}} & \gpfit{\lemptyset{};\evalCtxSpace{};\evalCtx{}\vdash_{\mathsf S} \lee{}^{\lpseudo} \xlongrightarrow{\lbn{0}^{\lstar} \loprod \lbn{1}} \glvop{GPINV}(\glvreshape{\leev{}}{\lsnv{1} \loprod \lR{}}{\lsnv{1}},\lenv{1},\lenv{0})\glresulttype{\lR{} \loprod \lsnv{1}}}\\
\gpfitbox{\textsc{S-E-Inner\allowbreak Std}} & \gpfit{\lDelta{};\lepsilon{};\evalCtx{}\vdash_{\mathsf S} \lee{}\xlongrightarrow{\lb{}}\leev{}\glresulttype{\ltyv{}}} & \gpfit{\lemptyset{};\evalCtxSpace{};\evalCtx{}\vdash_{\mathsf S} \linner{}_{\ls{}} \xlongrightarrow{\lbn{0}^{\lstar} \loprod \lbn{0}^{\lstar}} \lidelta{\lsv{}}\glresulttype{\lsv{} \loprod \lsv{}}}\\
\gpfitbox{\textsc{S-E-Embed\allowbreak Ker}} & \gpfit{\lDelta{};\lepsilon{};\evalCtx{}\vdash_{\mathsf S} \lee{}\xlongrightarrow{\lb{}}\leev{}\glresulttype{\ltyv{}}} & \gpfit{\lemptyset{};\evalCtxSpace{};\evalCtx{}\vdash_{\mathsf S} \lembd{}_{\Kernel(\lee{})} \xlongrightarrow{\pBasisC{\Kernel(\lee{})}{\lbn{0}}{\lbn{1}}^{\lstar} \loprod \lbn{0}} \Trans(\glvop{GSVDUI}(\leev{},\lenv{0},\lenv{1}))\glresulttype{\lR^{\glvreshape{\leev{}}{\lsnv{0} \loprod \lsnv{1}}{\lsnv{0} \loprod \lsnv{1}},\Kernel{},\lsnv{0} \loprod \lsnv{1}} \loprod \lsnv{0}}}\\
\gpfitbox{\textsc{S-E-Embed\allowbreak Coker}} & \gpfit{\lDelta{};\lepsilon{};\evalCtx{}\vdash_{\mathsf S} \lee{}\xlongrightarrow{\lb{}}\leev{}\glresulttype{\ltyv{}}} & \gpfit{\lemptyset{};\evalCtxSpace{};\evalCtx{}\vdash_{\mathsf S} \lembd{}_{\CoKernel(\lee{})} \xlongrightarrow{\pBasisC{\CoKernel(\lee{})}{\lbn{0}}{\lbn{1}}^{\lstar} \loprod \lbn{0}} \Trans(\glvop{GSVDU}(\leev{},\lenv{0},\lenv{1}))\glresulttype{\lR^{\glvreshape{\leev{}}{\lsnv{0} \loprod \lsnv{1}}{\lsnv{0} \loprod \lsnv{1}},\CoKernel{},\lsnv{0} \loprod \lsnv{1}} \loprod \lsnv{0}}}\\
\gpfitbox{\textsc{S-E-Embed\allowbreak Img}} & \gpfit{\lDelta{};\lepsilon{};\evalCtx{}\vdash_{\mathsf S} \lee{}\xlongrightarrow{\lb{}}\leev{}\glresulttype{\ltyv{}}} & \gpfit{\lemptyset{};\evalCtxSpace{};\evalCtx{}\vdash_{\mathsf S} \lembd{}_{\Image(\lee{})} \xlongrightarrow{\pBasisC{\Image(\lee{})}{\lbn{0}}{\lbn{1}}^{\lstar} \loprod \lbn{1}} \Trans(\glvop{GSVDV}(\leev{},\lenv{0},\lenv{1}))\glresulttype{\lR^{\glvreshape{\leev{}}{\lsnv{0} \loprod \lsnv{1}}{\lsnv{0} \loprod \lsnv{1}},\Image{},\lsnv{0} \loprod \lsnv{1}} \loprod \lsnv{1}}}\\
\gpfitbox{\textsc{S-E-Embed\allowbreak Coimg}} & \gpfit{\lDelta{};\lepsilon{};\evalCtx{}\vdash_{\mathsf S} \lee{}\xlongrightarrow{\lb{}}\leev{}\glresulttype{\ltyv{}}} & \gpfit{\lemptyset{};\evalCtxSpace{};\evalCtx{}\vdash_{\mathsf S} \lembd{}_{\CoImage(\lee{})} \xlongrightarrow{\pBasisC{\CoImage(\lee{})}{\lbn{0}}{\lbn{1}}^{\lstar} \loprod \lbn{1}} \Trans(\glvop{GSVDVI}(\leev{},\lenv{0},\lenv{1}))\glresulttype{\lR^{\glvreshape{\leev{}}{\lsnv{0} \loprod \lsnv{1}}{\lsnv{0} \loprod \lsnv{1}},\CoImage{},\lsnv{0} \loprod \lsnv{1}} \loprod \lsnv{1}}}\\
\gpfitbox{\textsc{S-E-App\allowbreak LLin}} & \gpfit{\lDelta{};\lepsilon{};\evalCtx{}\vdash_{\mathsf S} \lee{}\xlongrightarrow{\lb{}}\leev{}\glresulttype{\ltyv{}}} & \gpfit{{\lDelta{}_{0}} \ldisun {\lDelta{}_{1}};\evalCtxSpace{};\evalCtx{}\vdash_{\mathsf S} \leval{\len{0}}{}{\len{1}} \xlongrightarrow{\lbn{1}} \glvreshape{\LTTVPP{\lsv{}}{\lenv{0}}{\lenv{1}}}{\lsnv{{\mathrm{svOut}}}}{\lsnv{{\mathrm{svOut}}}}\glresulttype{\lsnv{{\mathrm{svOut}}}}}\\
\gpfitbox{\textsc{S-E-App\allowbreak ILin}} & \gpfit{\lDelta{};\lepsilon{};\evalCtx{}\vdash_{\mathsf S} \lee{}\xlongrightarrow{\lb{}}\leev{}\glresulttype{\ltyv{}}} & \gpfit{{\lDelta{}_{0}} \ldisun {\lDelta{}_{1}};\evalCtxSpace{};\evalCtx{}\vdash_{\mathsf S} \leval{\len{0}}{}{\len{1}} \xlongrightarrow{\lbone{}} \LTTVPP{\lsv{}}{\glvreshape{\lenv{0}}{\lsv{} \loprod \lR{}}{\lsv{}}}{\lenv{1}}\glresulttype{\lR{}}}\\
\gpfitbox{\textsc{S-E-Nllam}} & \gpfit{\lDelta{};\lepsilon{};\evalCtx{}\vdash_{\mathsf S} \lee{}\xlongrightarrow{\lb{}}\leev{}\glresulttype{\ltyv{}}} & \gpfit{\lemptyset{};\evalCtxSpace{};\evalCtx{}\vdash_{\mathsf S} (\nllambda \llpl x\colon \ltyn{0}\llpr \ldotsp \lee{}) \xlongrightarrow{\frameB{x}{\lbn{0}}{\lbn{1}}} \llambda(x\colon \ltynv{0})\lrh \ltynv{1}\colon \glvreshape{\leev{}}{\ltynv{1}}{\ltynv{1}}\glresulttype{(x\colon \ltynv{0}\fnto \ltynv{1})}}\\
\gpfitbox{\textsc{S-E-App\allowbreak NVal}} & \gpfit{\lDelta{};\lepsilon{};\evalCtx{}\vdash_{\mathsf S} \lee{}\xlongrightarrow{\lb{}}\leev{}\glresulttype{\ltyv{}}} & \gpfit{\lemptyset{};\evalCtxSpace{};\evalCtx{}\vdash_{\mathsf S} \leval{\len{0}}{}{\len{1}} \xlongrightarrow{\lbn{1}\subst{x}{\len{1}}} \glvreshape{\glvapp{\lenv{0}}{\lenv{2}}}{{\ltyv{}_{\mathsf{out}}}}{\ltynv{{\mathrm{C}}}\subst{x}{\lenv{2}}}\glresulttype{{\ltyv{}_{\mathsf{out}}}}}\\
\gpfitbox{\textsc{S-E-Frame\allowbreak Reanchor}} & \gpfit{\lDelta{};\lepsilon{};\evalCtx{}\vdash_{\mathsf S} \lee{}\xlongrightarrow{\lb{}}\leev{}\glresulttype{\ltyv{}}} & \gpfit{\lemptyset{};\evalCtxSpace{};\evalCtx{}\vdash_{\mathsf S} \len{0} \xlongrightarrow{\frameB{x}{\lbn{0}}{\lbn{1}}} \llambda(x\colon \ltynv{0})\lrh \ltynv{1}'\colon \glvreshape{\glvapp{\leev{}}{\lenv{2}}}{\ltynv{1}'}{\ltynv{1}'}\glresulttype{(x\colon \ltynv{0}\fnto \ltynv{1}')}}\\
\gpfitbox{\textsc{S-E-Frame\allowbreak Eta}} & \gpfit{\lDelta{};\lepsilon{};\evalCtx{}\vdash_{\mathsf S} \lee{}\xlongrightarrow{\lb{}}\leev{}\glresulttype{\ltyv{}}} & \gpfit{\lemptyset{};\evalCtxSpace{};\evalCtx{}\vdash_{\mathsf S} \len{0} \xlongrightarrow{\frameB{x}{\lbn{0}}{\lbn{1}}} \llambda(x\colon \ltynv{0})\lrh \ltynv{1}\colon \glvreshape{\leev{}}{\ltynv{1}}{\ltynv{1}}\glresulttype{(x\colon \ltynv{0}\fnto \ltynv{1})}}\\
\gpfitbox{\textsc{S-E-Llam\allowbreak Assemble}} & \gpfit{\lDelta{};\lepsilon{};\evalCtx{}\vdash_{\mathsf S} \lee{}\xlongrightarrow{\lb{}}\leev{}\glresulttype{\ltyv{}}} & \gpfit{\lDelta{};\evalCtxSpace{};\evalCtx{}\vdash_{\mathsf S} (\lllambda \llpl x\colon \sof{\lb{}'}\llpr \ldotsp \lee{}) \xlongrightarrow{\lb{}'^{\lstar} \loprod \lb{}} \glvreshape{\lsum^{\lsv{}'}_{\mathit{i}<\overline{\operatorname{dimOf}}(\lsv{}')}\left(\ldelta{\mathit{i}}{\lsv{}'}\loprod \glvreshape{\leev{}_{\mathit{i}}}{\lsv{}}{\lsv{}}\right)}{\lsv{}' \loprod \lsv{}}{\begin{cases}\lsempty{}&\overline{\operatorname{dimOf}}(\lsv{}')=0\\ \lsv{}' \loprod \lsv{}&\text{otherwise}\end{cases}}\glresulttype{\lsv{}' \loprod \lsv{}}}\\
\gpfitbox{\textsc{S-E-Llam\allowbreak Assemble\allowbreak Real}} & \gpfit{\lDelta{};\lepsilon{};\evalCtx{}\vdash_{\mathsf S} \lee{}\xlongrightarrow{\lb{}}\leev{}\glresulttype{\ltyv{}}} & \gpfit{\lDelta{};\evalCtxSpace{};\evalCtx{}\vdash_{\mathsf S} (\lllambda \llpl x\colon \sof{\lb{}'}\llpr \ldotsp \lee{}) \xlongrightarrow{\lb{}'^{\lstar}} \glvreshape{\lsum^{\lsv{}'}_{\mathit{i}<\overline{\operatorname{dimOf}}(\lsv{}')}\left(\glvreshape{\leev{}_{\mathit{i}}}{\lR{}}{\lR{}} \lvscale \ldelta{\mathit{i}}{\lsv{}'}\right)}{\lsv{}'}{\begin{cases}\lsempty{}&\overline{\operatorname{dimOf}}(\lsv{}')=0\\ \lsv{}'&\text{otherwise}\end{cases}}\glresulttype{\lsv{}'}}\\
\gpfitbox{\textsc{S-E-Zero}} & \gpfit{\lDelta{};\lepsilon{};\evalCtx{}\vdash_{\mathsf S} \lee{}\xlongrightarrow{\lb{}}\leev{}\glresulttype{\ltyv{}}} & \gpfit{\lDelta{};\evalCtxSpace{};\evalCtx{}\vdash_{\mathsf S} \lvzero{}_{\sof{\lb{}}} \xlongrightarrow{\lb{}} \lvzero{}_{\lsv{}}\glresulttype{\lsv{}}}\\
\gpfitbox{\textsc{S-E-Rlit}} & \gpfit{\lDelta{};\lepsilon{};\evalCtx{}\vdash_{\mathsf S} \lee{}\xlongrightarrow{\lb{}}\leev{}\glresulttype{\ltyv{}}} & \gpfit{\lemptyset{};\evalCtxSpace{};\evalCtx{}\vdash_{\mathsf S} \mathit{r} \xlongrightarrow{\lbone{}} \mathit{r}\glresulttype{\lR{}}}\\
\gpfitbox{\textsc{C-S-Real}} & \gpfit{\lepsilon{};\evalCtx{}\vdash_{\mathsf C} \ls{}\evalArrow{}\lsv{}} & \gpfit{\evalCtxSpace{};\evalCtx{}\vdash_{\mathsf C} \lR{} \evalArrow{} \lR{}}\\
\gpfitbox{\textsc{C-S-Empty}} & \gpfit{\lepsilon{};\evalCtx{}\vdash_{\mathsf C} \ls{}\evalArrow{}\lsv{}} & \gpfit{\evalCtxSpace{};\evalCtx{}\vdash_{\mathsf C} \lsempty{} \evalArrow{} \lsempty{}}\\
\gpfitbox{\textsc{C-S-Dual}} & \gpfit{\lepsilon{};\evalCtx{}\vdash_{\mathsf C} \ls{}\evalArrow{}\lsv{}} & \gpfit{\evalCtxSpace{};\evalCtx{}\vdash_{\mathsf C} \ls{}^{\lstar} \evalArrow{} \lsv{}}\\
\gpfitbox{\textsc{C-S-Ker}} & \gpfit{\lepsilon{};\evalCtx{}\vdash_{\mathsf C} \ls{}\evalArrow{}\lsv{}} & \gpfit{\evalCtxSpace{};\evalCtx{}\vdash_{\mathsf C} \Kernel(\lee{}) \evalArrow{} \lR^{\glvreshape{\leev{}}{\lsnv{{\mathrm{svb}}}}{\lsnv{{\mathrm{svb}}}},\Kernel{},\lsnv{{\mathrm{svb}}}}}\\
\gpfitbox{\textsc{C-S-Coker}} & \gpfit{\lepsilon{};\evalCtx{}\vdash_{\mathsf C} \ls{}\evalArrow{}\lsv{}} & \gpfit{\evalCtxSpace{};\evalCtx{}\vdash_{\mathsf C} \CoKernel(\lee{}) \evalArrow{} \lR^{\glvreshape{\leev{}}{\lsnv{{\mathrm{svb}}}}{\lsnv{{\mathrm{svb}}}},\CoKernel{},\lsnv{{\mathrm{svb}}}}}\\
\gpfitbox{\textsc{C-S-Img}} & \gpfit{\lepsilon{};\evalCtx{}\vdash_{\mathsf C} \ls{}\evalArrow{}\lsv{}} & \gpfit{\evalCtxSpace{};\evalCtx{}\vdash_{\mathsf C} \Image(\lee{}) \evalArrow{} \lR^{\glvreshape{\leev{}}{\lsnv{{\mathrm{svb}}}}{\lsnv{{\mathrm{svb}}}},\Image{},\lsnv{{\mathrm{svb}}}}}\\
\gpfitbox{\textsc{C-S-Coimg}} & \gpfit{\lepsilon{};\evalCtx{}\vdash_{\mathsf C} \ls{}\evalArrow{}\lsv{}} & \gpfit{\evalCtxSpace{};\evalCtx{}\vdash_{\mathsf C} \CoImage(\lee{}) \evalArrow{} \lR^{\glvreshape{\leev{}}{\lsnv{{\mathrm{svb}}}}{\lsnv{{\mathrm{svb}}}},\CoImage{},\lsnv{{\mathrm{svb}}}}}\\
\gpfitbox{\textsc{C-S-Tprod}} & \gpfit{\lepsilon{};\evalCtx{}\vdash_{\mathsf C} \ls{}\evalArrow{}\lsv{}} & \gpfit{\evalCtxSpace{};\evalCtx{}\vdash_{\mathsf C} \lsn{0} \loprod \lsn{1} \evalArrow{} \lsnv{0} \loprod \lsnv{1}}\\
\gpfitbox{\textsc{C-S-Dsum}} & \gpfit{\lepsilon{};\evalCtx{}\vdash_{\mathsf C} \ls{}\evalArrow{}\lsv{}} & \gpfit{\evalCtxSpace{};\evalCtx{}\vdash_{\mathsf C} \lsn{0} \losum \lsn{1} \evalArrow{} \lsnv{0} \losum \lsnv{1}}\\
\gpfitbox{\textsc{C-T-Space}} & \gpfit{\lepsilon{};\evalCtx{}\vdash_{\mathsf C} \lty{}\evalArrow{}\ltyv{}} & \gpfit{\evalCtxSpace{};\evalCtx{}\vdash_{\mathsf C} \ls{} \evalArrow{} \lsv{}}\\
\gpfitbox{\textsc{C-T-Pi\allowbreak Ty}} & \gpfit{\lepsilon{};\evalCtx{}\vdash_{\mathsf C} \lty{}\evalArrow{}\ltyv{}} & \gpfit{\evalCtxSpace{};\evalCtx{}\vdash_{\mathsf C} (x\colon \ltyn{0}\fnto \ltyn{1}) \evalArrow{} (x\colon \ltynv{0}\fnto \ltynv{1})}\\
\gpfitbox{\textsc{C-TA-Space}} & \gpfit{\lepsilon{};\evalCtx{}\vdash_{\mathsf C} \lb{}\evalArrow{}_{\mathsf{basis}}\ltyv{}} & \gpfit{\evalCtxSpace{};\evalCtx{}\vdash_{\mathsf C} \lb{} \evalArrow{}_{\mathsf{basis}} \lsv{}}\\
\gpfitbox{\textsc{C-TA-Frame}} & \gpfit{\lepsilon{};\evalCtx{}\vdash_{\mathsf C} \lb{}\evalArrow{}_{\mathsf{basis}}\ltyv{}} & \gpfit{\evalCtxSpace{};\evalCtx{}\vdash_{\mathsf C} \frameB{x}{\lbn{0}}{\lbn{1}} \evalArrow{}_{\mathsf{basis}} (x\colon \ltynv{0}\fnto \ltynv{1})}\\
\gpfitbox{\textsc{C-C-Inverse}} & \gpfit{\lepsilon{};\evalCtx{}\vdash_{\mathsf C} \lbn{0}\xlongleftrightarrow{\leev{}}\lbn{1}} & \gpfit{\evalCtxSpace{};\evalCtx{}\vdash_{\mathsf C} \lbn{0} \xlongleftrightarrow{\leev{}^{-1}} \lbn{1}}\\
\gpfitbox{\textsc{C-C-Trans}} & \gpfit{\lepsilon{};\evalCtx{}\vdash_{\mathsf C} \lbn{0}\xlongleftrightarrow{\leev{}}\lbn{1}} & \gpfit{\evalCtxSpace{};\evalCtx{}\vdash_{\mathsf C} \lbn{0} \xlongleftrightarrow{\glvmtm{\lsv{}}{\lsv{}}{\lenv{1}}{\glvreshape{\lenv{0}}{\lsv{} \loprod \lsv{}}{\lsv{} \loprod \lsv{}}}} \lbn{2}}\\
\gpfitbox{\textsc{C-C-Tprod}} & \gpfit{\lepsilon{};\evalCtx{}\vdash_{\mathsf C} \lbn{0}\xlongleftrightarrow{\leev{}}\lbn{1}} & \gpfit{\evalCtxSpace{};\evalCtx{}\vdash_{\mathsf C} \lbn{0}' \loprod \lbn{1}' \xlongleftrightarrow{\vkrn(\lenv{0}, \lenv{1})} \lbn{0} \loprod \lbn{1}}\\
\gpfitbox{\textsc{C-C-Dsum}} & \gpfit{\lepsilon{};\evalCtx{}\vdash_{\mathsf C} \lbn{0}\xlongleftrightarrow{\leev{}}\lbn{1}} & \gpfit{\evalCtxSpace{};\evalCtx{}\vdash_{\mathsf C} \lbn{0}' \losum \lbn{1}' \xlongleftrightarrow{\vblk(\lenv{0}, \lenv{1})} \lbn{0} \losum \lbn{1}}\\
\gpfitbox{\textsc{C-C-Dual\allowbreak Std}} & \gpfit{\lepsilon{};\evalCtx{}\vdash_{\mathsf C} \lbn{0}\xlongleftrightarrow{\leev{}}\lbn{1}} & \gpfit{\evalCtxSpace{};\evalCtx{}\vdash_{\mathsf C} \lbn{0}^{\lstar} \xlongleftrightarrow{\Trans(\leev{}^{-1})} \stdBasisDCtx{\tof{\lbn{0}^{\lstar}}}}\\
\gpfitbox{\textsc{C-C-Image}} & \gpfit{\lepsilon{};\evalCtx{}\vdash_{\mathsf C} \lbn{0}\xlongleftrightarrow{\leev{}}\lbn{1}} & \gpfit{\evalCtxSpace{};\evalCtx{}\vdash_{\mathsf C} \iBasisC{\lbn{0}} \xlongleftrightarrow{\leev{}} \iBasisC{\lbn{0}'}}\\
\gpfitbox{\textsc{C-C-Image\allowbreak Svds}} & \gpfit{\lepsilon{};\evalCtx{}\vdash_{\mathsf C} \lbn{0}\xlongleftrightarrow{\leev{}}\lbn{1}} & \gpfit{\evalCtxSpace{};\evalCtx{}\vdash_{\mathsf C} \iBasisC{\pBasisC{\CoKernel(\lee{})}{\lbn{0}}{\lbn{1}}} \xlongleftrightarrow{\glvreshape{\lenv{{\mathrm{P}}}}{\lR^{\lenv{{\mathrm{M}}},\CoKernel{},\lsnv{0} \loprod \lsnv{1}} \loprod \lR^{\lenv{{\mathrm{M}}},\Image{},\lsnv{0} \loprod \lsnv{1}}}{\operatorname{inferSpace}(\lenv{{\mathrm{P}}})}} \pBasisC{\Image(\lee{})}{\lbn{0}}{\lbn{1}}}\\
\gpfitbox{\textsc{C-C-Proj\allowbreak Ker\allowbreak Coker}} & \gpfit{\lepsilon{};\evalCtx{}\vdash_{\mathsf C} \lbn{0}\xlongleftrightarrow{\leev{}}\lbn{1}} & \gpfit{\evalCtxSpace{};\evalCtx{}\vdash_{\mathsf C} \pBasisC{\ls{}}{\lbn{0}}{\lbn{1}} \xlongleftrightarrow{\glvmtm{\lsnv{0}'}{\lsnv{0}'}{\glvreshape{\lenv{1}}{\lsnv{0}' \loprod \lsv{}'}{\lsnv{0}' \loprod \lsv{}'}}{\glvmtm{\lsnv{0}}{\lsnv{0}}{\glvreshape{\lenv{2}}{\lsnv{0} \loprod \lsnv{0}'}{\lsnv{0} \loprod \lsnv{0}'}}{\glvreshape{\lenv{3}}{\lsv{} \loprod \lsnv{0}}{\lsv{} \loprod \lsnv{0}}}}} \pBasisC{\ls{}}{\lbn{0}'}{\lbn{1}}}\\
\gpfitbox{\textsc{C-C-Proj\allowbreak Img\allowbreak Coimg}} & \gpfit{\lepsilon{};\evalCtx{}\vdash_{\mathsf C} \lbn{0}\xlongleftrightarrow{\leev{}}\lbn{1}} & \gpfit{\evalCtxSpace{};\evalCtx{}\vdash_{\mathsf C} \pBasisC{\ls{}}{\lbn{0}}{\lbn{1}} \xlongleftrightarrow{\glvmtm{\lsnv{1}'}{\lsnv{1}'}{\glvreshape{\lenv{1}}{\lsnv{1}' \loprod \lsv{}'}{\lsnv{1}' \loprod \lsv{}'}}{\glvmtm{\lsnv{1}}{\lsnv{1}}{\glvreshape{\lenv{2}}{\lsnv{1} \loprod \lsnv{1}'}{\lsnv{1} \loprod \lsnv{1}'}}{\glvreshape{\lenv{3}}{\lsv{} \loprod \lsnv{1}}{\lsv{} \loprod \lsnv{1}}}}} \pBasisC{\ls{}}{\lbn{0}}{\lbn{1}'}}\\
\gpfitbox{\textsc{C-C-Terms\allowbreak Rref}} & \gpfit{\lepsilon{};\evalCtx{}\vdash_{\mathsf C} \lbn{0}\xlongleftrightarrow{\leev{}}\lbn{1}} & \gpfit{\evalCtxSpace{};\evalCtx{}\vdash_{\mathsf C} \lb{} \xlongleftrightarrow{\glvreshape{\glvop{RREFRep}(\lenv{0},\glvreshape{\lenv{1}}{\lsv{} \loprod \lsv{}}{\lsv{} \loprod \lsv{}})}{\lsv{} \loprod \lsv{}}{\lsv{} \loprod \lsv{}}} \llpl{} \lee{} \mathbin{:\!:} \mathit{es}\llpr{}_{\lb{}'}}\\
\gpfitbox{\textsc{C-C-Refl}} & \gpfit{\lepsilon{};\evalCtx{}\vdash_{\mathsf C} \lbn{0}\xlongleftrightarrow{\leev{}}\lbn{1}} & \gpfit{\evalCtxSpace{};\evalCtx{}\vdash_{\mathsf C} \lb{} \xlongleftrightarrow{\glvreshape{\lidelta{\lsv{}}}{\lsv{} \loprod \lsv{}}{\lsv{} \loprod \lsv{}}} \lb{}}\\
\gpfitbox{\textsc{C-C-Proj\allowbreak Ker\allowbreak Coker\allowbreak Cod}} & \gpfit{\lepsilon{};\evalCtx{}\vdash_{\mathsf C} \lbn{0}\xlongleftrightarrow{\leev{}}\lbn{1}} & \gpfit{\evalCtxSpace{};\evalCtx{}\vdash_{\mathsf C} \pBasisC{\ls{}}{\lbn{0}}{\lbn{1}} \xlongleftrightarrow{\glvmtm{\lsnv{0}}{\lsnv{0}}{\glvreshape{\lenv{1}}{\lsnv{0} \loprod \lsv{}'}{\lsnv{0} \loprod \lsv{}'}}{\glvreshape{\lenv{3}}{\lsv{} \loprod \lsnv{0}}{\lsv{} \loprod \lsnv{0}}}} \pBasisC{\ls{}}{\lbn{0}}{\lbn{1}'}}\\
\gpfitbox{\textsc{C-C-Proj\allowbreak Img\allowbreak Coimg\allowbreak Dom}} & \gpfit{\lepsilon{};\evalCtx{}\vdash_{\mathsf C} \lbn{0}\xlongleftrightarrow{\leev{}}\lbn{1}} & \gpfit{\evalCtxSpace{};\evalCtx{}\vdash_{\mathsf C} \pBasisC{\ls{}}{\lbn{0}}{\lbn{1}} \xlongleftrightarrow{\glvmtm{\lsnv{1}}{\lsnv{1}}{\glvreshape{\lenv{1}}{\lsnv{1} \loprod \lsv{}'}{\lsnv{1} \loprod \lsv{}'}}{\glvreshape{\lenv{3}}{\lsv{} \loprod \lsnv{1}}{\lsv{} \loprod \lsnv{1}}}} \pBasisC{\ls{}}{\lbn{0}'}{\lbn{1}}}\\
\gpfitbox{\textsc{C-E-Var\allowbreak Lin}} & \gpfit{\lDelta{};\lepsilon{};\evalCtx{}\vdash_{\mathsf C} \lee{}\xlongrightarrow{\lb{}}\leev{}\glresulttype{\ltyv{}}} & \gpfit{\{(\mathit{x}, \ls{})\};\evalCtxSpace{};\evalCtx{}\vdash_{\mathsf C} \mathit{x} \xlongrightarrow{\lb{}} \leev{}\glresulttype{\ltyv{}}}\\
\gpfitbox{\textsc{C-E-Var\allowbreak Nlin}} & \gpfit{\lDelta{};\lepsilon{};\evalCtx{}\vdash_{\mathsf C} \lee{}\xlongrightarrow{\lb{}}\leev{}\glresulttype{\ltyv{}}} & \gpfit{\lemptyset{};\evalCtxSpace{};\evalCtx{}\vdash_{\mathsf C} \mathit{x} \xlongrightarrow{\lb{}} \leev{}\glresulttype{\ltyv{}}}\\
\gpfitbox{\textsc{C-E-Promote}} & \gpfit{\lDelta{};\lepsilon{};\evalCtx{}\vdash_{\mathsf C} \lee{}\xlongrightarrow{\lb{}}\leev{}\glresulttype{\ltyv{}}} & \gpfit{\lDelta{};\mathit{lin} \ldisun \evalCtxSpace{}'';\evalCtx{}\vdash_{\mathsf C} \lee{} \xlongrightarrow{\lb{}} \leev{}\glresulttype{\ltyv{}}}\\
\gpfitbox{\textsc{C-E-Cob}} & \gpfit{\lDelta{};\lepsilon{};\evalCtx{}\vdash_{\mathsf C} \lee{}\xlongrightarrow{\lb{}}\leev{}\glresulttype{\ltyv{}}} & \gpfit{\lDelta{};\evalCtxSpace{};\evalCtx{}\vdash_{\mathsf C} \lee{} \xlongrightarrow{\lbn{0}} \LTTVPP{\lsv{}}{\glvreshape{\lenv{0}}{\lsv{} \loprod \lsv{}}{\lsv{} \loprod \lsv{}}}{\glvreshape{\leev{}}{\lsv{}}{\lsv{}}}\glresulttype{\lsv{}}}\\
\gpfitbox{\textsc{C-E-Reshape\allowbreak Cob}} & \gpfit{\lDelta{};\lepsilon{};\evalCtx{}\vdash_{\mathsf C} \lee{}\xlongrightarrow{\lb{}}\leev{}\glresulttype{\ltyv{}}} & \gpfit{\lDelta{};\evalCtxSpace{};\evalCtx{}\vdash_{\mathsf C} \lee{} \xlongrightarrow{\lb{}} \glvreshape{\leev{}}{\ltyv{}'}{\ltyv{}}\glresulttype{\ltyv{}'}}\\
\gpfitbox{\textsc{C-E-Scale}} & \gpfit{\lDelta{};\lepsilon{};\evalCtx{}\vdash_{\mathsf C} \lee{}\xlongrightarrow{\lb{}}\leev{}\glresulttype{\ltyv{}}} & \gpfit{{\lDelta{}_{0}} \ldisun {\lDelta{}_{1}};\evalCtxSpace{};\evalCtx{}\vdash_{\mathsf C} \len{0} \lvscale \len{1} \xlongrightarrow{\lb{}} \lenv{0} \lvscale \lenv{1}\glresulttype{\lsv{}}}\\
\gpfitbox{\textsc{C-E-Add}} & \gpfit{\lDelta{};\lepsilon{};\evalCtx{}\vdash_{\mathsf C} \lee{}\xlongrightarrow{\lb{}}\leev{}\glresulttype{\ltyv{}}} & \gpfit{\lDelta{};\evalCtxSpace{};\evalCtx{}\vdash_{\mathsf C} \len{0} \lvadd \len{1} \xlongrightarrow{\lb{}} \lenv{0} \lvadd_{\lsv{}} \lenv{1}\glresulttype{\lsv{}}}\\
\gpfitbox{\textsc{C-E-Dsum\allowbreak E}} & \gpfit{\lDelta{};\lepsilon{};\evalCtx{}\vdash_{\mathsf C} \lee{}\xlongrightarrow{\lb{}}\leev{}\glresulttype{\ltyv{}}} & \gpfit{\lDelta{};\evalCtxSpace{};\evalCtx{}\vdash_{\mathsf C} \len{0}\losum \len{1} \xlongrightarrow{\lbn{0} \losum \lbn{1}} \lenv{0}\losum \lenv{1}\glresulttype{\lsnv{0} \losum \lsnv{1}}}\\
\gpfitbox{\textsc{C-E-Tprod\allowbreak E}} & \gpfit{\lDelta{};\lepsilon{};\evalCtx{}\vdash_{\mathsf C} \lee{}\xlongrightarrow{\lb{}}\leev{}\glresulttype{\ltyv{}}} & \gpfit{{\lDelta{}_{0}} \ldisun {\lDelta{}_{1}};\evalCtxSpace{};\evalCtx{}\vdash_{\mathsf C} \len{0}\loprod \len{1} \xlongrightarrow{\lbn{0} \loprod \lbn{1}} \lenv{0}\loprod \lenv{1}\glresulttype{\lsnv{0} \loprod \lsnv{1}}}\\
\gpfitbox{\textsc{C-E-Pinv\allowbreak TStd}} & \gpfit{\lDelta{};\lepsilon{};\evalCtx{}\vdash_{\mathsf C} \lee{}\xlongrightarrow{\lb{}}\leev{}\glresulttype{\ltyv{}}} & \gpfit{\lemptyset{};\evalCtxSpace{};\evalCtx{}\vdash_{\mathsf C} \lee{}^{\lpseudo} \xlongrightarrow{\lbn{0}^{\lstar} \loprod \lbn{1}} \leev{}^{+}\glresulttype{\lsnv{0} \loprod \lsnv{1}}}\\
\gpfitbox{\textsc{C-E-Pinv\allowbreak TGpinv}} & \gpfit{\lDelta{};\lepsilon{};\evalCtx{}\vdash_{\mathsf C} \lee{}\xlongrightarrow{\lb{}}\leev{}\glresulttype{\ltyv{}}} & \gpfit{\lemptyset{};\evalCtxSpace{};\evalCtx{}\vdash_{\mathsf C} \lee{}^{\lpseudo} \xlongrightarrow{\lbn{0}^{\lstar} \loprod \lbn{1}} \glvop{GPINV}(\leev{},\lenv{1},\lenv{0})\glresulttype{\lsnv{0} \loprod \lsnv{1}}}\\
\gpfitbox{\textsc{C-E-Pinv\allowbreak DStd}} & \gpfit{\lDelta{};\lepsilon{};\evalCtx{}\vdash_{\mathsf C} \lee{}\xlongrightarrow{\lb{}}\leev{}\glresulttype{\ltyv{}}} & \gpfit{\lemptyset{};\evalCtxSpace{};\evalCtx{}\vdash_{\mathsf C} \lee{}^{\lpseudo} \xlongrightarrow{\lbn{0}^{\lstar} \loprod \lbn{1}} \glvreshape{\leev{}}{\lsnv{1} \loprod \lR{}}{\lsnv{1}}^{+}\glresulttype{\lR{} \loprod \lsnv{1}}}\\
\gpfitbox{\textsc{C-E-Pinv\allowbreak DGpinv}} & \gpfit{\lDelta{};\lepsilon{};\evalCtx{}\vdash_{\mathsf C} \lee{}\xlongrightarrow{\lb{}}\leev{}\glresulttype{\ltyv{}}} & \gpfit{\lemptyset{};\evalCtxSpace{};\evalCtx{}\vdash_{\mathsf C} \lee{}^{\lpseudo} \xlongrightarrow{\lbn{0}^{\lstar} \loprod \lbn{1}} \glvop{GPINV}(\glvreshape{\leev{}}{\lsnv{1} \loprod \lR{}}{\lsnv{1}},\lenv{1},\lenv{0})\glresulttype{\lR{} \loprod \lsnv{1}}}\\
\gpfitbox{\textsc{C-E-Inner\allowbreak Std}} & \gpfit{\lDelta{};\lepsilon{};\evalCtx{}\vdash_{\mathsf C} \lee{}\xlongrightarrow{\lb{}}\leev{}\glresulttype{\ltyv{}}} & \gpfit{\lemptyset{};\evalCtxSpace{};\evalCtx{}\vdash_{\mathsf C} \linner{}_{\ls{}} \xlongrightarrow{\lbn{0}^{\lstar} \loprod \lbn{0}^{\lstar}} \lidelta{\lsv{}}\glresulttype{\lsv{} \loprod \lsv{}}}\\
\gpfitbox{\textsc{C-E-Embed\allowbreak Ker}} & \gpfit{\lDelta{};\lepsilon{};\evalCtx{}\vdash_{\mathsf C} \lee{}\xlongrightarrow{\lb{}}\leev{}\glresulttype{\ltyv{}}} & \gpfit{\lemptyset{};\evalCtxSpace{};\evalCtx{}\vdash_{\mathsf C} \lembd{}_{\Kernel(\lee{})} \xlongrightarrow{\pBasisC{\Kernel(\lee{})}{\lbn{0}}{\lbn{1}}^{\lstar} \loprod \lbn{0}} \Trans(\glvop{GSVDUI}(\leev{},\lenv{0},\lenv{1}))\glresulttype{\lR^{\glvreshape{\leev{}}{\lsnv{0} \loprod \lsnv{1}}{\lsnv{0} \loprod \lsnv{1}},\Kernel{},\lsnv{0} \loprod \lsnv{1}} \loprod \lsnv{0}}}\\
\gpfitbox{\textsc{C-E-Embed\allowbreak Coker}} & \gpfit{\lDelta{};\lepsilon{};\evalCtx{}\vdash_{\mathsf C} \lee{}\xlongrightarrow{\lb{}}\leev{}\glresulttype{\ltyv{}}} & \gpfit{\lemptyset{};\evalCtxSpace{};\evalCtx{}\vdash_{\mathsf C} \lembd{}_{\CoKernel(\lee{})} \xlongrightarrow{\pBasisC{\CoKernel(\lee{})}{\lbn{0}}{\lbn{1}}^{\lstar} \loprod \lbn{0}} \Trans(\glvop{GSVDU}(\leev{},\lenv{0},\lenv{1}))\glresulttype{\lR^{\glvreshape{\leev{}}{\lsnv{0} \loprod \lsnv{1}}{\lsnv{0} \loprod \lsnv{1}},\CoKernel{},\lsnv{0} \loprod \lsnv{1}} \loprod \lsnv{0}}}\\
\gpfitbox{\textsc{C-E-Embed\allowbreak Img}} & \gpfit{\lDelta{};\lepsilon{};\evalCtx{}\vdash_{\mathsf C} \lee{}\xlongrightarrow{\lb{}}\leev{}\glresulttype{\ltyv{}}} & \gpfit{\lemptyset{};\evalCtxSpace{};\evalCtx{}\vdash_{\mathsf C} \lembd{}_{\Image(\lee{})} \xlongrightarrow{\pBasisC{\Image(\lee{})}{\lbn{0}}{\lbn{1}}^{\lstar} \loprod \lbn{1}} \Trans(\glvop{GSVDV}(\leev{},\lenv{0},\lenv{1}))\glresulttype{\lR^{\glvreshape{\leev{}}{\lsnv{0} \loprod \lsnv{1}}{\lsnv{0} \loprod \lsnv{1}},\Image{},\lsnv{0} \loprod \lsnv{1}} \loprod \lsnv{1}}}\\
\gpfitbox{\textsc{C-E-Embed\allowbreak Coimg}} & \gpfit{\lDelta{};\lepsilon{};\evalCtx{}\vdash_{\mathsf C} \lee{}\xlongrightarrow{\lb{}}\leev{}\glresulttype{\ltyv{}}} & \gpfit{\lemptyset{};\evalCtxSpace{};\evalCtx{}\vdash_{\mathsf C} \lembd{}_{\CoImage(\lee{})} \xlongrightarrow{\pBasisC{\CoImage(\lee{})}{\lbn{0}}{\lbn{1}}^{\lstar} \loprod \lbn{1}} \Trans(\glvop{GSVDVI}(\leev{},\lenv{0},\lenv{1}))\glresulttype{\lR^{\glvreshape{\leev{}}{\lsnv{0} \loprod \lsnv{1}}{\lsnv{0} \loprod \lsnv{1}},\CoImage{},\lsnv{0} \loprod \lsnv{1}} \loprod \lsnv{1}}}\\
\gpfitbox{\textsc{C-E-App\allowbreak LLin}} & \gpfit{\lDelta{};\lepsilon{};\evalCtx{}\vdash_{\mathsf C} \lee{}\xlongrightarrow{\lb{}}\leev{}\glresulttype{\ltyv{}}} & \gpfit{{\lDelta{}_{0}} \ldisun {\lDelta{}_{1}};\evalCtxSpace{};\evalCtx{}\vdash_{\mathsf C} \leval{\len{0}}{}{\len{1}} \xlongrightarrow{\lbn{1}} \glvreshape{\LTTVPP{\lsv{}}{\lenv{0}}{\lenv{1}}}{\lsnv{{\mathrm{svOut}}}}{\lsnv{{\mathrm{svOut}}}}\glresulttype{\lsnv{{\mathrm{svOut}}}}}\\
\gpfitbox{\textsc{C-E-App\allowbreak ILin}} & \gpfit{\lDelta{};\lepsilon{};\evalCtx{}\vdash_{\mathsf C} \lee{}\xlongrightarrow{\lb{}}\leev{}\glresulttype{\ltyv{}}} & \gpfit{{\lDelta{}_{0}} \ldisun {\lDelta{}_{1}};\evalCtxSpace{};\evalCtx{}\vdash_{\mathsf C} \leval{\len{0}}{}{\len{1}} \xlongrightarrow{\lbone{}} \LTTVPP{\lsv{}}{\glvreshape{\lenv{0}}{\lsv{} \loprod \lR{}}{\lsv{}}}{\lenv{1}}\glresulttype{\lR{}}}\\
\gpfitbox{\textsc{C-E-Nllam}} & \gpfit{\lDelta{};\lepsilon{};\evalCtx{}\vdash_{\mathsf C} \lee{}\xlongrightarrow{\lb{}}\leev{}\glresulttype{\ltyv{}}} & \gpfit{\lemptyset{};\evalCtxSpace{};\evalCtx{}\vdash_{\mathsf C} (\nllambda \llpl x\colon \ltyn{0}\llpr \ldotsp \lee{}) \xlongrightarrow{\frameB{x}{\lbn{0}}{\lbn{1}}} \llambda(x\colon \ltynv{0})\lrh \ltynv{1}\colon \glvreshape{\leev{}}{\ltynv{1}}{\ltynv{1}}\glresulttype{(x\colon \ltynv{0}\fnto \ltynv{1})}}\\
\gpfitbox{\textsc{C-E-App\allowbreak NVal}} & \gpfit{\lDelta{};\lepsilon{};\evalCtx{}\vdash_{\mathsf C} \lee{}\xlongrightarrow{\lb{}}\leev{}\glresulttype{\ltyv{}}} & \gpfit{\lemptyset{};\evalCtxSpace{};\evalCtx{}\vdash_{\mathsf C} \leval{\len{0}}{}{\len{1}} \xlongrightarrow{\lbn{1}\subst{x}{\len{1}}} \glvreshape{\glvapp{\lenv{0}}{\lenv{2}}}{{\ltyv{}_{\mathsf{out}}}}{\ltynv{{\mathrm{C}}}\subst{x}{\lenv{2}}}\glresulttype{{\ltyv{}_{\mathsf{out}}}}}\\
\gpfitbox{\textsc{C-E-Frame\allowbreak Reanchor}} & \gpfit{\lDelta{};\lepsilon{};\evalCtx{}\vdash_{\mathsf C} \lee{}\xlongrightarrow{\lb{}}\leev{}\glresulttype{\ltyv{}}} & \gpfit{\lemptyset{};\evalCtxSpace{};\evalCtx{}\vdash_{\mathsf C} \len{0} \xlongrightarrow{\frameB{x}{\lbn{0}}{\lbn{1}}} \llambda(x\colon \ltynv{0})\lrh \ltynv{1}'\colon \glvreshape{\glvapp{\leev{}}{\lenv{2}}}{\ltynv{1}'}{\ltynv{1}'}\glresulttype{(x\colon \ltynv{0}\fnto \ltynv{1}')}}\\
\gpfitbox{\textsc{C-E-Frame\allowbreak Eta}} & \gpfit{\lDelta{};\lepsilon{};\evalCtx{}\vdash_{\mathsf C} \lee{}\xlongrightarrow{\lb{}}\leev{}\glresulttype{\ltyv{}}} & \gpfit{\lemptyset{};\evalCtxSpace{};\evalCtx{}\vdash_{\mathsf C} \len{0} \xlongrightarrow{\frameB{x}{\lbn{0}}{\lbn{1}}} \llambda(x\colon \ltynv{0})\lrh \ltynv{1}\colon \glvreshape{\leev{}}{\ltynv{1}}{\ltynv{1}}\glresulttype{(x\colon \ltynv{0}\fnto \ltynv{1})}}\\
\gpfitbox{\textsc{C-E-Llam\allowbreak Assemble}} & \gpfit{\lDelta{};\lepsilon{};\evalCtx{}\vdash_{\mathsf C} \lee{}\xlongrightarrow{\lb{}}\leev{}\glresulttype{\ltyv{}}} & \gpfit{\lDelta{};\evalCtxSpace{};\evalCtx{}\vdash_{\mathsf C} (\lllambda \llpl x\colon \sof{\lb{}'}\llpr \ldotsp \lee{}) \xlongrightarrow{\lb{}'^{\lstar} \loprod \lb{}} \glvreshape{\lsum^{\lsv{}'}_{\mathit{i}<\overline{\operatorname{dimOf}}(\lsv{}')}\left(\ldelta{\mathit{i}}{\lsv{}'}\loprod \glvreshape{\leev{}_{\mathit{i}}}{\lsv{}}{\lsv{}}\right)}{\lsv{}' \loprod \lsv{}}{\begin{cases}\lsempty{}&\overline{\operatorname{dimOf}}(\lsv{}')=0\\ \lsv{}' \loprod \lsv{}&\text{otherwise}\end{cases}}\glresulttype{\lsv{}' \loprod \lsv{}}}\\
\gpfitbox{\textsc{C-E-Llam\allowbreak Assemble\allowbreak Real}} & \gpfit{\lDelta{};\lepsilon{};\evalCtx{}\vdash_{\mathsf C} \lee{}\xlongrightarrow{\lb{}}\leev{}\glresulttype{\ltyv{}}} & \gpfit{\lDelta{};\evalCtxSpace{};\evalCtx{}\vdash_{\mathsf C} (\lllambda \llpl x\colon \sof{\lb{}'}\llpr \ldotsp \lee{}) \xlongrightarrow{\lb{}'^{\lstar}} \glvreshape{\lsum^{\lsv{}'}_{\mathit{i}<\overline{\operatorname{dimOf}}(\lsv{}')}\left(\glvreshape{\leev{}_{\mathit{i}}}{\lR{}}{\lR{}} \lvscale \ldelta{\mathit{i}}{\lsv{}'}\right)}{\lsv{}'}{\begin{cases}\lsempty{}&\overline{\operatorname{dimOf}}(\lsv{}')=0\\ \lsv{}'&\text{otherwise}\end{cases}}\glresulttype{\lsv{}'}}\\
\gpfitbox{\textsc{C-E-Zero}} & \gpfit{\lDelta{};\lepsilon{};\evalCtx{}\vdash_{\mathsf C} \lee{}\xlongrightarrow{\lb{}}\leev{}\glresulttype{\ltyv{}}} & \gpfit{\lDelta{};\evalCtxSpace{};\evalCtx{}\vdash_{\mathsf C} \lvzero{}_{\sof{\lb{}}} \xlongrightarrow{\lb{}} \lvzero{}_{\lsv{}}\glresulttype{\lsv{}}}\\
\gpfitbox{\textsc{C-E-Rlit}} & \gpfit{\lDelta{};\lepsilon{};\evalCtx{}\vdash_{\mathsf C} \lee{}\xlongrightarrow{\lb{}}\leev{}\glresulttype{\ltyv{}}} & \gpfit{\lemptyset{};\evalCtxSpace{};\evalCtx{}\vdash_{\mathsf C} \mathit{r} \xlongrightarrow{\lbone{}} \mathit{r}\glresulttype{\lR{}}}\\
}
\gldefcmdrows{syntax}{
\gpfit{\lsempty{}} & Empty Space & \gpfitbox{\texttt{\textbackslash glsyn\{space.empty\}}}\\
\gpfit{\lR{}} & Scalar Space & \gpfitbox{\texttt{\textbackslash glsyn\{space.real\}}}\\
\gpfit{\lsn{0}\losum\lsn{1}} & Direct Sum Space & \gpfitbox{\texttt{\textbackslash glsyn\{space.dsum\}}}\\
\gpfit{\ls{}^{\lstar}} & Dual Space & \gpfitbox{\texttt{\textbackslash glsyn\{space.dual\}}}\\
\gpfit{\lsn{0}\loprod\lsn{1}} & Tensor Product Space & \gpfitbox{\texttt{\textbackslash glsyn\{space.tprod\}}}\\
\gpfit{\Kernel{}(\lee)} & Kernel of a Linear Operator & \gpfitbox{\texttt{\textbackslash glsyn\{space.ker\}}}\\
\gpfit{\CoKernel{}(\lee)} & Cokernel of a Linear Operator & \gpfitbox{\texttt{\textbackslash glsyn\{space.coker\}}}\\
\gpfit{\Image{}(\lee)} & Image of a Linear Operator & \gpfitbox{\texttt{\textbackslash glsyn\{space.img\}}}\\
\gpfit{\CoImage{}(\lee)} & Coimage of a Linear Operator & \gpfitbox{\texttt{\textbackslash glsyn\{space.coimg\}}}\\
\gpfit{\ls{}} & Space as a Type & \gpfitbox{\texttt{\textbackslash glsyn\{ty.space\}}}\\
\gpfit{\fntypep{x}{\ltyn{0}}{\ltyn{1}}} & Dependent Function Type & \gpfitbox{\texttt{\textbackslash glsyn\{ty.piTy\}}}\\
\gpfit{\lkbase{}} & Space Kind & \gpfitbox{\texttt{\textbackslash glsyn\{kind.base\}}}\\
\gpfit{\lknto{\lkn{0}}{\lkn{1}}} & Function Kind & \gpfitbox{\texttt{\textbackslash glsyn\{kind.arrow\}}}\\
\gpfit{\lvzero_{\els{}}} & Zero Vector & \gpfitbox{\texttt{\textbackslash glsyn\{term.zero\}}}\\
\gpfit{\len{0} \lvscale \len{1}} & Scalar-Vector Product & \gpfitbox{\texttt{\textbackslash glsyn\{term.smul\}}}\\
\gpfit{r} & Real Number & \gpfitbox{\texttt{\textbackslash glsyn\{term.real\}}}\\
\gpfit{\len{0} \lvadd \len{1}} & Vector Sum & \gpfitbox{\texttt{\textbackslash glsyn\{term.add\}}}\\
\gpfit{\linner{}_{\els{}}} & Inner Product & \gpfitbox{\texttt{\textbackslash glsyn\{term.inner\}}}\\
\gpfit{\leval{\len{0}}{}{\len{1}}} & Linear Application & \gpfitbox{\texttt{\textbackslash glsyn\{term.appL\}}}\\
\gpfit{\leval{\len{0}}{}{\len{1}}} & Inner-Product Application & \gpfitbox{\texttt{\textbackslash glsyn\{term.appI\}}}\\
\gpfit{\leval{\len{0}}{}{\len{1}}} & Non-linear Application & \gpfitbox{\texttt{\textbackslash glsyn\{term.appN\}}}\\
\gpfit{\lllambdaArg{x\ecolon{}\els{}}{\lee}} & Multi-linear Function & \gpfitbox{\texttt{\textbackslash glsyn\{term.llam\}}}\\
\gpfit{\nllambdaArg{x\ecolon{}\elty{}}{\lee}} & Non-linear Function & \gpfitbox{\texttt{\textbackslash glsyn\{term.nllam\}}}\\
\gpfit{x} & Variable & \gpfitbox{\texttt{\textbackslash glsyn\{term.var\}}}\\
\gpfit{\lee{}^{\lpseudo{}}} & Linear Pseudo-Inverse (tensor) & \gpfitbox{\texttt{\textbackslash glsyn\{term.pinvT\}}}\\
\gpfit{\lee{}^{\lpseudo{}}} & Linear Pseudo-Inverse (dual) & \gpfitbox{\texttt{\textbackslash glsyn\{term.pinvD\}}}\\
\gpfit{\len{0}\losum\len{1}} & Direct Sum & \gpfitbox{\texttt{\textbackslash glsyn\{term.dsumE\}}}\\
\gpfit{\len{0}\loprod\len{1}} & Tensor Product & \gpfitbox{\texttt{\textbackslash glsyn\{term.tprodE\}}}\\
\gpfit{\lembd{}_{\els}} & Embedding & \gpfitbox{\texttt{\textbackslash glsyn\{term.embed\}}}\\
\gpfit{\lbzero{}} & Basis of $\lsempty$ & \gpfitbox{\texttt{\textbackslash glsyn\{basis.empty\}}}\\
\gpfit{\lbone{}} & Basis of $\lR$ & \gpfitbox{\texttt{\textbackslash glsyn\{basis.one\}}}\\
\gpfit{\lb{}^{\lbstar}} & Dual Basis & \gpfitbox{\texttt{\textbackslash glsyn\{basis.dual\}}}\\
\gpfit{\lbn{0}\lbosum\lbn{1}} & Direct Sum Basis & \gpfitbox{\texttt{\textbackslash glsyn\{basis.dsum\}}}\\
\gpfit{\lbn{0}\lboprod\lbn{1}} & Tensor Product Basis & \gpfitbox{\texttt{\textbackslash glsyn\{basis.tprod\}}}\\
\gpfit{\pBasisC{\ls}{\lbn{0}}{\lbn{1}}} & Projected Basis & \gpfitbox{\texttt{\textbackslash glsyn\{basis.proj\}}}\\
\gpfit{\iBasisC{\lb{}}} & Image Basis & \gpfitbox{\texttt{\textbackslash glsyn\{basis.image\}}}\\
\gpfit{\llpl{}\len{0},\dots,\len{n}\llpr{}_{\lb{}}} & Term-sequence Basis & \gpfitbox{\texttt{\textbackslash glsyn\{basis.terms\}}}\\
\gpfit{\frameB{x}{\lbn{0}}{\lbn{1}}} & Frame (Dependent Basis) & \gpfitbox{\texttt{\textbackslash glsyn\{basis.frame\}}}\\
\gpfit{\Kernel{}} & \texttt{kernel} & \gpfitbox{\texttt{\textbackslash glsyn\{valFFSProp.kernel\}}}\\
\gpfit{\CoKernel{}} & \texttt{cokernel} & \gpfitbox{\texttt{\textbackslash glsyn\{valFFSProp.cokernel\}}}\\
\gpfit{\Image{}} & \texttt{image} & \gpfitbox{\texttt{\textbackslash glsyn\{valFFSProp.image\}}}\\
\gpfit{\CoImage{}} & \texttt{coimage} & \gpfitbox{\texttt{\textbackslash glsyn\{valFFSProp.coimage\}}}\\
\gpfit{\lsempty{}} & \texttt{empty} & \gpfitbox{\texttt{\textbackslash glsyn\{valSpace.empty\}}}\\
\gpfit{\lR{}} & \texttt{real} & \gpfitbox{\texttt{\textbackslash glsyn\{valSpace.real\}}}\\
\gpfit{\lsnv{0} \losum \lsnv{1}} & \texttt{dsum} & \gpfitbox{\texttt{\textbackslash glsyn\{valSpace.dsum\}}}\\
\gpfit{\lsnv{0} \loprod \lsnv{1}} & \texttt{tprod} & \gpfitbox{\texttt{\textbackslash glsyn\{valSpace.tprod\}}}\\
\gpfit{\lR^{\leev{},\lprop{},\lsv{}}} & Four fundamental subspaces of a value linear map & \gpfitbox{\texttt{\textbackslash glsyn\{valSpace.ffs\}}}\\
\gpfit{\lsv{}} & \texttt{space} & \gpfitbox{\texttt{\textbackslash glsyn\{valTy.space\}}}\\
\gpfit{(x\colon \ltynv{0}\fnto \ltynv{1})} & \texttt{pi\allowbreak Ty} & \gpfitbox{\texttt{\textbackslash glsyn\{valTy.piTy\}}}\\
\gpfit{\lvzero{}_{\lsv{}}} & Vector-space generic & \gpfitbox{\texttt{\textbackslash glsyn\{valExpr.zero\}}}\\
\gpfit{r} & \texttt{rlit} & \gpfitbox{\texttt{\textbackslash glsyn\{valExpr.rlit\}}}\\
\gpfit{\lenv{0} \lvscale \lenv{1}} & \texttt{scale} & \gpfitbox{\texttt{\textbackslash glsyn\{valExpr.scale\}}}\\
\gpfit{\lenv{0} \lvadd_{\lsv{}} \lenv{1}} & \texttt{vadd} & \gpfitbox{\texttt{\textbackslash glsyn\{valExpr.vadd\}}}\\
\gpfit{\lenv{0}\losum \lenv{1}} & \texttt{dsum\allowbreak E} & \gpfitbox{\texttt{\textbackslash glsyn\{valExpr.dsumE\}}}\\
\gpfit{\lenv{0}\loprod \lenv{1}} & \texttt{tprod\allowbreak E} & \gpfitbox{\texttt{\textbackslash glsyn\{valExpr.tprodE\}}}\\
\gpfit{\ldelta{n}{\lsv{}}} & Basis vector / identity tensor & \gpfitbox{\texttt{\textbackslash glsyn\{valExpr.delta\}}}\\
\gpfit{\lidelta{\lsv{}}} & \texttt{idelta} & \gpfitbox{\texttt{\textbackslash glsyn\{valExpr.idelta\}}}\\
\gpfit{\lsum^{\lsv{}}_{v\in \mathit{vs}}v} & Sum over a value-space index & \gpfitbox{\texttt{\textbackslash glsyn\{valExpr.vsum\}}}\\
\gpfit{\llambda(x\colon \ltynv{0})\lrh \ltynv{1}\colon \leev{}} & Lambdas & \gpfitbox{\texttt{\textbackslash glsyn\{valExpr.vlam\}}}\\
\gpfit{x_{\ltyv{}}} & Free variable & \gpfitbox{\texttt{\textbackslash glsyn\{valExpr.var\}}}\\
\gpfit{\LTTVPP{\lsv{}}{\lenv{0}}{\lenv{1}}} & Tensor-vector partial contraction (paper: & \gpfitbox{\texttt{\textbackslash glsyn\{valExpr.ttv\}}}\\
\gpfit{\glvmtm{\lsnv{c}}{\lsnv{1}}{\lenv{0}}{\lenv{1}}} & Matrix-tensor-matrix / "MTM" (paper: & \gpfitbox{\texttt{\textbackslash glsyn\{valExpr.mtm\}}}\\
\gpfit{\glvapp{\lenv{0}}{\lenv{1}}} & Application (linear / non-linear share one constructor in the value layer; typing distinguishes) & \gpfitbox{\texttt{\textbackslash glsyn\{valExpr.appV\}}}\\
\gpfit{\vkrn(\lenv{0}, \lenv{1})} & Kronecker / block products on values & \gpfitbox{\texttt{\textbackslash glsyn\{valExpr.vkrn\}}}\\
\gpfit{\vblk(\lenv{0}, \lenv{1})} & \texttt{vblk} & \gpfitbox{\texttt{\textbackslash glsyn\{valExpr.vblk\}}}\\
\gpfit{\glvreshape{\leev{}}{\ltynv{t}}{\ltynv{s}}} & Reshape between value types & \gpfitbox{\texttt{\textbackslash glsyn\{valExpr.reshap\}}}\\
\gpfit{\Trans(\leev{})} & Transpose of a tensor-shaped value & \gpfitbox{\texttt{\textbackslash glsyn\{valExpr.vtrans\}}}\\
\gpfit{\glvop{GSVDS}(\lenv{0},\lenv{1},\lenv{2})} & (G)SVD & \gpfitbox{\texttt{\textbackslash glsyn\{valExpr.gsvdS\}}}\\
\gpfit{\glvop{GSVDU}(\lenv{0},\lenv{1},\lenv{2})} & \texttt{gsvd\allowbreak U} & \gpfitbox{\texttt{\textbackslash glsyn\{valExpr.gsvdU\}}}\\
\gpfit{\glvop{GSVDUI}(\lenv{0},\lenv{1},\lenv{2})} & \texttt{gsvd\allowbreak UI} & \gpfitbox{\texttt{\textbackslash glsyn\{valExpr.gsvdUI\}}}\\
\gpfit{\glvop{GSVDV}(\lenv{0},\lenv{1},\lenv{2})} & \texttt{gsvd\allowbreak V} & \gpfitbox{\texttt{\textbackslash glsyn\{valExpr.gsvdV\}}}\\
\gpfit{\glvop{GSVDVI}(\lenv{0},\lenv{1},\lenv{2})} & \texttt{gsvd\allowbreak VI} & \gpfitbox{\texttt{\textbackslash glsyn\{valExpr.gsvdVI\}}}\\
\gpfit{\leev{}^{+}} & Pseudo-inverse of a value (paper writes & \gpfitbox{\texttt{\textbackslash glsyn\{valExpr.pinv\}}}\\
\gpfit{\glvop{GPINV}(\lenv{0},\lenv{1},\lenv{2})} & Generalized pseudo-inverse with explicit inner-product operands & \gpfitbox{\texttt{\textbackslash glsyn\{valExpr.gpinv\}}}\\
\gpfit{\glvop{RREFRep}(\lenv{0},\lenv{1})} & RREF representation & \gpfitbox{\texttt{\textbackslash glsyn\{valExpr.rrefRep\}}}\\
\gpfit{\leev{}^{-1}} & Inverse-form & \gpfitbox{\texttt{\textbackslash glsyn\{valExpr.inv\}}}\\
\gpfit{\glvop{SVDS}(\leev{})} & Standard-basis singular-value block & \gpfitbox{\texttt{\textbackslash glsyn\{valExpr.svds\}}}\\
\gpfit{\lkbase{}} & \texttt{base} & \gpfitbox{\texttt{\textbackslash glsyn\{valKind.base\}}}\\
\gpfit{\lkn{0} \fnto \lkn{1}} & \texttt{arrow} & \gpfitbox{\texttt{\textbackslash glsyn\{valKind.arrow\}}}\\
}
\gldefcmdrows{meta}{
\gpfit{\lDelta{}} & Linear context & \gpfitbox{\texttt{\textbackslash glsyn\{meta.lctx\}}}\\
\gpfit{\lepsilon{}} & Non-linear context & \gpfitbox{\texttt{\textbackslash glsyn\{meta.nlctx\}}}\\
\gpfit{\evalCtx{}} & Evaluation context (the store) & \gpfitbox{\texttt{\textbackslash glsyn\{meta.evalCtx\}}}\\
\gpfit{\sigma} & Denotational environment & \gpfitbox{\texttt{\textbackslash glsyn\{meta.denv\}}}\\
\gpfit{\valCheckCtx{}} & Value typing context & \gpfitbox{\texttt{\textbackslash glsyn\{meta.valCtx\}}}\\
\gpfit{\glvEnv} & Value semantic environment & \gpfitbox{\texttt{\textbackslash glsyn\{meta.valEnv\}}}\\
\gpfit{\mathcal{O}(\glvEnv;\Gamma;\Delta)} & Constructive decomposition oracle at a named environment and local type/value stacks & \gpfitbox{\texttt{\textbackslash glsyn\{meta.directOracle\}}}\\
\gpfit{\tof{\lb{}}} & Type of a basis & \gpfitbox{\texttt{\textbackslash glsyn\{meta.tof\}}}\\
\gpfit{\sof{\lb{}}} & Space of a basis & \gpfitbox{\texttt{\textbackslash glsyn\{meta.sof\}}}\\
\gpfit{\stdBasisCtx{\lty{}}{\lctxnle}} & Syntax-directed standard basis of a source type & \gpfitbox{\texttt{\textbackslash glsyn\{meta.stdBasis\}}}\\
\gpfit{\stdBasisCtx{\ls{}}{\lctxnle}} & Space-level recursion used by the standard-basis construction & \gpfitbox{\texttt{\textbackslash glsyn\{meta.stdBasisSpace\}}}\\
\gpfit{\iof{\ls{}}} & Derived source term denoting the canonical inner product on a space & \gpfitbox{\texttt{\textbackslash glsyn\{meta.iof\}}}\\
\gpfit{\dof{\lsv{}}{\glvEnv}} & Degrees of freedom of a value space & \gpfitbox{\texttt{\textbackslash glsyn\{meta.dof\}}}\\
\gpfit{\shpOf{\lsv{}}{\glvEnv}} & Shape of a value space & \gpfitbox{\texttt{\textbackslash glsyn\{meta.shpOf\}}}\\
\gpfit{\operatorname{inferType}(\leev{})} & Inferred type of a value & \gpfitbox{\texttt{\textbackslash glsyn\{meta.inferType\}}}\\
\gpfit{\operatorname{inferSpace}(\leev{})} & Inferred result space of a value & \gpfitbox{\texttt{\textbackslash glsyn\{meta.inferSpace\}}}\\
\gpfit{\Sem{\lee{}}{\sigma}} & Denotation of a source term & \gpfitbox{\texttt{\textbackslash glsyn\{meta.denotTerm\}}}\\
\gpfit{\SemT{\lty{}}{\sigma}} & Denotation of a source type & \gpfitbox{\texttt{\textbackslash glsyn\{meta.denotTy\}}}\\
\gpfit{\SemBB{\lb{}}{\sigma}} & Denotation of a basis & \gpfitbox{\texttt{\textbackslash glsyn\{meta.denotBasis\}}}\\
\gpfit{\SemA{\leev{}}{\glvEnv}{}} & Denotation of a value expression & \gpfitbox{\texttt{\textbackslash glsyn\{meta.denotVal\}}}\\
}
\gldefcmdrows{rules}{
\gpfitbox{\textsc{Ctx-Empty}} & \gpfitbox{\texttt{\textbackslash glrule\{inductiveCtxWF.empty\}}} & \gpfitbox{\texttt{\footnotesize GLang/Source\allowbreak Context\allowbreak Inductive.lean:29}}\\
\gpfitbox{\textsc{Ctx-Linear}} & \gpfitbox{\texttt{\textbackslash glrule\{inductiveCtxWF.linear\}}} & \gpfitbox{\texttt{\footnotesize GLang/Source\allowbreak Context\allowbreak Inductive.lean:30}}\\
\gpfitbox{\textsc{Ctx-Nonlinear}} & \gpfitbox{\texttt{\textbackslash glrule\{inductiveCtxWF.nonlinear\}}} & \gpfitbox{\texttt{\footnotesize GLang/Source\allowbreak Context\allowbreak Inductive.lean:36}}\\
\gpfitbox{\textsc{S-Empty}} & \gpfitbox{\texttt{\textbackslash glrule\{sValid.empty\}}} & \gpfitbox{\texttt{\footnotesize GLang/Typing.lean:133}}\\
\gpfitbox{\textsc{S-Real}} & \gpfitbox{\texttt{\textbackslash glrule\{sValid.real\}}} & \gpfitbox{\texttt{\footnotesize GLang/Typing.lean:135}}\\
\gpfitbox{\textsc{S-Dual}} & \gpfitbox{\texttt{\textbackslash glrule\{sValid.dual\}}} & \gpfitbox{\texttt{\footnotesize GLang/Typing.lean:137}}\\
\gpfitbox{\textsc{S-Dsum}} & \gpfitbox{\texttt{\textbackslash glrule\{sValid.dsum\}}} & \gpfitbox{\texttt{\footnotesize GLang/Typing.lean:139}}\\
\gpfitbox{\textsc{S-Tprod}} & \gpfitbox{\texttt{\textbackslash glrule\{sValid.tprod\}}} & \gpfitbox{\texttt{\footnotesize GLang/Typing.lean:141}}\\
\gpfitbox{\textsc{S-Ker}} & \gpfitbox{\texttt{\textbackslash glrule\{sValid.ker\}}} & \gpfitbox{\texttt{\footnotesize GLang/Typing.lean:149}}\\
\gpfitbox{\textsc{S-Coker}} & \gpfitbox{\texttt{\textbackslash glrule\{sValid.coker\}}} & \gpfitbox{\texttt{\footnotesize GLang/Typing.lean:152}}\\
\gpfitbox{\textsc{S-Img}} & \gpfitbox{\texttt{\textbackslash glrule\{sValid.img\}}} & \gpfitbox{\texttt{\footnotesize GLang/Typing.lean:155}}\\
\gpfitbox{\textsc{S-Coimg}} & \gpfitbox{\texttt{\textbackslash glrule\{sValid.coimg\}}} & \gpfitbox{\texttt{\footnotesize GLang/Typing.lean:158}}\\
\gpfitbox{\textsc{K-Space}} & \gpfitbox{\texttt{\textbackslash glrule\{tValid.space\}}} & \gpfitbox{\texttt{\footnotesize GLang/Typing.lean:165}}\\
\gpfitbox{\textsc{K-Pi\allowbreak Ty}} & \gpfitbox{\texttt{\textbackslash glrule\{tValid.piTy\}}} & \gpfitbox{\texttt{\footnotesize GLang/Typing.lean:170}}\\
\gpfitbox{\textsc{Sub-Sub\allowbreak Ker}} & \gpfitbox{\texttt{\textbackslash glrule\{subType.subKer\}}} & \gpfitbox{\texttt{\footnotesize GLang/Typing.lean:181}}\\
\gpfitbox{\textsc{Sub-Sub\allowbreak Coker}} & \gpfitbox{\texttt{\textbackslash glrule\{subType.subCoker\}}} & \gpfitbox{\texttt{\footnotesize GLang/Typing.lean:184}}\\
\gpfitbox{\textsc{Sub-Sub\allowbreak Img}} & \gpfitbox{\texttt{\textbackslash glrule\{subType.subImg\}}} & \gpfitbox{\texttt{\footnotesize GLang/Typing.lean:187}}\\
\gpfitbox{\textsc{Sub-Sub\allowbreak Coimg}} & \gpfitbox{\texttt{\textbackslash glrule\{subType.subCoimg\}}} & \gpfitbox{\texttt{\footnotesize GLang/Typing.lean:190}}\\
\gpfitbox{\textsc{T-Zero}} & \gpfitbox{\texttt{\textbackslash glrule\{hasType.zero\}}} & \gpfitbox{\texttt{\footnotesize GLang/Typing.lean:198}}\\
\gpfitbox{\textsc{T-Smul}} & \gpfitbox{\texttt{\textbackslash glrule\{hasType.smul\}}} & \gpfitbox{\texttt{\footnotesize GLang/Typing.lean:205}}\\
\gpfitbox{\textsc{T-Add}} & \gpfitbox{\texttt{\textbackslash glrule\{hasType.add\}}} & \gpfitbox{\texttt{\footnotesize GLang/Typing.lean:214}}\\
\gpfitbox{\textsc{T-Inner}} & \gpfitbox{\texttt{\textbackslash glrule\{hasType.inner\}}} & \gpfitbox{\texttt{\footnotesize GLang/Typing.lean:221}}\\
\gpfitbox{\textsc{T-Real\allowbreak Lit}} & \gpfitbox{\texttt{\textbackslash glrule\{hasType.realLit\}}} & \gpfitbox{\texttt{\footnotesize GLang/Typing.lean:226}}\\
\gpfitbox{\textsc{T-App\allowbreak Linear}} & \gpfitbox{\texttt{\textbackslash glrule\{hasType.appLinear\}}} & \gpfitbox{\texttt{\footnotesize GLang/Typing.lean:233}}\\
\gpfitbox{\textsc{T-App\allowbreak Inner}} & \gpfitbox{\texttt{\textbackslash glrule\{hasType.appInner\}}} & \gpfitbox{\texttt{\footnotesize GLang/Typing.lean:247}}\\
\gpfitbox{\textsc{T-App\allowbreak Nonlinear}} & \gpfitbox{\texttt{\textbackslash glrule\{hasType.appNonlinear\}}} & \gpfitbox{\texttt{\footnotesize GLang/Typing.lean:260}}\\
\gpfitbox{\textsc{T-Var\allowbreak Lin}} & \gpfitbox{\texttt{\textbackslash glrule\{hasType.varLin\}}} & \gpfitbox{\texttt{\footnotesize GLang/Typing.lean:267}}\\
\gpfitbox{\textsc{T-Var\allowbreak Nlin}} & \gpfitbox{\texttt{\textbackslash glrule\{hasType.varNlin\}}} & \gpfitbox{\texttt{\footnotesize GLang/Typing.lean:274}}\\
\gpfitbox{\textsc{T-Pinv\allowbreak Tensor}} & \gpfitbox{\texttt{\textbackslash glrule\{hasType.pinvTensor\}}} & \gpfitbox{\texttt{\footnotesize GLang/Typing.lean:281}}\\
\gpfitbox{\textsc{T-Pinv\allowbreak Dual}} & \gpfitbox{\texttt{\textbackslash glrule\{hasType.pinvDual\}}} & \gpfitbox{\texttt{\footnotesize GLang/Typing.lean:288}}\\
\gpfitbox{\textsc{T-Promote}} & \gpfitbox{\texttt{\textbackslash glrule\{hasType.promote\}}} & \gpfitbox{\texttt{\footnotesize GLang/Typing.lean:303}}\\
\gpfitbox{\textsc{T-Dsum\allowbreak E}} & \gpfitbox{\texttt{\textbackslash glrule\{hasType.dsumE\}}} & \gpfitbox{\texttt{\footnotesize GLang/Typing.lean:318}}\\
\gpfitbox{\textsc{T-Tprod\allowbreak E}} & \gpfitbox{\texttt{\textbackslash glrule\{hasType.tprodE\}}} & \gpfitbox{\texttt{\footnotesize GLang/Typing.lean:327}}\\
\gpfitbox{\textsc{T-Llam\allowbreak Real}} & \gpfitbox{\texttt{\textbackslash glrule\{hasType.llamReal\}}} & \gpfitbox{\texttt{\footnotesize GLang/Typing.lean:340}}\\
\gpfitbox{\textsc{T-Llam\allowbreak Tensor}} & \gpfitbox{\texttt{\textbackslash glrule\{hasType.llamTensor\}}} & \gpfitbox{\texttt{\footnotesize GLang/Typing.lean:347}}\\
\gpfitbox{\textsc{T-Embed}} & \gpfitbox{\texttt{\textbackslash glrule\{hasType.embed\}}} & \gpfitbox{\texttt{\footnotesize GLang/Typing.lean:355}}\\
\gpfitbox{\textsc{T-Nllam}} & \gpfitbox{\texttt{\textbackslash glrule\{hasType.nllam\}}} & \gpfitbox{\texttt{\footnotesize GLang/Typing.lean:364}}\\
\gpfitbox{\textsc{B-Empty}} & \gpfitbox{\texttt{\textbackslash glrule\{basisValid.empty\}}} & \gpfitbox{\texttt{\footnotesize GLang/Basis.lean:162}}\\
\gpfitbox{\textsc{B-One}} & \gpfitbox{\texttt{\textbackslash glrule\{basisValid.one\}}} & \gpfitbox{\texttt{\footnotesize GLang/Basis.lean:163}}\\
\gpfitbox{\textsc{B-Dual}} & \gpfitbox{\texttt{\textbackslash glrule\{basisValid.dual\}}} & \gpfitbox{\texttt{\footnotesize GLang/Basis.lean:164}}\\
\gpfitbox{\textsc{B-Dsum}} & \gpfitbox{\texttt{\textbackslash glrule\{basisValid.dsum\}}} & \gpfitbox{\texttt{\footnotesize GLang/Basis.lean:167}}\\
\gpfitbox{\textsc{B-Tprod}} & \gpfitbox{\texttt{\textbackslash glrule\{basisValid.tprod\}}} & \gpfitbox{\texttt{\footnotesize GLang/Basis.lean:171}}\\
\gpfitbox{\textsc{B-Proj\allowbreak Ker}} & \gpfitbox{\texttt{\textbackslash glrule\{basisValid.projKer\}}} & \gpfitbox{\texttt{\footnotesize GLang/Basis.lean:180}}\\
\gpfitbox{\textsc{B-Proj\allowbreak Coker}} & \gpfitbox{\texttt{\textbackslash glrule\{basisValid.projCoker\}}} & \gpfitbox{\texttt{\footnotesize GLang/Basis.lean:184}}\\
\gpfitbox{\textsc{B-Proj\allowbreak Img}} & \gpfitbox{\texttt{\textbackslash glrule\{basisValid.projImg\}}} & \gpfitbox{\texttt{\footnotesize GLang/Basis.lean:188}}\\
\gpfitbox{\textsc{B-Proj\allowbreak Coimg}} & \gpfitbox{\texttt{\textbackslash glrule\{basisValid.projCoimg\}}} & \gpfitbox{\texttt{\footnotesize GLang/Basis.lean:192}}\\
\gpfitbox{\textsc{B-Image}} & \gpfitbox{\texttt{\textbackslash glrule\{basisValid.image\}}} & \gpfitbox{\texttt{\footnotesize GLang/Basis.lean:198}}\\
\gpfitbox{\textsc{B-Terms}} & \gpfitbox{\texttt{\textbackslash glrule\{basisValid.terms\}}} & \gpfitbox{\texttt{\footnotesize GLang/Basis.lean:204}}\\
\gpfitbox{\textsc{B-Frame}} & \gpfitbox{\texttt{\textbackslash glrule\{basisValid.frame\}}} & \gpfitbox{\texttt{\footnotesize GLang/Basis.lean:213}}\\
\gpfitbox{\textsc{LC-S-Empty}} & \gpfitbox{\texttt{\textbackslash glrule\{lc.space.empty\}}} & \gpfitbox{\texttt{\footnotesize GLang/Source\allowbreak LN.lean:179}}\\
\gpfitbox{\textsc{LC-S-Real}} & \gpfitbox{\texttt{\textbackslash glrule\{lc.space.real\}}} & \gpfitbox{\texttt{\footnotesize GLang/Source\allowbreak LN.lean:180}}\\
\gpfitbox{\textsc{LC-S-Dual}} & \gpfitbox{\texttt{\textbackslash glrule\{lc.space.dual\}}} & \gpfitbox{\texttt{\footnotesize GLang/Source\allowbreak LN.lean:181}}\\
\gpfitbox{\textsc{LC-S-Dsum}} & \gpfitbox{\texttt{\textbackslash glrule\{lc.space.dsum\}}} & \gpfitbox{\texttt{\footnotesize GLang/Source\allowbreak LN.lean:182}}\\
\gpfitbox{\textsc{LC-S-Tprod}} & \gpfitbox{\texttt{\textbackslash glrule\{lc.space.tprod\}}} & \gpfitbox{\texttt{\footnotesize GLang/Source\allowbreak LN.lean:183}}\\
\gpfitbox{\textsc{LC-S-Ker}} & \gpfitbox{\texttt{\textbackslash glrule\{lc.space.ker\}}} & \gpfitbox{\texttt{\footnotesize GLang/Source\allowbreak LN.lean:184}}\\
\gpfitbox{\textsc{LC-S-Coker}} & \gpfitbox{\texttt{\textbackslash glrule\{lc.space.coker\}}} & \gpfitbox{\texttt{\footnotesize GLang/Source\allowbreak LN.lean:185}}\\
\gpfitbox{\textsc{LC-S-Img}} & \gpfitbox{\texttt{\textbackslash glrule\{lc.space.img\}}} & \gpfitbox{\texttt{\footnotesize GLang/Source\allowbreak LN.lean:186}}\\
\gpfitbox{\textsc{LC-S-Coimg}} & \gpfitbox{\texttt{\textbackslash glrule\{lc.space.coimg\}}} & \gpfitbox{\texttt{\footnotesize GLang/Source\allowbreak LN.lean:187}}\\
\gpfitbox{\textsc{LC-T-Space}} & \gpfitbox{\texttt{\textbackslash glrule\{lc.ty.space\}}} & \gpfitbox{\texttt{\footnotesize GLang/Source\allowbreak LN.lean:190}}\\
\gpfitbox{\textsc{LC-T-Pi\allowbreak Ty}} & \gpfitbox{\texttt{\textbackslash glrule\{lc.ty.piTy\}}} & \gpfitbox{\texttt{\footnotesize GLang/Source\allowbreak LN.lean:191}}\\
\gpfitbox{\textsc{LC-E-Var}} & \gpfitbox{\texttt{\textbackslash glrule\{lc.term.var\}}} & \gpfitbox{\texttt{\footnotesize GLang/Source\allowbreak LN.lean:195}}\\
\gpfitbox{\textsc{LC-E-Real}} & \gpfitbox{\texttt{\textbackslash glrule\{lc.term.real\}}} & \gpfitbox{\texttt{\footnotesize GLang/Source\allowbreak LN.lean:196}}\\
\gpfitbox{\textsc{LC-E-Zero}} & \gpfitbox{\texttt{\textbackslash glrule\{lc.term.zero\}}} & \gpfitbox{\texttt{\footnotesize GLang/Source\allowbreak LN.lean:197}}\\
\gpfitbox{\textsc{LC-E-Smul}} & \gpfitbox{\texttt{\textbackslash glrule\{lc.term.smul\}}} & \gpfitbox{\texttt{\footnotesize GLang/Source\allowbreak LN.lean:198}}\\
\gpfitbox{\textsc{LC-E-Add}} & \gpfitbox{\texttt{\textbackslash glrule\{lc.term.add\}}} & \gpfitbox{\texttt{\footnotesize GLang/Source\allowbreak LN.lean:199}}\\
\gpfitbox{\textsc{LC-E-Dsum\allowbreak E}} & \gpfitbox{\texttt{\textbackslash glrule\{lc.term.dsumE\}}} & \gpfitbox{\texttt{\footnotesize GLang/Source\allowbreak LN.lean:200}}\\
\gpfitbox{\textsc{LC-E-Tprod\allowbreak E}} & \gpfitbox{\texttt{\textbackslash glrule\{lc.term.tprodE\}}} & \gpfitbox{\texttt{\footnotesize GLang/Source\allowbreak LN.lean:201}}\\
\gpfitbox{\textsc{LC-E-Inner}} & \gpfitbox{\texttt{\textbackslash glrule\{lc.term.inner\}}} & \gpfitbox{\texttt{\footnotesize GLang/Source\allowbreak LN.lean:202}}\\
\gpfitbox{\textsc{LC-E-Embed}} & \gpfitbox{\texttt{\textbackslash glrule\{lc.term.embed\}}} & \gpfitbox{\texttt{\footnotesize GLang/Source\allowbreak LN.lean:203}}\\
\gpfitbox{\textsc{LC-E-Pinv\allowbreak T}} & \gpfitbox{\texttt{\textbackslash glrule\{lc.term.pinvT\}}} & \gpfitbox{\texttt{\footnotesize GLang/Source\allowbreak LN.lean:204}}\\
\gpfitbox{\textsc{LC-E-Pinv\allowbreak D}} & \gpfitbox{\texttt{\textbackslash glrule\{lc.term.pinvD\}}} & \gpfitbox{\texttt{\footnotesize GLang/Source\allowbreak LN.lean:205}}\\
\gpfitbox{\textsc{LC-E-Llam}} & \gpfitbox{\texttt{\textbackslash glrule\{lc.term.llam\}}} & \gpfitbox{\texttt{\footnotesize GLang/Source\allowbreak LN.lean:206}}\\
\gpfitbox{\textsc{LC-E-Nllam}} & \gpfitbox{\texttt{\textbackslash glrule\{lc.term.nllam\}}} & \gpfitbox{\texttt{\footnotesize GLang/Source\allowbreak LN.lean:209}}\\
\gpfitbox{\textsc{LC-E-App\allowbreak L}} & \gpfitbox{\texttt{\textbackslash glrule\{lc.term.appL\}}} & \gpfitbox{\texttt{\footnotesize GLang/Source\allowbreak LN.lean:211}}\\
\gpfitbox{\textsc{LC-E-App\allowbreak I}} & \gpfitbox{\texttt{\textbackslash glrule\{lc.term.appI\}}} & \gpfitbox{\texttt{\footnotesize GLang/Source\allowbreak LN.lean:213}}\\
\gpfitbox{\textsc{LC-E-App\allowbreak N}} & \gpfitbox{\texttt{\textbackslash glrule\{lc.term.appN\}}} & \gpfitbox{\texttt{\footnotesize GLang/Source\allowbreak LN.lean:214}}\\
\gpfitbox{\textsc{LC-B-Empty}} & \gpfitbox{\texttt{\textbackslash glrule\{lc.basis.empty\}}} & \gpfitbox{\texttt{\footnotesize GLang/Source\allowbreak LN.lean:219}}\\
\gpfitbox{\textsc{LC-B-One}} & \gpfitbox{\texttt{\textbackslash glrule\{lc.basis.one\}}} & \gpfitbox{\texttt{\footnotesize GLang/Source\allowbreak LN.lean:220}}\\
\gpfitbox{\textsc{LC-B-Dual}} & \gpfitbox{\texttt{\textbackslash glrule\{lc.basis.dual\}}} & \gpfitbox{\texttt{\footnotesize GLang/Source\allowbreak LN.lean:221}}\\
\gpfitbox{\textsc{LC-B-Dsum}} & \gpfitbox{\texttt{\textbackslash glrule\{lc.basis.dsum\}}} & \gpfitbox{\texttt{\footnotesize GLang/Source\allowbreak LN.lean:222}}\\
\gpfitbox{\textsc{LC-B-Tprod}} & \gpfitbox{\texttt{\textbackslash glrule\{lc.basis.tprod\}}} & \gpfitbox{\texttt{\footnotesize GLang/Source\allowbreak LN.lean:223}}\\
\gpfitbox{\textsc{LC-B-Proj}} & \gpfitbox{\texttt{\textbackslash glrule\{lc.basis.proj\}}} & \gpfitbox{\texttt{\footnotesize GLang/Source\allowbreak LN.lean:224}}\\
\gpfitbox{\textsc{LC-B-Image}} & \gpfitbox{\texttt{\textbackslash glrule\{lc.basis.image\}}} & \gpfitbox{\texttt{\footnotesize GLang/Source\allowbreak LN.lean:225}}\\
\gpfitbox{\textsc{LC-B-Terms}} & \gpfitbox{\texttt{\textbackslash glrule\{lc.basis.terms\}}} & \gpfitbox{\texttt{\footnotesize GLang/Source\allowbreak LN.lean:226}}\\
\gpfitbox{\textsc{LC-B-Frame}} & \gpfitbox{\texttt{\textbackslash glrule\{lc.basis.frame\}}} & \gpfitbox{\texttt{\footnotesize GLang/Source\allowbreak LN.lean:227}}\\
\gpfitbox{\textsc{LC-VS-Empty}} & \gpfitbox{\texttt{\textbackslash glrule\{lc.valSpace.empty\}}} & \gpfitbox{\texttt{\footnotesize GLang/Value\allowbreak LN.lean:528}}\\
\gpfitbox{\textsc{LC-VS-Real}} & \gpfitbox{\texttt{\textbackslash glrule\{lc.valSpace.real\}}} & \gpfitbox{\texttt{\footnotesize GLang/Value\allowbreak LN.lean:529}}\\
\gpfitbox{\textsc{LC-VS-Dsum}} & \gpfitbox{\texttt{\textbackslash glrule\{lc.valSpace.dsum\}}} & \gpfitbox{\texttt{\footnotesize GLang/Value\allowbreak LN.lean:530}}\\
\gpfitbox{\textsc{LC-VS-Tprod}} & \gpfitbox{\texttt{\textbackslash glrule\{lc.valSpace.tprod\}}} & \gpfitbox{\texttt{\footnotesize GLang/Value\allowbreak LN.lean:531}}\\
\gpfitbox{\textsc{LC-VS-Ffs}} & \gpfitbox{\texttt{\textbackslash glrule\{lc.valSpace.ffs\}}} & \gpfitbox{\texttt{\footnotesize GLang/Value\allowbreak LN.lean:532}}\\
\gpfitbox{\textsc{LC-VT-Space}} & \gpfitbox{\texttt{\textbackslash glrule\{lc.valTy.space\}}} & \gpfitbox{\texttt{\footnotesize GLang/Value\allowbreak LN.lean:535}}\\
\gpfitbox{\textsc{LC-VT-Pi\allowbreak Ty}} & \gpfitbox{\texttt{\textbackslash glrule\{lc.valTy.piTy\}}} & \gpfitbox{\texttt{\footnotesize GLang/Value\allowbreak LN.lean:536}}\\
\gpfitbox{\textsc{LC-VE-Var}} & \gpfitbox{\texttt{\textbackslash glrule\{lc.valExpr.var\}}} & \gpfitbox{\texttt{\footnotesize GLang/Value\allowbreak LN.lean:541}}\\
\gpfitbox{\textsc{LC-VE-Vlam}} & \gpfitbox{\texttt{\textbackslash glrule\{lc.valExpr.vlam\}}} & \gpfitbox{\texttt{\footnotesize GLang/Value\allowbreak LN.lean:542}}\\
\gpfitbox{\textsc{LC-VE-Zero}} & \gpfitbox{\texttt{\textbackslash glrule\{lc.valExpr.zero\}}} & \gpfitbox{\texttt{\footnotesize GLang/Value\allowbreak LN.lean:547}}\\
\gpfitbox{\textsc{LC-VE-Rlit}} & \gpfitbox{\texttt{\textbackslash glrule\{lc.valExpr.rlit\}}} & \gpfitbox{\texttt{\footnotesize GLang/Value\allowbreak LN.lean:548}}\\
\gpfitbox{\textsc{LC-VE-Scale}} & \gpfitbox{\texttt{\textbackslash glrule\{lc.valExpr.scale\}}} & \gpfitbox{\texttt{\footnotesize GLang/Value\allowbreak LN.lean:549}}\\
\gpfitbox{\textsc{LC-VE-Vadd}} & \gpfitbox{\texttt{\textbackslash glrule\{lc.valExpr.vadd\}}} & \gpfitbox{\texttt{\footnotesize GLang/Value\allowbreak LN.lean:550}}\\
\gpfitbox{\textsc{LC-VE-Dsum\allowbreak E}} & \gpfitbox{\texttt{\textbackslash glrule\{lc.valExpr.dsumE\}}} & \gpfitbox{\texttt{\footnotesize GLang/Value\allowbreak LN.lean:552}}\\
\gpfitbox{\textsc{LC-VE-Tprod\allowbreak E}} & \gpfitbox{\texttt{\textbackslash glrule\{lc.valExpr.tprodE\}}} & \gpfitbox{\texttt{\footnotesize GLang/Value\allowbreak LN.lean:553}}\\
\gpfitbox{\textsc{LC-VE-Delta}} & \gpfitbox{\texttt{\textbackslash glrule\{lc.valExpr.delta\}}} & \gpfitbox{\texttt{\footnotesize GLang/Value\allowbreak LN.lean:554}}\\
\gpfitbox{\textsc{LC-VE-Idelta}} & \gpfitbox{\texttt{\textbackslash glrule\{lc.valExpr.idelta\}}} & \gpfitbox{\texttt{\footnotesize GLang/Value\allowbreak LN.lean:555}}\\
\gpfitbox{\textsc{LC-VE-Vsum}} & \gpfitbox{\texttt{\textbackslash glrule\{lc.valExpr.vsum\}}} & \gpfitbox{\texttt{\footnotesize GLang/Value\allowbreak LN.lean:556}}\\
\gpfitbox{\textsc{LC-VE-Ttv}} & \gpfitbox{\texttt{\textbackslash glrule\{lc.valExpr.ttv\}}} & \gpfitbox{\texttt{\footnotesize GLang/Value\allowbreak LN.lean:557}}\\
\gpfitbox{\textsc{LC-VE-Mtm}} & \gpfitbox{\texttt{\textbackslash glrule\{lc.valExpr.mtm\}}} & \gpfitbox{\texttt{\footnotesize GLang/Value\allowbreak LN.lean:558}}\\
\gpfitbox{\textsc{LC-VE-App\allowbreak V}} & \gpfitbox{\texttt{\textbackslash glrule\{lc.valExpr.appV\}}} & \gpfitbox{\texttt{\footnotesize GLang/Value\allowbreak LN.lean:560}}\\
\gpfitbox{\textsc{LC-VE-Vkrn}} & \gpfitbox{\texttt{\textbackslash glrule\{lc.valExpr.vkrn\}}} & \gpfitbox{\texttt{\footnotesize GLang/Value\allowbreak LN.lean:561}}\\
\gpfitbox{\textsc{LC-VE-Vblk}} & \gpfitbox{\texttt{\textbackslash glrule\{lc.valExpr.vblk\}}} & \gpfitbox{\texttt{\footnotesize GLang/Value\allowbreak LN.lean:562}}\\
\gpfitbox{\textsc{LC-VE-Reshap}} & \gpfitbox{\texttt{\textbackslash glrule\{lc.valExpr.reshap\}}} & \gpfitbox{\texttt{\footnotesize GLang/Value\allowbreak LN.lean:563}}\\
\gpfitbox{\textsc{LC-VE-Vtrans}} & \gpfitbox{\texttt{\textbackslash glrule\{lc.valExpr.vtrans\}}} & \gpfitbox{\texttt{\footnotesize GLang/Value\allowbreak LN.lean:565}}\\
\gpfitbox{\textsc{LC-VE-Gsvd\allowbreak S}} & \gpfitbox{\texttt{\textbackslash glrule\{lc.valExpr.gsvdS\}}} & \gpfitbox{\texttt{\footnotesize GLang/Value\allowbreak LN.lean:566}}\\
\gpfitbox{\textsc{LC-VE-Gsvd\allowbreak U}} & \gpfitbox{\texttt{\textbackslash glrule\{lc.valExpr.gsvdU\}}} & \gpfitbox{\texttt{\footnotesize GLang/Value\allowbreak LN.lean:567}}\\
\gpfitbox{\textsc{LC-VE-Gsvd\allowbreak UI}} & \gpfitbox{\texttt{\textbackslash glrule\{lc.valExpr.gsvdUI\}}} & \gpfitbox{\texttt{\footnotesize GLang/Value\allowbreak LN.lean:568}}\\
\gpfitbox{\textsc{LC-VE-Gsvd\allowbreak V}} & \gpfitbox{\texttt{\textbackslash glrule\{lc.valExpr.gsvdV\}}} & \gpfitbox{\texttt{\footnotesize GLang/Value\allowbreak LN.lean:569}}\\
\gpfitbox{\textsc{LC-VE-Gsvd\allowbreak VI}} & \gpfitbox{\texttt{\textbackslash glrule\{lc.valExpr.gsvdVI\}}} & \gpfitbox{\texttt{\footnotesize GLang/Value\allowbreak LN.lean:570}}\\
\gpfitbox{\textsc{LC-VE-Pinv}} & \gpfitbox{\texttt{\textbackslash glrule\{lc.valExpr.pinv\}}} & \gpfitbox{\texttt{\footnotesize GLang/Value\allowbreak LN.lean:571}}\\
\gpfitbox{\textsc{LC-VE-Gpinv}} & \gpfitbox{\texttt{\textbackslash glrule\{lc.valExpr.gpinv\}}} & \gpfitbox{\texttt{\footnotesize GLang/Value\allowbreak LN.lean:572}}\\
\gpfitbox{\textsc{LC-VE-Rref\allowbreak Rep}} & \gpfitbox{\texttt{\textbackslash glrule\{lc.valExpr.rrefRep\}}} & \gpfitbox{\texttt{\footnotesize GLang/Value\allowbreak LN.lean:573}}\\
\gpfitbox{\textsc{LC-VE-Inv}} & \gpfitbox{\texttt{\textbackslash glrule\{lc.valExpr.inv\}}} & \gpfitbox{\texttt{\footnotesize GLang/Value\allowbreak LN.lean:574}}\\
\gpfitbox{\textsc{LC-VE-Svds}} & \gpfitbox{\texttt{\textbackslash glrule\{lc.valExpr.svds\}}} & \gpfitbox{\texttt{\footnotesize GLang/Value\allowbreak LN.lean:575}}\\
\gpfitbox{\textsc{VS-Real}} & \gpfitbox{\texttt{\textbackslash glrule\{valSValid.real\}}} & \gpfitbox{\texttt{\footnotesize GLang/Runtime\allowbreak Value\allowbreak Typing.lean:12}}\\
\gpfitbox{\textsc{VS-Empty}} & \gpfitbox{\texttt{\textbackslash glrule\{valSValid.empty\}}} & \gpfitbox{\texttt{\footnotesize GLang/Runtime\allowbreak Value\allowbreak Typing.lean:13}}\\
\gpfitbox{\textsc{VS-Dsum}} & \gpfitbox{\texttt{\textbackslash glrule\{valSValid.dsum\}}} & \gpfitbox{\texttt{\footnotesize GLang/Runtime\allowbreak Value\allowbreak Typing.lean:14}}\\
\gpfitbox{\textsc{VS-Tprod}} & \gpfitbox{\texttt{\textbackslash glrule\{valSValid.tprod\}}} & \gpfitbox{\texttt{\footnotesize GLang/Runtime\allowbreak Value\allowbreak Typing.lean:16}}\\
\gpfitbox{\textsc{VS-Ffs}} & \gpfitbox{\texttt{\textbackslash glrule\{valSValid.ffs\}}} & \gpfitbox{\texttt{\footnotesize GLang/Runtime\allowbreak Value\allowbreak Typing.lean:18}}\\
\gpfitbox{\textsc{VK-Space}} & \gpfitbox{\texttt{\textbackslash glrule\{valTValid.space\}}} & \gpfitbox{\texttt{\footnotesize GLang/Runtime\allowbreak Value\allowbreak Typing.lean:25}}\\
\gpfitbox{\textsc{VK-Pi\allowbreak Ty}} & \gpfitbox{\texttt{\textbackslash glrule\{valTValid.piTy\}}} & \gpfitbox{\texttt{\footnotesize GLang/Runtime\allowbreak Value\allowbreak Typing.lean:26}}\\
\gpfitbox{\textsc{VT-Var}} & \gpfitbox{\texttt{\textbackslash glrule\{valHasType.var\}}} & \gpfitbox{\texttt{\footnotesize GLang/Runtime\allowbreak Value\allowbreak Typing.lean:33}}\\
\gpfitbox{\textsc{VT-Zero}} & \gpfitbox{\texttt{\textbackslash glrule\{valHasType.zero\}}} & \gpfitbox{\texttt{\footnotesize GLang/Runtime\allowbreak Value\allowbreak Typing.lean:34}}\\
\gpfitbox{\textsc{VT-Rlit}} & \gpfitbox{\texttt{\textbackslash glrule\{valHasType.rlit\}}} & \gpfitbox{\texttt{\footnotesize GLang/Runtime\allowbreak Value\allowbreak Typing.lean:35}}\\
\gpfitbox{\textsc{VT-Scale}} & \gpfitbox{\texttt{\textbackslash glrule\{valHasType.scale\}}} & \gpfitbox{\texttt{\footnotesize GLang/Runtime\allowbreak Value\allowbreak Typing.lean:36}}\\
\gpfitbox{\textsc{VT-Vadd}} & \gpfitbox{\texttt{\textbackslash glrule\{valHasType.vadd\}}} & \gpfitbox{\texttt{\footnotesize GLang/Runtime\allowbreak Value\allowbreak Typing.lean:40}}\\
\gpfitbox{\textsc{VT-Dsum\allowbreak E}} & \gpfitbox{\texttt{\textbackslash glrule\{valHasType.dsumE\}}} & \gpfitbox{\texttt{\footnotesize GLang/Runtime\allowbreak Value\allowbreak Typing.lean:45}}\\
\gpfitbox{\textsc{VT-Tprod\allowbreak E}} & \gpfitbox{\texttt{\textbackslash glrule\{valHasType.tprodE\}}} & \gpfitbox{\texttt{\footnotesize GLang/Runtime\allowbreak Value\allowbreak Typing.lean:49}}\\
\gpfitbox{\textsc{VT-Delta}} & \gpfitbox{\texttt{\textbackslash glrule\{valHasType.delta\}}} & \gpfitbox{\texttt{\footnotesize GLang/Runtime\allowbreak Value\allowbreak Typing.lean:53}}\\
\gpfitbox{\textsc{VT-Idelta}} & \gpfitbox{\texttt{\textbackslash glrule\{valHasType.idelta\}}} & \gpfitbox{\texttt{\footnotesize GLang/Runtime\allowbreak Value\allowbreak Typing.lean:54}}\\
\gpfitbox{\textsc{VT-Ttv\allowbreak Space}} & \gpfitbox{\texttt{\textbackslash glrule\{valHasType.ttvSpace\}}} & \gpfitbox{\texttt{\footnotesize GLang/Runtime\allowbreak Value\allowbreak Typing.lean:57}}\\
\gpfitbox{\textsc{VT-Ttv\allowbreak Real}} & \gpfitbox{\texttt{\textbackslash glrule\{valHasType.ttvReal\}}} & \gpfitbox{\texttt{\footnotesize GLang/Runtime\allowbreak Value\allowbreak Typing.lean:62}}\\
\gpfitbox{\textsc{VT-Mtm}} & \gpfitbox{\texttt{\textbackslash glrule\{valHasType.mtm\}}} & \gpfitbox{\texttt{\footnotesize GLang/Runtime\allowbreak Value\allowbreak Typing.lean:68}}\\
\gpfitbox{\textsc{VT-Vsum}} & \gpfitbox{\texttt{\textbackslash glrule\{valHasType.vsum\}}} & \gpfitbox{\texttt{\footnotesize GLang/Runtime\allowbreak Value\allowbreak Typing.lean:75}}\\
\gpfitbox{\textsc{VT-Vlam}} & \gpfitbox{\texttt{\textbackslash glrule\{valHasType.vlam\}}} & \gpfitbox{\texttt{\footnotesize GLang/Runtime\allowbreak Value\allowbreak Typing.lean:81}}\\
\gpfitbox{\textsc{VT-App\allowbreak V}} & \gpfitbox{\texttt{\textbackslash glrule\{valHasType.appV\}}} & \gpfitbox{\texttt{\footnotesize GLang/Runtime\allowbreak Value\allowbreak Typing.lean:86}}\\
\gpfitbox{\textsc{VT-Vkrn}} & \gpfitbox{\texttt{\textbackslash glrule\{valHasType.vkrn\}}} & \gpfitbox{\texttt{\footnotesize GLang/Runtime\allowbreak Value\allowbreak Typing.lean:92}}\\
\gpfitbox{\textsc{VT-Vblk}} & \gpfitbox{\texttt{\textbackslash glrule\{valHasType.vblk\}}} & \gpfitbox{\texttt{\footnotesize GLang/Runtime\allowbreak Value\allowbreak Typing.lean:97}}\\
\gpfitbox{\textsc{VT-Reshap}} & \gpfitbox{\texttt{\textbackslash glrule\{valHasType.reshap\}}} & \gpfitbox{\texttt{\footnotesize GLang/Runtime\allowbreak Value\allowbreak Typing.lean:102}}\\
\gpfitbox{\textsc{VT-Vtrans}} & \gpfitbox{\texttt{\textbackslash glrule\{valHasType.vtrans\}}} & \gpfitbox{\texttt{\footnotesize GLang/Runtime\allowbreak Value\allowbreak Typing.lean:107}}\\
\gpfitbox{\textsc{VT-Gsvd\allowbreak S}} & \gpfitbox{\texttt{\textbackslash glrule\{valHasType.gsvdS\}}} & \gpfitbox{\texttt{\footnotesize GLang/Runtime\allowbreak Value\allowbreak Typing.lean:110}}\\
\gpfitbox{\textsc{VT-Gsvd\allowbreak U}} & \gpfitbox{\texttt{\textbackslash glrule\{valHasType.gsvdU\}}} & \gpfitbox{\texttt{\footnotesize GLang/Runtime\allowbreak Value\allowbreak Typing.lean:118}}\\
\gpfitbox{\textsc{VT-Gsvd\allowbreak UI}} & \gpfitbox{\texttt{\textbackslash glrule\{valHasType.gsvdUI\}}} & \gpfitbox{\texttt{\footnotesize GLang/Runtime\allowbreak Value\allowbreak Typing.lean:125}}\\
\gpfitbox{\textsc{VT-Gsvd\allowbreak V}} & \gpfitbox{\texttt{\textbackslash glrule\{valHasType.gsvdV\}}} & \gpfitbox{\texttt{\footnotesize GLang/Runtime\allowbreak Value\allowbreak Typing.lean:131}}\\
\gpfitbox{\textsc{VT-Gsvd\allowbreak VI}} & \gpfitbox{\texttt{\textbackslash glrule\{valHasType.gsvdVI\}}} & \gpfitbox{\texttt{\footnotesize GLang/Runtime\allowbreak Value\allowbreak Typing.lean:138}}\\
\gpfitbox{\textsc{VT-Pinv\allowbreak Tensor}} & \gpfitbox{\texttt{\textbackslash glrule\{valHasType.pinvTensor\}}} & \gpfitbox{\texttt{\footnotesize GLang/Runtime\allowbreak Value\allowbreak Typing.lean:144}}\\
\gpfitbox{\textsc{VT-Pinv\allowbreak Raw}} & \gpfitbox{\texttt{\textbackslash glrule\{valHasType.pinvRaw\}}} & \gpfitbox{\texttt{\footnotesize GLang/Runtime\allowbreak Value\allowbreak Typing.lean:147}}\\
\gpfitbox{\textsc{VT-Gpinv}} & \gpfitbox{\texttt{\textbackslash glrule\{valHasType.gpinv\}}} & \gpfitbox{\texttt{\footnotesize GLang/Runtime\allowbreak Value\allowbreak Typing.lean:151}}\\
\gpfitbox{\textsc{VT-Rref\allowbreak Rep}} & \gpfitbox{\texttt{\textbackslash glrule\{valHasType.rrefRep\}}} & \gpfitbox{\texttt{\footnotesize GLang/Runtime\allowbreak Value\allowbreak Typing.lean:156}}\\
\gpfitbox{\textsc{VT-Inv\allowbreak Tensor}} & \gpfitbox{\texttt{\textbackslash glrule\{valHasType.invTensor\}}} & \gpfitbox{\texttt{\footnotesize GLang/Runtime\allowbreak Value\allowbreak Typing.lean:160}}\\
\gpfitbox{\textsc{VT-Inv\allowbreak Raw}} & \gpfitbox{\texttt{\textbackslash glrule\{valHasType.invRaw\}}} & \gpfitbox{\texttt{\footnotesize GLang/Runtime\allowbreak Value\allowbreak Typing.lean:163}}\\
\gpfitbox{\textsc{VT-Svds}} & \gpfitbox{\texttt{\textbackslash glrule\{valHasType.svds\}}} & \gpfitbox{\texttt{\footnotesize GLang/Runtime\allowbreak Value\allowbreak Typing.lean:167}}\\
\gpfitbox{\textsc{ES-Real}} & \gpfitbox{\texttt{\textbackslash glrule\{evalS.real\}}} & \gpfitbox{\texttt{\footnotesize GLang/Value\allowbreak Operational.lean:289}}\\
\gpfitbox{\textsc{ES-Empty}} & \gpfitbox{\texttt{\textbackslash glrule\{evalS.empty\}}} & \gpfitbox{\texttt{\footnotesize GLang/Value\allowbreak Operational.lean:291}}\\
\gpfitbox{\textsc{ES-Dual}} & \gpfitbox{\texttt{\textbackslash glrule\{evalS.dual\}}} & \gpfitbox{\texttt{\footnotesize GLang/Value\allowbreak Operational.lean:293}}\\
\gpfitbox{\textsc{ES-Ker}} & \gpfitbox{\texttt{\textbackslash glrule\{evalS.ker\}}} & \gpfitbox{\texttt{\footnotesize GLang/Value\allowbreak Operational.lean:297}}\\
\gpfitbox{\textsc{ES-Coker}} & \gpfitbox{\texttt{\textbackslash glrule\{evalS.coker\}}} & \gpfitbox{\texttt{\footnotesize GLang/Value\allowbreak Operational.lean:312}}\\
\gpfitbox{\textsc{ES-Img}} & \gpfitbox{\texttt{\textbackslash glrule\{evalS.img\}}} & \gpfitbox{\texttt{\footnotesize GLang/Value\allowbreak Operational.lean:321}}\\
\gpfitbox{\textsc{ES-Coimg}} & \gpfitbox{\texttt{\textbackslash glrule\{evalS.coimg\}}} & \gpfitbox{\texttt{\footnotesize GLang/Value\allowbreak Operational.lean:330}}\\
\gpfitbox{\textsc{ES-Tprod}} & \gpfitbox{\texttt{\textbackslash glrule\{evalS.tprod\}}} & \gpfitbox{\texttt{\footnotesize GLang/Value\allowbreak Operational.lean:339}}\\
\gpfitbox{\textsc{ES-Dsum}} & \gpfitbox{\texttt{\textbackslash glrule\{evalS.dsum\}}} & \gpfitbox{\texttt{\footnotesize GLang/Value\allowbreak Operational.lean:344}}\\
\gpfitbox{\textsc{ET-Space}} & \gpfitbox{\texttt{\textbackslash glrule\{evalT.space\}}} & \gpfitbox{\texttt{\footnotesize GLang/Value\allowbreak Operational.lean:358}}\\
\gpfitbox{\textsc{ET-Pi\allowbreak Ty}} & \gpfitbox{\texttt{\textbackslash glrule\{evalT.piTy\}}} & \gpfitbox{\texttt{\footnotesize GLang/Value\allowbreak Operational.lean:362}}\\
\gpfitbox{\textsc{EB-Inverse}} & \gpfitbox{\texttt{\textbackslash glrule\{evalBB.inverse\}}} & \gpfitbox{\texttt{\footnotesize GLang/Value\allowbreak Operational.lean:408}}\\
\gpfitbox{\textsc{EB-Trans}} & \gpfitbox{\texttt{\textbackslash glrule\{evalBB.trans\}}} & \gpfitbox{\texttt{\footnotesize GLang/Value\allowbreak Operational.lean:413}}\\
\gpfitbox{\textsc{EB-Tprod}} & \gpfitbox{\texttt{\textbackslash glrule\{evalBB.tprod\}}} & \gpfitbox{\texttt{\footnotesize GLang/Value\allowbreak Operational.lean:432}}\\
\gpfitbox{\textsc{EB-Dsum}} & \gpfitbox{\texttt{\textbackslash glrule\{evalBB.dsum\}}} & \gpfitbox{\texttt{\footnotesize GLang/Value\allowbreak Operational.lean:445}}\\
\gpfitbox{\textsc{EB-Dual\allowbreak Std}} & \gpfitbox{\texttt{\textbackslash glrule\{evalBB.dualStd\}}} & \gpfitbox{\texttt{\footnotesize GLang/Value\allowbreak Operational.lean:451}}\\
\gpfitbox{\textsc{EB-Image}} & \gpfitbox{\texttt{\textbackslash glrule\{evalBB.image\}}} & \gpfitbox{\texttt{\footnotesize GLang/Value\allowbreak Operational.lean:463}}\\
\gpfitbox{\textsc{EB-Image\allowbreak Svds}} & \gpfitbox{\texttt{\textbackslash glrule\{evalBB.imageSvds\}}} & \gpfitbox{\texttt{\footnotesize GLang/Value\allowbreak Operational.lean:474}}\\
\gpfitbox{\textsc{EB-Proj\allowbreak Ker\allowbreak Coker}} & \gpfitbox{\texttt{\textbackslash glrule\{evalBB.projKerCoker\}}} & \gpfitbox{\texttt{\footnotesize GLang/Value\allowbreak Operational.lean:496}}\\
\gpfitbox{\textsc{EB-Proj\allowbreak Img\allowbreak Coimg}} & \gpfitbox{\texttt{\textbackslash glrule\{evalBB.projImgCoimg\}}} & \gpfitbox{\texttt{\footnotesize GLang/Value\allowbreak Operational.lean:542}}\\
\gpfitbox{\textsc{EB-Terms\allowbreak Rref}} & \gpfitbox{\texttt{\textbackslash glrule\{evalBB.termsRref\}}} & \gpfitbox{\texttt{\footnotesize GLang/Value\allowbreak Operational.lean:575}}\\
\gpfitbox{\textsc{EB-Refl}} & \gpfitbox{\texttt{\textbackslash glrule\{evalBB.refl\}}} & \gpfitbox{\texttt{\footnotesize GLang/Value\allowbreak Operational.lean:601}}\\
\gpfitbox{\textsc{EB-Proj\allowbreak Ker\allowbreak Coker\allowbreak Cod}} & \gpfitbox{\texttt{\textbackslash glrule\{evalBB.projKerCokerCod\}}} & \gpfitbox{\texttt{\footnotesize GLang/Value\allowbreak Operational.lean:617}}\\
\gpfitbox{\textsc{EB-Proj\allowbreak Img\allowbreak Coimg\allowbreak Dom}} & \gpfitbox{\texttt{\textbackslash glrule\{evalBB.projImgCoimgDom\}}} & \gpfitbox{\texttt{\footnotesize GLang/Value\allowbreak Operational.lean:655}}\\
\gpfitbox{\textsc{E-Var}} & \gpfitbox{\texttt{\textbackslash glrule\{opStep.var\}}} & \gpfitbox{\texttt{\footnotesize GLang/Value\allowbreak Operational.lean:708}}\\
\gpfitbox{\textsc{E-Cob}} & \gpfitbox{\texttt{\textbackslash glrule\{opStep.cob\}}} & \gpfitbox{\texttt{\footnotesize GLang/Value\allowbreak Operational.lean:719}}\\
\gpfitbox{\textsc{E-Scale}} & \gpfitbox{\texttt{\textbackslash glrule\{opStep.scale\}}} & \gpfitbox{\texttt{\footnotesize GLang/Value\allowbreak Operational.lean:747}}\\
\gpfitbox{\textsc{E-Add}} & \gpfitbox{\texttt{\textbackslash glrule\{opStep.add\}}} & \gpfitbox{\texttt{\footnotesize GLang/Value\allowbreak Operational.lean:759}}\\
\gpfitbox{\textsc{E-Dsum\allowbreak E}} & \gpfitbox{\texttt{\textbackslash glrule\{opStep.dsumE\}}} & \gpfitbox{\texttt{\footnotesize GLang/Value\allowbreak Operational.lean:772}}\\
\gpfitbox{\textsc{E-Tprod\allowbreak E}} & \gpfitbox{\texttt{\textbackslash glrule\{opStep.tprodE\}}} & \gpfitbox{\texttt{\footnotesize GLang/Value\allowbreak Operational.lean:789}}\\
\gpfitbox{\textsc{E-Pinv\allowbreak TStd}} & \gpfitbox{\texttt{\textbackslash glrule\{opStep.pinvTStd\}}} & \gpfitbox{\texttt{\footnotesize GLang/Value\allowbreak Operational.lean:801}}\\
\gpfitbox{\textsc{E-Pinv\allowbreak TGpinv}} & \gpfitbox{\texttt{\textbackslash glrule\{opStep.pinvTGpinv\}}} & \gpfitbox{\texttt{\footnotesize GLang/Value\allowbreak Operational.lean:818}}\\
\gpfitbox{\textsc{E-Pinv\allowbreak DStd}} & \gpfitbox{\texttt{\textbackslash glrule\{opStep.pinvDStd\}}} & \gpfitbox{\texttt{\footnotesize GLang/Value\allowbreak Operational.lean:854}}\\
\gpfitbox{\textsc{E-Pinv\allowbreak DGpinv}} & \gpfitbox{\texttt{\textbackslash glrule\{opStep.pinvDGpinv\}}} & \gpfitbox{\texttt{\footnotesize GLang/Value\allowbreak Operational.lean:881}}\\
\gpfitbox{\textsc{E-Inner\allowbreak Std}} & \gpfitbox{\texttt{\textbackslash glrule\{opStep.innerStd\}}} & \gpfitbox{\texttt{\footnotesize GLang/Value\allowbreak Operational.lean:911}}\\
\gpfitbox{\textsc{E-Embed\allowbreak Ker}} & \gpfitbox{\texttt{\textbackslash glrule\{opStep.embedKer\}}} & \gpfitbox{\texttt{\footnotesize GLang/Value\allowbreak Operational.lean:926}}\\
\gpfitbox{\textsc{E-Embed\allowbreak Coker}} & \gpfitbox{\texttt{\textbackslash glrule\{opStep.embedCoker\}}} & \gpfitbox{\texttt{\footnotesize GLang/Value\allowbreak Operational.lean:963}}\\
\gpfitbox{\textsc{E-Embed\allowbreak Img}} & \gpfitbox{\texttt{\textbackslash glrule\{opStep.embedImg\}}} & \gpfitbox{\texttt{\footnotesize GLang/Value\allowbreak Operational.lean:982}}\\
\gpfitbox{\textsc{E-Embed\allowbreak Coimg}} & \gpfitbox{\texttt{\textbackslash glrule\{opStep.embedCoimg\}}} & \gpfitbox{\texttt{\footnotesize GLang/Value\allowbreak Operational.lean:1002}}\\
\gpfitbox{\textsc{E-App\allowbreak LLin}} & \gpfitbox{\texttt{\textbackslash glrule\{opStep.appLLin\}}} & \gpfitbox{\texttt{\footnotesize GLang/Value\allowbreak Operational.lean:1020}}\\
\gpfitbox{\textsc{E-App\allowbreak ILin}} & \gpfitbox{\texttt{\textbackslash glrule\{opStep.appILin\}}} & \gpfitbox{\texttt{\footnotesize GLang/Value\allowbreak Operational.lean:1038}}\\
\gpfitbox{\textsc{E-Nllam}} & \gpfitbox{\texttt{\textbackslash glrule\{opStep.nllam\}}} & \gpfitbox{\texttt{\footnotesize GLang/Value\allowbreak Operational.lean:1058}}\\
\gpfitbox{\textsc{E-App\allowbreak NVal}} & \gpfitbox{\texttt{\textbackslash glrule\{opStep.appNVal\}}} & \gpfitbox{\texttt{\footnotesize GLang/Value\allowbreak Operational.lean:1084}}\\
\gpfitbox{\textsc{E-Frame\allowbreak Reanchor}} & \gpfitbox{\texttt{\textbackslash glrule\{opStep.frameReanchor\}}} & \gpfitbox{\texttt{\footnotesize GLang/Value\allowbreak Operational.lean:1097}}\\
\gpfitbox{\textsc{E-Frame\allowbreak Eta}} & \gpfitbox{\texttt{\textbackslash glrule\{opStep.frameEta\}}} & \gpfitbox{\texttt{\footnotesize GLang/Value\allowbreak Operational.lean:1136}}\\
\gpfitbox{\textsc{E-Reshape\allowbreak Cob}} & \gpfitbox{\texttt{\textbackslash glrule\{opStep.reshapeCob\}}} & \gpfitbox{\texttt{\footnotesize GLang/Value\allowbreak Operational.lean:1161}}\\
\gpfitbox{\textsc{E-Llam\allowbreak Assemble}} & \gpfitbox{\texttt{\textbackslash glrule\{opStep.llamAssemble\}}} & \gpfitbox{\texttt{\footnotesize GLang/Value\allowbreak Operational.lean:1188}}\\
\gpfitbox{\textsc{E-Llam\allowbreak Assemble\allowbreak Real}} & \gpfitbox{\texttt{\textbackslash glrule\{opStep.llamAssembleReal\}}} & \gpfitbox{\texttt{\footnotesize GLang/Value\allowbreak Operational.lean:1233}}\\
\gpfitbox{\textsc{E-Zero}} & \gpfitbox{\texttt{\textbackslash glrule\{opStep.zero\}}} & \gpfitbox{\texttt{\footnotesize GLang/Value\allowbreak Operational.lean:1258}}\\
\gpfitbox{\textsc{E-Rlit}} & \gpfitbox{\texttt{\textbackslash glrule\{opStep.rlit\}}} & \gpfitbox{\texttt{\footnotesize GLang/Value\allowbreak Operational.lean:1274}}\\
\gpfitbox{\textsc{S-S-Real}} & \gpfitbox{\texttt{\textbackslash glrule\{s.space.real\}}} & \gpfitbox{\texttt{\footnotesize GLang/Adequacy/Strengthened.lean:74}}\\
\gpfitbox{\textsc{S-S-Empty}} & \gpfitbox{\texttt{\textbackslash glrule\{s.space.empty\}}} & \gpfitbox{\texttt{\footnotesize GLang/Adequacy/Strengthened.lean:75}}\\
\gpfitbox{\textsc{S-S-Dual}} & \gpfitbox{\texttt{\textbackslash glrule\{s.space.dual\}}} & \gpfitbox{\texttt{\footnotesize GLang/Adequacy/Strengthened.lean:76}}\\
\gpfitbox{\textsc{S-S-Ker}} & \gpfitbox{\texttt{\textbackslash glrule\{s.space.ker\}}} & \gpfitbox{\texttt{\footnotesize GLang/Adequacy/Strengthened.lean:80}}\\
\gpfitbox{\textsc{S-S-Coker}} & \gpfitbox{\texttt{\textbackslash glrule\{s.space.coker\}}} & \gpfitbox{\texttt{\footnotesize GLang/Adequacy/Strengthened.lean:89}}\\
\gpfitbox{\textsc{S-S-Img}} & \gpfitbox{\texttt{\textbackslash glrule\{s.space.img\}}} & \gpfitbox{\texttt{\footnotesize GLang/Adequacy/Strengthened.lean:97}}\\
\gpfitbox{\textsc{S-S-Coimg}} & \gpfitbox{\texttt{\textbackslash glrule\{s.space.coimg\}}} & \gpfitbox{\texttt{\footnotesize GLang/Adequacy/Strengthened.lean:105}}\\
\gpfitbox{\textsc{S-S-Tprod}} & \gpfitbox{\texttt{\textbackslash glrule\{s.space.tprod\}}} & \gpfitbox{\texttt{\footnotesize GLang/Adequacy/Strengthened.lean:113}}\\
\gpfitbox{\textsc{S-S-Dsum}} & \gpfitbox{\texttt{\textbackslash glrule\{s.space.dsum\}}} & \gpfitbox{\texttt{\footnotesize GLang/Adequacy/Strengthened.lean:117}}\\
\gpfitbox{\textsc{S-T-Space}} & \gpfitbox{\texttt{\textbackslash glrule\{s.ty.space\}}} & \gpfitbox{\texttt{\footnotesize GLang/Adequacy/Strengthened.lean:126}}\\
\gpfitbox{\textsc{S-T-Pi\allowbreak Ty}} & \gpfitbox{\texttt{\textbackslash glrule\{s.ty.piTy\}}} & \gpfitbox{\texttt{\footnotesize GLang/Adequacy/Strengthened.lean:129}}\\
\gpfitbox{\textsc{S-TA-Space}} & \gpfitbox{\texttt{\textbackslash glrule\{s.tyAt.space\}}} & \gpfitbox{\texttt{\footnotesize GLang/Adequacy/Strengthened.lean:147}}\\
\gpfitbox{\textsc{S-TA-Frame}} & \gpfitbox{\texttt{\textbackslash glrule\{s.tyAt.frame\}}} & \gpfitbox{\texttt{\footnotesize GLang/Adequacy/Strengthened.lean:152}}\\
\gpfitbox{\textsc{S-C-Inverse}} & \gpfitbox{\texttt{\textbackslash glrule\{s.cob.inverse\}}} & \gpfitbox{\texttt{\footnotesize GLang/Adequacy/Strengthened.lean:167}}\\
\gpfitbox{\textsc{S-C-Trans}} & \gpfitbox{\texttt{\textbackslash glrule\{s.cob.trans\}}} & \gpfitbox{\texttt{\footnotesize GLang/Adequacy/Strengthened.lean:172}}\\
\gpfitbox{\textsc{S-C-Tprod}} & \gpfitbox{\texttt{\textbackslash glrule\{s.cob.tprod\}}} & \gpfitbox{\texttt{\footnotesize GLang/Adequacy/Strengthened.lean:180}}\\
\gpfitbox{\textsc{S-C-Dsum}} & \gpfitbox{\texttt{\textbackslash glrule\{s.cob.dsum\}}} & \gpfitbox{\texttt{\footnotesize GLang/Adequacy/Strengthened.lean:187}}\\
\gpfitbox{\textsc{S-C-Dual\allowbreak Std}} & \gpfitbox{\texttt{\textbackslash glrule\{s.cob.dualStd\}}} & \gpfitbox{\texttt{\footnotesize GLang/Adequacy/Strengthened.lean:193}}\\
\gpfitbox{\textsc{S-C-Image}} & \gpfitbox{\texttt{\textbackslash glrule\{s.cob.image\}}} & \gpfitbox{\texttt{\footnotesize GLang/Adequacy/Strengthened.lean:199}}\\
\gpfitbox{\textsc{S-C-Image\allowbreak Svds}} & \gpfitbox{\texttt{\textbackslash glrule\{s.cob.imageSvds\}}} & \gpfitbox{\texttt{\footnotesize GLang/Adequacy/Strengthened.lean:205}}\\
\gpfitbox{\textsc{S-C-Proj\allowbreak Ker\allowbreak Coker}} & \gpfitbox{\texttt{\textbackslash glrule\{s.cob.projKerCoker\}}} & \gpfitbox{\texttt{\footnotesize GLang/Adequacy/Strengthened.lean:221}}\\
\gpfitbox{\textsc{S-C-Proj\allowbreak Img\allowbreak Coimg}} & \gpfitbox{\texttt{\textbackslash glrule\{s.cob.projImgCoimg\}}} & \gpfitbox{\texttt{\footnotesize GLang/Adequacy/Strengthened.lean:242}}\\
\gpfitbox{\textsc{S-C-Terms\allowbreak Rref}} & \gpfitbox{\texttt{\textbackslash glrule\{s.cob.termsRref\}}} & \gpfitbox{\texttt{\footnotesize GLang/Adequacy/Strengthened.lean:263}}\\
\gpfitbox{\textsc{S-C-Refl}} & \gpfitbox{\texttt{\textbackslash glrule\{s.cob.refl\}}} & \gpfitbox{\texttt{\footnotesize GLang/Adequacy/Strengthened.lean:276}}\\
\gpfitbox{\textsc{S-C-Proj\allowbreak Ker\allowbreak Coker\allowbreak Cod}} & \gpfitbox{\texttt{\textbackslash glrule\{s.cob.projKerCokerCod\}}} & \gpfitbox{\texttt{\footnotesize GLang/Adequacy/Strengthened.lean:283}}\\
\gpfitbox{\textsc{S-C-Proj\allowbreak Img\allowbreak Coimg\allowbreak Dom}} & \gpfitbox{\texttt{\textbackslash glrule\{s.cob.projImgCoimgDom\}}} & \gpfitbox{\texttt{\footnotesize GLang/Adequacy/Strengthened.lean:300}}\\
\gpfitbox{\textsc{S-E-Var\allowbreak Lin}} & \gpfitbox{\texttt{\textbackslash glrule\{s.term.varLin\}}} & \gpfitbox{\texttt{\footnotesize GLang/Adequacy/Strengthened.lean:325}}\\
\gpfitbox{\textsc{S-E-Var\allowbreak Nlin}} & \gpfitbox{\texttt{\textbackslash glrule\{s.term.varNlin\}}} & \gpfitbox{\texttt{\footnotesize GLang/Adequacy/Strengthened.lean:332}}\\
\gpfitbox{\textsc{S-E-Promote}} & \gpfitbox{\texttt{\textbackslash glrule\{s.term.promote\}}} & \gpfitbox{\texttt{\footnotesize GLang/Adequacy/Strengthened.lean:339}}\\
\gpfitbox{\textsc{S-E-Cob}} & \gpfitbox{\texttt{\textbackslash glrule\{s.term.cob\}}} & \gpfitbox{\texttt{\footnotesize GLang/Adequacy/Strengthened.lean:353}}\\
\gpfitbox{\textsc{S-E-Reshape\allowbreak Cob}} & \gpfitbox{\texttt{\textbackslash glrule\{s.term.reshapeCob\}}} & \gpfitbox{\texttt{\footnotesize GLang/Adequacy/Strengthened.lean:366}}\\
\gpfitbox{\textsc{S-E-Scale}} & \gpfitbox{\texttt{\textbackslash glrule\{s.term.scale\}}} & \gpfitbox{\texttt{\footnotesize GLang/Adequacy/Strengthened.lean:373}}\\
\gpfitbox{\textsc{S-E-Add}} & \gpfitbox{\texttt{\textbackslash glrule\{s.term.add\}}} & \gpfitbox{\texttt{\footnotesize GLang/Adequacy/Strengthened.lean:383}}\\
\gpfitbox{\textsc{S-E-Dsum\allowbreak E}} & \gpfitbox{\texttt{\textbackslash glrule\{s.term.dsumE\}}} & \gpfitbox{\texttt{\footnotesize GLang/Adequacy/Strengthened.lean:397}}\\
\gpfitbox{\textsc{S-E-Tprod\allowbreak E}} & \gpfitbox{\texttt{\textbackslash glrule\{s.term.tprodE\}}} & \gpfitbox{\texttt{\footnotesize GLang/Adequacy/Strengthened.lean:407}}\\
\gpfitbox{\textsc{S-E-Pinv\allowbreak TStd}} & \gpfitbox{\texttt{\textbackslash glrule\{s.term.pinvTStd\}}} & \gpfitbox{\texttt{\footnotesize GLang/Adequacy/Strengthened.lean:419}}\\
\gpfitbox{\textsc{S-E-Pinv\allowbreak TGpinv}} & \gpfitbox{\texttt{\textbackslash glrule\{s.term.pinvTGpinv\}}} & \gpfitbox{\texttt{\footnotesize GLang/Adequacy/Strengthened.lean:429}}\\
\gpfitbox{\textsc{S-E-Pinv\allowbreak DStd}} & \gpfitbox{\texttt{\textbackslash glrule\{s.term.pinvDStd\}}} & \gpfitbox{\texttt{\footnotesize GLang/Adequacy/Strengthened.lean:443}}\\
\gpfitbox{\textsc{S-E-Pinv\allowbreak DGpinv}} & \gpfitbox{\texttt{\textbackslash glrule\{s.term.pinvDGpinv\}}} & \gpfitbox{\texttt{\footnotesize GLang/Adequacy/Strengthened.lean:455}}\\
\gpfitbox{\textsc{S-E-Inner\allowbreak Std}} & \gpfitbox{\texttt{\textbackslash glrule\{s.term.innerStd\}}} & \gpfitbox{\texttt{\footnotesize GLang/Adequacy/Strengthened.lean:472}}\\
\gpfitbox{\textsc{S-E-Embed\allowbreak Ker}} & \gpfitbox{\texttt{\textbackslash glrule\{s.term.embedKer\}}} & \gpfitbox{\texttt{\footnotesize GLang/Adequacy/Strengthened.lean:482}}\\
\gpfitbox{\textsc{S-E-Embed\allowbreak Coker}} & \gpfitbox{\texttt{\textbackslash glrule\{s.term.embedCoker\}}} & \gpfitbox{\texttt{\footnotesize GLang/Adequacy/Strengthened.lean:503}}\\
\gpfitbox{\textsc{S-E-Embed\allowbreak Img}} & \gpfitbox{\texttt{\textbackslash glrule\{s.term.embedImg\}}} & \gpfitbox{\texttt{\footnotesize GLang/Adequacy/Strengthened.lean:523}}\\
\gpfitbox{\textsc{S-E-Embed\allowbreak Coimg}} & \gpfitbox{\texttt{\textbackslash glrule\{s.term.embedCoimg\}}} & \gpfitbox{\texttt{\footnotesize GLang/Adequacy/Strengthened.lean:543}}\\
\gpfitbox{\textsc{S-E-App\allowbreak LLin}} & \gpfitbox{\texttt{\textbackslash glrule\{s.term.appLLin\}}} & \gpfitbox{\texttt{\footnotesize GLang/Adequacy/Strengthened.lean:563}}\\
\gpfitbox{\textsc{S-E-App\allowbreak ILin}} & \gpfitbox{\texttt{\textbackslash glrule\{s.term.appILin\}}} & \gpfitbox{\texttt{\footnotesize GLang/Adequacy/Strengthened.lean:583}}\\
\gpfitbox{\textsc{S-E-Nllam}} & \gpfitbox{\texttt{\textbackslash glrule\{s.term.nllam\}}} & \gpfitbox{\texttt{\footnotesize GLang/Adequacy/Strengthened.lean:594}}\\
\gpfitbox{\textsc{S-E-App\allowbreak NVal}} & \gpfitbox{\texttt{\textbackslash glrule\{s.term.appNVal\}}} & \gpfitbox{\texttt{\footnotesize GLang/Adequacy/Strengthened.lean:612}}\\
\gpfitbox{\textsc{S-E-Frame\allowbreak Reanchor}} & \gpfitbox{\texttt{\textbackslash glrule\{s.term.frameReanchor\}}} & \gpfitbox{\texttt{\footnotesize GLang/Adequacy/Strengthened.lean:627}}\\
\gpfitbox{\textsc{S-E-Frame\allowbreak Eta}} & \gpfitbox{\texttt{\textbackslash glrule\{s.term.frameEta\}}} & \gpfitbox{\texttt{\footnotesize GLang/Adequacy/Strengthened.lean:652}}\\
\gpfitbox{\textsc{S-E-Llam\allowbreak Assemble}} & \gpfitbox{\texttt{\textbackslash glrule\{s.term.llamAssemble\}}} & \gpfitbox{\texttt{\footnotesize GLang/Adequacy/Strengthened.lean:671}}\\
\gpfitbox{\textsc{S-E-Llam\allowbreak Assemble\allowbreak Real}} & \gpfitbox{\texttt{\textbackslash glrule\{s.term.llamAssembleReal\}}} & \gpfitbox{\texttt{\footnotesize GLang/Adequacy/Strengthened.lean:701}}\\
\gpfitbox{\textsc{S-E-Zero}} & \gpfitbox{\texttt{\textbackslash glrule\{s.term.zero\}}} & \gpfitbox{\texttt{\footnotesize GLang/Adequacy/Strengthened.lean:720}}\\
\gpfitbox{\textsc{S-E-Rlit}} & \gpfitbox{\texttt{\textbackslash glrule\{s.term.rlit\}}} & \gpfitbox{\texttt{\footnotesize GLang/Adequacy/Strengthened.lean:728}}\\
\gpfitbox{\textsc{C-S-Real}} & \gpfitbox{\texttt{\textbackslash glrule\{c.space.real\}}} & \gpfitbox{\texttt{\footnotesize GLang/Adequacy/Raw\allowbreak Certified\allowbreak Operational.lean:32}}\\
\gpfitbox{\textsc{C-S-Empty}} & \gpfitbox{\texttt{\textbackslash glrule\{c.space.empty\}}} & \gpfitbox{\texttt{\footnotesize GLang/Adequacy/Raw\allowbreak Certified\allowbreak Operational.lean:33}}\\
\gpfitbox{\textsc{C-S-Dual}} & \gpfitbox{\texttt{\textbackslash glrule\{c.space.dual\}}} & \gpfitbox{\texttt{\footnotesize GLang/Adequacy/Raw\allowbreak Certified\allowbreak Operational.lean:34}}\\
\gpfitbox{\textsc{C-S-Ker}} & \gpfitbox{\texttt{\textbackslash glrule\{c.space.ker\}}} & \gpfitbox{\texttt{\footnotesize GLang/Adequacy/Raw\allowbreak Certified\allowbreak Operational.lean:38}}\\
\gpfitbox{\textsc{C-S-Coker}} & \gpfitbox{\texttt{\textbackslash glrule\{c.space.coker\}}} & \gpfitbox{\texttt{\footnotesize GLang/Adequacy/Raw\allowbreak Certified\allowbreak Operational.lean:47}}\\
\gpfitbox{\textsc{C-S-Img}} & \gpfitbox{\texttt{\textbackslash glrule\{c.space.img\}}} & \gpfitbox{\texttt{\footnotesize GLang/Adequacy/Raw\allowbreak Certified\allowbreak Operational.lean:55}}\\
\gpfitbox{\textsc{C-S-Coimg}} & \gpfitbox{\texttt{\textbackslash glrule\{c.space.coimg\}}} & \gpfitbox{\texttt{\footnotesize GLang/Adequacy/Raw\allowbreak Certified\allowbreak Operational.lean:63}}\\
\gpfitbox{\textsc{C-S-Tprod}} & \gpfitbox{\texttt{\textbackslash glrule\{c.space.tprod\}}} & \gpfitbox{\texttt{\footnotesize GLang/Adequacy/Raw\allowbreak Certified\allowbreak Operational.lean:71}}\\
\gpfitbox{\textsc{C-S-Dsum}} & \gpfitbox{\texttt{\textbackslash glrule\{c.space.dsum\}}} & \gpfitbox{\texttt{\footnotesize GLang/Adequacy/Raw\allowbreak Certified\allowbreak Operational.lean:75}}\\
\gpfitbox{\textsc{C-T-Space}} & \gpfitbox{\texttt{\textbackslash glrule\{c.ty.space\}}} & \gpfitbox{\texttt{\footnotesize GLang/Adequacy/Raw\allowbreak Certified\allowbreak Operational.lean:84}}\\
\gpfitbox{\textsc{C-T-Pi\allowbreak Ty}} & \gpfitbox{\texttt{\textbackslash glrule\{c.ty.piTy\}}} & \gpfitbox{\texttt{\footnotesize GLang/Adequacy/Raw\allowbreak Certified\allowbreak Operational.lean:87}}\\
\gpfitbox{\textsc{C-TA-Space}} & \gpfitbox{\texttt{\textbackslash glrule\{c.tyAt.space\}}} & \gpfitbox{\texttt{\footnotesize GLang/Adequacy/Raw\allowbreak Certified\allowbreak Operational.lean:105}}\\
\gpfitbox{\textsc{C-TA-Frame}} & \gpfitbox{\texttt{\textbackslash glrule\{c.tyAt.frame\}}} & \gpfitbox{\texttt{\footnotesize GLang/Adequacy/Raw\allowbreak Certified\allowbreak Operational.lean:110}}\\
\gpfitbox{\textsc{C-C-Inverse}} & \gpfitbox{\texttt{\textbackslash glrule\{c.cob.inverse\}}} & \gpfitbox{\texttt{\footnotesize GLang/Adequacy/Raw\allowbreak Certified\allowbreak Operational.lean:125}}\\
\gpfitbox{\textsc{C-C-Trans}} & \gpfitbox{\texttt{\textbackslash glrule\{c.cob.trans\}}} & \gpfitbox{\texttt{\footnotesize GLang/Adequacy/Raw\allowbreak Certified\allowbreak Operational.lean:130}}\\
\gpfitbox{\textsc{C-C-Tprod}} & \gpfitbox{\texttt{\textbackslash glrule\{c.cob.tprod\}}} & \gpfitbox{\texttt{\footnotesize GLang/Adequacy/Raw\allowbreak Certified\allowbreak Operational.lean:138}}\\
\gpfitbox{\textsc{C-C-Dsum}} & \gpfitbox{\texttt{\textbackslash glrule\{c.cob.dsum\}}} & \gpfitbox{\texttt{\footnotesize GLang/Adequacy/Raw\allowbreak Certified\allowbreak Operational.lean:145}}\\
\gpfitbox{\textsc{C-C-Dual\allowbreak Std}} & \gpfitbox{\texttt{\textbackslash glrule\{c.cob.dualStd\}}} & \gpfitbox{\texttt{\footnotesize GLang/Adequacy/Raw\allowbreak Certified\allowbreak Operational.lean:151}}\\
\gpfitbox{\textsc{C-C-Image}} & \gpfitbox{\texttt{\textbackslash glrule\{c.cob.image\}}} & \gpfitbox{\texttt{\footnotesize GLang/Adequacy/Raw\allowbreak Certified\allowbreak Operational.lean:157}}\\
\gpfitbox{\textsc{C-C-Image\allowbreak Svds}} & \gpfitbox{\texttt{\textbackslash glrule\{c.cob.imageSvds\}}} & \gpfitbox{\texttt{\footnotesize GLang/Adequacy/Raw\allowbreak Certified\allowbreak Operational.lean:163}}\\
\gpfitbox{\textsc{C-C-Proj\allowbreak Ker\allowbreak Coker}} & \gpfitbox{\texttt{\textbackslash glrule\{c.cob.projKerCoker\}}} & \gpfitbox{\texttt{\footnotesize GLang/Adequacy/Raw\allowbreak Certified\allowbreak Operational.lean:179}}\\
\gpfitbox{\textsc{C-C-Proj\allowbreak Img\allowbreak Coimg}} & \gpfitbox{\texttt{\textbackslash glrule\{c.cob.projImgCoimg\}}} & \gpfitbox{\texttt{\footnotesize GLang/Adequacy/Raw\allowbreak Certified\allowbreak Operational.lean:200}}\\
\gpfitbox{\textsc{C-C-Terms\allowbreak Rref}} & \gpfitbox{\texttt{\textbackslash glrule\{c.cob.termsRref\}}} & \gpfitbox{\texttt{\footnotesize GLang/Adequacy/Raw\allowbreak Certified\allowbreak Operational.lean:221}}\\
\gpfitbox{\textsc{C-C-Refl}} & \gpfitbox{\texttt{\textbackslash glrule\{c.cob.refl\}}} & \gpfitbox{\texttt{\footnotesize GLang/Adequacy/Raw\allowbreak Certified\allowbreak Operational.lean:234}}\\
\gpfitbox{\textsc{C-C-Proj\allowbreak Ker\allowbreak Coker\allowbreak Cod}} & \gpfitbox{\texttt{\textbackslash glrule\{c.cob.projKerCokerCod\}}} & \gpfitbox{\texttt{\footnotesize GLang/Adequacy/Raw\allowbreak Certified\allowbreak Operational.lean:241}}\\
\gpfitbox{\textsc{C-C-Proj\allowbreak Img\allowbreak Coimg\allowbreak Dom}} & \gpfitbox{\texttt{\textbackslash glrule\{c.cob.projImgCoimgDom\}}} & \gpfitbox{\texttt{\footnotesize GLang/Adequacy/Raw\allowbreak Certified\allowbreak Operational.lean:258}}\\
\gpfitbox{\textsc{C-E-Var\allowbreak Lin}} & \gpfitbox{\texttt{\textbackslash glrule\{c.term.varLin\}}} & \gpfitbox{\texttt{\footnotesize GLang/Adequacy/Raw\allowbreak Certified\allowbreak Operational.lean:283}}\\
\gpfitbox{\textsc{C-E-Var\allowbreak Nlin}} & \gpfitbox{\texttt{\textbackslash glrule\{c.term.varNlin\}}} & \gpfitbox{\texttt{\footnotesize GLang/Adequacy/Raw\allowbreak Certified\allowbreak Operational.lean:290}}\\
\gpfitbox{\textsc{C-E-Promote}} & \gpfitbox{\texttt{\textbackslash glrule\{c.term.promote\}}} & \gpfitbox{\texttt{\footnotesize GLang/Adequacy/Raw\allowbreak Certified\allowbreak Operational.lean:297}}\\
\gpfitbox{\textsc{C-E-Cob}} & \gpfitbox{\texttt{\textbackslash glrule\{c.term.cob\}}} & \gpfitbox{\texttt{\footnotesize GLang/Adequacy/Raw\allowbreak Certified\allowbreak Operational.lean:315}}\\
\gpfitbox{\textsc{C-E-Reshape\allowbreak Cob}} & \gpfitbox{\texttt{\textbackslash glrule\{c.term.reshapeCob\}}} & \gpfitbox{\texttt{\footnotesize GLang/Adequacy/Raw\allowbreak Certified\allowbreak Operational.lean:328}}\\
\gpfitbox{\textsc{C-E-Scale}} & \gpfitbox{\texttt{\textbackslash glrule\{c.term.scale\}}} & \gpfitbox{\texttt{\footnotesize GLang/Adequacy/Raw\allowbreak Certified\allowbreak Operational.lean:335}}\\
\gpfitbox{\textsc{C-E-Add}} & \gpfitbox{\texttt{\textbackslash glrule\{c.term.add\}}} & \gpfitbox{\texttt{\footnotesize GLang/Adequacy/Raw\allowbreak Certified\allowbreak Operational.lean:345}}\\
\gpfitbox{\textsc{C-E-Dsum\allowbreak E}} & \gpfitbox{\texttt{\textbackslash glrule\{c.term.dsumE\}}} & \gpfitbox{\texttt{\footnotesize GLang/Adequacy/Raw\allowbreak Certified\allowbreak Operational.lean:359}}\\
\gpfitbox{\textsc{C-E-Tprod\allowbreak E}} & \gpfitbox{\texttt{\textbackslash glrule\{c.term.tprodE\}}} & \gpfitbox{\texttt{\footnotesize GLang/Adequacy/Raw\allowbreak Certified\allowbreak Operational.lean:369}}\\
\gpfitbox{\textsc{C-E-Pinv\allowbreak TStd}} & \gpfitbox{\texttt{\textbackslash glrule\{c.term.pinvTStd\}}} & \gpfitbox{\texttt{\footnotesize GLang/Adequacy/Raw\allowbreak Certified\allowbreak Operational.lean:381}}\\
\gpfitbox{\textsc{C-E-Pinv\allowbreak TGpinv}} & \gpfitbox{\texttt{\textbackslash glrule\{c.term.pinvTGpinv\}}} & \gpfitbox{\texttt{\footnotesize GLang/Adequacy/Raw\allowbreak Certified\allowbreak Operational.lean:391}}\\
\gpfitbox{\textsc{C-E-Pinv\allowbreak DStd}} & \gpfitbox{\texttt{\textbackslash glrule\{c.term.pinvDStd\}}} & \gpfitbox{\texttt{\footnotesize GLang/Adequacy/Raw\allowbreak Certified\allowbreak Operational.lean:405}}\\
\gpfitbox{\textsc{C-E-Pinv\allowbreak DGpinv}} & \gpfitbox{\texttt{\textbackslash glrule\{c.term.pinvDGpinv\}}} & \gpfitbox{\texttt{\footnotesize GLang/Adequacy/Raw\allowbreak Certified\allowbreak Operational.lean:417}}\\
\gpfitbox{\textsc{C-E-Inner\allowbreak Std}} & \gpfitbox{\texttt{\textbackslash glrule\{c.term.innerStd\}}} & \gpfitbox{\texttt{\footnotesize GLang/Adequacy/Raw\allowbreak Certified\allowbreak Operational.lean:434}}\\
\gpfitbox{\textsc{C-E-Embed\allowbreak Ker}} & \gpfitbox{\texttt{\textbackslash glrule\{c.term.embedKer\}}} & \gpfitbox{\texttt{\footnotesize GLang/Adequacy/Raw\allowbreak Certified\allowbreak Operational.lean:444}}\\
\gpfitbox{\textsc{C-E-Embed\allowbreak Coker}} & \gpfitbox{\texttt{\textbackslash glrule\{c.term.embedCoker\}}} & \gpfitbox{\texttt{\footnotesize GLang/Adequacy/Raw\allowbreak Certified\allowbreak Operational.lean:465}}\\
\gpfitbox{\textsc{C-E-Embed\allowbreak Img}} & \gpfitbox{\texttt{\textbackslash glrule\{c.term.embedImg\}}} & \gpfitbox{\texttt{\footnotesize GLang/Adequacy/Raw\allowbreak Certified\allowbreak Operational.lean:485}}\\
\gpfitbox{\textsc{C-E-Embed\allowbreak Coimg}} & \gpfitbox{\texttt{\textbackslash glrule\{c.term.embedCoimg\}}} & \gpfitbox{\texttt{\footnotesize GLang/Adequacy/Raw\allowbreak Certified\allowbreak Operational.lean:505}}\\
\gpfitbox{\textsc{C-E-App\allowbreak LLin}} & \gpfitbox{\texttt{\textbackslash glrule\{c.term.appLLin\}}} & \gpfitbox{\texttt{\footnotesize GLang/Adequacy/Raw\allowbreak Certified\allowbreak Operational.lean:525}}\\
\gpfitbox{\textsc{C-E-App\allowbreak ILin}} & \gpfitbox{\texttt{\textbackslash glrule\{c.term.appILin\}}} & \gpfitbox{\texttt{\footnotesize GLang/Adequacy/Raw\allowbreak Certified\allowbreak Operational.lean:544}}\\
\gpfitbox{\textsc{C-E-Nllam}} & \gpfitbox{\texttt{\textbackslash glrule\{c.term.nllam\}}} & \gpfitbox{\texttt{\footnotesize GLang/Adequacy/Raw\allowbreak Certified\allowbreak Operational.lean:555}}\\
\gpfitbox{\textsc{C-E-App\allowbreak NVal}} & \gpfitbox{\texttt{\textbackslash glrule\{c.term.appNVal\}}} & \gpfitbox{\texttt{\footnotesize GLang/Adequacy/Raw\allowbreak Certified\allowbreak Operational.lean:573}}\\
\gpfitbox{\textsc{C-E-Frame\allowbreak Reanchor}} & \gpfitbox{\texttt{\textbackslash glrule\{c.term.frameReanchor\}}} & \gpfitbox{\texttt{\footnotesize GLang/Adequacy/Raw\allowbreak Certified\allowbreak Operational.lean:583}}\\
\gpfitbox{\textsc{C-E-Frame\allowbreak Eta}} & \gpfitbox{\texttt{\textbackslash glrule\{c.term.frameEta\}}} & \gpfitbox{\texttt{\footnotesize GLang/Adequacy/Raw\allowbreak Certified\allowbreak Operational.lean:608}}\\
\gpfitbox{\textsc{C-E-Llam\allowbreak Assemble}} & \gpfitbox{\texttt{\textbackslash glrule\{c.term.llamAssemble\}}} & \gpfitbox{\texttt{\footnotesize GLang/Adequacy/Raw\allowbreak Certified\allowbreak Operational.lean:627}}\\
\gpfitbox{\textsc{C-E-Llam\allowbreak Assemble\allowbreak Real}} & \gpfitbox{\texttt{\textbackslash glrule\{c.term.llamAssembleReal\}}} & \gpfitbox{\texttt{\footnotesize GLang/Adequacy/Raw\allowbreak Certified\allowbreak Operational.lean:657}}\\
\gpfitbox{\textsc{C-E-Zero}} & \gpfitbox{\texttt{\textbackslash glrule\{c.term.zero\}}} & \gpfitbox{\texttt{\footnotesize GLang/Adequacy/Raw\allowbreak Certified\allowbreak Operational.lean:676}}\\
\gpfitbox{\textsc{C-E-Rlit}} & \gpfitbox{\texttt{\textbackslash glrule\{c.term.rlit\}}} & \gpfitbox{\texttt{\footnotesize GLang/Adequacy/Raw\allowbreak Certified\allowbreak Operational.lean:684}}\\
}
\gldefcmdrows{defs}{
\gpfit{\mathsf{Agree}(\mathcal{O}, \sigma{}, \glvEnv{}, \mathit{a}, \Gamma_{\mathrm{a}}, \mathit{a}, \mathit{a}, \ltynv{{\mathrm{a}}}, \mathit{a}, \mathit{a}, \mathit{a})} & \gpfitbox{\texttt{\textbackslash gldefn\{partialAdequacy.agree\}}} & \gpfitbox{\texttt{\footnotesize GLang/Adequacy/Partial\allowbreak Source\allowbreak Agreement.lean:9}}\\
\gpfit{\mathsf{AgreeDims}} & \gpfitbox{\texttt{\textbackslash gldefn\{agreeDims\}}} & \gpfitbox{\texttt{\footnotesize GLang/Adequacy/Raw\allowbreak Reification.lean:75}}\\
\gpfit{\mathsf{AgreeDims}} & \gpfitbox{\texttt{\textbackslash gldefn\{adequacy.partialDirect.agreeDims\}}} & \gpfitbox{\texttt{\footnotesize GLang/Adequacy/Partial\allowbreak Source\allowbreak Public.lean:10}}\\
\gpfit{\SemT{\lty{}}{\sigma{}} \equiv_{\mathit{B}} \SemT{\ltyv{}}{\glvEnv{}}} & \gpfitbox{\texttt{\textbackslash gldefn\{adequacy.direct.agreeDims\}}} & \gpfitbox{\texttt{\footnotesize GLang/Adequacy/Direct\allowbreak Public.lean:55}}\\
\gpfit{\SemT{\lty{}}{\sigma{}} \equiv_{\mathit{B}} \SemT{\ltyv{}}{\glvEnv{}}} & \gpfitbox{\texttt{\textbackslash gldefn\{agreeDimsAt\}}} & \gpfitbox{\texttt{\footnotesize GLang/Adequacy/Raw\allowbreak Reification.lean:62}}\\
\gpfit{{\sigma{}_{1}} =_{\mathit{S}} {\sigma{}_{2}}} & \gpfitbox{\texttt{\textbackslash gldefn\{dEnv.agreeOn\}}} & \gpfitbox{\texttt{\footnotesize GLang/FVIrrelevance.lean:25}}\\
\gpfit{\mathsf{BasisAt}} & \gpfitbox{\texttt{\textbackslash gldefn\{adequacy.basisAt\}}} & \gpfitbox{\texttt{\footnotesize GLang/Adequacy/Public.lean:36}}\\
\gpfit{\mathit{B} \;\text{is a basis for}\; \SemT{\lty{}}{\sigma{}}} & \gpfitbox{\texttt{\textbackslash gldefn\{adequacy.direct.basisAt\}}} & \gpfitbox{\texttt{\footnotesize GLang/Adequacy/Direct\allowbreak Public.lean:42}}\\
\gpfit{\mathsf{Bounded}(\mathit{x}, \mathit{x})} & \gpfitbox{\texttt{\textbackslash gldefn\{valueQueryDependencies.bounded\}}} & \gpfitbox{\texttt{\footnotesize GLang/Value\allowbreak Query\allowbreak Dependencies.lean:45}}\\
\gpfit{\mathsf{Cob}} & \gpfitbox{\texttt{\textbackslash gldefn\{operational.cob\}}} & \gpfitbox{\texttt{\footnotesize GLang/Operational.lean:38}}\\
\gpfit{\mathsf{CobAgree}} & \gpfitbox{\texttt{\textbackslash gldefn\{adequacy.partialDirect.cobAgree\}}} & \gpfitbox{\texttt{\footnotesize GLang/Adequacy/Partial\allowbreak Source\allowbreak Public.lean:11}}\\
\gpfit{\SemBB{\textcolor{lang}{{\lb{}_{\mathrm{src}}}}}{\sigma{}} = \SemA{\leev{}}{\glvEnv{}}{} \cdot \SemBB{\textcolor{lang}{{\lb{}_{\mathrm{tgt}}}}}{\sigma{}}} & \gpfitbox{\texttt{\textbackslash gldefn\{cobAgree\}}} & \gpfitbox{\texttt{\footnotesize GLang/Adequacy/Agree.lean:253}}\\
\gpfit{\SemBB{\textcolor{lang}{{\lb{}_{\mathrm{src}}}}}{\sigma{}} = \SemA{\leev{}}{\glvEnv{}}{} \cdot \SemBB{\textcolor{lang}{{\lb{}_{\mathrm{tgt}}}}}{\sigma{}}} & \gpfitbox{\texttt{\textbackslash gldefn\{adequacy.direct.cobAgree\}}} & \gpfitbox{\texttt{\footnotesize GLang/Adequacy/Direct\allowbreak Public.lean:61}}\\
\gpfit{\mathsf{CobAgree}(\mathcal{O}, \sigma{}, \glvEnv{}, \textcolor{lang}{{\lb{}_{\mathrm{src}}}}, \textcolor{lang}{{\lb{}_{\mathrm{tgt}}}}, \leev{})} & \gpfitbox{\texttt{\textbackslash gldefn\{partialAdequacy.cobAgree\}}} & \gpfitbox{\texttt{\footnotesize GLang/Adequacy/Partial\allowbreak Source\allowbreak Cob.lean:20}}\\
\gpfit{\mathsf{CobResult}(\mathcal{O}, \glvEnv{}, \Gamma{}, \lDelta{}, \mathit{V}, \mathit{source}, \mathit{target}, \leev{})} & \gpfitbox{\texttt{\textbackslash gldefn\{partialAdequacy.cobResult\}}} & \gpfitbox{\texttt{\footnotesize GLang/Adequacy/Partial\allowbreak Source\allowbreak Cob.lean:15}}\\
\gpfit{\mathsf{CobRows}(\mathit{V}, \mathit{source}, \mathit{target}, \mathit{a})} & \gpfitbox{\texttt{\textbackslash gldefn\{partialAdequacy.cobRows\}}} & \gpfitbox{\texttt{\footnotesize GLang/Adequacy/Partial\allowbreak Source\allowbreak Cob.lean:9}}\\
\gpfit{\mathsf{CodomainLaw}(\mathcal{O})} & \gpfitbox{\texttt{\textbackslash gldefn\{runtimeValue.totalSubstitution.codomainLaw\}}} & \gpfitbox{\texttt{\footnotesize GLang/Runtime\allowbreak Value\allowbreak Total\allowbreak Semantic\allowbreak Substitution.lean:506}}\\
\gpfit{\mathsf{Consistent}} & \gpfitbox{\texttt{\textbackslash gldefn\{adequacy.partialDirect.consistent\}}} & \gpfitbox{\texttt{\footnotesize GLang/Adequacy/Partial\allowbreak Source\allowbreak Public.lean:8}}\\
\gpfit{(\lDelta{}, \evalCtxSpace{}); \evalCtx{}; \valCheckCtx{}\vdash \sigma{}, \glvEnv{}} & \gpfitbox{\texttt{\textbackslash gldefn\{adequacy.direct.consistent\}}} & \gpfitbox{\texttt{\footnotesize GLang/Adequacy/Direct\allowbreak Public.lean:26}}\\
\gpfit{\mathsf{Consistent}(\mathcal{O}, \lDelta{}, \evalCtxSpace{}, \evalCtx{}, \valCheckCtx{}, \sigma{}, \glvEnv{})} & \gpfitbox{\texttt{\textbackslash gldefn\{partialAdequacy.consistent\}}} & \gpfitbox{\texttt{\footnotesize GLang/Adequacy/Partial\allowbreak Source\allowbreak History.lean:1379}}\\
\gpfit{\mathsf{Consistent}} & \gpfitbox{\texttt{\textbackslash gldefn\{consistent\}}} & \gpfitbox{\texttt{\footnotesize GLang/Adequacy/Raw\allowbreak Typed\allowbreak Consistent.lean:20}}\\
\gpfit{\mathsf{ContextHistory}} & \gpfitbox{\texttt{\textbackslash gldefn\{operational.contextHistory\}}} & \gpfitbox{\texttt{\footnotesize GLang/Operational.lean:87}}\\
\gpfit{\mathsf{CtxTele}} & \gpfitbox{\texttt{\textbackslash gldefn\{v2.ctxTele\}}} & \gpfitbox{\texttt{\footnotesize GLang/Value\allowbreak Progress.lean:380}}\\
\gpfit{\mathsf{Scoped}(\lDelta{}, \lepsilon{})} & \gpfitbox{\texttt{\textbackslash gldefn\{ctxWF\}}} & \gpfitbox{\texttt{\footnotesize GLang/Wellformed.lean:31}}\\
\gpfit{\mathsf{CtxWF'}(\lDelta{}, \evalCtxSpace{})} & \gpfitbox{\texttt{\textbackslash gldefn\{v2.ctxWF'\}}} & \gpfitbox{\texttt{\footnotesize GLang/Value\allowbreak Preservation\allowbreak V2.lean:57}}\\
\gpfit{\mathsf{DimsAt}(\mathcal{O}, \sigma{}, \glvEnv{}, \mathit{a}, \Gamma_{\mathrm{a}}, \mathit{a}, \mathit{a}, \ltynv{{\mathrm{a}}}, \mathit{a})} & \gpfitbox{\texttt{\textbackslash gldefn\{partialAdequacy.dimsAt\}}} & \gpfitbox{\texttt{\footnotesize GLang/Adequacy/Partial\allowbreak Source\allowbreak Agreement.lean:41}}\\
\gpfit{\lDelta{}, \lepsilon{}\vdash \sigma{}} & \gpfitbox{\texttt{\textbackslash gldefn\{envWF\}}} & \gpfitbox{\texttt{\footnotesize GLang/Wellformed.lean:50}}\\
\gpfit{\mathsf{EvalBB}} & \gpfitbox{\texttt{\textbackslash gldefn\{operational.unchecked.evalBB\}}} & \gpfitbox{\texttt{\footnotesize GLang/Operational.lean:209}}\\
\gpfit{\mathsf{EvalCtxCreationRank}(\evalCtx{})} & \gpfitbox{\texttt{\textbackslash gldefn\{rawConsistentRank.evalCtxCreationRank\}}} & \gpfitbox{\texttt{\footnotesize GLang/Adequacy/Raw\allowbreak Consistent\allowbreak Rank.lean:30}}\\
\gpfit{\mathsf{EvalCtxLamAnchored}(\evalCtx{})} & \gpfitbox{\texttt{\textbackslash gldefn\{v2.evalCtxLamAnchored\}}} & \gpfitbox{\texttt{\footnotesize GLang/Value\allowbreak Preservation\allowbreak V2.lean:75}}\\
\gpfit{\mathsf{EvalCtxLc}(\evalCtx{})} & \gpfitbox{\texttt{\textbackslash gldefn\{evalCtxLc\}}} & \gpfitbox{\texttt{\footnotesize GLang/Value\allowbreak Realize\allowbreak Lc.lean:21}}\\
\gpfit{\vdash \evalCtx{}} & \gpfitbox{\texttt{\textbackslash gldefn\{v2.evalCtxValid\}}} & \gpfitbox{\texttt{\footnotesize GLang/Value\allowbreak Preservation\allowbreak V2.lean:67}}\\
\gpfit{\vdash \evalCtx{}} & \gpfitbox{\texttt{\textbackslash gldefn\{v2.evalCtxValidPlus\}}} & \gpfitbox{\texttt{\footnotesize GLang/Value\allowbreak Preservation\allowbreak V2.lean:92}}\\
\gpfit{\mathsf{EvalEntryCreation}(\evalCtx{}, \mathit{x}, \lb{}, \leev{})} & \gpfitbox{\texttt{\textbackslash gldefn\{v2.evalEntryCreation\}}} & \gpfitbox{\texttt{\footnotesize GLang/Value\allowbreak Progress.lean:358}}\\
\gpfit{\mathsf{EvalEntry}(\evalCtx{}, \mathit{x}, \lb{}, \leev{})} & \gpfitbox{\texttt{\textbackslash gldefn\{v2.evalEntryValid\}}} & \gpfitbox{\texttt{\footnotesize GLang/Value\allowbreak Progress.lean:349}}\\
\gpfit{\mathsf{FrameAnchor}(\evalCtx{}, \lb{}, \leev{})} & \gpfitbox{\texttt{\textbackslash gldefn\{v2.evalFrameAnchored\}}} & \gpfitbox{\texttt{\footnotesize GLang/Value\allowbreak Progress.lean:339}}\\
\gpfit{\mathsf{EvalS}} & \gpfitbox{\texttt{\textbackslash gldefn\{operational.unchecked.evalS\}}} & \gpfitbox{\texttt{\footnotesize GLang/Operational.lean:203}}\\
\gpfit{\mathsf{Finite}(\mathit{root})} & \gpfitbox{\texttt{\textbackslash gldefn\{valueQueryDependencies.finite\}}} & \gpfitbox{\texttt{\footnotesize GLang/Value\allowbreak Query\allowbreak Dependencies.lean:51}}\\
\gpfit{\mathsf{GSVDInputsShape}({\mathit{d}_{0}}, {\mathit{d}_{1}}, \mathit{map}, \mathit{leftMetric}, \mathit{rightMetric})} & \gpfitbox{\texttt{\textbackslash gldefn\{gSVDInputsShape\}}} & \gpfitbox{\texttt{\footnotesize GLang/Value\allowbreak Numerical\allowbreak Inputs.lean:12}}\\
\gpfit{\mathsf{GSVDecomposition}(\mathit{m}, \mathit{n})} & \gpfitbox{\texttt{\textbackslash gldefn\{gSVDecomposition\}}} & \gpfitbox{\texttt{\footnotesize GLang/Value\allowbreak SVD.lean:104}}\\
\gpfit{\mathsf{HOLinear}(\sigma{}; \mathit{x}; \ls{}; \lee{}; \mathit{x})} & \gpfitbox{\texttt{\textbackslash gldefn\{higherOrderLinearAtVar\}}} & \gpfitbox{\texttt{\footnotesize GLang/Source\allowbreak Soundness\allowbreak Strong.lean:36}}\\
\gpfit{\mathsf{History}(\mathcal{O})} & \gpfitbox{\texttt{\textbackslash gldefn\{partialAdequacy.history\}}} & \gpfitbox{\texttt{\footnotesize GLang/Adequacy/Partial\allowbreak Source\allowbreak History.lean:338}}\\
\gpfit{\vdash (\lDelta{}, \lepsilon{})} & \gpfitbox{\texttt{\textbackslash gldefn\{inductiveCtxWF\}}} & \gpfitbox{\texttt{\footnotesize GLang/Source\allowbreak Context\allowbreak Inductive.lean:27}}\\
\gpfit{\mathsf{InferSpaceMatchesBase}(\leev{}, \ltynv{{\mathrm{x}}})} & \gpfitbox{\texttt{\textbackslash gldefn\{valExpr.inferSpaceMatchesBase\}}} & \gpfitbox{\texttt{\footnotesize GLang/Adequacy/Raw\allowbreak Consistent.lean:354}}\\
\gpfit{\mathsf{iofFresh}(\lepsilon{})} & \gpfitbox{\texttt{\textbackslash gldefn\{iofFresh\}}} & \gpfitbox{\texttt{\footnotesize GLang/Inner\allowbreak Product.lean:294}}\\
\gpfit{\mathsf{IsInnerProduct}(\mathit{V}, \mathit{inst}, \mathit{inst}, \mathit{B})} & \gpfitbox{\texttt{\textbackslash gldefn\{isInnerProduct\}}} & \gpfitbox{\texttt{\footnotesize GLang/Denot.lean:52}}\\
\gpfit{\mathit{vs} \;\text{is an ordered basis of}\; \mathit{V}} & \gpfitbox{\texttt{\textbackslash gldefn\{isListBasis\}}} & \gpfitbox{\texttt{\footnotesize GLang/Basis\allowbreak FSVGSVD.lean:57}}\\
\gpfit{\mathrm{dom}(\evalCtx{}) \subseteq \mathcal{S}} & \gpfitbox{\texttt{\textbackslash gldefn\{v2.keyScope\}}} & \gpfitbox{\texttt{\footnotesize GLang/Value\allowbreak Preservation\allowbreak V2.lean:143}}\\
\gpfit{\mathsf{LegacyAgree}(\mathcal{O}, \sigma{}, \glvEnv{}, \mathit{x}, \Gamma{}, \Delta{}, \mathit{x}, \ltynv{{\mathrm{x}}}, \mathit{x}, \mathit{x}, \mathit{x})} & \gpfitbox{\texttt{\textbackslash gldefn\{legacyAgree\}}} & \gpfitbox{\texttt{\footnotesize GLang/Adequacy/Agree.lean:105}}\\
\gpfit{\mathsf{LinearAtVar}(\sigma{}, \mathit{x}, \ls{}, \lee{}, \mathit{x})} & \gpfitbox{\texttt{\textbackslash gldefn\{linearAtVar\}}} & \gpfitbox{\texttt{\footnotesize GLang/Soundness.lean:42}}\\
\gpfit{\mathsf{OpStep}} & \gpfitbox{\texttt{\textbackslash gldefn\{operational.unchecked.opStep\}}} & \gpfitbox{\texttt{\footnotesize GLang/Operational.lean:212}}\\
\gpfit{\mathsf{OracleSpec}(\mathcal{O})} & \gpfitbox{\texttt{\textbackslash gldefn\{oracleSpec\}}} & \gpfitbox{\texttt{\footnotesize GLang/Value\allowbreak Denot\allowbreak Sem.lean:2102}}\\
\gpfit{\mathsf{PartialCoherent}(\ltynv{{\mathrm{x}}}, \mathit{x}, \mathit{x})} & \gpfitbox{\texttt{\textbackslash gldefn\{partialValueDomains.partialCoherent\}}} & \gpfitbox{\texttt{\footnotesize GLang/Partial\allowbreak Value\allowbreak Relation.lean:20}}\\
\gpfit{\mathsf{PartialEnvWF}(\mathcal{O}, \valCheckCtx{}, \glvEnv{})} & \gpfitbox{\texttt{\textbackslash gldefn\{runtimeValue.partialEnvWF\}}} & \gpfitbox{\texttt{\footnotesize GLang/Runtime\allowbreak Value\allowbreak Correspondence.lean:11}}\\
\gpfit{\mathsf{PartialEnvWF}(\mathcal{O}, \valCheckCtx{}, \glvEnv{})} & \gpfitbox{\texttt{\textbackslash gldefn\{partialValueDomains.partialEnvWF\}}} & \gpfitbox{\texttt{\footnotesize GLang/Partial\allowbreak Value\allowbreak Runtime\allowbreak Relation.lean:29}}\\
\gpfit{\mathsf{PartialEnvironmentsRelated}(\mathcal{O}, \valCheckCtx{}, \glvEnv{}, \glvEnv{}')} & \gpfitbox{\texttt{\textbackslash gldefn\{partialValueDomains.partialEnvironmentsRelated\}}} & \gpfitbox{\texttt{\footnotesize GLang/Partial\allowbreak Value\allowbreak Runtime\allowbreak Relation.lean:21}}\\
\gpfit{\mathsf{PartialRel}(\mathit{A}, \mathit{B}, \mathit{R}, \mathit{x}, \mathit{x})} & \gpfitbox{\texttt{\textbackslash gldefn\{paperValueDomains.partialRel\}}} & \gpfitbox{\texttt{\footnotesize GLang/Paper\allowbreak Value\allowbreak Extensionality.lean:10}}\\
\gpfit{\mathsf{PartialRelated}(\ltynv{{\mathrm{x}}}, \mathit{x}, \mathit{x}, \mathit{x}, \mathit{x})} & \gpfitbox{\texttt{\textbackslash gldefn\{partialValueDomains.partialRelated\}}} & \gpfitbox{\texttt{\footnotesize GLang/Partial\allowbreak Value\allowbreak Relation.lean:10}}\\
\gpfit{\mathsf{PartialRuntimeRelated}(\ltynv{{\mathrm{x}}}, \mathit{x}, \mathit{x}, \mathit{x}, \mathit{x})} & \gpfitbox{\texttt{\textbackslash gldefn\{partialValueDomains.partialRuntimeRelated\}}} & \gpfitbox{\texttt{\footnotesize GLang/Partial\allowbreak Value\allowbreak Runtime\allowbreak Relation.lean:10}}\\
\gpfit{\mathsf{PointwiseLinear}(\mathit{V}; \mathit{x}; \mathit{x})} & \gpfitbox{\texttt{\textbackslash gldefn\{hType.pointwiseLinear\}}} & \gpfitbox{\texttt{\footnotesize GLang/Source\allowbreak Soundness\allowbreak Strong.lean:25}}\\
\gpfit{\mathsf{RawAdequacyOracleSpec}(\mathcal{O})} & \gpfitbox{\texttt{\textbackslash gldefn\{rawAdequacyOracleSpec\}}} & \gpfitbox{\texttt{\footnotesize GLang/Adequacy/Raw\allowbreak Agreement.lean:247}}\\
\gpfit{\mathsf{RawAgree}(\mathcal{O}, \sigma{}, \glvEnv{}, \mathit{a}, \Gamma_{\mathrm{a}}, \mathit{a}, \mathit{a}, \ltynv{{\mathrm{a}}}, \mathit{a}, \mathit{a}, \mathit{a})} & \gpfitbox{\texttt{\textbackslash gldefn\{rawAgree\}}} & \gpfitbox{\texttt{\footnotesize GLang/Adequacy/Raw\allowbreak Agreement.lean:21}}\\
\gpfit{\SemT{\mathit{a}}{\sigma{}} \equiv_{\mathit{a}} \SemT{\ltynv{{\mathrm{a}}}}{\glvEnv{}}} & \gpfitbox{\texttt{\textbackslash gldefn\{rawAgreeDimsAt\}}} & \gpfitbox{\texttt{\footnotesize GLang/Adequacy/Raw\allowbreak Agreement.lean:92}}\\
\gpfit{\mathsf{RawApplicationOracleSpec}(\mathcal{O})} & \gpfitbox{\texttt{\textbackslash gldefn\{rawApplicationOracleSpec\}}} & \gpfitbox{\texttt{\footnotesize GLang/Adequacy/Raw\allowbreak Application\allowbreak Oracle\allowbreak Spec.lean:6}}\\
\gpfit{\mathit{a} \;\text{is a basis for}\; \SemT{\mathit{a}}{\sigma{}}} & \gpfitbox{\texttt{\textbackslash gldefn\{rawBasisAt\}}} & \gpfitbox{\texttt{\footnotesize GLang/Adequacy/Raw\allowbreak Reification.lean:23}}\\
\gpfit{\mathsf{RawConsistent}(\mathcal{O})} & \gpfitbox{\texttt{\textbackslash gldefn\{rawConsistent\}}} & \gpfitbox{\texttt{\footnotesize GLang/Adequacy/Raw\allowbreak Consistent.lean:377}}\\
\gpfit{\mathsf{RawReifiableAt}(\mathcal{O}, \sigma{}, \glvEnv{}, {\lDelta{}_{s}}, \Gamma_{\mathrm{Gv}}, {\lDelta{}_{v}}, \lty{}, \ltyv{}, \mathit{B})} & \gpfitbox{\texttt{\textbackslash gldefn\{rawReifiableAt\}}} & \gpfitbox{\texttt{\footnotesize GLang/Adequacy/Raw\allowbreak Reification.lean:39}}\\
\gpfit{\mathit{x} \equiv^{\ltyv{}}_{\mathit{B}} \mathit{r}} & \gpfitbox{\texttt{\textbackslash gldefn\{rawResultAgree\}}} & \gpfitbox{\texttt{\footnotesize GLang/Adequacy/Raw\allowbreak Agreement.lean:47}}\\
\gpfit{\mathsf{RawSourceInjectiveAt}(\mathcal{O}, \sigma{}, \glvEnv{}, {\lDelta{}_{s}}, \Gamma_{\mathrm{Gv}}, {\lDelta{}_{v}}, \lty{}, \ltyv{}, \mathit{B})} & \gpfitbox{\texttt{\textbackslash gldefn\{rawSourceInjectiveAt\}}} & \gpfitbox{\texttt{\footnotesize GLang/Adequacy/Raw\allowbreak Reification.lean:48}}\\
\gpfit{\mathsf{RawTypePresentation}(\mathcal{O}, \ltynv{{\mathrm{x}}}, \ltynv{{\mathrm{x}}}, \mathit{x}, \mathit{x})} & \gpfitbox{\texttt{\textbackslash gldefn\{paperValueDomains.rawTypePresentation\}}} & \gpfitbox{\texttt{\footnotesize GLang/Paper\allowbreak Value\allowbreak Raw\allowbreak Presentation.lean:12}}\\
\gpfit{\mathsf{RawTypedConsistent}(\mathcal{O}, \lDelta{}, \evalCtxSpace{}, \evalCtx{}, \valCheckCtx{}, \sigma{}, \glvEnv{})} & \gpfitbox{\texttt{\textbackslash gldefn\{rawTypedConsistent\}}} & \gpfitbox{\texttt{\footnotesize GLang/Adequacy/Raw\allowbreak Typed\allowbreak Consistent.lean:11}}\\
\gpfit{\mathsf{RawValueOpenPushOracleCoherent}(\mathcal{O})} & \gpfitbox{\texttt{\textbackslash gldefn\{rawValueOpenPushOracleCoherent\}}} & \gpfitbox{\texttt{\footnotesize GLang/Adequacy/Raw\allowbreak Agreement.lean:205}}\\
\gpfit{\mathit{x} \;\text{is recursively orthonormal}} & \gpfitbox{\texttt{\textbackslash gldefn\{basisDenot.recursiveOrthonormal\}}} & \gpfitbox{\texttt{\footnotesize GLang/Basis\allowbreak Soundness.lean:318}}\\
\gpfit{\mathsf{ResultAgree}(\mathcal{O}, \sigma{}, \glvEnv{}, {\lDelta{}_{s}}, \Gamma{}, \lDelta{}, \lty{}, \ltynv{{\mathrm{tv}}}, \mathit{B}, \mathit{x}, \mathit{r})} & \gpfitbox{\texttt{\textbackslash gldefn\{partialAdequacy.resultAgree\}}} & \gpfitbox{\texttt{\footnotesize GLang/Adequacy/Partial\allowbreak Source\allowbreak Agreement.lean:35}}\\
\gpfit{\mathsf{ResultAgree}} & \gpfitbox{\texttt{\textbackslash gldefn\{adequacy.partialDirect.resultAgree\}}} & \gpfitbox{\texttt{\footnotesize GLang/Adequacy/Partial\allowbreak Source\allowbreak Public.lean:9}}\\
\gpfit{\mathsf{ResultAgree}} & \gpfitbox{\texttt{\textbackslash gldefn\{resultAgree\}}} & \gpfitbox{\texttt{\footnotesize GLang/Adequacy/Raw\allowbreak Agreement.lean:131}}\\
\gpfit{\mathit{source} \equiv^{\ltyv{}}_{\mathit{B}} \mathit{result}} & \gpfitbox{\texttt{\textbackslash gldefn\{adequacy.direct.resultAgree\}}} & \gpfitbox{\texttt{\footnotesize GLang/Adequacy/Direct\allowbreak Public.lean:48}}\\
\gpfit{\mathsf{ResultMatrix}} & \gpfitbox{\texttt{\textbackslash gldefn\{valueCanonicalFSV.resultMatrix\}}} & \gpfitbox{\texttt{\footnotesize GLang/Value\allowbreak Canonical\allowbreak FSV.lean:17}}\\
\gpfit{\lDelta{}\vdash \Sem{\lee{}}{\sigma{}} \in \SemT{\lty{}}{\sigma{}}} & \gpfitbox{\texttt{\textbackslash gldefn\{soundClaim\}}} & \gpfitbox{\texttt{\footnotesize GLang/Soundness.lean:58}}\\
\gpfit{\mathsf{ContextConditions}(\lDelta{}, \lepsilon{})} & \gpfitbox{\texttt{\textbackslash gldefn\{metatheory.sourceContextWF\}}} & \gpfitbox{\texttt{\footnotesize GLang/Source\allowbreak Context\allowbreak Inductive.lean:17}}\\
\gpfit{\mathsf{SourceCtxValid}(\lDelta{}, \evalCtxSpace{})} & \gpfitbox{\texttt{\textbackslash gldefn\{sourceCtxValid\}}} & \gpfitbox{\texttt{\footnotesize GLang/Source\allowbreak Soundness\allowbreak Strong.lean:54}}\\
\gpfit{\mathsf{SourceEntryLinked}(\lDelta{}, \evalCtxSpace{}, \mathit{x}, \lb{})} & \gpfitbox{\texttt{\textbackslash gldefn\{v2.sourceEntryLinked\}}} & \gpfitbox{\texttt{\footnotesize GLang/Value\allowbreak Progress.lean:325}}\\
\gpfit{\mathsf{SourceEntry}(\lDelta{}, \evalCtxSpace{}, \mathit{x})} & \gpfitbox{\texttt{\textbackslash gldefn\{v2.sourceEntryValid\}}} & \gpfitbox{\texttt{\footnotesize GLang/Value\allowbreak Progress.lean:250}}\\
\gpfit{\mathsf{Space}} & \gpfitbox{\texttt{\textbackslash gldefn\{operational.space\}}} & \gpfitbox{\texttt{\footnotesize GLang/Operational.lean:29}}\\
\gpfit{\mathsf{StdAnchored}} & \gpfitbox{\texttt{\textbackslash gldefn\{v2.stdAnchored\}}} & \gpfitbox{\texttt{\footnotesize GLang/Value\allowbreak Progress.lean:86}}\\
\gpfit{\mathsf{StdComplete}(\evalCtx{})} & \gpfitbox{\texttt{\textbackslash gldefn\{v2.stdComplete\}}} & \gpfitbox{\texttt{\footnotesize GLang/Value\allowbreak Progress.lean:233}}\\
\gpfit{\mathsf{StdPiAnchored}(\evalCtx{}, \ltyn{0}, \ltyn{1})} & \gpfitbox{\texttt{\textbackslash gldefn\{v2.stdPiAnchored\}}} & \gpfitbox{\texttt{\footnotesize GLang/Value\allowbreak Progress.lean:127}}\\
\gpfit{\mathsf{StrongSourceSound}(\sigma{}; \lDelta{}; \evalCtxSpace{}; \lee{}; \lty{})} & \gpfitbox{\texttt{\textbackslash gldefn\{strongSourceSoundProp\}}} & \gpfitbox{\texttt{\footnotesize GLang/Source\allowbreak Soundness\allowbreak Strong.lean:124}}\\
\gpfit{\mathsf{Term}} & \gpfitbox{\texttt{\textbackslash gldefn\{operational.term\}}} & \gpfitbox{\texttt{\footnotesize GLang/Operational.lean:41}}\\
\gpfit{\mathsf{Ty}} & \gpfitbox{\texttt{\textbackslash gldefn\{operational.ty\}}} & \gpfitbox{\texttt{\footnotesize GLang/Operational.lean:32}}\\
\gpfit{\mathsf{TyAt}} & \gpfitbox{\texttt{\textbackslash gldefn\{operational.tyAt\}}} & \gpfitbox{\texttt{\footnotesize GLang/Operational.lean:35}}\\
\gpfit{\mathsf{TypedContext}} & \gpfitbox{\texttt{\textbackslash gldefn\{operational.typedContext\}}} & \gpfitbox{\texttt{\footnotesize GLang/Operational\allowbreak Typed\allowbreak Context.lean:11}}\\
\gpfit{\mathsf{TypingImplyEval}(\lDelta{}, \evalCtxSpace{}, \evalCtx{})} & \gpfitbox{\texttt{\textbackslash gldefn\{v2.typingImplyEval\}}} & \gpfitbox{\texttt{\footnotesize GLang/Value\allowbreak Preservation\allowbreak V2.lean:129}}\\
\gpfit{\glvEnv{}; \Gamma{}; \lDelta{}\vdash_{\mathsf{canonical}} \mathit{r}\colon \ltyv{}} & \gpfitbox{\texttt{\textbackslash gldefn\{valCanonicalResultSoundOfDenotSemS\}}} & \gpfitbox{\texttt{\footnotesize GLang/Adequacy/Value\allowbreak Soundness\allowbreak Sem\allowbreak S.lean:70}}\\
\gpfit{\glvEnv{}; \Gamma{}; \Delta{}\vdash_{\mathsf{canonical}} \mathit{x}\colon \ltynv{{\mathrm{x}}}} & \gpfitbox{\texttt{\textbackslash gldefn\{valCanonicalSoundOfDenotSemS\}}} & \gpfitbox{\texttt{\footnotesize GLang/Adequacy/Value\allowbreak Soundness\allowbreak Sem\allowbreak S.lean:49}}\\
\gpfit{\mathsf{validValCtx}(\valCheckCtx{})} & \gpfitbox{\texttt{\textbackslash gldefn\{valCtxValid\}}} & \gpfitbox{\texttt{\footnotesize GLang/Value\allowbreak Sound\allowbreak Vals.lean:85}}\\
\gpfit{\glvEnv{}; \Gamma{}; \lDelta{}\vdash \mathit{r}\colon \ltyv{}} & \gpfitbox{\texttt{\textbackslash gldefn\{valResultSoundOfDenotSemS\}}} & \gpfitbox{\texttt{\footnotesize GLang/Adequacy/Value\allowbreak Soundness\allowbreak Sem\allowbreak S.lean:64}}\\
\gpfit{\glvEnv{}; \Gamma{}; \Delta{}\vdash \mathit{x}\colon \ltynv{{\mathrm{x}}}} & \gpfitbox{\texttt{\textbackslash gldefn\{valSoundOfDenotSemS\}}} & \gpfitbox{\texttt{\footnotesize GLang/Adequacy/Value\allowbreak Soundness\allowbreak Sem\allowbreak S.lean:33}}\\
\gpfit{\vdash \ltyv{}\colon \tvalid{}} & \gpfitbox{\texttt{\textbackslash gldefn\{valTyOk\}}} & \gpfitbox{\texttt{\footnotesize GLang/Value\allowbreak Valid\allowbreak C.lean:52}}\\
\gpfit{\mathsf{World}} & \gpfitbox{\texttt{\textbackslash gldefn\{paperValueDomains.world\}}} & \gpfitbox{\texttt{\footnotesize GLang/Paper\allowbreak Value\allowbreak Extensionality.lean:34}}\\
\gpfit{\mathsf{WorldRuntimeRelated}(\mathcal{O}, \ltyv{}, \mathit{w}, \mathit{w}')} & \gpfitbox{\texttt{\textbackslash gldefn\{partialValueDomains.worldRuntimeRelated\}}} & \gpfitbox{\texttt{\footnotesize GLang/Partial\allowbreak Value\allowbreak Family\allowbreak Binders.lean:9}}\\
\gpfit{({\mathit{x}}^{\mathit{x}\rightarrow \lky{}})} & \gpfitbox{\texttt{\textbackslash gldefn\{space.closeRec\}}} & \gpfitbox{\texttt{\footnotesize GLang/Source\allowbreak LN.lean:105}}\\
\gpfit{({\mathit{x}}^{\mathit{x}\rightarrow \lky{}})} & \gpfitbox{\texttt{\textbackslash gldefn\{ty.closeRec\}}} & \gpfitbox{\texttt{\footnotesize GLang/Source\allowbreak LN.lean:116}}\\
\gpfit{({\mathit{x}}^{\mathit{x}\rightarrow \lky{}})} & \gpfitbox{\texttt{\textbackslash gldefn\{term.closeRec\}}} & \gpfitbox{\texttt{\footnotesize GLang/Source\allowbreak LN.lean:120}}\\
\gpfit{({\leev{}}^{\mathit{x}\rightarrow \lky{}})} & \gpfitbox{\texttt{\textbackslash gldefn\{valExpr.closeRec\}}} & \gpfitbox{\texttt{\footnotesize GLang/Value\allowbreak LN.lean:363}}\\
\gpfit{({\textcolor{lang}{\mathit{x}}}^{\mathit{x}\rightarrow \lky{}})} & \gpfitbox{\texttt{\textbackslash gldefn\{basis.closeRec\}}} & \gpfitbox{\texttt{\footnotesize GLang/Source\allowbreak LN.lean:147}}\\
\gpfit{({\ltynv{{\mathrm{x}}}}^{\mathit{x}\rightarrow \lky{}})} & \gpfitbox{\texttt{\textbackslash gldefn\{valTy.closeRec\}}} & \gpfitbox{\texttt{\footnotesize GLang/Value\allowbreak LN.lean:359}}\\
\gpfit{({\lsnv{{\mathrm{x}}}}^{\mathit{x}\rightarrow \lky{}})} & \gpfitbox{\texttt{\textbackslash gldefn\{valSpace.closeRec\}}} & \gpfitbox{\texttt{\footnotesize GLang/Value\allowbreak LN.lean:352}}\\
\gpfit{\mathsf{closeRecList}(\lky{}, \mathit{x}, \mathit{es})} & \gpfitbox{\texttt{\textbackslash gldefn\{term.closeRecList\}}} & \gpfitbox{\texttt{\footnotesize GLang/Source\allowbreak LN.lean:143}}\\
\gpfit{\mathsf{closeRecList}(\lky{}, \mathit{x}, \mathit{x})} & \gpfitbox{\texttt{\textbackslash gldefn\{valExpr.closeRecList\}}} & \gpfitbox{\texttt{\footnotesize GLang/Value\allowbreak LN.lean:401}}\\
\gpfit{\SemBB{\textcolor{lang}{\mathit{a}}}{\sigma{}}} & \gpfitbox{\texttt{\textbackslash gleqn\{denotBasis.empty\}}} & \gpfitbox{\texttt{\footnotesize GLang/Basis.lean:886}}\\
\gpfit{\SemBB{\textcolor{lang}{\mathit{a}}}{\sigma{}}} & \gpfitbox{\texttt{\textbackslash gleqn\{denotBasis.one\}}} & \gpfitbox{\texttt{\footnotesize GLang/Basis.lean:886}}\\
\gpfit{\SemBB{\textcolor{lang}{\mathit{a}}}{\sigma{}}} & \gpfitbox{\texttt{\textbackslash gleqn\{denotBasis.dual\}}} & \gpfitbox{\texttt{\footnotesize GLang/Basis.lean:886}}\\
\gpfit{\SemBB{\textcolor{lang}{\mathit{a}}}{\sigma{}}} & \gpfitbox{\texttt{\textbackslash gleqn\{denotBasis.dsum\}}} & \gpfitbox{\texttt{\footnotesize GLang/Basis.lean:886}}\\
\gpfit{\SemBB{\textcolor{lang}{\mathit{a}}}{\sigma{}}} & \gpfitbox{\texttt{\textbackslash gleqn\{denotBasis.tprod\}}} & \gpfitbox{\texttt{\footnotesize GLang/Basis.lean:886}}\\
\gpfit{\SemBB{\textcolor{lang}{\mathit{a}}}{\sigma{}}} & \gpfitbox{\texttt{\textbackslash gleqn\{denotBasis.proj.ker\}}} & \gpfitbox{\texttt{\footnotesize GLang/Basis.lean:886}}\\
\gpfit{\SemBB{\textcolor{lang}{\mathit{a}}}{\sigma{}}} & \gpfitbox{\texttt{\textbackslash gleqn\{denotBasis.proj.coker\}}} & \gpfitbox{\texttt{\footnotesize GLang/Basis.lean:886}}\\
\gpfit{\SemBB{\textcolor{lang}{\mathit{a}}}{\sigma{}}} & \gpfitbox{\texttt{\textbackslash gleqn\{denotBasis.proj.img\}}} & \gpfitbox{\texttt{\footnotesize GLang/Basis.lean:886}}\\
\gpfit{\SemBB{\textcolor{lang}{\mathit{a}}}{\sigma{}}} & \gpfitbox{\texttt{\textbackslash gleqn\{denotBasis.proj.coimg\}}} & \gpfitbox{\texttt{\footnotesize GLang/Basis.lean:886}}\\
\gpfit{\SemBB{\textcolor{lang}{\mathit{a}}}{\sigma{}}} & \gpfitbox{\texttt{\textbackslash gleqn\{denotBasis.image\}}} & \gpfitbox{\texttt{\footnotesize GLang/Basis.lean:886}}\\
\gpfit{\SemBB{\textcolor{lang}{\mathit{a}}}{\sigma{}}} & \gpfitbox{\texttt{\textbackslash gleqn\{denotBasis.terms\}}} & \gpfitbox{\texttt{\footnotesize GLang/Basis.lean:886}}\\
\gpfit{\SemBB{\textcolor{lang}{\mathit{a}}}{\sigma{}}} & \gpfitbox{\texttt{\textbackslash gleqn\{denotBasis.frame\}}} & \gpfitbox{\texttt{\footnotesize GLang/Basis.lean:886}}\\
\gpfit{\SemBB{\textcolor{lang}{\mathit{a}}}{\sigma{}}} & \gpfitbox{\texttt{\textbackslash gldefn\{denotBasis\}}} & \gpfitbox{\texttt{\footnotesize GLang/Basis.lean:886}}\\
\gpfit{\SemT{\ls{}}{\sigma{}}} & \gpfitbox{\texttt{\textbackslash gleqn\{denotSpaceAux.empty\}}} & \gpfitbox{\texttt{\footnotesize GLang/Denot.lean:714}}\\
\gpfit{\SemT{\ls{}}{\sigma{}}} & \gpfitbox{\texttt{\textbackslash gleqn\{denotSpaceAux.real\}}} & \gpfitbox{\texttt{\footnotesize GLang/Denot.lean:714}}\\
\gpfit{\SemT{\ls{}}{\sigma{}}} & \gpfitbox{\texttt{\textbackslash gleqn\{denotSpaceAux.dual\}}} & \gpfitbox{\texttt{\footnotesize GLang/Denot.lean:714}}\\
\gpfit{\SemT{\ls{}}{\sigma{}}} & \gpfitbox{\texttt{\textbackslash gleqn\{denotSpaceAux.dsum\}}} & \gpfitbox{\texttt{\footnotesize GLang/Denot.lean:714}}\\
\gpfit{\SemT{\ls{}}{\sigma{}}} & \gpfitbox{\texttt{\textbackslash gleqn\{denotSpaceAux.tprod\}}} & \gpfitbox{\texttt{\footnotesize GLang/Denot.lean:714}}\\
\gpfit{\SemT{\ls{}}{\sigma{}}} & \gpfitbox{\texttt{\textbackslash gleqn\{denotSpaceAux.ker\}}} & \gpfitbox{\texttt{\footnotesize GLang/Denot.lean:714}}\\
\gpfit{\SemT{\ls{}}{\sigma{}}} & \gpfitbox{\texttt{\textbackslash gleqn\{denotSpaceAux.coker\}}} & \gpfitbox{\texttt{\footnotesize GLang/Denot.lean:714}}\\
\gpfit{\SemT{\ls{}}{\sigma{}}} & \gpfitbox{\texttt{\textbackslash gleqn\{denotSpaceAux.img\}}} & \gpfitbox{\texttt{\footnotesize GLang/Denot.lean:714}}\\
\gpfit{\SemT{\ls{}}{\sigma{}}} & \gpfitbox{\texttt{\textbackslash gleqn\{denotSpaceAux.coimg\}}} & \gpfitbox{\texttt{\footnotesize GLang/Denot.lean:714}}\\
\gpfit{\SemT{\ls{}}{\sigma{}}} & \gpfitbox{\texttt{\textbackslash gldefn\{denotSpaceAux\}}} & \gpfitbox{\texttt{\footnotesize GLang/Denot.lean:714}}\\
\gpfit{\Sem{\mathit{tm}}{\sigma{}}} & \gpfitbox{\texttt{\textbackslash gleqn\{denotTermAux.zero\}}} & \gpfitbox{\texttt{\footnotesize GLang/Denot.lean:761}}\\
\gpfit{\Sem{\mathit{tm}}{\sigma{}}} & \gpfitbox{\texttt{\textbackslash gleqn\{denotTermAux.real\}}} & \gpfitbox{\texttt{\footnotesize GLang/Denot.lean:761}}\\
\gpfit{\Sem{\mathit{tm}}{\sigma{}}} & \gpfitbox{\texttt{\textbackslash gleqn\{denotTermAux.smul\}}} & \gpfitbox{\texttt{\footnotesize GLang/Denot.lean:761}}\\
\gpfit{\Sem{\mathit{tm}}{\sigma{}}} & \gpfitbox{\texttt{\textbackslash gleqn\{denotTermAux.add\}}} & \gpfitbox{\texttt{\footnotesize GLang/Denot.lean:761}}\\
\gpfit{\Sem{\mathit{tm}}{\sigma{}}} & \gpfitbox{\texttt{\textbackslash gleqn\{denotTermAux.inner\}}} & \gpfitbox{\texttt{\footnotesize GLang/Denot.lean:761}}\\
\gpfit{\Sem{\mathit{tm}}{\sigma{}}} & \gpfitbox{\texttt{\textbackslash gleqn\{denotTermAux.appL\}}} & \gpfitbox{\texttt{\footnotesize GLang/Denot.lean:761}}\\
\gpfit{\Sem{\mathit{tm}}{\sigma{}}} & \gpfitbox{\texttt{\textbackslash gleqn\{denotTermAux.appI\}}} & \gpfitbox{\texttt{\footnotesize GLang/Denot.lean:761}}\\
\gpfit{\Sem{\mathit{tm}}{\sigma{}}} & \gpfitbox{\texttt{\textbackslash gleqn\{denotTermAux.appN\}}} & \gpfitbox{\texttt{\footnotesize GLang/Denot.lean:761}}\\
\gpfit{\Sem{\mathit{tm}}{\sigma{}}} & \gpfitbox{\texttt{\textbackslash gleqn\{denotTermAux.var\}}} & \gpfitbox{\texttt{\footnotesize GLang/Denot.lean:761}}\\
\gpfit{\Sem{\mathit{tm}}{\sigma{}}} & \gpfitbox{\texttt{\textbackslash gleqn\{denotTermAux.pinvT\}}} & \gpfitbox{\texttt{\footnotesize GLang/Denot.lean:761}}\\
\gpfit{\Sem{\mathit{tm}}{\sigma{}}} & \gpfitbox{\texttt{\textbackslash gleqn\{denotTermAux.pinvD\}}} & \gpfitbox{\texttt{\footnotesize GLang/Denot.lean:761}}\\
\gpfit{\Sem{\mathit{tm}}{\sigma{}}} & \gpfitbox{\texttt{\textbackslash gleqn\{denotTermAux.dsumE\}}} & \gpfitbox{\texttt{\footnotesize GLang/Denot.lean:761}}\\
\gpfit{\Sem{\mathit{tm}}{\sigma{}}} & \gpfitbox{\texttt{\textbackslash gleqn\{denotTermAux.tprodE\}}} & \gpfitbox{\texttt{\footnotesize GLang/Denot.lean:761}}\\
\gpfit{\Sem{\mathit{tm}}{\sigma{}}} & \gpfitbox{\texttt{\textbackslash gleqn\{denotTermAux.llam\}}} & \gpfitbox{\texttt{\footnotesize GLang/Denot.lean:761}}\\
\gpfit{\Sem{\mathit{tm}}{\sigma{}}} & \gpfitbox{\texttt{\textbackslash gleqn\{denotTermAux.nllam\}}} & \gpfitbox{\texttt{\footnotesize GLang/Denot.lean:761}}\\
\gpfit{\Sem{\mathit{tm}}{\sigma{}}} & \gpfitbox{\texttt{\textbackslash gleqn\{denotTermAux.embed.ker\}}} & \gpfitbox{\texttt{\footnotesize GLang/Denot.lean:761}}\\
\gpfit{\Sem{\mathit{tm}}{\sigma{}}} & \gpfitbox{\texttt{\textbackslash gleqn\{denotTermAux.embed.coker\}}} & \gpfitbox{\texttt{\footnotesize GLang/Denot.lean:761}}\\
\gpfit{\Sem{\mathit{tm}}{\sigma{}}} & \gpfitbox{\texttt{\textbackslash gleqn\{denotTermAux.embed.img\}}} & \gpfitbox{\texttt{\footnotesize GLang/Denot.lean:761}}\\
\gpfit{\Sem{\mathit{tm}}{\sigma{}}} & \gpfitbox{\texttt{\textbackslash gleqn\{denotTermAux.embed.coimg\}}} & \gpfitbox{\texttt{\footnotesize GLang/Denot.lean:761}}\\
\gpfit{\Sem{\mathit{tm}}{\sigma{}}} & \gpfitbox{\texttt{\textbackslash gldefn\{denotTermAux\}}} & \gpfitbox{\texttt{\footnotesize GLang/Denot.lean:761}}\\
\gpfit{\SemT{\lty{}}{\sigma{}}} & \gpfitbox{\texttt{\textbackslash gleqn\{denotTyAux.piTy\}}} & \gpfitbox{\texttt{\footnotesize GLang/Denot.lean:751}}\\
\gpfit{\SemT{\lty{}}{\sigma{}}} & \gpfitbox{\texttt{\textbackslash gldefn\{denotTyAux\}}} & \gpfitbox{\texttt{\footnotesize GLang/Denot.lean:751}}\\
\gpfit{\mathcal{E}\llbracket \lenv{{\mathrm{a}}}\rrbracket_{\Gamma_{\mathrm{a}};\mathit{a}}} & \gpfitbox{\texttt{\textbackslash gldefn\{denotValBSemS\}}} & \gpfitbox{\texttt{\footnotesize GLang/Value\allowbreak Denot\allowbreak Sem.lean:811}}\\
\gpfit{\mathcal{T}\llbracket \ltynv{{\mathrm{x}}}\rrbracket_{\Gamma{};\Delta{}}} & \gpfitbox{\texttt{\textbackslash gldefn\{denotValTyS\}}} & \gpfitbox{\texttt{\footnotesize GLang/Value\allowbreak Denot\allowbreak Sem.lean:511}}\\
\gpfit{\mathsf{dof}(\mathcal{O}, \lsv{})} & \gpfitbox{\texttt{\textbackslash gldefn\{valSpace.dof\}}} & \gpfitbox{\texttt{\footnotesize GLang/Value\allowbreak Denot.lean:135}}\\
\gpfit{|\mathit{arrays}|=\mathit{count} \land \forall a\in \mathit{arrays}.\; |a|=\mathit{width}} & \gpfitbox{\texttt{\textbackslash gldefn\{valueShape.family\}}} & \gpfitbox{\texttt{\footnotesize GLang/Value\allowbreak Primitive\allowbreak Shapes.lean:41}}\\
\gpfit{\mathsf{fv}(\mathit{x})} & \gpfitbox{\texttt{\textbackslash gldefn\{space.fv\}}} & \gpfitbox{\texttt{\footnotesize GLang/Syntax.lean:271}}\\
\gpfit{\mathsf{fv}(\ltynv{{\mathrm{x}}})} & \gpfitbox{\texttt{\textbackslash gldefn\{valTy.fv\}}} & \gpfitbox{\texttt{\footnotesize GLang/Values.lean:376}}\\
\gpfit{\mathsf{fv}(\leev{})} & \gpfitbox{\texttt{\textbackslash gldefn\{valExpr.fv\}}} & \gpfitbox{\texttt{\footnotesize GLang/Values.lean:381}}\\
\gpfit{\mathsf{fv}(\textcolor{lang}{\mathit{x}})} & \gpfitbox{\texttt{\textbackslash gldefn\{basis.fv\}}} & \gpfitbox{\texttt{\footnotesize GLang/Syntax.lean:327}}\\
\gpfit{\mathsf{fv}(\lsnv{{\mathrm{x}}})} & \gpfitbox{\texttt{\textbackslash gldefn\{valSpace.fv\}}} & \gpfitbox{\texttt{\footnotesize GLang/Values.lean:369}}\\
\gpfit{\mathsf{fv}(\mathit{x})} & \gpfitbox{\texttt{\textbackslash gldefn\{ty.fv\}}} & \gpfitbox{\texttt{\footnotesize GLang/Syntax.lean:282}}\\
\gpfit{\mathsf{fv}(\mathit{x})} & \gpfitbox{\texttt{\textbackslash gldefn\{term.fv\}}} & \gpfitbox{\texttt{\footnotesize GLang/Syntax.lean:287}}\\
\gpfit{\mathsf{fvList}(\mathit{x})} & \gpfitbox{\texttt{\textbackslash gldefn\{valExpr.fvList\}}} & \gpfitbox{\texttt{\footnotesize GLang/Values.lean:420}}\\
\gpfit{\mathsf{fvList}(\mathit{x})} & \gpfitbox{\texttt{\textbackslash gldefn\{term.fvList\}}} & \gpfitbox{\texttt{\footnotesize GLang/Syntax.lean:307}}\\
\gpfit{\operatorname{inferSpace}(\leev{})} & \gpfitbox{\texttt{\textbackslash gldefn\{valExpr.inferSpace\}}} & \gpfitbox{\texttt{\footnotesize GLang/Value\allowbreak Denot.lean:1020}}\\
\gpfit{\operatorname{inferSpaceCtx}(\Gamma{}, \leev{})} & \gpfitbox{\texttt{\textbackslash gldefn\{valExpr.inferSpaceCtx\}}} & \gpfitbox{\texttt{\footnotesize GLang/Value\allowbreak Denot.lean:1139}}\\
\gpfit{\operatorname{inferType}(\leev{})} & \gpfitbox{\texttt{\textbackslash gldefn\{valExpr.inferType\}}} & \gpfitbox{\texttt{\footnotesize GLang/Value\allowbreak Denot.lean:790}}\\
\gpfit{\operatorname{inferTypeCtx}(\Gamma{}, \leev{})} & \gpfitbox{\texttt{\textbackslash gldefn\{valExpr.inferTypeCtx\}}} & \gpfitbox{\texttt{\footnotesize GLang/Value\allowbreak Denot.lean:918}}\\
\gpfit{\leev{}\subst{x}{\lenv{{\mathrm{u}}}}} & \gpfitbox{\texttt{\textbackslash gldefn\{valExpr.instantiate\}}} & \gpfitbox{\texttt{\footnotesize GLang/Value\allowbreak LN.lean:509}}\\
\gpfit{\lty{}\subst{x}{\mathit{u}}} & \gpfitbox{\texttt{\textbackslash gldefn\{ty.instantiate\}}} & \gpfitbox{\texttt{\footnotesize GLang/Source\allowbreak LN.lean:162}}\\
\gpfit{\ltyv{}\subst{x}{\lenv{{\mathrm{u}}}}} & \gpfitbox{\texttt{\textbackslash gldefn\{valTy.instantiate\}}} & \gpfitbox{\texttt{\footnotesize GLang/Value\allowbreak LN.lean:511}}\\
\gpfit{\lb{}\subst{x}{\mathit{u}}} & \gpfitbox{\texttt{\textbackslash gldefn\{basis.instantiate\}}} & \gpfitbox{\texttt{\footnotesize GLang/Source\allowbreak LN.lean:163}}\\
\gpfit{\lee{}\subst{x}{\mathit{u}}} & \gpfitbox{\texttt{\textbackslash gldefn\{term.instantiate\}}} & \gpfitbox{\texttt{\footnotesize GLang/Source\allowbreak LN.lean:160}}\\
\gpfit{\isStdBasis{\textcolor{lang}{\mathit{x}}}} & \gpfitbox{\texttt{\textbackslash gleqn\{basis.isStd.empty\}}} & \gpfitbox{\texttt{\footnotesize GLang/Value\allowbreak Operational.lean:259}}\\
\gpfit{\isStdBasis{\textcolor{lang}{\mathit{x}}}} & \gpfitbox{\texttt{\textbackslash gleqn\{basis.isStd.one\}}} & \gpfitbox{\texttt{\footnotesize GLang/Value\allowbreak Operational.lean:259}}\\
\gpfit{\isStdBasis{\textcolor{lang}{\mathit{x}}}} & \gpfitbox{\texttt{\textbackslash gleqn\{basis.isStd.dual\}}} & \gpfitbox{\texttt{\footnotesize GLang/Value\allowbreak Operational.lean:259}}\\
\gpfit{\isStdBasis{\textcolor{lang}{\mathit{x}}}} & \gpfitbox{\texttt{\textbackslash gleqn\{basis.isStd.dsum\}}} & \gpfitbox{\texttt{\footnotesize GLang/Value\allowbreak Operational.lean:259}}\\
\gpfit{\isStdBasis{\textcolor{lang}{\mathit{x}}}} & \gpfitbox{\texttt{\textbackslash gleqn\{basis.isStd.tprod\}}} & \gpfitbox{\texttt{\footnotesize GLang/Value\allowbreak Operational.lean:259}}\\
\gpfit{\isStdBasis{\textcolor{lang}{\mathit{x}}}} & \gpfitbox{\texttt{\textbackslash gleqn\{basis.isStd.proj\}}} & \gpfitbox{\texttt{\footnotesize GLang/Value\allowbreak Operational.lean:259}}\\
\gpfit{\isStdBasis{\textcolor{lang}{\mathit{x}}}} & \gpfitbox{\texttt{\textbackslash gleqn\{basis.isStd.image\}}} & \gpfitbox{\texttt{\footnotesize GLang/Value\allowbreak Operational.lean:259}}\\
\gpfit{\isStdBasis{\textcolor{lang}{\mathit{x}}}} & \gpfitbox{\texttt{\textbackslash gleqn\{basis.isStd.terms\}}} & \gpfitbox{\texttt{\footnotesize GLang/Value\allowbreak Operational.lean:259}}\\
\gpfit{\isStdBasis{\textcolor{lang}{\mathit{x}}}} & \gpfitbox{\texttt{\textbackslash gleqn\{basis.isStd.frame\}}} & \gpfitbox{\texttt{\footnotesize GLang/Value\allowbreak Operational.lean:259}}\\
\gpfit{\isStdBasis{\textcolor{lang}{\mathit{x}}}} & \gpfitbox{\texttt{\textbackslash gldefn\{basis.isStd\}}} & \gpfitbox{\texttt{\footnotesize GLang/Value\allowbreak Operational.lean:259}}\\
\gpfit{\operatorname{Mat}_{\mathit{q}}(\lsv{},\mathit{a}) \land \operatorname{Mat}_{\mathit{q}}(\lsnv{{\mathrm{t}}},\lb{})} & \gpfitbox{\texttt{\textbackslash gldefn\{valueShape.matrices\}}} & \gpfitbox{\texttt{\footnotesize GLang/Value\allowbreak Primitive\allowbreak Shapes.lean:21}}\\
\gpfit{\exists s_0,s_1,t_0,t_1.\; \lsv{}=s_0\otimes s_1 \land \lsnv{{\mathrm{t}}}=t_0\otimes t_1 \land |\mathit{a}|=d_{\mathit{q}}(s_0)d_{\mathit{q}}(s_1) \land |\lb{}|=d_{\mathit{q}}(t_0)d_{\mathit{q}}(t_1) \land d_{\mathit{q}}(s_0)=d_{\mathit{q}}(t_1)=d_{\mathit{q}}(\lsnv{{\mathrm{contract}}})} & \gpfitbox{\texttt{\textbackslash gldefn\{valueShape.mtm\}}} & \gpfitbox{\texttt{\footnotesize GLang/Value\allowbreak Primitive\allowbreak Shapes.lean:33}}\\
\gpfit{({\textcolor{lang}{\mathit{x}}}^{\lky{}\leftarrow \mathit{u}})} & \gpfitbox{\texttt{\textbackslash gldefn\{basis.openRec\}}} & \gpfitbox{\texttt{\footnotesize GLang/Source\allowbreak LN.lean:89}}\\
\gpfit{({\leev{}}^{\lky{}\leftarrow \lenv{{\mathrm{u}}}})} & \gpfitbox{\texttt{\textbackslash gldefn\{valExpr.openRec\}}} & \gpfitbox{\texttt{\footnotesize GLang/Value\allowbreak LN.lean:44}}\\
\gpfit{({\mathit{x}}^{\lky{}\leftarrow \mathit{u}})} & \gpfitbox{\texttt{\textbackslash gldefn\{space.openRec\}}} & \gpfitbox{\texttt{\footnotesize GLang/Source\allowbreak LN.lean:42}}\\
\gpfit{({\lsnv{{\mathrm{x}}}}^{\lky{}\leftarrow \lenv{{\mathrm{u}}}})} & \gpfitbox{\texttt{\textbackslash gldefn\{valSpace.openRec\}}} & \gpfitbox{\texttt{\footnotesize GLang/Value\allowbreak LN.lean:32}}\\
\gpfit{({\ltynv{{\mathrm{x}}}}^{\lky{}\leftarrow \lenv{{\mathrm{u}}}})} & \gpfitbox{\texttt{\textbackslash gldefn\{valTy.openRec\}}} & \gpfitbox{\texttt{\footnotesize GLang/Value\allowbreak LN.lean:39}}\\
\gpfit{({\mathit{x}}^{\lky{}\leftarrow \mathit{u}})} & \gpfitbox{\texttt{\textbackslash gldefn\{ty.openRec\}}} & \gpfitbox{\texttt{\footnotesize GLang/Source\allowbreak LN.lean:53}}\\
\gpfit{({\mathit{x}}^{\lky{}\leftarrow \mathit{u}})} & \gpfitbox{\texttt{\textbackslash gldefn\{term.openRec\}}} & \gpfitbox{\texttt{\footnotesize GLang/Source\allowbreak LN.lean:58}}\\
\gpfit{\mathsf{openRecList}(\lky{}, \mathit{u}, \mathit{es})} & \gpfitbox{\texttt{\textbackslash gldefn\{term.openRecList\}}} & \gpfitbox{\texttt{\footnotesize GLang/Source\allowbreak LN.lean:85}}\\
\gpfit{\mathsf{openRecList}(\lky{}, \lenv{{\mathrm{u}}}, \mathit{x})} & \gpfitbox{\texttt{\textbackslash gldefn\{valExpr.openRecList\}}} & \gpfitbox{\texttt{\footnotesize GLang/Value\allowbreak LN.lean:82}}\\
\gpfit{\leev{}} & \gpfitbox{\texttt{\textbackslash gldefn\{valExpr.openVar\}}} & \gpfitbox{\texttt{\footnotesize GLang/Value\allowbreak LN.lean:513}}\\
\gpfit{\ltynv{{\mathrm{sig}}}} & \gpfitbox{\texttt{\textbackslash gldefn\{valTy.openVar\}}} & \gpfitbox{\texttt{\footnotesize GLang/Value\allowbreak LN.lean:516}}\\
\gpfit{\lb{}} & \gpfitbox{\texttt{\textbackslash gldefn\{basis.openVar\}}} & \gpfitbox{\texttt{\footnotesize GLang/Source\allowbreak LN.lean:168}}\\
\gpfit{\sigma{}} & \gpfitbox{\texttt{\textbackslash gldefn\{ty.openVar\}}} & \gpfitbox{\texttt{\footnotesize GLang/Source\allowbreak LN.lean:167}}\\
\gpfit{\lee{}} & \gpfitbox{\texttt{\textbackslash gldefn\{term.openVar\}}} & \gpfitbox{\texttt{\footnotesize GLang/Source\allowbreak LN.lean:165}}\\
\gpfit{|\mathit{a}|=d_{\mathit{q}}(\lsv{}) \;\land\; |\lb{}|=d_{\mathit{q}}(\lsnv{{\mathrm{t}}})} & \gpfitbox{\texttt{\textbackslash gldefn\{valueShape.pair\}}} & \gpfitbox{\texttt{\footnotesize GLang/Value\allowbreak Primitive\allowbreak Shapes.lean:13}}\\
\gpfit{\mathsf{reshapArrayBetweenS}(\mathcal{O}, \glvEnv{}, \Gamma_{\mathrm{Gt}}, \mathit{t}, \Gamma_{\mathrm{Gs}}, {\lDelta{}_{s}}, \mathit{a}, \lsnv{{\mathrm{st}}}, {\ltyv{}_{\mathsf{src}}})} & \gpfitbox{\texttt{\textbackslash gldefn\{reshapArrayBetweenS\}}} & \gpfitbox{\texttt{\footnotesize GLang/Value\allowbreak Denot\allowbreak Sem.lean:662}}\\
\gpfit{\mathsf{reshapDenotSemBetweenS}(\mathcal{O}, \glvEnv{}, \Gamma_{\mathrm{Gt}}, \mathit{t}, \Gamma_{\mathrm{Gs}}, {\lDelta{}_{s}}, \mathit{a}, \ltynv{{\mathrm{a}}}, \ltynv{{\mathrm{a}}})} & \gpfitbox{\texttt{\textbackslash gldefn\{reshapDenotSemBetweenS\}}} & \gpfitbox{\texttt{\footnotesize GLang/Value\allowbreak Denot\allowbreak Sem.lean:681}}\\
\gpfit{\mathsf{reshapDenotSemS}(\mathcal{O}, \glvEnv{}, \Gamma{}, \lDelta{}, \mathit{x}, \ltynv{{\mathrm{x}}}, \ltynv{{\mathrm{x}}})} & \gpfitbox{\texttt{\textbackslash gldefn\{reshapDenotSemS\}}} & \gpfitbox{\texttt{\footnotesize GLang/Value\allowbreak Denot\allowbreak Sem.lean:726}}\\
\gpfit{|\mathit{a}|=1 \;\land\; |\lb{}|=d_{\mathit{q}}(\lsv{})} & \gpfitbox{\texttt{\textbackslash gldefn\{valueShape.scale\}}} & \gpfitbox{\texttt{\footnotesize GLang/Value\allowbreak Primitive\allowbreak Shapes.lean:10}}\\
\gpfit{\mathit{x} \in \operatorname{idx}(\lsnv{{\mathrm{x}}})} & \gpfitbox{\texttt{\textbackslash gldefn\{valSpace.shapeInBounds\}}} & \gpfitbox{\texttt{\footnotesize GLang/Value\allowbreak Denot.lean:250}}\\
\gpfit{\shpOf{\lsnv{{\mathrm{a}}}}{\mathcal{O}}} & \gpfitbox{\texttt{\textbackslash gldefn\{valSpace.shpOf\}}} & \gpfitbox{\texttt{\footnotesize GLang/Value\allowbreak Denot.lean:116}}\\
\gpfit{\sof{\textcolor{lang}{\mathit{x}}}} & \gpfitbox{\texttt{\textbackslash gleqn\{basis.sof.empty\}}} & \gpfitbox{\texttt{\footnotesize GLang/Basis.lean:60}}\\
\gpfit{\sof{\textcolor{lang}{\mathit{x}}}} & \gpfitbox{\texttt{\textbackslash gleqn\{basis.sof.one\}}} & \gpfitbox{\texttt{\footnotesize GLang/Basis.lean:60}}\\
\gpfit{\sof{\textcolor{lang}{\mathit{x}}}} & \gpfitbox{\texttt{\textbackslash gleqn\{basis.sof.dual\}}} & \gpfitbox{\texttt{\footnotesize GLang/Basis.lean:60}}\\
\gpfit{\sof{\textcolor{lang}{\mathit{x}}}} & \gpfitbox{\texttt{\textbackslash gleqn\{basis.sof.dsum\}}} & \gpfitbox{\texttt{\footnotesize GLang/Basis.lean:60}}\\
\gpfit{\sof{\textcolor{lang}{\mathit{x}}}} & \gpfitbox{\texttt{\textbackslash gleqn\{basis.sof.tprod\}}} & \gpfitbox{\texttt{\footnotesize GLang/Basis.lean:60}}\\
\gpfit{\sof{\textcolor{lang}{\mathit{x}}}} & \gpfitbox{\texttt{\textbackslash gleqn\{basis.sof.proj\}}} & \gpfitbox{\texttt{\footnotesize GLang/Basis.lean:60}}\\
\gpfit{\sof{\textcolor{lang}{\mathit{x}}}} & \gpfitbox{\texttt{\textbackslash gleqn\{basis.sof.image\}}} & \gpfitbox{\texttt{\footnotesize GLang/Basis.lean:60}}\\
\gpfit{\sof{\textcolor{lang}{\mathit{x}}}} & \gpfitbox{\texttt{\textbackslash gleqn\{basis.sof.terms\}}} & \gpfitbox{\texttt{\footnotesize GLang/Basis.lean:60}}\\
\gpfit{\sof{\textcolor{lang}{\mathit{x}}}} & \gpfitbox{\texttt{\textbackslash gldefn\{basis.sof\}}} & \gpfitbox{\texttt{\footnotesize GLang/Basis.lean:60}}\\
\gpfit{\stdBasisDCtx{\mathit{x}}} & \gpfitbox{\texttt{\textbackslash gleqn\{basis.std.piTy\}}} & \gpfitbox{\texttt{\footnotesize GLang/Basis.lean:523}}\\
\gpfit{\stdBasisDCtx{\mathit{x}}} & \gpfitbox{\texttt{\textbackslash gldefn\{basis.std\}}} & \gpfitbox{\texttt{\footnotesize GLang/Basis.lean:523}}\\
\gpfit{\stdBasisDCtx{\mathit{x}}} & \gpfitbox{\texttt{\textbackslash gleqn\{basis.stdSpace.empty\}}} & \gpfitbox{\texttt{\footnotesize GLang/Basis.lean:506}}\\
\gpfit{\stdBasisDCtx{\mathit{x}}} & \gpfitbox{\texttt{\textbackslash gleqn\{basis.stdSpace.real\}}} & \gpfitbox{\texttt{\footnotesize GLang/Basis.lean:506}}\\
\gpfit{\stdBasisDCtx{\mathit{x}}} & \gpfitbox{\texttt{\textbackslash gleqn\{basis.stdSpace.dual\}}} & \gpfitbox{\texttt{\footnotesize GLang/Basis.lean:506}}\\
\gpfit{\stdBasisDCtx{\mathit{x}}} & \gpfitbox{\texttt{\textbackslash gleqn\{basis.stdSpace.dsum\}}} & \gpfitbox{\texttt{\footnotesize GLang/Basis.lean:506}}\\
\gpfit{\stdBasisDCtx{\mathit{x}}} & \gpfitbox{\texttt{\textbackslash gleqn\{basis.stdSpace.tprod\}}} & \gpfitbox{\texttt{\footnotesize GLang/Basis.lean:506}}\\
\gpfit{\stdBasisDCtx{\mathit{x}}} & \gpfitbox{\texttt{\textbackslash gleqn\{basis.stdSpace.ker\}}} & \gpfitbox{\texttt{\footnotesize GLang/Basis.lean:506}}\\
\gpfit{\stdBasisDCtx{\mathit{x}}} & \gpfitbox{\texttt{\textbackslash gleqn\{basis.stdSpace.coker\}}} & \gpfitbox{\texttt{\footnotesize GLang/Basis.lean:506}}\\
\gpfit{\stdBasisDCtx{\mathit{x}}} & \gpfitbox{\texttt{\textbackslash gleqn\{basis.stdSpace.img\}}} & \gpfitbox{\texttt{\footnotesize GLang/Basis.lean:506}}\\
\gpfit{\stdBasisDCtx{\mathit{x}}} & \gpfitbox{\texttt{\textbackslash gleqn\{basis.stdSpace.coimg\}}} & \gpfitbox{\texttt{\footnotesize GLang/Basis.lean:506}}\\
\gpfit{\stdBasisDCtx{\mathit{x}}} & \gpfitbox{\texttt{\textbackslash gleqn\{basis.stdSpace.ker.formula\}}} & \gpfitbox{\texttt{\footnotesize GLang/Basis.lean:506}}\\
\gpfit{\stdBasisDCtx{\mathit{x}}} & \gpfitbox{\texttt{\textbackslash gleqn\{basis.stdSpace.ker.condition\}}} & \gpfitbox{\texttt{\footnotesize GLang/Basis.lean:506}}\\
\gpfit{\stdBasisDCtx{\mathit{x}}} & \gpfitbox{\texttt{\textbackslash gleqn\{basis.stdSpace.coker.formula\}}} & \gpfitbox{\texttt{\footnotesize GLang/Basis.lean:506}}\\
\gpfit{\stdBasisDCtx{\mathit{x}}} & \gpfitbox{\texttt{\textbackslash gleqn\{basis.stdSpace.coker.condition\}}} & \gpfitbox{\texttt{\footnotesize GLang/Basis.lean:506}}\\
\gpfit{\stdBasisDCtx{\mathit{x}}} & \gpfitbox{\texttt{\textbackslash gleqn\{basis.stdSpace.img.formula\}}} & \gpfitbox{\texttt{\footnotesize GLang/Basis.lean:506}}\\
\gpfit{\stdBasisDCtx{\mathit{x}}} & \gpfitbox{\texttt{\textbackslash gleqn\{basis.stdSpace.img.condition\}}} & \gpfitbox{\texttt{\footnotesize GLang/Basis.lean:506}}\\
\gpfit{\stdBasisDCtx{\mathit{x}}} & \gpfitbox{\texttt{\textbackslash gleqn\{basis.stdSpace.coimg.formula\}}} & \gpfitbox{\texttt{\footnotesize GLang/Basis.lean:506}}\\
\gpfit{\stdBasisDCtx{\mathit{x}}} & \gpfitbox{\texttt{\textbackslash gleqn\{basis.stdSpace.coimg.condition\}}} & \gpfitbox{\texttt{\footnotesize GLang/Basis.lean:506}}\\
\gpfit{\stdBasisDCtx{\mathit{x}}} & \gpfitbox{\texttt{\textbackslash gldefn\{basis.stdSpace\}}} & \gpfitbox{\texttt{\footnotesize GLang/Basis.lean:506}}\\
\gpfit{\mathsf{subst}(\mathit{x}, \lee{}, \mathit{x})} & \gpfitbox{\texttt{\textbackslash gldefn\{term.subst\}}} & \gpfitbox{\texttt{\footnotesize GLang/Syntax.lean:215}}\\
\gpfit{\mathsf{subst}(\mathit{x}, \lee{}, \mathit{x})} & \gpfitbox{\texttt{\textbackslash gldefn\{space.subst\}}} & \gpfitbox{\texttt{\footnotesize GLang/Syntax.lean:194}}\\
\gpfit{\mathsf{subst}(\mathit{x}, \lenv{{\mathrm{u}}}, \lsnv{{\mathrm{x}}})} & \gpfitbox{\texttt{\textbackslash gldefn\{valSpace.subst\}}} & \gpfitbox{\texttt{\footnotesize GLang/Value\allowbreak LN.lean:418}}\\
\gpfit{\mathsf{subst}(\mathit{x}, \lenv{{\mathrm{u}}}, \leev{})} & \gpfitbox{\texttt{\textbackslash gldefn\{valExpr.subst\}}} & \gpfitbox{\texttt{\footnotesize GLang/Value\allowbreak LN.lean:429}}\\
\gpfit{\mathsf{subst}(\mathit{x}, \lee{}, \textcolor{lang}{\mathit{x}})} & \gpfitbox{\texttt{\textbackslash gldefn\{basis.subst\}}} & \gpfitbox{\texttt{\footnotesize GLang/Syntax.lean:249}}\\
\gpfit{\mathsf{subst}(\mathit{x}, \lee{}, \mathit{x})} & \gpfitbox{\texttt{\textbackslash gldefn\{ty.subst\}}} & \gpfitbox{\texttt{\footnotesize GLang/Syntax.lean:209}}\\
\gpfit{\mathsf{subst}(\mathit{x}, \lenv{{\mathrm{u}}}, \ltynv{{\mathrm{x}}})} & \gpfitbox{\texttt{\textbackslash gldefn\{valTy.subst\}}} & \gpfitbox{\texttt{\footnotesize GLang/Value\allowbreak LN.lean:425}}\\
\gpfit{\mathsf{substList}(\mathit{x}, \lenv{{\mathrm{u}}}, \mathit{x})} & \gpfitbox{\texttt{\textbackslash gldefn\{valExpr.substList\}}} & \gpfitbox{\texttt{\footnotesize GLang/Value\allowbreak LN.lean:466}}\\
\gpfit{\operatorname{Mat}_{\mathit{q}}(\lsv{},\mathit{a})} & \gpfitbox{\texttt{\textbackslash gldefn\{valueShape.tensor\}}} & \gpfitbox{\texttt{\footnotesize GLang/Value\allowbreak Primitive\allowbreak Shapes.lean:16}}\\
\gpfit{\tof{\textcolor{lang}{\mathit{x}}}} & \gpfitbox{\texttt{\textbackslash gleqn\{basis.tof.frame\}}} & \gpfitbox{\texttt{\footnotesize GLang/Basis.lean:74}}\\
\gpfit{\tof{\textcolor{lang}{\mathit{x}}}} & \gpfitbox{\texttt{\textbackslash gleqn\{basis.tof.space\}}} & \gpfitbox{\texttt{\footnotesize GLang/Basis.lean:74}}\\
\gpfit{\tof{\textcolor{lang}{\mathit{x}}}} & \gpfitbox{\texttt{\textbackslash gldefn\{basis.tof\}}} & \gpfitbox{\texttt{\footnotesize GLang/Basis.lean:74}}\\
\gpfit{|\mathit{a}|=d_{\mathit{q}}(\lsv{}) \land |\lb{}|=d_{\mathit{q}}(\lsnv{{\mathrm{t}}}) \land d_{\mathit{q}}(\partial(\lsv{}))=d_{\mathit{q}}(\lsnv{{\mathrm{contract}}})=d_{\mathit{q}}(\lsnv{{\mathrm{t}}})} & \gpfitbox{\texttt{\textbackslash gldefn\{valueShape.ttv\}}} & \gpfitbox{\texttt{\footnotesize GLang/Value\allowbreak Primitive\allowbreak Shapes.lean:24}}\\
Value domains & \gpfitbox{\texttt{\textbackslash gldefn\{value.domains\}}} & \gpfitbox{\texttt{\footnotesize GLang/Paper\allowbreak Value\allowbreak Domains.lean:16}}\\
Fundamental-space dimensions from SVD & \gpfitbox{\texttt{\textbackslash gldefn\{value.ffs.svd\}}} & \gpfitbox{\texttt{\footnotesize GLang/Partial\allowbreak Value\allowbreak SVDDimensions.lean:52}}\\
Inference auxiliaries & \gpfitbox{\texttt{\textbackslash gldefn\{value.inference.helpers\}}} & \gpfitbox{\texttt{\footnotesize GLang/Value\allowbreak Denot.lean:62}}\\
Environments and arrays & \gpfitbox{\texttt{\textbackslash gldefn\{value.denot.conventions\}}} & \gpfitbox{\texttt{\footnotesize GLang/Value\allowbreak Denot\allowbreak Sem.lean:811}}\\
Array shape conditions & \gpfitbox{\texttt{\textbackslash gldefn\{value.denot.guards\}}} & \gpfitbox{\texttt{\footnotesize GLang/Value\allowbreak Primitive\allowbreak Shapes.lean:16}}\\
Value-type representation & \gpfitbox{\texttt{\textbackslash gldefn\{value.denot.types\}}} & \gpfitbox{\texttt{\footnotesize GLang/Value\allowbreak Denot\allowbreak Sem.lean:511}}\\
Value representation & \gpfitbox{\texttt{\textbackslash gldefn\{value.denot.representation\}}} & \gpfitbox{\texttt{\footnotesize GLang/Value\allowbreak Denot\allowbreak Sem.lean:248}}\\
Variables and scalar operations & \gpfitbox{\texttt{\textbackslash gldefn\{value.denot.scalars\}}} & \gpfitbox{\texttt{\footnotesize GLang/Value\allowbreak Denot\allowbreak Sem.lean:811}}\\
Coordinates, tensor products, and direct sums & \gpfitbox{\texttt{\textbackslash gldefn\{value.denot.coordinates\}}} & \gpfitbox{\texttt{\footnotesize GLang/Value\allowbreak Denot\allowbreak Sem.lean:811}}\\
Transpose & \gpfitbox{\texttt{\textbackslash gldefn\{value.denot.transpose\}}} & \gpfitbox{\texttt{\footnotesize GLang/Value\allowbreak Denot\allowbreak Sem.lean:811}}\\
Finite list sums & \gpfitbox{\texttt{\textbackslash gldefn\{value.denot.sum\}}} & \gpfitbox{\texttt{\footnotesize GLang/Value\allowbreak Denot\allowbreak Sem.lean:811}}\\
Tensor contractions & \gpfitbox{\texttt{\textbackslash gldefn\{value.denot.contractions\}}} & \gpfitbox{\texttt{\footnotesize GLang/Value\allowbreak Denot\allowbreak Sem.lean:811}}\\
Kronecker and block-diagonal assembly & \gpfitbox{\texttt{\textbackslash gldefn\{value.denot.assembly\}}} & \gpfitbox{\texttt{\footnotesize GLang/Value\allowbreak Denot\allowbreak Sem.lean:811}}\\
Matrix conversion and output lengths & \gpfitbox{\texttt{\textbackslash gldefn\{value.denot.matrix-conventions\}}} & \gpfitbox{\texttt{\footnotesize GLang/Value\allowbreak Denot\allowbreak Sem.lean:811}}\\
Inverse operations & \gpfitbox{\texttt{\textbackslash gldefn\{value.denot.inverse\}}} & \gpfitbox{\texttt{\footnotesize GLang/Value\allowbreak Denot\allowbreak Sem.lean:811}}\\
Generalized SVD and weighted pseudoinverse & \gpfitbox{\texttt{\textbackslash gldefn\{value.denot.gsvd\}}} & \gpfitbox{\texttt{\footnotesize GLang/Value\allowbreak Denot\allowbreak Sem.lean:811}}\\
RREF representatives & \gpfitbox{\texttt{\textbackslash gldefn\{value.denot.rref\}}} & \gpfitbox{\texttt{\footnotesize GLang/Value\allowbreak Denot\allowbreak Sem.lean:811}}\\
Lambda abstraction & \gpfitbox{\texttt{\textbackslash gldefn\{value.denot.lambda\}}} & \gpfitbox{\texttt{\footnotesize GLang/Value\allowbreak Denot\allowbreak Sem.lean:811}}\\
Function application & \gpfitbox{\texttt{\textbackslash gldefn\{value.denot.application\}}} & \gpfitbox{\texttt{\footnotesize GLang/Partial\allowbreak Value\allowbreak Eval.lean:77}}\\
Reshape & \gpfitbox{\texttt{\textbackslash gldefn\{value.denot.reshape\}}} & \gpfitbox{\texttt{\footnotesize GLang/Partial\allowbreak Value\allowbreak Reshape\allowbreak Runtime.lean:50}}\\
Raw evaluation and checked soundness & \gpfitbox{\texttt{\textbackslash gldefn\{value.denot.soundness\}}} & \gpfitbox{\texttt{\footnotesize GLang/Metatheory.lean:266}}\\
Value domains & \gpfitbox{\texttt{\textbackslash gldefn\{preview.valueDomains\}}} & \gpfitbox{\texttt{\footnotesize GLang/Paper\allowbreak Value\allowbreak Domains.lean:16}}\\
Value environments & \gpfitbox{\texttt{\textbackslash gldefn\{preview.valueEnvWF\}}} & \gpfitbox{\texttt{\footnotesize GLang/Paper\allowbreak Value\allowbreak Environment.lean:9}}\\
Operational entries & \gpfitbox{\texttt{\textbackslash gldefn\{preview.operationalEntries\}}} & \gpfitbox{\texttt{\footnotesize GLang/Value\allowbreak Progress.lean:358}}\\
Operational context formation & \gpfitbox{\texttt{\textbackslash gldefn\{preview.operationalContextWF\}}} & \gpfitbox{\texttt{\footnotesize GLang/Value\allowbreak Progress.lean:380}}\\
Value dimensions and shapes & \gpfitbox{\texttt{\textbackslash gldefn\{value.shapes.compact\}}} & \gpfitbox{\texttt{\footnotesize GLang/Value\allowbreak Denot.lean:116}}\\
Value formation and kinding & \gpfitbox{\texttt{\textbackslash gldefn\{preview.valueKinding\}}} & \gpfitbox{\texttt{\footnotesize GLang/Value\allowbreak Valid\allowbreak C.lean:40}}\\
Creation histories & \gpfitbox{\texttt{\textbackslash gldefn\{preview.op.history\}}} & \gpfitbox{\texttt{\footnotesize GLang/Operational.lean:87}}\\
Checked operational semantics & \gpfitbox{\texttt{\textbackslash gldefn\{preview.op.checked\}}} & \gpfitbox{\texttt{\footnotesize GLang/Adequacy/Raw\allowbreak Certified\allowbreak Operational.lean:276}}\\
Value-annotation regularity & \gpfitbox{\texttt{\textbackslash gldefn\{preview.op.regularity\}}} & \gpfitbox{\texttt{\footnotesize GLang/Value\allowbreak Valid\allowbreak C.lean:40}}\\
Agreement at an exact basis & \gpfitbox{\texttt{\textbackslash gldefn\{preview.adequacy.agreement\}}} & \gpfitbox{\texttt{\footnotesize GLang/Adequacy/Partial\allowbreak Source\allowbreak Agreement.lean:9}}\\
Basis-indexed type agreement & \gpfitbox{\texttt{\textbackslash gldefn\{preview.adequacy.dimensions\}}} & \gpfitbox{\texttt{\footnotesize GLang/Adequacy/Partial\allowbreak Source\allowbreak Agreement.lean:41}}\\
Consistent environments & \gpfitbox{\texttt{\textbackslash gldefn\{preview.adequacy.consistency\}}} & \gpfitbox{\texttt{\footnotesize GLang/Adequacy/Partial\allowbreak Source\allowbreak Public.lean:8}}\\
Inner-product notation & \gpfitbox{\texttt{\textbackslash gldefn\{iof.conventions\}}} & \gpfitbox{\texttt{\footnotesize GLang/Inner\allowbreak Product.lean:193}}\\
Zero space & \gpfitbox{\texttt{\textbackslash gleqn\{iof.empty\}}} & \gpfitbox{\texttt{\footnotesize GLang/Inner\allowbreak Product.lean:193}}\\
Real scalars & \gpfitbox{\texttt{\textbackslash gleqn\{iof.real\}}} & \gpfitbox{\texttt{\footnotesize GLang/Inner\allowbreak Product.lean:193}}\\
Dual space & \gpfitbox{\texttt{\textbackslash gleqn\{iof.dual\}}} & \gpfitbox{\texttt{\footnotesize GLang/Inner\allowbreak Product.lean:193}}\\
Direct sum & \gpfitbox{\texttt{\textbackslash gleqn\{iof.dsum\}}} & \gpfitbox{\texttt{\footnotesize GLang/Inner\allowbreak Product.lean:193}}\\
Tensor product & \gpfitbox{\texttt{\textbackslash gleqn\{iof.tprod\}}} & \gpfitbox{\texttt{\footnotesize GLang/Inner\allowbreak Product.lean:193}}\\
Kernel & \gpfitbox{\texttt{\textbackslash gleqn\{iof.ker\}}} & \gpfitbox{\texttt{\footnotesize GLang/Inner\allowbreak Product.lean:193}}\\
Cokernel & \gpfitbox{\texttt{\textbackslash gleqn\{iof.coker\}}} & \gpfitbox{\texttt{\footnotesize GLang/Inner\allowbreak Product.lean:193}}\\
Image & \gpfitbox{\texttt{\textbackslash gleqn\{iof.img\}}} & \gpfitbox{\texttt{\footnotesize GLang/Inner\allowbreak Product.lean:193}}\\
Coimage & \gpfitbox{\texttt{\textbackslash gleqn\{iof.coimg\}}} & \gpfitbox{\texttt{\footnotesize GLang/Inner\allowbreak Product.lean:193}}\\
Source denotational soundness & \gpfitbox{\texttt{\textbackslash glthm\{preview.sourceSoundness\}}} & \gpfitbox{\texttt{\footnotesize GLang/Paper\allowbreak Source\allowbreak Results.lean:21}}\\
Inner-product choice is sound & \gpfitbox{\texttt{\textbackslash glthm\{preview.innerProduct\}}} & \gpfitbox{\texttt{\footnotesize GLang/Paper\allowbreak Source\allowbreak Results.lean:35}}\\
Soundness for values & \gpfitbox{\texttt{\textbackslash glthm\{preview.valueSoundness\}}} & \gpfitbox{\texttt{\footnotesize GLang/Paper\allowbreak Value\allowbreak Results.lean:10}}\\
Basis denotational soundness & \gpfitbox{\texttt{\textbackslash glthm\{preview.basisSoundness\}}} & \gpfitbox{\texttt{\footnotesize GLang/Paper\allowbreak Basis\allowbreak Results.lean:17}}\\
Standard basis soundness & \gpfitbox{\texttt{\textbackslash glthm\{preview.standardBasisSoundness\}}} & \gpfitbox{\texttt{\footnotesize GLang/Paper\allowbreak Basis\allowbreak Results.lean:25}}\\
Term preservation & \gpfitbox{\texttt{\textbackslash glthm\{preview.op.preserveTerm\}}} & \gpfitbox{\texttt{\footnotesize GLang/Paper\allowbreak Operational\allowbreak Previews.lean:9}}\\
Type preservation & \gpfitbox{\texttt{\textbackslash glthm\{preview.op.preserveType\}}} & \gpfitbox{\texttt{\footnotesize GLang/Paper\allowbreak Operational\allowbreak Previews.lean:30}}\\
Space preservation & \gpfitbox{\texttt{\textbackslash glthm\{preview.op.preserveSpace\}}} & \gpfitbox{\texttt{\footnotesize GLang/Metatheory.lean:474}}\\
Change-of-basis preservation & \gpfitbox{\texttt{\textbackslash glthm\{preview.op.preserveCob\}}} & \gpfitbox{\texttt{\footnotesize GLang/Metatheory.lean:519}}\\
Checked progress & \gpfitbox{\texttt{\textbackslash glthm\{preview.op.progress\}}} & \gpfitbox{\texttt{\footnotesize GLang/Metatheory.lean:553}}\\
Standard-basis progress & \gpfitbox{\texttt{\textbackslash glthm\{preview.op.progressStd\}}} & \gpfitbox{\texttt{\footnotesize GLang/Metatheory.lean:553}}\\
Checked space progress & \gpfitbox{\texttt{\textbackslash glthm\{preview.op.progressSpace\}}} & \gpfitbox{\texttt{\footnotesize GLang/Metatheory.lean:626}}\\
Checked change-of-basis progress & \gpfitbox{\texttt{\textbackslash glthm\{preview.op.progressCob\}}} & \gpfitbox{\texttt{\footnotesize GLang/Metatheory.lean:665}}\\
Checked term adequacy & \gpfitbox{\texttt{\textbackslash glthm\{preview.adequacy.term\}}} & \gpfitbox{\texttt{\footnotesize GLang/Metatheory.lean:682}}\\
Checked space adequacy & \gpfitbox{\texttt{\textbackslash glthm\{preview.adequacy.space\}}} & \gpfitbox{\texttt{\footnotesize GLang/Metatheory.lean:698}}\\
Exact-basis type adequacy & \gpfitbox{\texttt{\textbackslash glthm\{preview.adequacy.type\}}} & \gpfitbox{\texttt{\footnotesize GLang/Metatheory.lean:849}}\\
Change-of-basis adequacy & \gpfitbox{\texttt{\textbackslash glthm\{preview.adequacy.cob\}}} & \gpfitbox{\texttt{\footnotesize GLang/Metatheory.lean:826}}\\
Progress and adequacy & \gpfitbox{\texttt{\textbackslash glthm\{preview.adequacy.endToEnd\}}} & \gpfitbox{\texttt{\footnotesize GLang/Metatheory.lean:719}}\\
Checked value evaluation at the retained type & \gpfitbox{\texttt{\textbackslash glthm\{preview.adequacy.checkedValue\}}} & \gpfitbox{\texttt{\footnotesize GLang/Adequacy/Direct\allowbreak Public.lean:113}}\\
}
\gldefcmdrows{thms}{
Source-space soundness & \gpfitbox{\texttt{\textbackslash glthm\{metatheory.sourceSpaceSoundness\}}} & \gpfitbox{\texttt{\footnotesize GLang/Metatheory.lean:111}}\\
Source-type soundness & \gpfitbox{\texttt{\textbackslash glthm\{metatheory.sourceTypeSoundness\}}} & \gpfitbox{\texttt{\footnotesize GLang/Metatheory.lean:137}}\\
Source-term soundness & \gpfitbox{\texttt{\textbackslash glthm\{metatheory.sourceTermSoundness\}}} & \gpfitbox{\texttt{\footnotesize GLang/Metatheory.lean:88}}\\
Value-shape soundness & \gpfitbox{\texttt{\textbackslash glthm\{metatheory.valueShapeSoundness\}}} & \gpfitbox{\texttt{\footnotesize GLang/Metatheory.lean:196}}\\
Value-type soundness & \gpfitbox{\texttt{\textbackslash glthm\{metatheory.valueTypeSoundness\}}} & \gpfitbox{\texttt{\footnotesize GLang/Metatheory.lean:204}}\\
Value-term soundness & \gpfitbox{\texttt{\textbackslash glthm\{metatheory.valueTermSoundness\}}} & \gpfitbox{\texttt{\footnotesize GLang/Metatheory.lean:284}}\\
Ordinary space preservation & \gpfitbox{\texttt{\textbackslash glthm\{metatheory.spacePreservation\}}} & \gpfitbox{\texttt{\footnotesize GLang/Metatheory.lean:474}}\\
Ordinary term preservation & \gpfitbox{\texttt{\textbackslash glthm\{metatheory.termPreservation\}}} & \gpfitbox{\texttt{\footnotesize GLang/Metatheory.lean:428}}\\
Ordinary change-of-basis preservation & \gpfitbox{\texttt{\textbackslash glthm\{metatheory.cobPreservation\}}} & \gpfitbox{\texttt{\footnotesize GLang/Metatheory.lean:519}}\\
Checked space progress & \gpfitbox{\texttt{\textbackslash glthm\{metatheory.checkedSpaceProgress\}}} & \gpfitbox{\texttt{\footnotesize GLang/Metatheory.lean:626}}\\
Checked term progress & \gpfitbox{\texttt{\textbackslash glthm\{metatheory.termProgress\}}} & \gpfitbox{\texttt{\footnotesize GLang/Metatheory.lean:568}}\\
General checked change-of-basis progress & \gpfitbox{\texttt{\textbackslash glthm\{metatheory.checkedCobProgress\}}} & \gpfitbox{\texttt{\footnotesize GLang/Metatheory.lean:665}}\\
Space adequacy & \gpfitbox{\texttt{\textbackslash glthm\{metatheory.spaceAdequacy\}}} & \gpfitbox{\texttt{\footnotesize GLang/Metatheory.lean:698}}\\
Term adequacy & \gpfitbox{\texttt{\textbackslash glthm\{metatheory.termAdequacy\}}} & \gpfitbox{\texttt{\footnotesize GLang/Metatheory.lean:682}}\\
Change-of-basis adequacy & \gpfitbox{\texttt{\textbackslash glthm\{metatheory.cobAdequacy\}}} & \gpfitbox{\texttt{\footnotesize GLang/Metatheory.lean:826}}\\
Source-basis soundness & \gpfitbox{\texttt{\textbackslash glthm\{metatheory.sourceBasisSoundness\}}} & \gpfitbox{\texttt{\footnotesize GLang/Metatheory.lean:150}}\\
Value-type regularity & \gpfitbox{\texttt{\textbackslash glthm\{runtimeValue.valHasType.regular\}}} & \gpfitbox{\texttt{\footnotesize GLang/Runtime\allowbreak Value\allowbreak Regular.lean:23}}\\
Exact inferred value type & \gpfitbox{\texttt{\textbackslash glthm\{runtimeValue.valHasType.inferTypeEqOfTyped\}}} & \gpfitbox{\texttt{\footnotesize GLang/Runtime\allowbreak Value\allowbreak Infer.lean:37}}\\
Exact inferred value space & \gpfitbox{\texttt{\textbackslash glthm\{valHasType.inferSpaceEqOfSpace\}}} & \gpfitbox{\texttt{\footnotesize GLang/Value\allowbreak Infer\allowbreak Space\allowbreak Sound.lean:356}}\\
Value shape and dimension totality & \gpfitbox{\texttt{\textbackslash glthm\{lemSoundValsMetricsTotal\}}} & \gpfitbox{\texttt{\footnotesize GLang/Value\allowbreak Sound\allowbreak Vals.lean:55}}\\
Standard basis has the requested type & \gpfitbox{\texttt{\textbackslash glthm\{basis.tofStd\}}} & \gpfitbox{\texttt{\footnotesize GLang/Basis.lean:579}}\\
Standard basis is valid & \gpfitbox{\texttt{\textbackslash glthm\{basis.stdValid\}}} & \gpfitbox{\texttt{\footnotesize GLang/Basis.lean:688}}\\
Standard basis passes the standardness check & \gpfitbox{\texttt{\textbackslash glthm\{basis.isStdStd\}}} & \gpfitbox{\texttt{\footnotesize GLang/Progress\allowbreak Std\allowbreak Basis.lean:56}}\\
Standard basis/frame is recursively complete and orthonormal & \gpfitbox{\texttt{\textbackslash glthm\{metatheory.standardBasisSoundness\}}} & \gpfitbox{\texttt{\footnotesize GLang/Metatheory.lean:161}}\\
Progress at the standard basis & \gpfitbox{\texttt{\textbackslash glthm\{v2.progressStd\}}} & \gpfitbox{\texttt{\footnotesize GLang/Progress\allowbreak Master.lean:65}}\\
Canonical inner-product term is well typed & \gpfitbox{\texttt{\textbackslash glthm\{space.iofHasType\}}} & \gpfitbox{\texttt{\footnotesize GLang/Inner\allowbreak Product.lean:1523}}\\
Canonical inner-product term denotes an inner product & \gpfitbox{\texttt{\textbackslash glthm\{space.iofDenotIsInnerProduct\}}} & \gpfitbox{\texttt{\footnotesize GLang/Inner\allowbreak Product.lean:6809}}\\
Historical source-soundness projection & \gpfitbox{\texttt{\textbackslash glthm\{metatheory.sourceSoundnessWeak\}}} & \gpfitbox{\texttt{\footnotesize GLang/Metatheory.lean:78}}\\
Space regularity & \gpfitbox{\texttt{\textbackslash glthm\{metatheory.sourceSpaceRegularity\}}} & \gpfitbox{\texttt{\footnotesize GLang/Metatheory.lean:98}}\\
Type regularity & \gpfitbox{\texttt{\textbackslash glthm\{metatheory.sourceTypeRegularity\}}} & \gpfitbox{\texttt{\footnotesize GLang/Metatheory.lean:124}}\\
Oracle-polymorphic canonical value soundness & \gpfitbox{\texttt{\textbackslash glthm\{metatheory.valueSoundnessOfOracle\}}} & \gpfitbox{\texttt{\footnotesize GLang/Metatheory.lean:215}}\\
Historical weak value-soundness projection & \gpfitbox{\texttt{\textbackslash glthm\{metatheory.valueSoundnessWeak\}}} & \gpfitbox{\texttt{\footnotesize GLang/Metatheory.lean:254}}\\
Internal checked exact-index certificate & \gpfitbox{\texttt{\textbackslash glthm\{metatheory.checkedPreservation\}}} & \gpfitbox{\texttt{\footnotesize GLang/Metatheory.lean:312}}\\
Checked base-shape preservation & \gpfitbox{\texttt{\textbackslash glthm\{metatheory.checkedBaseShapePreservation\}}} & \gpfitbox{\texttt{\footnotesize GLang/Metatheory.lean:444}}\\
Checked space preservation & \gpfitbox{\texttt{\textbackslash glthm\{metatheory.checkedSpacePreservation\}}} & \gpfitbox{\texttt{\footnotesize GLang/Metatheory.lean:489}}\\
Type preservation & \gpfitbox{\texttt{\textbackslash glthm\{metatheory.typePreservation\}}} & \gpfitbox{\texttt{\footnotesize GLang/Metatheory.lean:504}}\\
Checked change-of-basis preservation & \gpfitbox{\texttt{\textbackslash glthm\{metatheory.checkedCobPreservation\}}} & \gpfitbox{\texttt{\footnotesize GLang/Metatheory.lean:535}}\\
Endpoint-aware change-of-basis preservation & \gpfitbox{\texttt{\textbackslash glthm\{metatheory.cobEndpointPreservation\}}} & \gpfitbox{\texttt{\footnotesize GLang/Cob\allowbreak Endpoint\allowbreak Preservation.lean:69}}\\
Oracle-polymorphic common COB dimension & \gpfitbox{\texttt{\textbackslash glthm\{metatheory.cobCommonDimensionAdequacyOfOracle\}}} & \gpfitbox{\texttt{\footnotesize GLang/Cob\allowbreak Endpoint\allowbreak Preservation.lean:98}}\\
Common COB dimension & \gpfitbox{\texttt{\textbackslash glthm\{metatheory.cobCommonDimensionAdequacy\}}} & \gpfitbox{\texttt{\footnotesize GLang/Cob\allowbreak Endpoint\allowbreak Preservation.lean:140}}\\
Oracle-polymorphic COB automorphism & \gpfitbox{\texttt{\textbackslash glthm\{metatheory.cobCoordinateMatrixBijectiveOfOracle\}}} & \gpfitbox{\texttt{\footnotesize GLang/Cob\allowbreak Endpoint\allowbreak Preservation.lean:245}}\\
COB coordinate-matrix automorphism & \gpfitbox{\texttt{\textbackslash glthm\{metatheory.cobCoordinateMatrixBijective\}}} & \gpfitbox{\texttt{\footnotesize GLang/Cob\allowbreak Endpoint\allowbreak Preservation.lean:280}}\\
Checked type progress & \gpfitbox{\texttt{\textbackslash glthm\{metatheory.checkedTypeProgress\}}} & \gpfitbox{\texttt{\footnotesize GLang/Metatheory.lean:635}}\\
Checked type progress at a basis & \gpfitbox{\texttt{\textbackslash glthm\{metatheory.checkedTypeProgressAtBasis\}}} & \gpfitbox{\texttt{\footnotesize GLang/Metatheory.lean:645}}\\
Checked change-of-basis progress & \gpfitbox{\texttt{\textbackslash glthm\{metatheory.checkedBasisProgress\}}} & \gpfitbox{\texttt{\footnotesize GLang/Metatheory.lean:655}}\\
Basis adequacy & \gpfitbox{\texttt{\textbackslash glthm\{metatheory.basisAdequacy\}}} & \gpfitbox{\texttt{\footnotesize GLang/Metatheory.lean:709}}\\
Type adequacy at a basis & \gpfitbox{\texttt{\textbackslash glthm\{metatheory.typeAdequacyAtBasis\}}} & \gpfitbox{\texttt{\footnotesize GLang/Metatheory.lean:849}}\\
Checked space progress and adequacy & \gpfitbox{\texttt{\textbackslash glthm\{metatheory.spaceProgressAdequacy\}}} & \gpfitbox{\texttt{\footnotesize GLang/Metatheory.lean:868}}\\
Checked type progress and adequacy at a basis & \gpfitbox{\texttt{\textbackslash glthm\{metatheory.typeProgressAdequacyAtBasis\}}} & \gpfitbox{\texttt{\footnotesize GLang/Metatheory.lean:882}}\\
Checked change-of-basis progress and adequacy & \gpfitbox{\texttt{\textbackslash glthm\{metatheory.basisProgressAdequacy\}}} & \gpfitbox{\texttt{\footnotesize GLang/Metatheory.lean:899}}\\
Multiple-representation checked-progress regression & \gpfitbox{\texttt{\textbackslash glthm\{rawMultipleRepresentationRegression.checkedProgressAfterExtendRep\}}} & \gpfitbox{\texttt{\footnotesize GLang/Adequacy/Raw\allowbreak Multiple\allowbreak Representation\allowbreak Regression.lean:16}}\\
Strengthened term adequacy (oracle-free) & \gpfitbox{\texttt{\textbackslash glthm\{adequacy.direct.term\}}} & \gpfitbox{\texttt{\footnotesize GLang/Adequacy/Direct\allowbreak Public.lean:97}}\\
Strengthened COB adequacy (oracle-free) & \gpfitbox{\texttt{\textbackslash glthm\{adequacy.direct.cob\}}} & \gpfitbox{\texttt{\footnotesize GLang/Adequacy/Direct\allowbreak Public.lean:132}}\\
Strengthened dimension adequacy (oracle-free) & \gpfitbox{\texttt{\textbackslash glthm\{adequacy.direct.dimsAt\}}} & \gpfitbox{\texttt{\footnotesize GLang/Adequacy/Direct\allowbreak Public.lean:142}}\\
}

\usepackage{longtable}

\begin{document}
\title{\GLang: The Coordinate-Independent Linear Algebra Programming Language}

\author{Theadora Fields Collin}
\email{teoc@mit.edu}
\affiliation{\institution{Massachusetts Institute of Technology}
  \city{Cambridge}\state{Massachusetts}\country{USA}}

\author{Gilbert Bernstein}
\email{gilbo@cs.washington.edu}
\affiliation{\institution{University of Washington}
  \city{Seattle}\state{Washington}\country{USA}}

\author{Oded Stein}
\email{ostein@usc.edu}
\affiliation{\institution{University of Southern California}
  \city{Los Angeles}\state{California}\country{USA}}
\affiliation{\institution{Technion---Israel Institute of Technology}
  \city{Haifa}\country{Israel}}

\author{Saman Amarasinghe}
\email{saman@csail.mit.edu}
\affiliation{\institution{Massachusetts Institute of Technology}
  \city{Cambridge}\state{Massachusetts}\country{USA}}

\renewcommand{\shortauthors}{Collin et al.}

\begin{abstract}
Scientists, engineers, and mathematicians often formulate algorithms using \textit{coordinate-independent} linear algebra because coordinate-independent tools are elegant, generative, conceptually clear, naturally composable, and harder to misuse.
However, for numerical processing, users must eventually translate  coordinate-independent ideas into coordinate-dependent array code.
Moreover, most practitioners prematurely translate to coordinate-dependent code because tools such as NumPy do not provide coordinate-independent representations.
In the few languages with coordinate-independent constructs, users must still translate prematurely because preexisting systems do not model enough linear algebra to capture computational situations where coordinate-independent ideas truly matter.
In this work, we propose a small, novel language, \GLang{}, to bridge the gap between coordinate-independent linear algebra and coordinate-dependent implementations.
\GLang{} uses dependent types, linear types, and partially (pseudo-)invertible computation to model coordinate-independent linear algebra in the situations that motivate it, such as exterior algebra and differential geometry.
\GLang{} is particularly targeted at the linear algebra that occurs at a point on a manifold, which is a prerequisite for many specifications in scientific computing and computer graphics.
Additionally, to provide users with the full freedom and power of coordinate-independent thinking, we augment \GLang{} with a language for coordinates.
We provide a denotational semantics that assigns meaning to \GLang{} without coordinates and we provide an operational semantics that translates \GLang{} to array code under a choice of coordinates.
In doing so, we model how practitioners translate linear algebra for computers.
 Our key meta-theoretic result shows that operational (coordinate-dependent) evaluations agree with the denotational (coordinate-independent) semantics, ensuring correctness under an arbitrary choice of coordinates and suggesting a new approach to the algorithmic optimization of linear algebra at a high level of abstraction.
 Our optimized implementation shows that our approach can generate code for exterior algebra operations with overheads no more than 10\%.

\end{abstract}



\maketitle

\section{Introduction}
Scientists, engineers, and mathematicians often prefer to express their ideas using \textit{coordinate-independent} linear algebra.
Coordinate-independent linear algebra allows one to express complex mathematical objects in a uniform, clear, and safe manner.
For example, given three vectors $p, v_{0}, v_{1}$, the area of the parallelogram with sides $(p, p+v_{0})$ and $(p, p+v_{1})$ has a variety of formulas for different situations with varying degrees of applicability (\textit{e.g.}, you can use the norm of the cross product in three dimensions or the singular value decomposition if you are careful to reason about the signs).
Further, you can easily misuse the formula for the cross product by using it on two vectors in different coordinate systems, as the object defined by the cross product does not transform like a vector or even like a covector.
In two dimensions, for example, the determinant of the coordinate vectors changes with the coordinate system; its signed value is preserved by changes of coordinates with determinant one.
Though some works do try to avoid coordinate system bugs with types and metadata~\cite{Geisler2020, Chen2024}, they don't supply the full power of coordinate-independent thinking.
For example, the wedge product (denoted $v_{0}\wedge v_{1}$) is unambiguously a generalized area and can be defined uniformly across a wide variety of situations (\textit{e.g.}, dimensions, input coordinate systems, and number of input vectors).
Further, the wedge product is hard to abuse as part of a system for geometric reasoning via algebra, called exterior algebra.
As a general rule of thumb, if a concept in linear algebra can be expressed without reference to coordinates, the concept is probably unambiguously defined and harder to abuse.
(``If it type checks, it runs'' for normal programs is equivalent to ``if it doesn't use coordinates it is probably well defined'' for linear algebra.) 

Scientists and engineers have few languages that provide the coordinate-independent algebra needed for these constructions directly.
For the universal goal of finding solutions to differential equations (DEs), practitioners set up equations that at each point relate various vector and tensor quantities via coordinate-independent linear algebra.
For exterior algebra in particular, the wedge product's unambiguous notion of area provides a way to cleanly define a uniform notion of integration and to relate rates of change of generalized geometric quantities, a fundamental aspect of mechanics.
However, if one surveys the many papers using exterior calculus to formulate and solve DEs, one will find table after table showing how (under certain assumptions) to translate the concepts into well-known, easy-to-abuse array primitives like cross or dot products. 
For examples, see~\cite[Tables~1--2]{10.1145/3592402},~\cite[Table IV]{1456316},~\cite[the~dictionary]{Frankel2011-cd},~\cite[Table~6.1]{Arnold2018}, or some of~\cite{BLAIRPEROT20141373,Gunn2019}.
Outside computer algebra systems, few programming languages express coordinate-independent linear algebra; examples include~\cite{Elliott2017,bernardy2025domain}.
These systems do not combine the wedge product, fundamental subspaces, and basis-dependent compilation needed for our differential-geometric constructions, which historically motivated coordinate-independent linear algebra~\cite{Crowe2003-rz}.

Now, we cannot fully discard coordinates for a few reasons.
Some algorithms, including many numerical methods, amount mostly to a choice of coordinates!
Other concepts are often modeled with coordinates (\textit{e.g.}, the orientation of a manifold).
However, we ought to be able to avoid coordinates where they are not necessary, and we ought to be able to ensure we use coordinates correctly and seamlessly when they are necessary.
We believe there are three reasons for this gap.
First, to actually express coordinate-independent linear algebra requires a specific combination of features that, to the best of our knowledge, have not been combined before in previous work on linear algebra (including linear types, a specific form of dependent or quotient types, and pseudo-invertible computation\footnote{Inversion of computation that is not necessarily injective via finding the norm-minimizing input that produces the desired output.}).
In particular, for the linear algebra that occurs at a point on a manifold where one is solving a differential equation, one needs to model how the linear algebra types depend on that point, necessitating dependent types.
Second, since we need to model coordinates too, we need much stronger soundness guarantees than just linearity: we need to show that coordinates are linearly independent and spanning. 
Third, we need semantics to seamlessly link the coordinate-independent framework and the coordinate-dependent framework, which ought to produce code in an array language with \NLang{}-like primitives~\cite{Harris2020} (\textit{i.e.}, what most programmers are currently familiar with).
In this work, we present a small extension to the lambda calculus called \GLang{} to address these issues.

We offer the following contributions:
\begin{enumerate}
    \item We present \GLang{}, a small novel language for coordinate-independent linear algebra and coordinates, with a unique combination of features that allow \GLang{} to express core features of coordinate-independent linear algebra, naturally giving rise to concepts such as exterior algebra and other algebraic tools, sufficient for doing the linear algebra at a point on a manifold.
    \item \GLang{} recovers projections, the musical isomorphism $\sharp$, linear uncurrying, and other operations with pseudo-inversion (\cref{sec:nas}).
    \item We provide a coordinate-independent (denotational) semantics for \GLang{}, which provides a clear meaning for terms and coordinates.
    \item We provide a coordinate-dependent (operational) semantics for \GLang{}, which models how practitioners translate coordinate-independent notation into array code given some choices of coordinates.
    \item We develop the basic metatheory for \GLang{}, proving that operational evaluation agrees with the denotational semantics (adequacy) under appropriate assumptions. Thus, our semantics preserves meaning under change of basis. The mechanized results and their Lean~\cite{deMoura2021} dependencies are listed in Appendix~\ref{app:lean-results}.
    \item We provide an implementation of a superset of \GLang{} and validating micro-benchmarks of exterior-algebra kernels: generated FLOP counts match or improve on the references, with runtimes no more than 10\% slower.
\end{enumerate}

We focus on a small core language, still leaving out many useful connections such as automatic differentiation or parametric polymorphism.
We focus narrowly on the core of coordinate-independent linear algebra and the relationship between the coordinate-independent and coordinate-dependent semantics.
We also restrict ourselves to real finite-dimensional vector spaces.

\section{Examples}
We consider a sequence of three examples to introduce \GLang{}'s capabilities and scope.
In particular, we will show enough \GLang{} programs to accurately specify some basics of differential geometry that are used at a point in sophisticated differential equations.
\subsection{The Area of a Parallelogram}

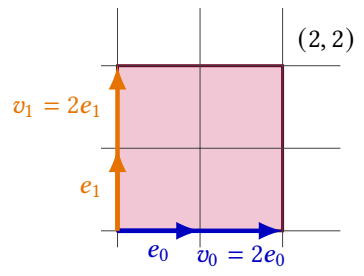
\begin{wrapfigure}{r}{0.18\columnwidth}
  \vspace{-2\baselineskip}
  \centering

  \scalebox{0.70}{%
  \begin{tikzpicture}[
      x=0.78cm,
      y=0.78cm,
      >=latex,
      vec/.style={->,very thick},
      every node/.style={font=\scriptsize}
    ]

    \filldraw[
      fill=purple!22,
      draw=purple!60!black,
      thick
    ]
      (0,0) rectangle (2,2);

        \draw[step=1,black!75,very thin]
      (-0.2,-0.2) grid (2.7,2.7);

    \draw[vec,blue!75!black]
      (0,0) -- (2,0)
      node[near end,below] {$v_0=2e_0$};

    \draw[vec,orange!90!black]
      (0,0) -- (0,1)
      node[midway,left] {$e_1$};

    \draw[vec,blue!75!black]
      (0,0) -- (1,0)
      node[midway,below] {$e_0$};

    \draw[vec,orange!90!black]
      (0,0) -- (0,2)
      node[near end,left] {$v_1=2e_1$};

    \node[above right] at (2,2)
      {$(2,2)$};

  \end{tikzpicture}%
  }

  \vspace{-2\baselineskip}
  \label{fig:ordinary-rectangle}
\end{wrapfigure}
Let's consider the area of a parallelogram defined by two vectors $v_{0}$ and $v_{1}$ at the origin in $\mathbb{R}^{2}$. 
Our perspective on it is outlined via the coordinate system, $\{e_{0}, e_{1}\}.$
Visually, the area of the parallelogram is four.
We can define the area of the parallelogram with the cross product of vectors computationally:
\begin{equation}
\label{eq:second}
    v_{0}\times v_{1} = v_{0}[0] v_{1}[1] - v_{0}[1] v_{1}[0].
\end{equation}
Indeed $v_{0}\times v_{1} = 2 * 2 - 0*0 = 4$.
We consider \eqref{eq:second} to illustrate the virtues of coordinate-independent thinking and to consider programming language requirements.
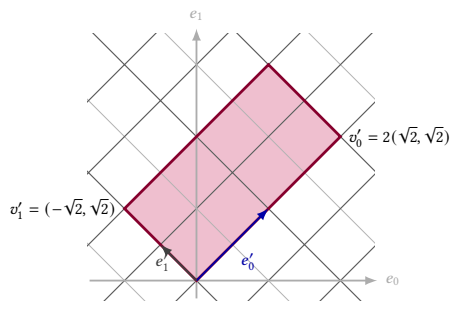
\begin{wrapfigure}[4]{r}{0.22\columnwidth}
  \vspace{-2\baselineskip}
  \centering

  \resizebox{0.22\columnwidth}{!}{%
  \begin{tikzpicture}[
  x=0.78cm, y=0.78cm,
  >=latex,
  every node/.style={font=\scriptsize}
]
  \coordinate (O)  at (0,0);
  \coordinate (E0) at ({sqrt(2)},{sqrt(2)});
  \coordinate (E1) at ({-1/sqrt(2)},{1/sqrt(2)});

  \coordinate (A) at (O);
  \coordinate (B) at ({2*sqrt(2)},{2*sqrt(2)});
  \coordinate (D) at ({-sqrt(2)},{sqrt(2)});
  \coordinate (C) at ({sqrt(2)},{3*sqrt(2)});

  \fill[purple!25] (A) -- (B) -- (C) -- (D) -- cycle;

  \begin{scope}
    \clip (-2.15,-0.4) rectangle (3.5,4.85);
    \begin{scope}[rotate=45]
      \draw[black!30,very thin,step=1]
        (-6,-6) grid (8,8);

      \draw[black!65,very thin,xstep=2,ystep=1]
        (-6,-6) grid (8,8);
    \end{scope}
  \end{scope}

  \draw[->,gray!65,thick]
    (-2.1,0) -- (3.55,0) node[right] {$e_0$};
  \draw[->,gray!65,thick]
    (0,-0.35) -- (0,4.95) node[above] {$e_1$};

  \draw[draw=purple!70!black,very thick]
    (A) -- (B) -- (C) -- (D) -- cycle;

  \draw[->,blue!65!black,thick]
    (O) -- (E0)
    node[midway,below right] {$e'_0$};
  \draw[->,black!75,thick]
    (O) -- (E1)
    node[midway,left] {$e'_1$};

  \node[right] at (B) {$v'_0=2(\sqrt{2},\sqrt{2})$};
  \node[left] at (D) {$v'_1=(-\sqrt{2},\sqrt{2})$};
\end{tikzpicture}
  }

  \vspace{-2\baselineskip}
  \label{fig:same-area-different-measurement}
\end{wrapfigure}
Though simple, \cref{eq:second} has a problem: we secretly assumed that $v_{i} = \sum_{j=0}^{1} v_{i}[j] e_{j}$.
If instead we picked $v^{\prime}_{i} = \sum_{j=0}^{1} v_{i}[j] e_{j}^{\prime}$ for some other vectors $e_{j}^{\prime}$, mistakenly using the array with a different basis, then we would end up with a different result.
We could imagine this as a distorting shift of our perspective.
Suppose that $e_{0}^{\prime} =R_{45^{\circ}}2 e_{0}$ and $e_{1}^{\prime}=R_{45^{\circ}}e_{1}$ where $R_{45^{\circ}}$ stands for anti-clockwise rotations by 45 degrees. 
Reading the same arrays in this basis gives a different parallelogram, so we get a different area: $v_{0}^{\prime}\times v_{1}^{\prime} = 8$ because $v_{0}^{\prime} = 2(\sqrt{2}, \sqrt{2})$ and $v_{1}^{\prime} = (-\sqrt{2},\sqrt{2})$.
Note that based on the visuals of $e_{i}^{\prime}$ and $v_{i}^{\prime}$, this makes sense: we see eight units of area in the convention $\times$  assumes, but four in the convention for $\times$ implied by using $e_{0}^{\prime}, e_{1}^{\prime}$.

In our language, we can avoid this mistake by requiring definitions independent of the choice of $e_{j}^{\prime}$ vs. $e_{j}$.
We can improve on \eqref{eq:second} with a simple translation. 
We can make a function that operates on two vectors to produce a scalar:
\begin{equation}
\label{eq:first}
    \lllambdaArg{v_{0}, v_{1}}{\lkprojpp{0}(v_{0})\lvscale{} \lkprojpp{1}(v_{1}) - \lkprojpp{1}(v_{0})\lvscale{} \lkprojpp{0}(v_{1})},
\end{equation}
 utilizing $\lkprojpp{j}$ to extract the $j$th coordinate of a vector and $\limp$ to indicate \textit{multi-linearity}; that is, linearity in each argument.
To use \eqref{eq:first}, our language requires that users separately specify coordinate systems for $v_{0}, v_{1}$. 
By hiding access to the coordinates behind $\lkprojpp{j}$, we avoid the pitfalls of \eqref{eq:second}!
Concretely, when we evaluate \eqref{eq:first}, we must specify the coordinates
in which we write $v_{0}, v_{1}$.
To correct our mistake and represent the
original parallelogram in the basis $e^{\prime}_{0}, e^{\prime}_{1}$, we must
communicate that $v_{0}$ is $(\tfrac{1}{\sqrt{2}}, -\sqrt{2})$ in that basis,
which relates to the original basis by
\begin{equation*}
    \begin{bmatrix} e^{\prime}_{0}\\ e^{\prime}_{1} \end{bmatrix}
    = R_{45^{\circ}}\begin{bmatrix} 2 & 0 \\ 0 & 1 \end{bmatrix}
      \begin{bmatrix} e_{0}\\ e_{1} \end{bmatrix}
    \quad\Rightarrow\quad
    \lkprojpp{0}(v_{0})
    = \Big(R_{45^{\circ}}\begin{bmatrix} 2 & 0 \\ 0 & 1 \end{bmatrix}
      \begin{bmatrix} 1/\sqrt{2} \\ -\sqrt{2} \end{bmatrix}\Big)[0] = 2.
\end{equation*}
With this bookkeeping, we select the parallelogram that we want; above, we keep the one with area four.
\GLang{} has an entire language for bases that allows coordinate transformations, such as the matrix above, to be automatically derived and appropriately inserted into code.
Thus, in \GLang{}, users \textbf{first}  write their notion of area with linear algebra independently of coordinates via \eqref{eq:first}. Then, as a \textbf{second} independent step, they choose their basis vectors, which allows them to turn \eqref{eq:first} into a number, area.
However, \eqref{eq:first} still has implicit choices that can lead to mistakes, such as the unit of area associated with the output.
Eq. \eqref{eq:first} only requires users to make the grid explicit via the choice of basis for the inputs.
Looking at our visuals, we ought to define area with a single unified choice of grid cell and unit.

To be able to declare the grid and unit area, we ultimately use a slightly more complex coordinate-independent representation.
Before we do this with our language, let's do it mathematically with a new coordinate-independent object that manages the grid and unit area.
To do this, we do a calculation involving the tensor product $v_{0}^{\prime}\otimes v_{1}^{\prime}$.
A basic property of the tensor product is that, for arbitrary vectors $w_{0}, w_{1}, w_{2}, w_{3}$ and scalars $a,b,c,d$,
\begin{equation}
\label{eq:tensor_product}
    (a w_{0} + b w_{1})\otimes (c w_{2} + d w_{3}) = ac w_{0}\otimes w_{2} + ad w_{0}\otimes w_{3} + bc w_{1}\otimes w_{2} + bd w_{1}\otimes w_{3}.
\end{equation}
The wedge (or exterior) product is defined using the tensor product (note the similarity to \eqref{eq:second}):
\begin{equation}
\label{eq:wedge_product}
    w_{0}\wedge w_{1} :=  (w_{0}\otimes w_{1} - w_{1}\otimes w_{0}).
\end{equation}
We can compute an area number for the wedge product in terms of another wedge product, using \eqref{eq:tensor_product} without actually expanding out the tensor product, but just regrouping back to a wedge product:
\begin{equation*}
    v_{0}^{\prime}\wedge v_{1}^{\prime} = (2 e_{0}^{\prime}\otimes 2e_{1}^{\prime} - 2e_{1}^{\prime}\otimes 2e_{0}^{\prime}) = 4(e_{0}^{\prime}\otimes e_{1}^{\prime}) - 4(e_{1}^{\prime}\otimes e_{0}^{\prime}) = 4 (e_{0}^{\prime}\otimes e_{1}^{\prime} - e_{1}^{\prime}\otimes e_{0}^{\prime}) = 4 (e_{0}^{\prime}\wedge e_{1}^{\prime}).
\end{equation*}
This looks similar to the ``mistake'' we allowed in \eqref{eq:second}: calculating the area to be four when it should be eight! But of course that is not what is happening: we simply have four units of $e_{0}^{\prime}\wedge e_{1}^{\prime}$.
Better yet, $e_{0}^{\prime}\wedge e_{1}^{\prime}$ is a unit of area; this is the object that is simply the area of the parallelogram (and analogous to the \emph{grid unit area} discussed earlier).
Our calculation has clarified the gap in  \eqref{eq:first}: we should be modeling the space of wedges of vectors (the space of bivectors) and using a basis to measure bivectors in a given unit system.

Now we want to translate this idea to our programming language and implement a truly coordinate-independent approach to this problem.
We want to compute the area in two steps, modeling and measurement, similar to \eqref{eq:first}.
\textbf{First}, we should model parallelogram areas via the space of bivectors without reference to a coordinate system or area units.
\textbf{Second}, we produce an area by choosing a basis for the space of bivectors and using it to convert a bivector into a real number.
The basis in this case will be one-dimensional, essentially a single choice for our unit area grid cell.

We now model the space of bivectors in \GLang{}.
To properly model the space of $w_{0}\wedge w_{1}$ objects
\GLang{} provides a \emph{novel} and \emph{powerful} linear algebraic modeling tool: an explicit linear quotient construction.
Given a linear map $A$, we can quotient the domain of $A$ to produce the image of a linear map.
We cannot use \eqref{eq:first} or \eqref{eq:wedge_product} as these are bilinear, but we can reformulate them.
We examine the type of \eqref{eq:first}: a linear map from the plane to the reals is written $(\lR{}^{2})^{\lstar}$ and a bilinear map can be written $(\lR{}^{2})^{\lstar} \loprod (\lR{}^{2})^{\lstar}$ where $\loprod$ builds the space of linear pairs.
Now we define the space of bivectors by defining the following multi-linear map, analogous to \eqref{eq:wedge_product}:
 \begin{equation}
 \label{eq:antisymmap}
    \operatorname{antiSymMap} =  (\luncurry_{0, 1, 2})(\lllambdaArg{x_{0}, x_{1}}{\frac{1}{2}(x_{0}\loprod x_{1} - x_{1}\loprod x_{0})})
\end{equation}
where $\luncurry_{0, 1, 2}$ is the uncurry\footnote{given $f(a,b)$, $g=\luncurry_{0, 1, 2}(f)$ is defined so that $g(a\otimes b) = f(a,b)$} operator for this bilinear map.
(Also note that we do not use $\lkprojpp{j}$ so this object is defined on any vector space, though we will still use $\mathbb{R}^{2}$ for the moment.)
Without uncurrying, we would type this object with $(\lR{}^{2})^{\lstar}\loprod (\lR{}^{2})^{\lstar}\loprod (\lR{}^{2}\loprod\lR{}^{2})$, but the uncurrying makes the object $(\lR{}^{2}\loprod \lR{}^{2})^{\lstar}\loprod (\lR{}^{2}\loprod\lR{}^{2})$, which is in the form of a linear map from a space to itself.
Thus, we can define our space of bivectors via $\operatorname{Img} = \Image(\operatorname{antiSymMap})$.
(Since this may feel a bit abstract, we point out for intuition that under appropriate bases $\operatorname{antiSymMap}$ is simply $\lambda A. (A-A^{T})/2$ and the $\operatorname{Img}$ can be similarly identified with anti-symmetric matrices.) 
Objects in $\operatorname{Img}$ correspond exactly to oriented parallelogram areas.
This completes our first step.

We now make a parallelogram and measure it with an explicit choice of basis, our second step.
Suppose we have $p$ a variable or expression in $\operatorname{Img}$ (which we can construct from $v_{0}, v_{1}$ if needed).
We can measure it via some basis.
Since our space of bivectors on $\mathbb{R}^{2}$ is one-dimensional\footnote{This can break down in higher dimensions, but continuing with the theory can dissolve this issue}, if we find a basis $b_{0}$ for the space, we can compute a real number $r$ such that $b_{0,0}r=p$ (\textit{i.e.}, $r$ represents $p$ with the only basis vector $b_{0,0}$).
To make the typical choices, we might select the unit square with a particular ordering of the sides to have area positive one:
$b_{0} = \llpl \lproj{}((1\lbosum 0 )\loprod( 0\lbosum 1 )) \llpr{}_{\lbn{1}}$,
where $x\lbosum y$ is essentially $(x,y)$, $\lproj{}$ projects values into subspaces such as $\operatorname{Img}$, and $\lbn{1}$ is a fallback\footnote{Our system ensures one exists for soundness, but we don't need it here.} basis for completing a list of expressions.
In $b_{0}$ we simply list the one vector that is our unit.
By developing the space $\operatorname{Img}$, we were able to completely specify the grid and unit area associated to it via a choice of a single basis vector, which we use to convert a bivector $p$  corresponding to a parallelogram to an area under the unit convention specified via the basis.
We have completed our second step.
Though the difference between the coordinate-free approach and coordinate-dependent approach (just use \cref{eq:second}) seems rather small with the former being slightly more complex, there are a few virtues that lead many scientists to prefer the former approach.
First, our approach here is far more general: this idea works for any $n$ vectors in $n \geq 2$-dimensional space as well as for lower-dimensional parallelepipeds in higher-dimensional space (see the next section).
Second, as stated earlier, we have saved ourselves from mistakes about coordinates; our formula for the cross product would not make sense if we mixed conventions.
(See~\cite{Geisler2020} for examples of similar bugs.)
Third, we have removed arbitrary choices from the core definition of a geometric object in linear algebra.
Fourth, we imposed a discipline via our two-step process: we separated modeling geometry with linear algebra from computing a number with linear algebra via a basis.
Lastly, this framing quickly reveals many other linear algebra operations that have a coordinate-independent meaning (\textit{e.g.}, adding classes of parallelograms, extending a class of parallelograms with another vector to a parallelepiped, determining a class of parallelograms of the same area in the plane orthogonal to another object, and many more~\cite{Gunn2019}); and we can get many of these for free.
\edef\MFrestoreat{\catcode`\noexpand\@=\the\catcode`\@\relax}
\catcode`\@=11\relax

\newif\ifmf@persp
\newif\ifmf@hideback

\pgfkeys{
  /manifold/.cd,
  azimuth/.store in=\mf@az,
  elevation/.store in=\mf@el,
  distance/.store in=\mf@D,
  target height/.store in=\mf@tz,
  perspective/.is if=mf@persp,
  hide back outline/.is if=mf@hideback,
  u cells/.store in=\mf@nu,
  v cells/.store in=\mf@nv,
  mesh step/.store in=\mf@mstep,
  surface color/.store in=\mf@fillc,
  shade base/.store in=\mf@sbase,
  shade gain/.store in=\mf@sgain,
  mesh style/.store in=\mf@meshstyle,
  outline style/.store in=\mf@outline,
  seam width/.store in=\mf@seam,
  curve samples/.store in=\mf@R,
  azimuth=18,
  elevation=42,
  distance=11,
  target height=0.7,
  perspective=true,
  hide back outline=false,
  u cells=24, v cells=20,
  mesh step=4,
  surface color=gray,
  shade base=8,
  shade gain=28,
  mesh style={gray!35, line width=.3pt, line cap=round, line join=round},
  outline style={black, thick, line cap=round, line join=round},
  seam width=.3pt,
  curve samples=6,
}

\newcommand\manifoldsetup[1]{\pgfkeys{/manifold/.cd,#1}}

\newcommand\manifoldheight[1]{\def\mf@expr{#1}}
\manifoldheight{%
  1.2*exp(-0.5*(((\mfu-0.15)*(\mfu-0.15))/(1.6*1.6)
               +((\mfv+0.1)*(\mfv+0.1))/(1.25*1.25)))
  +0.10*sin(deg(0.9*\mfu+0.6))*cos(deg(0.8*\mfv))
  +0.05*\mfu}

\newcommand\manifolddomain[2]{\def\mf@dmu{#1}\def\mf@dmv{#2}}
\manifolddomain
  {2.9*\mfp*(1+0.08*\mfq)+0.14*\mfq*\mfq}
  {2.0*\mfq*(1+0.08*\mfp)+0.20*\mfp*\mfp+0.15*\mfp*\mfp*\mfp
   +0.8*exp(-\mfp*\mfp/0.3)*(1-\mfq)/2-0.3}

\def\mf@z#1#2{%
  \pgfmathsetmacro\mfu{#1}%
  \pgfmathsetmacro\mfv{#2}%
  \pgfmathsetmacro\mfz{\mf@expr}}

\def\mf@eval#1#2{%
  \pgfmathsetmacro\mf@eu{#1}%
  \pgfmathsetmacro\mf@ev{#2}%
  \mf@z{\mf@eu+0.01}{\mf@ev}\let\mf@za\mfz
  \mf@z{\mf@eu-0.01}{\mf@ev}\let\mf@zb\mfz
  \mf@z{\mf@eu}{\mf@ev+0.01}\let\mf@zc\mfz
  \mf@z{\mf@eu}{\mf@ev-0.01}\let\mf@zd\mfz
  \mf@z{\mf@eu}{\mf@ev}%
  \pgfmathsetmacro\mffu{(\mf@za-\mf@zb)/0.02}%
  \pgfmathsetmacro\mffv{(\mf@zc-\mf@zd)/0.02}}

\def\mf@camera{%
  \pgfmathsetmacro\mf@sa{sin(\mf@az)}\pgfmathsetmacro\mf@ca{cos(\mf@az)}%
  \pgfmathsetmacro\mf@se{sin(\mf@el)}\pgfmathsetmacro\mf@ce{cos(\mf@el)}%
  \pgfmathsetmacro\mf@Cx{\mf@D*\mf@sa*\mf@ce}%
  \pgfmathsetmacro\mf@Cy{-\mf@D*\mf@ca*\mf@ce}%
  \pgfmathsetmacro\mf@Cz{\mf@tz+\mf@D*\mf@se}}

\def\mf@proj#1#2#3{%
  \pgfmathsetmacro\mf@dx{#1-\mf@Cx}%
  \pgfmathsetmacro\mf@dy{#2-\mf@Cy}%
  \pgfmathsetmacro\mf@dz{#3-\mf@Cz}%
  \pgfmathsetmacro\mf@dr{\mf@dx*\mf@ca+\mf@dy*\mf@sa}%
  \pgfmathsetmacro\mf@dup{-\mf@dx*\mf@sa*\mf@se+\mf@dy*\mf@ca*\mf@se+\mf@dz*\mf@ce}%
  \ifmf@persp
    \pgfmathsetmacro\mf@df{-\mf@dx*\mf@sa*\mf@ce+\mf@dy*\mf@ca*\mf@ce-\mf@dz*\mf@se}%
    \pgfmathsetmacro\mfsx{\mf@D*\mf@dr/\mf@df}%
    \pgfmathsetmacro\mfsy{\mf@D*\mf@dup/\mf@df}%
  \else
    \let\mfsx\mf@dr
    \let\mfsy\mf@dup
  \fi}

\def\mf@stpt#1#2{%
  \pgfmathsetmacro\mfs{#1}\pgfmathsetmacro\mft{#2}%
  \pgfmathsetmacro\mfp{\mfs*sqrt(1-\mft*\mft/2)}%
  \pgfmathsetmacro\mfq{\mft*sqrt(1-\mfs*\mfs/2)}%
  \pgfmathsetmacro\mf@uu{\mf@dmu}%
  \pgfmathsetmacro\mf@vv{\mf@dmv}%
  \mf@z{\mf@uu}{\mf@vv}%
  \mf@proj{\mf@uu}{\mf@vv}{\mfz}}

\def\mf@buildgrid{%
  \mf@camera
  \foreach \mf@i in {0,...,\mf@nu}{%
    \foreach \mf@j in {0,...,\mf@nv}{%
      \pgfmathsetmacro\mfs{-1+2*\mf@i/\mf@nu}%
      \pgfmathsetmacro\mft{-1+2*\mf@j/\mf@nv}%
      \pgfmathsetmacro\mfp{\mfs*sqrt(1-\mft*\mft/2)}%
      \pgfmathsetmacro\mfq{\mft*sqrt(1-\mfs*\mfs/2)}%
      \pgfmathsetmacro\mf@uu{\mf@dmu}%
      \pgfmathsetmacro\mf@vv{\mf@dmv}%
      \mf@eval{\mf@uu}{\mf@vv}%
      \mf@proj{\mf@uu}{\mf@vv}{\mfz}%
      \pgfmathsetmacro\mf@k{(0.45*\mffu+0.35*\mffv+0.82)/sqrt(\mffu*\mffu+\mffv*\mffv+1)}%
      \pgfmathsetmacro\mf@s{max(2,min(24,\mf@sbase+\mf@sgain*(\mf@k-0.82)))}%
      \global\expandafter\edef\csname mf@P@\mf@i @\mf@j\endcsname{\mfsx,\mfsy}%
      \global\expandafter\edef\csname mf@S@\mf@i @\mf@j\endcsname{\mf@s}%
      \global\expandafter\edef\csname mf@U@\mf@i @\mf@j\endcsname{\mf@uu}%
      \global\expandafter\edef\csname mf@V@\mf@i @\mf@j\endcsname{\mf@vv}%
    }}%
  \mf@buildedges}

\def\mf@buildedges{%
  \pgfmathtruncatemacro\mf@Rm{\mf@R-1}%
  \pgfmathtruncatemacro\mf@numax{\mf@nu-1}%
  \pgfmathtruncatemacro\mf@nvmax{\mf@nv-1}%
  \foreach \mf@i in {0,...,\mf@numax}{%
    \foreach \mf@j in {0,...,\mf@nv}{%
      \pgfmathtruncatemacro\mf@need{ifthenelse(or(or(\mf@j==0,\mf@j==\mf@nv),and(\mf@mstep>0,mod(\mf@j,max(\mf@mstep,1))==0)),1,0)}%
      \ifnum\mf@need=1
        \gdef\mf@pl{}%
        \ifnum\mf@Rm>0
          \pgfmathsetmacro\mf@ea{-1+2*\mf@i/\mf@nu}%
          \pgfmathsetmacro\mf@eb{-1+2*(\mf@i+1)/\mf@nu}%
          \pgfmathsetmacro\mf@ec{-1+2*\mf@j/\mf@nv}%
          \foreach \mf@k in {1,...,\mf@Rm}{%
            \pgfmathsetmacro\mf@sk{\mf@ea+(\mf@eb-\mf@ea)*\mf@k/\mf@R}%
            \mf@stpt{\mf@sk}{\mf@ec}%
            \xdef\mf@pl{\mf@pl--(\mfsx,\mfsy)}}%
        \fi
        \global\expandafter\let\csname mf@EH@\mf@i @\mf@j\endcsname\mf@pl
      \fi}}%
  \foreach \mf@i in {0,...,\mf@nu}{%
    \foreach \mf@j in {0,...,\mf@nvmax}{%
      \pgfmathtruncatemacro\mf@need{ifthenelse(or(or(\mf@i==0,\mf@i==\mf@nu),and(\mf@mstep>0,mod(\mf@i,max(\mf@mstep,1))==0)),1,0)}%
      \ifnum\mf@need=1
        \gdef\mf@pl{}%
        \ifnum\mf@Rm>0
          \pgfmathsetmacro\mf@ea{-1+2*\mf@j/\mf@nv}%
          \pgfmathsetmacro\mf@eb{-1+2*(\mf@j+1)/\mf@nv}%
          \pgfmathsetmacro\mf@ec{-1+2*\mf@i/\mf@nu}%
          \foreach \mf@k in {1,...,\mf@Rm}{%
            \pgfmathsetmacro\mf@sk{\mf@ea+(\mf@eb-\mf@ea)*\mf@k/\mf@R}%
            \mf@stpt{\mf@ec}{\mf@sk}%
            \xdef\mf@pl{\mf@pl--(\mfsx,\mfsy)}}%
        \fi
        \global\expandafter\let\csname mf@EV@\mf@i @\mf@j\endcsname\mf@pl
      \fi}}}

\def\mf@getP#1#2#3{\expandafter\let\expandafter#1\csname mf@P@#2@#3\endcsname}
\def\mf@getE#1#2#3#4{\expandafter\let\expandafter#1\csname mf@#2@#3@#4\endcsname}
\def\mf@getS#1#2#3{\expandafter\let\expandafter#1\csname mf@S@#2@#3\endcsname}

\def\mf@drawmesh{\expandafter\draw\expandafter[\mf@meshstyle]}
\def\mf@drawoutline{\expandafter\draw\expandafter[\mf@outline]}

\def\mf@frontlist{}
\newif\ifmf@frontedge
\def\mf@edge#1#2#3#4#5#6#7{%
  \mf@getP\mf@eA{#1}{#2}\mf@getP\mf@eB{#3}{#4}\mf@getE\mf@epl{#5}{#6}{#7}%
  \pgfmathsetmacro\mf@mu{(\csname mf@U@#1@#2\endcsname+\csname mf@U@#3@#4\endcsname)/2}%
  \pgfmathsetmacro\mf@mv{(\csname mf@V@#1@#2\endcsname+\csname mf@V@#3@#4\endcsname)/2}%
  \pgfmathparse{(\mf@mu-\mf@ccu)*(\mf@Cx-\mf@mu)+(\mf@mv-\mf@ccv)*(\mf@Cy-\mf@mv)}%
  \ifdim\pgfmathresult pt>0pt\relax\mf@frontedgetrue\else\ifmf@hideback\mf@frontedgefalse\else\mf@frontedgetrue\fi\fi
  \ifmf@frontedge
    \let\mf@oldf\mf@frontlist
    \xdef\mf@frontlist{\unexpanded\expandafter{\mf@oldf}\noexpand\mf@drawoutline(\mf@eA)\mf@epl--(\mf@eB);}%
  \else
    \mf@drawoutline (\mf@eA) \mf@epl -- (\mf@eB);%
  \fi}

\def\mf@cell#1#2{%
  \pgfmathtruncatemacro\mf@ip{#1+1}%
  \pgfmathtruncatemacro\mf@jp{#2+1}%
  \mf@getP\mf@pA{#1}{#2}\mf@getP\mf@pB{\mf@ip}{#2}%
  \mf@getP\mf@pC{\mf@ip}{\mf@jp}\mf@getP\mf@pD{#1}{\mf@jp}%
  \mf@getS\mf@sA{#1}{#2}\mf@getS\mf@sB{\mf@ip}{#2}%
  \mf@getS\mf@sC{\mf@ip}{\mf@jp}\mf@getS\mf@sD{#1}{\mf@jp}%
  \pgfmathsetmacro\mf@pct{(\mf@sA+\mf@sB+\mf@sC+\mf@sD)/4}%
  \filldraw[fill=\mf@fillc!\mf@pct, draw=\mf@fillc!\mf@pct, line width=\mf@seam]
    (\mf@pA) -- (\mf@pB) -- (\mf@pC) -- (\mf@pD) -- cycle;
  \ifnum\mf@mstep>0
    \pgfmathtruncatemacro\mf@m{mod(#1,\mf@mstep)}%
    \ifnum\mf@m=0 \mf@getE\mf@epl{EV}{#1}{#2}\mf@drawmesh (\mf@pA) \mf@epl -- (\mf@pD);\fi
    \pgfmathtruncatemacro\mf@m{mod(\mf@ip,\mf@mstep)}%
    \ifnum\mf@m=0 \mf@getE\mf@epl{EV}{\mf@ip}{#2}\mf@drawmesh (\mf@pB) \mf@epl -- (\mf@pC);\fi
    \pgfmathtruncatemacro\mf@m{mod(#2,\mf@mstep)}%
    \ifnum\mf@m=0 \mf@getE\mf@epl{EH}{#1}{#2}\mf@drawmesh (\mf@pA) \mf@epl -- (\mf@pB);\fi
    \pgfmathtruncatemacro\mf@m{mod(\mf@jp,\mf@mstep)}%
    \ifnum\mf@m=0 \mf@getE\mf@epl{EH}{#1}{\mf@jp}\mf@drawmesh (\mf@pD) \mf@epl -- (\mf@pC);\fi
  \fi
  \pgfmathsetmacro\mf@ccu{(\csname mf@U@#1@#2\endcsname+\csname mf@U@\mf@ip @#2\endcsname
    +\csname mf@U@\mf@ip @\mf@jp\endcsname+\csname mf@U@#1@\mf@jp\endcsname)/4}%
  \pgfmathsetmacro\mf@ccv{(\csname mf@V@#1@#2\endcsname+\csname mf@V@\mf@ip @#2\endcsname
    +\csname mf@V@\mf@ip @\mf@jp\endcsname+\csname mf@V@#1@\mf@jp\endcsname)/4}%
  \ifnum#1=0 \mf@edge{#1}{#2}{#1}{\mf@jp}{EV}{#1}{#2}\fi
  \ifnum\mf@ip=\mf@nu \mf@edge{\mf@ip}{#2}{\mf@ip}{\mf@jp}{EV}{\mf@ip}{#2}\fi
  \ifnum#2=0 \mf@edge{#1}{#2}{\mf@ip}{#2}{EH}{#1}{#2}\fi
  \ifnum\mf@jp=\mf@nv \mf@edge{#1}{\mf@jp}{\mf@ip}{\mf@jp}{EH}{#1}{\mf@jp}\fi}

\def\mf@NB{3600}
\newcommand\manifoldSurface[1][]{%
  \begingroup
  \pgfkeys{/manifold/.cd,#1}%
  \mf@buildgrid
  \gdef\mf@frontlist{}%
  \pgfmathsetmacro\mf@H{veclen(\mf@Cx,\mf@Cy)}%
  \foreach \mf@b in {0,...,\mf@NB}{%
    \global\expandafter\def\csname mf@B@\mf@b\endcsname{}}%
  \pgfmathtruncatemacro\mf@numax{\mf@nu-1}%
  \pgfmathtruncatemacro\mf@nvmax{\mf@nv-1}%
  \foreach \mf@ci in {0,...,\mf@numax}{%
    \foreach \mf@cj in {0,...,\mf@nvmax}{%
      \pgfmathtruncatemacro\mf@ci@p{\mf@ci+1}%
      \pgfmathtruncatemacro\mf@cj@p{\mf@cj+1}%
      \pgfmathsetmacro\mf@cu{(\csname mf@U@\mf@ci @\mf@cj\endcsname
        +\csname mf@U@\mf@ci@p @\mf@cj\endcsname
        +\csname mf@U@\mf@ci@p @\mf@cj@p\endcsname
        +\csname mf@U@\mf@ci @\mf@cj@p\endcsname)/4}%
      \pgfmathsetmacro\mf@cv{(\csname mf@V@\mf@ci @\mf@cj\endcsname
        +\csname mf@V@\mf@ci@p @\mf@cj\endcsname
        +\csname mf@V@\mf@ci@p @\mf@cj@p\endcsname
        +\csname mf@V@\mf@ci @\mf@cj@p\endcsname)/4}%
      \pgfmathtruncatemacro\mf@b{max(0,min(\mf@NB,floor((\mf@H+6-veclen(\mf@cu-\mf@Cx,\mf@cv-\mf@Cy))*300)))}%
      \expandafter\let\expandafter\mf@old\csname mf@B@\mf@b\endcsname
      \global\expandafter\edef\csname mf@B@\mf@b\endcsname{%
        \unexpanded\expandafter{\mf@old}\noexpand\mf@cell{\mf@ci}{\mf@cj}}%
    }}%
  \foreach \mf@b in {0,...,\mf@NB}{%
    \csname mf@B@\mf@b\endcsname}%
  \mf@frontlist
  \endgroup}

\newcommand\manifoldFrame[4]{%
  \pgfmathsetmacro\mfx@tmp{#1}\xdef\mfVa{\mfx@tmp}%
  \pgfmathsetmacro\mfx@tmp{#2}\xdef\mfVb{\mfx@tmp}%
  \pgfmathsetmacro\mfx@tmp{#3}\xdef\mfWa{\mfx@tmp}%
  \pgfmathsetmacro\mfx@tmp{#4}\xdef\mfWb{\mfx@tmp}}

\newcommand\manifoldOrthoFrame[4]{%
  \pgfmathsetmacro\mfx@ou{#1}\pgfmathsetmacro\mfx@ov{#2}%
  \mf@eval{\mfx@ou}{\mfx@ov}%
  \pgfmathsetmacro\mfx@a{#3/sqrt(1+\mffu*\mffu)}%
  \xdef\mfVa{\mfx@a}\xdef\mfVb{0}%
  \pgfmathsetmacro\mfx@wa{-\mffu*\mffv/(1+\mffu*\mffu)}%
  \pgfmathsetmacro\mfx@nrm{sqrt(\mfx@wa*\mfx@wa+1+(\mffu*\mfx@wa+\mffv)*(\mffu*\mfx@wa+\mffv))}%
  \pgfmathsetmacro\mfx@sc{#4/\mfx@nrm}%
  \pgfmathsetmacro\mfx@wa{\mfx@wa*\mfx@sc}%
  \xdef\mfWa{\mfx@wa}\xdef\mfWb{\mfx@sc}}

\def\mf@derivs#1#2{%
  \pgfmathsetmacro\mfd@u{#1}\pgfmathsetmacro\mfd@v{#2}%
  \pgfmathsetmacro\mfd@h{0.1}%
  \mf@z{\mfd@u}{\mfd@v}\let\mfd@fC\mfz
  \mf@z{\mfd@u+\mfd@h}{\mfd@v}\let\mfd@fE\mfz
  \mf@z{\mfd@u-\mfd@h}{\mfd@v}\let\mfd@fW\mfz
  \mf@z{\mfd@u}{\mfd@v+\mfd@h}\let\mfd@fN\mfz
  \mf@z{\mfd@u}{\mfd@v-\mfd@h}\let\mfd@fS\mfz
  \mf@z{\mfd@u+\mfd@h}{\mfd@v+\mfd@h}\let\mfd@fNE\mfz
  \mf@z{\mfd@u-\mfd@h}{\mfd@v+\mfd@h}\let\mfd@fNW\mfz
  \mf@z{\mfd@u+\mfd@h}{\mfd@v-\mfd@h}\let\mfd@fSE\mfz
  \mf@z{\mfd@u-\mfd@h}{\mfd@v-\mfd@h}\let\mfd@fSW\mfz
  \pgfmathsetmacro\mfd@fu{(\mfd@fE-\mfd@fW)/(2*\mfd@h)}%
  \pgfmathsetmacro\mfd@fv{(\mfd@fN-\mfd@fS)/(2*\mfd@h)}%
  \pgfmathsetmacro\mfd@fuu{(\mfd@fE-2*\mfd@fC+\mfd@fW)/(\mfd@h*\mfd@h)}%
  \pgfmathsetmacro\mfd@fvv{(\mfd@fN-2*\mfd@fC+\mfd@fS)/(\mfd@h*\mfd@h)}%
  \pgfmathsetmacro\mfd@fuv{(\mfd@fNE-\mfd@fNW-\mfd@fSE+\mfd@fSW)/(4*\mfd@h*\mfd@h)}}

\def\mf@rhs#1#2{%
  \pgfmathsetmacro\mfr@H{\mfd@fuu*(#1)*\mfx@cu+\mfd@fuv*((#1)*\mfx@cv+(#2)*\mfx@cu)+\mfd@fvv*(#2)*\mfx@cv}%
  \pgfmathsetmacro\mfr@D{\mfr@H/(1+\mfd@fu*\mfd@fu+\mfd@fv*\mfd@fv)}%
  \pgfmathsetmacro\mfr@a{-\mfr@D*\mfd@fu}%
  \pgfmathsetmacro\mfr@b{-\mfr@D*\mfd@fv}}

\newcommand\manifoldTransportSeg[5]{%
  \pgfmathsetmacro\mfx@pua{#1}\pgfmathsetmacro\mfx@pva{#2}%
  \pgfmathsetmacro\mfx@cu{#3-(#1)}\pgfmathsetmacro\mfx@cv{#4-(#2)}%
  \pgfmathsetmacro\mfx@dt{1/(#5)}%
  \foreach \mfx@st in {1,...,#5}{%
    \pgfmathsetmacro\mfx@tau{(\mfx@st-1)*\mfx@dt}%
    \pgfmathsetmacro\mfx@pu{\mfx@pua+\mfx@tau*\mfx@cu}%
    \pgfmathsetmacro\mfx@pv{\mfx@pva+\mfx@tau*\mfx@cv}%
    \mf@derivs{\mfx@pu}{\mfx@pv}%
    \mf@rhs{\mfVa}{\mfVb}\let\mfx@kVa\mfr@a\let\mfx@kVb\mfr@b
    \mf@rhs{\mfWa}{\mfWb}\let\mfx@kWa\mfr@a\let\mfx@kWb\mfr@b
    \pgfmathsetmacro\mfx@mVa{\mfVa+0.5*\mfx@dt*\mfx@kVa}%
    \pgfmathsetmacro\mfx@mVb{\mfVb+0.5*\mfx@dt*\mfx@kVb}%
    \pgfmathsetmacro\mfx@mWa{\mfWa+0.5*\mfx@dt*\mfx@kWa}%
    \pgfmathsetmacro\mfx@mWb{\mfWb+0.5*\mfx@dt*\mfx@kWb}%
    \pgfmathsetmacro\mfx@pu{\mfx@pua+(\mfx@tau+0.5*\mfx@dt)*\mfx@cu}%
    \pgfmathsetmacro\mfx@pv{\mfx@pva+(\mfx@tau+0.5*\mfx@dt)*\mfx@cv}%
    \mf@derivs{\mfx@pu}{\mfx@pv}%
    \mf@rhs{\mfx@mVa}{\mfx@mVb}%
    \pgfmathsetmacro\mfx@nVa{\mfVa+\mfx@dt*\mfr@a}%
    \pgfmathsetmacro\mfx@nVb{\mfVb+\mfx@dt*\mfr@b}%
    \mf@rhs{\mfx@mWa}{\mfx@mWb}%
    \pgfmathsetmacro\mfx@nWa{\mfWa+\mfx@dt*\mfr@a}%
    \pgfmathsetmacro\mfx@nWb{\mfWb+\mfx@dt*\mfr@b}%
    \xdef\mfVa{\mfx@nVa}\xdef\mfVb{\mfx@nVb}\xdef\mfWa{\mfx@nWa}\xdef\mfWb{\mfx@nWb}%
  }}

\newcommand\manifoldPoint[3]{%
  \mf@camera
  \pgfmathsetmacro\mf@pu{#2}%
  \pgfmathsetmacro\mf@pv{#3}%
  \mf@eval{\mf@pu}{\mf@pv}%
  \mf@proj{\mf@pu}{\mf@pv}{\mfz}%
  \global\expandafter\edef\csname mf@pt@#1@u\endcsname{\mf@pu}%
  \global\expandafter\edef\csname mf@pt@#1@v\endcsname{\mf@pv}%
  \global\expandafter\edef\csname mf@pt@#1@z\endcsname{\mfz}%
  \global\expandafter\edef\csname mf@pt@#1@fu\endcsname{\mffu}%
  \global\expandafter\edef\csname mf@pt@#1@fv\endcsname{\mffv}%
  \coordinate (#1) at (\mfsx,\mfsy);}

\newcommand\manifoldTangent[4]{%
  \mf@camera
  \edef\mf@bu{\csname mf@pt@#2@u\endcsname}%
  \edef\mf@bv{\csname mf@pt@#2@v\endcsname}%
  \edef\mf@bz{\csname mf@pt@#2@z\endcsname}%
  \edef\mf@bfu{\csname mf@pt@#2@fu\endcsname}%
  \edef\mf@bfv{\csname mf@pt@#2@fv\endcsname}%
  \pgfmathsetmacro\mf@ta{#3}%
  \pgfmathsetmacro\mf@tb{#4}%
  \pgfmathsetmacro\mf@Pu{\mf@bu+\mf@ta}%
  \pgfmathsetmacro\mf@Pv{\mf@bv+\mf@tb}%
  \pgfmathsetmacro\mf@Pz{\mf@bz+\mf@ta*\mf@bfu+\mf@tb*\mf@bfv}%
  \mf@proj{\mf@Pu}{\mf@Pv}{\mf@Pz}%
  \coordinate (#1) at (\mfsx,\mfsy);}

\let\manifoldPlaneCorner\manifoldTangent

\NewDocumentCommand\manifoldTangentPlane{O{} m m m m m}{%
  \manifoldTangent{mf@tp@1}{#2}{#3}{#5}%
  \manifoldTangent{mf@tp@2}{#2}{#4}{#5}%
  \manifoldTangent{mf@tp@3}{#2}{#4}{#6}%
  \manifoldTangent{mf@tp@4}{#2}{#3}{#6}%
  \filldraw[fill=blue!12, fill opacity=.65, draw=blue!55!black, thick, #1]
    (mf@tp@1) -- (mf@tp@2) -- (mf@tp@3) -- (mf@tp@4) -- cycle;}

\newcommand\manifoldSpace[4]{%
  \mf@camera
  \mf@proj{#2}{#3}{#4}%
  \coordinate (#1) at (\mfsx,\mfsy);}

\MFrestoreat

\subsection{The Tangent Space and Linear Algebra at a Point on a Manifold}
\label{sec:tangentSpaceOpt}

Another reason to prefer the coordinate-independent system is that we can extend our linear algebra tools to trickier situations where vectors depend on a point on a surface.
For example, if we wish to analyze parallelograms on a surface immersed in Euclidean space, we would normally encounter some pain because a vector space associated to each point on the surface (the tangent space) would use $n$ (say $2$) dimensions when the ambient Euclidean space allows such vectors to be $m$-dimensional (say $3$).
Both methods \eqref{eq:second} and \eqref{eq:first} fail in this case because they would only compute the portion of the parallelogram's area in the $xy$ plane.

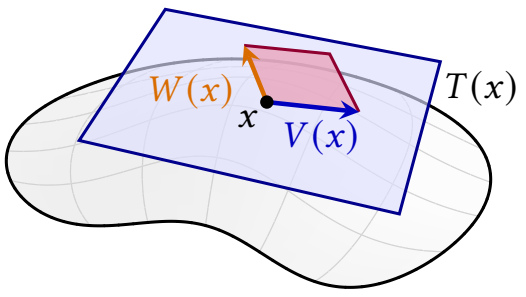
\begingroup
\setlength{\intextsep}{3pt}
\begin{wrapfigure}[4]{r}{0.25\textwidth}
\centering
\resizebox{\linewidth}{!}{\input{tex/figs/manifold1-new}}
\end{wrapfigure}
To solve this, we must model a situation where we have a surface, a tangent space $T(x)$ at a point $x$ on the surface, and two vectors, $V(x), W(x)$ modeling the parallelogram in the tangent space.
Note that $V,W$ are really vector fields and together constitute a moving parallelogram in the moving tangent space $T$.

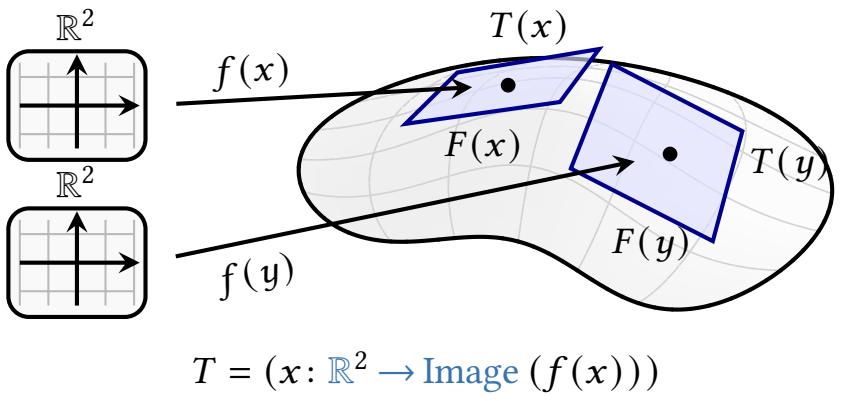
\begin{wrapfigure}{r}{0.4\textwidth}
\centering
\resizebox{\linewidth}{!}{\input{tex/figs/manifold-new}}
\par{\scriptsize
\(T=\fntype{x}{\lR{}^{2}}{\Image{}(f(x))}\)}
\end{wrapfigure}
To model this situation in our system and ultimately produce a moving area of our moving parallelogram, we imagine a user provides some parameterization $F\colon \mathbb{R}^{2}\to\mathbb{R}^{3}$ of a surface in $\mathbb{R}^{3}$ and differentiates it to define $f\colon \lR{}^{2}\to (\lR{}^{2})^{\lstar{}}\loprod\lR{}^{3}.$
In doing so, the user provides a parameterization of the tangent space, the plane tangent to a point on the surface.
We can model this with dependent types.
We can define $T$ as a dependent function type for the tangent space at a point;
this models the common pattern in scientific computing and computer graphics.
Now via our space of bivectors, we can then compute the area of a moving parallelogram in $T(x)$ defined by two moving vectors $V, W\colon T$ where $\fntype{x}{\ltyn{0}}{\ltyn{1}}$ is a dependent function type and we use the arrow superscript to indicate a function is non-linear:
\begingroup
\setlength{\abovedisplayskip}{3pt}
\setlength{\abovedisplayshortskip}{0pt}
\setlength{\belowdisplayskip}{3pt}
\setlength{\belowdisplayshortskip}{3pt}
\begin{equation}
\label{eq:movingarea}
\operatorname{area} \colon \fntype{x}{\lR{}^{2}}{\Image(\operatorname{antiSymMap})}
=
\nllambdaArg{x}{
  \lproj{}\bigl(V(x)\loprod W(x)\bigr)
}.
\end{equation}
\endgroup
Here $\operatorname{antiSymMap}$ abbreviates the term of \eqref{eq:antisymmap} instantiated at $T(x)$ instead of $\lR^2$, with its linear abstraction inside the outer $\nllambda(x)$. 
Proper inference here is outside the scope of this paper.

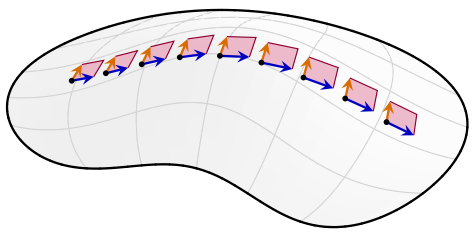
\begin{wrapfigure}[4]{r}{0.22\textwidth}
\centering
\resizebox{\linewidth}{!}{\input{tex/figs/manifold2-new}}
\end{wrapfigure}
We can now move on to calculus, as we can use this to define the area form, the basic tool for computing integrals on manifolds.
Roughly, by realizing the $\operatorname{area}$ operation at many points, we can sum infinitesimal areas together to compute an integral~\cite{Lee2012}.
Note that this example roughly delineates our scope: \GLang{} handles the linear algebra that we do at a few potentially generic points on a manifold.
Similarly to stencil languages, we work on a neighborhood, but we don't handle the global definitions (\textit{e.g.}, an abstract integral or global differential operator turned into a sparse linear algebra operator that goes through a complex direct solver).
See Appendix~\ref{app:ufl-comparison} for a comparison with the Unified Form Language, which has a similar situated component.
\par\endgroup

\subsection{The Tangent Space and Algorithmic Optimizations via Choice of Coordinates}
The example \eqref{eq:movingarea} also leads us to one last advantage for coordinate-independent models:  optimization of representations via choice of basis.
We will show two examples of this in this section based on the following idea:
in scientific applications, it is common to represent vectors in $T$ with vectors in $\lR{}^{2}$ that are implicitly parametrizing a vector in $T(x)$ via $f(x)$.
We can model this situation via a discrete vector frame (a function that supplies a basis at every point), and a special basis construction, an $\iBasis$, which maps a basis for the domain of a map to a basis for $\Image (\lee)$, by roughly\footnote{Technically, the basis needs to be for the cokernel but we omit this here for simplicity.} applying $\lee$ to every basis element.
Given a basis $\lb{}$ for the domain and noting the dependence of $\lee$ on $x$, we can write this as:
\begin{equation}
\label{eq:ftb}
T_{B} = \frameB{x}{(\lbone{}^{2})}{  \iBasis\llpl{}f(x), \lb{}\llpr{}}.
\end{equation}

This reads as: given a point $x$ in the standard basis for $\mathbb{R}^{2}$ (written $\lbone^{2}$), we yield a basis for the tangent space at $x$, $T(x)$, given via the action of $f(x)$ on a basis for the domain.
This can serve as a suitable frame for $V$ and $W$ in the situation when there exist functions $V^{0},V^{1}, W^{0}, W^{1}$ that parametrize the $V$ and $W$ in terms of $x$ (as opposed to the corresponding $y$ on the surface).
Additionally, we can use this to build a frame (a moving basis) for functions of type  
\begin{equation}
\label{eq:ftbb}
\fntype{x}{\lR{}^{2}}{\Image(\operatorname{antiSymMap})} \mbox{ via }
\operatorname{area}_{B} = \frameB{x}{(\lbone{}^{2})}{\llpl \lproj{}(T_{B}(x) \loprod T_{B}(x)) \llpr{}_{\lbn{2}(x)}}.
\end{equation}
where $\lbn{2}(x)$ is a fallback basis, which can be automatically provided.

\begingroup
\setlength{\intextsep}{0pt}
\begin{wrapfigure}[7]{r}{0.4\textwidth}
\centering
\resizebox{\linewidth}{!}{\input{tex/figs/manifold3-new}}
\par{\scriptsize
\(\begin{aligned}
V(x)&=V^0(x)f(x)(e_0)+V^1(x)f(x)(e_1),\\
W(x)&=W^0(x)f(x)(e_0)+W^1(x)f(x)(e_1).
\end{aligned}\)}
\end{wrapfigure}
Our first example is simply to measure $\operatorname{area}$ with $\operatorname{area}_{B}$.
In lots of scientific applications, we actually want to measure area in terms of $\operatorname{area}_{B}$\footnote{\textit{e.g.}, in Firedrake~\cite{Rathgeber2016}, this corresponds to the ``DG L2'' option} because it is only an intermediate quantity.
And critically, if we use $T_{B}$ for $V,W$ and measure under the basis $\operatorname{area}_{B}$, then the measured area is the Euclidean area of the parallelogram $(V^{0}(x), V^{1}(x)), (W^{0}(x), W^{1}(x))$.
Thus, we can reuse this calculation when $x, V^{i}, W^{i}$ are shared between surfaces but the parameterizations are not.
Our semantics captures this idea by producing a value in this situation that is independent of $y$ or $f(x)$ modulo a potential dimension calculation (which could be optimized away).
This is a critical optimization for a class of PDE operators that are sometimes referred to as universal element operators, basically block-diagonal matrices where each block  is the same, corresponding to roughly a different part of the surface parameterized in a different way~\cite{Silvester1982}.
With this example, we captured a simple version of an optimization in~\cite{Rognes2010} and demonstrated how certain mesh computations therein sometimes deliver highly structured sparse matrices (\textit{e.g.}, matrices found in~\cite[section~9.4]{desbrun2005discrete}).
\par\endgroup

We provide one more similar and simpler example.
Let's consider:
\begin{equation}
\label{eq:eqtaneval}
ev\colon \fntype{x}{\lR{}^{2}}{T(x)^{\lstar\lstar}\loprod T(x)^{\lstar}} = \nllambdaArg{x}{\lllambdaArg{y, z}{y(z)}}.
\end{equation}
Given a point, a cotangent vector (note the dual of the dual), and a tangent vector, this program produces the action of the cotangent vector against the vector.
We can use $T_{B}$ to build a frame for $\fntype{x}{\lR{}^{2}}{T(x)^{\lstar\lstar}\loprod T(x)^{\lstar}}$ via 
$\frameB{x}{(\lbone{}^{2})}{T_{B}(x)^{\lstar\lstar}\loprod T_{B}(x)^{\lstar}}$ where $(-)^{\lstar}$ creates a basis for $\ls{}^{\lstar}$ from a basis for $\ls{}$.
If we combine this basis with our term in \cref{eq:eqtaneval}, our program will operationally produce essentially the identity matrix as a constant function of $x$, due to the duality and despite the fact that our basis is being transformed by some expression $\lee$ that depends on $x$.
We say essentially because we will again need to emit code to compute or check dimensions, but again these can be potentially omitted or optimized away. 
By specifying a term without coordinates and then picking a good basis, we got a very favorable program.

\section{Defining \GLang}
\label{sec:defgrassman}

\subsection{Syntax for Spaces and Tensors}
\label{sec:syntaxIntro}
\begin{figure}[!htbp]
\scriptsize
\setlength{\tabcolsep}{4pt}
\centering
\begin{minipage}[t]{0.48\textwidth}
\centering
\begin{tabular}{rclp{3.8cm}}
\toprule
$x$ & $\in$ & String & Variable Names\\
$n$ & $\in$ & $\mathbb{N}$ & Constant Natural Number\\
$r$ & $\in$ & $\mathbb{R}$ & Constant Real\\
\midrule
$\ls{}$ & $::=$ & $\lsempty$ & Empty Space\\
 & $|$ & $\lR$ & Scalar Space\\
 & $|$ & $\lsn{0}\losum\lsn{1}$ & Direct Sum Space\\
 & $|$ & $\ls{}^{\lstar}$ & Dual Space\\
 & $|$ & $\lsn{0}\loprod\lsn{1}$ & Tensor Product Space\\
 & $|$ & $\Kernel{}(\lee)$ & Kernel of a Linear Operator\\
 & $|$ & $\CoKernel{}(\lee)$ & Cokernel of a Linear Operator\\
 & $|$ & $\Image{}(\lee)$ & Image of a Linear Operator\\
 & $|$ & $\CoImage{}(\lee)$ & Coimage of a Linear Operator\\
\midrule
$\lty{}$ & $::=$ & $\ls{}$ & Space as a Type \\
 & $|$ & $\fntypep{x}{\ltyn{0}}{\ltynx{1}}$ & Dependent Function Type \\
\bottomrule
\end{tabular}
\end{minipage}
\hfill
\begin{minipage}[t]{0.48\textwidth}
\centering
\begin{tabular}{rclp{3.8cm}}
\toprule
\lee & $::=$ & $\lvzero_{\els{}}$ & Zero Vector\\
 & $|$ & $\len{0} \lvscale \len{1}$ & Scalar-Vector Product\\
 & $|$ & $r$ & Real Number\\
 & $|$ & $\len{0} \lvadd \len{1}$ & Vector Sum\\
 & $|$ & $\linner{}_{\els{}}$ & Inner Product\\
 & $|$ & $\leval{\len{0}}{}{\len{1}}$ & Evaluation\\
 & $|$ & $\lllambdaArg{x\ecolon{}\els{}}{\lee}$ & Multi-linear Function\\
 & $|$ & $\nllambdaArg{x\ecolon{}\elty{}}{\lee}$ & Non-linear Function\\
 & $|$ & $x$ & Variable\\
 & $|$ & $\lee{}^{\lpseudo{}}$ & Linear-Pseudo-Inverse\\
 & $|$ & $\len{0}\losum\len{1}$ & Direct Sum\\
 & $|$ & $\len{0}\loprod\len{1}$ & Tensor Product\\
 & $|$ & $\lembd{}_{\els}$ & Embedding\\
\midrule
$\lky{}$ & $::=$ & $\lkbase$ & Space Kind\\
 & $|$ & $\lknto{\lkn{0}}{\lkn{1}}$ & Function Kind\\
\bottomrule
\end{tabular}
\end{minipage}

\caption{Syntax of \GLang{} for spaces, types, kinds, and tensors.}
\label{fig:grassDef}
\end{figure}

We now discuss the syntax of \GLang{}, which is depicted in \cref{fig:grassDef}.

\paragraph{Types:}\GLang{} has two kinds of types, $\lty$, a (vector) space $\ls$ and a non-linear function type where the output type is allowed to depend on the input.
The former contains the core spaces such as the empty space, $\lR{}$ for the real numbers, direct sum, dual, tensor product, and the four fundamental subspaces of a linear map ($\lee{}$):
\begin{defn}[Four Fundamental Subspaces~\cite{Strang2025}]

\label{def:ffs}
    Assume $W$ and $V$ are inner product spaces with inner products $\langle \cdot, \cdot\rangle_{W}$ and $\langle\cdot, \cdot\rangle_{V}$.
    Let $L$ be a linear map from $W$ to $V$.\footnote{Our names follow the four orthogonal subspaces: $\operatorname{coker}(L)=\ker(L)^\perp$ in the domain and $\operatorname{coimage}(L)=\operatorname{image}(L)^\perp$ in the codomain. These represent what are conventionally called the coimage and cokernel, respectively.}
     Two subspaces form the domain $W$:
     \begin{equation*}
         \operatorname{ker}(L) = \{x\in W\mid L(x) = 0\}\quad \operatorname{coker}(L) = \{x\in W\mid \forall x^{\prime}\in\operatorname{ker}(L).\;\langle x,x^{\prime} \rangle_{W} = 0\}.
     \end{equation*}
     Further $W\cong\operatorname{ker}(L)\oplus \operatorname{coker}(L)$.
     We recall that $\oplus$ is the direct sum, a special pair value (see~\cite{Dummit2003-np}).
     Similarly, we have $V\cong\operatorname{image}(L)\oplus \operatorname{coimage}(L)$ where:
     \begin{equation*}
         \operatorname{image}(L) = \{y\in V\mid \exists x\in W.\; L(x)=y\}\quad 
         \operatorname{coimage}(L) = \{y\in V\mid \forall y^{\prime}\in\operatorname{image}(L).\; \langle y,y^{\prime}\rangle_{V} = 0\}.
     \end{equation*}

\end{defn}
Function types represent dependent functions: at each input, the function produces not only a value, but a type that the value inhabits. 
The $x$ in our dependent function type  is a placeholder for a term that will be used in one of the four fundamental subspaces (as opposed to a type variable).
For this type, we should keep in mind the idea of a function (vector field) producing values in the tangent space, as the tangent space changes at each point.

\paragraph{Terms:} Most of \GLang{}'s terms can be grouped by space/type, with relatively few terms associated to each space.
Via the vector space axioms, every space has access to a zero, scaling, scalars, and sums.
Dual spaces, formally the space of linear maps from a space to the real numbers, give us access to linear evaluation and abstraction whereas a dependent function type requires non-linear evaluation and non-linear function abstraction.
Since these operate via dual spaces and tensor products instead of linear maps, we utilize the following idea:
\begin{prop}[Linear maps and contraction]

\label{prop:corelin}
Let $V_0$ and $V_1$ be finite-dimensional vector spaces.
Then there exists a canonical isomorphism
$\phi:V_0^{\star}\otimes V_1
\longrightarrow \operatorname{Hom}(V_0,V_1)$.
For $A\in V_0^{\star}\otimes V_1$ and $v\in V_0$, let $\contractN{}{A}{v}$ represent evaluation via this isomorphism: $ \contractN{}{A}{v}=\phi(A)(v)\in V_1$.

\end{prop}
From the four fundamental subspaces, we feature $\lembd{}_{\ls}$, a subspace embedding map from the space to the domain or codomain of the linear map.
These spaces are implicitly related to the pseudo-inverse of a linear map $L$:
\begin{prop}[First Isomorphism Theorem, Pseudo-inverse]

\label{prop:core}
    For $L\in W^{\star}\otimes V$:
    \begin{enumerate}
    \item $L$ is an isomorphism from $\operatorname{coker}(L)$ to $\operatorname{image}(L)$.
    \item Since any subspace $A$ of an inner product space $B$ has a unique least-squares projection, denoted $P_{A}\in B^{\star}\otimes A$, there is a unique pseudo-inverse for any linear map given by $L^{\dagger} = L^{-1}\circ P_{\operatorname{image}(L)}$. Additionally,
    \begin{equation}
        L^{\dagger} = \lambda v\colon V.\; \argmin_{w\in W}\left\{\norm{w}_{W}\colon \norm{\contractN{}{L}{w} - v }_{V} = \min_{w^{\prime}\in W}\norm{\contractN{}{L}{w^{\prime}} - v}_{V}\right\}
    \end{equation}

\end{enumerate}

\end{prop}
Via this, we can avoid many additional constructions (\textit{e.g.}, a tuple/direct sum accessor or various symmetries associated to tensor products).
We discuss this in more detail in \cref{sec:nas}.
Similarly, we also provide each space with a distinguished inner product; though this inner product can be constructed in the language (\cref{sec:inner}), we provide it because the inner product is key to the semantics of the pseudo-inverse and subspaces.

\paragraph{Terms and Types:} Several terms in \cref{fig:grassDef} feature spaces with an alternative color: $\els{}$. 
The alternative coloration marks that these terms morally ought not to need these and we omitted them in our examples. 
To handle dependent types and polymorphism, we could follow dependent ML~\cite{dependentML}.
We reserve polymorphism for later work.

\paragraph{Kinds} We have two kinds to distinguish space values, $\lkbase{}$, from dependent functions $\lknto{\lkn{0}}{\lkn{1}}$.

\subsubsection{Natural Additional Syntax}
\label{sec:nas}
To further introduce \GLang{} and facilitate succinct auxiliary definitions, we define some additional syntax.
We note that many of these definitions are often built into other languages, which illustrates that \GLang{} is small.
We use a basic pattern for these definitions, which we introduce by asking how to destruct a direct sum (pair value). The answer is to define an embedding and then pseudo-invert it.
Given spaces $\lsn{0}$ and $\lsn{1}$, our typing rules admit direct sum embedding functions:
\begin{equation}
\label{eq:emtupl}
    \lkembedpp{0} := \lllambdaArg{x}{x \losum \lvzero}\quad\text{and}\quad \lkembedpp{1} := \lllambdaArg{x}{\lvzero \losum x}.
\end{equation}
Then we can define:
\begin{equation*}
    \lkprojpp{i} := {\lkembedpp{i}}^{\lpseudo{}}.
\end{equation*}
If we consider $\lR{}\losum{}\lR{}$, then $\lkembedpp{0}$ is identifiable with the vector $\begin{bmatrix} 1 \\ 0 \\ \end{bmatrix}$ (via appropriate choice of basis) and then the pseudo-inverse of this in the usual inner product is $\begin{bmatrix} 1 & 0 \end{bmatrix}$, the action of which extracts the first element of a vector in $\mathbb{R}^{2}$.
We omitted many common operations from our syntax because they can be recovered in this manner.
For example, we can define projection to one of the four fundamental subspaces via:
\begin{equation*}
    \lproj := \lembd{}^{\lpseudo{}}.
\end{equation*}
Similarly, we can define the musical isomorphisms that map primal values to dual values and vice versa via:
\begin{equation*}
  \lflat{} := \lllambdaArg{x}{\linner{}(x)} \quad\text{and}\quad
  \lsharp{} := \lflat{}^{\lpseudo{}}.
\end{equation*}
We could also recover the various mappings required by symmetric monoidal categories for more general tensor product spaces, but in the presence of duality, we are especially interested in defining curry and uncurry operations.
We start by assuming syntactic right associativity of tensor products: $\lsn{0}\loprod\lsn{1}\loprod\lsn{2}:= \lsn{0}\loprod(\lsn{1}\loprod\lsn{2})$. Similarly, we define multi-variable linear lambdas: $\lllambdaArg{x_{0},\dots, x_{n}}{\lee} := \lllambdaArg{x_{0}}{\lllambdaArg{x_{1}}{\cdots\lllambdaArg{x_{n}}{\lee}}}$ and successive linear applications via: $\lee(\len{0},\dots, \len{n}) := \lee(\len{0})(\len{1})\cdots{}(\len{n})$.
Lastly, we need linear evaluation at argument $k$: $\len{0}^{(k, n)}(\len{1}) := \lllambdaArg{x_{0},\dots, x_{k-1}, x_{k+1},\dots, x_{n}}{\len{0}(x_{0},\dots, x_{k-1}, \len{1}, x_{k+1},\dots,x_{n})}.$
Given a tensor product $\lsn{0}\loprod\cdots\loprod\lsn{k}\loprod\cdots\loprod\lsn{m}$, we can promote an input $\lsn{k} = (\lsnp{0}\loprod\cdots\lsnp{n})^{\lstar{}}$ to ${\lsnp{0}}^{\lstar{}}\loprod\cdots{\loprod\lsnp{n}}^{\lstar{}}\loprod\lsn{0}\loprod\cdots\loprod{}\lsn{k-1}\loprod\lsn{k+1}\loprod\cdots\loprod\lsn{m}$ via a curry map and we can undo this with the inverse
\begin{equation*}
    \lcurry_{k, n, m} :=\lllambdaArg{x}{\lllambdaArg{x_{0},\dots, x_{n}}{x^{k, m}(x_{0}\loprod\cdots\loprod x_{n})}}\quad\text{and}\quad \luncurry_{k, n, m} := {\lcurry_{k, n, m}}^{\lpseudo{}}.
\end{equation*}
The inverse allows us to turn a multi-linear function into a linear function.
We note that this development of $\lcurry$ and $\luncurry$ mirrors that found in proofs: the former is directly from the definition and inverting it to form $\luncurry$ is the core work of the proofs of the key properties (symmetry, associativity, and other common, often built-in, tensor operations)\footnote{See~\cite{Dummit2003-np} or \url{https://kconrad.math.uconn.edu/blurbs/linmultialg/tensorprod.pdf}}.
An immediate application is that we can define operations like Kronecker products, block matrices, and trace, shown here:
\begingroup
\setlength{\belowdisplayskip}{0pt}
\setlength{\belowdisplayshortskip}{0pt}
\begin{equation*}
    \ltr{} := (\luncurry_{0, 1, 1})(\lllambdaArg{x_{0}, x_{1}}{x_{0}(x_{1})}).
\end{equation*}
\endgroup
\begingroup
\setlength{\intextsep}{6pt}
\begin{table}[htb]
\centering
\caption{Derived operators and notation in \cref{sec:nas}.}
\label{tab:natural-additional-syntax}
\small
\begin{tabular}{@{}ll@{}}
\toprule
Notation & Meaning \\
\midrule
$\lkembedpp{i},\lkprojpp{i}$ ($i=0,1$); $\lproj$ & Sum injection/projection; subspace projection. \\
$\lflat,\lsharp$ & Convert primal to dual / dual to primal. \\
$\lcurry_{k,n,m},\luncurry_{k,n,m}$ & Split / merge tensor arguments. \\
$\lllambdaArg{x_0,\ldots,x_n}{\lee}$; $\lee(\len{0},\ldots,\len{n})$; $\len{0}^{(k,n)}(\len{1})$ & Construct; evaluate; evaluate slot $k$. \\
$\ltr$ & Endomorphism trace. \\
\bottomrule
\end{tabular}
\end{table}
\endgroup

\subsubsection{Inner Products}
\label{sec:inner}
In our semantics, we will not explicitly select an inner product.
However, given the obvious choices for the real numbers and the empty space, an inner product follows for the rest of the spaces.
We formalize this choice in \cref{fig:inner} via defining the inner products with our syntax.
For example, $\gleqn{iof.real}$. For a dual space, we use the musical isomorphism to recover primal vectors: $\gleqn{iof.dual}$. The remaining clauses in \cref{fig:inner} proceed inductively, using the destructors in \cref{sec:nas}.
This demonstrates the expressivity of our language and we use it as a proof tool (\textit{e.g.}, we can prove soundness of our semantics without this specific choice of inner product and we can easily show some bases are orthonormal in our choice of inner products).

\subsection{Types and Typing}
Our judgments use a typing context $\lctx$, split into linear variables
$\lDelta$ and non-linear variables $\lepsilon$ to make the following judgments:
\[
\begin{array}{ll}
\lctx\vdash\lty{}\colon\tvalid{}
& \text{$\lty{}$ is a well-formed type with kind $\tvalid{}$} \\[3pt]
\lctx\vdash\lee\colon\lty
& \text{$\lee$ has type $\lty$ and is linear in $\lDelta$} \\[3pt]
\lctx\vdash\lsn{0}\leq\lsn{1}
& \text{$\lsn{0}$ is a subspace of $\lsn{1}$.}
\end{array}
\]

Kinding judgments validate spaces and separate spaces from dependent functions:
\begin{center}
{\tiny
\(\begin{array}{@{}c@{\qquad}c@{\qquad}c@{}}
\glrule{sValid.empty} & \glrule{sValid.coimg} & \glrule[2]{tValid.piTy}
\end{array}\)
}
\end{center}

For a kernel, for example, we need that $\lee$ is a linear map: $\lctxnle\vdash\lee\colon\lsn{0}^{\lstar{}}\loprod\lsn{1}$.
The complete kinding judgments are in \cref{fig:typingSpace}.

Our term typing judgments seek to ensure basic typing properties, the meaning of special terms, and linearity.
For example, the following judgments ensure inner product, inclusion, and dependent functions behave correctly: 
\begin{center}
{\tiny
\(\begin{gathered}
\begin{array}{@{}c@{\qquad}c@{\qquad}c@{}}
\glrule{hasType.inner} & \glrule{hasType.embed} & \glrule{subType.subKer}
\end{array}\\[3pt]
\begin{array}{@{}c@{\qquad}c@{}}
\glrule[2]{hasType.nllam} & \glrule[2]{hasType.appNonlinear}
\end{array}
\end{gathered}\)
}
\end{center}

Up to now, we have avoided situations where linear dependence was appropriate, leaving our linear context empty.
We are able to find linearity in terms via rules that partition the linear context, terminating in a single resource as the environment for a linear variable.
We use $\ldisun$ to assert a disjoint union of environments, following the patterns of other linear type-checking systems~\cite{diaz2024linear,Bernardy2021,valiron2014finite,velez2026compiling}:
\begin{center}
{\tiny
\(\begin{gathered}
\begin{array}{@{}c@{\qquad}c@{\qquad}c@{\qquad}c@{}}
\glrule[3]{hasType.llamReal} & \glrule[2]{hasType.appLinear} &
\glrule{hasType.varLin} & \glrule{hasType.varNlin}
\end{array}
\end{gathered}\)
}
\end{center}

However, three rules vary a bit from the linear typing literature and the linear types for linear algebra literature. The promotion rule requires we check that variables used in space constructions are not promoted. The direct sum and zero rules allow for the definition in \eqref{eq:emtupl} and together represent the additive conjunction ($\mathbin{\&}$) of linear logic with its unit $\top$~\cite{Girard1987}:
\begin{center}
{\tiny
\(\begin{gathered}
\begin{array}{@{}c@{\qquad\qquad}c@{\qquad\qquad}c@{}}
{\let\displaybigcup\bigcup\renewcommand{\bigcup}{\textstyle\displaybigcup\nolimits}\glrule[2]{hasType.promote}}
& \glrule[2]{hasType.dsumE}
& \glrule{hasType.zero}
\end{array}
\end{gathered}\)
}
\end{center}

The complete typing and subspace judgments are in \cref{fig:typing} and \cref{fig:sub}.
As a final matter, we must ensure our contexts are  well formed, which we denote as $\vdash \lctx$.
\begin{defn}[$\gldefnsig{inductiveCtxWF}$, well-formed typing context]
\label{def:wfc}
We have $\vdash(\emptyset,\emptyset)$. Given $\vdash(\lDelta,\lepsilon)$ and a name $x$ fresh in both components,
$\vdash(\lDelta\ldisun\{x\colon\ls{}\},\lepsilon)$ if $\fv{\ls{}}\subseteq\operatorname{dom}(\lepsilon)$ and $(\emptyset,\lepsilon)\vdash\ls{}\colon\svalid$;
and $\vdash(\lDelta,\lepsilon\ldisun\{x\colon\lty{}\})$ if $\fv{\lty{}}\subseteq\operatorname{dom}(\lepsilon)$ and $(\emptyset,\lepsilon)\vdash\lty{}\colon\lkn{}$ for some kind $\lkn{}$.
The corresponding construction rules are in \cref{app:context-formation}.

\end{defn}

\subsection{Semantics}
To define our semantics, we ought to take a term $\lee$ with a type $\lty$ and produce a tuple of the set-theoretic vector, the real Hilbert space or dependent function space, and in the former case the data for the Hilbert space (\textit{e.g.}, the definition of the addition and inner product operations).
However, we will describe only the vector and the set corresponding to our type, as most of the rest is clear from context.
We note that the semantics here  are parametric in the choice of inner product, which later is fixed via \cref{sec:inner}.
Given a semantic environment $\sigma$ mapping names to vectors or dependent functions, we define two mutually recursive semantic functions, $\SE$ and $\ST$, on kinding and typing derivations:
{\setlength{\arraycolsep}{3pt}
\[
\begin{array}{ll}
\ST(-,\sigma)\colon
(\lctx\vdash\ls{}\colon\svalid{})\to
\mathsf{Hilb}_{\mathbb R}^{\mathrm{fd}}
& \text{interprets a space as a finite-dimensional real Hilbert space}
\\[3pt]
\ST(-,\sigma)\colon
(\lctx\vdash\lty{}\colon\tvalid{})\to\mathsf{Set}
& \text{gives the vectors or dependent functions inhabiting $\lty$.}
\\[3pt]
\SE(-,\sigma)\colon
(\lctx\vdash\lee\colon\lty)\to\SemT{\lty}{\sigma}
& \text{interprets a term as an element of its type's interpretation.}
\end{array}
\]
}

We provide representative cases for $\ST$, to show how $\ST$ uses the derivation in the case of the four fundamental subspaces and the handling of dependent types.

\par\noindent
{\scriptsize
\setlength{\tabcolsep}{5pt}
\renewcommand{\arraystretch}{1.2}%
\begin{tabular}{@{}c@{}}
\makebox[\linewidth]{\hfill$\gleqn{denotSpaceAux.real}$\hfill$\gleqn{denotSpaceAux.ker}$\hfill$\gleqn{denotSpaceAux.coker}$\hfill}\\[3pt]
\makebox[\linewidth]{\hfill$\gleqn{denotSpaceAux.dual}$\hfill$\gleqn{denotSpaceAux.tprod}$\hfill$\gleqn{denotTyAux.piTy}$\hfill}
\end{tabular}\par
}\smallskip
The complete semantics for $\ST$ are  in \cref{fig:spaceSemantics}.
For $\SE$, we wish to illustrate a few representative cases:
\par\noindent
{\scriptsize
\setlength{\tabcolsep}{5pt}
\renewcommand{\arraystretch}{1.2}%
\begin{tabular}{@{}c@{}}
\makebox[\linewidth]{\hfill$\gleqn{denotTermAux.var}$\hfill$\gleqn{denotTermAux.real}$\hfill$\gleqn{denotTermAux.zero}$\hfill}\\[3pt]
\makebox[\linewidth]{\hfill$\gleqn{denotTermAux.inner}$\hfill$\gleqn{denotTermAux.embed.ker}$\hfill$\gleqn{denotTermAux.pinvT}$\hfill}\\[3pt]
\makebox[\linewidth]{\hfill$\gleqn{denotTermAux.add}$\hfill$\gleqn{denotTermAux.smul}$\hfill}
\end{tabular}\par
}\smallskip

These emphasize our use of the external inner product, basic mathematical operations of the Hilbert space, and the inclusion operation via a simple linear embedding (recall the judgment $\lctxnle\vdash\lsn{0}\leq \lsn{1}$ and \cref{def:ffs}) and that we need access to the pseudo-inverse (\cref{prop:core}).
Finally, we provide some (non-)linear evaluation and abstraction semantics:
\begin{center}
{\scriptsize
\setlength{\tabcolsep}{5pt}
\renewcommand{\arraystretch}{1.2}%
\begin{tabular}{@{}c@{\qquad}c@{}}
$\gleqn{denotTermAux.llam}$ & $\gleqn{denotTermAux.nllam}$\\[6pt]
\multicolumn{2}{c}{$\gleqn{denotTermAux.appL}$}\\[3pt]
\multicolumn{2}{c}{$\gleqn{denotTermAux.appN}$}
\end{tabular}
}
\end{center}
The main contrast to observe is that we use $\contractN{}{}{}$ via \cref{prop:corelin} to identify spaces like $V^{\star}\otimes W$  with linear maps that can be evaluated.
We provide a complete definition of $\SE$ in  \cref{fig:vectorSemantics}.

Before proceeding, we need to link $\sigma$ to $\lctx$.

\begin{defn}[$\lctx\vdash\sigma$]
\label{def:lctxsigma}
We write $\lctx\vdash\sigma$ iff, for every
$(x,\lty{})\in\lDelta\ldisun\lepsilon$,
$\sigma(x)$ is defined and $\sigma(x)\in\SemT{\lty{}}{\sigma}$.
\end{defn}

\section{Meta Theory: Part 1}
\label{sec:meta1}
Our denotational semantics is sound and we prove this in two steps. 
First, we assume a choice of inner product in \cref{lem:soundLemma} and then we prove that our choice of inner product is sound in \cref{lem:innerLema}.
For the first lemma, we wrap up several results together that are mostly proved by mutual induction on the typing derivation and validity derivation.
\begin{lem}[Soundness up to inner product]
\label{lem:soundLemma}
Suppose $\vdash\lctx$, $\lctx\vdash\lee{}\colon\lty{}$, and $\lctx\vdash\sigma$.
Then $(\emptyset,\lepsilon)\vdash\lty{}\colon\lkn{}$ for some $\lkn{}$.
For a fixed choice of inner products, $\Sem{\lee{}}{\sigma}$ and $\SemT{\lty{}}{\sigma}$
are independent of the typing derivation.
Moreover:
\begin{enumerate}
\item $\Sem{\lee{}}{\sigma}\in\SemT{\lty{}}{\sigma}$ is a well-defined vector or dependent function;
\item for every $(x,\ls{})\in\lDelta$, $y\mapsto\Sem{\lee{}}{\sigma[x\mapsto y]}$
is linear on $\SemT{\ls{}}{\sigma}$, pointwise for dependent-function results;
\item $\SemT{\lty{}}{\sigma}$ is a well-defined Hilbert space or set of dependent functions.
\end{enumerate}

\end{lem}
\begin{proof}
See proof of \cref{lem:soundLemmaA}.
\end{proof}

To make our semantics sound, we now only need to verify that our choice of inner product (\cref{sec:inner} for definition of $\iof{\ls}$) is sound.

\begin{lem}[Inner product choice is sound]
\label{lem:innerLema}
Suppose $\vdash\lctxnle$, $\lctxnle\vdash\ls{}\colon\svalid$,
and $\lctxnle\vdash\sigma$.
Then:
\begin{enumerate}
\item $\lctxnle\vdash\iof{\ls{}}\colon\ls{}^{\lstar}\loprod\ls{}^{\lstar}$;
\item $\Sem{\iof{\ls{}}}{\sigma}$ is an inner product on $\SemT{\ls{}}{\sigma}$.
\end{enumerate}

\end{lem}
\begin{proof}
See \cref{lem:innerLemap}.
\end{proof}

\section{Defining \GLang{} Part 2: Bases and Operational Semantics}
To define our operational semantics, we require a notion of basis.
Our denotational semantics did not need such a notion, but it is also very far from anything a computer would implement.
When users get to the point of using a matrix multiplication or Einstein index notation\footnote{admittedly, this depends on how one defines index notation; some definitions are considered coordinate-independent but most languages do not enforce this}, they have selected a basis and used it to operationalize linear algebra.
Thus, to define our operational (basis-dependent) semantics, we first define bases (\cref{sec:syntaxbasis}), basis type checking (\cref{sec:basisCheck}), and basis denotational semantics (\cref{sec:basisDenote}).
Then we define our operational semantics (\cref{sec:operational}).
\subsection{Bases}
\label{sec:def2b1}
\subsubsection{Syntax for Bases}
\label{sec:syntaxbasis}
\begin{figure}[H]
    \centering
    \scriptsize
    \setlength{\tabcolsep}{4pt}
    \begin{tabular}{@{}ll@{\qquad}ll@{}}
    \toprule
    $n\in\mathbb{N}$ & Constant Number
      & $x\in\mathrm{String}$ & Variable\\
    \midrule
    $\lb{}::=\lbzero{}$ & Empty Basis
      & $\mid\pBasisC{\ls{}}{\lbn{0}}{\lbn{1}}$ & Projected Basis\\
    $\mid\lbone{}$ & Scalar Basis
      & $\mid\iBasisC{\lb{}}$ & Image basis\\
    $\mid\lb{}^{\lbstar}$ & Dual Basis
      & $\mid\llpl{}\len{0},\dots,\len{n}\llpr{}_{\lb{}}$ & Basis from expressions for a space\\
    $\mid\lbn{0}\lbosum\lbn{1}$ & Direct Sum of Bases
      & $\mid\frameB{x}{\lbn{0}}{\lbn{1}}$ & Abstracted basis or Frame\\
    $\mid\lbn{0}\lboprod\lbn{1}$ & Tensor Product of Bases
      & & \\
    \bottomrule
    \end{tabular}
    \caption{Syntax of \GLang{} for bases, a sequence of linearly independent vectors that span the space.}
    \label{fig:basisDef}
\end{figure}

By basis, we mean a sequence (ordered) of linearly independent vectors that span our space.
We provide a special syntax in part to isolate these properties, to avoid writing sequences of vectors, and to provide a separation between coordinates and vectors.
The syntax for bases is depicted in \cref{fig:basisDef}; the syntax in some respects mirrors that for spaces, as we have a basis for the empty space, the space $\lR{}$, the dual space, the tensor product space, and the direct sum spaces.
The $\pBasisC{\ls{}}{\lbn{0}}{\lbn{1}}$ constructs a basis for the four fundamental subspaces in terms of bases for the domain and codomain, as we will explain further in the semantics for bases.
The $\frameB{x}{\lbn{0}}{\lbn{1}}$ defines a discrete frame: given a basis $\lbn{0}$, we construct a basis $\lbn{1}$ using the variable $x$. 
The basis $\lbn{0}$ indicates the space of $x$ and the basis that will be used to represent $x$.
To properly model these, we need a fixed-point construction of the class of higher-order discrete frames:
\begin{defn}[Discrete (Vector) Frame]

\label{def:bunframe}
Suppose $f$ is a mapping from points in a set $X$ to real finite-dimensional vector spaces $V_{x}$, \textit{i.e.}, the codomain is the set $\{V_{x}\colon x\in X\}$.
We could  write $f\colon \Pi_{x\colon X} V_{x}$.
Given a finite-dimensional vector space $V$, let  $V^{\circ }$ be the set of bases for $V$.
Then a \textbf{discrete vector frame } is a mapping $g\colon (\Pi_{x\in X} V_{x}^{\circ})$ such that for all $x\in X$, $g(x)$ is a basis for $f(x)$.
We could denote this as $g\colon (\Pi_{x\in X} V_{x})^{\circ}$.
A more general discrete frame emerges naturally.
Given $f\colon \Pi_{y\colon Y}\Pi_{x\colon X} V_{y,x}$ where $V_{y,x}$ is a vector space, then the corresponding \textbf{discrete frame} is defined recursively as a mapping $g\colon \Pi_{y\in Y} (\Pi_{x\colon X} V_{y,x})^{\circ}$ where given $y\in Y$, $g(y)$ is a discrete vector frame for $f(y)$.
Then given some $f\colon \Pi_{y\colon Y}B_{y}$, we say that $g$ is a discrete frame for $f$, if $g\colon \Pi_{y\colon Y}B_{y}^{\circ}$.

\end{defn}
Our syntax treats the frame as a basis but the typing judgments prevent it from being used in any of the other constructors.
Thus we have, for each space, at least one corresponding basis constructor, and these are sufficient to define our operational semantics.
However, to examine possibilities for optimizing code or computer algebra rewriting in an exterior algebra context, we include two other useful constructs.
First, we include $\iBasisC{\lb{}}$ to represent a basis for an image space produced from a basis for a corresponding cokernel space.
Second, we include a construction to build a basis from a collection of terms and another basis to draw from in case the terms are not a basis: $\llpl{} \len{0},\dots, \len{n}\llpr{}_{ \lb{}}$.
These were essential in our examples.

\subsubsection{Typing for Bases}
\label{sec:basisCheck}
Our typing rules for bases  exist to ensure that bases are in the correct space, which requires functions to relate bases to types.
These are straightforward and in the appendix (\cref{fig:spbasis} and \cref{fig:basisValid}):
\[
\begin{array}{ll}
\lctxnle\vdash\lb{}\colon\bvalid
& \text{$\lb{}$ is a valid basis or discrete frame for $\tof{\lb{}}$}
\\[3pt]
\tof{-}\colon\mathsf{Basis}\to\mathsf{Ty}
& \text{returns the type represented by a basis or discrete frame}
\\[3pt]
\sof{-}\colon\mathsf{Basis}\to\mathsf{Space}
& \text{returns the space represented by a space basis.}
\end{array}
\]
Here $\mathsf{Basis}$, $\mathsf{Ty}$, and $\mathsf{Space}$ denote
the syntactic categories.
We show the following examples:
\begin{center}
{\scriptsize
\expandafter\patchcmd\csname gl@rule@prem@basisValid.terms\endcsname
  {\\ n \geq 0}{}{}{\PackageError{GLang}{Cannot hide the term-basis index bound}{Check the generated premise list.}}
\begin{tabular}{@{}c@{\quad}c@{}}
$\glrule[2,1]{basisValid.projKer}$ & $\glrule[2,1]{basisValid.image}$\\[6pt]
$\glrule[2,2]{basisValid.terms}$ & $\glrule[2]{basisValid.frame}$
\end{tabular}}
\end{center}

\subsubsection{Mathematical Preliminaries for Bases}
\label{sec:mpb}
To define our semantics for bases, we need several auxiliary functions to utilize matrix factorization in our semantics without ambiguities.
In particular, the GSVD, \cref{def:gsvd}, gives infinitely many choices of bases for the four fundamental subspaces, which is not acceptable for our semantics.
Even routines such as orthogonalization have vagueness in their definitions due to how one handles linearly dependent vectors (\textit{i.e.}, pivoting).

We first define the lexicographically minimal largest independent subsequence:
\begin{defn}[$\LexInd{}$]
\label{def:lex}
Given a finite sequence $(v_i)_{i<n}$, define
$\LexInd((v_i)_{i<n})=(v_{i_j})_{j<r}$ by choosing the lexicographically least
increasing index tuple $(i_j)_{j<r}$ whose selected vectors are linearly independent, with $r$ maximal .

\end{defn}

Next, we constrain how we will use orthogonalization:
\begin{defn}[$\GramSchdmit{}$]

\label{def:gs}
    We define $\GramSchdmit{}(v_{0},\dots, v_{n})$ as the normalized Gram--Schmidt orthogonalization of the sequence provided that $v_{0},\dots, v_{n}$ are linearly independent in a known inner product space.

\end{defn}

Now, to define an algorithm that picks out specific subsequences of singular vectors, allowing us to select an SVD, we need an auxiliary result that we immediately use to define our choice.
\begin{prop}[Spectral Projector]
\label{prop:sp}
Let a positive semi-definite linear map $F\in V^{\star}\otimes V$ be given. Suppose that $\lambda_{0},\dots,\lambda_{m}$ are the eigenvalues with duplicates discarded. Let $p_{\lambda_{j}}(A) = \prod_{i=0, i\neq j}^{m} \frac{A-\lambda_{i}I}{\lambda_{j} - \lambda_{i}}$.  Then $p_{\lambda_{j}}(F)\in V^{\star}\otimes V$ and projects $V$ onto $V_{\lambda_{j}} = \{v\colon F(v) = \lambda_{j}v\}$.
\end{prop}
\begin{proof}
See \cref{prop:spA}.
\end{proof}

\begin{defn}[Filtered singular vectors]
\label{def:fsv}
For $A\in V^\star\otimes W$, ordered bases $v=(v_i)_{i<n}$ of $V$
and $w=(w_i)_{i<m}$ of $W$, $f:\mathbb R\to\operatorname{bool}$,
and $d\in\{\operatorname{Left},\operatorname{Right}\}$, put
$S_L=A^\star\circ A$ and $S_R=A\circ A^\star$.
Here $A^\star=\sharp_V\circ(\lambda\phi.\,\phi\circ A)\circ\flat_W$.
For a positive semidefinite operator $S$, write
$P_\lambda(S)=
\prod_{\mu\in\operatorname{spec}(S)\setminus\{\lambda\}}
(S-\mu I)/(\lambda-\mu)$ when $\lambda\in\operatorname{spec}(S)$, the eigenvalues of $S$,
and $P_\lambda(S)=0$ otherwise.
Define
$B_{\operatorname{Left},\lambda}
=\GramSchdmit\bigl(\LexInd((P_\lambda(S_L)(v_i))_{i<n})\bigr)$.
For $\lambda>0$, put
$B_{\operatorname{Right},\lambda}
=(A(u)/\sqrt{\lambda})_{u\in B_{\operatorname{Left},\lambda}}$;
at zero, put
$B_{\operatorname{Right},0}
=\GramSchdmit\bigl(\LexInd((P_0(S_R)(w_i))_{i<m})\bigr)$.
Let $(\lambda_j)_{j<r}$ list the distinct elements of
$\operatorname{spec}(S_L)\cup\operatorname{spec}(S_R)$
in decreasing order. Then
\[
 \FSV(A,f,d,v,w)
 :=\mathop{++}\nolimits_{j<r,\,f(\lambda_j)=\operatorname{true}}
 B_{d,\lambda_j},
\]
where $++$ concatenates the selected blocks.
\end{defn}
The validity is in the proof of \cref{lem:denoteBasisSound}.

\subsubsection{Basis-Independent Semantics for Bases}
\label{sec:basisDenote}

A basis is nothing but a linearly independent and spanning sequence of vectors.
And in the case of dependent functions, a basis is a discrete (vector) frame (\cref{def:bunframe}).
Thus, an appropriate denotational semantics, $\SB$, should, analogously to $\ST$, map a derivation $\lctxnle\vdash\lb{}\colon\bvalid$ to a basis or discrete vector frame given an environment $\sigma$.
Basis validity entails
$\lctxnle\vdash\tof{\lb{}}\colon\tvalid$
(\cref{lem:denoteBasisSound}). Thus:
\[
\SB(-,\sigma)\colon
(\lctxnle\vdash\lb{}\colon\bvalid)
\to\bigl(\SemLUS{\tof{\lb{}}}{\sigma}\bigr)^{\circ}.
\]
Bases are ordered sequences; $+\!\!+$ denotes sequence concatenation, and comprehensions follow lexicographic order: 
\begin{center}
{\scriptsize
\begin{tabular}{@{}c@{}}
$\gleqn{denotBasis.dsum}$\\[3pt]
\makebox[\linewidth]{\hfill$\gleqn{denotBasis.one}$\hfill$\gleqn{denotBasis.tprod}$\hfill}
\end{tabular}}
\end{center}

Semantics for other bases rely more closely on some of the underlying mathematical ideas, starting with the explicit duality relation for dual bases:
\begin{center}
{\scriptsize $\gleqn{denotBasis.dual}$}
\end{center}
Then for image bases and frames we need to use \cref{prop:core} and \cref{def:bunframe}:
\begin{center}
{\scriptsize
\begin{tabular}{@{}c@{}}
$\gleqn{denotBasis.image}$\\[3pt]
$\gleqn{denotBasis.frame}$
\end{tabular}}
\end{center}
We note that for the frame, many might expect $x'\colon\SemBB{\lbn{0}}{\sigma}$, but we have no basis variables; we use terms. Operationally, however, a non-linear function will need a basis for the domain and a different choice of basis will change the operational result.

Finally, for the four fundamental subspaces' projected bases and the term basis, we need the machinery of \cref{def:fsv} and \cref{def:lex}:
\begin{center}
{\scriptsize
\begin{tabular}{@{}c@{}}
$\gleqn{denotBasis.proj.ker}$\\[3pt]
$\gleqn{denotBasis.terms}$
\end{tabular}}
\end{center}
All basis semantics are gathered together in \cref{fig:basisSemantics}.
We utilize these observations to prove \cref{lem:denoteBasisSound}:
\begin{lem}[Basis Denotational Semantics is Sound]
\label{lem:denoteBasisSound}
{\raggedright Suppose $\vdash\lctxnle$ and $\lctxnle\vdash\lb{}\colon\bvalid$.
\begin{enumerate}
\item $\lctxnle\vdash\tof{\lb{}}\colon\lkn{}$ for some $\lkn{}$.
\item If also $\lctxnle\vdash\sigma$, then $\SemBB{\lb{}}{\sigma}$ is a basis or discrete frame
for $\SemT{\tof{\lb{}}}{\sigma}$. Its denotation is independent of the typing derivation.
\end{enumerate}\par}

\end{lem}
\begin{proof}
See proof of \cref{lem:denoteBasisSoundA}.
\end{proof}

\subsubsection{Standard Basis}
\label{sec:stdbasis}

Before we move on to define our operational semantics, we need a standard orthogonal basis for every space under our choice of inner product,~\cref{fig:inner}.
We emphasize that this is a convenience notion and not a canonical choice. 
We define this in \cref{fig:stdBasis}, with representative clauses for real, tensor-product, and projected bases:
\begin{center}
{\scriptsize
\begin{tabular}{@{}c@{}}
\makebox[\linewidth]{\hfill$\gleqn{basis.stdSpace.real}$\hfill$\gleqn{basis.stdSpace.tprod}$\hfill}\\[3pt]
$\gleqn{basis.stdSpace.ker}$
\end{tabular}}
\end{center}
For our operational semantics, though, the standard basis requiring the typing context is inconvenient, but thankfully we can equivalently check if a basis is the standard basis, which we can do without the typing derivation via $\isStdBasis{}$ defined in \cref{fig:isStdBasis}. %
Via a simple induction on the structure of the standard basis, one can prove that this basis is orthonormal with respect to our chosen inner product.
We can use this to prove that our standard basis detection works via $\isStdBasis{}$.
\begin{lem}[Standard Basis is orthonormal]
\label{lem:stdBasisOrthonormal}
{\raggedright Suppose $\vdash\lctxnle$ and $\lctxnle\vdash\lty{}\colon\lkn{}$. Then:
\begin{enumerate}
\item $\lctxnle\vdash\stdBasisDCtx{\lty{}}\colon\bvalid$,
$\tof{\stdBasisDCtx{\lty{}}}=\lty{}$, and $\isStdBasis{\stdBasisDCtx{\lty{}}}$.
\item If also $\lctxnle\vdash\sigma$, then $\SemBB{\stdBasisDCtx{\lty{}}}{\sigma}$ is an orthonormal basis or
discrete frame for $\SemT{\lty{}}{\sigma}$ under the chosen inner products.
\item If $\lctxnle\vdash\lb{}\colon\bvalid$, $\tof{\lb{}}=\lty{}$, and
$\isStdBasis{\lb{}}$, then $\stdBasisDCtx{\lty{}}=\lb{}$.
\end{enumerate}\par}

\end{lem}
\begin{proof}
See \cref{lem:stdBasisOrthonormalA}.
\end{proof}

\subsection{Basis-Dependent (Operational) Semantics}
\label{sec:operational}
We can now introduce our operational semantics.
Though our denotational semantics is much simpler, our operational semantics will capture how introducing coordinates allows mathematicians and scientists to implement linear algebra  with something like \NLang{} or PyTorch.
Thus, we call these semantics our coordinate-dependent semantics.
To make this precise, we will need to define an expansive notion of values that captures a subset of reasonable linear algebra machine primitives. 

\subsubsection{Values}
\begin{figure}[!htbp]
\scriptsize
\centering
\begin{minipage}[t]{0.48\textwidth}
\centering
\begin{tabular}[t]{rcll}
\toprule
$n$ & $\in$ & $\mathbb{N}$ & Constant Number\\
$x$ & $\in$ & String & Variable\\
\midrule
\lsv{} & $::=$ & $\glsyn{valSpace.empty}$ & Empty Space \\
      & $|$ & $\glsyn{valSpace.real}$ & Real Numbers\\
      & $|$ & $\glsyn{valSpace.dsum}$ & Direct Sum Space\\
      & $|$ & $\glsyn{valSpace.tprod}$ & Tensor Product Space\\
      & $|$ & $\glsyn{valSpace.ffs}$ & Masked Vector Subspace\\
\midrule
\lprop{} & $::=$ & $\glsyn{valFFSProp.kernel}\mid\glsyn{valFFSProp.cokernel}$ & Domain Subspace Labels\\
        & $|$ & $\glsyn{valFFSProp.image}\mid\glsyn{valFFSProp.coimage}$& Codomain Subspace Labels\\
\midrule
$\ltyv{}$ & $::=$ & $\glsyn{valTy.space}$ & Space \\
        & $|$ & $\glsyn{valTy.piTy}$ & Function \\
\midrule
$\lky{}$ & $::=$ & $\glsyn{valKind.base}$ & Space Kind\\
       & $|$ & $\glsyn{valKind.arrow}$ & Function Kind\\
\midrule
\leev{} & $::=$ & $\glsyn{valExpr.zero}$ & Zero Vector\\
& $|$ & $\glsyn{valExpr.scale}$ & Scalar-Vector Product\\
& $|$ & $\glsyn{valExpr.rlit}$ & Real Number\\
& $|$ & $\glsyn{valExpr.vadd}$ & Vector Sum\\
& $|$ & $\glsyn{valExpr.dsumE}$ & Direct Sum\\
\bottomrule
\end{tabular}
\end{minipage}
\hfill
\begin{minipage}[t]{0.50\textwidth}
\centering
\begin{tabular}[t]{rcll}
\toprule
\leev{} & $|$ & $\glsyn{valExpr.tprodE}$ & Tensor Product\\
& $|$ & $\glsyn{valExpr.delta}$ & One hot\\
& $|$ & $\glsyn{valExpr.idelta}$ & Identity matrix\\
& $|$ & $\glsyn{valExpr.ttv}$ & Tensor Times Vector \\
& $|$ & $\glsyn{valExpr.vtrans}$ & Transpose \\
& $|$ & $\glsyn{valExpr.vkrn}$ & Kronecker Product \\
& $|$ & $\glsyn{valExpr.vblk}$ & Block Diagonal Matrix \\
& $|$ & $\glsyn{valExpr.reshap}$ & Reshape\\
& $|$ & $\glsyn{valExpr.mtm}$ & Matrix Times Matrix \\
& $|$ & $\glsyn{valExpr.vsum}$ & Finite Family Sum\\
& $|$ & $\glsyn{valExpr.vlam}$ & Non-Linear Function\\
& $|$ & $\glsyn{valExpr.appV}$ & Function Eval\\
& $|$ & $\glsyn{valExpr.var}$ & Variable\\
& $|$ & $\glsyn{valExpr.pinv}$ & Pseudoinverse\\
& $|$ & $\glsyn{valExpr.gpinv}$ & Generalized Pseudoinverse\\
& $|$ & $\glsyn{valExpr.inv}$ & Inverse Form\\
& $|$ & $\glsyn{valExpr.svds}$ & Singular Values\\
& $|$ & $\glvop{GSVD}(X)(\lenv{0},\lenv{1},\lenv{2})$ & Generalized SVD\\
& & $X\in\{\mathrm{S},\mathrm{U},\mathrm{UI},\mathrm{V},\mathrm{VI}\}$ & Component\\
& $|$ & $\glsyn{valExpr.rrefRep}$ & RREF Representation\\
\bottomrule
\end{tabular}
\end{minipage}

\caption{Values for operational semantics.}
\label{fig:Values}
\end{figure}

 Our values are depicted in \cref{fig:Values}, built on the principle that we should lower to a machine that operates on arrays of numbers with \NLang{}-like primitives.
Due to the presence of non-linear operations and dynamic constructions,  we end up with a language that is roughly the lambda calculus with errors and \NLang{} features. 
In this language, spaces should correspond to array types (shapes), function types to dynamic shapes, functions to functions, and values in spaces to arrays.
Value variables $x_{\ltyv}$ and constructors retain their type and space annotations (Appendix~\ref{app:gen-values}).
To support our four fundamental subspaces or the pseudo-inverse in our value language, we will use the generalized singular value decomposition (G)SVD, which we provide here:
\begin{defn}[(G)SVD,~\cite{golub2013matrix,Stewart1990-mt}]

\label{def:gsvd}
Let $L\in W^{\star}\otimes V$ be given, where
$\dim W=m$ and $\dim V=n$, and let $\operatorname{rank}(L)=r\le \min(m,n)$.
After choosing bases for $W$ and $V$, we may represent $L$ as a matrix in
$\mathbb{R}^{n\times m}$. 
Let $M_W$ and $M_V$ denote (positive definite) matrices of the inner products on $W$ and $V$, respectively.
Then we may compute matrices
\[
\begin{gathered}
\GSVDU(L,M_W,M_V)\in\mathbb{R}^{m\times r},\qquad
\GSVDUI(L,M_W,M_V)\in\mathbb{R}^{m\times (m-r)},\\
\GSVDV(L,M_W,M_V)\in\mathbb{R}^{n\times r},\qquad
\GSVDVI(L,M_W,M_V)\in\mathbb{R}^{n\times (n-r)},
\end{gathered}
\]
whose columns are mutually orthonormal in the inner products on $W$ and $V$,
respectively, together with a diagonal matrix
$\GSVDS(L,M_W,M_V)=\operatorname{diag}(\sigma_1,\dots,\sigma_r)\in\mathbb{R}^{r\times r}$,
where each $\sigma_i>0$ and $\sigma_i\ge \sigma_{i+1}$.

These satisfy
\[
L=\GSVDV(L,M_W,M_V)\,
\GSVDS(L,M_W,M_V)\,
\GSVDU(L,M_W,M_V)^T M_W.
\]

Moreover,
\[
\begin{gathered}
\operatorname{image}(L)=\operatorname{span}(\GSVDV(L,M_W,M_V)),\qquad
\operatorname{coimage}(L)=\operatorname{span}(\GSVDVI(L,M_W,M_V)),\\
\operatorname{kernel}(L)=\operatorname{span}(\GSVDUI(L,M_W,M_V)),\qquad
\operatorname{cokernel}(L)=\operatorname{span}(\GSVDU(L,M_W,M_V)).
\end{gathered}
\]

We fix the choices of singular vectors by \cref{def:fsv}.
If the inner products are represented by identity matrices, we drop the prefix
$G$ and write simply SVD.
Finally, the Moore--Penrose pseudo-inverse is
\begin{equation}
\label{eq:gsvdpseudo}
L^\dagger
=
\GSVDU(L,M_W,M_V)\,
\GSVDS(L,M_W,M_V)^{-1}\,
\GSVDV(L,M_W,M_V)^T M_V.
\end{equation}

\end{defn}
\paragraph{Finite-array operations.}
Since we target array operations, specifying our semantics requires a few syntactic bits.
 We write $\mathsf{arr}_n[f]=(f(0),\ldots,f(n-1))$
for array construction and $a[i]$ for checked access.
We write $\mathsf{mat}_{m,n}(a), \mathsf{flat}(B)$, and $\mathsf{out}_n(B)$ for a C-order matrix view, flattening, and a length-$n$ array containing the entries of $B$, respectively; the required dimensions and array lengths are checked.
Failed checks or accesses yield $\glverr$, which propagates.
The full definitions appear in Appendix~\ref{app:gen-defs:valdenot}.

\paragraph{Value Environments, domains, and denotation functions} We will use $\glvEnv$ to map value variables to their semantics, arrays of finite size or  higher-order dependent functions.
We will allow errors represented via $\glverr$.
Semantic dimensions and shapes range over $\mathbb N\sqcup\{\glverr\}$ and $\mathbb N^{<\omega}\sqcup\{\glverr\}$, respectively.  
For a fixed value environment $\glvEnv$, we will define the following denotation functions:
\[
\begin{array}{@{}l@{\quad}l@{}}
\mathcal V_{\glvEnv}\colon
\mathsf{ValTy}\to\mathsf{Set}
& \text{gives the domain of a value type}
\\[3pt]
\Sem{-}{\glvEnv}\colon
\mathsf{ValExpr}\to\mathsf{Value}\sqcup\{\glverr\}
& \text{evaluates a value expression}
\\[3pt]
\dof{-}{\glvEnv}\colon
\mathsf{ValSpace}\to\mathbb N\sqcup\{\glverr\}
& \text{computes the number of scalar entries}
\\[3pt]
\shpOf{-}{\glvEnv}\colon
\mathsf{ValSpace}\to\mathbb N^{<\omega}\sqcup\{\glverr\}
& \text{computes the array dimensions.}
\end{array}
\]
Here $\mathsf{Value}$ consists of finite arrays and (higher-order) functions on arrays. We define our evaluation domain as follows:
\begingroup\small
\expandafter\patchcmd\csname gl@defn@body@value.domains\endcsname
  {,\\}{,\quad}{}{\PackageError{GLang}{Cannot arrange value domains inline}{Check the generated definition.}}
\gldefn{value.domains}
\endgroup
All semantic functions propagate $\glverr$.

\paragraph{Values, Kinding, and Type Checking}
We use the following judgments and inference functions:
\[
\begin{array}{@{}l@{\quad}l@{}}
\valCheckCtx\vdash\lsv\colon\svalid
& \text{$\lsv$ is a well-formed value space}
\\[3pt]
\valCheckCtx\vdash\ltyv\colon\lkn{}
& \text{$\ltyv$ is a value type of kind $\lkn{}$}
\\[3pt]
\valCheckCtx\vdash\leev\colon\ltyv
& \text{$\leev$ has value type $\ltyv$}
\\[3pt]
\operatorname{inferType}\colon
\mathsf{ValExpr}\to\mathsf{ValTy}
& \text{infers a value type from syntax}
\\[3pt]
\operatorname{inferSpace}\colon
\mathsf{ValExpr}\to\mathsf{ValSpace}
& \text{infers a value space from syntax.}
\end{array}
\]
Here $\valCheckCtx$ assigns types to value variables.
See  \cref{sec:vals} for judgments and \cref{app:value-inference} for functions.
Free-variable functions are given in \cref{fig:fvVals}.

\paragraph{Space Values: Dimension and Shape} The oddest construct in \cref{fig:Values} is $\lR{}^{\leev{}, \lprop{}, \lsv}$, which handles dynamic dimensions induced by an SVD corresponding to the $\lprop{}$ fundamental space of $\leev{}$, which has a shape $\lsv{}$. 
 We evaluate the matrix and obtain the selected fundamental subspace's dimension from its SVD.
 Thus, value semantics and dimension semantics are defined recursively when you include the SVD case:

\begin{center}
{\scriptsize\begin{tabular}{@{}c@{}}
\makebox[\linewidth]{\hfill$\dof{\lsv}{\glvEnv}=\prod\shpOf{\lsv}{\glvEnv}$\hfill$\shpOf{\lsnv{0}\losum\lsnv{1}}{\glvEnv}=(\dof{\lsnv{0}}{\glvEnv}+\dof{\lsnv{1}}{\glvEnv},)$\hfill}\\
\makebox[\linewidth]{\hfill$\gleqn{value.shapeOf.real}$\hfill$\shpOf{\lsnv{0}\loprod\lsnv{1}}{\glvEnv}=(\dof{\lsnv{0}}{\glvEnv},\dof{\lsnv{1}}{\glvEnv})$\hfill}
\end{tabular}}
\end{center}
\begingroup\scriptsize
\newcommand{\compactFFS}[2]{%
  \expandafter\patchcmd\csname gl@defn@body@value.ffs.svd\endcsname
    {#1}{#2}{}{\PackageError{GLang}{Cannot compact the SVD dimension clauses}{Check the generated definition.}}}
\compactFFS{\begin{gather*}}{$}
\compactFFS{,\\}{$, so $}
\compactFFS{\end{gather*}}{$}
\compactFFS{Failed shape or value evaluation, non-array values, and incorrect lengths give $\glverr$.}{}
\compactFFS{,\qquad r=|\{k\mid\Sigma_{kk}>0\}|}{}
\compactFFS{d_{\Kernel}=d_0-r,\qquad d_{\CoKernel}=d_{\Image}=r,\qquad d_{\CoImage}=d_1-r.}
  {\shpOf{\lR^{\leev,\Image,\lsv}}{\glvEnv}=(|\{k\mid\Sigma_{kk}>0\}|,).}
\gldefn{value.ffs.svd}\par
\endgroup

\paragraph{Values and their semantics}

At value-space type $\lsv=\operatorname{inferSpace}(\leev)$, a successful denotation
$a=\Sem{\leev}{\glvEnv}$ is a flat array of length $\dof{\lsv}{\glvEnv}$, associated to a shape $\shpOf{\lsv{}}{\glvEnv}$.
Scalars use one entry; vectors use $\SemA{\leev}{\glvEnv}{i}=a[i]$;
matrices of shape $(m,n)$ use
$\SemA{\leev}{\glvEnv}{i,j}=a[ni+j]$ for $0\leq i<m$ and $0\leq j<n$
(C-order), though we also support flattened access to matrices via a single index.
Zero-dimensional spaces use empty arrays; out-of-bounds accesses yield $\glverr$.

Some of our values are drawn from a  subset of those in our original language that can be easily interpreted as real array/lambda calculus operations while others are special constructs defined to handle dynamically determined dimensions.
For example, in the former category, we have tensor product while in the latter we have a one hot vector:

\begin{center}
{\scriptsize\begin{tabular}{@{}c@{}}
$\gleqn{value.denot.tprod}$\\
$\gleqn{value.denot.delta}$
\end{tabular}}
\end{center}
where $\lsnv{j}=\operatorname{inferSpace}(\lenv{j})$ and $d_j=\dof{\lsnv{j}}{\glvEnv}$.
These array semantics and their type-checking rules are not novel or interesting so we relegate their complete description to the appendix (\cref{sec:vals}).
We do include the matrix times matrix typing and semantics below because (a) the conventions are necessary to understand the idea of the operational semantics and (b) they reveal something important about typing in our array language.
Write $d_j=\dof{\lsnv{j}}{\glvEnv}$, carrying primes from spaces to their dimensions likewise:
\begin{center}
{\scriptsize
\newcommand{\compactMTM}[2]{%
  \expandafter\patchcmd\csname gl@equation@body@value.denot.mtm.compact\endcsname
    {#1}{#2}{}{\PackageError{GLang}{Cannot compact the matrix-product equation}{Check the generated equation.}}}
\compactMTM{\sum}{\sum\nolimits}
\compactMTM{,\\ 0\leq r<d_2}{,\quad 0\leq r<d_2}
\compactMTM{,\quad |\SemA{\lenv{0}}{\glvEnv}{}|}{,\\|\SemA{\lenv{0}}{\glvEnv}{}|}
\compactMTM{\\ \operatorname{inferSpace}}{,\quad\operatorname{inferSpace}}
\compactMTM{,\\d_1'=d_1=d_1''}{,\quad d_1'=d_1=d_1''}
\compactMTM{,\quad d_j=\dof{\lsnv{j}}{\glvEnv}\ (j=0,1,2),\quad d_1'=\dof{\lsnv{1}'}{\glvEnv},\quad d_1''=\dof{\lsnv{1}''}{\glvEnv}}{}
\(\begin{gathered}
\glrule[4]{valHasType.mtm}\\[3pt]
\gleqn{value.denot.mtm.compact}
\end{gathered}\)
}
\end{center}

Typing identifies $\lsnv{1}$ with the right operand's second factor; evaluation checks its dimension against the contraction space $\lsnv{1}''$ and the inferred factors, propagating $\glverr$ on failure.
Value typing alone cannot ensure that these array operations represent linear operators; information about bases supplies this guarantee and, as we later show, prevents errors.
We find this analysis particularly compelling as the utility of our value typing is within the scope of that provided by other treatments of array languages~\cite{Henriksen2021, 10.5555/AAI28088718,paszke2021getting, trojahner2009dependently,Berkling1991,sinkarovs2017lambda}.
For this work, we need a consistency definition:
\begin{defn}[$\valCheckCtx\vdash\glvEnv$]
\label{def:vrho}
$\valCheckCtx\vdash\glvEnv\Leftrightarrow
\forall(x\colon\ltyv)\in\valCheckCtx.\;
 (\exists\lkn{}.\;\valCheckCtx\vdash\ltyv\colon\lkn{})
 \land\glvEnv(x)\in\mathcal V_\glvEnv(\ltyv)\setminus\{\glverr\}$.

\end{defn}

Write $\SemCheck{v}{\glvEnv}$ for denotation checked against $\operatorname{inferType}(v)$ (Appendix~\ref{app:preview-checked}). This allows us to prove:
\begin{lem}[Soundness for values]
\label{lem:soundVals}
If $\valCheckCtx\vdash\glvEnv$ and $\valCheckCtx\vdash\leev\colon\ltyv$,
then:
\begin{enumerate}
\item $\exists\lkn{}.\;\valCheckCtx\vdash\ltyv\colon\lkn{}$;
\item $\operatorname{inferType}(\leev)=\ltyv$;
\item $\SemCheck{\leev}{\glvEnv}\in\mathcal V_\glvEnv(\ltyv)$.
\end{enumerate}

\end{lem}

\subsubsection{Operational Semantics}
\label{sec:opersem}

Our operational semantics uses an environment $\evalCtx$ where $\evalCtxVars$ maps variable-basis pairs to values that represent the variable in the given basis.
We use four judgments for our operational semantics:
\[
\setlength{\arraycolsep}{4pt}
\begin{array}{ll}
\evalSJudgeK{\ls{}}{\lsv{}} 
& \text{a space becomes a value space or function type} \\[3pt]

\evalSJudgeK{\lty{}}{\ltyv{}} & \text{a source type becomes a value type} \\[3pt]
\evalJudgeK{\lee}{\lb{}}{\leev{}} 
& \text{$\lee$ becomes value $\leev{}$ in the basis $\lb{}$} \\[3pt]

\evalBBJudgeK{\lbn{0}}{\leev{}}{\lbn{1}} 
& \text{value $\leev{}$ is a matrix converting $\lbn{0}$ to $\lbn{1}$.}
\end{array}
\]
We note that the first judgment uses bases too, but the choice of basis does not impact $\dof{\lsv}{\glvEnv}$ so we de-emphasize this.
We examine some of these rules in turn, but the complete operational semantics are split into four parts in the appendix: space rules (\cref{fig:operationalSemanticsSpace}), base term rules (\cref{fig:operationalSemantics1}), function and evaluation rules (\cref{fig:operationalSemanticsEvals}), and change of basis rules (\cref{fig:operationalSemanticsCOB}). 

Throughout these rules, binders are chosen fresh and paired indices range over $i\in\{0,1\}$.
For our space rules, most cases replicate the space/type structure:
\begin{center}
{\scriptsize
\makebox[\linewidth]{\hfill$\glrule{evalS.real}$\hfill$\glrule{evalS.tprod}$\hfill$\glrule[3]{evalT.piTy}$\hfill}
}
\end{center}

However, computers have no equivalent for dual spaces or the four fundamental subspaces.
In the case of a dual space, our rules forget the dual space structure while for the fundamental spaces our rules use a dynamic dimension:
\begin{center}
{\scriptsize
\makebox[\linewidth]{\hfill$\glrule{evalS.dual}$\hfill$\glrule[4]{evalS.ker}$\hfill}
}
\end{center}

 Similarly to the space rules, many term rules replicate structure where possible, often corresponding to cases where coordinate-independent concepts are easy to translate under any combination of bases:
\begin{center}
{\scriptsize
\makebox[\linewidth]{\hfill$\glrule{opStep.var}$\hfill$\glrule[3]{opStep.add}$\hfill}\par\smallskip
\makebox[\linewidth]{\hfill$\glrule[2]{opStep.scale}$\hfill$\glrule[2]{opStep.tprodE}$\hfill}
}
\end{center}

However, for concepts that must ultimately relate to \cref{def:ffs}, our rules are more complex, but also offer opportunities (\textit{e.g.}, pseudo-inverse in the standard basis vs. pseudo-inverse in a general basis can lead to calling a simpler function):
\begin{center}
{\scriptsize
\makebox[\linewidth]{\hfill$\glrule[3]{opStep.innerStd}$\hfill$\glrule[2]{opStep.pinvTGpinv}$\hfill}\par\smallskip
\makebox[\linewidth]{\hfill$\glrule[2]{opStep.pinvTStd}$\hfill$\glrule[3]{opStep.embedKer}$\hfill}
}
\end{center}

Most of the above rules don't provide much choice for the target basis, but our linear and non-linear evaluation rules can trigger the above optimization opportunities because they introduce choice of basis:
\begin{center}
{\scriptsize
\makebox[\linewidth]{\hfill$\glrule[2,2]{opStep.appLLin}$\hfill$\glrule[3,2]{opStep.appNVal}$\hfill}
}
\end{center}
Note that we need to reshape arguments and results to get a well-typed result though this is unnecessary under our final result; this issue reoccurs for non-linear abstraction:
\begin{center}
{\scriptsize
$\glrule[4]{opStep.nllam}$
}
\end{center}

In contrast, our rules handle linear abstraction via matrix assembly~\cite{kirby2006compiler}, which is similar to~\cite{velez2026compiling}:
\begin{center}
{\scriptsize
$\gpfit{\glrule[4]{opStep.llamAssemble}}$
}
\end{center}
The function $\overline{\operatorname{dimOf}}(\lsv)$ provides an upper bound on dimensions (\cref{def:dimension-upper-bound}).

Finally, we use the change of basis judgment $\evalBBJudgeK{\lbn{0}}{\leev{}}{\lbn{1}}$ explicitly within the evaluation judgment in one way, but implicitly we use it to change the basis of inputs and outputs to frames:
\begin{center}
{\scriptsize
$\glrule[4]{opStep.cob}$\par\smallskip
$\glrule[4,2,2]{opStep.frameReanchor}$
}
\end{center}

Many of the change of basis rules themselves are fairly standard; moreover, we might hope that we could express the change of basis operations with our terms and the evaluation rules. However, doing so would lead our rules to an infinite recursion, preventing progress (\cref{thm:progA}), which is why we need \NLang{} operations for block matrices and  Kronecker products:
\begin{center}
{\scriptsize
\makebox[\linewidth]{\hfill$\glrule{evalBB.tprod}$\hfill$\glrule{evalBB.dsum}$\hfill}
}
\end{center}

Naturally, we allow one to use the inverse change of basis rule and to use change of basis rules transitively:
\begin{center}
{\scriptsize
\makebox[\linewidth]{\hfill$\glrule{evalBB.inverse}$\hfill$\glrule[3]{evalBB.trans}$\hfill}
}
\end{center}

For more complex bases, we often have to invoke special functions like the inverse for dual bases
\begin{center}
{\scriptsize
$\glrule[2]{evalBB.dualStd}$
}
\end{center}

For projected bases, we change a basis in the super space of the subspace and then evaluate projection/embedding to turn this into a change of basis in the subspace:
\begin{center}
{\scriptsize
\newcommand{\compactProjOperand}[4]{%
  \expandafter\patchcmd\csname gl@rule@concl@evalBB.projKerCoker\endcsname
    {\glvreshape{#2}{#3}{#4}}{\widehat{#2}}
    {}{\PackageError{GLang}{Cannot factor projected-basis operand #1}{Check the generated rule.}}}
\compactProjOperand{1}{\lenv{1}}{\lsnv{0}' \loprod \lsv{}'}{\lsnv{0}' \loprod \lsv{}'}
\compactProjOperand{2}{\lenv{2}}{\lsnv{0} \loprod \lsnv{0}'}{\lsnv{0} \loprod \lsnv{0}'}
\compactProjOperand{3}{\lenv{3}}{\lsv{} \loprod \lsnv{0}}{\lsv{} \loprod \lsnv{0}}
\def\compactProjConclusion\begin#1\evalBBJudge#2#3#4#5\\[#6]#7=#8\end#9\stopProj{%
  \expandafter\def\csname gl@rule@concl@evalBB.projKerCoker\endcsname{\evalBBJudge{#2}{#3}{#8}{#5}}}
\expandafter\let\expandafter\projConclusion\csname gl@rule@concl@evalBB.projKerCoker\endcsname
\expandafter\compactProjConclusion\projConclusion\stopProj
\(\begin{gathered}
\glrule[5,2,2]{evalBB.projKerCoker}\\
\widehat{\lenv{1}}=\glvreshape{\lenv{1}}{\lsnv{0}'\loprod\lsv'}{\lsnv{0}'\loprod\lsv'},\quad
\widehat{\lenv{2}}=\glvreshape{\lenv{2}}{\lsnv{0}\loprod\lsnv{0}'}{\lsnv{0}\loprod\lsnv{0}'},\quad
\widehat{\lenv{3}}=\glvreshape{\lenv{3}}{\lsv\loprod\lsnv{0}}{\lsv\loprod\lsnv{0}}.
\end{gathered}\)
}
\end{center}
We have two rules to convert between different image bases, using the source or destination basis:
\begin{center}
{\scriptsize
$\glrule[2]{evalBB.image}$\par\smallskip
$\glrule[4,3,1]{evalBB.imageSvds}$
}
\end{center}

Finally, for a basis of terms, we exploit that the row-reduced echelon form of a matrix of vectors and some basis tells one how to write the pivot columns (the term basis) in terms of the other given basis. We summarize this process with just $RREFRep$, happily avoiding some of the details of extracting this from the actual row-reduced echelon form:
\begin{center}
{\scriptsize
$\glrule[3]{evalBB.termsRref}$
}
\end{center}

Before continuing, we need to relate operational and typing environments:
\begin{defn}[$\lctx;\valCheckCtx\vdash\evalCtx$]
\label{def:nonrecurseeval}
Let $\operatorname{names}(\Psi)=\{x\mid\exists b,v.\;((x,b)\mapsto v)\in\Psi\}$. Write $\Psi\vdash(x,\lb{})\mapsto\leev$ when
$\evalSJudge{\Psi}{\tof{\lb{}}}{\operatorname{inferType}(\leev)}$ and
$\operatorname{fv}(\leev)\subseteq\operatorname{names}(\Psi)\cup\{x\}$.
For $\lb{}=\frameB{y}{b_0}{b_1}$ and
$\leev=\llambda(y\colon\ltynv{0})\lrh\ltynv{1}\colon v$, additionally require
$\evalSJudge{\Psi[(y,b_0)\mapsto y_{\ltynv{0}}]}{\tof{b_1}}{\ltynv{1}}$ for fresh $y$.

Assuming $\vdash(\Delta,\epsilon)$ throughout (\cref{def:wfc}), define
$(\Delta,\epsilon);\valCheckCtx\vdash\Psi$ inductively, with
$(\emptyset,\emptyset);\emptyset\vdash\emptyset$.
Suppose $(\Delta,\epsilon);\valCheckCtx\vdash\Psi$,
$(\emptyset,\epsilon)\vdash\lb{}\colon\bvalid$,
$(x,\lb{})\notin\Psi$, and $\Psi\vdash(x,\lb{})\mapsto\leev$.
Put $T=\tof{\lb{}}$ and $\ltyv=\operatorname{inferType}(\leev)$.
Form $(\Delta',\epsilon';\valCheckCtx')$ by adding $x:T$ to
$(\Delta,\epsilon)$ and $x:\ltyv$ to $\valCheckCtx$ if $x$ is undeclared;
otherwise leave the contexts unchanged, with $(x:T)\in\Delta\sqcup\epsilon$.
If $\valCheckCtx\vdash\ltyv:\lkn{}$ and
$\valCheckCtx'\vdash\leev:\ltyv$, then
$(\Delta',\epsilon');\valCheckCtx'\vdash\Psi[(x,\lb{})\mapsto\leev]$.

\end{defn}

\subsubsection{Basic Meta Theory of the Operational Semantics}

Now we can state our preservation result.
\begin{thm}[Preservation]
\label{thm:preserve}
Suppose $(\Delta,\epsilon);\valCheckCtx\vdash\Psi$.
\begin{enumerate}
\item If $(\emptyset,\epsilon)\vdash\lty:\lkn{}$ and
$\evalSJudge{\Psi}{\lty}{\ltyv{}}$, then
$\valCheckCtx\vdash\ltyv{}:\lkn{}$ for some $\lkn{}$ and
$\operatorname{fv}(\ltyv{})\subseteq\operatorname{names}(\Psi)$.
\item If $(\Delta,\epsilon)\vdash t:\tof{b}$,
$(\emptyset,\epsilon)\vdash b:\bvalid$, and $\evalJudge{\Psi}{t}{b}{v}$, then
$\valCheckCtx\vdash v:\operatorname{inferType}(v)$,
$\valCheckCtx\vdash\operatorname{inferType}(v):\lkn{}$ for some $\lkn{}$, and
$\operatorname{fv}(\operatorname{inferType}(v))\cup\operatorname{fv}(v)\subseteq\operatorname{names}(\Psi)$.
\item If $(\emptyset,\epsilon)\vdash b_0:\bvalid$ and
$\evalBBJudge{\Psi}{b_0}{v}{b_1}$, then
$\valCheckCtx\vdash v:\lsnv{0}\loprod \lsnv{1}$ and
$\operatorname{inferType}(v)=\lsnv{0}\loprod \lsnv{1}$ for some $\lsn{0},\lsn{1},\lsnv{0},\lsnv{1}$ with
$\evalSJudge{\Psi}{\lsn{i}}{\lsnv{i}}$ $(i=0,1)$ and
$\operatorname{fv}(v)\subseteq\operatorname{names}(\Psi)$.
\end{enumerate}

\end{thm}
\begin{proof}
A fairly mechanical proof; see \cref{thm:preserveA}. \end{proof}
To state a strong version of progress that feeds adequacy, we need an auxiliary definition:
\begin{defn}[Checked operational evaluation]
\label{def:envimply}
Assume $(\Delta,\epsilon);\valCheckCtx\vdash\Psi$.
The subscript $\mathsf C$ marks an operational derivation checked at every step:
$\glCty{\epsilon}{\Psi}{\lty}{\ltyv}$,
$\glCterm{\Delta}{\epsilon}{\Psi}{\lee}{\lb{}}{\leev}{\ltyv}$, and
$\glCcob{\epsilon}{\Psi}{\lbn{0}}{\leev}{\lbn{1}}$.
At each term step, for example, these require
$(\Delta,\epsilon)\vdash t:\tof{b}$,
$(\{\},\epsilon)\vdash b:\bvalid$,
$\glCat{\epsilon}{\Psi}{b}{\ltyv{}}$,
$\evalJudge{\Psi}{t}{b}{v}$, and
$\evalSJudge{\Psi}{\tof{b}}{\ltyv{}}$.
The judgment $\glCat{\epsilon}{\Psi}{\lb{}}{\ltyv}$ simply shows what basis was used in $\glCty{\epsilon}{\Psi}{\lty}{\ltyv}$, which is needed to call dynamic dimensions properly.
The rather verbose rules are in \cref{app:gen-rules:c-space,app:gen-rules:c-ty,app:gen-rules:c-tyAt,app:gen-rules:c-cob,app:gen-rules:c-term}.

\end{defn}
For progress, we construct evaluations at standard bases and then use change of basis (\cref{thm:progA}).
\begin{thm}[Progress]
\label{thm:prog}
Suppose $(\Delta,\epsilon);\valCheckCtx\vdash\Psi$.
\begin{enumerate}
\item If $(\emptyset,\epsilon)\vdash\ls{}:\svalid$,
then $\glCspace{\epsilon}{\Psi}{\ls{}}{\lsv}$ for some $\lsv$.
\item If $(\Delta,\epsilon)\vdash t:\tof{b}$ and
$(\emptyset,\epsilon)\vdash b:\bvalid$, then
$\glCterm{\Delta}{\epsilon}{\Psi}{t}{b}{v}{\ltyv{}}$ for some $v,\ltyv{}$ with
$\operatorname{inferType}(v)=\ltyv{}$.
\item If $(\emptyset,\epsilon)\vdash b_i:\bvalid$ and
$\tof{b_i}=\ls{}$ $(i=0,1)$, then
$\glCcob{\epsilon}{\Psi}{b_0}{v}{b_1}$ for some $v$.
\end{enumerate}

\end{thm}

\section{Meta Theory Part 2: Adequacy}
\label{sec:meta2}
Adequacy expresses the idea that a vector and its array representation
have the same meaning when produced via an appropriate evaluation.
For a space-valued term $\lee$, a checked evaluation at basis $\lb$ produces $\leev$ and records the particular space realization $\lsv$ selected at that basis.
Under consistent environments, $\Sem{\leev}{\glvEnv}=w\ne\glverr$ and,
with $B=\SemBB{\lb}{\sigma}$ and $N=\dof{\lsv}{\glvEnv}$,
\begin{equation}
\label{eq:compp}
\Sem{\lee}{\sigma}=\sum_{i=0}^{N-1}w_iB_i.
\end{equation}
The index $i$ selects both a basis vector and an entry of the flattened array.

The following definition provides the appropriate higher-order definition for this intuitive idea:
\begin{defn}[Agreement]
\label{def:agree}
Write $a\equiv_{B;\sigma,\glvEnv}^{\lty,\ltyv}w$ for agreement between a source value
and a runtime value at basis or frame $B$, interpreting $\lty$ under $\sigma$
and $\ltyv$ under $\glvEnv$.
At spaces, put $N=\dof{\lsv}{\glvEnv}$; then
$a\equiv_{B;\sigma,\glvEnv}^{\ls,\lsv}w$ iff
$|B|=|w|=N$ and $a=\sum_{i=0}^{N-1}w_iB_i$.
At functions $\lty=x:\ltyn{0}\fnto\ltyn{1}$ and
$\ltyv=x:\ltynv{0}\fnto\ltynv{1}$, let $B$ be a frame with domain
component $B_0$ and fiber family $a\mapsto B_1(a)$.
Then $f\equiv_{B;\sigma,\glvEnv}^{\lty,\ltyv}g$ iff, for every
$a\equiv_{B_0;\sigma,\glvEnv}^{\ltyn{0},\ltynv{0}}w$, there is $z$ such that
$g(w)=z\ne\glverr$ and
$f(a)\equiv_{B_1(a);\sigma',\glvEnv'}^{\ltyn{1},\ltynv{1}}z$, where
$\sigma'=\sigma[x\mapsto a]$ and $\glvEnv'=\glvEnv[x\mapsto w]$.
Here $g$ has the runtime domain prescribed by $\ltynv{0}$.

\end{defn}
We also define type agreement $\lty\simeq_{B;\sigma,\glvEnv}\ltyv$ and its shape variant $\lty\sim_{B;\sigma,\glvEnv}\ltyv$; see \cref{def:agreedims}. We need a consistency definition to wrap up all our other consistency definitions and ensure agreement is satisfied in an existing environment.
\begin{defn}[Consistent environments]
\label{def:allCtxAgree}
We write $(\Delta,\epsilon);\Psi;\valCheckCtx\vdash(\sigma,\glvEnv)$ when
$(\Delta,\epsilon);\valCheckCtx\vdash\Psi$, $\valCheckCtx\vdash\glvEnv$,
$(\Delta,\epsilon)\vdash\sigma$ at every step, and
$(\Delta,\epsilon,\Psi,\valCheckCtx,\sigma,\glvEnv)$ is constructed as follows.
The tuple $(\emptyset,\emptyset,\emptyset,\emptyset,\emptyset,\glvEnv)$ is constructed for arbitrary $\glvEnv$.
Suppose $(\Delta,\epsilon,\Psi,\valCheckCtx,\sigma,\glvEnv)$ has been constructed.
Let $\Psi'=\Psi[(x,\lb{})\mapsto\leev{}]$ and update
$(\Delta,\epsilon,\valCheckCtx)$ to $(\Delta',\epsilon',\valCheckCtx')$
as in \cref{def:nonrecurseeval}.
Put $T=\tof{\lb{}}$,
$\ltyv{}=\operatorname{inferType}(\leev{})$, and
$B=\SemBB{\lb{}}{\sigma}$, and require
$\Sem{\leev{}}{\glvEnv}=w\ne\glverr$,
$a\equiv_{B;\sigma,\glvEnv}^{T,\ltyv{}}w$, and
$T\sim_{B;\sigma,\glvEnv}\ltyv{}$.
If $x\notin\operatorname{dom}(\Delta\sqcup\epsilon)$,
require $(x\notin\operatorname{fv}(\leev{})\ \lor\
\leev{}=x_{\ltyv{}})$ and put
$\sigma'=\sigma[x\mapsto a]$, and
$\glvEnv'=\glvEnv[x\mapsto w]$.
Otherwise require $a=\sigma(x)$ and put
$(\sigma',\glvEnv')=(\sigma,\glvEnv)$.
This constructs $(\Delta',\epsilon',\Psi',\valCheckCtx',\sigma',\glvEnv')$.

\end{defn}

We can now state our adequacy theorem. Intuitively, this theorem is nothing but the core idea that the coordinate-independent meaning of a vector should be the same as the coordinate-dependent meaning when interpreted via the coordinates as coordinate-independent objects. Our results imply invariance of our semantics under change of basis.
\begin{thm}[Adequacy]
\label{thm:main}
Suppose $(\Delta,\epsilon);\Psi;\valCheckCtx\vdash(\sigma,\glvEnv)$.
\begin{enumerate}
\item If $\glCat{\epsilon}{\Psi}{b}{\ltyv{}}$,
then $\tof{b}\simeq_{\SemBB{b}{\sigma};\sigma,\glvEnv}\ltyv{}$.
\item If $\glCterm{\Delta}{\epsilon}{\Psi}{t}{b}{v}{\ltyv{}}$,
then $\Sem{v}{\glvEnv}=w\ne\glverr$ for some $w$,
$\Sem{t}{\sigma}\equiv_{\SemBB{b}{\sigma};\sigma,\glvEnv}^{\tof{b},\ltyv{}}w$,
and $\tof{b}\simeq_{\SemBB{b}{\sigma};\sigma,\glvEnv}\ltyv{}$.
\item If $\glCcob{\epsilon}{\Psi}{b_0}{v}{b_1}$,
put $B_i=\SemBB{b_i}{\sigma}$ and $N=|B_0|=|B_1|$.
Then $\Sem{v}{\glvEnv}=A\ne\glverr$ for some array $A$ of length $N^2$,
and $B_0=B_1A^{\mathsf T}$, reading $A$ as a row-major $N\times N$ matrix
and the basis vectors as columns.
\end{enumerate}

\end{thm}
\begin{proof}[Proof sketch]
A sketch of the mutual induction is in \cref{thm:mainA}; we give one case here.
By the denotation $\gleqn{denotTermAux.embed.ker}$, it suffices to show that the returned
program evaluates without error and represents a linear map sending each
kernel vector to itself in the ambient space. Consider $\glrulename{c.term.embedKer}$
(Appendix~\ref{app:gen-rules:c-term}):
\begingroup\small
\newcommand{\factorEmbedBasis}[1]{%
  \expandafter\patchcmd\csname gl@rule@#1@c.term.embedKer\endcsname
    {\pBasisC{\Kernel(\lee{})}{\lbn{0}}{\lbn{1}}^{\lstar} \loprod \lbn{0}}
    {\lb{}}{}{%
      \PackageError{GLang}{Cannot factor the embedding basis}{Check the generated rule.}}}
\factorEmbedBasis{prem}
\factorEmbedBasis{prem}
\factorEmbedBasis{prem}
\factorEmbedBasis{concl}
\expandafter\pretocmd\csname gl@rule@prem@c.term.embedKer\endcsname
  {\lb{}=\pBasisC{\Kernel(\lee{})}{\lbn{0}}{\lbn{1}}^{\lstar}\loprod\lbn{0}\\}
  {}{\PackageError{GLang}{Cannot add the embedding basis equality}{Check the generated rule.}}
\glrulesplit[none]{c.term.embedKer}{3,2,2}
\endgroup
Put $V=\SemT{\lsn{0}}{\sigma}$, $W=\SemT{\lsn{1}}{\sigma}$,
$L=\Sem{\lee}{\sigma}:V\to W$; thus
$\SemT{\Kernel(\lee)}{\sigma}=\operatorname{ker}L$.
Basis validity gives bases
$E=\SemBB{\lbn{0}}{\sigma}=(E_i)_{i<N}$ of $V$ and
$F=\SemBB{\lbn{1}}{\sigma}=(F_\ell)_{\ell<M}$ of $W$.
The kernel basis is
$K=\SemBB{\pBasisC{\Kernel(\lee)}{\lbn{0}}{\lbn{1}}}{\sigma}=(K_j)_{j<k}$.
By induction, $N=\dof{\lsnv{0}}{\glvEnv}$ and $M=\dof{\lsnv{1}}{\glvEnv}$.
The arrays denoting $\leev,\lenv{0},\lenv{1}$ succeed with lengths
$NM,N^2,M^2$; read them in input-first order as
\[
 A_{\ell i}=\SemA{\leev}{\glvEnv}{iM+\ell},\qquad
 G_{ii'}=\SemA{\lenv{0}}{\glvEnv}{iN+i'},\qquad
 H_{\ell\ell'}=\SemA{\lenv{1}}{\glvEnv}{\ell M+\ell'}.
\]
By the term and inner-product induction hypotheses,
$L(E_i)=\sum_{\ell<M}A_{\ell i}F_\ell$,
$G_{ii'}=\langle E_i,E_{i'}\rangle$ and
$H_{\ell\ell'}=\langle F_\ell,F_{\ell'}\rangle$.
Basis validity makes $G,H$ positive definite. Thus the SVD denotation's
dimension and inner product checks (\cref{fig:valValSemantics}) succeed.

Put $k=N-\operatorname{rank}A=N-\operatorname{rank}L$ and
$Q_{ij}=\SemA{\glvop{GSVDUI}(\leev,\lenv{0},\lenv{1})}{\glvEnv}{ik+j}$.
By the definition of the SVD (\cref{def:gsvd}),
$K_j=\sum_{i<N}Q_{ij}E_i$.
The first inductive hypothesis gives output dimensions $k,N$;
the kernel block has $Nk$ entries, so transposition and the output-length
check also succeed. Thus, by \cref{fig:valValSemantics},
$a=\Sem{\Trans(\glvop{GSVDUI}(\leev,\lenv{0},\lenv{1}))}{\glvEnv}\ne\glverr$
has length $kN$ and $a[jN+i]=Q_{ij}$.
At the result basis $K^\star\otimes E$, this array represents the linear map
$J=\sum_{j<k,i<N}a[jN+i]K_j^\star\otimes E_i:\operatorname{ker}L\to V$.
For any $x=\sum_{j<k}u_jK_j\in\operatorname{ker}L$, calculate
\[
 J(x)=\sum\nolimits_{j<k,i<N}a[jN+i]K_j^\star(x)E_i
 =\sum\nolimits_{j<k}u_j\Bigl(\sum\nolimits_{i<N}Q_{ij}E_i\Bigr)
 =\sum\nolimits_{j<k}u_jK_j=x\quad\text{in }V.
\]
Hence $J=\iota_{\operatorname{ker}L}$, as required.
This gives value agreement; basis validity and the dimensions give type agreement.

\end{proof}

Combining progress with this theorem gives \cref{cor:progressAdequacy} in the appendix.

Before continuing to the next section, we wish to note that, though these ideas
were originally developed by hand, we used Anthropic's Claude and OpenAI's
Codex to assist in developing the Lean formalization, including generating
proof code. We also used Claude to generate the paper's rules and definitions
from the Lean development, check their correspondence with the formalization,
and translate between informal proof sketches and Lean proofs, including the
one above.

\section{Implementation and Performance}

We have implemented a superset of \GLang{}, adding built-in operations, map properties such as injectivity and surjectivity, basis selection strategies, and more.
 We will provide the implementation and a tutorial as an artifact. 
 Our initial, hand-written implementation had all capabilities described here; we later rewrote it with Claude.
 Here we provide some validation via basic performance data as well as intuition on how to implement \GLang{}.

\begingroup
\setlength{\intextsep}{6pt}
\begin{figure}[!htb]
\centering
\begingroup
\definecolor{codekeyword}{HTML}{245A9B}
\definecolor{codefunction}{HTML}{76529B}
\definecolor{codecomment}{HTML}{55745D}
\definecolor{codestring}{HTML}{A34D27}
\lstset{language=Python,basicstyle=\ttfamily\scriptsize,
  keywordstyle=\bfseries\color{codekeyword},
  commentstyle=\itshape\color{codecomment},
  morekeywords={match,case},
  emph={lower_inner,lower_apply,dimension,is_orthonormal,orthonormal_basis,
    change_of_basis,emit_identity,emit_assert,emit_matmul,emit_einsum,lower,convert,
    Tensor,Dual,Space,Basis,Context,Term,Program},
  emphstyle=\color{codefunction},stringstyle=\color{codestring},
  showstringspaces=false,columns=fullflexible,keepspaces=true,
  breaklines=false,frame=none,aboveskip=4pt,belowskip=0pt}
\begin{lstlisting}[aboveskip=0pt]
def lower(t: Term, ctx: Context) -> tuple[Program, Basis]: ...
\end{lstlisting}
\smallskip
\begin{minipage}[t]{0.48\linewidth}
\small\textbf{(a) Inner product in the target basis.}
\begin{lstlisting}
def lower_inner(V: Space, target: Basis,
                ctx: Context) -> tuple[Program, Basis]:
    n = dimension(V, ctx)  # Dynamic OK.
    I = ctx.emit_identity(n)
    match target:
        case Tensor(Dual(b), Dual(c)):
            if b == c and is_orthonormal(b, ctx):
                return I, target
    q = orthonormal_basis(V, ctx)
    source = Tensor(Dual(q), Dual(q))
    return convert(I, source, target, ctx), target
\end{lstlisting}
\end{minipage}\hfill
\begin{minipage}[t]{0.50\linewidth}
\small\textbf{(b) Apply $f\in A^\star\otimes B$ to $x\in A$.}
\begin{lstlisting}
def lower_apply(f: Term, x: Term, target: Basis,
                ctx: Context) -> tuple[Program, Basis]:
    (F, bf), (X, bx) = lower(f, ctx), lower(x, ctx)
    match bf:
        case Tensor(Dual(a), b) if b == target:
            if bx != a: X = convert(X, bx, a, ctx)
        case _:
            want = Tensor(Dual(bx), target)
            F = convert(F, bf, want, ctx)
    ctx.emit_assert(F.shape[0] == X.shape[0])
    return ctx.emit_einsum("ij,i->j", F, X), target
\end{lstlisting}
\end{minipage}
\endgroup
\caption{Representative lowering rules (schematic Python).}
\label{fig:implementation-rules}
\end{figure}
\endgroup

 \paragraph{Intuition for Implementation}
 Building on existing NumPy-like tensor-contraction interfaces, we can implement \GLang{}'s value language using a small subset of NumPy and Python's basic control operators.
\Cref{fig:implementation-rules} sketches two cases of a lowering function that builds programs from our operational rules for inner products and linear application.
Here, shape expressions are symbolic: $\texttt{emit\_assert}$ emits a runtime assertion that the matrix and vector contraction dimensions agree after conversion.
Additionally, these rules follow our progress proof: they fall back on changes of basis when no  rule fires.

Where our operational rules and schematics fall short of our implementation is that we take considerably more care in choosing bases.
For example, whenever a basis is existential, we actually run the operational semantics from the leaves up trying to pick bases that produce constant outputs.
 Similarly, we try to pick the same source and target bases for multi-linear maps and we try to prefer image bases over other bases. 
 Since a good choice of bases tends to lead to SVDs of constant operators, our target IR has aggressive constant propagation and simplification operations on small-scale linear algebra, designed to find SVDs of constants and simplify programs in a corresponding manner, often reducing them to pure rational functions of the inputs.
We also note that our implementation's version of \GLang{} is considerably more complex (\textit{e.g.}, more built-in operations, utilities, properties of maps such as injectivity and surjectivity).
Our goal in this work is to analyze the idea.

\paragraph{Performance}

We compare exterior-algebra kernels with hand-written PyTorch~\cite{Ansel2024} implementations of their known coordinate formulas~\cite{Frankel2011-cd,Arnold2018}. Appendix~\ref{app:implementation-performance-correctness} gives the setup and reference programs; the artifact contains the benchmark programs.
 See \cref{tab:operator-benchmarks} for the results.
  Our results show that a language like \GLang{} can be compiled, run, and also optimized to produce fragments of code that users would otherwise have looked up in textbooks. 
  These fragments may be far from optimal, but match what users would write by hand (we also had Claude attempt further improvements).
  We don't aim at improvement here, but to show that this idea works.

  \begin{table*}[t]
    \centering
    \caption{FLOP counts and runtimes for \GLang{}-generated code vs.\ a reference
      implementation of each operator's known form. Runtimes are mean ns per cell over a
      batch of $2^{16}$ cells, single-threaded, compiled with CPU Inductor.
      Here $\star:\Lambda^k\mathbb R^n\to\Lambda^{n-k}\mathbb R^n$ is the Hodge star. \emph{Parity} marks generated runtime no more than 10\% above the reference. Standard deviations are 2–7\% of the mean (16–19\% for $a\wedge b$ in $\mathbb{R}^3$).
      FLOPs are counted per cell in Inductor's post-grad graph. The two
      $\langle a \wedge b, c \wedge d\rangle$ rows compute the same value: once by pairing
      $\Lambda^2$ coordinates, once written via the Cauchy--Binet identity.   See \cref{app:implementation-performance-correctness} for details. }
    \label{tab:operator-benchmarks}
    \scriptsize
    \setlength{\tabcolsep}{3.5pt}
    \renewcommand{\arraystretch}{1.12}
    \setlength{\arrayrulewidth}{0.4pt}
    \resizebox{\textwidth}{!}{%
    \begin{tabular}{@{}c@{\hspace{7pt}}|@{\hspace{7pt}}c@{}}
        \begin{tabular}{|l|l|l|r|r|r|r|c|}
          \hline
          Operator & Space & Known form & \multicolumn{2}{c|}{FLOPs} & \multicolumn{2}{c|}{Time (ns)} & Par. \\
          & & & Gen. & Ref. & Gen. & Ref. & \\
          \hline
          $a \wedge b$ & $\mathbb{R}^{2}$ & determinant & 4 & 4 & 0.87 & 0.86 & \checkmark \\ \hline
          $\star|_{\Lambda^{1}}$ & $\mathbb{R}^{2}$ & $90^\circ$ rotation & 2 & 2 & 0.72 & 1.69 & \checkmark \\ \hline
          $a \wedge b$ & $\mathbb{R}^{3}$ & cross-product coordinates & 12 & 12 & 1.35 & 1.39 & \checkmark \\ \hline
          $\star|_{\Lambda^{1}}$ & $\mathbb{R}^{3}$ & cross-product matrix & 3 & 3 & 0.94 & 2.82 & \checkmark \\ \hline
          $\star(a \wedge b)$ & $\mathbb{R}^{3}$ & cross product & 12 & 12 & 1.57 & 1.58 & \checkmark \\ \hline
          $\langle a \wedge b,c \wedge d\rangle$ & $\mathbb{R}^{3}$ & pairing of minors & 24 & 24 & 2.16 & 2.21 & \checkmark \\ \hline
          $\langle a \wedge b,c \wedge d\rangle$ (C--B) & $\mathbb{R}^{3}$ & Lagrange identity & 24 & 24 & 2.16 & 2.19 & \checkmark \\ \hline
        \end{tabular}
        &
        \begin{tabular}{|l|l|l|r|r|r|r|c|}
          \hline
          Operator & Space & Known form & \multicolumn{2}{c|}{FLOPs} & \multicolumn{2}{c|}{Time (ns)} & Par. \\
          & & & Gen. & Ref. & Gen. & Ref. & \\
          \hline
          $\star(a \wedge b \wedge c)$ & $\mathbb{R}^{3}$ & scalar triple product & 15 & 15 & 1.63 & 1.62 & \checkmark \\ \hline
          $a \wedge b$ & $\mathbb{R}^{4}$ & $2 \times 2$ minors & 24 & 24 & 2.66 & 2.63 & \checkmark \\ \hline
          $\star|_{\Lambda^{1}}$ & $\mathbb{R}^{4}$ & dual trivector & 4 & 4 & 1.02 & 4.85 & \checkmark \\ \hline
          $\star(a \wedge b)$ & $\mathbb{R}^{4}$ & dual bivector & 24 & 24 & 2.38 & 2.46 & \checkmark \\ \hline
          $\langle a \wedge b,c \wedge d\rangle$ & $\mathbb{R}^{4}$ & pairing of minors & 48 & 48 & 4.42 & 4.40 & \checkmark \\ \hline
          $\langle a \wedge b,c \wedge d\rangle$ (C--B) & $\mathbb{R}^{4}$ & Lagrange identity & 32 & 32 & 3.57 & 3.60 & \checkmark \\ \hline
          $\star(a \wedge b \wedge c)$ & $\mathbb{R}^{4}$ & generalized cross product & 42 & 60 & 4.67 & 5.58 & \checkmark \\ \hline
        \end{tabular}
    \end{tabular}%
    }
  \end{table*}

\section{Related Work}
Due to the fundamental importance of linear algebra, programming languages and software systems have explored categorical semantics and compilation~\cite{Elliott2017,Bernardy2021,Hasegawa,Selinger2011,DazCaro2020,abramsky2004categorical},
vector-space models and algebraic lambda calculi~\cite{valiron2014finite,Arrighi2017,diaz2024linear,vaux2009algebraic,Girard1999,ehrhard2003differential},
linear types and resource management~\cite{makwana2019numlin,10.1145/3158093,Bernardy2021,diaz2024linear,fu2020linear},
array languages and tensor notation~\cite{Slepak2014,paszke2021getting,Henriksen2021,futhark2017,accelerate2011,bernardy2025domain,breuer2025einsum1},
algebraic polymorphism and semirings~\cite{shaikhha2019finally,kepner2016mathematical,shaikhha2022functional},
automatic differentiation~\cite{elliott2009beautiful,Elliott2018,Radul2023,Vkr2022,Giusti2026,paszke2021getting},
neural-network compilation~\cite{velez2026compiling},
quantum programming and equational reasoning~\cite{qwerty2024,adams2025asdf,Xu2025,Por2023,fu2020linear,Carette2024,Fang2026,Matsushita2026,qunity2023},
explicit bases, coordinates, and spatial representations~\cite{Geisler2020,Chen2024,qwerty2024,adams2025asdf,bernardy2025domain,collin2019point,kindlmann2005nrrd,unite:2025},
scientific operators and solver composition~\cite{alnaes2014unified,kirby2006compiler,kirby2008benchmarking,Betcke2021,Burns2020,gibson2020slate,devito2017opt,brown2012composable,Zhang2025},
linear-algebra compilation and optimization~\cite{Baumgartner2005,breuer2025einsum1,franchetti2009operator,taco2017,ahrens2025finch,10.1145/3446632,Psarras2022,halide2013},
graphics and visualization~\cite{chiw2012diderot,collin2019point,devito2017opt},
and mathematical notation and authoring~\cite{Li2021,bernardy2025domain,Xu2025}.
In this discussion, we narrow our focus to coordinates vs. coordinate-free as well as language features.
\citet{Padula2009} provide a coordinate-free linear algebra software interface for Hilbert spaces that can express linear solvers, but do not give a formal language definition or operational and denotational semantics for that interface.
Of the languages above, several utilize the idea of coordinate-free linear algebra, often via a categorical foundation:~\cite{Elliott2017, bernardy2025domain, Carette2024,valiron2014finite, diaz2024linear,vaux2009algebraic,Girard1999}.
On the other side, several also emphasize coordinates or a specific choice of coordinates as part of the types, semantics, or even a core part of the language, especially~\cite{Chen2024, Geisler2020, adams2025asdf,Xu2025p, Xu2025, Carette2024, Por2023, Matsushita2026}.
Many of the coordinate-free languages and coordinate-dependent ones share similarities with our language (\textit{e.g.}, dual spaces plus tensor products in~\cite{bernardy2025domain}, tensor product bases in~\cite{Xu2025}, direct sums and tensor products in~\cite{Carette2024}, inner products in~\cite{Giusti2026}, frames in~\cite{Geisler2020}, and linear abstraction evaluation via bases in~\cite{velez2026compiling}), but we don't know of any that rely on the specific interactions of inverses, linearity, inner products, the four fundamental subspaces,  coordinates, and basis representation as compilation.
We do note that~\citet{Geisler2020} and~\citet{Chen2024} do propose optimizing changes of basis or removing unnecessary ones.
Additionally, we are unaware of any formal PL work that builds exterior algebra from first principles. 
Geometric algebra has also been formalized in Lean using a construction independent of a choice of basis~\cite{wieser2022formalizing}.
Several recent languages engage with invertible computation in other contexts~\cite{Matsuda2020,Choudhury2022}. Algeo~\cite{Henglein2022} gives every function a generalized inverse in a language based on infinite-dimensional modules and the weak right inverse concept is also similar~\cite{morita2007automatic}; reversible quantum computing involves invertible linear algebra~\cite{Fang2026}  and incremental linear algebra has been considered~\cite{10.1145/3385398}.
We take inspiration from various computer algebra projects for expressing exterior and geometric algebra~\cite{Fontijne2006,Sangwine2016,eid2009gmac,Gevorkyan2023,Velasco2025,Ablamowicz2005,Dubinsky2019,MacCallum2018} though most are specialized to just geometric algebra; we are not aware of any that use our combinations of features or any that pose a formal semantics.
Another line of related work comes from examining how to target and use operations such as the SVD and other factorizations as part of the linear algebra mapping problem~\cite{Psarras2022, 10.1145/3446632, 2511.20198}. 
This is closely related to solver DSLs and libraries~\cite{gibson2020slate,devito2017opt, brown2012composable,Zhang2025, Padula2009}.
Our compilation procedure examines this problem as well (\textit{e.g.}, by choosing when to change basis or compute an SVD).
For a broader survey of performance and architecture, see~\cite{10.1145/3743135}; systems such as those in~\cite{ahrens2025finch,shaikhha2022functional} explore these issues, and the work in~\cite{unite:2025} relates them to spatial representations.
Our compilation procedure is heavily inspired by languages for scientific computation where compilation is placing a multi-linear operator in a given basis~\cite{alnaes2014unified, kirby2006compiler, kirby2008benchmarking, Betcke2021, Burns2020}.
Lastly, \GLang{} models a small part of the finite element language described in Collin's master's thesis~\cite{Collin2025Automated}, which sketches a similar core without metatheory.

\section{Discussion and Future Work}
We have defined a core language for linear algebra via coordinate-independent principles, utilizing a few concepts that allow us to express situations where the utility of coordinate independence comes through.
We also define a language to describe coordinate systems along similar principles.
We have defined two semantics, a coordinate-independent denotational semantics, and a coordinate-dependent operational semantics.
We argued that the operational semantics captures how one turns linear algebra into array programs via a choice of coordinates.
Our metatheory proves that checked evaluation exists at every valid basis and agrees with the coordinate-independent denotation.

We see many future directions for our work, along four broad axes. 
First, our language can become more expressive in a programming language sense; we are particularly interested in the details of parametric polymorphism as they relate to coordinate independence and vector spaces defined via recursion or limiting procedures.
Second, our language can become more mathematically expressive, taking on infinite-dimensional vector spaces, function spaces, which means taking on integrals,  derivatives, and other higher-order constructs in their proper setting along with the functional analytic issues therein.
Within this axis, we are also interested in the complex field and the $n$th roots of operators as an approach to quantum programming languages.
At the intersection of the first two axes, we could also consider spaces defined via general spans of vectors or other dynamic construction of spaces/inner products.
A third axis is optimization, which can take many forms.
We would like to consider the rewriting theory of our language, as a tool for proofs and optimization.
We could further extend this by considering how some optimizations can be modeled at the highest level (in our language strength reduction can often be captured via beta-reduction vs. assembling a lambda).
We can also consider the quality of results produced via our basis-dependent semantics in terms of their suitability for other tasks, such as usage in computer algebra systems.
We are particularly interested in invertible computation via either more general pseudo-invertible computation or regular invertible computation in \GLang{}.
Finally, we are interested in showing that this truly works for real applications. 
Collin's master's thesis describes a finite element system underlaid via a language similar to the one described here~\cite{Collin2025Automated}.
In theory, we believe much of the work could be completely done in a system like \GLang{} though it requires us to subsume other mathematical concepts.
Currently we only subsume certain core ideas (\textit{e.g.}, our handling of lambdas captures a high-level commitment of several scientific DSLs~\cite{alnaes2014unified,kirby2006compiler,Burns2020,Betcke2021}), but we are optimistic about a broad scientific project to develop constructive scientific computing.

\section*{Acknowledgments}
We acknowledge support from Intel; the National Science Foundation through PPoSS grant CCF-2217064 and SHF grant CCF-2107244; DARPA through PROWESS award HR0011-23-C-0101 and SBIR award HR001123C0139; and the U.S. Department of Energy through PSAAP center award DE-NA0003965.
Theadora Fields Collin acknowledges support from the NSF Graduate Research Fellowship Program.
We also acknowledge support from the MIT Center for Exascale Simulation of Coupled High Enthalpy Fluid-Solid Interactions (CHEFSI), sponsored by the U.S. Department of Energy's National Nuclear Security Administration through the Predictive Science Academic Alliance Program.
Oded Stein acknowledges support from NSF grant 2335493 and a gift from Adobe Inc.
We wish to acknowledge Sofia Chandler-Freed, Eleanor Mast, and Christopher McNally for their extensive editing. We particular wish to acknowledge Christopher McNally for many substantive conversations about this work. 

\section*{Data Availability Statement}
We offer two artifacts. First, we provide a \href{https://github.com/wraith1995/Grassmann/tree/2200d0d05e7100ec3c1fca17b5f4e9c4eec442dc}{Python implementation} of a larger language that includes \GLang{}.
Though the languages and semantics do not exactly match, this implementation is effectively a superset of the work presented here.
Further, this implementation includes our benchmarks and a tutorial on how to implement a highly sophisticated exterior calculus, extending our examples to actual differential equations.
Second, we offer the \href{https://github.com/wraith1995/GlangLean/tree/b4116e7e45bb4ca24d929f12d304c8433e34685d}{Lean formalization} of the results, explained in Appendix~\ref{app:lean-results}.
The formalization closely follows the paper; its use of classical mathematics makes it a verification of our metatheory rather than an executable implementation.
For all of these artifacts, Anthropic and OpenAI agents were used in the implementation.
The implementation has been extensively tested and read and we've carefully audited the definitions and reported statements of the Lean proofs as well as their correspondence to the material reported here.

\bibliographystyle{ACM-Reference-Format}
\bibliography{sample-base}

\clearpage
\appendix
\counterwithin{figure}{section}
\renewcommand{\theHfigure}{\theHsection.\arabic{figure}}
\begingroup
\input{tex/appendices/helpers}
\section{Catalogue}\label{app:catalogue}
 Appendices B--D collect syntax and semantics; E collects binding operations; F collects definitions, theorem statements, and proof sketches. Appendices~\ref{app:implementation-performance-correctness} and~\ref{app:ufl-comparison} cover implementation details and the comparison with UFL.
 
\subsection{Syntactic categories}
\begin{center}\begin{tabular}{ll@{\qquad}ll}
$\lkn{}$ & source kind & $\glsynlhs{valKind.base}$ & value kind\\
$\ls{}$ & source space & $\lsv{}$ & value space\\
$\lty{}$ & source type & $\ltyv{}$ & value type\\
$\lee{}$ & source term & $\leev{}$ & value expression\\
$\lb{}$ & basis or frame & $x,y$ & variable names\\
$r\in\mathbb R$ & scalar & $i,j,k\in\mathbb N$ & indices
\end{tabular}\end{center}
\subsection{Names and environments}\label{tab:metaList}\label{tab:envs}
\begin{longtable}{@{}p{.36\linewidth}p{.59\linewidth}@{}}
\textbf{Notation} & \textbf{Meaning}\\\hline\endhead
$\glsyn{meta.lctx}$ & \glsyndesc{meta.lctx}\\
$\glsyn{meta.nlctx}$ & \glsyndesc{meta.nlctx}\\
$\glsyn{meta.evalCtx}$ & \glsyndesc{meta.evalCtx}\\
$\glsyn{meta.denv}$ & \glsyndesc{meta.denv}\\
$\glsyn{meta.valCtx}$ & \glsyndesc{meta.valCtx}\\
$\glsyn{meta.valEnv}$ & \glsyndesc{meta.valEnv}\\
$\glsyn{meta.tof}$ & \glsyndesc{meta.tof}\\
$\glsyn{meta.sof}$ & \glsyndesc{meta.sof}\\
$\glsyn{meta.stdBasis}$ & \glsyndesc{meta.stdBasis}\\
$\glsyn{meta.stdBasisSpace}$ & \glsyndesc{meta.stdBasisSpace}\\
$\glsyn{meta.iof}$ & \glsyndesc{meta.iof}\\
$\glsyn{meta.dof}$ & \glsyndesc{meta.dof}\\
$\glsyn{meta.shpOf}$ & \glsyndesc{meta.shpOf}\\
$\glsyn{meta.inferType}$ & \glsyndesc{meta.inferType}\\
$\glsyn{meta.inferSpace}$ & \glsyndesc{meta.inferSpace}\\
$\glsyn{meta.denotTerm}$ & \glsyndesc{meta.denotTerm}\\
$\glsyn{meta.denotTy}$ & \glsyndesc{meta.denotTy}\\
$\glsyn{meta.denotBasis}$ & \glsyndesc{meta.denotBasis}\\
$\glsyn{meta.denotVal}$ & \glsyndesc{meta.denotVal}\\
\end{longtable}
\clearpage
\subsection{Types of the catalogue functions}
\label{app:semantic-signature-review}

\begingroup\small
\begin{longtable}{@{}p{.30\linewidth}p{.65\linewidth}@{}}
\textbf{Function / notation} & \textbf{Type or dependent result}\\\hline\endhead
$\glsyn{meta.tof}$ & $\operatorname{typeOf}:\mathsf{Basis}\to\mathsf{Ty}$\\[3pt]
$\glsyn{meta.sof}$ & $\operatorname{spaceOf}:\mathsf{Basis}\to\mathsf{Space}$\\[3pt]
$\glsyn{meta.stdBasis}$ & $\operatorname{stdBasis}(-,\lctxnle):\mathsf{Ty}\to\mathsf{Basis}$\\[3pt]
$\operatorname{isStdBasis}(\lb)$ & $\operatorname{isStdBasis}:\mathsf{Basis}\to\mathsf{Prop}$\\[3pt]
$\glsyn{meta.stdBasisSpace}$ & $\operatorname{stdBasis}(-,\lctxnle):\mathsf{Space}\to\mathsf{Basis}$\\[3pt]
$\glsyn{meta.iof}$ & $\operatorname{innerOf}:\mathsf{Space}\to\mathsf{Term}$;
$\SemT{\ls}{\sigma}\times\SemT{\ls}{\sigma}\to\mathbb R$ after interpretation\\[3pt]
$\glsyn{meta.dof}$ & $\operatorname{dimOf}(-,\glvEnv):$
$\mathsf{ValSpace}\to\mathsf{Result}(\mathbb N)$\\[3pt]
$\glsyn{meta.shpOf}$ & $\operatorname{shapeOf}(-,\glvEnv):$
$\mathsf{ValSpace}\to\mathsf{Result}(\mathbb N^{<\omega})$\\[3pt]
$\glsyn{meta.inferType}$ & $\operatorname{inferType}:\mathsf{ValExpr}\to\mathsf{ValTy}$\\[3pt]
$\glsyn{meta.inferSpace}$ & $\operatorname{inferSpace}:$
$\mathsf{ValExpr}\to\mathsf{ValSpace}$\\[3pt]
$\SemT{\ls}{\sigma}$ (space) &
$\ST(-,\sigma):\mathsf{Space}\to\mathsf{Hilb}_{\mathbb R}^{\mathrm{fd}}$\\[3pt]
$\glsyn{meta.denotTy}$ (type) &
$\ST(-,\sigma):\mathsf{Ty}\to\mathsf{Set}$\\[3pt]
$\glsyn{meta.denotTerm}$ &
$\SE(-,\sigma):\{t\in\mathsf{Term}\mid\lctx\vdash t:\lty\}$
$\to\SemT{\lty}{\sigma}$\\[3pt]
$\glsyn{meta.denotBasis}$ (space) &
$\SemBB{\lb}{\sigma}\in\SemT{\ls}{\sigma}^{\circ}$;
$V^\circ$ is the set of ordered bases of $V$\\[3pt]
$\glsyn{meta.denotBasis}$ (function) &
A domain basis/frame $B_0$ and a family $B_1(a)$ of codomain bases/frames,
for $a\in\SemT{T_0}{\sigma}$, interpreted under $\sigma[x\mapsto a]$
at type $T_1$\\[3pt]
$\glsyn{meta.denotVal}$ &
$\SemA{-}{\glvEnv}{}:\mathsf{ValExpr}\to\mathsf{Result}(\mathsf{Value})$\\[3pt]
$\SemCheck{\leev}{\glvEnv}$ &
$\SemCheck{\leev}{\glvEnv}\in\mathcal V_{\glvEnv}(\ltyv)$ under
value typing and environment validity\\[3pt]
$\mathcal V_{\glvEnv}$ & $\mathsf{ValTy}\to\mathsf{Set}$\\
\end{longtable}
\endgroup

\subsection{Judgments}\label{tab:myjudges}

\begin{longtable}{@{}p{.50\linewidth}p{.39\linewidth}r@{}}
\textbf{Judgment} & \textbf{Meaning} & \textbf{App.}\\\hline\endhead
$\gljudghead{inductiveCtxWF}$ & Inductive well-formed source contexts. & B\\
$\gljudghead{sValid}$ & Well-formed spaces. & B\\
$\gljudghead{tValid}$ & Well-formed types. & B\\
$\gljudghead{subType}$ & Subspace relation. & B\\
$\gljudghead{hasType}$ & Term typing. & B\\
$\gljudghead{basisValid}$ & Basis validity. & C\\
$\gljudghead{valSValid}$ & Value-space validity. & D\\
$\gljudghead{valTValid}$ & Value-type validity. & D\\
$\gljudghead{valHasType}$ & Value typing. & D\\
$\gljudghead{evalS}$ & Space evaluation. & C\\
$\gljudghead{evalT}$ & Type evaluation. & C\\
$\gljudghead{evalBB}$ & Change-of-basis evaluation. & C\\
$\gljudghead{opStep}$ & Term evaluation. & C\\
$\gljudghead{c.space}$ & Checked space evaluation. & C\\
$\gljudghead{c.ty}$ & Checked type evaluation. & C\\
$\gljudghead{c.tyAt}$ & Checked exact-basis type evaluation. & C\\
$\gljudghead{c.cob}$ & Checked change of basis. & C\\
$\gljudghead{c.term}$ & Checked term evaluation. & C\\
\end{longtable}

\subsection{Definition and theorem index}
\begin{longtable}{@{}p{.23\linewidth}p{.72\linewidth}@{}}
\textbf{Statement} & \textbf{Name}\\\hline\endhead
Definition~\ref{def:ffs} & Four Fundamental Subspaces\\
Proposition~\ref{prop:corelin} & Linear maps and contraction\\
Proposition~\ref{prop:core} & First Isomorphism Theorem, Pseudo-inverse\\
Definition~\ref{def:wfc} & $\gldefnsig{inductiveCtxWF}$, well-formed typing context\\
Definition~\ref{def:lctxsigma} & $\lctx\vdash\sigma$\\
Lemma~\ref{lem:soundLemma} & Soundness up to inner product\\
Lemma~\ref{lem:innerLema} & Inner product choice is sound\\
Definition~\ref{def:bunframe} & Discrete (Vector) Frame\\
Definition~\ref{def:lex} & $\LexInd{}$\\
Definition~\ref{def:gs} & $\GramSchdmit{}$\\
Proposition~\ref{prop:sp} & Spectral Projector\\
Definition~\ref{def:fsv} & Filtered singular vectors\\
Lemma~\ref{lem:denoteBasisSound} & Basis Denotational Semantics is Sound\\
Lemma~\ref{lem:stdBasisOrthonormal} & Standard Basis is orthonormal\\
Definition~\ref{def:gsvd} & (G)SVD,~\cite{golub2013matrix,Stewart1990-mt}\\
Definition~\ref{def:vrho} & $\valCheckCtx\vdash\glvEnv$\\
Lemma~\ref{lem:soundVals} & Soundness for values\\
Definition~\ref{def:nonrecurseeval} & $\lctx;\valCheckCtx\vdash\evalCtx$\\
Theorem~\ref{thm:preserve} & Preservation\\
Definition~\ref{def:envimply} & Checked operational evaluation\\
Theorem~\ref{thm:prog} & Progress\\
Definition~\ref{def:agree} & Agreement\\
Definition~\ref{def:allCtxAgree} & Consistent environments\\
Theorem~\ref{thm:main} & Adequacy\\
Corollary~\ref{cor:progressAdequacy} & Progress and adequacy\\
\end{longtable}
\subsection{Public Lean results}\label{app:lean-results}
The table lists the formal counterparts of the paper's language metatheory.
The development uses Lean~4.29.1~\cite{deMoura2021} and
mathlib~\cite{mathlib2020} at commit \texttt{5e932f97dd25}.
The public source, value, basis, preservation, progress, and adequacy results
have no project axioms or \texttt{sorry} dependencies: their axiom closures contain
only \texttt{propext}, \texttt{Quot.sound}, and \texttt{Classical.choice}.
Spectral projection and the paired SVD are proved in Lean, including agreement
between the operational SVD selections and the source filtered singular vectors.
The polynomial projector of \cref{prop:sp} is identified with the eigenspace
projector by \texttt{GLang.BasisSpectral.projectorPolynomial\_eq}.

\begin{longtable}{@{}p{.18\linewidth}p{.76\linewidth}@{}}
\textbf{Statement} & \textbf{Lean declaration}\\\hline\endhead
\ref{lem:soundLemma} & \glthmlean{metatheory.sourceSpaceSoundness}\\
\ref{lem:soundLemma} & \glthmlean{metatheory.sourceTypeSoundness}\\
\ref{lem:soundLemma} & \glthmlean{metatheory.sourceTermSoundness}\\
\ref{lem:innerLema} & \glthmlean{space.iofHasType}\\
\ref{lem:innerLema} & \glthmlean{space.iofDenotIsInnerProduct}\\
\ref{lem:denoteBasisSound} & \glthmlean{metatheory.sourceBasisSoundness}\\
\ref{lem:stdBasisOrthonormal} & \glthmlean{metatheory.standardBasisSoundness}\\
\ref{lem:soundVals} & \glthmlean{metatheory.valueTypeSoundness}\\
\ref{lem:soundVals} & \glthmlean{metatheory.valueTermSoundness}\\
\ref{thm:preserve} & \glthmlean{metatheory.spacePreservation}\\
\ref{thm:preserve} & \glthmlean{metatheory.termPreservation}\\
\ref{thm:preserve} & \glthmlean{metatheory.cobPreservation}\\
\ref{thm:prog} & \glthmlean{metatheory.checkedSpaceProgress}\\
\ref{thm:prog} & \glthmlean{metatheory.termProgress}\\
\ref{thm:main} & \glthmlean{metatheory.spaceAdequacy}\\
\ref{thm:main} & \glthmlean{metatheory.termAdequacy}\\
\ref{thm:main} & \glthmlean{metatheory.cobAdequacy}\\
\ref{cor:progressAdequacy} & \glthmlean{preview.adequacy.endToEnd}\\
\end{longtable}

\clearpage
\section{Source language}\label{app:gen-syntax}\label{app:gen-rules}\label{app:gen-defs}
\appgrammar{kind}{Kinds}\appgrammar{space}{Spaces}\appgrammar{ty}{Types}\appgrammar{term}{Terms}
\subsection{Formation and typing}\label{app:gen-rules:context}\label{app:gen-rules:sValid}\label{app:gen-rules:tValid}\label{app:gen-rules:hasType}\label{app:gen-rules:subType}
\subsubsection{Space and type formation}\label{fig:typingSpace}\leavevmode\par
\appRule{sValid.empty}
\appRule{sValid.real}
\appRule{sValid.dual}
\appRule{sValid.dsum}
\appRule{sValid.tprod}
\appRule{sValid.ker}
\appRule{sValid.coker}
\appRule{sValid.img}
\appRule{sValid.coimg}
\appRule{tValid.space}
\appRule{tValid.piTy}
\subsubsection{Context formation}\label{app:context-formation}\leavevmode\par
These rules present the context construction of \cref{def:wfc}.
\appRule{inductiveCtxWF.empty}
\appRule{inductiveCtxWF.linear}
\appRule{inductiveCtxWF.nonlinear}

\subsubsection{Term typing}\label{fig:typing}\leavevmode\par
\appRule{hasType.zero}
\appRule{hasType.smul}
\appRule{hasType.add}
\appRule{hasType.inner}
\appRule{hasType.realLit}
\appRule{hasType.appLinear}
\appRule{hasType.appInner}
\appRule{hasType.appNonlinear}
\appRule{hasType.varLin}
\appRule{hasType.varNlin}
\appRule{hasType.pinvTensor}
\appRule{hasType.pinvDual}
\appRule{hasType.promote}
\appRule{hasType.dsumE}
\appRule{hasType.tprodE}
\appRule{hasType.llamReal}
\appRule{hasType.llamTensor}
\appRule{hasType.embed}
\appRule{hasType.nllam}
\subsubsection{Subspace rules}\label{fig:sub}\leavevmode\par
\appRule{subType.subKer}
\appRule{subType.subCoker}
\appRule{subType.subImg}
\appRule{subType.subCoimg}

\clearpage
\subsection{Denotation and inner products}\label{app:gen-defs:denot}
\begin{figure}[!htbp]
{\scriptsize
\begin{mdframed}[nobreak=true,everyline=true,topline=true,frametitle=$\SemLUS{\ltyn{0}}{\sigma}$]
\setlength{\tabcolsep}{5pt}
\renewcommand{\arraystretch}{1.2}
\begin{tabular}{@{}c@{}}
\makebox[\linewidth]{\hfill$\gleqn{denotSpaceAux.empty}$\hfill$\gleqn{denotSpaceAux.ker}$\hfill}\\[3pt]
\makebox[\linewidth]{\hfill$\gleqn{denotSpaceAux.real}$\hfill$\gleqn{denotSpaceAux.coker}$\hfill}\\[3pt]
\makebox[\linewidth]{\hfill$\gleqn{denotSpaceAux.dual}$\hfill$\gleqn{denotSpaceAux.img}$\hfill}\\[3pt]
\makebox[\linewidth]{\hfill$\gleqn{denotSpaceAux.dsum}$\hfill$\gleqn{denotSpaceAux.coimg}$\hfill}\\[3pt]
\makebox[\linewidth]{\hfill$\gleqn{denotSpaceAux.tprod}$\hfill$\gleqn{denotTyAux.piTy}$\hfill}
\end{tabular}
\end{mdframed}
}
\caption{(Hilbert) Space and Function Space Semantics}
\label{fig:spaceSemantics}
\end{figure}
\begin{figure}[!htbp]
{\scriptsize
\begin{mdframed}[nobreak=true,everyline=true,topline=true,frametitle=$\SemLK{\lee}{\sigma}{\lty}$]
\setlength{\tabcolsep}{5pt}
\renewcommand{\arraystretch}{1.2}
\begin{tabular}{@{}c@{}}
\makebox[\linewidth]{\hfill$\gleqn{denotTermAux.zero}$\hfill$\gleqn{denotTermAux.inner}$\hfill}\\[3pt]
\makebox[\linewidth]{\hfill$\gleqn{denotTermAux.real}$\hfill$\gleqn{denotTermAux.smul}$\hfill}\\[3pt]
\makebox[\linewidth]{\hfill$\gleqn{denotTermAux.var}$\hfill$\gleqn{denotTermAux.add}$\hfill}\\[3pt]
\makebox[\linewidth]{\hfill$\gleqn{denotTermAux.pinvT}$\hfill$\gleqn{denotTermAux.appI}$\hfill}\\[3pt]
\makebox[\linewidth]{\hfill$\gleqn{denotTermAux.dsumE}$\hfill$\gleqn{denotTermAux.appL}$\hfill}\\[3pt]
\makebox[\linewidth]{\hfill$\gleqn{denotTermAux.tprodE}$\hfill$\gleqn{denotTermAux.appN}$\hfill}\\[3pt]
\makebox[\linewidth]{\hfill$\gleqn{denotTermAux.llam}$\hfill$\gleqn{denotTermAux.nllam}$\hfill}\\[3pt]
\makebox[\linewidth]{\hfill$\gleqn{denotTermAux.embed.ker}$\hfill$\gleqn{denotTermAux.embed.coker}$\hfill}\\[3pt]
\makebox[\linewidth]{\hfill$\gleqn{denotTermAux.embed.img}$\hfill$\gleqn{denotTermAux.embed.coimg}$\hfill}
\end{tabular}
\end{mdframed}
}
\caption{Vector Semantics}
\label{fig:vectorSemantics}
\end{figure}
\begin{figure}[!htbp]
{\scriptsize
\begin{mdframed}[nobreak=true,everyline=true,topline=true,frametitle=$\iof{\ls{}}$]
\begin{gather*}
\gleqn{iof.empty}\\[3pt]
\gleqn{iof.real}\\[3pt]
\gleqn{iof.dual}\\[3pt]
\gleqn{iof.dsum}\\[5pt]
\gleqn{iof.tprod}\\[5pt]
\gleqn{iof.ker}\\[3pt]
\gleqn{iof.coker}\\[3pt]
\gleqn{iof.img}\\[3pt]
\gleqn{iof.coimg}\\[3pt]
\bigl(\lctxnle\vdash\lee{}\colon\lsn{0}^{\lstar}\loprod\lsn{1}\bigr)
\end{gather*}
\end{mdframed}
}
\caption{Inner products for spaces in \GLang{}}
\label{fig:inner}
\end{figure}

\clearpage

\section{Bases and operational semantics}
\appgrammar{basis}{Basis syntax}
\subsection{Basis functions and denotation}\label{app:gen-defs:basis}
\begin{figure}[!htbp]
\begin{mdframed}[frametitle={$\sof{\lb{}},\;\tof{\lb{}}$}]
{\scriptsize\gldefn{basis.sof}\gldefn{basis.tof}}
\end{mdframed}
\caption{Space/Type of a basis}
\label{fig:spbasis}
\end{figure}
\begin{figure}[t]
\begin{mdframed}[frametitle=$\stdBasisDCtx{\lty}$]
{\scriptsize%
\begin{gather*}
\gleqn{basis.stdSpace.empty}\\[3pt]
\gleqn{basis.stdSpace.real}\\[3pt]
\gleqn{basis.stdSpace.dual}\\[3pt]
\gleqn{basis.stdSpace.dsum}\\[3pt]
\gleqn{basis.stdSpace.tprod}\\[3pt]
\gleqn{basis.stdSpace.ker.formula}\\[3pt]
\gleqn{basis.stdSpace.coker.formula}\\[3pt]
\gleqn{basis.stdSpace.img.formula}\\[3pt]
\gleqn{basis.stdSpace.coimg.formula}\\[3pt]
\gleqn{basis.stdSpace.ker.condition}\\[3pt]
\gleqn{basis.std.piTy}
\end{gather*}}
\end{mdframed}
\caption{StandardBasis}
\label{fig:stdBasis}
\end{figure}
\begin{figure}[t]
\begin{mdframed}[frametitle=$\isStdBasis{\lb}$]
{\scriptsize\gldefn{basis.isStd}}
\end{mdframed}
\caption{IsStandardBasis}
\label{fig:isStdBasis}
\end{figure}
\begin{figure}[t]
\begin{mdframed}[frametitle=$\SemBB{\lb{}}{\sigma}$]
{\scriptsize%
\begin{gather*}
\gleqn{denotBasis.empty}\\[3pt]
\gleqn{denotBasis.one}\\[3pt]
\gleqn{denotBasis.dual}\\[3pt]
\gleqn{denotBasis.dsum}\\[3pt]
\gleqn{denotBasis.tprod}\\[3pt]
\gleqn{denotBasis.proj.ker}\\[3pt]
\gleqn{denotBasis.proj.coker}\\[3pt]
\gleqn{denotBasis.proj.img}\\[3pt]
\gleqn{denotBasis.proj.coimg}\\[3pt]
\gleqn{denotBasis.image}\\[3pt]
\gleqn{denotBasis.terms}\\[3pt]
\gleqn{denotBasis.frame}
\end{gather*}}
\end{mdframed}
\caption{Basis Semantics}
\label{fig:basisSemantics}
\end{figure}

\clearpage
\subsection{Basis validity}\label{app:gen-rules:basisValid}
\label{fig:basisValid}
\appRule{basisValid.empty}
\appRule{basisValid.one}
\appRule{basisValid.dual}
\appRule{basisValid.dsum}
\appRule{basisValid.tprod}
\appRule{basisValid.projKer}
\appRule{basisValid.projCoker}
\appRule{basisValid.projImg}
\appRule{basisValid.projCoimg}
\appRule{basisValid.image}
\appRule{basisValid.terms}
\appRule{basisValid.frame}
\clearpage
\subsection{Ordinary evaluation}\label{app:gen-rules:evalS}\label{app:gen-rules:evalT}\label{app:gen-rules:evalBB}\label{app:gen-rules:opStep}
\subsubsection{Ordinary spaces and types}\label{fig:operationalSemanticsSpace}\leavevmode\par
\appRule{evalS.real}
\appRule{evalS.empty}
\appRule{evalS.tprod}
\appRule{evalS.dsum}
\appRule{evalT.piTy}
\appRule{evalS.dual}
\appRule{evalT.space}
\appRule{evalS.ker}
\appRule{evalS.coker}
\appRule{evalS.img}
\appRule{evalS.coimg}

\subsubsection{Ordinary terms}\label{fig:operationalSemantics1}\leavevmode\par
\appRule{opStep.var}
\appRule{opStep.rlit}
\appRule{opStep.zero}
\appRule{opStep.scale}
\appRule{opStep.add}
\appRule{opStep.dsumE}
\appRule{opStep.tprodE}
\appRule{opStep.cob}
\appRule{opStep.reshapeCob}
\appRule{opStep.pinvTStd}
\appRule{opStep.innerStd}
\appRule{opStep.pinvTGpinv}
\appRule{opStep.pinvDStd}
\appRule{opStep.pinvDGpinv}
\appRule{opStep.embedKer}
\appRule{opStep.embedCoker}
\appRule{opStep.embedImg}
\appRule{opStep.embedCoimg}

\subsubsection{Ordinary functions and application}\label{fig:operationalSemanticsEvals}\leavevmode\par
\appRule{opStep.appLLin}
\appRule{opStep.appILin}
\appRule{opStep.appNVal}
\appRule{opStep.nllam}
\appTermRule{opStep.llamAssemble}
\appTermRule{opStep.llamAssembleReal}
\appRule{opStep.frameReanchor}
\appRule{opStep.frameEta}

\subsubsection{Ordinary changes of basis}\label{fig:operationalSemanticsCOB}\leavevmode\par
\appRule{evalBB.inverse}
\appRule{evalBB.refl}
\appRule{evalBB.trans}
\appRule{evalBB.tprod}
\appRule{evalBB.dsum}
\appRule{evalBB.dualStd}
\appRule{evalBB.image}
\appRule{evalBB.termsRref}
{\small\appRule{evalBB.imageSvds}}
\begingroup\small
\renewcommand{\LTTT}[3]{\textcolor{lang}{\MTM{}_{#1}}%
  \left\langle\begin{gathered}#2,\\[3pt]#3\end{gathered}\right\rangle}
\appRule{evalBB.projKerCoker}
\appRule{evalBB.projImgCoimg}
\appRule{evalBB.projKerCokerCod}
\appRule{evalBB.projImgCoimgDom}
\endgroup
\clearpage
\subsection{Checked evaluation}
\subsubsection{Checked spaces}\label{app:gen-rules:c-space}
\noindent\emph{Judgment:}\enspace $\gljudghead{c.space}$\quad(\,9 rules\,)\par
\glrulev{c.space.real}
\glrulev{c.space.empty}
\glrulev{c.space.dual}
\glrulev{c.space.ker}
\glrulev{c.space.coker}
\glrulev{c.space.img}
\glrulev{c.space.coimg}
\glrulev{c.space.tprod}
\glrulev{c.space.dsum}
\subsubsection{Checked types}\label{app:gen-rules:c-ty}
\noindent\emph{Judgment:}\enspace $\gljudghead{c.ty}$\quad(\,2 rules\,)\par
\glrulev{c.ty.space}
\glrulev{c.ty.piTy}
\subsubsection{Checked types at a basis}\label{app:gen-rules:c-tyAt}
\noindent\emph{Judgment:}\enspace $\gljudghead{c.tyAt}$\quad(\,2 rules\,)\par
\glrulev{c.tyAt.space}
\glrulev{c.tyAt.frame}
\subsubsection{Checked change of basis}\label{app:gen-rules:c-cob}
\noindent\emph{Judgment:}\enspace $\gljudghead{c.cob}$\quad(\,13 rules\,)\par
\glrulev{c.cob.inverse}
\glrulev{c.cob.trans}
\glrulev{c.cob.tprod}
\glrulev{c.cob.dsum}
\glrulev{c.cob.dualStd}
\glrulev{c.cob.image}
\glcobrulepage{c.cob.imageSvds}
\glcobrulepage{c.cob.projKerCoker}
\glcobrulepage{c.cob.projImgCoimg}
\glrulev{c.cob.termsRref}
\glrulev{c.cob.refl}
\glcobrulepage{c.cob.projKerCokerCod}
\glcobrulepage{c.cob.projImgCoimgDom}
\subsubsection{Checked terms}\label{app:gen-rules:c-term}
\noindent\emph{Judgment:}\enspace $\gljudghead{c.term}$\quad(\,28 rules\,)\par
\glrulev{c.term.varLin}
\glrulev{c.term.varNlin}
\glrulev{c.term.promote}
\glrulev{c.term.cob}
\glrulev{c.term.reshapeCob}
\glrulev{c.term.scale}
\glrulev{c.term.add}
\glrulev{c.term.dsumE}
\glrulev{c.term.tprodE}
\glrulev{c.term.pinvTStd}
\glrulev{c.term.pinvTGpinv}
\glrulev{c.term.pinvDStd}
\glrulev{c.term.pinvDGpinv}
\glrulev{c.term.innerStd}
{\small\glrulev{c.term.embedKer}}
{\small\glrulev{c.term.embedCoker}}
{\small\glrulev{c.term.embedImg}}
{\small\glrulev{c.term.embedCoimg}}
\glrulev{c.term.appLLin}
\glrulev{c.term.appILin}
\glrulev{c.term.nllam}
\glrulev{c.term.appNVal}
\gltermrule[checked]{c.term.frameReanchor}
\glrulev{c.term.frameEta}
\gltermrule[checked]{c.term.llamAssemble}
\gltermrule[checked]{c.term.llamAssembleReal}
\glrulev{c.term.zero}
\glrulev{c.term.rlit}

\clearpage\section{Value language}\label{sec:vals}\label{app:gen-values}
\subsection{Syntax}
\label{app:gen-syntax:valKind}\label{app:gen-syntax:valSpace}\label{app:gen-syntax:valTy}\label{app:gen-syntax:valExpr}\label{app:gen-syntax:valFFSProp}
\begin{center}\scriptsize

\end{center}
\subsection{Formation and typing}\label{app:gen-rules:valSValid}\label{app:gen-rules:valTValid}\label{app:gen-rules:valHasType}

\subsubsection{Value-space formation}\label{fig:valChecking}\leavevmode\par
\appRule{valSValid.real}
\appRule{valSValid.empty}
\appRule{valSValid.dsum}
\appRule{valSValid.tprod}
\appRule{valSValid.ffs}

\subsubsection{Value-type formation}\leavevmode\par
\appRule{valTValid.space}
\appRule{valTValid.piTy}

\subsubsection{Value typing}\leavevmode\par
\appRule{valHasType.var}
\appRule{valHasType.zero}
\appRule{valHasType.rlit}
\appRule{valHasType.scale}
\appRule{valHasType.vadd}
\appRule{valHasType.dsumE}
\appRule{valHasType.tprodE}
\appRule{valHasType.delta}
\appRule{valHasType.idelta}
\appRule{valHasType.ttvSpace}
\appRule{valHasType.ttvReal}
\appRule{valHasType.mtm}
\appRule{valHasType.vsum}
\appRule{valHasType.vlam}
\appRule{valHasType.appV}
\appRule{valHasType.vkrn}
\appRule{valHasType.vblk}
\appRule{valHasType.reshap}
\appRule{valHasType.vtrans}
\appRule{valHasType.gsvdS}
\appRule{valHasType.gsvdU}
\appRule{valHasType.gsvdUI}
\appRule{valHasType.gsvdV}
\appRule{valHasType.gsvdVI}
\appRule{valHasType.pinvTensor}
\appRule{valHasType.pinvRaw}
\appRule{valHasType.gpinv}
\appRule{valHasType.rrefRep}
\appRule{valHasType.invTensor}
\appRule{valHasType.invRaw}
\appRule{valHasType.svds}
\clearpage
\subsection{Shapes and denotation}\label{app:gen-defs:valdenot}\label{app:gen-defs:valaux}\label{app:gen-defs:reshap}
\par\noindent\begin{minipage}{\linewidth}
\paragraph{Finite-array operations.}
 The denotation uses finite real arrays
$a\in\mathbb R^n$, with $|a|=n$, and the following operations.
These are semantic operations used to interpret value expressions.
Array layout is C-order; failed checks and reads produce $\glverr$,
which all operations propagate.
\begin{center}\small
\begin{tabular}{@{}p{.30\linewidth}p{.65\linewidth}@{}}
\textbf{Notation} & \textbf{Meaning}\\\hline
$\mathsf{arr}_n[f]$ & Construct $(f(0),\ldots,f(n-1))\in\mathbb R^n$
from $f:\{0,\ldots,n-1\}\to\mathbb R$.
For $n=0$, this is the empty array.\\[3pt]
$a[i]$ & Read entry $i$ if $0\leq i<|a|$; otherwise return $\glverr$.\\[3pt]
$a[r,c]$ & Read $a[nr+c]$ in the matrix view with $n$ columns,
for matrix indices $0\leq r<m$, $0\leq c<n$.\\[3pt]
$\mathsf{mat}_{m,n}(a)$ & View an array of length $mn$ as an
$m\times n$ matrix: $\mathsf{mat}_{m,n}(a)_{r,c}=a[nr+c]$.
The surrounding operation checks the shape and length.\\[3pt]
$\mathsf{flat}(B)[i]$ & For $B\in\mathbb R^{m\times n}$, read
$B_{i\mathbin{\mathrm{div}}n,\,i\bmod n}$ when $0\leq i<mn$,
and $0$ otherwise. This is a zero-extended mathematical view.\\[3pt]
$\mathsf{out}_n(B)$ & Construct
$\mathsf{arr}_n[i\mapsto\mathsf{flat}(B)[i]]$.
Each rule supplies the output length $n$ and its shape checks.\\
\end{tabular}
\end{center}
\end{minipage}\par

Dimensions and shapes range over $\mathbb N\sqcup\{\glverr\}$ and
$\mathbb N^{<\omega}\sqcup\{\glverr\}$, respectively.
Write $\mathbb R^n$ for flat arrays of length $n$; all clauses propagate $\glverr$.
The runtime domain $\mathsf{Value}$ contains finite real arrays and functions
that record their argument representation and may return $\glverr$.
The value domains $\mathcal V_{\glvEnv}$ are given in \cref{fig:valSpaceSemantics}.
\begin{defn}[Syntactic dimension bound]\label{def:dimension-upper-bound}
The function $\overline{\operatorname{dimOf}}:\mathsf{ValSpace}\to\mathbb N$
is defined recursively. Writing $D=\overline{\operatorname{dimOf}}$,
\[
\begin{gathered}
D(\lsempty{})=0,\qquad D(\lR{})=1,\\
D(\lsnv{0}\losum\lsnv{1})=D(\lsnv{0})+D(\lsnv{1}),\qquad
D(\lsnv{0}\loprod\lsnv{1})=D(\lsnv{0})D(\lsnv{1}),\\
D(\lR^{\leev{},\lprop{},\lsnv{0}\loprod\lsnv{1}})=
\begin{cases}
D(\lsnv{0}) & \lprop{}\in\{\glsyn{valFFSProp.kernel},\glsyn{valFFSProp.cokernel}\},\\
D(\lsnv{1}) & \lprop{}\in\{\glsyn{valFFSProp.image},\glsyn{valFFSProp.coimage}\},
\end{cases}\\
D(\lR^{\leev{},\lprop{},\lsv{}})=D(\lsv{})\quad
\text{if $\lsv{}$ is not a tensor-product constructor.}
\end{gathered}
\]

\end{defn}
\begin{figure}[!htbp]
{\scriptsize
\gldefn{value.shapes.compact}
\gldefn{value.denot.conventions}
\gldefn{value.domains}
If dimension evaluation fails, the domain is $\{\glverr\}$. Errors propagate before a result fibre is selected.
}
\caption{Finite shapes and value domains}
\label{fig:valSpaceSemantics}
\end{figure}
\begin{figure}[!htbp]
{\scriptsize
\gldefn{value.denot.scalars}
\gldefn{value.denot.coordinates}
}
\caption{Value denotation}
\label{fig:valValSemantics}
\end{figure}
\begin{figure}[!htbp]
\ContinuedFloat
{\scriptsize
\gldefn{value.denot.transpose}
\gldefn{value.denot.sum}
\gldefn{value.denot.contractions}
}
\caption{Value denotation (continued)}
\end{figure}
\begin{figure}[!htbp]
\ContinuedFloat
{\scriptsize
\gldefn{value.denot.assembly}
}
\caption{Value denotation (continued)}
\end{figure}
\begin{figure}[!htbp]
\ContinuedFloat
{\scriptsize
\gldefn{value.denot.matrix-conventions}
\gldefn{value.denot.inverse}
}
\caption{Value denotation (continued)}
\end{figure}
\begin{figure}[!htbp]
\ContinuedFloat
{\scriptsize
\gldefn{value.denot.gsvd}
\gldefn{value.denot.rref}
}
\caption{Value denotation (continued)}
\end{figure}
\begin{figure}[!htbp]
\ContinuedFloat
{\scriptsize
\gldefn{value.denot.lambda}
\gldefn{value.denot.application}
\gldefn{value.denot.reshape}
}
\caption{Value denotation (continued)}
\end{figure}
\clearpage
\subsection{Value inference}\label{app:value-inference}
The functions $\operatorname{inferType}$ and $\operatorname{inferSpace}$
read the annotated value syntax. Variables carry their declared types;
binders use the named-variable convention of the main text.
\gldefn{value.inference.helpers}
\gldefnblock{valExpr.inferSpace}
\gldefnblock{valExpr.inferType}

\clearpage\section{Free variables and binding}\label{sec:fv}\label{app:gen-defs:fv}
These equations use named binders, up to renaming. Substitution replaces free names and avoids capture; bound names are renamed to satisfy the displayed freshness conditions. Value annotations use the same arguments and type ascriptions as the main text. Free-variable collection and substitution traverse every annotation, including value expressions inside dynamic spaces. Bound names scope over function codomains and bodies, and over the codomain of a basis frame, but not over domains or the result space of a linear lambda.
\subsection{Free variables}
For source spaces, types, terms, and bases, $\fv{A}$ is defined on a kinding,
typing, or basis-validity derivation of $A$, left implicit in the notation.
The derivation supplies omitted annotations; their free variables are included,
but unrelated entries of the context are not.
In the four fundamental subspaces, the derivation supplies
$\lepsilon\vdash\lee:\lsn{0}^{\lstar}\loprod\lsn{1}$.
For a linear lambda, $\ls$ is the body's result space; for a term basis,
$\ls=\sof{\lb}$.
Rule names distinguish the application and pseudoinverse cases.
Figures~\ref{fig:fv}, \ref{fig:fvbasis}, and~\ref{fig:fvVals} give the free-variable functions for source syntax, bases, and values.
\begin{figure}[!htbp]
{\scriptsize
\gldefn{space.fv}
\gldefn{ty.fv}
\gldefn{term.fv}
}
\caption{Free variables on source kinding and typing derivations}
\label{fig:fv}
\end{figure}
\begin{figure}[!htbp]
{\scriptsize
\gldefn{basis.fv}
}
\caption{Free variables on basis-validity derivations}
\label{fig:fvbasis}
\end{figure}
\begin{figure}[!htbp]
{\scriptsize
\gldefn{valSpace.fv}
\gldefn{valTy.fv}
\gldefn{valExpr.fv}
}
\caption{Free variables of value syntax, including annotations}
\label{fig:fvVals}
\end{figure}
\clearpage
\subsection{Substitution}\label{app:substitution}
As with free variables, source substitution $A\subst{x}{\lee}$ is defined on
an implicit kinding, typing, or basis-validity derivation of $A$.
It also substitutes in the inferred annotations supplied by that derivation;
these remain omitted on both sides of the equations.
\gldefnblock{space.subst}
\gldefnblock{ty.subst}
\gldefnblock{term.subst}
\gldefnblock{basis.subst}
\gldefnblock{valSpace.subst}
\gldefnblock{valTy.subst}
\gldefnblock{valExpr.subst}
\subsection{Free variables of lists}
\gldefnblock{term.fvList}
\gldefnblock{valExpr.fvList}

\clearpage\section{Definitions and metatheory}\label{app:preview-source-results}
The statements follow the order of the main text and reuse its definition and theorem sources. Each proof sketch follows its statement.
\subsection{Source language (\cref{sec:defgrassman})}
\begin{defn}[Four Fundamental Subspaces (\cref{def:ffs})]

\label{def:ffs}
    Assume $W$ and $V$ are inner product spaces with inner products $\langle \cdot, \cdot\rangle_{W}$ and $\langle\cdot, \cdot\rangle_{V}$.
    Let $L$ be a linear map from $W$ to $V$.\footnote{Our names follow the four orthogonal subspaces: $\operatorname{coker}(L)=\ker(L)^\perp$ in the domain and $\operatorname{coimage}(L)=\operatorname{image}(L)^\perp$ in the codomain. These represent what are conventionally called the coimage and cokernel, respectively.}
     Two subspaces form the domain $W$:
     \begin{equation*}
         \operatorname{ker}(L) = \{x\in W\mid L(x) = 0\}\quad \operatorname{coker}(L) = \{x\in W\mid \forall x^{\prime}\in\operatorname{ker}(L).\;\langle x,x^{\prime} \rangle_{W} = 0\}.
     \end{equation*}
     Further $W\cong\operatorname{ker}(L)\oplus \operatorname{coker}(L)$.
     We recall that $\oplus$ is the direct sum, a special pair value (see~\cite{Dummit2003-np}).
     Similarly, we have $V\cong\operatorname{image}(L)\oplus \operatorname{coimage}(L)$ where:
     \begin{equation*}
         \operatorname{image}(L) = \{y\in V\mid \exists x\in W.\; L(x)=y\}\quad 
         \operatorname{coimage}(L) = \{y\in V\mid \forall y^{\prime}\in\operatorname{image}(L).\; \langle y,y^{\prime}\rangle_{V} = 0\}.
     \end{equation*}

\end{defn}
\begin{prop}[Linear maps and contraction (\cref{prop:corelin})]

\label{prop:corelin}
Let $V_0$ and $V_1$ be finite-dimensional vector spaces.
Then there exists a canonical isomorphism
$\phi:V_0^{\star}\otimes V_1
\longrightarrow \operatorname{Hom}(V_0,V_1)$.
For $A\in V_0^{\star}\otimes V_1$ and $v\in V_0$, let $\contractN{}{A}{v}$ represent evaluation via this isomorphism: $ \contractN{}{A}{v}=\phi(A)(v)\in V_1$.

\end{prop}
\begin{prop}[First Isomorphism Theorem, Pseudo-inverse (\cref{prop:core})]

\label{prop:core}
    For $L\in W^{\star}\otimes V$:
    \begin{enumerate}
    \item $L$ is an isomorphism from $\operatorname{coker}(L)$ to $\operatorname{image}(L)$.
    \item Since any subspace $A$ of an inner product space $B$ has a unique least-squares projection, denoted $P_{A}\in B^{\star}\otimes A$, there is a unique pseudo-inverse for any linear map given by $L^{\dagger} = L^{-1}\circ P_{\operatorname{image}(L)}$. Additionally,
    \begin{equation}
        L^{\dagger} = \lambda v\colon V.\; \argmin_{w\in W}\left\{\norm{w}_{W}\colon \norm{\contractN{}{L}{w} - v }_{V} = \min_{w^{\prime}\in W}\norm{\contractN{}{L}{w^{\prime}} - v}_{V}\right\}
    \end{equation}

\end{enumerate}

\end{prop}
\begin{defn}[$\gldefnsig{inductiveCtxWF}$, well-formed typing context (\cref{def:wfc})]
\label{def:wfc-appendix}
We have $\vdash(\emptyset,\emptyset)$. Given $\vdash(\lDelta,\lepsilon)$ and a name $x$ fresh in both components,
$\vdash(\lDelta\ldisun\{x\colon\ls{}\},\lepsilon)$ if $\fv{\ls{}}\subseteq\operatorname{dom}(\lepsilon)$ and $(\emptyset,\lepsilon)\vdash\ls{}\colon\svalid$;
and $\vdash(\lDelta,\lepsilon\ldisun\{x\colon\lty{}\})$ if $\fv{\lty{}}\subseteq\operatorname{dom}(\lepsilon)$ and $(\emptyset,\lepsilon)\vdash\lty{}\colon\lkn{}$ for some kind $\lkn{}$.
The corresponding construction rules are in \cref{app:context-formation}.

\end{defn}
\begin{defn}[$\lctx\vdash\sigma$ (\cref{def:lctxsigma})]
\label{def:lctxsigma-appendix}
We write $\lctx\vdash\sigma$ iff, for every
$(x,\lty{})\in\lDelta\ldisun\lepsilon$,
$\sigma(x)$ is defined and $\sigma(x)\in\SemT{\lty{}}{\sigma}$.
\end{defn}
\subsection{Source metatheory (\cref{sec:meta1})}
\begin{lem}[Soundness up to inner product (\cref{lem:soundLemma})]
\label{lem:soundLemmaA}
Suppose $\vdash\lctx$, $\lctx\vdash\lee{}\colon\lty{}$, and $\lctx\vdash\sigma$.
Then $(\emptyset,\lepsilon)\vdash\lty{}\colon\lkn{}$ for some $\lkn{}$.
For a fixed choice of inner products, $\Sem{\lee{}}{\sigma}$ and $\SemT{\lty{}}{\sigma}$
are independent of the typing derivation.
Moreover:
\begin{enumerate}
\item $\Sem{\lee{}}{\sigma}\in\SemT{\lty{}}{\sigma}$ is a well-defined vector or dependent function;
\item for every $(x,\ls{})\in\lDelta$, $y\mapsto\Sem{\lee{}}{\sigma[x\mapsto y]}$
is linear on $\SemT{\ls{}}{\sigma}$, pointwise for dependent-function results;
\item $\SemT{\lty{}}{\sigma}$ is a well-defined Hilbert space or set of dependent functions.
\end{enumerate}

\end{lem}
\begin{proof}
To prove $(\emptyset,\lepsilon)\vdash\lty\colon\lkn{}$ for some $\lkn{}$, we induct on $\lctx\vdash\lee\colon\lty$. In many cases (zero, real numbers, and abstractions), the required kinding follows from the premises. For variables, we use well-formedness of the context; for addition and scaling, we apply the inductive hypothesis. In the inclusion case, \glrulename{subType.subKer}, \glrulename{subType.subCoker}, \glrulename{subType.subImg}, and \glrulename{subType.subCoimg} supply the typing premises from which we recover validity of the ambient space.
For dependent application, \glrulename{hasType.appNonlinear}, we also use substitution for kinding: if $(\emptyset,\lepsilon)\vdash\len{1}\colon\ltyn{0}$ and $(\emptyset,\lepsilon,x\colon\ltyn{0})\vdash\ltyn{1}\colon\lkn{}$, then $(\emptyset,\lepsilon)\vdash\ltyn{1}\subst{x}{\len{1}}\colon\lkn{}$. Here $x$ is fresh for $\lepsilon$, and substitution is capture avoiding. This follows by mutual induction on validity and typing, using substitution as defined in \cref{app:substitution}.

To show that the semantics is independent of the typing derivation, we can argue that a syntax annotated with appropriate typing information can be produced from a well-typed term, which can be erased. We annotated the syntax with the info needed for each place the semantics uses the typing derivation (\textit{e.g.}, domain and codomain for anywhere the semantics expects a function; each direct sub-term's type for everywhere the derivation needs to use an operation like zero or times). We can show by induction that the typing derivation fixes these types uniquely, which is sufficient.

It remains to show that these denotations have the stated mathematical properties. The arguments for semantic membership, linearity, and type interpretation depend on one another, so we prove them simultaneously by mutual induction on typing and validity, generalizing over environments satisfying the context. We consider a few interesting cases.

\paragraph{$\Sem{\lee}{\sigma}$ is a well-defined vector or dependent function.}
Most algebraic cases follow from the vector space operations and the inductive hypotheses. We consider addition, abstraction, application, inclusion, and the pseudo-inverse.
\begin{enumerate}
    \item For \glrulename{hasType.add}, we use the rule
    \begin{equation*}
        \glrule{hasType.add},
    \end{equation*}
    and the corresponding denotation

    \begin{equation*}
        \gleqn{denotTermAux.add}.
    \end{equation*}
    The inductive hypotheses give a Hilbert space $\SemT{\ls}{\sigma}$ and two vectors in it. Their sum is therefore a vector in the same space. 

    \item For linear abstraction, \glrulename{hasType.llamReal} and \glrulename{hasType.llamTensor}, we use the rules
    \begin{equation*}
        \glrule{hasType.llamReal}.
    \end{equation*}
    \begin{equation*}
        \glrule{hasType.llamTensor},
    \end{equation*}
    and their denotation
    \begin{equation*}
        \gleqn{denotTermAux.llam}.
    \end{equation*}
    The body satisfies (1), and (2) makes this function linear in $y$. The finite-dimensional identifications in \cref{prop:corelin} then give an element of the indicated dual or tensor space.

    \item For linear evaluation, \glrulename{hasType.appLinear} and \glrulename{hasType.appInner}, we invoke \cref{prop:corelin}: contraction represents application of the linear map to its argument. For \glrulename{hasType.pinvTensor} and \glrulename{hasType.pinvDual}, the corresponding claim follows from \cref{prop:core}.
    \begin{equation*}
        \glrule{hasType.appLinear}.
    \end{equation*}
    \begin{equation*}
        \glrule{hasType.appInner}.
    \end{equation*}
    \begin{equation*}
        \glrule{hasType.pinvTensor}.
    \end{equation*}
    \begin{equation*}
        \glrule{hasType.pinvDual}.
    \end{equation*}

    \item For inclusion, \glrulename{hasType.embed}, the subspace rules and \cref{def:ffs} show that $\lctxnle\vdash\lsn{0}\leq\lsn{1}$ implies that $\SemT{\lsn{0}}{\sigma}$ is a subspace of $\SemT{\lsn{1}}{\sigma}$. The inclusion therefore has the required codomain.
    \begin{equation*}
        \glrule{hasType.embed}.
    \end{equation*}
    \begin{equation*}
        \glrule{subType.subKer}.
    \end{equation*}
    \begin{equation*}
        \glrule{subType.subCoker}.
    \end{equation*}
    \begin{equation*}
        \glrule{subType.subImg}.
    \end{equation*}
    \begin{equation*}
        \glrule{subType.subCoimg}.
    \end{equation*}

    \item For dependent abstraction, \glrulename{hasType.nllam}, we use the rule
    \begin{equation*}
        \glrule{hasType.nllam},
    \end{equation*}
    and its denotation

    \begin{equation*}
        \gleqn{denotTermAux.nllam}.
    \end{equation*}
    For every $y\in\lvert\SemT{\ltyn{0}}{\sigma}\rvert$, the extended environment satisfies the body's context. By (1), its denotation belongs to $\SemT{\ltyn{1}}{\sigma[x\mapsto y]}$. Thus the function belongs to the dependent product
    \begin{equation}\label{eq:source-proof-pi-denotation}
        \gleqn{denotTyAux.piTy}.
    \end{equation}

    \item For dependent application, we use the rule
    \begin{equation*}
        \glrule{hasType.appNonlinear}.
    \end{equation*}
    By the inductive hypotheses and \cref{eq:source-proof-pi-denotation}, applying the denoted function to $\Sem{\len{1}}{\sigma}$ gives an element of the corresponding fiber. Semantic substitution, proved mutually for spaces, types, and terms, identifies this fiber with the conclusion's type:
    \begin{equation*}
        \SemT{\ltyn{1}\subst{x}{\len{1}}}{\sigma}
        = \SemT{\ltyn{1}}{\sigma[x\mapsto\Sem{\len{1}}{\sigma}]}.
    \end{equation*}
\end{enumerate}

\paragraph{$y\mapsto\Sem{\lee}{\sigma[x\mapsto y]}$ is linear for each $x\in\lDelta$.}
\begin{enumerate}
    \item When the linear context is empty, the claim is vacuous. For \glrulename{hasType.zero}, the zero function is linear in every variable; for \glrulename{hasType.varLin}, the denotation is the identity.
    \begin{equation*}
        \glrule{hasType.zero}.
    \end{equation*}
    \begin{equation*}
        \glrule{hasType.varLin}.
    \end{equation*}

    \item For addition, both operands are linear in each variable by induction, so their sum is linear.
    \item For scalar multiplication, linear application, and tensor product, the premises split the linear context using $\ldisun$.
    \begin{equation*}
        \glrule{hasType.smul}.
    \end{equation*}
    \begin{equation*}
        \glrule{hasType.tprodE}.
    \end{equation*}
    Each variable therefore occurs on only one side of the bilinear operation. Holding the other side fixed gives a linear function. This is why these rules admit $xy$ with distinct linear variables, but not $x^{2}$.
    \item For direct sum, both premises of \glrulename{hasType.dsumE} use the same linear context:
    \begin{equation*}
        \glrule{hasType.dsumE}.
    \end{equation*}
    Both components are linear in each variable by induction, so their direct sum is linear.  Since the zero case doesn't care about the linear variables, we are able to define a linear inclusion into a direct sum.
    \item For linear abstraction, the inductive hypothesis gives linearity pointwise in every remaining linear variable. The identification with a dual or tensor space preserves the vector space operations. Inclusion is itself a linear map; the remaining rules requiring an empty linear context contribute no further cases.
\end{enumerate}

\paragraph{$\SemT{\lty}{\sigma}$ is a Hilbert space or set of dependent functions.}
The ordinary space constructors preserve finite-dimensional Hilbert spaces. We consider the fundamental subspaces and dependent products.
\begin{enumerate}
    \item For the kernel, we use the validity rule and denotation:
    \begin{equation*}
        \glrule{sValid.ker},
        \qquad \gleqn{denotSpaceAux.ker}.
    \end{equation*}
    By (1), $A=\Sem{\lee}{\sigma}$ represents a linear map from $\SemT{\lsn{0}}{\sigma}$ to $\SemT{\lsn{1}}{\sigma}$. Its kernel is a subspace of the former: $A(0)=0$, and $A(\alpha u+\beta v)=\alpha A(u)+\beta A(v)=0$ whenever $u,v\in\ker A$. Note at this stage, we need not argue this is a Hilbert space as an inner product is assumed and the space is finite dimensional.
    \item For the dependent type, \glrulename{tValid.piTy} supplies validity of the base and of each fiber in an extended environment.
    \begin{equation*}
        \glrule{tValid.piTy}.
    \end{equation*}
    By induction these interpretations exist, and \cref{eq:source-proof-pi-denotation} gives their dependent product. To recover the section interpretation, write $X=\lvert\SemT{\ltyn{0}}{\sigma}\rvert$ and $F_y=\lvert\SemT{\ltyn{1}}{\sigma[x\mapsto y]}\rvert$. Then
    \begin{equation*}
        E=\bigsqcup_{y\in X}\{y\}\times F_y,
        \qquad \pi\colon E\to X,
        \qquad \pi(y,z)=y.
    \end{equation*}
    A dependent function $f$ determines a section $s\colon X\to E$ by $s(y)=(y,f(y))$, with $\pi\circ s=\mathrm{id}_X$. Conversely, each section selects one element of every fiber. Thus the dependent product is precisely the set of sections of this family over the discrete base.
\end{enumerate}
\end{proof}

\begin{lem}[Inner product choice is sound (\cref{lem:innerLema})]
\label{lem:innerLemap}
Suppose $\vdash\lctxnle$, $\lctxnle\vdash\ls{}\colon\svalid$,
and $\lctxnle\vdash\sigma$.
Then:
\begin{enumerate}
\item $\lctxnle\vdash\iof{\ls{}}\colon\ls{}^{\lstar}\loprod\ls{}^{\lstar}$;
\item $\Sem{\iof{\ls{}}}{\sigma}$ is an inner product on $\SemT{\ls{}}{\sigma}$.
\end{enumerate}

\end{lem}
\begin{proof}

We induct on $\lctxnle\vdash\ls\colon\svalid$. The typing claim follows from the corresponding definition of $\iof{\ls}$ in \cref{fig:inner} and the typing rules. For the semantic claim, \cref{lem:soundLemma} gives bilinearity. We check symmetry and positive definiteness by cases.
Write $B_{\ls}$ for $\Sem{\iof{\ls}}{\sigma}$ viewed as a bilinear form via \cref{prop:corelin}.
\begin{enumerate}
    \item For a direct sum, the definition is
    \begin{equation*}
        \gleqn{iof.dsum}.
    \end{equation*}
    Write $x=x_0\oplus x_1$ and $y=y_0\oplus y_1$, where $x_i,y_i\in\SemT{\lsn{i}}{\sigma}$. The projections return the corresponding components, so abstraction, application, and addition give
    \begin{equation*}
        B_{\lsn{0}\losum\lsn{1}}(x,y)
        =B_{\lsn{0}}(x_0,y_0)+B_{\lsn{1}}(x_1,y_1).
    \end{equation*}
    By induction, this is the direct-sum inner product and is symmetric. Both summands in $B_{\lsn{0}\losum\lsn{1}}(x,x)$ are nonnegative. If $x\ne0$, some $x_i\ne0$, and hence
    \begin{equation*}
        B_{\lsn{0}\losum\lsn{1}}(x,x)
        =\sum_{j=0}^{1}B_{\lsn{j}}(x_j,x_j)
        \ge B_{\lsn{i}}(x_i,x_i)>0.
    \end{equation*}

    \item For a dual space, we use
    \begin{equation*}
        \gleqn{iof.dual}.
    \end{equation*}
    Thus $B_{\ls^{\lstar}}(f,g)=B_{\ls}(\sharp f,\sharp g)$, the dual inner product. The musical map $\sharp$ is a linear isomorphism. Symmetry follows from the inductive hypothesis, and $f\ne0$ implies $\sharp f\ne0$, giving positive definiteness.

    \item For a tensor product, the uncurrying operations in \cref{fig:inner} give the bilinear extension of
    \begin{equation*}
        B_{\lsn{0}\loprod\lsn{1}}(u_0\otimes u_1,v_0\otimes v_1)
        =B_{\lsn{0}}(u_0,v_0)B_{\lsn{1}}(u_1,v_1).
    \end{equation*}
    This is symmetric by induction. Choose orthonormal bases $(a_i)$ and $(b_j)$ for the two interpreted spaces. Writing $z=\sum_{i,j}c_{ij}a_i\otimes b_j$ gives $B_{\lsn{0}\loprod\lsn{1}}(z,z)=\sum_{i,j}c_{ij}^{2}$, which is positive for $z\ne0$. 

    \item For the fundamental subspaces, the definitions restrict the ambient inner product along inclusion; for example,
    \begin{equation*}
        \gleqn{iof.ker}.
    \end{equation*}
    Thus $B_{\Kernel(\lee)}(u,v)=B_{\lsn{0}}(\iota u,\iota v)$, where $\iota$ is the inclusion into $\SemT{\lsn{0}}{\sigma}$. This is the inherited inner product: symmetry follows from the ambient form, and injectivity of $\iota$ preserves positive definiteness. The other three subspaces are identical arguments in their respective ambient spaces.

    \item The scalar definition gives $B_{\lR}(x,y)=xy$. The empty-space definition gives the zero form on the zero-dimensional space, where positive definiteness is vacuous.
\end{enumerate}
\end{proof}

\subsection{Bases (\cref{sec:def2b1})}
\begin{defn}[Discrete (Vector) Frame (\cref{def:bunframe})]

\label{def:bunframe}
Suppose $f$ is a mapping from points in a set $X$ to real finite-dimensional vector spaces $V_{x}$, \textit{i.e.}, the codomain is the set $\{V_{x}\colon x\in X\}$.
We could  write $f\colon \Pi_{x\colon X} V_{x}$.
Given a finite-dimensional vector space $V$, let  $V^{\circ }$ be the set of bases for $V$.
Then a \textbf{discrete vector frame } is a mapping $g\colon (\Pi_{x\in X} V_{x}^{\circ})$ such that for all $x\in X$, $g(x)$ is a basis for $f(x)$.
We could denote this as $g\colon (\Pi_{x\in X} V_{x})^{\circ}$.
A more general discrete frame emerges naturally.
Given $f\colon \Pi_{y\colon Y}\Pi_{x\colon X} V_{y,x}$ where $V_{y,x}$ is a vector space, then the corresponding \textbf{discrete frame} is defined recursively as a mapping $g\colon \Pi_{y\in Y} (\Pi_{x\colon X} V_{y,x})^{\circ}$ where given $y\in Y$, $g(y)$ is a discrete vector frame for $f(y)$.
Then given some $f\colon \Pi_{y\colon Y}B_{y}$, we say that $g$ is a discrete frame for $f$, if $g\colon \Pi_{y\colon Y}B_{y}^{\circ}$.

\end{defn}
\begin{defn}[$\LexInd{}$ (\cref{def:lex})]
\label{def:lex}
Given a finite sequence $(v_i)_{i<n}$, define
$\LexInd((v_i)_{i<n})=(v_{i_j})_{j<r}$ by choosing the lexicographically least
increasing index tuple $(i_j)_{j<r}$ whose selected vectors are linearly independent, with $r$ maximal .

\end{defn}
\begin{defn}[$\GramSchdmit{}$ (\cref{def:gs})]

\label{def:gs}
    We define $\GramSchdmit{}(v_{0},\dots, v_{n})$ as the normalized Gram--Schmidt orthogonalization of the sequence provided that $v_{0},\dots, v_{n}$ are linearly independent in a known inner product space.

\end{defn}
\begin{prop}[Spectral Projector (\cref{prop:sp})]
\label{prop:spA}
Let a positive semi-definite linear map $F\in V^{\star}\otimes V$ be given. Suppose that $\lambda_{0},\dots,\lambda_{m}$ are the eigenvalues with duplicates discarded. Let $p_{\lambda_{j}}(A) = \prod_{i=0, i\neq j}^{m} \frac{A-\lambda_{i}I}{\lambda_{j} - \lambda_{i}}$.  Then $p_{\lambda_{j}}(F)\in V^{\star}\otimes V$ and projects $V$ onto $V_{\lambda_{j}} = \{v\colon F(v) = \lambda_{j}v\}$.
\end{prop}
\begin{proof}
The proof follows almost immediately from the spectral theorem. If $w\in V_{\lambda_{j}}^{\perp}$ then it can be represented as a sum of vectors in the other eigenspaces. Every factor $F-I\lambda_k$ preserves these eigenspaces, and one factor annihilates each such component. If $w\in  V_{\lambda_{j}}$ then it will be preserved by all product operands. To see that it will be preserved exactly, we can show that $\sum_{i=0}^{m} p_{\lambda_{i}}(F) = I$ via partition of unity for Lagrange polynomials so $\sum_{i=0, i\neq j}^{m} p_{\lambda_{i}}(F) = I - p_{\lambda_{j}}(F)$. Thus, if $w\in  V_{\lambda_{j}}$, then $w$ will be annihilated by $p_{\lambda_{i}}(F)$ for all $i\neq j$ so $Iw = p_{\lambda_{j}}(F)w$.
\end{proof}

\begin{defn}[Filtered singular vectors (\cref{def:fsv})]
\label{def:fsv}
For $A\in V^\star\otimes W$, ordered bases $v=(v_i)_{i<n}$ of $V$
and $w=(w_i)_{i<m}$ of $W$, $f:\mathbb R\to\operatorname{bool}$,
and $d\in\{\operatorname{Left},\operatorname{Right}\}$, put
$S_L=A^\star\circ A$ and $S_R=A\circ A^\star$.
Here $A^\star=\sharp_V\circ(\lambda\phi.\,\phi\circ A)\circ\flat_W$.
For a positive semidefinite operator $S$, write
$P_\lambda(S)=
\prod_{\mu\in\operatorname{spec}(S)\setminus\{\lambda\}}
(S-\mu I)/(\lambda-\mu)$ when $\lambda\in\operatorname{spec}(S)$, the eigenvalues of $S$,
and $P_\lambda(S)=0$ otherwise.
Define
$B_{\operatorname{Left},\lambda}
=\GramSchdmit\bigl(\LexInd((P_\lambda(S_L)(v_i))_{i<n})\bigr)$.
For $\lambda>0$, put
$B_{\operatorname{Right},\lambda}
=(A(u)/\sqrt{\lambda})_{u\in B_{\operatorname{Left},\lambda}}$;
at zero, put
$B_{\operatorname{Right},0}
=\GramSchdmit\bigl(\LexInd((P_0(S_R)(w_i))_{i<m})\bigr)$.
Let $(\lambda_j)_{j<r}$ list the distinct elements of
$\operatorname{spec}(S_L)\cup\operatorname{spec}(S_R)$
in decreasing order. Then
\[
 \FSV(A,f,d,v,w)
 :=\mathop{++}\nolimits_{j<r,\,f(\lambda_j)=\operatorname{true}}
 B_{d,\lambda_j},
\]
where $++$ concatenates the selected blocks.
\end{defn}
\begin{lem}[Basis Denotational Semantics is Sound (\cref{lem:denoteBasisSound})]
\label{lem:denoteBasisSoundA}
{\raggedright Suppose $\vdash\lctxnle$ and $\lctxnle\vdash\lb{}\colon\bvalid$.
\begin{enumerate}
\item $\lctxnle\vdash\tof{\lb{}}\colon\lkn{}$ for some $\lkn{}$.
\item If also $\lctxnle\vdash\sigma$, then $\SemBB{\lb{}}{\sigma}$ is a basis or discrete frame
for $\SemT{\tof{\lb{}}}{\sigma}$. Its denotation is independent of the typing derivation.
\end{enumerate}\par}

\end{lem}
\begin{proof}

We induct on $\lctxnle\vdash\lb\colon\bvalid$, generalizing over environments satisfying the context. The validity premises give $\lctxnle\vdash\tof{\lb}\colon\lkn{}$ for some $\lkn{}$ immediately.
For uniqueness of denotation, we can argue immediately from the uniqueness of the denotation of terms, \cref{lem:soundLemma}, as these determine the only denotations that depend on derivation.
For denotation's correctness, the empty, scalar, direct-sum, and tensor-product cases follow from their definitions in \cref{fig:basisSemantics} and the inductive hypotheses. We consider the remaining cases.
\begin{enumerate}
    \item For a dual basis, the denotation is
    \begin{equation*}
        \gleqn{denotBasis.dual}.
    \end{equation*}
    The inductive hypothesis gives a basis $(v_i)$ of $\SemT{\sof{\lb}}{\sigma}$. Its coordinate functionals are the unique $v_i'$ satisfying $v_i'(v_j)=\delta_{ij}$, and form a basis of the dual space.

    \item For an image basis, \glrulename{basisValid.image} supplies a basis of the cokernel and a well-typed map. The denotation is
    \begin{equation*}
        \gleqn{denotBasis.image}.
    \end{equation*}
    Write $A=\Sem{\lee}{\sigma}$. By \cref{lem:soundLemma,def:ffs}, the restriction $A\colon\operatorname{coker}(A)\to\operatorname{image}(A)$ is an isomorphism: it is injective on the orthogonal complement of its kernel, and every image vector has a preimage there. Thus it sends the basis supplied by induction to a basis of the image.

    \item For projected bases, the four denotations in \cref{fig:basisSemantics} use $\FSV$ (\cref{def:fsv}). Let $A=\Sem{\lee}{\sigma}$ and let $S=A^\star A$ for the left construction, or $S=AA^\star$ for the right construction. By \cref{lem:soundLemma,lem:innerLema}, these are well-defined self-adjoint positive-semidefinite operators. For example, $\langle A^\star Au,u\rangle=\langle Au,Au\rangle\ge0$.
    For each selected eigenvalue $\lambda$, \cref{prop:sp} gives the spectral projector $P_\lambda(S)$ onto its eigenspace. On the left, projecting the prescribed input basis gives a spanning sequence; $\LexInd$ extracts a basis and $\GramSchdmit$ makes it orthonormal. At zero on the right, we apply the same construction to $S_R=AA^\star$ and the right input basis.
    For $\lambda>0$, the right block is paired with the left block by $u\mapsto A(u)/\sqrt\lambda$. If $u,u'$ lie in the left $\lambda$-eigenspace, then
    \[
      \left\langle\frac{A(u)}{\sqrt\lambda},\frac{A(u')}{\sqrt\lambda}\right\rangle
      =\lambda^{-1}\langle u,A^\star A(u')\rangle=\langle u,u'\rangle.
    \]
    This map is an isomorphism onto the right $\lambda$-eigenspace, with inverse $w\mapsto A^\star(w)/\sqrt\lambda$, so it sends the left orthonormal block to an orthonormal basis of that right eigenspace. Distinct eigenspaces are orthogonal, and concatenating the selected blocks gives a basis of their sum.
    The zero and positive eigenspaces of $A^\star A$ give $\operatorname{ker}(A)$ and $\operatorname{coker}(A)$; those of $AA^\star$ give $\operatorname{coimage}(A)$ and $\operatorname{image}(A)$. These are exactly the filters in the four clauses. As in \cref{def:fsv}, the distinct eigenvalues in the union of the two spectra are processed in decreasing order, with the prescribed input order fixing each projected block.

    \item For a term basis, \glrulename{basisValid.terms} and \cref{lem:soundLemma} put all supplied vectors in the space of the fallback basis. We use
    \begin{equation*}
        \gleqn{denotBasis.terms}.
    \end{equation*}
    The appended basis spans by induction, so the concatenated sequence spans even when the supplied terms are linearly dependent or zero. $\LexInd$ retains a linearly independent subsequence with the same span, hence a basis. Its prescribed scan order makes this choice unique.

    \item For a frame, \glrulename{basisValid.frame} gives validity of the domain basis and of the fiber basis under the extended context. Its denotation is
    \begin{equation*}
        \gleqn{denotBasis.frame}.
    \end{equation*}
    Every argument $x'\in\SemT{\tof{\lbn{0}}}{\sigma}$ gives an environment satisfying that extended context. The inductive hypothesis therefore supplies a basis or frame in each fiber. Together with the domain basis, this is the discrete frame required by \cref{def:bunframe}.
\end{enumerate}
\end{proof}

\begin{lem}[Standard Basis is orthonormal (\cref{lem:stdBasisOrthonormal})]
\label{lem:stdBasisOrthonormalA}
{\raggedright Suppose $\vdash\lctxnle$ and $\lctxnle\vdash\lty{}\colon\lkn{}$. Then:
\begin{enumerate}
\item $\lctxnle\vdash\stdBasisDCtx{\lty{}}\colon\bvalid$,
$\tof{\stdBasisDCtx{\lty{}}}=\lty{}$, and $\isStdBasis{\stdBasisDCtx{\lty{}}}$.
\item If also $\lctxnle\vdash\sigma$, then $\SemBB{\stdBasisDCtx{\lty{}}}{\sigma}$ is an orthonormal basis or
discrete frame for $\SemT{\lty{}}{\sigma}$ under the chosen inner products.
\item If $\lctxnle\vdash\lb{}\colon\bvalid$, $\tof{\lb{}}=\lty{}$, and
$\isStdBasis{\lb{}}$, then $\stdBasisDCtx{\lty{}}=\lb{}$.
\end{enumerate}\par}

\end{lem}
\begin{proof}

For validity, the type equation, and standardness, we induct on type validity using the clauses in \cref{fig:stdBasis,fig:isStdBasis}. Each space constructor uses the corresponding basis-validity rule and the inductive hypotheses. For a dependent type, \glrulename{basisValid.frame} applies to the standard basis of the domain and the standard basis of the codomain under the extended context. Thus $\stdBasisDCtx{\lty}$ is valid, has type $\lty$, and satisfies $\isStdBasis{\stdBasisDCtx{\lty}}$.
By \cref{lem:denoteBasisSound}, its denotation is a basis or discrete frame. To show orthonormality, use the same induction and the inner products established in \cref{lem:innerLema}. We consider the cases.
\begin{enumerate}
    \item The empty basis is orthonormal vacuously, and the scalar basis consists of the unit vector $1$. For a direct sum, the component bases are orthonormal by induction and the two summands are orthogonal.

    \item For a dual space, the defining clause is
    \begin{equation*}
        \gleqn{basis.stdSpace.dual}.
    \end{equation*}
    If $(a_i)$ is the orthonormal basis supplied by induction and $(a^i)$ its dual basis, then $\sharp a^i=a_i$. The dual inner product therefore gives $\langle a^i,a^j\rangle=\langle a_i,a_j\rangle=\delta_{ij}$.

    \item For a tensor product, we use
    \begin{equation*}
        \gleqn{basis.stdSpace.tprod}.
    \end{equation*}
    By the tensor-basis denotation in \cref{fig:basisSemantics}, its vectors are $a_i\otimes b_j$, where $(a_i)$ and $(b_j)$ are the component standard bases. The tensor inner product calculated in \cref{lem:innerLemap} gives
    \begin{equation*}
        \langle a_i\otimes b_j,a_k\otimes b_\ell\rangle
        =\langle a_i,a_k\rangle\langle b_j,b_\ell\rangle
        =\delta_{ik}\delta_{j\ell}.
    \end{equation*}
    Thus the tensor basis is orthonormal.

    \item For the four fundamental subspaces, the standard-basis clauses select projected bases. The $\FSV$ construction in \cref{def:fsv} normalizes the basis within each selected eigenspace, and distinct eigenspaces are orthogonal. Their concatenation is therefore orthonormal in the inherited inner product.

    \item For a dependent type, the frame clause in \cref{fig:stdBasis} applies the construction to the domain and recursively to each fiber. The inductive hypotheses give an orthonormal domain basis or frame and an orthonormal basis or frame in every fiber. 
\end{enumerate}

For the final claim, induct on $\lctxnle\vdash\lb\colon\bvalid$, assuming $\isStdBasis{\lb}$. The clauses in \cref{fig:isStdBasis} exclude image and term bases. In every remaining case, they assert that the component bases are standard; the inductive hypotheses identify these with the standard bases of their types. The corresponding clause in \cref{fig:stdBasis} then reconstructs $\lb$ with the appropriate type.
\end{proof}

\subsection{Values (\cref{sec:operational})}\label{app:preview-checked}
\begin{defn}[(G)SVD (\cref{def:gsvd})]

\label{def:gsvd}
Let $L\in W^{\star}\otimes V$ be given, where
$\dim W=m$ and $\dim V=n$, and let $\operatorname{rank}(L)=r\le \min(m,n)$.
After choosing bases for $W$ and $V$, we may represent $L$ as a matrix in
$\mathbb{R}^{n\times m}$. 
Let $M_W$ and $M_V$ denote (positive definite) matrices of the inner products on $W$ and $V$, respectively.
Then we may compute matrices
\[
\begin{gathered}
\GSVDU(L,M_W,M_V)\in\mathbb{R}^{m\times r},\qquad
\GSVDUI(L,M_W,M_V)\in\mathbb{R}^{m\times (m-r)},\\
\GSVDV(L,M_W,M_V)\in\mathbb{R}^{n\times r},\qquad
\GSVDVI(L,M_W,M_V)\in\mathbb{R}^{n\times (n-r)},
\end{gathered}
\]
whose columns are mutually orthonormal in the inner products on $W$ and $V$,
respectively, together with a diagonal matrix
$\GSVDS(L,M_W,M_V)=\operatorname{diag}(\sigma_1,\dots,\sigma_r)\in\mathbb{R}^{r\times r}$,
where each $\sigma_i>0$ and $\sigma_i\ge \sigma_{i+1}$.

These satisfy
\[
L=\GSVDV(L,M_W,M_V)\,
\GSVDS(L,M_W,M_V)\,
\GSVDU(L,M_W,M_V)^T M_W.
\]

Moreover,
\[
\begin{gathered}
\operatorname{image}(L)=\operatorname{span}(\GSVDV(L,M_W,M_V)),\qquad
\operatorname{coimage}(L)=\operatorname{span}(\GSVDVI(L,M_W,M_V)),\\
\operatorname{kernel}(L)=\operatorname{span}(\GSVDUI(L,M_W,M_V)),\qquad
\operatorname{cokernel}(L)=\operatorname{span}(\GSVDU(L,M_W,M_V)).
\end{gathered}
\]

We fix the choices of singular vectors by \cref{def:fsv}.
If the inner products are represented by identity matrices, we drop the prefix
$G$ and write simply SVD.
Finally, the Moore--Penrose pseudo-inverse is
\begin{equation}
\label{eq:gsvdpseudo}
L^\dagger
=
\GSVDU(L,M_W,M_V)\,
\GSVDS(L,M_W,M_V)^{-1}\,
\GSVDV(L,M_W,M_V)^T M_V.
\end{equation}

\end{defn}
\paragraph{Value domains.}
\gldefn{value.domains}
If a space dimension fails, its domain is $\{\glverr\}$; all semantic functions propagate $\glverr$.
\paragraph{Checked value denotation.}
Write $\SemCheck{v}{\glvEnv}:=\mathcal E^{\mathrm{check}}_{\operatorname{inferType}(v)}(v,\glvEnv)$ for value denotation checked against its inferred type.
\begin{defn}[$\valCheckCtx\vdash\glvEnv$ (\cref{def:vrho})]
$\valCheckCtx\vdash\glvEnv\Leftrightarrow
\forall(x\colon\ltyv)\in\valCheckCtx.\;
 (\exists\lkn{}.\;\valCheckCtx\vdash\ltyv\colon\lkn{})
 \land\glvEnv(x)\in\mathcal V_\glvEnv(\ltyv)\setminus\{\glverr\}$.

\end{defn}
\begin{lem}[Soundness for values (\cref{lem:soundVals})]
\label{lem:soundValsA}
If $\valCheckCtx\vdash\glvEnv$ and $\valCheckCtx\vdash\leev\colon\ltyv$,
then:
\begin{enumerate}
\item $\exists\lkn{}.\;\valCheckCtx\vdash\ltyv\colon\lkn{}$;
\item $\operatorname{inferType}(\leev)=\ltyv$;
\item $\SemCheck{\leev}{\glvEnv}\in\mathcal V_\glvEnv(\ltyv)$.
\end{enumerate}

\end{lem}
\begin{proof}
We prove the three conclusions separately. For regularity, induct on
$\valCheckCtx\vdash\leev\colon\ltyv$. The premise
$\valCheckCtx\vdash\glvEnv$ of \cref{def:vrho} supplies formation of every
declared value type and recursive membership of its runtime binding.
In particular, the value context is valid. Each typing rule either supplies formation of its result
type explicitly, or constructs that type from the types supplied by its
premises. For example, \glrulename{valHasType.vlam} supplies formation of the
whole function type, and \glrulename{valHasType.appV} supplies formation of the
substituted result type; both rules are displayed below. Thus
$\exists\lkn{}.\;\valCheckCtx\vdash\ltyv\colon\lkn{}$, proving (1).

For (2), induct on the same typing derivation and unfold
$\operatorname{inferType}$ using \cref{app:value-inference}. The induction hypotheses identify the types of the
operands; the annotations and equalities in the typing rule then identify the
result. In particular, application substitutes its argument in the inferred
function codomain, exactly as its typing rule does. This proves the
syntactic equality $\operatorname{inferType}(\leev)=\ltyv$ before evaluating
$\leev$; it does not require evaluation to succeed. We give the corresponding
inference calculation in each case below.

For (3), induct simultaneously on space formation, type formation, and value
typing, generalizing to related environments and arguments. The stronger
hypothesis relates equal arrays and functions that take related arguments to
related results. It also relates the dependent result types in the two extended
environments, which is needed in the abstraction, application, and reshape
cases. Specializing the two environments to $\glvEnv$ gives membership in the
value domain. Write
$\mathcal V^+_\glvEnv(\ltyv)=\mathcal V_\glvEnv(\ltyv)\setminus\{\glverr\}$.
The domain clauses used below are
\gldefn{value.domains}
We consider a few cases, spelling out successful results
and errors separately.
\begin{enumerate}
 \item For variables, the typing rule is
 \[\glrulenamed{valHasType.var}.\]
 Its premise selects the binding $(x\colon\ltyv)\in\valCheckCtx$.
 The environment assumption therefore gives
 $\glvEnv(x)\in\mathcal V^+_\glvEnv(\ltyv)$. Lookup and type inference give
 \[
   \Sem{x_{\ltyv}}{\glvEnv}=\glvEnv(x),\qquad
   \operatorname{inferType}(x_{\ltyv})=\ltyv.
 \]
 The stored value already belongs to the prescribed domain, so recursive
 checking retains a representative of that same value. Consequently
 $\SemCheck{x_{\ltyv}}{\glvEnv}\in\mathcal V_\glvEnv(\ltyv)$.
 Scalar constants and zero arrays follow from their denotations in
 \cref{fig:valValSemantics}.

 \item For matrix multiplication, use
 \[\glrulenamed{valHasType.mtm}.\]
 The operand hypotheses give their displayed tensor types; hence inference
 returns
 \[
 \operatorname{inferType}(\glvmtm{\lsnv{1}''}{\lsnv{1}}{\lenv{0}}{\lenv{1}})
   =\lsnv{2}\loprod\lsnv{0}.
 \]
 Put $d_c=\dof{\lsnv{1}''}{\glvEnv}$. On successful dimension and operand
 evaluation, the denotation clause is
 {\small\[\gleqn{value.denot.mtm.compact}.\]}
 The side conditions check both operand lengths and equality of the three
 contraction dimensions. These are runtime checks: the typing rule alone does
 not assert that the contraction dimensions evaluate to the same natural
 number. When they pass, each summand reads within the stored arrays, since
 \[
 \begin{gathered}
 0\leq k<d_c,\quad 0\leq c<d_0,\quad 0\leq r<d_2,\\
 0\leq d_0k+c<d_cd_0,\qquad 0\leq d_cr+k<d_2d_c.
 \end{gathered}
 \]
 Thus every output coordinate is a real number. The clause constructs exactly
 one coordinate for each $(r,c)$, so the resulting flat array $C$ has length
 $d_2d_0$. By the space-domain clause,
 \[
 C\in\mathbb R^{d_2d_0}\subseteq
 \mathcal V_\glvEnv(\lsnv{2}\loprod\lsnv{0}).
 \]
 Recursive checking accepts this length. If a dimension, operand, or shape
 check fails, the result is instead $\glverr$, which belongs to the same
 domain. In either case the checked multiplication has the required
 membership. The other array clauses use their own bounds and output-length
 checks in the same way.

 \item For abstraction, the typing and denotation clauses are
 \[\glrulenamed{valHasType.vlam}.\]
 \gldefn{value.denot.lambda}
 The first typing premise supplies formation of the function type, and
 inference reads the two annotations:
 \[
 \operatorname{inferType}(\llambda(x\colon\ltynv{0})\lrh\ltynv{1}\colon\leev)
   =(x\colon\ltynv{0}\fnto\ltynv{1}).
 \]
 If evaluation of the argument type fails, abstraction returns $\glverr$,
 already in the required domain. Otherwise fix any
 $a\in\mathcal V^+_\glvEnv(\ltynv{0})$. Extending the related environments by
 this argument makes the body induction hypothesis applicable. Its membership
 specialization gives
 \[
 f(a):=\SemCheck{\leev}{\glvEnv[x\mapsto a]}
    \in\mathcal V_{\glvEnv[x\mapsto a]}(\ltynv{1}).
 \]
 This holds for every argument in the domain, including when evaluation of a
 dependent result type or of the body returns $\glverr$. Therefore
 \[
 f\in\prod_{a\in\mathcal V^+_\glvEnv(\ltynv{0})}
       \mathcal V_{\glvEnv[x\mapsto a]}(\ltynv{1})
   \subseteq\mathcal V_\glvEnv(x\colon\ltynv{0}\fnto\ltynv{1}).
 \]
 The checked abstraction is precisely this recursively checked function, so
 this inclusion is its required conclusion.

 \item For application, use
 \[\glrulenamed{valHasType.appV}.\]
 \gldefn{value.denot.application}
 Write $\operatorname{inferType}(\lenv{0})=x:\ltynv{0}\fnto\ltynv{1}$.
 The operand induction hypotheses and the typing rule give
 \[
 \operatorname{inferType}(\lenv{1})=\ltynv{0},\qquad
 \operatorname{inferType}(\glvapp{\lenv{0}}{\lenv{1}})
   =\ltynv{1}\subst{x}{\lenv{1}}=\ltynv{\rm out}.
 \]
 The last premise supplies formation of this substituted type.
 If either operand or the domain check fails, error propagation gives
 $\glverr\in\mathcal V_\glvEnv(\ltynv{\rm out})$.
 Otherwise, put $f=\Sem{\lenv{0}}{\glvEnv}$ and
 $a=\Sem{\lenv{1}}{\glvEnv}$. The induction hypotheses give
 \[
 a\in\mathcal V^+_\glvEnv(\ltynv{0}),\qquad
 f(a)\in\mathcal V_{\glvEnv[x\mapsto a]}(\ltynv{1}).
 \]
 Dependent semantic substitution identifies this last domain with
 $\mathcal V_\glvEnv(\ltynv{1}\subst{x}{\lenv{1}})$.
 The family induction also covers higher-order arguments and dependent
 results. Thus the checked application has the required membership.

 \item For reshape, the typing rule is
 \[\glrulenamed{valHasType.reshap}.\]
 Write $\ltynv{\rm in}=\operatorname{inferType}(\leev)$.
 The rule types the operand at $\ltynv{\rm in}$ and supplies formation of
 the stored source annotation and the target annotation at the same kind.
 Their dimensions need not coincide.
 Inference returns the target annotation. The recursive denotation is
 \gldefn{value.denot.reshape}
 For a space source annotation, the displayed source type is that annotation.
 For a function source annotation, it is the actual inferred function type
 $\ltynv{\rm in}$. An inferred space in this latter case gives $\glverr$.
 Thus the source function hypothesis below is taken at its inferred type;
 it does not identify that type with a separately stored annotation.
 For space types, a successful input array is returned only after checking
 $|a|=\dof{\lsnv{s}}{\glvEnv^s}=\dof{\lsnv{t}}{\glvEnv^t}$.
 It therefore belongs to
 $\mathbb R^{\dof{\lsnv{t}}{\glvEnv^t}}\subseteq
 \mathcal V_{\glvEnv^t}(\lsnv{t})$; a failed check returns $\glverr$ in that
 same target domain. A function where an array is required also gives
 $\glverr$. Likewise, a successful array cannot be converted to a function.

 For function types, fix a successful target argument
 $a\in\mathcal V^+_{\glvEnv^t}(\ltynv{t0})$.
 The first recursive conversion in the displayed clause converts it to the
 source argument type. If that conversion succeeds, its induction conclusion
 is $c\in\mathcal V^+_{\glvEnv^s}(\ltynv{s0})$. The source function hypothesis
 then gives
 $f(c)\in\mathcal V_{\glvEnv^s[y\mapsto c]}(\ltynv{s1})$.
 The second recursive conversion uses these two extended environments and
 gives
 \[
 F(a)\in\mathcal V_{\glvEnv^t[x\mapsto a]}(\ltynv{t1}).
 \]
 If either conversion or the source function fails, $F(a)=\glverr$ satisfies
 this same conclusion. Since it holds for every successful target argument,
 \[
 (\lambda a.\,F(a))\in
 \mathcal V_{\glvEnv^t}(x\colon\ltynv{t0}\fnto\ltynv{t1}).
 \]
 This is the membership required for the reshaped function. The argument is
 converted from target to source, and the result from source to target;
 retaining the two extended environments is what handles dependent fibers.
\end{enumerate}
The relation induction thus gives membership after recursive checking at
$\ltyv$. The correspondence with finite arrays and dependent functions
identifies the raw and checked representatives in that domain, proving
$\SemCheck{\leev}{\glvEnv}\in\mathcal V_\glvEnv(\ltyv)$, which is (3).
Together with regularity and exact inference, this establishes all three
conclusions.
\end{proof}

\subsection{Operational semantics (\cref{sec:operational})}
\begin{defn}[$\lctx;\valCheckCtx\vdash\evalCtx$ (\cref{def:nonrecurseeval})]
Let $\operatorname{names}(\Psi)=\{x\mid\exists b,v.\;((x,b)\mapsto v)\in\Psi\}$. Write $\Psi\vdash(x,\lb{})\mapsto\leev$ when
$\evalSJudge{\Psi}{\tof{\lb{}}}{\operatorname{inferType}(\leev)}$ and
$\operatorname{fv}(\leev)\subseteq\operatorname{names}(\Psi)\cup\{x\}$.
For $\lb{}=\frameB{y}{b_0}{b_1}$ and
$\leev=\llambda(y\colon\ltynv{0})\lrh\ltynv{1}\colon v$, additionally require
$\evalSJudge{\Psi[(y,b_0)\mapsto y_{\ltynv{0}}]}{\tof{b_1}}{\ltynv{1}}$ for fresh $y$.

Assuming $\vdash(\Delta,\epsilon)$ throughout (\cref{def:wfc}), define
$(\Delta,\epsilon);\valCheckCtx\vdash\Psi$ inductively, with
$(\emptyset,\emptyset);\emptyset\vdash\emptyset$.
Suppose $(\Delta,\epsilon);\valCheckCtx\vdash\Psi$,
$(\emptyset,\epsilon)\vdash\lb{}\colon\bvalid$,
$(x,\lb{})\notin\Psi$, and $\Psi\vdash(x,\lb{})\mapsto\leev$.
Put $T=\tof{\lb{}}$ and $\ltyv=\operatorname{inferType}(\leev)$.
Form $(\Delta',\epsilon';\valCheckCtx')$ by adding $x:T$ to
$(\Delta,\epsilon)$ and $x:\ltyv$ to $\valCheckCtx$ if $x$ is undeclared;
otherwise leave the contexts unchanged, with $(x:T)\in\Delta\sqcup\epsilon$.
If $\valCheckCtx\vdash\ltyv:\lkn{}$ and
$\valCheckCtx'\vdash\leev:\ltyv$, then
$(\Delta',\epsilon');\valCheckCtx'\vdash\Psi[(x,\lb{})\mapsto\leev]$.

\end{defn}
\begin{thm}[Preservation (\cref{thm:preserve})]
\label{thm:preserveA}
Suppose $(\Delta,\epsilon);\valCheckCtx\vdash\Psi$.
\begin{enumerate}
\item If $(\emptyset,\epsilon)\vdash\lty:\lkn{}$ and
$\evalSJudge{\Psi}{\lty}{\ltyv{}}$, then
$\valCheckCtx\vdash\ltyv{}:\lkn{}$ for some $\lkn{}$ and
$\operatorname{fv}(\ltyv{})\subseteq\operatorname{names}(\Psi)$.
\item If $(\Delta,\epsilon)\vdash t:\tof{b}$,
$(\emptyset,\epsilon)\vdash b:\bvalid$, and $\evalJudge{\Psi}{t}{b}{v}$, then
$\valCheckCtx\vdash v:\operatorname{inferType}(v)$,
$\valCheckCtx\vdash\operatorname{inferType}(v):\lkn{}$ for some $\lkn{}$, and
$\operatorname{fv}(\operatorname{inferType}(v))\cup\operatorname{fv}(v)\subseteq\operatorname{names}(\Psi)$.
\item If $(\emptyset,\epsilon)\vdash b_0:\bvalid$ and
$\evalBBJudge{\Psi}{b_0}{v}{b_1}$, then
$\valCheckCtx\vdash v:\lsnv{0}\loprod \lsnv{1}$ and
$\operatorname{inferType}(v)=\lsnv{0}\loprod \lsnv{1}$ for some $\lsn{0},\lsn{1},\lsnv{0},\lsnv{1}$ with
$\evalSJudge{\Psi}{\lsn{i}}{\lsnv{i}}$ $(i=0,1)$ and
$\operatorname{fv}(v)\subseteq\operatorname{names}(\Psi)$.
\end{enumerate}

\end{thm}
\begin{proof}

We use mutual induction on the supplied ordinary space, type,
change-of-basis, and term evaluations. Put
$N=\operatorname{names}(\Psi)$.
The condition $(\Delta,\epsilon);\valCheckCtx\vdash\Psi$ of
\cref{def:nonrecurseeval} supplies a valid value context and
$\valCheckCtx\vdash v:\operatorname{inferType}(v)$ for every stored
$(x,b)\mapsto v$, as well as the recorded realizations and scope conditions.
The simultaneous induction threads these properties of $\Psi$ and
$\valCheckCtx$. A fresh nonlinear argument extends both contexts by the
same annotated variable. A linear column substitutes a basis vector into
$\Psi$ and needs no new value-variable declaration.

For a type evaluation $\evalSJudge{\Psi}{\lty}{\ltyv}$, the induction conclusion
is $\valCheckCtx\vdash\ltyv:\lkn{}$ for some $\lkn{}$, together with
$\operatorname{fv}(\ltyv)\subseteq N$. For a term evaluation
$\evalJudge{\Psi}{t}{b}{v}$, retain the stronger conclusion
\[
\begin{gathered}
 \exists\ltyv,\lkn{}.\;\evalSJudge{\Psi}{\tof{b}}{\ltyv},\qquad
 \valCheckCtx\vdash v:\operatorname{inferType}(v),\\
 \valCheckCtx\vdash\operatorname{inferType}(v):\lkn{},\quad\valCheckCtx\vdash\ltyv:\lkn{},\\
 \operatorname{fv}(\operatorname{inferType}(v))\cup\operatorname{fv}(v)
 \subseteq N.
\end{gathered}
\]
Thus we retain a realization of the source type and a common kind, without
requiring the inferred type to equal that realization.
We also use the kind determined by a type's space/function structure:
$\kappa(\ls)=\kappa(\lsv)=\lkbase$ and
$\kappa(x:T_0\fnto T_1)=\kappa(T_0)\fnto\kappa(T_1)$ in either language.
Induction on ordinary evaluation gives
\[
 \evalJudge{\Psi}{t}{b}{v}
 \quad\Longrightarrow\quad
 \kappa(\operatorname{inferType}(v))=\kappa(\tof{b}).
\]
The stored type realizations establish the variable case. The linear-application rule
\[\glrulenamed{opStep.appLLin}\]
requires $\tof{\lbn{1}}=\sof{\lbn{1}}$, so both its requested type and its
array result have kind $\lkbase$. The frame rules produce function types;
the remaining cases use their fixed array results or their operand hypotheses.
In particular, every recursive evaluation at a frame infers a function type.
For a change-of-basis
witness, also retain $\valCheckCtx\vdash v:s_{v0}\loprod s_{v1}$, the two
realized matrix factors, and scope as in (3).
For array operations, simultaneous induction also preserves the array
constructor structure: source tensors give inferred value tensors, with
possibly different dynamic dimensions. In particular, change-of-basis
witnesses infer matrix types. This supplies the tensor premises of matrix
multiplication and inverse typing without equating their dimensions.
The inference equations below use \cref{app:value-inference}; the scope
calculations use the free-variable definitions in \cref{sec:fv}, including
the annotations suppressed by the displayed syntax. We give representative
cases as follows.
\begin{enumerate}
 \item For a variable, the operational rule is
 \[\glrulenamed{opStep.var}.\]
 The lookup premise selects an entry $(x,b)\mapsto v$ in $\Psi$. Write
 $\Psi_0$ for the context immediately before this entry was added, and
 $\valCheckCtx_0$ for its value context. The entry condition
 $\Psi_0\vdash(x,b)\mapsto v$ in the definition of
 $(\Delta,\epsilon);\valCheckCtx\vdash\Psi$ supplies
 \[
 \evalSJudge{\Psi_0}{\tof{b}}{\operatorname{inferType}(v)},\qquad
 \operatorname{fv}(v)\subseteq\operatorname{names}(\Psi_0)\cup\{x\}.
 \]
 Formation and scope of the evaluated type give kinding of
 $\operatorname{inferType}(v)$ and bound its free variables by
 $\operatorname{names}(\Psi_0)$. The recorded evaluation persists under
 extension from $\Psi_0$ to $\Psi$, and $\operatorname{names}(\Psi_0)\cup\{x\}\subseteq N$.
 Choose $\ltyv=\operatorname{inferType}(v)$. Consequently, in the current context,
 \[
 \begin{gathered}
 \evalSJudge{\Psi}{\tof{b}}{\ltyv},\qquad
 \valCheckCtx\vdash\operatorname{inferType}(v):\lkn{},\quad\valCheckCtx\vdash\ltyv:\lkn{},\\
 \operatorname{fv}(\operatorname{inferType}(v))\cup\operatorname{fv}(v)
 \subseteq N.
 \end{gathered}
 \]
 Finally,
 \cref{def:nonrecurseeval} supplies $\valCheckCtx\vdash v:\ltyv$ after weakening.
 Together these are all the required term conclusions.

 \item For change-of-basis application, use
 \[\glrulenamed{opStep.cob}.\]
 The space premise and its induction hypothesis give
 $\valCheckCtx\vdash\lsv:\lkbase$ and $\operatorname{fv}(\lsv)\subseteq N$.
 The term and conversion hypotheses give
 $\operatorname{fv}(\leev)\subseteq N$ and
 $\operatorname{fv}(\lenv{0})\subseteq N$.
 Write the output expression as
 \[
 w=\LTTVPP{\lsv}{\glvreshape{\lenv{0}}{\lsv\loprod\lsv}{\lsv\loprod\lsv}}
                    {\glvreshape{\leev}{\lsv}{\lsv}}.
 \]
 Unfolding inference first at the two reshapes and then at the contraction
 gives
 \[
 \begin{aligned}
 \operatorname{inferType}(\glvreshape{\lenv{0}}{\lsv\loprod\lsv}{\lsv\loprod\lsv})
   &=\lsv\loprod\lsv,\\
 \operatorname{inferType}(\glvreshape{\leev}{\lsv}{\lsv})&=\lsv,\\
 \operatorname{inferType}(w)&=\lsv.
 \end{aligned}
 \]
 To establish typing as well, use
 \[\glrulenamed{valHasType.reshap},\qquad
   \glrulenamed{valHasType.ttvSpace}.\]
 The operand induction hypotheses type $\lenv{0}$ and $\leev$ at their
 inferred types; formation of $\lsv$ types both displayed reshapes.
 Contraction therefore gives $\valCheckCtx\vdash w:\lsv$.
 Regularity gives $\valCheckCtx\vdash\operatorname{inferType}(w):\lkbase$.
 The rule's equality $\tof{\lbn{0}}=\sof{\lbn{0}}$ turns its last premise
 into the required realization
 $\evalSJudge{\Psi}{\tof{\lbn{0}}}{\lsv}$; choose $\ltyv=\lsv$ and
 $\lkn{}=\lkbase$ in the stronger conclusion.
 The free-variable clauses for contraction and reshape also give
 \[
 \begin{aligned}
 \operatorname{fv}(\operatorname{inferType}(w))\cup\operatorname{fv}(w)
 &\subseteq\operatorname{fv}(\lenv{0})\cup\operatorname{fv}(\leev)
                \cup\operatorname{fv}(\lsv)\\
 &\subseteq N.
 \end{aligned}
 \]
 Thus this case supplies the type realization, both kinding judgments, and
 scope.

 \item For abstraction, the operational rule is
 \[\glrulenamed{opStep.nllam}.\]
 Choose $x$ fresh for the contexts and put
 \[
 \epsilon'=\epsilon[x\mapsto\ltyn{0}],\qquad
 \valCheckCtx'=\valCheckCtx[x\mapsto\ltynv{0}],\qquad
 \Psi'=\Psi[(x,\lbn{0})\mapsto x_{\ltynv{0}}].
 \]
 The domain evaluation supplies the realization for this entry, and its
 induction hypothesis gives $\valCheckCtx\vdash\ltynv{0}:\lkn{0}$.
 The variable rule
 \[\glrulenamed{valHasType.var}\]
 gives $\valCheckCtx'\vdash x_{\ltynv{0}}:\ltynv{0}$, with
 $\operatorname{inferType}(x_{\ltynv{0}})=\ltynv{0}$.
 The fresh-entry clause of \cref{def:nonrecurseeval} therefore gives
 $(\Delta,\epsilon');\valCheckCtx'\vdash\Psi'$.
 Apply the body and codomain induction hypotheses with this extended value context.
 Together with the domain hypothesis, they give
 \[
 \begin{gathered}
 \valCheckCtx\vdash\ltynv{0}:\lkn{0},\qquad
 \valCheckCtx'\vdash\ltynv{1}:\lkn{1},\qquad\valCheckCtx'\vdash\leev:\operatorname{inferType}(\leev),\\
 \operatorname{fv}(\ltynv{0})\subseteq N,\qquad
 \operatorname{fv}(\ltynv{1})\cup\operatorname{fv}(\leev)
     \subseteq N\cup\{x\}.
 \end{gathered}
 \]
 The codomain and body here are read with the fresh argument in scope.
 The output lambda has the explicit result annotation $\ltynv{1}$ and a
 body reshaped to that annotation. The reshape rule gives
 $\valCheckCtx'\vdash\glvreshape{\leev}{\ltynv{1}}{\ltynv{1}}:\ltynv{1}$.
 Applying the abstraction typing rule
 \[\glrulenamed{valHasType.vlam}\]
 gives $\valCheckCtx\vdash w:(x\colon\ltynv{0}\fnto\ltynv{1})$.
 Thus, for its output $w$,
 \[
 \operatorname{inferType}(w)=(x\colon\ltynv{0}\fnto\ltynv{1}),\qquad
 \valCheckCtx\vdash\operatorname{inferType}(w):\lkn{0}\fnto\lkn{1}.
 \]
 The type realization is rebuilt from the same domain and codomain
 evaluations using
 \[\glrulenamed{evalT.piTy}.\]
 Since $\tof{\frameB{x}{\lbn{0}}{\lbn{1}}}
 =(x\colon\ltyn{0}\fnto\ltyn{1})$, this produces exactly the realization
 needed by the stronger term conclusion. Binding $x$ removes its free
 occurrences in both the codomain annotation and the body:
 \[
 \begin{aligned}
 \operatorname{fv}(\operatorname{inferType}(w))\cup\operatorname{fv}(w)
 &\subseteq\operatorname{fv}(\ltynv{0})\cup
       ((\operatorname{fv}(\ltynv{1})\cup\operatorname{fv}(\leev))
          \setminus\{x\})\\
 &\subseteq N.
 \end{aligned}
 \]
 This establishes kinding and scope for the abstraction as well as its
 source-type realization.

 For dependent application, the rule is
 \[\glrulenamed{opStep.appNVal}.\]
 Its type-evaluation premise evaluates the instantiated codomain to $\ltynv{\rm out}$.
 The operand induction hypotheses give value typing. The function is evaluated
 at $\frameB{x}{\lbn{0}}{\lbn{1}}$, so the kind invariant gives
 \[
 \operatorname{inferType}(\lenv{0})=(x:D\fnto C),\qquad
 \kappa(D)=\kappa(\tof{\lbn{0}}),\quad
 \kappa(C)=\kappa(\tof{\lbn{1}}).
 \]
 The argument is evaluated at $\lbn{0}$; hence
 $\kappa(\operatorname{inferType}(\lenv{1}))=\kappa(D)$. Write
 \[
 u=\glvreshape{\lenv{1}}{D}{\operatorname{inferType}(\lenv{1})},\qquad
 w=\glvreshape{\glvapp{\lenv{0}}{u}}{\ltynv{\rm out}}{\operatorname{inferType}(\glvapp{\lenv{0}}{u})}.
 \]
 Type preservation for the codomain premise gives
 $\valCheckCtx\vdash\ltynv{\rm out}:\kappa(C)$ and
 $\operatorname{fv}(\ltynv{\rm out})\subseteq N$.
 The relevant value rules are
 \[\glrulenamed{valHasType.appV}\qquad\glrulenamed{valHasType.reshap}.\]
 The argument reshape gives $\valCheckCtx\vdash u:D$.
 Substitution gives formation of $C\subst{x}{u}$, and application gives
 $\valCheckCtx\vdash\glvapp{\lenv{0}}{u}:C\subst{x}{u}$.
 Inference assigns this call the same substituted codomain,
 whose kind is $\kappa(C\subst{x}{u})=\kappa(C)$.
 Thus both reshapes have source and target annotations of the same kind.
 The outer reshape gives
 $\valCheckCtx\vdash w:\ltynv{\rm out}$ and
 $\operatorname{inferType}(w)=\ltynv{\rm out}$.
 The reshapes check agreement of the independently realized dimensions
 during evaluation. Scope of inference and substitution gives
 \[
 \operatorname{fv}(\operatorname{inferType}(w))\cup\operatorname{fv}(w)
 \subseteq\operatorname{fv}(\lenv{0})\cup\operatorname{fv}(\lenv{1})
       \cup\operatorname{fv}(\ltynv{\rm out})\subseteq N.
 \]

 Choosing the rule's own $\ltynv{\rm out}$ supplies the type realization and
 the common kind. No identification with a separately computed result type
 is needed in this case.

 \item For type evaluation, the space case uses
 \[\glrulenamed{evalT.space}.\]
 The space induction hypothesis supplies $\valCheckCtx\vdash\lsv:\lkbase$ and
 $\operatorname{fv}(\lsv)\subseteq N$, which are precisely the type
 conclusion in this case. For a dependent function type, use
 \glrulename{evalT.piTy}, displayed above. Its domain hypothesis gives
 $\valCheckCtx\vdash\ltynv{0}:\lkn{0}$ and scope in $N$; its codomain hypothesis gives
 formation at $\lkn{1}$ with the fresh argument in scope.
 Function-type formation and binding that argument therefore give
 \[
 \valCheckCtx\vdash(x\colon\ltynv{0}\fnto\ltynv{1}):\lkn{0}\fnto\lkn{1},
 \qquad
 \operatorname{fv}(x\colon\ltynv{0}\fnto\ltynv{1})\subseteq N.
 \]
 These are the two conclusions of (1). Dynamic-space formation uses the term hypothesis to type its defining
 matrix, the space hypothesis to form its shape annotation, and the reshape
 already in the operational conclusion to give the exact annotated matrix
 type. The value-space formation rule then applies.

 For change of basis, the induction hypothesis for a witness $c$ retains
 \[
 \operatorname{inferType}(c)=s_{v0}\loprod s_{v1},\qquad
 \evalSJudge{\Psi}{s_i}{s_{vi}}\ (i=0,1),\qquad
 \operatorname{fv}(c)\subseteq N.
 \]
 To see how these conclusions are propagated, the inverse rule is
 \[\glrulenamed{evalBB.inverse}.\]
 Inference gives
 $\operatorname{inferType}(c^{-1})=s_{v1}\loprod s_{v0}$.
 The value inverse rule
 \[\glrulenamed{valHasType.invTensor}\]
 gives $\valCheckCtx\vdash c^{-1}:s_{v1}\loprod s_{v0}$.
 Exchange the two realization witnesses, and use
 $\operatorname{fv}(c^{-1})=\operatorname{fv}(c)\subseteq N$.
 These supply all three components of the change-of-basis conclusion.
 For composition, use
 \[\glrulenamed{evalBB.trans}.\]
 Write $\operatorname{inferType}(\lenv{1})=s_{v0}\loprod s_{v1}$ and
 $\evalSJudge{\Psi}{s_i}{s_{vi}}$ for the witnesses from its second conversion
 premise. The last premise supplies a realization $\lsv$ of
 $\sof{\lbn{0}}$. The reshape in the conclusion has inferred type
 $\lsv\loprod\lsv$; matrix-multiplication inference consequently gives
 \[
 \operatorname{inferType}
   (\glvmtm{\lsv}{\lsv}{\lenv{1}}{\glvreshape{\lenv{0}}{\lsv\loprod\lsv}{\lsv\loprod\lsv}})
       =\lsv\loprod s_{v1}.
 \]
 The reshape and matrix-multiplication typing rules
 \[\glrulenamed{valHasType.mtm}\]
 applied to the operand induction hypotheses give typing at exactly this
 inferred result type. Its two realization witnesses are therefore the rule's last premise and
 the second-factor witness for $\lenv{1}$.
 Finally its free variables are contained in
 $\operatorname{fv}(\lenv{0})\cup\operatorname{fv}(\lenv{1})
 \cup\operatorname{fv}(\lsv)\subseteq N$.
 This proves the required factorization, both realizations, and scope for
 composition.
\end{enumerate}
The type cases establish (1). The stronger term conclusion supplies value typing, kinding
of $\operatorname{inferType}(v)$ and the stated union of free-variable bounds,
which proves (2) after forgetting the extra realization witness. The
change-of-basis cases supply value typing, the two realized factors, and the free-variable
bound in (3). 
\end{proof}

\begin{defn}[Checked operational evaluation (\cref{def:envimply})]
\label{def:envimply}
Assume $(\Delta,\epsilon);\valCheckCtx\vdash\Psi$.
The subscript $\mathsf C$ marks an operational derivation checked at every step:
$\glCty{\epsilon}{\Psi}{\lty}{\ltyv}$,
$\glCterm{\Delta}{\epsilon}{\Psi}{\lee}{\lb{}}{\leev}{\ltyv}$, and
$\glCcob{\epsilon}{\Psi}{\lbn{0}}{\leev}{\lbn{1}}$.
At each term step, for example, these require
$(\Delta,\epsilon)\vdash t:\tof{b}$,
$(\{\},\epsilon)\vdash b:\bvalid$,
$\glCat{\epsilon}{\Psi}{b}{\ltyv{}}$,
$\evalJudge{\Psi}{t}{b}{v}$, and
$\evalSJudge{\Psi}{\tof{b}}{\ltyv{}}$.
The judgment $\glCat{\epsilon}{\Psi}{\lb{}}{\ltyv}$ simply shows what basis was used in $\glCty{\epsilon}{\Psi}{\lty}{\ltyv}$, which is needed to call dynamic dimensions properly.
The rather verbose rules are in \cref{app:gen-rules:c-space,app:gen-rules:c-ty,app:gen-rules:c-tyAt,app:gen-rules:c-cob,app:gen-rules:c-term}.

\end{defn}
\begin{thm}[Progress (\cref{thm:prog})]
\label{thm:progA}
Suppose $(\Delta,\epsilon);\valCheckCtx\vdash\Psi$.
\begin{enumerate}
\item If $(\emptyset,\epsilon)\vdash\ls{}:\svalid$,
then $\glCspace{\epsilon}{\Psi}{\ls{}}{\lsv}$ for some $\lsv$.
\item If $(\Delta,\epsilon)\vdash t:\tof{b}$ and
$(\emptyset,\epsilon)\vdash b:\bvalid$, then
$\glCterm{\Delta}{\epsilon}{\Psi}{t}{b}{v}{\ltyv{}}$ for some $v,\ltyv{}$ with
$\operatorname{inferType}(v)=\ltyv{}$.
\item If $(\emptyset,\epsilon)\vdash b_i:\bvalid$ and
$\tof{b_i}=\ls{}$ $(i=0,1)$, then
$\glCcob{\epsilon}{\Psi}{b_0}{v}{b_1}$ for some $v$.
\end{enumerate}

\end{thm}
\begingroup
\renewcommand{\evalCtxSpace}{\epsilon}
\begin{proof}

We construct term evaluations first at the standard basis of their source
type. Change of basis converts a stored variable representation to that
standard basis and, at the end, converts the standard-basis result to the
basis requested by the theorem; function frames use argument and fiber
transport for the same purpose. Since constructing a conversion can itself
require evaluating source terms, we first define context extensions and
certificates bounding these mutual calls. We prove standard-basis progress
and conversion together at every finite bound, then use
$(\Delta,\epsilon);\valCheckCtx\vdash\Psi$ to obtain a bound for the given inputs.
Finally, we transport the resulting checked evaluation to the requested
basis and reshape its result so that its inferred type is exactly the
realized type.

The construction alternates between source derivations and basis derivations.
For example, evaluating $x$ may select $(x,b)\mapsto v$ and require a
conversion of $b$ to $\stdBasisDCtx{\tof{b}}$; converting a projected basis
requires evaluating its defining map. We make the induction parameter precise
by giving separate bounds for these two kinds of recursive call.

\paragraph{Admissible context extensions.}
First specify the contexts in which a recursive call is allowed. Write
$\Psi_0\preceq_*\Psi$ when all lookups in $\Psi_0$ are preserved and every
new entry has a fresh name and a standard basis, with only one entry for each
new name. Explicitly,
\[
\begin{aligned}
 &\Psi_0(x,b)=v\ \Longrightarrow\ \Psi(x,b)=v,\\
 &(x,b)\in\operatorname{dom}(\Psi)\setminus\operatorname{dom}(\Psi_0)
 \ \Longrightarrow\
 \begin{cases}
 b=\stdBasisDCtx{\tof{b}},\\
 x\notin\operatorname{names}(\Psi_0),\\
 (x,b')\in\operatorname{dom}(\Psi)\ \Longrightarrow\ b'=b.
 \end{cases}
\end{aligned}
\]
The theorem assumes $(\Delta,\epsilon);\valCheckCtx\vdash\Psi$.
The construction uses its source-variable coverage, valid stored bases,
recorded realizations, and scope conditions. Its value-typing component
additionally gives $\valCheckCtx\vdash v:\operatorname{inferType}(v)$
for each stored representation. 

\paragraph{Certificates for the recursive calls.}
With the allowed extensions fixed, we define the two certificate families
that determine which recursive calls are available at a given bound.
For a source typing or formation derivation $D$, write
$\mathcal S_n^{\Psi_0}(D)$ for the following recursively defined condition.
Every immediate source premise of $D$ must satisfy the same condition at
index $n$. At each variable leaf $x:T$, in its local context $\epsilon$,
require, uniformly in the extensions $\Psi_0\preceq_*\Psi$ covering that
context,
\[
\begin{aligned}
 \mathcal S_0^{\Psi_0}(x:T)&=\bot,\\
 \mathcal S_{n+1}^{\Psi_0}(x:T)&\Longleftrightarrow
 \exists b,v.\;
 \Psi(x,b)=v\ \land\ \tof{b}=T\ \land\
 (\{\},\epsilon)\vdash b:\bvalid\ \land\ \mathcal B_n^{\Psi_0}(b).
\end{aligned}
\]

Here $\mathcal B_n^{\Psi_0}(b)$ is a condition on a valid basis and the
source derivations used in its validity proof. At index zero its grammar is
\[
 b::=\lbzero{}\mid\lbone{}\mid b^{\lbstar}
       \mid b\lbosum b\mid b\lboprod b.
\]
In particular, $\mathcal B_0$ excludes projected, image, term, and frame
bases. At index $n+1$, each immediate basis component must satisfy
$\mathcal B_{n+1}$, while each embedded source derivation must satisfy
$\mathcal S_n$. Schematically, for a basis constructor $K$,
\[
 \mathcal B_{n+1}^{\Psi_0}(K(b_1,\ldots,b_k;D_1,\ldots,D_m))
 \Longleftrightarrow
 \bigwedge_{i=1}^k\mathcal B_{n+1}^{\Psi_0}(b_i)
 \ \land\ \bigwedge_{j=1}^m\mathcal S_n^{\Psi_0}(D_j).
\]
The $D_j$ are the derivations needed to validate the choice of basis: the
map-typing derivation for a projected or image basis, the space-formation and
vector-typing derivations for a term basis, and the dependent-type formation
derivation for a frame. The component bases retain their validity and
space-equality premises. For example, with
$D_t:(\{\},\epsilon)\vdash t:s_0^\star\loprod s_1$ and
$\sof{b_i}=s_i$, the projected-kernel clause is
\[
 \mathcal B_{n+1}^{\Psi_0}(\pBasisC{\Kernel(t)}{b_0}{b_1})
 \Longleftrightarrow
 \mathcal B_{n+1}^{\Psi_0}(b_0)\land
 \mathcal B_{n+1}^{\Psi_0}(b_1)\land
 \mathcal S_n^{\Psi_0}(D_t).
\]
For a frame $\frameB{x}{b_0}{b_1}$, the component obligations also hold after
introducing a fresh argument: for every $\Psi_0\preceq_*\Psi$, fresh $y$,
and valid argument basis $a$ with $\tof{a}=\tof{b_0}$, put
$\Psi'=\Psi[(y,a)\mapsto u]$. Require the same-index conditions on $b_0$
and $b_1[y/x]$ in this extended context, uniformly in $u$. This allows
conversion between different frames, as well as the standard argument used
by abstraction. These binder obligations are uniform in the fresh name;
renaming does not increase the index.

Consequently the recursive calls have the following indices:
\[
 \mathcal S_{n+1}\ \xrightarrow{\text{selected variable basis}}\ \mathcal B_n,
 \qquad
 \mathcal B_{n+1}\ \xrightarrow{\text{embedded source derivation}}\ \mathcal S_n.
\]
Recursion to a source premise or basis component keeps the index and decreases
the corresponding derivation or syntax size. Thus the induction is
lexicographic: first the displayed index, then structural size. 

\paragraph{Mutual induction on the bound.}
The certificates specify the recursive calls; we now prove that those calls
produce checked evaluations and conversions. Below, quantification includes
all roots $\Psi_0$, their extensions $\Psi_0\preceq_*\Psi$, and value contexts $\valCheckCtx$
satisfying $(\Delta,\epsilon);\valCheckCtx\vdash\Psi$. The source assertion is
\[
\begin{aligned}
 P_n:\quad &D:(\Delta,\epsilon)\vdash t:T,\quad
              \mathcal S_n^{\Psi_0}(D)\\
 &\Longrightarrow\quad
 \exists v,\ltyv.\;
 \glCterm{\Delta}{\epsilon}{\Psi}{t}{\stdBasisDCtx{T}}{v}{\ltyv}.
\end{aligned}
\]
The simultaneous source assertions give checked space and type evaluations
from their formation derivations. For conversion of space bases, put
\[
\begin{aligned}
 C_n:\quad &(\{\},\epsilon)\vdash b:\bvalid,\quad
             \tof{b}=s,\quad \mathcal B_n^{\Psi_0}(b)\\
 &\Longrightarrow\quad
 \exists c.\;\glCcob{\epsilon}{\Psi}{b}{c}{\stdBasisDCtx{s}}.
\end{aligned}
\]
These are the context-formation specializations of the induction assertions;
the Lean proof retains the separate store and scope conditions so it can
also use them under the frame extensions above.
The induction steps are
\[
 P_0\land C_0,\qquad
 C_n\Longrightarrow P_{n+1},\qquad
 P_n\Longrightarrow C_{n+1},\qquad
 \therefore\ \forall n.\;P_n\land C_n.
\]
\paragraph{Base cases: $P_0\land C_0$.}
For $P_0$, induct structurally on the source derivation carrying
$\mathcal S_0$. A variable case would require
$\mathcal S_0(x:T)=\bot$, so no lookup or basis conversion is needed.
The remaining cases use the structural induction hypotheses and the checked
source rules, including the simultaneous space- and type-formation cases.
This constructs the standard-basis evaluations asserted by $P_0$.

For $C_0$, induct on the grammar defining $\mathcal B_0$: empty, scalar,
dual, direct sum, and tensor product. Empty and scalar bases use the identity
conversion; the other constructors combine the conversions supplied by their
component hypotheses. There is no embedded source derivation, so this
construction requires no instance of $P_n$. It gives the checked conversions
asserted by $C_0$, establishing $P_0\land C_0$.

\paragraph{Source step: $C_n\Longrightarrow P_{n+1}$.}
Fix $n$ and assume $C_n$ together with its depth-$n$ realization and transport
results. We prove $P_{n+1}$ by structural induction on the source derivation:
calls on proper source premises keep index $n+1$, whereas a selected variable
basis uses the already-proved depth-$n$ results. The following are cases of
this source induction.

\begin{enumerate}
    \item In the variable case of $P_{n+1}$, the certificate
    $\mathcal S_{n+1}$ selects an entry $(x,b)\mapsto v$ with
    $\mathcal B_n(b)$. Context formation and the depth-$n$ realization result
    supply its checked basis realization. For a linear variable, apply
    {\small\begin{equation*}
        \glrulenamed{c.term.varLin}.
    \end{equation*}}
    For a nonlinear variable, the selected entry instead gives
    $(x:\tof{b})\in\epsilon$. Together with basis validity and the exact
    realization supplied at depth $n$, these are the four premises of
    \[\glrulenamed{c.term.varNlin}.\]
    Thus, in either variable case, we first obtain a checked evaluation at
    the selected stored basis $b$. It remains to put that evaluation at
    $b_* = \stdBasisDCtx{\tof{b}}$ to prove the variable conclusion of
    $P_{n+1}$.

    If $b$ is a space basis, $C_n$ supplies
    $\glCcob{\epsilon}{\Psi}{b}{c}{b_*}$. The simultaneous realization
    results supply $\glCat{\epsilon}{\Psi}{b_*}{\lsv}$ and
    $\glCspace{\epsilon}{\Psi}{\sof{b_*}}{\lsv}$. The stored-variable
    evaluation and these three judgments instantiate
    {\small\[\glrulenamed{c.term.cob}.\]}
    with source basis $\lbn{1}=b$, target basis $\lbn{0}=b_*$, and source
    term $x$. Source typing and standard-basis validity supply its remaining
    premises. Its conclusion is the required checked evaluation of $x$ at
    $b_*$, with realized type $\lsv$.

    If $b$ is a frame, write
    \[
     b=\frameB{z}{b_0}{b_1},\qquad
     b_* =\frameB{z}{a_0}{a_1},\qquad
     \tof{a_0}=\tof{b_0}=T_0.
    \]
    The depth-$n$ transport construction uses eta expansion. It converts a
    fresh argument from $a_0$ to $b_0$, applies the stored function, and
    converts the result from $b_1$ to $a_1$. These recursive transports are
    at the smaller domain and codomain types; they do not invoke
    $P_{n+1}$ again on the original variable.

    First obtain the exact frame realizations. The rule used for each is
    \[\glrulenamed{c.tyAt.frame}.\]
    In particular, the target frame gives a domain realization
    $\glCat{\epsilon}{\Psi}{a_0}{\ltynv{t0}}$. For fresh $y$, put
    \[
     \epsilon'=\epsilon[y\mapsto T_0],\qquad
     \valCheckCtx'=\valCheckCtx[y\mapsto\ltynv{t0}],\qquad
     \Psi'=\Psi[(y,a_0)\mapsto y_{\ltynv{t0}}].
    \]
    Erase the domain realization and apply \cref{thm:preserve}(1) to obtain
    $\valCheckCtx\vdash\ltynv{t0}:\lkn{}$ for some $\lkn{}$.
    Lookup in $\valCheckCtx'$ types the fresh annotated variable.
    Hence \cref{def:nonrecurseeval} gives
    $(\Delta,\epsilon');\valCheckCtx'\vdash\Psi'$.
    Its codomain premise supplies
    $\glCat{\epsilon'}{\Psi'}{a_1[y/z]}{\ltynv{t1}}$.
    Here $\ltynv{t1}$ denotes the target codomain with $y$ in scope.
    The entry condition for the selected function and the certified frame
    realizations supply the corresponding source-domain and source-codomain
    judgments.

    Apply \glrulename{c.term.varNlin}, displayed above, to the new entry:
    lookup holds by construction, $(y:T_0)\in\epsilon'$, and the domain
    validity and realization are already available. Its conclusion is
    \[
     \glCterm{\emptyset}{\epsilon'}{\Psi'}{y}{a_0}
       {y_{\ltynv{t0}}}{\ltynv{t0}}.
    \]
    Domain transport, by the smaller-type induction hypothesis, then gives
    \[
     \glCterm{\emptyset}{\epsilon'}{\Psi'}{y}{b_0}{u}{\ltynv{a}}.
    \]
    For a space domain this uses \glrulename{c.term.cob}; for a function
    domain it uses this same frame construction recursively at the smaller
    type. Weakening preserves the original checked lookup of $x$ in $\Psi'$.
    Checked substitution into the source frame's codomain realization,
    using the checked argument $u$, gives
    $\glCat{\epsilon'}{\Psi'}{b_1[y/z]}{\ltynv{r}}$.
    This substitution may first reshape the argument to the source-domain
    realization; write $u$ for the resulting checked argument, retaining the
    displayed judgment at $b_0$.
    The function lookup, converted argument, and this result realization
    now supply the evaluation premises of
    {\small\[\glrulenamed{c.term.appNVal}.\]}
    Source application typing and validity of $b_1[y/z]$ supply the other
    two premises. With the source function instantiated by $x$ and its
    argument by $y$, put
    $\operatorname{inferType}(v)=(z:D\fnto C)$ and
    $u'=\glvreshape{u}{D}{\operatorname{inferType}(u)}$.
    The conclusion is
    \[
     \glCterm{\emptyset}{\epsilon'}{\Psi'}{\leval{x}{}{y}}
       {b_1[y/z]}{\glvreshape{\glvapp{v}{u'}}{\ltynv{r}}{C[u'/z]}}{\ltynv{r}}.
    \]

    Next apply the smaller-type transport hypothesis to the result fiber.
    This gives a checked evaluation of $x(y)$ at $a_1[y/z]$ with some
    realized type $\ltynv{w}$. Align it with the already chosen target
    codomain realization using
    \[\glrulenamed{c.term.reshapeCob}.\]
    Specifically, instantiate its target realization by $\ltynv{t1}$,
    its input realization by $\ltynv{w}$, and its source term by $x(y)$.
    Its conclusion supplies a body $w'$ satisfying
    \[
     \glCterm{\emptyset}{\epsilon'}{\Psi'}{\leval{x}{}{y}}
       {a_1[y/z]}{w'}{\ltynv{t1}}.
    \]
    This construction is uniform in fresh $y$. Together with source typing
    of $x$, validity of the target frame, its whole-frame realization, its
    domain realization, and its codomain realization, it supplies every
    premise of
    {\small\[\glrulenamed{c.term.frameEta}.\]}
    Instantiate the displayed source term $\len{0}$ by $x$ and the target
    frame by $\frameB{z}{a_0}{a_1}$. The rule closes the fresh argument and
    returns a lambda whose body is the reshaped $w'$. Hence
    \[
     \exists w,\ltyv.\;
     \glCterm{\emptyset}{\epsilon}{\Psi}{x}{b_*}{w}{\ltyv}.
    \]
    This is exactly the nonlinear-variable conclusion of $P_{n+1}$.
    The conversion and frame-transport services used throughout are the
    depth-$n$ results supplied with $C_n$.

    \item For algebraic terms in $P_{n+1}$, the structural induction
    hypotheses at the same index $n+1$ give checked operands. The simultaneous
    formation hypotheses supply their basis realizations, and the corresponding
    checked rule constructs the standard-basis result. These calls decrease
    derivation size, not the index. For example:
    {\small\begin{equation*}
        \glrule[2,2,1]{c.term.zero}.
    \end{equation*}}
    Linear application retains both the matrix basis and the argument basis (\glrulename{c.term.appLLin}); its conclusion reshapes the contraction to the realized output space. Thus the construction records the required intermediate basis realizations as well as the term evaluation.

    \item For abstraction in $P_{n+1}$, put
    $T=(x\colon\ltyn{0}\fnto\ltyn{1})$ and
    $b_* = \stdBasisDCtx{T}=\frameB{x}{b_0}{b_1}$.
    The domain basis $b_0$ is standard for $\ltyn{0}$, and the fiber
    $b_1$ is standard for $\ltyn{1}$ with $x$ in scope.
    Given the certified source derivation of
    $\nllambda(x\colon\ltyn{0})\ldotsp t:T$, our goal is
    \[
     \exists w,\ltyv.\;
     \glCterm{\emptyset}{\epsilon}{\Psi}
       {\nllambda(x\colon\ltyn{0})\ldotsp t}{b_*}{w}{\ltyv}.
    \]
    We construct the function's annotation first, compile its body with a
    symbolic argument, and then bind that argument in the resulting value.
    Both recursive calls use smaller source premises at the same index $n+1$.

    \emph{First, realize the function type.}
    The abstraction's type-formation premise is a smaller source derivation,
    with its $\mathcal S_{n+1}$ certificate supplied by the certificate for
    the abstraction. Apply the simultaneous type induction hypothesis to get
    \[
     \glCat{\epsilon}{\Psi}{b_*}
       {(x\colon\ltynv{0}\fnto\ltynv{1})}.
    \]
    Its frame clause is
    {\small\[\glrulenamed{c.tyAt.frame}.\]}
    Inverting this clause gives
    $\glCat{\epsilon}{\Psi}{b_0}{\ltynv{0}}$ and the codomain
    realization under each fresh argument. This fixes the domain annotation
    $\ltynv{0}$ and dependent result annotation $\ltynv{1}$ before the
    body is compiled.

    \emph{Second, supply the symbolic argument.}
    Choose $y$ fresh for the contexts, terms, and annotations, and put
    \[
    \begin{gathered}
     \epsilon'=\epsilon[y\mapsto\ltyn{0}],\qquad
     \valCheckCtx'=\valCheckCtx[y\mapsto\ltynv{0}],\\
     \Psi'=\Psi[(y,b_0)\mapsto y_{\ltynv{0}}].
    \end{gathered}
    \]
    The new entry represents the future source argument by an annotated
    value variable, so the body can be compiled before an argument is supplied.
    Erasing the domain realization and applying \cref{thm:preserve}(1)
    gives $\valCheckCtx\vdash\ltynv{0}:\lkn{}$ for some $\lkn{}$.
    The variable rule
    \[\glrulenamed{valHasType.var}\]
    gives $\valCheckCtx'\vdash y_{\ltynv{0}}:\ltynv{0}$.
    These facts, domain-basis validity, and freshness give
    $(\emptyset,\epsilon');\valCheckCtx'\vdash\Psi'$ by
    \cref{def:nonrecurseeval}. Since $b_0$ is standard,
    $\Psi\preceq_*\Psi'$ and hence $\Psi_0\preceq_*\Psi'$.
    Thus the extended context is one of the contexts permitted by the
    induction hypothesis. The codomain conclusion retained in the first
    step is now
    \[
     \glCat{\epsilon'}{\Psi'}{b_1[y/x]}{\ltynv{1}[y_{\ltynv{0}}/x]}.
    \]

    \emph{Third, compile the body.}
    Let $D_y$ be the source premise
    $(\emptyset,\epsilon')\vdash t[y/x]:\ltyn{1}[y/x]$.
    The binder clause of the abstraction's certificate supplies
    $\mathcal S_{n+1}^{\Psi_0}(D_y)$.
    Apply the structural body induction hypothesis in the context just
    constructed. It produces
    \[
     \glCterm{\emptyset}{\epsilon'}{\Psi'}
       {t[y/x]}{b_1[y/x]}{u_y}{\ltynv{\rm body}}
    \]
    for some $u_y,\ltynv{\rm body}$.
    The basis is $b_1[y/x]$ because standard-basis construction commutes
    with replacing the bound argument by a fresh variable.
    This call decreases the source derivation size. The annotation
    $\ltynv{\rm body}$ it chooses need not equal $\ltynv{1}[y_{\ltynv{0}}/x]$.

    \emph{Fourth, align the body's result annotation.}
    Apply the checked reshape rule at the same fiber basis:
    {\small\[\glrulenamed{c.term.reshapeCob}.\]}
    Its term premise is the body result, and its target-realization premise
    is the codomain judgment retained from the first step. Source typing
    and validity of the standard fiber supply its other premises. Hence,
    putting $v_y=\glvreshape{u_y}{\ltynv{1}[y_{\ltynv{0}}/x]}{\ltynv{\rm body}}$, we obtain
    \[
     \glCterm{\emptyset}{\epsilon'}{\Psi'}
       {t[y/x]}{b_1[y/x]}{v_y}{\ltynv{1}[y_{\ltynv{0}}/x]}.
    \]
    This gives the body evaluation at exactly the annotation fixed by the
    function-type realization. It uses the two constructed derivations,
    with no further progress call.

    \emph{Finally, bind the argument and apply the abstraction rule.}
    Fresh-variable renaming transports these body and fiber derivations
    to every fresh argument name. Write $v_1$ for the value body with that
    argument bound as $x$, so $v_1[y_{\ltynv{0}}/x]=v_y$. The domain and uniform fiber
    derivations give the whole-frame realization by
    \glrulename{c.tyAt.frame}, displayed above. Together with the given
    source typing, standard-frame validity, and aligned body derivation,
    they supply all premises of
    {\small\[\glrulenamed{c.term.nllam}.\]}
    Its output and conclusion are
    \[
    \begin{gathered}
     w=\llambda(x\colon\ltynv{0})\lrh\ltynv{1}\colon
          \glvreshape{v_1}{\ltynv{1}}{\ltynv{1}},\\
     \glCterm{\emptyset}{\epsilon}{\Psi}
       {\nllambda(x\colon\ltyn{0})\ldotsp t}{b_*}{w}
       {(x\colon\ltynv{0}\fnto\ltynv{1})}.
    \end{gathered}
    \]
    This is the required instance of $P_{n+1}$. The depth-$n$ services
    supplied by $C_n$ support any variable or basis transport inside the
    smaller type and body calls; this case combines their results using
    the frame, reshape, and abstraction rules.

    \item For dependent application in $P_{n+1}$, the structural
    induction hypotheses at index $n+1$ first give checked evaluations of the
    function and argument. Use the accompanying transport results to put the
    function at its actual frame and the argument at that frame's domain. Substituting the checked argument into the opened codomain realization gives the realization of $b_1\subst{x}{\len{1}}$. Apply the rule:
    {\small\begin{equation*}
        \glrulenamed{c.term.appNVal}.
    \end{equation*}}
    The instantiated codomain premise is obtained by checked substitution; it is not a new progress assumption. Transport the result to the standard basis of the instantiated codomain to obtain the conclusion of $P_{n+1}$. The argument and function realizations need not have syntactically identical value types.
\end{enumerate}
The space component of the same source induction provides $\exists\lsv.\;\glCspace{\epsilon}{\Psi}{\ls}{\lsv}$. Scalar, sum, and tensor cases use the structural space rules. A fundamental-subspace case compiles its defining map at the required matrix basis and then forms the corresponding dynamic value space.

This completes the source cases of $C_n\Longrightarrow P_{n+1}$.

\paragraph{Basis step: $P_n\Longrightarrow C_{n+1}$.}
Fix $n$ and assume $P_n$ with its simultaneous formation conclusions.
We prove $C_{n+1}$ by structural induction on the basis carrying
$\mathcal B_{n+1}$. Calls on basis components keep index $n+1$; calls on
embedded source derivations use $P_n$. Write
$b_i^*=\stdBasisDCtx{\tof{b_i}}$. We give three cases, distinguishing
structural conversion from the calls to source progress.
\begin{enumerate}
 \item For a dual basis $b_0^{\lbstar}$, the structural induction hypothesis
 gives $\glCcob{\epsilon}{\Psi}{b_0}{c_0}{b_0^*}$ at the same index
 $n+1$. Apply
 \[\glrulenamed{c.cob.dualStd}.\]
 Basis validity gives $\tof{b_0}=\sof{b_0}$ and validity of the dual;
 validity of its standard basis gives the remaining premise. The conclusion
 is explicitly
 \[
 \glCcob{\epsilon}{\Psi}{b_0^{\lbstar}}{\Trans(c_0^{-1})}
   {\stdBasisDCtx{\tof{b_0^{\lbstar}}}}.
 \]
 This proves the dual case of $C_{n+1}$ without a call to $P_n$.

 \item For a projected kernel, put $s=\Kernel(t)$ and
 \[
 \begin{aligned}
 B&=\pBasisC{s}{b_0}{b_1},\\
 M&=\pBasisC{s}{b_0}{b_1^*},\\
 B_*&=\pBasisC{s}{b_0^*}{b_1^*}
       =\stdBasisDCtx{s}.
 \end{aligned}
 \]
 The basis certificate supplies $\mathcal S_n(D_t)$ for the defining map.
 Thus $P_n$ gives its checked evaluation at the standard matrix basis.
 The structural induction hypotheses give
 $\glCcob{\epsilon}{\Psi}{b_i}{c_i}{b_i^*}$ for $i=0,1$.
 Their dual and tensor conversions transport the map evaluation to the
 matrix bases required by $B$, $M$, and $B_*$. The tensor rule used with
 the displayed dual rule is
 \[\glrulenamed{c.cob.tprod}.\]
 The simultaneous formation construction supplies the ambient and kernel
 space evaluations. Using these map evaluations, the embedding and projection
 constructions supply the checked embedding from $M$ into $b_0$ and the
 checked projection from $b_0^*$ into $B_*$. Consequently all premises of
 the following rule are available, with its unprimed right basis set to
 $b_1^*$:
 {\footnotesize
 \renewcommand{\LTTT}[3]{\textcolor{lang}{\MTM{}_{#1}}%
   \left\langle\begin{gathered}#2,\\#3\end{gathered}\right\rangle}
 \[\gpfit{\glrulenamed{c.cob.projKerCoker}}.\]}
 It constructs
 \[
  \glCcob{\epsilon}{\Psi}{M}{c_{M*}}{B_*}.
 \]
 In this instance the witness is the displayed composition of the embedding
 into $b_0$, the component conversion $c_0$, and the projection into $B_*$.
 The two kernel realizations may differ syntactically; the reshapes in the
 rule retain both annotations.

 This conversion is constructed first. It permits transport of the
 projection into $M$ from its standard matrix basis. Together with the
 checked embedding from $B$ into $b_0$, that projection supplies the last
 two premises of
 {\small
 \renewcommand{\LTTT}[3]{\textcolor{lang}{\MTM{}_{#1}}%
   \left\langle\begin{gathered}#2,\\#3\end{gathered}\right\rangle}
 \[\gpfit{\glrulenamed{c.cob.projKerCokerCod}}.\]}
 The other premises are the validity, ambient-space equalities, and space
 evaluations already obtained. Instantiating the changed right basis by
 $b_1^*$ gives
 \[
  \glCcob{\epsilon}{\Psi}{B}{c_{BM}}{M}.
 \]
 Finally compose these two conversions using
 \[\glrulenamed{c.cob.trans}.\]
 If $\glCspace{\epsilon}{\Psi}{s}{\lsv}$ is the source kernel realization,
 the conclusion is
 \[
 \glCcob{\epsilon}{\Psi}{B}
   {\glvmtm{\lsv}{\lsv}{c_{M*}}{\glvreshape{c_{BM}}{\lsv\loprod\lsv}{\lsv\loprod\lsv}}}{B_*}.
 \]
 This proves the projected-kernel case of $C_{n+1}$. The use of $P_n$
 evaluates the embedded source map and supplies the source evaluations used
 in these constructions; the conversions of $b_0$ and $b_1$ come from
 structural induction at index $n+1$. In particular, neither of the two
 displayed projected-basis conversions is assumed in order to construct
 itself.

 \item For a nonempty term basis, put
 \[
 B=(t_0,\ldots,t_{m-1})_{b'},\qquad
 s=\sof{b'},\qquad b_* =\stdBasisDCtx{s}.
 \]
 The structural induction
 hypothesis gives
 $\glCcob{\epsilon}{\Psi}{b'}{c}{b_*}$.
 The certificate supplies the vector-typing derivations at depth $n$.
 Apply $P_n$ to these derivations and combine their checked evaluations to
 obtain the direct-sum value $v_{\rm list}$ at the repeated basis $b'$.
 The required conversions of the repeated basis are built from $c$ by
 \[\glrulenamed{c.cob.dsum}.\]
 The simultaneous space conclusion gives
 $\glCspace{\epsilon}{\Psi}{s}{\lsv}$.

 The RREF rule needs a conversion in the opposite direction, from $b_*$
 to the fallback basis $b'$. Obtain it with
 \[\glrulenamed{c.cob.inverse}.\]
 This gives $\glCcob{\epsilon}{\Psi}{b_*}{c^{-1}}{b'}$.
 Together with the list evaluation, space evaluation, and validity of
 $b_*$ and $B$, it supplies all premises of
 {\small\[\glrulenamed{c.cob.termsRref}.\]}
 Instantiate the rule's $\lb$ by $b_*$ and $\lb'$ by $b'$. Its conclusion
 is a conversion from the standard basis to the term basis:
 \[
 \begin{gathered}
 \glCcob{\epsilon}{\Psi}{b_*}{r}{B},\\
 r=\glvreshape{\glvop{RREFRep}(v_{\rm list},
       \glvreshape{c^{-1}}{\lsv\loprod\lsv}{\lsv\loprod\lsv})}{\lsv\loprod\lsv}{\lsv\loprod\lsv}.
 \end{gathered}
 \]
 Apply \glrulename{c.cob.inverse} once more to conclude
 \[
  \glCcob{\epsilon}{\Psi}{B}{r^{-1}}{b_*},
 \]
 which is the required term-basis instance of $C_{n+1}$.
 Here $P_n$ supplies the vector evaluations, while structural induction
 supplies the fallback-basis conversion; the RREF rule and inversion assemble
 the final witness.
\end{enumerate}
Together with the base cases and the simultaneous realization and transport
results, these constructions establish $P_n\land C_n$ for every $n$.

\paragraph{Obtaining a bound from context formation.}
We have proved that $\mathcal S_n^{\Psi}(D)$ gives checked evaluation at
 the standard basis, and that $\mathcal B_n^{\Psi}(b)$ permits transport to
 $b$. Here $D$ is the given derivation of
 $(\Delta,\epsilon)\vdash t:\tof{b}$: $\mathcal S_n^{\Psi}(D)$ bounds its
 selected variable representations, while $\mathcal B_n^{\Psi}(b)$ bounds
 the component bases and embedded source derivations of the requested basis.
 It remains to derive
 \[
  \exists n.\;\mathcal S_n^{\Psi}(D)\land\mathcal B_n^{\Psi}(b)
 \]
 from $(\Delta,\epsilon);\valCheckCtx\vdash\Psi$, $D$, and validity of $b$.
 We induct on the derivation of
$(\Delta,\epsilon);\valCheckCtx\vdash\Psi$ in \cref{def:nonrecurseeval}.
The empty-context case has no stored entries and hence no selected-basis
obligations. For an extension, write
\[
 \Psi_1=\Psi_0[(x,b)\mapsto v].
\]
Here $\Psi_0$ is the context \emph{before} this entry is added. The definition
requires the following premises in that earlier context:
\[
\begin{gathered}
 (\Delta_0,\epsilon_0);\valCheckCtx_0\vdash\Psi_0,\qquad
 (\{\},\epsilon_0)\vdash b:\bvalid,\\
 \evalSJudge{\Psi_0}{\tof{b}}{\operatorname{inferType}(v)},\qquad
 \operatorname{fv}(v)\subseteq\operatorname{names}(\Psi_0)\cup\{x\}.
\end{gathered}
\]
Writing $\ltyv=\operatorname{inferType}(v)$, a fresh declaration also supplies
$\valCheckCtx_0\vdash\ltyv:\lkn{}$ and
$\valCheckCtx_1\vdash v:\ltyv$, where
$\valCheckCtx_1=\valCheckCtx_0[x\mapsto\ltyv]$.
For another representation, $\valCheckCtx_1=\valCheckCtx_0$ and
$\valCheckCtx_0\vdash v:\ltyv$ is supplied directly.
These typing premises are retained by context formation; the finite-bound
argument uses the validity, realization, and scope premises displayed above.
Thus validity of $b$ and the evaluation of its type are established before
installing $(x,b)\mapsto v$. For each source derivation $D_j$ needed by
validity of $b$, the induction hypothesis for $\Psi_0$ supplies a finite
bound for $\mathcal S_{m_j}^{\Psi_0}(D_j)$. Its variable leaves select
representations already available in $\Psi_0$, with the corresponding
lower-index $\mathcal B$ obligations. Combining these $\mathcal S$ conditions
with the component-basis $\mathcal B$ conditions gives
$\mathcal B_m^{\Psi_0}(b)$ for a sufficiently large finite $m$.
At a later use of the new entry, the variable clause of $\mathcal S_{m+1}$
uses this $\mathcal B_m$ evidence to select $(x,b)\mapsto v$.
The previously established $\mathcal S$ and $\mathcal B$ conditions persist
under the permitted extensions; increasing the bound accommodates the new
selection. When $x$ is new, $b$ is checked in $\epsilon_0$, which does not
contain $x$; when this adds another representation of an existing $x$, its
validity may use a representation already present in $\Psi_0$. Neither case
uses the entry being installed to justify its own basis.

For a stored function, the definition also gives an explicit condition.
If $b=\frameB{z}{b_0}{b_1}$ and
$v=\llambda(z\colon\ltynv{0})\lrh\ltynv{1}\colon u$, then, for fresh $y$,
\[
 \evalSJudge{\Psi_0[(y,b_0)\mapsto y_{\ltynv{0}}]}
   {\tof{b_1[y/z]}}{\ltynv{1}[y_{\ltynv{0}}/z]}.
\]
This is the codomain realization in the environment containing the fresh
argument at the function's domain basis. It supplies the additional
fresh-argument condition used in frame transport above. The frame clause
of $\mathcal B_m$ retains the corresponding component-basis and source
obligations uniformly in the fresh name.

Repeating the extension argument through the finite derivation of
$(\Delta,\epsilon);\valCheckCtx\vdash\Psi$, and combining the resulting
$\mathcal S$ and $\mathcal B$ conditions for the given typing and
basis-validity derivations, supplies a
common bound:
\[
\begin{gathered}
 (\Delta,\epsilon);\valCheckCtx\vdash\Psi,\quad
 D:(\Delta,\epsilon)\vdash t:\tof{b},\quad
 (\{\},\epsilon)\vdash b:\bvalid\\
 \Longrightarrow\quad
 \exists n.\;\mathcal S_n^{\Psi}(D)\land\mathcal B_n^{\Psi}(b).
\end{gathered}
\]
Choose such an $n$ and name the two conclusions
$H_S:\mathcal S_n^{\Psi}(D)$ and $H_B:\mathcal B_n^{\Psi}(b)$.
Write $b_* = \stdBasisDCtx{\tof{b}}$.
Applying $P_n$ to $D$ and $H_S$ gives the checked evaluation at $b_*$
on the left below. Applying the depth-$n$ transport result with $H_B$
then gives the evaluation at the requested basis $b$ on the right:
\begin{equation*}
    \glCterm{\Delta}{\epsilon}{\Psi}{t}{b_*}{v_*}{\ltynv{*}}
    \quad\Longrightarrow\quad
    \exists v_0,\ltyv.\;
    \glCterm{\Delta}{\epsilon}{\Psi}{t}{b}{v_0}{\ltyv}.
\end{equation*}
For a space basis, $H_B$ supplies the hypothesis of $C_n$ used in this
transport; for a frame, it supplies the domain and fiber obligations of the
accompanying frame-transport result. Thus both $H_S$ and $H_B$ have been
used, and their common bound $n$ was obtained from the theorem's premises.

\paragraph{Matching the inferred result type.}
We have obtained the required checked space evaluation and a checked term
evaluation at $b$. The remaining term conclusion is the equality of its
inferred type with the realized type.
The constructed trace retains $\glCat{\epsilon}{\Psi}{b}{\ltyv}$. Apply
{\small\begin{equation*}
    \glrule[2,1,1]{c.term.reshapeCob}
\end{equation*}}
with the target realization equal to $\ltyv$, and choose $v=\glvreshape{v_0}{\ltyv}{\ltyv}$. The type-inference clause then gives
\begin{equation*}
    \operatorname{inferType}(v)=\ltyv.
\end{equation*}
This is the selected checked evaluation in the conclusion. Erase it to an
ordinary evaluation and apply \cref{thm:preserve}(2):
\[
 \valCheckCtx\vdash v:\ltyv.
\]
If also $\valCheckCtx\vdash\glvEnv$, \cref{lem:soundVals} gives
$\SemCheck{v}{\glvEnv}\in\mathcal V_\glvEnv(\ltyv)$.
Successful raw denotation and source agreement are established by adequacy
under $(\Delta,\epsilon);\Psi;\valCheckCtx\vdash(\sigma,\glvEnv)$.
\end{proof}
\endgroup

\subsection{Adequacy (\cref{sec:meta2})}
\begin{defn}[Agreement (\cref{def:agree})]
Write $a\equiv_{B;\sigma,\glvEnv}^{\lty,\ltyv}w$ for agreement between a source value
and a runtime value at basis or frame $B$, interpreting $\lty$ under $\sigma$
and $\ltyv$ under $\glvEnv$.
At spaces, put $N=\dof{\lsv}{\glvEnv}$; then
$a\equiv_{B;\sigma,\glvEnv}^{\ls,\lsv}w$ iff
$|B|=|w|=N$ and $a=\sum_{i=0}^{N-1}w_iB_i$.
At functions $\lty=x:\ltyn{0}\fnto\ltyn{1}$ and
$\ltyv=x:\ltynv{0}\fnto\ltynv{1}$, let $B$ be a frame with domain
component $B_0$ and fiber family $a\mapsto B_1(a)$.
Then $f\equiv_{B;\sigma,\glvEnv}^{\lty,\ltyv}g$ iff, for every
$a\equiv_{B_0;\sigma,\glvEnv}^{\ltyn{0},\ltynv{0}}w$, there is $z$ such that
$g(w)=z\ne\glverr$ and
$f(a)\equiv_{B_1(a);\sigma',\glvEnv'}^{\ltyn{1},\ltynv{1}}z$, where
$\sigma'=\sigma[x\mapsto a]$ and $\glvEnv'=\glvEnv[x\mapsto w]$.
Here $g$ has the runtime domain prescribed by $\ltynv{0}$.

\end{defn}
\begin{defn}[Type and dimension agreement]
\label{def:agreedims}
Shape agreement $\lty\sim_{B;\sigma,\glvEnv}\ltyv$ is defined recursively.
At spaces, $\ls\sim_{B;\sigma,\glvEnv}\lsv$ iff
$\shpOf{\lsv}{\glvEnv}=\operatorname{shape}_{\sigma}(\ls)$.
At function types $\lty=x:\ltyn{0}\fnto\ltyn{1}$ and
$\ltyv=x:\ltynv{0}\fnto\ltynv{1}$, let $B$ be a frame with domain
component $B_0$ and fiber family $a\mapsto B_1(a)$. Require
$\ltyn{0}\sim_{B_0;\sigma,\glvEnv}\ltynv{0}$ and, for every
$a\equiv_{B_0;\sigma,\glvEnv}^{\ltyn{0},\ltynv{0}}w$,
$\ltyn{1}\sim_{B_1(a);\sigma[x\mapsto a],\glvEnv[x\mapsto w]}\ltynv{1}$.
The source shape $\operatorname{shape}_{\sigma}(\ls)$ follows the source-space
constructors: $(0,)$, $(1,)$, dual unchanged, direct-sum dimension,
tensor-factor dimensions, and fundamental-subspace dimension.
Type agreement $\lty\simeq_{B;\sigma,\glvEnv}\ltyv$ holds when
$\lty\sim_{B;\sigma,\glvEnv}\ltyv$ and $B$ is recursively a basis or discrete frame
for $\lty$ under $\sigma$.

\end{defn}
\begin{defn}[Consistent environments (\cref{def:allCtxAgree})]
We write $(\Delta,\epsilon);\Psi;\valCheckCtx\vdash(\sigma,\glvEnv)$ when
$(\Delta,\epsilon);\valCheckCtx\vdash\Psi$, $\valCheckCtx\vdash\glvEnv$,
$(\Delta,\epsilon)\vdash\sigma$ at every step, and
$(\Delta,\epsilon,\Psi,\valCheckCtx,\sigma,\glvEnv)$ is constructed as follows.
The tuple $(\emptyset,\emptyset,\emptyset,\emptyset,\emptyset,\glvEnv)$ is constructed for arbitrary $\glvEnv$.
Suppose $(\Delta,\epsilon,\Psi,\valCheckCtx,\sigma,\glvEnv)$ has been constructed.
Let $\Psi'=\Psi[(x,\lb{})\mapsto\leev{}]$ and update
$(\Delta,\epsilon,\valCheckCtx)$ to $(\Delta',\epsilon',\valCheckCtx')$
as in \cref{def:nonrecurseeval}.
Put $T=\tof{\lb{}}$,
$\ltyv{}=\operatorname{inferType}(\leev{})$, and
$B=\SemBB{\lb{}}{\sigma}$, and require
$\Sem{\leev{}}{\glvEnv}=w\ne\glverr$,
$a\equiv_{B;\sigma,\glvEnv}^{T,\ltyv{}}w$, and
$T\sim_{B;\sigma,\glvEnv}\ltyv{}$.
If $x\notin\operatorname{dom}(\Delta\sqcup\epsilon)$,
require $(x\notin\operatorname{fv}(\leev{})\ \lor\
\leev{}=x_{\ltyv{}})$ and put
$\sigma'=\sigma[x\mapsto a]$, and
$\glvEnv'=\glvEnv[x\mapsto w]$.
Otherwise require $a=\sigma(x)$ and put
$(\sigma',\glvEnv')=(\sigma,\glvEnv)$.
This constructs $(\Delta',\epsilon',\Psi',\valCheckCtx',\sigma',\glvEnv')$.

\end{defn}
\begin{lem}[Reshape identity under agreement]
\label{lem:reshape-agreement}
Let $R$ be recursive reshape from \cref{app:gen-defs:reshap}, with the
target type and environment first.
\begin{enumerate}
\item If $w$ is an array and
$|w|=\dof{\lsnv{s}}{\glvEnv^s}=\dof{\lsnv{t}}{\glvEnv^t}$, then
\[
 R_{\glvEnv^t,\glvEnv^s}(w;\lsnv{t},\lsnv{s})=w.
\]
\item Suppose
$\lty\simeq_{B;\sigma,\glvEnv^s}\ltynv{s}$ and
$\lty\simeq_{B;\sigma,\glvEnv^t}\ltynv{t}$.
If $a\equiv_{B;\sigma,\glvEnv^s}^{\lty,\ltynv{s}}w$, then for some $w'$,
\[
 R_{\glvEnv^t,\glvEnv^s}(w;\ltynv{t},\ltynv{s})=w'\ne\glverr,
 \qquad a\equiv_{B;\sigma,\glvEnv^t}^{\lty,\ltynv{t}}w'.
\]
\end{enumerate}
Thus reshape is the identity on arrays and preserves the represented
source value at function types.

\end{lem}
\begin{proof}
Part (1) is the array clause of $R$: both dimension computations succeed
and equal $|w|$, so its length tests succeed and it returns $w$.

For (2), induct on the sum of the sizes of the two value types.
When the types and environments are fixed, abbreviate agreement by
$a\equiv_B w$. At spaces, the hypotheses give
\[
 |w|=\dof{\lsnv{s}}{\glvEnv^s}
     =\dof{\lsnv{t}}{\glvEnv^t}=|B|.
\]
By (1), $R$ returns $w$ itself. Both its entries and
$a=\sum_i w_iB_i$ are unchanged, proving agreement at the target type.

At function types, write the source value as $f$ and its runtime
representation as $g_s$. Let $B$ have domain component $B_0$ and fiber
family $a\mapsto B_1(a)$, and write the value types as
$x:\ltynv{s0}\fnto\ltynv{s1}$ and
$x:\ltynv{t0}\fnto\ltynv{t1}$.
The domain clauses of type agreement make both runtime domains defined,
so $R$ returns a function $g_t$. To prove that it agrees with $f$, take
any $w_t$ in its prescribed runtime domain agreeing with $a$ at $B_0$.
The domain induction hypothesis gives
\[
 w_s=R_{\glvEnv^s,\glvEnv^t}
       (w_t;\ltynv{s0},\ltynv{t0})\ne\glverr,
 \qquad a\equiv_{B_0}w_s.
\]
Function agreement for $g_s$ then gives $g_s(w_s)=z_s\ne\glverr$ and
$f(a)\equiv_{B_1(a)}z_s$. Set
$\glvEnv^{s'}=\glvEnv^s[x\mapsto w_s]$ and
$\glvEnv^{t'}=\glvEnv^t[x\mapsto w_t]$.
Both arguments represent $a$, so the fiber clauses of the two type
agreements apply at the same source environment $\sigma[x\mapsto a]$
and the same fiber $B_1(a)$. The codomain induction hypothesis gives
\[
 g_t(w_t)=R_{\glvEnv^{t'},\glvEnv^{s'}}
     (z_s;\ltynv{t1},\ltynv{s1})=z_t\ne\glverr,
 \qquad f(a)\equiv_{B_1(a)}z_t.
\]
Thus $g_t$ recursively converts its argument to the stored domain,
calls $g_s$, and recursively converts the result to the target codomain.
The displayed agreement for every agreeing argument pair is precisely
function agreement for $f$ and $g_t$, proving (2).
\end{proof}

\begin{lem}[Pseudoinverses and projected bases in coordinates]
\label{lem:pseudoinverse-projected-bases}
Let $L:V\to W$ be linear between finite-dimensional real inner-product
spaces, with ordered bases $E=(E_i)_{i<N}$ and $F=(F_j)_{j<M}$ and
Gram matrices $G,H$. Write $r_B(c)=\sum_i c_iB_i$ for reconstruction
in a basis $B$, applied columnwise to matrices, and let
$A\in\mathbb R^{M\times N}$ satisfy
$L(r_E(c))=r_F(Ac)$. Put $r=\operatorname{rank}L$ and, using the
conventions of \cref{def:fsv} and \cref{def:gsvd},
\[
\begin{aligned}
 Q_K&=\GSVDUI(A,G,H),& Q_C&=\GSVDU(A,G,H),\\
 Q_I&=\GSVDV(A,G,H),& Q_O&=\GSVDVI(A,G,H),
\end{aligned}
\qquad \Sigma=\GSVDS(A,G,H).
\]
\begin{enumerate}
\item The columns of $r_E(Q_K)$, $r_E(Q_C)$, $r_F(Q_I)$, and
$r_F(Q_O)$ are exactly the ordered projected bases of
$\operatorname{ker}(L)$, $\operatorname{coker}(L)$, $\operatorname{image}(L)$, and $\operatorname{coimage}(L)$,
respectively, viewed in their ambient spaces. They are orthonormal
and have lengths $N-r$, $r$, $r$, and $M-r$.
\item The metric pseudo-inverse $A^+_{G,H}=Q_C\Sigma^{-1}Q_I^{\mathsf T}H$
satisfies
\[
 L^+(r_F(y))=r_E(A^+_{G,H}y),\qquad
 L^+=\sum_{j<M,\,i<N}(A^+_{G,H})_{ij}\,F^j\otimes E_i.
\]
For orthonormal $E,F$, $G=I_N$, $H=I_M$, and this is the ordinary
pseudoinverse.
\item More generally, let $K=(K_j)_{j<k}$ be an orthonormal basis of
$S\leq U$, and $B=(B_i)_{i<n}$ an ordered basis of the ambient inner-product space
$U$, with Gram matrix $M_B$. For the inclusion
$\iota_S:S\hookrightarrow U$ and orthogonal projection $\pi_S:U\to S$,
define $Q\in\mathbb R^{n\times k}$ by
$\iota_S(K_j)=\sum_iQ_{ij}B_i$. Then
\[
\begin{gathered}
 Q^{\mathsf T}M_BQ=I_k,\qquad Q^+_{I_k,M_B}=Q^{\mathsf T}M_B,\\
 \iota_S(r_K(z))=r_B(Qz),\qquad
 \pi_S(r_B(u))=r_K(Q^{\mathsf T}M_Bu),\qquad
 \pi_S\iota_S=\operatorname{id}_S.
\end{gathered}
\]
In particular, the four blocks satisfy
\[
 Q_KQ_K^{\mathsf T}G+Q_CQ_C^{\mathsf T}G=I_N,\qquad
 Q_IQ_I^{\mathsf T}H+Q_OQ_O^{\mathsf T}H=I_M.
\]
\end{enumerate}

\end{lem}
\begin{proof}
\emph{Ordered projected bases.}
For (1), the numerical columns must match the source lists,
including their signs and choices within repeated eigenspaces.
Both use the paired convention of
\cref{def:fsv}. The GSVD gives complete orthonormal spectral families
for $L^\star L$ and $LL^\star$ in the supplied Gram metrics.
For either self-adjoint operator $S$, group the vectors $(u_i)$ by
eigenvalue. The projector computed from a block is
\[
 P_\lambda(x)=\sum_{i:\lambda_i=\lambda}\langle u_i,x\rangle u_i.
\]
By \cref{prop:sp}, this is $p_\lambda(S)$. On the domain side,
projecting the prescribed ordered basis, selecting $\LexInd$, and
applying normalized Gram--Schmidt therefore gives
$B_{\operatorname{Left},\lambda}$ from \cref{def:fsv}.
For $\lambda>0$, set $z_i=\lambda^{-1/2}L(u_i)$. Then
\[
 \langle z_i,z_j\rangle
 =\lambda^{-1}\langle u_i,L^\star L(u_j)\rangle=\delta_{ij},
 \qquad LL^\star(z_i)=\lambda z_i,
 \qquad L(u_i)=\sqrt\lambda\,z_i.
\]
Thus these vectors form the paired codomain block. At zero, use the
projector construction independently on $\ker L$ and $\ker L^\star$.
Concatenation in decreasing eigenvalue order, filtered by the same
predicate, gives the two ordered $\FSV$ lists. Projector-based domain
selection fixes signs and repeated-eigenvalue choices; the positive
codomain blocks then follow by pairing.

To identify the numerical columns, the coordinate and Gram identities are
\[
 L(r_E(c))=r_F(Ac),\qquad
 \langle r_E(c),r_E(c')\rangle=c^{\mathsf T}Gc',\qquad
 \langle r_F(d),r_F(d')\rangle=d^{\mathsf T}Hd'.
\]
In particular, $c^{\mathsf T}Gc=\|r_E(c)\|^2>0$ for $c\ne0$, since
$E$ is a basis; similarly $H\succ0$. The adjoint identity gives
\[
 L^\star(r_F(d))=r_E(G^{-1}A^{\mathsf T}Hd),\qquad
 L^\star L(r_E(c))=r_E(G^{-1}A^{\mathsf T}HAc).
\]
Consequently the spectral projector satisfies
\[
 r_E\bigl(p_\lambda(G^{-1}A^{\mathsf T}HA)e_i\bigr)
 =p_\lambda(L^\star L)(E_i),
\]
where $e_i$ is the $i$th coordinate vector. Because $r_E$ is
invertible, the two ordered lists have the same linear dependencies,
so $\LexInd$ selects the same indices. The Gram identity makes every
inner product and normalization in Gram--Schmidt equal on the two
sides. Thus every computed column represents the corresponding source
block vector. On the codomain side, use $r_F$ and
$AG^{-1}A^{\mathsf T}H$; positive blocks also satisfy
$r_F(\lambda^{-1/2}Au)=\lambda^{-1/2}L(r_E(u))$.
This proves (1), including repeated eigenvalues and empty blocks.
The block sizes are $N\times(N-r)$, $N\times r$, $M\times r$, and
$M\times(M-r)$, respectively, as in \cref{def:gsvd}.

\emph{Pseudoinverse.}
For (2), the numerical result is
$A^+_{G,H}=Q_C\Sigma^{-1}Q_I^{\mathsf T}H$.
For each positive singular value $\sigma_j$, the paired-vector
identities give
\[
 L(r_E((Q_C)_j))=\sigma_j r_F((Q_I)_j),\qquad
 A^+_{G,H}(Q_I)_j=\sigma_j^{-1}(Q_C)_j.
\]
If $r_F(y)\perp\operatorname{image}(L)$, then $Q_I^{\mathsf T}Hy=0$, so
$A^+_{G,H}y=0$. Thus the computed map is the inverse of $L$ from its
image to its cokernel and zero on the orthogonal complement of its
image. These restrictions characterize the source Moore--Penrose map
$L^+$ of \cref{prop:core}, which satisfies
\[
 LL^+L=L,\quad L^+LL^+=L^+,\quad
 (LL^+)^\star=LL^+,\quad (L^+L)^\star=L^+L.
\]
Hence $L^+(r_F(y))=r_E(A^+_{G,H}y)$. Taking $y$ to be each
coordinate vector gives the stated tensor coefficients. In the
paper's row-array convention, entry $(i,j)$ is stored at $jN+i$,
so the length-$MN$ array represents $L^+$ in $F^\star\otimes E$.
For orthonormal bases the Gram matrices are identities, giving the
ordinary SVD pseudo-inverse. This also covers rank zero, with empty
positive blocks and the zero pseudo-inverse.

\emph{Inclusions and projections.}
For (3), orthonormality gives
$(Q^{\mathsf T}M_BQ)_{ij}=\langle K_i,K_j\rangle=\delta_{ij}$.
Put $P=Q^{\mathsf T}M_B$. Then $PQ=I_k$ and, for every $u$,
\[
 r_B(QPu)\in S,\qquad
 \langle K_j,r_B(u)-r_B(QPu)\rangle
 =(Pu)_j-(PQPu)_j=0.
\]
Thus $QP$ is the orthogonal projection in ambient coordinates.
Moreover, $P$ is the inverse of the inclusion on $S$ and vanishes on
$S^\perp$, so it is the inclusion's pseudo-inverse. This proves all
the formulas in (3). Finally,
$V=\operatorname{ker}(L)\oplus\operatorname{coker}(L)$ and
$W=\operatorname{image}(L)\oplus\operatorname{coimage}(L)$ give the two complementary
projection identities.
\end{proof}

\begin{thm}[Adequacy (\cref{thm:main})]
\label{thm:mainA}
Suppose $(\Delta,\epsilon);\Psi;\valCheckCtx\vdash(\sigma,\glvEnv)$.
\begin{enumerate}
\item If $\glCat{\epsilon}{\Psi}{b}{\ltyv{}}$,
then $\tof{b}\simeq_{\SemBB{b}{\sigma};\sigma,\glvEnv}\ltyv{}$.
\item If $\glCterm{\Delta}{\epsilon}{\Psi}{t}{b}{v}{\ltyv{}}$,
then $\Sem{v}{\glvEnv}=w\ne\glverr$ for some $w$,
$\Sem{t}{\sigma}\equiv_{\SemBB{b}{\sigma};\sigma,\glvEnv}^{\tof{b},\ltyv{}}w$,
and $\tof{b}\simeq_{\SemBB{b}{\sigma};\sigma,\glvEnv}\ltyv{}$.
\item If $\glCcob{\epsilon}{\Psi}{b_0}{v}{b_1}$,
put $B_i=\SemBB{b_i}{\sigma}$ and $N=|B_0|=|B_1|$.
Then $\Sem{v}{\glvEnv}=A\ne\glverr$ for some array $A$ of length $N^2$,
and $B_0=B_1A^{\mathsf T}$, reading $A$ as a row-major $N\times N$ matrix
and the basis vectors as columns.
\end{enumerate}

\end{thm}
\begingroup
\renewcommand{\evalCtxSpace}{\epsilon}
\begin{proof}

We induct simultaneously on the checked term, space, type-at-basis, and
change-of-basis derivations. The hypotheses assume the semantic clauses
of \cref{def:allCtxAgree}, which we extend by agreeing arguments under
binders.

We give the principal cases, writing sums over the finite index sets of the displayed bases. When the source and value types and the two environments are fixed, we abbreviate the agreement judgment by $a\equiv_B w$.
In each term case, put $B=\SemBB{b}{\sigma}$. Assuming the induction
hypotheses for the premises of its last rule, we must establish
\[
\begin{gathered}
 \Sem{v}{\glvEnv}=w\ne\glverr,\qquad
 \Sem{t}{\sigma}\equiv_{B;\sigma,\glvEnv}^{\tof{b},\ltyv}w,\\
 \tof{b}\simeq_{B;\sigma,\glvEnv}\ltyv.
\end{gathered}
\]
We also establish these agreements with $\ltyv$ replaced by
$\operatorname{inferType}(v)$. 

\paragraph{Term evaluation}
We give representative cases, each for one last rule of the checked
term derivation. Each case proves successful denotation and agreement
for that rule's conclusion.

\begin{enumerate}
 \item\label{proof:adequacy-variable} In the nonlinear-variable case, the last rule is
 {\small\[\glrulenamed{c.term.varNlin}.\]}
 Let $(x,b)\mapsto v$ be its selected entry and put
 $\ltynv{e}=\operatorname{inferType}(v)$ and $B=\SemBB{b}{\sigma}$.
 The entry conditions of \cref{def:allCtxAgree} give
 \[
 \Sem{v}{\glvEnv}=w\ne\glverr,\qquad
 \sigma(x)\equiv_{B;\sigma,\glvEnv}^{\tof{b},\ltynv{e}}w.
 \]
 The induction hypothesis for the premise $\glCat{\epsilon}{\Psi}{b}{\ltyv}$ gives
 $\tof{b}\simeq_{B;\sigma,\glvEnv}\ltyv$.
 At a space, combining this with agreement at the inferred type gives
 \[
 |w|=|B|=\dof{\ltynv{e}}{\glvEnv}=\dof{\ltyv}{\glvEnv},\qquad
 \sigma(x)=\sum_{i<|B|}w_iB_i.
 \]
 At a function, the recursive domain and fiber agreements give the
 corresponding function clause. Thus the returned $w$ agrees with
 $\sigma(x)$ at the requested $\ltyv$.

 \item In the addition case, the last rule is
 {\small\[\glrulenamed{c.term.add}.\]}
 Put $B=\SemBB{\lb{}}{\sigma}$. The operand induction hypotheses give
 \[
 \begin{aligned}
 u&=\Sem{\lenv{0}}{\glvEnv}\ne\glverr,
 &a&=\Sem{\len{0}}{\sigma}=\sum_i u_iB_i,\\
 w&=\Sem{\lenv{1}}{\glvEnv}\ne\glverr,
 &a'&=\Sem{\len{1}}{\sigma}=\sum_i w_iB_i.
 \end{aligned}
 \]
 If their annotations are $\lsnv{0},\lsnv{1}$ and the result annotation
 is $\lsv$, the operand and basis hypotheses give
 \[
 |u|=|w|=|B|=\dof{\lsnv{0}}{\glvEnv}
 =\dof{\lsnv{1}}{\glvEnv}=\dof{\lsv}{\glvEnv},\qquad
 a+a'=\sum_i(u_i+w_i)B_i.
 \]
 These length equalities make the addition checks succeed even when the
 annotations differ. Thus
 $\Sem{\lenv{0}\lvadd_{\lsv}\lenv{1}}{\glvEnv}_i=u_i+w_i$;
 the returned array has
 length $|B|$, so the displayed sum proves agreement with $a+a'$ at
 the result annotation.

 \item In the standard-inner-product case, the last rule is
 {\small\begin{equation*}\glrule[2,2,2]{c.term.innerStd}.\end{equation*}}
 Put $B=\SemBB{\lbn{0}}{\sigma}=(B_i)_{i<N}$ and
 $(B^i)_{i<N}=\SemBB{\lbn{0}^{\lstar}}{\sigma}$.
 By \cref{lem:stdBasisOrthonormal},
 \begin{equation*}
 \langle B_i,B_j\rangle=\delta_{ij},\qquad
 \Sem{\linner{}_{\ls}}{\sigma}=\sum_i B^i\otimes B^i.
 \end{equation*}
 The space hypothesis gives $\dof{\lsv}{\glvEnv}=N$, so
 $a=\Sem{\lidelta{\lsv}}{\glvEnv}\ne\glverr$ has
 $|a|=N^2$ and $a[iN+j]=\delta_{ij}$, representing the displayed tensor in
 $B^\star\otimes B^\star$. 

 \item\label{proof:adequacy-linear-application} In the linear-application case, the last rule is
 {\small\begin{equation*}\glrule[2,2,1,1,1]{c.term.appLLin}.\end{equation*}}
 The source denotation is
 \begin{equation*}\gleqn{denotTermAux.appL}.\end{equation*}
 Put $E=\SemBB{\lbn{0}}{\sigma}=(E_k)_{k<N}$ and $F=\SemBB{\lbn{1}}{\sigma}=(F_j)_{j<M}$. Let $A=\Sem{\lenv{0}}{\glvEnv}$ and $u=\Sem{\lenv{1}}{\glvEnv}$. The term hypotheses give successful arrays with
 \begin{equation*}
 \begin{gathered}
 |A|=NM,\quad |u|=N,\quad
 L:=\Sem{\len{0}}{\sigma}=\sum_{k<N,j<M}A_{kM+j}\,E^k\otimes F_j,\\
 x:=\Sem{\len{1}}{\sigma}=\sum_{\ell<N}u_\ell E_\ell.
 \end{gathered}
 \end{equation*}
 All realizations of the input space appearing in the rule have dimension $N$; the output and matrix-output realizations have dimension $M$. They need not be identical value-space expressions. Using $E^k(E_\ell)=\delta_{k\ell}$, calculate
 \begin{align*}
 L(x)
 &=\sum_{k<N,j<M,\ell<N}A_{kM+j}u_\ell E^k(E_\ell)F_j\\
 &=\sum_{j<M}\left(\sum_{k<N}A_{kM+j}u_k\right)F_j.
 \end{align*}
 The generated contraction clauses use the evaluation-state notation of \cref{fig:valValSemantics}: $d_q$ gives dimensions, $s_q$ gives inferred spaces, $n_q$ their dimensions, and $w$ denotes the full expression. They are
 {\small\gldefn{value.denot.contractions}}
 In the tensor-vector clause, the contracted dimensions both equal $N$,
 the matrix and argument lengths are $NM$ and $N$, and the output dimension
 is $M$. These are exactly its dimension and length tests. Every read is in
 bounds: for $k<N$ and $j<M$, $kM+j<NM$ and $k<N$.
 Hence
 \begin{equation*}
 \begin{gathered}
 w=\Sem{\LTTVPP{\lsv}{\lenv{0}}{\lenv{1}}}{\glvEnv}\ne\glverr,\qquad |w|=M,\\
 w_j=\sum_{k<N}A_{kM+j}u_k,\qquad
 L(x)=\sum_{j<M}w_jF_j.
 \end{gathered}
 \end{equation*}
 The final reshape has source and target dimensions $M$, so \cref{lem:reshape-agreement}(1) makes it succeed and preserve these coordinates. This proves agreement in $F$, rather than an equality between the source vector and the array itself. Empty dimensions give empty arrays or empty sums, with no out-of-bounds reads.

 \item\label{proof:adequacy-change-of-basis} In the change-of-basis application case, the last rule is
 {\small\[\glrulenamed{c.term.cob}.\]}
 Put $B_s=\SemBB{\lbn{1}}{\sigma}$ and
 $B_t=\SemBB{\lbn{0}}{\sigma}$. The induction hypotheses give
 \begin{equation*}
 \begin{gathered}
 u=\Sem{\leev{}}{\glvEnv}\ne\glverr,\qquad
 a=\Sem{\lee{}}{\sigma}=\sum_j u_j(B_s)_j,\\
 A_{ji}:=\Sem{\lenv{0}}{\glvEnv}_{jN+i},\qquad
 (B_s)_j=\sum_i A_{ji}(B_t)_i.
 \end{gathered}
 \end{equation*}
 Therefore
 \begin{equation*}
 a=\sum_i\left(\sum_j A_{ji}u_j\right)(B_t)_i.
 \end{equation*}
 The dimension hypotheses give $|B_s|=|B_t|=N$,
 $|\Sem{\lenv{0}}{\glvEnv}|=N^2$, and $|u|=N$.
 By \cref{lem:reshape-agreement}(1), these matching lengths make both
 reshapes leave all entries unchanged. Matrix-vector evaluation gives
 \[
 \begin{gathered}
 w=\Sem{\LTTVPP{\lsv}{\glvreshape{\lenv{0}}{\lsv\loprod\lsv}{\lsv\loprod\lsv}}{\glvreshape{\leev{}}{\lsv}{\lsv}}}{\glvEnv}
   \ne\glverr,\\
 |w|=N,\qquad w_i=(A^{\mathsf T}u)_i,\qquad
 a=\sum_iw_i(B_t)_i.
 \end{gathered}
 \]

 \item In the nonlinear-abstraction case, the last rule is
 {\small\begin{equation*}\glrulenamed{c.term.nllam}.\end{equation*}}
 Rename the bound variable to a fresh $y$ throughout this case. Put
 \[
 \begin{gathered}
 B=\SemBB{\frameB{y}{\lbn{0}}{\lbn{1}}}{\sigma},\qquad
 B_0=\SemBB{\lbn{0}}{\sigma},\qquad
 B_1(a)=\SemBB{\lbn{1}}{\sigma[y\mapsto a]},\\
 f=\Sem{\nllambda\llpl y\colon\ltyn{0}\llpr\ldotsp\lee{}}{\sigma},\\
 g=\Sem{\llambda(y\colon\ltynv{0})\lrh\ltynv{1}\colon
       \glvreshape{\leev{}}{\ltynv{1}}{\ltynv{1}}}{\glvEnv}.
 \end{gathered}
 \]
 To prove function agreement, take any
 $a\equiv_{B_0;\sigma,\glvEnv}^{\ltyn{0},\ltynv{0}}w$.
 Extend the environments by
 \begin{equation*}
 \begin{gathered}
 \epsilon'=\epsilon[y\mapsto\ltyn{0}],\qquad
 \valCheckCtx'=\valCheckCtx[y\mapsto\ltynv{0}],\\
 \sigma'=\sigma[y\mapsto a],\qquad
 \glvEnv'=\glvEnv[y\mapsto w],\qquad
 \Psi'=\Psi[(y,b_0)\mapsto y_{\ltynv{0}}].
 \end{gathered}
 \end{equation*}
 For the new entry, $\Sem{y_{\ltynv{0}}}{\glvEnv'}=w$,
 $\operatorname{inferType}(y_{\ltynv{0}})=\ltynv{0}$, and $\sigma'(y)=a$.
 Argument agreement and the domain induction hypothesis give its value and
 shape agreement. Together with freshness, these are the semantic extension
 conditions in \cref{def:allCtxAgree}. Apply the body and fiber induction
 hypotheses under these extended environments:
 \begin{equation*}
 \begin{gathered}
 \Sem{\leev{}}{\glvEnv'}=z\ne\glverr,\qquad
 \Sem{\lee{}}{\sigma'}
   \equiv_{B_1(a);\sigma',\glvEnv'}^{\ltyn{1},\ltynv{1}}z,\\
 \ltyn{1}\simeq_{B_1(a);\sigma',\glvEnv'}\ltynv{1}.
 \end{gathered}
 \end{equation*}
 By the lambda
 denotation in \cref{fig:valValSemantics}, calling the produced function
 evaluates this body under $\glvEnv'$. Its final reshape targets
 $\ltynv{1}$ under this same environment. At a space-valued fiber,
 $|z|=|B_1(a)|=\dof{\ltynv{1}}{\glvEnv'}$, so the array is unchanged.
 At a function-valued fiber, the body's inferred type is the source
 realization and $\ltynv{1}$ is the target; their agreement at $B_1(a)$
 satisfies the hypotheses of \cref{lem:reshape-agreement}(2).
 In either case the reshape returns $z'$ with
 $f(a)\equiv_{B_1(a)}z'$.
 Thus $g(w)=z'\ne\glverr$ for every agreeing argument pair,
 exactly as required by \cref{def:agree}.

\item\label{proof:adequacy-linear-assembly}
\emph{Linear assembly.}
The last rule is
\gltermrule[checked]{c.term.llamAssemble}
Write $V=\SemT{\sof{b'}}{\sigma}$ and
$W=\SemT{\sof{b}}{\sigma}$, and let
$E=\SemBB{b'}{\sigma}=(E_i)_{i<N}$ and
$F=\SemBB{b}{\sigma}=(F_j)_{j<M}$.
The source clause is
\[\gleqn{denotTermAux.llam}.\]
Rename the bound variable to a fresh $y$ throughout this case. Put $m=\overline{\operatorname{dimOf}}(\lsv')$ and
\[
\begin{aligned}
 f&=\Sem{\lllambda\llpl y\colon\sof{b'}\llpr\ldotsp\lee{}}{\sigma},\\
 q&=\glvreshape{\lsum^{\lsv'}_{i<m}
       \left(\ldelta{i}{\lsv'}\loprod\glvreshape{v_i}{\lsv}{\lsv}\right)}{
       \lsv'\loprod\lsv}{\lsnv{s}},\\
 \lsnv{s}&=\begin{cases}\lsempty&m=0,\\\lsv'\loprod\lsv&m>0.\end{cases}
\end{aligned}
\]
We must show that
$\Sem{q}{\glvEnv}$ succeeds and gives the coefficients of $f$ in
$E^\star\otimes F$, with the result type $\lsv'\loprod\lsv$.
In particular, the goal is agreement with a tensor, rather than function
agreement with a runtime closure.

\emph{Dimensions and bases.}
The two space premises, by their induction hypotheses, give
\[
 \dof{\lsv'}{\glvEnv}=\dim V=N=|E|,\qquad
 \dof{\lsv}{\glvEnv}=\dim W=M=|F|.
\]
The basis-validity premise and \cref{lem:denoteBasisSound} make $E$ and
$F$ bases. The result-basis premise, by its induction hypothesis, gives
type agreement at $E^\star\otimes F$ and result dimension $NM$.
Thus it remains to evaluate $q$ and prove its coefficient equation.

\emph{Apply the body induction hypothesis at each input basis vector.}
For $i<N$, the dimension equality above gives
\[
 \delta_i=\Sem{\ldelta{i}{\lsv'}}{\glvEnv}\ne\glverr,
 \qquad |\delta_i|=N,\qquad (\delta_i)_k=\delta_{ik},
 \qquad E_i=\sum_{k<N}\delta_{ik}E_k.
\]
The family premise is the body's derivation with its bound
argument named $y$ and represented by $\ldelta{i}{\lsv'}$.
For this instance, extend the contexts and environments by
\[
\begin{gathered}
 \Delta_i=\Delta[y\mapsto\sof{b'}],\qquad
 \Psi_i=\Psi[(y,b')\mapsto\ldelta{i}{\lsv'}],\qquad
 \valCheckCtx_i=\valCheckCtx[y\mapsto\lsv'],\\
 \sigma_i=\sigma[y\mapsto E_i],\qquad
 \glvEnv_i=\glvEnv[y\mapsto\delta_i].
\end{gathered}
\]
The new entry evaluates to $\delta_i$, its inferred type is $\lsv'$,
and it agrees with $E_i$ in $E$ by the displayed equation. The domain
space hypothesis gives its type agreement. Freshness gives the context
and free-variable conditions. These verify
$(\Delta_i,\epsilon);\Psi_i;\valCheckCtx_i
 \vdash(\sigma_i,\glvEnv_i)$, so the body induction hypothesis applies.
It gives an array $c_i$ such that
\[
 \Sem{v_i}{\glvEnv_i}=c_i\ne\glverr,
 \qquad |c_i|=M,
 \qquad f(E_i)=\Sem{\lee{}}{\sigma_i}=\sum_{j<M}(c_i)_jF_j.
\]
Freshness of $y$ for $b$ and $\lsv$ leaves the output basis $F$
and output dimension $M$ unchanged.
Since the family premise supplies the same $v_i$ for every fresh $y$,
scoping gives $y\notin\operatorname{fv}(v_i)$ and hence
$\Sem{v_i}{\glvEnv}=\Sem{v_i}{\glvEnv_i}=c_i\ne\glverr$.

\emph{Evaluate each summand, including its reshape.}
Put $c_{ij}=(c_i)_j$. The inner $\glvreshape{v_i}{\lsv}{\lsv}$ has source length $|c_i|=M$ and target
 dimension $\dof{\lsv}{\glvEnv}=M$. By \cref{lem:reshape-agreement}(1),
 it succeeds and returns $c_i$ unchanged.
The generated tensor denotation, with $d_0=N$ and $d_1=M$, is
{\small\[\gleqn{value.denot.tprod}.\]}
Thus
\[
\begin{gathered}
 r_i=\Sem{\ldelta{i}{\lsv'}\loprod\glvreshape{v_i}{\lsv}{\lsv}}{\glvEnv}
 \ne\glverr,\qquad |r_i|=NM,\\
 r_i[kM+j]=\delta_{ik}c_{ij}\qquad(k<N,\ j<M).
\end{gathered}
\]
The two operands have lengths $N$ and $M$, exactly the dimensions of
$\lsv'$ and $\lsv$. Their tensor has dimension $NM$, so its shape tests
succeed. Every entry used in this calculation is in bounds:
$k<N$, $j<M$, and $kM+j<NM$.

\emph{Sum the rows and evaluate the outer reshape.}
The finite-sum clause is
{\small\gldefn{value.denot.sum}}
Here $d=N$ and the output dimension is $NM$. The family supplies a
summand for every active index $i<N$; entries beyond that range are not
evaluated. All active summands have length $NM$, so
\[
 A=\Sem{\lsum^{\lsv'}_{i<m}
       \left(\ldelta{i}{\lsv'}\loprod\glvreshape{v_i}{\lsv}{\lsv}\right)}{\glvEnv}
 \ne\glverr,\qquad |A|=NM.
\]
Its entries satisfy
\[
 A[kM+j]=\sum_{i<N}r_i[kM+j]
         =\sum_{i<N}\delta_{ik}c_{ij}=c_{kj}.
\]
The final target $\lsv'\loprod\lsv$ also has dimension $NM$.
Its reshape therefore returns this same array, proving
$\Sem{q}{\glvEnv}=A\ne\glverr$.
If $N=0$ or $M=0$, the result is the empty array of length $NM$;
the coefficient equation has no indices and no array entry is read.

\emph{Identify the assembled tensor with the source denotation.}
Source typing makes the body linear in its bound argument. Put
$(E^k)_{k<N}=\SemBB{b'^{\lstar}}{\sigma}$, so
$E^k(E_i)=\delta_{ki}$. The tensor represented
by $A$ is
\[
 \sum_{k<N,j<M} A[kM+j]E^k\otimes F_j
 =\sum_{k<N}E^k\otimes\left(\sum_{j<M}c_{kj}F_j\right)
 =\sum_{k<N}E^k\otimes f(E_k).
\]
To see that this is precisely $f$, take an arbitrary
$u=\sum_{i<N}u_iE_i$. Its action on $u$ is
\[
 \sum_{k<N}E^k(u)f(E_k)
 =\sum_{k<N}u_kf(E_k)
 =f\left(\sum_{k<N}u_kE_k\right)=f(u).
\]
Here $E^k(u)=u_k$ by duality, and
$\sum_k u_kf(E_k)=f(\sum_k u_kE_k)$ by linearity;
orthonormality of $E$ is not needed. The length equality $|A|=NM$ and
this tensor equality are exactly value agreement at $E^\star\otimes F$.
Together with the result-basis induction hypothesis, they prove all
parts of the rule's conclusion.

\item In the dependent-application case, the last rule is
{\small\[\glrulenamed{c.term.appNVal}.\]}
Put
\[
\begin{gathered}
 f=\Sem{\len{0}}{\sigma},\quad a=\Sem{\len{1}}{\sigma},\quad
 g=\Sem{\lenv{0}}{\glvEnv},\quad w=\Sem{\lenv{1}}{\glvEnv},\\
 B=\SemBB{\frameB{x}{\lbn{0}}{\lbn{1}}}{\sigma},\qquad
 B_0=\SemBB{\lbn{0}}{\sigma},\qquad
 B_1(a)=\SemBB{\lbn{1}}{\sigma[x\mapsto a]}.
\end{gathered}
\]
The operand induction hypotheses give agreeing
values and types, supplying the premises of \cref{lem:reshape-agreement}(2)
for the argument conversion to the function's inferred domain.
Function agreement then gives a successful result representing $f(a)$.
Substitution identifies $B_1(a)$ with the conclusion's output basis;
the output-basis induction hypothesis supplies the target type agreement
for a second application of \cref{lem:reshape-agreement}(2).
Specifically, put
\[
\begin{gathered}
 \operatorname{inferType}(\lenv{0})=(x:D\fnto C),\qquad
 u=\glvreshape{\lenv{1}}{D}{\operatorname{inferType}(\lenv{1})},\\
 v=\glvreshape{\glvapp{\lenv{0}}{u}}{\ltynv{\rm out}}{C[u/x]}.
\end{gathered}
\]
Then $\Sem{v}{\glvEnv}\ne\glverr$ agrees with $f(a)$ at
$\ltynv{\rm out}$, as required by the conclusion.

\item In the reshape case, the last rule is
{\small\[\glrulenamed{c.term.reshapeCob}.\]}
Write $\ltyv$ for the source realization carried by the rule. Put $B=\SemBB{\lb{}}{\sigma}$, $a=\Sem{\lee{}}{\sigma}$, and
$w=\Sem{\leev{}}{\glvEnv}$. The operand induction hypothesis at
$\operatorname{inferType}(\leev{})$ and the basis-evaluation hypothesis supply value
agreement and both type agreements required by
\cref{lem:reshape-agreement}(2). Hence
$w'=\Sem{\glvreshape{\leev{}}{\ltyv{}'}{\ltyv}}{\glvEnv}\ne\glverr$ and
$a\equiv_{B;\sigma,\glvEnv}^{\tof{\lb{}},\ltyv{}'}w'$.

 \item\label{proof:adequacy-embedding} In the kernel-embedding case, the last rule is
 {\small\glrulev{c.term.embedKer}}
 Its source denotation is
 \[\gleqn{denotTermAux.embed.ker}.\]
 Put $L=\Sem{\lee{}}{\sigma}$, $S=\operatorname{ker}(L)$, and
 \[
 E=\SemBB{\lbn{0}}{\sigma}=(E_i)_{i<N},\qquad
 K=\SemBB{\pBasisC{\Kernel(\lee{})}{\lbn{0}}{\lbn{1}}}{\sigma}
   =(K_j)_{j<k}.
 \]
 The operand hypotheses and \cref{lem:pseudoinverse-projected-bases}(1) give
 \[
 Q_{ij}:=\Sem{\glvop{GSVDUI}(\leev{},\lenv{0},\lenv{1})}{\glvEnv}_{ik+j},
 \qquad \iota_S(K_j)=\sum_iQ_{ij}E_i.
 \]
 Hence $a=\Sem{\Trans(\glvop{GSVDUI}(\leev{},\lenv{0},\lenv{1}))}{\glvEnv}$ satisfies
 \[
 a[jN+i]=Q_{ij},\qquad
 \sum_{j<k,\,i<N}a[jN+i]K_j^{\star}\otimes E_i
 =\sum_{j<k}K_j^{\star}\otimes\iota_S(K_j)=\iota_S.
 \]
 Here $Q$ is the zero-eigenvalue block on the domain side. It has
 $Nk$ entries and the result dimension is $kN$, so the transpose and
 output-length checks succeed.

\item\label{proof:adequacy-pseudoinverse-rule}
In the general pseudo-inverse case, the last rule is
{\small\[\glrulenamed{c.term.pinvTGpinv}.\]}
Put
\[
 L=\Sem{\lee{}}{\sigma},\qquad
 E=\SemBB{\lbn{1}}{\sigma}=(E_j)_{j<N},\qquad
 F=\SemBB{\lbn{0}}{\sigma}=(F_i)_{i<M}.
\]
The operand and inner-product induction hypotheses give
\[
\begin{aligned}
 A_{ij}&:=\Sem{\leev{}}{\glvEnv}_{jM+i},
 &L(E_j)&=\sum_{i<M}A_{ij}F_i,\\
 G_{jk}&:=\Sem{\lenv{1}}{\glvEnv}_{jN+k},
 &G_{jk}&=\langle E_j,E_k\rangle,\\
 H_{i\ell}&:=\Sem{\lenv{0}}{\glvEnv}_{iM+\ell},
 &H_{i\ell}&=\langle F_i,F_\ell\rangle.
\end{aligned}
\]
\Cref{lem:pseudoinverse-projected-bases}(2) gives
\[
\Sem{\lee{}^{\lpseudo}}{\sigma}=L^+
=\sum_{j<M,i<N}(A^+_{G,H})_{ij}F^j\otimes E_i.
\]
Hence $p=\Sem{\glvop{GPINV}(\leev{},\lenv{1},\lenv{0})}{\glvEnv}$
has $p[jN+i]=(A^+_{G,H})_{ij}$, the required coefficients.
The operand arrays have lengths $NM,N^2,M^2$ and the result has
length $MN$, so the matrix, metric, and output-length checks succeed.

\end{enumerate}

\paragraph{Change of basis}
In each change-of-basis case, put $B_i=\SemBB{\lbn{i}}{\sigma}$.
For the conclusion $\lbn{0}\xlongleftrightarrow{\leev{}}\lbn{1}$,
we must prove
\begin{equation*}
\begin{gathered}
 a=\Sem{\leev{}}{\glvEnv}\ne\glverr,\qquad
 |a|=N^2,\qquad N=|B_0|=|B_1|,\\
 A_{ji}:=a[jN+i],\qquad
 (B_0)_j=\sum_i A_{ji}(B_1)_i.
\end{gathered}
\end{equation*}
Thus $A^{\mathsf T}$ maps column coordinates from $B_0$ to $B_1$.
\begin{enumerate}
 \item In the reflexivity case, the last rule is
 {\small\[\glrulenamed{c.cob.refl}.\]}
 Put $B=\SemBB{\lb{}}{\sigma}$ and $a=\Sem{\lidelta{\lsv}}{\glvEnv}$.
 The space induction hypothesis gives $\dof{\lsv}{\glvEnv}=N=|B|$.
 Thus $a\ne\glverr$, $|a|=N^2$, and $a[jN+i]=\delta_{ji}$;
 \cref{lem:reshape-agreement}(1) leaves it unchanged
 at the target dimension $N^2$. Since
 $B_j=\sum_i\delta_{ji}B_i$, it satisfies the conclusion's coefficient
 equation.

 \item In the inverse case, the last rule is
 {\small\[\glrulenamed{c.cob.inverse}.\]}
 Put $A_{ij}:=\Sem{\leev{}}{\glvEnv}_{iN+j}$.
 The induction hypothesis gives
 $\Sem{\leev{}}{\glvEnv}\ne\glverr$,
 $|\Sem{\leev{}}{\glvEnv}|=N^2$, and
 $(B_1)_i=\sum_j A_{ij}(B_0)_j$.
 Both endpoints are bases of the same space, so $A$ is invertible.
 The value inverse therefore satisfies
 \[
 \begin{gathered}
 a=\Sem{\leev{}^{-1}}{\glvEnv}\ne\glverr,\qquad
 |a|=N^2,\qquad a[jN+i]=(A^{-1})_{ji},\\
 (B_0)_j=\sum_i(A^{-1})_{ji}(B_1)_i,
 \end{gathered}
 \]
 which is the required conclusion.

 \item In the composition case, the last rule is
 {\small\[\glrulenamed{c.cob.trans}.\]}
 Put $A_{ji}:=\Sem{\lenv{0}}{\glvEnv}_{jN+i}$ and
 $D_{ik}:=\Sem{\lenv{1}}{\glvEnv}_{iN+k}$.
 The two conversion induction hypotheses give
 \begin{equation*}
 (B_0)_j=\sum_i A_{ji}(B_1)_i,\qquad
 (B_1)_i=\sum_k D_{ik}(B_2)_k.
 \end{equation*}
 Then
 \begin{equation*}
 (B_0)_j=\sum_k(AD)_{jk}(B_2)_k.
 \end{equation*}
 The generated matrix contraction clause fixes the operational order:
 {\small\begin{equation*}\gleqn{value.denot.mtm.compact}.\end{equation*}}
 All three bases have the same length $N$, so
 $|\Sem{\lenv{0}}{\glvEnv}|=|\Sem{\lenv{1}}{\glvEnv}|=N^2$.
 The space induction hypothesis gives target dimension $N^2$ for the
 reshaped operand, so the array reshape keeps its entries. Contraction
 gives
 \[
 \begin{gathered}
 a=\Sem{\glvmtm{\lsv}{\lsv}{\lenv{1}}{\glvreshape{\lenv{0}}{\lsv\loprod\lsv}{\lsv\loprod\lsv}}}{\glvEnv}
   \ne\glverr,\qquad |a|=N^2,\\
 a[jN+k]=(AD)_{jk}=\sum_iA_{ji}D_{ik}.
 \end{gathered}
 \]
 This is successful denotation and the
 coefficient equation for the conclusion's conversion from $B_0$ to $B_2$.

 \item In the direct-sum case, the last rule is
 {\small\[\glrulenamed{c.cob.dsum}.\]}
 Put $B_i'=\SemBB{\lbn{i}'}{\sigma}$, $N=|B_0|$, and $M=|B_1|$.
 The component hypotheses give
 \[
 \begin{aligned}
 A_{ji}&:=\Sem{\lenv{0}}{\glvEnv}_{jN+i},
 &(B_0')_j&=\sum_i A_{ji}(B_0)_i,\\
 D_{k\ell}&:=\Sem{\lenv{1}}{\glvEnv}_{kM+\ell},
 &(B_1')_k&=\sum_\ell D_{k\ell}(B_1)_\ell.
 \end{aligned}
 \]
 Thus
 \[
 a=\Sem{\vblk(\lenv{0},\lenv{1})}{\glvEnv}\ne\glverr,
 \qquad |a|=(N+M)^2,\qquad
 a[p(N+M)+q]=\begin{pmatrix}A&0\\0&D\end{pmatrix}_{pq}.
 \]
 The two diagonal blocks give the two component coefficient equations,
 and its off-diagonal blocks are zero. It therefore converts the
 concatenated bases, as required by the conclusion.

 \item In the tensor-product case, the last rule is
 {\small\[\glrulenamed{c.cob.tprod}.\]}
 Put
 \[
 \begin{aligned}
 E&=\SemBB{\lbn{0}}{\sigma},&E'&=\SemBB{\lbn{0}'}{\sigma},&N&=|E|,\\
 F&=\SemBB{\lbn{1}}{\sigma},&F'&=\SemBB{\lbn{1}'}{\sigma},&M&=|F|,\\
 A_{ji}&:=\Sem{\lenv{0}}{\glvEnv}_{jN+i},&
 D_{k\ell}&:=\Sem{\lenv{1}}{\glvEnv}_{kM+\ell}.
 \end{aligned}
 \]
 The component hypotheses give $E'_j=\sum_i A_{ji}E_i$ and
 $F'_k=\sum_\ell D_{k\ell}F_\ell$. Hence
 \begin{equation*}
 E'_j\otimes F'_k=\sum_{i,\ell}A_{ji}D_{k\ell}(E_i\otimes F_\ell).
 \end{equation*}
 The Kronecker clause gives the same coefficients in the flattened order:
 \[
 \begin{gathered}
 a=\Sem{\vkrn(\lenv{0},\lenv{1})}{\glvEnv}\ne\glverr,
 \qquad |a|=(NM)^2,\\
 a[(jM+k)NM+(iM+\ell)]=A_{ji}D_{k\ell}.
 \end{gathered}
 \]
 The displayed
 coefficients prove the required agreement for the tensor bases.

 \item In the dual-to-standard-basis case, the last rule is
 {\small\[\glrulenamed{c.cob.dualStd}.\]}
 Put
 \[
 \begin{gathered}
 E=\SemBB{\lbn{0}}{\sigma},\qquad
 E'=\SemBB{\stdBasisDCtx{\tof{\lbn{0}}}}{\sigma},\qquad N=|E|,\\
 A_{ji}:=\Sem{\leev{}}{\glvEnv}_{jN+i}.
 \end{gathered}
 \]
 The conversion hypothesis gives $E_j=\sum_iA_{ji}E'_i$, and
 $\SemBB{\stdBasisDCtx{\tof{\lbn{0}^{\lstar}}}}{\sigma}=(E')^\star$.
 Evaluation on basis vectors gives
 \begin{equation*}
 E^j=\sum_i(A^{-\mathsf T})_{ji}(E')^i.
 \end{equation*}
 Indeed, applying the right-hand side to $E_k$ gives
 $\sum_i(A^{-\mathsf T})_{ji}A_{ki}=(AA^{-1})_{kj}=\delta_{kj}$.
 Thus it is the $j$th dual basis vector. Since $A$ is an invertible
 $N\times N$ basis conversion, inverse and transpose succeed and
 give
 \[
 a=\Sem{\Trans(\leev{}^{-1})}{\glvEnv}\ne\glverr,\qquad
 |a|=N^2,\qquad a[jN+i]=(A^{-\mathsf T})_{ji}.
 \]
 This proves the conclusion.

 \item In the projected-kernel-basis case, the last rule is
 \glcobrule{c.cob.projKerCoker}
 Take $\ls=\Kernel(\lee{})$ in the rule and put
 $S=\operatorname{ker}(\Sem{\lee{}}{\sigma})$.
 Then
 \[
 \iota_S=\Sem{\lembd{}_{\ls}}{\sigma},\qquad
 \pi_S=\Sem{\lproj{}_{\ls,\sof{\lbn{0}'}}}{\sigma}.
 \]
 Put
 \[
 \begin{aligned}
 K&=\SemBB{\pBasisC{\ls}{\lbn{0}}{\lbn{1}}}{\sigma}=(K_j)_{j<r},\\
 K'&=\SemBB{\pBasisC{\ls}{\lbn{0}'}{\lbn{1}}}{\sigma}=(K'_\ell)_{\ell<r},
 \end{aligned}
 \]
 \[
 E=\SemBB{\lbn{0}}{\sigma}=(E_i)_{i<N},\qquad
 E'=\SemBB{\lbn{0}'}{\sigma}=(E'_k)_{k<N}.
 \]
 We must prove that the output converts $K$ to $K'$.
 The three operand induction hypotheses give successful arrays with
 \begin{equation*}
 \begin{aligned}
 I_{ij}&:=\Sem{\lenv{3}}{\glvEnv}_{jN+i},
 &\iota_S(K_j)&=\sum_{i<N}I_{ij}E_i,\\
 C_{ki}&:=\Sem{\lenv{2}}{\glvEnv}_{iN+k},
 &E_i&=\sum_{k<N}C_{ki}E'_k,\\
 P_{\ell k}&:=\Sem{\lenv{1}}{\glvEnv}_{kr+\ell},
 &\pi_S(E'_k)&=\sum_{\ell<r}P_{\ell k}K'_\ell.
 \end{aligned}
 \end{equation*}
 Since \cref{lem:pseudoinverse-projected-bases}(3) gives
 $\pi_S\iota_S=\operatorname{id}_S$,
 \begin{equation*}
 \begin{aligned}
 K_j=\pi_S(\iota_S(K_j))
 &=\sum_{i,k<N,\,\ell<r}P_{\ell k}C_{ki}I_{ij}K'_\ell\\
 &=\sum_{\ell<r}(PCI)_{\ell j}K'_\ell.
 \end{aligned}
 \end{equation*}
 The first matrix operation computes $CI$ and the second computes $P(CI)$;
 the resulting row array is $(PCI)^{\mathsf T}$. The hypotheses give
 $I\in\mathbb R^{N\times r}$, $C\in\mathbb R^{N\times N}$, and
 $P\in\mathbb R^{r\times N}$. Thus the contracted dimensions match and
 the result has $r^2$ entries. The input reshapes have dimensions
 $rN$, $N^2$, and $Nr$, respectively, exactly the lengths supplied by
 the hypotheses. \Cref{lem:reshape-agreement}(1) therefore keeps
 $I^{\mathsf T}$, $C^{\mathsf T}$, and $P^{\mathsf T}$ unchanged.
 Thus both contractions give $a=\Sem{\leev_{\mathrm{out}}}{\glvEnv}\ne\glverr$ with
 $|a|=r^2$ and $a[jr+\ell]=(PCI)_{\ell j}$. The displayed expansion
 of $K_j$ is exactly the conclusion's coefficient equation.

 \item In the case converting the image of the canonical cokernel basis to
 the canonical image basis, the last rule is
 \glcobrule{c.cob.imageSvds}
 Put
 \[
 \begin{gathered}
 L=\Sem{\lee{}}{\sigma},\qquad
 E=\SemBB{\lbn{0}}{\sigma}=(E_j)_{j<N},\qquad
 F=\SemBB{\lbn{1}}{\sigma}=(F_i)_{i<M},\\
 A_{ij}:=\Sem{\leev{}}{\glvEnv}_{jM+i}.
 \end{gathered}
 \]
 The term hypothesis gives $L(E_j)=\sum_iA_{ij}F_i$.
 The standard-basis premises give identity Gram matrices. Thus, putting
 \[
 P=\GSVDU(A,I_N,I_M),\qquad Q=\GSVDV(A,I_N,I_M),\qquad
 \Sigma=\GSVDS(A,I_N,I_M),
 \]
 \cref{lem:pseudoinverse-projected-bases}(1) gives
 \[
 \begin{aligned}
 K&=\SemBB{\pBasisC{\CoKernel(\lee{})}{\lbn{0}}{\lbn{1}}}{\sigma},
 &K_j&=\sum_iP_{ij}E_i,\\
 J&=\SemBB{\pBasisC{\Image(\lee{})}{\lbn{0}}{\lbn{1}}}{\sigma},
 &J_j&=\sum_iQ_{ij}F_i.
 \end{aligned}
 \]
 With $r=|K|=|J|$, we have $Q^{\mathsf T}Q=I_r$, and the columns of
 $AP$ lie in the span of $Q$. Hence
 \begin{equation*}
 C=Q^{\mathsf T}AP,\qquad QC=AP,\qquad
 L(K_j)=\sum_i(AP)_{ij}F_i=\sum_\ell C_{\ell j}J_\ell.
 \end{equation*}
 Thus column $j$ of $C$ gives the coordinates of $L(K_j)$ in $J$.
 The rule's two matrix contractions compute $(Q^{\mathsf T}AP)^{\mathsf T}$,
 using the displayed $\glvop{GSVDU}$ and $\glvop{GSVDV}$ operations.
 By the paired convention of \cref{def:fsv}, $AP=Q\Sigma$, and therefore
 \[
 C=Q^{\mathsf T}AP=Q^{\mathsf T}Q\Sigma=\Sigma.
 \]
 Here $\Sigma$ contains the positive singular values in decreasing order.
 The returned array therefore satisfies
 \[
 a=\Sem{\leev_{\mathrm{out}}}{\glvEnv}\ne\glverr,\qquad
 |a|=r^2,\qquad a[jr+i]=C_{ij}=\sigma_j\delta_{ij},
 \]
 the required coordinates.
 The inner reshape of the map has the matrix dimension given by its
 space hypothesis. The outer reshape has target dimension $r\cdot r$,
 since cokernel and image each have dimension $r$. Both are array
 reshapes with matching lengths, so both preserve these computed entries.
 This proves success and the coefficient equation for the conclusion.
 \item In the term-basis case, the last rule is
 \glcobrule{c.cob.termsRref}
 Put
 \[
 \begin{gathered}
 B=\SemBB{\lb{}}{\sigma}=(B_j)_{j<m},\qquad
 B'=\SemBB{\lb{}'}{\sigma}=(B'_i)_{i<m},\\
 (t_j)_{j<k}=\lee{}\mathbin{:\!:}\mathit{es}.
 \end{gathered}
 \]
 The term and conversion hypotheses give
 \[
 \begin{aligned}
 Z_{ij}&:=\Sem{\lenv{0}}{\glvEnv}_{jm+i},
 &\Sem{t_j}{\sigma}&=\sum_iZ_{ij}B'_i,\\
 X_{ji}&:=\Sem{\lenv{1}}{\glvEnv}_{jm+i},
 &B_j&=\sum_iX_{ji}B'_i.
 \end{aligned}
 \]
 The candidate matrix is
 \begin{equation*}
 Q=[Z\mid I_m].
 \end{equation*}
 Its columns represent the terms followed by the fallback basis, in exactly the order used by $\LexInd$. Row reduction preserves every linear dependence among columns. Therefore its increasing pivot-column indices select precisely the first columns which enlarge the span. These are the $\LexInd$ columns, giving the denotation
 \begin{equation*}\gleqn{denotBasis.terms}.\end{equation*}
 Put $R=\operatorname{RREF}(Q)$. Because $Q$ includes $I_m$, it has full
 row rank. If $p_0<\cdots<p_{m-1}$ are its pivot-column indices, then
 \[
 E:=(Q_{i,p_j})_{i,j<m},\qquad
 (R_{i,p_j})_{i,j<m}=I_m,\qquad R=E^{-1}Q.
 \]
 Consequently,
 \[
 \begin{gathered}
 K=\SemBB{\llpl{}\lee{}\mathbin{:\!:}\mathit{es}\llpr{}_{\lb{}'}}{\sigma},
 \qquad K_j=\sum_iE_{ij}B'_i,\\
 B_j=\sum_i(E^{-1}X^{\mathsf T})_{ij}K_i.
 \end{gathered}
 \]
 Thus the required row coefficients are
 \begin{equation*}
 (E^{-1}X^{\mathsf T})^{\mathsf T}=XE^{-\mathsf T}.
 \end{equation*}
 This is the returned $\operatorname{RREFRep}$ array. Since
 $R=[E^{-1}Z\mid E^{-1}]$, the last $m$ columns of the unique reduced
 matrix determine $E^{-1}$, regardless of the row operations used.
 The term array has length $mk$, the input conversion array has length
 $m^2$, and the returned matrix is $m\times m$. The inner and outer
 reshapes both have target dimension $m^2$. By
 \cref{lem:reshape-agreement}(1), they leave their
 respective entries unchanged. Thus
 \[
 a=\Sem{\leev_{\mathrm{out}}}{\glvEnv}\ne\glverr,\qquad
 |a|=m^2,\qquad a[jm+i]=(XE^{-\mathsf T})_{ji},
 \]
 the required conversion into $K$.

\end{enumerate}

\paragraph{Shapes, dimensions, and frames}
The remaining simultaneous induction cases establish shape agreement.
For structural spaces, the shape and dimension equations applied to the
premise induction hypotheses give the required shape directly. For a
fundamental subspace of $\lee{}$, with
$(\emptyset,\epsilon)\vdash\lee{}\colon\lsn{0}^{\lstar}\loprod\lsn{1}$, put
\[
\begin{gathered}
 L=\Sem{\lee{}}{\sigma},\qquad
 V=\SemT{\lsn{0}}{\sigma},\qquad W=\SemT{\lsn{1}}{\sigma},\\
 r=\operatorname{rank}L,\qquad N=\dim V,\qquad M=\dim W.
\end{gathered}
\]
The term hypothesis and \cref{lem:pseudoinverse-projected-bases}(1)
identify the selected subspace. Its computed dimension is therefore $N-r$, $r$, $r$, or $M-r$
for kernel, cokernel, image, or coimage, respectively.
At a frame, the domain hypothesis and the fiber hypothesis for every
agreeing argument pair give exactly the recursive function clause of
shape agreement in \cref{def:agreedims}.
Finally, \cref{lem:denoteBasisSound} supplies a basis at spaces and a
valid domain component and fiber family at frames. Combined with shape
agreement, this proves the required type agreement.

These simultaneous conclusions give successful value denotation, the required coefficient or function agreement, and agreement at the actual basis realization for the supplied checked evaluation. Applying them to the particular derivation selected by \cref{thm:prog} proves \cref{cor:progressAdequacy}; both conclusions concern the same produced value.
\end{proof}
\endgroup

\begin{cor}[Progress and adequacy]
\label{cor:progressAdequacy}
If $(\Delta,\epsilon);\Psi;\valCheckCtx\vdash(\sigma,\glvEnv)$,
$(\Delta,\epsilon)\vdash t:\tof{b}$, and $(\{\},\epsilon)\vdash b:\bvalid$, then
\[
\begin{gathered}
\exists v,\ltyv{},w.\quad\glCterm{\Delta}{\epsilon}{\Psi}{t}{b}{v}{\ltyv{}},\quad
\operatorname{inferType}(v)=\ltyv{},\\
\Sem{v}{\glvEnv}=w\ne\glverr,\qquad
\Sem{t}{\sigma}\equiv_{\SemBB{b}{\sigma};\sigma,\glvEnv}^{\tof{b},\ltyv{}}w,\qquad
\tof{b}\simeq_{\SemBB{b}{\sigma};\sigma,\glvEnv}\ltyv{}.
\end{gathered}
\]

\end{cor}
\begin{proof}[Proof sketch]
From $(\Delta,\epsilon);\Psi;\valCheckCtx\vdash(\sigma,\glvEnv)$,
\cref{def:allCtxAgree} gives $(\Delta,\epsilon);\valCheckCtx\vdash\Psi$.
Apply \cref{thm:prog} with this context to obtain a checked
evaluation with $\operatorname{inferType}(v)=\ltyv$.
Then apply \cref{thm:main} to that same evaluation under the given
consistent environments. Its ordinary evaluation and \cref{thm:preserve} give
$\valCheckCtx\vdash v:\ltyv$; combining this with
$\valCheckCtx\vdash\glvEnv$ and \cref{lem:soundVals} places its checked
value denotation in $\mathcal V_\glvEnv(\ltyv)$.
\end{proof}

\clearpage
\section{Implementation Details: Performance and Correctness}
\label{app:implementation-performance-correctness}
\paragraph{Correctness Details}
We checked correctness of our implementation in the following ways:
\begin{enumerate}
    \item We generated Einstein summation notation expressions randomly, converted them into \GLang{} , and checked the results.
    \item We implemented and checked the properties of exterior calculus (wedge products, interior products, Hodge stars, exterior derivatives, co-derivatives). 
    We also used this in our large finite element system to implement Hodge decompositions in 2D, 2D in 3D, and 3D. 
    \item We have tested pullbacks for a wide variety of tensor spaces, including exterior and symmetric algebra spaces. 
    \item We implemented a source to source automatic covariant differentiation for \GLang{} and used it to fuzz \GLang{} against a finite difference implementation. 
\end{enumerate}
\paragraph{Benchmarking Details}
These measurements concern the Python implementation, including its basis selection and simplification passes; the timed code is not extracted from Lean.
We ran all benchmarks on an Apple M3 Max (10 performance and 4 efficiency cores, 36\,GB) under macOS~26.5, using Python~3.13 and PyTorch~2.13 in double precision on a single thread.
  Both the \GLang{}-generated kernel and the reference are compiled with
  \texttt{torch.compile} (CPU Inductor backend, static shapes) and applied to a batch of
  $N = 2^{16}$ inputs. 
  Timings of successive compilations within one process drift, so every
  measurement runs in a fresh process: after 20 warm-up calls we time 30 runs of 100 consecutive calls each, and repeat this in 6 processes per kernel, alternating the order of
  generated and reference processes. 
  We report the mean time per cell over all $6 \times 30 = 180$ runs (18000 calls), dividing each batched-call time by $N$ to obtain time per cell.
  We count FLOPs per cell as the scalar arithmetic operations (add, subtract, multiply, negate, and reduction adds) in Inductor's post-grad graph. 
  The reference implementations are found in \cref{lst:bench-refs}.
  \begingroup
\setlength{\intextsep}{6pt}
\begin{figure}[!htb]
\centering
\definecolor{codekeyword}{HTML}{245A9B}
\definecolor{codefunction}{HTML}{76529B}
\definecolor{codecomment}{HTML}{55745D}
\definecolor{codestring}{HTML}{A34D27}
\lstset{
  language=Python,
  basicstyle=\ttfamily\scriptsize,
  keywordstyle=\bfseries\color{codekeyword},
  commentstyle=\itshape\color{codecomment},
  stringstyle=\color{codestring},
  emph={dot,minors,triples,sqrt,stack},
  emphstyle=\color{codefunction},
  showstringspaces=false,
  columns=fullflexible,
  keepspaces=true,
  breaklines=false,
  frame=none,
  aboveskip=0pt,
  belowskip=0pt
}
\begin{lstlisting}
from itertools import combinations
from math import sqrt
from torch import stack

def dot(x, y):
    return (x * y).sum(-1)

def minors(a, b, pairs):        # a_i b_j - a_j b_i for each (i, j)
    i, j = (list(t) for t in zip(*pairs))
    return a[..., i] * b[..., j] - a[..., j] * b[..., i]

def triples(a, b, c, rows):     # det[a,b,c] on coordinates (i,j,k)
    return stack([dot(a[..., [i, j, k]],
                      minors(b, c, [(j, k), (k, i), (i, j)]))
                  for i, j, k in rows], -1)

CROSS = [(1, 2), (2, 0), (0, 1)]
PAIRS = lambda n: list(combinations(range(n), 2))

wedge_R2   = lambda a, b: sqrt(2) * minors(a, b, [(0, 1)])
star1_R2   = lambda x: stack([-x[..., 1], x[..., 0]], -1) / sqrt(2)
wedge_R3   = lambda a, b: sqrt(2) * minors(a, b, [(0, 1), (0, 2), (1, 2)])
star1_R3   = lambda x: stack([x[..., 2], -x[..., 1], x[..., 0]], -1) / sqrt(3)
star_ab_R3 = lambda a, b: sqrt(2/3) * minors(a, b, CROSS)
triple_R3  = lambda a, b, c: sqrt(6) * dot(a, minors(b, c, CROSS))
wedge_R4   = lambda a, b: sqrt(2) * minors(a, b, PAIRS(4))
star1_R4   = lambda x: stack([-x[..., 3], x[..., 2],
                              -x[..., 1], x[..., 0]], -1) / 2
star_ab_R4 = lambda a, b: minors(a, b, [(2, 3), (3, 1), (1, 2),
                                        (0, 3), (2, 0), (0, 1)]) / sqrt(3)
triple_R4  = lambda a, b, c: sqrt(3/2) * triples(a, b, c,
                 [(2, 1, 3), (0, 2, 3), (1, 0, 3), (0, 1, 2)])

# <a^b, c^d>: pairing of Lambda^2 coordinates, and the Lagrange identity
gram_coords = lambda a, b, c, d: 2 * dot(minors(a, b, PAIRS(a.shape[-1])),
                                         minors(c, d, PAIRS(a.shape[-1])))
gram        = lambda a, b, c, d: 2 * (dot(a, c) * dot(b, d)
                                      - dot(a, d) * dot(b, c))
\end{lstlisting}
\caption{Reference implementations. 
Inputs carry a leading batch axis.
We choose to use simple formulas sometimes in place of builtin sum operations because calling LAPACK or BLAS on these kernels unfairly disadvantaged the reference code (as these codes perform better as vectorized loops).
We tried a few ways of writing the reference and found these were the most competitive (at least in terms of how PyTorch lowers them).}
\label{lst:bench-refs}
\end{figure}
\endgroup
\clearpage
\section{Scope relative to the Unified Form Language}
\label{app:ufl-comparison}
Even more concretely for those familiar with the Unified Form Language (UFL)~\cite{alnaes2014unified}, we handle tables 4 through 9 of their syntax modulo derivatives (though automatic differentiation can be input) and special functions (\textit{e.g.}, trig, absolute value, step functions), which is a simple extension.
However, our non-linear functions, $\nllambdaArg{a}{b}$, are not in vector spaces that they model (their table 1) and we don't model integrals on a mesh (their table 10).
We note, though, that these can be recovered by systems around \GLang{} without any loss as systems using UFL often convert integrals to quadrature, reducing some of the problem to what happens at each quadrature point.

\endgroup

\end{document}